%% file: astroWP_00main.tex
\documentclass[a4paper,11pt] {article}

\usepackage[utf8]{inputenc}
\usepackage[left=2.5cm,right=2.5cm,top=1cm,bottom=4cm]{geometry}
\usepackage{color}
\usepackage{colortbl}
\usepackage{fancyhdr}
\usepackage{graphicx}
\usepackage{eurosym}
\usepackage{enumitem}
\usepackage{lastpage}
\usepackage{booktabs}
\usepackage[yyyymmdd]{datetime}
\usepackage{multirow}
\usepackage{pifont}
\usepackage[colorlinks=true, urlcolor=blue!50!black, linkcolor=blue!50!black, citecolor=blue!50!black]{hyperref}
\usepackage{eurosym}
\usepackage{enumitem}
\usepackage{lastpage}
\usepackage{booktabs}
\usepackage[yyyymmdd]{datetime}
\usepackage{multirow}
\usepackage{pifont}
\usepackage{amssymb} 
\usepackage{amsmath}
\usepackage[color]{changebar}
\usepackage{listings}
\usepackage{threeparttable}
\usepackage{aas_macros_prc}
\usepackage{LatexShortcuts}
\usepackage{xcolor}
\usepackage{natbib}
\usepackage{soul}
\usepackage{longtable}

\usepackage{makecell}

\usepackage{lscape}
\usepackage[T1]{fontenc}
\usepackage[normalem]{ulem}
\usepackage{comment}

\definecolor{pink}{rgb}{0.55,0,0.52}
\definecolor{mygreen}{rgb}{0.19,0.55,0.11}
\definecolor{dkgreen}{rgb}{0,0.6,0}
\definecolor{gray}{rgb}{0.5,0.5,0.5}
\definecolor{mauve}{rgb}{0.58,0,0.82}
\definecolor{verbgray}{gray}{0.9}
\definecolor{lightblue}{rgb}{0.85,0.9,1}
\definecolor{lightgreen}{rgb}{0.85,1,0.85}
\definecolor{lightorange}{rgb}{1,0.94,0.8}
\definecolor{forestgreen}{rgb}{0.1,0.49,0.07}
\definecolor{burgundy}{rgb}{0.5, 0.0, 0.13}

\newcommand{\contr}[1]{\textbf{\color{black}Contributors: #1}}
\newcommand{\coord}[1]{\textbf{\color{black}Coordinators: #1}}

\def\institute#1{\gdef\@institute{#1}}

\begin{document}

\newcommand{\ReportTitle}{Astrophysics with the Laser Interferometer Space Antenna}           
\newcommand{\ReportShortTitle}{Astrophysics with the Laser Interferometer Space Antenna}    
\newcommand{\ReportIssue}{0}                                          
\newcommand{\ReportRevision}{1}                                             
\newcommand{\ReportRef}{LISA-}               
\newcommand{\ReportCurrentAuthor}{\href{authors.tex}{[full author details at the end of the article]}}
\newcommand{\ReportCurrentAuthorInstitute}{}           


\setcounter{secnumdepth}{4}
\setcounter{tocdepth}{3}

\lstnewenvironment{rawtxt}{%
  \lstset{backgroundcolor=\color{verbgray},
  frame=single,
  framerule=0pt,
  basicstyle=\ttfamily,
  columns=fullflexible}}{}

\newcommand{\delay}[1]{\mathcal{D}_{#1}}
\newtheorem{app}{Approximation}[section]
\pagenumbering{arabic}

\noindent
\resizebox{\textwidth}{!}{\textbf{\Huge \ReportTitle}}\\[2em]
\input{authors_abridged}\\[2em]

\input{abstract}

\input{AllAcronyms}

\tableofcontents

\input{intro}


\newpage
\input{astroWP_01ucb}

\input{astroWP_02mbh}

\input{astroWP_03emri}

\clearpage
\input{conclusions}
\clearpage

\clearpage
\input{acknowledgements}

\cleardoublepage
\phantomsection
\addcontentsline{toc}{section}{\protect\numberline{}References}
\bibliographystyle{spbasic-FS-PRC}
\bibliography{ucb,mbh,emri}

\input{authors}

\end{document}

%% file: authors_abridged.tex
\noindent
\textbf{
Amaro Seoane, Pau; 
Andrews, Jeff; 
Arca Sedda, Manuel; 
Askar, Abbas; 
Baghi, Quentin;
Balasov, Razvan; 
\hyperref[sec:author-details]{[full author details at the end of the article]}
}

%% file: abstract.tex
\section*{Abstract}

The Laser Interferometer Space Antenna (LISA) will be a transformative experiment for gravitational wave astronomy, and, as such, it will offer unique opportunities to address many key astrophysical questions in a completely novel way. The synergy with ground-based and space-born instruments in the electromagnetic domain, by enabling multi-messenger observations, will add further to the discovery potential of LISA. The next decade is crucial to prepare the astrophysical community for LISA’s first observations. This review outlines the extensive landscape of astrophysical theory, numerical simulations, and astronomical observations that are instrumental for modeling and interpreting the upcoming LISA datastream. To this aim, the current knowledge in three main source classes for LISA is reviewed; ultra-compact stellar-mass binaries, massive black hole binaries, and extreme or interme-diate mass ratio inspirals. The relevant astrophysical processes and the established modeling techniques are summarized. Likewise, open issues and gaps in our understanding of these sources are highlighted, along with an indication of how LISA could help making progress in the different areas. New research avenues that LISA itself, or its joint exploitation with upcoming studies in the electromagnetic domain, will enable, are also illustrated. Improvements in modeling and analysis approaches, such as the combination of numerical simulations and modern data science techniques, are discussed. This review is intended to be a starting point for using LISA as a new discovery tool for understanding our Universe.

%% file: AllAcronyms.tex
\section*{List of Acronyms}

\begin{longtable}{ll}
AGN   &  Active galactic nucleus/nuclei \\
AM~CVn & AM  Canum  Venaticorum\\
AU & Astronomical Unit \\
BD & brown dwarf (plural: BDs)\\
BH  &  black hole (plural: BHs) \\
BH+BH  &  (stellar-mass) binary black hole (plural: BH+BHs) \\
CDM   &  cold dark matter \\
CE & common envelope\\
CO & compact object \\
COM & centre-of-mass\\
CV & cataclysmic variable\\
DM    &  dark matter \\
ELM & extremely low-mass\\
EM & electromagnetic\\
EMRI  &  extreme mass ratio inspiral \\
EOS & equation of state\\
GR    &  general relativity/relativistic \\
GW & gravitational wave (plural: GWs)\\
HMXB & high-mass X-ray binary\\
IMBH  &  intermediate-mass black hole \\
IMF & initial mass function\\
IMRI  &  intermediate mass-ratio inspiral \\
IR & infra-red\\
ISCO &  innermost stable circular orbit \\
LMXB & low-mass X-ray binary\\
MBH   &  massive black hole \\
MBHB  &  massive black hole binary \\
MHD   &  magnetohydrodynamics/magnetohydrodynamic \\
MSP & millisecond radio pulsar \\
MW & Milky Way\\
MS & main sequence \\
NFW & Navarro-Frenk-and-White \\
NS & neutron star (plural: NSs)\\
NS+NS & double neutron star (plural: NS+NSs)\\
NSC  &  nuclear star cluster \\
PN & post-Newtonian \\
Pop~III & population~III\\
PTA   &  pulsar timing array \\
RF & radiative feedback \\
RLO & Roche-lobe overflow\\
SFH & star-formation history\\
SGWB & stochastic gravitational wave background\\
SMBH  &  supermassive black hole \\
SMS  &  supermassive star \\
SN    &  Supernova  (plural: SNe) \\
SNR  & signal-to-noise ratio \\
SPH   &  smoothed-particle hydrodynamics \\
SSO & substellar object \\
TDE & tidal disruption event\\
UCB & ultra-compact binary\\
UCXB & ultra-compact X-ray binary\\
UV & ultra-violet \\
WD & white dwarf (plural: WDs) \\
WD+WD & double white dwarf (plural: WD+WDs)\\
XMRB &  extremely large mass-ratio burst \\
XMRI &  extremely large mass-ratio inspiral \\
ZKL & von Zeipel--Kozai--Lidov\\
b-EMRI & binary-extreme mass ratio inspiral \\
\end{longtable}


%% file: intro.tex
\section*{General Introduction}
\addcontentsline{toc}{section}{\protect\numberline{}General Introduction}

Gravitational wave (GW) observations have opened a new way to observe and characterize compact objects
throughout the Universe and at all cosmic epochs. The Laser Interferometer Space Antenna \citep[LISA][]{2017arXiv170200786A}, with its low-frequency band coverage spanning nearly
three decades, will allow the detection and study of signals from a strikingly large variety of sources, ranging from stellar-mass binaries in our own galaxy to mergers between nascent massive black holes (MBH), called black hole (BH) \emph{seeds}, at high redshift.
LISA is expected to revolutionize our understanding of these astrophysical sources by allowing reconstruction of
their demographics and dynamical evolution, as well as discovery of new types of sources, including some that have
been theorized but not yet detected by conventional means. Since the first detection of GWs by the Laser Interferometer GW Observatory (LIGO)/Virgo collaboration in 2015 \citep{2016PhRvL.116f1102A}, ground-based GW observations already had a remarkable impact on astrophysics.
For instance, the gravitational-wave facilities LIGO and Virgo have
observed the mergers of stellar BHs in the range $\sim$6--95~M$_{\odot}$ \citep{2020arXiv201014527A}, greatly expanding our knowledge of the mass spectrum of BHs.
Recently, the first intermediate mass black holes (IMBH), with masses of $\sim$142~M$_{\odot}$, has been discovered \citep[GW190521, see][]{2020PhRvL.125j1102A}.
The existence of stellar-mass BHs with masses higher than observed before, as well as the discovery of an IMBH, have fostered new exciting developments in theoretical models for
the formation and evolution of stellar-origin black holes. The discovery of the double neutron star (NS+NS) merger GW170817 with accompanying electromagnetic observations \citep{2017ApJ...848L..12A} has had a great impact on our understanding of dense matter and the origin of heavy elements. These discoveries showcase the huge potential that gravitational
wave astronomy has to revolutionize our understanding of astrophysical objects and processes.

At the lower frequencies in LISA's observing band, the stellar-mass systems, in binaries or multiples, provide a very rich source population. The population in the Milky Way is expected to consist of millions of double white dwarf (WD+WD) binaries, with a smaller population of neutron star (NS)/BH binaries, and possibly some of the heavy BHs that LIGO/Virgo have already detected. LISA observations of the BH populations will capture a snapshot of BH systems when their orbital periods are tens of minutes, a few years before their coalescence at the high frequencies observed by LIGO-Virgo. Overall, LISA observations of Galactic binaries will address many open questions in stellar astrophysics, such as the evolution of binary star systems, the origin of different transient phenomena, the origin of the elements and even the structure of the Galaxy. It should be noted that among the stellar-mass binaries in the Milky Way, a few are already known from electromagnetic observing campaigns, and can be used as LISA verification sources. While the vast majority of the stellar-mass binaries are expected to be too dim to be detected by electromagnetic instruments, there will be a substantive number that will be excellent targets for electromagnetic follow-up after LISA discovers them. 

The observed BH mass spectrum spans ten orders of magnitude, ranging from a few $\rm M_{\odot}$ for
stellar-mass BHs to up to $10^{11} \,\rm M_{\odot}$ for the most extreme MBHs.  Many of the most massive MBHs,
with $M_{\rm BH} \gtrsim 10^{8}\,\rm M_{\odot}$, have been discovered in the high-redshift Universe, at $z > 6$,
powering some of the brightest quasars 
\citep{2003AJ....125.1649F, 2011Natur.474..616M, 2020ApJ...897L..14Y}.  
LISA will open up a wide discovery space for BHs. BH systems that merge at the millihertz frequencies, where LISA is most sensitive, are typically hosted in the most common type of galaxies, namely dwarf and massive spiral galaxies. The fact that LISA observations straddle the frequency bands of merging IMBHs and MBHs suggests that the potential impact on many fields of extragalactic astrophysics is huge. Such foreseen impact,
however, relies heavily on our understanding of the astrophysical processes preceding and accompanying  the
evolution of the binaries during inspiral and into merger \citep{2019NewAR..8601525D}.
For MBHs, this knowledge is
tightly entangled with the knowledge of the environments in which they evolve, namely their host
galaxies and galactic nuclei.
It follows then that LISA sources associated with MBH binaries
cannot be understood without a robust knowledge of the landscape of galaxy formation and evolution,
and in particular without  a detailed knowledge of stellar dynamical processes and  the interstellar medium inside
galactic nuclei. There is thus an inherent multi-disciplinarity in the approach needed to understand
these sources, which will naturally bring together various fields of galactic and extra-galactic astrophysics.
Furthermore, since LISA will be able to detect MBH binary sources up to
very high redshift ($z \sim 10-15$), one also needs ancillary knowledge of cosmic structure formation, 
as galaxies, and thus their relevant environments, evolve significantly from high to low redshift \citep{2019PASA...36...27W}.
The endeavour then extends into cosmology, and hints at great possibilities for derivative knowledge, some already
expected and others not, coming from the future discovery and characterization of LISA MBH binaries.

The stellar dynamics of the central
cluster of stars at the galactic centre (the S-stars, or S0-stars), provides compelling evidence for the existence of
a MBH of mass $\sim 4\times 10^6\,M_{\odot}$, Sgr A*
(see for a review \citealt{2010RvMP...82.3121G}, and references therein).
The stars in the centres of galaxies have the potential to interact with MBHs, but only if their pericentres are small enough.
LISA will be able to observe the inspiral of a compact object such
as a stellar-mass BH, a NS or a WD onto a (light) MBH, i.e.,
one with a mass between $\sim 10^4\,M_{\odot}$ and $\sim 10^7\,M_{\odot}$. Because of the difference in mass between the MBH and the $\lesssim{\rm few}$--${\rm tens}$ of solar 
masses of the compact object, we call these \textit{extreme mass-ratio inspirals} (EMRIs)---where the mass ratio is $10^{-8}\lesssim q \lesssim 10^{-5}$
\citep{2007CQGra..24R.113A}.
There is also a potential population of IMBHs
with masses between $10^2\,M_{\odot}$ and $10^4\,M_{\odot}$, which, through inspiral onto the central MBH,
would generate GWs detectable by LISA, this being a class of sources dubbed \textit{intermediate mass-ratio inspirals} (IMRIs) \citep{2007CQGra..24R.113A}.
In principle, EMRIs and IMRIs could occur in the nuclei of any galaxy hosting a central MBH. They should be
ubiquitous, since most galaxies host a central MBH and undergo a variety of merger events with other galaxies throughout their lives. IMRIs might also occur outside galactic nuclei, for example in a star cluster cannibalizing its own population of compact objects.
For EMRIs and IMRIs, astrophysical modelling of their origin are in their earliest theoretical stages; in recent years
a number of new astrophysical scenarios have been proposed in which they could form even outside the conventional stellar-dynamical scenarios in the galactic centre or in star clusters. These scenarios have been, for the most part, detached from the notion that their host galaxies are highly dynamical systems with a diverse range of properties, at large scales as well as at the level of galactic nuclei and star clusters. From the astrophysical perspective, this is thus the least explored, albeit potentially most exciting, class of sources in the LISA band. An assessment of the current knowledge and upcoming developments in this area is of paramount importance, to propel new research on the astrophysical impact that the discovery of EMRIs and IMRIs by LISA can have.

The joint exploitation of LISA data with data from terrestrial GW detectors and electromagnetic observations across essentially all possible wavelengths, from infrared and radio to X-ray and gamma-rays, will further enhance its astrophysical impact \citep{2020PhRvD.102h4056M}.
Indeed, essentially all of LISA's individual sources have potential electromagnetic counterparts. Achieving a quantitative characterization of such counterparts,
determining the feasibility of detecting them in one or more wavebands, and assessing the stage at which they would be detectable, relative to the inspiral and/or merger stage of the corresponding GW signal, are the main objectives ahead for current and upcoming research. An assessment of the current knowledge in this area is another important task.

The challenge to  bring all these different pieces of knowledge into a coherent, robust picture within the next decade is huge, perhaps the most ambitious that the astrophysical community has ever faced.
This review attempts to aid this ambitious, community-wide effort by assessing the status of knowledge in the modelling of LISA sources, and it summarizes our understanding of the astrophysical processes and environments relevant for the interpretation of the  LISA data.
Furthermore, it discusses the most important challenges ahead of us in the research of galactic binaries/multiples, massive and intermediate-mass black hole binaries, and EMRIs/IMRIs. Among these are the quest for identifying the different astrophysical formation
channels for these various sources, including how these might be encoded in the LISA data stream, and the daunting multi-scale modelling needed to reconstruct the
full dynamical history of such sources, from their emergence to the final inspiral phase and merger driven
by GW radiation. The review material presented will help foster a critical discussion
of the major gaps in our knowledge that need to be filled in the next decade, highlighting where disagreement exists
between results, and what should be done next to reach beyond the current state of the art.
This brings the discussion to important methodological tasks for the immediate future, from exploiting electromagnetic (EM) observations in the next decade, to improving simulation and semi-analytical techniques employed to build astrophysical models for the sources, and to refurbishing analysis and interpretation techniques for the models, for example by employing machine learning, neural networks and other modern inference strategies.

%% file: astroWP_01ucb.tex

\section{Stellar Compact Binaries and Multiples}

\textbf{Section coordinators: Silvia Toonen, Tassos Fragos, Thomas Kupfer, Thomas Tauris}
\newpage
\subsection	{Introduction and Summary}
\noindent \contr{Silvia Toonen, Tassos Fragos, Thomas Kupfer, Thomas Tauris}\\

Detection of GW emission from binary compact stars is one of the key drivers for the LISA mission. There are already at the time of writing (July~2022) about two dozen known Galactic sources, most of which are guaranteed to be detectable with LISA within a few years of its operation (Section~\ref{subsec:classes-binaries}). These are tight binaries (typically with orbital periods of $P_{\rm orb}\simeq 5-30\;{\rm min}$) of WD+WDs which give rise to continuous emission of GWs. 
Unlike binaries consisting of NSs and BHs, WD binaries (with their larger radii and thus lower orbital frequencies at merger) are not readily detectable by ground-based high-frequency (Hz--kHz) GW observatories, such as LIGO/Virgo/KAGRA, nor by the planned third generation of such detectors. These high-frequency detectors can observe the final a few - a few thousands orbits of inspiral (lasting a fraction of a second - minutes) and the merger event itself for NSs and BHs. Such merger events, however, are rare (of order a dozen events ${\rm Myr}^{-1}$ for a Milky~Way equivalent galaxy) and therefore they are only anticipated to be detected as extra-galactic sources, across volumes that encompass large numbers of galaxies. 
A major advantage of LISA is that the inspiral phase (due to orbital GW damping in the compact binaries) of the vast population of tight Galactic double WDs, NSs and BHs is in the low-frequency ($\sim$~mHz) GW window  for up to $\sim 10^6\;{\rm yr}$ prior to their merger event. Thus a significant number of such local sources are anticipated to be detected by LISA, even though their emitted GW luminosity is relatively small compared to that of the final merger process. The possibility that LISA can measure sky locations of its sources will allow for EM follow-up observations which may result in much more precise compact object component masses, e.g. compared to high-frequency GW mergers.

\begin{figure*}[h!]
    \begin{center}
   \includegraphics[width=0.60\textwidth]{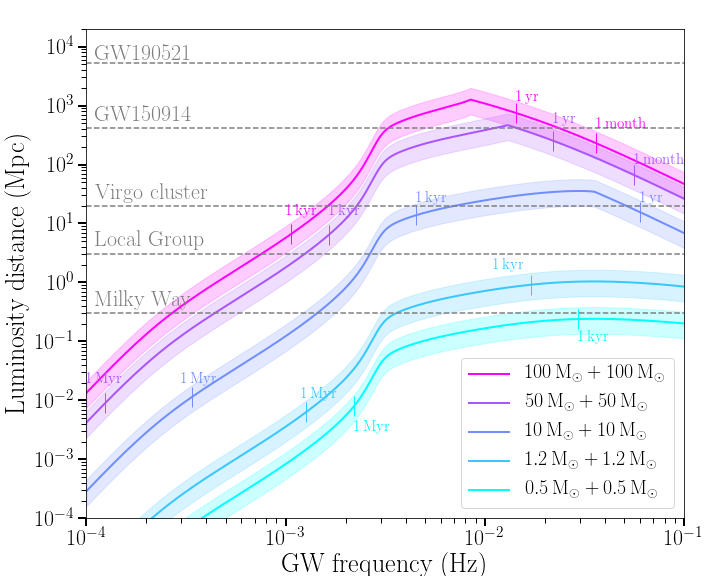}
\caption{Distance to which LISA binaries can be detected as a function of GW frequency. The coloured lines represent the SNR threshold of 7 (here computed assuming a mission duration of 4~yr with 100\% duty cycle) for (quasi-)stationary equal-mass circular binaries of different total masses in the distance--GW frequency parameter space. The shaded range represents angle-averaged curve limits for the optimal and worst binary orientation. The ticks on the curves represent binary merger times: for merger times $\gg 4\;{\rm yr}$ the binary will be seen by LISA as a monochromatic GW source, whereas for merger times $< 4\;{\rm yr}$ the binary will be seen as evolving. Note in particular that evolving sources like GW190521 and GW150914 remain within the LISA band for less than the mission lifetime. Figure credit: Antoine Klein \& Valeriya Korol.}
 \label{fig:horizons}
  \end{center}
\end{figure*}

Binary population synthesis studies and early data-analysis work predicts of order $10^4$ resolved Galactic WD+WD may be detected with LISA. This population includes both detached WD+WD and those undergoing mass transfer (the so-called AM Canum Venaticorum binaries or AM~CVns, see Section~\ref{paragr:AM-CVns}). NS+NS systems are also expected to be detected by LISA. Based on the known Galactic population of tight-orbit radio pulsar binaries in combination with population synthesis predictions, an estimated number of $10^1-10^2$ NS+NS systems with a significant signal-to-noise ratio (SNR) may be detected by LISA within a 4-year mission (Section~\ref{paragr:DNS}). An even larger number of NS+WD systems is expected to be detected too, including ultra-compact X-ray binaries (UCXBs, see Section~\ref{paragr:UCXBs}, a sub-class of low-mass X-ray binaries, LMXBs). Binary BHs (BH+BH) detectable by LISA are strong candidates to become the first discoveries of such systems in the Milky Way.  
Given that LISA’s volume sensitivity for a constant SNR scales with chirp mass to the fifth power, $\mathcal{M}^5_{\rm chirp}$, BH+BH sources may be detected in distant galaxies, located several hundreds of megaparsecs away (see examples in Fig.~\ref{fig:horizons}). Interestingly enough, this fortuitous condition will therefore allow LISA to discover extra-galactic BH+BHs several years before the final merger events that LIGO/Virgo/KAGRA or Einstein Telescope/Cosmic Explorer will detect. Finally, LISA is also expected to detect rare Galactic systems such as (see Section~\ref{subsubsec:other_sources}): triple stellar systems, tight systems of WDs with exoplanets, or helium star binaries.

The LISA mission will provide  opportunities to learn new physics and answer key scientific questions related to formation and evolutionary processes of tight binary and multiple stellar systems containing compact objects. This includes questions related to the stability and efficiency of mass transfer, common envelopes (CEs), tides and stellar angular momentum transport, irradiation effects, as well as details of their formation and destruction in core-collapse supernovae (SNe) and Type~Ia SNe (and related transients), respectively. Furthermore, information about the environments of these sources will be available too, and the number and Galactic distribution of LISA sources are excellent probes to gain new knowledge on the star formation history and the structure of the Milky~Way. Finally, the sheer numbers of LISA sources will provide crucial knowledge concerning their formation and evolution processes and help to place constraints on key physical parameters related to binary (and triple-star) interactions.

The current catalogue of known LISA ``guaranteed sources'' consists of detached WD+WDs, accreting AM~CVn binaries, a hot subdwarf binary, and an UCXB. Although the sample is still small and inhomogeneous, binary population synthesis predicts a large population of multi-messenger sources that are EM bright and also detectable by LISA. This includes up to a few thousand detectable WD+WDs as well as a few tens of NS
or BH binaries, with a population strongly peaking towards the Galactic Plane/Bulge. Many sources will be detected across different EM bands. Detached WD+WDs and NS+WDs are typically seen in optical and UV bands, whereas AM~CVn systems and UCXBs are also seen in X-rays. NSs in compact binaries can potentially be detected as pulsars in the radio band. Therefore, in parallel with the LISA mission, we expect an EM bright future of thousands of resolved Galactic LISA binaries.

Systems with orbital periods $<20\;{\rm min}$ will be the strongest Galactic LISA sources and will be detected by LISA within weeks after science operations begin. These verification binaries, as well as other so far unknown loud sources, are crucial in facilitating the functional tests of the instrument and maximize LISAs scientific output. Combined GW and EM multi-messenger studies of UCXBs will allow us to derive population properties of these systems with unprecedented quality including for the first time the effects of tides compared to GW  radiation. Tides are predicted to contribute up to $10\,\%$ of the orbital decay. For accreting WDs as well as NS binaries, multi-messenger observations give us the possibility to study the angular momentum transport due to mass transfer. In particular for monochromatic GW sources, EM observations are required to break degeneracies in the GW data (e.g. between masses and distance).

\newpage

\subsection	{Classes of LISA binaries}\label{subsec:classes-binaries} %
\subsubsection{Known binaries --- LISA verification sources}\label{subsec:verification sources}

\noindent \coord{Thomas Kupfer, Thomas Tauris}\\
\noindent \contr{Thomas Kupfer, Thomas Tauris, Silvia Toonen, Tassos Fragos}\\

The most abundant sources in the LISA band will be binary stars with orbital periods $<60$\,min, so-called ultra-compact binaries (UCBs). They are a class of binary stars with ultrashort orbital periods, consisting of a WD or NS primary and a compact helium-star/WD/NS secondary. A subset of the known UCBs have predicted GW strains high enough that they will be individually detected due to their strong GW signals \citep[e.g.][]{2020ApJ...905...32B}. These LISA guaranteed sources are termed {\it verification binaries} with some being expected to be detected on a timescale of weeks or a few months \citep{2006CQGra..23S.809S}. Currently, we know of only about two dozen of these systems although hundreds are predicted by theory to be detectable in our Galaxy \citep[e.g.][]{2004MNRAS.349..181N,2006PhRvD..73l2001T,2013MNRAS.429.2361L,2017MNRAS.470.1894K, 2017ApJ...846...95K, 2018MNRAS.480..302K, 2019MNRAS.490.5888L}.

\begin{figure*}[h!]
    \begin{center}
  \includegraphics[width=0.73\textwidth,]{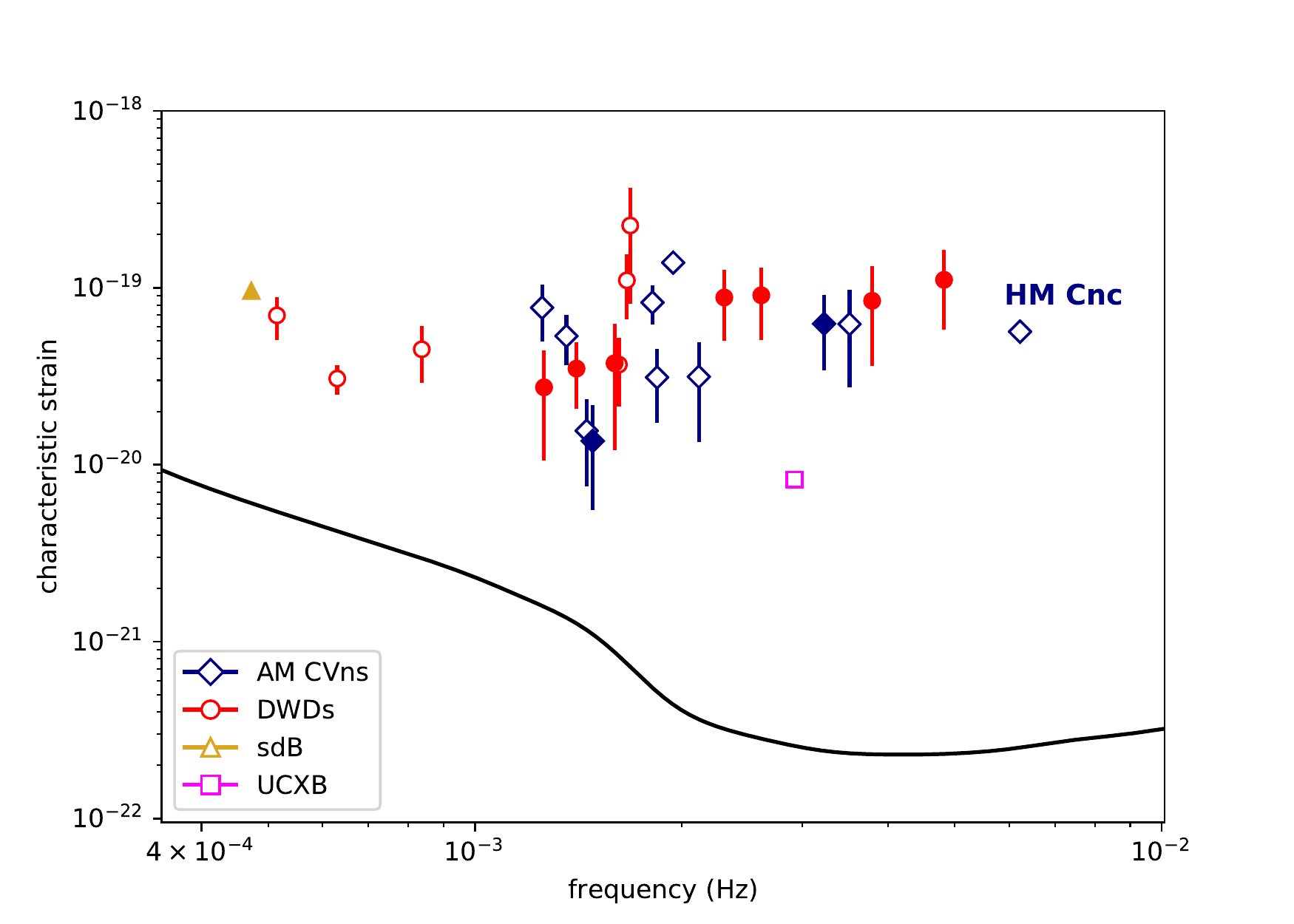}
\caption{Sensitivity plot for LISA assuming 10~yr of observation with an 80\% duty cycle showing the known binaries which reach a ${\rm SNR}\geq5$. Filled symbols represent eclipsing sources and open symbols represent non-eclipsing sources from \citet{2018MNRAS.480..302K}. The black lack solid line represents the LISA sensitivity curve. Acronyms for binaries: AM Canum Venaticorum (AM~CVns), WD+WD (DWDs), subdwarf B-star (sdB) and ultracompact X-ray binary (UCXB). Figure credit: Thomas Kupfer.}
 \label{fig:strain}
  \end{center}
\end{figure*}

\begin{figure*}[h!]
    \begin{center}
  \includegraphics[width=0.95\textwidth,]{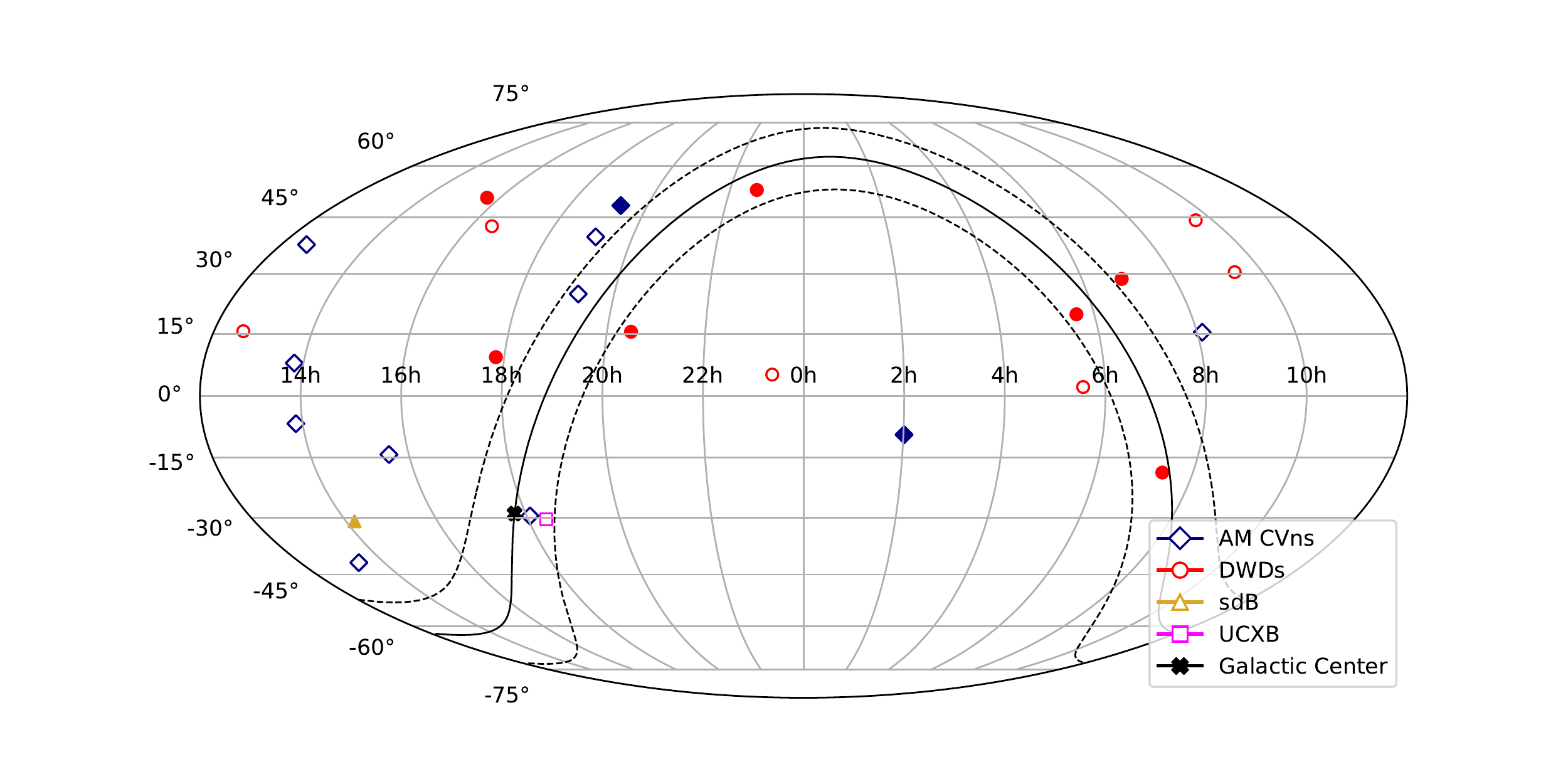}
\caption{Sky position of the verification binaries. The sky positions show a clear bias towards the northern hemisphere and to higher Galactic latitudes. The black line indicates the Galactic equator and $|b| = 10$\,deg, with the Galactic Centre located at the black cross. See caption in Fig.~\ref{fig:strain} for explanation of acronyms. Figure credit: Thomas Kupfer.}
 \label{fig:skylocation}
  \end{center}
\end{figure*}

At present, the catalogue of verification binaries include 13 WD+WDs, 11 semi-detached accreting WDs (AM\,CVn binaries, a subclass of Cataclysmic Variables, CVs), one hot subdwarf star with a WD companion, and one semi-detached UCXB. Table\,\ref{tab:verif_systems} and \ref{tab:verif_systems1} present an overview of the known systems with observed EM properties. Figure\,\ref{fig:strain} shows the characteristic strain of the known verification binaries which reach a predicted ${\rm SNR}\geq5$ in LISA assuming an optimistic 10~year mission with an 80\% duty cycle. So far large-scale searches for verification binaries have been conducted almost exclusively in the northern hemisphere, because large-scale survey instruments (e.g. the Sloan Digital Sky Survey, SDSS, and the Zwicky Transient Facility, ZTF) are located in the Northern Hemisphere, and mostly at high Galactic latitudes, to avoid stellar crowding. Fig.\,\ref{fig:skylocation} shows the sky location of the known verification systems which presents the strong bias towards sources in the Northern Hemisphere.

In 2018, Gaia data release 2 \citep{2018A&A...616A...1G} announced parallaxes for $\approx1.3$ billion sources. The Gaia catalogue contains the distances of many of the known LISA verification binaries, allowing accurate prediction of their GW strains. Using the Gaia distances, \citet{2018MNRAS.480..302K} found 13 sources will exceed an SNR of 5 after 4~yr of LISA observations. This  sample consists of 13 verification binaries from the current, known list; it is strongly biased and incomplete. It includes AM\,CVn, CR~Boo, V803~Cen and ES~Cet, which were all found as outliers in surveys for blue, high-Galactic latitude stars. HM~Cnc and V407~Vul are the most compact known AM\,CVn systems and were discovered during the course of the ROSAT All-Sky Survey showing an on/off X-ray profile modulated on a period of 321 and 569~s respectively. The known WD+WD verification binaries, such as SDSS J0651 and SDSS J0935, were found as part of the extremely low-mass (ELM) WD survey (\citealt{2020ApJ...889...49B} and references therein).

More recently, more systematic searches for UCBs were performed. UCBs show up in lightcurves with variations on timescales of the orbital period (e.g.\, due to eclipses or tidal deformation of the components). Therefore, photometric surveys are well suited to identify UCBs in a homogeneous way. A number of fast cadence ground-based surveys, including the Rapid Temporal Survey (RATS; \citealt{2005MNRAS.360..314R,2011MNRAS.413.2696B}), OmegaWhite (\citealt{2015MNRAS.454..507M}) survey as well as the ZTF high-cadence Galactic plane survey \citep{2019PASP..131a8003M,2021MNRAS.505.1254K}, have been executed to study the variable sky down to a few minute period aiming to find UCBs and increase the number of known verification binaries. The ELM survey targets a colour-selected sample of B-type hypervelocity candidates from SDSS \citep[][]{2005AJ....130.2230A,2007MNRAS.382..685R}, which are being followed up systematically (\citealt{2020ApJ...889...49B} and references therein). ELM~WDs can be separated efficiently from the bulk of WDs with a colour selection \citep{2010ApJ...723.1072B}.

Over the last few years the number of known verification binaries has almost doubled thanks to these large scale surveys. The two most significant contributors were the ELM survey (\citealt{2020ApJ...889...49B} and references therein) and ZTF \citep{2019Natur.571..528B, 2020ApJ...905...32B,2020arXiv201003555B}. The ELM survey discovered six WD+WD verification binaries including SDSS\,J0651: a detached eclipsing system with an orbital period of 12~min. Most recently ZTF released seven new WD+WD verification binaries, five systems found as eclipsing sources. Remarkably, one of the first ZTF discoveries was the shortest orbital period eclipsing WD+WD known to date, ZTF~J1539+5027, with an orbital period of just 6.91~min \citep{2019Natur.571..528B}. 

\begin{table*}[ht]
\begin{center}
\caption{Physical properties (orbital periods, component masses, inclination angles) of the known verification binaries which reach a ${\rm SNR}>5$ after a 10~yr LISA mission with 80\% duty cycle. Masses and inclination angles in brackets are assumed and based on evolutionary stage and mass ratio estimations.}
\begin{tabular}{lrllll}
\hline
\hline
Source         & $P_{\rm orb}$ &  $M_1$ & $M_2$ & $\iota$ & Refs.  \\
             &   (s)          &  (M$_\odot$) &    (M$_\odot$)  &  ($\deg$)    \\
                      \hline
\multicolumn{2}{l}{{\bf AM\,CVn type}}      &     \\
HM Cnc                     &   321.5  &   0.55     &  0.27  & $\approx$38 & 1, 2  \\
V407 Vul                   &   569.4  &   [0.8$\pm$0.1]   & [0.177$\pm$0.071]   & [60]  & 3  \\
ES Cet                     &   620.2  & [0.8$\pm$0.1]  &  [0.161$\pm$0.064] &  [60] & 4  \\ 
SDSS J135154.46$-$064309.0 &   943.8  &    [0.8$\pm$0.1]     &  [0.100$\pm$0.040]  & [60]  & 5  \\
AM CVn                     &  1028.7  &  0.68$\pm$0.06 &  0.125$\pm$0.012  & 43$\pm$2 & 6, 7  \\
SDSS J190817.07+394036.4   &  1085.7  & [0.8$\pm$0.1]  & [0.085$\pm$0.034]  & 10 - 20 & 8, 9  \\
HP Lib                     &  1102.7  &  0.49-0.80 & 0.048-0.088 & 26-34 & 10,11  \\
PTF1 J191905.19+481506.2   &  1347.3  & [0.8$\pm$0.1] & [0.066$\pm$0.026] & [60] & 12  \\
CXOGBS J175107.6$-$294037  &  1375.0  & [0.8$\pm$0.1] &  [0.064$\pm$0.026] & [60]  & 13  \\
CR Boo                     &  1471.3  & 0.67-1.10 &  0.044-0.088 & 30  & 10,14  \\
V803 Cen                   &  1596.4  &  0.78-1.17 & 0.059-0.109 & 12 - 15 & 11  \\
     \noalign{\smallskip}
 \multicolumn{3}{l}{{\bf Detached double WDs}}         & \\
 ZTF J1539+5027            &  414.8  &    $0.610^{+0.017}_{-0.022}$ &	$0.210\pm0.015$	& $84.15^{+0.64}_{-0.57}$   &   15    \\
 ZTF J2243+5242            &  527.9  &   $0.349^{+0.093}_{-0.074}$ & $0.384^{+0.114}_{-0.074}$ & $81.88^{+1.31}_{-0.69}$ &   16   \\
SDSS J065133.34+284423.4   &  765.5  &  0.247$\pm$0.015 &  0.49$\pm$0.02 & ${86.9^{+1.6}_{-1.0}}$  & 17, 18  \\
 ZTF J0538+1953            &  866.6  &    $0.45\pm0.05$  &	$0.32\pm0.03$   &  $85.43^{+0.07}_{-0.09}$  &   19 \\
SDSS J093506.92+441107.0   & 1188.0  &  0.312$\pm$0.019 &  0.75$\pm$0.24 &  [60]  & 20, 21  \\
SDSS J2322+0509            & 1201.4  &  $0.34\pm0.02$   & $>0.17$ &  [60]    & 22 \\
 PTF J0533+0209            & 1234.0  &   $0.652^{+0.037}_{-0.040}$ &  $0.167\pm0.030$  & $72.8^{+0.8}_{-1.4}$ & 19, 23 \\
 ZTF J2029+1534            & 1252.0  &   $0.32\pm0.04$  &   $0.3\pm0.04$   &  $86.64^{+0.70}_{-0.40}$ &   19 \\
 ZTF J0722$-$1839          & 1422.5  &   $0.38\pm0.04$  & $0.33\pm0.03$   & $89.66\pm0.22$   &  19  \\
 ZTF J1749+0924            & 1586.0  &  $0.40^{+0.07}_{-0.05}$  &  $0.28^{+0.05}_{-0.04}$  &	$85.45^{+1.40}_{-1.15}$  &  19   \\
SDSS J163030.58+423305.7   & 2389.8  &  0.298$\pm$0.019 &  0.76$\pm$0.24 & [60]  & 20, 24 \\
SDSS J1235+1543            & 3172.6  &  $0.35\pm0.01$ &  $0.27^{+0.06}_{-0.02}$  &	$27\pm3.8$ &  25, 26  \\
SDSS J092345.59+302805.0   & 3883.7  &  0.275$\pm$0.015 &  0.76$\pm$0.23 & [60] & 20, 27 \\
      \noalign{\smallskip}
  \multicolumn{3}{l}{{\bf Hot subdwarf binaries}}      & \\
  CD$-$30$^\circ$11223       & 4231.8   &   0.54$\pm$0.02     &  0.79$\pm$0.01  & 82.9$\pm$0.4 & 28 \\
        \noalign{\smallskip}
  \multicolumn{3}{l}{{\bf Ultracompact X-ray binaries}}      & \\
  4U1820$-$30                &  685.0  &   [1.4]	&  [0.069]	&  [60]  &  29, 30 \\  
\hline
 \hline
 \end{tabular}
\begin{flushleft}
[1] \citet{2005ApJ...627..920S}, [2] \citet{2010ApJ...711L.138R}, [3] \citet{2002MNRAS.331L...7M}, [4] \citet{2005PASP..117..189E}, [5] \citet{2018MNRAS.477.5646G}, [6] \citet{1999PASP..111.1281S}, [7] \citet{2006MNRAS.371.1231R}, [8] \citet{2011ApJ...726...92F}, [9] \citet{2015MNRAS.453..483K}, [10] \citet{2007MNRAS.379..176R}, [11] \citet{2021MNRAS.500.1222S}, [12] \citet{2014ApJ...785..114L}, [13] \citet{2016MNRAS.462L.106W}, [14] \citet{1997ApJ...480..383P}, [15] \citet{2019Natur.571..528B}, [16] \citet{2020arXiv201003555B}, [17] \citet{2011ApJ...737L..23B}, [18] \citet{2012ApJ...757L..21H}, [19] \citet{2020ApJ...905...32B}, [20] \citet{2016ApJ...824...46B},  [21] \citet{2014MNRAS.444L...1K}, [22] \citet{2020ApJ...892L..35B}, [23] \citet{2019ApJ...886L..12B}, [24] \citet{2011MNRAS.418L.157K}, [25] \citet{2017MNRAS.468.2910B}, [26] \citet{2017MNRAS.471.4218K}, [27] \citet{2010ApJ...723.1072B}, [28] \citet{2013A&A...554A..54G}, [29] \citet{1987vgex.conf..157S}, [30] \citet{2020ApJ...900L...8C}
 \label{tab:verif_systems}
\end{flushleft}
\end{center}
\end{table*}

\begin{table*}[ht]
\begin{center}
\caption{Measured EM properties (Galactic coordinates, GW frequency, magnitudes and parallaxes from Gaia early data  release 3 \citep{2020arXiv201201533G} of the known verification binaries which reach a ${\rm SNR}>5$ after a 10~yr LISA mission with 80\% duty cycle.}
\begin{tabular}{lrrrrc}
\hline
\hline
Source              & $f_{\rm GW}$    &  $l_{\rm Gal}$   &  $b_{\rm Gal}$ &  Gaia G &  $\varpi$   \\
                   & (mHz)    & (deg)  & (deg) & mag    &   (mas)      \\
                      \hline
\multicolumn{3}{l}{{\bf AM\,CVn type}}      &     \\
HM Cnc                     &  6.22  &  206.9246 &	23.3952 &	20.92 &  ---  \\
V407 Vul                   &  3.51  &  57.7281	&  6.4006   &  19.36  & 0.0978$\pm$0.2384 \\
ES Cet                     &  3.22  &  168.9684 & $-$65.8632	 & 16.80 & 0.5606$\pm$0.0677  \\ 
SDSS J135154.46$-$064309.0 &  2.12  &  328.5021 & 53.1240 &	18.72  & 0.6584$\pm$0.2197 \\
AM CVn                     &  1.94  &  140.2343 & 78.9382  & 14.06  & 3.3106$\pm$0.0303 \\
SDSS J190817.07+394036.4   &  1.84  &  70.6664 & 13.9349 & 16.22 & 1.0232$\pm$0.0335 \\
HP Lib                     &  1.81  & 352.0561 & 32.5467 & 13.60 & 3.5674$\pm$0.0313 \\
PTF1 J191905.19+481506.2.  &  1.48  & 79.5945 & 15.5977 & 19.75 & 0.6229$\pm$0.2385 \\
CXOGBS J175107.6$-$294037  &  1.45  &  359.9849 & $-$1.4108 & 16.27 & 0.8591$\pm$0.1733 \\
CR Boo                     &  1.34  & 340.9671 & 66.4884 &  15.47 & 2.8438$\pm$0.0367  \\
V803 Cen                   &  1.25  & 309.3671 & 20.7262 & 15.73 & 3.4885$\pm$0.0599 \\
     \noalign{\smallskip}
 \multicolumn{3}{l}{{\bf Detached double WDs}}         & \\
 ZTF J1539+5027            &  4.82  & 80.7746 & 50.5819 & 20.40 & $-$0.4926$\pm$0.5726 \\
 ZTF J2243+5242            &  3.79  & 104.1514 & $-$5.4496 & 20.55 & $-$1.2372$\pm$0.6578 \\
SDSS J065133.34+284423.4   &  2.61  & 186.9277 & 12.6886 & 19.28 & 1.0071$\pm$0.3091    \\
 ZTF J0538+1953            &  2.31  & 186.8104 & $-$6.2213 & 18.80 & 0.9617$\pm$0.2866 \\
SDSS J093506.92+441107.0   &  1.68  & 176.0796 & 47.3776 & 17.80 & 2.7034$\pm$0.6648  \\
 SDSS J2322+0509           &  1.66  & 85.9507 & $-$51.2104 & 18.75 & 1.1558$\pm$0.2244 \\
 PTF J0533+0209            &  1.62  & 201.8012 & $-$16.2238 & 19.05 & 0.7902$\pm$0.2396  \\
 ZTF J2029+1534            &  1.60  & 58.5836 & $-$13.4655 & 20.47 & 0.1240$\pm$0.9893   \\
 ZTF J0722$-$1839          &  1.40  & 232.9930 & $-$1.8604 & 19.05 & 0.6996$\pm$0.2457 \\
 ZTF J1749+0924            &  1.26  & 34.5093 & 17.9025 & 20.47 & $-$0.2961$\pm$0.8222 \\
SDSS J163030.58+423305.7   &  0.84  & 67.0760 & 43.3604 & 19.18 & 1.1748$\pm$0.1952 \\
SDSS J1235+1543            &  0.63  & 284.5186 & 78.0320 & 17.52 &  2.2504$\pm$0.1389  \\
SDSS J092345.59+302805.0   &  0.51  & 195.8199 & 44.7754 & 15.92 & 3.4795$\pm$0.0648 \\
      \noalign{\smallskip}
  \multicolumn{3}{l}{{\bf Hot subdwarf binaries}}      & \\
  CD--30$^\circ$11223       &   0.47  & 322.4875 & 28.9379 & 12.30 & 2.8198$\pm$0.0516 \\
        \noalign{\smallskip}
  \multicolumn{3}{l}{{\bf Ultracompact X-ray binaries}}      & \\
  4U1820$-$30                &  2.92  & 2.7896 & $-$7.9144 & 15.41 & $-$0.7676$\pm$0.2164 \\
\hline
 \hline
 \end{tabular}
\begin{flushleft}
 \label{tab:verif_systems1}
\end{flushleft}
\end{center}
\end{table*}

\clearpage
\bigskip

\subsubsection{Detached binaries}\label{subsubsec:detached-binaries}

\noindent \coord{Ashley Ruiter, Ross Church}\\
\noindent \contr{Ashley Ruiter (1.2.2.1), Thomas Tauris (1.2.2.2--4), Jeff Andrews (1.2.2.3), Simone Bavera (1.2.2.4), Ross Church (1.2.2.2), Tassos Fragos (1.2.2.4), Gijs Nelemans (1.2.2.1), Milton Ruiz (1.2.2.3), Alberto Sesana (1.2.2.4), Antonios Tsokaros (1.2.2.3), Shenghua Yu (1.2.2.3)}

\paragraph{WD+WD systems}\label{paragr:DWD-detached} %

\phantom{text}

For over three decades it has been known that WD+WD binaries will be the dominant contributor to signals detected by a space-based GW observatory \cite[][]{1990ApJ...360...75H}. While most extra-galactic sources \citep{2003MNRAS.346.1197F} as well as a significant fraction of those in the Galactic halo \citep{2009ApJ...705L.128R} are likely too distant to be individually detected by LISA, a large portion of the frequency band observed by LISA will be swamped with GWs from millions of WD+WDs existing in the Galactic disc and bulge. At low frequencies, the combined signal of these millions of WD+WDs will populate just a few frequency bins and merge to form an unresolved confusion foreground (often referred to as the galactic foreground or the galactic confusion noise), with louder resolvable sources standing out above the confusion (see also Section~\ref{subsubsec:GWforeground}). Together with high-frequency sources a large number these form $\sim 10^4$, (e. g., \citealt{2001A&A...365..491N,2003MNRAS.346.1197F,2010ApJ...717.1006R,2017MNRAS.470.1894K}) of resolved WD+WDs and we now discuss these key sources in more detail.

WD+WDs were discovered in the late 1980s and initially were dominated by low-mass ($\lesssim 0.4\;M_\odot$) helium-core (He-core)~WDs that cannot be formed in single-star evolution within a Hubble time, and, thus, were the targets for radial velocity searches for binarity amongst known WDs \citep{1995MNRAS.275..828M}. Later, (more) unbiased surveys were done, e.g. the Supernova Ia Progenitor surveY \citep[SPY, ][and references therein]{2020A&A...638A.131N} and studies using SDSS,
discovering also more massive WDs. Over the last decade, it has become more clear that previously-undetected WD+WD systems (and their progenitors, e.g. double-core planetary nebulae) are more easily detectable with today's sophisticated instrumentation \citep{2018MNRAS.480.4589W}. A dramatic increase in the number of WD+WDs has come from the ELM WD survey \citep{2012ApJ...751..141K,2017ApJ...835..180B,2020ApJ...889...49B} that targets a part of the parameter space in colour--colour diagrams that is occupied by (subdwarf) B-stars, but also by very low mass (below $\sim 0.3\;M_\odot$) proto-WDs that are still approaching the cooling track and are thus relatively large and bright \citep{2014A&A...571L...3I,2016A&A...595A..35I}. In total, the ELM survey alone has discovered 98 WD+WDs so far \citep{2020ApJ...889...49B}. 

Over the past couple of years, ZTF \citep[[]{2019PASP..131a8002B} has facilitated a rapid growth in the population of known WD+WDs with orbital periods under an hour. Three of the sources discovered by ZTF so far \citep{2020arXiv201003555B}, the eclipsing WD+WDs: ZTF~J1539+5027 ($P_b = 6.91\;{\rm min}$), ZTF~J2243+5242 ($P_b = 8.80\;{\rm min}$) and ZTF~J0538+1953 ($P_b = 14.4\;{\rm min}$),
should all be detected by LISA with a high SNR, enabling precise parameter estimation using GWs \citep{2019ApJ...881L..43L}.

The detached WD+WDs may consist of a pair of He-core WDs, carbon/oxygen-core (C/O-core) WDs, oxygen/neon/magnesium-core (O/Ne/Mg-core) WDs, or any mixed combination thereof. For some systems, LISA measurements of the orbital-decay rate will yield the chirp mass for a given system, which can be combined with EM observations to reveal individual WD component masses. The distribution of WD masses (and their mass ratios), along with the number of detectable sources in a local volume, will provide important information to help understand their formation history (see Section~\ref{subsec:formation-binaries}). 
With enough detached WD+WDs in a sample, it may even be possible to set limits on mass-transfer efficiencies and CE physics (Section~\ref{subsec:scientificQ}) through characterisation of chirp mass distributions \citep{2019MNRAS.484..698R}. 
Furthermore, the detected WD+WDs will provide unique information on the formation of progenitors of R~Coronae Borealis stars, thought to be formed by the merger of two WDs, e.g. \citep{2020A&A...635A..14T}, massive carbon-enhanced WDs \citep{2020MNRAS.491L..40K}, Type~Ia SNe (see Section~\ref{paragr:SNeIa}), and other transients. 
 
Over the last three decades, several works have made predictions about the scientific impact of LISA detections of Galactic WD+WDs. Different binary evolution population synthesis studies have uncovered how the WD+WD population will look to LISA in terms of characteristic strain amplitude \citep{2001A&A...375..890N,2004MNRAS.349..181N,2010A&A...521A..85Y,2019MNRAS.490.5888L,2020A&A...638A.153K}, spectral density \citep{2020ApJ...901....4B}, as well as how different populations of WD+WDs contribute to the spectral amplitude signal \citep{2009ApJ...705L.128R,2010ApJ...717.1006R}.

Detached WD+WD binaries that are resolvable with LISA are expected to be on par with or slightly outnumber the resolvable interacting WD+WD binaries \citep{2001A&A...368..939N,2004MNRAS.349..181N,2010ApJ...717.1006R}, and will be the sole WD+WD contributers to the LISA signal at GW frequencies below $\sim 2 \times 10^{-4}\;{\rm Hz}$. \cite{2017ApJ...846...95K} found that a number of {\em mass-transferring} WD+WDs (Section~\ref{subsubsec:interacting-binaries}) will be resolvable with LISA ($\sim 200-3000$ for SNRs between 10 and 5, respectively), many of which are likely to exhibit a negative chirp (caused by orbital widening) --- a diagnostic not applicable for detached WD+WDs. 
Finally, we expect that the number of detached WD+WDs, composed of a light He-core~WD and a more massive C/O-core~WD (or possibly an O/Ne/Mg-core~WD) detected by LISA must be in accordance with the number of similar interacting AM~CVn systems that LISA will detect, given their evolutionary connection (the detached systems being the precursors of the interacting WD+WDs, see Fig.~\ref{fig:DWD}). The transitional GW frequency between these two populations (detached and interacting) depends on the mass and temperature of the lighter (last-formed) WD, see examples in Figs.~\ref{fig:UCXB_strain+data} and \ref{fig:UCXB-AMCVn-detailed}.

\paragraph{NS+WD and BH+WD systems}\label{paragr:WD+N} 

\phantom{text}

The known population of Galactic NS+WD systems can be divided into two classes. Systems with: i) massive WDs (O/Ne/Mg-core or C/O-core WDs, typically more massive than $0.7\,M_\odot$), and ii) low-mass He-core~WDs (typically less massive than $0.3\,M_\odot$). The massive NS+WD systems can again be subdivided into two populations, depending on the formation order of the WD and the NS. The NS+WD systems are observed as binary radio pulsars and the formation order can be clearly distinguished from the properties of the pulsar: if the pulsar has a strong B-field and an eccentric orbit \citep[e.g.][]{2000A&A...355..236T,2006MNRAS.372..715C}, it is the last-formed compact object, whereas if the pulsar is (mildly) recycled with a low-B-field and a fairly rapid spin, and in a near-circular orbit, it is the first-formed compact object \citep[][]{2012MNRAS.425.1601T}. 
For LISA detections, the formation order is irrelevant and among both types of systems examples are known to merge within a Hubble time, thus producing a bright LISA source well before their final merger.

Among the detached low-mass He-core~WDs with NS companions, the systems in relatively tight orbits are completely dominated by millisecond radio pulsars. 
According to the ATNF Pulsar Catalogue \citep{2005AJ....129.1993M}, there are about 120 such systems known in the MW disc, a handful of which will merge within a Hubble time, producing a bright detached LISA source (depending on their distance) for approximately the last several tens of Myr of the inspiral, before an UCXB is formed (Section~\ref{paragr:UCXBs}).
Based on the observed population of radio pulsars and their selection effects, \citet{2018PhRvL.121m1105T} argue that LISA could detect about 50 of these systems while still detached, before they become UCXBs and widen their orbits again, resulting in a negative chirp of the GW signal (see Figs.~\ref{fig:UCXB_strain+data} and \ref{fig:UCXB-AMCVn-detailed}). 

At present, we do not know of any detached BH+WD binaries. However, this is probably due to observational selection effects since the only EM radiation we would expect from such detached systems would be from the cooling of the WD companion --- unlike the situation for semi-detached systems or systems containing NSs, which can be detected in X-rays and radio waves, respectively. Nevertheless, several Galactic LMXBs are known with low-mass donor stars and BH accretors \citep{2006csxs.book..157M} and thus we expect many of these systems to leave detached BH+WD systems, possibly (although still to be proven) in tight orbits that LISA will detect. A more viable formation channel for more massive WDs in tight orbits with BHs is formation via a CE.
Optical follow-up observations of the WD companion, in combination with the measured chirp mass, will constrain the BH mass in these systems. Early simulations \citep{2001A&A...375..890N} predict a Galactic merger rate of BH+WD binaries of order $\sim 100\;{\rm Myr}^{-1}$ and thus roughly $\sim 100$ such systems detectable by LISA.

\paragraph{NS+NS systems}\label{paragr:DNS} 

\phantom{text}

The known population NS+NS systems so far only manifest themselves as radio pulsars. The first one of these (PSR~B1913$+$16, the {\it Hulse--Taylor Pulsar}) was discovered in 1974 \citep{1975ApJ...195L..51H}. According to the ATNF Pulsar Catalogue \citep{2005AJ....129.1993M}, there are currently about 20 NS+NS systems detected in our Galaxy. Except for one case, the {\it double pulsar} PSR~J0737$-$3039 \citep{2004Sci...303.1153L}, only one of the two NSs is detected --- usually the recycled pulsar \citep{2017ApJ...846..170T}. The other NS, is either not an active radio pulsar anymore or it is not beaming in our direction. 

Given the small merger rate of NS+NS systems in our Galaxy\footnote{The empirical rate and its uncertainty measured by the LIGO network of detectors is currently based on only two events (GW170817 and GW190425) but is anticipated to improve substantially in the coming decade.} (most likely somewhere in the range from a few events up to a hundred events per~Myr), it is statistically highly improbable that ground-based high-frequency detectors (LIGO--Virgo--KAGRA) will detect a NS+NS merger in the Local Group earlier than the LISA era. 
The advantage of LISA is that it can follow the inspiral of Galactic NS+NS systems up to $\sim 10^6\;{\rm yr}$ prior to their merger event, and thus a significant number of NS+NS sources are anticipated to be detected in GWs by LISA.

About half of the 20 known NS+NS systems have orbital periods small enough (or eccentricities sufficiently large) to merge within a Hubble time. As an example, a ``standard'' NS+NS system with NS masses of $1.35\;M_\odot$ and e.g. an orbital period of 16~hr will merge in: 11.8~Gyr, 4.4~Gyr or 0.35~Gyr for an initial eccentricity, $e_0$ of 0.1, 0.5 or 0.8, respectively. 
The number of NS+NS sources that LISA will detect can be evaluated, approximately to first order, from a combination of the Galactic NS+NS merger rate and the distribution of these sources within the Milky Way. The above three standard NS+NS systems will have a remaining lifetime of between $\sim 247\;{\rm kyr}$ ($\sim 243\;{\rm kyr}$ for $e_0=0.8$) and $1.57\;{\rm Myr}$ ($1.48\;{\rm Myr}$ for $e_0=0.8$) by the time they enter the LISA band, if this occurs at a GW frequency of about 2~mHz and 1~mHz, respectively. Thus, if the Galactic merger rate is, say, $10\;{\rm Myr}^{-1}$, we can roughly expect to detect between a few and a dozen LISA sources. Of course, the details depend on the Galactic distribution of these sources, the SNR required for a detection, and the duration of the LISA mission.
The merger rate can be estimated from population synthesis, but its value is uncertain by, at least, one or two orders of magnitude \citep{2010CQGra..27q3001A}. The merger rate derived from an extrapolation of the LIGO/Virgo empirical merger rate of NS+NSs still has very large error bars due to small number statistics.

Recent works by \citet{2020MNRAS.492.3061L,2020ApJ...892L...9A} suggest that LISA may even detect up to $\sim 50-200$ Galactic NS+NS sources with a SNR greater than 7 within a 4~yr mission. Given that LISA’s volume sensitivity for a constant SNR scales with $\mathcal{M}^5_{\rm chirp}$, unlike double BH sources, very few NS+NS sources are anticipated to be detected outside the Milky Way, although a few such binaries may be found in both the LMC and M31 \citep{2019MNRAS.489.4513S}. 
Applying a more conservative number, however, for the merger rate of Galactic NS+NS system of about $3-14\;{\rm Myr}^{-1}$ \citep{2018MNRAS.481.1908K} would lead to a substantial reduction in the predicted number of LISA detections. The Galactic merger rate is expected to be significantly better constrained in the coming decade such that we will have a clear idea about the expected number of NS+NS sources detected by LISA prior to its operation. Finally, an expected reduction in LISA SNR for detecting eccentric NS+NS systems, compared to circular NS+NS systems with similar orbital period and NS masses, should be noticed \citep{2021arXiv210316030R}.

LISA may give us the opportunity to probe a hidden subpopulation of NS+NS systems with different properties compared to those of the well-known radio pulsar NS+NSs.
The nature of GW190425, a presumed NS+NS merger detected by the LIGO/Virgo network with a total mass of $3.4\;M_\odot$ \citep{2020ApJ...892L...3A}, is still a mystery.
With such a large total mass, GW190425 stands at five standard deviations away from the total mass distribution of Galactic NS+NSs detected in the Milky Way as radio pulsars \citep{2019ApJ...876...18F}. If a subpopulation of heavy GW190425-like NS+NSs exists in our Galaxy, it is not yet clear why it should be radio-quiet \citep[e.g.][]{2020ApJ...900...13S}. Thus, LISA may actually be the most suited instrument for detecting the population of GW190425-like binaries \citep{2021ApJ...909L..19G, 2021MNRAS.502.5576K}. In particular, \citet{2021MNRAS.502.5576K} demonstrated that if GW190425-like binaries constitute a fraction larger than 10\% of the total Galactic population, LISA should be able to recover this fraction with better than $\sim 15$\% accuracy, assuming the merger rate of $42\;{\rm Myr}^{-1}$.

Additional recent investigations \citep{2020MNRAS.493.5408T,2019MNRAS.483.2615K} have discussed the importance of sky-localization on LISA NS+NS sources for multi-messenger follow-ups that may allow to impose constraints on the equation-of-state of NSs by measuring the Lense--Thirring precession \citep{2020MNRAS.493.5408T} or test general relativity through the detection of radio pulses from Galactic NS+NS binaries in a very tight orbit with the period shorter than 10~min \citep{2019MNRAS.483.2615K}. Sky-localization may also help disentangle NS+NS systems from others sources by either knowing their position in the Milky Way, or in nearby galaxies, thus enhancing the possibility of EM follow-ups~\cite[e.g.][]{2020MNRAS.492.3061L}. In particular, in addition to differences in chirp masses, it will allow us to distinguish between eccentric NS+NS and eccentric WD+WD systems --- the latter only expected to be formed in globular clusters and ejected into the Galactic halo via dynamical interactions, while the former systems have an eccentricity encoded from the last SN explosion.

\paragraph{BH+NS and BH+BH systems} 

\phantom{text}

For several decades, a number of high-mass X-ray binaries (HMXBs) containing BH accretors have been identified in the Milky Way and nearby galaxies \citep[e.g.\, Cyg~X-1, LMC~X-1, LMC~X-3, MCW~656, M33~X-7, see][]{2019IAUS..346....1V}.
It has been shown that a fraction of these known wind-accreting HMXBs may eventually form BH+BHs or BH+NS systems \citep[][see also Fig.~\ref{fig:BHNS}]{2012arXiv1208.2422B,2013ApJ...764...96B}, while others will merge in an upcoming CE phase (Section~\ref{paragr:CE}), once the companion star evolves to a giant-star size and possibly initiates dynamically unstable Roche-lobe overflow (RLO), depending on its stellar structure and the mass ratio between the two binary components. 
The masses of compact objects in X-ray binaries can be estimated with astrometry and the Galactic stellar-mass BHs are found to have masses between roughly $5-21\;M_{\odot}$ \citep{2019MNRAS.485.2642G,2021MNRAS.502.5455A,2019IAUS..346....1V,2021Sci...371.1046M}. The astrometric satellite Gaia, can also be used to detect optical emission from the HMXB companion stars  \citep{2014arXiv1407.6163B,2017IAUS..324...41K,2017MNRAS.470.2611M,2017ApJ...850L..13B,2018ApJ...861...21Y}. 
Finally, non-interacting binaries with a BH component have also been discovered by combining radial velocity measurements with photometric variability data \citep{2017ApJ...850L..13B,2019Sci...366..637T,2019Natur.575..618L}, although their interpretations can, in some cases, be subject to alternative explanations \citep{2020Sci...368.3282V}. 

The LIGO--Virgo GW detectors have detected BH+BH mergers in distant galaxies, out to $\sim 5\;{\rm Gpc}$. \citep{2020arXiv201014527A}. The inferred BH masses\footnote{Notice, the $2.6\;M_\odot$ compact object in GW190814 might well have been a massive NS rather than a BH.} of their inspiralling BH components potentially range all the way from $\sim 2.6\pm{0.1}\;M_\odot$ \citep{2020ApJ...896L..44A} to $\sim 95\pm{10}\;M_\odot$ \citep{2020arXiv201014527A}, thus significantly more massive than the known Galactic stellar-mass BHs. This difference is mainly attributed to the relatively high metallicity content of the Galaxy \citep{2016Natur.534..512B,2018MNRAS.481.1908K}.
The LIGO--Virgo GW detectors have also identified two sources in close proximity or within the mass ranges expected for BH+NS binaries: GW190426 and GW190814. 
The first event has marginal significance (i.e. a high false-alarm rate, ${\rm FAR}=1.4\;{\rm yr}^{-1}$) and the second is likely to be a BH+BH, not a BH+NS. Nevertheless, 
LISA like LIGO is much more sensitive to the masses (as opposed to matter content) of the binaries, so the presence of similar binaries suggests LISA will copiously find similar sources.

Interestingly enough, although the LIGO--Virgo GW sources are located in distant galaxies at Gpc distances, their low-frequency GWs during the last few years of inspiral prior to the merger event is often so luminous that it allows for detection with LISA \citep{2016PhRvL.116w1102S}. For example, the very first GW source (GW150914) would have been observable by LISA several years before its merger (see Fig.~\ref{fig:UCXB_strain+data}). Similarly, the extreme event GW190521 \citep{2020PhRvL.125j1102A,2020arXiv201006056T}, with a total stellar mass of $\sim 160\;M_\odot$ and located at a distance of about 5~Gpc, would also have been detected during its inspiral in the LISA band. 

\paragraph{Stochastic background}\label{subsec:SGWB}

\phantom{text}

As discussed above, stellar-mass compact binaries (BH+BH, BH+NS, NS+NS) are one of the primary targets for LISA, with expected detection rates of between a few and a few thousands per year, as summarized in Sec\,\ref{subsec:rates}. Nevertheless, many of these sources will not be detected, either because they are too distant and thus have a low SNR, or because the signals from multiple long-lived sources will overlap in time and will be difficult to disentangle. These unresolved signals will combine incoherently and produce a stochastic GW background (SGWB). Current predictions of the amplitude of this background typically rely on the merger rates measured in the local Universe by LIGO-Virgo \citep{2016PhRvL.116m1102A,2018PhRvL.120i1101A}, but since most of the unresolved sources reside at higher redshifts, these predictions depend on the detailed population synthesis and galaxy evolution models. 

The expected amplitude of the SGWB from BH+BH binaries in the LISA band (without source subtraction) varies in the range $\Omega_{\rm GW}(f)\sim 10^{-13}-10^{-11}$ at $f=3$ mHz \citep{2016PhRvL.116w1102S,2016PhRvD..94j3011D,2020MNRAS.493L...1C,2020arXiv200804890P}. Up to a few thousands of these binaries (likely responsible for about $10\%$ of the background) will be individually detected by LISA \citep{2016PhRvL.116w1102S,2020arXiv200804890P}.  The prediction for the background from NS+NS is significantly lower at $\Omega_{\rm GW}(f)\sim 10^{-14}$ at $f=3$ mHz \citep{2020arXiv200804890P}, and the background from BH+NS in the model of \citet{2020arXiv200804890P} is slightly higher but similar to that of NS+NS. The uncertainties on these rates will be significantly reduced in the coming years with more detections of stellar-mass binaries by ground-based detectors, and improved modelling of source formation and evolution. Detection of this background and in particular its shape, will provide important information about the population at periods too short to be directly observed, but before the merger phase probed by ground-based detectors (see Section~\ref{paragr:highFreqGW}).

\bigskip

\subsubsection{Interacting binaries}\label{subsubsec:interacting-binaries}

\noindent \coord{Shenghua Yu, Thomas Tauris} \\
\noindent \contr{Ashley Ruiter(1.2.3.1), Thomas Kupfer(1.2.3.1), Thomas Tauris(1.2.3.1--2), Gijs Nelemans (1.2.3.1), Shenghua Yu (1.2.3.2)}

\paragraph{AM~CVn binaries (AM Canum Venaticorum binaries --- accreting WDs)}\label{paragr:AM-CVns}

\phantom{text}

AM~CVn binaries consist of a WD accreting from a hydrogen-deficient star (or WD) companion \citep{1995Ap&SS.225..249W,2010PASP..122.1133S}. In their formation history (Fig.~\ref{fig:DWD} and Section~\ref{paragr:DWD-formation}), AM~CVns form after at least one CE phase of their progenitor system. 
The current RLO is initiated, due to orbital damping caused by GW radiation, at orbital periods of typically $5-20\;{\rm min}$ (depending on the nature and the temperature of the companion star), and the mass-transfer rate is determined by a competition between orbital angular momentum loss through emission of GWs and orbital widening due to RLO from the less-massive donor star to the more-massive WD accretor. 

If the system survives the onset of the semi-detached phase, a stable accreting AM~CVn binary is formed in which the orbital separation widens shortly after onset of RLO (Fig.~\ref{fig:UCXB-AMCVn-detailed}), and the system evolves to longer orbital periods (see Fig.~\ref{fig:UCXB-AMCVn-detailed}). When they reach an orbital period of $\sim 60\;{\rm min.}$ (after a few Gyr), the donor star has been stripped down to about 5~Jupiter~masses \citep[$5~M_{\rm J}$,][]{2018PhRvL.121m1105T}. These systems have been hypothesised to be possible progenitors of faint thermonuclear explosions  \citep[flashes, or Type~``.Ia'' SNe,][]{2001A&A...368..939N,2007ApJ...662L..95B}.

\begin{figure*}
    \begin{center}
  \includegraphics[width=0.60\textwidth, angle=-90]{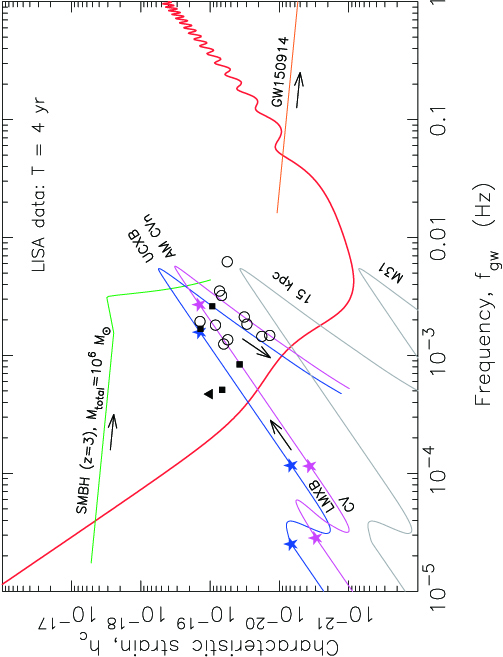}
\caption{Characteristic strain amplitude vs GW frequency for LISA. Evolutionary tracks are for an UCXB (blue) and an AM~CVn system (magenta) at a distance of $d_{\rm L} = 1\;{\rm kpc}$. Their slope on the inspiral leg is $\propto\, f_{\rm gw}^{7/6}$. The stars along the tracks represent (with increasing GW frequency) onset LMXB/CV stage, termination LMXB/CV stage, and onset UCXB/AM CVn stage. The evolutionary timescales along these tracks are shown in Fig.~\ref{fig:UCXB-AMCVn-detailed}. The LISA sensitivity curve (red line, ${\rm SNR}=1$) is based on four years of observations. The grey curves are for the UCXB at $d_{\rm L} = 15\;{\rm kpc}$ and $780\;{\rm kpc}$ (M31), respectively. Comparison tracks are shown for an MBH merger (green) and GW150914 (orange). Their inspiral slopes are $\propto\, f_{\rm gw}^{-1/6}$. Data from LISA verification sources \citep{2018MNRAS.480..302K} include detached double WD binaries (solid squares), AM~CVn systems (open circles), and a hot subdwarf binary (solid triangle). Figure from \citet{2018PhRvL.121m1105T}.}
 \label{fig:UCXB_strain+data}
  \end{center}
\end{figure*}

\begin{figure*}
    \begin{center}
  \includegraphics[width=0.60\textwidth, angle=-90]{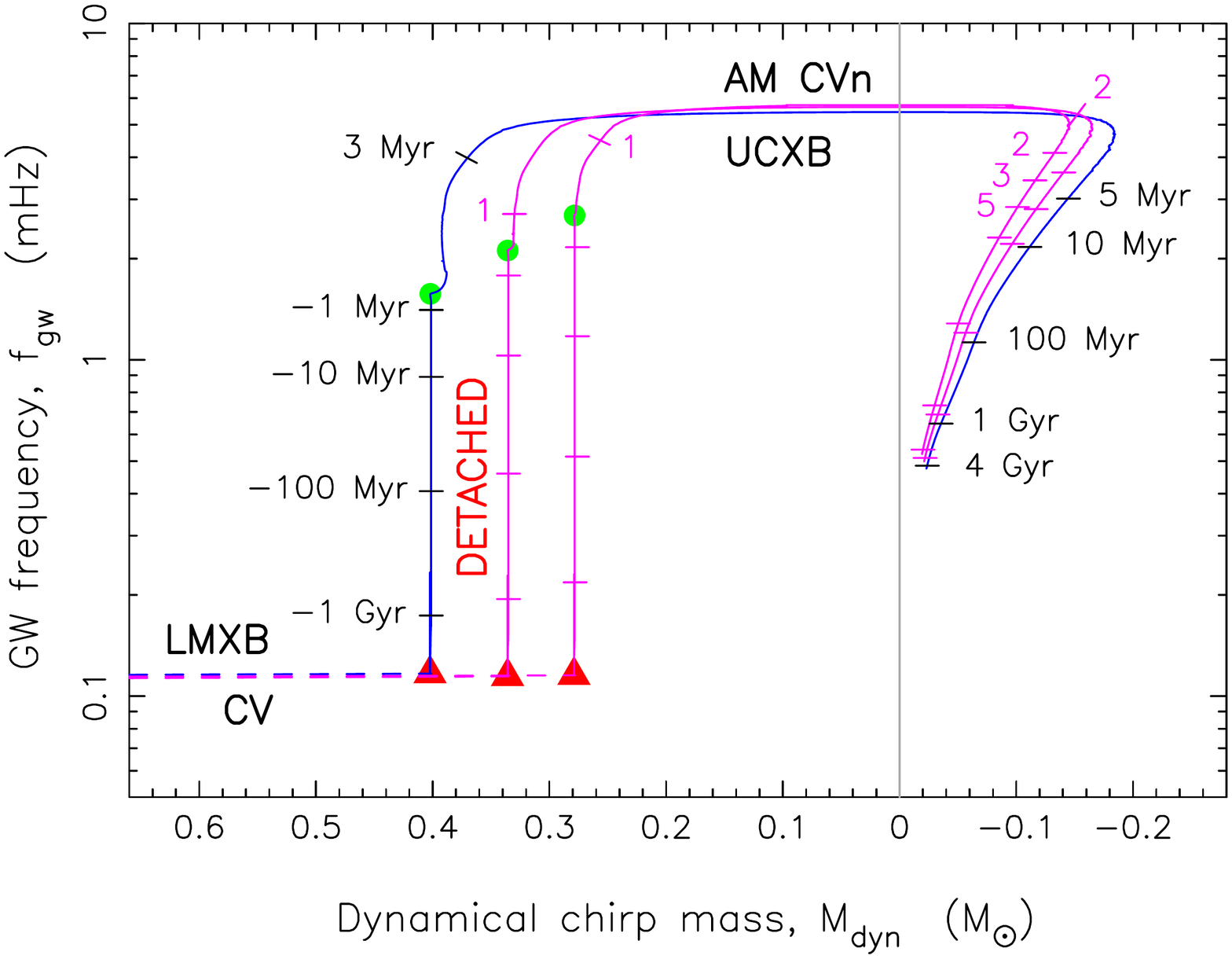}
\caption{GW frequency vs dynamical chirp mass for an UCXB and two AM~CVn systems, based on detailed mass transfer calculations (including finite-temperature effects of the WD donor star) using the MESA code \citep{2018PhRvL.121m1105T}. The end points of the first mass-transfer phases (LMXB and CV) are indicated by red triangles; the starting points of the second mass-transfer phases (UCXB and AM~CVn) are indicated by green circles. The time marks along the AM~CVn tracks are for the same values as indicated for the UCXB system, unless stated otherwise (in Myr). Time zero is defined at the onset of the second mass-transfer phase. The maximum GW frequencies (strongest LISA signal) in these three examples are 5.45~mHz (UCXB), 5.64~mHz (AM~CVn1), and 5.72~mHz (AM~CVn2) corresponding orbital periods of 6.12~min, 5.91~min and 5.83~min, respectively. The frequency at the onset of the RLO (green circles) depends on the temperature of the low-mass He~WD donor ($T_{\rm eff}$\,=\,10\,850~K, 9\,965~K and 8\,999~K, respectively).}
 \label{fig:UCXB-AMCVn-detailed}
  \end{center}
\end{figure*}

Figure~\ref{fig:UCXB_strain+data} shows examples of computed evolutionary tracks of AM~CVn (and UCXB) systems in the characteristic strain amplitude vs GW frequency diagram. As can be seen, AM~CVn systems are indeed anticipated to be detected by LISA --- in some cases even with a ${\rm SNR}>100$ (for the sources located within 1~kpc).

Though there are currently $\sim 65$ AM~CVn binaries known in the Galaxy \citep{2018A&A...620A.141R}, their formation pathways are still a puzzle \citep{2018MNRAS.476.1663G}. Their compact orbits and the lack of hydrogen in their spectra, led to three different proposed formation channels: i) the donor is a low-mass (likely He-core) WD \citep{1967AcA....17..287P}; ii) the donor is a semi-degenerate hydrogen-stripped, helium-burning star (e.g. main-sequence helium star, or hot subdwarf); or iii) the donor is a helium-rich core of a main-sequence star that has not undergone helium-burning since it had a rather low mass to begin with \citep[see][]{2010PASP..122.1133S}. Further discussions on their formation is given in Section~\ref{paragr:DWD-formation}. 

Only for eclipsing AM~CVns is it possible to fully determine all binary parameters and put constraints on the donor type. Recent results from eclipsing systems revealed that the donor stars are likely larger and more massive than previously assumed \citep{2011MNRAS.410.1113C,2018A&A...620A.141R}, implying that a semi-degenerate donor is more likely for such systems, unless the donor star is a low-mass He-core~WD which can remain bloated on a Gyr timescale \citep{2014A&A...571L...3I}. If that is the norm rather than the exception, it will lead to more AM~CVn GW sources than previously predicted.

Based on binary population synthesis, \citet{2001A&A...368..939N} predicted a space density of AM~CVn stars in a range of $0.4-1.7\times10^{-4}\;{\rm pc}^{-3}$ and a number of resolvable AM~CVn systems for LISA roughly equal to the number of detached WD+WDs \citep{2004MNRAS.349..181N}. More recently, \cite{2017ApJ...846...95K} predicts that $\sim 2700$ systems will be observable by LISA with a negative chirp of $0.1\;{\rm yr}^{-2}$ (i.e. $\dot{f}_{\rm gw}<0$, resulting from orbital expansion due to mass transfer, see Figs.~\ref{fig:UCXB_strain+data} and \ref{fig:UCXB-AMCVn-detailed}). Until very recently, when ZTF reported a large number of eclipsing WD+WDs \citep{2019Natur.571..528B, 2020arXiv201003555B}, the majority of known LISA verification binaries was dominated by AM~CVn systems \citep{2007ApJ...666.1174R,2010ApJ...711L.138R,2015MNRAS.453..483K,2018MNRAS.477.5646G,2018MNRAS.480..302K}. However, observational space density estimates from SDSS data are in strong disagreement with theoretical predictions from these binary population studies. \citet{2007MNRAS.382..685R,2013MNRAS.429.2143C} derived an observed space density about an order of magnitude below the prediction by \cite{2001A&A...368..939N,2017ApJ...846...95K} which would result in only $\lesssim1000$ resolvable systems in the LISA band. The discrepancy could be real with the population synthesis predicting too many systems, related to assumptions in binary evolution physics (especially the treatment of mass transfer in close binaries), and/or possibly because some of the systems that are predicted to evolve into AM~CVn binaries in fact merge in a CE shortly after the less-massive star fills its Roche lobe. On the other hand, it could also be that AM~CVn stars are more difficult to find than expected (when not in outburst) or that they are distributed with relative high concentration in the thick disc \citep{2012ApJ...758..131N}. \citet{2018A&A...620A.141R} argued, based on Gaia data release 2 parallaxes, that a significant fraction of AM~CVn systems, even within 100~pc, could still be undiscovered. Future transient sky surveys, such as LSST using the Vera~C.~Rubin Observatory, could have great success in detecting short-period binary systems with the implementation of appropriate cadence intervals (e.g. very short, $\sim 15\;{\rm s}$ sub-exposures). Indeed already some AM~CVn systems are thought (or known) to be eclipsing \citep{2020ApJ...905...32B}. 

Nonetheless, AM~CVn binaries are expected to be extremely useful for LISA because they simultaneously provide EM information across different wavelengths, as well as being observable in LISA's GW frequency range. For this reason, AM~CVn systems have been cited as being important verification sources for LISA \citep[e.g.][]{2018MNRAS.480..302K}. In other words, AM~CVn binaries will be multi-messenger sources once LISA flies. See Section~\ref{subsec:synergies} for further discussion on the multi-messenger opportunities for LISA.

\paragraph{UCXBs (ultra-compact X-ray binaries)}\label{paragr:UCXBs} 

\phantom{text}

It has been known for many years that tight-orbit post-LMXB systems, leaving behind a NS+WD binary that spirals-in due to GW radiation, may avoid a catastrophic event, once the WD fills its Roche lobe. The outcome is expected to be a long-lived UCXB \citep[][]{1979ApJ...227..178W,1986ApJ...304..231N,2002ApJ...565.1107P,2010MNRAS.401.1347N,2012A&A...537A.104V,2013ApJ...768..184H}.      
These sources are tight X-ray binaries observed with an accreting NS and a typical orbital period of less than 60~min. 
Because of the compactness of UCXBs, the donor stars are constrained to be either a WD, a semi-degenerate dwarf or a helium star \citep{1982ApJ...254..616R}.
UCXBs are not only excellent laboratories for testing binary-star evolution, but also important GW sources. Studies of their orbital parameters (mass, orbital period and eccentricity), chemical composition and spatial distribution may provide important information and clues to understand both the accretion processes of compact binaries (including spin-orbit and tidal interactions) and the long-term evolution of double compact object binaries.

Depending on the mass-transfer rate, the UCXBs are classified in two categories: persistent and transient sources \citep[e.g.][]{2013ApJ...768..184H}. Only about 14~UCXBs have been confirmed so far (9~persistent, 5~transient), and an additional $\sim 14$~candidates are known. 
Thus UCXBs are difficult to detect or represent a rare population. Earlier studies \citep[e.g.][]{2014A&A...571A..45I} have suggested the need for extreme fine tuning of initial parameters (stellar mass and orbital period of the LMXB progenitor systems) in order to produce an UCXB from an LMXB system.
UCXBs are detected with different chemical compositions in the spectra of their accretion discs \citep[e.g. H, He, C, O and Ne, see][]{2010MNRAS.401.1347N}. To explain this diversity requires donor stars which have evolved to different levels of nuclear 
burning and interior degeneracy, and therefore to different scenarios for the formation of UCXBs.
Since a large fraction of the UCXBs are found in globular clusters, some of these UCXB systems could also have formed via tidal captures, direct collisions or stellar exchange interactions \citep{1975MNRAS.172P..15F,1975A&A....44..227S,1983ApJ...268..319H}. 

Figure~\ref{fig:UCXB-AMCVn-detailed} displays the evolution of AM~CVns and UCXBs in the GW frequency vs dynamical chirp mass diagram. These systems undergo stable RLO and will start to widen their orbits again within a few Myr after the onset of the mass transfer. LISA will detect such Galactic systems continuously both during the inspiral phase for a few tens of Myr, while the systems are still detached, and after the onset of RLO on a timescale of up to 100~Myr (depending on their distance). 

The long-term stability of UCXBs has been a topic of debate. From an analytical investigation, \citet{2012A&A...537A.104V} argued that for a $1.4\;M_{\odot}$ NS accretor, only C/O-core~WDs with a mass of $\lesssim 0.4\;M_{\odot}$ lead to stable UCXB configurations. Subsequent hydrodynamical simulations suggested that this critical WD mass limit could be lower \citep{2017MNRAS.467.3556B}. The first successful numerical calculations of RLO from a WD to a NS were presented by \citet{2017MNRAS.470L...6S}, and they were able to follow the entire evolution until the low-mass He-core~WD donor star has become a $\sim 0.005\;M_\odot$ planet-like companion. These systems were further evolved \citep{2018PhRvL.121m1105T}, including the finite-temperature (entropy) effects of the WD donor stars, and the first evolutionary tracks of such sources across the LISA GW band were produced, see e.g. Figs~\ref{fig:UCXB_strain+data} and \ref{fig:UCXB-AMCVn-detailed}. Further independent studies on the detectability of UCXBs as LISA sources have been provided by e.g. \citet{2020ApJ...900L...8C,2021MNRAS.503.2776Y}.

It is also anticipated that LISA may detect interacting BH+WD systems \citep{2017MNRAS.467.2199B,2020arXiv201005974S}. It has been estimated in some studies \citep{2006A&A...454..559Y} that the Galaxy contains some $10^4$ of these systems. However, their formation process (especially those with low-mass WD companions) remains uncertain \citep{2003MNRAS.341..385P}.

\bigskip

\subsubsection{Other potential sources}\label{subsubsec:other_sources}

\noindent \contr{Camilla Danielski(1.2.4.1-2), Silvia Toonen(1.2.4.2), Thomas Kupfer(1.2.4.4), Jan van Roestel (1.2.4.3), Nicola Tamanini (1.2.4.1), Valeriya Korol (1.2.4.1)}

\paragraph{Helium-star binaries}

\phantom{text}

Subdwarf B~stars (sdBs) are stars of spectral type B with luminosities below that of main-sequence stars. The formation mechanism and evolution of sdBs are still debated, although most sdBs are likely He--burning stars with masses $\sim 0.5\;M_\odot$, radii as small as $\sim 0.1\;R_\odot$ and thin hydrogen envelopes \citep{2016PASP..128h2001H}. A large fraction are found in binary systems and, due to their compact nature, the most compact ones have orbital periods $\lesssim\,1$\,hr \citep{2012ApJ...759L..25V,2013A&A...554A..54G,2017ApJ...851...28K,2020ApJ...891...45K,2020ApJ...898L..25K}, making them potentially detectable sources for LISA. 

The most compact systems have WD companions and as such they are prime progenitor systems for double detonation Type~Ia~SNe. In this scenario a WD is orbited by a core He-burning sdB~star in an ultra-compact orbit ($P_{\rm orb}<80\;{\rm min}$). Due to the emission of GWs, the binary shrinks until the sdB~star fills its Roche lobe and starts mass transfer. He-rich material is then transferred to the C/O-core WD companion which will lead to the accumulation of a He-layer on top of the WD. After accreting about $0.1\;M_\odot$, He-burning is predicted to be ignited in this shell. This in turn triggers the ignition of carbon in the core, even if the WD~mass is significantly lower than the Chandrasekhar limit \citep{2010A&A...514A..53F}. So far, the only known candidate for this scenario is the ultra-compact sdB+WD binary CD$-$30$^\circ$11223 with an orbital period of $P=71\;{\rm min}$ \citep{2012ApJ...759L..25V,2013A&A...554A..54G}. 
This system was also found to be detectable for LISA with an expected SNR of $\sim 5$ after four years of LISA observations \citep{2018MNRAS.480..302K}. More recently, the first members of ultra-compact sdB binaries which have started to transfer material to the WD companion were discovered. The most compact system, ZTF\,J2130+4420, consisting of a low-mass sdB~star with $M_{\rm sdB}=0.337\;M_\odot$ has an orbital period of 39~min. The system has well measured properties from EM observations and is expected to have a SNR of $\sim 3$ after four years of LISA observations, adding to the growing number of LISA detectable He-burning stars.

\citet{2020arXiv200607382G} modelled the Galactic population of stripped stars, which contain the low-mass sdB~stars as well as more massive He-core burning stars, in tight orbits with compact companions, focusing on those that will be detectable by LISA. Their analysis predicts up to 100 stripped star + WD binaries and up to 4 stripped star + NS binaries with ${\rm SNR}>5$ after 10 years of observations with LISA. Although the expected numbers are significantly smaller than for WD+WDs or AM~CVns, \citet{2020arXiv200607382G} finds that all of the LISA detectable sources are within 1~kpc and therefore bright in EM flux which makes them ideal targets for multi-messenger studies (see Section~\ref{subsec:synergies} for more details on multi-messenger opportunities). 

\paragraph{Period bouncing CVs} 

\phantom{text}

Period bouncing CVs are highly evolved cataclysmic variables where the donor has lost almost all of its mass and has become degenerate. These systems have reached the minimum orbital period for a hydrogen donor (70~min) and are evolving to longer orbital periods (up to 100~min). Model predictions are that 40--70\% of all CVs are period bouncers \citep{1993A&A...271..149K,2011ApJS..194...28K}. However, only a few have been identified so far because of the low accretion rate and low temperature of the WD and donor \citep[e.g.][]{2018MNRAS.481.2523P}. 
While the donor-mass is low and the orbital periods are relatively long, nearby period bouncers are detectable with LISA. Given their high space density, a dozen of these systems are close enough to be detected by LISA.

\paragraph{Exoplanets, brown dwarfs and substellar companions} %

\phantom{text}

In the Galaxy, due to the slope of the Salpeter-like initial mass function (IMF), more than 97\% of all stars will terminate their lives as a WD, meaning that the vast majority of the known 4000+ planet-hosting stars will end their lives as WDs. 
In the last couple of decades, most of the attention in exoplanetary searches has been focused on the formation and characterisation of exoplanets orbiting host stars on the main sequence, but very little is known on planetary systems in which the host star evolves off the main sequence, to become a red giant. 
Theoretical models indicate that, if planets avoid engulfment and evaporation throughout the red-giant or/and the asymptotic-giant branch phases of the host star, they can survive \citep[see e.g.][]{1984MNRAS.208..763L,1998Icar..134..303D,1998A&A...335L..85N}. This is expected to be the fate of the planet Mars, and other planets orbiting further out in our Solar System \citep{2008MNRAS.386..155S}.
Observational evidence, in the form of photospheric contamination by the accreted debris \citep{2010ApJ...722..725Z,2014A&A...566A..34K}, dusty \citep{2009ApJ...694..805F} and gaseous circumstellar discs \citep{2006Sci...314.1908G,2016MNRAS.455.4467M}, supports the existence of dynamically active planetary systems around WDs. Up to very recently, only two planetesimals had been observed orbiting a WD \citep{2015Natur.526..546V,2019Sci...364...66M}. However, within the last years, two giant planets have been detected orbiting single WDs \citep{2019Natur.576...61G,2020Natur.585..363V}, 
showing that planets can survive single host-star evolution. 

Today, over 1000 brown dwarfs (BDs) have been detected in the Solar neighbourhood \citep{2018haex.bookE.118B}. 
Some of them have been discovered also around single WDs, and examples of BDs orbiting at distances beyond the tidal radius of the asymptotic-giant branch progenitor \citep[but also within it, e.g.\,WD~0137$-$349~B,][]{2006Natur.442..543M}, show that BDs can survive stellar evolution of their host star, whether or not they are engulfed by its expanding envelope. \citet{2005ApJS..161..394F} predicted that few tenths of percent of Galactic single WDs hosts a BD.

The most straightforward way with which LISA could detect sub-stellar objects, such as planets or BDs, would be the direct detection of GWs emitted by a binary system composed of a sub-stellar object in an tight orbit around a single star. 
However, the absolute orbital period minimum for a hydrogen-rich body (i.e. a star, BD or a gas giant planet) in a binary system is about $P_{\rm orb}\simeq 37\;{\rm min}$ \citep{2021arXiv210412083R}. This corresponds to a GW frequency of at most $f_{\rm GW}\simeq 0.9\;{\rm mHz}$.
Such a system could be detected only at close distances (say, within 1~kpc) and only for relatively high sub-stellar masses ($M\gtrsim 13\;M_{\rm J}$), possibly excluding all exoplanets.  Furthermore, the mass of the sub-stellar object cannot be directly inferred from direct detection, and at best only the chirp mass of the binary system can be retrieved.
Further investigations and EM observations are necessary to better understand the detectability and the rates of these sub-stellar objects, although at the moment it seems unlikely that a large number of these systems will be observed by LISA \citep{2019MNRAS.483L..33W}.

Another option is to search for circumbinary planets around WD+WD through a modulation of the WD+WD signal \citep{2019NatAs...3..858T}, that can probe regions of parameter space not probed by EM observations (far away and not towards the Galactic Centre).
The discovery of evolved planetary systems will statistically increase the current sample of post-main-sequence planets, filling an area of the planetary Hertzsprung--Russell diagram that is currently not explored  \citep{2019NatAs...3..858T}. LISA will provide observational constraints on both planets that can survive two CE stellar evolution phases and on a possible second-generation planet population produced from CE ejecta material \citep{2014A&A...563A..61S}. 
Even in the case where LISA will prove no detection anywhere in the Milky Way, it will be possible to set strong unbiased constraints on planetary evolution and dynamical theories, and in particular on the fate of exoplanets bound to a binary that undergoes two CE phases.

\paragraph{Triples and multiples}\label{paragr:triples} 

\phantom{text}

LISA's stellar sources will also contain multiple body systems, such as triples and quadruples. 
Hierarchical systems that consist of nested orbits represent stable configurations that can remain intact for several Gyr and throughout (despite) the evolution of the stellar components, as evidenced by observations.  
Within 20~pc of the Sun, there are already two such systems that harbor close WD+WDs. These are 
WD~0326$-$273 \citep{1949ApJ...109..528L,1994RMxAA..28...43P,2005A&A...440.1087N,2012ApJS..199...29G,2017A&A...602A..16T} and 
WD~0101+048 \citep{1998ApJ...502..394S,2000MNRAS.319..305M,2009A&A...507..251C,2012ApJS..199...29G,2017A&A...602A..16T}. 
The former is a triple that consists of a close WD+WD with a period of $\sim 1.8\;{\rm d}$, and an M5 star in a wide orbit. The latter is a quadruple consisting of a close WD+WD (with a period of $\sim1.2\;{\rm d}$, but see \citep{2000MNRAS.319..305M} and a MS+MS binary. 
Two triple systems with \textit{three} WD components are known as well, J1953$-$1019 \citep{2019MNRAS.483..901P} and WD~1704+481 \citep{2000MNRAS.314..334M}. The inner binary of the latter system has a period of $\sim0.15\;{\rm d}$, just inside the LISA frequency range.  
Even millisecond pulsars have been found to be part of triple-architectures; 
the PSR~J0337+1715 system harbors a compact NS+WD (1.6~d orbital period) inner binary which is orbited by another (tertiary) WD every 327~d \citep{2014Natur.505..520R,2014ApJ...781L..13T}. The globular cluster (M4) pulsar B1620$-$26 has a WD companion in a half-year orbit, and a planetary companion in a 100-yr orbit \citep{1999ApJ...523..763T,2003Sci...301..193S}. 

Theory suggests that exoplanets (and BDs) also exist around WD+WDs in the Galactic disc, and that such objects are more likely to survive around evolving close binary stars than around evolving single stars \citep{2016ApJ...832..183K}. The eclipse timing variation technique allowed the detection of a few post-CE systems (that is WD+low-mass star), and a BD companion(s) \citep[see e.g.][]{2015MNRAS.448.1118G, 2012A&A...543A.138B, 2019AJ....157..150A}, nevertheless today no BDs or exoplanets orbiting WD+WDs have been observed yet. 

LISA will be able to detect outer companions to compact (inner) binaries when they impose eccentricity oscillations in the inner orbit due to three-body dynamics \citep{1910AN....183..345V,1962P&SS....9..719L,1962AJ.....67..591K,2016ARA&A..54..441N}. In particular \citet{2019ApJ...875L..31H} showed such oscillations would be observable with LISA to distances up to a few Mpc for compact binaries near supermassive BHs, which can also be considered a three-body system.  Furthermore, LISA can detect outer companions by exploiting the Doppler frequency modulation on the GW waveform due to their gravitational pull  \citep{2018PhRvD..98f4012R}. 
The acceleration imparted by the hierarchical companions can be detected in the GW signal for outer periods as large as 100~yr \citep{2018PhRvD..98f4012R, 2020PhRvD.101f3002T}. 
For systems with orbital periods that are shorter than, or comparable to, the mission lifetime, the perturbation allows for the determination of the orbital period, eccentricity, initial orbital phase and radial velocity parameter of the companion \citep{2018PhRvD..98f4012R,2019NatAs...3..858T}. 
On a general level, the sensitivity of LISA will be able to detect WD+WDs companions with masses down to $\sim M_{\rm J}$
\citep{2019A&A...632A.113D}, and therefore allow not only for the detection of stellar companion and compact objects, but also BDs and exoplanets. This being an indirect detection, i.e.\,the observation of a periodic Doppler shift modulation of an existing strong binary GW signal, we are able to probe a wider mass range, whose inferior limit also covers the giant planets range. The novelty of using LISA for the detection of planetary/low-mass companions is that GWs provide a much larger spatial coverage than the one provided by EM techniques, enabling us to probe regions of our Galaxy currently not accessible to other methods.
More specifically, \cite{2019A&A...632A.113D} showed that during a 4~yr nominal mission LISA will detect from 3 to 83 exoplanets, and from 14 to 2218 BDs everywhere in the Milky Way. The sensitivity of LISA is such that in the most optimistic cases exoplanets could be detected orbiting WD+WDs in the Milky Way's satellites, in particular in the Large Magellanic Cloud \citep[LMC,][]{2020arXiv200707010D}. Such an observation could represent the first detection of an extra-galactic bound exoplanetary system.

\paragraph{Capturing the inspiral of a CE system}\label{paragr:other-interacting} 

\phantom{text}

It has been suggested by \citet{2021arXiv210200078R} that LISA may be able to detect the inspiral of binaries undergoing a CE phase. Depending on various assumptions, they anticipate that LISA could detect between 0.1 and 100 such GW sources in the Galaxy during the mission duration. Detecting this GW signal would provide direct insight into the gas-driven physics of CE evolution.

\subsection	{Formation of LISA binaries }\label{subsec:formation-binaries} 
\noindent \coord{Katie Breivik} 

\noindent \contr{Michela Mapelli (1.3.2-3), Simone Bavera, Katie Breivik, Martyna Chruslinska, Gijs Nelemans,  Pau Amaro Seoane (1.3.2), Manuel Arca Sedda (1.3.2-3), Thomas Tauris,  Silvia Toonen (1.3.2-3), Jeff Andrews, Tassos Fragos (1.3.1), Luca Graziani, Daryl Haggard (1.3.2), Melvyn B. Davies (1.3.2-3)}

\vspace{1.0cm}
In the following section, we discuss the formation of LISA binaries. For a review and broader description of the many physical aspects of stellar evolution and binary star interactions that are referred to below in the context of the formation of compact object binaries, we refer to the textbooks by, for example, \citet{slv94,2001icbs.book.....H,2006epbm.book.....E,10.1088/2514-3433/ac595f,TvdH23}.

\subsubsection{Isolated binaries}\label{subsubsec:isolated-binaries}
The formation pathways of isolated binaries observable by LISA are marked with several phases of mass loss or exchange. In the following Section, we refer to the initially more massive star as the primary and the initially less massive star as the secondary. Stable mass transfer can occur either through wind mass loss/accretion or RLO. Wind mass loss is generally assumed to be non-conservative across all phases of stellar evolution, with mass accretion efficiency ranging from $\lesssim 10\%$ for the Bondi--Hoyle--Lyttleton mechanism and $20\text{--}50\%$ if the accretion is focused \citep{2017MNRAS.468.3408D}. In this case, the orbit widens as the mass lost from the system causes an increase of the remaining specific angular momentum. In practice, the dynamics are  complicated and dependent on physics related to the geometry and structure of the wind, tidal effects, orbital characteristics and in some cases magnetic fields and radiation transport, thus calling for three-dimensional, multi-physics, hydrodynamical simulations \citep[e.g.][]{2018A&A...618A..50S, 2019A&A...626A..68S}. 

RLO occurs when the donor star expands to the point that its radius exceeds the Roche radius \citep{1983ApJ...268..368E}. RLO mass transfer can proceed in a dynamically stable or unstable fashion, depending on the structure of the donor and accretor as well as their mass ratio. The stability of RLO mass transfer is commonly described using the Webbink radius--mass exponents \citep{1985ibs..book...39W} that determine the timescales on which mass transfer will become unstable. In the case of dynamically stable mass transfer, the orbital evolution depends strongly on the mass ratio of the binary: if $M_{\rm donor}/M_{\rm accretor} > 1$,  the orbit tightens (for fully conservative mass-transfer), while in the converse case the orbit widens. Mass-transfer efficiency, as well as the assumed specific angular momentum carried away from mass lost from the system, play a crucial role in mass-transfer stability and the orbital evolution of the binary.

Dynamically unstable mass transfer is believed to generate a CE phase where the donor star's core and companion are enshrouded in the donor's envelope \citep[for a review see][]{2013A&ARv..21...59I}. The precise dynamics of how CE proceeds are still not fully understood. In the context of compact object binary formation and population synthesis studies, energy budget arguments are most often employed to estimate the post-CE properties of a binary. In the ``$\alpha_{\rm CE}$'' prescription, it is assumed that a fraction $\alpha_{\rm CE}$ of the released orbital energy is used to unbind the donor's envelope and eject it from the system \citep{1976IAUS...73...35V, 1984ApJ...277..355W}. Several recent studies have suggested that other sources of energy may be needed to successfully eject the envelope, including recombination energy \citep[e.g.][]{2014A&A...568A..68Z, 2016MNRAS.460.3992N} or jets launched by the companion \citep[e.g.][]{2019MNRAS.488.5615S}. Each of these will change the overall energy budget of the CE evolution and lead to differences in the final orbital separation \citep[e.g.][]{2018MNRAS.477.2349I}. Alternatively, in the ``$\gamma_{\rm CE}$'' prescription, angular momentum conservation arguments, which lead to less dramatic inspiral, have been considered to explain the orbital period distribution of WD+WDs \citep{2000A&A...360.1011N}.

\paragraph{WD+WD systems and AM~CVn binaries }\label{paragr:DWD-formation} 

\phantom{text}

\begin{figure*}
    \begin{center}
  \includegraphics[width=0.60\textwidth, angle=0]{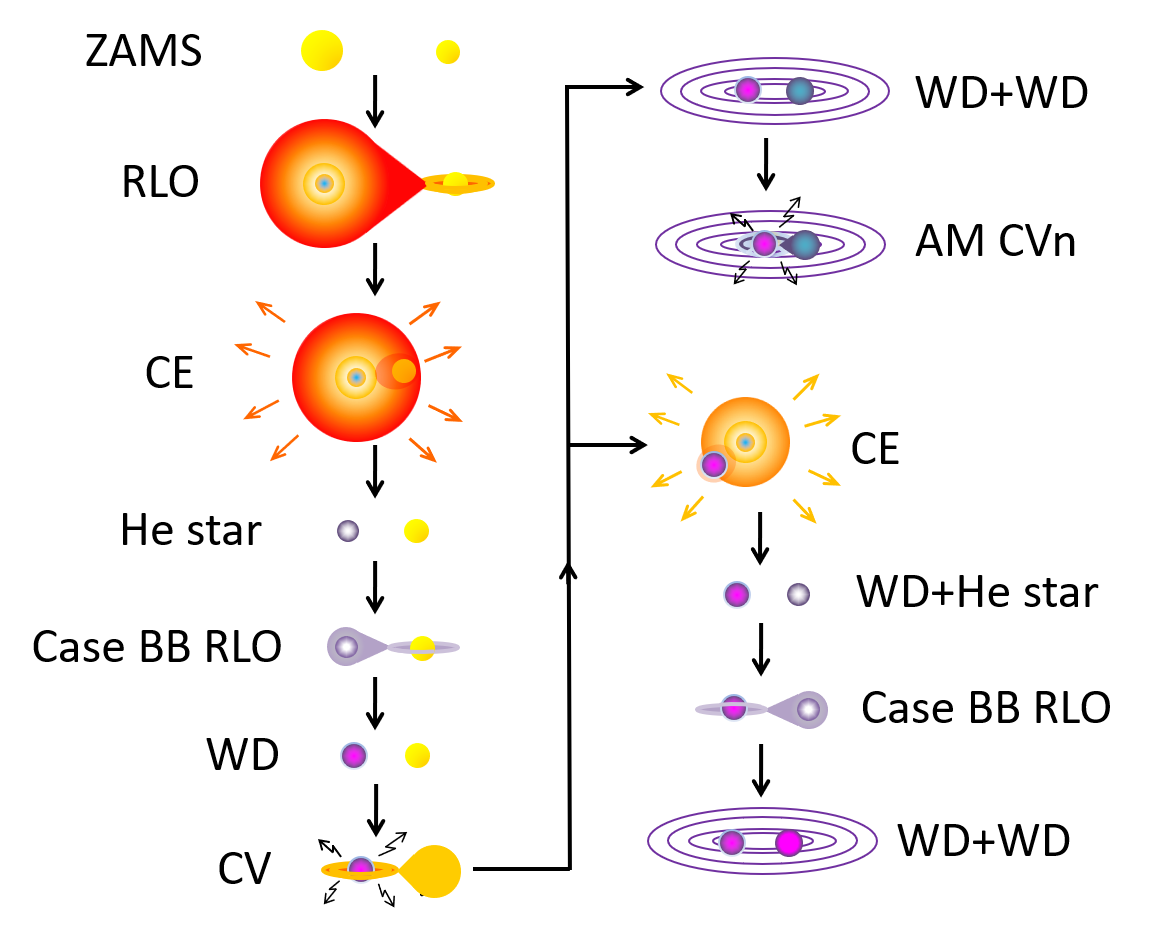}
\caption{Illustration of the formation of an AM~CVn system and a detached WD+WD binary. LISA sources are indicated with waves. Acronyms. ZAMS: zero-age main sequence; RLO: Roche-lobe overflow (mass transfer); CE: common envelope; He star: helium star; WD: white dwarf; CV: cataclysmic variable; AM~CVn: AM Canum Venaticorum binary \citep[Figure from][]{TvdH23}.
}
 \label{fig:DWD}
  \end{center}
\end{figure*}

The progenitors of isolated WD+WD and AM~CVn binaries begin with zero age main sequence stars with masses below $8-10\,M_{\odot}$. The formation pathways of close WD+WD and AM~CVn binaries contain several stages of stable and unstable mass transfer, or CE (see Fig.\,\ref{fig:DWD}). The uncertain outcomes of these interactions determine whether the progenitor binary continues on in its evolutionary path toward becoming a LISA source or if it merges with its companion. Conversely, LISA observations of the populations of WD+WD and AM~CVn sources will constrain these interactions.

Virtually all close WD+WD and AM~CVn progenitors experience an interaction as the primary star advances off the main sequence and fills its Roche lobe. This interaction can either proceed stably or unstably on dynamical timescales. In either case, the orbit will shrink because of the donor's higher mass relative to the accretor. For systems with late red giant and asymptotic giant branch donors, initially dynamically stable but thermally unstable mass transfer can produce mass loss from the L2 Lagrange point, which leads to a delayed dynamical instability and a CE phase \citep{2020ApJS..249....9G, 2020A&A...642A.174M}. In the rare case of close WD+WDs where the more massive WD forms second (converse to stellar lifetime expectations), a phase of stable mass transfer, followed by a CE generated by the initially lower mass star could be necessary \citep{2012ApJ...744...12W}. 

After each interaction, the star which donates mass becomes stripped leaving behind a He core that can have varying structure depending on the evolutionary phase at which the donor filled its Roche lobe. Such stripped stars orbiting main sequence companions have been widely observed throughout the Galaxy \citep{2007MNRAS.382.1377R} and been used to constrain CE ejection efficiencies \citep{2010A&A...520A..86Z, 2013A&A...557A..87T}. 

Since the previous interaction brings the two stars together, further interactions are likely. Interactions can occur while the secondary is still on the main sequence. In this case, if the mass transfer is stable, a cataclysmic variable is formed \citep{2011A&A...536A..42Z}. Conversely, mass transfer can also occur as the secondary advances up the giant branch. Due to previous envelope stripping, the mass ratio of the secondary donor to the WD accretor can be either greater or less than one. If the secondary is already less massive than the WD, stable mass transfer will occur and widen the orbit, removing the possibility of detection by LISA. However, if the secondary is more massive than the WD companion, the mass exchange will lead to orbital tightening. If the mass transfer is unstable, another CE phase takes place, potentially bringing the stars even closer together and leaving behind a WD with a stripped He core companion. The structure of the He core again depends on the evolutionary phase at which the secondary overflows its Roche lobe. At this point, a close WD+WD binary is assured and the slow evolution due to GW  emission brings the WD+WD toward the LISA band.

A key uncertainty in the formation pathways of AM~CVn binaries is the nature of the donor star. AM~CVn binaries consist of a WD accreting He-rich material originating from a WD, semi-degenerate helium star, or evolved MS donor  \citep{2010PASP..122.1133S}. Indeed, it could be the case that the observed AM~CVn population is a combination of all three with different relative contributions \citep{2004MNRAS.349..181N}. 
If the donor star is a non-degenerate evolved star, magnetic braking is required, along with GW emission, to maintain the ultra-short periods of observed AM~CVn systems \citep{2005A&A...440..973V}. 
Magnetic braking is a process in which orbital angular momentum in a tight synchronized binary is converted into spin angular momentum via a magnetic stellar wind (a process that therefore requires a low-mass stellar component with a convective envelope).
The ultra-compact orbital configuration is less problematic for semi-degenerate and fully-degenerate donors which originate from the ejection of a second CE, with tighter orbits allowed by more degenerate donors \citep{2008AstL...34..620Y}. In the case of fully-degenerate He-core~WD donors, the orbit can become so small that the mass lost from the donor directly impacts the accretor leading to a rapid decrease in orbital size followed by a long-lived phase of accretion which widens the orbit \citep{2004MNRAS.349..181N,2004MNRAS.350..113M, 2006ApJ...649L..99D, 2017ApJ...846...95K}. Regardless of the donor, a significant uncertainty still remains in how much He-rich material the accretor can handle until novae erupt on the WD's surface. While detailed binary evolution calculations \citep[e.g.][]{2018PhRvL.121m1105T} have shown that RLO mass transfer in WD+WD can be stable, it has been suggested that interactions of the donor star with the expanding nova shells will likely lead to a rapid orbit shrinkage and eventually a merger \citep{2015ApJ...805L...6S}.

\paragraph{White-dwarf binaries with neutron-star or black-hole companions}

\phantom{text}

\begin{figure*}
    \begin{center}
  \includegraphics[width=0.60\textwidth, angle=0]{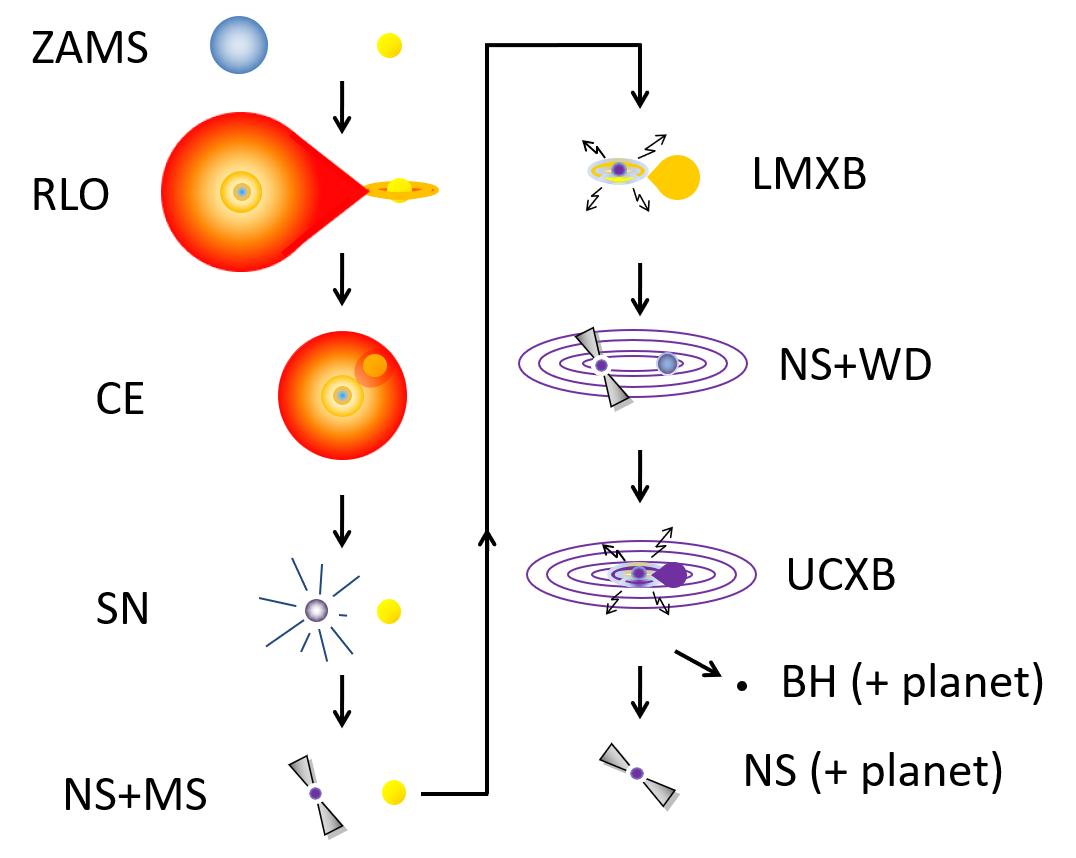}
\caption{Illustration of the formation of a detached NS+WD binary and an UCXB system. See Fig.~\ref{fig:DWD} for details. Additional acronyms: SN: supernova; NS: neutron star; LMXB: low-mass X-ray binary; BH: black hole \citep[Figure from][]{TvdH23}. }
 \label{fig:UCXB}
  \end{center}
\end{figure*}

\begin{figure*}
    \begin{center}
   \includegraphics[width=0.60\textwidth]{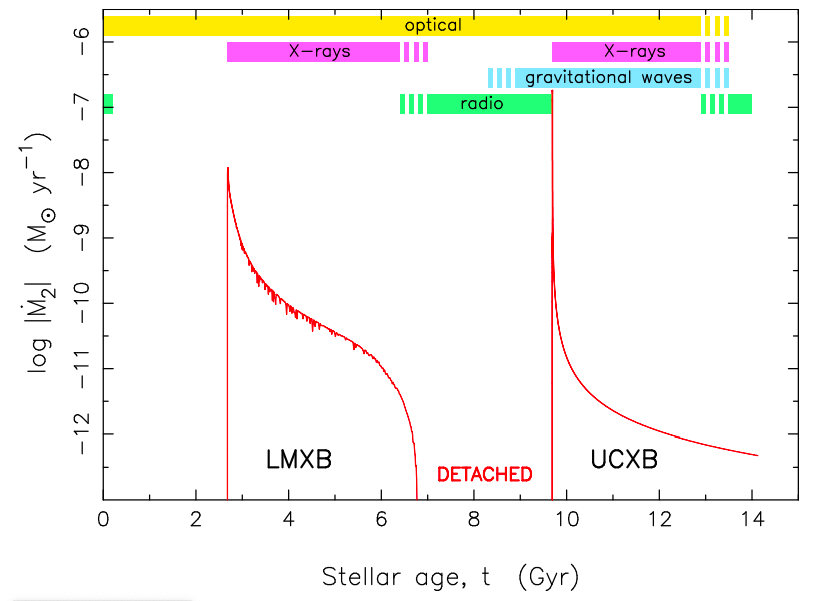}
\caption{Evolutionary sequence showing how ultra-compact X-ray binaries (prime LISA source candidates) are formed from merging NS+WD binaries, descending from LMXBs. Plotted here is mass-transfer rate of the donor star as a function of stellar age. The initial MS star\,+\,NS binary has components of $1.40\;M_\odot$ and $1.30\;M_\odot$, respectively. The system evolves through two observable stages of mass transfer:
an LMXB for 4~Gyr, followed by a detached phase lasting about 3~Gyr where the system is detectable as a radio millisecond pulsar orbiting the helium WD remnant of the donor star, until GW radiation brings the system into contact again, producing a UCXB. The colour bars indicate detectability in different regimes resulting in synergies between LISA and EM detectors \citep[Figure from][]{2018PhRvL.121m1105T}.}
 \label{fig:LMXB-UCXB-link}
  \end{center}
\end{figure*}

Compared to WD+WDs, the formation of detached binaries with WD and NS or BH companions occur in binaries with stars that are massive enough to explode in a SN (see Fig.~\ref{fig:UCXB}). Similar to WD+WD formation, the more massive primary evolves first and, because of the relatively large mass ratio, begins RLO mass transfer that is  often expected to be unstable and lead to a CE. Soon after, the primary evolves to become a compact object, through either a supernova explosion (NS or BH) or  direct collapse (BH only). Since a NS is thought to receive a kick during its formation, there is a significant  probability that the binary disrupts at this point.

The subsequent evolution, of a lower-mass non-degenerate star with a NS or BH, will typically go through a phase of stable mass transfer in which the binary becomes observable as X-ray binary, due to the strong heating of the accretion disc in the deep potential of the NS or BH. When the onset of the mass-transfer occurs after the secondary star has evolved past its main sequence, the core of the star has already contracted. Thus after the X-ray binary phase, when the envelope of the expanding star has been completely transferred, the NS or BH is left with a WD companion that was the core of the donor star.

In some cases, the NS/BH+WD binary is tight enough that angular momentum loss due to GW emmission will bring the two objects together as LISA sources (Fig.~\ref{fig:LMXB-UCXB-link}). At periods of $\sim 10\text{--}20\;{\rm min}$, i.e. within the LISA band, the WD will start to transfer mass to the NS/BH, forming an X-ray binary again, but now of ultra-short period, called an UCXB. Detailed numerical calculations, including finite-temperature (entropy) effects, have shown that UCXBs can indeed form via stable RLO from post-LMXBs systems \citep{2017MNRAS.470L...6S,2018PhRvL.121m1105T}.

\paragraph{Double neutron star/black hole binaries } \label{subsec:1313}

\phantom{text}

\begin{figure*}
    \begin{center} \label{fig:1.9}
  \includegraphics[width=0.60\textwidth, angle=0]{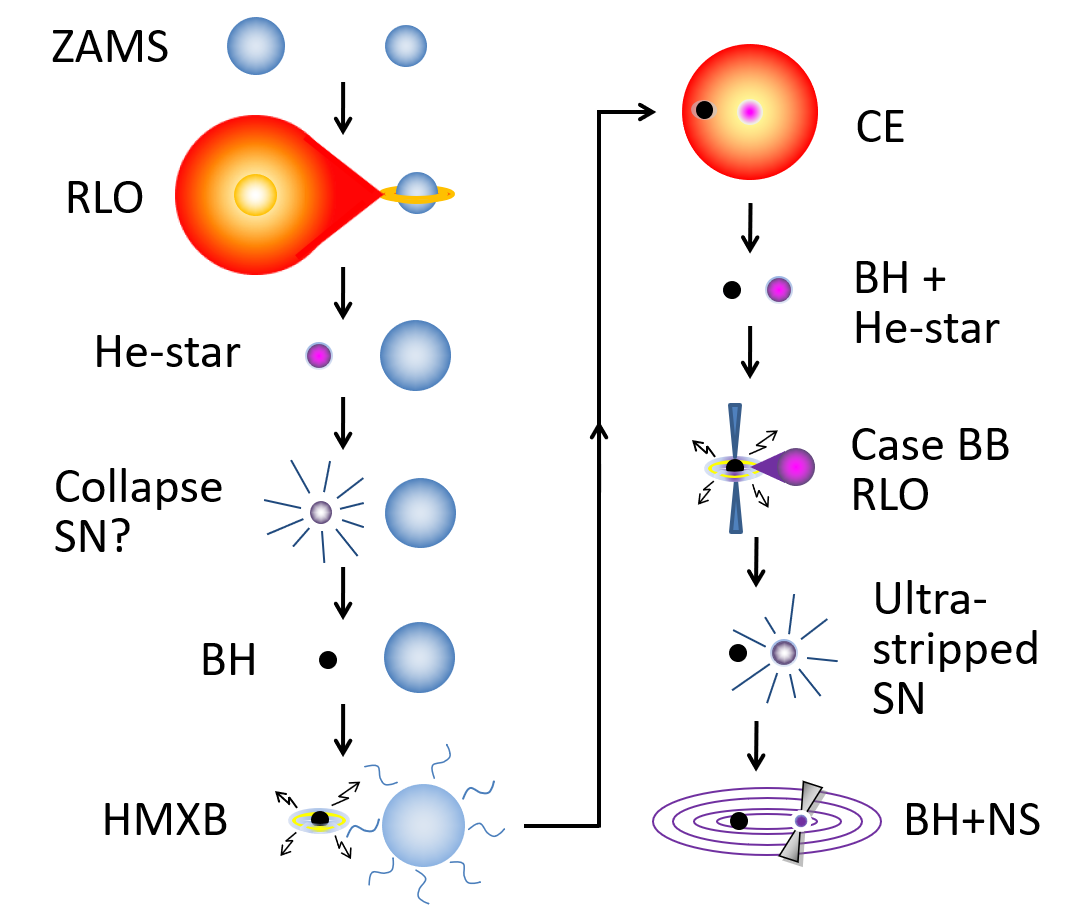}
\caption{Illustration of the formation of a tight BH+NS binary that evolves towards a merger. See Figs.~\ref{fig:DWD} and \ref{fig:UCXB} for explanation of acronyms \citep[Figure from][]{TvdH23}.}
 \label{fig:BHNS}
  \end{center}
\end{figure*}

NS+NS formation has been extensively discussed in the literature \citep[see][for a review]{2017ApJ...846..170T}. The standard scenario (see Fig.~\ref{fig:BHNS} for a schematic diagram) involves several phases of interaction, starting with a stable RLO, during which the primary loses part of its envelope before it undergoes a SN to form a NS. The newly formed NS HMXB is likely too dim to be detectable in X-rays, as the orbital separation is still large and the NS may only be able to capture an appreciable fraction of the companion's stellar wind when the latter evolves to the giant phase. During the subsequent evolution the orbital separation needs to decrease from $\sim 10^3\;R_{\odot}$ to a few $R_\odot$ for the final binary to merge within the Hubble time.
Significant tightening is typically achieved through a CE phase that occurs when the secondary this time fills its Roche lobe. The post-CE binary is expected to encounter another phase of mass transfer, initiated by a stripped helium-burning secondary \citep[i.e. so-called Case~BB mass transfer, often initiated when a $\sim 2.5-3.5\;M_\odot$ helium star expands during shell-helium burning,][]{1986A&A...165...95H}. This leads to further orbital tightening, stripping of the secondary's envelope, and NS spin-up. If this last mass-transfer episode is unstable and leads to a second CE phase, a fast merging NS+NS will be formed; a scenario invoked to explain r-process
element enrichment observed in some stellar systems \citep{2019ApJ...872..105S,2019ApJ...886....4Z}.
Such NS+NSs would be  effectively unobservable with current radio surveys, and if they exist within the Galaxy, their presence will be revealed by LISA \citep{2019MNRAS.483.2615K,2020ApJ...892L...9A}. However, recent detailed binary evolution calculations have shown that this last phase of Case~BB mass transfer is expected to be stable \citep{2015MNRAS.451.2123T,2018MNRAS.481.4009V} and do not support the existence of the aforementioned fast-merging channel \citep[in contrast to earlier works][]{2003ApJ...592..475I,2003MNRAS.344..629D}.

Besides the pre-HMXB evolution, the most important and uncertain aspects of our current understanding of NS+NS and mixed BH+NS formation are related to: 
i) CE evolution and spiral-in of the NS, ii) momentum kicks (magnitude and direction) imparted onto newborn NSs, and iii) the mass distribution of NSs. 

From an energetics point of view, it has been shown that an inspiralling NS may indeed be able to eject the
envelope of its massive star companion \citep[e.g.][]{ 2010ApJ...716..114X,2011ApJ...743...49L, 2016RAA....16..126W,  2016A&A...596A..58K}. However, predicting the final post-CE
separation is difficult for several reasons, including: estimating the location of the bifurcation point  within the massive star \citep{2001A&A...369..170T}, separating the remaining core from the ejected envelope \citep{2001A&A...369..170T, 2019ApJ...883L..45F}, additional energy sources such as accretion energy \citep{2015ApJ...798L..19M, 2020ApJ...901L..24M}, energy and radiation transport during the CE inspiral \citep{2019ApJ...883L..45F} and the effect of an inflated envelope of the exposed naked helium core \citep[][see also Section~\ref{paragr:CE}]{2015A&A...580A..20S}.

Newly formed NSs gain velocity \citep[natal kicks; e.g.][] {1970ApJ...160..979G,2005MNRAS.360..974H,2017A&A...608A..57V} due to asymmetries arising during their formation  \citep[e.g.][]{2012ARNPS..62..407J}. The properties of the modelled NS+NS population (e.g. number of systems formed, orbital parameters, merger locations relative to formation site) are highly dependent on the adopted natal kick prescription  \cite[e.g.][]{1998A&A...332..173P,1999MNRAS.305..763B, 2018MNRAS.474.2937C,2018MNRAS.480.2011G,2019MNRAS.486.3213A}. To match the current observational constraints on the NS+NS merger rate and parameters of several of the observed systems \cite[e.g.][]{2007AIPC..924..598V}, it is necessary to assume that a fraction of NS forms with natal kicks smaller than typically found for young single pulsars. Some scenarios involve low-mass NS progenitors and electron-capture triggered explosions \cite[e.g.][]{2006ApJ...644.1063D,2013ApJ...772..150J}. Others postulate a link between the natal kick magnitude and the mass of the NS progenitor and SN ejecta  \citep[e.g.][]{2016MNRAS.456.4089B,2016MNRAS.461.3747B,2017ApJ...837...84J}. 
These claims have been supported by 3D NS simulations of ultra-stripped stars \citep{2019MNRAS.484.3307M}.
In fact, it has been demonstrated that close-orbit, low-eccentricity NS+NS and BH+NS systems most likely form via ultra-stripped SNe when the last star explodes \citep{2013ApJ...778L..23T,2015MNRAS.451.2123T}. The reason being that the last Case~BB RLO mass-transfer phase causes the NS to significantly strip its evolved helium-star companion, almost to a naked metal core prior to its explosion, and thus there is very little SN ejecta (see also Section~\ref{paragr:ccSNe}).

Finally, a clear correlation has been predicted between the spin period of the recycled pulsar and the orbital period of the system after the second SN \citep{2017ApJ...846..170T}. This correlation can be tested in LISA binaries, if the spin period is measured, since only short orbit systems will enter the LISA band within a Hubble time, and these binaries should therefore contain the most rapidly spinning NSs of this population. Another hypothesis that can be tested by LISA, is the resulting mass distribution among NS+NS systems \cite[e.g.][and references therein]{2016ARA&A..54..401O}.

Merging BH+BHs and NS+BHs in the field are thought to occur under some specific binary interactions which either (i) bring the parent stars closer together during their evolution or (ii) prevent stars in close obits from expanding. 

The former one (i) occurs in a similar manner to the formation of NS+NSs described above, and involves many of the main uncertainties. In contrast to NS+NS formation, BH+BHs and to a lesser degree BH+NSs are sensitive to the metallicity of the progenitor stars, and they favor low-metallicity environments. In addition the second mass transfer episode, after the first compact object formation, can be either dynamically stable \citep[e.g.][]{2017MNRAS.471.4256V,2017MNRAS.468.5020I,2019MNRAS.490.3740N} or unstable \citep[e.g.,][]{1976ApJ...207..574S,1976IAUS...73...35V,1993MNRAS.260..675T,2007PhR...442...75K,2014LRR....17....3P,2016Natur.534..512B}. In the latter case this leads to a CE phase. The resulting tight system composed of a compact object and a Wolf--Rayet star can eventually undergo a tidal spin up of the star \citep{2018A&A...616A..28Q,2020A&A...635A..97B}. On the other end, if the second mass transfer is stable the binary will result in wider orbits compared to the evolution through CE and avoid a subsequent tidal spin up phase \citep{2020arXiv201016333B}. Eventually, following wind-driven mass loss, the secondary will collapse to a compact object. This leave us with either a BH+BH system or a NS+BH system with either a first- or second-born NS. 

The latter possibility (ii) occurs when two massive stars are born in a tight orbit (orbital periods less than 4 days) in low-metallicity environments which due to their tidal interactions can maintain the stars at almost critical rotation. Such rapidly rotating stars develop a temperature gradient between the poles and the equator leading to chemical homogeneous evolution \citep[e.g.,][]{2009A&A...497..243D,2016MNRAS.458.2634M,2016A&A...588A..50M,2020arXiv200211630D}. In these stars meridional circulation transport hydrogen from the surface into the core and helium out into the envelope until nearly all the hydrogen in the star is fused into helium. At the end of their main sequence these stars are essentially Wolf--Rayet stars and do not expand, hence, avoiding any additional mass-transfer phase.

LISA may answer whether or not mixed binaries of BHs and NSs, in which the NS formed first, are produced in the Galaxy. It is possible that the in-spiralling NS is unable to eject the envelope of the relatively massive BH progenitor star \citep{2018MNRAS.481.1908K}. 
Since we currently do not know of the existence of mixed BH+NS systems in the Galacy, any LISA detections of such systems, as well as double BH systems, will provide crucial information about their formation process.

\subsubsection{Sources in clusters}\label{subsubsec:cluster-sources} 

\begin{figure*}
    \begin{center}
  \includegraphics[width=0.60\textwidth, angle=0]{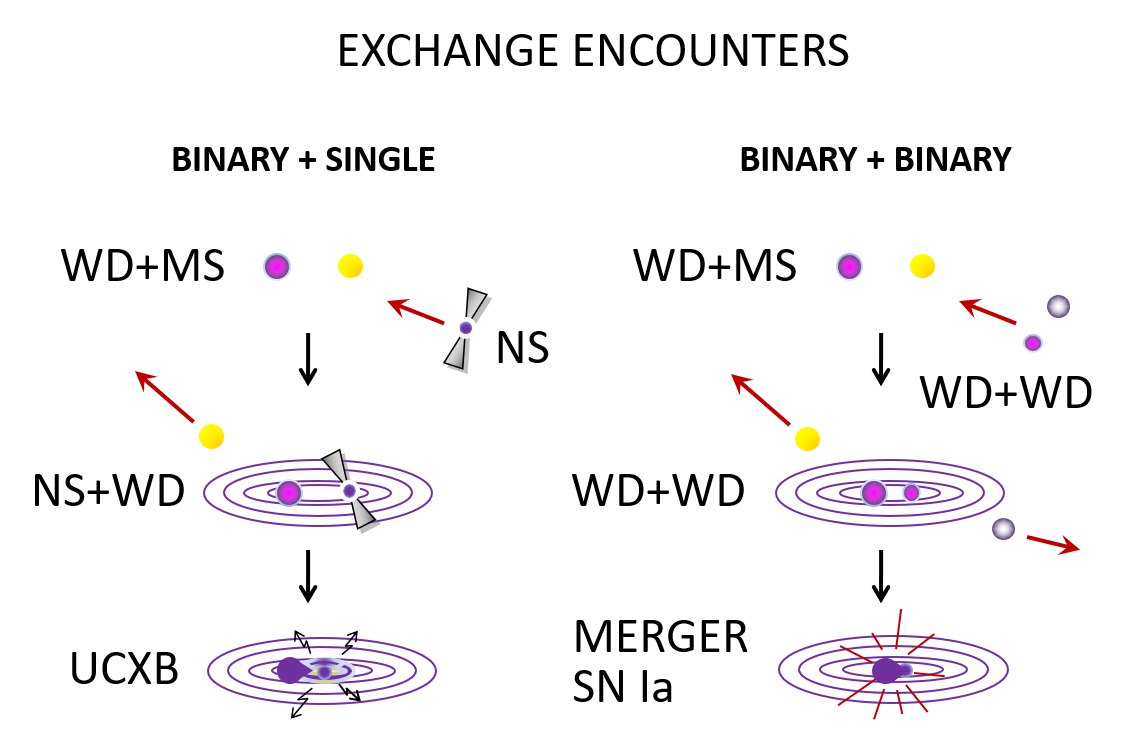}
\caption{Illustration of the formation of LISA sources via examples of exchange encounters. See Figs.~\ref{fig:DWD} and \ref{fig:UCXB} for explanation of acronyms \citep[Figure from][]{TvdH23}.}
 \label{fig:dynamical}
  \end{center}
\end{figure*}

The inner regions of stellar clusters are cosmic factories of compact binaries (i.e.\,binaries containing two compact objects; BHs,  NSs or WDs), owing to the dominant role played by stellar dynamics in such environments. Massive stars in stellar clusters lose kinetic energy to lighter stars and accumulate into the cluster centre. In just a few Myr, these stars evolve into stellar BHs and NSs. 

The production of compact binaries can take one of two routes. If only a small fraction of BHs are retained within a cluster, encounters between BHs and binary stars  lead to dynamical exchanges, where the BH  replaces a less massive star within the binary \citep{1980AJ.....85.1281H}. Globular clusters have a considerably-enhanced population of X-ray binaries \citep{2003ApJ...598..501H}, which might have formed when a NS or a BH exchanges into a binary star \citep[e.g.,][]{1976MNRAS.175P...1H}. After the first exchange, evolution of the stellar companion (which might also become a BH or a NS)  or a second dynamical exchange can produce a compact binary.  Binary-single interactions represent an efficient mechanism to harden these binaries  \citep{1975MNRAS.173..729H} to the point where they can merge via the emission of gravitational radiation (see Fig.~\ref{fig:dynamical} for a schematic representation).  

Alternatively, if a large fraction of
BHs are retained in the cluster, they can form a BH subsystem 
\citep[][but see \citealt{2013MNRAS.432.2779B}]{1969ApJ...158L.139S,2007MNRAS.379L..40M,2008MNRAS.386...65M,2018MNRAS.479.4652A,2020ApJS..247...48K}. BHs in the subsystem tend to strongly interact with each other, undergoing frequent pairing, exchanges, and ejections. The most efficient mechanism driving binary formation in globular clusters is via three-body scatterings \citep[e.g.,][]{2015ApJ...800....9M}. After formation, a binary can undergo dozens of interactions with passing stars and binaries, which can lead to the production of  very hard binaries, capable of merging via the action of GW  emission \citep{2000ApJ...528L..17P}. 
If the star cluster centre harbours a BH subsystem, the BHs dominate the dynamics, quenching mass segregation and preventing the formation of binaries containing other compact objects \citep[see e.g.][]{2020ApJ...888L..10Y}. However, dynamically evolved clusters can lose a substantial fraction of the BHs. In these BH-poor clusters, binary-single interactions can allow the formation of binary NS and BH+NS binaries. Furthermore, it has been proposed that a parabolic encounter between two compact objects could potentially lead to the formation of a binary due to an abrupt loss of energy emitted as gravitational radiation \cite[e.g.][]{1972PhRvD...5.1021H,1989ApJ...343..725Q,2006ApJ...648..411K,2015MNRAS.448..754H}. However, the event rate of this mechanism, which is often referred to as the ``gravitational brake'' capture, is very likely to be negligible due to the small cross-section \citep{1990ApJ...358...81K}.

Old BH-poor clusters may also be ideal for dynamical formation of WD+WD binaries as well as BH/NS+WD binaries \citep{2020ApJS..247...48K}, see Fig.~\ref{fig:dynamical}. In old globular clusters, WDs are by far the most abundant type of compact object (roughly $10^5$ WDs are expected in a $10^6\,M_{\odot}$ cluster). A number of analyses have studied ways WD binaries,  both accreting and detached, may be dynamically assembled in stellar clusters \citep{1995ApJ...455L..47G,2006MNRAS.372.1043I,2016MNRAS.462.2950B,2018ApJ...852...29K}. Furthermore, a handful of the stellar-mass BH binary candidates observed in Galactic globular clusters are suspected to be ultra-compact accreting BH+WD binaries \citep{2012Natur.490...71S,2017MNRAS.467.2199B,2017ApJ...851L...4C}. Overall, up to a few dozen dynamically formed WD binaries are expected to be resolved by LISA in the Galactic globular clusters, likely constituting the largest class of dynamically-formed LISA sources in the Galaxy \citep{Willems2007,2019PhRvD..99f3003K}. Currently two candidates to AM CVns in globular clusters have been identified \citep{ 2016MNRAS.460.3660Z,2018MNRAS.475.4841R} and several more are expected to be discovered in upcoming globular cluster surveys.

Comparing nuclear stellar clusters with globular clusters, the former tend to have somewhat larger escape speeds \citep[due in part to the presence of a central massive BH, MBH:][]{2009MNRAS.397.2148G}. This means that a larger fraction of BHs are likely to be retained \cite[e.g.][]{2009ApJ...692..917M}, while the higher dispersion velocity inhibits both exchange encounters and the dynamical formation of binaries \citep[e.g.][]{2003gmbp.book.....H}. The presence of a dense nuclear cluster surrounding the MBH can significantly affect the formation process of compact binaries in a number of ways. Dynamical three body encounters can form at least one compact BH+BH if the nuclear cluster-to-MBH mass ratio exceeds 10, whereas at lower values the reservoir of compact binaries might be replenished via star cluster inspiral \citep[e.g.][]{2020ApJ...891...47A}. The presence of an MBH can leave significant imprints on the BH+BH evolution, owing to the possible development of von Zeipel--Kozai--Lidov cycles \citep{1910AN....183..345V,1962AJ.....67..591K,1962P&SS....9..719L}, which can boost the rate of BH+BH mergers \citep{2002ApJ...578..775B,2012ApJ...757...27A,2018ApJ...856..140H,2019MNRAS.488...47F,2020ApJ...891...47A} and significantly affect the BH mass spectrum in these extreme environments \citep[e.g.][]{2020ApJ...891...47A}.

Young star clusters and open clusters, because of their relatively low total masses ($10^2\text{--}10^5$~M$_\odot$), host a smaller population of BHs with respect to globular and nuclear clusters \citep[e.g.,][]{2000ApJ...528L..17P,2010MNRAS.402..371B,2017MNRAS.467..524B,2021MNRAS.500.3002B}. BH+BHs in young/open star clusters mostly originate from dynamical exchanges or even from the evolution and hardening of primordial binaries \citep{2014MNRAS.441.3703Z,2019MNRAS.487.2947D,2020MNRAS.498..495D,2019MNRAS.486.3942K,2020MNRAS.495.4268K}. Furthermore, dynamical exchanges favour the formation of BH+NS binaries in young and open clusters \citep{2020MNRAS.497.1563R}.

Finally, hierarchical mergers in globular/nuclear clusters \citep[e.g.][]{2002ApJ...576..894M,2019PhRvD.100d3027R,2019MNRAS.486.5008A,2020ApJ...891...47A,2020ApJ...894..133A} or runaway collisions of massive stars \citep[e.g.][]{2004Natur.428..724P,2015MNRAS.454.3150G,2016MNRAS.459.3432M,2020arXiv200809571R} and binary star mergers in young star clusters \citep{2020MNRAS.497.1043D} might even lead to the formation of intermediate-mass BHs \citep{2019MNRAS.484..814G} and BHs with mass in the pair-instability gap \citep[e.g.][]{2020ApJ...894..133A}, similar to GW190521 \citep[][]{2020PhRvL.125j1102A,2020ApJ...900L..13A}.

\subsubsection{Triple stellar systems}\label{subsubsec:triples_formation}

\begin{figure*}
    \begin{center}
  \includegraphics[width=0.60\textwidth, angle=0]{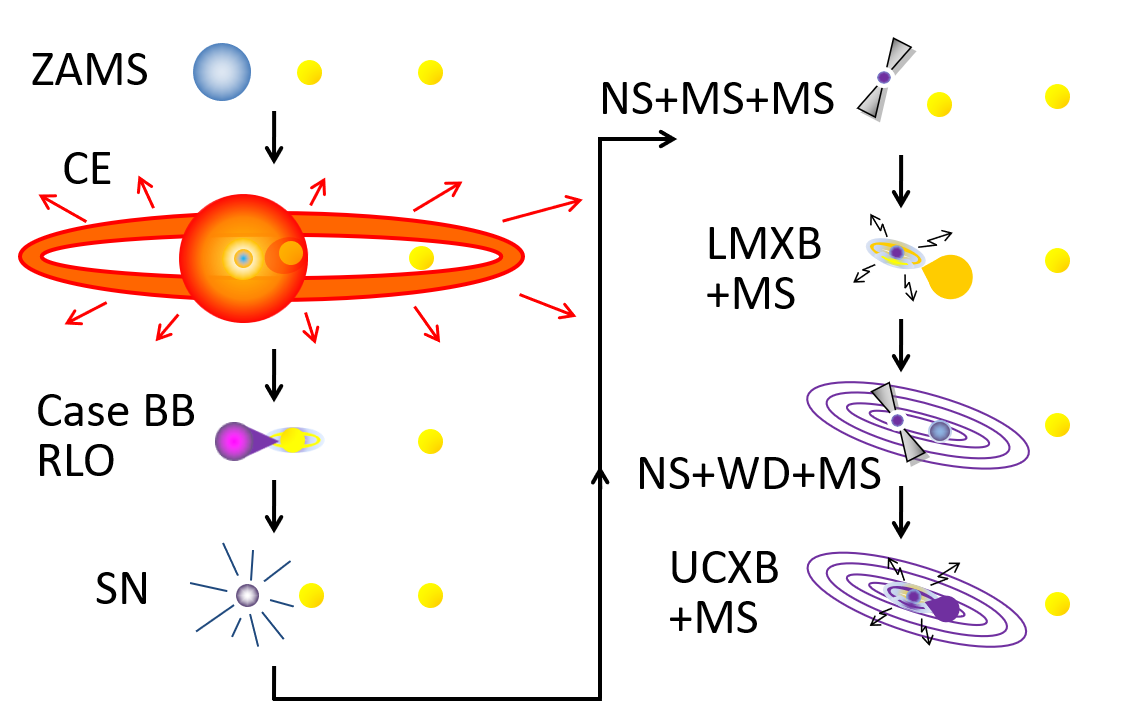}
\caption{Illustration of the formation of a LISA source in a triple system. See Figs.~\ref{fig:DWD} and \ref{fig:UCXB} for explanation of acronyms \citep[Figure from][]{TvdH23}.}
 \label{fig:triple}
  \end{center}
\end{figure*}

Some of the LISA sources may form as part of triples and higher-order multiples. This includes sources in the Galactic disc (formed through e.g.\,isolated triple evolution) as well as those in dense environments. Three-body (or more) interactions are important in the formation of compact sources in two ways: during short-lived dynamical interactions and in hierarchical triple systems.

Hierarchical triples, in which two bodies orbit each other, and a third body orbits the centre of mass of the inner orbit, can remain stable for secular timescales, and therefore stay intact for Hubble times \citep{1994MNRAS.270..936K,1999ASIC..522..385M,2008CeMDA.100..151G,2018MNRAS.474...20H}. They may form in clusters (where they may interact with interloper stars in the densest environments) or exist in the Galactic disc (and evolve in pure isolation). Their evolution differs from that of isolated binaries due to three-body effects. Hence, triples that live their lives in isolation bridge the gap between classically isolated LISA sources and dynamically-evolving sources (often used to mean cluster sources). The importance of three-body interactions in hierarchical systems has been recognised for the evolution of stellar triples \citep{2011ApJ...741...82T,2013MNRAS.430.2262H,2017ApJ...836...39S,2017ApJ...841...77A,2017ApJ...846L..11L,2018A&A...610A..22T,2019MNRAS.486.4443F,2019MNRAS.486.4781F,2020A&A...640A..16T}, triples that consists of a combinations of stars and planets \citep{2016MNRAS.462L..84H,2017MNRAS.466.4107H,2018MNRAS.481.2180V,2018AJ....156..128S,2020ApJ...889...45S}, as well as stellar binaries in dense environments \citep{2012ApJ...757...27A,2016ApJ...816...65A,2017ApJ...846..146P,2016MNRAS.460.3494S,2019ApJ...878...58S, 2019MNRAS.488.5489H, 2019ApJ...881L..13H,2020ApJ...900...16F,2020arXiv200908468M,2020ApJ...900...16F}. 

The general formation of binaries and multiples (compact and wide) in clusters is boosted during the  collapse of the dense cluster core, which is halted by frequent stellar interactions \citep{1987degc.book.....S,1992PASP..104..981H}. The formation of the first binaries takes place most likely via three-body scatterings, involving three initially unbound objects \citep{1993ApJ...403..271G,1995MNRAS.272..605L}. As soon as binaries start forming, binary--single \citep{1983ApJ...268..319H,1993ApJ...415..631S} and binary-binary \citep{1983MNRAS.203.1107M,2002ApJ...576..894M} interactions take over and become the dominant dynamical processes at play. Even for relatively low triple fractions, dynamical interactions involving triples occur roughly as often as encounters involving either single or binary stars alone, particularly in low-mass star clusters \citep{2013MNRAS.432.2474L}. When the objects involved in the interaction cross their mutual sphere of influence, a strong interaction can trigger the formation of a short-lived bound triple system \citep{1993ApJ...403..271G}. During these chaotic resonances, a pair of objects has a non-negligible probability of experiencing a very close passage, triggering the formation of a compact binary and subsequent merger \citep{2014ApJ...784...71S}. Depending on the cluster structure, binary mergers developing through resonant interactions can be highly eccentric at LISA frequencies and even still when entering the frequency range typical of ground-based detectors \citep{2018PhRvD..97j3014S,2018MNRAS.481.5445S,2018arXiv180506458A}. Binary--binary interactions \citep{1984MNRAS.207..115M,1991ApJ...372..111M,2002ApJ...576..894M} represent another  efficient mechanism to form triples, either 
in the form of short lived, resonant unstable triples \citep{1983ApJ...268..319H,2019ApJ...871...91Z,2018arXiv180506458A}, or in a hierarchical configuration \citep{2016ApJ...816...65A,2019ApJ...871...91Z,2018arXiv180506458A,2020arXiv200908468M, 2020ApJ...900...16F}.

The best known manifestation of three-body dynamics are the von Zeipel--Kozai--Lidov cycles \citep{1910AN....183..345V, 1962P&SS....9..719L,1962AJ.....67..591K} regime in which the inner orbit eccentricity and the inclination between the two orbits vary periodically. Strictly speaking this applies to the (inner) test particle regime, an axisymmetric outer potential and the lowest-order expansion of the Hamiltonian (i.e.\,quadrupole). Relaxing either one of these assumptions leads to qualitative different dynamical evolution, which include extreme eccentricity variations and orbital flips \citep[see][for a review]{2016ARA&A..54..441N}. The high eccentricities can lead to close passages between the bodies, mass transfer, and enhancement of dissipative processes such as from tides or by GW  emission. Over time, the latter can lead to a significant reduction of the inner orbital separation during the nuclear burning stages or during the compact object phase of the stars \citep{1979A&A....77..145M,1998MNRAS.300..292K,2007ApJ...669.1298F,2011ApJ...741...82T}. Through the internal stellar evolution, a triple may transition from one dynamical regime to another, enhancing (or diminishing) the three-body effects \citep{2013ApJ...766...64S,2014ApJ...794..122M,2016ComAC...3....6T}.

\subsection	{Expected LISA observations: numbers and rates}\label{subsec:rates}
\noindent \coord{Abbas Askar,Simone Bavera}\\
\noindent \contr{Abbas Askar,  Quentin Baghi, Simone Bavera, Tassos Fragos, Valeriya Korol, Kyle Kremer,  Manuel Arca Sedda}\\

\subsubsection{Binary's detectability}

The detectability of Galactic stellar binaries with LISA
primarily depends on the parameters involved in the GW amplitude:
\begin{equation} \label{eqn:amp}
    {\cal A} = \frac{2(G{\cal M})^{5/3}(\pi f)^{2/3}}{c^4 d},
\end{equation}
where $f$ is the binary's GW frequency, ${\cal M}={(m_1 m_2)^{3/5}}/{(m_1+m_2)^{1/5}}$ is the chirp mass (with $m_1$ and $m_2$ being the primary and secondary masses), $d$ is the luminosity distance, $G$ and $c$ are respectively the gravitational constant and the speed of light. This follows from the fact that the signal-to-noise ratio scales linearly with amplitude $\rho \propto {\cal A} \sqrt{T}$ with $T$ being the observation time \citep{2021arXiv210801167B}.
Besides, in contrast with electromagnetic observations, the observed GW signal scales as $1/d$, rather than $1/d^2$.

The GW frequency, defined as $f=2/P$ with $P$ being the binary's orbital period, has the strongest impact on the signal's detectability. Binaries emitting at $f>3\,$mHz fall in the most sensitive part of the LISA frequency band \citep{2017arXiv170200786A}. As a result, these high-frequency binaries -- even if consisting of the lowest mass WD components -- can be detected across the Milky Way also reaching satellite galaxies (cf. Figures~\ref{fig:horizons} and \ref{fig:MW}). At a set frequency, the maximum distance at which the binary can be detected increases with its chirp mass, as shown in Figure~1.1. In addition, the chirp mass defines how fast the GW frequency changes during the in-spiral phase. This so-called chirp is given by
\begin{equation} \label{eq:fdot}
\dot{f}=\frac{96}{5}  \pi^{8/3} \left( \frac{G{\cal M}}{c^3} \right)^{5/3} f^{11/3}.
\end{equation}
The chirping rate for stellar binaries in the LISA band is generally very slow ($\dot{f} < 10^{-15}$\,Hz$^2$ for typical WD+WD and NS+NS binaries emitting at frequencies lower than a few mHz).
Note that the frequency derivative (not the chirp mass) can be measured directly from GW data. The measurements is possible for binaries emitting at sufficiently large frequencies ($f > 3\,$mHz, \cite{2020ApJ...894L..15R}), such that during the observation time ($T$) with LISA the binary sweeps through at least a few frequency bins ($1/T$) . If $\dot{f}$ is measured and assuming that the inspiral is driven by radiation reaction only, the luminosity distance can be recovered by plugging in the measured value of the chirp mass into Eq.\eqref{eqn:amp}. At lower frequencies, binaries will be effectively ``seen'' by LISA as monochromatic. In which case, only measurements of $f$ and ${\cal A}$ will be possible, while the chirp mass estimate will be degenerate with that of the distance (cf. Eq.~\ref{eqn:amp}). 

Unlike circular binaries, eccentric ones emit GWs at multiple harmonics. Each signal can be thought as a collection of $n$ signals from circular binaries emitting at $f_n = n f/2$ and the amplitude ${\cal A}_n={\cal A}(2/n)^{5/3} g(n,e)^{1/2}$, where $g(n,e)$ is given in Peters \& Mathews (1963). The total signal-to-noise ratio can be estimated as the quadrature sum of the individual harmonics' signal-to-noise ratios. 
Thus, to measure the chirp mass in case of the eccentric binary, in addition to $\dot{f}$, one needs to simultaneously measure the eccentricity. This is possible when at least two harmonics are detected \citep[e.g.][]{2016MNRAS.460L...1S}, otherwise only an upper limit on the chirp mass can be derived.

Another aspect of Galactic binaries' detectability is that they will be observed continuously during the course of the mission. However, it is likely that the measurements will undergo occasional interruptions due to communication antenna re-pointing or spacecraft housekeeping. Additionally, spurious disturbances may affect the interferometer signals, such as the transient glitches observed in LISA Pathfinder \citep{2022arXiv220511938L, 2022PhRvD.105d2002B}, leading to corrupted data spans. These operating conditions may impact the duty cycle of the mission, with a minimum requirement of 89\% at the time of writing. If the masking events are frequent, they could impact the detection of low-frequency sources (around 0.1 mHz and below). However, mitigation techniques have been developed and show promising results for restoring the optimal detectability \citep{2019PhRvD.100b2003B,2022MNRAS.509.5902B}.

\subsubsection{Detection and parameter estimation expectations}
It is expected that the Galactic stellar binaries will dominate the number of individually detected GW sources at mHz frequencies (Table~\ref{tab:absolute_n_MWG}). Out of $\sim\mathcal{O}(10^7)$ stellar binaries emitting in the LISA frequency band, LISA is expected to deliver $\sim\mathcal{O}(10^4)$ individual detections \citep[e.g.][]{2020PhRvD.101l3021L, 2021PhRvD.104d3019K}.
Current estimates suggest that at frequencies $>3\,$mHz, Galactic WD+WD detectable by LISA will be counted in thousands, NS+NSs in few up to hundreds and BH+BHs in a few (see Tables \ref{tab:absolute_n_MWG} and \ref{tab:absolute_n_MWG_GC}). At frequencies $<3\,$mHz the number of stellar binaries is so large, that only a small fraction - the closest and more massive ones - will be individually detected, while the rest of the population will form an unresolved stochastic foreground (see Section~\ref{subsubsec:GWforeground}).

Amongst the resolvable systems, population synthesis simulations suggest that hundreds will be exceptionally strong LISA sources \citep[e.g.][]{2021PhRvD.104d3019K}. These binaries will be detectable within weeks from the start of mission operations, and over the full mission lifetime can accumulate SNRs up to $10^3$.
Based on an SNR evaluation via an iterative scheme for the estimate of the confusion foreground generated by Milky Way’s GW sources \citep[e.g.][]{2020PhRvD.101l3021L,2021PhRvD.104d3019K}, synthetic population analysis \citep[e.g.][]{2017MNRAS.470.1894K, 2019MNRAS.490.5888L} yields about up to $10^4$ binaries detectable with SNR$>20$, several $10^3$ with SNR$>100$, a few with SNR$>1000$. For all binaries, frequencies predicted be measured very accurately with $\sigma_{\rm f}/f < 10^{-5}$, which corresponds to $\sigma_{\rm P}<0.025$\,ms for a typical mHz WD+WD binary. Frequency derivative is estimated to be measured to better than 30\% for up to $10^4$ binaries, leading to the measurement of the chirp mass via Eq.~\eqref{eq:fdot}. Consequently, also distances can be derived to better than 30\% for several $10^3$ binaries. The sky localisation uncertainty depends strongly on the SNR and GW frequency ($\propto \rho^{-2}f^{-2}$). Additionally, it is also dependent on the ecliptic latitude: a source on the ecliptic has an order of magnitude more uncertainty than a source at the poles \citep{2020ApJ...894L..15R}. On average, sky localisation error for Galactic WD+WD systems is expected to be of several deg$^2$, improving with increasing GW frequency and SNR down to sub-square degree precision and below. These expectation are well supported by Bayesian-based data analysis exercises \citep{2020PhRvD.101l3021L} with mock data simulating observations over one year, and including 30 millions of injected Galactic sources. Consolidation of these results are expected once more realistic data simulations featuring mixed source types are analysed (see https://lisa-ldc.lal.in2p3.fr).

Beyond the Milky Way, the Local Group galaxies are expected to harbour from a few to a few hundreds LISA sources (mainly WD+WDs and some NS+NSs) depending on the total mass and the distance of the galaxy \citep{2019MNRAS.489.4513S, 2020ApJ...892L...9A,2020A&A...638A.153K,2020MNRAS.492.3061L} (see Fig.~\ref{fig:horizons}). For instance, the number of WD+WDs and NS+NS in the largest Milky Way satellites, the Large and Small Magellanic Clouds, will be high enough to overcome Galactic foreground and to unambiguously identify these galaxies in the LISA data \citep{2020ApJ...894L..15R}. Even further away, with the total mass comparable to the Milky Way's mass, the Andromeda galaxy could also be visible on the LISA sky as a group of GW sources \citep{2018ApJ...866L..20K}. 

The outside limits of the Local Group, LISA can access distances up to $\sim 1\,$Gpc through GW signal from stellar-mass BH+BHs, which will ultimately be observable during merger to ground-based detectors (see Section~\ref{paragr:highFreqGW}).
Studies based on cosmological simulations of galaxies at $z=0$ find that the present-day dwarf galaxies can accommodate a larger BH+BH merger rate compared to massive galaxies \citep{2017MNRAS.464.2831O}. Specifically, for massive BH+BHs similar to GW150914, about $40\%$ of mergers are expected to be in galaxy progenitors of Milky Way-like systems and the rest in smaller satellite or isolated dwarf galaxies \citep{2019MNRAS.484.3219M}. This translates into a large number of potential hosts within the estimated LISA horizon distance. Not only BH+BHs formed from the evolution of isolated binaries, but also BH+BHs formed in extragalactic globular clusters may  be detectable by LISA, with initial studies predicting the number of such sources to be in the range of 1 -- $10^2$ \citep{2018PhRvL.120s1103K}.

Given LISA's selection effects, extra-galactic LIGO-like BH+BH are expected to have quite low signal-to-noise ratios ($\sim 10$). For this class of stellar binaries, \cite{2021PhRvD.104d4065B} report sky localisation errors of a few tens of squared degrees and constrains on the detector-frame chirp mass down to $\pm 0.01$\,M$_\odot$. They also show
that at 10\,mHz the eccentricity for these binaries can be measured down to $10^{-3}$, while the merger time can be determined within a time window of 1\,hour.

For sources that form through isolated binary evolution, the rates (such as those mentioned above and in Table~\ref{tab:absolute_n_MWG}) are often estimated with the population synthesis approach.
There are many uncertainties that affect the rate calculations. These uncertainties can be divided in five broad categories: (i) binary evolution physics (e.g., CE evolution, mass transfer stability, mass-accretion efficiency, etc.), (ii) stellar evolution physics (e.g., metallicity dependent stellar winds, core-collapse mechanisms, natal kicks, pair-instability SN and pulsational pair instability, etc.), (iii) initial stellar and binary properties (e.g., initial mass function, binary fraction, initial distribution of separation, mass ratio, eccentricity, etc.), (iv) different assumptions for the Galactic spacial distribution (thin/thick disc and bulge) and star-formation history  and (v) LISA selection effects (e.g., GW foreground, mission length, sensitivity curve and SNR detection threshold).

As discussed in Section~\ref{subsubsec:cluster-sources}, LISA sources may also form dynamically in dense stellar environments such as globular clusters (and may have markedly different features compared to sources that form through isolated binary evolution, Section~\ref{subsubsec:isolated-binaries}). In Table \ref{tab:absolute_n_MWG_GC}, we show the estimated number of sources expected to be found in globular clusters in a Milky Way-like galaxy \citep{2018PhRvL.120s1103K}. The number of detectable LISA BH+BHs originating from young massive clusters and open clusters is expected to be several tens to $\sim$100 \citep[or about $0.5\text{--}3$ times the density of those clusters in the local volume in units of Mpc$^{-3}$;][]{Banerjee2020}. Merger rate estimates for GW sources produced in dynamical environments from cluster simulations, can also be influenced by many uncertain physical processes, some of which are common to the ones outlined for sources formed via isolated binary evolution. For instance, natal kicks for compact objects have a direct impact on retention of those objects in stellar clusters \citep{2013ApJ...763L..15M,2018MNRAS.479.4652A,2018MNRAS.474.3835W,2018A&A...617A..69P,2020A&A...639A..41B} which influences the number of dynamically formed binary systems. The number of compact objects that form in dense environments also depends on their metallicity and the initial mass function of their stars which may vary with their formation environment  \citep{2009MNRAS.394.1529D,2013ApJ...771...29G,2014PhR...539...49K,2020A&A...636A..10C}. Additionally, dissolved or tidally disrupted open and globular clusters can also contribute to the number of GW sources that may have dynamically formed \citep{2010ApJ...718.1266M,2018ApJ...856...92F,2019MNRAS.487.2412G}.

Apart from population synthesis-like simulations, another approach to predict the expected LISA rates is to derive empirical estimates from the already observed population of sources. For NS+NSs for instance, one can use the inferred merger rate coming from the known Galactic NS+NS population, and accounting for survey selection effects \citep{1991ApJ...380L..17P, 2003ApJ...584..985K}, or the inferred merger rate from LIGO--Virgo \citep{2020arXiv201014533T}, to predict that LISA should be able to detect $50\text{--}300$ NS+NSs in the Milky Way \citep{2020ApJ...892L...9A}. In a similar manner, based on O1 LIGO--Virgo detections, it was estimated that LISA maybe able to detect up to $\sim 50$ BH+BHs \citep{2016PhRvL.116w1102S}, but the inferred BH+BH merger rate density decreased in the most recent O3 run by a factor of $\sim 2$. Empirical estimates of the Galactic NS+WD population have been derived, based on the observed pulsar population, with $100-150$ being predicted to be observable by LISA \citep{2018PhRvL.121m1105T}.

\begin{table}
\centering                          
\begin{tabular}{c | c c }        
\hline
Source & $N$ & $N^\mathrm{detected}$ \\
\hline
    WD+WD & $\sim 10^8$  & 6,000$-$10,000  \\
    NS+WD & $\sim 10^7$  & 100$-$300  \\
    BH+WD & $ \sim 10^6$  & 0$-$3  \\
    NS+NS & $\sim 10^5$  & 2$-$100  \\
    BH+NS & $\sim 10^4 - 10^5$  & 0$-$20  \\
    BH+BH & $\sim 10^6$  & 0$-$70  \\
\hline
\end{tabular}
\caption{Estimated absolute number of compact binaries from \textit{isolated binary evolution} in the Milky Way. The columns show the source type, the total number of binaries in the galaxy at any frequency and the total number of estimated sources detected by LISA. We report values from indipendent studies which assume different LISA mission lifetimes and SNR. WD+WD models assume a frequency range $0.5-10$~mHz while models for the other sources assume a frequency range $0.1-10$~mHz. At lower frequencies the total number of LISA sources is so high that it might become impossible to distinguish individual sources from the GW foreground. The ranges of the expected intrinsic number of each binary type are extracted from \cite{2012ApJ...758..131N, 2020ApJ...898...71B, 2010ApJ...725..816B, 2018MNRAS.481.1908K, 2001A&A...365..491N, 2014A&A...569A..42V, 2018MNRAS.480.2704L} while the ranges of the expected number of LISA sources are extracted from \cite{2001A&A...375..890N, 2017MNRAS.470.1894K, 2019MNRAS.490.5888L, 2018ApJ...866L..20K, 2010ApJ...719.1546L, 2010ApJ...717.1006R, 2018PhRvL.121m1105T, 2020ApJ...898...71B, 2010ApJ...725..816B, 2018MNRAS.481.1908K, 2020MNRAS.492.3061L, 2020ApJ...892L...9A, 2020MNRAS.494L..75S}.}
\label{tab:absolute_n_MWG}
\end{table}

\begin{table}
\centering                          
\begin{tabular}{c | c c }        
\hline
 Source & $N$ & $N^\mathrm{detected}$ \\
\hline
    WD+WD & $\sim 2 \times 10^4$ & 4$-$20 \\
    NS+WD & $\sim 10^3$ & 3$-$6 \\
    BH+WD & $\sim 10^2$ & 2$-$4 \\
    NS+NS & $\sim 40$ & 1  \\
    BH+NS & $\sim 4$ & 0  \\
    BH+BH & $\sim 2 \times 10^2$ & 4$-$7 \\
\hline
\end{tabular}
 \caption{Estimated absolute number of compact binaries in \textit{globular clusters} in the Milky Way. The columns show the source type, the total number of binaries in the galaxy at any frequency and the total number of estimated sources detected by LISA assuming a frequency range of $10^{-5}-1$~Hz and a mission lifetime of 4 years with a SNR ranging between 2 and 7 \citep{2018PhRvL.120s1103K}. }
\label{tab:absolute_n_MWG_GC}
\end{table}

\subsection	{Synergies}

\subsubsection{Synergies with EM observations}\label{subsec:synergies}

\noindent \coord{Thomas Kupfer}\\
\noindent \contr{Thomas Kupfer, John Tomsick, Nicole Lloyd-Ronning, John Quenby, Thomas Tauris, Thomas J. Maccarone}


Many new verification binaries are expected to be discovered before the launch of LISA, in particular with wide-field optical surveys such as ZTF, BlackGEM and LSST on the Rubin Observatory. Once flying, LISA will be complemented with surveys across different frequency bands (radio to gamma-rays). New sources discovered by LISA can be studied with the next generation of follow-up facilities such as ESO/ELT, CTA, SKA, ngVLA or even Athena as well as smaller space missions which could be approved, built and launched in the early 2030s (eXTP or STROBE-X). This provides a plethora of large-scale follow-up resources for detailed multi-messenger studies but it requires a well planned follow-up strategy to generate the most useful results. 

The large number of facilities running in the 2020s and 2030s will provide a bright future for the research on compact Galactic binaries, with hundreds of additional EM discovered systems ready to be studied in detail through EM+GW observations as soon as LISA is operational. A significant sample of binaries, observed with EM+GW observations, will open up possibilities to explore and study astrophysical phenomena which are crucial to our understanding of the Universe. This includes the long-standing questions of the progenitors of SNe Ia, the formation and evolution of compact objects in binaries and accretion physics under extreme conditions.

\paragraph{UV/Optical/IR observations}

\phantom{text}

Previous studies predict that we will be able to observe several thousand Galactic binaries in both GW and optical emission \citep{2013MNRAS.429.2361L,2017MNRAS.470.1894K}. A subset of the known UCBs have orbital periods that lie in the LISA band and these will be individually detected by LISA due to their strong GW signals, some on a timescale of weeks or a few months. Combined GW and EM multi-messenger studies of UCBs will allow us to derive population properties of these systems such as masses, radii, orbital separations, and inclination angles but in many cases EM observations are required to complement GWs and break degeneracies in the GW data. \citet{2012A&A...544A.153S, 2013A&A...553A..82S, 2014ApJ...791...76S, 2018MNRAS.480..302K} present several studies on how EM observations can complement LISA GW data and vice versa. The GW amplitude and inclination is strongly correlated, but the GW amplitude can be improved by a factor of six when including EM constraints on the inclination and the sky position and inclination can reduce the uncertainty in amplitude by up to a factor of 60 \citep{2014ApJ...791...76S}. 

Additionally, knowing the distance from EM, e.g. from parallaxes measured by Gaia, will help to derive a chirp mass from GWs even for non-chirping sources because for many LISA sources only the frequency and amplitude will be measured. This leaves a degeneracy between chirp mass and distance and inclination, since a more massive binary further away can have the same amplitude as a lower mass one that is closer by and inclined systems closer by look like face on systems further away \citep{2012A&A...544A.153S,2014ApJ...790..161S}. 

\begin{figure*}
    \begin{center}
   \includegraphics[width=0.60\textwidth]{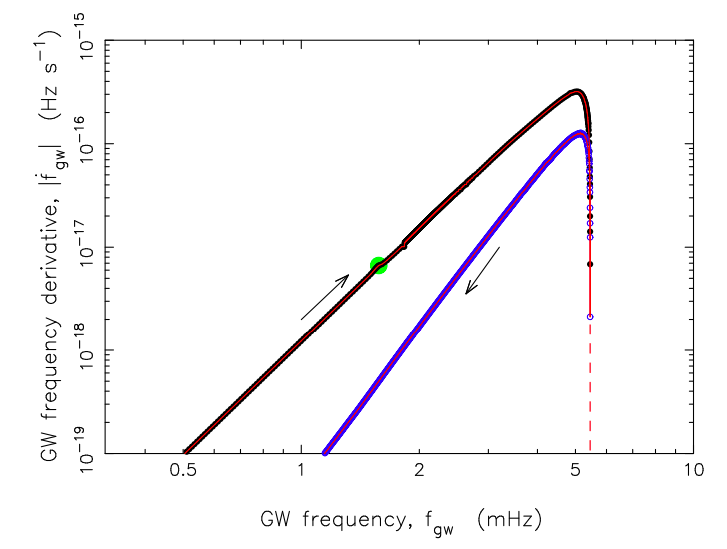}
   \includegraphics[width=0.60\textwidth]{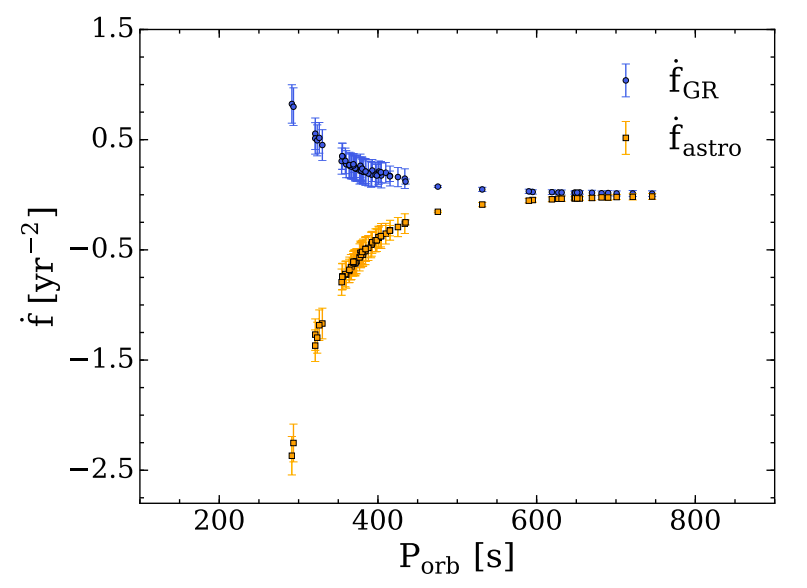}
\caption{{\bf Upper panel}: GW frequency derivative, $|\dot{f}_{\rm GW}|$ as a function of GW frequency, $f_{\rm GW}$ for the UCXB shown in Figs.~\ref{fig:UCXB_strain+data} and \ref{fig:UCXB-AMCVn-detailed}. The black coloured points correspond to the inbound leg (orbital shrinking) and the blue points correspond to the outbound leg (orbital expansion, after reaching the orbital period minimum) including mass transfer/loss from the system and finite-temperature effects of the WD donor. Each point represents a binary stellar MESA model. The green solid circle indicates the onset of the UCXB stage. (Before this point, the system is a detached NS+WD binary). Figure from \citet{2018PhRvL.121m1105T}. 
{\bf Bottom panel}: pure general relativistic (GR, blue) and astrophysical (orange) chirps as a function of GW frequency for WD+WD (AM~CVn) systems \citet [Figure from][]{2018ApJ...854L...1B}. The astrophysical contribution from mass transfer and tides leads to a significant deviation from the contribution from GR alone and will cause the systems to widen their orbit upon mass transfer. The decoupling of the astrophysical chirp from the GR chirp will be possible with {\em combined measurements} from LISA and Gaia. The error bars show the anticipated 1$\sigma$ measurement errors. On average, \citet{2018ApJ...854L...1B} find that about 50 AM~CVn systems in the Mikly Way with $P_{\rm orb}<800\;{\rm s}$ have resolvable GR and astrophysically driven chirps.} 
 \label{fig:f-dot}
  \end{center}
\end{figure*}

If the chirp mass has been measured from GWs using e.g.\,Gaia parallaxes and the mass ratio has been measured from EM, through radial velocities or ellipsoidal deformation of one component, both measurements can be combined to the measure the masses of both components with a few percent precision. Comparing the measured orbital decay with the predicted orbital decay from general relativity will allow us to measure the effects of tides compared to GWs. Tides are predicted to contribute up to $10\,\%$ of the orbital decay \citep{2011ApJ...740L..53P} but has not been measured so far and is very difficult with GW and EM alone. Some of the known AM\,CVn and detached WD+WD binaries have well constrained distances from Gaia \citep{2018MNRAS.480..302K, 2018A&A...620A.141R}. Measuring the distance from EM constrains the uncertainty in chirp mass to 20\%, whereas adding the period derivative $\dot{P}$ reduces it to 0.1\% (see e.g.\,Fig.~\ref{fig:f-dot}). A GW chirp mass measurement would provide the first detection of tidal heating in a merging pair of WDs from the deviations in predicted $\dot P$ \citep{2014ApJ...791...76S}. With a large enough sample of WD+WD binaries whose chirp masses can be determined, we can plausibly extract constraining information about CE phase evolution physics. In studying massive WD+WD binaries that are likely progenitors of merger-induced collapse NSs \citep{2019MNRAS.484..698R}, chirp mass distributions had different shapes depending on the adopted CE phase prescription in the binary evolution population synthesis model.

Several studies have shown that spectral and photometric analysis of detached WD+WD can provide precise sky positions, orbital periods and in some cases mass ratios, inclinations, and the rate of orbital period decay \citep[e.g.][]{2002MNRAS.332..745M,2011ApJ...737L..23B, 2012ApJ...757L..21H, 2019Natur.571..528B}. All of the known verification binaries have precise sky positions and orbital periods and five systems have a measured orbital decay from photometric monitoring \citep{2018MNRAS.480..302K}.  SDSSJ0651, a 12 min orbital period detached WD+WD \citep{2011ApJ...737L..23B}, and ZTFJ1539, a 7 min orbital period detached WD+WD \citep{2019Natur.571..528B}, are prime example of what can be accomplished. Using only one year of eclipse timing measurements, \citet{2012ApJ...757L..21H} found an orbital decay of $\dot{P} = (-9.8\pm 2.8)\times 10^{-12}$\,s\,s$^{-1}$ in SDSSJ0651 which has not been updated since then. \citet{2019Natur.571..528B} used photometric data from PTF/iPTF and ZTF covering a total of ten years and measured a very precise orbital decay of  $\dot P = (-2.373\pm0.005)\times 10^{-11}$\,s\,s$^{-1}$. However, neither of the two systems has a precision of WD component masses good enough from optical observations to see if their $\dot{P}$ differs from the predictions of the theory of General Relativity (GR). GW observations can solve that. Tidal theory predicts a 10\% deviation from GR if the WDs are tidally heating up.  Which means that combining EM+GW observations from LISA will allow a measurement of the amount of tidal heating in these merging pairs of WDs for the first time.

Combined EM+GW observations of the Galactic WD+WD population will help to solve another major problem in astrophysics: the SNe\,Ia progenitor problem. Although only the thermonuclear explosion of a WD following the interaction with a binary companion can explain the observed features in the SN light, much less is known about their progenitors. Recent results have shown that SN\,Iae show a large range of explosion energies and decay times, photometric and spectroscopic signatures indicating different progenitor systems \citep{2019NatAs...3..706J}. Several different explosion scenarios are under discussion, including the merger and subsequent explosion of an ultra-compact WD+WD system (double-degenerate model) or the explosion triggered by ignition of an helium shell accreted from a helium star in a UCB (double-detonation scenario). However, the number of known progenitor systems is limited. \citet{2019MNRAS.482.3656R} studied the probability of finding WD+WD progenitors of SNe\,Ia using a binary population synthesis approach, and found that the chance of identifying such progenitors purely in EM data is $\sim10^{-5}$. These include both double-lined spectroscopic binaries and the eclipsing systems. Even with the next generation of 30\,m class telescopes, the probability for detection only goes up by a factor of $\sim10$. \citet{2017MNRAS.470.1894K} predicts that LISA will individually resolve $\sim25,000$ detached WD+WD systems including the most massive systems. EM follow-up observations in combination with GW measurements will allow us to measure masses of individual systems and find and characterize the population of double degenerate SN\,Ia progenitors. 

Kilonovae \citep[see][]{2019LRR....23....1M} are optical/IR emission accompanying the merger of NS+NS, possibly NS+BH, and in special cases, WD+WD mergers \citep{2019JCAP...03..044R}. Although LISA is not sensitive to the actual merger that can produce a kilonova, it is sensitive to the GW emission from their progenitors. Therefore it is worthwhile to consider kilonova events in the nearby universe because they allow constraints on these degenerate stellar populations. 

\paragraph{X-ray observations}

\phantom{text}

Many of the LISA sources also emit X-rays, thus allowing for a number of joint LISA + X-ray investigations. The donor stars in UCXBs appear to be a mixture of C/O-core and He-core WDs and abundance measurements can help identify their formation scenario \citep{2010MNRAS.401.1347N}. In the oxygen-rich systems, oxygen is the dominant coolant in the accretion discs, and the iron emission lines are suppressed; the strength of the iron lines is broadly in agreement with the thermonuclear burst properties of the sources, strengthening the case that this donor classification process works reasonably well even in its simplest form \citep{2020arXiv200100716K}. However, in some cases, the abundances of the WD inferred from X-ray data are at odds with those inferred from Type I burst properties. In a few cases, the inference has been made from neon-to-oxygen rations \citep{2001ApJ...560L..59J}, and for this scenario, it has been shown that there is a channel of binary evolution that allows a He-core WD to have such neon-to-oxygen ratios \citep{2005A&A...441..675I}.  Alternatively, \citet{1992ApJ...384..143B} show that spallation of CNO elements in a NS atmosphere is quite likely. The combination of LISA measurements with X-ray (and optical) abundance measurements thus opens a window to determining which of these scenarios is correct.  If the apparent C/O-core WD donors are paired with high mass NSs, then the spallation scenario is strongly preferred. 

More detailed X-ray spectroscopy should potentially be able to make detailed abundance estimates for the donor stars, allowing, e.g., identification of systems in which the CNO processing may not have gone to completion, and perhaps estimating the time of formation through estimates of the abundances of non-CNO elements, which could yield the initial metallicity of the star. The X-ray measurements are essential given that a large fraction of the UCXBs are located in globular clusters, or deep in the Galactic Plane where ultraviolet and optical spectroscopy are more challenging. Substantive work has also been done in the optical wavebands \citep[see e.g.][]{2004MNRAS.348L...7N, 2006MNRAS.370..255N}.  Furthermore, when combined with LISA data the UCBs then may provide a means of testing how conservative accretion onto NSs is; C/O-core WD donors must start at masses of at least $\approx0.5 M_\odot$, but are generally observed with masses of $0.1\,M_\odot$ or less.  If the early stages of mass transfer in these systems are conservative, the NSs should typically be about $1.8\,M_\odot$, while if they show much lower masses, this implies that the mass transfer was strongly non-conservative.

Additionally, for 4U~1820$-$30, X-ray measurements provide a straightforward way to monitor the source's period derivative \citep{1991ApJ...374..291T}, as the source is deep in the potential well of a globular cluster whose gravitational acceleration leads to its negative period derivative.  It is easy to track the source's orbital period in the X-ray band \citep{1987ApJ...312L..17S,1991ApJ...374..291T}, and hard in other bands, due to the crowding in the cluster. With more intensive X-ray data, the period derivative of 47~Tuc~X-9, the best candidate BH+WD binary in the Milky Way \citep{2017MNRAS.467.2199B}, could be tracked.  Imposing these constraints, which are likely to lie outside the range of normal templates, can help improve the quality of the LISA GW detection.  Globular clusters are likely good hosts for LISA sources, but the globular cluster sources' periods have come from ultraviolet photometry \citep{2005ApJ...634L.105D,2009ApJ...699.1113Z} and the best non-cluster source's period comes from optical spectroscopy \citep{2013MNRAS.429.2986M}. Additional intensive monitoring campaigns would be required for the period derivative to be estimated.

For the AM~CVn systems, X-ray emission is also valuable.  The same abundance issues can be studied in the X-ray band in AM~CVn systems, although primarily from the emission lines from the boundary layer of the accretor, rather than disc reflection.  Relatively short period AM~CVn systems will be detectable to large distances, where reddening may be important, and in these cases, restricting the set of sources to those which are in a reasonable range of fluxes.  For the faintest AM~CVn sources with periods less than half an hour, the X-ray luminosities are typically about $10^{31}$ erg\,s$^{-1}$ \citep[e.g.][]{2004MNRAS.349..181N,2004ApJ...614..358S, 2005A&A...440..675R, 2006A&A...457..623R, 2016MNRAS.460.3660Z}, meaning that eROSITA should detect them to distances of about 3~kpc.  Combining with radio surveys to remove background Active Galactic Nuclei (AGN), and optical surveys to remove foreground stars and CVs should then yield a much more manageable list of candidates for high time resolution optical follow-up (which usually has limited fields of view) than without the X-ray data and  potentially add more LISA verification sources.

X-rays are also likely to provide the best EM distance estimators for many of these UCXBs (see Fig.\,\ref{fig:LMXB-UCXB-link})  complementing LISA data. Some are located in globular clusters, where the cluster distance can be used. None of the Galactic field UCXBs is bright enough for Gaia in the optical, and most are also too faint for radio parallaxes with current facilities.  Thermonuclear bursts with radius expansion can be used to estimate the Eddington luminosities \citep{2003A&A...399..663K}, and these can then be used in conjunction with the GW estimates of the masses to establish self-consistent properties for the sources.  For the persistent sources that burst in an appropriate manner, these data are already in hand, but obtaining such data for transients would require instruments with large collecting area and small deadtime \citep[e.g.\,NICER, STROBE-X, eXTP][]{2016SPIE.9905E..1HG, 2019arXiv190303035R, 2019SCPMA..6229502Z}.\footnote{most imaging X-ray telescopes will not be able to make such measurements because of pile-up or deadtime issues.}  The other approach that can be used to estimate distances is that of dust-scattering halos \citep{1973A&A....25..445T}, something that requires good angular resolution, good collecting area, and the ability to observe bright sources; while Chandra has done some work in this area, Athena should be able to help dramatically \citep{2019ApJ...874..155C}.

Some UCXBs may contain BHs as well. The first strong globular cluster BH candidate \citep{2007Natur.445..183M} in NGC~4472 is an ultracompact system \citep{2008ApJ...683L.139Z}, probably with an orbital period near 5 minutes, and 47~Tuc~X-9 \citep{2015MNRAS.453.3918M,2017MNRAS.467.2199B} is also a strong candidate ultracompact BH X-ray binary.  At the shortest orbital periods, BH+WD binaries should be detectable by LISA to distances of a few megaparsecs.  For these cases, imaging X-ray data would be needed, along with follow-up optical spectroscopy to look for [O III] nebulae as well as hydrogen emission similar to that in the globular cluster RZ~2109 in NGC~4472 \citep{2008ApJ...683L.139Z,2014ApJ...785..147S,2019MNRAS.489.4783D}.  In the Milky Way, UCXBs with BH accretors at relatively long orbital periods could be quite faint X-ray sources in quiescence (being considerably fainter than accreting NSs at the same mass transfer rate due to advection dominated accretion, \citealt{1994ApJ...428L..13N}). They could also exhibit only rare outbursts, meaning that sensitive X-ray observations would be needed to detect their counterparts.  Furthermore, these objects may be preferentially in globular clusters, meaning that excellent angular resolution, from Chandra or a Chandra successor mission like Lynx or AXIS would be needed to find their counterparts.  If some new BH UCXBs are discovered with X-ray outbursts, they may become bright enough to make BH spin estimates using reflection and/or disc continuum modelling.

For most of the topics related to accretion, there is a need for developing better spectral models that treat unusual abundances.  Development of reflection models that include both the reflection from the surface of the WD, and discs made from hydrogen-poor material is thus vital. It has already been found that details of how atomic physics is incorporated into the disc models can affect inferred spins and abundances \citep{2016MNRAS.462..751G,2018ApJ...855....3T}.

X-ray observations are also vital for understanding the detached systems with NS members.  In most of these systems, the older NS will have experienced significant spin-up, and will be a millisecond pulsar.  Pulsar beam opening angles are larger at high energy than at radio wavelengths (a phenomenon exhibited by objects like Geminga \citep{1992Natur.357..222H} and many of the Fermi-discovered pulsars), meaning that some fraction of these objects will be radio-quiet pulsars \citep{2015ApJ...802...78M}.   X-ray observations will then provide the most comprehensive means for estimating the spins of these systems and determining what fraction of them have become recycled.  Furthermore, in combination with radio searches for pulsations, having a gravitationally-selected sample will allow a clean determination of the ratio of pulsars with radio and X-ray emission, allowing an important constraint for developing a full picture of the pulsar beam geometry.  In an ideal case we may identify an object with thermal cap emission from the NS, such that pulse-profile measurement and modelling could be done to estimate its radius.  If this comes in conjunction with sufficiently good LISA GW measurements to provide an independent, precise, estimate of the NS's mass, this would give a constraint on the equation of state for NSs, even from a single object \citep{2019AIPC.2127b0008W}.

\paragraph{Radio observations}

\phantom{text}

The synergies from joint, multi-messenger observations of radio pulsar binaries entering the LISA band are very promising and will provide significantly more information than observations in the EM or GW bands alone (see Fig.\,\ref{fig:LMXB-UCXB-link}). Such benefits include better measuring the orbital inclination angle \citep{2012A&A...544A.153S} and sky position \citep{2013A&A...553A..82S} and potentially even constraining the NS mass-radius relation to within $\sim 0.2$\% \citep{2020MNRAS.493.5408T}. Additionally, radio astrometry can give parallax distances out to a few kpc already, and with ngVLA \citep{2018ASPC..517....3M}, that should increase dramatically. Most LISA sources would be nearby enough that with ngVLA, one would be able to get 10\% or better geometric parallaxes over about 3/4 of the sky. 

Binary NSs in NS+NS and NS+WD systems enter the LISA band at a GW frequencey of order $1\;{\rm mHz}$ (depending on their distance), corresponding to orbital periods of about $30\;{\rm min}$.
Doppler smearing of radio pulsations from pulsars in such tight binary orbits \citep{2013MNRAS.431..292E} could cause a selection bias against detection of e.g., rapidly spinning millisecond radio pulsars in many previous and present day acceleration searches (at least for dispersion measures, $< 100\;{\rm cm}^{-3}\,{\rm pc}$). However, using neural networks, \citet{2020arXiv201004151P} develop accurate modelling of the observed binary pulsar population and argue for a $\sim 50$--$80$\% chance of detecting at least one of these systems with $P_{\rm orb}\le 15\;{\rm min}$ using data from surveys with the Arecibo radio telescope, and
$\sim 80$-$95$\% using optimal integration times of $\sim 50\;{\rm s}$ in the next several years. The chances of a radio detection of a binary pulsar in the LISA GW band is expected to be significantly enhanced by the completion of the Square-Kilometre-Array \citep{2015aska.confE..40K}. 

It has been argued \citep{2020arXiv201004151P} that unequal mass NS+WD systems are easier to detect compared to the usually near-equal mass NS+NS systems. It should be kept in mind that RLO from these (often bloated) WD companions begins when $P_{\rm orb}$ has decreased to $25-15\;{\rm min}$, depending on their temperature \citep{2018PhRvL.121m1105T} (see also Figs.~\ref{fig:UCXB_strain+data} and \ref{fig:UCXB-AMCVn-detailed}). This will exclude radio detection of such pulsar binaries once accreted plasma enters the NS magnetosphere.

\paragraph{Particle observations }

\phantom{text}

For high-frequency GW detections, there are prospects for detection of neutrino's or cosmic rays, from jets produced in mergers or from supernovae \citep{2016PhRvD..9312010A}. For LISA, the prospects are not so clear, even though associations of LISA GW sources with AGN jets and tidal disruption events could be possible.

\subsubsection{Synergies with other GW  detectors}

\noindent \coord{Lijing Shao; Paul Groot} \\
\noindent \contr{Ilya Mandel (1.5.2.1), Alberto Sesana, Emanuele Berti, Lijing Shao (1.5.2.4), Davide Gerosa (1.5.2.1), Pau Amaro Seoane, Paul Groot, Thomas Tauris (1.5.2.2), Valeriya Korol (1.5.2.3)}

\paragraph{High-frequency GW merger precursors seen by LISA \label{paragr:highFreqGW}}

\phantom{text}

LISA has a unique capability of covering the full frequency spectrum for stellar-mass binary BHs and NSs when combined with the higher-frequency ground-based GW  detectors 
advanced LIGO \citep{AdvLIGO} and Virgo \citep{AdvVirgo}, and their third generation successors such  as the proposed Einstein Telescope \citep{2010CQGra..27h4007P} and Cosmic Explorer \citep{2017CQGra..34d4001A}.

Some individual sources can be tracked on human timescales from the LISA band to the $\gtrsim$ few Hz ground-based detector sensitive frequency band \citep{2016PhRvL.116w1102S}.  The GW  driven merger timescale for a circular binary with components of equal mass $m$ from a starting frequency $f$ is~\citep{1964PhRv..136.1224P}
\begin{equation}
\tau_\mathrm{GW} \simeq 5 \left( \frac{f}{0.01\ \mathrm{Hz}} \right)^{-8/3} \left( \frac{m}{68\ M_\odot} \right)^{-8/3}\ \mathrm{yr}.
\end{equation}
Thus, a signal like GW190521 \citep{2020PhRvL.125j1102A} could be tracked from 10 mHz to merger across the the full range of frequencies with the combination of LISA and ground-based detectors. Rate estimates have been presented by \citet[][]{2015Sci...349.1522S,2016MNRAS.462.2177K,2016PhRvL.116w1102S,2017JPhCS.840a2018S,2019PhRvD..99j3004G,2019MNRAS.488L..94M}, with predictions ranging from  0 to roughly a dozen detections during the LISA mission.

The high mass and correspondingly rapid orbital evolution of IMBHs with masses in the $100$--$1000\ M_\odot$ range makes them particularly appealing targets for tracking across the LISA and ground-based detector frequency bands.  IMBHs with these masses are a challenge for EM observations: their dynamical signature is relatively insignificant, while their X-ray emission can be confused with that of super-Eddington accretors \citep[e.g.,][]{MillerColbert:2004,2011NewAR..55..166F, 2019arXiv191109678G}.  On the other hand, IMBH mergers have been proposed in the context of both isolated binary evolution of very massive stars \citep{Belczynski:2014VMS} and globular cluster dynamics \citep{Amaro:2006imbh}.  The latter can also be responsible for intermediate-mass ratio inspirals of stellar-mass compact objects into IMBHs \citep{Mandel:2008,2016ApJ...832..192H}.  Meanwhile, hierarchical mergers of few-hundred $M_\odot$ seeds have been proposed as seeds of today's massive BHs \citep[][see Chapter~\ref{MBH_chapter}]{2003ApJ...582..559V}. Joint observations with LISA and third-generation ground-based detectors \citep{2011PhRvD..83d4036S,2011GReGr..43..485G} would provide the perfect tools for studying these elusive IMBHs.   

Observations of the same individual source across a broad range of frequencies can improve the accuracy of source parameter measurement. Some parameters are likely to be best measured at low frequencies.  For example, sky localisation accuracy depends on timing precision \citep{Fairhurst:2009,2010PhRvD..81h2001W,Grover:2013}.  The sky localization accuracy can be estimated as the timing accuracy divided by the light travel time across the detector baseline \citep{Mandel:2017}, which is $\sim$ astronomical unit (AU) for LISA, yielding a relative position error of:
\begin{equation}
\sigma_\theta \sim 0.025 \;\left(\frac{0.01\ \mathrm{Hz}}{f}\right) \,\left(\frac{8}{\rho}\right),
\end{equation}
where $\rho$ is the detection signal-to-noise ratio. For heavier sources that can evolve faster than the LISA observing duration, the LISA frequency bandwidth $f_\mathrm{bandwidth}$ should be used in place of $f$.  The angular resolution scales inversely with baseline.  Therefore, despite the lower observing frequency (lower bandwidth), for high SNR sources, LISA sky localisation is likely to be superior to the capabilities of ground-based detectors, whose baseline, even in a network, is limited by the size of the Earth (unless the signal is sufficiently long-lived to allow the effective baseline to be extended by the detector motion over the duration of the observation).

On the other hand, some source parameters will be better measured at higher frequencies, allowing ground-based detectors to provide complementary information to LISA observations.   These include measurements of the ringdown of the post-merger BH, which yield the final mass and spin, and the tidal effects for NSs.  

Yet other measurements could benefit from the joint constraints placed by low-frequency and high-frequency observations.  These include measurements of spin magnitudes and spin-orbit misalignment angles, which could carry information about formation scenarios \citep[e.g.,][]{2013PhRvD..87j4028G,Vitale:2015,Stevenson:2017spin,Zevin:2017,Farr:2017,2018PhRvD..98h4036G}.   Spin-orbit and spin-spin coupling enter the waveform at higher post-Newtonian orders in an expansion in the orbital frequency \citep{PoissonWill:1995}, and so may be better measured at higher frequencies by ground-based detectors.  On the other hand, massive binaries like GW190521 may spend a million cycles in the LISA band (only 4 cycles were observed in the LIGO band when this signal was detected in 2019; \citealt{2020PhRvL.125j1102A}).  Further analysis is necessary to explore the quantitative benefits of LISA for parameter estimation of such signals \citep[but see e.g.][]{2016PhRvL.117e1102V,2019MNRAS.488L..94M,2019PhRvD..99f4056M,2019BAAS...51c.109C}

Lower-mass GW  sources such as double NSs \citep{2020MNRAS.492.3061L,2020ApJ...892L...9A} will not be individually trackable on a human timescale from the LISA band to the band of ground-based detectors.  However, they may still benefit from tracking the entire population of sources as the sources evolve from the LISA frequency band to the frequency band of ground-based detectors.  For example, binaries circularise through GW emission (very roughly, the eccentricity scales inversely with the increase in frequency), making eccentricity challenging to observe with ground-based detectors \citep[e.g.,][]{RomeroShaw:2019,2020MNRAS.497.1966L}.  Thus, LISA observations at lower frequencies, where eccentricities are still significant, could help to distinguish compact binary formation scenarios \citep{2016ApJ...830L..18B,2016PhRvD..94f4020N,2017MNRAS.465.4375N}.  

This can be further aided by the detection of a stochastic  background from a superposition of GWs emitted by multiple individually unresolvable binaries (see Section~\ref{subsec:SGWB}) .  For circular binaries, the stochastic background should be a simple power-law in frequency, and any deviations from that could indicate the emergence of new binaries, particularly eccentric binaries, at high frequencies.  Moreover, the combined low-frequency and high-frequency stochastic background observations may make it easier to subtract the astrophysical background and reveal a possible GW  background of cosmological origin  \citep[e.g.][]{Mandic:2012,Lasky:2016,Callister:2016}.

LISA precursors to ground-based detector mergers could also have important repercussions in fundamental physics, allowing stringent tests of the BH no-hair theorems, as well as more stringent bounds on low-Post-Newtonian deviations from GR \citep{2020PhRvD.101j4038T,2016PhRvL.116x1104B,2019PhRvD..99l4043T,2019PhRvD.100f4024G,2020CQGra..37bLT01C,2017PhRvX...7d1025S}.

\paragraph{Dual-line GW sources}

\phantom{text}

A possibility in upcoming GW astronomy will be the potential discovery of a {\em dual-line} Galactic GW source \citep{2018PhRvL.121m1105T}, where ground-based detectors detect the continuous high-frequency GW emission from the rapid spinning (recycled) NS \citep{2019gwa..book.....A} and LISA simultaneously detects the gravitational damping of the system's orbital motion via continuous low-frequency GW emission. Such a system could very well be a UCXB \citep[e.g.][]{2012A&A...537A.104V,2013ApJ...768..184H}.
Combining the expressions for the strain amplitudes of the ground-based and LISA observations ($h_{\rm spin}$ and $h_{\rm orb}$ respectively) yields \citep{2018PhRvL.121m1105T}:
\begin{equation}
    I_{zz}\,\varepsilon = \sqrt{\frac{2}{5}}\,\left(\frac{\sqrt{G}}{2\pi}\right)^{4/3} \left(\frac{f_{\rm orb}^{1/3}}{f_{\rm spin}}\right)^2\,\mathcal{M}^{5/3}\,\left(\frac{h_{\rm spin}}{h_{\rm orb}}\right).
\end{equation}
Once the right-hand-side of this equation is determined observationally, and assuming that the NS mass, $M_{\rm NS}$ can be determined from the chirp mass, $\mathcal{M}$ \citep[see required assumptions on component mass determinations in][]{2018PhRvL.121m1105T}, constraints can be made on the NS moment of inertia, $I_{zz}$, and thus the NS radius \citep{1994ApJ...424..846R}. \citet{2021MNRAS.503.5495S} recently examined the dual-line detectability of tight Galactic binaries, and found that at least two of the known systems (4U 1820-30 and 4U 1728-34) may be visible to both ground-based and space-based instruments simultaneously. Although only measuring the moment of inertia in combination with the ellipticity, $\varepsilon$, it will still help in pinning down the long sought-after equation of state (EOS) of NS matter. The maximum spin rate and $\varepsilon$ for accreting NSs \citep[][]{2019gwa..book.....A} remain to be constrained firmly.

\paragraph{TianQin}

\phantom{text}

TianQin is a space-based GW observatory conceived as an equilateral triangle constellation of three drag-free satellites with frequency sensitivity at $10^{-3} - 10^{-1}\,$Hz \citep{2016CQGra..33c5010L}, between LISA and DECIGO. 
Unlike LISA, TianQin will follow a geocentric orbit with a radius of about $10^5$\,km \citep{2018CQGra..35i5008H,2019IJMPD..2850121Y}. Its constellation plane will be nearly perpendicular to the ecliptic plane and will have a fix orientation pointing toward RX J0806.3+1527 \citep{2005ApJ...627..920S}, a $5\,$min orbital period Galactic binary that is expected to be the strongest GW source among currently known systems \citep{2018MNRAS.480..302K}. Planned for the launch around 2035, TinQin will see the same GW sources as LISA \citep{2020arXiv200810332M}. Consequently, many synergies can be envisioned between the two missions. For instance, TianQin and LISA will simultaneously detect several thousand Galactic WD+WD binaries, which will improve the parameters estimation including the amplitude, inclination and sky localisation for these binaries \citep{2020arXiv200507889H}. 

\paragraph{Mid-band observatories, e.g. DECIGO}

\phantom{text}

After the discovery of GW150914~\citep{2016PhRvL.116f1102A}, it was realized that massive stellar-mass BHs will be detectable in both LISA and LIGO/Virgo bands~\citep{2016PhRvL.116w1102S,2010ApJ...722.1197A}. However, the SNRs are not expected to be large in the mHz band, and because of the use of template bank searching, in order to claim a confident detection, BH+BH signals in LISA require a larger SNR threshold than 15~\citep{2019MNRAS.488L..94M}. Fortunately, these sources will have large SNRs if they are seen in the decihertz band~\citep{2018PTEP.2018g3E01I, 2019arXiv190811375A, 2020MNRAS.496..182L}. DECihertz laser Interferometer Gravitational wave Observatory (DECIGO) is a representative GW detector in the relevant frequency band~\citep{2011PhRvD..83d4011Y, 2011CQGra..28i4011K, 2020arXiv200613545K}. Studies showed that, not only will observations in the decihertz band provide profound insights to astrophysics \citep[see][for a comprehensive discussion]{2019arXiv190811375A}, they will also provide unprecedented playgrounds for fundamental physics (e.g. testing the dipolar radiation~\citep{2016PhRvL.116x1104B, 2020MNRAS.496..182L}. The mid-band observations of decihertz frequency are natural means to bridge the gap between LISA and LIGO/Virgo observatories.

\bigskip
\subsection	{Technical aspects}

\noindent \coord{Irina Dvorkin}\\
\noindent \contr{Emanuele Berti, Sylvain Chaty, Astrid Lamberts, Alberto Sesana, Kinwah Wu (1.6.3), Shenghua Yu, Shane Larson, Irina Dvorkin (1.6.1, 1.6.3, 1.6.5), Pau Amaro Seoane, Giuseppe Lodato, Xian Chen, Valeriya Korol (1.6.2), Silvia Toonen (1.6.5)}

\subsubsection{How to distinguish between different compact binaries? }

One of the outstanding challenges of LISA will be to analyze a datastream that consists of multiple overlapping signals from astrophysical and possibly cosmological sources as well as instrumental noise. Since many LISA sources will remain in band for multiple orbits, from several days or months up to the entire duration of the mission, there will be an overlap between multiple sources in any given data stretch. Data analysis techniques suitable for this unique problem are currently under development, including in the context of the LISA Data Challenge \citep{2010CQGra..27h4009B,2020PhRvD.101l4008C,2020PhRvD.101l3021L}.
A standard procedure that allows us to extract WD+WD signals from a noisy datastream uses waveform templates that span a large parameter space \citep{1996PhRvD..53.6749O}. The most studied case (and the only class detected so far by LIGO-Virgo) is that of isolated compact binaries (see Section~\ref{subsubsec:detached-binaries}), which are characterized by the component masses, the distance to the binary, its position on the sky and the orbital eccentricity, as well as the orbital frequency. Contrary to the case of ground-based interferometers, these binaries are long-lived LISA sources. In other words their orbital evolution timescale due to the emission of GW is very slow compared to the mission duration. This will allow us to collect data from many cycles of each binary, increasing the SNR, but also puts stringent requirements on the accuracy of the template waveform. The waveform of quasi-monochromatic sources, such as WD+WD binaries, is relatively simple and can be quite accurately described by the leading order terms in the orbital dynamics \citep{2020PhRvD.101l3021L}. On the other hand, binaries that evolve in the LISA band (such as BH+BH) require a more detailed computation to higher Post-Newtonian order  \citep{2019PhRvD..99f4056M}. Accurate waveform templates are crucial for measuring the source parameters and distinguishing between various source classes since any error in the predicted phase of the template waveform will accumulate over the many cycles the binary stays in band.

The main parameter that can help to distinguish the different classes of isolated compact binaries is the chirp mass of the system. In order to establish the class of a quasi-monochromatic source one may use the fact that the chirp mass distribution of WD+WD binaries peaks around $\simeq 0.25M_{\odot}$ \citep{2017MNRAS.470.1894K} with the tail up to 1 M$_\odot$, while the chirp mass of NS+NS systems is expected to lie around $\simeq 1.2 M_{\odot}$. 
BH+NS and BH+BH systems will have higher chirp masses. However, high-mass WD+WD systems at the tail of the distribution with chirp masses of up to $\simeq 1.2M_{\odot}$ may be confused for a NS+NS or a NS+WD binary. Similarly, BH+NS binaries may be confused with NS+NS if the NS has an extremely high mass, or the BH has an extremely low mass. Indeed, the discovery by LIGO-Virgo of GW190814, a binary consisting of a $23M_{\odot}$ BH and a $2.6M_{\odot}$ compact object \citep{2020ApJ...896L..44A} is very difficult to interpret: the secondary component is either the lightest BH or the heaviest NS discovered to date. 

Additional clues as to the identity of the source are somewhat model-dependent, although priors from copious ground-based observations will help with NS+NS, NS+BH, BH+BH scale events. Thus, it may be possible to use eccentricity measurements to distinguish between WD+WDs and NS+NSs \citep{2020MNRAS.492.3061L}. Since WD+WDs are expected to have  formed  via  isolated  binary  evolution, their progenitors are expected to have circularised via multiple mass transfer episodes (see Sec.~\ref{paragr:DWD-formation}). This assumption is supported by the lack of observed eccentric galactic WD+WD binaries. On the other hand, NS+NSs in the LISA band could have measurable eccentricities: e.g. in the fiducial model by \citet{2020MNRAS.492.3061L} half of LISA NS+NS sources have eccentricities $e>0.1$. Thus, a detection of an eccentric source with chirp mass of around $\simeq 1.2M_{\odot}$ can be interpreted as a likely NS+NS. Nevertheless,  some rare eccentric WD+WD can be produced in globular clusters of the Milky Way, or via triple interactions \citep[e.g.][]{2018PhRvL.120s1103K}.

The case of interacting binaries (see Sec.\,\ref{subsubsec:interacting-binaries}) is potentially even more complex, since their orbital evolution is influenced not only by gravity, but also mass transfer and magnetic braking, which lead to  qualitatively different waveforms (such as anti-chirping phases) depending on the evolution stage of the binary \citep[e.g.][]{2017ApJ...846...95K, 2018PhRvL.121m1105T}. For example, as discussed in Sec.\,\ref{paragr:UCXBs} and \ref{paragr:AM-CVns}, AM CVns and UCXBs can be detectable by LISA either during their inspiral phase (when the binary components are detached and the orbits shrinks) or during mass transfer (when the orbit expands). The upside of this complexity is that anti-chirping signals are easier to distinguish from isolated compact binaries.

Clearly, an EM counterpart to a GW detection will help to identify the source. Indeed, as discussed in Sec.\,\ref{subsec:synergies}, EM observations can help in distinguishing between interacting and isolated sources, as well as identifying NS+NS or BH+NS binaries. For the technical aspects of EM synergies, see Sec.\,\ref{subsec:synergies}.

\subsubsection{Foreground sources}\label{paragr:foreground} 

\label{subsubsec:GWforeground}

    The Milky Way hosts a large variety of stellar binaries (Section~\ref{subsec:classes-binaries}), numbering in the millions below mHz frequencies (Table~\ref{tab:absolute_n_MWG}). 
    They will appear as nearly monochromatic (constant frequency) sources emitting over the whole duration of the mission (continuous GW sources). Up to tens of thousands - those with frequencies larger than a few mHz and/or located closer than a few kpc -  will be individually resolvable. The rest of Galactic binaries will blend together into the confusion-limited foreground that is expected to affect the LISA data stream at frequencies below 3 mHz \citep[e.g.][]{1997CQGra..14.1439B,2005PhRvD..71l2003E,2010ApJ...717.1006R,2017JPhCS.840a2024C}. 
    The optimal detection, characterization, and subtraction of Galactic binaries from the data stream has been recognized as one of the fundamental tasks for the LISA analysis. Over-fitting the population of Galactic binaries can result in a large contamination fraction in the catalogue of detected sources, while under-fitting it can degrade the analyses of extra-galactic GW sources in the data due to the excess residual.
    
    The waveforms for Galactic binaries are well predicted using only leading order terms for the orbital dynamics of the binary \citep{1963PhRv..131..435P} and can be computed at low computational cost using a fast/slow decomposition of the waveform combined with the instrument response \citep{2007PhRvD..76h3006C}. Nevertheless, their identification in the LISA data will be laborious due to the sheer number of sources expected to be in the measurement band ($\sim 10^4$) and the large number of parameters required to model each source (between 5 and 10, depending on if the source is chirping and if spins are important in the modelling of one or both of the components). In addition, the high density of Galactic binaries in the LISA band (to the extent that sources are overlapping) and the modulation effects caused by LISA’s orbital motion, which spreads a source’s spectral power across multiple frequency bins, makes the true number of signals at a given frequency difficult to identify. Several techniques have been developed to address this challenge.
    
    {\it A hierarchichal/iterative scheme}. The detectable binaries can be identified by using an iterative process that utilizes a median smoothing of the power spectrum to estimate the effective noise level at each iteration, regresses binaries from the data with signal-to-noise ratios above the established threshold as detected sources, repeating the process until the convergence \citep[]{2003PhRvD..67j3001C,2006PhRvD..73l2001T,2012ApJ...758..131N}.
    However, each iteration can leave behind some residual due to ``imperfect subtraction'' that can affect further analysis and bias parameter estimation.  In practice, the stochastic signal is not actually subtracted but rather included in the covariance matrix during the likelihood calculation. 
     
    {\it Global fit}. A number of studies show that a global fit to the resolvable binaries, while simultaneously fitting a model for the residual confusion or instrument noise and using Bayesian model selection to optimize the number of detectable sources can provide an effective solution to the Galactic binaries challenge 
    \citep{2005PhRvD..72d3005C,2005PhRvD..72b2001U}. These global fit methods have been demonstrated on the Galactic binaries using data from the LISA Data Challenges \citep{2020PhRvD.101l3021L}, but work still needs to be done to extend this into a fully-developed analysis pipeline with the full variety of overlapping LISA sources.

It should be noted that there will possibly be SGWBs of unresolved extragalactic sources or of cosmological origin (see Section~\ref{subsec:SGWB}). Such backgrounds will similarly be a broadband confusion signal, and similar considerations to identify and characterize SGWB in LISA data may be needed in order to reveal some of the fainter signals from astrophysical and cosmological sources. Techniques to identify and subtract the SGWB in LISA data are currently being developed by several groups  \citep{2019arXiv190609027K,2019JCAP...11..017C,2020JCAP...07..021P}.

\subsubsection{Tools}
\label{subsubsec:codes}

\paragraph{Modelling isolated binary evolution and populations}

\phantom{text}

The long-term evolution of stars and binaries is typically modelled in either of two methods; by solving the stellar structure equations, i.e. referred to as detailed calculations, or by faster approximate methods typically aimed at the population synthesis approach.
The latter either interpolates in a grid of detailed calculations or uses parametrised stellar evolution tracks which are fitted to detailed calculations. The advantage of this method is the highly boosted  computational speed (the simulation of the evolution of a single binary takes a fraction of a second in stead of hours or days), at the cost of detail; one only has access to those parameters included in the grid of tracks. Due to the speed,  the effect of different assumptions for poorly understood stellar physics (e.g., stellar mass-loss, interaction physics, supernova kick physics) can be tested in a statistical way, which leads to a deeper understanding of the underlying physical processes involved. The population synthesis approach has proven to work  well in retrieving the general characteristics of large binary populations \citep{2014A&A...562A..14T} and has led to many insights in binary evolution.

Population synthesis codes are crucial for LISA science, both in order to make forecasts for source rates, but also to develop data analysis pipelines. Indeed, ongoing work on building fast and reliable waveforms relies on the knowledge of the expected properties of the sources (masses, spins, eccentricities) and the accuracies required to detect them and measure these properties. The codes currently in use by the LISA community are listed in Table\,\ref{table:pop-synth}. 
 Further development of these codes, in particular the inclusion of additional physical processes as well as cross-checks between the codes will help to prepare for LISA observations.

\begin{table}
\caption{Population synthesis codes used by the community at the time of writing. Binary\_c, COSMIC, MOBSE are based on the BSE code \citep{2002MNRAS.329..897H}. \label{tbl:table:pop-synth}}

\begin{tabular}{ |c|c|c| } 
 \hline
 Code name & Reference & Publicly available  \\ 
 \hline
Binary\_c & \citet{2004PhDT........45I,2006AA...460..565I,2009AA...508.1359I}  & No  \\
 BSE & \citet{2002MNRAS.329..897H} & Yes  \\ 
 BPASS & \citet{2018MNRAS.479...75S}  & No \\
 ComBinE & \citet{2018MNRAS.481.1908K} & No \\
 COMPAS & \citet{2017NatCo...814906S} & Yes  \\
  COSMIC & \citet{2020ApJ...898...71B} & Yes  \\ 
 MOBSE & \citet{2018MNRAS.474.2959G} & No  \\ 
 POSYDON & \citet{2022arXiv220205892F} & Yes \\
Scenario Machine & \citet{1996smbs.book.....L,2009ARep...53..915L} & No \\
 SEVN & \citet{2015MNRAS.451.4086S} & Yes  \\ 
SeBa & \citet{1996AA...309..179P,2012AA...546A..70T} & Yes  \\
 
 StarTrack & \citet{2008ApJS..174..223B} & No   \\ 
TRES & \citet{2016ComAC...3....6T} & Yes   \\  
 \hline
\end{tabular}
\label{table:pop-synth}
\end{table}

\paragraph{Modelling binary evolution in dense environments}

\phantom{text}

Stellar-origin LISA GW sources can also be formed in dense stellar environments through dynamical interactions. Hence simulation codes that follow the evolution of dense stellar systems, either by direct integration or using Monte-Carlo techniques, are crucial for their study. Many of these codes follow simultaneously the evolution of single and binary stars within the dense stellar system, using one of the tools described in the previous Section. A list of stellar dynamics codes currently used in the community for the study of the formation of stellar-origin LISA GW sources can be fount in Table~\ref{table:dynamics-codes}.

\begin{table}
\caption{N-body and few-body dynamics codes used by the community at the time of writing. }

\begin{tabular}{ |c|p{4.5cm}|c| } 
 \hline
 Code name & Reference & Publicly available  \\ 
 \hline
NBODY6/NBODY6++GPU/NBODY7 & \citet{2012MNRAS.422..841A, 2012MNRAS.424..545N, 2015MNRAS.450.4070W}  & Yes  \\
phiGRAPE/phiGPU & \citet{2013hpc..conf...52B, 2011hpc..conf....8B} & No \\
HiGPU & \citet{2013JCoPh.236..580C} & No \\
CMC & \citet{2020ApJS..247...48K, 2019PhRvD..99f3003K, 2018PhRvL.120s1103K} & Yes \\
MOCCA & \citet{2013MNRAS.429.1221H, 2013MNRAS.431.2184G} & No\\
clusterBH & \citet{2020MNRAS.492.2936A} & No \\
 \hline
\end{tabular}
\label{table:dynamics-codes}
\end{table}

\paragraph{GW signal tools}

\phantom{text}

In order to calculate the GW signal of Galactic binary systems there are a number of approaches that can be used, ranging from detailed TDI based methods, e.g. via codes from the LISA Data Challenges (LDC, \url{https://lisa-ldc.lal.in2p3.fr}) to (more) analytic methods to calculate the signal and SNR of specific objects  \citep[e.g.][]{2003PhRvD..67j3001C,2019CQGra..36j5011R,2017MNRAS.470.1894K, 2018MNRAS.480..302K,2019PhRvD.100j4055S}. There are also some web-based tools to explore sensitivity of different GW detectors, including LISA and also detectability of sources in LISA, such as \url{https://gwplotter.com} \citep{2015CQGra..32a5014M} and the Gravitational Wave Universe Toolbox (\url{https://www.gw-universe.org}, \citealt{2021arXiv210613662Y})

\subsection	{Scientific objectives} 
\label{subsec:scientificQ}

\subsubsection{Constraining stellar and binary interaction physics}

\noindent \coord{Fritz Roepke, Alina Istrate}\\
\contr{Karel Temmink (1.7.1.1), Mike Lau (1.7.1.1, 1.7.1.7-8), Stephan Rosswog (1.7.1.2, 1.7.1.6), Vasileios Paschalidis (1.7.1.2), Alina Istrate (1.7.1.2), Fritz Roepke (1.7.1.3,1.7.1.6), Kinwah Wu (1.7.1.4-5), St\'ephane Mathis (1.7.1.4), Thomas Tauris (1.7.1.5), Stéphane Blondin (1.7.1.6), Ashley Ruiter(1.7.1.6, 1.7.1.8), Chris Fryer (1.7.1.7), Thierry Foglizzo (1.7.1.7) Manuel Arca Sedda (1.7.1.8), Kyle Kremer (1.7.1.8), Simone Bavera (1.7.1.8), Abbas Askar (1.7.1.8), Silvia Toonen (1.7.1.8), Gijs Nelemans (refs for 1.7.1.8),  Irina Dvorkin (refs for 1.7.1.8),  Valerya Korol (refs for 1.7.1.8), Astrid Lamberts (refs for 1.7.1.8)}\\

Throughout this Section we will highlight science questions related to LISA that can/should be addressed before the launch of the mission with the label \textbf{pre-LISA-launch objective}, while science questions that can only be addressed by using LISA data will be highlighted with the label \textbf{post-LISA-launch objective}.
It should be emphasized, that in most cases LISA detection of individual binaries is limited to GW frequency and a somewhat poor sky location. For a fraction of these thousands of systems, however, high SNR and/or high frequency will enable the measurement of the GW frequency derivative, a good sky localization, and possibly constraining the eccentricity of the binary system. The GW frequency derivative will reveal the chirp mass of the binary and thus the distance. Although these cases will be exceptional systems in the overall global perspective, they will be the key science drivers that deliver deep insight and new breakthroughs in our understanding of binary compact object systems --- often because of the enhanced chances of finding the EM counterpart of these well-localized systems.  

\paragraph{Dynamical stability and efficiency of mass transfer in the formation of LISA sources}

\phantom{text}

The formation of compact binary systems with two compact objects is still relatively poorly constrained. Typically, at least two phases of mass transfer are required to form a general observable stellar LISA system: one for each component star to lose their hydrogen envelope, and additional phases are possible to remove the helium-rich envelope. 
To explain the compactness of the orbit, typically one or more of these mass-transfer phases are considered to proceed in an unstable fashion in order to get the necessary amount of orbital shrinkage i.e. a CE phase \citep[e.g.][]{1972AcA....22...73P,1976IAUS...73...75P,1984ApJ...277..355W}. Hence, it is crucial to understand for which binary configurations mass transfer proceed stably, and for which it will be unstable.

\textbf{Pre-LISA-launch objective:} The precise value of the stability boundary (i.e. a critical mass ratio, $q$ between the two stellar components above which  no stable mass transfer is possible) remains under debate. Theoretical work has shown that mass transfer can proceed  significantly more stable than classical results  have previously implied \citep[e.g.][]{1987ApJ...318..794H,2008MNRAS.387.1416C, 2011ApJ...739L..48W, 2012ApJ...760...90P, 2015MNRAS.449.4415P, 2017MNRAS.471.4256V,2020A&A...642A.174M}. 
Similarly, the mass-retention fraction of the accreting companion remains relatively poorly understood \citep[e.g.][]{1972AcA....22...73P, 1999ApJ...513L..41K, 1999ApJ...522..487H, 2000ApJ...530L..93T, 2002MNRAS.329..897H, 2007ApJ...663.1269N, 2020MNRAS.498.4705V}.
These issues severely affect the predicted formation details of compact binaries, and determine which evolutionary pathways are dominating in the formation rate of LISA sources (Section~\ref{subsec:rates}), and hence would leave characteristic imprints on the numbers and properties (orbital periods, masses) of the resulting LISA population \citep[e.g.][]{2017MNRAS.470.1894K,2019MNRAS.484..698R}. 
For instance, whether or not the first phase of mass transfer in the formation of a WD+WD leads to a shrinkage or widening of the orbit (i.e. unstable or stable mass transfer) determines to which size the secondary star can evolve before filling its Roche lobe, which dictates the mass of its core at RLO, i.e. the mass of the resulting WD, and hence the mass ratio of the WD+WD \citep{2000A&A...360.1011N,2001A&A...365..491N,2006A&A...460..209V,2012AA...546A..70T}. 

\begin{figure*}
    \begin{center}
   \includegraphics[width=0.60\textwidth]{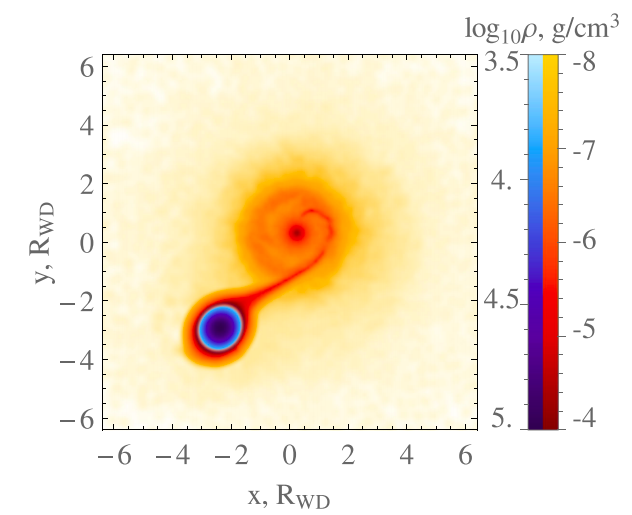}
\caption{Hydrodynamical investigation of dynamical stability in a UCXB with a $0.15\;M_\odot$ WD donor star and a $1.4\;M_\odot$ accreting NS, in an orbit with an initial eccentricity of 0.04. Plotted here is the mass density in the orbital plane after roughly 13 orbits of RLO. The density plot shows eccentric structures in the accretion disc, the complex character of the flow near the circularization radius and a strong density cusp near the NS. The envelope surrounding the binary is sparse but its total mass is significant compared to the disc. Figure from \citet{2017MNRAS.467.3556B}.}
 \label{fig:UCXB_hydro}
  \end{center}
\end{figure*}

Stability of mass transfer depends rather sensitively not only on the intricate details of the structure of the donor star, but also on the transfer and potential loss of mass and angular momentum \citep{1997A&A...327..620S}. 
\textbf{Pre-LISA-launch objective:} Hydrodynamical simulations can help settle the question of dynamical stability (see Fig.~\ref{fig:UCXB_hydro}).
The mass that is transferred to the companion can not always be fully accreted by the companion star: spin-up to critical rotation rates and/or strong optically thick winds or outflows (including jets) can result in significant fractions of transferred material being lost from the accreting star. Since mass that is not accreted leaves the binary system, carrying with it an amount of orbital angular momentum, the efficiency of accretion and the stability of mass transfer are linked. In the case of compact object accretors, it is expected that less conservative mass transfer is typically more stable than mass transfer where all mass is accreted \citep{1997A&A...327..620S}.

\textbf{Post-LISA-launch objective:} Observed samples are required to reverse-engineer the progenitor evolution, and constrain the evolutionary pathways \citep[e.g.][]{2000A&A...360.1011N, 2006A&A...460..209V,2010A&A...520A..86Z,2011MNRAS.411.2277D,2013MNRAS.429L..45P}.
On the EM side, only relatively small samples exist currently, with relatively large biases towards the lower-mass and hotter WDs, since they have longer (observational) lifetimes and are brighter. 
However, LISA will be sensitive to WD+WDs throughout the whole Galaxy and will be able to provide properties of the entire WD+WD population with relatively few selection biases. This will allow for stronger and more meaningful statistical analyses. Additionally, LISA will be able to almost directly measure the Galactic rate and masses of merging WD+WDs, which is another useful tool in constraining progenitor evolution.

The detection of NS+NSs (and potentially NS+BH systems) with LISA allow for a direct view of their formation pathways through their eccentricity, that is induced by the supernova kick associated with the formation of the second NS \citep{2018MNRAS.481.4009V,2020MNRAS.492.3061L}. 
The population of eccentric LISA NS+NS binaries originate from NS+NS binaries born right in or near the sensitivity window of LISA, so that there is little time for GWs to circularise the orbit.  The tight orbit prior to the second SN, leading to NS+NS formation, is characterized by the last phase of mass transfer (see Section~\ref{subsec:1313} and Fig.~\ref{fig:1.9}). This is Case~BB mass RLO initiated by the expansion of the naked helium-star after core-helium depletion. The Case~BB mass transfer episode is believed to be predominantly stable from detailed simulations \citep{2015MNRAS.451.2123T} and in order to match the observed period distribution of Galactic NS+NS systems \citep{2018MNRAS.481.4009V}. However, unstable Case~BB RLO would lead to an additional CE phase that produces NS+NSs with sub-hour periods. Yet, because such NS+NSs that have gone through unstable Case~BB RLO prior to the second SN are formed with higher GW frequencies, they also have a more rapid GW frequency evolution ($f_{\rm GW}/\dot{f}_{\rm GW} \propto f_{\rm GW}^{-8/3}$), which disfavours their detection by LISA \citep{2020MNRAS.492.3061L,2020ApJ...892L...9A}. 
\textbf{Pre-LISA-launch objective:} A deeper understanding of whether or not mass transfer is stable or unstable in Case~BB RLO is needed, and should be investigated further.

\paragraph{Dynamical stability and efficiency of mass transfer in accreting LISA sources}

\phantom{text}

A remarkable property of Roche-lobe filling stars is that, for mass ratios $< 0.8$, their
average density is related to their orbital period \citep[see e.g.][]{2002apa..book.....F}. For mass ratios, $M_{\rm donor}/M_{\rm accretor}<0.8$ the relation is given by:
\begin{equation}
P_{\rm orb}= 10.5\;{\rm hr}\,\left( \frac{\bar{\rho}}{{\rm g\,cm}^{-3}}\right)^{-1/2},
\label{eq:P_rho}
\end{equation}
As an example, Roche-lobe filling stars with orbital frequencies of $0.1\;{\rm mHz}$ ($f_{\rm GW}=0.2\;{\rm mHz}$)  have
$\bar{\rho}\sim 10\;{\rm g\,cm}^{-3}$, while those at 1~Hz possess an average
density of $\bar{\rho}\sim 10^9\;{\rm g\,cm}^{-3}$.
In other words, mass-transferring WDs are located right inside LISA's frequency band. For NS donors, mass transfer only sets in at much smaller separations when the frequencies are already close to the kHz-regime \citep{2011LRR....14....6S}. 

\begin{figure*}
    \begin{center}
   \includegraphics[width=0.60\textwidth]{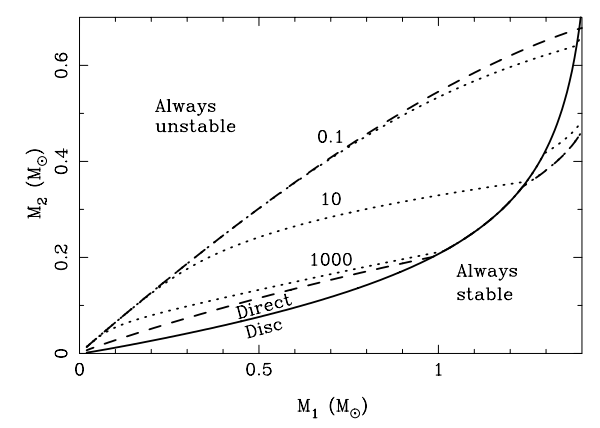}
   \includegraphics[width=0.60\textwidth]{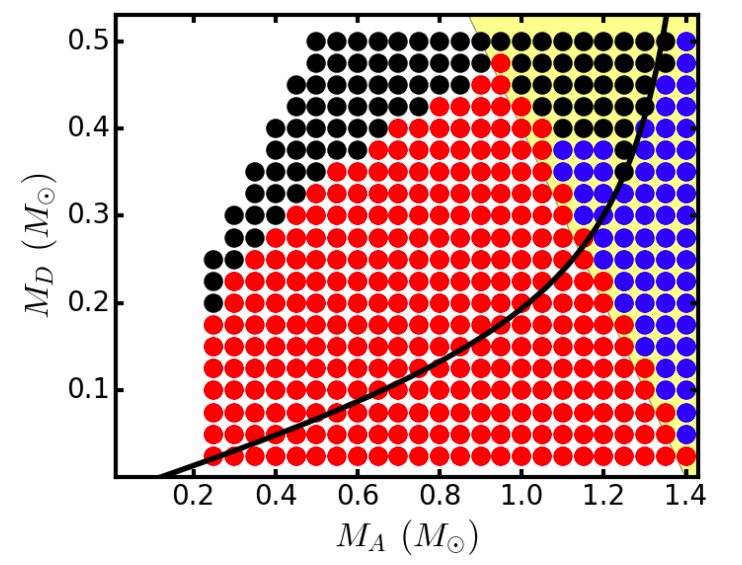}
\caption{Stability regions in the donor mass---accretor mass plane of WD+WDs and UCXBs. Donor star masses are on the vertical axes, accretor star mass on the horizontal axes. {\bf Upper panel}: analytical results of \citet{2004MNRAS.350..113M} where the solid line marks the transition between disc and direct impact accretion, and the other lines show how the strict stability limit of \citet{2001A&A...368..939N} is relaxed when dissipative torques feed angular momentum from the accretor back to the orbit. 
{\bf Bottom panel}: ballistic calculation approach by \citet{2017ApJ...846...95K} using zero-temperature mass--radius relations for the WD donor. Red systems are stable throughout their lifetimes, through stages of both direct-impact and disc accretion; black systems are unstable; and blue systems have an accretor that exceeds the Chandrasekhar limit during their evolution. The solid black line marks again the boundary between disc and direct-impact accretion for initially synchronous and circular binaries. The yellow region indicates systems with a total mass in excess of the Chandrasekhar limit, i.e. potential SN~Ia progenitors. 
It is evident from these figures that UCXBs with low-mass ($\lesssim 0.2\;M_\odot$) He~WD donor stars, and NS accretors, are always dynamically stable.} 
 \label{fig:DWD-stability}
  \end{center}
\end{figure*}

For a LISA stellar source, there are at least two possible scenarios for the subsequent evolution after the onset of mass transfer: a) after an initial brief phase of continued orbital shrinkage after RLO is initiated \citep{2018PhRvL.121m1105T}, mass flows on a much longer timescale from the WD toward the accretor star while the binary separation increases. This process is commonly referred to as a form of stable mass transfer (see Fig.~\ref{fig:DWD-stability}); b) The WD becomes tidally disrupted by the accretor, resulting in the binary merger. This process occurs on a dynamical (orbital) timescale. \textbf{Pre-LISA-launch objective:} Whether the binary undergoes stable  mass transfer vs a merger may have important implications on the type of GW templates that are necessary to detect these binaries with LISA. 

The above discussion makes it clear that for WD+WD, NS+WD, and BH+WD binaries the onset of mass transfer marks a turning point
since the stability of mass transfer decides whether the binary
can survive or will inevitably merge. As mentioned in the previous Sections, its fate depends sensitively on
the internal structure of the mass-donating star, on the mass ratio
and on the involved angular momentum exchange mechanisms,  which here depend primarily on whether mass transfer takes place with or without an accretion disc around the accretor \citep{1982ApJ...254..616R,1984ApJ...284..675H,2004MNRAS.350..113M,2007ApJ...655.1010G,2007ApJ...670.1314M,2009PhRvD..80b4006P,2011ApJ...737...89D,2015ApJ...805L...6S}.
Since fully degenerate WDs possess an inverted mass-radius relation, i.e.\,they grow in size as mass is removed, the mass-donating star will expand
and thereby tend to speed up the mass transfer. However, since the
mass is transferred to the heavier star, the orbit will tend to widen, and therefore stabilize the mass transfer \citep{1997A&A...327..620S,1999A&A...350..928T}.

The way the transferred mass settles onto the accretor star has a
decisive impact on the orbital evolution. If the circularization radius
of the transferred matter is smaller than the radius of the accretor, it will
directly impact onto the stellar surface and spin up the accreting
star -- this scenario is referred to as direct impact accretion. That means that orbital angular momentum is not fed back into the orbit, and therefore the orbital separation shrinks and
mass transfer accelerates. If instead the circularization radius is
larger than the radius of the heavier WD accretor, a disc can form
and --- via its large lever arm --- the disc can feed back angular
momentum into the orbital motion, increase the orbital separation, and thus stabilize the binary system
\citep{1998ApJ...503..344I,2011ApJ...740L..53P,2009PhRvD..80b4006P}.

The majority of studies of mass-transfer stability to date assume that the mass-transferring WD is tidally locked. However, as pointed out by~\cite{WebbinkIben1988}, spinning up a WD while being tidally locked from some initial separation down to the Roche limit is accompanied by tremendous energy release that, depending on the dissipation mechanism, could potentially lift the degeneracy throughout the star. Given that the dissipation mechanisms in WD interiors are not well understood, this makes things even more complex: The tidal interaction between the stars can substantially heat up the
mass-donating WD, change its internal structure, and thereby its response response to mass loss.

For LISA this means that the measured chirp of a mass-transferring binary is not only set by the decay of the orbit due to GW, but also due to mass transfer and tidal interactions, and therefore the chirp mass can not be directly measured as in the case of  detached (chirping) binaries. 
However, for an assumed cold equation-of-state mass-radius relation of the donor star, both the mass of the donor star and the model mass-transfer rate can be derived, as these are fully set by the orbital period \citep[][see also Eq.\,\ref{eq:P_rho}]{1971ApJ...170L..99F, 1971ApJ...168..217V}. 
\textbf{Pre-LISA-launch objective:}  Future work should aim at including finite-temperature effects in the WD EOS consistently, e.g.\,by using detailed stellar structure calculations following the formation and evolution of the system --- from the detachment of the CV/LMXB phase until onset of the AM~CVn/UCXB phase (Fig.~\ref{fig:LMXB-UCXB-link}).

\textbf{Post-LISA-launch objective:}  We do not expect the second derivative ($\ddot{f}_{\rm GW}$) of the GW frequency an AM~CVn system to be measurable with LISA, as the low mass ratios required from mass transfer stability considerations ($M_{\rm donor}\ll M_{\rm accretor}$), imply a low-mass WD donor, whose large radius prevents the AM~CVn system from penetrating into the highest frequency range where the second derivate is large \citep{2004MNRAS.349..181N}.
The combination of LISA with EM surveys, such as Gaia, is particularly promising for AM~CVn sources. If their distance is known, the chirp mass can be constrained, which allows for the orbital chirp to be decoupled into its different components \citep{2018ApJ...854L...1B}.

Apart from determining the orbital evolution, mass transfer in a WD+WD also
leads to an accumulation of helium (possibly also carbon and oxygen) on the surface of the accreting WD. If (or when) nuclear fusion commences in this layer, rapid burning follows that causes a nova outburst or, in case of persistent fusion, X-ray emission may be observed as for supersoft X-ray sources \citep{1997ARA&A..35...69K}. Rapid accretion during the last tens of orbits before a merger and the
interaction with the incoming accretion stream can trigger surface
detonations that cause weak SN~Ia-like transients \citep{2010ApJ...709L..64G}. A (tidal) disruption of a WD by a NS or BH could potentially lead to nuclear-dominated accretion flows \citep{2012MNRAS.419..827M,2013ApJ...763..108F}. 

\textbf{Post-LISA-launch objective:} The probability of detecting a Galactic WD+WD merger during the LISA mission is small (since the Galactic WD+WD merger rate is of order one per century \citep{2001A&A...375..890N}, let alone that this number includes all the WD+WD ``mergers'' giving rise to stable RLO after contact, i.e. the AM~CVn systems). Yet, the case of merging WD+WD binaries deserves a separate mention, because of the exciting possibility of multiband and/or multiwavelength observations: the inspiral phase would be detectable by LISA or a similar mission, and the merger phase by a future observatory such as DECIGO \citep{2017JPhCS.840a2010S}. If the merger gives rise to an optical transient (e.g. a SN of Type~Ia/Iax), these can be observed by EM transients surveys such as ZTF and the Vera Rubin Observatory. For NS+WD, the post-merger phase would be detectable by ground-based high-frequency GW observatories, such as LIGO, Virgo, KAGRA and future facilities like Cosmic Explorer \citep{ 2017CQGra..34d4001A} and the Einstein Telescope \citep{ 2010CQGra..27h4007P}, due to either the eventual collapse of the NS core or post-merger oscillations of the remnant \citep{2017LRR....20....7P}. Therefore, NS+WD systems offer the unique opportunity not only to study the dynamics/stability of mass-transfer, but also the potential to place constraints on the nuclear equation of state. Nevertheless, once again, we emphasize the small probability for a Galactic WD+WD merger event during the lifetime of the LISA mission.

\textbf{Post-LISA-launch objective:} In summary, mass transfer is crucial for determining the final fate of interacting close binaries, but many details and many questions still remain unanswered related to e.g. formation and evolution of AM~CVn and UCXB systems and their ultimate fates, and for related questions such as the progenitor systems of Type~Ia/Iax supernovae. Observations with LISA may therefore bring a major leap forward in our understanding of the physics of this crucial evolutionary phase of close-orbit stellar binaries with compact objects.

\paragraph{Common envelopes \label{paragr:CE}}

\phantom{text}

CE phases are one of the greatest uncertainties in binary stellar evolution theory \citep{2020cee..book.....I} and LISA will provide important measurements.
\textbf{Pre-LISA-launch objective:}  The inspiral of the secondary star into the envelope of the primary lacks obvious symmetries and is therefore not accessible to classical one-dimensional stellar evolution modelling approaches. At least some part of CE interaction takes place on a dynamical timescale. Therefore, parametrized prescriptions for CE evolution are used in stellar evolution modelling. Three-dimensional hydrodynamic simulations have been employed to study the process in more detail \citep{2012ApJ...744...52P,2018MNRAS.477.2349I, 2000ApJ...533..984S, 2012ApJ...746...74R,2015MNRAS.450L..39N,2016MNRAS.460.3992N,2016MNRAS.461..486K,2016ApJ...816L...9O,2016MNRAS.462L.121O,2018MNRAS.480.1898C,2019MNRAS.484..631R,1996ApJ...471..366R,2019MNRAS.486.5809P,2020arXiv200700019K, 2020arXiv200711000S,2020arXiv201106630L}, but they are numerically challenging due to the large dynamical range of spatial and temporal scales of the problem. Furthermore, most often three-dimensional hydrodynamic simulations do not incorporate modelling of physical processes like convection and radiation transport, which are thought to be important especially in the later phases of the inspiral. Complementing the global three-dimensional CE simulations with local, wind-tunnel type, simulations that study the details of the flow around the inspiraling object \cite[e.g.][]{2017ApJ...838...56M, 2020ApJ...897..130D, 2020ApJ...899...77E}, as well as one-dimensional but multi-physics hydrodynamic simulations \cite[e.g.][]{2017MNRAS.470.1788C, 2019ApJ...883L..45F} is a promising avenue to more physically accurate predictions of post-CE binary properties. Lastly, studies of post-CE binaries that are found observationally will provide insights into the CE phase. Valuable constraints on the CE mechanism have come from this method previously \citep{2000A&A...360.1011N,2006A&A...460..209V,2010A&A...520A..86Z,2013A&A...557A..87T}.

Because the actual interaction is short (up to about $10^3$ years), direct observations in optical astronomy are difficult. Some of the fainter optical transients  \citep[luminous red novae;][]{2003ApJ...582L.105S,2011A&A...528A.114T, 2007Natur.447..458K,2020MNRAS.492.3229H,2020A&A...639A.104S} have  been associated with CE events. The two fundamental, and to date unanswered, questions are: i) Which systems manage to eject the CE? ii) What is the final orbital separation of the two stellar cores in this case? \textbf{Post-LISA-launch objective:} The LISA mission is instrumental for clarifying many aspects of the physics of CE phases in two main directions:

\begin{enumerate}
\item \textbf{Direct observations of events related to CE interaction.}\\
The secondary star may be a compact object, but the primary (donor) is typically in a giant phase. Therefore, a generation of sufficiently strong GW signals during inspiral can only be expected if the core of the primary star is also relatively compact, and if the secondary (a compact object) comes close enough during the inspiral.

The prospects to observe GW signals from the inspiral phase, however, do not seem very promising in current models. While based on a parametrized description of CE inspiral, \cite{2020MNRAS.493.4861G} predict about one detection in a few centuries with LISA. 
The full three-dimensional hydrodynamic CE simulations of \cite{2016PhDT........74O} find weaker signals. The rates for the slower self-regulated phase that proceeds on a thermal timescale are more promising, with a rate of $\approx 0.1-100$ events in the Galaxy during the LISA mission duration \citep{2021arXiv210200078R}.

The detectability, however, depends on how close the stellar cores come to one another during the evolution. This is uncertain and depends on the above questions i) and ii). Due to the inspiral of the secondary into the primary star's envelope, orbital energy and angular momentum are transferred to this (now) CE. Some material becomes unbound and is ejected from the system. Simulations, however, show that this process alone is inefficient and other energy sources (such as the ionization of envelope material, \citealt{2015MNRAS.450L..39N, 2019MNRAS.486.5809P,2020arXiv200711000S}) have to be tapped to achieve full envelope removal. The exact parameters allowing for such a successful CE ejection are still unknown, but it is likely that most initial configurations may fail. Such failed events, however, may produce stronger GW signals. Moreover, more exotic cases  in which, for instance, triple systems enter CE evolution \citep{2020MNRAS.498.2957C,2021MNRAS.500.1921G}, may potentially also be sources of detectable GW signals.

\item \textbf{Indirect information from detecting post-CE sources.}\\
Since all stellar mass LISA sources have presumably gone through a CE phase (disregarding here sources produced in dense clusters via dynamical interactions), comparison of LISA populations with model predictions naturally test CE physics. 
Specifically, the occurrence (and detection) rates of these binary populations depend critically on the orbital separation of the stellar cores after the CE phase. In this sense, the LISA mission will statistically sample the outcome of CE events and the results provide valuable information for answering the above question ii). Moreover, if CEs produce binaries with sufficiently short orbital periods, such that they are within the LISA band immediately at envelope ejection, they will act as a source term for the population of LISA binaries. In the absence of this injection of sources, evolution of the orbits due to GWs produce a predictable expectation for the orbital period distribution of binaries, any deviations from that expectation could be ascribed to an injected population of post-CE systems. By simultaneously modelling the LISA noise curve, the GW foreground from unresolved Galactic sources, and the effect of GWs on a binary population, the initial post-CE separations for the shortest period binaries can, in principle, be derived, and thereby deliver important insight into CE physics.
\end{enumerate}

Contributing to these two aspects, the LISA mission will help to constrain physics of the mysterious, and yet crucial, CE phase in binary stellar evolution. Its results will be used to validate hydrodynamic simulations and to develop new efficient prescriptions of CE interactions that are to be used in binary stellar evolution and population synthesis studies.

\paragraph{Tides and angular momentum transport}

\phantom{text}

The orbital evolution of detached binaries is practically determined by the loss of orbital angular momentum and the exchanges of angular momentum between the stars and the orbit, and also in some situations, the mass loss from the system. 
For binaries with only BH or NS components, tidal effects are completely insignificant (except for the few final orbits before a merger event) and the orbital evolution is entirely determined by GWs alone. 
An example of such a system is the radio pulsar binary PSR~B1913+16 \citep[see][]{2016ApJ...829...55W}.  
For binaries with non-degenerate stars, or tight systems with WDs, tides can be important, leading to measurable differences in the LISA signal. 
The orbital dynamics of the wide-orbit systems is relatively simple, as the orbital angular momentum is decoupled from the spin of the two stars. 
The orbital evolution of these binaries is therefore simply driven by the wind-mass loss of the stars and, in principle, GWs, which nevertheless is negligible in such wide systems.
The situation is very different for the systems with a sufficiently short orbital period. 
First of all, the timescale for the loss of orbital angular momentum via GWs \citep{1963PhRv..131..435P} could be comparable to or shorter than the evolution of the stars and the tidal evolution of the system \citep{1992ApJ...400..175B}. 
Secondly, the sizes of the stars are no longer negligible compared with the orbital separation, and torques can be exerted on the stars effectively \citep{1994ApJ...437..742L,1981A&A....99..126H}.  
Hence, the structures and the hydrodynamical properties of the stars would play an important role in determining the orbital dynamics and the orbital evolution of these systems \citep{2011ApJ...740L..54B,2012MNRAS.421..426F,2014ApJ...790..161S}. 
These have, at least, two immediate consequences on the LISA science: 
i) the number density of persistent GW sources associated with these binaries observable within the LISA band, and 
ii) the event rate of burst sources associated with coalescence or merging, resulting from the orbital decay of these binaries, although these bursts are very rare in our Galaxy.
  
The exchange of angular momentum between the binary orbit and the spin of the stars can be facilitated by the viscous torque \citep[][]{1977A&A....57..383Z}, but in the compact star binaries this coupling is less efficient than the coupling caused by a stellar bulge cause by the tidal deformation of the star \citep{1992ApJ...400..175B,2013MNRAS.428..518D,1992ApJ...398..234K}.  
The degree of tidal deformation of a star depends on its internal structure and dynamics \citep[e.g.][]{2008PhRvD..77b1502F,2014ARA&A..52..171O,2019EAS....82....5M}. 
Thus, even in the absolute absence of viscosity, WDs and NSs would respond differently to tidal deformations \cite[cf. the studies of][]{2019PhRvD.100f3001V,1995ApJ...443..705L,2013MNRAS.428..518D,1992ApJ...400..175B,2013MNRAS.428..518D,1992ApJ...398..234K}. 
The presence of the close companion will trigger a large-scale flow induced by the hydrostatic adjustment of the studied primary to the tidal perturbation, the equilibrium tide \citep[e.g.][]{1966AnAp...29..313Z,2012A&A...544A.132R,2013MNRAS.429..613O}, and a broad diversity of tidal waves (i.e.\,gravity waves, inertial waves, gravito-inertial waves), and the dynamical tide \citep[e.g.][]{2017PhRvD..96h3005X,2020MNRAS.496.5482Y}. 
Their dissipation and the quadrupolar moment they induce modify the inspiral and cause changes in orbital frequency and phase shifts \citep[e.g.][]{1992ApJ...400..175B,2020arXiv200908300W,2020MNRAS.491.3000M}.     
Thus, WD binaries and NS binaries will behave and evolve differently, which will manifest in the LISA GW background. 
Their orbital evolution will also imprint signatures in the GWs that they emit \citep{2019PhRvD.100f3001V,2020MNRAS.491.3000M,2013MNRAS.428..518D,2014MNRAS.443.1057D,2014ApJ...791...76S}. 
The situation can be more complicated  if the compact stars have a large magnetic moment, which is not uncommon among magnetic WDs \citep{2015SSRv..191..111F}. The magnetic moments of NSs may not be as large as those of WDs since NSs are more compact. 
Direct magnetic interactions between two magnetised components should be taken into account and may compete with tidal interactions in LISA sources. 

\textbf{Post-LISA-launch objective:} Our understanding of the interplay between tidal interaction, feedback magnetic-field amplification and orbital angular momentum extraction by GWs is currently very primitive. 
The observations of GWs from such systems will surely expand our knowledge on this subject substantially \citep[e.g.][]{1990MNRAS.244..731K,2012ApJ...755...80P,2002MNRAS.331..221W,2018ApJ...868...19W}.

\paragraph{Irradiation of companion star }

\phantom{text}

Feedback irradiation effects on companion stars caused by the intense X-ray flux emitted from accreting compact objects may influence the evolution 
 of the orbits of binary stars \citep{1991Natur.350..136P,2014ApJ...786L...7B}. 
In detached systems, energetic millisecond pulsars  \citep[MSPs, recycled to high spin frequencies from a previous recycling phase;][]{1991PhR...203....1B} may irradiate their companion star with a pulsar wind of relativistic particles and hard photons \citep{1992Natur.356..320T}.
Observations have revealed a growing number of such MSPs with a non- or semi-degenerate companion star which is being ablated by the pulsar wind, the so-called black widows and redbacks \citep{2013IAUS..291..127R}. This is evidenced by the radio signal from the pulsar being eclipsed for some fraction of the orbit \citep{1988Natur.333..237F}.
Tidal dissipation of energy in the donor star envelope \citep{1994ApJ...436..312A} may cause the companion star to be thermally bloated and thereby evaporate more easily.
For LISA binaries, such mass loss via ablation/evaporation will modify their orbital evolution \citep[e.g.][]{2013ApJ...775...27C,2018ApJ...864...30H}, which is otherwise dictated by GWs and tides. For this reason, we may gain new insight on irradiation efficiency from LISA detections of such systems and precise measurements of their orbital frequency. \textbf{Pre-LISA-launch objective:} The impact and the modelling of this effect, often leading to cyclic accretion, is still unclear needs to be improved before LISA flies.

\textbf{Post-LISA-launch objective:} For accreting LISA sources, the irradiation will lead to disturbance of the thermal equilibrium of the companion star \citep{2004A&A...423..281B} and, in the extreme situation, geometrical deformation \citep{2002MNRAS.337..431P}, thereby affecting its mass-transfer rate and thus the orbital evolution of the binary. 
Such an effect may indeed be measured by LISA via its impact on the orbital frequency derivative, and thus the chirp mass of the system. 
Hence, detection of a number of mass-transferring UCXBs and AM~CVn systems by LISA could provide us with unique ways of probing the physics governing close compact object binaries \citep{2016ApJ...830..153J,2017ApJ...846...95K}.

\paragraph{Type~Ia supernovae and other transients }\label{paragr:SNeIa}

\phantom{text}

Stellar interactions in binary systems containing at least one WD are thought to trigger Type~Ia supernovae  (SN~Ia)
and likely a variety of other transients \citep[see e.g.][for a review]{2012NewAR..56..122W}. SN~Ia were of paramount importance for the discovery of the accelerated expansion of the Universe 
and they significantly contribute to cosmic nucleosynthesis, but the lack of a clear observational connection between a progenitor
system and the observable phenomenon has made their understanding difficult. Without proper initial conditions their
modelling remains uncertain.

The properties of the ensuing explosion  are determined by the pre-explosion state of the WD, but is it triggered
when approaching the Chandrasekhar-mass limit, or well before? The occurrence rate, the delay time between binary 
formation and SN explosion, and the ignition process are all determined by the nature of the progenitor system, 
and they have a strong impact on the contribution of thermonuclear SNe to galactic chemical evolution. A traditional 
broad classification is to distinguish between single-degenerate systems, where the companion of the exploding WD is a 
non-degenerate star --- and the double degenerate systems --- where the interaction of two WDs (mergers, or in rare cases
collisions) triggers the SN explosion \cite[see][for a breakdown of binary star progenitor configurations]{2020IAUS..357....1R}. 

None of the progenitors and explosion mechanisms is established beyond doubt. The single-degenerate 
Chandrasekhar-mass model, that served as a reference for a long time, has several shortcomings, but it seems 
to be needed to explain observed abundance trends \citep{2013A&A...559L...5S}. 
However, both population synthesis models and observations indicate that single degenerate explosions fall short of explaining the observed SN~Ia rate.
GW signals were derived from explosion simulations of near-Chandrasekhar mass WDs \citep{2011PhRvL.106t1103F, 2015PhRvD..92l4013S}, but the prospect of measuring individual events is low.

The major competing double-degenerate scenario received increased attention over the past years 
and is of particular interest in the context of LISA. In this scenario, however, the process initiating the 
actual thermonuclear explosion is unclear. For massive WDs, the remnant can reach or exceed the 
Chandrasekhar-mass, but the explosion could also be triggered in the merger process itself while the more 
massive WD is well below the Chandrasekhar mass limit \citep{2010Natur.463...61P, 2012ApJ...747L..10P}. 
Apart from GW-driven (close-to-circular) mergers, {\em collisions} can also (likely to a much smaller extent) contribute to the SN~Ia rate \citep{2009ApJ...705L.128R, 2009MNRAS.399L.156R}. They may occur in locations with large
stellar number densities such as globular cluster cores or galactic centres, but they
are generally thought to occur too infrequently to explain the bulk of SN~Ia \citep[][but see \citealt{2013ApJ...778L..37K} for more optimistic claims]{2018A&A...610A..22T}.
Such collisions, however, have the advantage of a very robust and physically understood explosion
mechanism: WDs of the most common type ($\sim 0.6\;M_\odot$),
that collide with velocities given by their mutual gravitational attraction, cause
strong shocks in the collision and nuclear burning occurs in the right density regime,
so that the resulting explosions appear as rather common Type~Ia SNe
\citep{2009ApJ...705L.128R}. Before the final collision causes a thermonuclear explosion,
the two WDs may undergo several close encounters causing a sequence of
GW bursts in the LISA band of increasing amplitude.

Further clarification of the double-degenerate progenitor channel of Type~Ia SNe 
requires the determination of the exact demographics of WD merger events --- what is their
occurrence frequency for different WD masses? It is further crucial to understand whether
the explosion is triggered during the merger itself or, maybe, already during the inspiral 
phase when mass transfer between both WDs sets in. Even if only small amounts of
mass are exchanged, the re-distribution of angular momentum can have a substantial
impact on the orbital dynamics and therefore on the GW signal \citep{2011ApJ...737...89D}.   
\textbf{Post-LISA-launch objective:}  The LISA mission has great potential to contribute here and to provide important clues
to the mechanism of Type~Ia SN explosions. \citet{2010ApJ...717.1006R} found that on the order of $\sim 500$ WD+WD pairs --- whose total mass exceeds the Chandrasekhar mass limit and will 
merge within a Hubble time --- could be resolvable by LISA in our own Galaxy. 
While most likely no such systems will merge and give rise to a SN~Ia during LISA's operation, 
much can be learned about SN~Ia (and more generally transient) demographics from 
detecting these plausible progenitor systems.

\paragraph{Core-collapse and supernova kicks }\label{paragr:ccSNe}  

\phantom{text}

Observations of compact objects, from pulsar proper motions \citep{2005MNRAS.360..974H} to compact binary properties \citep{2005MNRAS.363L..71D, 2017NewAR..78....1M}, argue that many NSs and some BHs receive natal kicks during the collapse and explosion of the massive star that forms them. 
Asymmetries in the explosion mechanism, manifested either through asymmetries in the mass ejecta \citep{2013A&A...552A.126W} or neutrino emission, have been studied as a source of these kicks. The different mechanisms \citep{2001ApJ...549.1111L} produce different predictions for the distribution of their magnitude \citep{2006A&A...457..963S}, their orientation with respect to the orbital angular momentum \citep{1961BAN....15..265B}, to the stellar spin \citep{2006ApJ...639.1007W, 2012MNRAS.423.2736N, 2013A&A...552A.126W}, and to the distribution of heavy elements \citep{2013A&A...552A.126W,2014Natur.506..339G}.
The asymmetries produced by strongly magnetized explosions are generally aligned with the angular momentum in the collapsing star \citep{2008ApJ...672..465S, 2020MNRAS.492.4613O, 2020ApJ...896..102K} and these mechanisms will produce kicks with directions aligned with the rotation axis, which typically is also aligned with the orbital angular momentum axis. Mechanisms produced by the large-scale convective eddies in the neutrino driven mechanism can produce kicks that are distributed more isotropically \citep{2013A&A...552A.126W, 2019MNRAS.484.3307M}. 

Different kick mechanisms also predict different kick magnitudes as a function of the compact remnant mass \citep[][and references therein]{2017ApJ...846..170T,2020arXiv200608360M}. These kick distributions, in turn, predict different properties in compact object binaries \citep{2003MNRAS.342.1169V, 2020MNRAS.492.3061L}. The NS kick properties can thus affect the number of NS+NS, BH+NS and BH+BH binaries detectable by LISA and LIGO/Virgo \citep{2003MNRAS.342.1169V,2016ApJ...819..108B,2018MNRAS.481.4009V,2018MNRAS.481.1908K,2020ApJ...891..141G,2020MNRAS.492.3061L}, as well as EMRIs \citep{2019MNRAS.485.2125B}.

For example, it has been demonstrated \citep{2013ApJ...778L..23T,2015MNRAS.451.2123T} that ultra-stripped SNe are at work in close-orbit NS+NS and BH+NS systems that LISA and LIGO will eventually detect. The reason being that extreme stripping of the companion star by the accreting NS or BH during the last mass-transfer stage (Case~BB RLO), produces an almost naked metal core prior to the second SN. This has an important effect on the magnitude of the kick added onto the newborn (second) NS, which affects the survival probabilities. It was argued qualitatively and quantitatively \citep{2017ApJ...846..170T} that the resulting kicks are often, but not always, small --- depending on the mass of the collapsing metal core and thus on the resulting NS mass --- which enhances the survival probability.

\textbf{Post-LISA-launch objective:} The overall detection rate of Galactic NS+NS systems by LISA is thus directly affected by the magnitude of the kick, since a large kick can disrupt the binary during the SN. A large kick may also produce moderately more eccentric LISA NS+NS sources \citep{2020MNRAS.492.3061L}. The systemic velocity imparted by the two SN kicks displaces a binary from its birth position in the thin Galactic disc. LISA’s ability to localise Galactic NS+NS sources on the sky to within a few degrees \citep{2019MNRAS.483.2615K, 2020MNRAS.492.3061L} may therefore constrain the kick distribution by measuring the Galactic NS+NS scale height.
By increasing the sample of observed compact binaries, LISA can thus be used to constrain the kick mechanism. In turn, this constrains the nature of SN explosions in binary system \cite{2004ApJ...612.1044P}.

 \paragraph{Neutron star equation of state}
  
\phantom{text}

Matter in the interior of a NS is compressed to densities exceeding those in the centre of atomic nuclei, providing a unique possibility to probe the nature of the strong interaction and to determine the NS composition.  Via the EOS, matter properties determine the star's radius for a given mass \citep{2016PhR...621..127L, 2016ARA&A..54..401O}.  Candidate EOSs can be tested by measuring the mass and radius for a NS or via the accurate measurement of a NS with a high mass because each EOS has a corresponding maximum allowed mass.  Thus, finding a NS with a mass above the maximum allowed for an EOS rules out that EOS, and the radio measurement that PSR J0740+6620 has a mass of 2.14 solar masses rules out many EOSs \citep{2020NatAs...4...72C}.  In addition, the Neutron Star Interior Composition Explorer (NICER) has enabled the measurement of the mass and radius for PSR 0030+0451. In particular, M/R is measured to 5\%, but M and R separately are known to $\approx10$\,\% \citep{2019ApJ...887L..21R, 2019ApJ...887L..24M}, and the uncertainties still do not allow for the determination of a unique EOS. STROBE-X and eXTP could do the same work as NICER to even to a larger distance. Masses provided by GW measurements would help dramatically, since  for the pulsars observed using NICER the pulse profile fitting is mostly sensitive to M/R, and having data points with M and M/R measured well is much more valuable than just having M/R.  With GW measurements, the determination of the tidal deformability for merging NSs is another measurable parameter that can constrain the EOS, as has been shown for GW170817 \citep{2018PhRvL.121p1101A}.  Given the small number of constraining measurements to date, it is clear that additional EM measurements of pulsars and GW measurements of merging NSs are both necessary \citep{2020ApJ...893L..21R} to obtain conclusions that will affect our understanding of fundamental physics.

\textbf{Post-LISA-launch objective:} Many binary systems with NSs produce GWs that will be detectable by LISA, leading to NS mass distributions for various binary populations, and some of these populations may have high mass NSs to further constrain the EOS.  While NS+NSs are somewhat rare, binaries with a WD and a NS are expected to be plentiful.  LISA will also detect binaries that are approaching mergers, and predicting mergers will allow for EM observations to be planned at the time of the merger.  UV, optical, and near-IR observations to determine the remnant type and to constrain the mass and velocity of the ejecta will be very powerful for constraining the EOS \citep{2018MNRAS.480.3871C, 2019ApJ...880L..15M}, especially with a facility like STROBE-X.

\paragraph{Disentangling formation environments based on LISA data}
\label{ucb_dis_form}

\phantom{text}

One of the exciting prospects of LISA observations is the possibility to disentangle the formation channels from compact sources in different environments based on their distinctive properties/demographics; most importantly isolated binary evolution in the Galactic disc or dynamical interactions in dense environments (e.g.\,open, globular and nuclear star clusters) or isolated triple evolution (Section\,\ref{subsec:formation-binaries}). 
Whereas the majority of LISA binaries are expected to form in isolation (Section\,\ref{subsec:rates}), several key properties, in particular orbital eccentricity and component masses, can reveal deviating birth environments.
If LISA is able to constrain these source properties from the GW signal for a given resolved system, the formation channel for that particular system may be inferred. Here, we describe briefly ways these properties may differ between different formation channels and describe applications to particular classes of binaries.

In general, BH and WD binaries that form as isolated systems through standard binary evolution processes are expected to be nearly circular by the time they enter the LISA frequency band. This is a consequence of the various dissipative forces expected to operate throughout the binary evolution that circularize the binary orbit, namely CE \citep{2013A&ARv..21...59I,2016A&A...596A..58K,2018MNRAS.480.2011G,2020arXiv200109829V} and tidal interactions \citep{1977A&A....57..383Z,2014LRR....17....3P,2020A&A...636A.104B}.
In contrast, LISA binaries that formed dynamically in dense stellar environments may have relatively high eccentricities. In the dense star clusters, frequent dynamical encounters impart large eccentricities to binaries \citep{2003gmbp.book.....H}, whereas the formation of triple systems with an inner double compact object can reach high eccentricities via von Zeipel-Kozai-Lidov cycles \citep{2016ApJ...816...65A,2018arXiv180506458A,2019MNRAS.483.1233R,2020arXiv200908468M}, which induce a secular variation of the inner binary eccentricity \citep{1910AN....183..345V,1962AJ.....67..591K,1962P&SS....9..719L}.

Dynamically formed binaries are expected to feature several distinct sub-classes of formation channels that may also be distinguished by their eccentricities. In order of increasing characteristic formation frequency, these dynamical sub channels include: binaries dynamically {\em ejected} from their host cluster that merge as isolated binaries ($f_{\rm{GW}}\approx 10^{-5}\;{\rm Hz}$), binaries that merge {\em in cluster} between strong dynamical encounters ($f_{\rm GW}\approx 10^{-3}\;{\rm Hz}$), and finally binaries that merge in cluster through {\em GW capture} during single--single ($f_{\rm GW}\approx 10^{-1}\;{\rm Hz}$) or few-body dynamical encounters encounters ($f_{\rm GW}\approx 1\;{\rm Hz}$) 
\citep{2016ApJ...830L..18B,Banerjee2018,2018PhRvL.120s1103K,2018MNRAS.481.5445S,2018MNRAS.481.4775D, 2018arXiv180506458A,Samsing2019a, 2019PhRvD..99f3003K, 2019ApJ...871...91Z, Banerjee2020, 2020ApJ...894..133A}. 
\textbf{Post-LISA-launch objective:}  Binaries formed through the ejected and in-cluster merger channels are expected to have eccentricities at GW frequencies of $10^{-2}\;{\rm Hz}$ of roughly $10^{-3}$ and $10^{-2}$, respectively, which are expected to be measurable by LISA \citep{2016PhRvD..94f4020N}. Furthermore, the likelihood of an eccentric merger is dependent on the eccentricity and orbital separation of the outer perturber's orbit and the mutual orientation of the outer and inner orbit \citep{2018ApJ...863...68L,2018arXiv180506458A}.  
Since the typical binary architecture can be connected with the cluster structure, in terms of either mass and radius or velocity dispersion, the detection of binaries with given orbital properties can carry insights on the type of cluster that harboured the merger.

In nuclear star clusters and, more in general, galactic nuclei, the formation and evolution of compact binaries can be substantially affected by the presence of an MBH \citep{1995MNRAS.272..605L,2002ApJ...578..775B,2009ApJ...692..917M,2020ApJ...891...47A}. MBHs are not only a common occurrence in nuclear star clusters \citep[e.g.,][]{2007ApJ...655...77G, 2008AJ....135..747G}, but their masses are correlated \citep{2009MNRAS.397.2148G, 2013ApJ...763...76S, 2016IAUS..312..269G}. The binary can develop ZKL oscillations as a result of secular perturbations exerted by the MBH tidal field \citep{2012ApJ...757...27A,2018ApJ...856..140H,2019MNRAS.488...47F,2020ApJ...891...47A}. Up to 40\% of binaries undergoing ZKL oscillations in galactic nuclei transit into the LISA band with an eccentricity $> 0.1$ \citep{2020ApJ...891...47A}. LISA has the potential to measure the eccentricity oscillations driven by an MBH onto a stellar BH+BH binary out to a few~Mpc \citep{2019ApJ...875L..31H} thus offering a unique way to probe the KL mechanism in galactic nuclei and to disentangle this sub-channel of the dynamical formation scenario. \textbf{Post-LISA-launch objective:} Thus, if measurable by LISA, eccentricities (or lack thereof) may serve as a strong fingerprint pointing toward the specific formation channel \citep{2016ApJ...830L..18B,2017MNRAS.465.4375N,2019arXiv190702283R,2019PhRvD..99f3003K}.

In the case of NS binaries, NS natal kicks \citep[e.g.,][]{2005MNRAS.360..974H} may result in high-eccentricities for binaries that form through isolated binary evolution. In this case, eccentricity may no longer be useful for distinguishing between the dynamical and isolated formation channels. However, even in this case, dynamical and disc binary NSs may still have distinguishable eccentricity distributions that can potentially be differentiated with LISA \citep{2020ApJ...892L...9A}.

Although rare, dynamically-formed NS+BH systems represent a class of GW sources that potentially offer the widest range of peculiarities compared to the isolated channel in terms of total mass, primary mass, and high eccentricity at mHz frequencies \citep{2020CmPhy...3...43A}. Isolated NS+BH systems  \citep{2018MNRAS.481.1908K} are mostly characterised by BHs with masses of $6-10\;M_\odot$ at high, Milky Way-like metallicity ($Z=0.0088$), or BH masses of $10-25\;M_\odot$ at low metallicity, ($Z=0.0002$), and nearly zero eccentricity at merger \citep{2018MNRAS.480.2011G}. Dynamical formation of these systems is not generally relevant for isolated LISA sources in the Milky Way, but in globular and nuclear clusters up to 50\% of dynamically formed compact NS+BH feature BH masses $>10\;M_\odot$, and a large probability ($\sim 50\%$) will have an eccentricity $> 0.1$ when transiting into the mHz~frequency band of LISA \citep{2020CmPhy...3...43A}.

In general, GW sources forming through dynamical channels may contain compact objects with masses that differ or that are even not expected to form at all from isolated binary evolution of Galactic disc sources. 
For instance, BHs with masses between $\sim 55\;M_\odot - 120\;M_\odot$ are not expected to form from the evolution of single massive stars due to pair-instability SNe \citep{2007Natur.450..390W,2012ApJ...749...91F,2016Natur.534..512B,2017MNRAS.470.4739S,2019ApJ...887...53F, 2021arXiv210307933W}. This range has been described as the upper-mass-gap for BHs.
It may be possible to form binaries containing BHs in this mass range through dynamical processes in stellar clusters \citep{2020MNRAS.497.1043D}. One channel to form such BHs could be through hierarchical mergers of stellar-mass BHs in nuclear and globular clusters \citep{2002MNRAS.330..232C,2019PhRvD.100d3027R,2020ApJ...894..133A,2020ApJ...891...47A,2020arXiv200609744S,2020arXiv200715022M}. Hierarchical mergers are most likely to happen in the densest stellar clusters with the highest escape velocities, such as nuclear star clusters, as these clusters can retain the binary despite the GW recoil kick from the merger \citep{2018ApJ...856...92F,2019MNRAS.486.5008A,2020MNRAS.tmp.2017F, 2020A&ARv..28....4N,2020ApJ...891...47A}.  In these extreme environments, binaries at formation are tighter, on average, than in normal clusters, and the interactions with flyby stars and the possible long-term effect of a central MBH can boost stellar collisions and BH mergers, thus possibly inducing a significant modification of the BH mass spectrum. Another possibility to form more massive BHs could be through collisional runaway mergers of BH progenitors in dense star clusters \citep{2002ApJ...576..899P,2004Natur.428..724P,2006MNRAS.368..121F,2006MNRAS.368..141F,2015MNRAS.454.3150G,2016MNRAS.459.3432M,2020MNRAS.497.1043D,2020arXiv200610771K}. 

\textbf{Post-LISA-launch objective:} Such runaway mergers and collisions in dense clusters can also lead to the formation of IMBHs in the mass range $10^2-10^4 \; M_\odot$ \citep{2001ApJ...562L..19E,2006ApJ...641..319P,2006ApJ...640L..39G,2007CQGra..24R.113A,2016ApJ...819...70M,2019arXiv190605864A,2020arXiv200604922A,2020MNRAS.tmp.2099H, 2020ApJ...894..133A, 2020arXiv200715022M}, and mergers of IMBH+IMBH with component masses in the range $10^3-10^4\;M_\odot$ can be observed with LISA up to redshift $z\lesssim 3$ \citep{2019arXiv190605864A,2019arXiv190811375A,2020arXiv200713746A,2020NatAs...4..260J}. 
The number and characteristics of BHs in the upper mass-gap could shed a light on the relative contribution of dynamically-formed or isolated sources to the overall population of BH+BH mergers. 

Lastly, the evolution of isolated triples can also lead to a mass distribution that deviates from that of isolated binary evolution. This is mainly due to two effects. Firstly, to form a LISA source, isolated binary evolution relies on one or more mass-transfer phases that reduce the orbital period (Section~\ref{subsec:formation-binaries}) down to the range observable by LISA. Mass transfer can also occur in triples \citep[even a larger fraction of triples experiences RLO;][]{2020A&A...640A..16T} (Sections~\ref{paragr:triples} and \ref{subsubsec:triples_formation}), however orbital shrinkage can also be achieved by the combination of three-body dynamics with dissipative processes. The increased eccentricities during von Zeipel-Kozai-Lidov cycles reduce the GW inspiral time \citep[e.g.][]{2011ApJ...741...82T,2017ApJ...841...77A,2018ApJ...863....7R,2019MNRAS.486.4443F}. Moreover, if a star does not fill its Roche lobe, and does not lose its envelope prematurely, it typically will form a more massive remnant compared to the case of RLO mass stripping. Such triples will on average contain stars that are more massive than those formed through isolated binary evolution \citep{2013MNRAS.430.2262H, 2018A&A...610A..22T}, and will not contain He-core WDs (which have masses $\lesssim 0.45M_{\odot}$ which can only be formed in a Hubble time through mass stripping). Secondly, similar to the evolution in star clusters, sequential mergers in multiples give rise to higher stellar masses \citep{2020ApJ...888L...3S, 2020ApJ...898...99H, 2020arXiv200910082L}. In addition, the effect of a tertiary perturber can induce precession of the spins and lead to spin misalignment \citep{2018MNRAS.480L..58A,2018ApJ...863...68L,2018ApJ...863....7R}. 

\bigskip

\subsubsection{LISA sources as galactic probes} 

\noindent \coord{Valeriya Korol}\\
\noindent \contr{Valeriya Korol, Raffaella Schneider, Luca Graziani, Astrid Lamberts, Samuel Boissier, Martyna Chruslinska, Alberto Sesana, Katie Breivik, Shane Larson, Michela Mapelli}\\

Stellar binaries detectable by LISA bear the imprint of the properties of their native stellar environments (galaxies and stellar clusters) such as the total stellar mass, IMF, star formation history (SFH), age and metallicity ($Z$). These properties can be investigated by combining binary population synthesis (BPS) models (Section~\ref{subsubsec:codes}) with models of galaxy formation and evolution.
Several methods have been developed to achieve this goal. The combination of BPS models with theoretical semi-analytic or observationally inferred cosmic star formation rate densities  provides a fast way of predicting the evolution of the overall birth and merger rates with redshift \citep{2001MNRAS.324..797S, 2011RAA....11..369R, 2011PhRvD..84l4037M,2013ApJ...779...72D,2016Natur.534..512B,2016MNRAS.461.3877D,2016MNRAS.463L..31L,2018MNRAS.473.1186E,2019MNRAS.488.5300C,2019ApJ...881..157B}.
In particular, observation-based approaches allow one to account for the current observational uncertainties on the birth metallicity distribution of stars forming over the cosmic history and evaluate the related uncertainty on the predicted properties of mergers \citep[e.g.][]{2019MNRAS.488.5300C}.
A detailed understanding of the properties of galaxies hosting GW sources can be gained from cosmological simulations, which provide a detailed description of the cosmic star formation in a more accurate context of the galaxy evolution. Galaxy catalogues from the Illustris \citep{2014MNRAS.444.1518V}, GASOLINE \citep{2001PhDT........21S,2004NewA....9..137W}, 
EAGLE \citep{2015MNRAS.446..521S} simulations have been used for predicting NS+NS and BH+BH mergers  \citep[e.g.][]{2017MNRAS.472.2422M,2017MNRAS.464.2831O,2019MNRAS.487.1675A}. Similarly, the Latte simulation of Milky Way-like galaxies of the FIRE hydrodynamical simulation project \citep{2014MNRAS.445..581H,2016ApJ...827L..23W} was adopted to study the properties of Galactic WD+WDs and BH+BHs accessible to LISA \citep{2018MNRAS.480.2704L,2019MNRAS.490.5888L}. Alternative hybrid pipelines  such as \texttt{GAMESH} \citep{2015MNRAS.449.3137G,2017MNRAS.469.1101G,2019Physi...1..412G}, combining a dark matter simulation with semi-analytic star formation, chemical enrichment and numerical radiative transfer, represent an advantageous alternative to study the redshift evolution of compact binaries along the assembly of a Milky Way-like galaxy and in its local volume dwarf satellites \citep{2017MNRAS.471L.105S,2019MNRAS.484.3219M,2020MNRAS.495L..81G}.

{\it Effect of the IMF.} The IMF is one of the key ingredients in the BPS that sets the distribution of initial masses and the relative proportions of stars forming in different mass ranges. Therefore it has a direct impact on the observed merger rates and properties of the LISA sources. Studies often adopt the IMF inferred from the observations of stars in the local Galactic neighborhood \citep[e.g.][]{2001MNRAS.322..231K,2003PASP..115..763C}. However the universality of this assumption is one of the fundamental open questions in astronomy and is still debated \citep[e.g.][]{2010ARA&A..48..339B}. Theoretical studies
show that with the assumption of the Milky Way-like IMF one may underestimate the number of WD and NS progenitors forming at redshifts $\lesssim$ 1, especially at low metallicities \citep{2020A&A...636A..10C}, and,
therefore,  underestimate  the  predicted  number  of  individual  LISA  detections  and  background/foreground  noise. 

\textbf{Post-LISA-launch objective:} LISA's observations of stellar remnants --- invisible to EM observatories --- will offer us an alternative way of probing the IMF. For instance, hundreds of Galactic WD+WDs with measured chirp mass \citep{2019MNRAS.482.3656R} can be used to constrain the low-mass end of the IMF in different Galactic habitats. In addition, numerous LISA detections in the Magellanic Clouds will enable the studies of the IMF with GWs in alternative environments \citep{2020A&A...638A.153K}.

{\it Effect of metallicity.}
The metallicity is another important assumption of the models that affects different types of stellar binaries in different ways \citep{2019MNRAS.482.5012C}. 
The predicted metallicity dependence of the formation efficiency of merging BH/NS binaries is a complex function of numerous poorly constrained phases of binary evolution. 
Specifically, BH+BH mergers resulting from isolated stellar evolution are typically found to form much more efficiently at low metallicity ($\lesssim 0.1-0.3\;$Z$_\odot$) than at solar metallicity \citep{2010ApJ...715L.138B, 2016MNRAS.462.3302E, 2017NatCo...814906S,2017MNRAS.471L.105S, 2018A&A...619A..77K, 2018MNRAS.474.2959G}. The differences in formation efficiency reach up to two orders of magnitude and consequently, the size of the observable BH+BH population is sensitive to the amount of star formation happening at low metallicity
\citep{2013ApJ...779...72D,2017MNRAS.472.2422M, 2019MNRAS.484.3219M, 2019MNRAS.482.5012C, 2019MNRAS.490.3740N,2020MNRAS.495L..81G, 2020ApJ...898..152S,2020arXiv200903911S}.
Furthermore, the most massive BH+BH are expected to form at low metallicity
and their mass distribution could potentially be linked to the metallicity distribution
of their progenitors.
Metallicity dependence of the formation efficiency of NS+NS mergers is typically found to be much weaker than for BH+BH, with the mixed systems falling in between.
For the case of WD+WD, the metallicity mainly changes the total number of WD+WDs by allowing lower masses for a star to reach the WD stage in a Hubble time with decreasing metallicity. This results in a moderate increase (few tens of percents) in the number of resolved LISA sources  \citep{2010A&A...521A..85Y,2020A&A...638A.153K}.

 {\it Effect of star formation histories.} The merger rate of compact stellar binaries across cosmic time is a direct consequence of the SFH \citep{2017ApJ...840...39M,2019MNRAS.487.1675A,2013ApJ...779...72D,2017MNRAS.472.2422M,2018MNRAS.479.4391M,2019ApJ...886L...1V,2019MNRAS.490.3740N,2020ApJ...898..152S}. Together with BPS models, SFH regulates the content of stellar binaries in the LISA band at a given time. Galactic WD+WDs can be used as a tool to study the SFHs of the MW components: due to the different timescales to reach the mHz frequencies, WD+WDs of different core composition dominate different parts of the Galaxy due to their distinct SFHs. Specifically, double He-core WDs with formation times that can exceed 10~Gyr populate the Galactic bulge, thick disc and stellar halo; double C/O-core WDs, typically form on timescales shorter than 2~Gyr and are associated with a much younger populations present in the thin disc; mixed He-C/O-core binaries present an intermediate distribution \citep{2010A&A...521A..85Y,2019MNRAS.490.5888L}. In addition, SFH has significant effects on the LISA detection rates in the Milky Way satellites \citep{2020A&A...638A.153K}.

{\it Structure of the Milky Way with resolved and unresolved sources.} It is expected that the Galactic GW population at mHz frequencies will be largely dominated by WD+WDs and will have two components in the LISA data: population of high-frequency individually resolved binaries and unresolved stochastic foreground from low-frequency binaries (Section~\ref{paragr:foreground}). Both resolved and unresolved WD+WDs encode global properties of Galactic stellar populations, and can thus be used as a tool to study the Milky Way’s stellar content and shape. 

\begin{figure*}
    \begin{center}
  \includegraphics[width=0.60\textwidth]{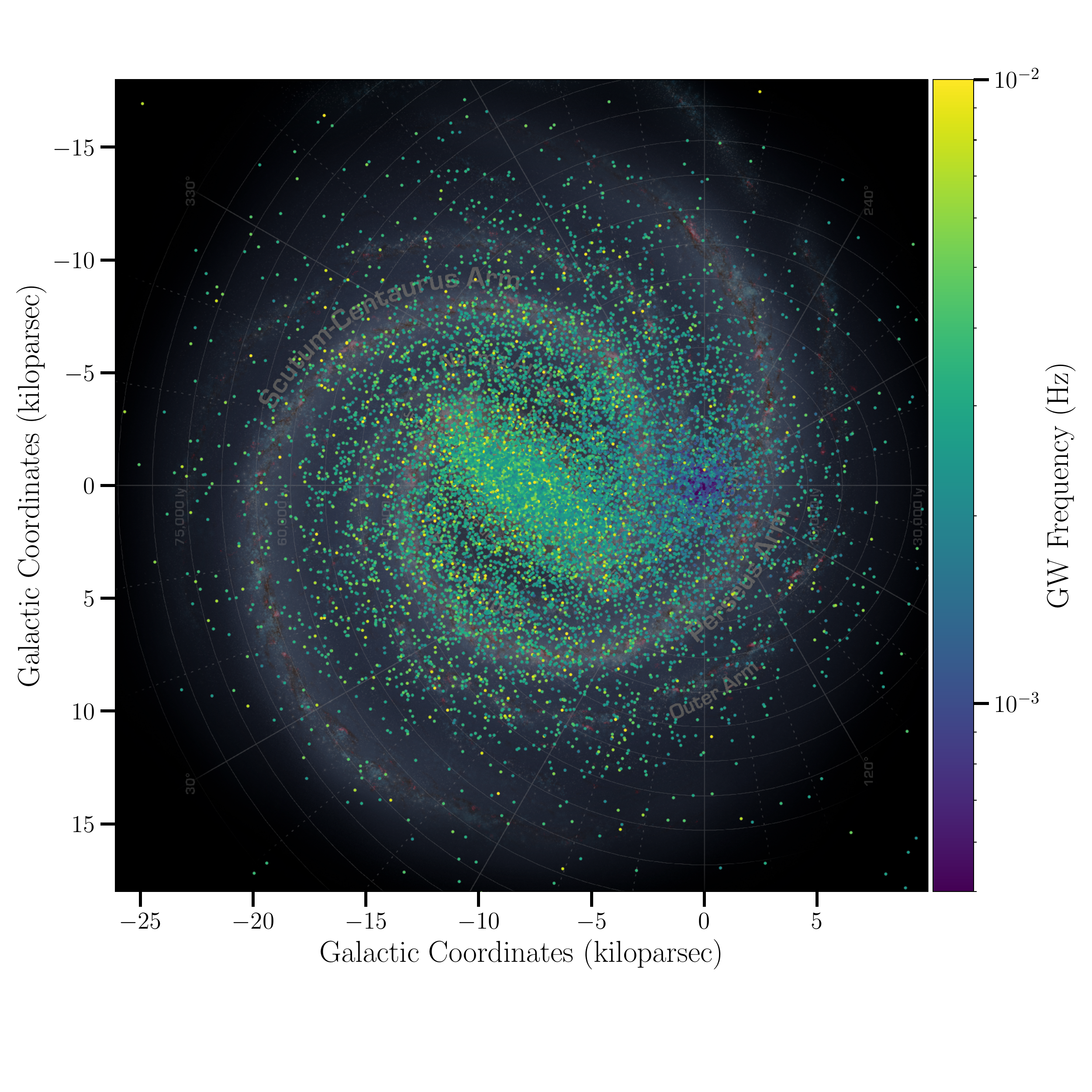}
\caption{Numerous WD+WDs detectable by LISA will enable the mapping of our Galaxy. In the background the artist impression of our current view of the MW. Over-plotted in colour WD+WD with SNR$>7$ from \citet{2020arXiv200311074W}. LISA's position is at $(0,0)$. The selection effect due to GW frequency in visible in colour. Figure credit: Valeriya Korol.}
 \label{fig:MW}
  \end{center}
\end{figure*}

\textbf{Post-LISA-launch objective:} Affected by different selection biases than EM observatories, LISA can probe the entire volume of the Milky Way and therefore will facilitate detailed studies of its the far side (Fig.\,~\ref{fig:MW}). Moreover, unaffected by the dust extinction and stellar crowding, LISA can also probe the inner Galaxy at all latitudes. For several thousands WD+WDs measurements of the sky positions and distances will enable the mapping of the Galaxy. Reconstructed density profiles of WD+WDs will provide unbiased constraints on the scale length parameters of Galactic bulge/bar and disc that are both accurate and precise, with statistical errors of a few \% to 10\% level \citep{2012PhRvD..86l4032A,2019MNRAS.483.5518K,2020arXiv200311074W}. The Galactic stellar halo is also expected to host up to a few thousand WD+WDs, and therefore can potentially be studied with WD+WDs in a similar way \citep{2009ApJ...693..383R,2010A&A...521A..85Y,2019MNRAS.490.5888L}. Furthermore, the LISA sample is found to be sufficient to disentangle between different commonly used disc density profiles, by well covering the disc out to sufficiently large radii. The stellar bar will also clearly appear in the GW map of the bulge. LISA's WD+WDs can accurately characterise the bar's physical parameters: length, axis ratio and orientation angle with respect to the Sun's position \citep{2020arXiv200311074W}. However, because of the low density contrast compared to the background disc, the spiral arms will be elusive to LISA. Finally, building upon the analogy with simple stellar population models used for inferring stellar masses of galaxies based on their total light, the total stellar mass of the Galaxy can be estimated from the number of LISA events. Using a simplified example of Milky Way satellites, \citet{2021MNRAS.502L..55K} showed that based on BPS models of LISA sources satellite masses can be recovered within 1) a factor two if the SFH of the satellite is known and 2) within an order of magnitude even when marginalising over alternative SFHs. When also accounting for the unresolved Galactic foreground, this method could be extended for measuring the total stellar mass of the Milky Way.

\textbf{Post-LISA-launch objective:} The power of constraining the overall properties of the Galactic potential will be significantly enhanced by using LISA detections in combination with EM observations of binaries motions. BPS studies forecast up to 150 detached and interacting WD+WDs detectable through both EM and GW radiation \citep[e.g.][see also Section~\ref{subsec:rates}]{2017MNRAS.470.1894K,2018ApJ...854L...1B}. For these multi-messenger binaries 3D positions provided by LISA can be combined with proper motions --- for example provided by Gaia or Vera Rubin Observatory --- into the rotation curve, which allows the derivation of the stellar masses of the Galactic baryonic components \citep{2019MNRAS.483.5518K}.

The unresolved Galactic foreground will provide complementary constraints on the Galactic structure. For example, the Galactic foreground will show whether the WD+WD population traces the spatial distribution of young, bright stars (and thus do experience significant kicks), or traces a vertically heated spatial distribution associated with Galaxy's oldest stellar populations.
This can be understood from the shape of Galactic power spectral density that depends on the characteristic scale height of the WD+WD population \citep{2006ApJ...645..589B}. 
\textbf{Post-LISA-launch objective:}  In addition, using the spherical harmonic decomposition of the LISA data streams, the structure of the disc population of Galactic WD+WDs can be constrained with an accuracy of 300~pc \citep{2020ApJ...901....4B}. The relative poor resolution compared with the resolved sources is a direct consequence of LISA's poor spatial resolution at low frequencies. Nevertheless, an independent measurement at low frequencies will either help to confirm the structure of the resolved sources or point to frequency-dependent Galactic structure.

%% file: astroWP_02mbh.tex
\section{Massive Black Hole Binaries}\label{MBH_chapter}

\input{mbh_chapter}

%% file: mbh_chapter.tex
\noindent {\bf \textcolor{black}{Section coordinators:} Elisa Bortolas, Pedro R. Capelo, Melanie Habouzit\\
\textcolor{black}{Section reviewers:} Laura Blecha, Massimo Dotti, Zoltan Haiman}

\newpage
\subsection{Introduction}
\label{MBHch_intro}
{\bf \textcolor{black}{Contributors:}
Elisa Bortolas,
Pedro R. Capelo,
Melanie Habouzit\\
}

The observed BH mass spectrum spans ten orders of magnitude, ranging from a few $\rm M_{\odot}$ of stellar-mass BHs to more than $10^{10} \,\rm M_{\odot}$ for the most extreme MBHs. Our knowledge of the mass spectrum has expanded over the past decade. On the low-mass end, the GW facilities Laser Interferometer Gravitational-wave Observatory (LIGO) and Virgo have observed the mergers of low-mass BHs in the range $\sim$6--80~M$_{\odot}$ \citep{2020PhRvL.125j1102A}.
At the high-mass end, we have discovered in the high-redshift Universe extremely bright objects, called quasars, powered by MBHs with masses similar to those of the most massive MBHs 
around us \citep[$M_{\rm BH}\geqslant 10^{8}\, \rm M_{\odot}$ at $z>6$, e.g.][]{2011Natur.474..616M,2018Natur.553..473B,2020ApJ...897L..14Y}. LISA has the ability to detect MBHs of $M_{\rm BH}=10^{3}-10^{7}\, \rm M_{\odot}$ through the last stages of in-spiral and merger up to $z\sim 20$, bridging these extremes of the mass spectrum.

The mass-redshift regime that LISA can probe is key to constraining the origin and growth of MBHs, and is one of LISA's main science goals. 
Considering the current state of observations, theory, and simulations, we still do not know how MBHs form and evolve in the early Universe. We do not know how they assemble with time and become present in almost all the galaxies in the local Universe, from dwarf galaxies \citep[with stellar masses of $\leqslant 10^{9.5}\, \msun$, e.g.][]{2020ApJ...898L..30M,2019arXiv191109678G,2018ApJ...863....1C,2018MNRAS.478.2576M,2015ApJ...809L..14B,2013ApJ...775..116R} to large ellipticals \citep[e.g.][]{1998AJ....115.2285M,2009ApJ...698..198G,2011Natur.480..215M,2013ARA&A..51..511K,2016ASSL..418..263G,2019ApJ...873...85D,2019ApJ...876..155S}. LISA observations will play a key role in addressing these enigmas.
In this Section, we only discuss MBHs, which we define as BHs with $\gtrsim 100\, \rm M_{\odot}$.  
Additionally, we do not make an explicit distinction within that range, i.e. we do not distinguish between intermediate-mass BHs, massive black holes, and supermassive BHs.

\begin{figure}
    \centering
    \includegraphics[width=\textwidth]{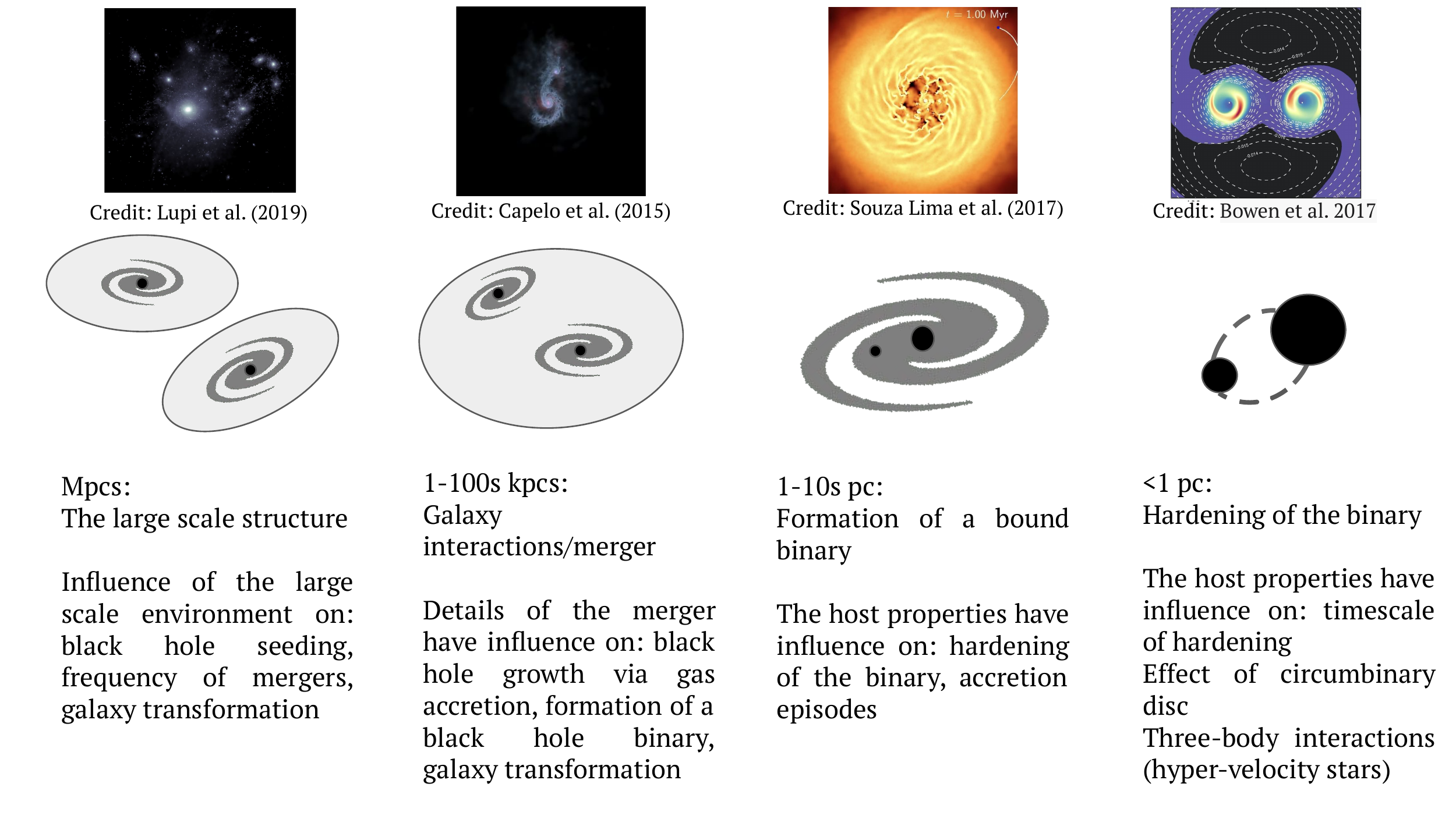}
    \caption{A schematic view of the complex and multi-scale processes affecting the formation of a hard MBH binary system. Figure credit: Silvia Bonoli and Alessandro Lupi.}
    \label{fig:big_picture}
\end{figure}

In the hierarchical paradigm of galaxy formation, we expect central MBHs to coalesce after the merger of their host galaxies. As shown in Fig.~\ref{fig:big_picture}, MBHs will have to cross an impressive range of scales, from when they are hosted in separate galaxies at early times to the end of their dance, when they coalesce with each other \citep{1980Natur.287..307B,2001ApJ...563...34M,2014MNRAS.444.2700D}.
Following a galaxy merger, while MBHs are still separated by kpc to tens of kpc scales, they will start losing orbital energy and angular momentum via gravitational drag from background gas and stars, causing them to sink to the centre of the remnant galaxy (a process referred to as dynamical friction). On $\lesssim$ pc scales, the MBHs will form a gravitationally-bound binary and evolve further via interactions with gas and individual stars (the so-called binary hardening phase). This may include interactions with a circumbinary disc on ${\sim}10^{-3}$ pc scales.
Finally, the MBHs enter the last stage of the dance, i.e. the GW regime ($\leqslant 10^{-5}\,\rm pc$ scale).

To maximize the scientific return of LISA, advances are needed in our theoretical understanding of MBH formation, dynamics, and evolution, a field of research that started in the 1980s \citep{1980Natur.287..307B}. Building powerful tools such as semi-analytical models, N-Body and hydrodynamic simulations is crucial to predict the MBH mergers that LISA will detect as a function of the intrinsic properties that describe both the MBH and galaxy. Currently, the predicted MBH merger rate spans more than one order of magnitude, from a few LISA detections per year to tens. 
Rate predictions depend on computational methods,  
and on the modelling of the relevant physics. In the coming years we will improve on the dynamical range of scales that we can resolve, and address how different mechanisms of MBH formation, galaxy environments, MBH growth models, and MBH dynamics can shape the merger rates of MBHs, thus paving the way for the interpretation of LISA data.

In the near future, several space missions will be launched with the goal of constraining the formation and evolution of MBHs and their environments. These missions will complement LISA in the EM domain, and will provide unprecedented constraints on the entire population of MBHs. The James Webb Space Telescope \citep[JWST;][]{2006SSRv..123..485G} and the Roman-Wide Field Infra-red Survey Telescope \citep{2015arXiv150303757S} will image the first galaxies \citep[e.g.][]{2018ApJS..236...33W},  the cradles of the first MBHs. The assembly of galaxies will also be witnessed by the new thirty-meter telescopes such as E-ELT, TMT, and GMT.
New X-ray facilities such as Athena \citep{2013arXiv1306.2307N}, as well as the LynX \citep{2018arXiv180909642T} and AXIS \citep{2018SPIE10699E..29M} concept missions, will aim at uncovering the population of accreting young MBHs at high redshift ($z>6$). 

With LISA and the aforementioned new instruments working in the EM domain, we will enter the new multimessenger era for MBHs. By
performing synergistic observations that combine low-frequency GW signals with EM signals from the same source, we will uncover previously unavailable information. These combined observations will precede, accompany, or follow, the MBH merger events, helping us to constrain MBH activity, understand their immediate surroundings (e.g. the nature of the accretion disc, jets, and the accreted/ejected material), and its relation with the host galaxy. One challenge of multimessenger observations is the localization of the sources, and the confirmation that they are indeed MBH binaries. Before the launch of LISA, we will have to better understand, among other aspects, how the different potential observational EM signatures of coalescing systems are
originated, and develop new analysis  tools to identify these GW source candidates in large datasets. 

LISA will also constitute a bridge between the two GW frequency regimes that are already being investigated: the highest GW frequencies (LIGO/Virgo/KAGRA), and the lowest GW frequencies which build the GW background (observed by Pulsar Timing Arrays; PTAs). In the coming years, we will have to fully exploit these missions to, e.g. select, monitor, confirm and characterise MBH binaries (MBHBs), but also understand their small-scale to galactic and large-scale environments, and how they fit within the full MBH population. 

Sections~\ref{sec:MassiveBlackHolesAndTheirPathToCoalescence} and~\ref{sec:MBHoriginandgrowthacrosscosmictime} review the theoretical background and highlight pre-launch objectives for the LISA community that can sharpen our preparation for the mission. Section~\ref{sec:StatisticsOnMBHMergers} distills the theoretical picture into LISA's observables and highlights uncertainties. The pre-launch objective is to compare different approaches to obtain realistic predictions that can be used, post-launch, to interpret LISA's data. The pre-LISA theoretical development is of paramount importance because the expectation is that LISA's event properties will be compared to theoretical models through a Bayesian framework in order to perform astrophysical inference \citep{2011PhRvD..83d4036S}. 
Section~\ref{sec:sec_2_4} focuses on EM signatures of MBHs, highlighting both pre-launch  (improve theoretical models, search for EM emission from MBHBs) and post-launch (devise strategies for searches of EM counterparts to MBHBs detected by LISA) objectives. Finally, Section~\ref{sec:sec_2_5} shows how LISA's results can be strengthened by complementary campaigns performed by different instruments and facilites, straddling pre-launch and post-launch objectives dependent on whether missions overlap or not.

\subsection{MBHs and their path to coalescence}
\label{sec:MassiveBlackHolesAndTheirPathToCoalescence}
{\bf \textcolor{black}{Coordinators:}
Matteo Bonetti,
Hugo Pfister\\
\textcolor{black}{Contributors:}
Emanuele Berti,
Tamara Bogdanovic,
Elisa Bortolas,
Pedro R. Capelo,
Monica Colpi,
Pratika Dayal,
Massimo Dotti,
Alessia Franchini,
Davide Gerosa,
Zoltan Haiman,
Peter Johansson,
Fazeel Mahmood Khan,
Giuseppe Lodato,
Lucio Mayer,
David Mota,
Vasileios Paschalidis,
Alberto Sesana,
Nick Stone,
Tomas Tamfal,
Marta Volonteri,
Lorenz Zwick\\
}
\begin{figure}[h]
    \centering
    \hspace*{-1.cm} 
    \includegraphics[width=1.1\textwidth]{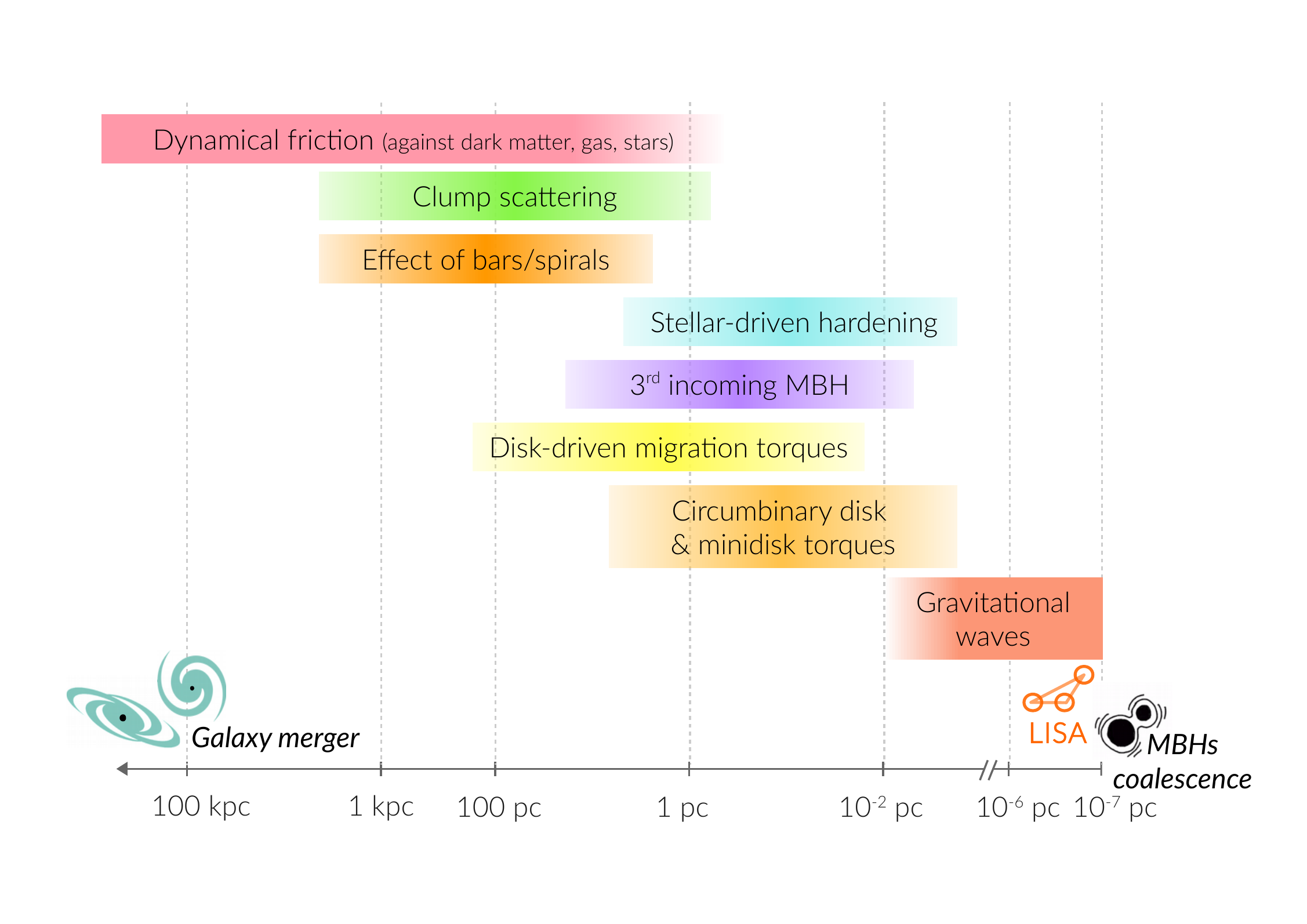}
    \caption{Illustration of the physical processes affecting two coalescing MBHs after the merger of their host galaxies. The cartoon reports the typical physical scales associated to each process for a nearly equal mass MBH pair of about $10^6 \msun$. Scales vary depending on the exact mass, mass ratio and host galaxy properties. These physical processes are described in Sections~\ref{sec:DynFriction},~\ref{sec:DFGasMedium} (dynamical friction); Section~\ref{sec:MoreComplexDensityProfiles} (clump scattering, effect of bars/spirals); Section~\ref{sec:hardening_star} (stellar-driven hardening); Section~\ref{sec:tripleMBHs} ($3^{rd}$ incoming MBH); Section~\ref{sec:hardening_gas} (disc-driven migration torques; circumbinary disc and minidisc torques); Section~\ref{sec:TheGWEmissionPhase} (gravitational waves). Figure credit: Elisa Bortolas.}
    \label{fig:cartoon_Elisa}
\end{figure}

There is observational evidence that a significant fraction of galaxies host MBHs in their centres \citep{2013ARA&A..51..511K}, and at least some of them harbour an MBH since the dawn of structure formation \citep[e.g.][]{2014AJ....148...14B, 2015Natur.518..512W, 2018Natur.553..473B}. This, combined with the notion that galaxies aggregate via repeated mergers of smaller structures \citep{2010MNRAS.406.2267F,2021MNRAS.501.3215O}, leads to the conclusion that a number of MBHBs should have formed across cosmic epochs, and that their ultimate coalescence phase could be observed by LISA \citep[e.g. ][]{2016PhRvD..93b4003K, 2019MNRAS.486.2336D, 2020ApJ...897...86C, 2020ApJ...904...16B, 2021MNRAS.500.4095V, 2019MNRAS.486.4044B}.

The exact number of detectable MBHB mergers and their properties (Section~\ref{sec:StatisticsOnMBHMergers}) will depend on still poorly understood parameters, such as the low-mass end of the MBH mass function and their seeding mechanism (Section~\ref{sec:MBHoriginandgrowthacrosscosmictime}), or the host galaxy structure and environment. However, as a start, we can try to address the following questions for MBHs in the LISA mass-redshift range: \textit{(i)} What are the mechanisms which bring two MBHs in distinct galaxies separated by tens of kpc close enough, so that they emit GWs and merge when they are at separations of the order of their gravitational radii, $\sim 10^{-6} \ (M_{\rm BH}/10^7\msun)$ pc? \textit{(ii)} Given the variety of galaxy types, MBH masses, and orbits they can have, are these mechanisms always efficient enough that a galaxy merger results in an MBH merger within the age of the Universe?
We begin by considering two galaxies with a range of properties hosting an MBH in their centre. We will then describe the different steps that may or may not lead to the MBH merger following the merger of the two galaxies.

In a seminal work, \cite{1980Natur.287..307B} were the first to explore the dynamics of MBH pairs in merging galaxies. In their study they highlighted the occurrence of three steps, which we will use as the foundation of this section: the initial {\it dynamical friction phase} (Section~\ref{sec:TheGalaxyMergerAndTheLargeScaleInspriral}; kpc scale), when MBHs and their hosts sink toward the centre of the remnant galaxy losing orbital energy and angular momentum until they are gravitationally bound and form a binary; the {\it binary hardening phase} (Section~\ref{sec:TheBinaryFormationAndInitialShrinking}; pc scale), when the binary mainly interacts with single stars and/or gas; and finally the {\it relativistic phase} (Section~\ref{sec:TheGWEmissionPhase}; mpc scale), when the dynamics is dominated by GW emission.

Throughout the years, this initial picture has been enriched by many different aspects, highlighting different astrophysical regimes
for MBH orbital decay determined by the nature of the galactic environment, with associated
different times-cales which depend  on the MBHs mass, mass ratio, mass distribution and thermodynamics in the galactic nucleus etc. (see Fig.~\ref{fig:cartoon_Elisa}). Below we  focus on the recent developments on the aforementioned stages of the  orbital decay, and we highlight the prospects for future research in the context of the science relevant to LISA. We refer the reader to the many existing reviews for a more complete presentation of the topic \citep[e.g.][]{2013CQGra..30x4008M,2014SSRv..183..189C,2012AdAst2012E...3D, 2019NewAR..8601525D}.

\subsubsection{The galaxy merger and the large-scale orbital decay at kpc scales}
\label{sec:TheGalaxyMergerAndTheLargeScaleInspriral}

In order to make forecasts for the LISA event rates the first step is to quantify robustly the range of decay time-scales at kpc scales, where the BH pair is expected to spend most of its time.

\paragraph{Dynamical friction in collisionless media}
\label{sec:DynFriction}

When a massive perturber, such as an MBH, with mass $M_{\rm BH}$ moves in a medium composed of collisionless particles (stars or dark matter, DM) with masses $m_\bigstar \ll M_{\rm BH}$, it deflects such particles from their unperturbed trajectories. As a result a trailing overdensity is generated, often referred to as ``wake'', which then pulls the perturber
towards it owing to its gravitational force, namely it causes a 
deceleration directed opposite to its motion. Such drag force
is the so-called dynamical friction \citep{1943ApJ....97..255C}. Under the assumption of an infinite homogeneous medium with density $\rho$, if the background is characterized by an isotropic Maxwellian velocity distribution with velocity dispersion $\sigma$, \cite{1943ApJ....97..255C} showed the force acting on the perturbing body is:

\begin{eqnarray}
\vec{F}_{\rm DF} \propto - M_{\rm BH}^2 \rho \mathcal{G}\left(\frac{v}{\sigma}\right) \ln \Lambda \frac{\vec{v}}{v^3} \, , \label{eq:force}
\end{eqnarray}

\noindent where $v$ is the perturber velocity relative to the surrounding background, $\ln \Lambda\sim 10$ is the Coulomb logarithm\footnote{The Coulomb term $\Lambda$ is the ratio of the maximum and minimum relevant impact parameters for encounters between
stars in the background, and the perturber.
$\Lambda$ is often estimated as the ratio of two global quantities characterizing the system: for example, the mass of the galaxy over the mass of the MBH, or the mass of the MBH over the mass of individual stars, with the first choice being more applicable to systems comprising both stars and dark matter.} and the function $\mathcal{G}(x), x = {v}/{\sigma}$ depends on the underlying velocity distribution; if the latter is Maxwellian, as typically assumed, $\mathcal{G}(x)$ scales as $x^3$ for $x \ll 1$ and as $\sim 1$ for $x\gtrsim 2$.  When Eq.~\ref{eq:force} is applied locally to the case of an MBH  moving on a circular orbit of radius $r$ in the stellar background of a singular isothermal sphere ($\rho \propto \sigma^2 r^{-2}$), a calculation \citep{1987gady.book.....B} shows that the orbital decay of $M_{\rm BH}$ occurs on a time-scale

\begin{eqnarray}
  \tau_{\rm DF} \approx \frac{8\, \rm {Gyr}}{\ln \Lambda} \left( \frac{r}{\rm {kpc}} \right)^2  \frac{\sigma}{200\, \rm {km/s}}  \frac{10^7\msun}{M_{\rm BH}}.
     \label{eq:tau_DF}
\end{eqnarray}

If we assume MBHs at kpc scale separations and a $10^6 \msun$  black hole in a galaxy with $\sigma=100\,
\rm {km/s}$, this calculation shows
that dynamical friction plays an important role in causing a rapid
sinking of MBHs with masses in the range accessible to LISA
as the process can take less than a Hubble time \citep[in the early stage of a galaxy merger, $M_{\rm BH}$ may be replaced by the mass of a residual galactic core embedding the MBHs, resulting in much shorter time-scales,][]{2002MNRAS.335..965Y}. Two $10^6 \, \msun$ black holes
are indeed expected to bind gravitationally and form a binary once
their separation is reduced to a few pc.
In the following we detail how this simplified picture is enriched when some of the assumptions made above are relaxed.  Overall, more complex dynamics lead to a much broader range of time-scales than expected based on the previous discussion, and render the formation of a binary a more uncertain outcome.\\

\noindent$\bullet$ {\bf Global asymmetries}

The description of dynamical friction given above implies that the drag is local, caused by the overdensity trailing the perturber, thus it neglects the global exchange of orbital angular momentum and energy between the MBH and the host system. Global asymmetries triggered in the mass distribution of the host system \citep[called \textit{modes}; see, e.g.][]{1984MNRAS.209..729T,1986ApJ...300...93W,1989MNRAS.239..549W} can give rise to global torques, and these can be enhanced at resonances between the perturber's orbital frequency and the orbital frequency of the background matter. Owing to new observational data \citep[][]{2018A&A...616A..11G}, as well as recent theoretical \citep{2020arXiv201114812H} and numerical work \citep[][]{2019ApJ...884...51G, 2020ApJ...898....4C, 2020arXiv200713763T, 2020arXiv201000816G}, the global halo mode theory has gained renewed attention. Accounting for the corrections to the dynamical friction time-scale introduced by global torques is likely important in order to provide robust estimates of the initial phase of black hole binary formation and sinking. Studies specifically for LISA MBHBs are required to ultimately assess the importance of these processes in the context of LISA's science.\\

\noindent$\bullet$ {\bf Power-law density profiles: cusps and cores}

The assumption of an isothermal sphere used to derive Eq.~\ref{eq:tau_DF} is also a simplification: all real galaxies feature much more complex profiles. Even referring only to the DM distribution, its inner density profile is typically believed to behave as a Navarro-Frenk-and-White (NFW) profile $\rho \propto r^{-\gamma}$ with $\gamma = 1$, or even shallower in low-mass dwarf galaxies \citep[see, e.g. the evidence on constant density cores in][]{2015AJ....149..180O}. This shallower core could be the result of baryonic feedback effects \citep{2010Natur.463..203G}, or of the phase-space density structure inherent to a specific DM model such as self-interacting DM or fuzzy DM \citep[][]{2017PhRvD..95d3541H}. LISA will be particularly sensitive to MBHBs in the range of masses $10^4$--$10^6$~M$_{\odot}$ mainly hosted in low-mass dwarf galaxies; since many dwarfs appear to be DM cored (at least at low redshift; see, e.g. \citealt{1994Natur.370..629M, 2018MNRAS.476.3124C, 2020MNRAS.493..320L}), this motivates a thorough study of the dynamics in shallow inner density profiles.

\cite{2018ApJ...864L..19T} modelled numerically the orbital dynamics of a pair of $10^5$~M$_{\odot}$ BHs during the equal-mass merger of two dwarf galaxies. They showed that, if the merging galaxies have kpc-sized cores, or at least a profile shallower than NFW (inner slope $\gamma \sim 0.6$ or lower), the pair of MBHs would stall at separations of 50--100~pc (i.e., when the bound binary is not formed yet) and the coalescence would be aborted. In a halo with an NFW profile, stalling does not occur, rather the MBHs sink very fast to sub-pc separations in less than a few $10^8$ yr after the two host galaxies have merged. In self-interacting DM models, in which cores can be $> 1$ kpc in size assuming a large specific cross section of interaction of $10$ cm$^2$/g, and which are under-dense compared with Cold Dark Matter (CDM) control cases, an analogous suppression of dynamical friction was found to occur at even larger scales, when galaxies are still in the process of merging, leading to many wandering MBH pairs with few kpc separation    \citep{2017MNRAS.469.2845D}. This opens the possibility that the event rate of MBH mergers detected by LISA could constrain the density profile of dark matter halos of their host galaxies, which in turn can shed light on the physical  properties of dark matter particles.

Gas dissipation and stellar feedback were not taken into account in the aforementioned studies; those  could delay the binary formation  even more. On the other hand, if at least one of the sinking MBHs is surrounded by a massive nuclear star cluster, as in the case of a captured ultra-compact dwarf galaxy, this may enhance the dynamical friction and aid the binary formation and shrinking even in cored DM profiles. These aspects should be investigated in detail in preparation for LISA.\\

\noindent$\bullet$ {\bf MBHBs with very unequal mass ratio}

When the mass enclosed within the binary orbit becomes of order the mass of the secondary, dynamical
friction is not effective anymore, and different processes are required to shrink the binary further (see \S\ref{sec:TheBinaryFormationAndInitialShrinking}). For  equal mass
binaries this critical separation  roughly corresponds  to the distance at which the binary becomes effectively bound, but for binaries in which the 
secondary MBH is much lighter than the primary, dynamical friction remains the main driver for the MBHB shrinking well below the separation at which the secondary becomes bound.
In this situation, the fact that
(Eq.~\ref{eq:force}) considers only the contribution of stars moving slower than the secondary MBH \citep{1943ApJ....97..255C} can be
a major limitation.  
\cite{2012ApJ...745...83A} found that, if the inner density profile scales as  $\rho \propto r^{-\gamma}$, the conventional application of Chandrasekhar's formula  works reasonably well if $\gamma>1$, but does not reproduce the inspiral of the secondary if the profile is very shallow ($\gamma \sim 0.6$). The reason is that, in the latter case, stars that move faster than the secondary MBH contribute to most of the force. As a consequence, conventional dynamical friction would predict  stalling of the secondary MBH, while the
orbit can keep shrinking, albeit at a much slower pace \citep{2017ApJ...840...31D}. However, this has only been verified
when  the secondary MBH is significantly smaller than the primary ($q\lesssim 0.01$, i.e. close to the IMRI regime) and orbits inside a nearly spherical and isotropic nucleus without net rotation. Assessing the outcome for a more realistic profile of the nucleus will be needed in the near future to prepare for LISA.

\paragraph{Dynamical friction in  a gaseous medium}
\label{sec:DFGasMedium}

In the previous sections, we considered orbital decay of a pair of MBHs in a collisionless background of stars and dark matter. However, gas constitutes a significant fraction of mass in many galaxies, especially at high-redshift  \citep[][]{2018ApJ...853..179T,2020ApJ...902..110D}. Similarly to stars, the gaseous wake lagging behind a massive perturber tends to slow it down but the details of the interaction depend on the geometry of the gas wake, which in the case of gas is subject to the additional effect of pressure.

For comparable densities gas-driven dynamical friction  is larger by a factor of $\sim$5 than that of stars in the transonic regime, i.e., when the Mach number of the perturber is around unity \citep{1999ApJ...513..252O}, while it is of similar order in the supersonic regime, i.e., for Mach numbers much larger than unity, and it is suppressed in the subsonic regime, i.e., below Mach numbers of order unity. However, the overall contribution of the gas-driven component to the total drag force suffered by an MBH in a galactic nucleus is still debated, as it depends sensitively on the dynamical and thermodynamical state of the medium as well as on its density and cooling properties in the vicinity of the perturber.

Hot, low-density gas in gas-poor galaxies is virialized and thus gives little contribution to the drag. However, the central cold and dense region may play an important role. Using semi-analytical models, \cite{2020ApJ...896..113L} find that galaxies with low gas fraction and a large stellar bulge favour the formation of binary MBHs, and that their dynamics is dominated by stellar dynamical friction. Hydrodynamical simulations find quite  a range of results, whose often large differences are likely driven by the different setups, astrophysical as well as numerical, considered by different simulations.

For example, \cite{2017MNRAS.471.3646P} also find that the contribution from gas  friction is negligible compared to that from stars as in their case \textit{(i)} the gas density is lower than stellar density; and \textit{(ii)} the stellar density profile is more regular than the gas one, so that stars act as a smooth background, which is conceptually consistent with the theory of dynamical friction. On the other hand, numerical simulations and observations also show that stellar morphology in merging systems is often highly disturbed and rapidly varying, which complicates this picture and suggests that galactic substructure might be important to take into account. \cite{2013MNRAS.429.3114C} find that  the collision between the two massive equal mass gas-rich galactic discs drives rapid sinking, primarily owing to gas-driven friction, of the MBHs that pair into a binary. Note that, in equal mass galaxy mergers, funnelling of gas to the centre of the merger remnant via gravitational torques and shocks is maximized relative to the unequal mass merger case considered by \cite{2017MNRAS.471.3646P}, and this leads to a much higher central gas density. It is therefore not surprising that the two studies reach different conclusions on the relative importance on gas-driven and stellar-driven friction.  All this shows that the processes leading to the formation of LISA binaries in realistic gaseous and stellar environments  deserves future investigations before LISA flies.

\paragraph{More complex mass distributions and additional physical phenomena}
\label{sec:MoreComplexDensityProfiles}

We expect galaxies to not be realistically represented by the spherical, power-law, and smooth density profiles. As already mentioned in the previous section, global asymmetries affect the sinking time-scale. In order to prepare for LISA, we need to investigate the effects of more complex structures onto the dynamics of MBHs.  We summarize below the recent results of several groups studying these effects and highlight some areas of particular interest for future study.\\

\noindent$\bullet$ {\bf Effects of discs}
    
\begin{figure*}[t]
\centering{
\includegraphics[trim=0 0 0 0, clip, scale=0.85,angle=0]{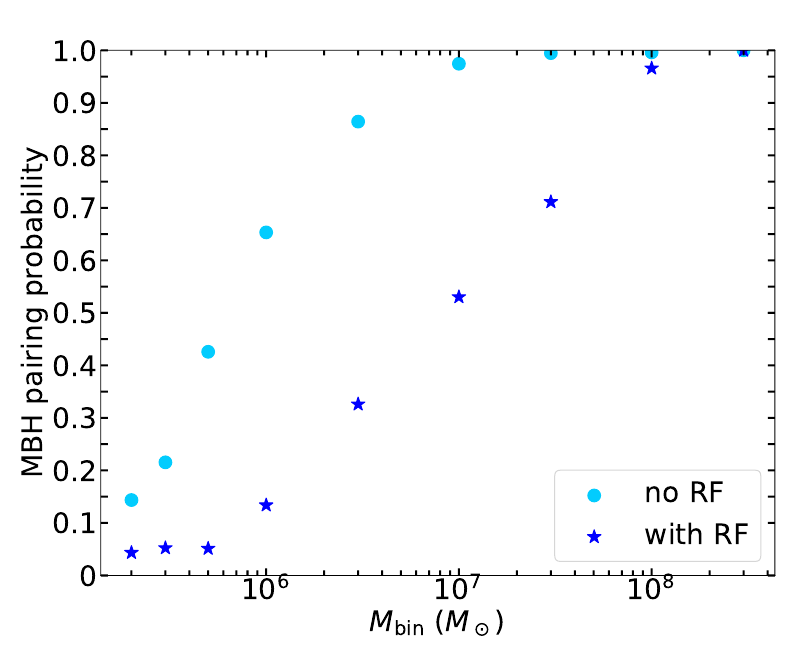} %
\includegraphics[trim=0 0 0 0, clip, scale=0.85,angle=0]{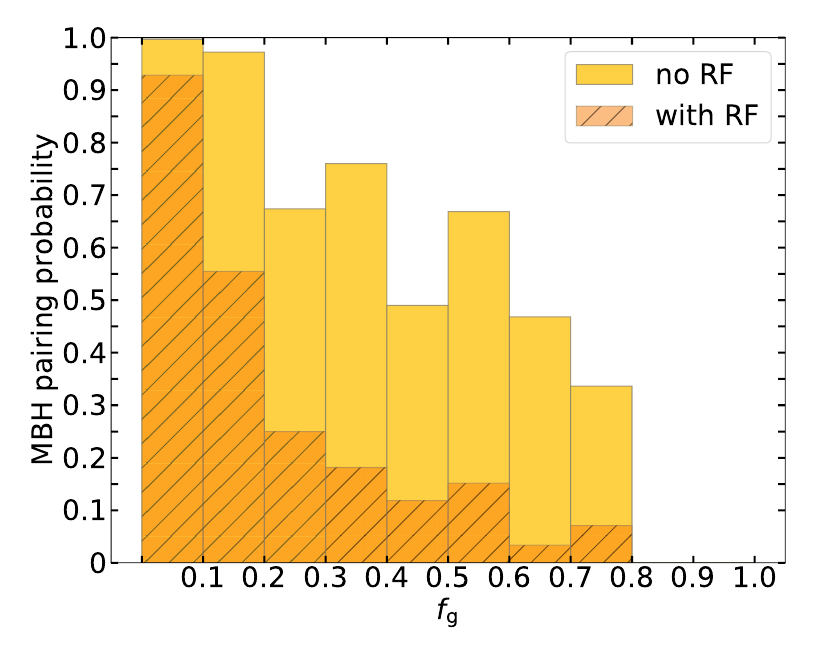}
}
\caption{MBH pairing probability as a function of the MBH pair mass (left)  and the gas fraction of the host galaxy (right) 
in models with and without radiative feedback (RF).
In the presence of radiative feedback, the suppression of MBH pairing is most severe in galaxies with MBH pairs with mass $< 10^8\,M_\odot$
and $f_{\rm g} \geq 0.1$.  
The pairing probability is calculated as a fraction of models in which the two MBHs reach a minimum separation of 1~pc within a Hubble time. Figure adapted from \citet{2020arXiv200702051L}.}
\label{fig:Li2020}      
\end{figure*}

The question of the effects of large-scale galactic discs  ($\sim$1--10 kpc) and circumnuclear discs ($\sim 100$ pc) 
on the formation of gravitationally bound MBH pairs is closely related to the question of what types of galaxies are the most likely progenitors of LISA sources. Simulations have already addressed the effect of dynamical friction in composite, rotationally supported environments; they suggested that, quite independently of whether the background is mainly stellar or gaseous \citep{2007MNRAS.379..956D}, dynamical friction acting in rotating discs usually induces the circularization of initially prograde and eccentric orbits, while it reverses the angular momentum of counter-rotating trajectories, then again promoting circularization \citep[see, e.g.][]{2006MNRAS.367..103D,2011ApJ...729...85C,2013ApJ...777L..14F, 2020MNRAS.494.3053B}.

Using a different approach, \citet{2020ApJ...896..113L} studied this aspect adopting a semi-analytical model to describe the orbital evolution of MBHs from separations of $\sim$1~kpc to $\sim$1~pc, under the influence of stellar and gaseous dynamical friction. Their study of the parameter space suggests that the dominant drivers of the MBH orbital evolution are the stellar bulge and galactic gas disc. They find that the 
chance of MBH pairing within a Hubble time is nearly 100 per cent in host galaxies with a gas fraction of $<0.2$, as shown in the right-hand panel of Fig.~\ref{fig:Li2020}. 
They also find that the orbital evolution 
sensitively depends on the relative speed between the gas disc and the MBHs. Semi-analytical models, however, are quite limited in their predictive
power when it comes to the effect of gas as they cannot account for the multi phase nature of the interstellar medium and the concurrent star formation
in the galactic nucleus, which are bound to have an impact on both the local drag and the global torques, motivating the investigation of this problem with various approaches in order to assess the impact on LISA's MBH mergers.\\

\noindent$\bullet$ {\bf Effects of feedback}

The distribution of gas in the host galaxy and its contribution to the total dynamical friction force on MBHs is  likely to be strongly impacted by radiative feedback \citep[e.g.][]{2011MNRAS.414.3656S, 2017ApJ...838...13S}. More specifically, it was recently shown for MBHs evolving in gas-rich backgrounds that ionizing radiation that emerges from the innermost parts of the MBHs’ accretion flows can strongly affect their gaseous dynamical friction wake and render gas  friction inefficient for a range of physical scenarios. Combined with the effect of radiation pressure, the radiative feedback
creates an ionized region larger than the characteristic size of the dynamical friction wake and a
dense shell of gas in front of the MBH, as a consequence of the snowplow effect. In this regime, the
dominant contribution to the MBH acceleration comes from the dense shell and such MBHs experience a
positive net force, meaning that they speed up, contrary to the expectations for gaseous dynamical
friction in absence of radiative feedback  \citep{2017ApJ...838..103P, 2020MNRAS.492.2755G, 2020MNRAS.496.1909T}. 
This effect was dubbed ``negative dynamical friction''.

If prevalent in real merging galaxies, negative gaseous dynamical friction can lengthen the inspiral time of MBHs and even offset the action of stellar dynamical friction. Its full implications for the formation and coalescence rate of MBHBs in galactic and cosmological settings for MBHs in the LISA mass range are however yet to be understood. Some early insights into this question are provided by \citet{2020arXiv200702051L}, who used a semi-analytical model to study the impact of negative dynamical friction on pairs of MBHs in merger remnant galaxies evolving under the combined influence of stellar and gaseous dynamical friction. They found that, for a wide range of galaxy mergers and MBH properties, negative dynamical friction reduces the MBH pairing probability to $\sim$50 per cent of that found in absence of radiative feedback (Fig.~\ref{fig:Li2020}, left). This effect is particularly prevalent in systems with a gas fraction above $0.1$, especially if the disc rotational velocity is comparable to the circular velocity (Fig.~\ref{fig:Li2020}, right). 
Importantly, the pairing probability in the presence of radiative feedback decreases five-fold (to $\lesssim 0.1$) for MBHBs with mass $\lesssim 10^6M_\odot$  \citep{2020arXiv200702051L}, implying that the pairing of the very population of MBHBs targeted by LISA may be greatly affected by it.

\citet{2019MNRAS.486.2336D} on the other hand point out that many MBHs in the LISA mass range ($\lesssim 10^6 M_\odot$) MBHs would reside in low mass haloes, in which SN feedback and radiation background due to reionization will expel and photo-evaporate most of the gas, thus curbing the growth of the MBHs and suppressing the effect of gas on their orbital evolution (see Section~ \ref{sec:accretion_vs_merger} for a related discussion). 
It is therefore crucial to understand how feedback affects gas dynamical friction in realistic merger remnants. If pairing and merger rates of MBHBs in gas-rich environments are  reduced, this has important implications for the likelihood of detection of multimessenger events with LISA and the contemporary EM observatories.
 For this reason,  a much wider range of scenarios needs to be explored to investigate the complex role of radiative feedback and gaseous dynamical friction for MBHs in the LISA mass/redshift range, exploring the impact of a different feedback geometry, energetics,  mass loading, momentum injection, etc.\\

\noindent$\bullet$ {\bf Effects of bars}
    
A large fraction of disc galaxies at low redshift show clear deviations from axisymmetry in their stellar distribution. At least at low redshift, about half of massive  ($M_*\gtrsim 10^{10} \, \msun$) discs  
\citep[e.g.][and references therein]{2016A&A...595A..67C}
host a prominent overdensity approximately symmetric with respect to the centre with constant phase, e.g. a bar, that can significantly affect the dynamical evolution of the different components of the host galaxies \citep{2002ASPC..275..141A, 2014RvMP...86....1S}. Quantifying the fraction of barred galaxies at high redshift is still challenging \citep[see, e.g.][]{2008ApJ...675.1141S, 2014MNRAS.438.2882M, 2014MNRAS.445.3466S}, but \textit{(a)} it has been observationally proposed that bars could be frequently hosted in massive galaxies at all redshifts, with an increasing mass threshold for entering in the bar unstable regime as a function of redshift \citep{2015A&A...580A.116G}, and \textit{(b)} bar formation has been found as early as at $z\sim7$ in cosmological zoom-in simulations \citep{2017MNRAS.467.4080F}.
    
Being such strong perturbations to the host potential, bars could significantly affect the pairing of MBHs during galaxy mergers. The occurrence of such effect could be increased when the actual merger is responsible for the triggering of a bar \citep{1986A&A...166...75B, 2004MNRAS.347..277M, 2008ApJ...687L..13R, 2016Galax...4....7M, 2018MNRAS.473.2608Z, 2018MNRAS.479.5214Z, 2019MNRAS.483.2721P}, even when this is short-lived and not sustained by the galactic potential, e.g. if the galaxy stellar disc is below the threshold for bar instability.
    
The effect of a forming and growing bar on the pairing of MBHs within LISA's reach has been recently explored  by \cite{2020MNRAS.498.3601B}. They populated a main-sequence $z\sim 7$ galaxy\footnote{The main galaxy forming in the cosmological zoom-in simulation Ponos \citep{2017MNRAS.467.4080F}.} with secondary MBHs at different radii and at different angles with respect to the forming bar, and found a stochastic behaviour in the pairing time-scales, with some of the secondary MBHs being pushed towards the centre of the main galaxy, and others being ejected by a slingshot with the bar. Noticeably, it was found that the orbital
decay of the secondary MBHs was dominated by the global torque provided by the bar rather than by the local effect of dynamical friction. This points  to the need
of including the effect of global torques in future recipes for sinking time-scales of MBH pairs   at $\sim$kpc distances.  A first semi-analytical attempt to explore the broad parameter space of MBH pairs/bar interaction is currently ongoing \citep{2022MNRAS.512.3365B}, in which a time-dependent bar potential has been added to the integrator of orbits in disc galaxy potentials presented in \cite{2020MNRAS.494.3053B,2021MNRAS.502.3554B}. 
A much more thorough analysis, considering (a) different galactic properties, different bar potentials and bar precession velocities, (b) the dependence of the fraction of barred discs as a function of redshift and in recent mergers, and (c) the host galaxy evolution  during the MBH pairing is needed to better evaluate the effect of bars on the population of MBH binaries in the LISA band.\\

\noindent$\bullet$ {\bf Effects of clumps}

If an MBH happens to get close to a massive interstellar cloud or dense star cluster, its orbit can be severely affected, and this effect is particularly strong in a clumpy interstellar medium. The typical masses of the perturbers depend on the background gas density which determines the conditions of fragmentation in the framework of the Toomre instability. These masses are  $\sim$ 10$^5$--$10^7$~M$_{\odot}$ for giant molecular clouds in present-day galaxies; and $10^7$--$10^8$~M$_{\odot}$ for giant star-forming clumps in galactic discs at higher redshift, which have a much larger gas fraction \citep{2015MNRAS.453.2490T,2017MNRAS.464.2952T,2017MNRAS.468.4792T}. 
As clumps have to be massive enough to have a dynamical impact on the MBH, this suggests that the effect of a clumpy medium is irrelevant for MBHs with masses $> 10^8$~M$_{\odot}$ \citep{2013ApJ...777L..14F}, but is likely relevant for the MBHs that are targeted by LISA. This is especially important considering that a significant fraction of galaxies  at $z\approx$ 1--3 (i.e. an epoch in which galaxy mergers are supposedly frequent, \citealt{2010MNRAS.406.2267F}) appears to be clumpy \citep{2010MNRAS.404.2151C, 2016ApJ...821...72S}.
Several numerical simulations show that the clumpiness of the gaseous medium renders the orbital decay highly stochastic \citep{2019NewAR..8601525D}: in some situations, the MBH separation does not shrink \citep{2015MNRAS.449..494R}, in others the decay is promoted \citep{2013ApJ...777L..14F,2015ApJ...811...59D}.

When the decay stalls, it is often because the lighter secondary MBH is scattered away from the disc plane (galactic or circumnuclear), ending up in a region of much lower stellar and gas density, where dynamical friction becomes inefficient. This effect is even more emphasized when stellar and AGN feedback are included, as they can open cavities of low density gas \citep{2017ApJ...838...13S}. Note that this is not a definitive effect, as a scattered MBH can eventually be dragged back to the disc: the net effect is to delay the formation of the binary which takes 10--100 times longer \citep{2015MNRAS.449..494R}, but this will contribute in shaping the redshift distribution of MBH coalescences \citep{2020MNRAS.498.2219V}.

In summary, while stalling of the MBH pair is an extreme outcome that cannot be verified due to the limited time-scales probed by current simulations, it is clear that the range of orbital decay time-scales of MBH pairs in a clumpy medium, from kpc scales to separations of order pc and below, can be widened by up to two orders of magnitude. In addition, the induced delay likely depends on the number and mass distribution of clumps within their hosts. Since LISA can detect MBHs up to high redshift, when clumpy galaxies were more common, future, better resolved observations of clumpy galaxies at $z>1$ would be beneficial for the community. The latter, aided by more accurate simulations of the same systems, will help in better constraining the effect of  clump-driven perturbations  on the orbital decay of MBHs, especially their impact on the rates and properties of the MBHBs that LISA will detect.

\begin{figure*}
\center
\includegraphics[width=0.48\columnwidth]{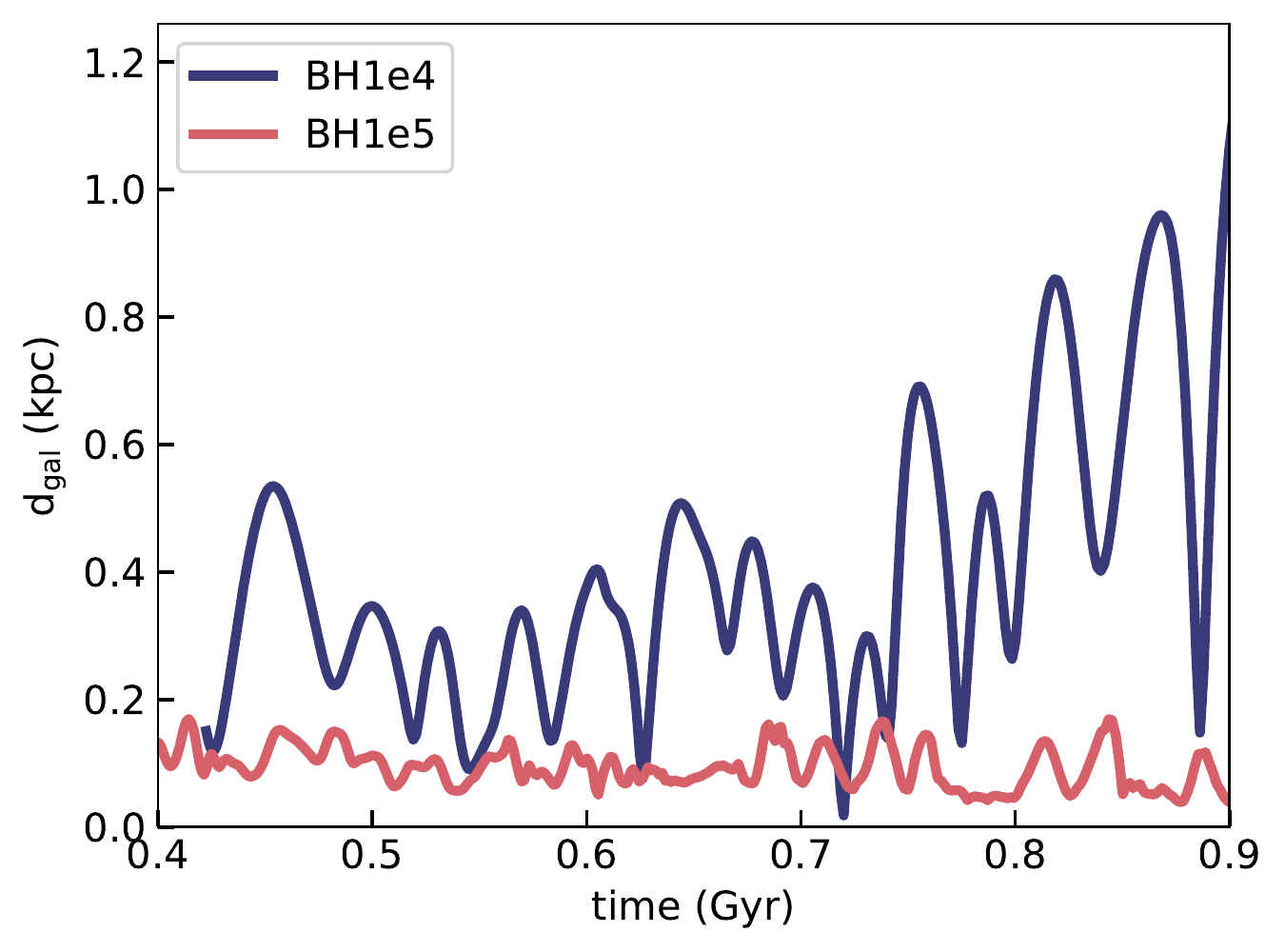}  \hfill \includegraphics[width=0.48\columnwidth]{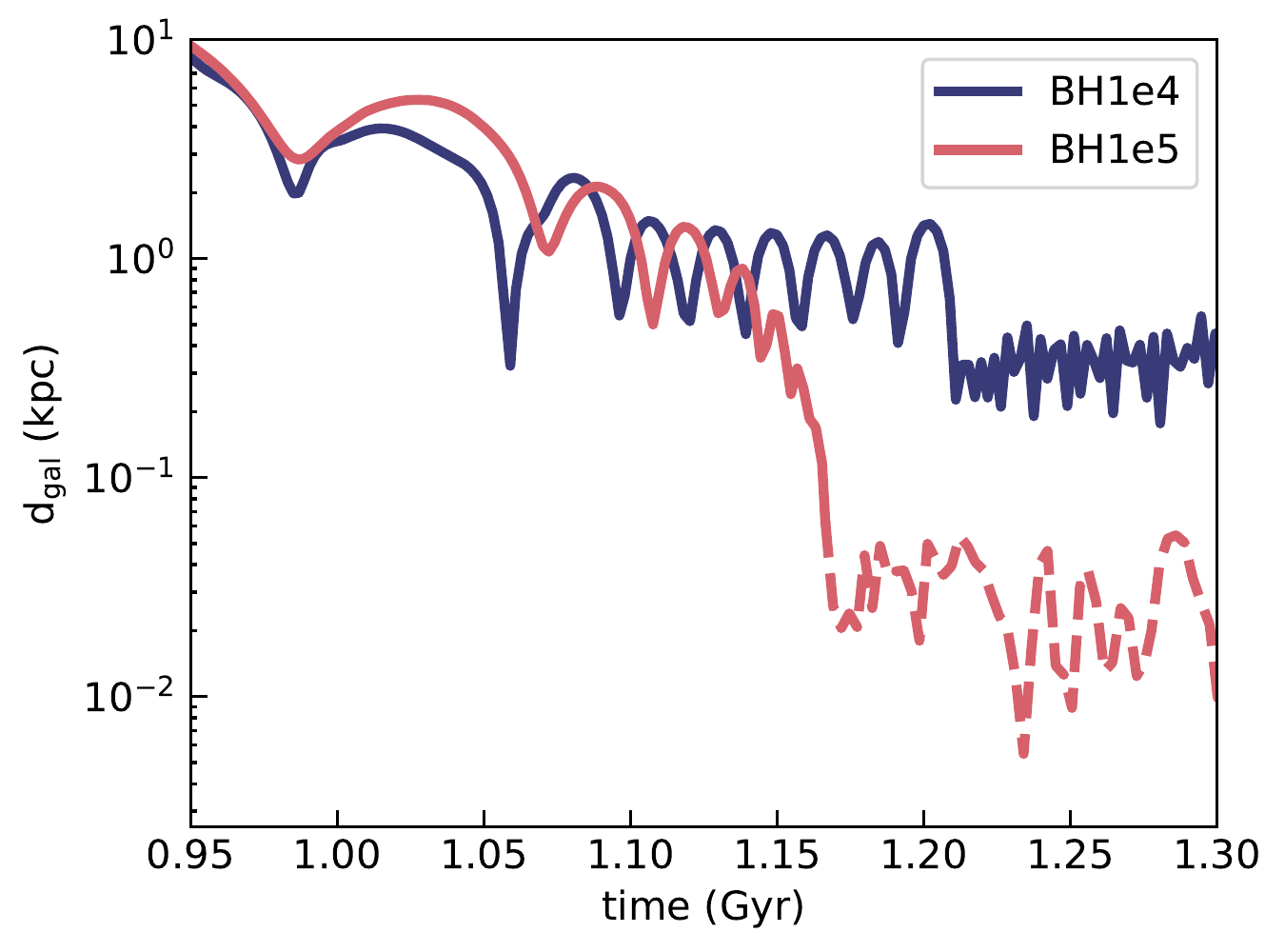}
\caption{Distance between an MBH and a galaxy as a function of time for two cosmological simulations which only differ by the mass of the MBHs ($10^4\msun$ in blue and $10^5\msun$ in pink). \textbf{Left:} We show the results for the central MBH of the galaxy. In both cases MBHs are smoothly off-centred due to inhomogeneities of the potential, but the more massive one remains in the centre, since dynamical friction is efficient, and the lighter one is instead displaced to kpc distances. \textbf{Right:} We show the results for the central MBH embedded in a satellite galaxy sinking towards the primary galaxy. The initial phase is similar in both cases as the whole satellite suffers dynamical friction, but as soon as material surrounding MBHs has been stripped, the light MBH stalls while the more massive one keeps sinking until it merges with the central MBH of the primary galaxy (subsequent evolution in dashed line). Adapted from \cite{2019MNRAS.486..101P}.}
\label{fig:EffectMass}      
\end{figure*}

\paragraph{Is there a final kpc problem?}\label{sec:LastkpcProblem}

In the previous sections we discussed several mechanisms that can cause complete stalling, or at least a significant delay of the orbital decay of a MBH pair. These mechanisms can essentially stifle the formation of an MBH binary in the first place, and the orbital evolution seems to be highly sensitive to the physical parameters involved.

To exemplify this, we show in Fig.~\ref{fig:EffectMass} the outcome of two cosmological simulations \citep{2019MNRAS.486..101P} which only differ by the mass of the MBHs. In the case of a light MBH ($10^4\msun$ in blue), not only the sinking MBH stalls at $\sim$kpc distances, similarly to \cite{2018ApJ...864L..19T}, but even the central one is smoothly off-centred due to inhomogeneties and never sinks back, as the dynamical friction time-scale is very long. While in this particular case, the massive MBH ($10^5\msun$ in pink) behaves smoothly in agreement with the classic picture of dynamical friction, we recall that \cite{2020MNRAS.498.3601B} have shown that massive MBHs can wander for a long time because of bars.
    
Perturbations, and the subsequent stalling, appear to be more relevant and likely to occur at higher redshift ($z > 1$), as host galaxies have clumpier, more turbulent, and more inhomogeneous gas and stellar density profiles \citep{2019MNRAS.486..101P}. At lower redshift, stalling is more likely to occur in low-mass/dwarf galaxies that have low background density \citep{2015MNRAS.451.1868T,2019MNRAS.482.2913B} or cored DM profiles \citep[][]{2018ApJ...864L..19T}, or in high-mass galaxies with core-S\'ersic profiles \citep{2003AJ....125.2951G}. Some isolated simulations of 1:4 massive spirals mergers \citep{2009ApJ...696L..89C, 2011ApJ...729...85C} also show a delayed inspiral with respect to the estimate of Eq.~\ref{eq:tau_DF}. However, the exact physics of the inspiral crucially depends on the details of gas accretion and star formation about the MBHs, as e.g. the formation of a dense stellar nucleus about the secondary may accelerate its orbital decay \citep{2014MNRAS.439..474V,2020MNRAS.493.3676O}. 

On the observational side, recent radio observations of AGN in local dwarf galaxies \citep{2020ApJ...888...36R}  have highlighted that at least some of these objects are not located in the centre of their host, which is often not easy to define, due to irregular galactic morphologies, in line with the results of the simulations presented above. Known offsets of nuclear star clusters offer further insight \citep{2000A&A...359..447B}. If the reason for the observed displacement is a failed inspiral (instead of e.g. the effect of a GW recoil following an MBH merger, or the interaction with a third MBH), it is easy to imagine that the formation of a bound MBH binary may become very unlikely. 
All the above suggests that the large-scale decay of MBHs is likely to be a stochastic process, highly dependent on the environmental conditions of the host galaxy nucleus, and on the orbital configuration of the MBH pair.

Modelling such stochasticity in a way simple enough that can be incorporated in population synthesis models for LISA MBHs,  but at the same time accurate enough to account for the relevant physical processes shaping the inspiral, is a key challenge ahead of us \citep[see][for an example investigating the effect on LISA's coalescences]{2020ApJ...904...16B}. 
Furthermore, it is practically hard to set the boundary between dynamical friction (thought as the response of the host to the perturber's passage) and different torquing mechanisms related to the galaxy mass distribution.
For this, an effort towards a detailed and realistic characterization of MBHs, along with the effect of their feedback, in a variety of systems at all redshifts is vital to properly model the MBH merger population that LISA is going to probe. In order to interpret LISA's data it is crucial to develop well-motivated models that can be compared with the detected events in order to extract astrophysical information.

\subsubsection{Orbital decay after binary formation at pc scales}\label{sec:TheBinaryFormationAndInitialShrinking}

As dynamical friction drives the orbital decay of the MBHs, they eventually  find themselves inside their mutual sphere of influence\footnote{The sphere of influence of an MBH of mass $M$ can be defined in different ways, but generally it is considered as the sphere with radius equal to $\sim G M / \sigma^2$, with $\sigma$ the velocity dispersion of the nearby stellar background.}, resulting in the formation of an MBHB \citep[][]{1980Natur.287..307B, 2001ApJ...563...34M}. The subsequent evolution of the newborn binary can be driven by several processes. In general, the efficiency of these processes is critically connected to the characteristics of the  environment surrounding the MBHB,  and every MBHB will follow its own different evolutionary path. 
As for the larger-scale dynamics at kiloparsec scales, we can broadly identify two classes of physical processes that shape the further shrinking of the binary separation in galactic nuclei: those that operate in gas-poor stellar environments and those that instead work when a consistent reservoir of gas is present. The boundary between the two classes is definitely not strict, and although most studies available today focus on only one of the two environments at a time, both stellar and gaseous hardening can operate at the same time (see e.g. \citealt{2017MNRAS.464.3131K,2021ApJ...918L..15B}). In the following we outline the key physical aspects featured by each shrinking mechanism, since they can all operate for LISA's MBHs, which are expected to dwell in environments rich in both gas and stars. 

\paragraph{Hardening in stellar environments}
\label{sec:hardening_star}

As soon as two MBHs form a bound binary system, dynamical friction gradually ceases to be effective since at such small scales, of order a parsec, the surrounding background mass is too low to generate a significant back-reaction to the perturbation induced by the MBHB itself. In a galactic nucleus whose density is dominated by stars, then, 
the prevalent mechanism that can continue to shrink the orbit of the binary is three-body encounters of individual stars with the MBHB \citep{1992MNRAS.259..115M,1996NewA....1...35Q,2006ApJ...651..392S}. After a first rapid  decay of the MBHB orbit in which dynamical friction and three-body encounters act in tandem, the binary shrinking starts to proceed at a slower but almost constant rate. The transition occurs around a separation commonly known as the hard binary separation, $a_{\rm h}$, corresponding to a semi-major axis \citep{2005LRR.....8....8M}

\begin{equation}
    a_{\rm h} \leq \frac{G\mu}{4\sigma^2},
\label{eq:a_hard}
\end{equation}

\noindent with $\mu$ denoting the reduced mass of the binary and $\sigma$ the local velocity dispersion of the surrounding stellar distribution. Physically, this scale approximately denotes the point at which the binary orbital velocity exceeds the characteristic speed of the stellar background, therefore representing a sort of decoupling length below which the dynamics is strongly dominated by the self-gravity of the two black holes. 
During the process stars are generally ejected out of galaxy centre with high velocities as a result of the interaction with the binary.
Therefore, ejections of stars by the MBHB result in a decrease of stellar density in the vicinity of the MBHB, with the damage extending typically up to a few influence radii \citep{1991Natur.354..212E, 2003ApJ...593..661V, 2012ApJ...749..147K} and effectively translating into less frequent stellar encounters. This mechanism possibly justifies the almost ubiquitous presence of stellar cores at the centre of the most massive galaxies (i.e. the ones
that likely experienced the largest number of mergers; \citealt{2018MNRAS.473L..94B}).\\

\begin{figure*}[t]
\centering
\includegraphics[width=0.85\textwidth]{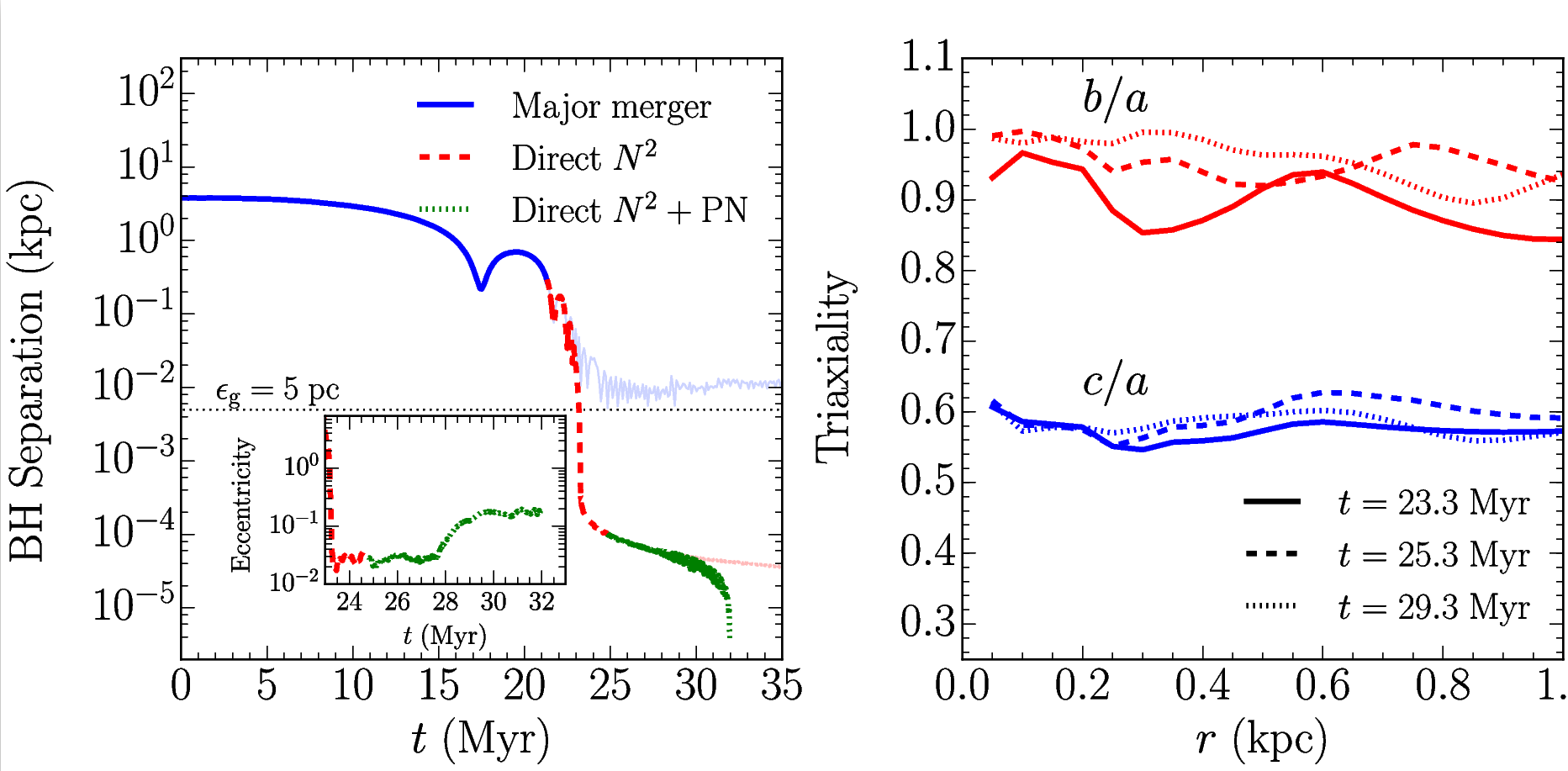} 
\caption{Evolution of an MBH pair in direct $N$-body simulations of a galaxy merger (obtained from cosmological simulations) at redshift $z \sim 3$. \textbf{Left:} MBH separation as a function of time; the small plot shows the time evolution of the Keplerian eccentricity past the binary formation. \textbf{Right:} Merger remnant axis ratios as a function of the radius for different simulation times. 
Figure adapted from \citet{2016ApJ...828...73K}.
}
\label{fig_MBBH_cos}      
\end{figure*}

\noindent$\bullet$ {\bf The final parsec problem}

As the MBHB enters in the hard binary regime, it is expected to shrink at a rate determined by

\begin{equation}
 \frac{d}{dt}\left(\frac{1}{a}\right)=\frac{G\rho}{\sigma}H, 
 \label{eq:hrate}
\end{equation}

\noindent where $\rho$ is the density of the stellar background, $\sigma$ is the velocity dispersion, $a$ is the binary Keplerian semi-major axis and $H\approx 15$--20 is a numerical coefficient weakly dependent on the properties of the binary (mass, mass ratio, and eccentricity; \citealt{1992MNRAS.259..115M}; \citealt{1996NewA....1...35Q}; \citealt{2006ApJ...651..392S}, but see \citealt{2020MNRAS.493.3676O}). The equation above shows that the shrinking rate would be constant for fixed $\rho$ and $\sigma$; however, the binary surroundings get perturbed by its scouring action, resulting typically in a mildly declining hardening rate  \citep{2015ApJ...810...49V, 2018MNRAS.477.2310B}. Eq.~\ref{eq:hrate} applies so long as
the MBHB loss cone (the region of phase space containing stars with angular momentum low enough to interact with the binary) remains populated with stars. 
However, the loss cone is generally depleted within a typical stellar orbital period at the beginning of the hardening phase, and  further MBHB shrinking crucially depends on the existence of processes able to repopulate it. In principle, the loss cone can be replenished by means of 
two-body relaxation. Unfortunately, this process acts on a time-scale much longer than a Hubble time if one considers the average properties of
galactic nuclei, assuming a spherically symmetric potential \citep[][although it may be short enough for dwarf galaxies hosting low mass MBHs in the LISA band]{1987gady.book.....B}. For this, the possibility of an MBHB stalling at pc scales has been put forward from both numerical \citep{2004ApJ...602...93M, 2005ApJ...633..680B} and theoretical grounds \citep{1980Natur.287..307B}, and has been referred to as the final parsec problem. \\

\noindent$\bullet$ {\bf The final parsec problem is not a problem}

Throughout the last decades, evidence has been building up that the final parsec problem would only occur 
in perfectly  spherical, idealized galaxies; in fact, in these systems stars are bound to conserve all components of their specific angular momentum. 
If the stellar bulge is triaxial --- as in real systems --- stellar orbits can be torqued by the asymmetric mass distribution and their angular momentum does not have to be conserved in time, meaning that the loss cone can be easily repopulated in a collisionless fashion 
\citep[][]{2002MNRAS.331..935Y, 2004ApJ...606..788M,2011ApJ...726...61M}. \cite{2006ApJ...642L..21B} first adopted numerical simulations to point out how rapid the MBHB coalescence can be in triaxial nuclei, which are themselves a natural outcome of the merger of two stellar bulges \citep[][see the right-hand panel in Fig.~\ref{fig_MBBH_cos}]{2011ApJ...732...89K,2011ApJ...732L..26P}. 
Those findings were confirmed and extended to systems with different MBHB mass ratios, galaxy density profiles, and orbits, and were generalized to galaxies with realistic two body relaxation rates \citep{2012ApJ...749..147K, 2015ApJ...810...49V,  2015MNRAS.454L..66S,2016ApJ...828...73K,2017MNRAS.464.2301G, 2018MNRAS.477.2310B}.

The general consensus is that MBHBs in realistic  merger remnants can reach the GW-driven coalescence through stellar hardening alone. The time-scale on which this happens, though, depends heavily on the details of the stellar density profile and the eccentricity growth of the binary, both of which are hard to pin down. For instance,   (galaxy-morphology)-dependent scaling relations that correlate the MBH mass with many host galaxy quantities \citep[e.g.][]{2009ApJ...698..198G,2011Natur.480..215M,2013ARA&A..51..511K,2015ApJ...813...82R,2015ApJ...798...54G,2018ApJ...869..113D}
can be used to probe the binary lifetimes.
\citet{2019MNRAS.487.4985B} found they can range between $10^{-2}$~Gyr to more than 10~Gyr for MBHs of $10^5$--$10^7 \msun$. When hosts are scaled to bulges of local galaxies, the  merger times of MBHBs derived from simulations are typically less than 500~Myr \citep{2018A&A...615A..71K}, but can reach 
$\sim 1$ Gyr depending on central density, which is varied in a realistic range \citep{2018ApJ...868...97K,2016ApJ...828...73K}.  
Time-scales, however, have also been shown to depend strongly on redshift because the scaling of mass density and velocity dispersion, which both affect hardening, is a rather steep function of redshift \citep{2017JPhCS.840a2025M}. This is the reason behind the extremely short MBH merging time-scales found in cosmological simulations of massive galaxies at redshift $\sim$3.3 (\citealt{2016ApJ...828...73K}, see Fig.~\ref{fig_MBBH_cos}), and is naturally explained if one considers the scaling of structural properties
of galaxies with respect to their host CDM halos as a function of redshift
\citep[][]{2017JPhCS.840a2025M}. A detailed knowledge of the  properties of stellar dominated galaxies hosting LISA MBHs  at different redshifts appears thus to be important in order to derive a realistic  distribution of hardening times for LISA MBHB evolving inside such hosts. 

An acceleration of the MBHB shrinking  can be induced by a non-zero orbital eccentricity during the hardening stage, as this would shorten the time-scale needed by the binary to enter the GW-dominated evolutionary stage \citep[][but see Section~\ref{sec:TheGWEmissionPhase}]{1964PhRv..136.1224P}. For example, \cite{2015MNRAS.454L..66S} find that the typical time spanning from the onset of the hardening to the GW-induced coalescence is  $\approx$30 times shorter for binaries with $e = 0.99$ compared to circular ones. Eccentricity evolution in the hardening phase depends on a fine balance between energy and angular momentum exchange, and it is sensitive to a number of factors. Three-body scattering experiments in a non-rotating stellar system find that eccentricity tends to grow as the binary shrinks
\citep{1996NewA....1...35Q}, and the growth is more prominent for binaries with moderately low mass ratio ($q \gtrsim 0.01$) residing in steeper stellar density cusps \citep{2008ApJ...686..432S}. For binaries with even lower mass ratio, the evolutionary trend is less clear, and below $q \sim 0.001$ the scattering process seems to even circularise binaries \citep{2019ApJ...878...17R,2020MNRAS.493L.114B}.
\citet{2012ApJ...749..147K} also noticed that mergers of  cuspy galaxies result in lower binary eccentricities at the time of binary formation, whereas galaxy mergers with shallower central density end up with higher eccentricity values. MBH mergers in the mass range $10^4$--$10^7 \msun$, around the peak of the LISA sensitivity window, are generally observed to be hosted by galactic nuclei with steep density profiles, which might favour relatively low eccentricities. However, the situation may greatly vary once the hosts rotation is taken into consideration, as discussed below. The expected eccentricity of binaries close to merger is not only important to estimate the merger time-scale, but it has repercussions on what type of waveforms should be developed for LISA data analysis.\\

\noindent$\bullet$ {\bf The host rotation}

Recent numerical studies investigated the impact of rotation in galactic nuclei on the evolution of MBHBs, showing that it can profoundly affect the orbital parameters of the bound binary  \citep{2017MNRAS.470..940M,2017ApJ...837..135R}. MBHs sink significantly faster in orbits co-rotating with galaxy rotation, because of the longer time for the encounter between the MBHs and the incoming stars, results in a more efficient extraction of energy from the orbit \citep{2015ApJ...810..139H}. Moreover, MBHs in co-rotating orbits circularise efficiently prior to binary formation, whereas those on counter-rotating ones tend to maintain their eccentricity, which starts to grow as the two MBHs approach the binary formation phase \citep{2011MNRAS.415L..35S,2020MNRAS.492..256K}. 
Before a hard binary forms, MBHBs in counter-rotating orbits attain very high values of the orbital eccentricity ($e \simeq 1$) and also flip their plane to align themselves with the orientation of the galactic angular momentum \citep{2012MNRAS.420L..38G, 2017ApJ...837..135R}. 
This means that, in principle, MBHBs evolving in rotating environments may typically end up being close to co-rotating with their background and having a very low eccentricity; however, more studies on the modelling of MBHBs in rotating systems is needed in this direction in order to get a more complete picture of the phenomenon before LISA flies.\\

\noindent$\bullet$ {\bf Substructure in galactic nuclei}

The evolution of an MBHB in its hardening stage can also be affected by the presence of perturbers near the galaxy nucleus. In the case of low-mass perturbers, such as the nearby stars, this results in Brownian wandering. The MBHB instantaneous centre-of-mass velocity gets continuously perturbed by gravitational three-body encounters with the nearby stars \citep{2001ApJ...556..245M}, and is balanced by dynamical friction, which acts as a restoring force. As a result, the MBHB centre of mass wanders about the centre of the galaxy. As the MBHB gets displaced from the centre, the MBHB loss-cone re-population can be enhanced, resulting in a possible boost of the MBHB hardening rate in spherically symmetric systems \citep{1997NewA....2..533Q, 2003ApJ...592...32C, 2003ApJ...596..860M}. For triaxial systems, where an almost full MBHB loss cone is usually expected \citep{2017MNRAS.464.2301G}, the MBHB shrinking rate can be impacted by the MBHB's wandering only if $M_{\rm BH}/m_\bigstar\lesssim 10^3$ \citep[being $m_\bigstar$ the typical mass of stars, see][]{2016MNRAS.461.1023B}. This suggests that the MBHB's wandering would not significantly affect the hardening rate of LISA MBHBs.

In the case of perturbers with mass much larger than the stellar one, the effects can be more significant (\citealt{2007ApJ...656..709P}; \citealt{2008ApJ...677..146P}). Massive perturbers may be in the form of star clusters,  giant molecular clouds or even a further inspiralling MBH, and may have masses up and above that of MBHs in the LISA band. Such objects can reach the galaxy centre and affect the binary inspiral in different ways. Due to their large mass, they scatter new stars into the loss cone,  thus indirectly enhancing the MBHB shrinking rate, somehow acting as boosters for two body relaxation \citep{1951ApJ...114..385S}. In addition, if they come close enough to the binary, they may displace it from the galaxy centre, thus again affecting the flux of stars in the loss cone; furthermore, if the massive perturber is a stellar cluster, once the object reaches the  binary, it  delivers new stars onto it, thus directly promoting its shrinking \citep{2018MNRAS.474.1054B, 2019MNRAS.484..520A}. 
Thus, in principle, massive perturbers may have a significant effect  on the orbital evolution of an MBHB. In order to properly model their impact on a population of LISA MBHBs, more studies are needed to pinpoint the rate at which different massive perturbers may interact with hardening binaries in different host environments. This
regime bears similarities with the clumpy medium regime in gas-rich galaxies and in circumnuclear discs as far as the dynamical interaction with the MBHB is concerned.

\paragraph{Hardening in gaseous environments}\label{sec:hardening_gas}

We now turn to the evolution of MBHBs embedded in a gas-dominated galactic nucleus. This is of great relevance for LISA, since it will detect low mass MBHs, which are expected to reside in high-redshift gas-rich galaxies. In Section~\ref{sec:TheGalaxyMergerAndTheLargeScaleInspriral} we  have already discussed extensively the dynamics of MBH pairs, until MBHB formation, in gas-rich galaxies, from kpc scales in the galactic disc, to pc scale separations in the circumnuclear disc, showing how various effects can both hamper or promote the sinking of the MBHs. We  now continue our investigation in gas-rich nuclei for smaller separations, namely following MBHB formation.

One first effect of gas which is relevant also at such small separations is related to accretion.
In the limiting case that the gas accreting on to the MBHs has zero angular momentum, the gas inflow will be purely radial and accretion on to the binary will be Bondi--Hoyle--Lyttleton-like~\citep[hereafter Bondi;][]{1939PCPS...35..405H,1944MNRAS.104..273B,1952MNRAS.112..195B}. 
Bondi accretion onto a binary has been studied in, e.g. \citet{2010PhRvD..81h4008F,2019ApJ...884...22A,2019MNRAS.490.5196C} and, with the inclusion of magnetic fields, in \citet{2012ApJ...752L..15G,2017PhRvD..96l3003K}. The main conclusion is that the sinking time-scale of the binary caused by their distorted wakes remains comparable to the usual gaseous Bondi drag time-scale for a single compact object, only a factor of few smaller, at least in the parameter ranges studied in the above papers.

\begin{figure}
    \centering
    \includegraphics[width=0.6 \textwidth]{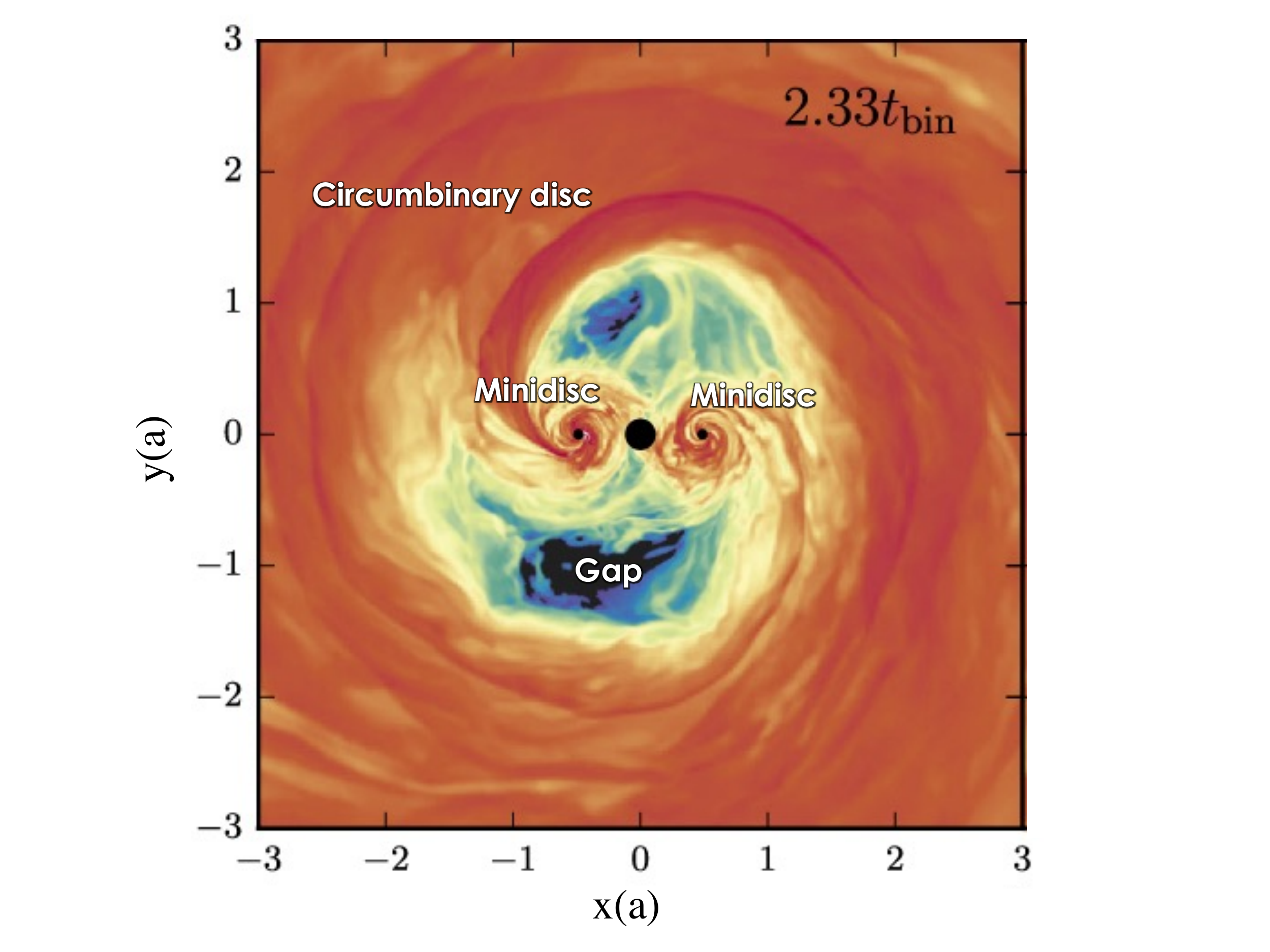}
    \caption{Illustration of the geometry of a circumbinary disc, with minidiscs surrounding each of the MBHs in the binary, a gap opened by the MBHs gravitational torques, and gas streams connecting the circumbinary disc to the minidiscs.  Adapted from \citet[][the central larger black circle at the coordinate origin marks a central excision and is not physical]{2018ApJ...853L..17B}. Image concept: Julian Krolik. Figure realization: Marta Volonteri.}
    \label{fig:CBD_mini}
\end{figure}

In reality, gas on large scales is likely to possess a specific angular momentum much larger than the one corresponding to the innermost stable circular orbit (ISCO) of the MBHs. Therefore, considerable loss of angular momentum is required to drive gas from kpc to sub-pc scales, and it is likely that some residual angular momentum remains on small scales, resulting in the formation of a disc surrounding the MBHB: the so-called circumbinary disc (for a single MBH the analogous structure is an accretion disc). An illustration of the geometry of a circumbinary disc is provided in Fig.~\ref{fig:CBD_mini}. In this section we focus on the effect of the disc onto the MBH binary dynamics, while we refer to Section \ref{sec:AGN_feedback} for the implications on accretion.\\

\noindent$\bullet$ {\bf The circumbinary disc}

Due to the computational burden, large scale simulations able to resolve $\sim$100 pc scales have not yet managed to fully and self-consistently determine the properties of such circumbinary discs, with few exceptions \citep{2020ApJ...899..126S}. To circumvent the limits of large scale simulations, \citet{2016MNRAS.455.1989G,2017MNRAS.472..514G,2018MNRAS.478.1726M,2018MNRAS.479.3438G} detailed the properties of circumbinary discs  through an extensive, though highly idealised, set of simulations, where the disc was built through a bombardment  of gas clouds towards a central MBHB. These studies demonstrated that the detailed properties of circumbinary discs depend on the dynamical properties of the infalling material.

When the MBHB reaches a critical small separation, its gravitational torque on the surrounding disc material becomes stronger than the angular momentum losses per unit of time due to the disc dissipative processes; at this point, depending on the mass ratio of the binary, either an annular gap centred on the secondary radius \citep{2008ApJ...672...83M}, or a large cavity encompassing both the MBHs can be opened \citep{2016MNRAS.459.2379D}. 
It was initially suggested that the creation of such cavity would inhibit gas accretion onto the pair; more recent and resolved simulations seem instead to suggest that accretion may remain sustained through the inner edge of the disc \citep[e.g.,][]{2015MNRAS.447L..80F,2020ApJ...899..126S}. 

If and when the binary reaches sufficiently small separations, the mass of the circumbinary disc enclosed within the MBHB orbit becomes much smaller than that of the binary itself,
making the disc gravitationally stable against fragmentation \citep{2003MNRAS.339..937G}. The simplest expectation in this regime is that the gas disc will cause the binary to harden on a time-scale comparable to the viscous time-scale \citep[in analogy with Type~II planetary migration;][]{1997Icar..126..261W} down to the decoupling radius where GWs start dominating the MBHB dynamics. For typical, thin \citeauthor{1973A&A....24..337S} discs (with ratio between the vertical length scale $H$ and the radial extent $R$ around $H/R \approx 0.05$) and close to equal mass MBHBs, this occurs at $\sim 100$ gravitational radii \citep{2014PhRvD..90j4030G}.

At such close separations, for close to equal mass MBHBs ($q\sim1$) the time-scale needed by the disc to refill the cavity would get longer than the GW-driven coalescence. If, on the other hand, $q \ll 1$, the ratio between the mass of the secondary MBH and the mass of the disc enclosed in the MBHB orbit, $q_{\rm 2, disc}\equiv M_2/M_{\rm disc}$, is a key parameter. A $q \ll 1$ binary is expected to harden on the viscous time-scale of the surrounding disc, up to the binary separation when $q_{2,\rm disc}>1$, afterwards, the migration rate falls below the viscous rate. The MBHs separation at which $q_{2,\rm disc}$ grows above unity  can occur outside the region where the disc is stable against self-gravity-driven fragmentation (see figures~3 and 4 in \citealt{2009ApJ...700.1952H} and Figure~6 in \citealt{Lodato09}). The conclusion is that, if $q_{2,\rm disc}\gg 1$ at large separations ($\gtrsim$0.1--1~pc), the ensuing slow-down  would preclude the merger \citep{Lodato09}, or else it would have to occur in a self-gravitating, clumpy disc.  At smaller separations, the viscous time is generally short, and rapid merger can be promoted by a stable disc, despite the slow-down occurring when $q_{2,\rm disc}\gg 1$. 

As commented earlier in the section,  simulations have observed that gas continues to cross the inner edge of the circumbinary disc \citep[e.g.][]{2016MNRAS.459.2379D}, but in an unstable and strongly fluctuating fashion, and the spatial symmetry of the circumbinary gas is lost, resulting in a strongly lopsided, precessing disc, preventing analytical modelling of these processes. 
In the simplest case of an equal-mass binary on a circular orbit, surrounded by a locally isothermal but warm disc (with a low Mach number, or a high aspect ratio $H/R = 0.1$), several  recent simulations have converged on the same conclusion: the disc causes the binary to outspiral \citep{2017MNRAS.469.4258T,2019ApJ...875...66M,2019ApJ...871...84M,2020ApJ...889..114M}. The outspiral rate  is quite rapid, for accretion rates comparable to those of bright, near-Eddington quasars. This of course could represent an important obstacle to binary mergers, but more recent work  
suggests that this conclusion is peculiar, and holds only for the above, idealized, specific configuration.  
In particular, the sign of the torques and migration changes from positive to negative when the mass ratio is below $q\lesssim 0.05$~\citep{2019arXiv191105506D}. This means that binaries below this mass ratio may migrate inwards -- at least until the secondary accretes a sufficient mass to increase $q$ above 0.05 (after which, in the absence of any other effects, the torque would change sign, causing the binary to migrate outwards). More importantly, the disc torque has been found to strongly depend on the disc temperature (or equivalently Mach number or aspect ratio). \citet{2020arXiv200509555T} have emphasized that real AGN discs in the inner regions are expected to be thinner/colder. They measured the torques in simulations of such cooler discs, and have found that outspiral changes to inspiral, at a comparable rate, when $H/R\lesssim 0.04$. They attributed this to the importance of direct gravitational torques of the gas accumulating near the binary with an asymmetric distribution (as opposed to accretion torques). 
The dependence on the disc temperature was later confirmed by \citet{2020A&A...641A..64H}, who found that binaries outspiral only for $H/R\gtrsim 0.2$. However, \citet{2022ApJ...929L..13F} showed that, using high-resolution simulations, the result does not depend only on the disc temperature, but also on viscosity and argue that there is no threshold for expansion in terms of disc aspect ratios. However, these recent papers  all agree on the same conclusion: binaries embedded in thin ($H/R \lesssim 0.05$) locally isothermal discs do inspiral as a result of the interaction with the gas. 

A good understanding of this gas-disc driven phase is important to better understand the properties of MBHBs when they enter in the LISA band. Assuming
for simplicity that one can neglect accretion flows towards
the binary, the viscous time-scale of the disc is the relevant evolution time-scale. This means that the MBHB will simply
shrink as the disc material itself shrinks due to internal viscous stresses. Then one can define a {\it decoupling radius} by equating the viscous time-scale in the disc with the GW inspiral time-scale of the binary. 

For an equal-mass binary at the decoupling radius, the GW observed frequency is given by 

\begin{equation}
 f_{\rm GW} \sim 10{-4} \frac{1}{1+z} {\rm Hz} \left(\frac{H/R}{0.05}\right)^{6/5} \left(\frac{\alpha}{0.1}\right)^{3/5} \left(\frac{M}{10^6{\rm \,M}_\odot}\right)^{-1},
\end{equation}

\noindent assuming the viscous time-scale $t_{\nu} \sim R^2/\nu$ follows the $\alpha$-disc scaling. This suggests that for typical values of the \citeauthor{1973A&A....24..337S} viscosity parameter $\alpha$ and sufficiently large $H/R$ ratios, 
binary-disc decoupling may occur just inside the LISA band. Gas interaction becomes even more relevant in the mHz regime for binaries with smaller component masses or lower mass ratios.
Also, the binary residual eccentricity when it enters the LISA band may be determined at binary-disc decoupling~\citep{2011MNRAS.415.3033R}, suggesting that residual eccentricities of up to $10^{-2}$ in the LISA band are possible~\citep[see also][]{2009MNRAS.393.1423C,2019ApJ...871...84M}.

It is unclear, however, how realistic this way of reasoning is. Fully relativistic 3D magnetohydrodynamic (MHD) simulations  \citep{2012PhRvL.109v1102F,2014PhRvD..90j4030G,2018PhRvD..97d4036K} find that accretion on to the binary proceeds all the way through the binary merger, albeit at progressively slower rate, suggesting that there is never a true decoupling between disc and MBHB.
These studies also showed that if the disc is cooler, then decoupling is more pronounced and the accretion on to the binary declines earlier than in hot discs. 
Recent 2D simulations \citep{2015MNRAS.447L..80F,2018MNRAS.476.2249T} are in agreement with the relativistic studies and suggest that angular momentum transport of the gas in the vicinity of the binary is driven by shocks, which enable it to flow inwards and follow the binary even well past the canonical decoupling radius. 

Before the launch of LISA a number of improvements to these models are needed in order to develop tools (e.g., include eccentricity in waveforms and data analysis) and use them as guide for EM searches (see Section~\ref{sec:sec_2_4}). Descriptions of the fuelling processes from large scale down to the central pc of galaxies, with a higher resolution than the one achieved in the current available studies, are needed to pin down the properties of circumbinary discs. Furthermore, current simulations of circumbinary discs are idealized in many ways (e.g. some simulations are in 2D rather than 3D, some do not include magneto-hydrodynamics, most have simplified equations of state and treatment of thermodynamics, all of them neglect radiative feedback, and disc self-gravity is rarely included except in \citealt{2009MNRAS.393.1423C,2011MNRAS.415.3033R,2012A&A...545A.127R,2014MNRAS.439.3476R,2021MNRAS.507.1458F}. 
Moreover, although it is expected that discs at the decoupling radius are gravitationally stable, except for binaries too massive to be detected by LISA \citep{2009ApJ...700.1952H}, eccentricity evolution would be different in a self-gravitating regime, as the disc would become strongly distorted in response to its own self-gravity.
Therefore, more sophisticated models of accretion in conjunction with future observations are necessary to properly predict the residual eccentricity of binaries and other aspects of their dynamics when they enter the LISA band.\\

\noindent$\bullet$ {\bf The formation and evolution of mini-discs}

The matter that crosses the gap/cavity region (as discussed in the previous section) can form mini-discs around each MBH (see illustration in Fig.~\ref{fig:CBD_mini}). 
Their presence may depend on the thermal state of the circumbinary disc, with colder and thinner discs producing lower-mass and shorter-lived mini-discs than those in hotter and thicker circumbinary discs \citep{2016MNRAS.460.1243R}. While their masses may be small \citep{2010MNRAS.407.2007C,2015MNRAS.449.1118T,2017MNRAS.468L..50F}, they mediate the rate of accretion on to the MBHs (and determine their spin evolution, see the discussion in Section~\ref{sec:spin}) and may play a role in the migration rate of the binary. Present 2D simulations find that the resulting disc torque that affects the binary evolution receives a dominant or significant contribution from the gas near the edge of the mini-discs, and, from the numerical point of view, therefore depends on the treatment of mini-discs and possibly even on the sink particles, and/or the inner boundary conditions that mimic MBHs in Newtonian simulations~\citep{2017MNRAS.469.4258T,2019ApJ...871...84M,2019ApJ...875...66M,2020arXiv200509555T}. The importance of the mini-discs torques has also been confirmed with 3D numerical simulations by \citet{2022ApJ...929L..13F}, therefore calling for comprehensive investigations about mini-discs modeling in Newtonian numerical simulations.

Calculations partially involving relativistic corrections \citep{2012ApJ...755...51N} or full GR \citep{2012PhRvL.109v1102F,2014PhRvD..90j4030G,2014PhRvD..89f4060G,2018PhRvD..97d4036K} did not find persistent mini-discs. The more recent studies of \cite{2017ApJ...838...42B,2018ApJ...853L..17B} initialized the simulations with mini-discs already in place, and found mini-discs which are more persistent, but also found that they undergo periods of depletion and replenishment. In \cite{2014PhRvD..90j4030G}, it was argued that the reason for the absence of persistent mini-discs in relativistic simulations at small orbital separations was due to the fact that the ISCO around the individual MBHs  is larger than the corresponding Hill spheres, thereby any matter that is gravitationally bound to one MBH is immediately accreted. This hypothesis was recently confirmed in fully GR simulations in \cite{2021arXiv210206712P}.

Despite the progress made so far, studies of all these topics are at an infant state at the moment, and more sophisticated models are necessary to understand how the presence of mini-discs and a circumbinary disc affects the MBH spins, and the binary orbit as it evolves toward the LISA band.

\paragraph{The effect of AGN feedback in the hardening phase}\label{sec:AGN_feedback}

Irrespective of the dominant hardening mechanism, AGN feedback, i.e. the energy injection from an accreting MBH, can affect the dynamics of an evolving MBHB, as it does in the phases before binary formation (see Section \ref{sec:MoreComplexDensityProfiles}). For a binary migrating in a circumbinary disc, the effect of AGN feedback has been explored, with smoothed-particle hydrodynamics (SPH) simulations, only for binaries with parsec separation, i.e. in the early stages of binary evolution \citep{2018MNRAS.480..439D}. The effect of feedback in the late binary evolution has not been investigated explicitly yet.  \citet{2018MNRAS.480..439D} consider the two main regimes of binary evolution, one where the binary opens a gap in the disc and one where a gap does not form. As said, if viscous torques are inefficient in redistributing the angular momentum extracted from the binary, a low density cavity (gap) forms around the MBHs. 
In this situation, very little gas flows towards the MBHs, which have low accretion rates and AGN feedback is characterized by outflows carrying little mass and escaping through the cavity.  They do not affect the binary orbital evolution which, however, is very slow exactly because of the presence of the cavities and inefficient torques. If the redistribution of angular momentum extracted from the binary is efficient, no gap forms, and the MBHs are embedded in a dense gas bath, leading to rapid migration. Under these conditions, however, gas accretion on the MBHs is also favoured. MBHs then produce mass-loaded winds that interact with the gas in the disc, shredding it and ejecting it in all directions. The ejection of gas leads to the formation of a hollow region (``feedback gap'') around the MBHs, and the binary migration is stalled by the lack of gas with which to exchange torques. In these simulations feedback was injected isotropically, but outflows could  be non-isotropic if launched by a disc, and the effect of a collimated outflow could be different and it would be worthwhile to explore this in future studies. These jets and collimated outflows could result in unique EM signatures, as discussed in Sections~\ref{sec:sec_2_4},~\ref{sec:sec_2_5}.
   
For a binary evolving instead by stellar hardening, the effect of AGN feedback has not been explicitly studied. We can speculate that thermal or kinetic energy injection should have little effect on the distribution of existing stars, however, it can affect, and even suppress, the formation of new stars. If binary evolution is slower than star formation, persistent AGN feedback would prevent the formation of new stars that can repopulate the loss cone and further the shrinking of the binary. If the amount of gas present is very little, this effect is likely limited. If gas is copious, then this effect can become important, but then one has to consider that the dynamics of the binary will occur in a ``mixed environment'', where both scattering with stars and gas torques contribute to the binary migration.  

In summary, this regime is still largely unexplored, and may have important consequences for the orbital evolution and the EM counterparts of LISA's detections. In the near future, both simulations including isotropic and collimated AGN feedback in the late evolution of MBHs in circumbinary discs, and simulations of the stellar hardening phase including gas, star formation and AGN feedback, will need to be developed to address/understand the impact of feedback on MBH coalescence.

\paragraph{The formation of triplets/multiplets of MBHs}
\label{sec:tripleMBHs}

In the high-redshift Universe, the environment in which MBHs live is highly dynamical, as halo interactions and mergers are far more frequent \citep[e.g. the Jackpot nebula, a system at $z\sim 2$ containing several AGN in the same 400 kpc-wide Ly-$\alpha$ nebula, see][]{2015Sci...348..779H}. The outcome of these encounters could be either the formation of an MBHB  or, at least temporarily, a wandering MBH, leading to multiple MBHs in the grown galaxy halo, each inherited from a different merger \citep{2019MNRAS.486..101P}.  Failures in the binary formation process affect the specific merger rate as a function of redshift, MBH mass, and mass ratio \citep{2016PhRvD..93b4003K,2019MNRAS.486.4044B,2020ApJ...904...16B}. 

In these situations, the formation of  MBH triplets or multiplets can arise, possibly triggering a richer and more complex range of few-body dynamics (\citealt{1990ApJ...348..412M,2001A&A...371..795H,2002ApJ...578..775B,2007MNRAS.377..957H,2010MNRAS.402.2308A,2012MNRAS.422.1306K,2017ApJ...840...53R,2018MNRAS.473.3410R,2018MNRAS.477.3910B}; \citealt{2021ApJ...912L..20M}. Triplets of MBHs generally start their evolution as spatially hierarchical systems, i.e. systems where the hierarchy of orbital separations (or semi-major axes) defines an inner ($a_{\rm in}$) and an outer binary ($a_{\rm out}$), the latter consisting of the newly arrived MBH coming from $\sim$kpc scales plus the former binary (viewed as an effective single body located at its centre of mass). Under certain circumstances, these MBH systems may undergo von Zeipel-Lidov-Kozai (ZKL) oscillations (\citealt{1910AN....183..345V}; \citealt{1962AJ.....67..591K,1962P&SS....9..719L}), in which secular exchanges of angular momentum between the two binaries periodically excite the inner binary's orbital eccentricity at the expense of the relative inclination (see also Sections~\ref{subsec:triples_formation} and~\ref{ucb_dis_form} for the same process in the context of stellar-mass compact objects), resulting in efficient GW emission (see Section~\ref{sec:TheGWEmissionPhase}).

Despite the ZKL mechanism's efficiency in increasing the orbital eccentricity, it has been shown that relativistic precession (or other types of precession) can interfere with it. In practice, if the apsidal precession time-scale is shorter than that of the ZKL oscillations, then precession destroys the coherent accumulation of secular torques, hindering eccentricity growth (see e.g. \citealt{2000ApJ...535..385F}; \citealt{2016ARA&A..54..441N}; \citealt{2020PhRvD.102f4033L}). In the context of MBH triplets, the ZKL mechanism can be further re-enhanced by the orbital decay of the intruder MBH, due to its interaction with the host galaxy environment, which tends to shrink its separation from the inner binary \citep[see e.g.][]{2018MNRAS.477.3910B}. This produces a shortening of the ZKL oscillation period and a strengthening of the perturbing force acting on the inner binary, again promoting the increase of eccentricity with subsequent GW emission and possible coalescence \citep{2018MNRAS.477.3910B}.
    
Although ZKL oscillations may sometimes lead to a direct merger of the inner binary, there are many initial conditions under which no merger can occur during the secular evolution phase of MBH triplets. For example, the mutual inclination may not be high enough, the perturber may be too light, or the binary may be too wide for efficient emission of GWs. In this case, the triplet is likely to become Hill-unstable as the perturber's shrinking orbit brings it closer to the inner binary. The final fate of many MBH triplets is thus dynamical instability, wherein the secular interaction gives way to chaotic dynamics characterized by strong encounters, exchanges, and ejections. Again, this may not represent the end of the story, since in fact an ejected MBH may leave on a wide but bound trajectory, in which case it may return back and perturb the inner binary, this time through close energetic encounters, depending on the galactic potential (spherical, axisymmetric or triaxial), the specific outgoing trajectory and also on the dynamical friction efficiency.
Repeated chaotic interactions between the ejected MBH and the leftover binary can increase the orbital eccentricity again, promoting coalescence in a non-negligible fraction of cases \citep[see, e.g.][]{2018MNRAS.477.3910B}. Still, since this is not always effective, a considerable number of ejected MBHs may keep wandering inside galaxies.
    
Finally, when the lifetime of hierarchical triplets is long enough, new galaxy mergers provide additional MBHs, forming hierarchical quadruplets and even higher-order multiplets. Considering quadruplets, a natural way in which they can form is when two merging galaxies each host MBHBs. In this particular case, the system can behave like a hierarchical triplet 
until the four-body nature of the system becomes manifest, leading again to chaotic dynamics. The dynamics of MBHs multiplets can be highly stochastic and largely non-predictable, requiring therefore numerical investigations. Still, a  likely signature of MBHB coalescence triggered by dynamical interaction is the very high acquired eccentricity, that will be retained (or at least, retained in residual form) well inside the GW-dominated phase \citep[][]{2018MNRAS.473.3410R,2019MNRAS.486.4044B}. 

In the context of LISA, pre-launch more work is needed to generally include triple/multiple interactions in models of MBH evolution \citep[as done in][]{2019MNRAS.486.4044B}, and to assess the consequences on the need of eccentric waveforms. Post-launch, detection of highly eccentric MBHBs would point to triple/multiple interactions as important drivers of MBHB coalescences.


\subsubsection{The GW-emission phase at mpc scale}\label{sec:TheGWEmissionPhase}

As the MBHB continues to efficiently interact with the surrounding environment, which continuously drains energy and angular momentum from the MBHB system \citep[e.g.][]{1980AJ.....85.1281H}, it eventually enters into the gravitational radiation dominated phase. During this phase, the main parameters driving the evolution are the masses and spins of the MBHs, as well as the binary separation and eccentricity. \\

\noindent$\bullet$ {\bf Relativistic evolution}

Although relativistic effects can influence the binary evolution also in the previous hardening phase, at this stage we must necessarily take them into account. Relativistic effects can be introduced through spin-dependent post-Newtonian (PN) corrections in the equations of motion of the MBHs.  
    
Schematically, the PN-corrected acceleration can be written as

\begin{equation}
\mathbf{a} = \mathbf{a}_{\rm N} + 
\mathbf{a}_{\rm 1\,PN} + 
\mathbf{a}_{\rm 2\,PN} + 
\mathbf{a}_{\rm 3\,PN} + 
\mathbf{a}_{\rm 2.5\,PN} + 
\mathbf{a}_{\rm 3.5\,PN} + ..., 
\end{equation}

\noindent where, in the case of $N$-body  numerical simulations, the Newtonian acceleration $\mathbf{a}_{\rm N}$ is usually computed including the surrounding stellar particles, whereas the PN-terms only include contributions from two MBHs (see, e.g. \citealt{2006LRR.....9....3W,2006MNRAS.371L..45K,2013MNRAS.434.2999B}; \citealt{2014LRR....17....2B};\citealt{2019ApJ...887...35M}). The PN-correction terms are labelled so that they are proportional to the corresponding power of the formal PN expansion parameter $\epsilon_{\rm PN}$, i.e.

\begin{equation}
|\mathbf{a}_{i \,\rm{PN}}| \propto \epsilon_{\rm PN}^{i} 
\sim \left(\frac{v}{c}\right)^{2i} 
\sim \left( \frac{r_{\rm g}}{R}\right)^i,
\end{equation}

\noindent where $v$ and $R$ are the relative velocity and separation of the MBHB, while $r_{\rm g} = GM/c^2$ is the gravitational radius, with  $c$  the speed of light in vacuum, $G$ the gravitational constant, and $M$ the binary total mass. The PN terms of integer order are conservative, whereas the half-integer order terms are dissipative radiation reaction terms related to the emission of gravitational radiation.

The PN corrections, and thus the GW emission, are still negligibly small when the binary separation is of the order $a \sim a_\mathrm{h}$ (see Eq.~\ref{eq:a_hard}). The PN radiative loss terms in the equations of motion start dominating the evolution when the binary separation drops down to $a \sim a_{\mathrm{GW}} \sim 0.01 \times a_\mathrm{h}$ (e.g. \citealt{1996NewA....1...35Q}; \citealt{2006ApJ...651..392S}; \citealt{2018ApJ...864..113R}). This corresponds to a typical physical separation of $a_{\mathrm{GW}} \sim 10^{-4}$--$10^{-3}$~pc for equal-mass binaries with individual MBH masses of $M_{\rm BH} \sim 10^{6}$--$10^{7}$~M$_{\odot}$, with the required separation being correspondingly smaller for lower-mass MBHs. However, it is also possible for the gas component to follow the binary essentially all the way down to merger, and thus gas can be present even in this GW-emission stage~\citep[see Section~\ref{sec:hardening_gas} and][]{2015MNRAS.447L..80F,2018MNRAS.476.2249T}. In a novel attempt to quantify the effects of environmental perturbations (such as gas friction and torques) with respect to those due to PN corrections, \citet{2021arXiv210200015Z} found simple analytical expressions for the regions of phase space wherein the two are comparable.\\

\noindent$\bullet$ {\bf The GW-driven inspiral}

If we assume that the evolution of the system is purely driven by GW emission, to leading order, the secular evolution of the Keplerian orbital parameters of the isolated MBHB can be approximated following the seminal work by  \cite{1964PhRv..136.1224P}. While the orbital period scales as $t_{\rm}\sim (a/r_{\rm g})^{3/2}$ (where $a$ is the semi-major axis), the radiation reaction time-scale scales instead as $t_{\rm RR}\sim (a/r_{\rm g})^{4}$. The inequality $a\gg r_{\rm g}$ implies that $t_{\rm orb} \ll t_{\rm RR}$: the binary is thus approximately Keplerian, and the orbital parameters $a$ and $e$ change slowly. Using angular brackets to denote orbit averaging, the evolution of the binary's semi-major axis is described by (\citealt{1963PhRv..131..435P}; \citealt{1964PhRv..136.1224P})

 \begin{equation} \label{eq:peters1}
\left<\frac{d a}{d t} \right>_{\rm GW} = - \frac{64}{5} \frac{G^3 m_1 m_2 M}{c^5 a^3 (1-e^2)^{7/2}} \biggl(1+\frac{73}{24} e^2 + \frac{37}{96}e^4 \biggr) = - \frac{64}{5} \frac{G^3 m_1 m_2 M}{c^5 a^3}f(e),
\end{equation} 
where $f(e) = (1+ 73 e^2 /24 + 37 e^4 /96) (1-e^2)^{-7/2}$ is the so-called eccentricity enhancement function, whereas $m_1,m_2$, and $M$ denote the masses of two bodies and the binary total mass, respectively. The evolution of the eccentricity $e$ is instead dictated by

\begin{equation} \label{eq:peters2}
\left<\frac{d e}{d t}\right>_{\rm GW} = -\frac{304}{15}  \frac{G^3 m_1 m_2 M}{c^5 a^4
(1-e^2)^{5/2}} e \biggl( 1+\frac{121}{304} e^2 \biggr).
\end{equation}

The overall minus sign ensures that both the semi-major axis and the eccentricity decrease as the binary evolves, resulting in an increasingly tighter and more circular binary orbit. 
 
For $e \ll 1$, Eq.s~\ref{eq:peters1} and \ref{eq:peters2} imply that the eccentricity decays faster than the orbital separation. This causes a fast circularization of initially eccentric systems and, unless the initial eccentricity is extremely high, binaries in this GW-driven regime would mostly be circular.

An important caveat to the rather simplistic discussion presented above is that when all PN corrections up to a given order (e.g. 3.5~PN) are included in the motion of the MBHB, the standard Keplerian elements are no longer constant over an orbit, but rather they oscillate, especially near the pericentre of an eccentric orbit \citep[e.g.][]{2006LRR.....9....3W,2019ApJ...887...35M,2004PhRvD..70j4011M}. 
When MBHs are spinning, their spins (both modulus and direction) also participate in shaping the dynamics of inspiraling binaries and profoundly affect the orbital motion \citep{1994PhRvD..49.2658C,1994PhRvD..49.6274A,1995PhRvD..52..821K,2015PhRvL.114h1103K,2015PhRvD..92f4016G}, as well the emitted GWs. 

Despite the inclusion of high PN order being able to describe very well the evolution down to a few gravitational radii, at binary separation of about $a\sim 6 r_{\rm g}$ and below, the strongly non-linear gravitational field makes the PN expansion to become unreliable, and full GR simulations are necessary \citep[see, e.g.][]{2005PhRvL..95l1101P,2006PhRvL..96k1101C,2006PhRvL..96k1102B}. \\

\noindent$\bullet$ {\bf The GW inspiral time-scale}

A reasonable question to ask is the following: if the binary enters the GW-driven phase of its evolution with certain initial orbital parameters (semi-major axis and eccentricity), how much time will it take to merge?
    
A proper answer requires the numerical integration of the evolution Equations \ref{eq:peters1} and \ref{eq:peters2}, as discussed above. Still, a reasonable analytical approximation, valid for mildly eccentric binaries, is given by the so-called Peters' time-scale \citep{1964PhDT........51P} i.e.

\begin{equation}
   t_{\rm P} = \frac{5c^5(1 + q)^2}{256 G^3 M^3q} \frac{a_0^4}{f(e_0)} \approx 0.32\frac{(1 + q)^2}{q f(e_0)} \left(\frac{a_0}{\rm{AU}}\right)^4 \left(\frac{M}{10^6 \rm{M}_{\odot}}\right)^{-3} \rm{yr}.
\end{equation}

\noindent where $a_0$ and $e_0$ are the initial semi-major axis and eccentricity, respectively. The interpretation of this time-scale is simple: the more massive and the more compact the binary is, the faster it will decay. Moreover, for a given semi-major axis, highly eccentric orbits decay much faster than circular ones, simply because the two MBHs at pericentre are closer to each other and the strong GW emission efficiently extracts a large amount of orbital energy. 
    
Because of its simplicity, this formula has been widely used to estimate the decay time-scale of compact binaries, as done in many preceding Sections when the efficiency of GW-induced decay must be compared against other factors that affect the orbital evolution. 

While Peters's formula often suffices as an order-of-magnitude estimate for the decay time-scale, it has two major limitations that are known but often overlooked. First of all, it is only a lower bound to the results of numerical integration, and it can underestimate the numerical time-scale by a factor of 1--8 \citep{1964PhRv..136.1224P}. In addition, Peters and Mathews' analysis assumes that the binary follows a Keplerian path, and that it only radiates according to the quadrupole formula: both of these assumptions are only true at the lowest order in the PN expansion. Corrections to the classic formula have been recently presented in \citet{2020MNRAS.495.2321Z} up to first order in PN theory. 
For orbits that are either eccentric or highly relativistic, one can expect errors of order ten to be accounted for by the correction factors. Recently, \citet{2021arXiv210200015Z} expanded further on those results,  obtaining a new spin-dependent correction. The corrected formula reads, for a given total mass and mass ratio, in the case of highly eccentric orbits:

\begin{equation}
 t_{\rm PN}(a_0,e_0,s_1) = \label{eq:corrpet} 
 \underbrace{\frac{5c^5(1 + q)^2}{256 G^3 M^3q} \frac{a_0^4}{f(e_0)}}_{\rm{Peters' \, formula}}
  \underbrace{R(e_0) \exp\left(\frac{2.8r_{\rm S}}{p_0} + s_1 \frac{0.3r_{\rm S}}{p_0} +|s_1|^{3/2} \left(\frac{1.1r_{\rm S}}{p_0}\right)^{5/2}\right)}_{\rm{eccentricity, \, spin \, and \, PN \, correction}},
\end{equation}

\noindent where $p_0 = a_0(1-e_0)$, $r_{\rm S} = 2G M/c^2$, $R(e_0) = 8^{1-\sqrt{1-e_0}}$, and $s_1 \equiv S_1 \cos\theta$, with $S_1$ being the magnitude of the spin of the most massive MBH and $\theta$ the angle between that MBH's spin vector and the orbital angular momentum vector. Adopting more accurate GW-emission time-scales in studies devoted to LISA would be beneficial to improve the investigations of MBHB dynamics and merger rates.\\

\noindent$\bullet$ {\bf MBH coalescence and kicks/recoils}
\label{MBH_kicks}

When MBHs finally reach coalescence, the emitted GWs are responsible for dissipating not only energy and angular momentum (causing the shrinking of the orbit), but also linear momentum~\citep{1961RSPSA.265..109B, 1962PhRv..128.2471P,1973ApJ...183..657B}. Conservation of linear momentum implies that the MBH left behind following a merger has a non-zero recoil velocity (or ``kick''), which is independent of the MBH mass and depends only on the mass ratio, spins, and eccentricity of the merging binary. While energy and angular momentum are dissipated more gradually during the inspiral, linear-momentum emission is strongly peaked during the last few orbits prior to and at merger (e.g. \citealt{2008PhRvD..77l4047B,2018PhRvD..97j4049G}). This implies that, although PN predictions are possible \citep{1983MNRAS.203.1049F,1995PhRvD..52..821K,2005ApJ...635..508B}, kicks can be modelled accurately only using numerical-relativity simulations (e.g. \citealt{2007PhRvL..98w1102C, 2007PhRvL..98w1101G,2007PhRvD..76f1502T, 2011PhRvL.107w1102L}). Such kind of (very expensive) simulations show that MBH recoils can reach velocities as large as $\sim$5000~km~s$^{-1}$ (the so-called ``superkicks''). A variety of tools, ranging from fitting formulae \citep{2007ApJ...659L...5C,2007PhRvL..98i1101G, 2008PhRvD..77d4028L,2013PhRvD..87h4027L,  2010ApJ...719.1427V, 2016PhRvD..93l4066G} to full surrogate models  \citep{2018PhRvD..97j4049G, 2019PhRvL.122a1101V} calibrated on numerical-relativity results are now available to quickly estimate MBH kicks for large parameter-space explorations. 

Kicks around 1000~km~s$^{-1}$ imply that MBH merger remnants might have velocities that exceed the escape speed of their galactic hosts~\citep{1989ComAp..14..165R,2004ApJ...607L...9M, 2015MNRAS.446...38G}. The astrophysical consequences of recoils are several. Amongst them, we find that energetic kicks can critically modify the merger rate of MBHs, induce scatter in the correlations between MBHs and galaxy hosts, deplete low-mass galaxies of MBHs, create cores in the central stellar distribution, hinder the formation of $>10^9 \msun$ MBHs powering $z>6$ quasars, and generate a population of wandering MBHs and AGN. These possibilities were explored by various authors \citep[e.g.][]{2004ApJ...613...36H,2004ApJ...613L..37B,2005MNRAS.358..913V,2007MNRAS.382L...6S,2008ApJ...678..780G,2007ApJ...663L...5V,2008ApJ...682..758S, 2008ApJ...686..829H,2008MNRAS.390.1311B,2011MNRAS.412.2154B,2020ApJ...896...72D,2020arXiv200606647S}. In the LISA context, the occurrence of kicks might have important consequences for the MBHB event rate, although the assessment of their impact depends very sensitively on the assumed spin directions that can be strongly affected by the interaction with the surrounding environment \citep{2007ApJ...667L.133S,2007ApJ...661L.147B,2010PhRvD..81h4054K,2010ApJ...715.1006K,2012PhRvD..85l4049B,2013ApJ...774...43M,2015MNRAS.451.3941G,2020MNRAS.496.3060G, 2010MNRAS.402..682D}. Furthermore, recoiling MBHs would produce a post-merger EM signature that can aid in the identification of the merged MBH \citep{2005ApJ...622L..93M,2007ApJ...662L..63S,2008ApJ...684..835S,2008ApJ...676L...5L,2010MNRAS.404..947C,2010MNRAS.401.2021R}.

Potential EM signatures of GW recoils are reviewed by  \citet{2012AdAst2012E..14K}. If the recoiling MBHs carry the bound gas as they recoil, they would shine as off-nuclear AGN \citep{2008MNRAS.390.1311B,2008ApJ...687L..57V}. The most characteristic signature is a set of broad emission lines, which led to the identification of several observational candidates 
\citep{2008ApJ...678L..81K, 2012ApJ...752...49C,
2011ApJ...738...20T, 2014MNRAS.445..515K, 2017A&A...600A..57C,
2017ApJ...840...71K, 2017ApJ...851L..15K} and the development of various detection strategies \citep{2014ApJ...795..146L, 2016MNRAS.455..484R,
2016MNRAS.456..961B}. 
Identification of such candidates is a particularly active field of research and is a difficult task (see Section~\ref{sec:sec_2_4}); some candidates cited above have already been disproved in the recent years. Detection or confirmation of some candidates would prove that indeed MBHBs merge in the Universe, supporting LISA's science case. 

Recoiling systems are also expected to present GW signatures. These include a relative Doppler shift between inspiral and ringdown \citep{2016PhRvL.117a1101G}, different higher-order mode content \citep{2018PhRvL.121s1102C}, and statistical correlation with the spin properties \citep{2020PhRvL.124j1104V}. \cite{2016PhRvL.117a1101G,2018PhRvL.121s1102C,2020PhRvL.124j1104V} all agree that the direct detectability of GW signatures from kicked MBHs is well within the reach of LISA. 


\subsection{MBH origin and growth across the cosmic time}
\label{sec:MBHoriginandgrowthacrosscosmictime}
{\bf \textcolor{black}{Coordinators:}
Pratika Dayal,
John Regan\\
\noindent \textcolor{black}{Contributors:}
Pau Amaro-Seoane,
Abbas Askar,
Razvan Balasov,
Emanuele Berti,
Pedro R. Capelo,
Laurentiu Caramete,
Monica Colpi,
Davide Gerosa,
Melanie Habouzit,
Daryl Haggard,
Peter Johansson,
Fabio Pacucci,
Raffaella Schneider,
Stuart L. Shapiro,
Caner Unal,
Rosa Valiante,
Marta Volonteri\\
}

MBHs are ubiquitous across space and time. Observations have revealed the likelihood that MBHs populate every massive galaxy in the Universe \citep[e.g.,][]{2013ARA&A..51..511K}, with MBHs of upwards of $10^4\, \rm M_{\odot}$ populating some, possibly large, fraction of dwarf galaxies (e.g. \citealt{2015ApJ...809L..14B,2018ApJ...863....1C,2018MNRAS.478.2576M,2019MNRAS.484..814G,2019arXiv191109678G}; \citealt{2020ApJ...898L...3B}). At the massive end of the MBH mass function, MBHs are remarkably well-centred in the cores of galaxy bulges, and their mass is tightly correlated with many properties of the galaxy host, as the stellar mass of the bulge. Luminous quasars, powered by $10^{8-9}\, \rm M_{\odot}$ MBHs, were identified when the Universe was less than a billion years old \citep[$z \sim 7.5$,][]{2019ApJ...881L..23B,2020ApJ...897L..14Y,2021arXiv210103179W}, evidence that MBH evolution started well before then.  
 
LISA will bring a wealth of new independent information on the population census and the ability of MBH mergers to contribute to the growth of MBHs, all the way to the realm of MBH "seeds" postulated by different formation models. No telescope can search for MBHs at redshifts as high as LISA can ($z>10$) , allowing us to observe an otherwise inaccessible region of the Universe. LISA will play a crucial role in, for instance, pinpointing the main formation channel of MBH seeds at high redshift, with light seeding models (numerous but low mass BH seeds) expected to drive a significantly higher merger rate at $z\gtrsim 10$ compared to heavy seeds (rare but massive BH seeds), provided that their dynamical decay is efficient (see Section~\ref{sec:stuntlight}). Several studies (e.g. \citealt{2008ApJ...684..822B, 2011PhRvD..83d4036S, 2012MNRAS.423.2533B, 2017arXiv170200786A, 2019MNRAS.486.2336D, 2019MNRAS.486.4044B, 2020ApJ...895...95P, 2020ApJ...904...16B, 2021MNRAS.500.4095V}) have shown that LISA will provide a unique view of the merger history of MBHs up to very high redshift ($z\sim 20$).

In this section, we first review our current understanding of the different seed MBH formation mechanisms, the resultant seed masses, and the outstanding questions in the field.
In the second part, we examine the growth of MBHs across cosmic time under the assumption that a population of seeds, formed at $z \sim 10$--30, grow over cosmic time via the two key mechanisms of gas accretion and coalescence.  Growth by accretion can typically occur by a stable influx of material, usually organised in a thin/thick accretion disc (e.g.~\citealt{1976MNRAS.175..613S,2014ApJ...796..106J}) or via chaotic accretion with cold gas raining from random directions (e.g.~\citealt{2013MNRAS.432.3401G,2015A&A...579A..62G,2017ApJ...845...80V}).
Coalescence between MBHs also contributes to their growth, with some small fraction of the total mass being radiated away via GWs. Through this section, we discuss how LISA will be crucial in shedding light on the origin and evolution of MBHs.

\subsubsection{MBH seeds: formation mechanisms}\label{sec:formation}

LISA will be  sensitive to the detection of MBHs with masses in excess of a few thousand solar masses out to high redshifts, and therefore uniquely able to probe how MBHs form in the first galaxies. Theoretical models show that seeding mechanisms are crucial to make detailed predictions for the number density of MBH mergers and hence for predicted detections by LISA (see detailed description in Section~\ref{sec:StatisticsOnMBHMergers}). Correctly modelling the seeding of MBHs thus becomes of paramount importance to prepare and then interpret LISA data. 

\begin{figure}
    \centering
    \hspace*{-1.cm}
    \includegraphics[scale=0.38]{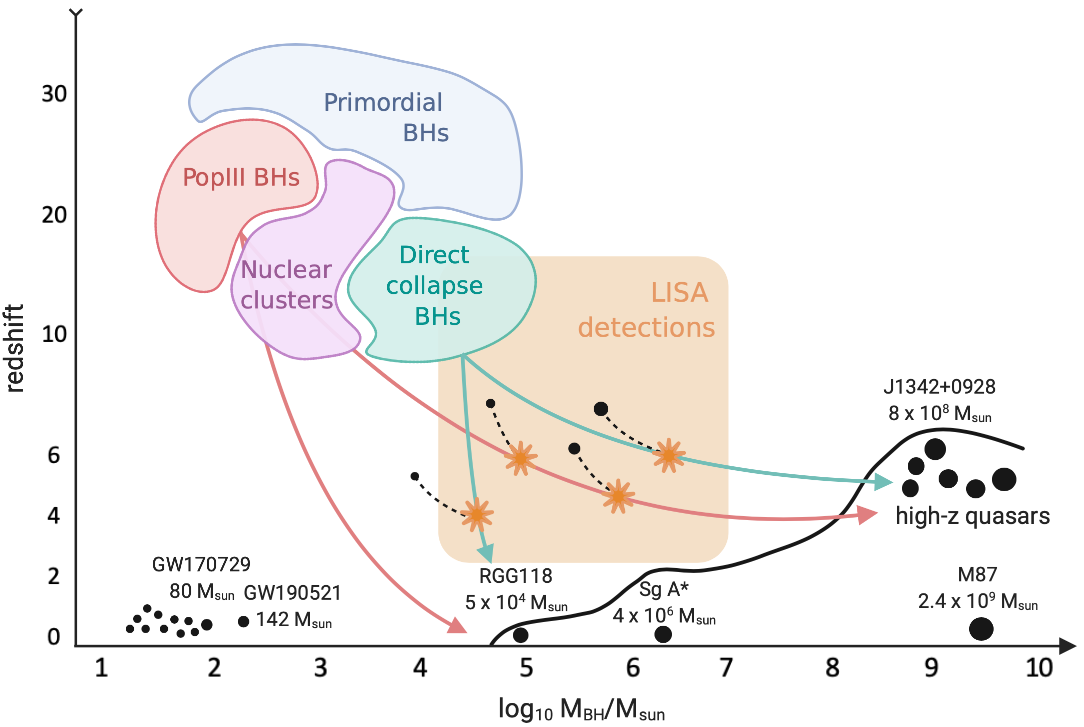}
    \caption{
    Pathways towards the formation of MBHs are numerous, and include the collapse of first-generation stars ({\it Pop~III BHs}, M$_{\rm BH} \lesssim 10^3  \msun$), the collapse and/or coalescence of massive stars formed in compact stellar clusters ({\it nuclear clusters}, $10^2 \msun \lesssim M_{\rm BH} \lesssim 10^4 \msun$), the collapse of SMS formed in primordial environment ({\it direct collapse}, $M_{\rm BH} \gtrsim 10^3 \msun$), and the collapse of cosmological density perturbations ({\it primordial BHs}, $1 \msun \lesssim M_{\rm BH} \lesssim 10^{10} \msun$). The shaded orange region shows the redshift and MBH mass ranges of LISA, and the orange starburst symbols the LISA detections. LISA will significantly extend the current MBH EM detections, shown below the curved solid black line (from the local Universe at $z\sim0$ to the high-redshift quasars at $z\geqslant 6$). Figure credit: Melanie Habouzit}
    \label{fig:formation_cartoon}
\end{figure}

In this section, we focus on formation pathways which can lead to the production of such seeds ($M_{\rm BH}=10^{2}$--$10^{6}\,\rm M_{\odot}$) - see Fig. \ref{fig:formation_cartoon}. This includes {\it (a)} seeds from metal-free Population~III (Pop~III) stars; {\it (b)} seeds originating from the dynamical processes in dense stellar clusters; {\it (c)} seeds born from the collapse of supermassive stars (SMSs); and {\it (d)} primordial MBH seeds. More detailed information on each of these scenarios is dealt with in other reviews (e.g. \citealt{2010A&ARv..18..279V, 2016PASA...33....7J, 2017PASA...34...31V, 2019arXiv191105791I}; \citealt{2021NatRP...3..732V}). Here we outline the mechanisms behind each pathway as well as underlining outstanding issues in the field, especially those pertinent to LISA. The consequences on the detection rate and properties of mergers identified by LISA will be discussed in Section~\ref{sec:SAMs}.\\

\noindent$\bullet$ {\bf Formation of MBHs as Pop~III remnants} [M$_{\rm BH} \lesssim 10^3  \msun$]

 One of the popular explanations behind the formation of high-redshift MBHs is related to Pop~III stars, the hypothesized first-generation stars. Pop~III stars are born in $\sim$10$^5$--$10^6 \msun$ DM ``minihaloes''. The primordial gas in these first haloes is cooled primarily by H$_2$, which allows the temperature of the gas to cool to approximately 200~K \citep{2002Sci...295...93A}. This inefficient cooling channel leads to a top-heavy initial mass IMF expected for Pop~III stars compared to present day star formation \citep{2009Sci...325..601T, 2011ApJ...727..110C, 2011Sci...331.1040C}, with mass values ranging from $10 \msun$ to $10^3 \msun$ \citep{2014ApJ...781...60H}.

\indent Pop~III stars with masses $M_{*} \gtrsim 260 \msun$ will directly collapse into BHs, losing very little of their progenitor mass in the process \citep{2003ApJ...591..288H}. The retention of a significant amount of the parent star mass is expected as a result of the weak stellar winds associated with metal-free stars. As a result, a large population of Pop~III remnant BHs is expected to be left behind in these first minihaloes that are ubiquitous at early times. 
Less massive Pop~III stars will explode as SNae, enriching their surroundings with metals. As metal enrichment is extended to nearby galaxies \citep[e.g.][]{2015MNRAS.452.2822S, 2020arXiv200905499H} through both winds and halo mergers, the formation of Pop~III stars declines severely and less massive Population-II stars begin to dominate the star formation history of the Universe \citep{2015ApJ...807L..12O, 2016ApJ...823..140X}. Nonetheless, this first generation of stars leaves in its wake a large number of Pop~III remnant BHs, which may act as the seeds to future MBHs \citep{2001ApJ...551L..27M,2014ApJ...781...60H}. A key open question is therefore whether these Pop~III remnants can grow into a population of MBHs, and under what conditions rapid growth can be achieved (this is 
particularly relevant for the high-z quasars) and their mergers be expected to be detected by LISA. We will explore research in this area and the significant challenges to their growth which must be overcome in Section~\ref{sec:growth}.\\

\noindent$\bullet$ {\bf Formation of MBHs in dense stellar environments} [$10^2 \msun \lesssim M_{\rm BH} \lesssim 10^4 \msun$]
 Seed MBHs of $10^{2}$--$10^{4} \msun$ can form in dense and massive stellar clusters of $\sim$10$^{5}\, \rm M_{\odot}$ through dynamical interactions (e.g. \citealt{2008ApJ...686..801O, 2009ApJ...694..302D, 2018A&A...614A..14R}; \citealt{2022MNRAS.512.6192S}). During the early evolution of star clusters with initial central densities $\gtrsim 10^{5} \msun \rm \ pc^{-3}$, massive stars segregate to the cluster centre due to dynamical friction, where they may undergo runaway collisions resulting in the formation of very massive stars with masses of approximately $10^{2}$--$10^{3} \msun$ \citep{2002ApJ...576..899P,2004Natur.428..724P,2004ApJ...604..632G,2006MNRAS.368..121F,2006MNRAS.368..141F}. In low-metallicity clusters, such massive stars 
may collapse into an MBH \citep{2015MNRAS.451.2352K,2015MNRAS.454.3150G,2016MNRAS.459.3432M,2017MNRAS.472.1677S, 2015MNRAS.454.3150G,2020arXiv200809571R}.

Another possibility of forming an MBH in stellar clusters is through runaway mergers of stellar-mass BHs, which are expected to form from the evolution of massive stars. If stellar-mass BHs form with low-velocity natal kicks or are embedded in a dense gaseous halo \citep{2002ApJ...572..407B,2016MNRAS.456..578M,2018MNRAS.480.2011G, 2011ApJ...740L..42D}, a significant fraction can be retained within the star cluster \citep{2013MNRAS.430L..30S,2013ApJ...763L..15M,2015ApJ...800....9M,2016MNRAS.458.1450W,2018MNRAS.479.4652A,2018MNRAS.478.1844A,2019ApJ...871...38K}. While the mergers of these stellar mass sized BHs will generate GWs, their frequency ranges put them outside of the sensitivity range of LISA - they may however be detectable by future GW detectors like the 
Einstein Telescope \citep{2021MNRAS.500.4095V}.
A potential barrier to this formation scenario is that the retention of any MBH will depend on the GW recoil kicks that they receive during the merger process. If recoil kick velocities are larger than the escape speed of the cluster, then the seed MBH may be ejected out of the cluster \citep{2008ApJ...686..829H,2011ApJ...740L..42D, 2012ApJ...755...81M, 2014ApJ...794..104S, 2018MNRAS.481.2168M}.

It may also be possible to grow stellar-mass BHs through gas accretion (rather or in conjunction with mergers) inside stellar clusters. Retained stellar-mass BHs could effectively grow and become MBHs by accreting the interstellar gas inside massive stellar clusters \citep{2013MNRAS.429.2997L,2020arXiv200909156N}.
Moreover, BHs of $\sim 100 \msun$ can become more massive by growing through tidal capture and disruption of stars in dense nuclear star clusters \citep{2017MNRAS.467.4180S}. Such runaway events can grow the mass of a BH from $10^{2-3} \, \msun$ to up to $10^{5} \, \msun$ \citep{2009ApJ...695..404R,2016ApJ...819....3M,2017NatAs...1E.147A,2017MNRAS.467.4180S,2018MNRAS.476..366B,2019MNRAS.484.4665S}. \\

\noindent$\bullet$  {\bf Formation of very massive seeds in atomic cooling haloes and primordial galaxies} [$\rm{M_{BH}} \gtrsim 10^3 \msun$]
 SMSs\footnote{While the term direct collapse is often used in the literature to describe the formation of an MBH seed, that terminology is ambiguous and the formation of the intermediate stellar stage is expected in the general case \citep{2019arXiv191105791I} except perhaps under extreme conditions \citep[e.g.][]{2010Natur.466.1082M}.} 
 were originally invoked to explain the existence of quasars prior to their origin being understood as the accretion of matter on to MBHs. More recently, SMSs have been ``reinvoked'' as potential seeds for MBHs. SMSs are thought to form through the rapid accumulation of gas during the early stages of stellar evolution. If gas can be rapidly accreted with accretion rates in excess of $10^{-3} \msun$~yr$^{-1}$ \citep{2018MNRAS.474.2757H, 2003ApJ...589..677O}, then the stellar envelope remains bloated and cool (with a temperature T$_{\rm{eff}} \sim 5000$~K). Detailed numerical simulations have shown that such objects do not provide enough negative (radiative) feedback to halt accretion and the end result is an SMS \citep{2016MNRAS.459.1137S, 2018MNRAS.475.4104C, 2020arXiv200902629S}. However, sustaining this accretion rate is nonetheless challenging, due to the complex dynamics between the gas and the stellar component \citep{2020MNRAS.494.2851C, 2020OJAp....3E...9R}.

The ideal environmental conditions for SMS formation can be achieved in so-called atomic cooling haloes \citep{2009ApJ...696.1798T}, where line-emission cooling due to neutral hydrogen allows the gas to cool and condense in a sufficiently massive halo (with a virial temperature T$_{\rm vir} \sim 8000$~K and a virial mass M$_{\rm vir} \sim 5 \times 10^7 \msun$ at $z \sim 15$). The larger mass of the atomic cooling halo, compared to the minihaloes in which Pop~III stars are typically born, provides a larger baryonic reservoir for (metal-free) star formation. The key requirement for the development of an SMS is that the gas inflow onto the stellar surface remains high. Fragmentation of the gas into a (dense) stellar cluster must also be avoided for a truly SMS to form\footnote{Current research is trending towards a likely overlap between the conditions necessary for a dense stellar cluster to form and for SMS formation and hence an overlap in mass scales between SMS formation and the formation of a dense stellar cluster.} \citep{2020OJAp....3E...9R}. Fragmentation may be avoided if the gas is sufficiently  metal-poor \citep{2020MNRAS.494.2851C, 2020ApJ...892...36T}, with metallicities not exceeding $Z \approx 10^{-3}$~Z$_{\odot}$, and perhaps also if the halo is not tidally disrupted \citep{2018MNRAS.475.4104C}. Given the difficulties in achieving monolithic SMS formation, the question of whether true SMS formation can be achieved remains an open and active research question.

Radiative feedback in the Lyman-Werner (LW) band (in the energy range 11.2--13.6~eV) allows for the dissociation of H$_2$, which suppresses Pop~III star formation, allowing a halo to remain star-free (and hence metal-free). An attractive scenario here is the synchronised pair \citep{2008MNRAS.391.1961D, 2017NatAs...1E..75R} mechanism, whereby a pair of halos closely separated in time and space evolve together. The first of these haloes that forms stars could then provide the second halo with a strong enough LW background \citep{2014MNRAS.445.1056V}; the key issue with this model is that the number density of such environments may be too rare to explain the number densities of expected MBHs.

Alternative scenarios for avoiding premature star formation are to dynamically heat the gas (rather than photo-dissociating H$_2$). In this scenario, the gas can be shock-heated either through galactic collisions \citep{2016MNRAS.459.3738I} or through a rapid succession of minor and major mergers \citep{2003ApJ...592..645Y, 2014MNRAS.439.3798F, 2019Natur.566...85W}. The appeal of this scenario is that it arises more naturally through 
the mechanisms of DM structure formation and that 
the number density of MBH seed formation looks promising \citep[e.g.][]{2020MNRAS.492.3021R}
though further work on the expected number density of MBH seeds is required. 

Finally, the collisions of massive galaxies at moderately high redshifts ($z \sim 8-10)$ can lead to the direct formation of an MBH without any intermediate stage \citep{2010Natur.466.1082M, 2015ApJ...810...51M}. In this scenario (which,
in stark contrast with the atomic cooling halo scenario,
can occur also at solar metallicities) major mergers between the rare, most massive high-z galaxies funnel gas to their 
centre at rates exceeding $1000 \,\rm M_{\odot}$/yr.  The resulting accumulation of billions of solar masses of gas in a nuclear region less than a parsec in size could 
either induce the formation of a very large  SMS, and hence a massive BH seed  by direct collapse, or even directly 
form a large MBH via the radial general-relativistic instability of a supermassive protostellar precursor. Recent models
show that an accreting SMS, owing to the much higher accretion rates occurring in the merger-driven scenario, can grow
in mass much more than in the atomic cooling halo case, namely to $> 10^7 \,\rm M_{\odot}$ in absence of rotation,  
before collapsing into an MBH seed \citep{Haemmerle_et_al_2021}. \\

\noindent$\bullet$  {\bf Primordial Black Holes} 

 Primordial BHs are another plausible way to explain the formation of MBHs. Their abundance is constrained at various mass scales \citep{2020arXiv200212778C}, but they can still form a considerable fraction of DM in mass ranges $1-10^2 \msun$ \citep{Bird:2016dcv,Sasaki:2016jop,Clesse:2016vqa} and $10^{-13}-10^{-11} \msun$ \citep{Saito:2008jc,Garcia-Bellido:2017aan,Domcke:2017fix,Bartolo:2018rku,Cai:2018dig,Unal:2018yaa}. Moreveor, primordial BHs of mass ${\cal O}(10-10^5)\msun$ formed in the early universe (before recombination) could be the seeds of MBHs \citep{Duechting:2004dk,2014MPLA...2940005B, 2015PhRvD..92b3524C,2016PhRvD..94j3522N,Garcia-Bellido:2016dkw}.  The tail of their mass function\footnote{It should be noted that primordial BH mass function can also peak at BH masses higher than solar mass to form SMBHs.} reaching a few hundred or thousand solar masses can grow many orders of magnitude (depending on formation mass) via accretion and mergers \citep{Mack:2006gz,Ali-Haimoud:2017rtz,Raidal:2018bbj,Inman:2019wvr,Serpico:2020ehh,DeLuca:2020fpg}. This claim has been studied and found to be consistent with the current cosmological probes of cosmological history.

Primordial BHs are formed by large density contrasts, and the most likely stage to produce these large perturbations is during inflation. Although cosmic microwave background-scale perturbations must be Gaussian and nearly scale invariant with a typical amplitude of $10^{-5}$, the fluctuations at smaller scales can be larger. There exist characteristic signatures of these enhanced fluctuations in various multimessenger probes, including cosmic microwave background distortions (\citealt{Chluba:2012we,2017PhRvD..95d3534A,Blum:2016cjs,2017PhRvD..95l3510I,Garcia-Bellido:2017aan,Nakama:2017xvq}; \citealt{2022ApJ...926..205C}) and secondary stochastic GWs resulting from
the enhanced perturbations that re-enter the horizon in the radiation (or matter) dominated era (in particular
enhanced inflationary perturbations that produce $1-10^4 \, \msun$ primordial BHs) which also produce stochastic GWs at 
Pulsar Timing Arrays (PTA) scales \citep{2017PhRvD..95l3510I,Garcia-Bellido:2017aan,2020arXiv200907832V,2020arXiv200911853K,2020arXiv200908268D}. The next generation PTAs,  which can constrain the stochastic GW Background, as well as the cosmic microwave background experiments using spectral distortions,  will probe inflationary fluctuations so sensitively that they could conclusively test the existence of primordial BHs from inflationary perturbations \citep{Byrnes:2018txb,Inomata:2018epa,Kalaja:2019uju,2020arXiv200803289G}. We refer the reader to \cite{CosWGwhitepaper2021} for more details on primordial BHs and LISA.

\bigskip

Research into the seeding of MBHs remains a highly active area of research. In an era where vigorous development of both semi-analytical models and full numerical calculations continues apace (see Section~\ref{sec:StatisticsOnMBHMergers}), understanding the mechanisms of MBH seeding becomes all the more important. A definitive pathway to forming MBHs remains an open question. An important metric of success for any formation model is to explain naturally the current abundance of MBHs of all masses in the nuclei of galaxies. These have a currently measured number density of n$_{\rm MBH} \sim (0.2-1.0)\times10^{-2}$ Mpc$^{-3}$ at $z=0$, depending on how far down in mass function is integrated \citep[e.g.][]{2007MNRAS.378..198G, 2009NewAR..53...57S, 2016ApJ...830L..12T}.  A key goal in any of the seeding models discussed above is therefore a calculation of the resulting number density of MBHs in the Universe as a function of redshift. So far, calculations within the community have varied significantly between approximately $10^{-3}$--10$^{-9} \ \rm{comoving\, Mpc}^{-3}$ for the very massive seeds formed in atomic cooling halos (\citealt{2008MNRAS.391.1961D}; \citealt{2012MNRAS.425.2854A, 2014MNRAS.442.2036D,  2014MNRAS.443..648A, 2016MNRAS.463..529H, 2019Natur.566...85W}), while we expect more seeds from e.g, the Pop~III remnant formation mechanism.  For reference, the number density of galaxies in the Universe today is ${\sim} 10^{-1}\, \rm{comoving\, Mpc}^{-3}$, and the number density of quasars at $z\sim 6$ is ${\sim}10^{-9} \ \rm{comoving\, Mpc}^{-3}$. MBH formation needs to explain both the population of high-redshift quasars, and the population of MBHs in the local Universe. 
  
A central challenge of models in the next decade leading up to LISA's launch will be to reduce the number density uncertainties associated with different models of MBH seed formation. A focal point of simulations in the next  decade will be to accurately model the assembly of galaxies including modelling the environments, 
in a cosmological context, in which MBH seeds can form. Given the large dynamical range of nonlinear physical processes required to form MBH seeds, this is challenging. The use of focused, high-resolution and relatively large-scale numerical simulations with detailed (Pop~III) star formation and MBH formation prescriptions will ultimately be required to break the current degeneracies between models which currently exist and provide models to be used for inference with LISA detections.

\subsubsection{MBH growth across time and space} \label{sec:growth}

LISA will  measure not only the masses but also the spins of massive black holes.
In this section, we discuss three compelling open questions: {\it how do MBH seeds grow across cosmic time? What is the impact of such growth on the spin of an MBH? What can the final spin reveal about its past accretion history?} In the discussion of these issues, throughout this section, we will differentiate between \textit{light} seeds and \textit{heavy} seeds. \textit{Light} seeds have masses of at most $10^3 M_{\odot}$ and are typically those
formed by the first generation of metal-free stars, while \textit{heavy}  seeds 
are those with higher masses that can result from stellar dynamical processes or from the direct collapse scenarios discussed above. 

\paragraph{How to grow light seeds}\label{sec:stuntlight}

 Pop~III remnants are predicted to have low mass, $\leqslant 10^{3}\, \rm M_{\odot}$. In order for these seeds to grow massive enough to be in the LISA band, or to grow massive enough to even become the extremely massive quasars that we observe at $z\sim$6--7, they would need to clear two main hurdles. If not formed in the centre of their galaxies, these seeds must sink efficiently to the centre, but also sustain efficient accretion for a significant fraction of their lifetime. To produce the population of high-redshift quasars, they have to sustain near-Eddington accretion rates for nearly a Gyr \citep{2001ApJ...552..459H}. In the conventional picture of a spherically symmetric accretion flow whose energy loss is only controlled by radiation
propagating isotropically, 
the Eddington limit expresses a maximum allowed accretion rate. Therefore, seeds would have to grow at near the maximum rate allowed for their entire lifetime, unless a mechanism for super-Eddington accretion is invoked by resorting to more complex configurations of the fluid flow and
radiation field, and to a different energy transport mechanism.

High accretion efficiency is challenging to explain physically, given that radiative feedback both from the surrounding stellar component and MBH growth can unbind gas in the vicinity of the seed BH, thus preventing further growth. These hurdles were first examined in the early 2000s \citep{2002MNRAS.332...59O, 2003MNRAS.346..456O, 2004ApJ...610...14W}, with each study finding that Pop~III BHs initially find themselves in low-density environments within the galaxy, where they are unable to grow. Expanding on earlier studies, \cite{2018MNRAS.480.3762S} investigated the growth of more than 15,000 light-seed BHs 
using the Renaissance simulations and found that none were able to grow by more than 10 percent for the 300~Myr for which their growth was followed. This time period represents a significant fraction of the Hubble time at this epoch. These predominantly numerical works have been confirmed by semi-analytical approaches which also find that light seeds struggle to achieve significant growth \citep[e.g.][and references therein]{2016MNRAS.457.3356V, 2017ApJ...850L..42P}.

A mechanism of rapid growth may be required in order to grow light seeds, both to help stabilise their orbits (see Section \ref{sec:MassiveBlackHolesAndTheirPathToCoalescence}) within the galactic centre and to allow them evolve into MBHs, as examined in Section~\ref{sec:formation}. A number of studies have also shown that light seeds can grow through super-Eddington accretion given the correct environmental conditions \citep{Alexander_2014, 2016MNRAS.459.3738I, 2019arXiv191105791I, 2016MNRAS.456.2993L, 2016MNRAS.458.3047P, 2017MNRAS.471..589P}.

In either case, a growing BH must reach a critical mass before it can sink to the centre of the potential and become a central MBH.  Recent investigations by \cite{2019MNRAS.486..101P} have shown that MBHs with masses M$_{\rm BH} \lesssim 10^5 \msun$ are unable to sink via dynamical friction as the stellar component of high-redshift galaxies tends to be too irregular. This leads to a population of wandering MBHs that cannot efficiently accrete gas or merge with other BHs. The idea of a population of wandering MBHs is not new; this population has been previously associated with galaxy mergers which result in off-nuclear MBHs from the failure of them to reach the centre of the merger remnant \citep[e.g.][]{2010A&ARv..18..279V}. Nonetheless, more 
recent, high resolution, simulations have shown that wandering MBHs may result from seeds 
with masses M$_{BH} \lesssim 10^5 \msun$ that are unable to settle to the centre of the 
galactic potential. 
This result has been confirmed by other high-resolution simulations 
which show that a large population of wandering MBHs with M$_{\rm BH} \lesssim 10^5 \msun$ is likely in most, if not all, galaxies \citep{2018ApJ...857L..22T, 2019MNRAS.482.2913B, 2020OJAp....3E...9R}. Interestingly, this result has also been tentatively confirmed by observations of off-center MBHs in galaxies \citep{2020ApJ...888...36R}. However, if associated with a compact massive star cluster, dynamical friction will be greater, and such off-centre MBHs may be transiting rather than stalled.  Once MBHs exceed M$_{\rm BH} \sim 10^6 \msun$, they are less prone to ``jittering'' \citep[but see][]{2021MNRAS.508.1973M}, although they still remain susceptible to ejections (via triple interactions and GW recoils), the velocities of which depend on MBH mass ratios rather than absolute mass, although of course retaining the MBHs depends on the potential well of the galaxy. 

\paragraph{Accretion versus MBH mergers}\label{sec:accretion_vs_merger}

\begin{figure*}[ht]
\centering
\includegraphics[angle=0,width=.9\textwidth]{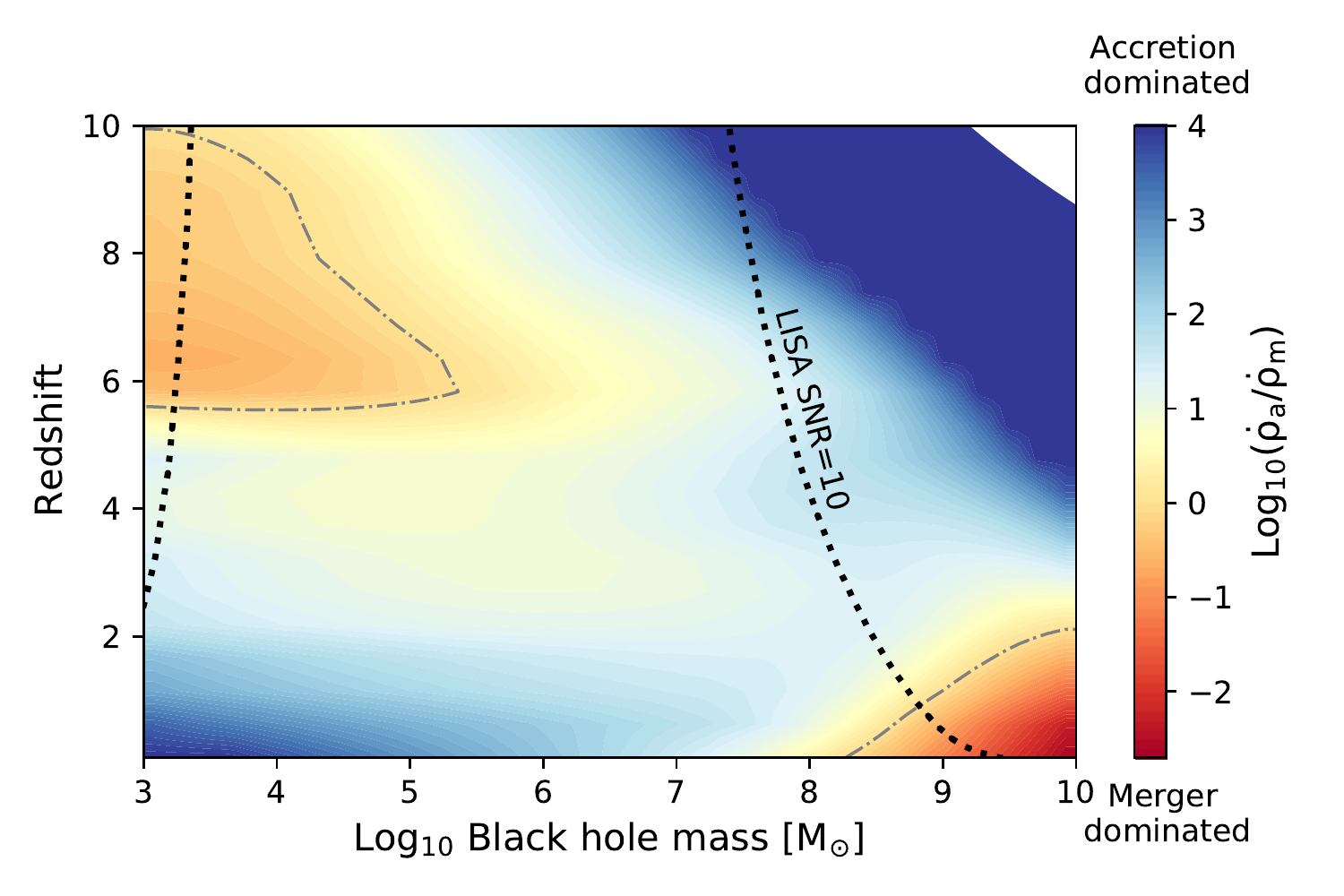} 
\caption{\label{fig:fabio}Predictions on the relative importance of MBH growth by gas accretion (blue shades) and mergers (red shades). $\dot{\rho}_a$ and $\dot{\rho}_m$ are the predicted mass growth rates by gas accretion and mergers, respectively. The contour where $\dot{\rho}_a = \dot{\rho}_m$ is represented with a dash-dotted line. The region corresponding to a LISA signal-to-noise ratio $\geq10$ for MBHBs with mass ratio 0.2  is delimited within the two thick, dotted lines. The whited-out area on the top right corner indicates the region of the parameter space where no MBHs should be present (adapted from \citealt{2020ApJ...895...95P}).} 
\label{fig:history}
\end{figure*}
Despite large uncertainties in the physical parameters that enable a mapping between luminosity and mass (e.g. duty cycle, matter-to-energy accretion efficiency, Eddington ratios, and bolometric corrections; see, e.g.  \citealt{2009ApJ...696.1798T}), a consistent picture is now emerging. Observations and theoretical models suggest that most of the mass growth over cosmic time occurred via gas accretion, and that more massive MBHs grew at earlier cosmic times, whereas lighter MBHs were still growing at $z \lesssim 1$ \citep{1982MNRAS.200..115S, 2004MNRAS.351..169M, 2008MNRAS.388.1011M, 2009ApJ...690...20S}. Assuming a combination of light and heavy MBH seeds at $z\sim 20$--30, recent studies have confirmed that growth by gas accretion is dominant for most MBH masses during a large fraction of the evolution of the Universe ($0\leq z \leq 9$--10), especially for $M_{\rm BH} > 10^6 \msun$ and $z < 8$ \citep{2020ApJ...895...95P, 2021MNRAS.500.2146P}.

Growth by mergers -- which we recall can at most double an MBH mass at each merger -- can become dominant for $M_{\rm BH} < 10^{4-5} \msun$ at $z > 6$ \citep{2019MNRAS.486.2336D, 2021MNRAS.500.2146P}, and for $M_{\rm BH} > 10^8 \msun$ at $z < 2$ \citep{2020ApJ...895...95P}. This is possible if one or more of the following conditions are met: {\it (a)}the number density of MBHs is large, and {\it(b)} the cold gas available for accretion is scarce given that the accreted mass fraction depends on the richness (over-density) of the environment \citep{2014MNRAS.440.1590D}. The first condition can be met at high redshifts if light seeding mechanisms are dominant, leading to a large number density of MBHs. This could in turn result in frequent mergers, although light seeds are unlikely to merge and sink to the centre as shown in Section \ref{sec:MassiveBlackHolesAndTheirPathToCoalescence}. The second condition can be verified at $z \lesssim 1$ \citep{2010MNRAS.406...43P}. Predictions on the contribution of mergers to the cosmic growth of MBHs strongly depend on a multitude of parameters, many of which are unknown or loosely constrained. For example, the number density of heavy MBH seeds can vary over $\sim$6 orders of magnitude (at a given redshift $z \gtrsim 8$) in modern cosmological simulations (see, e.g. \citealt{2016MNRAS.463..529H, 2019PASA...36...27W}), with huge uncertainties being introduced by the time-scale on which MBHs can actually merge \citep[e.g.][]{2019MNRAS.486.2336D,2020ApJ...904...16B}.  Nonetheless, the presence of partially-depleted cores in massive galaxies offers the promise of a substantial number of MBH mergers at least at late cosmic times \citep{1980Natur.287..307B, 2004ApJ...613L..33G, 2013ApJ...768...36D}.

 Despite significant unknowns, e.g. the contribution of obscured accretion, which is invisible in all bands apart from X-rays and higher energies, \citep[e.g.][]{2005MNRAS.357.1281W, 2009ApJ...693..447F, 2015A&A...574L..10C}, we now have a clear picture of growth by accretion (see Fig.~\ref{fig:fabio}). LISA, along with future third-generation GW observatories (e.g. the Einstein Telescope and/or Cosmic Explorer), is however the only way to actually measure the merger history of the full MBH mass spectrum, and this is what is expected to be delivered after its launch. However, as stressed already, to accurately assess the role that LISA will play in constraining the relative role of MBH mergers and accretion in MBH growth, theoretical models have to be refined in order to allow for inference on the astrophysical picture by comparing the data stream to predictions.

\paragraph{Feedback as a barrier to MBH growth}\label{sec:feedback_barrier}

As detailed in Section~\ref{sec:MassiveBlackHolesAndTheirPathToCoalescence}, the ionizing radiation that emerges from the innermost parts of the MBHs’ accretion flows can render gas dynamical friction inefficient for a range of physical scenarios \citep{2017ApJ...838..103P, 2019ApJ...883..209P}, although this depends on the surrounding gas 
environment \citep{2020MNRAS.496.1909T}. This can lengthen the inspiral time of MBHs and reduce the MBH pairing probability \citep{2020ApJ...896..113L}. The suppression of MBH pairing is most severe in galaxies with MBH pairs with mass $< 10^8$~M$_\odot$ and low mass ratio, which are direct progenitors of the merging binaries targeted by LISA. See Section~\ref{sec:MoreComplexDensityProfiles} for additional details.

Secondly, both hydrodynamic cosmological simulations with a physical model for light seed MBH formation \citep{2017MNRAS.468.3935H} and semi-analytical models \citep{2020ApJ...904...16B} converge on the fact that the number of MBHs growing enough to enter the LISA band depends on the strength of SN feedback. In case of strong feedback, SN winds can expel gas from the nuclear region of relatively low-mass galaxies ($M_{\star}\leqslant 10^{10}\, \rm M_{\odot}$), depleting the gas reservoir of the MBHs \citep{2015MNRAS.452.1502D}. This prevents the MBHs to grow in mass until the gravitational potential well of their host galaxies is deep enough to confine again the cold gas close to the central region. The MBHs may remain too light to be detected by LISA. 

In addition to affecting the merger rates, strong feedback generated by the MBHs themselves can significantly slow down the growth of MBH seeds. As shown, e.g. in \citet{2019MNRAS.486.3892R} the strong outflows generated by the jets are able to deplete a region of $\sim 0.1$ pc around the seed. Although the outflow generally does not reach the escape velocity from the host galaxy, it does suppress the growth for a time-scale comparable to the dynamical time. A super-Eddington ($\dot{M}_{\rm BH} > \dot{M}_{\rm Edd} = L_{\rm Edd}/c^2$)
accretion rate would then translate into a time-weighted, effective accretion rate of 0.1--0.5 the Eddington rate, significantly slowing down the growth of the MBH over $\sim 0.5$ Gyr by factors $\sim 30-3000$, when compared to the growth required to match the observations of $z\sim 7$ quasars. While
this is potentially an issue to explain the brightest high-z quasars, it could act to increase event rates in the LISA band at the highest redshifts as heavy BH seeds could remain longer within the mass range where LISA is most sensitive.

Finally, galaxies can also experience external radiative feedback due to the heating background created by reionization photo-evaporating gas from the outskirts of low-mass galaxies in ionized regions \citep{2010MNRAS.408.1139V, 2018PhR...780....1D}. However, this feedback has almost no effect on the mass build-up of MBHs in the early Universe since the MBHs of such reionization feedback affected galaxies are already accretion-starved due to SN feedback \citep{2019MNRAS.486.2336D}.

\bigskip

The variety of astrophysical processes involved in modelling MBH growth, described in this Section, highlights that one of the challenges ahead of us to prepare for LISA is to assess degeneracies that can affect the interpretation of LISA's data. Overall, the large number of parameters and scales involved makes this a complex problem -- at the same level of galaxy formation. Progress in delivering realistic models that can be compared to LISA's detection will require on the one hand more detailed investigations in all the subfields, and on the other hand a way to consolidate these results into coherent models.

\paragraph{Spin evolution of MBHs under accretion and mergers}\label{sec:spin}

LISA has a unique potential in providing measurements of the spins of merging MBHs: this means a theoretical understanding of how MBH spins evolve is necessary in order to be able to interpret LISA's results. Accretion and mergers establish profound links between the spin and the mass of the MBHs, which therefore have to be studied jointly.
In the accretion process, the spin is a critical physical parameter, as it determines the radiative efficiency. For a geometrically thin accretion disc, the efficiency of converting mass into light  varies from 0.057 for a non-spinning MBH to 0.43 for a maximally spinning MBH \citep[e.g.][]{1973blho.conf..343N}. This has a direct impact on the rate of MBH mass growth, on the amount of radiated energy, and on the spin magnitude and orientation at the end of an accretion episode. Also, a key manifestation of the spin when an MBH is accreting from a magnetized plasma is the launch of a collimated jet of matter and radiation which directly tracks its orientation \citep{1977MNRAS.179..433B}.  The link between spin, accretion and jet power/efficiency has started being compared to observations \citep{Unal:2020add} and being used to set lower bounds on AGN spins \citep{2020arXiv201212790U}. 

Spins determine how efficiently the accreted matter is transformed into energy, but in turn the way in which MBHs accrete gas has a crucial bearing on their spins: depending on the accretion geometry, the resulting MBH spin's magnitude and direction can vary widely. Taking the limiting case of prolonged coherent accretion from a viscous disc, the spin can increase up to its limiting value of 0.998 (\citealt{1974ApJ...191..507T}; see also \citealt{1998ApJ...504..419P}; \citealt{2004ApJ...602..312G}) after the MBH has accreted an amount of gas comparable to its initial mass, regardless of the flow being initially prograde or retrograde \citep[][]{1970Natur.226...64B}.  The spin in this case gets aligned with the angular momentum of the disc from which it is fed \citep{1975ApJ...195L..65B}, and the time-scale for the alignment is short  ($10^5$~yr) compared to the typical time for mass growth. At the other extreme, chaotic accretion, made up of randomly oriented small-mass accretion events, results instead in an erratic orientation of the spin and, in general, in a spun-down MBH \citep[][]{2006MNRAS.373L..90K, 2008MNRAS.385.1621K}. Several semi-analytical models of MBH evolution have included either one \citep[e.g.][]{2005ApJ...620...69V} or both \citep[e.g.][]{2007ApJ...667..704V,2008ApJ...684..822B,2012MNRAS.423.2533B} of these two limiting cases.  More recent semi-analytical models \citep[][]{2014ApJ...794..104S} have included accretion flows that are neither perfectly coherent nor perfectly isotropic depending on the fuelling geometry \citep[][]{2013ApJ...762...68D}. These studies, together with numerical works that follow the evolution of the spin in relation to the dynamics of the accreting gas \citep[e.g.][]{2013ApJ...767...37M,2014MNRAS.440.1590D,2015MNRAS.452.1502D,2020arXiv200910578D, 2020arXiv200606647S}, have shown that the distribution of MBH spins depends on several quantities, such as host morphology, MBH mass, mass ratios, and redshift.

In principle in a binary all spin orientations and all spin magnitudes allowed by GR are possible. However, when an MBH binary, in its latest stages of evolution,  is  surrounded by a circumbinary disc, the interaction with the external gas leads both the binary orbital axis and the individual MBH spins to reorient their directions into a configuration of minimum energy where the two spins are aligned to a large degree with the orbital angular momentum axis, as discussed  in \citet{2007ApJ...661L.147B} \citep[see also][]{2010MNRAS.402..682D,2013ApJ...774...43M}. This has a strong impact on the final spin of the new MBH  and on the magnitude  of the velocity acquired by gravitational recoil, which  depends sensibly not only on the mass ratio, but also on the magnitudes and orientations of the spins \citep[][]{2010ApJ...715.1006K,2012PhRvD..85h4015L,2012PhRvD..85l4049B}. Extrapolation of MBH coalescences with large initial spins (larger than $\sim$0.9) exactly aligned with the orbital angular momentum yields a final spin as large as $\sim 0.95$ \citep{2008PhRvD..77f4010M,2008ApJ...684..822B,2010PhRvD..81h4054K,2011PhRvD..83b4010L}. 

In gas-poor conditions, the potential lack of a massive circumbinary disc leads MBH binaries to have spins randomly oriented at the time of their coalescence relative to the orbital plane, with  magnitudes determined by the previous accretion history.
Statistically, when spins are equally distributed in all directions relative to the orbital axis, the remnant MBH spin depends on the binary's mass ratio: if an MBH merges with many lower-mass MBHs it tends to spin down, as the final spin is dominated by the orbit at plunge and retrograde accretion at larger radii reduces (on average) the spin of the larger MBH \citep[][]{2003ApJ...585L.101H}.
If instead the mergers involve MBHs of comparable mass, on average the remant will have a spin $\sim 0.7$ \citep[][]{2008ApJ...684..822B}, consistent with the value  
of the final spin resulting from the merger of  two equal-mass, nonspinning MBHs \citep[][]{2009PhRvD..79b4003S}.

\cite{2008ApJ...684..822B}  studied the co-evolution of MBH masses and spins in a cosmological context, showing that in general accretion dominates over mergers in determining the spin evolution of the whole MBH population.  While in prolonged accretion episodes spin-up is very efficient, with a large fraction of MBHs having individual spins in excess of 0.9,  isotropic mergers reduce the fraction of high-spin MBHs and create a roughly uniform distribution. If accretion is chaotic, most MBHs have spins below 0.1 prior to merging. This demonstrates  how spins offer the best diagnostics on whether MBHs before  coalescence have experienced either coherent or chaotic accretion. These studies are 
important preparation for LISA as they provide insight for modelling realistic spin distributions to be used as priors during the analysis of waveforms to extract source parameters.

Indeed, LISA will measure not only the MBH individual masses, but also a mass-weighted combination of the individual spins projected along the orbital angular momentum (the so-called ``effective spin'' $\chi_{\rm eff}= (M_1 \chi_{1z} + M_2 \chi_{2z})/(M_1+M_2)$, where $M_1$ and $M_2$ are the MBH masses, $\chi_{1z}$ and $\chi_{2z}$ are the components of the spins along the orbital angular momentum) and possibly their precessional dynamics, which is encoded in the amplitude and phase of the waveform. A measurement of $\chi_{\rm eff}$  alone does not constrain the individual spins.  For example, a small $\chi_{\rm eff}$  could result from both MBHs having small spins; from each MBH having significant spins in the angular momentum direction, but anti-aligned with each other; or from nonzero spins oriented along the orbital plane.
Through parameter estimation of precessing binaries, however, it is possible to infer posterior distributions for both spins. Preliminary work on simulated MBH populations  has shown that the spin of the primary MBH can be measured by LISA with an exquisite accuracy ($\sim 1-10\%$) for nearby, loud events. This precision in the measurement mirrors the fact that the primary MBH leaves a bigger imprint in the waveform through the mass-weighted $\chi_{\rm eff}$. The measurement is more problematic for the spin of the secondary, that can be either determined to an accuracy of 0.1, or can remain completely undetermined, depending on the mass ratio and  spin magnitude  \citep{2016PhRvD..93b4003K}. 
  
If LISA's detection rates will be at the high end of current estimates, it may be possible to learn about the statistical distribution of the spins, and therefore constrain the relative importance of  mergers and accretion in shaping the MBH spin population in the mass range below $10^6\,\rm M_\odot$, which is poorly constrained by EM observations (see Section~\ref{sec:sec_2_5}).  Having a comparison between spin measurements from LISA and EM observations, which are tracing different populations,  will be of paramount importance \citep{2014ApJ...794..104S}.

In preparation for LISA, further improvements in numerical simulations are needed to make use of novel techniques to model physical processes below the resolution limits \citep[e.g.][]{2014MNRAS.440.1590D,2018MNRAS.477.3807F}, and to include changes in the spin directions that affect feedback \citep[][]{2019MNRAS.490.4133B,2020MNRAS.500.3719C,2020MNRAS.tmp.3351S,2020arXiv200910578D}. Semi-analytical models are also needed to understand whether the interaction between MBHs and their accretion discs can lead to spin alignment (see e.g. \cite{2013ApJ...774...43M,2013MNRAS.429L..30L,2015MNRAS.451.3941G,2020MNRAS.496.3060G}).
Finally, an interesting possible outcome of MBH mergers is that in non-aligned conditions, the direction of the remnant's spin can flip with respect to those of the progenitors: this would leave an observed imprint in the surrounding medium in the form of a particular shape \citep[X-shaped radio galaxies;][]{2010CQGra..27s4009G}, and observational searches for such systems \citep{2015ApJ...810L...6R} can provide information on the spin properties of merging MBHBs complementary to those obtained from theoretical models. 

\subsection{Statistics on MBH mergers}
\label{sec:StatisticsOnMBHMergers}
{\bf \textcolor{black}{Coordinators:}
Silvia Bonoli,
Alessandro Lupi\\
\noindent\textcolor{black}{Contributors:}
Monica Colpi,
Pratika Dayal,
Massimo Gaspari,
Melanie Habouzit,
Chung-Pei Ma,
Lucio Mayer,
Sean McGee,
Hugo Pfister,
Raffaella Schneider,
Alberto Sesana,
Rosa Valiante,
Marta Volonteri\\
}

MBHs are not born nor evolve in isolation. The physical properties of the host galaxies are key not only to set the MBH initial mass (see Section~\ref{sec:MBHoriginandgrowthacrosscosmictime}), but also to modulate the subsequent growth and mergers.  Indeed, most of the mass of today's MBHs is likely the result of multiple accretion episodes throughout  their entire lifetime \citep{1982MNRAS.200..115S}, likely triggered by secular processes or during violent events, such as galaxy interactions. For this reason there are various aspects
of galaxy formation and evolution that are indirectly very relevant to LISA, and which need to be well understood in order to 
enable predictions for observable MBH merger event rates as a function of key parameters, such as masses, mass ratios and spins of the MBHs, as well as their dependence on redshift. Likewise, the same deep understanding is required for post-launch interpretation
of the LISA datastream.
In the currently accepted cosmological framework, the $\Lambda$CDM model, galaxies  are expected to experience a large number of interactions and mergers during their lifetimes \citep{1993MNRAS.262..627L}. Galaxy interactions not only likely foster the activation of accretion episodes \citep[e.g.][]{2000MNRAS.311..576K, 2005Natur.433..604D,2015MNRAS.447.2123C}, but also lead to the formation of  binary MBH systems \citep{2007Sci...316.1874M,2017MNRAS.470.1121T,2020MNRAS.498.2219V}. The creation of triplets and multiple MBH systems is also possible \citep{2019MNRAS.486.4044B}, in particular for galaxies in dense environments which generally experience more frequent mergers. 
The formation time-scale of an MBHB is, however, dependent on the properties of the host galaxies. As discussed in Section~\ref{sec:formation}, the formation of a bound system is subject to the ability of the secondary MBH to sink towards the centre of the 
merger remnant, where the primary MBH is expected to reside. A substantial amount of orbital angular momentum needs to be transported away, with the distance between the two MBHs having to decrease by several orders of magnitude (from kpc to pc scales, see Fig.~\ref{fig:big_picture}).  The sinking process, driven by dynamical friction and global torques, depends non-trivially on the properties of the host, such as the overall galaxy structure, the  gas fraction, the presence of clumps or stellar clusters and structures such as discs or bars (see Section~\ref{sec:MassiveBlackHolesAndTheirPathToCoalescence}). Once a bound binary forms, its ability to harden still depends on the properties of the surrounding medium \citep[e.g.][]{2015MNRAS.454L..66S,2019MNRAS.487.4985B}. The hardening time-scale is shorter in galaxies with a large amount of stars that can cross the 
binary ``loss-cone'' and/or with enough gas in the centre to create a circumnuclear disc \citep[e.g. ][]{2004ApJ...606..788M, 2007MNRAS.379..956D}. 

Given that galaxy properties are tightly connected to the large-scale environment, the frequency of MBH mergers that LISA will detect depends on the global cosmological evolution of the host galaxies. All these physical processes, highly non-linear, can only be studied via sophisticated theoretical models, either analytical, semi-analytical, or fully numerical.  The main difficulty resides in the extremely wide range of physical processes and scales that need to be resolved simultaneously, from the Mpc cosmological scales to the sub-pc scales where GWs become dominant
(see Fig.~\ref{fig:big_picture}). We are currently unable to resolve the full dynamical range that would be required to predict the number and properties of MBH mergers for LISA.

Despite such modelling difficulties, building the infrastructure for interpreting the LISA data-stream in the context of structure formation and evolution is one of the key tasks for the LISA Consortium and the astrophysical community at large. In fact, LISA will provide a catalogue of MBHBs with posterior distribution of the parameters of each source, including masses, sky localization, distance, magnitude and orientation of individual MBH spins, and eccentricity of the orbit. The degree of precision of these measurements will obviously depend on the specific source. 

Individual binary parameters and parameter distribution across the detected population encode important information about the physics underlying MBHB formation. For example, the mass function and redshift distribution of detected events strongly depend on the nature and efficiency of the seeding mechanism. The spins of individual MBHBs are strongly affected by their main accretion channel, whether this is accretion of cold gas, tidal disruptions of stars or capture of compact objects, or previous mergers with other MBHs. 

The information encoded in LISA's catalogue of events has the potential to revolutionize our understanding of MBH formation and evolution, and the degree to which such potential can be exploited depends on the sophistication of the astrophysical inference models and pipelines at hand. \citet{2011PhRvD..83d4036S} conducted a pilot study demonstrating the power of inference on LISA data. They considered a number of different MBH cosmic evolution scenarios, encompassing different seeding models (Pop~III versus direct collapse), accretion efficiency (Eddington versus sub-Eddington) and geometry (coherent versus chaotic), demonstrating that LISA will be able to discriminate among them with just a handful of detections. They also considered mixed models in which, for example, different seed populations were combined, and found that LISA could correctly recover the presence of multiple sub-populations and their relative abundance. 

Although this was a successful first step, ideally, the community should employ all the arsenal of analytical and numerical models to distill a meaningful mapping of key astrophysical processes into MBHB parameter distributions. The LISA catalogue can then be used to tackle the `inverse problem' of reconstructing the cosmic history of MBHs from GW observations. A proof of concept example of such process can be found in \citet{2020JCAP...11..055P}. They used a parametric toy model connecting the MBH properties to the host DM haloes, to demonstrate that LISA would be able to constrain the halo occupation fraction and the MBH-halo relation. 

As we summarized above, the properties of MBHs are shaped by a number of physical ingredients that go beyond the host DM halo and involve a number of gas and stellar dynamical processes, and likely involve a non-negligible degree of stochasticity. In this respect, exascale numerical simulations combined with neural network techniques for model emulation, as we outline below, can be used to anchor and inform flexible semi-analytical models that can efficiently map a vast physical parameter space into a likelihood function of the MBHB population. Ideally, those models would be flexible enough to include information coming from future observations across the EM spectrum, including Rubin, JWST, Athena, and SKA, to name few notable examples, to enhance their constraining power (see Section~\ref{sec:sec_2_5}). Last but not least, the ultimate LISA MBHB astrophysical inference pipeline will also take advantage of any EM counterpart to individual LISA sources, which will provide additional information about the environment of the merging binary.

In this section, we first review the state of the art of the models that attempt to connect the small-scale processes of MBH formation, growth, and dynamics with the broader cosmological context of galaxy evolution. We then discuss the current estimates for the number of events that LISA will be able to detect as predicted by both numerical and semi-analytical models. At the end of the section, we provide an outlook on the need of pushing these models to be progressively  more and more  accurate and flexible, taking advantage of both the progress in computational power and new computational and statistical techniques. The theoretical framework connecting MBH mergers with the broader cosmological picture will play a fundamental role in the data analysis and the physical interpretation of LISA events.

\subsubsection{Modelling MBH evolution in a cosmological context } 
	
MBH assembly is considered an essential component of  galaxy formation \citep[e.g.][]{2013ARA&A..51..511K}. As anticipated,
while not specific to LISA science, this is a central topic to enable pre-launch studies, such as to inform models of LISA event rates,
as well as instruct post-launch studies by setting the framework for the interpretation of the data, allowing to generate
astrophysical inference work.
The inclusion of physical processes related to MBH growth into simulations of galaxy formation has been initially driven by the need of understanding the role of MBHs in shaping the host galaxies via processes such as AGN feedback.  In particular, feedback from AGN has been invoked to explain the observed quenching of massive galaxies, 
which
could not be explained by stellar feedback alone
 \citep[e.g.][]{2005ApJ...620L..79S,2006MNRAS.365...11C,2006MNRAS.370..645B}.

Models of galaxy formation and MBH assembly can be grouped into two categories: cosmological hydrodynamical simulations and semi-analytical models.  In the former, the DM and baryonic components of the Universe are followed simultaneously, starting from given initial conditions set by the chosen cosmological model. These simulations are computationally expensive and, in their set-up, a trade-off has to be made between the size of the cosmological volume that needs to be probed and the mass and spatial
resolution desired. Thanks to the fast advance of computational power, state-of-the-art simulations are able to encompass large volumes of $\sim 100^{3}\,\rm cMpc^3$ with kpc resolution (e.g. \citealt{2014MNRAS.444.1453D,2014MNRAS.444.1518V}; \citealt{2014MNRAS.442.2304H}; \citealt{2015MNRAS.446..521S, 2019MNRAS.490.3196P,2019MNRAS.486.2827D}). 
While allowing to study the evolution of DM and baryons in a self-consistent way down to the resolved physical scales, astrophysical processes that act at smaller scales (e.g. gas cooling, star formation, stellar feedback, MBH seeding, MBH accretion, MBH feedback) need to be included via sub-grid recipes.

Semi-analytical models, instead, follow the evolution of the baryonic component of the Universe through a series of differential equations  which link the time-evolution of the baryons to that of the underlying DM haloes. While losing some level of self-consistency, this approach has the advantage of being able to statistically explore how different physical assumptions affect the global galaxy population or targeted sub-samples \citep[see, e.g. the seminal paper by][]{1993MNRAS.264..201K}. The merger trees can either be derived via the Press-Schechter formalism \citep{1974ApJ...187..425P} 
or using the outputs of $N$-body simulations. While the first approach is computationally less expensive and merger trees with a broad range of masses can be resolved (e.g. down to the mass of the first star-forming haloes), $N$-body simulations offer the advantage that the 3D spatial distribution of galaxies  is fully tracked, as the dynamical evolution of the underlying DM haloes is properly followed. This allows the modelling and studying of the complex link between physical non-linear processes and the large-scale environment. 

Independently of the adopted technique, the models that track the evolution of the MBH population need to use sub-grid assumptions derived from higher-resolution simulations or analytical derivations, whose parameters are calibrated using observed properties of MBHs and their host galaxies. 
The local stellar mass function \citep[e.g.][]{2012MNRAS.421..621B}, the distribution of galaxy colours  \citep[e.g.][]{2004ApJ...600..681B}, and the evolution of the star formation rate density \citep[e.g.][]{2014ARA&A..52..415M} are some of the key observables used to calibrate the parameters regulating galaxy evolution, e.g., the galaxy and MBH sub-grid physics.

Models including the growth and evolution of MBHs are typically anchored to local relationships between the MBH masses and host properties, such as stellar mass and velocity dispersion  \citep{1995ARA&A..33..581K,1998AJ....115.2285M,2000ApJ...539L...9F,2000ApJ...539L..13G,2002ApJ...574..740T,2015ApJ...798...54G, 2019ApJ...876..155S}. 
Besides the calibration, the validation of the models is done by comparing the resulting MBH population to observational constraints. Typically, the AGN luminosity function from the local Universe to high redshifts \citep[up to $z\sim 4$,][]{2007ApJ...654..731H,2015ApJ...802..102L,2020MNRAS.495.3252S}, which constrains MBH accretion rates over cosmic time, is often used as diagnostics of the simulation or semi-analytical subgrid models.
Additional diagnostics include the Eddington-ratio distribution of AGN \citep[e.g.][]{2009ApJ...696..891H,2018MNRAS.474.1225A}, the number density of the highest-redshift quasars \citep[e.g.][]{2006ARA&A..44..415F,2011Natur.474..616M,2018ApJ...854...97D}, and the clustering of active and luminous MBHs \citep[e.g.][]{2005A&A...430..811G, 2009ApJ...697.1634R}. 

Here, we briefly review the different modelling aspects of MBHs -- seeding, fuelling, feedback, and dynamics -- and discuss different implementations and uncertainties among models, highlighting in particular those that are relevant for LISA. 

\paragraph{MBH seeding}

The first aspect that is crucial to determine the MBH occupation fraction in galaxies, and therefore MBH formation efficiency, is MBH seeding. Despite the strong effort by the community and the variety of proposed models 
(see Section~\ref{sec:MBHoriginandgrowthacrosscosmictime} for a detailed review), MBH seeding mechanisms are still unconstrained. 
As in Section~\ref{sec:MBHoriginandgrowthacrosscosmictime}, we consider models assuming ``heavy'' seeding, resulting in massive ($10^{4-6}$~M$_\odot$) but rare seeds, as well as models
assuming ``light'' seeding, which results in less massive ($\leqslant 10^3$~M$_\odot$) but more abundant seeds. 
LISA, by being sensitive to the mass of the merging MBHs that generate the GW signal, has the potential to constrain these models at statistical level through  Bayesian analysis,  because detection rates in the various mass intervals depend upon the seeding mechanism \citep{2011PhRvD..83d4036S,2016PhRvD..93b4003K,2019MNRAS.486.4044B, 2020ApJ...904...16B}.  Of course, right after their emergence, the mass growth
of the seeds through various mechanisms will also affect that MBH merger mass distribution that LISA can detect at any given time, which may complicate the interpretation
of the statistics (see next section).

In state-of-the-art cosmological hydrodynamical simulations of $\sim 100^{3}\, \rm cMpc^{3}$, the typical mass of DM particles is $M_{\rm DM}\sim 10^{6-8}\,\rm M_{\odot}$. 
This is not enough to resolve the haloes where we expect MBH formation to happen, for example the atomic cooling haloes where we expect direct-collapse MBH seeds to form. These simulations also do not have enough resolution to model self-consistently some key physical processes required by the different channels of MBH formation described in Section~\ref{sec:MBHoriginandgrowthacrosscosmictime}. 
Instead, MBHs are commonly inserted as sink particles ``by hand'', either in massive haloes of $M_{\rm h}\geqslant 10^{10}\, \rm M_{\odot}$ 
\citep[e.g.][]{2005ApJ...620L..79S,2014MNRAS.442.2304H, 2015MNRAS.452..575S, 2015MNRAS.446..521S, 2019MNRAS.486.2827D}, or in regions of the volume depending on the local properties of the medium, such as gas density (e.g.\citealt{2014MNRAS.442.2751T}; \citealt{2016MNRAS.459.2603B,2017MNRAS.468.3935H,2014MNRAS.444.1453D, 2020arXiv200910578D}).

In smaller-volume cosmological simulations with higher resolution, it has become possible to start implementing more physical MBH formation recipes.
For example, several simulations formed heavy seeds according to the local gas properties in high-redshift haloes 
\citep{2017MNRAS.470.1121T,2019MNRAS.482.2913B}.
A model by \cite{2018ApJ...861...39D} additionally includes Lyman-Werner flux as a seed formation criterion, most closely mimicking the direct collapse model.  Other realistic seed formation mechanisms forming lighter MBHs in cosmological simulations were explored in
\cite{2017MNRAS.468.3935H}: seed MBHs were formed in dense metal-free collapsing regions
to mimic the collapse of the first generation of stars or of dense nuclear star clusters.  

Leveraging the ability to efficiently probe larger effective volumes and smaller halo masses, semi-analytical models remain valuable for testing seeding models and statistically exploring the impact of seeding on multiple observables across cosmic times.  Using a model connecting heavy MBH seeding to halo properties, \citet{2006MNRAS.371.1813L,2008MNRAS.383.1079V,2009MNRAS.400.1911V} explore, for example, the observational consequences of light seeding models compared to direct collapse models with varying efficiencies. These and other works \citep[e.g.][]{2014MNRAS.437.1576B} provided novel quantitative predictions on how seeding reflects on the galaxy-MBH correlation, e.g., $M_\mathrm{BH}-\sigma$, with $\sigma$ the galaxy velocity dispersion. Other works have instead focused on the high-redshift universe, analyzing the ability of different seeding scenarios to lead to a population of $z>6$ quasars consistent with current observational data \citep[see the review of][]{2017PASA...34...31V}. 
As discussed in subsequent sections, semi-analytical models also predict clear seeding signatures in the mass function of GW event rates detectable by LISA.

 However, our current knowledge of the MBH population and their hosts across redshift is not sufficient to put tight constraints on seeding models, leaving predictions for the signatures of seeding on LISA events largely degenerate with other physical assumptions on MBH growth and dynamical evolution. 
This underlines on the one hand the importance of improving observational constraints of the key measurements in the seed mass regime to constrain models. This is needed to create reliable models to compare with LISA's event properties. On the other hand,  LISA's results will likely provide the most stringent constraints on MBH seeds, since it can explore redshifts closer to seed formation ($z\gtrsim 10$) than any EM facility can do. 

\paragraph{MBH fuelling}\label{sec:MBH_fuelling}

After MBHs have formed, their growth is mainly driven by gas accretion, whose rate is determined by the efficiency of the fuelling process onto the MBH from galactic (kpc) scales down to the nuclear region.  Because of the limited resolution, and the inability to properly track the angular momentum evolution of the inflowing gas and the formation of the accretion disc,  cosmological simulations almost always describe the accretion process through some version of  Bondi--Hoyle--Lyttleton accretion \citep[hereafter Bondi;][accretion rate $\propto M_{\rm BH}^{2}$]{1939PCPS...35..405H,1944MNRAS.104..273B,1952MNRAS.112..195B}, 
which assumes spherical symmetry and can be inaccurate in most realistic physical scenarios \citep[e.g.][]{2010ApJ...716.1386L, 2012MNRAS.421.3443H, 2017MNRAS.466..677G,2017MNRAS.467.3475N}.
In case of significant angular momentum of the MBH accreting material, the accretion rate may not be well represented by the Bondi model. As such, the EAGLE simulation employs a modified Bondi model that takes into account the circularization and subsequent viscous transport of the infalling material \citep{2015MNRAS.454.1038R,2016MNRAS.462..190R}.
The gravitational torque-driven model \citep{2010MNRAS.407.1529H}, implemented in some recent cosmological simulations \citep[e.g.][]{2017MNRAS.464.2840A,2019MNRAS.486.2827D}, takes a different approach and since the accretion rate $\propto M_{\rm BH}^{1/6}$, low-mass MBHs initially grow more than in the Bondi model \citep[e.g.][]{2020arXiv200712185C}. This emphasizes the need for progress in bridging the gap between galactic and accretion disc scales.
Indeed, zoom-in high-resolution simulations (from galactic down to sub-pc scales) show that the actual accretion flow often proceeds in the form of chaotic cold accretion  \citep{2013MNRAS.432.3401G,2015A&A...579A..62G}, in which fractal clouds condense out of the turbulent hot halo, rain on to the nuclear region and, via frequent inelastic collisions, boost the accretion rate by $100\times$ over the simple Bondi rate (see also Section~\ref{s5:host_galaxies}). 

Semi-analytical models often tie the growth of MBHs to galaxy mergers \citep[e.g.][]{2000MNRAS.311..576K, 2008MNRAS.385.1846M},  or events of starbursts or bulge growth, assuming some form of MBH-galaxy co-evolution model \citep[e.g.][]{2008MNRAS.391..481S,2019MNRAS.482.4846S}.  Models which do not track the full evolution of the galaxy population, have modelled similar co-evolution with the velocity dispersion derived directly from the DM halo \citep{2003ApJ...582..559V}  or estimating the velocity dispersion of the galaxy based on empirical relations \citep{2018MNRAS.474.1995R}.  In this way, it is assumed that some combination of fuelling and feedback produces $M_\mathrm{BH}$-host relations. Other models directly relate the growth of MBHs to the evolution of different gas phases or different dynamical processes, such as disc instabilities \citep{2017MNRAS.466..677G,2019MNRAS.486.2336D,2020MNRAS.495.4681I}.  The gravitational torque-driven model introduced by \citet{2010MNRAS.407.1529H} has also been implemented in semi-analytical models, together with analytic models for disc instabilities \citep{2014A&A...569A..37M,2015A&A...576A..32G}.

On accretion disc scales, one of the most important physical ingredients is the Eddington limit, the accretion rate at which radiation pressure balances gravity for a spherical accretor (as defined in Section~\ref{sec:feedback_barrier}). Typically, MBH accretion rates are capped at Eddington, and indeed the overall quasar population appears to obey this limit \citep[e.g.][]{2015Natur.518..512W}. However, there are several theoretical motivations to consider relaxing this assumption.  First, MBH accretion does not occur spherically, but rather through an accretion disc.  State-of-the-art radiative MHD simulations have demonstrated that Super-Eddington flow regimes can be sustained for  many disc orbits \citep{2014ApJ...796..106J,2015MNRAS.454L...6M,2016MNRAS.456.3929S,2018ApJ...859L..20D}. In addition, the existence of $10^{9-10}$~M$_\odot$ quasars at $z\sim 6$ requires optimistic duty cycles to grow from the seed mass if an Eddington rate cap is assumed, even under a heavy seeding scenario (see the discussion in Section~\ref{sec:growth}).  

Being able to resolve the full journey of the gas inflow from galaxy scales down to the nuclear region is essential not only to properly address MBH accretion, but also to study formation of circumbinary discs (see Section~\ref{sec:hardening_gas}, and below). Spin evolution is also connected to the frequency and properties of the accretion process. We refer the reader to Section~\ref{sec:spin} for a discussion of the physical approaches, and we only recall here that coherent accretion leads to maximally spinning MBHs, whereas randomly oriented accretion can spin MBHs down.
MBH spins also evolve during  coalescence, depending on the combination of the orbital angular momentum and the two initial spins.  The latter
part of the evolution is included in cosmological simulations by adopting
 fitting formulae to GR simulations \citep{2008PhRvD..78d4002R,2010CQGra..27k4006L,2009ApJ...704L..40B,2016ApJ...825L..19H}. 
 Only few semi-analytical models \citep[e.g.][]{2005ApJ...620...69V,2012MNRAS.423.2533B,2013ApJ...775...94V,  2020MNRAS.495.4681I} and cosmological simulations \citep{2014MNRAS.440.1590D,2019MNRAS.490.4133B,2020arXiv200204045T,2020arXiv200910578D} follow MBH spin self-consistently with a sub-grid model. 

 Improvements on the modelling of MBH fuelling, tighter observational constraints on MBH accretion rates across a wide range of masses and redshift, and direct estimates of MBH spins \citep[see the review of ][]{2019NatAs...3...41R} will help discriminating between different spin evolution models on the run up to LISA to sharpen predictions. 
The spin of MBHs also bears a relation to the energy that can be released through AGN feedback. MBHs with high spins are predicted to release more specific energy than MBHs that have low spins or are non rotating  \citep{2014MNRAS.440.1590D,2019MNRAS.490.4133B}. 
Constraining the spin distribution of MBHs with LISA could help us to better constrain AGN feedback models, which we discuss below.

\paragraph{MBH feedback}

Feedback from AGN is arguably one of the most important and still open aspects of MBH-galaxy co-evolution. AGN are short-lived phases of MBHs evolution, and release substantial amounts of energy in their surroundings. The feedback drives winds, outflows, and jets, which create large-scale X-ray cavities in clusters and groups of galaxies. However, even if there is observational evidence for the role of AGN in quenching star formation 
and cooling flows in the host haloes, the exact mechanisms of such energetic processes are still unclear today \citep[e.g.][]{2012ARA&A..50..455F,2020NatAs...4...10G}.  
Indeed, the range of physical scales ($\sim$9 orders of magnitude) tied to gas inflows and negative/positive feedback processes in the host make this a very challenging problem to solve (see also Section~\ref{s5:host_galaxies}). 

Cosmological simulations often partition feedback into two modes, depending on the efficiency of the accretion on to the MBHs \citep[e.g.][]{2008MNRAS.388.1011M}. During accretion phases characterized by high Eddington ratios and thin accretion discs, often called `quasar' mode, AGN feedback is strongly ejective and radiatively driven, 
whereas during phases of lower accretion rates, often called `radio' mode and characterized by a thick disc or chaotic cold accretion, AGN feedback is mostly driven via radio jets or sub-relativistic outflows that maintain the macro-scale gaseous haloes in quasi-thermal equilibrium for several billion years.
The transition between the two states likely occurs around an Eddington ratio of $\sim$0.01--0.1
\citep{2014ARA&A..52..529Y}. Based on analytic arguments, the MBH mass-velocity dispersion scaling relation ($M_\mathrm{BH}$-$\sigma$) may be a direct consequence of how these wind powers ought to scale with the host properties to curtail further accretion \citep{1998MNRAS.300..817H,2003ApJ...596L..27K,2012MNRAS.426.2751Z}. 

The modelling of AGN feedback is one of the aspects that needs to be improved in cosmological simulations. Bridging the scale gap is unfeasible, and thus many sub-grid models often invoke simple direct heating mechanisms (either central or as pairs of hot bubbles) to quench local star formation \citep[e.g.][]{2005MNRAS.361..776S,2007MNRAS.380..877S,2014MNRAS.444.1518V,2015MNRAS.452.1502D,2015MNRAS.446..521S}.
Other approaches have included injection of  kinetic energy via jets or winds (\citealt{2012MNRAS.420.2662D,2012ApJ...754..125C,2012ApJ...746...94G,2016MNRAS.461.1548B,2017MNRAS.472.4707B,2017MNRAS.465.3291W,2020MNRAS.498.4983W}) instead of or jointly with heating.

While sub-grid models are rooted in physical insight, they are still not able to follow the full range of processes related to MBH accretion and star formation in the interstellar/circumgalactic medium. Therefore, future improvements should take into account magnetic fields, employ  at least some approximate radiative transfer, 
and consider more realistic models of stellar feedback and of the clumpy, multi-phase interstellar medium \citep[e.g.][]{2018MNRAS.480..800H,2019MNRAS.489.4233M}. 
One key aspect is the connection between AGN feedback and the spin of MBHs. Given that LISA will provide us with spin distribution of the merging systems, this is a direction that we need to address in the coming years. The strength of AGN feedback scales with the radiative efficiency, which is closely tied to MBH spin. Therefore, a self-consistent treatment of AGN feedback should account for the effect of spin on radiative efficiency \citep[as in][for example]{2020arXiv200204045T,2020arXiv200910578D}, which could become
feasible if the spin distribution will be robustly constrained by future GW datasets.

\paragraph{MBH dynamics}

MBH dynamics is key to model LISA's MBH mergers. Between when a galaxy merger begins and the final MBHs coalesce, an MBH must complete a journey of many 
orders of magnitude in spatial scales. We refer the reader to Section~\ref{sec:MassiveBlackHolesAndTheirPathToCoalescence} for a detailed account of the orbital decay mechanisms acting on
different scales, and in varying astrophysical environments.
However, most cosmological simulations are unable to follow the dynamics of the infalling MBH down to the scale where the MBH binary system can form, because of the trade-off between the maximum resolution achievable and the simulation volume. 
For example, large-scale cosmological simulations like Horizon-AGN (\citealt{2014MNRAS.444.1453D}; \citealt{2016MNRAS.460.2979V}), Illustris \citep{2015MNRAS.452..575S}, and Eagle \citep{2015MNRAS.446..521S}, with spatial resolutions of about 1~kpc, cannot follow MBH dynamics down to the centre of galaxies.  The kpc-scale regime can now be directly probed with smaller volume cosmological simulations, in which multiple MBHs are allowed to co-exist within the same galaxy, although reaching the required resolution is very challenging.  Taking care to correct the dynamical friction force onto MBHs lost due to gravitational softening, \citet{2018MNRAS.475.4967T} find a wide range of delay times between galaxy merger and MBH pairing in the {\sc Romulus} cosmological simulation, which can impact GW event rates \citep{2020ApJ...904...16B}. Even higher resolutions can be instead achieved by means of zoom-in simulations or isolated galaxy mergers, that allow to resolve the dynamics down to a few tens of pc \citep{2014MNRAS.439..474V,2019MNRAS.482.2913B,2019MNRAS.486..101P}. 

When MBHs become gravitationally bound, their orbit must still shrink, the hardening phase, before GWs can act to bring about coalescence (see Section~\ref{sec:TheBinaryFormationAndInitialShrinking} for a description of the physical processes).  
The binary hardening phase can not be resolved in cosmological simulations, thus  assumptions for  the estimate  of the hardening time-scale need to be adopted in the post-processing analysis of hydro-simulations or in semi-analytical models. Delay times can be assumed to be fixed \citep[][]{2020arXiv201200775D} or to depend on some simplified way on the properties of the host galaxy  \citep[e.g.][]{2020MNRAS.495.4681I} and/or of the circumbinary disc \citep[e.g.][]{2017MNRAS.471.4508K, 2020MNRAS.498.2219V, 2020arXiv200606647S}. Finally, in any situation characterised by moderately long binary hardening times, triple or even multiple MBH systems are also likely to form in galaxies experiencing frequent mergers, and cosmological models should also take those into account \citep[see, e.g.][]{2017ApJ...840...53R,2018MNRAS.473.3410R,2019MNRAS.486.4044B}.
It is clear that in preparation for LISA, significant developments are needed to both semi-analytical models and simulations to improve how dynamics is treated and obtain convergence in the predicted rates and MBHB properties.
  
\subsubsection{State of the art on MBH merger rates from cosmological simulations}\label{sec:merger_rate}

In what follows, we discuss how the modelling and assumptions for the processes mentioned above affect the predicted rate and properties of LISA events. The range of predictions for the merger rate of MBHs that LISA can detect currently spans a wide range, from about one to several hundreds per year. 
The reason for this large span lies both in different physical assumptions and in the different techniques used. 
To give a rapid overview, in terms of physical modelling the merger rate is high when MBHs are abundant in galaxies, hinging on the efficiency of the MBH formation model adopted, and when the dynamical evolution is fast. Then, the rate decreases as one or the other of these assumptions is relaxed. The rate of mass growth also is important, as it determines the redshift at which MBHBs enter and exit the frequency range accessible by LISA. There are also other subtleties that enter the models. For instance, the spins and mass ratios of the binaries at the time of coalescence, which are determined by formation, growth, and dynamical evolution all together, influence the speed of recoil kicks \citep[e.g.][]{1962PhRv..128.2471P,2006PhRvD..73l4006D}, 
which in turn modulate further the merger rate by ejecting MBHs from galaxies. Thus, not only the merger rates but also mass and spin distributions of merging MBHs depend sensitively on the specific assumptions of the models in terms of seeding, accretion, feedback, and spin evolution, which are largely unconstrained by available observational data-sets. The details of these physical aspects are described in Section~\ref{sec:MassiveBlackHolesAndTheirPathToCoalescence} and Section~\ref{sec:MBHoriginandgrowthacrosscosmictime} of this paper and below we discuss how they influence the statistics of LISA's merging MBHs also in dependence of the technique used.  

The techniques adopted also have a bearing on the resulting merger rate. The main parameter in this context is the mass resolution of the cosmological simulation or the DM merger tree used to build a model Universe. LISA’s MBHs have masses in the range $10^3$--$10^7 \msun$ and can be hosted in haloes with stellar masses as low as $10^6 \msun$ \citep{2019MNRAS.482.2913B,2020MNRAS.498.2219V}. 
If the resolution of the model does not allow to resolve low-mass haloes, the merger history of MBHs in these haloes cannot be tracked and therefore the MBH merger rate obtained will be a lower limit to the real merger rate. 
The volume of the simulation also matters, and ideally a large diversity of environments is needed to accurately derive reliable MBH merger rates that can provide sensible predictions in preparation for LISA. 
Obviously, both high resolution and large volume requirements increase the computational cost. Therefore, models are currently a compromise and not an ideal set-up. 

In the next two sections, we review the existing predictions for LISA and critically discuss their physical assumptions and technical approaches.  

\paragraph{Cosmological hydrodynamical simulations}
 
Cosmological simulations are a recent addition to predictions for LISA's merger rates and properties of merging MBHs, which started with analytical and semi-analytical models about 15--20 years ago \citep{1994MNRAS.269..199H,2005ApJ...623...23S}. 
For LISA, these simulations are, in principle, the best tool, since they incorporate, to some extent, all the processes regarding
MBH formation and evolution, and this in the context of an evolving population of host galaxies computed from high to low redshift.
As a result, cosmological hydrodynamical simulations have the  advantage over isolated merger simulations in that they naturally include a variety of mass ratios, orbital configurations, and galaxy structures. For instance, in isolated merger simulations the minimum mass ratio of a galaxy for MBHs to bind within $\sim$1~Gyr seems to be $>0.25$ \citep{2009ApJ...696L..89C,2014MNRAS.439..474V,2015MNRAS.447.2123C}, but cosmological simulations show (i) that also galaxy mergers with lower mass ratio contribute to the MBH merger rate \citep{2018MNRAS.475.4967T,2020MNRAS.498.2219V} and also (ii) the effect of irregular  potentials in high-redshift and dwarf galaxies  \citep{2019MNRAS.486..101P,2019MNRAS.482.2913B,2020MNRAS.498.3601B}.

Cosmological hydrodynamical simulations have for the most part focused on the merger rates  and mass ratio distributions of the merging events \citep{2016MNRAS.463..870S,2020MNRAS.491.2301K,2020MNRAS.491.4973D,2020MNRAS.498.2219V}. Spin has been for now included in post-processing in GW-related studies  \citep{2020arXiv200606647S}: although some simulations with spin evolution exist \citep{2014MNRAS.440.1590D,2019MNRAS.490.4133B,2020arXiv200204045T,2020arXiv200910578D}, for the moment the spins of merging MBHs has only been investigated in post-processing \citep{2020arXiv200606647S}. 
 
Most cosmological simulations used to investigate statistically merging MBHs are large-volume ($ \geq 100^3~ {\rm Mpc}^3$), low-resolution (DM particle mass $\sim$10$^7$--$10^8 \msun$, with the proviso that $\sim$50--100 particles are required to identify a halo; star particle mass $\sim$10$^6 \msun$, spatial resolution 0.4--1~kpc) simulations \citep{2016MNRAS.463..870S,2020MNRAS.491.2301K,2020MNRAS.491.4973D,2020MNRAS.498.2219V}. However, such simulations are not suited for studying MBHs in the LISA mass range. Large volume is a positive aspect, improving statistics and capturing various environments in the large-scale structure. Mass resolution, as noted above, is a key point. LISA's MBHs have masses in the range $10^3$--$10^7 \msun$, 
with some MBH formation models predicting MBHs with mass $\sim$10$^4 \msun$ in haloes with mass as low as $10^8 \msun$ (see Section~\ref{sec:growth}). Therefore, such low-mass haloes must be resolved in order to capture the full merger rate of LISA's MBHs. Most of the MBHs in well-resolved galaxies in low-resolution simulations are simply {\it too massive} and therefore merge outside the LISA band, at lower frequencies \citep[they are better suited for PTA experiments,][]{2017MNRAS.471.4508K}. This means that we have to be aware that the merger rates predicted by current simulations -- generally $<1$ per year -- 
could be a lower limit.
 
\cite{2020MNRAS.498.2219V} present the first analysis of the merger rate and merging MBH properties in a high-resolution simulation (``NewHorizon'', DM particle mass $\sim 10^6 \msun$, star particle mass $\sim 10^4 \msun$,  spatial resolution 0.04~kpc) with a sufficiently large volume (a sphere of radius $\sim 10$~cMpc) to have some statistics, while \cite{2019MNRAS.482.2913B} simulate a number of isolated dwarf galaxies at somewhat lower spatial resolution but higher mass resolution and \cite{2016ApJ...828...73K} simulate one single galaxy at similar resolution in a sphere with radius 13.5~kpc (they then extract and resimulate further the central nucleus of the galaxy at higher resolution, but without hydrodynamics).  \cite{2020MNRAS.498.2219V} analyze in the same way a  high-resolution, small-volume simulation and a low-resolution large-volume simulation and show explicitly that indeed the merger rate of the former is higher. This is because dwarf galaxies are resolved, and there are more dwarf galaxies than high-mass galaxies in the Universe. Furthermore, a significant fraction of dwarf galaxies host MBHs: in NewHorizon at $z \sim 0.5$, about 10 per cent of galaxies with mass $10^6 \msun$ host an MBH, increasing to 100 per cent at  $10^9 \msun$. Observationally, between 10 and 100 per cent of galaxies with mass $\sim$10$^9$--$10^{10} \msun$ at $z=0$ appear to host an MBH \citep{2019arXiv191109678G}, implying that MBH mergers in dwarf galaxies are indeed crucial for the low-mass MBHs relevant for LISA. Similar results have also been found by \citet{2019MNRAS.482.2913B}, where, in addition, MBHs typically appear off-centred relative to the host.

\begin{figure}
    \centering
    \includegraphics[scale=0.6,trim= 0.5cm 0cm 0.5cm 0.5cm,clip]{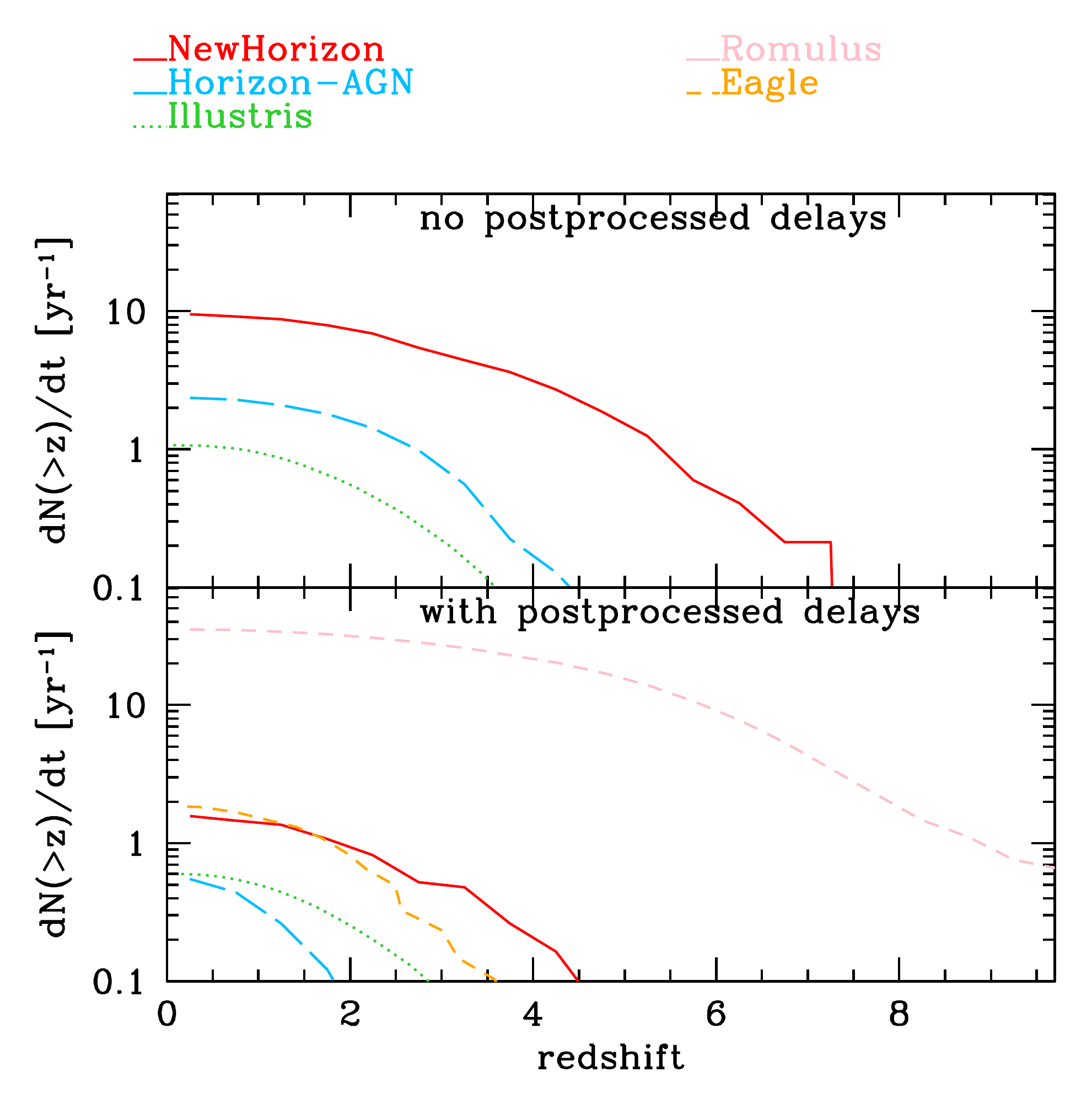}
    \caption{Comparison of merger rates from different cosmological hydrodynamical simulations, with (bottom) and without (top) the addition of a delay in post-processing. No SNR LISA cut has been applied: this is the merger rate of all MBHs independent of whether LISA can detect them or not, e.g. most MBHs in low-resolution simulations are too massive to enter the LISA band.  NewHorizon and Horizon-AGN \citep{2020MNRAS.498.2219V} include intrinsic delays from dynamical friction from gas, and additional below-resolution delays (bottom panels), and model MBHs above $10^4$ and $10^5\rm\, M_\odot$, respectively.  Illustris \citep{2020MNRAS.491.2301K}, with $M_{\rm BH}\sim 10^5\rm\, M_\odot$, does not implement any intrinsic delay and adds (bottom panel) physically motivated delays in post-processing.  Romulus \citep{2018MNRAS.475.4967T}, where $M_{\rm BH}>10^6\rm\, M_\odot$, includes intrinsic delays from dynamical friction from particles, and a fixed 0.1~Gyr  below-resolution delay. Eagle \citep{2016MNRAS.463..870S} seeds MBHs with $M_{\rm BH
    }\sim 10^5\rm\, M_\odot$ and does not include any intrinsic delay, adding in post-processing fixed delays of 0.1~Gyr for gas-rich galaxies and 5~Gyr for gas-poor galaxies. Figure credit: Marta Volonteri.
    }
    \label{fig:predictions_sims}
\end{figure}

Besides mass resolution, how MBH dynamics is treated in simulations is also important, and this circles back to spatial resolution. Typically, large-scale cosmological simulations do not have the sub-pc resolution needed to resolve MBH dynamics. Two approaches have been used in the literature. 

The first is to not treat explicitly MBH dynamics: MBHs are repositioned at each timestep at the position of the lowest potential (gas) particle near the MBH. 
In a merger, an MBH would very rapidly be moved to the centre of the potential well, on time-scales much shorter than in reality. As a consequence, the merger rate of MBHs is increased. Some studies do not consider delays \citep{2020MNRAS.491.4973D}, in others delays have been added in post-processing, either using a fixed value \citep{2016MNRAS.463..870S} or adopting a physical approach \citep{2020MNRAS.491.2301K}. 
\citep{2016MNRAS.463..870S,2020MNRAS.491.2301K}.

The second approach is to include sub-grid dynamical friction from gas \citep{2013MNRAS.428.2885D}, from stars and DM \citep{2015MNRAS.451.1868T}, or all of the above \citep{2019MNRAS.486..101P}. Adding dynamical friction in the code, however, poses an additional challenge: the ratio of MBH mass to the mass of star particles (or unsmoothed DM particles) must be $>10$ to avoid spurious oscillations. This means a very challenging computational task when MBHs have mass $<10^5 \msun$. On-the-fly dynamical friction helps in having realistic dynamics down to the resolution of the simulations: the force acting on the MBHs captures the inhomogeneous, time-varying density distribution and irregular potential wells where MBHs, especially at high redshift, evolve. Still, this approach operates only down to the spatial resolution of the simulation, which is $\sim$10--50~pc in high-resolution cosmological simulations and 0.3--1~kpc in low-resolution simulations. Below this scale, statistical studies of MBH mergers can only rely on adding additional time-scales of binary evolution (stellar hardening, torques in circumnuclear discs and circumbinary discs; see Section~\ref{sec:formation}) in post-processing \citep{2020MNRAS.491.2301K,2020MNRAS.498.2219V,2020arXiv200606647S}, although there are prospects for a full on-the-fly treatment \citep{2017ApJ...840...53R}. 

An important point is that for the moment the mass ratio of merging binaries is based either on information obtained long before the MBH mergers (before including the dynamical delays) or on specific choices applied in post-processing \citep{2020arXiv200606647S}, which may or may not capture how each of the MBHs grows in mass during the final phase of dynamical friction and during the hardening and circumbinary disc phase. Moreover, the limited resolution limits the ability to self-consistently follow the tidal stripping of the galaxy nucleus during the dynamical friction phase, and this affects the orbital decay. 
A comparison of the predictions obtained by different state-of-the-art simulations is reported in Fig.~\ref{fig:predictions_sims}, with (bottom panel) and without (top panel) the inclusion of a post-processed delay between the time when MBHs merge in the simulation and the estimate of the coalescence time taking into account the expected, but unresolved, physical processes.

\paragraph{Analytical and semi-analytical models}\label{sec:SAMs}
 
Several studies have developed analytical and semi-analytical models to predict merger rates and chirp masses for LISA, with various assumptions on the main seeding mechanism for MBHs. Most of these studies pre-date the use of cosmological hydrodynamical
simulations in the context of LISA, and have paved the ground for the latter.
The predictions of these models can vary significantly, mostly because the physics of the formation and of the orbital shrinking of the MBHBs are thus far loosely constrained, although some advancements have been recently put forward. 
Analytical and semi-analytical models suggest that different seed populations have a different impact on the total number and mass distribution of potential LISA sources at different cosmic epochs \citep[see, e.g.][]{2003ApJ...582..559V,2011PhRvD..83d4036S,2012MNRAS.423.2533B,2016PhRvD..93b4003K,2019MNRAS.486.4044B,2019MNRAS.486.2336D,2020ApJ...904...16B,2020MNRAS.491.2301K,2021MNRAS.500.4095V}.
Generally speaking, all the models converge on predicting that the merger rates are significantly higher if seeding occurs mainly with light seeding mechanisms, e.g. MBHs are formed as remnants of Pop~III stars, with a typical mass $\lesssim 10^3\, \msun$  \citep[see, e.g.][for a description of initial mass functions for light and heavy seeds]{2014MNRAS.443.2410F, 2016MNRAS.457.3356V, 2018ApJ...864L...6P}. Specifically, \cite{2018MNRAS.481.3278R} predict that LISA will observe $\sim$20 times more events if seeding occurred mainly from light seeds, with an upper limit of $\sim$300 events (over a 4-year mission duration) with a typical mass $\sim$10$^3 \, \msun$ in the most optimistic scenario. Similarly, \cite{2019MNRAS.486.2336D} predict that light-seeding scenarios will drive the merger rates up, ending with a more conservative prediction of 12--20 mergers during a 4-year mission duration. Even when light and heavy seeds are combined in the same cosmological evolution history, as in \citet{2019MNRAS.486.2336D} and \citet{2021MNRAS.500.4095V}, the number of predicted LISA events is dominated by (growing) light seed binary mergers, although the impact of feedback (reionization, SNae, AGN) by suppressing MBH growth or hindering dynamical friction, reduces the importance of the mergers of light and heavy seeds \citep{2019MNRAS.486.2336D,2020ApJ...904...16B,2020arXiv200702051L}. Notably, what drives significant differences in predictions is the probability that MBHs actually coalesce, once their host galaxies have merged (see a broad description of the issue in, e.g. \citealt{2019arXiv191105791I} and in Section~\ref{sec:MassiveBlackHolesAndTheirPathToCoalescence}). \citet[][]{2019MNRAS.486.4044B} predict a rate of $\sim$25 and $\sim$75 LISA events per year, respectively, in heavy and light seeding models, which is reduced to $\sim$10--20~yr$^{-1}$ if MBHB mergers are efficiently driven only via triple interactions (i.e. if gas/stellar-driven shrinking mechanisms were to fail in driving the binary to coalescence). 
In addition, as the GWs emitted during the coalescence phase carry linear momentum, also the inclusion of gravitational recoil can  impact the  halo occupation fraction, hence the merger rates \citep[see, e.g.][]{2004ApJ...613...36H, 2009ApJ...696.1798T, 2019arXiv191105791I,2020MNRAS.495.4681I}. 

\begin{figure}
    \centering
    \includegraphics[scale=0.6,trim= 0.5cm 5cm 1cm 2.5cm,clip]{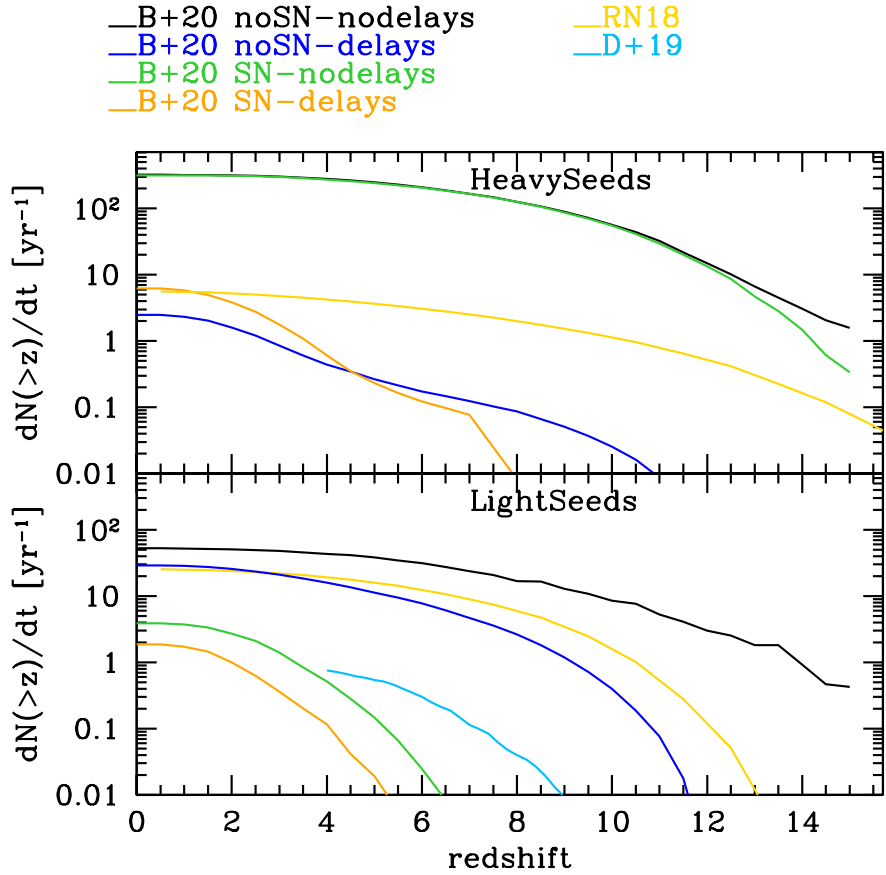}
    \caption{Comparison of merger rates from different semi-analytical models, assuming heavy seeds (top panel) and light seeds (bottom panel). For all models, we employed the Science Requirement curve \citep{2021arXiv210801167B} applying an SNR cut of 8. Different assumptions for models by \citet{2020ApJ...904...16B} are shown, with or without SN feedback, and including or not delays. \citet{2019MNRAS.486.2336D} include reionisation feedback and delays, whereas \citet{2018MNRAS.481.3278R} do not include delays. The still large uncertainties in the modelling result in significant variations, up to two orders of magnitude, with mergers between light seeds typically dominating the event rate, but for the case when SN feedback is included, as in \citet{2020ApJ...904...16B}. Figure credit: Marta Volonteri.}
    \label{fig:predictions_sams}
\end{figure}

A comparison of the prediction by different semi-analytical models is reported in Fig.~\ref{fig:predictions_sams}, for light seeds (bottom panel) and heavy seeds (top panel). In general, the predicted event rates span a wide range, from no detection to a few hundred events, depending on the adopted description of the multi-scale and complex processes regulating seed MBH formation, mergers, accretion, and dynamics, which are far from being fully understood. It is therefore important, to reliably predict the rate of MBH coalescences alongside the hierarchical assembly of galaxies, to get full control of the assumptions made to describe these processes, on different scales/times \citep[see, e.g.][and Sections~\ref{sec:MassiveBlackHolesAndTheirPathToCoalescence} and \ref{sec:StatisticsOnMBHMergers}]{2005ARAOJ...7...34E,2011PhRvD..83d4036S,2016PhRvD..93b4003K,2016JCAP...04..002T,2018MNRAS.474.1995R,2018MNRAS.481.3278R, 2019MNRAS.486.4044B,2019MNRAS.486.2336D,2020MNRAS.498.2219V,2020ApJ...904...16B,2021MNRAS.500.4095V}.

From a statistical point of view, LISA detections (or non-detections) may reflect more the dynamical properties and evolution of binary MBHs (i.e. their ability to form and merge) rather than their origin. For instance, heavy seeds are expected to form binaries more efficiently than the more common light seeds. Therefore, a low number (or even the lack) of detections of high-redshift sources in the LISA band may indicate that heavy seeds are very rare and/or that they are not able to merge, after binding in binaries (because of inefficient hardening mechanisms in their host galaxies).

\subsubsection{How to advance and optimize the scientific return of LISA}

As we have seen above, predictions for LISA events depend in a complicated  way on a large number of assumptions, from the seed mass to spin evolution and the dynamics of binary systems. In turn,  these aspects are  tightly linked to the properties of the host galaxies and of the environment.

The interplay between all these different non-linear physical processes leads to predictions for the merger rates that are highly degenerate.

Learning about spin evolution, merger time-scales, accretion physics, and seed masses from the merger rates of LISA requires a  data analysis process where the multi-dimensional parameter space can be quickly explored. By the time LISA launches, the community needs to be ready with a comprehensive and flexible set of theoretical models that can be efficiently confronted with the data.

Numerical simulations of small-scale physical processes will need to be connected to simulations that trace the full cosmological evolution of structures and they will need to inform analytical or semi-anaytical models, that can scan and test the parameter space efficiently. This is central to quantify robustly the mapping between galaxy mergers and MBH mergers. 
Furthermore, we will need to understand which classes of galaxies, and in which environment,  LISA events are most likely hosted. This aspect will be vital if an EM counterpart of a GW merger is to be discovered. Given the current capabilities of LISA to localize  nearby sources \citep[][and section \ref{sec:sec_2_4_2}]{2020PhRvD.102h4056M}, 
from a few to a few hundreds of galaxies could be in the field of view of, e.g. X-ray telescopes, depending on the loudness of the source. Thus, anticipating the characteristic properties of the host galaxies from simulations will help identifying the host and its redshift.

In the remaining of this section, we will give a brief outlook on the expected advances  in ``traditional'' techniques and on the possibility of using new statistical methods, and how those will be used to inform the LISA data processing. 

\paragraph{Improvements on current techniques}

An important role in building the theoretical framework will be played by the transition to the exascale computing, that will allow us to develop simulations of much larger portions of the observable Universe, of order comoving Gpc$^3$ (either as single simulations or as several smaller-volume ones targeting different environments), compared to the current ones (limited to a few hundreds of comoving Mpc), and to further increase the resolution in order to resolve ever smaller scales currently achieved only via dedicated idealised studies. 
The combination of ultra-large simulated cosmological volumes and very high resolution is the best strategy to enable
astrophysical inference studies with the LISA datastream because the properties of the MBH binaries that enter the LISA
band are determined by both large scale and small scale processes, which demands both large volumes and high resolution, and because LISA is an all-sky instrument that probes the Universe from low  to very high redshift, which again calls for very large volumes.
However, exascale computing is not a guaranteed solution, since even the best cosmological hydrodynamic codes available today are far from being able to scale on the billion-core platforms that will characterize such a  computing phase over the next decade, when the resolution is increased beyond a certain threshold. This is because in simulations with very high resolution, reaching below tens of parsecs, load balancing on large core counts becomes a computational bottleneck which, unfortunately, gets worse as the number of computing cores is increased. 
This is an intrinsic challenge with modelling the non-linear process of gravitational collapse. Thus, unless a quantum leap occurs in the parallel architecture of simulation codes, for example owing to improvements in task-based parallelism, the larger cosmological volumes will still be limited in their ability to capture the small-scale stellar and gaseous processes (at sub-pc scales) that drive the hardening phase. This is why simulations will always need to be complemented by other techniques. Different techniques will also require different improvements, and should be combined together to exploit the respective advantages, i.e. the speed and the easy parameter exploration of semi-analytical models and the spatial information of hydrodynamic simulations.

On the side of semi-analytical models, more sophisticated and comprehensive assumptions for MBH seeding and the dynamics of binaries and multiplets will have to be included. High-resolution small-scale numerical simulations covering a wide range in the parameter space will be needed to create new parametric prescriptions for these physical processes. Moreover, semi-analytical models can be combined together to offer a wide dynamic range: Press-Schechter-based models can be combined with models based on $N$-body simulations. Also, $N$-body simulations of different mass resolution and cosmological volume can be combined together.  

For hydrodynamic simulations -- including the so-called ``zoom-in'' cosmological simulations which probe small volumes, often a single galaxy -- the larger computational power will also allow to increase the resolution, reaching scales currently achievable only by dedicated simulations with idealized boundary conditions (down to sub-pc and AU scales).  Since it is already clear from the vendors' strategic plans that exascale platforms will have ``fat nodes'' with at least 128~cores, these simulations, which are much smaller for number of compute elements compared to cosmological volumes, could fit on just a few nodes, partially resolving the load balancing issue mainly caused by communication {\it between} nodes. This also means that, with exascale computing, many more zoom-in simulations could be run with significantly less resources than today, allowing to probe a larger fraction of the parameter space.

For these improvements to be effective, a strong effort aimed at improving current sub-grid models of MBH formation, growth, and dynamics, and including new physical processes (e.g. magnetic fields, cosmic rays, non-equilibrium chemistry, radiation transport, and GR effects) is required. Furthermore, increasing resolution would ease the need for simplified prescriptions or post-processing models, but resolution cannot be increased \emph{ad libitum} and a  treatment of small-scale phases needs nevertheless to be added to simulations, based on the detailed results of smaller-scale simulations. This combination of different scales will be key to properly estimate the MBH spins, masses, and dynamics of MBHBs, and therefore the subsequent cosmic evolution of MBHs and sharpen predictions for, and intepretation of, LISA detections. 

\paragraph{New methodologies: artificial intelligence integrated with simulations}
\label{Sec:New methodologies: artificial intelligence[...]}

Despite the foreseen progress in simulations with the advent of exascale computing, the parameter space potentially probed by LISA will always be too large to be explored at the resolution needed to capture all the effects that determine the time-scales and occurrence rate of MBH mergers. Such effects represent a truly daunting computational challenge of global models. Stochastic processes play a role throughout, which implies that to derive truly robust quantitative models for LISA predictions, e.g. on the mapping between galaxy and MBH mergers, one would need to run a very large sample of simulations. Stochasticity applies both to scales that might be directly resolved in the next generation of cosmological hydrodynamic simulations (10--100~pc) and to scales that will not be resolved for long (parsec scales and below, into the circumbinary disc regime). While semi-analytical models could be used in complement, the complex dependencies of torques/drag regimes on the interstellar medium properties and the stochastic nature of the processes themselves conceptually speak against the use of deterministic phenomenological recipes, which is instead the standard approach of semi-analytical models.

An alternative, promising avenue, which is gaining increasing momentum in observational cosmology and in the analysis of large-scale structure statistics, is artificial intelligence \citep[e.g.][]{2019PhRvD.100f3514F,2020MNRAS.495.1311T,2020arXiv201006608L}. This often entails using neural networks of varying complexity to recognize correlations and patterns, and subsequently produce many realizations of a given model (model emulator technique). One particular interesting class of such neural networks are generative adversarial networks. Such networks are at the base of modern facial recognition algorithms, which are becoming increasingly sophisticated, and are thus designed to work with an extremely large parameter space (each facial feature can be cast as a parameter, essentially). The networks are designed in such a way that they can be continuously updated to recognize deeper features and patterns without retraining, thus essentially allowing  to tune the response based on the needs (namely based on the target, which would be determined by the scientific application). One can imagine training such algorithms to identify complex interstellar medium patterns and to correlate them with orbital decay regimes/time-scales for MBHs. Training would of course have to be done on small-scale simulations (non-cosmological, galactic and nuclear scale). For example, a first application of such techniques to galaxy dynamics is the morphological identification of merging versus isolated galaxies \citep{2018PASJ...70S..37G,2019ApJ...872...76N,2019MNRAS.486.3702S,2020MNRAS.493..922P}, which is becoming increasingly common in these years. 
An emulator of the ``small-scale dynamics'' could be then be designed by integrating the sub-grid model computed via neural networks within a large-scale simulation, using a zoom-in simulation as intermediate step, to encapsulate their trends and results, and implant them in simulations of large cosmological volumes.

\paragraph{Summary of LISA measurable quantities and how it will inform us on MBH physics}

\begin{itemize}

    \item LISA can determine the mass of merging MBHs at any time. We expect LISA to discover MBHs closest to the redshift of their formation. At such high redshift, the emitted radiation of these MBHs (that are likely low-mass objects) is too faint to be detected by current EM missions. On single events, the detection of MBHs with $M_{\rm MBH}\leqslant 10^{5}\,\rm M_{\odot}$ would confirm the existence of light MBH seeds, while not ruling out the existence of heavy ones (see Section~\ref{sec:formation} for a description of MBH formation channels). The detections of MBHs with $M_{\rm MBH}\geqslant 10^{5}\,\rm M_{\odot}$ cannot, however, validate the existence of heavy seeds as the MBHs could be grown light seeds (except if many of these detections take place at very high redshift). In case of a sufficiently large number of events, LISA will provide us with constraints on the most likely dominant MBH formation channels, as well as the first constraints on the low-mass end of the MBH mass function from low to high redshift.
    
    \item LISA can measure the effective spin of merging MBHs (see Section~\ref{sec:spin}). Posterior distributions of the spins will be used to determine the spins of the two merging MBHs. For single events, it provides us with information on the dominant nature of the growth of these MBHs, i.e. whether their accretion histories were chaotic or coherent (in other words, whether MBH growth is accretion- or merger-dominated). Spin distributions for the population of MBHs detected by LISA will constrain the relative contributions of MBH growth channels as a function of MBH mass and redshift (see Sections~\ref{sec:accretion_vs_merger}, \ref{sec:feedback_barrier}, and \ref{sec:spin}). In particular, spin measurements are likely going to be possible up to very high redshift ($z\sim 10$) with per-cent precision for nearly a third of the detections {\citep{2016PhRvD..93b4003K}}.
    
    \item LISA will measure on the full sky the merger rate of MBHs in the mass range $10^{4}$--$10^{7}\, \rm M_{\odot}$. First, the observations of MBH mergers would be the evidence that these BHs dynamically pair and merge within relatively short time-scales, especially if observed at high redshift. Second, the merger rate of LISA will constrain a combination of MBH physical characteristics (MBH seeding, MBH dynamics, efficiency of MBHs to sink to galaxy centers, MBH growth) and characteristics of their host galaxies (Section~\ref{sec:merger_rate}). LISA will constrain the number density of merging MBHs, independently of their activity. As such, LISA could enable new investigations of the fraction of obscured AGN (by e.g. comparing LISA results to current and future AGN surveys).
    
    \item Localization of the LISA events on the sky will be crucial to enable multi-messenger science towards a full characterization of MBH physics and demographic evolution of MBHs. Among many new potential directions, LISA could open a new window on the origins of gamma-ray bursts (Section~\ref{sec:supermassive_stars}), jet (Sections~\ref{sec:supermassive_stars} and \ref{sec:astroparticle}) and cosmic ray astrophysics (Section~\ref{sec:astroparticle}), and MBH accretion (Section~\ref{sec:MBH_fuelling}). Localization of the events in space and time could also help linking merging MBHs to their galactic and larger-scale environments and further disentangle MBH formation and growth channels as well as MBH and galaxy co-evolution (e.g. Section~\ref{s5:host_galaxies}).
    
\end{itemize}

\subsection{Multimessenger on single events: What do we learn about BH physics from the multimessenger view of the coalescence of MBHs? }
\label{sec:sec_2_4}
{\bf \noindent \textcolor{black}{Coordinators:}
Ioana Du\c{t}an,
Delphine Porquet\\
\textcolor{black}{Contributors:}
Imre Bartos,
Tamara Bogdanovic,
Federico Cattorini,
Maria Charisi,
Monica Colpi,
Alessandra De Rosa,
Daniel D'Orazio,
Massimo Dotti,
Massimo Gaspari,
Alberto Mangiagli,
Sean McGee,
Vasileios Paschalidis,
John Quenby,
Milton Ruiz,
Jessie Runnoe,
Antonios Tsokaros,
Rosa Valiante,
Maurice van Putten,
Silvia Zane\\
}

The scientific exploitation of the LISA mission would be greatly increased by performing synergistic, multimessenger observations; that is, combining low-frequency GW observations by LISA with contemporary, prior, or follow-up observations of the same source by EM and astroparticle  messengers. The overall goal of this section is to highlight the multimessenger view of single collisions between MBHs detected by LISA in the astrophysical environment posed by their host galaxies.
We start this section by presenting the expected multimessenger signatures of coalescing MBHs (precursor, coincident, and afterglows observations). We then elaborate on the best observational strategies to maximize the multimessenger observations.
Moreover, we present different inputs on what we need to prepare to improve estimations of the source parameters (e.g. sky position, luminosity distance, chirp mass, and mass ratio). Finally, at the end of this section, we present what is needed in the near future to maximize the scientific returns of LISA.  In this section, particular attention is also given to the synergy between the LISA and Athena\footnote{\href{https://sci.esa.int/web/athena}{https://sci.esa.int/web/athena}} missions, both of which will operate at the same time. 

\subsubsection{The expected multimessenger signatures}
 
The stages which precede and follow the merger of an MBHB feature different spacetime geometries, and the ability to simultaneously detect both the GW and EM signals during each step differs as well. We distinguish between the \textit{pre-merger} (late inspiral) phase, that could lead to the detection of an \textit{X-ray precursor signal}, and the \textit{post-merger} phase, that could lead to \textit{disc rebrightening,} the formation of an \textit{X-ray corona}, and that of an \textit{incipient jet}. This subsection covers first \textit{pre-merger}  signatures, and subsequently
 possible signatures {\it during merger} and  \textit{post-merger}.  There are then additional opportunities of  multi-messenger observations
 associated with potential precursor objects of MBHs themselves, such as SMSs.

\paragraph{Expected EM signatures of MBHB in-spirals at sub-pc scales} 
\label{Sec-4.1.1} 

In order to maximize the synergy between contemporaneous LISA and EM observations on single MBHB coalescence events, an understanding of the pre-merger population of MBHBs at sub-pc scales using EM  observations is crucial. This section focuses on searches for MBHBs that are being carried out at the present time, and that can inform us of the expected LISA merger rate and possibly the expected orbital parameter distributions at merger time. The power of these predictions, however, will rely on how close to merger we can probe an EM identifiable MBHB population. After LISA detects MBHBs, interpretation of formation and evolution channels will rely on the characterization of EM identified populations. In such a case, population samples over the widest possible range of MBHB orbital parameters will be useful in piecing together the entire life stories of MBHBs.
 
While multiple methods for EM identification of a population of MBHBs have been proposed and practiced over the last two decades (for more details, see Section~\ref{sec:sec_2_5}), there is currently no definitive observational evidence for MBHBs with separations of order one parsec or smaller. 
Hydrodynamical simulations of circumbinary accretion show that the accretion rate onto an MBHB can be strongly modulated at multiples of the orbital periods \citep{2009ApJ...700.1952H, 2008ApJ...672...83M, 2013MNRAS.436.2997D}. This has led to searches for sub-pc separation MBHBs manifesting as $\mathcal{O}(yr)$ time-scale periodicity in quasar light curves. Of order 100 such candidates exist to date \citep{2015MNRAS.453.1562G, 2016MNRAS.463.2145C, 2019ApJ...884...36L}. However, distinguishing the periodicities from the noise processes intrinsic to AGN variability remains a significant challenge \citep[e.g.][]{2016MNRAS.461.3145V, 2020ApJ...900..117Z}.
Signatures unique to MBHBs, with which to vet these periodic quasar candidates, have been proposed through the relativistic Doppler boost and binary self-lensing models for periodic variability and flares \citep{2015Natur.525..351D, 2018MNRAS.474.2975D, 2020MNRAS.495.4061H,2018MNRAS.476.4617C}.   

Most of the effort has been focused on the exploration of large optical spectroscopic surveys (e.g. SDSS) using several approaches,  searching for:
 \begin{itemize}
 \item large velocity differences between the narrow and broad emission lines, tracing the host galaxy and at least one of the two MBHs, respectively \citep{2011ApJ...738...20T, 2012ApJS..201...23E, 2013MNRAS.433.1492D, 2014ApJ...789..140L,
 2015ApJS..221....7R,
 2017MNRAS.468.1683R}, 
 \item a time varying shift of the broad emission lines, tracing the highly accelerated motion of one of the two MBHs in a binary \citep{2013ApJ...777...44J, 2013ApJ...775...49S,
 2017ApJ...834..129W, 
 2019MNRAS.482.3288G}, or
 \item peculiar ratios between broad emission lines with different ionizing potentials due to the tidal effect of the other component of the candidate binary \citep{2011MNRAS.412...26M, 2012MNRAS.425.1633M}. 
 \end{itemize}
 
If any of these systems are true MBHBs, then the modelling of their broad optical emission lines can in principle yield the properties of the binary, such as the minimum mass, separation, and mass ratio \citep[e.g.][]{2016ApJ...828...68N, 2016ApJS..225...29B, 2017MNRAS.468.1683R, 2019ApJ...870...16N, 2020ApJ...894..105N}. It is important to mention, however, that emission-line features mentioned above are not unique to MBHBs. As a consequence, searches like this can generate relatively large samples of MBHB {\it candidates} whose nature must be tested through continued follow-up or with help of other complementary observational techniques. For example, in the case of SDSS J0927+2943, multi-wavelength follow-up observations disproved both the binary and recoiling-MBH hypotheses \citep{2014MNRAS.445.1558D}. 

Because they rely on the existing EM spectroscopic surveys, searches of this type are generally biased toward active MBHB candidates with masses $\gtrsim 10^{6-7} \msun$ and orbital separations $\gtrsim 0.01$ pc (\citealt{2018ApJ...861...59P}; \citealt{2021MNRAS.506.2408X}). Similarly, they are sensitive to MBHBs at redshifts $z\lesssim 1$--2 \citep{2011MNRAS.412...26M, 2012MNRAS.425.1633M, 2016ApJ...828...68N}. Therefore, because of the observational selection effects, these widely used techniques may uncover a fraction of MBHB systems that are {\it progenitors} to binaries in the LISA frequency band but will not be detected by LISA because the coalescence time-scale is too long. These same techniques will miss low-mass and high-redshift systems, as well as the systems that do not show AGN activity intense enough to allow for a proper modelling of the broad emission lines. 

More promising from the standpoint of the coincidental multimessenger detections are MBHBs with smaller orbital separations than those discovered by optical spectroscopic searches (the latter provides signatures too weak to detect if separation distances are smaller than the typical scales of the broad line regions).
One can in principle search for such MBHBs using the broad iron fluorescence emission lines, observed at about 6.4\,keV in the X-ray spectra of many individual AGN with masses as low as $\sim$10$^6 \msun$ \citep[e.g.][]{2014SSRv..183..277R}. The broad iron emission lines are emitted by the parts of the accretion flow in the close proximity of the MBH (within $\sim$10--1000 gravitational radii). They can therefore trace the relative motion of the two MBHs even when they are well within the LISA band \citep{2012MNRAS.420..860S, 2015MNRAS.452L...1M, 2018MNRAS.479.3804S}, as long as at least one of the MBHs exhibits observable AGN activity (see the discussion in Section~\ref{sec:MassiveBlackHolesAndTheirPathToCoalescence}). However, their current observations are limited to redshifts significantly smaller than those of the expected bulk of LISA MBHB coalescences, indicating a need for a high-sensitivity X-ray detector that will be able to measure broad iron emission lines in the spectra of AGN at higher redshift, such as the Athena mission. Additional signatures include modification of the disc emission caused by the presence of a gap, ``suppressing'' emission from an annulus in the multi-colour black-body model, and shocks caused by matter hitting the minidiscs detectable in X-rays   \citep{2012MNRAS.420..860S,2014ApJ...785..115R}. In all cases, large enough signal-to-noise ratios (necessary in order to discriminate between single and double MBHs) will require long integration times, comparable to or longer than the orbital period of binaries in the LISA band. 

What about directly imaging and tracking the orbits of many MBHBs at sub-pc separation in the near future?
Advances in Very Long Base Inteferometry (VLBI) at mm-wavelengths should make the direct imaging possible, and this would definitely be very complementary to the indirect methods described above. For example, the Event Horizon Telescope \citep[EHT;][]{2019ApJ...875L...2E} has the angular resolution and sensitivity to astrometrically track the orbits of MBHBs separated by $0.01$~pc at Gpc distances.
Simple MBHB population models suggest that near-future mm-VLBI experiments could directly image and track the orbits of many such MBHB in, and possibly before, the LISA era \citep{2019BAAS...51g.235J,2018ApJ...863..185D}. 

The next generation Very Large Array (ngVLA) will be added to this effort, with the ability to resolve MBHB pairs down to sub-10 pc separations and also track binary orbits through changing pc-scale jet morphology \citep{2018arXiv180804368B}.

In summary, to prepare for LISA we must invest in theoretical understanding of accretion flows around MBHBs:\begin{itemize}
    \item to better understand what drives MBHB orbital evolution, and hence generate more accurate predictions for MBHB populations,
    \item to more accurately predict observational signatures generated before, during, and after merger, that we can reliably disentangle from AGN variability associated with single MBHs -- this partly requires understanding such intrinsic variability better as well, 
    \item to better model mm-wavelength emission from MBHB accretion for direct EM detection prospects.
\end{itemize}
On the observational side, we must:
\begin{itemize}
    \item continue to extend time-domain surveys to longer baselines, in order to mitigate false-periodicity detections due to AGN red noise, 
    \item improve our understanding of intrinsic AGN noise processes and quasi-periodic oscillations,
    \item  improve methods for detecting non-standard periodicity (e.g. variable accretion and self-lensing induced periodic flares),
    \item advance MBHB-related science goals for VLBI experiments that could directly image MBHB orbits.
\end{itemize}

\paragraph{Expected EM counterparts during the late inspiral and merger stages}\label{sec:EMlatestage}
 
The multimessenger detection of MBHB inspirals and mergers will certainly establish unique breakthroughs in various fields of physics and astrophysics; yet, one shall be mindful of a series of caveats that make the concurrent observation of this class of events uncertain. Besides lacking firm predictions on the EM light-curves and spectra of coalescing MBHBs under a variety of conditions (see Section \ref{TheoObsImprove}), the structure and properties of the astrophysical environment around MBHBs are uncertain as well and largely depend on the supply of gas for accretion in the aftermath of a galactic merger. 

Generically, MBHBs can be surrounded by a circumbinary disc, and mini-discs can form around the two black holes  (see Section~\ref{sec:hardening_gas}).
The accretion of mini-disc gas onto each MBH is expected to produce copious amounts of X-ray radiation. Analogously to what  was discussed in Section~\ref{Sec-4.1.1}, the orbital motion of the binary may imprint a modulation to the expected X-ray emission thanks to Doppler boosting or modulations in the accretion rate. The modulation is expected to be in phase with the GW incoming signal, allowing the correct identification of the host galaxy in the relatively large area provided by LISA \citep{2017PhRvD..96b3004H,2018MNRAS.476.2249T,2019ApJ...886..146D}. After the identification, alerts could be sent to other facilities in order to observe the very prompt emission. 

Dynamical GR simulations of MBHBs in the force-free limit, which assumes that the plasma around the BHs is tenuous, suggest that two separate jets during the inspiral, one around each BH, could emerge from these systems \citep{2010Sci...329..927P, 2012ApJ...749L..32M}, providing a complementary way to search for MBHBs in the late inspiral phase. 

What happens at the time of merger is an active subject of research. The natal kick imparted by the GW recoil affects the properties of the accretion disc leading to modifications in the spectrum and light curve that can be non-universal depending on the orientations of the kick relative to the orbital plane pre-merger \citep{2008ApJ...684..835S,2010MNRAS.401.2021R}. The birth or rebrightening of a jet \citep{2014PhRvD..90j4030G,2018PhRvD..97d4036K} are also possible outcomes. These studies have revealed new possibilities for EM counterparts from binary BHs that arise from jets in binary AGN. As a result, both non-thermal X-ray and gamma-ray signatures from these systems are expected, which would be of interest to Athena as well as other X-ray and gamma-ray satellites. 

\paragraph{Possible GW and EM signatures of MBH formation from the collapse of supermassive stars}\label{sec:supermassive_stars}

A widely accepted model of long gamma-ray bursts, with a typical duration of $\sim 30$ s, is the so-called
collapsar scenario. In this model, a BH accretion disc system forms after the core-collapse of a massive low-metallicity star, and launches a relativistic jet. The jet breaks through the stellar debris producing gamma-rays \citep{1992ApJ...395L..83N,1986ApJ...308L..43P,1993ApJ...405..273W}.
 
Supermassive stars can be responsible for the formation of MBH seeds (see Section~\ref{sec:formation}), hence they could have played a crucial role in generating the population of MBH binaries that LISA can detect out to very high redshift. If they were common at $z > 10$, then a direct-collapse population of high-redshift MBH binaries would have been prominent, leading to a very different population of GW sources detectable with LISA at high redshift relative to the case of light MBH seeds originating from Pop III stars (see Section~\ref{sec:formation}). Hydrodynamic simulations in~\cite{2016PhRvD..94b1501S} found that the collapse of a $\gtrsim 10^5\, \rm M_\odot$  massive star at redshift $z=3$ emits GWs, with a peak amplitude of  $5\times 10^{-21}$ at a frequency of $\sim 5\,\rm mHz$. These GWs may be detectable by LISA \citep[see also][]{Liu:2007cf,Sun:2017voo,Sun:2018puk}. Simulations also found that after~$\Delta t\approx 2000(M_{\rm MBH}/10^6\, \rm M_\odot)$\,s following the MBH formation a magnetically-driven jet is launched. The jet has a lifetime $\Delta t\sim 10^5(M_{\rm MBH}/10^6\, \rm M_\odot)$ s, and the outgoing Poynting luminosity is $L_{\rm EM}\sim 10^{51-52}$ erg~s$^{-1}$~\citep{Sun:2017voo,Sun:2018puk}. These engines can shine for very long times compared to standard gamma-ray bursts and could be detected as ultra long gamma-ray bursts. The combination of GW and EM signals could help us constrain the origins of GRBs and MBHs. 

\paragraph{Expectations from astroparticle observations}\label{sec:astroparticle}

The most likely origin of cosmic rays of energy above $10^{15}$ eV is
in the jets of AGN. 
Shock acceleration is the popular explanation of the power law relativistic
proton and electron energy spectra. 
A number of phenomena occurring in the dense hot plasma
surrounding an MBH binary can affect the jet production and evolution, hence
opening the possibility to use LISA sources, specifically merging MBH binaries,
as novel laboratories for jet and cosmic ray astrophysics.
A favourable observational window would seem to be during a merger where
the separate MBH jets tend to co-align.
The possibility of a spin flip turning two misaligned
jets into one where a single enhanced jet is pointing close to the direction of the spin axis of the
more massive of the two MBHs
has been discussed in \cite{ApJ.697.1621}
and applies to mass ratios $\geq 0.1$. X-rays from the accretion disc relate to the seed particles for the accelerator and the source of p-nucleon or p$\gamma$
neutrino production.
The emergence of a gap in the circumbinary disc or lack of stars to be swallowed in the MBH 
could cause observable EM emission to cease. This situation is suggested
from the lack of EM emission in the observation of the merger of stellar mass BHs.
Correlated observation of GWs with those of the Athena X-ray mission and the
Square Kilometer Array (SKA) in the radio bands could help in 
understanding cosmic ray origin. Specifically, data from Athena would reveal the amount of accreted gas available during the merger to power the jet, while simultaneous radio information would both yield the strength of the magnetic field  associated with the jet and determine the spectrum of the accelerated relativistic 
electrons, which could then be directly related to the acceleration of protons.

For IceCube to detect neutrinos from p-nucleon collision, in the favourable case of a jet boosted flux and if 3 per cent of
the accretion energy is available, requires that $M_{8}m_{e}\gamma_{10}^{4}D_{4}^{-2} \geq 3$
 where $M_{8}$ is mass in units of $10^{8}M_{\odot}$, $m_{e}$ is the ratio of the
accretion rate to the Eddington rate, $\gamma_{10}$ is the jet Lorentz factor
in units of 10 and $D_{4}$ is luminosity distance in units of $10^{4}$ Mpc.
However, the
chance of seeing such a favourable geometry for the dominant jet in a merger is only $10^{-3}\gamma_{10}^{-2}$.
Successful co-observation of neutrinos and GWs is yet to occur \citep{2016PhRvD..9312010A}.

\subsubsection{Multimessenger observation strategy for MBHB mergers with LISA}
\label{sec:sec_2_4_2}

\begin{figure}[ht]
    \centering
    \includegraphics[scale=0.18]{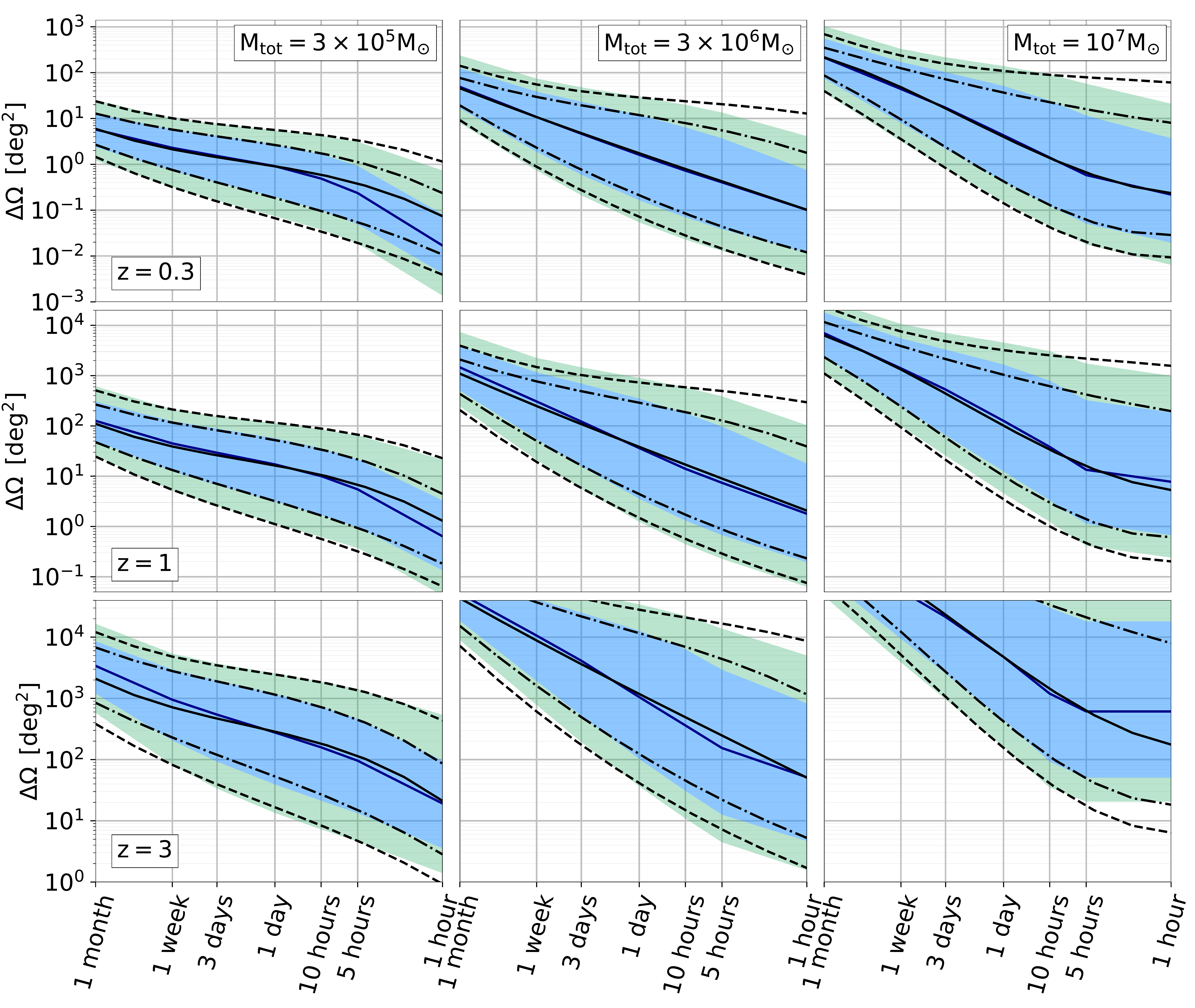}
    \caption{Time evolution of sky position uncertainties from Fisher Matrix simulations, from \citet{2020PhRvD.102h4056M}, for different source-frame MBHB total mass and redshift. Blue lines correspond to the median of distribution, whereas blue and green areas correspond to the 68 and 95 percentiles, respectively.
    Overall, lower-mass systems are localized better than more massive MBHBs. At $z=1$, systems with total mass of $3 \times 10^5 \msun$ are localized within $10 \rm \,deg ^2$ ten hours before merger. The same accuracy is reached for MBHB with $10^7 \msun$ total mass only 1 hour before merger.
    }
    \label{fig:skyloc_and_fit}
\end{figure}

\citet{2020PhRvD.102h4056M} recently demonstrated that overall the parameter estimation {\it on the fly} of light systems at $z\sim 1$ and with total intrinsic mass $\sim 10^5 \ \mathrm{M}_{\odot}$ shows  smaller uncertainties than in heavy systems ($10^7 \ \mathrm{M}_{\odot}$). The chirp mass and mass ratio are well constrained prior to the merger proper, with errors at the per cent level. In Fig.~\ref{fig:skyloc_and_fit}, we report LISA's abilities to constrain the sky position of the source. At $z \approx 1$, MBHBs with a total mass of $3 \times 10^5 \ \mathrm{M}_{\odot}$ can be localized with a median precision of $\sim 100$~deg$^2$ (1~deg$^2$) at 1~month (1~hour) before merger, whereas the sky position of $10^7 \msun$ MBHBs can be determined to within 10~deg$^2$ only 1~hour before merger.  Thus, only light and nearby sources can be traced during the inspiral phase. If the MBHs are embedded in a circumbinary disc, optical emission is predicted from the inner ring of the circumbinary disc and soft and hard X-rays from the mini-discs and the shock heated cavity  \citep{2018MNRAS.476.2249T, 2018ApJ...865..140D}. Modulation of the light curve is expected at frequencies commensurate to the fluid patterns \citep{2017ApJ...838...42B, 2018ApJ...865..140D, 2018MNRAS.476.2249T}. Thus, observatories such as the Vera Rubin large synoptic telescope could detect the optical signal when the sky localization uncertainty falls below $\sim$10~deg$^2$. Athena can strategically tile the optical field of view and then narrow down the sky position to detect a potential modulated X-ray chirp. This is possible for MBHBs in the near Universe ($z \lesssim 0.3$). 

At merger, the sky localization improves down to $\sim$10$^{-1}$~deg$^2$ for all masses, giving us the chance to detect the post-merger emission by staring at the source for a sufficiently long time, from weeks to months, and witness a re-brightening  of an AGN.  Again, no definite spectral template exists to identify the source within the narrower error box indicated by LISA \citep[but see][]{2008ApJ...684..835S,2010MNRAS.401.2021R}, and work in this direction should be performed before LISA flies.  
These multimessenger observations will be unique as for the first time and {\it in real time} it will be possible to correlate the masses and spins of the merging BHs with the  EM emission by the surrounding gas to give quantitative estimates on the efficiency of the emission under extraordinary conditions, such as during the violence of a merger and from gas bound to a moving BH.

The exposure time needed to detect an Eddington-limited system varies with MBH mass, redshift, and can be more efficient in the soft or hard X-ray band depending on the obscuration of the source.
For example, unobscured systems with $M$ $\sim$ 10$^{6-7}$~M$_\odot$ require an Athena exposure time of less than 1~kilosecond (i.e., a single pointing) up to $z=1.5$ in the soft band, whereas systems of $M$ $\approx$ 10$^{5}$~M$_\odot$ can be detected in a kilosecond up to $z = 0.4$.
Similarly, systems of $M$ $>$ 10$^7$ M$_\odot$ require less than kilosecond exposures at redshifts of $z < 4.5$ (\citealt{2020NatAs...4...26M}; \citealt{2021arXiv211015677P}). For super-Eddington sources, shorter exposure times are expected, 
possibly through gas squeezing \citep{2002ApJ...567L...9A, 2016MNRAS.457..939C}. 
However, in the case of obscured sources \citep[whose fraction remains poorly constrained, and could increase with redshift]{2022arXiv220603508G}, whose detections would be more efficient in the hard X-ray band, or objects accreting at low rates below the Eddington limit, the required exposure time of Athena can increase significantly, as described in \citet{2020NatAs...4...26M,2021arXiv211015677P}. A system with $\sim 10^6\, \rm M_\odot$ and a luminosity of $L=0.1 L_{\rm Edd}$ at $z \sim 1$ would require an exposure of more than 100 ks, against less than 10 ks for the same system with $L=L_{\rm Edd}$.

Identifying the best observational strategies to maximize the synergy between LISA and other missions such as Athena is a very recent and active field of research. 
Besides the detectability of the emission, in fact, matching the GW source to its EM counterpart requires the ability of identifying the host among a large number of potential candidates within the LISA error box. This aspect has recently been investigated by \citet{2022arXiv220710683L}, who considered the synergy between LISA and the future X-ray observatories LynX \citep{2018arXiv180909642T} and Athena (see below for the description of the missions). Assuming an active binary at merger, they found that most LISA sources with masses in the range $10^5-10^7\msun$ at $z<2$ will be detectable by those instruments within kiloseconds in a single pointing. However, the number of contaminating AGN unrelated to the GW event can be up to thousands for high-redshift signals, making it hard to pinpoint the correct host. Identification strategies need to be developed but require a better theoretical understanding of the peculiar features associated with the EM counterpart. For example, \citet{2018MNRAS.476.2249T} find that the EM luminosity of a merging binary is suppressed in the last cycles prior to merger and enhanced after coalescence; if this is the case, a viable identification strategy would be to perform sequential pointings and search for a source displaying a monotonically increasing flux; in this case, the exposure time for each pointing might depend on the sky localization posterior distribution provided by LISA with longer exposure times for regions with higher probability to host the MBHB event. Identification of newborn jets powered by an highly spinning merger remnant (mentioned in Sec.~\ref{sec:EMlatestage}) might offer another possibility for unambiguous counterpart identification. The best way to suppress the  number of contaminants would be to improve the GW localization, which would be possible if LISA is joined by a second space-borne detector such as Taiji \citep{2018arXiv180709495R}
or TianQin \citep{2016CQGra..33c5010L}. In particular, several works \citep{2021Resea202114164R, 2021PhRvD.104b4012W, 2022PhRvD.105f4055S}  showed that LISA-Taiji joint observations would improve the precision of the sky localization by three orders of magnitude. With this assumption,  \citet{2022arXiv220710683L} demonstrated that unambiguous identification of the active AGN related to the binary would be possible up to $z=2$.

In the context of sources identification, it is also important to mention that the astrophysical uncertainties on the population of merging MBHBs and on the type of EM emission strongly affect the number of expected EM counterparts. Recently, \citet{2022arXiv220710678M} computed the number of expected EM counterparts, starting from catalogs of merging MBHBs. Combining the information from radio, optical, and X-ray emission with the information from LISA sky localization, they estimated between 7 and 20 counterparts in 4 yr of LISA time mission. However, in the case of obscuration or collimated radio emission, the number of EM counterparts reduces to 2 or 3. This implies that a better understanding and modeling of the galaxies hosting MBHBs mergers are necessary to be ready for the LISA mission.

In general, it is clear that, in order to best prepare for LISA, we need to investigate better these multimessenger aspects, especially in view of forthcoming missions that could be operational when LISA will fly. AXIS \citep{2018SPIE10699E..29M} and LynX \citep{2018arXiv180909642T} are two NASA concept X-ray observatories that could also fly simultaneously with LISA. 
Their flux sensitivity will be at least one order of magnitude better than Athena (while having smaller fields of view, see Section~\ref{sec:sec_2_5}), making it possible to observe the X-ray emission from fainter AGN than achievable by current missions or Athena. Compared to the 5-10 arcsecond angular resolution of Athena, 
the high angular resolution of AXIS (sub-arcsecond resolution compared to the 5-10 arcsecond resolution of Athena), its fast slew rate and ToO response could be key for monitoring MBH binaries until coalescence.  As mentioned above, further investigations are required in the near future to determine whether the sky position uncertainties of merging MBHB systems would be compatible with the characteristics of AXIS and LynX, and particularly their small fields of view.

\begin{figure*}[ht]
    \centering
    \includegraphics[width=0.9\textwidth]{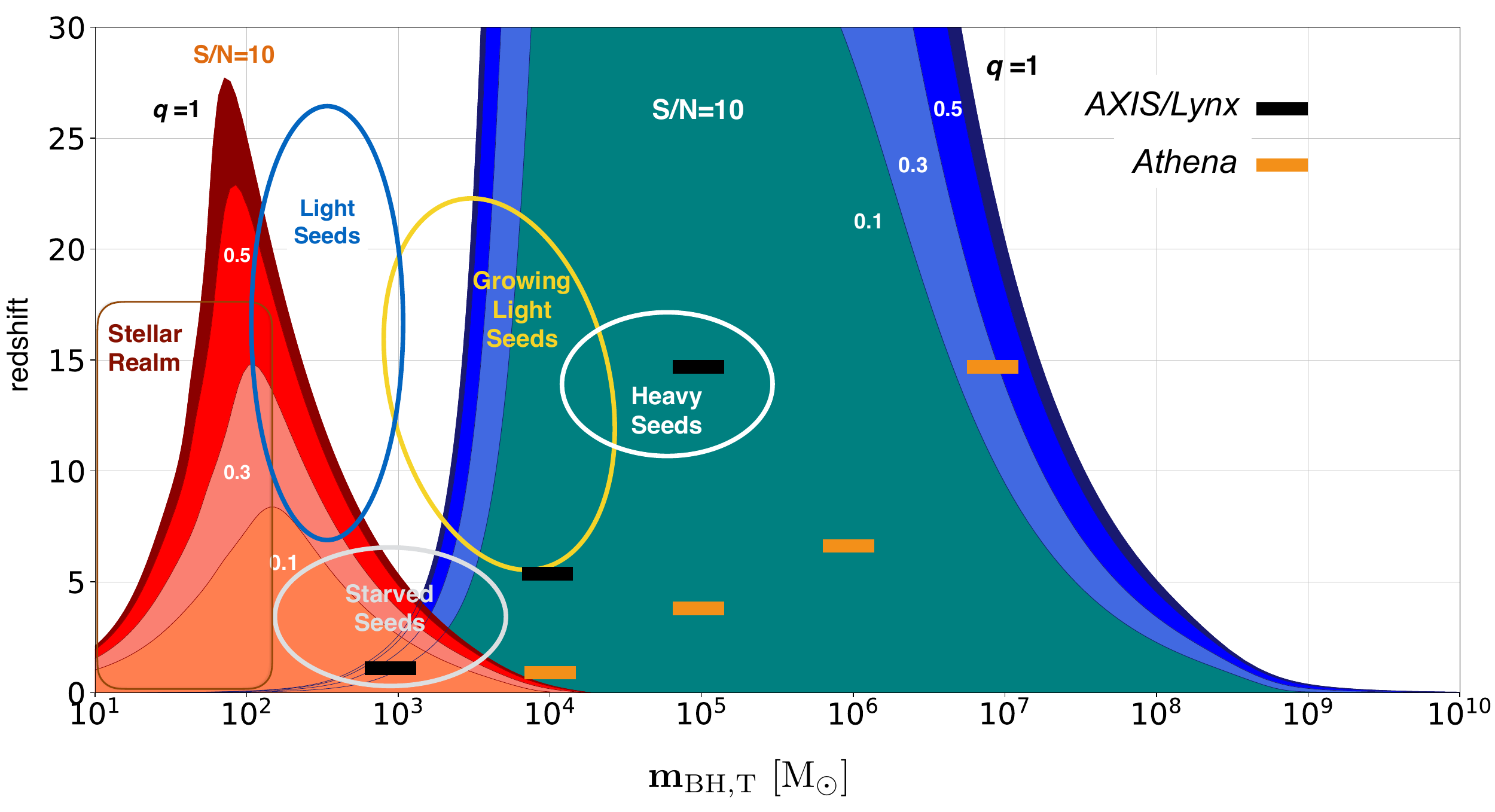}
    \vspace{-0.4cm}
    \caption{LISA will be complemented by the X-ray mission Athena (launch expected in the early 2030s), and potentially by the NASA concept missions LynX and AXIS. These missions are shown in orange and black horizontal symbols, which indicate the sensitivity of the deepest pointing, in the [0.5--2]~keV observed band, by {\it Athena} (orange) and {\it LynX/AXIS} (black).
    Waterfall plots show the average GW horizon computed for signal-to-noise ratio $\rm SNR = 10$ and different BH mass ratios for the Einstein Telescope (red) and LISA (blue/green) bandwidth \citep{Santamaria10,Hild2011,2019CQGra..36j5011R}. 
    For reference, main MBH formation mechanisms are shown with ellipses. The growth of some of  MBH seeds could be stunted by several processes and could be detectable only at late times when merging with other MBHs at $z\leq 5$ \citep[``starved MBHs'' in the white bottom ellipse,][]{2021MNRAS.500.4095V}.
    Figure taken from \citet{2021MNRAS.500.4095V}.}
    \label{fig:GWsummary}
\end{figure*}

We show in Fig.~\ref{fig:GWsummary} how these X-ray missions will complement LISA by partially covering the same MBH mass and redshift ranges.
The figure also illustrates the possible synergy between LISA and the Einstein Telescope (ET). As developed in Section~\ref{sec:MBHoriginandgrowthacrosscosmictime}, there are a lot of hurdles to grow light seeds efficiently in the high redshift Universe, and a population of long-living ``starved'' (i.e.g, ungrown) merging MBH seeds  could exist   \citep{2021MNRAS.500.4095V}. In this mass range coordinated multi-band GW observations are possible, with LISA having the capability to first follow the early inspiral of MBHBs, and tracking the merger phase. This unique combination will revolutionise our ability to carry out precise measurements of the source parameters also at $z\sim 5$ \citep{2020NatAs...4..260J}. This mass and redshift range would also be covered by the X-ray missions LynX and AXIS.

In this section, we mainly discussed multimessenger observations with X-ray observatories. However, multimessenger observations with LISA span a large range of wavelengths. 
For example, we can learn about the physics of jets using radio and optical observations on LISA systems. Radio astronomy, for instance ngVLA as well as SKA, will allow us to observe MBH jets turning on. SKA should also be capable of detecting very
luminous flares in the radio emitted by an equal mass MBH binary at merger time, and, through that, help with their
sky localization \citep{2016JCAP...04..002T}. This can be complemented by observations of optical flares from e.g., the Roman Space Telescope or the Rubin Observatory (see Section~\ref{sec:sec_2_5} for a more complete descriptions of relevant instruments and  space missions).
Observational and theoretical constraints on these EM flares are still very poor, e.g., on the frequency of the peak emissions.

\subsubsection{The path towards LISA}

In this subsection, we present several ideas for what we need to prepare to exploit the unique characteristics of LISA in the context of multimessenger study of MBHBs from the perspectives of theory, observations, and artificial intelligence.

\paragraph{Theoretical and observational improvements in the multimessenger study of MBHBs}\label{TheoObsImprove}

On the theoretical front,  work is necessary to understand accretion on to binary MBHs and their EM signatures at various wavelengths; on the observational front, efforts are necessary to find and understand more MBHB candidates. Strengthening the collaborative studies between the EM and GW scientific communities is thus very important for scientific utilization of LISA data products.\\

\noindent$\bullet$  {\bf Numerical simulations  of  EM counterparts to MBHB inspirals and mergers}

Over the last decade, several theoretical groups studied MBHBs in a circumbinary disc or more tenuous gas clouds (see Section~\ref{sec:EMlatestage}), systematically adding the layers of physics necessary to investigate potential mechanisms for EM counterpart signals emerging during MBHB inspiral and merger.
Newtonian viscous hydrodynamics \citep{2015MNRAS.446L..36F, 2017MNRAS.469.4258T} and MHD \citep{2016ApJ...832...22S} simulations investigated the dynamics of the gas streams being stripped off the inner edge of circumbinary discs. MHD simulations over a post-Newtonian background spacetime explored the first stages of the strongly relativistic behaviour of MBHBs in circumbinary discs in the form of the disc's response to binary orbital evolution by GW emission \citep{2012ApJ...755...51N} and, more recently, examined the mass-feeding mechanisms onto the individual mini-discs around the BHs \citep{2017ApJ...838...42B, 2018ApJ...853L..17B} and the systems' radiative properties in the stage immediately prior to merger adopting ray-tracing techniques \citep{2018ApJ...865..140D}.

The first simulations in full, dynamical GR with resolved BH horizons and the MHD plasma from a circumbinary disc were performed in~\cite{2012PhRvL.109v1102F} and \cite{2014PhRvD..89f4060G}, modelling the binary-disc pre-decoupling epoch, and in~\cite{2014PhRvD..90j4030G}, modelling the post-decoupling, merger, and post-merger epochs. The inclusion of the BH horizons in these studies showed that powerful outflows and jets are launched from these systems even when the BHs are  non-spinning. The more recent study in~\cite{2018PhRvD..97d4036K} found that accretion rates, temperatures, and jet launching from the interactions of the horizons with the magnetized medium exhibit modest dependence on the initial disc thickness. Jets in binary AGN would produce both non-thermal X-ray and gamma-ray signatures, which would be of interest to Athena as well as other X-ray and gamma-ray satellites. However, modelling from first principles of such EM signals is currently absent. Therefore, efforts must be made towards adding radiation transport in dynamical-spacetime general relativistic MHD  (GRMHD) simulations of accreting MBHBs.

In addition to GRMHD simulations in dynamical spacetime, dynamical GR simulations of MBHBs have been performed in the force-free limit, which assumes that the plasma around the BHs is tenuous~\citep{2009PhRvL.103h1101P, 2010PhRvD..81h4007P, 2010PhRvD..82d4045P,2010Sci...329..927P,2010PhRvD..81f4017M,  2012ApJ...749L..32M}. These studies showed how the orbital motion of the BHs alters magnetic and electric fields and leads to possible EM emissions. In particular, it was suggested that two separate jets during the inspiral, one around each BH, could emerge from these systems \citep{2010Sci...329..927P, 2012ApJ...749L..32M}. 

Dynamical-spacetime simulations of MBHBs have also been performed in moderately magnetized clouds in \cite{2012ApJ...752L..15G}, showing a rapid amplification of the magnetic field over the last few orbits, leading to the creation of a post-merger magnetically dominated funnel aligned with the spin axis of the final BH, with properties relatively insensitive to aspects of the initial configuration \citep{2017PhRvD..96l3003K} 

The future of numerical simulations will require an improved insight into the fuelling rate and the MHD properties of plasma accreting on to the BHs to sharpen the EM predictions. Furthermore, it will be necessary to match a range of different spatial scales in order to more properly address the evolution of the accreting gas during the early inspiral up to merger. These simulations will also need to account for radiation processes in order to correctly estimate EM light curves/spectra, as well as other modes of accretions \citep[such as chaotic cold accretion][Section~\ref{s5:host_galaxies}]{2013MNRAS.432.3401G,2015A&A...579A..62G} and radiation feedback (e.g. \citealt{2017MNRAS.468.1398S}). The development of reliable radiation transport schemes in dynamical spacetime is therefore a high priority. 

\paragraph{Artificial intelligence: Deep learning methods to identify GW source candidates and to estimate LISA source parameters}

In recent years, artificial intelligence has been intensively applied in astronomy for a wide variety of tasks. As a sub-field of artificial intelligence, machine learning\footnote{
Any machine learning algorithm aims to learn from data. 
The data is divided in a {\it training} data set used for the algorithm learning process, a {\it validation} data set (optional) to evaluate the progress of learning, and a {\it test} data set to evaluate the algorithm performance. 
The level of accuracy of an algorithm depends on the available data, its complexity, and more specifically on the size of the training data.
} has gained increasing popularity among astronomers, especially through utilization of one of its sub-sets, namely deep learning, when big data is involved. The most widely used deep learning algorithms are neural networks\footnote{Neural networks were inspired by the structure and the function of the brain, and they can be thought of as networks of neurons organised in layers: predictors (or inputs) form the bottom layer, forecasts (or outputs) form the top layer, and there may also be intermediate layers containing hidden neurons.} containing multiple hidden layers that progressively extract higher-level features from input data.\\

\begin{figure}
    \centering
    \includegraphics[width=0.8\textwidth]{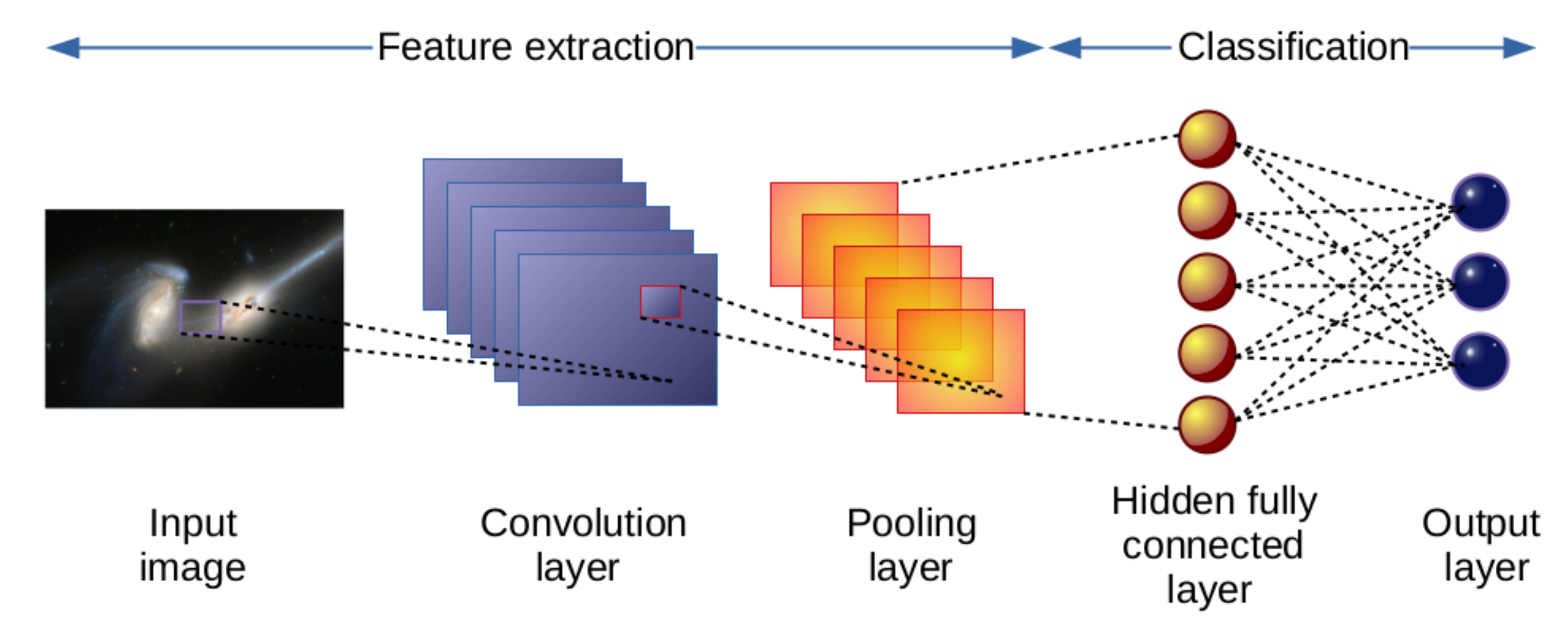}
    \caption{Schematic representation of a basic convolutional neural network architecture. Such numerical network can be trained on simulations, and later apply to observations to systematically identify MBHB candidates. Figure credit: Ioana Dutan}
    \label{fig:SchemeCNN}
\end{figure}

\noindent$\bullet$  {\bf Deep learning methods to identify GW source candidates from EM observations}
An important issue in astronomy is to find astronomical sources in survey images in order to build source catalogues. These catalogues are valuable tools used for testing theories and numerical simulations against observational data. Convolutional neural networks are deep learning neural networks designed for processing structured arrays of data such as images (see Fig.~\ref{fig:SchemeCNN}). Convolutional neural networks are very good at learning features from images by hierarchical convolutional and pooling operations. 

More precisely, convolutional neural networks algorithms have been already applied for image classification in order to find sources/objects in different EM wavebands. Among the uses in relation to LISA science we can mention detection and classification of quasars from light curves and identification of galaxy mergers  \citep[e.g.,][]{2019A&A...626A..49P, 2018MNRAS.479..415A}. Image classification employed in observational cosmology and in the analysis of large-scale structure statistics can set the stage for improving the estimations of time-scales and occurrence rate of MBH mergers via integration of artificial intelligence with simulations (see discussions in Section~\ref{Sec:New methodologies: artificial intelligence[...]}). 

More importantly, the time spent to generate catalogues decreases dramatically when using deep learning algorithms instead of standard approaches. 
However, to reach a desired level of accuracy in image classification, training a deep learning algorithm can be costly in terms of duration and computational resources. Nevertheless, once properly trained, the algorithm can quickly classify thousands of GW source candidates (e.g. \citealt{2019A&A...626A..49P}). Here, the training and validation/test samples can be either observational data from ongoing and upcoming galaxy imaging surveys or simulated data. Using such an approach, some possible biases in observations or additional requirements in simulations might be identified. In spite of current achievements, we need to further design algorithms that are able to learn representative features faster and achieve higher performance in image classification in order to better understand the AGN population and to find binarity signatures in the observational data.\\

\noindent$\bullet$  {\bf Deep learning algorithms for estimation of LISA source parameters}
Standard algorithms used for estimation of the physical parameters that govern GW signals are effective, but the computations are time-consuming and they can take up to a few days. For example, in the case of synergistic observations with both LISA and Athena, a poor sky localisation of the source during the inspiral phase limits the possibility of performing concurrent observations. Moreover, the Athena capability of carrying out a {\it target of opportunity} is about 4 hours; that is, a low-latency alert should be released in less than 4 hours in order for Athena to be able to watch the merger phase. Therefore, reducing the computational time of source parameters is crucial for multimessenger studies. Over the past few years,  deep learning algorithms have been employed for classification of glitches (non-Gaussian noise transients) in Advanced LIGO data
(data referring here to the ``BH coalescence signal + noises'', e.g. \citealt{2018PhRvD..97j1501G,2018CQGra..35i5016R}). This allowed the identification of signals in Advanced LIGO data, where the training of the algorithm is performed on simulated stellar-mass BH merger signals in synthetic Gaussian noise representative to LIGO sensitivity (e.g. \citealt{2018PhRvL.120n1103G}), and for estimation of source parameters (e.g. \citealt{2020PhRvL.124d1102C,2020arXiv200803312G}). 
Such models may have limited capacity as they do not currently account for an holistic approach to a quasi-realistic GW data analysis specific to LISA, where tens of thousands of signals overlap with many gaps and glitches. Nevertheless, the current models represent a starting point from which novel architectures can be trained on non-stationary, non-Gaussian noise LISA-like data to conduct parameter estimation. Such development can allow us to perform time-sensitive multimessenger searches to greatly increase the science return of the LISA and other (future) experiments and observatories.

\clearpage

\setlength{\headheight}{75pt}

\subsection{Multimessenger view of MBH populations}

\label{sec:sec_2_5}
{\bf \noindent \textcolor{black}{Coordinators:}
Maria Charisi,
Alessandra De Rosa\\
\textcolor{black}{Contributors:}
Stefano Bianchi,
Tamara Bogdanovic,
Monica Colpi,
Pratika Dayal,
Ioana Dutan,
Saavik Ford,
Massimo Gaspari,
Melanie Habouzit,
Albert Kong,
Sean McGee,
Barry McKernan,
Francesca Panessa,
Delphine Porquet,
Raffaella Schneider,
Stuart Shapiro,
Rosa Valiante,
Maurice van Putten,
Cristian Vignali,
Marta Volonteri,
Silvia Zane\\
}

LISA will bring crucial constraints on mass, redshift, and spin of merging BHs in the mass range $\sim 10^4-10^7 \,\rm M_\odot$. To achieve a complete understanding of the population of MBHs, from high redshift to the local Universe, from low to high mass, single and in binaries, the synergy of LISA with other missions will be key. In this section, we provide a global view on the facilities that complement LISA, or will complement it in the near future. We discuss how these  missions will address different aspects of MBH physics and populations, but also how they will help us to understand the galactic and large-scale environments in which MBHs assemble, which is a major question in modern astrophysics.

\subsubsection{A landscape of new missions to understand MBH formation, growth, and environment}

\begin{figure}[h]
    \centering
    \includegraphics[scale=0.44,angle=270]{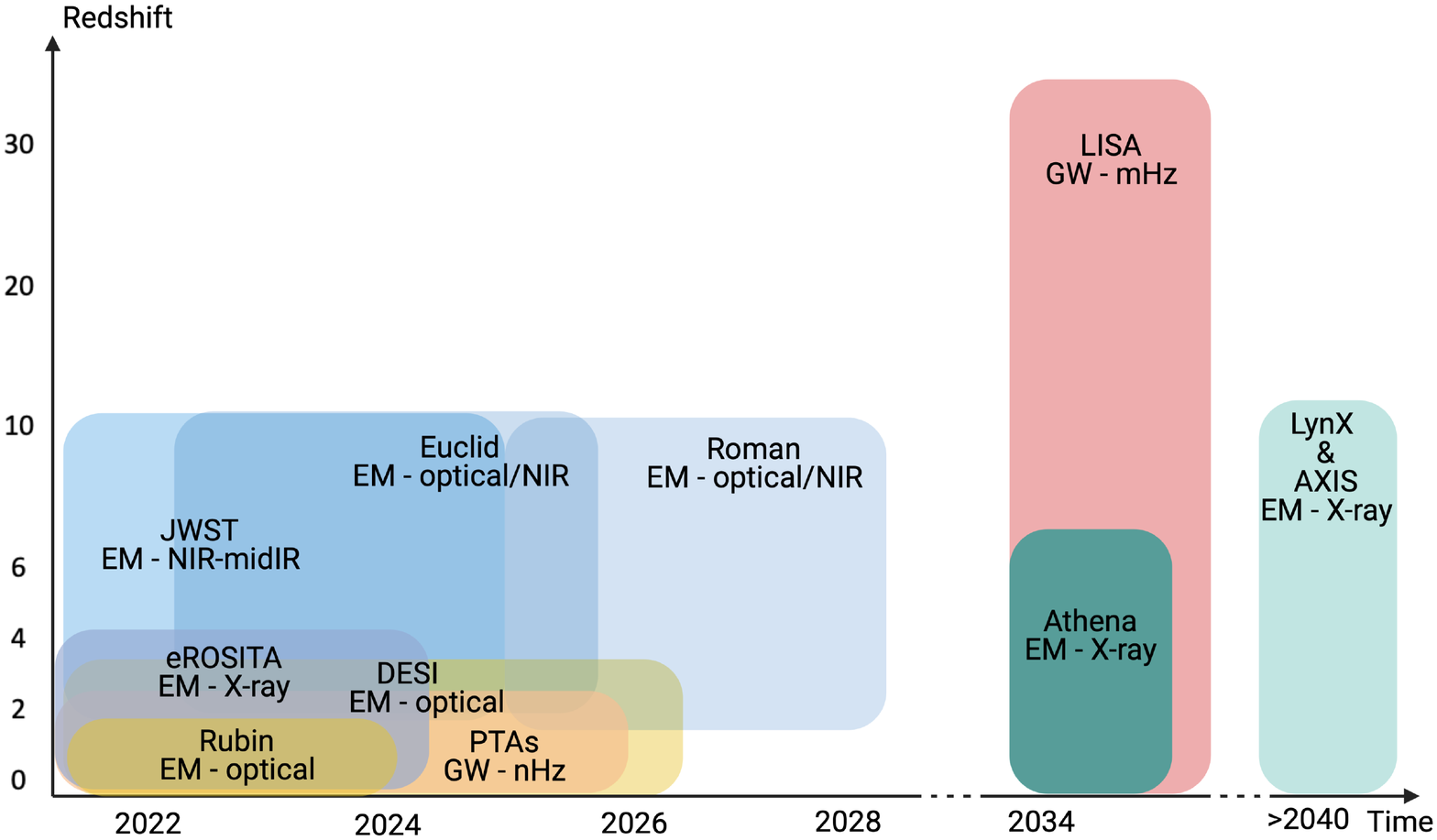}
    \caption{Landscape of the upcoming and concept missions aiming at constraining the population of MBHs and their host galaxies, from the local to the high-redshift Universe. These missions will significantly increase current EM detections towards high redshifts ($z\sim 10$), while LISA will reach redshifts (e.g., $z\geqslant 30$) that will not be available with EM observations. We caution that the timelines reported in the figure are only indicative as delays in the launch of any of the missions, especially those a few years away from the time of writing, are always possible. In addition, at the time of writing the likely launch time-frame for ATHENA is set to slightly earlier than that of LISA. 
    Characteristics of these and other missions are listed in Table~\ref{Table:table_summary_missions}. Figure credit: Melanie Habouzit.
    }
    \label{fig:figure_allmissions}
\end{figure}

\paragraph{A diversity of missions to complement LISA}

By exploring the ``light'' MBHs (the low end of the mass distribution), LISA will open a new window on the GW spectrum, bridging the gap between high-frequency ground-based GW observations (e.g. by LIGO and Virgo), and the nano-hertz frequency observations by PTAs \citep{2016MNRAS.458.3341D,2019BAAS...51g.195R,2019MNRAS.490.4666P,2020PASA...37...20K}. 
PTAs are currently building up constraints on the GW background, generated by tight binaries of MBHs at the high-mass end ($M_{\rm BH}\sim 10^{7-9}\, \rm M_{\odot}$) in the low-redshift Universe \citep{2019A&ARv..27....5B}.  
LISA will be deaf to the population of even lower-mass seed mergers with $10^2$--$10^3 \,\rm M_\odot$, whose signal falls below the detection threshold (although some
portion of the in-spiral might be 
accessible).
These are potential sources for ground-based GW observatories, such as the Einstein Telescope (ET) and Cosmic Explorer. Space- and ground-based missions together will provide a complete census of MBHs, from the earliest seed BH mergers to the largest MBHs today.

The GW detections will be complemented by new observations of MBHBs from space- and ground-based facilities across the EM spectrum, as shown in Fig. \ref{fig:figure_allmissions}.
The ESA L2 mission Athena \citep{2015JPhCS.610a2008B}, the proposed NASA missions AXIS \citep{2019BAAS...51g.107M} and LynX (\citealt{2018arXiv180909642T}; \citealt{2019JATIS...5b1001G}), and the ongoing eROSITA mission \citep{2010SPIE.7732E..0UP} will probe the accretion properties of AGN in X-rays, while surveys like the Dark Energy Spectroscopic Instrument (DESI; \citealt{2016arXiv161100036D}), the James Webb Space Telescope (JWST; \citealt{2006SSRv..123..485G}), the Nancy Grace Roman Space Telescope \citep{2012arXiv1208.4012G} and Euclid \citep{2013LRR....16....6A} in the optical and IR band will investigate galaxy hosts up to the highest redshifts. 
The next-generation ground-based optical telescopes, like the Extremely Large Telescope (ELT; \citealt{2016SPIE.9906E..0WT}) and the Thirty-Meter Telescope (TMT; \citealt{2013JApA...34...81S}), will reveal the assembly of the first galaxies, and large-area photometric and spectroscopic surveys, like the Rubin Observatory Legacy Survey of Space and Time (LSST; \citealt{2019ApJ...873..111I}) and the Sloan Digital Sky Survey-V (SDSS-V; \citealt{2017arXiv171103234K}), are expected to discover a treasure trove of binary candidates. 
Wide-area and deep radio surveys will be available with radio interferometry provided by the Square Kilometre Array (SKA, \citealt{2009IEEEP..97.1482D}).

In Table~\ref{Table:table_summary_missions}, we summarize some of the existing and upcoming space- and ground-based telescopes, with their sky coverage and key science that they will address both leading up to LISA's launch and concurrently with LISA. We discuss in detail the role of all these missions in the following.

\paragraph{The synergy of LISA and EM missions to answer major questions on MBHs and their host galaxies}\label{s5:host_galaxies}

In the next decades, EM and GW messengers will work in concert, providing new knowledge of the galaxy and MBH assembly processes, as well as of the interplay between dynamic gravity and the relativistic plasma. Multimessenger observations are an emerging research domain of modern astrophysics. In the following, we detail how several missions can work in synergy with LISA to answer key scientific questions.\\

\begin{landscape}
\begin{table}
\caption{Landscape of the upcoming and concept missions that will be complementary to LISA, and will provide us with crucial constraints on the population of MBHs and their host galaxies. See the text for references.
} 

{\footnotesize
\label{Table:table_summary_missions}
\begin{tabular}{|l|l|l|l|l|l|}
\hline
Missions &Wavelength & Types & Sky coverage  & Launch  &Goals \\
\hline
LISA & GW mHz & laser interferometery & all sky & mid 2030s & MBH mergers with $M_{\rm BH}=$ up to $z=20$,\\
&&&&&constraints on BH mass, redshift, spin.\\

\hline
PTAs & GW nHz & pulsed radio emission & all sky & current & GW background powered by low-redshift\\
&&&&&MBHs of $M_{\rm BH}\sim 10^{7-9}\, \rm M_{\odot}$.\\

\hline
eROSITA & X-ray (0.3--10 {\rm keV})& spectroscopy & all sky & 2019 & 3 million AGN, of which several tens of thousands at z$\ge$ 3 \\
&& imaging &&&\\

\hline
Athena & X-ray (0.3--10 {\rm keV})& spectroscopy & 2.4/30 $\rm deg^{2}$  & 2030s & AGN with $L_{2-10\, \rm keV}\geqslant 10^{41-43}\, \rm erg/s$ depending on redshift.\\
&& imaging &WFI deep/shallow &&\\

\hline
AXIS & X-ray (0.3--10 {\rm keV})& spectroscopy &  2.5/50  $\rm deg^{2}$ & concept  & AGN with $L_{2-10\, \rm keV}\geqslant 10^{40-43}\, \rm erg/s$  depending on redshift.\\
&& imaging & Medium/Wide &mission&\\

\hline
Lynx & X-ray (0.3--10 {\rm keV}) & spectroscopy & 2 $\rm deg^{2}$ & concept  &
AGN with $L_{2-10\, \rm keV}\geqslant 10^{39-41}\, \rm erg/s$ depending on redshift, \\
&&imaging&&mission& potentially reaching MBHs of $M_{\rm BH}\sim 2\times 10^{4}\rm M_{\odot}$ at $z=7$-10.\\

\hline
 IXPE/  & X-ray (2-8 \rm{keV}) & polarimetry &  pointed observations, & 2022  & MBH accretion in star forming galaxies\\
 XL-Calibur/eXTP & & & limited survey capability  & & MBH spin and mass, astrophysical environments \\
 &&&&&  of the MBHs, AGN outflows  \\

\hline
JWST & NIR-midIR & spectroscopy & 46/190 $\rm arcmin^2$   & 2021 & High-redshift galaxies up to $z\sim 10$, \\
&($0.6-28\mu m$)&imaging& Deep/Medium &&high-redshift quasars, spectrum of young MBHs, \\
&&&&&constraints on MBH formation mechanisms.\\

\hline
Roman & optical/NIR & imaging & 2000 $\rm deg^{2}$  & $\sim 2025$ & Mapping high-redshift galaxies,\\
&($0.5-2 {\rm \mu m}$) && WFI HLS &&detection of massive quasars of $\sim$ 10$^9$ M$_\odot$ up to $z \sim 10$\\

\hline
Euclid & optical/NIR & spectroscopy &  40/15000 $\rm deg^{2}$  & $\sim 2022$ &Mapping high-redshift galaxies,\\
&($550-2000 nm$) &imaging& Deep/Wide && detection of massive quasars of $\sim$10$^9$~M$_\odot$ up to $z \sim 10$\\
\hline 
DESI & 360-980 $nm$ & spectroscopy & 14000 $\rm deg^{2}$ & 2021 & Mapping high-redshift galaxies and quasars\\
\hline
E-ELT &$0.35-14\mu m$& imaging &  & 2025 & confirmation of high-redshift quasar candidates\\
& & spectroscopy & & & \\
\hline
SKA & $0.01-4m$ & radio interferometry & 10-20  $\rm deg^{2}$  & 2027 & duty cycle of jet launching in AGN \\
&&& SKA1-MID deep &  & provide detailed insights on feedback/feeding loop in AGN\\

\hline
Rubin & 320–1050 $nm$ & photometry & 18000 deg$^2$ & 2023 & detection of sub-pc MBHBs through photometry\\
LSST &&&&  &variability study on 10$^{4-5}$ AGN\\

\hline
SDSS-V & $380-920 nm$ & spectroscopy & all sky & 2020 & detection of sub-pc MBHBs through spectroscopy \\
     & &&&& spectroscopic identification and redshift of quasar/AGN\\
     & &&&& MBH mass at $z = 0.1$--4.5\\
\hline
\end{tabular}
}
\end{table}
\end{landscape}

\noindent$\bullet$ {\bf The formation of the first MBHs}

Thanks to its distant horizon and vast volume probed, LISA will be able to detect the first coalescing massive seeds of $10^4$--$10^5 \,\rm M_\odot$, witnessing the dawn of MBHBs at redshifts that are not reachable with EM observations. LISA will not, however, detect all the first MBHs, but only those that form binaries and merge, or those that form from the collapse of SMSs that provide a sufficiently high GW signal. EM observations, targeting a complementary population, supplement LISA's detections to provide an improved understanding of the fundamental question of MBH formation. 

Currently, the only EM observational insights on low-mass MBHs can be obtained from relatively local (up to $z\sim 2$) dwarf galaxies of total stellar mass $M_{\star}=10^{7}-10^{9.5}\, \rm M_{\odot}$ \citep[e.g.][]{2015ApJ...813...82R,2015ApJ...809L..14B,2016ApJ...817...20M, 2018MNRAS.478.2576M}.  Unlike massive local galaxies, which have experienced significant mass growth, local dwarfs have experienced less growth through cosmic history \citep{2017MNRAS.468.3935H}. Their MBHs are also expected to have experienced a similar limited growth. Properties of MBH formation could have been preserved in local low-mass galaxies (\citealt{2008MNRAS.383.1079V}; \citealt{2012NatCo...3.1304G}).

To directly probe the properties of seed MBHs, before they  grow  significantly in mass via gas accretion \citep{2018MNRAS.474.3825V}, it is necessary to search for such sources at high redshifts ($z>10$). Theoretical models, including spectral-synthesis, predict that the emission from accreting heavy seeds (e.g., direct collapse MBH seeds), will be strong mainly in the IR-submm  and X-ray bands, and thus could be detected by JWST up to $z\sim 15$ and Athena up to $z>6$. By contrast, the emission from lighter accreting seeds ($\sim$10$^4 \,\rm M_\odot$) is expected to be weaker and difficult to observe with EM facilities at high redshifts \citep{2015MNRAS.454.3771P, 2017ApJ...838..117N, 2018MNRAS.476..407V,2018NatAs...2..987B}. The NASA concept mission Lynx, and to a lesser extend AXIS, which has lower sensitivity, directly aim at detecting these young, faint and faraway AGN. 

Given the small fraction of the sky that EM missions such as JWST, Athena, and Lynx will cover (see Table~\ref{Table:table_summary_missions}), an optimized observational strategy is crucial to detect as many low-mass MBHs as possible. 
The MBH community is currently leading an effort to build up target selection criteria designed on the basis of the combined analysis of IR colours (colour-colour cuts), X-rays-to-optical flux ratios (rest frame), IR excess, and UV continuum slopes to efficiently detect and distinguish candidates \citep{2017ApJ...838..117N, 2018MNRAS.476..407V}. \\

\noindent$\bullet$  {\bf The growth of MBHs}

MBH growth is one of the major open questions in astrophysics, and because of this, constraining it is one of the main goals for several of the upcoming EM surveys. Understanding the processes that determine the growth of MBHs from low-mass seeds to MBHs with mass $10^8$--$10^{10}$~M$_{\odot}$ requires observations of MBHs at different evolutionary stages over cosmic history. With the large samples of AGN/quasars discovered by these surveys, we expect significant developments before LISA flies.  Having a better understanding of MBH growth will help us refine the theoretical models that will be confronted with LISA data and sharpen the astrophysical interpretation of LISA's detections.

The current population of rare bright high-redshift quasars \citep[$z \sim 6$--7,][]{2011Natur.474..616M,2016ApJS..227...11B,2018ApJ...856L..25B,2019ApJ...883..183M,2020ApJ...897L..14Y} powered by MBHs of $10^8$--$10^{10}$~M$_{\odot}$ will be extended in the coming decade by several EM space and ground-based missions, and will provide us with a unique snapshot in the MBH growth timeline, offering a complementary view to LISA probing the low-mass end of the MBH mass spectrum. The Nancy Grace Roman Space Telescope \citep{2019BAAS...51c.121F} and the Euclid space telescope \citep{2019A&A...631A..85E} are  set to increase tenfold the number of high-redshift quasars discovered in the near-IR. By mapping large fractions of the sky, these surveys will identify quasar candidates, which will be confirmed with spectroscopic follow-up. At lower redshift, the  SDSS-V Black Hole mapper program will deliver MBH mass measurements for about 1000--1500 quasars/AGN between redshift 0.1 and 4.5 \citep{2017arXiv171103234K}.

X-ray observatories will also greatly enlarge the population of known AGN, in particular at high redshift. eROSITA, an all sky survey strong of an expected sample of about 3 million AGN, will study the accretion history of MBHs by measuring the luminosity-dependent fraction of obscured objects; studying the clustering properties of X-ray selected AGN at least up to $z\sim$2; and identifying rare AGN sub-populations such as high redshift, possibly highly obscured nuclei. Athena, with higher sensitivity than current missions, aims at detecting over 400,000 AGN, several hundred of which at $z\ge$6. With even higher sensitivity, the concept missions Lynx and AXIS aim to push the quest for faint AGN by two orders of magnitude in intrinsic luminosity. Besides probing high-redshift quasars, X-ray telescopes will complement the census of MBHs by discovering obscured AGN at the peak of the accreting MBH activity ($z\sim 2-4$), which are inaccessible with optical/near-IR facilities but are crucial in order to obtain a complete census of MBH growth.  We will be able to build the mass and spin distributions of a large population of MBHs  that will provide  crucial information on the growth process (e.g. merger versus accretion,  \citealt{2008ApJ...684..822B},  see also Section~\ref{sec:spin}).\\

\noindent$\bullet$ {\bf The co-evolution of MBHs and cosmic structures}

Over twenty years of observations have unveiled fundamental correlations between the properties of galaxies and the mass of their central MBH. This promoted important advancements in extragalactic astronomy, suggesting that the central MBHs and the host galaxies co-evolve from high to low redshift. 
Notable correlations are the $M_{\rm BH}$-$\sigma$ and $M_{\rm BH}$- $M_{\rm bulge}$ relations, which link the stellar velocity dispersion $\sigma$ and the mass of the stellar bulge $M_{\rm bulge}$ with the mass of the MBH (see \citealt{2013ARA&A..51..511K}, \citealt{2016ASSL..418..263G} for reviews). The correlation extends to haloes of galaxies, relating the MBH mass to the hot plasma halo temperature or luminosity (\citealt{2019ApJ...884..169G,2019A&A...630A.144B}). These correlations indicate that the MBH, albeit tiny compared to the entire galaxy, is linked to the stellar component and the surrounding intracluster/intragroup medium (up to $\sim10$ per cent of the virial radius). The co-evolution between MBHs and galaxies/haloes is possible due to the self-regulation between feeding and feedback processes, from near the MBHs' horizon to the edge of the bound stellar and dark matter structure
(see \citealt{2020NatAs...4...10G} for a review). 
One way LISA will contribute to investigating the link between MBHs and their hosts is by offering completely independent mass measurements, allowing us to better calibrate the known correlations with galaxy host properties since currently mass measurements suffer from biases introduced by EM observations. Advanced optical/IR facilities will be instrumental in constraining the related stellar properties and evolution of the hosts, both for LISA sources and for MBHs with more uncertain mass measurements.
Future galaxy surveys of JWST, Euclid, and Roman will include the host galaxies of LISA-band MBHs, i.e. lower-mass galaxies than possible to detect today. Respectively, these telescopes should uncover galaxies with stellar mass of $\geqslant 10^{7}, 10^{9.5}, 10^{8}\, \rm M_{\odot} $ at high redshift. Among others, the PRIMER JWST Treasury Program should detect about 120 000 galaxies out to $z \sim 12$ \citep{2021jwst.prop.1837D}, the FRESCO Program $\sim 1200$ galaxies at $z\sim 5$--6.5 \citep{2021jwst.prop.1895O} and $~\sim 300$ galaxies at $z\sim7$--9), and the WDEEP Program $\geqslant 1000$ mostly low-mass galaxies with $10^{6-7}\, \rm M_{\odot}$ \citep{2021jwst.prop.2079F}. Further investigations in the community are required to assess whether the galaxies of these surveys could be later matched to LISA events.
To connect MBH and galaxy mass with the feeding and feedback physics expected to establish their self-regulation, next-generation X-ray telescopes (Athena, LynX and XRISM; \citealt{2018SPIE10699E..22T}) will constrain the inner hot accretion flows and surrounding plasma haloes, in particular by leveraging IFU instruments with high spectral and spatial resolution. Radio-mm telescopes (such as ALMA and LOFAR/SKA) will provide constraints on relativistic jets (especially at low frequencies), their launching mode, and duty cycle of AGN kinetic feedback, thus providing us with a comprehensive view of the role of feeding, feedback and self-regulation in the co-evolution of galaxies and MBHs.\\

\noindent$\bullet$  {\bf The impact of the cosmic large-scale structure on the MBH merger rate}

Observational studies of MBH scaling relations (eg, $M_{\rm BH}$-$\sigma$ and $M_{\rm BH}$-$M_{\mathrm{bulge}}$) have shown evidence for a dependence on large scale environment. Both central and satellite galaxies in galaxy groups and brightest cluster galaxies  appear to have larger MBH masses given their galaxy velocity dispersion \citep{2013ApJ...764..184M, 2013MNRAS.436.2708M, 2019ApJ...886...80D,2012ApJ...753..140B, 2013MNRAS.436.2708M}.  These departures in the scaling relations remain controversial and could be due to selection effects in the observational samples. If the results are confirmed, the cause could be an enhanced galaxy and MBH merger rate in dense environment, but alternatively tidal effects from the host group/cluster could strip stellar material from the host galaxy \citep{2008MNRAS.384.1387V,2019MNRAS.484..794G, 2019MNRAS.485..396V} or an additional channel of MBH growth could result from the host galaxy's interaction with the hot intragroup/cluster medium \citep{2017Natur.548..304P, 2020ApJ...895L...8R}.

In group/cluster environments it is difficult to disentangle mergers and interactions from the abundant projection effects, so observational results on their relative rate are not firmly established \citep{2012MNRAS.425..287E, 2014ApJ...797..127P}. The low-surface brightness and wide-field capabilities of the Rubin Observatory will enable the identification of diffuse merger and tidal features which should narrow the observational uncertainty \citep{2020arXiv200111067B}. The identification of large samples of stripped galaxies \citep[e.g.,][]{2010AJ....140.1814Y} combined with AGN measures from time-varying photometric analysis or X-ray measurements will allow a robust quantification of this growth channel. The improvements in the understanding of this physics prior to the launch of LISA should enable stronger predictions for the environmental dependence of the MBH merger rate. If MBH mergers are more common in biased overdense regions , this will also help focusing the efforts for finding the EM counterpart of LISA sources.\\

\noindent$\bullet$ {\bf Matter behaviour in the strong field gravity regime }

Astrophysical BHs span 10 orders of magnitude in mass, allowing for unique tests of the scale invariance of gravitational effects. LISA will detect in-spiraling and merging MBHs.
LISA's ability to perform tests of gravity through BH coalescences and EMRIs as well as to
provide spin measurements will be complemented by EM and GW observations that probe the behaviour of matter in different gravity regimes. Electromagnetic 
signatures of  the inspiral and merger phases in the X-ray domain are produced so close to BHs that relativity enters into modelling their production. 
Spin measurements using EM observations include relativistic effects in modelling and data analysis.  

The motion of accreting plasma near BHs provides a powerful diagnostic to study the very deep potential well generated by the central object. The infalling matter forms an accretion disc that may extend down to the ISCO, in the vicinity of which the bulk of the X-ray radiation is emitted. X-ray timing, spectroscopic and polarimetric techniques for probing matter flows into the strong gravity regime have been developed and, the first two have already been applied to real data, allowing us to infer the mass and spin of MBHs \citep{2000PASP..112.1145F,2006ARA&A..44...49R,2011CQGra..28k4009M,2014SSRv..183..277R}. Moreover, observations of matter orbiting a BH can be used to verify some of the key predictions of GR in a stationary spacetime metric, i.e a very different - and complementary - setting to that probed using GW measurements of merging BHs.

The relativistically broadened Fe lines observed in accreting BHs are direct diagnostics of matter behaviour in the strong-field gravity regime. In the standard scenario, the hot gas in the `corona' produces thermal Comptonized emission that is reflected by the inner regions of the accretion disc, resulting in the Fe K$\alpha$ emission line.
Special relativity (Doppler boost and relativistic aberration) and GR (gravitational redshift, light bending) affect the shape of the Fe line \citep{2000PASP..112.1145F}. When line profile templates are fit to real data in both stellar mass BHs and MBHs, the disc inner radius and inclination can be measured. If the inner disc radius is assumed to be the ISCO, then the spin of BHs, which depends directly on the spin magnitude and whether the accretion disc is prograde or retrograde with respect to the BH rotation, can be inferred  \citep{2006ApJ...652.1028B,2007ARA&A..45..441M,2014SSRv..183..277R}. In the near future, the X-ray polarimetry mission IXPE (launched in December 2021, \citealt{2016SPIE.9905E..17W}) will offer an independent method to measure inclinations and BH spins, mainly in stellar mass BH \citep{1980ApJ...235..224C,2009ApJ...691..847L,2009ApJ...701.1175S}. Larger effective area X-ray polarimetry missions, such as eXTP \citep{2019SCPMA..6229502Z}, will extend the technique to the weakest AGN.

In order to further advance our understanding of the behaviour of matter around BHs, we need higher sensitivity (i.e. large effective area) and higher energy resolution, allowing a better characterization of the BH environment through the study of the narrow emission/absorption features in the X-ray spectrum. 
This will be achieved with the next generation of X-ray telescopes such as Athena, AXIS, Lynx, STROBE-X \citep{2018SPIE10699E..19R}, eXTP, and XRISM, which are expected to produce unprecedented quality spectra with short exposures. 
Observations with such telescopes will minimize the modelling uncertainties (e.g. due to disc inclination, absorption properties, geometry), since these facilities will use different techniques (combining spectral-timing, and spectral-timing-polarimetry information) to measure the distinct physical quantities such as BH spin, accretion geometry, and BH mass \citep{2008MNRAS.391...32D,2013arXiv1306.2331D,2019SCPMA..6229504D}. The sample of EM-measured MBH spins will also be enlarged. \\

\noindent$\bullet$ {\bf Signatures of MBHB arising from circumbinary discs}

In contrast to earlier studies, recent simulations of GW-driven, nearly equal-mass binaries all the way to coalescence
\citep{2015MNRAS.447L..80F,2018MNRAS.476.2249T,2018ApJ...853L..17B,2019ApJ...879...76B,2014ApJ...785..115R} have shown that the gas is able to accrete on to the BHs all the way to the merger, despite the rapid contraction of the binary orbit and the formation of a central cavity (see Section~\ref{sec:hardening_gas}).
In the case of an optically thick flow, coronal emission around the two MBHs may give rise to hard X-ray emission at the mini-disc scales and soft X-ray emission from the inner rim of the circumbinary disc. 
Simulations suggest that the binaries can be very bright in hard X-rays when the spatial separation of the two MBHs is below about 100 gravitational radii, with thermal emission from the minidiscs dominating.
The modulation  of the X-ray emission might depend  on the orientation of the binary orbital plane relative to the line of sight, Doppler beaming, and gravitational lensing (see Sections~\ref{sec:hardening_gas} and \ref{sec:EMlatestage}).

eROSITA can detect candidates through the hard X-ray binary signature of shocks in mini-discs \citep{2019ApJ...879..110K}, while notch signatures (i.e. the lower thermal output at the frequencies that would have been radiated from the radii in the cavity)
can be detected in optical surveys.  For instance, the plan for SDSS-V is to acquire spectra for eROSITA's AGN, then joint signatures (notch and shock) can be looked for in the same sources. eROSITA and SDSS-V, however, have relatively shallow flux limits, therefore only rapidly accreting MBHs at the upper end of the masses of interest for LISA can be detected. 

As discussed in Section~\ref{sec:sec_2_4_2}, for sources with mass $\sim 3\times 10^5 \,M_{\odot}$ at $z<0.5$ LISA's sky localization can be of a few square degrees weeks to months prior to merger. This will allow wide-field X-ray (and possibly optical) instruments to observe the EM precursor signal. Such sources, however, are expected to be few. For most sources a few square degrees in the sky localization uncertainty can be generally obtained only days/hours prior to merger, making pre-merger EM observations extremely challenging (if not impossible).   
In the post-merger phase (with $\sim$ 0.1-10 square degrees sky  localization)  we will have the chance to observe the disc re-brightening, the formation of an X-ray corona, and that of an incipient jet. 
In fact, in the post-merger phase, a relativistic jet may be launched by the newly born MBH, with production of gamma-ray emission and afterglow emission in its impact with the interstellar medium \citep{2014PhRvD..90j4030G}. \\

\noindent$\bullet$ \textbf{Astrophysical neutrinos from MBHBs}

Astrophysical neutrinos may originate in AGN jets, as supported by the detection of $10^{15}$ eV neutrinos possibly associated with the blazar TXS 0506+056 \citep{Science.361.6398}. Since MBHBs are also likely associated with AGN, they may be promising sources for GWs+neutrinos observations. Coincident detections of GWs+neutrinos may be facilitated by the fact that neutrino observatories are all-sky detectors (like LISA) and do not need to be pointed towards a specific direction. If neutrinos are detected from a sizeable sample of LISA MBHBs, this will provide invaluable insights on the currently unexplored mechanisms of jet launching and acceleration in the presence of an MBHB.

\subsubsection{Preparing LISA using prior knowledge on MBHBs from current and upcoming missions}

LISA detection rates are uncertain, varying between several to few hundreds over the planned 4-yr mission lifetime. Although the bulk of these events will involve MBHBs with $M_{\rm BH} < 10^5\msun$ at $z>5$, 
more massive sources with $M_{\rm BH}>10^6\msun$ at lower redshift ($z<3$) might be detected at a rate of a few per year (see Section~\ref{sec:StatisticsOnMBHMergers}). A number of EM and GW facilities (already operating or upcoming) are expected to deliver significant results even before LISA flies, allowing us to tackle a number of key questions that will prepare the way for LISA. Detections by LISA will then complement these findings either through multimessenger observations or by opening a new window in the GW spectrum. Some of the main questions are as follows: \textit{How do MBHs pair following a galaxy merger? What role does the gas play during the MBH mergers and on which time-scale does coalescence occur? Can the merging MBHs shine down to the final coalescence?}
In this section, we summarize the EM and GW facilities that will operate before LISA, and discuss their main contribution to understanding MBHBs. LISA will bring unique and invaluable insights on this topic, enhancing the importance of the upcoming discoveries.

\paragraph{Multi-band gravitational waves}  

\noindent$\bullet$ {\bf {Exploring the nHz GW band with Pulsar Timing Arrays to uncover the most massive MBH binaries in the local Universe}}

The detection of GWs with LISA will expand the GW spectrum in the mHz regime, thus enhancing the discoveries of PTAs in the nHz window.
PTAs systematically monitor stable milli-second pulsars over a long period of time, currently spanning almost two decades. 
GWs passing between a pulsar and the Earth change the time required by successive pulses
to travel the path from the pulsar to the Earth. If such deviations in the travel time pulses are correlated over multiple pulsars in the array showing a characteristic quadrupolar correlation signature (Hellings \& Downs curve; \citealt{1983ApJ...265L..39H}), then GWs can be detected. PTAs are sensitive to nHz GWs and thus target MBHBs with masses of $10^8$--$10^{10}$~M$_{\odot}$ at $z\sim 1-2$.  PTAs are expected to detect primarily two signals: (1) the stochastic GW background from the superposition of many unresolved signals, and (2) continuous (monochromatic) GWs from individual sources that stand above the background. The former is expected to be detectable within the next few years, whereas GWs from individual binaries likely will follow soon after \citep{2015MNRAS.451.2417R,2016ApJ...819L...6T,2017NatAs...1..886M,2018MNRAS.477..964K,2020arXiv200904496A}. 
Recently, the North American Nanohertz Observatory for Gravitational Waves (NANOgrav) collaboration, based on their 12.5 years data release with a total of 47 pulsars studied with the
Arecibo Observatory and Green Bank Telescope, showed that the stochastic GW background is consistent with predictions
for the  spectrum produced by SMBHs in the accessible frequency bands \citep{NANOgrav:2020bcs}.
However uncertainties remain large, and admit alternative
explanations such as cosmic strings.  which result in a slightly different spectral slope. Using a larger number of pulsars,
longer observations time, and reducing systematic errors, will be needed to improve upon this latest result.

Because of large theoretical uncertainties in binary evolution, models with similar amplitudes for the GW background predict different merger rates for LISA.
The amplitude of the GW background depends on how often galaxy mergers deliver sub-pc binaries, which in turn depends on how often galaxies merge and on how rapidly bound binaries to reach the GW regime. It further depends on the mass of the MBHBs in galaxies \citep{2019ApJ...876..155S}, with recent upper limits placing constraints on the MBH scaling relationships \citep{2016ApJ...826...11S}.  It is expected that, prior to LISA's launch, PTAs will constrain not only the GW background amplitude, but also the shape of the background spectrum, which encodes information about the binary eccentricities and/or environmental coupling \citep{2009MNRAS.394.2255S,2017PhRvL.118r1102T,2017MNRAS.471.4508K,2019BAAS...51c.336T}. 

Connecting the binary population at two cosmic epochs (i.e. the lower-mass binaries at higher redshifts observed with LISA and the higher-mass local binaries that dominate the GW background in the PTA band) will constrain the processes that drive the binary evolution following a galaxy merger, which have remained highly uncertain for several decades \citep{1980Natur.287..307B}, significantly improving our understanding of galaxy evolution, one of the most fundamental open questions in astronomy. Last but not least, individual MBHBs  in the nHz band are candidates for multimessenger observations \citep{2019BAAS...51c.490K,2020ApJ...900..102A}: the sky localization for PTA detections will be very poor (of order $\sim 100 \, \rm{deg}^2$), making the identification of the host galaxy challenging.  Therefore, PTAs will develop and refine strategies for follow-up observations that will be invaluable for LISA.\\

\noindent$\bullet$  {\bf Prospects from astrometry to reduce the gap between PTAs and LISA}

Low-frequency GWs can also be detected with precise astrometry. GWs passing through the MW can alter the apparent position of the stars on the sky, resulting in a characteristic oscillatory pattern. This requires long-term monitoring of the precise position of a large sample of stars. Fortunately, this can be achieved in the near future by Gaia, which at the end of its planned 5-year mission will provide precise astrometric measurements for billions of stars. 
It has been suggested that Gaia observations will provide high-quality data that would complement data from PTAs.
This because, while the frequency domain is similar to that of PTAs, sensitivity is somewhat higher towards
the high frequency tail accessible of the latter, around 300nHz \citep{2017PhRvL.119z1102M}.  The sensitivity of Gaia at those frequencies, which corresponds
to a strain of order $10^{-14}$, could perhaps allow to detect individual loud sources, such as a supermassive black hole binary with a mass of a few $10^8 M_{\odot}$,  namely straddling between the typical PTA and typical LISA range, provided that the source is very nearby. Yet, even if detections would mainly occur for supermassive black hole binaries in the same range of masses of PTAs, the fact that the detection technique is completely different will, by itself, represent an important step forward.
The three experiments (PTAs, Gaia, and LISA) together will consolidate our knowledge of the evolution of MBHBs through cosmic time.

\paragraph{Multimessenger astrophysics}

\noindent$\bullet$  {\bf The Legacy Survey of Space and Time of Vera Rubin observatory to detect AGN binaries through photometric variability} 

The Vera C. Rubin Observatory will perform the Legacy Survey of Space and Time (LSST), which will provide time-domain observations of unprecedented quality and quantity \citep{2009arXiv0912.0201L}. This is particularly significant for EM searches of MBHBs, since they can be detected as AGN with periodic variability. LSST will monitor a large number of quasars (of order one million) providing multi-band observations with high cadence, and long baselines, extending up to 10 yr. Therefore, it is perfectly suited to detect the relatively short-lived and short-period MBHBs emitting GWs in the LISA band.

Already large numbers of binary candidates are identified in existing photometric datasets from the Catalina Real-time Transient Survey (CRTS; \citealt{2015MNRAS.453.1562G}), the Palomar Transient Factory (\citealt{2016MNRAS.463.2145C}), the Panoramic Survey Telescope and Rapid Response System (PanSTARRS; \citealt{2019ApJ...884...36L}), and the Dark Energy Survey (DES; \citealt{2020arXiv200812329C}). However, currently it is extremely challenging to distinguish the sources with genuinely periodic variability from the typical AGN that show intrinsic red noise variability \citep{2016MNRAS.461.3145V}. 
LSST will also facilitate binary searches from that perspective. The vast sample of AGN will allow an improved statistical description of the red noise properties of AGN, thus minimizing the false periodic detections.

The upcoming detections with LSST , along with the current candidates, will illuminate the accretion processes in the presence of a binary, paving the way for multimessenger observations with LISA. More importantly, LSST will constrain the demographics of the population of GW-emitting binaries, the distribution of periods, masses, and mass ratios. Additionally, LISA will provide independent measurements for the binary parameters, allowing us to examine potential biases in EM searches for binaries.

Another exciting possibility arises from the expectation that LSST will detect thousands of tidal disruption events (TDEs). The rate of TDEs depends on \textit{(i)} the dynamics of stars surrounding MBHs; and \textit{(ii)} the density surrounding MBHs. As orbits of stars can be perturbed by MBHB, it is expected that bound MBHB have a different rate than single MBHs, and N-body simulations actually find that galaxies hosting an MBHB should have a significantly higher rate of TDEs \citep{2017ApJ...834..195L}. Therefore this is a possible explanation to the over-representation of TDEs in galaxies which undergone a starburst $\sim$1~Gyr ago but currently exhibit no sign of star formation \citep[E+A galaxies; ][]{2016ApJ...818L..21F}. However, theoretical works have not converged on the origin of these post-starburst galaxies: galaxy mergers triggering nuclear star formation and enhancing the central stellar density \citep{2016ApJ...825L..14S,2019MNRAS.488L..29P,2021MNRAS.500.3944P} provides a possible explanation, but the higher TDE rate could also be due anisotropy in the nuclear star cluster produced caused by the starburst \citep{2016ApJ...831...84L} or a merger \citep{2018MNRAS.480.5060S}. In any case, the galaxies in which TDEs are detected may be more promising hosts of MBHBs or MBH pairs, and may serve as signposts for binary follow-up observations.\\

\noindent$\bullet$ {\bf Spectroscopic search of MBHBs with the fifth Sloan Digital Sky Survey (SDDS-V)}

Spectroscopic surveys, like SDSS, provide another potential route to detect sub-pc MBHBs with EM observations. If one of the MBHs in a binary system is surrounded by enough gas to produce a prominent broad line region, the motion of the MBH will result in detectable Doppler shifts in the broad emission lines (\citealt{2020ApJ...894..105N} and references therein). The spectroscopic database of SDSS has already provided significant samples for spectroscopic searches of MBHBs and dozens of binary candidates have been identified from their large broad-line offsets \citep{2012ApJS..201...23E}. However, these are not unique signatures for binaries, and long-term spectroscopic follow-up is necessary in order to observe coherent changes in the broad emission lines and confirm the binary nature of the sources \citep{2017MNRAS.468.1683R}.

SDSS-V will provide a promising sample for this type of search, since the BH mapper program will spectroscopically monitor thousands of AGN over multiple epochs \citep{2017arXiv171103234K}. This time-domain component to the spectroscopic survey will allow the detection of several more candidates. These candidates are likely progenitors of LISA sources before entering the GW-dominated phase of their evolution, since at mpc separations, the broad line region around individual MBHs cannot be that prominent. However, they can bridge the gap in our understanding of binary evolution in the sub-pc regime.\\

\noindent$\bullet$ {\bf Identifying MBHBs through morphological and spectral investigations at radio wavelength}

Radio emission in galaxies can directly mark the location of the MBH, since it is typically associated with active MBHs. In case of binary systems, if both nuclei are active, then a double radio core can be resolved. However, such systems are rarely found \citep{2006ApJ...646...49R}. Sometimes, jets are produced 
and their associated synchrotron emission can help in tracing the past and current dynamics of MBHs in a merging system. Radio observations are crucial in identifying pairs via a morphological, spectral, and variability investigation. 

Nowadays the highest spatial resolutions on ground are achieved by Global VLBI (Very Long Baseline Interferometry) network observations, that combine radio telescopes all over the world to synthesize an equivalent Earth size instrument, being able to reach angular resolution at milli-arcsec scales, allowing to map the nuclear sub-pc scales for nearby sources \citep{2020arXiv200702347V}.

Future radio observatories such as Next Generation Very Large Array (ngVLA) and Square Kilometre Array (SKA) will work in excellent synergy with LISA on several grounds. On the one hand, they will be able to identify the radio EM counterparts to GWs due to MBHBs mergers, thanks to their high-resolution, sensitivity, and fast-mapping capabilities. On the other hand, the large-scale surveys on wide areas and with nearly  $\mu$Jy/beam sensitivity will significantly increase the dual AGN population at sub-kpc separation,  by several orders of magnitude (see \citealt{2015aska.confE.143P}). For instance,  SKA1-MID is expected to detect a few hundreds of dual AGN per square degree and probe scales of 1--100~kpc  \citep{2014Natur.511...57D}. 
In addition, precise measurements of AGN core positions will allow the investigation of offset MBH predicted by gravitational recoil. A combination of long baselines and high frequencies can ideally map and identify cores from MBHBs at sub-pc separations. The ngVLA and SKA with a VLBI expansion will allow to resolve the sub-mJy source population, tracking the orbital motions of radio cores for the most nearby GW candidates \citep{2017ApJ...843...14B,2018arXiv180804368B}. 
Jet precession might be due to the presence of an MBHB, potentially producing an X-shape morphology \citep{2020MNRAS.493.3911H} that can be traced by high-sensitivity low surface brightness observations as offered by SKA. In addition, radio light curves of AGN can show periodic activity that can be associated to orbital precession. 
Candidate binaries, dual and offset MBH can be cross-matched with multi-frequency observations to confirm their nature (redshifts from the optical spectra, X-ray emission, e.g. from Athena, etc.).\\

\subsubsection{The path towards LISA}

In the following, we inventory the different steps, studies, surveys, and developments, which to us seem crucial in view of LISA, and which are based on current and upcoming observational facilities.

\begin{itemize}
\item In the near future, the eROSITA X-ray survey will dramatically improve constraints on the MBH population at the upper end of the LISA band and beyond, up to high redshift.
While waiting for the new X-ray missions with better sensitivity and spatial resolution, such as Athena, AXIS, and Lynx, we should aim at exploiting the best capabilities of Chandra and XMM in order to characterize and confirm the candidate MBHB selected through optical variability (photometric variability).
Moreover, we should improve modelling of intrinsic emission related to disc-corona in AGN, in order to reduce systematic uncertainties on the estimates of MBH spin and mass. This goal requires the use of spectral-timing (and spectral polarimetry in the future with, e.g. the eXTP mission) techniques that need deep observations of specific targets to investigate the variability properties.

\item The upcoming surveys from DESI, JWST, Euclid, ROMAN, and the next phase of SDSS will provide massive catalogues of galaxies. It is imperative to enhance these catalogues with measurements of their MBHs (e.g. MBH mass) which will facilitate the identification of host galaxies for a large sample of LISA MBHB coalescences.  

\item The optical photometric surveys offer the possibility to select large samples of MBHB candidates (see previous section). These candidates should be further monitored with highly rewarding, albeit risky, X-ray observations in order to confirm or reject their binary nature. This will constrain the expected event rate for LISA. Moreover, X-ray observations with their ability to arbitrate between genuine MBHs and false positives will allow us to validate and refine the selection techniques. Such techniques will be widely used and improved in future facilities such as LSST. 

\item It is also crucial to improve the numerical simulations of inspiralling MBHBs embedded in gaseous discs, considering accretion properties and detailed thermodynamics of single MBHs and including radiative feedback. These studies will provide more reliable EM signatures that will allow the detection of LISA EM counterparts. They will also facilitate the efficient discovery (with low contamination of false detections) of binary candidates populations with current and future EM facilities.
\end{itemize}

%% file: astroWP_03emri.tex
\section{Extreme and Intermediate Mass-Ratio Inspirals}
\noindent {\bf \textcolor{black}{Chapter coordinators:} Pau Amaro Seoane, Saavik Ford, Alejandro Torres-Orjuela, Martina Toscani, Lorenz Zwick}\\

\subsection{Introduction}

\noindent
{\bf Coordinators: Pau Amaro Seoane, Saavik Ford and Cole Miller

\noindent
Contributors: Pau Amaro Seoane, Christopher Berry, Alvin Chua, Saavik Ford, Barry McKernan, Cole Miller, Carlos F. Sopuerta, Alejandro Torres-Orjuela and Veronica Vazquez-Aceves}\\

Thanks to high-resolution observations of the kinematics of stars and gas we
know that most nearby bright galaxies host a dark, massive, compact
object  \citep[e.g.,][]{2004cbhg.symp....1K,2010RvMP...82.3121G,2013ARA&A..51..511K,2016ASSL..418..263G}. 
One of the most impressive cases is our own Galactic Centre. 
The stellar dynamics of the central cluster of stars (the S-stars, or S0-stars),
provides compelling evidence for the existence of a massive BH (MBH) of mass $\sim 4\times 10^6\,M_{\odot}$, Sgr A*  \citep[see for a review][]{2010RvMP...82.3121G}.
In particular, the star S4714 in this cluster has an orbital eccentricity of $0.985$ and moves at about $8\%$ the speed of light at pericentre, with an orbital
period of $9.9$ years around the MBH in our Galactic centre. 
Another extreme case, S62, comes within 16 AU of Sgr A* \citep{2020ApJ...899...50P}.
Also very recently, \cite{2020A&A...636L...5G} have presented the detection of pericentre precession in the orbit of the star S2.
The best fit to a relativistic orbit yields a precession rate between 0.196 and 0.272 degrees per orbit, which is consistent with GR predictions at the $\sim 17$\% level.
Further compelling evidence for a MBH comes from the Event Horizon Telescope (EHT) observations of the centre of the galaxy M87. 
EHT measured the mass of its central MBH to be $\sim 6.5\times10^9\,M_{\odot}$, with an event horizon diameter of about $0.0013~\mathrm{pc}$ \citep{2019ApJ...875L...1E}. 

More generally, it is believed that most large galaxies host a MBH. 
The currently highest inferred mass is $6.6\times 10^{10}\,M_{\odot}$ \citep{2004ApJ...614..547S}.  
The stars in the centres of such galaxies have the potential to interact with MBHs, but only if their pericentres are small enough. 
These orbits typically lead to the tidal disruption of an extended star; when these orbits are represented in phase-space by their energy and angular momentum, the section of phase space that leads to tidally disrupted systems is called the loss cone \citep{1976MNRAS.176..633F}.

The range of frequencies that LISA will cover is ideally matched to the inspiral of a compact object such
as a stellar-mass BH, a NS or a WD on to a (light) MBH; i.e.,
one with a mass between $\sim 10^4\,M_{\odot}$ and $\sim 10^7\,M_{\odot}$. Because of the  difference in mass between the MBH and the $\sim{\rm few}$--${\rm tens}$ of solar masses of the compact object,
we call these EMRIs---where the mass ratio $q$ is $10^{-8}<q<10^{-5}$. There is also a potential population of BHs
with masses between $10^2\,M_{\odot}$ and $10^4\,M_{\odot}$, which are called intermediate-mass BHs (IMBHs). 
In principle, such BHs can be involved in intermediate mass-ratio inspiral (IMRI; $10^{-5}<q<10^{-2}$) systems, either with a compact object inspiralling into them, or with them inspiralling into a MBH. These would fill the gap between EMRIs and comparable-mass binaries. IMRI observations are naturally complementary to EMRI observations, providing further insight into the development of MBHs and their surroundings and the possible evolutionary link between IMBHs and MBHs.
Table~(\ref{tab.acronyms}) introduces the different acronyms used throughout this chapter, their meaning, mass ratio ranges, and configurations.

\begin{table}[ht]
\centering
    \begin{tabular}{|m{2.5cm}|m{4.5cm}|m{2.5cm}|m{4cm}|}
        \hline
         \textbf{Acronym} & \textbf{Meaning} & \textbf{Mass ratio} & \textbf{Constituents} \\
         \hline
         \textbf{light IMRI} & light intermediate mass-ratio inspiral & \centering{$10^{-5}$--$10^{-2}$} & IMBH \& stellar-mass compact object\\
         \hline
         \textbf{heavy IMRI} & heavy intermediate mass-ratio inspiral & \centering{$10^{-5}$--$10^{-2}$} & MBH \& IMBH \\
         \hline
         \textbf{EMRI} & extreme mass-ratio inspiral & \centering{$10^{-8}$--$10^{-5}$} & MBH \& stellar-mass compact object\\
         \hline
         \textbf{b-EMRI} & binary-extreme mass-ratio inspiral & \centering{$10^{-8}$--$10^{-5}$} & MBH \& binary stellar-mass compact object\\
         \hline
         \textbf{XMRI} & extremely large mass-ratio inspiral & \centering{$\lesssim10^{-8}$} & MBH \& sub-stellar object\\
         \hline

    \end{tabular}
             \caption{Nomenclature for the different types of mass ratio inspirals used in this chapter.}
         \label{tab.acronyms}
\end{table}

The frequencies of EMRIs are inaccessible to ground-based GW observatories, as are all but the lightest ($\lesssim10^3\msun$) IMRIs \citep{2011GReGr..43..485G,2013A&A...557A.135K,2016MNRAS.457.4499H,2018PhRvD..98f3018A}, yet their astrophysical production may be related to stellar-mass binary BHs (BH+BHs). Indeed, astrophysical mechanisms for generating EMRIs and IMRIs touch on an extremely diverse range of topics (see below), and LISA will lead the way in distinguishing between various formation channels, and furthering our knowledge in all of these areas.
Both EMRIs and IMRIs have been reviewed in \citet{2007CQGra..24R.113A} and more recently in \citet{2020arXiv201103059A,2018LRR....21....4A,2019BAAS...51c..42B}. Substellar objects, in particular brown dwarfs, with masses around $10^{-2}\,M_{\odot}$ form a third class of inspirals potentially observable in the LISA frequency range. These objects can last as many as $10^8$ cycles before crossing the event horizon, due to their extremely large mass ratio, which is why they have been dubbed extremely large mass-ratio inspirals (XMRIs; $q<10^{-8}$).  XMRIs are particularly important in our own Galactic Centre, where a few of them should be present when LISA observations begin \citep{2006ApJ...649L..25R,2019PhRvD..99l3025A}. 

Ordinary stars that are similar to our Sun would only complete a single periapsis cycle around a 
MBH before being tidally disrupted (for a close enough passage to enter the LISA frequency range). In contrast, compact stars can revolve around an MBH thousands or even hundreds of thousands of times, with the number of cycles roughly inversely proportional to the mass ratio.\footnote{Even compact objects may not always avoid tidal disruptions during the inspiral process; for example, a WD could be disrupted outside the event horizon of an IMBH \citep[e.g.,][]{2008NewAR..51..884M,2008MNRAS.391..718S,2008ApJ...679.1385R,2020arXiv200512528R,2020SSRv..216...39M}.} 
Although the system is constantly emitting GWs, it is at the periapsis when the EMRI emits a strong burst of gravitational radiation. Since the orbit shrinks and precesses, we can envisage this as a probe taking pictures of spacetime around a MBH in the strong regime. 

Observing the large number of cycles of gravitational radiation emitted before disappearing into the event horizon has three main benefits:
\begin{itemize}

\item As an EMRI can spend up to hundreds of thousands of orbits within a few gravitational radii of the MBH, careful analysis promises to provide exceptionally precise tests of the nature of strong-field gravity and the Kerr nature of MBHs.

\item Tracking the complicated orbit through many thousands of cycles yields outstanding measurements of parameters including the redshifted mass and spin of the MBH, without any modelling other than general relativity. In turn, this will give us hints about the evolution of MBHs, in a population that is typically inaccessible to EM observations, except, perhaps, for a limited number of local MBHs, e.g., with the Next Generation Event Horizon Telescope, but nevertheless with much less precision on mass and spin measurements compared to what LISA will do.

\item If there is a correlation between the environment and the parameters of the
EMRI or IMRI, we could reverse-engineer these to extract unique astrophysical information.

\end{itemize} 
However, these exciting prospects must be earned: the weakness of the signal from individual EMRI orbits means that detection, let alone parameter estimation, will require highly accurate computation of thousands of waveform cycles. 
EMRI waveform templates are challenging to model. 
Traditional computation techniques are not suitable because the post-Newtonian (PN) approximation \citep{2014LRR....17....2B} is inapplicable to these highly relativistic systems and numerical relativity calculations \citep{2019RPPh...82a6902D} are infeasibly computationally expensive because of the large difference in scales between the two binary components.  Instead, templates can be calculated by treating the effects of the compact object as a perturbation to the background spacetime of the more massive MBH \citep{2011LRR....14....7P,2019RPPh...82a6904B}. 
For IMRIs, the systems lie at the boundary of where perturbative methods may apply, where PN approximations may be used for the inspiral, and where numerical relativity simulations may be possible. 
Therefore, a combination of techniques will be needed to simulate IMRI templates.
For EMRI and IMRI science, it will be essential to accurately compute these long waveforms in order to sift out these multi-year signals from the LISA data stream. 

In this chapter we first give a summary about the formation
mechanisms for EMRIs, which have received more detailed study than IMRIs or XMRIs, as well as the many different physical scenarios
that play a crucial role in their event rate estimation. The fundamental theory on which these estimates rely, relaxation theory, is robustly understood and has yielded theoretical results which have been corroborated observationally. There remain astrophysical uncertainties which impact the EMRI rate, and there may be subtle effects that leave the theory incomplete. However, the theory has received extensive and detailed investigation in the context of EMRI formation and evolution in galactic nuclei over the last few decades.
The narrative is complex. Here we will briefly summarise EMRI formation in the context of relaxation theory; for a detailed review see \citet{2020arXiv201103059A,2018LRR....21....4A}.
\textit{Assuming} that at least one EMRI will be detected during the LISA mission, we lay out the anticipated science that is guaranteed, plausible, and speculative.

\subsubsection{Guaranteed science with the detection of EMRIs} 

EMRIs are essentially guaranteed to happen in our Universe. The expected rates span a wide enough range that we cannot absolutely guarantee an observed EMRI in a 5-year mission \citep{2012A&A...542A.102M}. 
However, the uncertainties are such that LISA might also see multiple EMRIs---and if even one is observed, it is guaranteed to be a direct probe of the MBH spin. Currently our best measurements of MBH spin comes from studies of the broad component of the FeK$\alpha$ line in X-rays. The FeK$\alpha$ line is broadened due to relativistic effects (special and general) within a few $10r_\mathrm{g}$ (here $r_\mathrm{g}=GM_{\rm MBH}/c^{2}$, the gravitational radius) from the MBH. 
By assuming a compact X-ray illumination source close to the MBH, the FeK$\alpha$ line shape strongly constrains MBH spin \citep[e.g.][]{2003PhR...377..389R,2011ApJ...736..103B}, but only for $O(20)$ nearby AGN at present  \citep{2016MNRAS.458.2012V}, and both the statistical uncertainties and the possible systematic errors on the spin measurements are substantial. 
If we assume the Blandford--Znajek mechanism powers jets in radio-loud sources, it may also be possible to put constraints on MBH spin in a larger number of radio galaxies by measuring jet power \citep{2011MNRAS.414.1253D}. MBH spins have the power to reveal their growth history: what is the contribution due to mergers with other MBH versus gas accretion? The answers have implications for our understanding of galaxy assembly and evolution. In particular, near maximal spin would indicate that the most recent significant mass increase occurred via gas accretion, predominantly from a thin disc with coherent direction of the angular momentum; other individual spin measurements lead to less clear-cut conclusions but can permit constraints on the accretion history (see Section~\ref{sec:spin} for details). 
If many EMRIs are observed, we will have the opportunity to probe the distribution of MBH spins. 
Ideally, this would also enable us to probe a couple of decades in MBH mass, informing the underlying astrophysics of the $M_\mathrm{BH}$--$\sigma$ relation. 
Second, GW inferences of MBH spin will allow us to test the assumptions underpinning EM measurements inferring spins \citep{2011MNRAS.414.1253D,2016MNRAS.458.2012V}.

EMRIs (and IMRIs) are also unique sources of GWs for studying fundamental physics with LISA---see more details in the white paper of the Fundamental Physics Working Group~\citep{2020GReGr..52...81B}---mainly because the small body spends a relatively high number of cycles (that scales with the inverse of the small mass ratio $q$) very close to the horizon of the MBH, where precessional effects (periastron shift and orbital plane precession) become
as strong as they can be~\citep{2017PhRvD..95j3012B,2019BAAS...51c..42B}. The orbital dynamics get imprinted in the GW signal by introducing the associated fundamental frequencies and their harmonics~\citep{2009PhRvD..79j4016D}. In the case of IMRIs there are extra timescales due to coupling of the small object spin with the orbital angular momentum and also with the MBH spin. These timescales evolve slowly due to gravitational backreaction of the small body gravitational field on its own trajectory. The orbital timescales depend on the MBH spacetime geometry (the Kerr geometry according to general relativity) and their evolution due to gravitational backreaction depends on the gravitational dynamics, which can be very sensitive to modifications to Einstein's equations of general relativity, such as modified gravity, extra fields, extra dimensions, etc.

It is expected \citep{2017PhRvD..95j3012B} that EMRI and IMRI waveforms will be sensitive to both the parameters that describe the MBH geometry (mass, spin, and other gravitational multipole moments) and the parameters that describe deviations from the general relativity paradigm (coupling constants, extra dimension length scales, etc.). 
Therefore, they are unique probes of the geometry of the MBHs in galactic centres and of the particular details of the gravitational theory (and other non-gravitational fields that may affect the dynamics of EMRI/IMRIs) responsible for GW generation~\citep{2019RPPh...82a6904B}.

However, in order to extract meaningful constraints from an EMRI (or IMRI) detection, it is essential to have reliable astrophysical predictions for the \emph{distribution} of the key parameters of these systems. 
The values of such parameters determine to what level we can test the no-hair conjecture and general relativity, and what kind of fundamental physics we can expect to carry out with LISA observations of EMRI/IMRIs. 
This leads us to discuss important plausible science, especially involving EMRIs in gas-rich environments.

\subsubsection{Plausible science with the detection of EMRIs}

Identifying EMRIs in gas-rich environments could be an important observational result, as the effects of gas could in some cases mimic the effect of alternative theories of gravity. 
For AGN-driven EMRIs, we expect the gas to circularize prograde orbiters, and for the merger to occur in the equatorial plane of the MBH. For retrograde orbiters, however, the eccentricity could be driven to extremely large values \citep{2019ApJ...878...85S}, and the interplay of gas- and GW-driven decay may be challenging to disentangle. 
By isolating the gas-driven mergers, we may also be able to directly probe important parameters of the gas, notably the density and viscosity---however, the detectability of the gas-driven phase shift requires substantial gas densities \citep{2008PhRvD..77j4027B,2011PhRvD..84b4032K,2011PhRvL.107q1103Y,2014PhRvD..89j4059B,2019MNRAS.486.2754D,2020arXiv200511333D}.

Although there are no known IMRI systems, there are multiple plausible channels for their formation. In particular:
\begin{itemize}
    \item Dwarf galaxies may contain an IMBH that could interact with stellar mass BHs \citep{2017A&A...601A..20K,2019MNRAS.484..794G,2019MNRAS.484..814G};
    \item Mergers of IMBH-bearing dwarf galaxies with larger MBH-bearing galaxies can produce systems with the relevant mass ratios;
    \item Globular clusters may contain IMBH that could (similar to dwarf galaxies) interact with stellar mass BH or decay into an MBH-bearing galactic nucleus;
    \item Finally, IMBHs could form through hierarchical mergers in a galactic nucleus---in particular in an AGN---which would continue to accrete stellar mass BH, while simultaneously creating a natural IMBH--MBH system.

\end{itemize}
We will discuss each of these formation scenarios in more detail in Section~\ref{sec.ForEMRI}, and we further separate IMRIs into 2 classes: \emph{light} IMRIs, where the primary mass is an IMBH and \emph{heavy} IMRIs, where the primary mass is an MBH. 
Broadly, in the scenarios where we have the most theoretical confidence, the expected rates are small enough that there is a low probability of an event in the lifetime of a LISA mission (though as with EMRI rates, some uncertainties remain).  
In other scenarios, our uncertainty on many input parameters is such that the rate per galactic nucleus within the LISA observational horizon could be anywhere from zero to one per few years. However, this provides an excellent opportunity, even in the case of non-detections, to place important constraints on nuclear star clusters (NSCs) and their formation mechanisms, as well as on the structure and lifetime of AGN discs. In effect, LISA will enable us to reverse engineer important properties of AGN discs, including their radial surface density profiles and lifetimes.

One theoretical concern regarding IMRIs has recently been addressed: IMBH ($10^{2}-10^{5}M_{\odot}$) \emph{do} exist in the relatively nearby universe. 
GW190521 demonstrated the formation of an IMBH ($>100M_{\odot})$ at $z<1$ \citep{2020ApJ...900L..13A}. 
At the other end, some dwarf galaxies may host central IMBHs, at least at the lower end of their mass estimates \citep{2014AJ....148..136M,2017A&A...601A..20K} and some of these may correspond to an extrapolation of the mass function and scaling relations towards low MBH mass ($10^{5}-10^{6}M_{\odot}$) \citep{2013ApJ...775..116R,2015ApJ...798...54G}.

XMRIs, with $q < 10^{-8}$ are systems containing brown dwarfs \citep{2019PhRvD..99l3025A}. 
On account of their mass ratio, XMRIs would not evolve appreciably over the course of the LISA mission \citep{2019A&A...627A..92G}. 
XMRIs are the GW event mostly likely to permit LISA to probe the Milky Way MBH and its nuclear star cluster.
However, XMRI GWs would only be detectable from the Milky Way, or perhaps a few nearby galaxies. Nevertheless, given their expected rate from Sgr A*, and their possible interactions with the stellar cusp as they inspiral, XMRIs are an exciting probe of our nearest MBH and its environs.

In addition to the direct effects of fundamental physics on waveform generation, there are other effects that are accumulated during the propagation of the GWs from the source to the detector, such as those due to possible high-energy effects beyond general relativity: breaking of Lorentz invariance or of the weak equivalence principle, extra polarizations, gravitational parity violation, etc.~\citep{2020GReGr..52...81B}. We can in principle detect those effects in EMRI/IMRI waveforms, but in the case of LISA, sources that are detectable at higher redshift, i.e.\ MBH binary coalescence, are more competitive in this regard. The test of the no-hair conjecture that EMRI/IMRIs provide are complementary to those that can be performed using quasinormal models excited in the ringdown of the final MBH after a MBH binary merger~\citep{Berti:2018vdi}. 

\subsubsection{Speculative science with the detection of EMRIs}

EMRI/IMRI observations can also have impact on two other important subjects in fundamental physics. The first is the search for primordial BHs~\citep{2020arXiv200212778C}.  Given the high precision expected for EMRI mass estimates, a detection would determine with confidence when the small compact object has a mass below what is reasonable from any astrophysical channel. 
This would be a strong indication of the primordial origin of that object. 
The other subject where EMRI/IMRIs can have an impact is in the understanding of dark matter.  This is not independent from the previous subject since primordial BHs have been proposed to constitute all the dark matter in the observable universe~\citep{2020ARNPS..7050520C}. An example of how EMRI/IMRIs may probe the nature of dark matter is in the case where it is made by bosons that can form clouds around MBHs~\citep[see for our Galactic Centre][]{2010JCAP...11..002A}, which would affect the orbital dynamics of EMRI/IMRIs and hence would leave an imprint in the waveforms (see, e.g. \cite{Hannuksela:2018izj}). 
Moreover, EMRI/IMRIs could also contribute to the understanding of dark energy by adding significantly to the knowledge of the expansion history of the universe, assuming that we are able to determine their redshift, either from direct EM counterparts or via correlation with galaxy surveys \citep{MacLeod:2007jd}.

Finally, the fundamental physics (also the cosmology) that we can investigate using some IMRI systems may be enhanced if we can do multiband GW astronomy, that is, by combining the information obtained with LISA with the information from detections with ground-based detectors, provided that the IMRI masses are such the system can enter the frequency band of the ground-detectors at a later time~\citep{2018PhRvD..98f3018A,2020arXiv200612137D}.

\subsubsection{Data analysis \& waveform modelling} 

To unlock the rich scientific potential of EMRI and IMRI observations, we must be able to extract these signals from the LISA data stream. 
EMRI detection and characterisation is one of the most challenging problems in LISA data analysis \citep{2007CQGra..24R.113A,2015JPhCS.610a2002A,2018LRR....21....4A,2020arXiv201103059A}. 
There are three main sources of challenge: 
\begin{itemize}
    \item The complexity of EMRI signals---EMRI orbits are generically eccentric and precessing over a large number of cycles. The waveforms are thus extremely sensitive to the source parameters, and there is a gargantuan space of potential signals to be searched.
    \item The length of EMRI signals---EMRIs need to be tracked for an extended time in order to accumulate sufficient signal-to-noise ratio (SNR) to be detectable. If the phase cannot be accurately tracked, either due to hard-to-model effects, like transient resonances \citep{2012PhRvL.109g1102F,2016PhRvD..94l4042B} or higher-order corrections due to nonlinear interactions of the compact object's gravitational field \citep{2019RPPh...82a6904B}, or unaccounted for environmental effects, such as viscous drag \citep{2008PhRvD..77j4027B,2011PhRvD..84b4032K,2014PhRvD..89j4059B} or perturbations from a nearby object \citep{2012ApJ...744L..20A,2019PhRvL.123j1103B}, this will impact detectability.
    \item The number of EMRI signals---EMRIs are long-lived and possibly numerous. Thus there may be many EMRI signals in the LISA data stream at any given time, overlapping with one another as well as with signals from the multitude of other sources. This means that any data analysis strategy for EMRIs must be part of a global fit that analyzes all signals concurrently.
\end{itemize}
The first challenge means that, unlike when searching for LIGO--Virgo signals \citep[e.g.,][]{2012PhRvD..85l2006A,2016PhRvD..93l2003A}, it is computationally infeasible to perform a matched-filter search with a regular grid of templates: it has been estimated that $\sim10^{40}$ templates may be needed for this goal \citep{2004CQGra..21S1595G}. 
Instead, we must trade search sensitivity for computational expediency. Multiple data analysis approaches have been explored following the initiation of the Mock LISA Data Challenges \citep{2008CQGra..25k4037B,2008CQGra..25r4026B,2010CQGra..27h4009B}. 
Techniques include identifying time--frequency tracks \citep{2005CQGra..22S.445W,2007CQGra..24.1145G,2008JPhCS.122a2037G,2008CQGra..25r4031G}, although this can be difficult in the presence of multiple signals, or using Monte Carlo techniques to stochastically search for signals, either using EMRI templates \citep{2006AIPC..873..444S,2008CQGra..25r4030G,2009CQGra..26m5004B,2011CQGra..28i4016C,2012CQGra..29n5014A} or phenomenological waveforms \citep{2012PhRvD..86j4050W}. 
These techniques still do not extend to the full scope of global EMRI search, which must ultimately be conducted in a hierarchical fashion \citep{2004CQGra..21S1595G,2017PhRvD..96d4005C}. 
Stochastically searching parameter space to fit for EMRI signals is especially challenging as there may be many regions in parameter space where there are good matches to the data, aside from the vicinity of the true parameters \citep{2008CQGra..25r4026B,2010CQGra..27h4009B}; 
the full extent and severity of such parameter degeneracy is difficult to determine due to the size of the parameter space and the lack of tractable waveforms, and is currently being investigated \citep{2021arXiv210914254C}.
Design of EMRI analyses hence remains an open area of research. 
IMRI detection is less well studied, but should be possible with a combination of the techniques designed for EMRIs and equal-mass binaries. 

Essential to measuring the properties of EMRIs and IMRIs is the modelling of the gravitational waveforms. 
Only with accurate models can the source properties be inferred. 
For EMRI waveforms, the highest accuracy waveforms are calculated using the self-force formalism \citep{2011LRR....14....7P,2019IJMPD..2830011B}:
the compact object's gravitational field is treated as a perturbation to the background spacetime of the larger black hole.
For full characterisation of EMRI signals, we will require calculation of self-force effects to second order in the mass ratio for generic orbits in Kerr spacetime \citep{2006PhRvD..73d4034R}. 
The self-force program is now well advanced \citep{2019RPPh...82a6904B}; the self-force has been calculated to first order for a generic orbit in Kerr spacetime \citep{2018PhRvD..97j4033V}, and the foundations have been laid for a second-order calculation \citep{2014PhRvD..89j4020P,2014PhRvD..90h4039P,2016PhRvD..94j4018M,2017PhRvD..95j4056P,2018PhRvD..97j5001M,2020PhRvL.124b1101P}. 
It is expected that concerted waveform development will lead to successful computation of EMRI waveforms ahead of LISA's launch. 
In the meantime, less accurate approximate waveforms are used for developing LISA data analysis. 
The most common approximations are the analytic kludge \citep{2004PhRvD..69h2005B,2017PhRvD..96d4005C}, based upon a Keplerian orbit augmented with relativistic corrections, and the numerical kludge \citep{2007PhRvD..75b4005B}, based upon Kerr geodesics mapped to flat spacetime for waveform generation.
The low computational cost of these models makes them suitable for early stages of LISA data analysis, where we are looking for EMRI-like signals, but are not concerned about the precise parameter values. 
IMRI waveforms present a challenge as they lie at the boundary of different techniques for waveform generation \citep{2008PhRvD..78f4028H,2014LRR....17....2B}. 
The self-force calculations may cover a large range of IMRI parameter space \citep{2020arXiv200612036V}.
These can be supplemented with calculations from PN theory \citep{2014LRR....17....2B,2009PhRvD..80h4043B} and effective one-body theory used for more equal mass binaries \citep{1999PhRvD..59h4006B,2000PhRvD..62f4015B,2014PhRvD..89f1502T,2020PhRvD.102d4055O}.
Finally, numerical relativity can model IMRIs. 
Numerical relativity should give exact numerical solutions to the Einstein equations \citep{2015CQGra..32l4011S}, but IMRIs require extremely high spatial and temporal resolution. 
Therefore, computation of high fidelity numerical relativity IMRI waveforms for LISA may require the development of a new generation of codes. 
The best IMRI waveform models should be produced through combining the strengths of each of these formalisms.
Both EMRIs and IMRIs provide a valuable chance to validate our waveform calculation theory in new regions of parameter space.

\subsection{Formation channels}
\label{sec.ForEMRI}

\noindent
{\bf Coordinators: Manuel Arca Sedda, Xian Chen, and Andrea Derdzinski  

\noindent
Contributors: Pau Amaro Seoane, Manuel Arca Sedda, Jillian Bellovary, Elisa Bortolas, Pedro R. Capelo, Xian Chen, Andrea Derdzinksi, Gaia Fabj, Saavik Ford, Jean-Baptiste Fouvry, Zoltan Haiman, Wen-Biao Han, Giuseppe Lodato, Barry McKernan, Syeda Nasim, Amy Secunda and Martina Toscani}\\

\subsubsection{Gas-poor dynamics: Galactic nuclei including dwarfs, and globular clusters}
Observations of galaxies and their nuclei have revealed close relationships between several galactic properties and the masses of their central MBHs \citep[][]{2008ApJ...678L..93S,2009ApJ...698..198G, 2013ARA&A..51..511K,2013ApJ...769..132B, 2015ApJ...813...82R, 2016ASSL..418..263G,2017MNRAS.471.2187D,2018ApJ...869..113D,2019ApJ...873...85D,2019ApJ...876..155S, 2019ApJ...887...10S,2019ApJ...877...64D,2020ApJ...903...97S}. Extrapolating these relations to the lower
mass end, one would expect $10^3$--$10^5 \msun$ IMBHs to exist in the centres of dwarf galaxies \citep{2017IJMPD..2630021M,2019arXiv191109678G}, as suggested by X-ray observations of low-mass AGN \citep[][]{2017A&A...601A..20K,2018MNRAS.478.2576M,2019MNRAS.484..794G,2019MNRAS.484..814G,2020ApJ...888...36R}. 
IMBHs may also form via collisions and mergers of stars and stellar-mass BHs in dense clusters \citep{2000ApJ...528L..17P,2002MNRAS.330..232C,2015MNRAS.454.3150G,2016MNRAS.459.3432M,2021MNRAS.507.5132D,2021MNRAS.501.5257R,2021ApJ...920..128A,2021ApJ...908L..29G,2022MNRAS.512..884R}. Dynamical friction can subsequently lead to the orbital decay of globular clusters into a galactic nucleus \citep{1975ApJ...196..407T,1976ApJ...203..345T,1993ApJ...415..616C}, allowing the formation of an IMBH--MBH system \citep{2001ApJ...562L..19E,2004ApJ...614..864M,2006ApJ...641..319P,2007ApJ...656..879M,2009ApJ...705..361G,2018MNRAS.477.4423A}.
Depending on the populations of stars and BHs in these environments, galactic nuclei (including dwarf nuclei) and globular clusters each provide plausible formation channels for EMRIs and IMRIs. 
We group these channels together because the underlying physical mechanisms for EMRI and IMRI formation is similar for all (gravitational interactions alone); in addition, these channels interact with one another astrophysically through mergers. 
There remain major astrophysical uncertainties in each formation channel, meaning that LISA observations (perhaps including non-detections) can crucially constrain the astrophysics that lead to their formation.

\subsubsection{Formation of EMRIs in gas-poor galactic nuclei}

\paragraph{Physics of EMRI formation I. Relaxation processes}

\begin{figure}[htbp!]
    \centering
   \includegraphics[width=0.75 \textwidth]{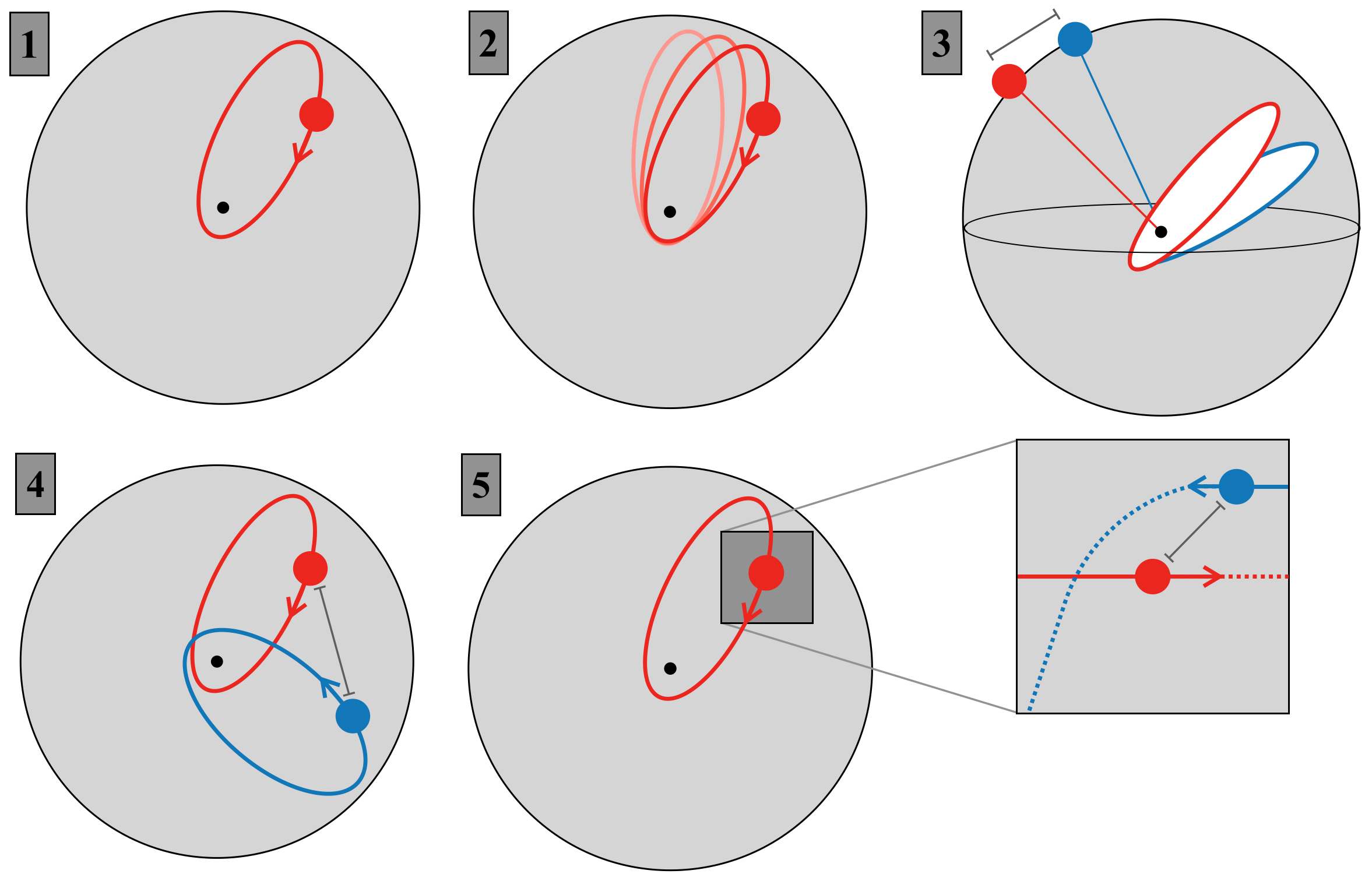}
   \caption{Illustration of the hierarchy of timescales in a NSC:
   (1) the dynamical time;
   (2) the precession time;
   (3) the vector resonant relaxation time;
   (4) the scalar resonant relaxation time;
   (5) the two-body relaxation time.
   }
   \label{fig:Timescales_NSC}
\end{figure}

MBHs, often surrounded by nuclear star clusters, seem ubiquitous in the centre of nearby galaxies \citep{2009MNRAS.397.2148G,2010RvMP...82.3121G,2013ARA&A..51..511K,2020A&ARv..28....4N}.
Yet, the details of MBH formation, and their impact on both their surrounding NSC and their host galaxy, remain uncertain \citep{2014ARA&A..52..589H}.
Owing to the overwhelming mass of the central MBH
and the steep potential well that it generates,
NSCs encompass a wide range of dynamical processes
that act on radically different timescales \citep{1996NewA....1..149R,2006ApJ...645.1152H,2017ARA&A..55...17A}.
The key dynamical processes that can generate an EMRI around a MBH hosted in an NSC
are briefly illustrated in Fig.~\ref{fig:Timescales_NSC}. Of these processes, the two-body relaxation time is the slowest, but also the main mechanism for producing EMRIs. We now discuss each timescale and their relevance for studying EMRIs with LISA. After that, we will give a description of how our understanding of EMRI formation has evolved in
the last decades, and how LISA can help in that respect.

\begin{itemize}
    \item[(1)] \textbf{Dynamical time}

    On account of its mass, the central MBH dominates the nucleus's mean potential,
    and imposes, at leading order, a Keplerian motion
    to any object orbiting within the MBH's sphere of influence.
    These motions take to leading order the form of closed ellipses,
    for example as currently monitored
    for the S cluster around Sgr A* (see, e.g.\@, Fig.~\ref{fig:Orbits_NSC}).
    \begin{figure}[htbp!]
    \begin{center}
    \raisebox{-0.5\height}{\includegraphics[width=0.45 \textwidth]{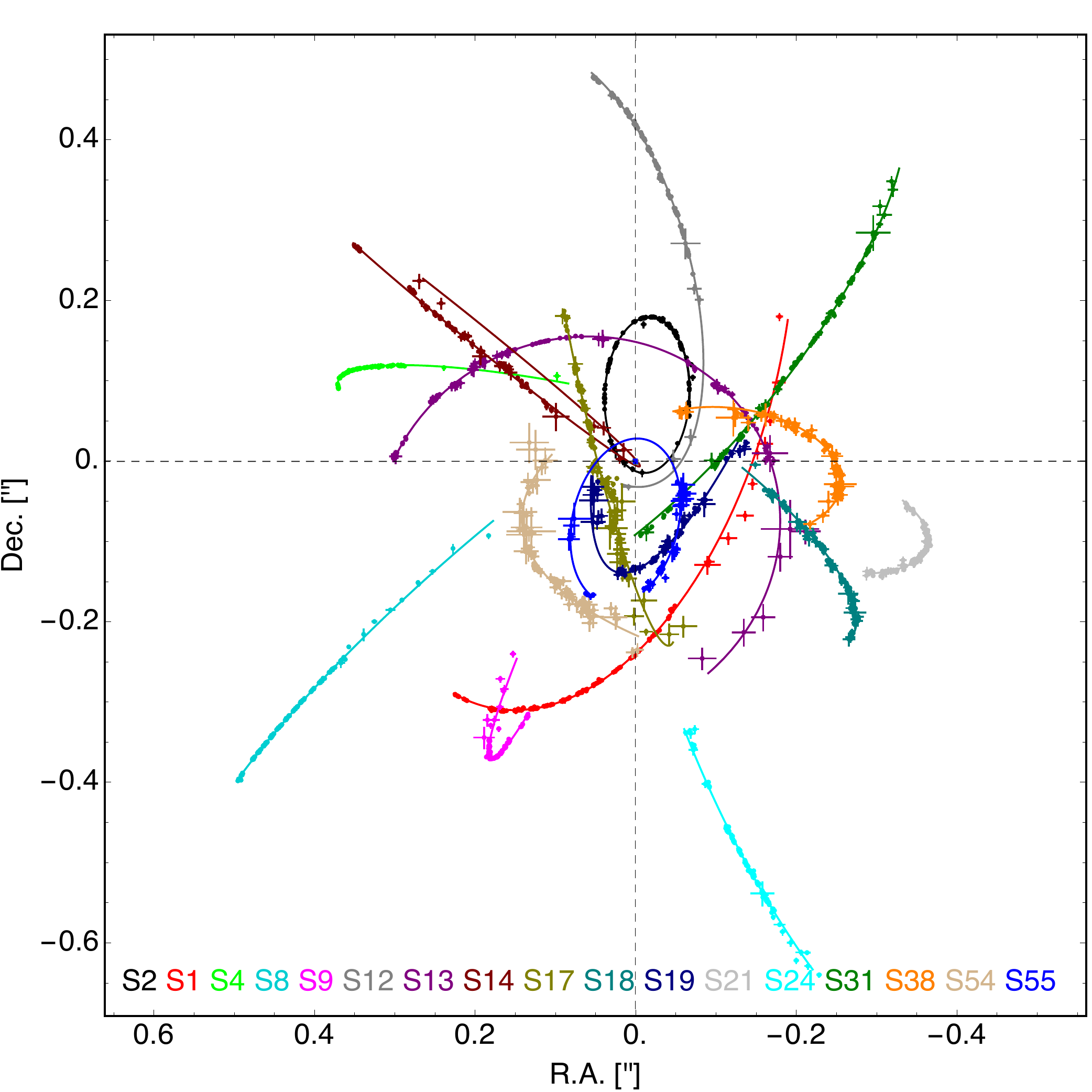}}
    \hspace{0.02\textwidth}
    \raisebox{-0.5\height}{\includegraphics[width=0.50\textwidth]{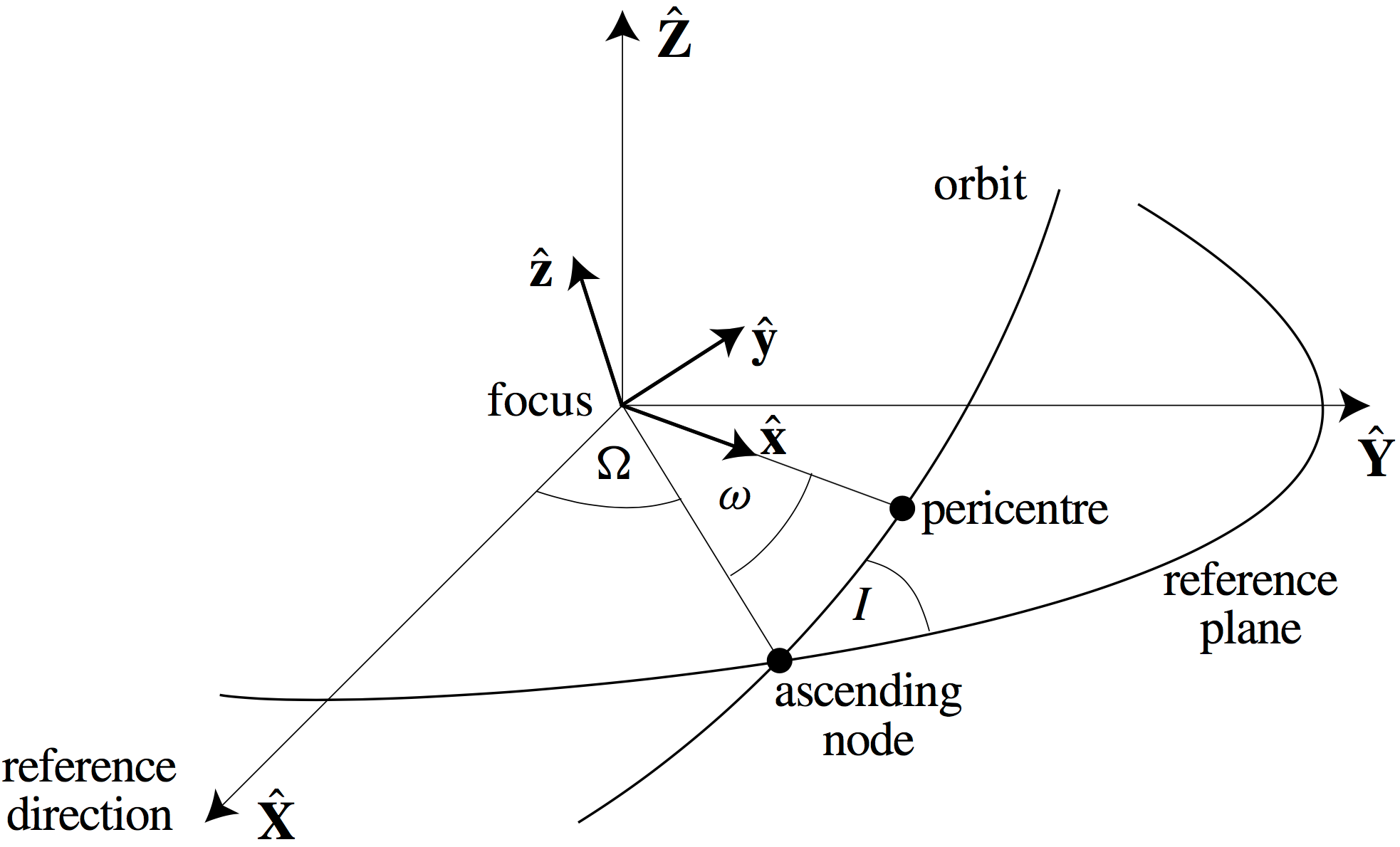}}
    \caption{Keplerian orbits around an MBH.
    Left: Detailed observations of the Keplerian dynamics of the S-stars in the vicinity of Sgr A* (${\sim 10~\mpc}$) from \cite{2017ApJ...837...30G}. At leading order, orbits take the form of closed ellipses, because the central MBH dominates the gravitational potential. Right: Illustration of the Keplerian orbital elements from \cite{1999ssd..book.....M}.
    }
    \label{fig:Orbits_NSC}
    \end{center}
    \end{figure}
    A Keplerian orbit can be described by its orbital elements \citep{1999ssd..book.....M},
    which we denote with ${ (M , \omega , \Omega , \Lc , L , \Lz) }$.
    Here, $M$ stands for the mean anomaly,
    $\omega$ is the argument of the pericentre,
    and $\Omega$ the longitude of the ascending node.
    An orbit is also characterised by three actions
    ${ (\Lc , L , \Lz) \!=\! (\sqrt{G \MBH a},\Lc \sqrt{1 \!-\! e^{2}},L \cos (I))}$.
    Here, $\MBH$ is the mass of the central MBH,
    $a$ the ellipse's semi major axis, $e$ its eccentricity,
    $I$ its inclination, $L$ the norm of the angular momentum,
    $\Lz$ its projection along a given axis,
    and finally $\Lc$ the circular angular momentum.
    Describing dynamics in NSCs amounts
    then to describing the dynamics of these particular orbital elements.
    The dynamical time is associated
    with the motion ${ \dot{M} = \sqrt{G\MBH/a^{3}} }$.
    This dynamical time being so short, one is naturally led to performing
    an orbit-average over it, i.e.\ smearing out the orbiting objects
    along their Keplerian ellipses \citep[see, e.g.\@,][]{2009MNRAS.394.1085T}.
    
    \item[(2)] \textbf{Precession time}

    On longer timescales, the gravitational potential self-consistently
    generated by the stellar cluster, as well as the relativistic corrections imposed by the MBH, namely the Schwarzschild precession \citep{2013degn.book.....M} cause the ellipses
    to precess in their planes.
    This drives the evolution of $\omega$.
    Importantly, one can note that the relativistic precession frequency
    diverges as orbits get more and more eccentric,
    which ultimately leads to the breakdown of the orbit-averaged assumption.
    Such a relativistic precession has recently been observed
    for the S2 star around Sgr A*
    by the {Gravity} interferometer \citep{2018A&A...615L..15G}. This is presently the best direct observational constraint on the metric in the vicinity of Sgr A*.
    
    \item[(3)] \textbf{Vector resonant relaxation}

    Subsequently, because of the non-spherical stellar fluctuations in the NSC, as well as the relativistic corrections induced by a spinning MBH, the Lense-Thirring precession \citep{2013degn.book.....M}, the ellipses' orbital orientations get reshuffled.
    This process is called vector resonant relaxation \citep{2015MNRAS.448.3265K}.
    In that limit, the orbits' angular momenta
    change in their orientations, ${ \hbL = \bL / |\bL| }$,
    without changing in magnitude $|\bL|$ (equivalently in $e$),
    nor in energy $\Lc$ (equivalently in $a$).
    
    The process of vector resonant relaxation has
    been studied, among others, through numerical simulations \citep{2009ApJ...698..641E},
    orbit-averaged simulations \citep{2015MNRAS.448.3265K},
    as well as thermodynamical \citep{2017ApJ...842...90R,2018ApJ...856..113T}
    and kinetic theories \citep{2019ApJ...883..161F}.
    Vector resonant relaxation is essential
    to describe the warping of accretion \citep{2012ApJ...748...63B}
    and stellar discs \citep{2015MNRAS.448.3265K},
    and can enhance the rate of binary mergers in NSCs (which could naturally produce an IMBH--MBH binary).
    Furthermore, it can also explain the possible presence of stellar discs \citep{2009ApJ...697.1741B,2014ApJ...783..131Y},
    or even strong anisotropic mass segregation
    of IMBHs discs \citep{2018PhRvL.121j1101S}.

    \item[(4)] \textbf{Scalar resonant relaxation}

    On longer timescales, resonant torques between in-plane
    precessing orbits lead to an efficient diffusion
    of the ellipses' eccentricity.
    This process is called scalar resonant relaxation \citep{1996NewA....1..149R}
    as the quantity that diffuses is the norm of the orbit's
    angular momentum.
    It is also said to be resonant,
    as only orbits that precess with matching in-plane precession frequencies
    will effectively and constructively interact with one another.
    
    This relaxation process has been investigated through ad hoc methods \citep{2006ApJ...645.1152H,2009ApJ...698..641E,2011ApJ...738...99M,2013ApJ...763L..10A,2015ApJ...804...52M},
    as well as $N$-body simulations \citep{2009ApJ...702..884P,2011PhRvD..84d4024M,2013ApJ...763L..10A,2014MNRAS.443..355H},
    and kinetic theories \citep{2014CQGra..31x4003B,2016MNRAS.458.4143S,2018ApJ...860L..23B}.
    Scalar resonant relaxation may be paramount
    to explain the thermal distribution of stellar eccentricities
    around Sgr A* \citep{2020ApJ...896..137G},
    while not necessarily mandatory~\citep{2014ApJ...786L..14C}.
    However, its efficiency drastically
    damps as orbits get very eccentric,
    an effect called the Schwarzschild barrier \citep{2011PhRvD..84d4024M}.
    This particular problem has been addressed in detail by \citet{2014CQGra..31x4003B}.
    They have shown that the divergence of the relativistic precession
    frequency as orbits get more and more eccentric is responsible
    for a drastic dampening of the efficiency of resonant relaxation.
    As such, the Schwarzschild barrier
    corresponds to the abrupt transition
    from a relaxation dominated by resonant interactions
    (for not too eccentric orbits) to a relaxation dominated
    by non-resonant two-body scatterings (for eccentric enough orbits).
    Given that mainly highly eccentric stellar orbits undergo EMRIs,
    the total contribution of scalar resonant relaxation
    to EMRI event rates is small
    \citep{2016ApJ...820..129B}.

    \item[(5)] \textbf{Two-body relaxation}

    Finally, on even longer timescales, rather than being driven
    by the interaction between Keplerian ellipses,
    an object will start to see its evolution being driven
    by nearby pairwise interactions, as a result of close
    two-body encounters.
    It is only through the cumulative contributions from these localised
    scatterings that objects can ultimately relax in their Keplerian energy
    (i.e.\ in $a$), through a process called two-body relaxation \citep{1976ApJ...209..214B,1978ApJ...226.1087C,1978ApJ...225..603S,2016ApJ...820..129B,2013ApJ...764...52B,2017ApJ...848...10V,2018LRR....21....4A,2020arXiv201103059A}.
    It is generically the slowest relaxation timescale in NSCs.
    The main mechanism for producing EMRI in NSCs is two-body relaxation. This is because it allows for the orbits to become highly eccentric, where other resonant processes significantly damp eccentricity \citep{2014CQGra..31x4003B}; the expected EMRI rates depend on the spin of the central MBH \citep{2013MNRAS.429.3155A}. We will elaborate on this later.
    
\end{itemize}

One of the first attempts to understand how to produce a successful orbit
in a galactic nucleus that will lead to the formation of an EMRI goes back to the work
of \cite{1997MNRAS.284..318S}. By using standard relaxation and loss-cone theory (see Sec.~\ref{sec.ForEMRI}), the authors derived the event rate for
compact objects to merge with a MBH in a galactic nucleus. It is important
to note that for MBHs above about a few $10^7\,M_{\odot}$, the timescales for relaxation exceed a Hubble time; LISA is going to observe MBHs below this threshold, down to $10^5\,M_{\odot}$. For 
lower masses, i.e.\ for IMBHs and hence IMRIs, we cannot further assume that
the MBH is fixed at the centre of the stellar system and
any analytical derivation becomes more difficult.

We define now a standard EMRI to consist of a stellar mass object of mass $10\,M_{\odot}$ and a MBH with a mass such as that of Sgr A*, the MBH in our own Galaxy. The event rate for this kind of EMRI, we obtain of the order of $10^{-5}$--$10^{-6}~\textrm{yr}^{-1}$ \citep[see e.g.][and references therein]{2018LRR....21....4A,2020arXiv201103059A}. This analytical result has been reproduced using numerical algorithms such as in the work of \cite{2001CQGra..18.4033F}. The properties of EMRIs formed via relaxation are such that they have large eccentricities at semi-major axis of about $0.1$--$1~\textrm{pc}$. They describe a random-walk-like evolution in phase-space, in particular in angular momentum, until one of two things happens: (i) the small mass deviates off the orbit which would evolved into an EMRI that inspirals into the MBH, or (ii) they cross a threshold in phase-space which separates orbital evolution dominated by dynamics into a regime where orbital evolution is driven only by the emission of GWs. When systems cross this line, which can be roughly derived by equating the relaxation timescale at pericenter to the associated timescale due to the emission of gravitational radiation, as derived by \cite{1964PhRv..136.1224P}, we can ignore any dynamical perturbation.

The increase in eccentricity during the EMRI formation can lead to
a situation in which the eccentricity is so high that the smaller BH falls radially on to the MBH, and there is only one or a handful of gravitational radiation bursts before the source is lost. This can be regarded as a head-on collision. This is what is commonly referred to as a direct plunge in the related literature (not to be confused with the plunge when the EMRI crosses the event horizon after hundreds of thousands of orbits). Direct plunge sources are basically lost, because we can extract little or no information from it \citep[but see][]{2007MNRAS.378..129H,2013MNRAS.429..589B,2013MNRAS.435.3521B}. It has been shown that the ratio between successful EMRIs and direct plunges could be of about $1$:$200$ respectively \citep[][and references therein]{2018LRR....21....4A,2020arXiv201103059A}.

This result led to an interesting new avenue in investigating the role of other
types of relaxation. By getting closer and closer to the MBH, the stellar
density drops, so that the danger of producing direct plunges due to the
accumulation of gravitational tugs of the orbit at apocentre is accordingly reduced. At the same time, the usual two-body relaxation time
increases more and more. In addition, the process of scalar resonant relaxation was found to be inefficient in this region of phase-space.

However, as we explain later in this section, direct plunges mostly occur in Schwarzschild MBHs. If the MBH has a spin, any direct-plunging
orbit turns out to be a successful EMRI, meaning that it spends tens and up to hundreds of thousands of cycles in the LISA band, depending on the inclination and the
spin of the MBH, as shown in \cite{2013MNRAS.429.3155A}. This has an impact on the event rate, because the ratio of 1:200 that we mentioned before 
increases in favour of successful EMRI orbits.

We note that recently, \cite{2020MNRAS.495.2321Z} derived an improvement to the pioneering work of \cite{1964PhRv..136.1224P}, extending the timescale to be accurate to first post-Newtonian order. By taking into account this modification,
the EMRI rates drop by at least one order of magnitude per nucleus. But then the role of the spin has another impact on the inspiraling timescale that might again enhance the event rate (\citealt{2021arXiv210200015Z}; Vazquez-Aceves et al in prep 2020b).

To sum up, relaxation is a robust theory which has been tested in observations in dense stellar systems. In particular, recent results show that theory, numerical simulations and observational data agree on the existence of a segregated stellar cusp at our Galactic Centre \citep{2018A&A...609A..28B,2018A&A...609A..27S,2018A&A...609A..26G,2019MNRAS.484.3279P}. This theory has been used in numerical simulations along with relativistic corrections (both PN and geodesic ones) to derive the event rate of EMRIs for a Milky~Way-like nucleus \citep{2011CQGra..28i4017A,2014MNRAS.437.1259B,2019MNRAS.483..152A}. 

\paragraph{Physics of EMRI formation II. Formation and disruption of binaries around a massive black hole} 

Besides stellar relaxation, binary separation is another way of delivering stellar BHs to the vicinity of a MBH and forming EMRIs \citep{2005ApJ...631L.117M}. In this model, a binary containing a
stellar-mass BH could form
relatively far away from the MBH and later be scattered by other stars to the vicinity of the MBH. If 
the periastron distance is smaller than the tidal-disruption radius of the binary, the most likely outcome is that the binary gets tidally disrupted, leaving the stellar-mass BH gravitationally 
bound to the MBH and the other binary component ejected from the system. The event rate is
difficult to estimate because of the uncertainties in the physical properties of nuclear
star clusters, but given the large cross section for tidal separation of binaries, it is considered that
this channel could make a significant contribution to the total EMRI population. Unlike the EMRIs
formed via stellar relaxation, the EMRIs produced by tidal separation have a much lower eccentricity
in the LISA band because the captured stellar-mass BHs initially have a larger binding energy. As a result,
these low-eccentricity EMRIs are less susceptible to the perturbation by the stars around the MBHs
and hence are more stable. 

\paragraph*{The contribution of LISA to the physics of EMRI formation}

As we have tried to convey in this section, there are a number of
different processes that leave the expected EMRI event rate detectable by LISA substantially uncertain. It seems likely that many competing effects produce unique signatures in the LISA observables--either for individual events or in the distribution of their properties over multiple events. Theory work in advance of launch will help determine which effects may be dominant, and what observables are correlated with which effects.
Hence, with detections in hand, we can hope to observe these phenomena and therefore address many open questions related to astrophysics, in addition to fundamental physics.

\begin{itemize}
\item LISA can determine the ratio of plunges (coalescence that involve tens or fewer orbits) to insprials (coalescence involving thousands or more orbits), since it can distinguish the waveforms of the two types of coalescences. Such a number can be used to test the predictions of stellar-dynamics models \citep[][and references therein]{2018LRR....21....4A,2020arXiv201103059A}.
\item LISA can measure the eccentricity of an EMRI to a precession of $10^{-6}$ \citep{2017PhRvD..95j3012B}. Such a measurement can reveal those EMRIs formed by the binary-separation channel, since they have relatively mild eccentricities \citep[$\sim$ 0.1,][]{2005ApJ...631L.117M}, while those produced by stellar relaxation preferentially have extreme eccentricities ($>$0.9). 
\item LISA can identify those EMRIs with zero eccentricities, and they are most likely produced in AGN accretion disks. Moreover, the small sky-localization error \citep[on average smaller than 1 deg$^2$,][]{2017PhRvD..95j3012B} of LISA may help us identify the host AGN of the EMRI and understand the condition favorable to the formation of such \textit{wet} EMRIs.   
    
\end{itemize}

\subsubsection{Physics of IMRI formation: Dwarf galaxies, galactic nuclei and globular clusters}

\paragraph{Heavy IMRIs from galaxy mergers} 
By providing the first access to the GW universe in the millihertz band, LISA will
reveal the population of high-redshift MBHs co-evolving with their host
galaxies to ultimately assemble the galaxies of today's Universe. While much of our astrophysical
understanding of BHs and galaxies through the cosmic dawn era, say $5 < z < 20$, has so far
focused on the largest structures including massive high-$z$ quasars and their hosts, these are not
the ancestors of today's typical galaxies. Galaxies like the MW were assembled from smaller,
mainly dwarf, galaxies, and our galaxy's MBH grew through some combination of
accretion and merger of smaller BHs hosted in these dwarfs, provided that orbital decay is efficient (see Section~\ref{sec:TheGalaxyMergerAndTheLargeScaleInspriral}). EM observations of
the smaller systems through this epoch will remain extremely challenging, making LISA's data
crucial for understanding our cosmic history.

Regardless of how IMBH seeds do form, it is likely that they form in
low-mass ($\sim 10^{8}$--$10^{9}\msun$) galaxies at high redshifts.  These
galaxies merge over time to build up more massive galaxies, and any
BHs they host will likely merge as well.  Further evidence in the
local universe supports this hypothesis: dwarf galaxies possibly hosting
IMBHs/MBHs have been discovered recently, lending further support that low-mass
galaxies can be IMBH hosts \citep{2013ApJ...775..116R,2014AJ....148..136M,2016ApJ...831..203P,Ahn:2017lzs, 2018ApJ...863....1C}.
When these low-mass galaxies merge with larger halos, they are tidally stripped and disrupted, with their remnants joining the greater halo population.  An IMBH orbiting in the halo will eventually spiral to the centre due to dynamical friction, and merge with the central MBH.

\begin{figure}
\includegraphics[width=7cm]{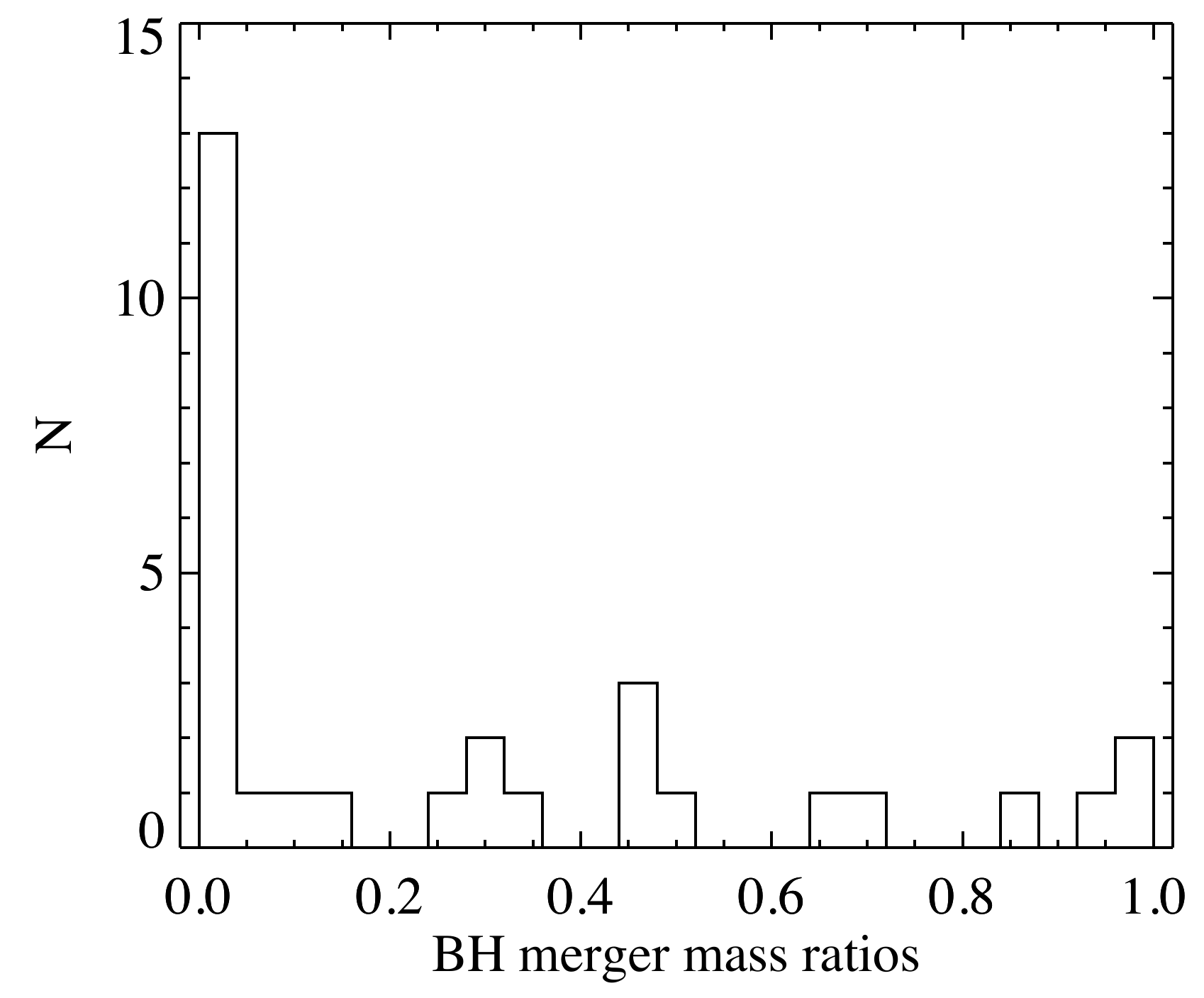}
\caption{Peak in IMBH/MBH mergers at mass ratios $\sim 20$:$1$ \citep{2019MNRAS.482.2913B}.
\label{massratios}}
\end{figure}

Cosmological simulations have shown that when dwarf galaxies hosting IMBHs merge with larger halos, they are tidally disrupted, leaving the IMBH to wander within the larger galaxy halo \citep{2010ApJ...721L.148B,2017MNRAS.470.1121T,2019MNRAS.482.2913B}.  Depending on their orbits and dynamical timescales, these IMBHs may spiral into the galactic centre and merge with the existing MBH. \citet{2019MNRAS.482.2913B} have shown that this event  has a characteristic mass ratio of $\sim 20$:$1$, which is reflected in the large peak at low mass ratios in Fig.~\ref{massratios}.

Dwarf galaxies are the most numerous in the Universe, and while the occupation fraction of BHs in dwarfs may be less than 1, because dwarf mergers with larger halos are common, this case cannot be ignored.  This type of BH--BH merger is the most common interaction in low-mass galaxy environments.

\paragraph{Heavy IMRIs from galactic nuclei assembly}

As discussed in the previous section, the mergers between dwarf satellites and their main galaxies
can lead to the formation of IMBH--MBH pairs (corresponding to heavy IMRIs) in galactic nuclei. However, the corresponding merger timescale is long due to the large mass ratio between the merging galaxies,
leaving this scenario unfavorable for the formation of IMRIs \citep{2007CQGra..24R.113A}. For example, theoretical models show that hierarchical galaxy mergers produce $<10\%$ of binary BHs with $q<0.01$
\citep{2003ApJ...582..559V,2020MNRAS.498.2219V}. 
For those IMRIs whose primary MBHs fall in the mass range of
$10^5$--$10^7\,M_\odot$ so that they can be detected by LISA, 
the event rate may be low because the host galaxies are relatively small, and low-mass galaxies merge much less frequently than those heavier ones.

Another viable route to the formation of an IMBH--MBH binary is related to the possible formation of IMBHs via stellar collisions and accretion onto stellar-mass BHs in the nuclei of dense stellar clusters, i.e.\ globular clusters \citep{2002ApJ...576..899P,2015MNRAS.454.3150G,2016MNRAS.459.3432M,2019arXiv190500902A,2020arXiv200809571R}. 

Clusters forming sufficiently close to the centre of their host galaxy can migrate toward the galactic centre via dynamical friction \citep{1975ApJ...196..407T,1993ApJ...415..616C}. This mechanism is thought to contribute to the growth of galactic nuclei \citep{1975ApJ...196..407T,1993ApJ...415..616C,2014ApJ...785...71G,2014MNRAS.444.3738A,2013ApJ...763...62A}. Clusters harboring an IMBH can bring the black hole to the galactic innermost regions and release it in the galactic centre. Such a mechanism might contribute to the seeding and growth of MBHs 
\citep{2001ApJ...562L..19E,2006ApJ...641..319P,2015ApJ...806..220A,2017MNRAS.464.3060A,2020arXiv200604922A}. If one or more IMBHs reach the galactic centre after the MBH is fully grown, the subsequent interaction between the MBH and the IMBH can trigger the formation of a massive binary that might undergo coalescence within a Hubble time. Given the typical range of mass of IMBHs ($10^2$--$10^5\msun$) and MBHs ($10^5$--$10^{10}\msun$) these merging binaries would have mass ratios in the range of $10^{-2}$--$10^{-5}$, typical of IMRIs. In massive elliptical galaxies, this mechanism can drive the formation of several IMRIs over a Hubble time, with an inferred rate of around $0.003-0.03~{\rm Gpc}^{-3}{\rm yr}^{-1}$ \citep[e.g.][]{2018MNRAS.477.4423A,2019MNRAS.483..152A}. Upon the simplest assumption that these mergers are distributed uniformly through space and are all detectable with LISA within a redshift $z<1$ \citep{2019arXiv190811391S}, we can derive an upper limit to the number of LISA detections of $2-20~{\rm yr}^{-1}$.

Massive ellipticals are not the only suitable nurseries for IMBH--MBH binaries. A number of IMBHs might be hiding in plain sight in our own Galaxy. 
The possible mass and location of such IMBHs can be constrained by the proper motion of Sgr A* and the kinematics of the S-stars close to it
\citep{2003ApJ...599.1129Y,2003ApJ...593L..77H,2004ApJ...616..872R}, as well as
the TDE rate in the Galactic Centre \citep{2013ApJ...762...95C}. According to these earlier
studies, the possibility of an IMBH with a mass of $\lesssim2000\,M_\odot$ and residing at a distance of $\lesssim10^{-3}$ pc from Sgr A* is not excluded.
However, the motion of S-stars orbiting Sgr A* suggests that 
if the MW MBH has a companion IMBH, its mass should be most likely smaller than $10^3\msun$ if the IMBH--MBH binary orbital period exceeds 5 yr, or up to $10^5\msun$  if the binary separation falls in the range $0.1$--$1~\mathrm{mpc}$ \citep{2009ApJ...705..361G,2018MNRAS.477.4423A}. The recent measurements of relativistic precession in the S2 star \citep{2018A&A...615L..15G} helped in further constraining the phase space allowed for an IMBH, ruling out companions with a mass $10^5\msun$ orbiting within $170~\mathrm{AU}$ ($0.8~\mathrm{mpc}$) from Sgr A* \citep{2020ApJ...888L...8N}. Nonetheless, there is growing suspicion of IMBH candidates orbiting farther away, around $1$--$10~\mathrm{pc}$ from the MBH. These putative IMBHs are supposedly harboured in a handful of compact gaseous clouds, whose measured velocity dispersion is so high to suggest the presence in their centres of point-like objects with masses in the range $10^4$--$10^5\msun$ \citep{2017NatAs...1..709O,2019ApJ...871L...1T,2020ApJ...890..167T}. However, depending on their orbital properties, a population of IMBHs lurking at the Galactic Centre would affect significantly the motion of S-stars \citep{2020ApJ...892..130D} and the structure of the nuclear star cluster \citep{2014ApJ...796...40M}.

The detection of such heavy IMRIs by LISA would have huge implications for our understandings of IMBH formation and evolution. 

\paragraph{Light IMRIs in stellar clusters and dwarf nuclei} 

We can also imagine the formation of light IMRIs, where the IMBH is the more massive partner in a merger with a stellar mass BH. At first glance, one might expect these to be simply scaled down versions of the EMRI problem with a central MBH. However, the challenge of understanding stellar dynamics with an IMBH is that we cannot assume that
the IMBH is fixed at the centre of the system, as we do with MBH in galactic nuclei. The
wandering of the IMBH makes it demanding, to say the least, to attempt an analytical
study, and hence we have to resort to numerical simulations to get an idea of what could be
the IMRI event rate and the characteristic properties. Globular clusters and dwarf nuclei may also present quite different dynamical scenarios (higher escape velocities due to dark matter but lower stellar densities in dwarf galaxies, for example).

The first dynamical simulation addressing the evolution of a globular cluster harbouring an IMBH
which successfully led to the formation of an IMRI was done by
\cite{2013A&A...557A.135K}. In this work they employed a direct-summation $N-$body code with
relativistic corrections as presented by  and a live treatment
of the relativistic recoil. They find that IMBHs with masses $500$--$1000\,M_{\odot}$ merge with stellar-mass
BHs and escape the host globular cluster due to the low escape velocity of the system. 
The IMBH is in a binary in almost all cases. The companion is a
stellar-mass BH of mass $\sim 20$--$26\,M_{\odot}$, and semi-major axis
of about $5$--$7~\mathrm{AU}$. Later, \cite{2014MNRAS.444...29L} found
similar results for this mass range. In their simulations, the
heaviest stellar-mass BH forms a tight binary with the IMBH in the
system. The work of \cite{2016ApJ...832..192H} is basically a reproduction of
\cite{2013A&A...557A.135K}, but using a different numerical scheme. This is
interesting because it validates of the findings of \cite{2013A&A...557A.135K}. Later,
\cite{2016ApJ...819...70M} explore lighter mass ranges for the IMBH, with masses at most of
$150\,M_{\odot}$. They confirm that the IMBH has a bound companion most of the time,
with the probability distribution function for the semi-major axis
maximised at $2~\mathrm{AU}$.

Recently, \citet{2021MNRAS.500.4628P} performed a series of Fokker-Planck simulations to explore the occurrence of light IMRIs around IMBHs of $10^5$~M$_{\odot}$ residing at the centre of massive star-forming clumps in high-redshift galaxies, finding event rates of $10^{-8}$--$10^{-7}$~yr$^{-1}$, depending on the assumptions for the initial inner density profile.

The IMRI rate from globular clusters and detectable by LISA depends intrinsically on a number of unknown quantities, namely the fraction of clusters capable of nursing the IMBH seed and growth \citep{2002ApJ...576..899P,2015MNRAS.454.3150G}, the amount of stellar-mass BHs and other compact objects lurking in the IMBH closest vicinity \citep{2016ApJ...819...70M,2019arXiv190500902A}, the number of times the same IMBH can pair in an IMRI \citep{2016ApJ...819...70M}, and the probability that upon merger an IMBH is ejected from the parent cluster due to anisotropic GW emission \citep{2008ApJ...686..829H, 2018ApJ...856...92F,2020arXiv200713746A}. 
Depending on all these quantities, LISA might be able to detect $0.01$--$60$ IMRIs from globular clusters per year out to redshift $z \sim 2$ \citep{2019arXiv190811375A,2020arXiv200713746A}. The number of light IMRIs from massive star-forming clumps detectable by LISA falls in the same ballpark: by integrating their computed IMRI event rate over $z = 1$--3, when clumpy galaxies are more numerous, \citet{2021MNRAS.500.4628P} computed that LISA should be able to detect $\sim$2 IMRIs per year, conservatively assuming that one star-forming clump per clumpy galaxy hosts a central IMBH. IMRIs with IMBHs in the mass range between $10^2 \, M_\odot$ and a few $10^3\, M_\odot$ might be detected with LISA and provide advanced warning to ground-based detectors with a precision up to a second \citep{2018PhRvD..98f3018A}.

\paragraph*{The contribution of LISA to the physics of heavy- and light-IMRI formation in gas-poor environments}

In the processes we have just described with regard to the formation
and evolution of heavy- and light IMRIs in gas-poor scenarios, our astrophysical theory so far only provides loose guidance about how the physics of the assembly
process connects to the characteristics of this population. Current models suggest, though, that
understanding heavy IMRIs will be an important discriminator
between hierarchical formation models.

IMRIs provoke more open questions than EMRIs, as the theoretical frameworks required to address IMRI formation are even more complex, due to the mobility of an IMBH, as compared to an MBH. Thus, many approximations used in relaxation theory to predict the gas-poor dynamics of EMRI formation cannot be applied to an otherwise similar IMBH system. In addition, there are several plausible channels of IMRI formation that involve theoretical frameworks that have far larger uncertainties than relaxation theory.

As for galactic nuclei assembly, several uncertainties might affect the formation of such heavy IMRIs, namely the number of clusters capable of reaching the galactic centre, the probability of IMBH formation, the relation between IMBH formation and the host cluster properties. 
Theoretical models suggest that star cluster infall and dispersal could bring 1-50 IMBHs in the MBH vicinity \citep{2006ApJ...641..319P,2014ApJ...796...40M,2018MNRAS.477.4423A,2019MNRAS.483..152A,2022MNRAS.514.5751L,2022arXiv220205618F}. If these models prove right, the delivery of IMBHs can give rise to up to 0.001-0.03 Gpc$^{-3}\,yr^{-1}$ \citep{2019MNRAS.483..152A,2022arXiv220205618F}, corresponding roughly to 1 event per year within redshift 0.5-3 \citep{2022arXiv220205618F}. Interestingly, the possible ejection of the IMRI product from the parent nucleus owing to GW radiation could account for up to $10^5$ Gpc$^{-3}$ MBH wandering outside their host galaxies at redshift < 1 \citep{2019MNRAS.483..152A}.


Finally, less consideration has been given to possible light IMRI rates in dwarf nuclei, although the presence of at least some IMBH in dwarf galaxies is more secure than in globular clusters. This is an important open question, especially since IMRI detection in dwarf nuclei could enable us to probe IMBH in \emph{quiescent} dwarf galaxies. This in turn leads to a better understanding of the formation of seed MBHs.
The detection of even a single light IMRI with LISA would incredibly improve our knowledge of how and where IMBHs form and grow.

\subsubsection{Formation of EMRIs and IMRIs in gas-rich galactic nuclei: AGN discs}

\paragraph{Gas-rich dynamics: Active galactic nuclei}

As discussed above, many galactic nuclei harbor a dense nuclear cluster of stellar-origin objects surrounding an MBH. Above we considered the state of the NSC as a result of stellar evolution, dynamical friction, secular evolution and minor mergers \citep{1993ApJ...408..496M,2014ApJ...794..106A,2018MNRAS.478.4030G}. Further complicating our picture, however, are the existence of AGN, which occur when low angular momentum gas forms a disc that accretes onto the MBH. As a result, a fraction of the nuclear cluster will end up embedded in the AGN disc via coincident orbits or capture \citep{1991MNRAS.250..505S,1993ApJ...409..592A}. While one might anticipate that AGN would be a subdominant mechanism for producing EMRIs or IMRIs, given that AGN represent only a fraction of all galactic nuclei (or, more likely, a relatively brief episode or series of episodes in the life of any given galactic nucleus), the presence of gas qualitatively changes the dynamics in the NSC.

Prograde orbiters embedded in a disc are expected to have their eccentricities rapidly damped \citep{1988Icar...73..330W,2004ApJ...602..388T,2007A&A...473..329C,2010A&A...523A..30B}, although there is a strong dependence on the details and resolution of gas-flow on horse-shoe orbits \citep{2010A&A...523A..30B}. For plausible AGN disc densities, gas dynamical cooling dominates over spherical component dynamical heating, so prograde orbiters should experience very rapid ($<0.1~\mathrm{Myr}$) eccentricity damping \citep{2012MNRAS.425..460M,Kennedy16,2020ApJ...889...94M}. 

The majority of NSC objects begin as inclined orbiters not coincident with the disc. Stars experience geometric drag and BHs experience dynamical drag as they pass through the disc, causing a significant portion to be captured within a plausible range of AGN disc lifetimes ($0.1$--$100~\mathrm{Myr}$). Stellar and BH orbiters not coincident with the disc experience geometric and dynamical drag forces, damping first the orbital eccentricity, followed by orbital inclination  \citep{2011A&A...530A..41B,2012ApJ...758...51J,Kennedy16,Bekdaulet2018, 2020ApJ...889...94M,2020arXiv200611229F}. \cite{2020arXiv200611229F} find $O(10\%)$ of prograde NSC BH are captured within a \citet{SG} type disc, for typical disc lifetimes, ignoring accretion. A much smaller fraction are captured by lower-density type \citet{TQM} AGN discs. Stellar objects are primarily captured by the disc at small  radii, losing around an order of magnitude in semi-major  axis, whereas BHs are captured across the full range of disc radii but rapidly get delivered to the innermost disc \citep{2020arXiv200611229F}. The possibility of the accumulation of BHs at small radii across disc models has significant implications for the LIGO--Virgo merger detection rate \citep{2020arXiv200611229F}, but also for the possible EMRI/IMRI production rate.

The accumulation of prograde BH from the NSC in the inner disc leads to high interaction cross sections at low relative velocity. All embedded stellar-origin objects on prograde orbits should undergo mass-dependent Type I migration due to torques from gas at Lindblad resonances and co-rotating gas \citep{2002ApJ...565.1257T}, enhancing pile-up of BHs in inner AGN discs, leading to a high merger rate \citep{2012MNRAS.425..460M,2014MNRAS.441..900M,2017ApJ...835..165B,2017MNRAS.464..946S}. 

\paragraph{Heavy IMRIs in AGN}
The rate of change of surface density in AGN disc models implies that we should expect the occurrence of locations in the discs where the outward and inward migration torques cancel \citep{2016ApJ...819L..17B}. At such so-called migration traps, the local merger rate is significantly enhanced and IMBHs with masses $\sim10^{3}~M_{\odot}$ can quickly ($1$~Myr) be produced \citep{2019ApJ...878...85S,2020arXiv200411936S,2019PhRvL.123r1101Y,2020MNRAS.494.1203M}. The IMBH-formation merger GW190521 detected by LIGO--Virgo  \citep{2020ApJ...900L..13A} consisted of two BHs in the upper mass gap with mis-aligned spins, suggestive of a merger in a dynamically rich, deep gravitational potential well. If we assume this merger happened at a migration trap then the rate of such mergers inferred by LIGO--Virgo is $\sim 0.7~\mathrm{Gpc^{-3}\,yr^{-1}}$ \citet{2020arXiv201014533T}. From \cite{2020MNRAS.498.4088M}, assuming $O(15)$ mergers at migration traps over a Myr disc lifetime, and an AGN fraction of $O(1\%)$ of galactic nuclei (quasars and the brightest Seyfert nuclei), we find $\sim 1~\mathrm{Gpc^{-3}\,yr^{-1}}$ mergers at migration traps, consistent with the observed rate of GW190521-like events.  
All of these effects taken together enhance stellar BH merger rates and encourage the formation of IMBH in AGN discs. IMBH formed in this manner automatically create an IMBH--MBH binary, which will typically decay due to GW emission on timescales of a few hundred Myr \citep{2016ApJ...819L..17B}. 

\paragraph{Light IMRIs in AGN}
A large IMBH sitting in an AGN disc migration trap is an excellent site for the creation of light IMRIs---less massive BH will be delivered to the IMBH from more distant regions of the disc via disc migration torques, with an approximate merger rate of $O(1)~\mathrm{Gpc^{-3}\,yr^{-1}}$ \citep{2020MNRAS.498.4088M}, assuming \citet{SG} type AGN disc models. LISA can detect IMBHs of several hundred $M_{\odot}$ out to $\sim 10~\mathrm{Gpc}$. So for migration trap mergers with an IMBH of few hundred $M_{\odot}$, this suggests LISA can detect $O(10^{3})~\mathrm{yr}^{-1}$ light IMRI mergers from AGN discs. If migration traps are less common in AGN discs, the maximum IMBH masses from bulk disc mergers are much smaller $\sim10^{2}~M_{\odot}$---and migration torques could drive objects rapidly onto the MBH, typically creating EMRIs with small eccentricities and inclination angles.\footnote{In providing fuel to the MBH, this process may contribute to explaining the $M_\mathrm{bh}$--$\sigma$ relation \citep{2005ApJ...619...30M}.}

\paragraph{EMRIs in AGN}
One likely exception to the small $(e,i)$ expectation due to AGN gas damping comes from retrograde orbiters. Approximately half of the initial (NSC) population that is geometrically coincident with the disc, should lie on retrograde orbits \citep{2015A&A...576A..29I}. Migration torques on retrograde orbiters in AGN discs are a small fraction of that on prograde orbiters \citep{2014MNRAS.441..900M}. However, retrograde orbiters experience eccentricity pumping at apocenter which rapidly drives them to very high eccentricities \citep{2013MNRAS.428.3072D,2016MNRAS.458.3221T,2019EAS....82..415T,2020ApJ...889...94M} and increases the decay rate of the semi-major axis  \citep{2020arXiv200903910S}. As a result we expect a significant population of retrograde orbiters in the innermost AGN disc. This population could yield a very high EMRI rate. The eccentricities of EMRIs from this population depend on the masses of the retrograde orbiters, since the GW circularization rate increases with mass \citep{1964PhRv..136.1224P} and the eccentricity driving of the gas decreases with mass \citep{2020arXiv200903910S}. For example, a $10\,M_{\odot}$ retrograde orbiter will likely have an eccentricity over $0.9$ when it inspirals, whereas a $50\,M_{\odot}$ orbiter will circularize before inspiraling, making it much easier for LISA to detect over many cycles. However, the higher rate of eccentricity driving of lower mass retrograde orbiters also implies that their EMRI rates will be higher. Therefore, unlike prograde orbiters whose orbits tend to be circularized by the gas disc, retrograde orbiters could commonly produce highly eccentric EMRIs, though this effect is strongest at high (AGN disc) gas density and for low mass BH. 

There may also be observable EM signals associated with smaller BHs in AGN discs. Stellar-mass BH binaries (BH+BHs) can merge at high rates in AGN discs \citep{2012MNRAS.425..460M,2014MNRAS.441..900M,2017ApJ...835..165B}. Under these circumstances, since the BH+BH is surrounded by gas, there will always be an EM counterpart. Indeed, a candidate EM counterpart to GW190521 has recently been suggested in an AGN \citep{2020PhRvL.124y1102G}.
Several key questions underpin the search for EM counterparts to BH+BH mergers in AGN discs: Is the EM counterpart detectable through a potentially large optical depth? Is the emission completely outshined by the AGN emission, and on what timescale? Does the radiation from the BH reduce the EM emission? 

At merger, a remnant stellar-mass BH formed from the BH+BHs recoils with a kick velocity $v_k$ depending on the mass asymmetry and spin orientations of the progenitors. In an AGN disc, gas at distance $R_{\rm bound} < GM_{\rm BH+BH}/v_k^{2}$
is bound to the merged BH+BH and attempts to follow the kicked merger product. In doing so, it collides with surrounding disc gas and a shock luminosity emerges on a time-scale of $\sim 20$ days, with a bound gas energy of about $10^{45}\, \rm erg$ (depending on the BH+BH and gas properties). After the kicked stellar-mass BH has shed the bound gas, the passage of the stellar-mass BH through the gas in the accretion disc produces a shocked Bondi drag tail \citep[e.g.][]{1999ApJ...513..252O}. This tail both decelerates the  stellar-mass BH and accretes onto it, generating a potentially high luminosity. 
In order for enough radiation to escape to make for a bright flare against the AGN, a jet or collimated outflow is required. Such jets may also produce detectable X-ray or gamma-ray signatures. 

Alternatively,  stellar-mass BHs in AGN discs may undergo substantial accretion even prior to merger \citep{2020ApJ...901L..34Y}, which can lead to e.g., longer term X-ray emission. This scenario largely depends on the poorly understood accretion efficiency and gap opening by  stellar-mass BHs in AGN discs.
We recommend that future simulations of hyper-Eddington accretion establish whether there is an upper limit to accretion which can choke off jet formation and launching. 
This will help establish  luminosity upper limits on any flares that originate from kicked stellar-mass BH mergers in AGN discs and can guide searches for EM counterparts from AGN discs. We also recommend simulations of lightcurves from false-positive flaring events such as SNe and TDEs breaking out from within the AGN disc, or lightcurves of micro-lensing events.\\

\paragraph{In situ formation of stars in AGN discs: a special population of EMRIs}

AGN discs are known to be prone to gravitational (Toomre) instability in the outer regions, and expected to form stars vigorously \citep{1989ApJ...341..685S,2003MNRAS.339..937G,2007MNRAS.374..515L,2007MNRAS.379...21N}. 
This expectation is supported by observations of nearby stellar discs in the nucleus of the MW \citep{2003ApJ...590L..33L} and M31 \citep{2005ApJ...631..280B} which, due to their large masses and orbital configurations, can be interpreted as remnants of a prior accretion episode from a gaseous disc \citep{2007MNRAS.374..515L}. 
Numerous observational studies also suggest a broader connection between AGN activity and nuclear starbursts \citep[e.g.,][]{2007ApJ...671.1388D, 2010MNRAS.405..933W,2016MNRAS.463.1291I}, although whether this connection is causal
remains uncertain, given that AGN feeding occurs on scales difficult to resolve ($\lesssim$ parsecs) and is often obscured \citep{2012NewAR..56...93A}. 
Theoretically, the stars formed in the outskirts of AGN discs are expected to be unusually massive because the disc material is much hotter and denser than star-forming regions in the galactic ISM.  Once formed, a population of stars embedded in a gas disc will undergo a stellar evolution that is notably altered by their environment (\citealt{2020arXiv200903936C}). Throughout their lifetime stars can grow by accretion, becoming even more massive \citep{2004ApJ...608..108G,2020MNRAS.tmp.2525D}. In addition to experiencing drag and dynamical friction, they will excite perturbations in the disc that will exert torques on their orbit, typically causing inward migration (as discussed above). Furthermore, we expect an increased number of binaries within the stellar population in the AGN disc \citep[e.g.,][]{2008MNRAS.389.1655A}, which can become harder due to disc-satellite interactions \citep[e.g.,][]{2011ApJ...726...28B,2020ApJ...891...47A}. All this makes fertile ground for forming BH remnants in the disc, which can subsequently encounter each other in migration traps (described above) or migrate to the inner regions of the disc where their evolution becomes GW dominated. Initial estimates show that this process may be an efficient method for feeding MBHs at early times in a way that is not Eddington-limited \citep{2020MNRAS.493.3732D}.

Future numerical studies can help us narrow down the vast parameter space of accretion disc structures as a function of relevant parameters such as MBH mass, accretion rate, gas supply or redshift.  As we improve our understanding of AGN discs, more investigations are needed to fully understand how embedded stars and BHs evolve over a range of system parameters and disc properties.  Including more detailed physics (such as disc instabilities and stochastic torques, radiation transport or feedback from accretion, to name a few) may change the evolutionary outcome of these sources, the predicted rates and characteristic properties, as well as the precise waveform signatures and whether or not they are distinguishable amongst formation channels.  
 At the same time, gas-embedded EMRIs/IMRIs also present a powerful opportunity to probe AGN properties in regions that are historically electromagnetically unresolvable, 
either with deviations in the GW waveforms that correlate with properties of the gas (see Sec.~\ref{sec:waveforms}) or with populations statistics, if multiple events are detected. This is in addition to the possibility for multimessenger astrophysics with associated EM counterparts (e.g. variability in emission, see Section~\ref{sec:MMA}).

As a distinguishing characteristic between various formation channels, here we quote approximate expectations of eccentricity and inclination at late stages of the inspiral, which we define as approaching the central MBH ISCO. Gas-driven, prograde EMRIs are expected to have low e ($e\lesssim0.01$) given that circularization by gas and GWs is efficient, but there remains a possibility of disc-driven eccentricity pumping for relatively massive secondaries or IMRIs \citep[e.g.,][]{2006ApJ...652.1698D}.
These sources are also likely fully embedded in the disc with low inclination---hence if the orientation of the disc aligns with the spin of the central MBH (which may be true to varying degrees, see \citealt{2013ApJ...775...94V}), we expect the spins of the binary components to be closely aligned. 
Gas driven, \emph{retro-}grade EMRIs, on the other hand, may reach very high eccentricities even at the ISCO \citep[$e\lesssim0.9$,][]{2020arXiv200903910S}. They should also retain a low inclination, although the spin alignment of the BHs will depend on the accretion history of the embedded BH which remains to be investigated across the full range of parameter space. 
These estimates will be further constrained with future work that includes more realistic disc modelling and treatment of gas dynamics and accretion onto embedded BHs.

\paragraph*{The  contribution  of  LISA  to  the  physics  of formation of EMRIs and IMRIs in gas-rich galactic nuclei}

From what we have presented, we can derive that depending on the (highly uncertain) duty cycle of AGN, an IMBH--MBH binary could correspond to a heavy IMRI in nearly every galactic nucleus. However, the absence of any such detection with LISA would seriously constrain the existence of a migration trap in a generic AGN disc. The absence of a migration trap implies the absence of strong changes in the disc surface density gradient and thus tightly constrains the transition between the radiation pressure and gas pressure dominated regions of AGN discs.

As we have mentioned, there is a clear correlation between the dynamical parameters of EMRIs formed in AGN discs and their detection. Thus the rate of EMRIs detected by LISA can put strong constraints on the populations and dynamics we expect to live in innermost AGN discs, a system lurking in a region inaccessible to spatially resolved EM observations. 

Prediction of rates and precise characteristics of disc-embedded EMRIs/IMRIs is a multi-faceted problem that relies on details of gravitational instability, stellar evolution, nonlinear gas dynamics, and accretion physics. Given that MBHs spend $1$--$10\%$ of their evolution in an AGN phase \citep{2013MNRAS.428..421S,2016ApJ...831..203P}
 we expect at minimum the same fraction of EMRIs to occur in gas-rich environments. 
In-situ star formation likely leads to a population of compact remnants \emph{in addition} to those that are captured from the nucleus. 
Thus we expect that dense accretion discs in near-Eddington AGN may not only boost EMRI rates, but also produce a population that is uniquely characteristic: with low eccentricity and (some degree of) spin alignment with the central MBH. These $(e,i)$ expectations are strongest for EMRIs from objects formed in-situ. A single EMRI with low eccentricity will indicate a current (or recent) interaction with gas, showing that the host galaxy harbors an active (or recently active) MBH. The precise parameters (e.g. eccentricity, secondary BH mass) are intimately connected to prior evolution of the accretion disk, and thus these  measurements will give constraints on the efficiency of gas-driven circularization and accretion disk structure. If a larger population of such EMRIs is detected, the distribution of orbital characteristics will tell us about the diversity of AGN disks. Rates and orbital characteristics will constrain various aspects of accretion onto MBHs -- for example, the number of detected events will inform how many nearby AGN host disk-embedded BHs. If the number is high, it will challenge accretion models.  Just as the migration rates of stars and compact remnants depend sensitively on characteristics of the disk \citep{2013LNP...861..201B,2015ApJ...806..182D}, the secondary masses will be a consequence of BH accretion or hierarchical mergers, all of which are tied to disk models. Overall, measurements that shed light on accretion disk structure and prevalence will improve our constraints on AGN duty cycles and MBH growth.

\subsubsection{Alternative formation scenarios}

\paragraph{XMRIs} 

The possibility of observing an EMRI at our own galactic centre when LISA flies is basically zero, since the rates for an EMRI formed via relaxation in a 
MW-like galaxy are at most about $10^{-6}~\mathrm{yr}^{-1}$ \citep[e.g.,][]{2018LRR....21....4A,2020arXiv201103059A}. 
This means that about once every million years a stellar-mass BH plunges through the event horizon of
our central MBH. Since the lifetime of LISA will be only a few years, the probability of detecting one at our
galactic centre is negligible. 

We can however find another class of EMRIs at our galactic centre.
It has been recently put forward \citep{2019PhRvD..99l3025A} that substellar objects, in particular brown dwarfs,
stand very high chances of being in band of the detector when it is launched. The reason for this is very simple: these substellar objects have mass ratios of about $q\sim 10^{-8}$ as compared to the central MBH, Sgr~A$^*$. Such XMRIs (extremely large mass ratio inspirals) can therefore
cover up to $\sim 10^8$ cycles before crossing the event horizon, since the number of cycles is roughly inversely proportional to the
mass ratio. This means that they stay in band for millions of years. About $2\times10^6~\mathrm{yr}$ before merger they have an
SNR at the galactic centre of $10$. Later, $\sim 10^4~\mathrm{yr}$ before merger, the SNR reaches several thousands, i.e.\ they are at the level of the loudest MBH mergers.
At the last stages of their evolution, some $\sim 10^3~\mathrm{yr}$ before the merger, they can reach SNR as high as a few $10^4$ \citep{2019PhRvD..99l3025A,2019A&A...627A..92G,2004PhRvD..69h2005B}.

The work of \cite{2019PhRvD..99l3025A} predicts that at any given moment there should be of the order of $\gtrsim5$ XMRI that are
highly eccentric and are located at higher frequencies, and about $\gtrsim15$
are circular and are at lower frequencies. The mass ratio for an XMRI is about
three orders of magnitude smaller than that of stellar-mass BH EMRIs. Since
backreaction depends on $q$, the orbit closely follows a standard geodesic,
which means that many approximations work better in the calculation of the orbit.
XMRIs can be sufficiently loud so as to track the systematic growth of their
SNR, which can be high enough to bury that of MBH binaries.

In addition, there are also plunge events during the formation of inspiralling sources. The GWs from low mass objects (brown dwarfs, primordial BHs, etc.) plunging into the central MBH are burst signals. For LISA, the SNRs of these bursts are quite high if they happen in our Galaxy. However, the event rates are estimated as $\sim 0.01~\mathrm{yr}^{-1}$ for the Galaxy. 
If we are lucky, this kind of very extreme mass-ratio burst will offer a unique chance to reveal the nearest MBH and nucleus dynamics. 
The event rate could be as large as $4$--$8~\mathrm{yr}^{-1}$ within $10~\mathrm{Mpc}$, and because the signal is strong enough for observations by space-borne detectors, there is a good chance of being able to use these events to probe the nature of neighbouring BHs \citep{2013MNRAS.433.3572B}. 
This kind of burst sources are called XMRBs (extreme mass ratio bursts) \citep{2020MNRAS.498L..61H}.

\paragraph{Binary and multiple EMRIs} 

Recent theoretical studies pointed out the existence of a new type of EMRI
in which the small body is a stellar-mass BH+BH.
Such a triple system could form either due to tidal capture of a BH+BH by a MBH
\citep{2019GReGr..51...38A,2018CmPhy...1...53C}
or the formation and migration of a BH+BH in the accretion disc of an AGN \citep{2019MNRAS.485L.141C}.
While the latter channel is considered to be more effective than the former one,
both channels could deliver BH+BHs to a distance as small as tens of gravitational radii of the
central MBH. As a result, the binary, as a single entity, spirals into the MBH due to GW radiation.
For this reason, the source is referred to as a binary-EMRI, or b-EMRI.

\begin{figure}[h]
\includegraphics[scale=0.35]{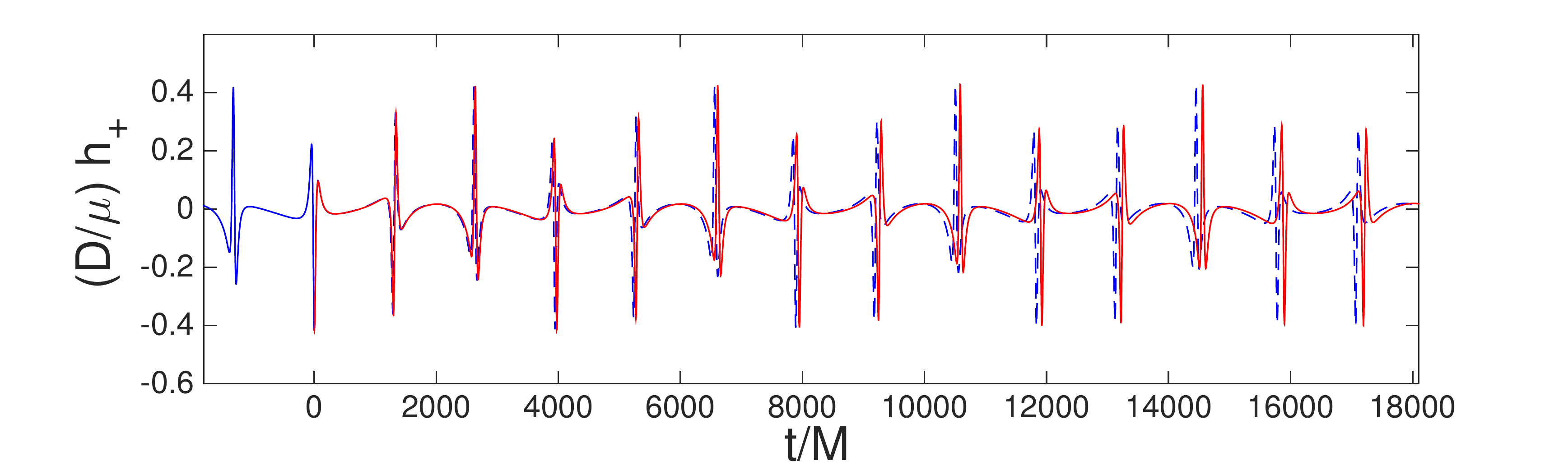}
\caption{Comparing the waveforms of the EMRIs with (red, solid) and without (blue, dashed) a glitch. 
The MBH has a mass of $M = 10^6\, M_\odot$ and a spin parameter of $0.9$. 
The total mass of the BH+BH is $m = 20\, M_\odot$, and $D$ refers to the luminosity distance. In this example, the centre of mass of the stellar-mass BH+BH initially is moving inside the equatorial plane of the MBH with an orbital eccentricity of $e = 0.7$ and a semilatus rectum of $p = R(1 - e^2) = 17 r_\mathrm{g}$. 
At the time $t = 0$ a kick to the centre-of-mass velocity of the binary happens, in the polar direction and with a magnitude of $1500~\kms$. 
As a result, the orbital parameters change to $p = 16.9990 r_\mathrm{g}$ and 
$e = 0.7019$, and the orbital plane of the EMRI becomes inclined by 
$\iota = 0.5233^\circ$ relative to the equatorial plane of the MBH.\label{bemriwave}}
\label{spectrum}
\end{figure}

The uniqueness of the b-EMRI lies in the
fact that it simultaneously emits two kinds of GWs. One in the LISA band, due
to the orbital motion of the BH+BH around the MBH, and the other in the
ground-based detector band, when the BH+BH coalesces due to the tidal perturbation by the MBH.
A coordinated observation by LISA and ground-based detectors would allow us to identify
such interesting sources and, more importantly, constrain
several aspects of fundamental physics to a precision more than one order of magnitude better than the current limit,
including the loss of rest mass due to GW radiation, the recoil velocity
of the merging BH+BH, and the dispersion of GWs of difference frequencies \citep{2019MNRAS.485L..29H}. Due to the merger of the BH+BH, the remnant will obtain a recoil velocity which may be up to a few thousand $\kms$. This sudden kick will induce a glitch on the waveform of a b-EMRI (see Fig.~\ref{bemriwave}). 

\paragraph{Supernova-driven EMRIs} 

In addition to the aforementioned mechanisms, an EMRI can also be generated via the supernova explosion that accompanies the formation of a CO. When this happens, the velocity of the compact object gets almost instantaneously significantly perturbed, so that the compact object settles on a brand new trajectory: the timescale for the CO to coalesce with the MBH via GWs on this new orbit may be shorter than the timescale for two-body relaxation to perturb it, so that the CO is bound to evolve into an EMRI. Focusing on the Galactic Centre environment, \citet{2019MNRAS.485.2125B} showed that one supernova out of $10^4-10^5$ occurring within the star forming structures present about the MW MBH will give rise to a supernova-driven EMRI. This result, coupled with the expected frequency of core-collapse supernovae explosions occurring in the Galactic Centre, implies a frequency of supernova-driven EMRIs up to $10^{-8}$ yr$^{-1}$, i.e. an EMRI rate that is comparable or only mildly lower than the one associated to the standard two-body relaxation process \citep{2019MNRAS.485.2125B}.

\paragraph*{The  contribution  of  LISA to relativistic stellar dynamics and supernovae rates}

What we have described about XMRIs allows us to understand that these might be envisaged as a  double-edged sword. From the one
side we have a promising and strong source of GWs from an extreme-mass ratio which is easy to model. On the other hand they might pose a problem
because their SNRs (as high as $10^4$ for one-year observation if one resides in the Galactic Center) are such that can bury binaries of MBHs. Also, if they are present in most nuclei harbouring
MBHs in the range of LISA, they might interact with EMRIs or even scatter them off from their inspiraling orbit. The
detection of XMRIs will allow us to infer information on relativistic astrodynamics impossible to obtain otherwise.

LISA can distinguish b-EMRIs from normal EMRIs by detecting the GWs from the small binary black hole months to years prior to its coalescence around the SMBH. If the frequency of the GWs from the small binary matches a fundamental frequency of the SMBH, the SMBH could be resonantly excited and the EMRI waveform could contain an enhanced quasi-normal mode \citep{2021PhRvD.103h1501C}. Moreover, the binarity of the small body also induces an addititional phase shift to the EMRI waveform, which can be used to identify b-EMRIs as well \citep{2022arXiv220608104C}.

 
Regarding the detection of EMRIs formed via supernova, this result calls for a more extended analysis, in order to investigate this process in a wider range of galaxy environments and star-formation rates, and exploring in more detail the waveform signatures associated to this kind of EMRI compared to other EMRI formation mechanisms. In this framework, LISA will thus help us shed light on the rates of supernovae near MBHs via the detection of EMRIs.

\subsection{Multimessenger prospects}
\label{sec:MMA}

\noindent
{\bf Coordinators: Giuseppe Lodato and Martina Toscani

\noindent
Contributors: Pau Amaro Seoane, Jillian Bellovary, Stefano Bianchi, Saavik Ford, Barry McKernan, Giuseppe Lodato, Tom Kimpson, Scott Noble, Martina Toscani, Kinwah Wu, Ziri Younsi and Silvia Zane}\\

We outline a variety of proposed EM counterparts of EMRIs and IMRIs, including TDE (in multiple configurations), AGN-related signatures, and pulsar EMRIs.

\subsubsection{Tidal disruption events}

TDEs \citep[][for a recent review]{1983A&A...121...97C,1988Natur.333..523R,1989IAUS..136..543P,2020arXiv200512528R} can be considered a particular type of EMRI, where the star is disrupted by the MBH tides during the first passage at the pericenter. For this to happen, the pericenter radius should be smaller \citep[cf.][]{2020ApJ...904...99R} than the \textit{tidal radius}  
\begin{equation}
    r_\mathrm{t} \approx R_*\left( \frac{M_{\bullet}}{M_*}\right)^{1/3},
    \label{eq:tidalradius}
\end{equation}
where $R_*$ and $M_*$ are the stellar radius and stellar mass respectively, while $M_{\bullet}$ is the BH mass; $r_\mathrm{t}$ is usually a factor of $10$--$20$ times the MBH Schwarzshild radius. For MBH masses larger than $\sim 10^8 \, M_{\odot}$ ($10^9\,M_{\odot}$ for rapidly spinning BHs), the tidal radius is within the event horizon and no TDEs can happen.

TDEs are very luminous events over a broad range of EM bands. The first EM observations occurred in 1990s thanks to the ROSAT survey \citep{1996A&A...309L..35B,1999A&A...349L..45K,1999A&A...350L..31G,2000A&A...362L..25G}, which detected some bright flares from the cores of non-AGN galaxies. Since then, the number of X-ray detections has incresed. These observations seem to be in agreement with the theoretical expectation of X-ray emission from an accretion disc \citep[e.g.,][]{1999ApJ...514..180U,2017ApJ...838..149A,2011MNRAS.410..359L}. Initially these flares are powered by a near-Eddington accretion, then the luminosity decreases over a period from months to years \citep[][ and references therein]{2020SSRv..216...85S}. Over the last decade, also a growing number of optical TDEs has been detected. It remains unclear what processes are at the origin of this optical emission. Some hypotheses concern the shocks from self-crossing debris \citep{2015ApJ...806..164P,2015ApJ...804...85S} or reprocessing in an outflow \citep[e.g.,][]{2009MNRAS.400.2070S,2011MNRAS.410..359L,2016MNRAS.461..948M}. Detailed reviews of optical TDEs are found in \citet{2019MNRAS.487.4136W} and \citet{2020SSRv..216..124V}. A small fraction of these jetted events has also shown significant radio emission \citep[][ and references therein]{2020SSRv..216...81A}.

Furthermore, TDEs emit GWs. Their GW emission is produced by three different varying quadrupoles: (a) the star--BH quadrupole \citep{2004ApJ...615..855K,2022MNRAS.510..992T}; (b) the stellar internal quadrupole \citep{2009ApJ...705..844G,2013MNRAS.435.1809S} and (c) the quadrupole of the compact torus formed after disruption \citep[e.g.,][]{2001PhRvL..87i1101V,2002ApJ...575L..71V,2011PhRvL.106y1102K,2019MNRAS.489..699T,2019EPJP..134..537V}. The dominant contribution is the first term, that could be well described as a GW burst with strain \citep{2004ApJ...615..855K,2022MNRAS.510..992T}
\begin{equation}
  h \approx \beta \times \frac{r_\mathrm{s}r_\mathrm{s*}}{r_\mathrm{t}d}  
  \approx \beta \times 2 \times 10^{-22} \left(\frac{M_*}{\mathrm{M_\odot}}\right)^{4/3} \left(\frac{M_{\bullet}}{10^6 {M}_{\odot}}\right)^{2/3} \left(\frac{R_*}{{R_{\odot}}}\right)^{-1}\left(\frac{d}{16\,\mathrm{Mpc}} \right)^{-1},
  \label{eq:TDE_strain}
\end{equation}   
and an associated Keplerian frequency of 
\begin{equation}
   f \approx \frac{\beta^{3/2}}{2\pi}\left( \frac{G M_{\bullet}}{r_\mathrm{t}^3}\right)^{1/2}
\approx \beta^{3/2}\times 10^{-4}\,\mathrm{Hz}\times \left(\frac{M_*}{{M_{\odot}}}\right)^{1/2}\left( \frac{R_*}{{R}_{\odot}}\right)^{-3/2}. 
\end{equation}
In the above formulas we have introduced the Schwarzschild radius of the BH, $r_\mathrm{s} = 2 r_\mathrm{g}$, the Schwarzschild radius of the star, $r_\mathrm{s*}$, and the penetration factor
\begin{equation}
    \beta=\frac{r_\mathrm{t}}{r_\mathrm{p}},
\end{equation}
where $r_\mathrm{p}$ is the pericentre distance. A library of gravitational waveforms from TDEs has been presented in \citet{2022MNRAS.510..992T}, generated using a general relativistic smoothed particle hydrodynamic code, \textsc{phantom} \citep{2019MNRAS.485..819L}. To date, this numerical study models the star as a polytropic sphere with index $\gamma=5/3$, but we expect the strain to have a dependence on the internal structure of the star. This dependence needs to be further investigated. 

The expected strain for a Sun-like star being disrupted by a $10^6 \, M_{\odot}$ MBH at $15~\mathrm{Mpc}$ distance (assuming $\beta$=1) is $h\sim 10^{-22}$, with a frequency $f\sim 10^{-4}~\mathrm{Hz}$. The two other contributions are expected to have similar frequency but are scaled down by some (two-five) orders of magnitude.

 \cite{2021arXiv210305883P} estimate the rate of TDEs which could be observed with different instruments. They find that for LISA it is unlikely to detect GWs from TDEs, unless BHs are surrounded by particularly massive stars. The next generation of detectors beyond LISA should however be able to detect GW from TDEs up to cosmological redshifts $z\geq 1$.

An interesting signal to study is the GW background from the entire cosmic population of TDEs. Details on this signal and its derivation may be found in Sec.~\ref{sec:TDE_background}.

\paragraph{TDEs outside galactic nuclei}
In the majority of TDE studies  
  the MBH that disrupts a star 
  is implicitly taken to be the nuclear BH of the galaxy. 
The Swift transient source AT2018cow has many characteristics resembling a TDE 
  \citep{2019MNRAS.487.2505K}, 
  but peculiarly the source is not located 
  at the nuclear region of its suspected host galaxy Z~137-068. 
This leads to consideration of whether it is a TDE, 
  with an MBH disrupting a WD,  
  i.e.\ WD-TDE \citep[][]{2018ApJ...856...82H},  
  or alternatively a violent stellar explosion. 
The question is now: Can TDE involving a wandering\footnote{See Sections~\ref{sec:DynFriction},~\ref{sec:tripleMBHs} and~\ref{MBH_kicks}.} MBH occur? 
This question would be answered if there are mechanisms 
  to populate a galaxy with massive BHs of non-stellar nature. 
Stellar systems are not stationary structures.  
A stellar cluster can dissolve on timescales 
  of $10$--$100~\mathrm{Myr}$ \citep[e.g.,][]{2008A&A...482..165G}. 
Globular clusters can survive longer 
  but they can also be disrupted 
  \citep[e.g.,][]{2006ApJ...637L..29B,2020Natur.583..768W} 
  or dissolved  \citep[see][]{2009gcgg.book..387B}. 
Similarly, dwarf galaxies can be disrupted and dissolved 
  \citep[e.g.,][]{2018ApJ...866...22L,2018MNRAS.478.3879S}
 when they encounter and are accreted by a larger galaxy. 
The nuclear BHs, if present  
  in these stellar systems,  
  would be dispersed into the interstellar space 
  of the cannibal galaxy.  
Stars are also carried along into the interstellar space 
  by these BHs, 
  and some of them will eventually spiral into their carrier BH 
  and become a TDE.

\subsubsection{Electromagnetic counterparts of light IMRIs in AGN discs}
\label{EM_IMRIs_AGNdisks}
Most BHs that merge in AGN discs (including IMBH--BH mergers) are expected to experience a GW recoil kick at the moment of merger with speeds $v_\mathrm{k}$ of up to a few hundred $\kms$. Such merger kicks would happen for any comparable system, with or without gas; however, the consequences of such kicks for a gas-embedded merger may include a detectable EM counterpart. In an AGN disc, gas within
\begin{equation}
R_\mathrm{bound} < \frac{GM_\mathrm{BH+BH}}{v_\mathrm{k}^{2}}
\end{equation}
is bound to the merged BH+BH and attempts to follow the kicked merger product. In doing so, it collides with surrounding disc gas and, as long as the disc is geometrically thin or optically thin, a shock luminosity can emerge on a timescale $t_\mathrm{bound}=R_\mathrm{bound}/v_\mathrm{k}=GM_\mathrm{BH+BH}/v_\mathrm{k}^{3}$ \citep{2019ApJ...884L..50M} or
\begin{equation}
    t_\mathrm{bound} \sim 20\left(\frac{M_\mathrm{BH+BH}}{100\, M_{\odot}}\right)\left(\frac{v_\mathrm{k}}{200~\mathrm{km}\, \mathrm{s}^{-1}}\right)^{-3}~\mathrm{day}.
\end{equation} 
The total energy delivered to the bound gas is $E_\mathrm{bound}=(1/2) M_\mathrm{bound} v_\mathrm{k}^{2}=(3/2) N k_\mathrm{B} T_\mathrm{bound}$ where $M_\mathrm{bound} = N m_{H}$ is the mass of the bound gas expressed as $N$ atoms of Hydrogen (mass $m_{H}$), $k_\mathrm{B}$ is the Boltzmann constant, and $T_\mathrm{bound}$ is the average temperature of the post-shock gas. This energy is
    \begin{eqnarray}
      E_\mathrm{bound}&=&3\times 10^{45}\left(\frac{\rho}{10^{-10}\, \mathrm{g}\, \mathrm{cm}^{-3}}\right)\left(\frac{M_\mathrm{BH+BH}}{100\, M_{\odot}}\right)^{3}  \left(\frac{v_k}{200\, \mathrm{km}\, \mathrm{s}^{-1}}\right)^{-4}~\mathrm{erg},
   \end{eqnarray}
 and the resulting average hot spot temperature is
 \begin{equation}
T_\mathrm{bound} \sim 1.8 \times 10^{6} \left( \frac{v_k}{200\, \mathrm{km}\, \mathrm{s}^{-1}}\right)^{2}~\mathrm{K}.
\end{equation}
The resulting UV/optical flare occurs between  $t=[0,t_\mathrm{ram}]$, has an average (low) luminosity $E_\mathrm{bound}/t_\mathrm{ram}$ and a shape given by $\sin^2\left( \pi t/ 2t_\mathrm{ram}\right)$. 

Once the kicked BH leaves behind originally bound gas, the disc gas it passes through is accelerated around the BH, producing an asymmetric low angular momentum Bondi tail inside the stagnation point \citep[e.g.,][]{1999ApJ...513..252O,2019ApJ...884...22A}. This tail both acts as a drag on the BH and accretes onto it. The Bondi--Hoyle--Lyttleton luminosity is $L_\mathrm{BHL}=\eta \dot{M}_\mathrm{BHL}c^{2}$ where $\eta$ is the radiative efficiency and
\begin{equation}
    \dot{M}_\mathrm{BHL}=\frac{4\pi G^{2}M_\mathrm{BH+BH}^{2} \rho}{v_\mathrm{rel}^{3}},
\end{equation}
with $v_\mathrm{rel}=v_{k}+c_{s}$ and $c_{s}$ is the gas sound speed. In principle, hyper-Eddington accretion is allowed by this process. This should cause trapping of emergent radiation, unless a collimated outflow allows radiation to escape. However, if the kicked merger product travels out of the dense disc midplane into a more tenuous disc atmosphere, such signatures may be bright enough to be detected even against bright AGN hosts \citep{2020PhRvL.124y1102G}.    

\subsubsection{FeK$\alpha$ lines (or other EM signatures) as probes of small separation MBH--IMBH binaries} 

The relativistically broadened component of the fluorescent FeK$\alpha$ line (centered around $6.4$--$7~\mathrm{keV}$ source-frame) is believed to be a probe of material in the innermost accretion disc \citep{1997ApJ...477..602N,2000PASP..112.1145F,2003PhR...377..389R}. 
EMRIs and heavy IMRIs pre-merger will disrupt the flow of gas in the innermost disc, yielding flicker in the innermost disc (EMRIs) or carving gaps or a central cavity (MBH--MBH, MBH--IMBH binaries) with minidiscs. 
The resulting re-arrangement of gas yields signatures prior to GW merger events which may be detectable in the broad FeK$\alpha$ with high-throughput X-ray telescopes like Athena. 
A gap-opening secondary IMBH close to the primary MBH will leave an imprint in the broad component of the FeK$\alpha$ emission line, which varies in a unique and predictable manner \citep{2013MNRAS.432.1468M}.

\subsubsection{EMRIs containing a pulsar}

Pulsars are spinning NSs, mainly identified by their radio observations.  
To date there are about 2800 known pulsars in the Galaxy 
  \citep[]{2020MNRAS.493.1063C}, 
  of which about $160$ are found to be associated 
  with globular clusters 
  (see the ATNF pulsar catalogue, \citet{2005AJ....129.1993M}). 
Among these radio pulsars, 
  more than $170$ are MSPs 
  \citep[e.g.,][]{2013MNRAS.434.1387L}, 
  and the majority of them actually reside in globular clusters. 
Although the number of known MSPs is growing, 
  most MSPs are yet to be discovered. 
It has been suggested that   
  the total population number of MSPs in the MW 
  could be about $100,000$ or even more \citep{2013MNRAS.434.1387L}.  
Radio pulsar timing is a relatively mature technique in astronomy,  
  as researchers have accumulated experience in pulsar research 
  over decades. 
The current developments 
  in instrumentation and search techniques 
  will enable us to detect radio pulsars outside the MW 
  \citep[][]{2015aska.confE..40K},  
  and the detection range will be further extended 
  by the time LISA is operating.  
  
EMRIs containing a radio pulsar are a special class of 
  GW sources 
  with a guaranteed EM counterpart, if they are close enough.  
The presence of a MSP provides researchers 
  with several advantages 
  to study these EMRI systems and their associated physics. 
NSs have a small mass range 
   centred around $1.4 \, \mathrm{M}_\odot$. 
Knowing the mass and the spin of one component in the EMRI system
  reduces the parameter space,  
  hence easing the computational demands 
  in the template matching and searching 
  for establishing their GW properties, 
  whereas other EMRI systems would require  
  the determination of the system parameters simultaneously,  
  relying solely on the GW signals. 
The availability of the EM signals  
  with measurements at high precision will give another advantage.  
Both pulsar timing observations 
  and GW experiments 
  can obtain measurements to high precision; 
  the accuracy and precision of pulsar timing  
  is among the highest achievable
  in astrophysical time-domain analysis 
  \citep{2011RvMP...83....1H}. 
Radio pulsar timing and GW experiments 
  employ different analysis techniques. 
As such, 
  the orbital and spin dynamics, as well as the system parameters 
  which they determine, will be independent, 
  thereby giving us a means 
  to understand certain systematic properties 
  in the statistical and data analyses.     

\paragraph*{The contribution of LISA to multimessenger science}
  
The event rate of the potential emission of GWs by extended stars approaching a MBH
will provide us with additional information about tidal disruption events. This combined with
EM detections will deliver much more precise catalogues of disruptions and, within some limits,
information about the star and MBH which is inaccessible via traditional telescopes.

 Regarding TDEs outside galactic nuclei, a WD-TDE system would emit GWs \citep{2018ApJ...856...82H}
 as well as bursts of EM radiation \citep{2019MNRAS.487.2505K}. 
With the additional constraints 
 provided by the GW observations, 
 it would easily resolve the dispute about certain candidate TDE sources, 
 such as AT\,2018cow  
 \citep[][]{2019MNRAS.487.2505K,2019MNRAS.484.1031P}.

 From what we have explained about EM counterparts of light IMRIs, we can conclude that a population of stellar-origin BH+BH in the LISA band that harden into the ground-based GW detector band in AGN discs can yield potentially detectable optical/UV counterparts. IMBH--BH and IMBH--IMBH binary mergers in AGN discs are likely to occur at migration traps in the inner disc \citep{2016ApJ...819L..17B}, so kicked merger products remain bound to the MBH. The kicked BH must splash back down into the AGN disc possibly yielding a repeat flare on half the orbital timescale. An off-center luminous flare should be detectable as an asymmetry in broad optical lines as the broad line region responds to non-central illumination \citep{2019ApJ...884L..50M}.

Double relativistic FeK$\alpha$ lines may be detectable from binary mini-disc emission, allowing us to localize LISA sources well before merger \citep{2012MNRAS.420..860S}. 
The barycenter of a MBH binary will lie outside the event horizon of the primary BH for modest values of mass ratio and binary separation. Analogous to the radial velocity method of planet detection, whereby the wobble of a star indicates the presence of a nearby Jupiter-sized planet, the radial velocity of the primary BH around the binary barycenter can leave a tell-tale oscillation in the broad component of FeK$\alpha$ emission \citep{2015MNRAS.452L...1M}. Such oscillations are detectable by Athena for binaries with mass ratios $q \geq 0.01$, at binary separations of up to $O(10^{2}r_\mathrm{g})$. Both the general-relativistic and Lense--Thirring precession of the periapse of the secondary orbit imprint a detectable modulation on these oscillations \citep{2015MNRAS.452L...1M}.
Athena is likely to detect $O(30)$ FeK$\alpha$ broad lines at sufficient statistical significance in local AGN to carry out tests for ripples and oscillations of such binaries. $O(100)$ AGN may have broad FeK$\alpha$ components that will allow us to search for double components \citep{2020NatAs...4...26M}. Hence, the input from LISA and Athena can be compounded to extract information about the separation of heavy IMRIs.

  LISA detection of EMRIs with a MSP will provide opportunities 
  to investigate a variety of fundamental issues in gravitational physics. 
This is rooted in the extreme mass ratio between 
  the MSP and the BH 
  (the MSP being a test mass) 
  and the ultra-fast rotation of the MSP 
  (the MSP being an extreme gyro and a stable time-keeper).  
More specifically, 
  the spin--spin, spin--orbit, and the spin--curvature interactions  
  \citep{2005PhLA..343....1C,2012GReGr..44..719I,2013MNRAS.430.1940R,2014MNRAS.441..800S}
  between the MSP and the BH 
  will manifest in the spin and orbital dynamics of the MSP 
  \citep{2019MNRAS.485.1053L,2020MNRAS.497.5421K}, 
  which will in turn modify the pulsar timing signals 
  via modification of the pulse period and the pulse arrival time 
  \citep{2019MNRAS.486..360K,2020MNRAS.495..600K}. 
Together 
  with the information extracted  
  from the GWs 
  generated by the EMRI, 
  researchers will be able to investigate 
  how EM waves propagate  
  in a non-vacuum space time \citep{2019MNRAS.484.2411K} 
  or in a slightly perturbed space time,  
  as well as having the opportunity to gain some understanding 
  of certain fundamental issues, such as the gravitational self-force 
  \cite[e.g.,][]{2019RPPh...82a6904B}
  in GW sources.   
  
\subsection{Environmental effects on waveforms}
\label{sec:waveforms}

\noindent
{\bf Coordinators: Alvin Chua, Alejandro Torres-Orjuela and Lorenz Zwick

\noindent
Contributors: Pau Amaro Seoane, Manuel Arca Sedda, Emanuele Berti, Xian Chen, Alvin Chua, Andrea Derdzinski, Kyriakos Destounis, Wen-Biao Han, Kostas Kokkotas, Cole Miller, Scott Noble, Arthur Suvorov, Alejandro Torres-Orjuela and Lorenz Zwick}\\

We know that EMRI/IMRI events can form in a variety of interesting astrophysical environments. Some of these environments may leave detectable imprints on the waveforms measured by LISA (though detecting the imprints may be challenging). We outline a wide variety of environmental effects, including gas-driven effects and many-body effects. Some effects may be degenerate with one another or with deviations from general relativity, even when an effect is detectable. Fortunately, complex dynamics sometimes lend themselves to breaking degeneracies through e.g., Doppler effects. We also use this section to address the possibility of detecting (and extracting astrophysical information from) the EMRI background, and possibly detecting the signatures of chaotic systems. Finally, we specifically consider the degeneracies between environmental effects and PN/self-force effects.

\subsubsection{Gas torques}  
For gas-embedded EMRIs/IMRIs, gas torques can speed up or slow down an inspiral while it is in the LISA band. The magnitude of the torques will scale with the disc density, and the precise value and direction of the torque will depend on the mass of the inspiralling CO and disc properties. 
The effect is several orders of magnitude weaker than GWs in this regime, but even a small dephasing over several thousand cycles may accumulate to a detectable phase shift (up to a few radians), depending on the density of the environment, as shown in analytic work \citep{2011PhRvL.107q1103Y, 2011PhRvD..84b4032K, 2014PhRvD..89j4059B} as well as more recently in 2D hydrodynamical simulations \citep{2019MNRAS.486.2754D, 2020arXiv200511333D}. 
Accretion discs in bright AGN are expected to be thin and dense, but their inner regions are hot and radiation pressure-dominated. Analytical estimates of densities in the inner regions of such discs from simple models predict surface densities varying from $\sim10$--$10^7~\mathrm{g\,cm}^{-2}$ \citep{1973A&A....24..337S,2002apa..book.....F}.
The wide range arises from our uncertainty on how viscosity scales with the (gas or total) pressure. State-of-the art 3D global magnetohydrodynamical disc simulations \citep{2016ApJ...827...10J,2019ApJ...885..144J,2020ApJ...900...25J} suggest that densities are between these values, somewhat closer to the lower end, which could make gas dephasing too small to detect, but this may change as we continue to explore the parameter space of MBH masses. 

Whether or not this effect is \emph{detectable} will depend on the density of the environment as well as the SNR of the source. 
An EMRI embedded in a Shakura--Sunyaev alpha-disc with $\Sigma \sim 10^2~\mathrm{g\,cm}^{-2}$ can accumulate a phase shift up to $\lesssim10^{-2}$ radians within 4 years, whereas if embedded in a beta-disc would dephase over $10^{1}$--$10^{2}$ radians over 4 years  \citep{2020arXiv200511333D}. IMRIs, due to their higher mass, accumulate higher SNR and also feel stronger torques (since they scale with secondary mass), making them ideal events for producing detectable gas signatures. 
    
Whether or not this effect is \emph{distinguishable} from other waveform deviations will depend on how well one can measure the phase shift and how this accumulates as the frequency evolves.
Simulations suggest that the torque can be approximated by simple analytical formulae---torques are within an order of magnitude of the Type I torque derived by \citet{2002ApJ...565.1257T}, although variability in the torque can arise for sufficiently massive secondaries ($q \gtrsim 10^{-3}$). This means that one can estimate how the deviation accumulates with frequency, assuming we know the disc density profile, and if $\ddot{f}$ can be measured from the GW data, degeneracies between the GW waveform distortions due to disc torques versus parameter variations or other effects can be disentangled in principle. 
In practice this may prove difficult given that the small mass ratio of these sources (and expected low eccentricity) means they will chirp slowly, and will more likely appear as near-continuous wave sources within a few year observation.

\subsubsection{Many-body interactions}\label{sec:many-body}
\paragraph{XMRIs and EMRIs}
As explained above, we expect a handful of XMRIs to be present in our own Galactic Centre \citep{2019PhRvD..99l3025A}. Since these systems are so loud, reaching SNRs of up to $\sim 20,000$, in principle we should be able to detect them in nearby galaxies harbouring MBHs in the mass range $10^5$--$10^7\,M_{\odot}$ \citep[i.e.\ nuclei for which the relaxation time is below a Hubble time][]{2020arXiv201103059A,2018LRR....21....4A,2010arXiv1005.4048A}. Farther away XMRI systems with much lower SNR will not be detected. However, XMRIs pose a problem for normal EMRI systems: Since XMRIs live in band for millions of years, and the estimated event rate for an EMRI is between $10^{-5}$--$10^{-6}~\mathrm{yr}^{-1}$, the possibility that an EMRI encounters an XMRI on its way to the MBH is non-negligible. 

\paragraph{EMRIs interacting with a perturbing star}    
Although unlikely, it is not ruled out that a star can be located close to an EMRI inspiraling towards the central MBH. In the work of \cite{2012ApJ...744L..20A} the authors derive the shortest radius from the MBH within which one might expect to have at least one star. Then they run direct-summation $N-$body simulations with
relativistic corrections following the first implementation as presented in \citet{2006MNRAS.371L..45K} and find that periapsis shift along with gravitational-radiation effects induce non-determinism in the evolution of the EMRI. This means that for two identical dynamical setups of an EMRI system with a perturbing star located at a distance of about $\sim 5 a_\mathrm{EMRI}$, with $a_\mathrm{EMRI}$ the semi-major axis of the EMRI, small changes of any dynamical parameter induces a different evolution of the EMRI. The presence of a perturbing star, therefore, can be misinterpreted as a deviation of general relativity, and this should be taken into account in the development of data analysis algorithms.

\paragraph{Binary-EMRIs}
     The formation rate of b-EMRIs for BH+BHs tidally captured by MBHs is equivalent to $(10^{-5}$--$10^{-4})~\mathrm{Gpc^{-3}\,yr^{-1}}$ in the pessimistic case and 0.1 $\mathrm{Gpc^{-3}\,yr^{-1}}$ in the most optimistic one. However, due to the non-negligible lifetime of b-EMRIs, within a spherical volume of $1~\mathrm{Gpc}^3$ (corresponding to a radial distance of about $600~\mathrm{Mpc}$), there are, on average, about $0.02$--$20$ b-EMRIs expected during LISA's mission duration.
    
    The most exciting property of b-EMRIs is that they are multi-band GW sources, i.e., radiate low and high frequency GWs synchronously \citep{2019GReGr..51...38A}. Though the time scale of BH+BH merger is much shorter than the inspiral of binary into the MBH, LISA and the ground-based detectors (LIGO etc.) may observe the b-EMRI at the same time. The high-frequency GWs could be redshifted because they are generated close to a MBH \citep{2019MNRAS.485L.141C}, providing an opportunity of studying the propagation of GWs in the regime of strong gravity.

\subsubsection{Moving sources}
\label{sec:waveformDoppler}

Almost all astrophysical objects are moving relative to us and GWs sources are no exception. The motion of the centre of mass of a source is often related to the properties of its environment, e.g., the orbital motion induced by the interaction with other bodies \citep{2003ApJ...598..419W,2012MNRAS.425..460M,2012ApJ...757...27A,2016ARA&A..54..441N,2020ApJ...891...47A,2017MNRAS.464..946S,2017ApJ...835..165B,2020ApJ...898...25T} and the motion of its host system like the peculiar velocity of galaxies \citep{1984ApJS...55...45Z,1988ARA&A..26..631B,1996ApJ...462...32C,2001MNRAS.328..726S,2016MNRAS.455..386S,2017MNRAS.471.1045C,1996ApJ...457...61G,2014ApJ...792...45R}. Therefore, the detection of velocity can provide valuable and versatile information about the sources environment and its host system.
    
In the case of a constant velocity the Doppler effect changes the observed GW frequency $f_{obs}$ by a factor \citep{2019MNRAS.485L.141C}
    \begin{equation}
        f_{obs} = f(1+z)^{-1},
    \end{equation}
    and its derivative, $\dot{f}_{obs}$, by the same factor squared
    \begin{equation}
        \dot{f}_{obs} = \dot{f}(1+z)^{-2}.
    \end{equation}
    Here, $f$ and $\dot{f}$ being, respectively, the GW frequency and its derivative in the source's rest frame. When only considering the dominant mode of GWs, these shifts lead to a wrong estimation for the sources actual chirp mass $\mathcal{M}$,
    \begin{equation}
        \mathcal{M}_{obs} = \mathcal{M}(1+z),
    \end{equation}
    and actual luminosity distance, $d_\mathrm{L}$,
    \begin{equation}
        d_{obs} = d_\mathrm{L}(1+z),
    \end{equation}
    thus fundamentally affecting our interpretation of the source. An analogous effect appears for the cosmological redshift of GWs. Although the latter one is considered in current GW models and detections, the same is in general not true for the effect of velocity \citep{2019PhRvX...9c1040A}.
    
    If the velocity of the source varies in time, the previous picture changes significantly. The Doppler effect induces a time-dependent phase shift, proportional to the velocity of the source along the line of sight, which can be detected when having accurate models of the evolution of the phase of the source and the velocity profile \citep{2017PhRvD..96f3014I,2017ApJ...834..200M,2019MNRAS.488.5665W}. For a LISA mission of $4$ years probably no accelerated sources could be detected. However, the number of detections could be increased to up to $3$ when conducting joint measurements with a ground-based detector. For a LISA mission of $10$ years up to $40$ accelerated sources could be detected by LISA alone and up to 103 when conducting joint measurements with an earth based detector \citep{2020PhRvD.101f3002T}. Moreover, the aberration of the GWs rays affects the line of sight thus inducing an additional phase shift \citep{2020PhRvD.101h3028T}. The magnitude of the aberrational phase shift is of the same order as the Doppler phase shift but proportional to the components of the velocity perpendicular to the line-of-sight. Therefore, considering the total phase shift, i.e.\ Doppler effect plus aberrational phase shift, the SNR required for the detection of the acceleration could be reduced by a factor of up to $1.8$, thus allowing the detection of up to $5.8$ times more sources \citep{2020PhRvD.101h3028T}.

\subsubsection{Dark matter as an environmental effect}
The density profile of dark matter halos has a cusp at the centre of galaxies because of the large potential well there \citep{2012PDU.....1...50K}. If a MBH resides at the centre of the galaxy, the strong gravity could lead to a significant increase of density in the central region and create a spike, which enhances the dark matter annihilation rate \citep{1999PhRvL..83.1719G,2013PhRvD..88f3522S}. Similarly, IMBHs may have a smaller dark matter spike \citep{2005PhRvL..95a1301Z,2005PhRvD..72j3517B}. 
The gravitational potential of the dark matter could impact the evolution of an EMRI/IMRI, particularly where enhancement of the density occurs due to spikes or a superradiant instability, leading to a detectable signature \citep{2013PhRvL.110v1101E,2018PhRvD..97f4003Y,2020PhRvD.102j3022H}.
However, dynamical events such as mergers of host galaxies can weaken the dark matter cusp \citep{2001PhRvD..64d3504U,2002PhRvL..88s1301M,2004PhRvL..92t1304M,2005PhRvD..72j3502B}, which makes its presence harder to detect.

\subsubsection{Astrophysical chaos} 

Owing to the huge mass disparity for the objects involved in an EMRI, the dynamics of the companion body can be modelled as a point particle traversing the gravitational field of the (super-massive) primary to high accuracy. Characteristics of GWs emitted during the inspiral are therefore dominated by the particulars of the metric geometry of the primary. Fundamental symmetries, or the absence thereof, associated with the spacetime geometry can therefore be probed by LISA. For a primary which is both stationary and axisymmetric---properties expected of astrophysically stable BHs---the energy and one component of the angular momentum of a companion are both constants of motion with respect to the orbital dynamics. The companion's Hamiltonian, $H \sim g_{\mu \nu} p^{\mu} p^{\nu}$ for momentum $\boldsymbol{p}$, provides a third constant of motion. In \emph{four} spacetime dimensions, however, having only \emph{three} conserved quantities implies that the equations of motion are not Liouville integrable \citep{2002ocda.book.....C}. In this context, a non-integrable system exhibits chaotic orbital phenomena, as is familiar from the three-body problem in (post-)Newtonian gravity \citep{2014PhRvD..89l4034H}. The Kerr spacetime, which uniquely represents stable BHs in general relativity, also however admits a rank-two Killing tensor, which provides a fourth constant of motion in the form of the Carter constant \citep{1968PhRv..174.1559C}. This implies the absence of astrophysical chaos in EMRIs within general relativity, at least for companions which are not themselves spinning rapidly \citep{2004PhRvD..70f4036K,2014PhRvD..90j4019L,2020PhRvD.102b4041P,2020PhRvD.101b4037Z}.

\begin{figure}[h]
\includegraphics[scale=0.55]{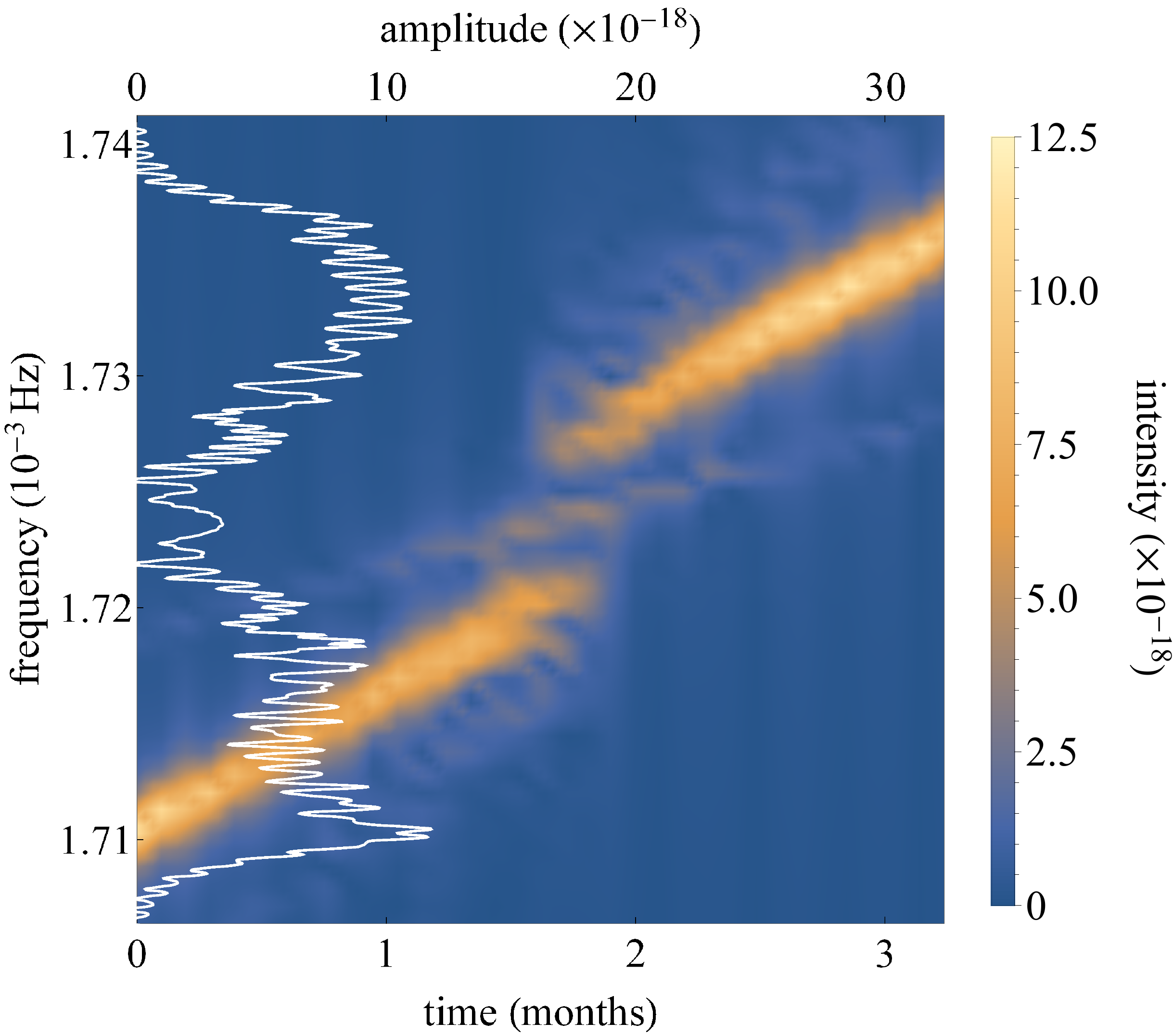}
\caption{Periodogram of the fundamental frequency of a gravitational waveform associated with an inspiral on a particular non-Kerr spacetime. A sudden jump in the frequency evolution appears when the inspiralling object crosses a Birkhoff island, which is associated with a valley in the amplitude of the signal's frequencies (white line).}
\label{birkhoff-spectrum}
\end{figure}

If, however, high-energy corrections to the field equations present themselves in nature, the particulars of gravitational collapse \citep[e.g.,][]{2012JCAP...04..021C} and accretion \citep[e.g.,][]{2010CQGra..27j5010H} may be such that a non-Kerr object resides within  galactic centres or elsewhere. There are many ways in which a hypothetical departure from a Kerr description may manifest within the spacetime metric, such as those described in \cite{2013PhRvD..88d4002J}. One interesting possibility is that the Carter symmetry is broken, thereby giving rise to a non-Kerr object, as opposed to a deformed-Kerr body which still forbids chaotic phenomena even if the spacetime is not exactly Kerr \citep{2018CQGra..35r5014P,PhysRevD.102.064041}.

In general, sections of the inspiral that behave as bound orbits can be characterised by both radial $(\omega_{r})$ and angular $(\omega_{\theta})$ libration frequencies, which describe the rate of transition from the periastron to the apastron of the orbit and longitudinal oscillations about the equatorial plane, respectively \citep{2002ocda.book.....C}. On the other hand, classical results from dynamical systems theory infer that small islands of stability (Birkhoff islands) form around periodic orbits in the phase space of non-integrable dynamical systems \citep{1978mmcm.book.....A}. When an inspiralling orbit crosses an island, the ratio $\omega_{r} / \omega_{\theta}$, which defines what is called the \emph{rotation curve}, remains constant, while otherwise it behaves monotonically as a function of radius. Absence of islands therefore implies an everywhere monotonic rotation curve, while the dynamics display transient plateau features for non-Kerr spacetimes when orbits intersect with an island \citep{2009PhRvL.103k1101A}.

Several studies have shown that these transient plateaus also introduce features into the gravitational waveforms which are, in principle, discernible from deformed-Kerr features \citep{2009PhRvL.103k1101A,2010PhRvD..81l4005L,2011IJBC...21.2261C,2018CQGra..35p5010C,PhysRevD.102.064041,2021hgwa.bookE..42L}.  Figure~\ref{birkhoff-spectrum} shows the fundamental frequency evolution of a gravitational waveform, associated with a particular non-Kerr spacetime (see \cite{PhysRevD.102.064041,2021PhRvL.126n1102D,2021PhRvD.104f4023D} for details). However, non-integrable perturbations in the \cite{1978mmcm.book.....A} sense of the particle Hamiltonian may also arise due to environmental effects within general relativity \citep{2022PhRvD.105f1501C}. For instance, $N>2$-body interactions (as in the Newtonian case; see also Sec.~\ref{sec:many-body}) \citep{2007PhRvD..75f4026B,2012ApJ...744L..20A} or significant internal spins in the companion \citep{2004PhRvD..70f4036K,2014PhRvD..90j4019L} can induce chaos. 
   
\subsubsection{Environment versus PN/self-force degeneracies} 

The secular evolution of EMRI orbital elements is intimately connected with the phase and the shape of the GWs that will be measured by LISA. In vacuum this evolution is fully described by general relativity, and it can in principle be computed to arbitrary precision with approximation schemes such as perturbation theory (an expansion in the small mass ratio $q$ of the binary) or the PN expansion (an expansion in the small parameter $v/c$, where $v$ is the orbital velocity and $c$ is the speed of light). At leading order in perturbation theory, the system can be described as a point-like particle moving in a geodesic orbit around the large BH. The energy flux can be computed either numerically or analytically to varying degrees of accuracy. For example, \citet{2020arXiv200810622M} computed the energy radiated from eccentric orbits around nonrotating BHs up to 19PN order. 
\citet{2015PTEP.2015c3E01F} computed the energy flux from a particle in \emph{circular} orbit around a rotating BH up to 11PN order. \citet{2015PTEP.2015g3E03S} computed the expansion for \emph{eccentric} orbits around rotating BHs up
to 4PN, and up to order $e^6$ in a small-eccentricity expansion. 
The PN expansion is an asymptotic series, and it is known to converge quite slowly for EMRIs \citep{2008PhRvD..77l4006Y,2011PhRvD..84b4029Z}.
At higher orders in $q$, interaction of the particle with its own gravitational perturbation gives rise to  gravitational self-force, which drives the radiative evolution of the orbit, and whose effects
can be accounted for order by order in $q$
\citep{2019RPPh...82a6904B}.

EMRIs are however by necessity embedded in astrophysical environments, and as such it is likely that their secular evolution will differ from the pure vacuum case.
Of all the environmental factors, gravitational torques from accretion flows are likely to be most the significant \citep{2019arXiv190905870C}.  Obtaining realistic estimates for the influence of accreting gas on the orbital evolution and phase of the binary is difficult because accretion dynamics is a largely unexplored 3D magnetohydrodynamics problem over a large dynamic range.  The best theoretical prediction for the impact of gas dynamics on IMRI/EMRI phase errors are from 2D viscous hydrodynamics  \citep{2019MNRAS.486.2754D,2020arXiv200511333D}.  Phase errors grow with the surface density of the accretion disc.  The sign in the tidal torque, i.e.\ whether the separation increases or decreases, depends on the mass ratio, the strength of the viscosity (the $\alpha$ parameter), and the rate of inspiral \citep{2020arXiv200511333D}. Some parameters even experience stochastic variability in the sign of the tidal torque for EMRIs; such variability would be in stark contrast with secular phase errors coming from truncating the PN expansion. Phase errors due to non-stochastic effects, however, will likely be comparable to PN errors for some disc configurations and orbital separations  \citep{2014PhRvD..89j4059B,2019arXiv190905870C,2020arXiv200703700A}. Understanding whether these situations are likely requires a combination of results from 2D viscous simulations \citep[cf.][]{2019MNRAS.486.2754D,2020arXiv200511333D} with EM surveys of AGN discs, population synthesis modelling of AGN/binary discs  \citep[e.g.,][]{2019ApJ...879..110K}, and estimates of the distribution of the observed binary parameters for LISA. Since all EMRI/IMRI simulations have used 2D Newtonian viscous hydrodynamics, it will be interesting to see how these results change when using more realistic 3D general-relativistic magnetohydrodynamical simulations. Unfortunately,  performing a series of simulations with $O(2000)$ orbits, which seems to be required to reach a steady-state in 2D simulations, is computationally prohibitive at present.
    
Another obvious effect that can spoil the vacuum evolution of an EMRI is the influence of a third gravitational body. In the case of a hierarchical triple, two effects can take place. First, the influence of the perturber can produce a shift in the binding energy of the inner binary.  \citet{2014CQGra..31x4001W} showed that this shift is constant even if the inner binary undergoes perihelion advance. Second, if the perturber is sufficiently inclined it can induce von Zeipel--Kozai--Lidov oscillations \citep{1910AN....183..345V,1962AJ.....67..591K,1962P&SS....9..719L} in the inner binary. This can in principle cause an enhancement in the eccentricity of the inner binary. In the case of EMRIs, however, we can expect gravitational perturbations in a hierarchical triple to be very weak. The pull of the third body essentially acts as a tidal force between the components of the inner binary. Therefore, it scales as $\sim a R^{-3}$, where $a$ is the typical separation of the inner binary and $R$ is the distance of the perturber  to its centre of mass. For EMRIs, $a$ will generally be very small ($10$ to $10^3$ Schwarzschild radii) and the third power of $R$ will strongly suppress tidal forces. As an example, one can compare the von Zeipel--Kozai--Lidov oscillation timescale $t_{\rm{KL}}$ with the gravitational radiation reaction timescale $t_{\rm{GW}}$:
\begin{equation}
t_{\rm{GW}} \sim t_{\rm{KL}} = 2 \pi \frac{\sqrt{G M}}{G m_3}\frac{R^3}{a^{3/2}}.
\end{equation}
By using Peters' formula \citep{1964PhRv..136.1224P} one can find the typical orbital separation at which GW emission and von Zeipel--Kozai--Lidov oscillations change the orbital elements on the same timescale. For circular orbits, this yields
\begin{equation}
R_{\rm{KL/GW}} \approx a \left( \frac{ m_3}{q M} \right)^{1/3}   \left(\frac{a}{r_{\rm S}} \right)^{5/6},
\end{equation}
where $q$ is the mass ratio and $M$ the total mass of the inner binary, while $r_{\rm S}$ is the Schwarzschild radius of the central MBH and $m_3$ the mass of the perturber. For EMRIs, it is clear that the two timescales can be comparable only for very massive or very close perturbers. Nonetheless, recent works have used this result to compute event rates for binaries that are affected by the Kozai-Lidov eccentricity enhancement \citep{2019arXiv190208604R,2020ApJ...901..125D}, and \cite{2011PhRvD..83d4030Y} showed that a sufficiently large perturber ($\sim 10^6 \, M_{\odot}$) at sub-parsec distances can dephase the GW signal of an EMRI by a detectable amount.

The astrophysical community has mostly focused on understanding and modelling the influence of the environment on vacuum sources. Claims of detectability for any given effect are therefore often based on simple phenomenological criteria rather than complete signal injections and parameter inference methods. The simplest and most ubiquitous criterion is based on the concept of the SNR of a deviation from a vacuum waveform.
While the SNR of a GW event can be estimated by the following formula:
\begin{align}
    \rm{SNR}=\sqrt{2 \cdot 4\int^{{f}_{\text{max}}}_{{f}_{\text{min}}}{\rm d}{f}^\prime\frac{h_{\rm{c}}^2({f}^\prime)}{S_{\rm{t}}({f}^\prime){f}^{\prime2}}}.
\end{align}
The SNR of a deviation can be found by replacing $h_{\rm c}$ with:
\begin{align}
    h_{\rm c}\exp\left(i \phi \right) \to h_{\rm c}\exp\left(i \phi \right) - h_{\rm c}'\exp\left(i \phi' \right),
\end{align}
where the primes denote the waveform of a source that is modified through the action of some environmental effect.
For GW sources in general, an SNR value of $\sim 8$ is chosen as a threshold required in order to claim detectability. The same is assumed to be true for a given deviation, $\delta h_{\rm c}$ which is deemed detectable whenever $\Delta$SNR > 20 (at least for EMRIs). While such criteria can serve as a first order approximation, they do not take into account many of the complications that will plague data analysis procedures required to extract signal from LISA's datastream.
Effects such as degeneracies, the lack of appropriate waveforms and subtleties of Bayesian analysis in parameter spaces with many dimensions are only a few of the many considerations that should in principle be taken into account when considering the detectability of environmental effects.
Degeneracies are especially important, since the influence of many environmental effects might be misinterpreted as sources with different intrinsic parameters.
As a simple example, consider the evolution of a BH binary in gas. The primary effect of gas would be to change the rate at which the binary chirps, i.e. $\dot{f}$.
However, in a blind waveform template search, the same variation in $\dot{f}$ could likely be accounted for by a slightly modified mass of the system, inducing a bias in parameter estimation. With sufficient SNR, it will be likely possible to break such degeneracies from leading order parameters such as chirp mass and distance, using e.g. the information contained in higher derivatives of the frequency, $\ddot{f}$. Further work should confirm that the same can be said for subtler GR vacuum effects, mainly the spin components and the eccentricity of the source.
A possible way forward would be an increased collaboration between the data analysis and astrophysics working groups: the former can provide more sophisticated phenomenological detectability criteria, while the latter could direct the data analysis efforts towards those effects that are expected to be relevant.

\paragraph*{The contribution of LISA to our understanding of the host environment}

The variety of gas torques suggests that, if chirping, EMRIs in gaseous environments will exhibit characteristic signatures that may allow us to probe the inner regions of AGN discs. If a phase shift is detected and confirmed to be of gas origin, its magnitude and evolution will be a direct consequence of interaction with local gas properties around the BHs.  However, detecting such deviations will require improving models of environmental effects such that we can include them properly in parameter estimation. A critical step is to assess for which regions of parameter space these effects will be unique or degenerate with system parameters. In the latter case, neglecting them may induce biases in parameter estimation. We expect such signatures to arise in only a subset of systems, whereas deviations from general relativity would arise in all EMRIs (depending on the observed frequency). Gas effects can also lead to additional waveform implications: for example prograde, disc-embedded sources will likely have low eccentricity and some degree of spin/inclination alignment with the central MBH.  If deep within the potential well of the MBH, these effects may be complemented by a phase-shifting from the Doppler effect (Sec.~\ref{sec:waveformDoppler}). 

Elaborating on what we have presented here but also in the previous sections, the presence of an XMRI can alter the evolution of an EMRI on its way to cross the event horizon of the MBH. This can lead to extreme situations in which the orbital dynamics of the EMRI is not just affected by the presence of the XMRI, but to the point of scattering off the EMRI from its inspiraling orbit towards the MBH (Vretinaris \& Amaro Seoane, in prep.). Since our Galactic Centre and MBH might be envisaged as a typical target for LISA, this means that many, if not all nuclei in the LISA observational volume are prone to this problem.
 
Our theory of how stars distribute around MBHs is more than four decades old and seems to be robust. However, at distances very close to the MBH, the power-law distribution of the stellar system (the ``cusp'') is ill-defined, because the number density drops significantly. If a star happened to be close to an EMRI, in principle one could reverse-engineer the modulation induced in the waveform, in particular in the phase, to recover information about such perturbing stars from a region which is too obscured to be accessible to EM telescopes.
 
The merger in a b-EMRI system induces a kick to the BH remnant \citep{2010RvMP...82.3069C}. This kick causes a glitch in the EMRI waveform, which, through a careful analysis, is discernible in the data stream \citep{2019MNRAS.485L..29H}. The b-EMRI can hence accurately weigh the mass loss due to the BH+BH merger, and offer an opportunity to test general-relativistic effects (in particular the dispersion relation of GWs and the weak equivalence principle).  
  
Since a significant fraction of EMRIs can be hosted in galaxies which move relative to us at very high speeds, the imprint of the aberration and beaming effects on the waveform can be crucial.  A constant drift of the centre of mass of a source also can affect the higher multipoles of the gravitational waveform. This, in turn, affects the frequency and amplitude of the wave as seen by a distant observer \citep{2008PhRvD..78d4024G,2020arXiv201015856T}. Therefore, higher modes can be used to break the aforementioned degeneracy between a constant velocity and the mass/distance of the source. Considering the change of the modes for EMRIs, LISA should be able to detect constant velocities of just $1000~\kms$ for an SNR of around $70$ \citep{2020arXiv201015842A}. Moreover, as mentioned in that work, we could use this information to obtain a detailed map of the relative speed distribution of galactic clusters out to distances unaccessible to EM observations.
  
Another interesting possibility is that dark matter minispikes could impact the gravitational waveform, inducing dephasings that could be detected by LISA \citep{2013PhRvL.110v1101E,2018PhRvD..97f4003Y,2020PhRvD.102j3022H,2020PhRvD.102h3006K}. 
Furthermore, the existence of dark matter halos around IMBHs could accelerate the formation of IMRIs \citep{2018PhRvD..97f4003Y}.
Therefore, the event rates of IMRIs may be much higher than previous estimates, which did not include a DM halo \citep{2019ApJ...874...34Y}.

As for astrophysical chaos, we can deduce that the detectability of the plateau scales with the magnitude of the non-Kerr parameters and the mass ratio of the EMRI: the ``islands'' presented above become larger for greater non-Kerr parameters, and the system requires more time to cross an island for greater mass disparity \citep{2010PhRvD..81l4005L}. Crossing into an island leads to a period of frequency modulation, which can, in principle, be detected by LISA \citep{PhysRevD.102.064041}. More careful data analysis (using, e.g., a Fisher matrix study) is therefore required to determine whether chaotic phenomena have a non-general-relativistic origin, since the frequency jumps described in Fig.~\ref{birkhoff-spectrum} may be mimicked by environmental effects.

The research carried out so far demonstrates the difficulty of distinguishing gas-driven environmental effects from poorly modelled GR effects. Work to date has explored a relatively narrow range of parameter space for possible environmental effects, and more work should be done to understand the dominant effects, even within currently available 2D Newtonian hydrodynamical models. As computational resources increase, closer to launch, it would also be helpful to expand theoretical efforts to include at least some 3D general relativistic magnetohydrodynamical simulations. Though challenging, further work must also be done to examine the degeneracies between, e.g., additional PN terms and gas-driven departures from GR. The most important efforts are finding effects that share the same frequency or mass dependence---for effects that do not share dependencies, we can hope to distinguish the source from the GW observations themselves. Gas-driven sources should be a subset of all sources (and should have eccentricities and inclinations which help to distinguish them), while higher-order corrections should apply to every system (though the corrections may be mass- or frequency-dependent). Assuming we overcome the challenges listed here, disentangling various environmental signatures from GWs will give us access to unique measurements of MBH environments purely through GWs. Such measurements are inaccessible via EM observations. Characteristic deviations (or even a lack thereof) will provide constraints on gas densities, dark matter profiles, or the presence of external perturbers.

\subsection{EMRI background} 
{\bf Coordinators: Pau Amaro Seoane, Andrea Derdzinski

\noindent
Contributors: Pau Amaro Seoane, Andrea Derdzinski, Giuseppe Lodato, Martina Toscani }\\

While the majority of this Chapter describes resolvable sources (as they are certainly the most interesting for guaranteed science),  most EMRIs/IMRIs throughout the Universe will not be individually detectable, particularly if they are too distant, 
at earlier stages of their inspiral,
or their GWs are too weak (which is moreso an issue for inspirals of WDs or NSs). 
The combined signal from the population of faint, unresolved sources will constitute a stochastic background.

The EMRI background may lie well below the LISA sensitivity or exceed it, contributing an additional confusion noise. Its amplitude scales with the EMRI rate, although not necessarily linearly \citep{2004PhRvD..70l2002B}, and its precise spectral shape will depend on the efficiency of various formation channels over cosmic time.
 Seminal predictions find that the background signal will only become comparable to the LISA noise if the EMRI rate is substantial: e.g., if the detection rate is as high as $\mathcal{O}(10^2)$ detections per year, the corresponding background may increase the LISA noise by a factor of nearly $\sim2$ \citep{2004PhRvD..70l2002B}. 
More recent estimates based on EMRI catalogues by \citet{2017PhRvD..95j3012B} 
use an updated version of the LISA sensitivity curve and  
find that, for a range of EMRI rates, the background may add considerable noise (attaining an SNR of a few to few-hundred) within the LISA sensitivity bucket around $f\sim3~\mathrm{mHz}$  \citep{2020arXiv200714403B}. 
Higher levels of confusion noise may compromise our ability to detect faint sources that fall into this frequency range, such as high redshift, low mass MBH mergers--although one could argue that this would be compensated for by the generous resolvable EMRI rate.

A detectable background may provide additional information on the cosmological EMRI population and the efficiency of various formation mechanisms, if there exists robust differences in the spectrum between formation channels. 
The trick to detecting (and then hopefully characterizing) a background signal is to distinguish it from the instrumental noise as well as other confusion sources 
(see \citealt{2017LRR....20....2R} for a comprehensive review). For LISA, the main contributor to confusion noise is expected to arise from Galactic binaries: while many will be individually resolvable, the rest will form a confusion foreground \citep[e.g.,][ but see Chapter 1]{2012ApJ...758..131N} that may overwhelm any extragalactic stochastic signal. Fortunately,
LISA's orbital motion around the Sun introduces an annual modulation in the anisotropic galactic foreground, and this makes it possible to distinguish the astrophysical signal from the instrument noise. With a prior understanding of the LISA noise, knowing the distinct spectral shape of an astrophysical foreground further helps us separate the two, so there is hope for detecting an underlying stochastic background \citep{2014PhRvD..89b2001A}. Such techniques were successfully applied in the LISA mock data challenge \citep{2008CQGra..25r4019R}.
At the moment, predictions for the EMRI background signal suffer from the same uncertainties as detection rates (see Table~\ref{tab.RatesSNR}), but these can be improved as we increase our understanding of formation mechanisms. Improving waveform modelling or finding other methods of accurate signal extraction will also be critical if we hope to detect an underlying signal. 

An important step in this analysis would be to distinguish the EMRI background from other possible background sources.  In addition to EMRIs, there may be characteristic background signals from 
extra-galactic WD, NS, or BH binaries  \citep[e.g.,][]{2019ApJ...871...97C,2018MNRAS.481.4775D}, TDEs (discussed below), 
phase transitions in the early universe
\citep{2000gr.qc.....8027M,2012JCAP...11..006G,2012JCAP...10..024L},  
or cosmic strings \citep{2007PhRvL..98k1101S}. 
If the spectrum is sufficiently constrained, then it is likely that the origin of the signal, whether from a large number of unresolved EMRIs or other extragalactic sources, can be determined \citep{2020GReGr..52...81B}. 
 
\subsubsection{TDE background}
\label{sec:TDE_background}
A particular type of EMRI background is the one generated by the unresolvale GW signal from the cosmic TDE population. The calculation of such background has been performed by \citet{2020MNRAS.498..507T} for both main-sequence stars being disrupted by MBHs in galactic nuclei and WDs being disrupted by IMBHs in globular clusters. The signal has a characteristic spectral shape $h_\mathrm{c} \propto f^{-1/2}$, due to the specific impulsive nature of these events. The predicted amplitude of the background is generally low, with WDs on IMBHs providing typical strains of $\approx 10^{-23}$--$10^{-21}$ and main-sequence stars on MBHs providing $\approx 10^{-22}$. 
\paragraph*{The contribution of LISA to our understanding of backgrounds of inspirals}

In summary, LISA will have the capability to detect a stochastic background signal, once the galactic foreground is subtracted.  
This measurement will improve throughout the mission lifetime as we constrain the instrument noise and resolve individual sources \citep{2014PhRvD..89b2001A}. 
If EMRIs provide the dominant contribution within some frequency range---e.g., around $3~\mathrm{mHz}$, as predicted by \citet{2020arXiv200714403B}---a measurement of the background spectrum can serve as an additional measurement of dynamics in galactic nuclei. For example, the amplitude and spectrum of the background are related to the number of EMRIs that are either at earlier inspiral stages or at higher redshifts, as well as the MBH mass function.  If the background is above the LISA sensitivity and not well-characterised, it will contribute to the noise budget, possibly complicating the detection of other weak signals. To avoid this, it is important that we improve our predictions on the EMRI rate.
Further work is also needed to constrain the expected spectrum of various background signals, in order to determine which will be dominant, distinguishable, and removable from the LISA data.
If the EMRI rate is low enough such that the background falls below the LISA sensitivity, then it becomes possible to detect other stochastic signals, such as those predicted from extragalactic binaries or signatures from the early Universe.

As for the background from TDEs, its detection could provide interesting insights both on the distribution of quiescent MBHs (for main sequence stars tidally disrupted) and on the occupation fraction of IMBHs in globular clusters (for TDEs of WDs), up to redshift $\approx 3$. Yet, this detection seems to be very difficult. Indeed, the background produced by WD TDEs will lie in a high part of the frequency window (deciHertz to a few Hertz), where LISA will be less sensitive (yet, more sensitive interforemeters in this frequency interval are planned for the future). Instead, the background from MS TDEs is expected at lower frequency ($10^{-4}-10^{-2}$ Hz), but will be still below the threshold LISA sensitivity. Hence, this detection seems unlikely (although some background signal below the strain sensitivity might still be visible in some cases, \citet[see][ for more details] {2016PhRvL.116w1102S}).

\subsection{Conclusions}
{\bf Contributor: Saavik Ford}\\
\renewcommand\theadalign{bc}
\renewcommand\theadfont{\bfseries}
\renewcommand\theadgape{\Gape[4pt]}
\renewcommand\cellgape{\Gape[4pt]}
\begin{table}
\begin{tabular}{|
>{\columncolor[HTML]{ECF4FF}}c |c|c|c|}
\hline
\cellcolor[HTML]{C0C0C0}Inspiral type & \cellcolor[HTML]{C0C0C0}Rate ($\mathrm{yr}^{-1}$)                                       & \cellcolor[HTML]{C0C0C0}SNR & \cellcolor[HTML]{C0C0C0}References \\ \hline
EMRI                                  & $10$--$10^3$                                                                     & $\sim 100$                  &    \footnotesize\cite{2018LRR....21....4A,2020arXiv201103059A},\cite{2017PhRvD..95j3012B}\normalsize                                            \\ \hline
light IMRI                            & $6$--$60$                                                                                & $10$--$10^3$                 &   \footnotesize\cite{2020arXiv200713746A,2020arXiv201103059A}\normalsize                                             \\ \hline
heavy IMRI                            &  2-20                                                                                        & $10$--$100$                  &    \footnotesize\makecell{\cite{2005ApJ...618..426M,2018MNRAS.477.4423A} \\ \cite{2019MNRAS.483..152A}}\normalsize                                            \\ \hline
XMRI                                  & \begin{tabular}[c]{@{}c@{}}$\sim\textrm{few~tens}$\\ (at any given moment)\end{tabular} & $10$--$10^4$                   &    \footnotesize\cite{2019PhRvD..99l3025A,2020arXiv201103059A}\normalsize                                            \\ \hline
\end{tabular}
\label{tab.RatesSNR}
\caption{Rates and SNRs for inspirals. Note that the rates for XMRIs are at any given moment in the MW and, possibly, nearby galaxies (see reference).}
\end{table}

To summarise: the basic physics of EMRI mergers has been known for a long
time. We can expect to find EMRIs in NSCs harboring an MBH, and
can predict the dynamics of their formation and evolution using
relaxation theory. The waveform modelling for EMRIs is also reasonably
advanced, such that the 
path to detectability of such signals is understood (if challenging). However, there are substantial uncertainties in the
astrophysical parameters that govern the \emph{rates} of EMRI production,
notably the spins of MBHs and, most critically, the radial mass
distribution of NSCs. These astrophysical unknowns
will change the ratio of plunging extreme mass ratio mergers to bona fide EMRIs -- in the
case of plunging mergers, an extreme mass ratio merger does occur,
but the interactions of the merging object with the inner edge of
the stellar cusp
alters the trajectory of the low mass object after only a few orbits, and produce a rapid
merger. Since LISA detection of EMRIs will depend on the buildup of a
sufficient SNR over many orbits, plunges are rendered undetectable.

However, assuming the inner edge of an NSC is typically sufficiently far from the MBH, and if MBH spins are typically non-zero, EMRIs can occur at a high enough rate that LISA would detect one or more over the lifetime of the mission. If an EMRI is detected, we will immediately obtain a wide variety of both fundamental physical and astrophysical information (including, implicitly, information about the mass distribution in NSCs and MBH spins). Because of the many-orbit nature of an EMRI, such events can provide a detailed map of the gravitational field in the vicintiy of the MBH, yielding exquisite measurements of the mass and spin of that MBH, and providing an opportunity to probe fundamental physics by testing for subtle departures from GR.

Given the current uncertainties on EMRI rates, it is most useful to proceed along multiple fronts:
\begin{itemize}

\item Theoretical work to understand higher-order dynamical effects which may preserve more EMRIs for sufficient cycles to allow detection by LISA (i.e. preventing plunges)
\item Observational work constraining the inner edge of the NSC cusp in nuclei other than the MW (M32 would be notably useful)
\item Theoretical and observational work constraining the binary fraction in typical NSCs (enabling better estimates of the EMRI rate due to binary tidal separation)
\item Theoretical work to develop non-standard EMRI channels, especially AGN and SNe routes, to constrain rates and parameter distributions, and permit reverse engineering of astrophysical parameters

\end{itemize}

In addition, there is groundwork to be done on the waveform, data analysis and coordination front:

\begin{itemize}

\item Self-force calculations to second order in mass ratio for generic orbits in a Kerr metric to enable high precision waveform calculations
\item Further data analysis work with updated waveforms to improve EMRI extraction from the LISA datastream
\item Coordination of data analysis with radio and ground-based GW observatories in case of pulsar or b-EMRI detection

\end{itemize}

IMRIs provide still more exciting science opportunities, but correspondingly more
challenging uncertainties. Due to their larger mass ratio, IMRIs cannot
be treated using the same theoretical mechanisms as EMRIs (i.e.\ as
small perturbations), yet they are also not sufficiently large to be
treated using the mechanisms that apply to near-equal-mass
binaries. This reality provokes difficulties in several directions -- we
cannot generally apply the same relaxation theory strategies to predict IMRI
formation dynamics, nor can we readily produce IMRI waveform models
using numerical relativity
without changing computational strategy. In addition, there are at
least 2 types of IMRIs to be considered: 1) light IMRIs, where the
more massive partner is an IMBH and the less massive partner is a
stellar mass BH; and 2) heavy IMRIs, where the more massive partner
is an MBH and the less massive partner is an IMBH. There are multiple
formation channels for each, and thus large astrophysical uncertainties in
predicting their rates.
One substantial uncertainty has recently been
removed: with the announcement of GW190521, we are certain that
low-mass IMBHs do exist. Though their formation environment remains
uncertain, work thus far points to some kind of dynamical origin,
encouraging expectations that there may be environments conducive to the formation of at least some light IMRIs.

Broadly, the channels for light IMRIs include: 1) formation in globular clusters (assuming the presence of an IMBH in the cluster); 2) formation in dwarf galaxies (assuming the presence of an IMBH in the galaxy); 3) formation in an AGN disc (assuming the formation of an IMBH at a disc migration trap). In each of these cases, the first uncertainty is the presence of an IMBH in the relevant environment. There are no universally accepted detections of IMBH in globular clusters. While there are some IMBH known in dwarf galaxies, their occupation fraction is not well-measured and depends on the still unknown physics of BH seed formation at high redshift. IMBH formation in AGN discs is expected to be nearly universal if such discs contain a migration trap and are sufficiently long-lived. Unfortunately neither condition is sufficiently theoretically or observationally well-constrained to make a confident statement on the rate of IMBH formation in AGN discs.

Sadly that is not the end of our uncertainties for the formation of these systems -- the dynamics in each case are difficult to model, as noted above, and as with EMRIs, the systems might produce a beautiful, detectable inspiral, or a rapid plunge. Among the important open questions are the observational presence or absence of IMBH in each formation environment, and observations and theoretical investigation of the (extremely different) dynamical environment around each IMBH.

For heavy IMRIs, we have a similar diversity of formation channels: 1) globular clusters containing an IMBH infalling into a galactic nucleus containing a MBH; 2) dwarf galaxies infalling into a galactic nucleus containing a MBH; 3) AGN-produced IMBH falling into their host MBH (likely in the post-AGN phase). Globular clusters, being denser than dwarf galaxies, will deposit their IMBHs in galactic nuclei more rapidly than dwarf galaxies, and in general, are expected to dominate the rate in this formation channel. However, if globulars do not contain IMBHs, we should consider dwarf galaxy mergers quite carefully, since at least some dwarfs are known to harbor IMBHs. For low-mass galaxy groups, dwarf mergers will be the most common type of merger and do lead to the formation of bound IMBH-MBH systems in less than a Hubble time. However, theoretical  astrophysical rate calculations for resulting IMRIs, over the volume probed by LISA, remain an important open question. Observations of dwarf galaxies and their evolution over cosmic time, as well as observations that inform the occupation fraction of globular clusters and dwarf galaxies are, consequently, critical unknowns. AGN production of IMBHs provide a potentially extremely high rate of IMRIs, given the formation location and expected GW inspiral time; such IMRI systems will not likely be disrupted by interactions with the NSC. However, as with the light IMRI channel, substantial uncertainties in the structure of AGN discs and their lifetimes lead to orders-of-magnitude uncertainties in the rate estimates for this channel.

IMRIs have received less attention in the literature to date, and consequently the tasks to complete before LISA launches tend to be larger. These include:
\begin{itemize}

\item Determining the occupation fraction of globular clusters and dwarf galaxies
\item Determining the contribution to the heavy IMRI rate from low mass galactic environments
\item Determining the formation environment of GW190521-like sources
\item IMRI waveform modelling and extraction (may require expensive numerical relativity modelling)
\item Modelling of IMRI formation (and rates) in both gas-poor and gas-rich environments

\end{itemize}

Finally, we consider the open questions related to the relatively new
class of unequal mass ratio inspirals, XMRIs. Here, the physics of
their formation and evolution is similar to that of EMRIs, but they
are labelled eXtreme due to the very small mass of the secondary -- for
LISA frequencies, the secondary would typically be a brown
dwarf. Similar to the situation for EMRIs, the uncertainties largely
relate to our astrophysical ignorance; however, we are able to limit
the locations we must investigate. Due to their low mass and
consequently small strains, XMRIs will only be detectable within roughly our Local
Group, meaning either from the MW or M31. In addition to
the relevant questions for EMRIs (especially NSC radial mass
distribution), we must understand the mass function of those NSCs.
How many brown dwarfs are there in the galactic centre? If the IMF is
top heavy, there may be very few -- however, if we assume a more
standard IMF, and the radial mass distribution and MBH spin are
favorable, LISA can expect to detect one or more XMRIs from Sgr A*
over the mission lifetime.

For XMRIs there are several useful items to work on as we proceed towards launch:
\begin{itemize}
    \item Determine the low end mass function in the galactic centre
    \item Investigate possible interactions between XMRIs and EMRIs (and find distinguishing observables between interacting EMRI-XMRI systems and departures from GR such as the chaotic behaviour introduced by non-Kerr objects)
    \item Data analysis work to properly characterize potential loud XMRIs
    
\end{itemize}

As we have discussed, various mechanisms for producing EMRIs and IMRIs
may have EM
counterparts -- this may enable independent rate constraints either
prior to or concurrent with LISA; further, if specific 
counterparts are reliably identified, we can use the complementary
information provided by each messenger to learn more about the
astrophysics of the emitting system. Notable work to be done includes:

\begin{itemize}
    \item Detailed hydrodynamical models of EM emission mechanisms for GW events in AGN discs
    \item Athena observations of candidate IMRI systems if Athena will not be flying simultaneously with LISA (or coordination if missions are concurrent)
\end{itemize}

Each of these types of inspirals present a large parameter-space
of possible waveforms, making detection itself a notable
challenge. Narrowing the theoretical uncertainties for each channel,
in advance of LISA, therefore also has implications for the
detectability of their signals. However, work may also be done on the
signal processing side to find new strategies for extracting the
signals, and doing reliable parameter estimation on them, from the
LISA data stream. In this context, there is currently substantial concern
over possible degeneracies between gas-induced phase shifts and
departures from general relativity; fortunately, if we have multiple events from
multiple populations, departures from general relativity should be universal, while
gas effects will impact only a subset of events. On the other hand,
some signals containing environmental effects will be clearly identifiable
(e.g., b-EMRIs). All of these channels may contribute to a
detectable EMRI background; however, disentangling multiple populations
from such an unresolved background will be extremely challenging and
fundamentally requires more theoretical development from each
contributing channel. From these areas, we would especially like to highlight the need for a thorough parameter space exploration with at least 3d Newtonian hydrodynamical simulations of the impact of gas on the inspiral waveforms.

Substantially unequal mass ratio inspirals of all types represent an
important class of sources, uniquely detectable by LISA. In order to
best exploit the astrophysical and fundamental physical science
achievable by LISA using these types of events, in the years preceding launch, we will need to
work primarily on developing
more detailed models for each formation channel, and on observationally
constraining the parameters used as inputs of those models, along the lines we have outlined above.

%% file: conclusions.tex
\section*{General Summary} \label{sec.generalConclusion}
\addcontentsline{toc}{section}{\protect\numberline{}General Summary}

The decade prior to LISA's launch will be an exciting one for the astrophysics community, presenting unique challenges and opportunities in preparing for LISA's first observations. This review outlines the extensive landscape of astrophysical theory, numerical simulations, and current astronomical observations that will influence preparations for the pipelines that will deliver LISA data, and guide our interpretations of the first LISA observations and catalogues.

This review describes the current state of knowledge regarding three main source classes for LISA: ultra-compact stellar-mass binaries, massive black hole binaries, and extreme or intermediate mass ratio inspirals. For each of these three source classes, our current understanding of the astrophysical processes that create them and guide their ongoing evolution is a rich tapestry formed from extant observations (usually electromagnetic), numerical simulations and modelling, and theoretical considerations. LISA data will be added to this, providing new independent information that will help constrain the physics governing these systems, and opening up new avenues of investigation for future observations, theory, and simulations.

Astronomy observations will continue to evolve and alter the scientific landscape prior to LISA's launch, and theory and modelling will become more refined. Such advances will inform our understanding of the ways in which LISA data can be used, and they can also sharpen the focus on the important ways in which gravitational wave data will expand and enhance our ability to understand astrophysical phenomena in many different environments and scales. This review endeavours to provide a framework within which to consider these possibilities, and should be a good starting point for those interested in using LISA as a new observational tool for understanding the Universe.

%% file: acknowledgements.tex
\section*{Acknowledgements}
\addcontentsline{toc}{section}{\protect\numberline{}Acknowledgements}

P. Dayal acknowledges support from the European Research council (ERC-717001) and from the Netherlands Research Council NWO (016.VIDI.189.162).\\
P.H. Johansson acknowledges the support from the European Research Council (ERC-818930).\\
S. Toonen acknowledges support from the Netherlands Research Council NWO (VENI 639.041.645 grant)\\
C. Unal is supported by European Structural and Investment Funds and the Czech Ministry of Education, Youth and Sports (Project CoGraDS - CZ.02.1.01/0.0/0.0/15\_003/0000437).\\
S. Chaty acknowledges the LabEx UnivEarthS for the funding of Interface project I10 ``From binary evolution towards merging of compact objects".\\
A. De Rosa acknowledges financial contribution from the agreement ASI-INAF n.2017-14-H.O \\
E.Berti is supported by NSF Grants No. PHY-1912550 and AST-2006538, NASA ATP Grants No. 17-ATP17-0225 and 19-ATP19-0051, NSF-XSEDE Grant No. PHY-090003, and NSF Grant PHY-20043.\\
D. Gerosa is supported by European Union's H2020  ERC Starting Grant No. 945155--GWmining, Leverhulme Trust Grant No. RPG-2019-350 and Royal Society Grant No. RGS-R2-202004.\\
T.Bogdanovic acknowledges support by the NASA award No. 80NSSC19K0319 and by the NSF award AST-1908042.
D. Porquet acknowledges funding support from CNES. \\  
C. Danielski  acknowledges financial support from the State Agency for Research of the Spanish MCIU through the "Center of Excellence Severo Ochoa" award to the Instituto de Astrofísica de Andalucía (SEV-2017-0709)\\
B. L. Davis acknowledges support from Tamkeen under the NYU Abu Dhabi Research Institute grant CAP3. \\
F. Pacucci acknowledges support from a Clay Fellowship by the SAO and from the Black Hole Initiative, which is funded by grants from the John Templeton Foundation and the Gordon and Betty Moore Foundation.\\
A. J. Ruiter acknowledges support from the Australian Research Council Future Fellowship grant FT170100243.\\
V. Paschalidis is supported by NSF grant  PHY-1912619 and NASA grant 80NSSC20K1542 to the University of Arizona, and NSF-XSEDE grant TG-PHY190020.\\
D. Haggard acknowledges support from the NSERC Discovery Grant and Canada Research Chairs programs, and the Bob Wares Science Innovation Prospectors Fund.\\
M. Toscani acknowledges European Union’s Horizon 2020 research and innovation program under the Marie Sklodowska-Curie grant agreement NO 823823 (RISE DUSTBUSTERS project) and COST Action CA16104 - Gravitational waves, black holes and fundamental physics, supported by COST (European Cooperation in Science and Technology).\\
M. Chruslinska, A. Istrate and G. Nelemans acknowledge support from Netherlands Research Council NWO. \\
T. Fragos and S. Bavera acknowledge support from a Swiss  National  Science  Foundation  Professorship  grant  (project  numbers  PP00P2\_176868 and PP00P2\_211006)\\
V. Korol is supported by the Netherlands Research Council NWO (Rubicon 019.183EN.015 grant).

%% file: authors.tex
\section*{Authors and affiliations} \label{sec:author-details}

Amaro Seoane, Pau; 
Andrews, Jeff; 
Arca Sedda, Manuel; 
Askar, Abbas; 
Baghi, Quentin;
Balasov, Razvan; 
Bartos, Imre; 
Bavera, Simone S.; 
Bellovary, Jillian; 
Berry, Christopher P. L.; 
Berti, Emanuele; 
Bianchi, Stefano; 
Blecha, Laura; 
Blondin, Ste\'phane; 
Bogdanovi\'c, Tamara; 
Boissier, Samuel; 
Bonetti, Matteo; 
Bonoli, Silvia; 
Bortolas, Elisa; 
Breivik, Katelyn; 
Capelo, Pedro R.; 
Caramete, Laurentiu; 
Cattorini, Federico; 
Charisi, Maria; 
Chaty, Sylvain; 
Chen, Xian; 
Chru{\'s}li{\'n}ska, Martyna; 
Chua, Alvin J. K.; 
Church, Ross; 
Colpi, Monica; 
D'Orazio, Daniel; 
Danielski, Camilla; 
Davies, Melvyn B.; 
Dayal, Pratika; 
De Rosa, Alessandra; 
Derdzinski, Andrea; 
Destounis, Kyriakos; 
Dotti, Massimo; 
Du\c{t}an, Ioana; 
Dvorkin, Irina; 
Fabj, Gaia; 
Foglizzo, Thierry; 
Ford, Saavik; 
Fouvry, Jean-Baptiste; 
Franchini, Alessia;
Fragos, Tassos; 
Fryer, Chris; 
Gaspari, Massimo; 
Gerosa, Davide; 
Graziani, Luca; 
Groot, Paul; 
Habouzit, Melanie; 
Haggard, Daryl; 
Haiman, Zoltan; 
Han, Wen-Biao; 
Istrate, Alina; 
Johansson, Peter H.; 
Khan, Fazeel Mahmood; 
Kimpson, Tomas; 
Kokkotas, Kostas; 
Kong, Albert; 
Korol, Valeriya; 
Kremer, Kyle; 
Kupfer, Thomas; 
Lamberts, Astrid; 
Larson, Shane; 
Lau, Mike; 
Liu, Dongliang; 
Lloyd-Ronning, Nicole; 
Lodato, Giuseppe; 
Lupi, Alessandro; 
Ma, Chung-Pei; 
Maccarone, Tomas; 
Mandel, Ilya; 
Mangiagli, Alberto; 
Mapelli, Michela; 
Mathis, St\'{e}phane;
Mayer, Lucio; 
McGee, Sean; 
McKernan, Berry; 
Miller, M. Coleman; 
Mota, David F.; 
Mumpower, Matthew; 
Nasim, Syeda S; 
Nelemans, Gijs; 
Noble, Scott; 
Pacucci, Fabio; 
Panessa, Francesca; 
Paschalidis, Vasileios; 
Pfister, Hugo; 
Porquet, Delphine; 
Quenby, John; 
Ricarte, Angelo;
R\"opke,  Friedrich K.; 
Regan, John; 
Rosswog, Stephan; 
Ruiter, Ashley; 
Ruiz, Milton; 
Runnoe, Jessie; 
Schneider, Raffaella; 
Schnittman, Jeremy; 
Secunda, Amy; 
Sesana, Alberto; 
Seto, Naoki; 
Shao, Lijing; 
Shapiro, Stuart; 
Sopuerta, Carlos; 
Stone, Nicholas C.; 
Suvorov, Arthur; 
Tamanini, Nicola; 
Tamfal, Tomas; 
Tauris, Thomas; 
Temmink, Karel; 
Tomsick, John; 
Toonen, Silvia; 
Torres-Orjuela, Alejandro; 
Toscani, Martina; 
Tsokaros, Antonios; 
Unal, Caner; 
V\'{a}zquez-Aceves, Ver\'{o}nica; 
Valiante, Rosa; 
van Putten, Maurice; 
van Roestel, Jan; 
Vignali, Christian; 
Volonteri, Marta; 
Wu, Kinwah; 
Younsi, Ziri; 
Yu, Shenghua; 
Zane, Silvia; 
Zwick, Lorenz; 
Antonini, Fabio; 
Baibhav, Vishal; 
Barausse, Enrico; 
Bonilla Rivera, Alexander; 
Branchesi, Marica; 
Branduardi-Raymont, Graziella; 
Burdge, Kevin; 
Chakraborty, Srija; 
Cuadra, Jorge; 
Dage, Kristen; 
Davis, Benjamin; 
de Mink, Selma E.; 
Decarli, Roberto; 
Doneva, Daniela; 
Escoffier, Stephanie; 
Fragione, Giacomo;
Gandhi, Poshak; 
Haardt, Francesco; 
Lousto, Carlos O.; 
Nissanke, Samaya; 
Nordhaus, Jason; 
O'Shaughnessy, Richard; 
Portegies Zwart, Simon; 
Pound, Adam; 
Schussler, Fabian; 
Sergijenko, Olga; 
Spallicci, Alessandro; 
Vernieri, Daniele; 
Vigna-G\'omez, Alejandro 

\clearpage
 P. Amaro Seoane:  Institute of Multidisciplinary Mathematics, Universitat Polit{\'e}cnica de Val{\'e}ncia, Spain;  Lanzhou Center for Theoretical Physics, Key Laboratory of Theoretical Physics of Gansu Province, School of Physical Science and Technology, Lanzhou University, Lanzhou 730000, People's Republic of China; Institute of Theoretical Physics \& Research Center of Gravitation,Lanzhou University, Lanzhou 730000, People's Republic of China; Kavli Institute for Astronomy and Astrophysics, Beijing, China;  {amaro@riseup.net};  
 
 J. Andrews:  Center for Interdisciplinary Exploration and Research in Astrophysics (CIERA); Department of Physics and Astronomy, Northwestern University, 1800 Sherman Ave, Evanston, IL 60201, USA;  {jeffrey.andrews@northwestern.edu};  
 
 M. Arca Sedda:  Astronomisches Rechen Instituy (University of Heidelberg);  {m.arcasedda@gmail.com};  
 
 A. Askar:  Lund Observatory, Department of Astronomy, and Theoretical Physics, Lund University, Box 43, SE-221 00 Lund, Sweden;  {askar@astro.lu.se};  

 Q. Baghi, IRFU, CEA, Universit\'{e} Paris-Saclay, F-91191 Gif-sur-Yvette, France; {quentin.baghi@cea.fr}
 
 R. Balasov:  {Institute of Space Science, Romania}; {Faculty of Physics, University of Bucharest, Romania};  {rabalasov@spacescience.ro};  
 
 I. Bartos:  Department of Physics, University of Florida, PO Box 118440, Gainesville, FL 32611-8440, USA;  {imrebartos@ufl.edu};  
 
 S. Bavera:  Geneva Observatory, University of Geneva, Chemin Pegasi 51, 1290 Versoix, Switzerland; Gravitational Wave Science Center (GWSC), Université de Genève, CH1211 Geneva, Switzerland; {simone.bavera@unige.ch};  
 
 J. Bellovary:  CUNY - Queensborough Community College; American Museum of Natural History; CUNY Graduate Center;  {jbellovary@amnh.org};  
 
 C. Berry:  Center for Interdisciplinary Exploration and Research in Astrophysics (CIERA), Department of Physics and Astronomy, Northwestern University, 1800 Sherman Ave, Evanston, IL 60201, USA; SUPA, School of Physics and Astronomy, University of Glasgow, Kelvin Building, University Ave, Glasgow G12 8QQ, UK;  {christopher.berry.2@glasgow.ac.uk};  
 
 E. Berti:  Johns Hopkins University;  {berti@jhu.edu};  
 
 S. Bianchi:  Dipartimento di Matematica e Fisica, Universit\`a degli Studi Roma Tre, via della Vasca Navale 84, I-00146 Roma, Italy;  {bianchi@fis.uniroma3.it};  
 
 L. Blecha:  Department of Physics, University of Florida;  {lblecha@ufl.edu};  
 
 S. Blondin:  Aix Marseille Univ, CNRS, CNES, LAM, Marseille, France;  {stephane.blondin@lam.fr};  
 
 T. Bogdanovi\'c:  School of Physics and Center for Relativistic Astrophysics, 837 State St NW, Georgia Institute of Technology, Atlanta, GA 30332, USA;  {tamarab@gatech.edu};  
 
 S. Boissier:  Aix Marseille Univ, CNRS, CNES, LAM, Marseille, France;  {samuel.boissier@lam.fr};  
 
 M. Bonetti:  Dipartimento di Fisica G. Occhialini, Universita' degli studi di Milano-Bicocca, Piazza della Scienza 3, 20126 Milano Italy;  {matteo.bonetti@unimib.it};  
 
 S. Bonoli:  Donostia International Physics Centre (DIPC), Paseo Manuel de Lardizabal 4, 20018 Donostia-San Sebastian, Spain; IKERBASQUE, Basque Foundation for Science, E-48013, Bilbao, Spain;  {silvia.bonoli@dipc.org};  
 
 E. Bortolas:  Dipartimento di Fisica ``G. Occhialini'', Universit\'a degli Studi di Milano-Bicocca, Piazza della Scienza 3, I-20126 Milano, Italy; INFN, Sezione di Milano-Bicocca, Piazza della Scienza 3, I-20126 Milano, Italy;  {elisa.bortolas@unimib.it};  
 
 P. Capelo:  Center for Theoretical Astrophysics and Cosmology, Institute for Computational Science, University of Zurich, Winterthurerstrasse 190, CH-8057 Z\"urich, Switzerland;  {pcapelo@physik.uzh.ch};  
 
 L. Caramete:  Institute of Space Science, Magurele, Romania;  {lcaramete@spacescience.ro};  
 
 F. Catorini:  DiSAT, Universit\'a degli studi dell'Insubria, Via Valleggio, 11, I-22100 Como, Italy; INFN, Sezione di Milano-Bicocca, Piazza della Scienza 3, I-20126 Milano, Italy;  {fcattorini@uninsubria.it};  
 
 M. Charisi:  Department of Physics \& Astronomy, Vanderbilt University, 2301 Vanderbilt Place, Nashville, TN 37235, USA;  {maria.charisi@nanograv.org};  
 
 S. Chaty:  Universit\'e de Paris, CNRS, AstroParticule et Cosmologie,  F-75013 Paris, France;  {sylvain.chaty@u-paris.fr};  
 
 X. Chen:  Astronomy Department, School of Physics, Peking University, Beijing 100871, P.R. China;  {xian.chen@pku.edu.cn};  
 
 M. Chru{\'s}li{\'n}ska:  Department of Astrophysics/IMAPP, Radboud University, PO Box 9010, 6500 GL, The Netherlands;  {mchruslinska@mpa-garching.mpg.de};  
 
 A. Chua:  Theoretical Astrophysics Group, California Institute of Technology, Pasadena, CA 91125, U.S.A.;  {achua@caltech.edu};  
 
 R. Church:  Lund Observatory;  {ross@astro.lu.se};  
 
 M. Colpi:  Department of Physics, University of Milano Bicocca, Milan, Italy;  {monica.colpi@unimib.it};  
 
 D. D'Orazio:  Niels Bohr International Academy, Niels Bohr Institute, Blegdamsvej 17, 2100 Copenhagen, Denmark;  {daniel.dorazio@nbi.ku.dk};  
 
 C. Danielski:  Instituto de Astrof\'isica de Andaluc\'ia (IAA-CSIC), Glorieta de la Astronom\'ia S/N, 18008 Granada, Spain;  {cdanielski@iaa.es};  
 
 M. Davies:  Centre for Mathematical Sciences, Lund University, Box 118, 221 00 Lund, Sweden;  {melvyn$\_$b.davies@math.lu.se};  
 
 P. Dayal:  Kapteyn Astronomical Institute, University of Groningen, P.O. Box 800, 9700 AV Groningen, The Netherlands;  {p.dayal@rug.nl};  
 
 A. De Rosa:  INAF - Istituto di Astrofisica e Planetologia Spaziali, via Fosso del Cavaliere, I-133 Roma, Italy;  {alessandra.derosa@inaf.it};  
 
 A. Derdzinski:  Center for Theoretical Astrophysics and Cosmology, Institute for Computational Science, University of Zurich, Winterthurerstrasse 190, 8057 Zurich, Switzerland;  {andrea@ics.uzh.ch};  
 
 K. Destounis:  Theoretical Astrophysics, IAAT, University of Tübingen, 72076 Tübingen, Germany;  {kyriakos.destounis@uni-tuebingen.de};  
 
 M. Dotti:  Dipartimento di Fisica ``G. Occhialini'', Universit\'a degli Studi di Milano-Bicocca, Piazza della Scienza 3, I-20126 Milano, Italy; INFN, Sezione di Milano-Bicocca, Piazza della Scienza 3, I-20126 Milano, Italy; INAF, Osservatorio Astronomico di Brera, Via E. Bianchi 46, I-23807, Merate, Italy;  {massimo.dotti@unimib.it};  
 
 I. Du\c{t}an:  Institute of Space Science, Atomi\c{s}tilor 409, RO-077125 M\u{a}gurele, Romania;  {idutan@spacescience.ro};  
 
 I. Dvorkin:  Institut d'Astrophysique de Paris, Sorbonne Universit\'e \& CNRS, UMR 7095, 98 bis bd Arago, F-75014 Paris, France;  {dvorkin@iap.fr};  
 
 G. Fabj:  Astronomisches Rechen-Institut, Zentrum f\"{u}r Astronomie, Universit\"{a}t Heidelberg, 69120 Heidelberg, Germany; Dept. of Astrophysics, American Museum of Natural History, New York, NY 10024 USA;  {gaia.fabj@stud.uni-heidelberg.de};  
 
 T. Foglizzo:  AIM, CEA, CNRS, Universit\'e Paris-Saclay, Universit\'e Paris Diderot, Sorbonne Paris Cit\'e, F-91191 Gif-sur-Yvette, France;  {thierry.foglizzo@cea.fr};  
 
 S. Ford:  Department of Astrophysics, American Museum of Natural History, New York, NY 10024, USA; Center for Computational Astrophysics, Flatiron Institute, New York, NY 10010, USA; Graduate Center, City University of New York, 365 5th Avenue, New York, NY 10016, USA; Department of Science, BMCC, City University of New York, New York, NY 10007, USA;  {sford@amnh.org};  
 
 J. Fouvry:  CNRS and Sorbonne Universit\'e, UMR 7095, Institut d'Astrophysique de Paris, 98 bis Boulevard Arago, F-75014 Paris, France;  {fouvry@iap.fr};  
 
 T. Fragkos:  Département d'Astronomie, Universit\'e de Geneve, Chemin Pegasi 51, CH-1290 Versoix, Switzerland; Gravitational Wave Science Center (GWSC), Université de Genève, CH1211 Geneva, Switzerland;  {anastasios.fragkos@unige.ch}; 
 
 A. Franchini:  Dipartimento di Fisica ‘G. Occhialini’, Universita degli Studi di Milano-Bicocca, Piazza della Scienza 3, I-20126 Milano, Italy; INFN – Sezione di Milano-Bicocca, Piazza della Scienza 3, I-20126 Milano, Italy; {alessia.franchini@unimib.it}
 
 C. Fryer:  Center for Theoretical Astrophysics, Los Alamos National Laboratory, Los Alamos, NM, 87545, USA;  {fryer@lanl.gov};  
 
 M. Gaspari:  INAF - Osservatorio di Astrofisica e Scienza dello Spazio, via P. Gobetti 93/3, I-40129 Bologna, Italy;  Department of Astrophysical Sciences, Princeton University, 4 Ivy Lane, Princeton, NJ 08544-1001, USA;  {massimo.gaspari@inaf.it};  
 
 D. Gerosa:  Dipartimento di Fisica ``G. Occhialini'', Universit\'a degli Studi di Milano-Bicocca, Piazza della Scienza 3, 20126 Milano, Italy INFN, Sezione di Milano-Bicocca, Piazza della Scienza 3, 20126 Milano, Italy School of Physics and Astronomy \& Institute for Gravitational Wave Astronomy, University of Birmingham, Birmingham, B15 2TT, United Kingdom;  {davide.gerosa@unimib.it};  
 
 L. Graziani:  Dipartimento di Fisica, Sapienza, Universit$\grave{a}$ di Roma, Piazzale Aldo Moro 5, 00185, Roma, Italy; INFN, Sezione di Roma I, P.le Aldo Moro 2, 00185 Roma, Italy; INAF/Osservatorio Astrofisico di Arcetri, Largo E. Femi 5, 50125 Firenze, Italy;  {luca.graziani@uniroma1.it};  
 
 P. Groot:   Department of Astrophysics/IMAPP, Radboud University, P.O. Box 9010, 6500 GL Nijmegen, The Netherlands; Department of Astronomy, University of Cape Town, Private Bag X3, Rondebosch, 7701, South Africa; South African Astronomical Observatory, P.O. Box 9, Observatory, 7935, South Africa; The Inter-University Institute for Data Intensive Astronomy, University of Cape Town, Private Bag X3, Rondebosch, 7701, South Africa;  {p.groot@astro.ru.nl};  
 
 M. Habouzit:  Max-Planck-Institut f\"ur Astronomie, K\"onigstuhl 17, D-69117 Heidelberg, Germany; Zentrum für Astronomie der Universit\"at Heidelberg, ITA, Albert-Ueberle-Str. 2, D-69120 Heidelberg, Germany;  {habouzit.astro@gmail.com};  
 
 D. Haggard:  McGill Space Institute and Department of Physics, McGill University, 3600 rue University, Montr\'eal, Qu\'ebec, H3A 2T8, Canada;  {daryl.haggard@mcgill.ca};  
 
 Z. Haiman:  Columbia University;  {zoltan@astro.columbia.edu};  
 
 W. Han:  Shanghai Astronomical Observatory, CAS, 80 Nandan Road, Shanghai, China 200030;  {wbhan@shao.ac.cn};  
 
 A. Istrate:  Department of Astrophysics/IMAPP, Radboud University, PO Box 9010, 6500 GL, The Netherlands;  {a.istrate@astro.ru.nl};  
 
 P. Johansson:  Department of Physics, Gustaf H\"allstr\"omin katu 2, FI-00014, University of Helsinki, Finland;  {Peter.Johansson@helsinki.fi};  
 
 F. Khan:  Department of Space Science, Institute of Space Technology, Islamabad 44000, Pakistan;  {khanfazeel.ist@gmail.com};  
 
 T. Kimpson:  Mullard Space Science Laboratory, University College London, Holmbury St. Mary, Dorking, Surrey, RH5 6NT, UK;  {tom.kimpson.16@ucl.ac.uk};  
 
 K. Kokkotas:  Theoretical Astrophysics, University of Tuebingen, Germany;  {kostas.kokkotas@uni-tuebingen.de};  
 
 A. Kong:  Institute of Astronomy, National Tsing Hua University, Hsinchu 30013, Taiwan;  {akong@gapp.nthu.edu.tw};  
 
 V. Korol:  Max-Planck-Institut f{\"u}r Astrophysik, Karl-Schwarzschild-Straße 1, 85741 Garching, Germany; Institute for Gravitational Wave Astronomy \& School of Physics and Astronomy, University of Birmingham, Birmingham, B15 2TT, UK;  {korol@mpa-garching.mpg.de};  
 
 K. Kremer:  TAPIR, California Institute of Technology, Pasadena, CA 91125, USA; The Observatories of the Carnegie Institution for Science, Pasadena, CA 91101, USA;  {kkremer@caltech.edu};  
 
 T. Kupfer:  Department of Physics and Astronomy, Texas Tech University, PO Box 41051, Lubbock, TX 79409, USA;  {tkupfer@ttu.edu};  
 
 A. Lamberts:  Universit\'e C\^ote d'Azur, Observatoire de la C\^ote d'Azur, CNRS, Laboratoire Lagrange, Laboratoire Art\'emis, Bd de l'Observatoire,CS 34229, 06304 Nice cedex 4, France.;  {astrid.lamberts@oca.eu};  
 
 S. Larson:  Center for Interdisciplinary Exploration and Research in Astrophysics (CIERA); Department of Physics and Astronomy, Northwestern University, 1800 Sherman Ave, Evanston, IL 60201, USA;  {s.larson@northwestern.edu};  
 
 M. Lau:  Monash Centre for Astrophysics, School of Physics and Astronomy, Monash University, Clayton, Victoria 3800, Australia; OzGrav, Australian Research Council Centre of Excellence for Gravitational Wave Discovery, Australia;  {};  
 
 D. Liu:  National Astronomical Observatories, Chinese Academy of Sciences;  {dlliu@bao.ac.cn};  
 
 N. Lloyd-Ronning:  Los Alamos National Lab and The University of New Mexico;  {lloyd-ronning@lanl.gov};  
 
 G. Lodato:  Universit\'a degli Studi di Milano;  {giuseppe.lodato@unimi.it};  
 
 A. Lupi:  Dipartimento di Fisica ``G. Occhialini'', Universit\`a degli Studi di Milano-Bicocca, Piazza della Scienza 3, I-20126 Milano, Italy; INFN – Sezione di Milano-Bicocca, Piazza della Scienza 3, I-20126 Milano, Italy;  {alessandro.lupi@unimib.it};  
 
 C. Ma :  Department of Astronomy, Department of Physics, University of California at Berkeley, CA 94720 USA;  {cpma@berkeley.edu};  
 
 T. Maccarone:  Department of Physics \& Astronomy, Texas Tech University, Box 41051, Lubbock TX 79409-1051;  {thomas.maccarone@ttu.edu};  
 
 I. Mandel:  Monash Centre for Astrophysics, School of Physics and Astronomy, Monash University, Clayton, Victoria 3800, Australia; OzGrav, Australian Research Council Centre of Excellence for Gravitational Wave Discovery, Australia; Institute of Gravitational Wave Astronomy and School of Physics and Astronomy, University of Birmingham, Birmingham, B15 2TT, United Kingdom;  {ilya.mandel@monash.edu};  
 
 A. Mangiagli:  APC, AstroParticule et Cosmologie, Universit\'e de Paris, CNRS, F-75013 Paris, France;  {mangiagli@apc.in2p3.fr};  
 
 M. Mapelli:  Physics and Astronomy Department Galileo Galilei, University of Padova, Vicolo dell'Osservatorio 3, I--35122, Padova, Italy; INFN--Padova, Via Marzolo 8, I--35131 Padova, Italy; INAF--Osservatorio Astronomico di Padova, Vicolo dell'Osservatorio 5, I--35122, Padova, Italy;  {michela.mapelli@unipd.it};  
 
 S. Mathis:  Département d'Astrophysique-AIM, CEA/DRF/IRFU, CNRS/INSU, Universit\'e Paris-Saclay, Universit\'e Paris-Diderot, Universit\'e De Paris, F-91191 Gif-sur-Yvette, France;  {stephane.mathis@cea.fr};  
 
 L. Mayer:  Center for Theoretical Astrophysics and Cosmology, Institute for Computational Science, University of Zurich, Winterthurerstrasse 190, 8057, Zurich, Switzerland;  {lmayer@physik.uzh.ch};  
 
 S. McGee:  School of Physics and Astronomy, University of Birmingham, Edgbaston, Birmingham B15 2TT, UK;  {smcgee@star.sr.bham.ac.uk}; 
 
 M. Miller:  University of Maryland, Department of Astronomy, College Park MD 20742-2421;  {miller@astro.umd.edu};  
 
 D. Mota:  Institute of Theoretical Astrophysics, University of Oslo, PO Box 1029, Blindern 0315, Oslo, Norway;  {d.f.mota@astro.uio.no};  
 
 M. Mumpower:  Los Alamos National Laboratory;  {mumpower@lanl.gov};
 
 S. Nasim:  Missouri University of Science and Technology \& Rolla, MO (USA); American Museum of Natural History \& New York, NY (USA);  {ssnkct@mst.edu};  
 
 G. Nelemans:  Department of Astrophysics/IMAPP, Radboud University, PO Box 9010, 6500 GL, The Netherlands; SRON, Netherlands Institute for Space Research, Sorbonnelaan 2, NL-3584 CA Utrecht, The Netherlands; Institute of Astronomy, KU Leuven, Celestijnenlaan 200D, B-3001 Leuven, Belgium;  {nelemans@astro.ru.nl};  
 
 S. Noble:  Gravitational Astrophysics Laboratory, NASA Goddard Space Flight Center, Greenbelt, MD 20771, USA;  {scott.c.noble@nasa.gov};  
 
F. Pacucci:  Center for Astrophysics $\vert$ Harvard \& Smithsonian, Cambridge, MA 02138, USA; Black Hole Initiative, Harvard University, Cambridge, MA 02138, USA;  {fabio.pacucci@cfa.harvard.edu};  
 
 F. Panessa:  INAF - Istituto di Astrofisica e Planetologia Spaziali, via Fosso del Cavaliere 100, I-00133 Roma, Italy;  {francesca.panessa@inaf.it};  
 
 V. Paschalidis:  Departments of Astronomy and Physics, University of Arizona, Tucson, AZ 85719, USA;  {vpaschal@email.arizona.edu};  
 
 H. Pfister:  Department of Physics, The University of Hong Kong, Pokfulam Road, Hong Kong, China; DARK, Niels Bohr Institute, University of Copenhagen, Jagtvej 128, 2200 København, Denmark;  {pfisterastro@gmail.com};  
 
 D. Porquet:  Aix Marseille Univ, CNRS, CNES, LAM, Marseille, France;  {delphine.porquet@lam.fr};  
 
 J. Quenby:  Imperial College London UK;  {j.quenby@imperial.ac.uk};  
 
 F. R\"opke:  Zentrum f{\"u}r Astronomie der Universit{\"a}t Heidelberg, Institut f{\"u}r Theoretische Astrophysik; Heidelberg Institute for Theoretical Studies;  {friedrich.roepke@h-its.org};  
 
 J. Regan:  Department of Theoretical Physics, Maynooth University, Maynooth, Ireland;  {john.regan@mu.ie};  
 
 S. Rosswog:  The Oskar Klein Centre, Department of Astronomy, Stockholm University;  {stephan.rosswog@astro.su.se};  
 
 A. Ruiter :  University of New South Wales (Canberra);  {ashley.ruiter@adfa.edu.au};  
 
 M. Ruiz:  Department of Physics, University of Illinois at Urbana-Champaign, Urbana, IL 61801;  {ruizm@illinois.edu};  
 
 J. Runnoe:  Vanderbilt University, Department of Physics \& Astronomy, 6301 Stevenson Center, Nashville, TN 37235, USA;  {jessie.c.runnoe@vanderbilt.edu};  
 
 R. Schneider:  Dipartimento di Fisica, Universita di Roma La Sapienza, Piazzale Aldo Moro 2, 00185 Roma Italy; Sapienza School for Advanced Studies, Viale Regina Elena 291, 00161 Roma Italy; Istituto Nazionale di Astrofisica/Osservatorio Astronomico di Roma, via Frascati 33, 00078 Monte Porzio Catone, Roma Italy Istituto Nazionale di Fisica Nucleare, Sezione di Roma1, Piazzale Aldo Moro 2, 00185 Roma Italy;  {raffaella.schneider@uniroma1.it};  
 
 J. Schnittman:  NASA Goddard Space Flight Center, 8800 Greenbelt Rd, Greenbelt, MD 20771;  {jeremy.schnittman@nasa.gov};  
 
 A. Secunda:  Department of Astrophysical Sciences, Princeton University, Peyton Hall, Princeton, NJ 08544, USA;  {asecunda@princeton.edu};  
 
 A. Sesana:  Dipartimento di Fisica “G. Occhialini", Universit\`a degli Studi di Milano-Bicocca, Piazza della Scienza 3, IT-20126 Milano, Italy;  {alberto.sesana@unimib.it};  
 
 N. Seto:  Department of Physics, Kyoto University, Kyoto 606-8502, Japan;  {seto@tap.scphys.kyoto-u.ac.jp};  
 
 L. Shao:  Kavli Institute for Astronomy and Astrophysics, Peking University, Beijing 100871, China;  {lshao@pku.edu.cn};  
 
 S. Shapiro:  University of Illinois at Urbana-Champaign;  {slshapir@illinois.edu};  
 
 C. Sopuerta:  Institut de Ci\'encies de l'Espai (ICE, CSIC), Campus UAB, Carrer de Can Magrans s/n, 08193 Cerdanyola del Vall\'es, Spain; Institut d'Estudis Espacials de Catalunya (IEEC), Edifici Nexus, Carrer del Gran Capit\'a 2-4, despatx 201, 08034 Barcelona, Spain;  {carlos.f.sopuerta@csic.es};  
 
 N. Stone:  The Hebrew University of Jerusalem;  {nicholas.stone@mail.huji.ac.il}; 
 
 A. Suvorov:  Theoretical Astrophysics, IAAT, University of T\"ubingen, T\"ubingen 72076, Germany;  {arthur.suvorov@tat.uni-tuebingen.de};  
 
 N. Tamanini:  Laboratoire des 2 Infinis - Toulouse (L2IT-IN2P3), Universit\'e de Toulouse, CNRS, UPS, F-31062 Toulouse Cedex 9, France;  {nicola.tamanini@l2it.in2p3.fr};  
 
 T. Tamfal:  Center for Theoretical Astrophysics and Cosmology, Institute for Computational Science, University of Zurich, Winterthurerstrasse 190, CH-8057 Z\"urich, Switzerland;  {tomas.tamfal@uzh.ch};  
 
 T. Tauris:  Department of Materials and Production, Aalborg University, Denmark; 
 {tauris@mp.aau.dk};  
 
 K. Temmink:  Department of Astrophysics/IMAPP, Radboud University Nijmegen;  
 {Karel.Temmink@ru.nl};  
 
 J. Tomsick:  Space Sciences Laboratory, 7 Gauss Way, University of California, Berkeley, CA 94720-7450, USA;  {jtomsick@berkeley.edu};  
 
 S. Toonen:  Anton Pannekoek Institute for Astronomy, University of Amsterdam, 1090 GE Amsterdam, The Netherlands;  {toonen@uva.nl};  
 
 A. Torres-Orjuela:  MOE Key Laboratory of TianQin Mission, TianQin Research Center for Gravitational Physics \& School of Physics and Astronomy, Frontiers Science Center for TianQin, CNSA Research Center for Gravitational Waves, Sun Yat-Sen University (Zhuhai Campus), Zhuhai 519082, China;  {atorreso@mail.sysu.edu.cn};  
 
 M. Toscani:  Laboratoire des 2 Infinis - Toulouse (L2IT-IN2P3), Université de Toulouse, CNRS, UPS, F-31062 Toulouse Cedex 9, France; Dipartimento di Fisica, Universit\'a Degli Studi di Milano, Via Celoria, 16, Milano, 20133, Italy;  {martina.toscani@l2it.in2p3.fr};  
 
 A. Tsokaros:  University of Illinois at Urbana-Champaign;  {tsokaros@illinois.edu};  
 
 C. Unal:  CEICO, Institute of Physics of the Czech Academy of Sciences, Na Slovance 1999/2, 182 21 Praha 8, Czechia;  {unalx005@umn.edu};  
 
 V. V\'{a}zquez-Aceves:  Institute of Applied Mathematics, Academy of Mathematics and Systems Science, Chinese Academy of Sciences, 100190 Beijing, China;  {veronica@nao.cas.cn};  
 
 R. Valiante:  INAF-Osservatorio Astronomico di Roma, via di Frascati 33, I-00078 Monteporzio Catone, Italy; INFN, Sezione di Roma I, P.le Aldo Moro 2, I-00185 Roma, Italy;  {rosa.valiante@inaf.it};  
 
 M. van Putten:  Physics and Astronomy, Sejong University, 209 Neungdong-ro, Gwangjin-gu. Seoul 143-747;  {mvp@sejong.ac.kr};  
 
 J. van Roestel:  Caltech, 1201 E. California Blvd, Pasadena, CA 91125, California (CA) United States;  {jvanroes@caltech.edu};  
 
 C. Vignali:  Dipartimento di Fisica e Astronomia ``Augusto Righi", Universit\'a degli Studi di Bologna, Via Gobetti 93/2, I-40129 Bologna, Italy;  INAF -- Osservatorio di Astrofisica e Scienza dello Spazio di Bologna (OAS), Via Gobetti 93/3, I-40129 Bologna, Italy;  {cristian.vignali@unibo.it};  
 
 M. Volonteri:  Sorbonne Universit\'{e}, CNRS, UMR 7095, Institut d'Astrophysique de Paris, 98 bis bd Arago, 75014 Paris, France;  {martav@iap.fr};  
 
 K. Wu:  Mullard Space Science Laboratory, University College London, Holmbury St Mary, Surrey RH5 6NT, United Kingdom;  {kinwah.wu@ucl.ac.uk};  
 
 Z. Younsi:  Mullard Space Science Laboratory, University College London, Holmbury St.~Mary, Dorking, Surrey, RH5 6NT, UK;  {z.younsi@ucl.ac.uk};  
 
 S. Yu:  National Astronomical Observatories, Chinese Academy of Sciences;  {shenghuayu@bao.ac.cn};  
 
 S. Zane:  Mullard Space Science Laboratory, University College London, Holmbury St Mary, Dorking, Surrey, RH5 6NT, UK;  {s.zane@ucl.ac.uk};  
 
 L. Zwick:  University of Z\"urich; Centre for Theoretical Astrophysics and Cosmology;  {zwicklo@ics.uzh.ch};  
 
 F. Antonini:  Gravity Exploration Institute, School of Physics and Astronomy, Cardiff University, Cardiff, CF24 3AA, United Kingdom;  {antoninif@cardiff.ac.uk};  
 
 V. Baibhav:  Department of Physics and Astronomy, Johns Hopkins University, 3400 N. Charles St, Baltimore, Maryland 21218, USA;  {baibhavv@gmail.com};  
 
 E. Barausse:  SISSA, Via Bonomea 265, 34136 Trieste, Italy and INFN Sezione di Trieste; IFPU - Institute for Fundamental Physics of the Universe, Via Beirut 2, 34014 Trieste, Italy;  {barausse@sissa.it};  
 
 A. Bonilla Rivera:  Departamento de F\'isica, Universidade Federal de Juiz de Fora, 36036-330, Juiz de Fora, MG, Brazil.;  {alex.acidjazz@gmail.com};  
 
 G. Branduardi-Raymont:  Mullard Space Science Laboratory - University College London;  {g.branduardi-raymont@ucl.ac.uk};  
 
 K. Burdge:  Division of Physics, Mathematics and Astronomy, California Institute of Technology, Pasadena, CA 91125, USA;  {kburdge@caltech.edu};  
  S. Chakraborty:  Scuola Normale Superiore;  {srija.chakraborty@sns.it};  
 
 J. Cuadra:  Departamento de Ciencias, Facultad de Artes Liberales, Universidad Adolfo Ib\'a\~nez, Padre Hurtado 750, Vi\~na del Mar, Chile;  {jcuadra@npf.cl};  
 
 K. Dage:  Department of Physics, McGill University, 3600 University Street, Montr\'eal, QC H3A 2T8, Canada; McGill Space Institute, McGill University, 3550 University Street, Montr\'eal, QC H3A 2A7, Canada;  {dagek@physics.mcgill.ca};  
 
 B. Davis:  Center for Astro, Particle, and Planetary Physics, New York University Abu Dhabi;  {ben.davis@nyu.edu};  
 
 S. de Mink:  Max-Planck-Institut für Astrophysik, Karl-Schwarzschild-Straße 1, 85741 Garching, Germany;  {sedemink@mpa-garching.mpg.de};  
 
 R. Decarli:  INAF -- Osservatorio di Astrofisica e Scienza dello Spazio di Bologna, via Gobetti 93/3, I-40129, Bologna, Italy;  {roberto.decarli@inaf.it};  
 
 D. Doneva:  University of Tuebingen;  {daniela.doneva@uni-tuebingen.de};  
 
 S. Escoffier:  Aix Marseille Univ, CNRS/IN2P3, CPPM, Marseille, France;  {escoffier@cppm.in2p3.fr};  

G. Fragione: Center for Interdisciplinary Exploration and Research in Astrophysics (CIERA); Department of Physics and Astronomy, Northwestern University, 1800 Sherman Ave, Evanston, IL 60201, USA;  {giacomo.fragione@northwestern.edu};
 
 P. Gandhi:  School of Physics \& Astronomy, University of Southampton, Southampton, SO17 1BJ, UK;  {poshak.gandhi@soton.ac.uk};  
 
 F. Haardt:  Dipartimento di Scienza e Alta Tecnologia, Universit\`a degli Studi dell'Insubria, via Valleggio 11, I-22100 Como, Italy;  {haardt@uninsubria.it};  
 
 C. Lousto:  CCRG, Rochester Institute of Technology;  {colsma@rit.edu};  
 
 J. Nordhaus:  Center for Computational Relativity and Gravitation, Rochester Institute of Technology, Rochester, NY 14623, USA;  {nordhaus@astro.rit.edu};  

A. Pound:  School of Mathematical Sciences and STAG Research Centre, University of Southampton, SO17 1BJ, United Kingdom;  {a.pound@soton.ac.uk};  
 
 R. O'Shaughnessy:  Center for Computational Relativity and Gravitation, Rochester Institute of Technology;  {richard.oshaughnessy@ligo.org};  
 
 S. Portegies Zwart:  Leiden Observatory, Leiden, the Netherlands;  {spz@strw.leidenuniv.nl};  
 
 F. Schussler:  IRFU, CEA, Universit\'e Paris-Saclay, F-91191 Gif-sur-Yvette, France;  {fabian.schussler@cea.fr};  
  O. Sergijenko:  Astronomical Observatory of Taras Shevchenko National University of Kyiv, Observatorna str., 3, Kyiv, 04053, Ukraine; Main Astronomical Observatory of the National Academy of Sciences of Ukraine, Zabolotnoho str., 27, Kyiv, 03680, Ukraine;  {olga.sergijenko.astro@gmail.com};  
 
 A. Spallicci:  Universit\'e d'Orl\'eans - Centre National de la Recherche Scientifique;  {spallicci@cnrs-orleans.fr};  
 
 D. Vernieri:  Dipartimento di Fisica ``E. Pancini'', Università di Napoli ``Federico II'' and INFN, Sezione di Napoli, Compl. Univ. di Monte S. Angelo, Edificio G, Via Cinthia, I-80126, Napoli, Italy.;  {daniele.vernieri@unina.it};  
 
 A. Vigna-G\'omez:  DARK, Niels Bohr Institute, University of Copenhagen, Jagtvej 128, 2200, Copenhagen, Denmark;  {avignagomez@nbi.ku.dk}

%% file: astroWP_00main.bbl
\hyphenation{Post-Script Sprin-ger}
\begin{thebibliography}{1923}
\expandafter\ifx\csname url\endcsname\relax
 \def\url#1{\burl{#1}}\fi
\expandafter\ifx\csname urlprefix\endcsname\relax\def\urlprefix{URL }\fi
\providecommand{\bibinfo}[2]{#2}
\providecommand{\eprint}[2][]{\url{#2}}
\providecommand{\doi}[1]{\urlstyle{rm}\url{https://doi.org/#1}}

\bibitem[{{Aarseth}(2012)}]{2012MNRAS.422..841A}
{Aarseth} SJ (2012) {Mergers and ejections of black holes in globular
  clusters}. \mnras 422(1):841--848. \doi{10.1111/j.1365-2966.2012.20666.x}.
  {\href{https://arxiv.org/abs/1202.4688}{{arXiv:1202.4688}}} {[astro-ph.SR]}

\bibitem[{Aartsen et~al.(2018)}]{Science.361.6398}
Aartsen M, et~al. (2018) Neutrino emission from the direction of the blazar txs
  0506+056 prior to the icecube-170922a alert. Science 361(6398):147--151.
  \doi{10.1126/science.aat2890},
  \urlprefix\url{https://science.sciencemag.org/content/361/6398/147}

\bibitem[{{Aasi} et~al.(2015){Aasi}, {Abbott}, {Abbott}, {Abbott}, {Abernathy},
  {Ackley}, {Adams}, {Adams}, {Addesso}, and et~al.}]{AdvLIGO}
{Aasi} J, {Abbott} BP, {Abbott} R, {Abbott} T, {Abernathy} MR, {Ackley} K,
  {Adams} C, {Adams} T, {Addesso} P, et~al (2015) {Advanced LIGO}. Classical
  and Quantum Gravity 32(7):074001. \doi{10.1088/0264-9381/32/7/074001}.
  {\href{https://arxiv.org/abs/1411.4547}{{arXiv:1411.4547}}} {[gr-qc]}

\bibitem[{{Abadie} et~al.(2010){Abadie}, {Abbott}, {Abbott}, {Abernathy},
  {Accadia}, {Acernese}, {Adams}, {Adhikari}, {Ajith}, {Allen}, and
  et~al.}]{2010CQGra..27q3001A}
{Abadie} J, {Abbott} BP, {Abbott} R, {Abernathy} M, {Accadia} T, {Acernese} F,
  {Adams} C, {Adhikari} R, {Ajith} P, {Allen} B, et~al. (2010) {TOPICAL REVIEW:
  Predictions for the rates of compact binary coalescences observable by
  ground-based gravitational-wave detectors}. Classical and Quantum Gravity
  27(17):173001. \doi{10.1088/0264-9381/27/17/173001}.
  {\href{https://arxiv.org/abs/1003.2480}{{arXiv:1003.2480}}} {[astro-ph.HE]}

\bibitem[{{Abbott} et~al.(2016{\natexlab{a}}){Abbott}, {Abbott}, {Abbott},
  {Abernathy}, {Acernese}, {Ackley}, {Adams}, {Adams}, {Addesso}, {Adhikari},
  {Adya}, {Affeldt}, {Agathos}, {Agatsuma}, {Aggarwal}, {Aguiar}, {Aiello},
  {Ain}, {Ajith}, {Allen}, {Allocca}, {Altin}, {Anderson}, {Anderson}, {Arai},
  {Araya}, {Arceneaux}, {Areeda}, {Arnaud}, {Arun}, {Ascenzi}, {Ashton}, {Ast},
  {Aston}, {Astone}, {Aufmuth}, {Aulbert}, {Babak}, {Bacon}, {Bader}, {Baker},
  {Baldaccini}, {Ballardin}, {Ballmer}, {Barayoga}, {Barclay}, {Barish},
  {Barker}, {Barone}, {Barr}, {Barsotti}, {Barsuglia}, {Barta}, {Bartlett},
  {Bartos}, {Bassiri}, {Basti}, {Batch}, {Baune}, {Bavigadda}, {Bazzan},
  {Behnke}, {Bejger}, {Bell}, {Bell}, {Berger}, {Bergman}, {Bergmann}, {Berry},
  {Bersanetti}, {Bertolini}, {Betzwieser}, {Bhagwat}, {Bhand are}, {Bilenko},
  {Billingsley}, {Birch}, {Birney}, {Biscans}, {Bisht}, {Bitossi}, {Biwer},
  {Bizouard}, {Blackburn}, {Blair}, {Blair}, {Blair}, {Bloemen}, {Bock},
  {Bodiya}, {Boer}, {Bogaert}, {Bogan}, {Bohe}, {Bojtos}, {Bond}, {Bondu},
  {Bonnand}, {Boom}, {Bork}, {Boschi}, {Bose}, {Bouffanais}, {Bozzi},
  {Bradaschia}, {Brady}, {Braginsky}, {Branchesi}, {Brau}, {Briant}, {Brillet},
  {Brinkmann}, {Brisson}, {Brockill}, {Brooks}, {Brown}, {Brown}, {Buchanan},
  {Buikema}, {Bulik}, {Bulten}, {Buonanno}, {Buskulic}, {Buy}, {Byer},
  {Cadonati}, {Cagnoli}, {Cahillane}, {Bustillo}, {Callister}, {Calloni},
  {Camp}, {Cannon}, {Cao}, {Capano}, {Capocasa}, {Carbognani}, {Caride},
  {Diaz}, {Casentini}, {Caudill}, {Cavagli{\`a}}, {Cavalier}, {Cavalieri},
  {Cella}, {Cepeda}, {Baiardi}, {Cerretani}, {Cesarini}, {Chakraborty},
  {Chalermsongsak}, {Chamberlin}, {Chan}, {Chao}, {Charlton},
  {Chassande-Mottin}, {Chen}, {Chen}, {Cheng}, {Chincarini}, {Chiummo}, {Cho},
  {Cho}, {Chow}, {Christensen}, {Chu}, {Chua}, {Chung}, {Ciani}, {Clara},
  {Clark}, {Cleva}, {Coccia}, {Cohadon}, {Colla}, {Collette}, {Cominsky},
  {Constancio}, {Conte}, {Conti}, {Cook}, {Corbitt}, {Cornish}, {Corsi},
  {Cortese}, {Costa}, {Coughlin}, {Coughlin}, {Coulon}, {Countryman},
  {Couvares}, {Cowan}, {Coward}, {Cowart}, {Coyne}, {Coyne}, {Craig},
  {Creighton}, {Cripe}, {Crowder}, {Cumming}, {Cunningham}, {Cuoco}, {Dal
  Canton}, {Danilishin}, {D'Antonio}, {Danzmann}, {Darman}, {Dattilo}, {Dave},
  {Daveloza}, {Davier}, {Davies}, {Daw}, {Day}, {DeBra}, {Debreczeni},
  {Degallaix}, {De Laurentis}, {Del{\'e}glise}, {Del Pozzo}, {Denker}, {Dent},
  {Dereli}, {Dergachev}, {DeRosa}, {De Rosa}, {DeSalvo}, {Dhurandhar},
  {D{\'\i}az}, {Di Fiore}, {Di Giovanni}, {Di Lieto}, {Di Pace}, {Di Palma},
  {Di Virgilio}, {Dojcinoski}, {Dolique}, {Donovan}, {Dooley}, {Doravari},
  {Douglas}, {Downes}, {Drago}, {Drever}, {Driggers}, {Du}, {Ducrot}, {Dwyer},
  {Edo}, {Edwards}, {Effler}, {Eggenstein}, {Ehrens}, {Eichholz}, {Eikenberry},
  {Engels}, {Essick}, {Etzel}, {Evans}, {Evans}, {Everett}, {Factourovich},
  {Fafone}, {Fair}, {Fairhurst}, {Fan}, {Fang}, {Farinon}, {Farr}, {Farr},
  {Favata}, {Fays}, {Fehrmann}, {Fejer}, {Ferrante}, {Ferreira}, {Ferrini},
  {Fidecaro}, {Fiori}, {Fiorucci}, {Fisher}, {Flaminio}, {Fletcher},
  {Fournier}, {Franco}, {Frasca}, {Frasconi}, {Frei}, {Freise}, {Frey}, {Frey},
  {Fricke}, {Fritschel}, {Frolov}, {Fulda}, {Fyffe}, {Gabbard}, {Gair},
  {Gammaitoni}, {Gaonkar}, {Garufi}, {Gatto}, {Gaur}, {Gehrels}, {Gemme},
  {Gendre}, {Genin}, {Gennai}, {George}, {Gergely}, {Germain}, {Ghosh},
  {Ghosh}, {Giaime}, {Giardina}, {Giazotto}, {Gill}, {Glaefke}, {Goetz},
  {Goetz}, {Gondan}, {Gonz{\'a}lez}, {Castro}, {Gopakumar}, {Gordon},
  {Gorodetsky}, {Gossan}, {Gosselin}, {Gouaty}, {Graef}, {Graff}, {Granata},
  {Grant}, {Gras}, {Gray}, {Greco}, {Green}, {Groot}, {Grote}, {Grunewald},
  {Guidi}, {Guo}, {Gupta}, {Gupta}, {Gushwa}, {Gustafson}, {Gustafson},
  {Hacker}, {Hall}, {Hall}, {Hammond}, {Haney}, {Hanke}, {Hanks}, {Hanna},
  {Hannam}, {Hanson}, {Hardwick}, {Haris}, {Harms}, {Harry}, {Harry}, {Hart},
  {Hartman}, {Haster}, {Haughian}, {Heidmann}, {Heintze}, {Heitmann}, {Hello},
  {Hemming}, {Hendry}, {Heng}, {Hennig}, {Heptonstall}, {Heurs}, {Hild},
  {Hoak}, {Hodge}, {Hofman}, {Hollitt}, {Holt}, {Holz}, {Hopkins}, {Hosken},
  {Hough}, {Houston}, {Howell}, {Hu}, {Huang}, {Huerta}, {Huet}, {Hughey},
  {Husa}, {Huttner}, {Huynh-Dinh}, {Idrisy}, {Indik}, {Ingram}, {Inta}, {Isa},
  {Isac}, {Isi}, {Islas}, {Isogai}, {Iyer}, {Izumi}, {Jacqmin}, {Jang}, {Jani},
  {Jaranowski}, {Jawahar}, {Jim{\'e}nez-Forteza}, {Johnson}, {Jones}, {Jones},
  {Jonker}, {Ju}, {Kalaghatgi}, {Kalogera}, {Kandhasamy}, {Kang}, {Kanner},
  {Karki}, {Kasprzack}, {Katsavounidis}, {Katzman}, {Kaufer}, {Kaur}, {Kawabe},
  {Kawazoe}, {K{\'e}f{\'e}lian}, {Kehl}, {Keitel}, {Kelley}, {Kells},
  {Kennedy}, {Key}, {Khalaidovski}, {Khalili}, {Khan}, {Khan}, {Khan},
  {Khazanov}, {Kijbunchoo}, {Kim}, {Kim}, {Kim}, {Kim}, {Kim}, {Kim}, {King},
  {King}, {Kinzel}, {Kissel}, {Kleybolte}, {Klimenko}, {Koehlenbeck},
  {Kokeyama}, {Koley}, {Kondrashov}, {Kontos}, {Korobko}, {Korth}, {Kowalska},
  {Kozak}, {Kringel}, {Kr{\'o}lak}, {Krueger}, {Kuehn}, {Kumar}, {Kuo},
  {Kutynia}, {Lackey}, {Land ry}, {Lange}, {Lantz}, {Lasky}, {Lazzarini},
  {Lazzaro}, {Leaci}, {Leavey}, {Lebigot}, {Lee}, {Lee}, {Lee}, {Lee}, {Lenon},
  {Leonardi}, {Leong}, {Leroy}, {Letendre}, {Levin}, {Levine}, {Li}, {Libson},
  {Littenberg}, {Lockerbie}, {Logue}, {Lombardi}, {Lord}, {Lorenzini},
  {Loriette}, {Lormand}, {Losurdo}, {Lough}, {L{\"u}ck}, {Lundgren}, {Luo},
  {Lynch}, {Ma}, {MacDonald}, {Machenschalk}, {MacInnis}, {Macleod},
  {Maga{\~n}a-Sandoval}, {Magee}, {Mageswaran}, {Majorana}, {Maksimovic},
  {Malvezzi}, {Man}, {Mandel}, {Mandic}, {Mangano}, {Mansell}, {Manske},
  {Mantovani}, {Marchesoni}, {Marion}, {M{\'a}rka}, {M{\'a}rka}, {Markosyan},
  {Maros}, {Martelli}, {Martellini}, {Martin}, {Martin}, {Martynov}, {Marx},
  {Mason}, {Masserot}, {Massinger}, {Masso-Reid}, {Matichard}, {Matone},
  {Mavalvala}, {Mazumder}, {Mazzolo}, {McCarthy}, {McClelland}, {McCormick},
  {McGuire}, {McIntyre}, {McIver}, {McManus}, {McWilliams}, {Meacher},
  {Meadors}, {Meidam}, {Melatos}, {Mendell}, {Mendoza-Gandara}, {Mercer},
  {Merilh}, {Merzougui}, {Meshkov}, {Messenger}, {Messick}, {Meyers},
  {Mezzani}, {Miao}, {Michel}, {Middleton}, {Mikhailov}, {Milano}, {Miller},
  {Millhouse}, {Minenkov}, {Ming}, {Mirshekari}, {Mishra}, {Mitra},
  {Mitrofanov}, {Mitselmakher}, {Mittleman}, {Moggi}, {Mohan}, {Mohapatra},
  {Montani}, {Moore}, {Moore}, {Moraru}, {Moreno}, {Morriss}, {Mossavi},
  {Mours}, {Mow-Lowry}, {Mueller}, {Mueller}, {Muir}, {Mukherjee}, {Mukherjee},
  {Mukherjee}, {Mukund}, {Mullavey}, {Munch}, {Murphy}, {Murray}, {Mytidis},
  {Nardecchia}, {Naticchioni}, {Nayak}, {Necula}, {Nedkova}, {Nelemans},
  {Neri}, {Neunzert}, {Newton}, {Nguyen}, {Nielsen}, {Nissanke}, {Nitz},
  {Nocera}, {Nolting}, {Normandin}, {Nuttall}, {Oberling}, {Ochsner}, {O'Dell},
  {Oelker}, {Ogin}, {Oh}, {Oh}, {Ohme}, {Oliver}, {Oppermann}, {Oram},
  {O'Reilly}, {O'Shaughnessy}, {Ottaway}, {Ottens}, {Overmier}, {Owen}, {Pai},
  {Pai}, {Palamos}, {Palashov}, {Palomba}, {Pal-Singh}, {Pan}, {Pankow},
  {Pannarale}, {Pant}, {Paoletti}, {Paoli}, {Papa}, {Paris}, {Parker},
  {Pascucci}, {Pasqualetti}, {Passaquieti}, {Passuello}, {Patricelli},
  {Patrick}, {Pearlstone}, {Pedraza}, {Pedurand}, {Pekowsky}, {Pele}, {Penn},
  {Perreca}, {Phelps}, {Piccinni}, {Pichot}, {Piergiovanni}, {Pierro},
  {Pillant}, {Pinard}, {Pinto}, {Pitkin}, {Poggiani}, {Popolizio}, {Post},
  {Powell}, {Prasad}, {Predoi}, {Premachandra}, {Prestegard}, {Price},
  {Prijatelj}, {Principe}, {Privitera}, {Prodi}, {Prokhorov}, {Puncken},
  {Punturo}, {Puppo}, {P{\"u}rrer}, {Qi}, {Qin}, {Quetschke}, {Quintero},
  {Quitzow-James}, {Raab}, {Rabeling}, {Radkins}, {Raffai}, {Raja},
  {Rakhmanov}, {Rapagnani}, {Raymond}, {Razzano}, {Re}, {Read}, {Reed},
  {Regimbau}, {Rei}, {Reid}, {Reitze}, {Rew}, {Reyes}, {Ricci}, {Riles},
  {Robertson}, {Robie}, {Robinet}, {Rocchi}, {Rolland}, {Rollins}, {Roma},
  {Romano}, {Romano}, {Romanov}, {Romie}, {Rosi{\'n}ska}, {Rowan},
  {R{\"u}diger}, {Ruggi}, {Ryan}, {Sachdev}, {Sadecki}, {Sadeghian}, {Salconi},
  {Saleem}, {Salemi}, {Samajdar}, {Sammut}, {Sanchez}, {Sandberg}, {Sandeen},
  {Sanders}, {Sassolas}, {Sathyaprakash}, {Saulson}, {Sauter}, {Savage},
  {Sawadsky}, {Schale}, {Schilling}, {Schmidt}, {Schmidt}, {Schnabel},
  {Schofield}, {Sch{\"o}nbeck}, {Schreiber}, {Schuette}, {Schutz}, {Scott},
  {Scott}, {Sellers}, {Sentenac}, {Sequino}, {Sergeev}, {Serna}, {Setyawati},
  {Sevigny}, {Shaddock}, {Shah}, {Shahriar}, {Shaltev}, {Shao}, {Shapiro},
  {Shawhan}, {Sheperd}, {Shoemaker}, {Shoemaker}, {Siellez}, {Siemens}, {Sigg},
  {Silva}, {Simakov}, {Singer}, {Singer}, {Singh}, {Singh}, {Singhal},
  {Sintes}, {Slagmolen}, {Smith}, {Smith}, {Smith}, {Son}, {Sorazu},
  {Sorrentino}, {Souradeep}, {Srivastava}, {Staley}, {Steinke}, {Steinlechner},
  {Steinlechner}, {Steinmeyer}, {Stephens}, {Stone}, {Strain}, {Straniero},
  {Stratta}, {Strauss}, {Strigin}, {Sturani}, {Stuver}, {Summerscales}, {Sun},
  {Sutton}, {Swinkels}, {Szczepa{\'n}czyk}, {Tacca}, {Talukder}, {Tanner},
  {T{\'a}pai}, {Tarabrin}, {Taracchini}, {Taylor}, {Theeg},
  {Thirugnanasambandam}, {Thomas}, {Thomas}, {Thomas}, {Thorne}, {Thorne},
  {Thrane}, {Tiwari}, {Tiwari}, {Tokmakov}, {Tomlinson}, {Tonelli}, {Torres},
  {Torrie}, {T{\"o}yr{\"a}}, {Travasso}, {Traylor}, {Trifir{\`o}}, {Tringali},
  {Trozzo}, {Tse}, {Turconi}, {Tuyenbayev}, {Ugolini}, {Unnikrishnan}, {Urban},
  {Usman}, {Vahlbruch}, {Vajente}, {Valdes}, {van Bakel}, {van Beuzekom}, {van
  den Brand}, {Van Den Broeck}, {Vander-Hyde}, {van der Schaaf}, {van
  Heijningen}, {van Veggel}, {Vardaro}, {Vass}, {Vas{\'u}th}, {Vaulin},
  {Vecchio}, {Vedovato}, {Veitch}, {Veitch}, {Venkateswara}, {Verkindt},
  {Vetrano}, {Vicer{\'e}}, {Vinciguerra}, {Vine}, {Vinet}, {Vitale}, {Vo},
  {Vocca}, {Vorvick}, {Voss}, {Vousden}, {Vyatchanin}, {Wade}, {Wade}, {Wade},
  {Walker}, {Wallace}, {Walsh}, {Wang}, {Wang}, {Wang}, {Wang}, {Wang}, {Ward},
  {Warner}, {Was}, {Weaver}, {Wei}, {Weinert}, {Weinstein}, {Weiss}, {Welborn},
  {Wen}, {We{\ss}els}, {Westphal}, {Wette}, {Whelan}, {White}, {Whiting},
  {Williams}, {Williamson}, {Willis}, {Willke}, {Wimmer}, {Winkler}, {Wipf},
  {Wittel}, {Woan}, {Worden}, {Wright}, {Wu}, {Yablon}, {Yam}, {Yamamoto},
  {Yancey}, {Yap}, {Yu}, {Yvert}, {Zadrozny}, {Zangrando}, {Zanolin}, {Zendri},
  {Zevin}, {Zhang}, {Zhang}, {Zhang}, {Zhang}, {Zhao}, {Zhou}, {Zhou}, {Zhu},
  {Zucker}, {Zuraw}, {Zweizig}, {LIGO Scientific Collaboration}, and {Virgo
  Collaboration}}]{2016PhRvL.116m1102A}
{Abbott} BP, {Abbott} R, {Abbott} TD, {Abernathy} MR, {Acernese} F, {Ackley} K,
  {Adams} C, {Adams} T, {Addesso} P, {Adhikari} RX, et~al. (2016{\natexlab{a}})
  {GW150914: Implications for the Stochastic Gravitational-Wave Background from
  Binary Black Holes}. \prl 116(13):131102.
  \doi{10.1103/PhysRevLett.116.131102}.
  {\href{https://arxiv.org/abs/1602.03847}{{arXiv:1602.03847}}} {[gr-qc]}

\bibitem[{{Abbott} et~al.(2016{\natexlab{b}}){Abbott}, {Abbott}, {Abbott},
  {Abernathy}, {Acernese}, {Ackley}, {Adams}, {Adams}, {Addesso}, {Adhikari},
  {Adya}, {Affeldt}, {Agathos}, {Agatsuma}, {Aggarwal}, {Aguiar}, {Aiello},
  {Ain}, {Ajith}, {Allen}, {Allocca}, {Altin}, {Anderson}, {Anderson}, {Arai},
  {Araya}, {Arceneaux}, {Areeda}, {Arnaud}, {Arun}, {Ascenzi}, {Ashton}, {Ast},
  {Aston}, {Astone}, {Aufmuth}, {Aulbert}, {Babak}, {Bacon}, {Bader}, {Baker},
  {Baldaccini}, {Ballardin}, {Ballmer}, {Barayoga}, {Barclay}, {Barish},
  {Barker}, {Barone}, {Barr}, {Barsotti}, {Barsuglia}, {Barta}, {Bartlett},
  {Bartos}, {Bassiri}, {Basti}, {Batch}, {Baune}, {Bavigadda}, {Bazzan},
  {Behnke}, {Bejger}, {Bell}, {Bell}, {Berger}, {Bergman}, {Bergmann}, {Berry},
  {Bersanetti}, {Bertolini}, {Betzwieser}, {Bhagwat}, {Bhand are}, {Bilenko},
  {Billingsley}, {Birch}, {Birney}, {Biscans}, {Bisht}, {Bitossi}, {Biwer},
  {Bizouard}, {Blackburn}, {Blair}, {Blair}, {Blair}, {Bloemen}, {Bock},
  {Bodiya}, {Boer}, {Bogaert}, {Bogan}, {Bohe}, {Boh{\'e}mier}, {Bojtos},
  {Bond}, {Bondu}, {Bonnand}, {Boom}, {Bork}, {Boschi}, {Bose}, {Bouffanais},
  {Bozzi}, {Bradaschia}, {Brady}, {Braginsky}, {Branchesi}, {Brau}, {Briant},
  {Brillet}, {Brinkmann}, {Brisson}, {Brockill}, {Brooks}, {Brown}, {Brown},
  {Brown}, {Buchanan}, {Buikema}, {Bulik}, {Bulten}, {Buonanno}, {Buskulic},
  {Buy}, {Byer}, {Cabero}, {Cadonati}, {Cagnoli}, {Cahillane}, {Calder{\'o}n
  Bustillo}, {Callister}, {Calloni}, {Camp}, {Cannon}, {Cao}, {Capano},
  {Capocasa}, {Carbognani}, {Caride}, {Casanueva Diaz}, {Casentini}, {Caudill},
  {Cavagli{\`a}}, {Cavalier}, {Cavalieri}, {Cella}, {Cepeda}, {Cerboni
  Baiardi}, {Cerretani}, {Cesarini}, {Chakraborty}, {Chalermsongsak},
  {Chamberlin}, {Chan}, {Chao}, {Charlton}, {Chassande-Mottin}, {Chen}, {Chen},
  {Cheng}, {Chincarini}, {Chiummo}, {Cho}, {Cho}, {Chow}, {Christensen}, {Chu},
  {Chua}, {Chung}, {Ciani}, {Clara}, {Clark}, {Clayton}, {Cleva}, {Coccia},
  {Cohadon}, {Cokelaer}, {Colla}, {Collette}, {Cominsky}, {Constancio},
  {Conte}, {Conti}, {Cook}, {Corbitt}, {Cornish}, {Corsi}, {Cortese}, {Costa},
  {Coughlin}, {Coughlin}, {Coulon}, {Countryman}, {Couvares}, {Cowan},
  {Coward}, {Cowart}, {Coyne}, {Coyne}, {Craig}, {Creighton}, {Creighton},
  {Cripe}, {Crowder}, {Cumming}, {Cunningham}, {Cuoco}, {Dal Canton},
  {Danilishin}, {D'Antonio}, {Danzmann}, {Darman}, {Dattilo}, {Dave},
  {Daveloza}, {Davier}, {Davies}, {Daw}, {Day}, {De}, {DeBra}, {Debreczeni},
  {Degallaix}, {De Laurentis}, {Del{\'e}glise}, {Del Pozzo}, {Denker}, {Dent},
  {Dereli}, {Dergachev}, {DeRosa}, {De Rosa}, {DeSalvo}, {Dhurand har},
  {D{\'\i}az}, {Dietz}, {Di Fiore}, {Di Giovanni}, {Di Lieto}, {Di Pace}, {Di
  Palma}, {Di Virgilio}, {Dojcinoski}, {Dolique}, {Donovan}, {Dooley},
  {Doravari}, {Douglas}, {Downes}, {Drago}, {Drever}, {Driggers}, {Du},
  {Ducrot}, {Dwyer}, {Edo}, {Edwards}, {Effler}, {Eggenstein}, {Ehrens},
  {Eichholz}, {Eikenberry}, {Engels}, {Essick}, {Etzel}, {Evans}, {Evans},
  {Everett}, {Factourovich}, {Fafone}, {Fair}, {Fairhurst}, {Fan}, {Fang},
  {Farinon}, {Farr}, {Farr}, {Favata}, {Fays}, {Fehrmann}, {Fejer}, {Ferrante},
  {Ferreira}, {Ferrini}, {Fidecaro}, {Fiori}, {Fiorucci}, {Fisher}, {Flaminio},
  {Fletcher}, {Fotopoulos}, {Fournier}, {Franco}, {Frasca}, {Frasconi}, {Frei},
  {Frei}, {Freise}, {Frey}, {Frey}, {Fricke}, {Fritschel}, {Frolov}, {Fulda},
  {Fyffe}, {Gabbard}, {Gair}, {Gammaitoni}, {Gaonkar}, {Garufi}, {Gatto},
  {Gaur}, {Gehrels}, {Gemme}, {Gendre}, {Genin}, {Gennai}, {George}, {Gergely},
  {Germain}, {Ghosh}, {Ghosh}, {Giaime}, {Giardina}, {Giazotto}, {Gill},
  {Glaefke}, {Goetz}, {Goetz}, {Goggin}, {Gondan}, {Gonz{\'a}lez}, {Gonzalez
  Castro}, {Gopakumar}, {Gordon}, {Gorodetsky}, {Gossan}, {Gosselin}, {Gouaty},
  {Graef}, {Graff}, {Granata}, {Grant}, {Gras}, {Gray}, {Greco}, {Green},
  {Groot}, {Grote}, {Grunewald}, {Guidi}, {Guo}, {Gupta}, {Gupta}, {Gushwa},
  {Gustafson}, {Gustafson}, {Hacker}, {Hall}, {Hall}, {Hammond}, {Haney},
  {Hanke}, {Hanks}, {Hanna}, {Hannam}, {Hanson}, {Hardwick}, {Harms}, {Harry},
  {Harry}, {Hart}, {Hartman}, {Haster}, {Haughian}, {Heidmann}, {Heintze},
  {Heitmann}, {Hello}, {Hemming}, {Hendry}, {Heng}, {Hennig}, {Heptonstall},
  {Heurs}, {Hild}, {Hoak}, {Hodge}, {Hofman}, {Hollitt}, {Holt}, {Holz},
  {Hopkins}, {Hosken}, {Hough}, {Houston}, {Howell}, {Hu}, {Huang}, {Huerta},
  {Huet}, {Hughey}, {Husa}, {Huttner}, {Huynh-Dinh}, {Idrisy}, {Indik},
  {Ingram}, {Inta}, {Isa}, {Isac}, {Isi}, {Islas}, {Isogai}, {Iyer}, {Izumi},
  {Jacqmin}, {Jang}, {Jani}, {Jaranowski}, {Jawahar}, {Jim{\'e}nez-Forteza},
  {Johnson}, {Jones}, {Jones}, {Jones}, {Jonker}, {Ju}, {Haris}, {Kalaghatgi},
  {Kalogera}, {Kandhasamy}, {Kang}, {Kanner}, {Karki}, {Kasprzack},
  {Katsavounidis}, {Katzman}, {Kaufer}, {Kaur}, {Kawabe}, {Kawazoe},
  {K{\'e}f{\'e}lian}, {Kehl}, {Keitel}, {Kelley}, {Kells}, {Keppel}, {Kennedy},
  {Key}, {Khalaidovski}, {Khalili}, {Khan}, {Khan}, {Khan}, {Khazanov},
  {Kijbunchoo}, {Kim}, {Kim}, {Kim}, {Kim}, {Kim}, {Kim}, {King}, {King},
  {Kinzel}, {Kissel}, {Kleybolte}, {Klimenko}, {Koehlenbeck}, {Kokeyama},
  {Koley}, {Kondrashov}, {Kontos}, {Korobko}, {Korth}, {Kowalska}, {Kozak},
  {Kringel}, {Krishnan}, {Kr{\'o}lak}, {Krueger}, {Kuehn}, {Kumar}, {Kuo},
  {Kutynia}, {Lackey}, {Landry}, {Lange}, {Lantz}, {Lasky}, {Lazzarini},
  {Lazzaro}, {Leaci}, {Leavey}, {Lebigot}, {Lee}, {Lee}, {Lee}, {Lee}, {Lenon},
  {Leonardi}, {Leong}, {Leroy}, {Letendre}, {Levin}, {Levine}, {Li}, {Libson},
  {Littenberg}, {Lockerbie}, {Logue}, {Lombardi}, {Lord}, {Lorenzini},
  {Loriette}, {Lormand}, {Losurdo}, {Lough}, {L{\"u}ck}, {Lundgren}, {Luo},
  {Lynch}, {Ma}, {MacDonald}, {Machenschalk}, {MacInnis}, {Macleod},
  {Maga{\~n}a-Sandoval}, {Magee}, {Mageswaran}, {Majorana}, {Maksimovic},
  {Malvezzi}, {Man}, {Mandel}, {Mandic}, {Mangano}, {Mansell}, {Manske},
  {Mantovani}, {Marchesoni}, {Marion}, {M{\'a}rka}, {M{\'a}rka}, {Markosyan},
  {Maros}, {Martelli}, {Martellini}, {Martin}, {Martin}, {Martynov}, {Marx},
  {Mason}, {Masserot}, {Massinger}, {Masso-Reid}, {Matichard}, {Matone},
  {Mavalvala}, {Mazumder}, {Mazzolo}, {McCarthy}, {McClelland}, {McCormick},
  {McGuire}, {McIntyre}, {McIver}, {McKechan}, {McManus}, {McWilliams},
  {Meacher}, {Meadors}, {Meidam}, {Melatos}, {Mendell}, {Mendoza-Gandara},
  {Mercer}, {Merilh}, {Merzougui}, {Meshkov}, {Messaritaki}, {Messenger},
  {Messick}, {Meyers}, {Mezzani}, {Miao}, {Michel}, {Middleton}, {Mikhailov},
  {Milano}, {Miller}, {Millhouse}, {Minenkov}, {Ming}, {Mirshekari}, {Mishra},
  {Mitra}, {Mitrofanov}, {Mitselmakher}, {Mittleman}, {Moggi}, {Mohan},
  {Mohapatra}, {Montani}, {Moore}, {Moore}, {Moraru}, {Moreno}, {Morriss},
  {Mossavi}, {Mours}, {Mow-Lowry}, {Mueller}, {Mueller}, {Muir}, {Mukherjee},
  {Mukherjee}, {Mukherjee}, {Mukund}, {Mullavey}, {Munch}, {Murphy}, {Murray},
  {Mytidis}, {Nardecchia}, {Naticchioni}, {Nayak}, {Necula}, {Nedkova},
  {Nelemans}, {Neri}, {Neunzert}, {Newton}, {Nguyen}, {Nielsen}, {Nissanke},
  {Nitz}, {Nocera}, {Nolting}, {Normandin}, {Nuttall}, {Oberling}, {Ochsner},
  {O'Dell}, {Oelker}, {Ogin}, {Oh}, {Oh}, {Ohme}, {Oliver}, {Oppermann},
  {Oram}, {O'Reilly}, {O'Shaughnessy}, {Ottaway}, {Ottens}, {Overmier}, {Owen},
  {Pai}, {Pai}, {Palamos}, {Palashov}, {Palomba}, {Pal-Singh}, {Pan}, {Pan},
  {Pankow}, {Pannarale}, {Pant}, {Paoletti}, {Paoli}, {Papa}, {Paris},
  {Parker}, {Pascucci}, {Pasqualetti}, {Passaquieti}, {Passuello},
  {Patricelli}, {Patrick}, {Pearlstone}, {Pedraza}, {Pedurand}, {Pekowsky},
  {Pele}, {Penn}, {Perreca}, {Phelps}, {Piccinni}, {Pichot}, {Piergiovanni},
  {Pierro}, {Pillant}, {Pinard}, {Pinto}, {Pitkin}, {Poggiani}, {Popolizio},
  {Post}, {Powell}, {Prasad}, {Predoi}, {Premachandra}, {Prestegard}, {Price},
  {Prijatelj}, {Principe}, {Privitera}, {Prodi}, {Prokhorov}, {Puncken},
  {Punturo}, {Puppo}, {P{\"u}rrer}, {Qi}, {Qin}, {Quetschke}, {Quintero},
  {Quitzow-James}, {Raab}, {Rabeling}, {Radkins}, {Raffai}, {Raja},
  {Rakhmanov}, {Rapagnani}, {Raymond}, {Razzano}, {Re}, {Read}, {Reed},
  {Regimbau}, {Rei}, {Reid}, {Reitze}, {Rew}, {Reyes}, {Ricci}, {Riles},
  {Robertson}, {Robie}, {Robinet}, {Robinson}, {Rocchi}, {Rodriguez},
  {Rolland}, {Rollins}, {Roma}, {Romano}, {Romanov}, {Romie}, {Rosi{\'n}ska},
  {Rowan}, {R{\"u}diger}, {Ruggi}, {Ryan}, {Sachdev}, {Sadecki}, {Sadeghian},
  {Salconi}, {Saleem}, {Salemi}, {Samajdar}, {Sammut}, {Sanchez}, {Sand berg},
  {Sandeen}, {Sanders}, {Santamar{\'\i}a}, {Sassolas}, {Sathyaprakash},
  {Saulson}, {Sauter}, {Savage}, {Sawadsky}, {Schale}, {Schilling}, {Schmidt},
  {Schmidt}, {Schnabel}, {Schofield}, {Sch{\"o}nbeck}, {Schreiber}, {Schuette},
  {Schutz}, {Scott}, {Scott}, {Sellers}, {Sengupta}, {Sentenac}, {Sequino},
  {Sergeev}, {Serna}, {Setyawati}, {Sevigny}, {Shaddock}, {Shah}, {Shahriar},
  {Shaltev}, {Shao}, {Shapiro}, {Shawhan}, {Sheperd}, {Shoemaker}, {Shoemaker},
  {Siellez}, {Siemens}, {Sigg}, {Silva}, {Simakov}, {Singer}, {Singer},
  {Singh}, {Singh}, {Singhal}, {Sintes}, {Slagmolen}, {Smith}, {Smith},
  {Smith}, {Son}, {Sorazu}, {Sorrentino}, {Souradeep}, {Srivastava}, {Staley},
  {Steinke}, {Steinlechner}, {Steinlechner}, {Steinmeyer}, {Stephens}, {Stone},
  {Strain}, {Straniero}, {Stratta}, {Strauss}, {Strigin}, {Sturani}, {Stuver},
  {Summerscales}, {Sun}, {Sutton}, {Swinkels}, {Szczepa{\'n}czyk}, {Tacca},
  {Talukder}, {Tanner}, {T{\'a}pai}, {Tarabrin}, {Taracchini}, {Taylor},
  {Theeg}, {Thirugnanasambandam}, {Thomas}, {Thomas}, {Thomas}, {Thorne},
  {Thorne}, {Thrane}, {Tiwari}, {Tiwari}, {Tokmakov}, {Tomlinson}, {Tonelli},
  {Torres}, {Torrie}, {T{\"o}yr{\"a}}, {Travasso}, {Traylor}, {Trifir{\`o}},
  {Tringali}, {Trozzo}, {Tse}, {Turconi}, {Tuyenbayev}, {Ugolini},
  {Unnikrishnan}, {Urban}, {Usman}, {Vahlbruch}, {Vajente}, {Valdes}, {van
  Bakel}, {van Beuzekom}, {van den Brand}, {Van Den Broeck}, {Vander-Hyde},
  {van der Schaaf}, {van Heijningen}, {van Veggel}, {Vardaro}, {Vass},
  {Vas{\'u}th}, {Vaulin}, {Vecchio}, {Vedovato}, {Veitch}, {Veitch},
  {Venkateswara}, {Verkindt}, {Vetrano}, {Vicer{\'e}}, {Vinciguerra}, {Vine},
  {Vinet}, {Vitale}, {Vo}, {Vocca}, {Vorvick}, {Voss}, {Vousden}, {Vyatchanin},
  {Wade}, {Wade}, {Wade}, {Walker}, {Wallace}, {Walsh}, {Wang}, {Wang}, {Wang},
  {Wang}, {Wang}, {Ward}, {Warner}, {Was}, {Weaver}, {Wei}, {Weinert},
  {Weinstein}, {Weiss}, {Welborn}, {Wen}, {We{\ss}els}, {West}, {Westphal},
  {Wette}, {Whelan}, {White}, {Whiting}, {Wiesner}, {Williams}, {Williamson},
  {Willis}, {Willke}, {Wimmer}, {Winkler}, {Wipf}, {Wiseman}, {Wittel}, {Woan},
  {Worden}, {Wright}, {Wu}, {Yablon}, {Yam}, {Yamamoto}, {Yancey}, {Yap}, {Yu},
  {Yvert}, {Zadrozny}, {Zangrando}, {Zanolin}, {Zendri}, {Zevin}, {Zhang},
  {Zhang}, {Zhang}, {Zhang}, {Zhao}, {Zhou}, {Zhou}, {Zhu}, {Zucker}, {Zuraw},
  {Zweizig}, {LIGO Scientific Collaboration}, and {Virgo
  Collaboration}}]{2016PhRvD..93l2003A}
{Abbott} BP, {Abbott} R, {Abbott} TD, {Abernathy} MR, {Acernese} F, {Ackley} K,
  {Adams} C, {Adams} T, {Addesso} P, {Adhikari} RX, et~al. (2016{\natexlab{b}})
  {GW150914: First results from the search for binary black hole coalescence
  with Advanced LIGO}. \prd 93(12):122003. \doi{10.1103/PhysRevD.93.122003}.
  {\href{https://arxiv.org/abs/1602.03839}{{arXiv:1602.03839}}} {[gr-qc]}

\bibitem[{{Abbott} et~al.(2016{\natexlab{c}}){Abbott}, {Abbott}, {Abbott},
  {Abernathy}, {Acernese}, {Ackley}, {Adams}, {Adams}, {Addesso}, {Adhikari},
  and et~al.}]{2016PhRvL.116f1102A}
{Abbott} BP, {Abbott} R, {Abbott} TD, {Abernathy} MR, {Acernese} F, {Ackley} K,
  {Adams} C, {Adams} T, {Addesso} P, {Adhikari} RX, et~al. (2016{\natexlab{c}})
  {Observation of Gravitational Waves from a Binary Black Hole Merger}. \prl
  116(6):061102. \doi{10.1103/PhysRevLett.116.061102}.
  {\href{https://arxiv.org/abs/1602.03837}{{arXiv:1602.03837}}} {[gr-qc]}

\bibitem[{{Abbott} et~al.(2017{\natexlab{a}}){Abbott}, {Abbott}, {Abbott},
  {Abernathy}, {Ackley}, {Adams}, {Addesso}, {Adhikari}, {Adya}, {Affeldt},
  {Aggarwal}, {Aguiar}, {Ain}, {Ajith}, {Allen}, {Altin}, {Anderson},
  {Anderson}, {Arai}, {Araya}, {Arceneaux}, {Areeda}, {Arun}, {Ashton}, {Ast},
  {Aston}, {Aufmuth}, {Aulbert}, {Babak}, {Baker}, {Ballmer}, {Barayoga},
  {Barclay}, {Barish}, {Barker}, {Barr}, {Barsotti}, {Bartlett}, {Bartos},
  {Bassiri}, {Batch}, {Baune}, {Bell}, {Berger}, {Bergmann}, {Berry},
  {Betzwieser}, {Bhagwat}, {Bhandare}, {Bilenko}, {Billingsley}, {Birch},
  {Birney}, {Biscans}, {Bisht}, {Biwer}, {Blackburn}, {Blair}, {Blair},
  {Blair}, {Bock}, {Bogan}, {Bohe}, {Bond}, {Bork}, {Bose}, {Brady},
  {Braginsky}, {Brau}, {Brinkmann}, {Brockill}, {Broida}, {Brooks}, {Brown},
  {Brown}, {Brown}, {Brunett}, {Buchanan}, {Buikema}, {Buonanno}, {Byer},
  {Cabero}, {Cadonati}, {Cahillane}, {Calder{\'o}n Bustillo}, {Callister},
  {Camp}, {Cannon}, {Cao}, {Capano}, {Caride}, {Caudill}, {Cavagli{\`a}},
  {Cepeda}, {Chamberlin}, {Chan}, {Chao}, {Charlton}, {Cheeseboro}, {Chen},
  {Chen}, {Cheng}, {Cho}, {Cho}, {Chow}, {Christensen}, {Chu}, {Chung},
  {Ciani}, {Clara}, {Clark}, {Collette}, {Cominsky}, {Constancio}, {Cook},
  {Corbitt}, {Cornish}, {Corsi}, {Costa}, {Coughlin}, {Coughlin}, {Countryman},
  {Couvares}, {Cowan}, {Coward}, {Cowart}, {Coyne}, {Coyne}, {Craig},
  {Creighton}, {Cripe}, {Crowder}, {Cumming}, {Cunningham}, {Dal Canton},
  {Danilishin}, {Danzmann}, {Darman}, {Dasgupta}, {Da Silva Costa}, {Dave},
  {Davies}, {Daw}, {De}, {DeBra}, {Del Pozzo}, {Denker}, {Dent}, {Dergachev},
  {DeRosa}, {DeSalvo}, {Devine}, {Dhurandhar}, {D{\'\i}az}, {Di Palma},
  {Donovan}, {Dooley}, {Doravari}, {Douglas}, {Downes}, {Drago}, {Drever},
  {Driggers}, {Dwyer}, {Edo}, {Edwards}, {Effler}, {Eggenstein}, {Ehrens},
  {Eichholz}, {Eikenberry}, {Engels}, {Essick}, {Etzel}, {Evans}, {Evans},
  {Everett}, {Factourovich}, {Fair}, {Fairhurst}, {Fan}, {Fang}, {Farr},
  {Farr}, {Favata}, {Fays}, {Fehrmann}, {Fejer}, {Fenyvesi}, {Ferreira},
  {Fisher}, {Fletcher}, {Frei}, {Freise}, {Frey}, {Fritschel}, {Frolov},
  {Fulda}, {Fyffe}, {Gabbard}, {Gair}, {Gaonkar}, {Gaur}, {Gehrels}, {Geng},
  {George}, {Gergely}, {Ghosh}, {Ghosh}, {Giaime}, {Giardina}, {Gill},
  {Glaefke}, {Goetz}, {Goetz}, {Gondan}, {Gonz{\'a}lez}, {Gopakumar}, {Gordon},
  {Gorodetsky}, {Gossan}, {Graef}, {Graff}, {Grant}, {Gras}, {Gray}, {Green},
  {Grote}, {Grunewald}, {Guo}, {Gupta}, {Gupta}, {Gushwa}, {Gustafson},
  {Gustafson}, {Hacker}, {Hall}, {Hall}, {Hammond}, {Haney}, {Hanke}, {Hanks},
  {Hanna}, {Hannam}, {Hanson}, {Hardwick}, {Harry}, {Harry}, {Hart}, {Hartman},
  {Haster}, {Haughian}, {Heintze}, {Hendry}, {Heng}, {Hennig}, {Henry},
  {Heptonstall}, {Heurs}, {Hild}, {Hoak}, {Holt}, {Holz}, {Hopkins}, {Hough},
  {Houston}, {Howell}, {Hu}, {Huang}, {Huerta}, {Hughey}, {Husa}, {Huttner},
  {Huynh-Dinh}, {Indik}, {Ingram}, {Inta}, {Isa}, {Isi}, {Isogai}, {Iyer},
  {Izumi}, {Jang}, {Jani}, {Jawahar}, {Jian}, {Jim{\'e}nez-Forteza}, {Johnson},
  {Jones}, {Jones}, {Ju}, {Haris}, {Kalaghatgi}, {Kalogera}, {Kandhasamy},
  {Kang}, {Kanner}, {Kapadia}, {Karki}, {Karvinen}, {Kasprzack},
  {Katsavounidis}, {Katzman}, {Kaufer}, {Kaur}, {Kawabe}, {Kehl}, {Keitel},
  {Kelley}, {Kells}, {Kennedy}, {Key}, {Khalili}, {Khan}, {Khan}, {Khazanov},
  {Kijbunchoo}, {Kim}, {Kim}, {Kim}, {Kim}, {Kim}, {Kim}, {Kim}, {Kimbrell},
  {King}, {King}, {Kissel}, {Klein}, {Kleybolte}, {Klimenko}, {Koehlenbeck},
  {Kondrashov}, {Kontos}, {Korobko}, {Korth}, {Kozak}, {Kringel}, {Krueger},
  {Kuehn}, {Kumar}, {Kumar}, {Kuo}, {Lackey}, {Landry}, {Lange}, {Lantz},
  {Lasky}, {Laxen}, {Lazzarini}, {Leavey}, {Lebigot}, {Lee}, {Lee}, {Lee},
  {Lee}, {Lenon}, {Leong}, {Levin}, {Lewis}, {Li}, {Libson}, {Littenberg},
  {Lockerbie}, {Lombardi}, {London}, {Lord}, {Lormand}, {Lough}, {L{\"u}ck},
  {Lundgren}, {Lynch}, {Ma}, {Machenschalk}, {MacInnis}, {Macleod},
  {Maga{\~n}a-Sandoval}, {Maga{\~n}a Zertuche}, {Magee}, {Mandic}, {Mangano},
  {Mansell}, {Manske}, {M{\'a}rka}, {M{\'a}rka}, {Markosyan}, {Maros},
  {Martin}, {Martynov}, {Mason}, {Massinger}, {Masso-Reid}, {Matichard},
  {Matone}, {Mavalvala}, {Mazumder}, {McCarthy}, {McClelland}, {McCormick},
  {McGuire}, {McIntyre}, {McIver}, {McManus}, {McRae}, {McWilliams}, {Meacher},
  {Meadors}, {Melatos}, {Mendell}, {Mercer}, {Merilh}, {Meshkov}, {Messenger},
  {Messick}, {Meyers}, {Miao}, {Middleton}, {Mikhailov}, {Miller}, {Miller},
  {Miller}, {Miller}, {Millhouse}, {Ming}, {Mirshekari}, {Mishra}, {Mitra},
  {Mitrofanov}, {Mitselmakher}, {Mittleman}, {Mohapatra}, {Moore}, {Moore},
  {Moraru}, {Moreno}, {Morriss}, {Mossavi}, {Mow-Lowry}, {Mueller}, {Muir},
  {Mukherjee}, {Mukherjee}, {Mukherjee}, {Mukund}, {Mullavey}, {Munch},
  {Murphy}, {Murray}, {Mytidis}, {Nayak}, {Nedkova}, {Nelson}, {Neunzert},
  {Newton}, {Nguyen}, {Nielsen}, {Nitz}, {Nolting}, {Normandin}, {Nuttall},
  {Oberling}, {Ochsner}, {O'Dell}, {Oelker}, {Ogin}, {Oh}, {Oh}, {Ohme},
  {Oliver}, {Oppermann}, {Oram}, {O'Reilly}, {O'Shaughnessy}, {Ottaway},
  {Overmier}, {Owen}, {Pai}, {Pai}, {Palamos}, {Palashov}, {Pal-Singh}, {Pan},
  {Pankow}, {Pannarale}, {Pant}, {Papa}, {Paris}, {Parker}, {Pascucci},
  {Patrick}, {Pearlstone}, {Pedraza}, {Pekowsky}, {Pele}, {Penn}, {Perreca},
  {Perri}, {Phelps}, {Pierro}, {Pinto}, {Pitkin}, {Poe}, {Post}, {Powell},
  {Prasad}, {Predoi}, {Prestegard}, {Price}, {Prijatelj}, {Principe},
  {Privitera}, {Prokhorov}, {Puncken}, {P{\"u}rrer}, {Qi}, {Qin}, {Qiu},
  {Quetschke}, {Quintero}, {Quitzow-James}, {Raab}, {Rabeling}, {Radkins},
  {Raffai}, {Raja}, {Rajan}, {Rakhmanov}, {Raymond}, {Read}, {Reed}, {Reid},
  {Reitze}, {Rew}, {Reyes}, {Riles}, {Rizzo}, {Robertson}, {Robie}, {Rollins},
  {Roma}, {Romanov}, {Romie}, {Rowan}, {R{\"u}diger}, {Ryan}, {Sachdev},
  {Sadecki}, {Sadeghian}, {Sakellariadou}, {Saleem}, {Salemi}, {Samajdar},
  {Sammut}, {Sanchez}, {Sandberg}, {Sandeen}, {Sanders}, {Sathyaprakash},
  {Saulson}, {Sauter}, {Savage}, {Sawadsky}, {Schale}, {Schilling}, {Schmidt},
  {Schmidt}, {Schnabel}, {Schofield}, {Sch{\"o}nbeck}, {Schreiber}, {Schuette},
  {Schutz}, {Scott}, {Scott}, {Sellers}, {Sengupta}, {Sergeev}, {Shaddock},
  {Shaffer}, {Shahriar}, {Shaltev}, {Shapiro}, {Shawhan}, {Sheperd},
  {Shoemaker}, {Shoemaker}, {Siellez}, {Siemens}, {Sigg}, {Silva}, {Singer},
  {Singer}, {Singh}, {Singh}, {Sintes}, {Slagmolen}, {Smith}, {Smith}, {Smith},
  {Son}, {Sorazu}, {Souradeep}, {Srivastava}, {Staley}, {Steinke},
  {Steinlechner}, {Steinlechner}, {Steinmeyer}, {Stephens}, {Stone}, {Strain},
  {Strauss}, {Strigin}, {Sturani}, {Stuver}, {Summerscales}, {Sun}, {Sunil},
  {Sutton}, {Szczepa{\'n}czyk}, {Talukder}, {Tanner}, {T{\'a}pai}, {Tarabrin},
  {Taracchini}, {Taylor}, {Theeg}, {Thirugnanasambandam}, {Thomas}, {Thomas},
  {Thomas}, {Thorne}, {Thrane}, {Tiwari}, {Tokmakov}, {Toland}, {Tomlinson},
  {Tornasi}, {Torres}, {Torrie}, {T{\"o}yr{\"a}}, {Traylor}, {Trifir{\`o}},
  {Tse}, {Tuyenbayev}, {Ugolini}, {Unnikrishnan}, {Urban}, {Usman},
  {Vahlbruch}, {Vajente}, {Valdes}, {Vander-Hyde}, {van Veggel}, {Vass},
  {Vaulin}, {Vecchio}, {Veitch}, {Veitch}, {Venkateswara}, {Vinciguerra},
  {Vine}, {Vitale}, {Vo}, {Vorvick}, {Voss}, {Vousden}, {Vyatchanin}, {Wade},
  {Wade}, {Wade}, {Walker}, {Wallace}, {Walsh}, {Wang}, {Wang}, {Wang}, {Wang},
  {Ward}, {Warner}, {Weaver}, {Weinert}, {Weinstein}, {Weiss}, {Wen},
  {We{\ss}els}, {Westphal}, {Wette}, {Whelan}, {Whiting}, {Williams},
  {Williamson}, {Willis}, {Willke}, {Wimmer}, {Winkler}, {Wipf}, {Wittel},
  {Woan}, {Woehler}, {Worden}, {Wright}, {Wu}, {Wu}, {Yablon}, {Yam},
  {Yamamoto}, {Yancey}, {Yu}, {Zanolin}, {Zevin}, {Zhang}, {Zhang}, {Zhang},
  {Zhao}, {Zhou}, {Zhou}, {Zhu}, {Zucker}, {Zuraw}, {Zweizig}, {(LIGO
  Scientific Collaboration}, and {Harms}}]{2017CQGra..34d4001A}
{Abbott} BP, {Abbott} R, {Abbott} TD, {Abernathy} MR, {Ackley} K, {Adams} C,
  {Addesso} P, {Adhikari} RX, {Adya} VB, {Affeldt} C, et~al.
  (2017{\natexlab{a}}) {Exploring the sensitivity of next generation
  gravitational wave detectors}. Classical and Quantum Gravity 34(4):044001.
  \doi{10.1088/1361-6382/aa51f4}.
  {\href{https://arxiv.org/abs/1607.08697}{{arXiv:1607.08697}}} {[astro-ph.IM]}

\bibitem[{{Abbott} et~al.(2017{\natexlab{b}}){Abbott}, {Abbott}, {Abbott},
  {Acernese}, {Ackley}, {Adams}, {Adams}, {Addesso}, {Adhikari}, {Adya},
  {Affeldt}, {Afrough}, {Agarwal}, {Agathos}, {Agatsuma}, {Aggarwal}, {Aguiar},
  {Aiello}, {Ain}, {Ajith}, {Allen}, {Allen}, {Allocca}, {Altin}, {Amato},
  {Ananyeva}, {Anderson}, {Anderson}, {Angelova}, {Antier}, {Appert}, {Arai},
  {Araya}, {Areeda}, {Arnaud}, {Arun}, {Ascenzi}, {Ashton}, {Ast}, {Aston},
  {Astone}, {Atallah}, {Aufmuth}, {Aulbert}, {AultONeal}, {Austin},
  {Avila-Alvarez}, {Babak}, {Bacon}, {Bader}, {Bae}, {Baker}, {Baldaccini},
  {Ballardin}, {Ballmer}, {Banagiri}, {Barayoga}, {Barclay}, {Barish},
  {Barker}, {Barkett}, {Barone}, {Barr}, {Barsotti}, {Barsuglia}, {Barta},
  {Barthelmy}, {Bartlett}, {Bartos}, {Bassiri}, {Basti}, {Batch}, {Bawaj},
  {Bayley}, {Bazzan}, {B{\'e}csy}, {Beer}, {Bejger}, {Belahcene}, {Bell},
  {Berger}, {Bergmann}, {Bero}, {Berry}, {Bersanetti}, {Bertolini},
  {Betzwieser}, {Bhagwat}, {Bhandare}, {Bilenko}, {Billingsley}, {Billman},
  {Birch}, {Birney}, {Birnholtz}, {Biscans}, {Biscoveanu}, {Bisht}, {Bitossi},
  {Biwer}, {Bizouard}, {Blackburn}, {Blackman}, {Blair}, {Blair}, {Blair},
  {Bloemen}, {Bock}, {Bode}, {Boer}, {Bogaert}, {Bohe}, {Bondu}, {Bonilla},
  {Bonnand}, {Boom}, {Bork}, {Boschi}, {Bose}, {Bossie}, {Bouffanais}, {Bozzi},
  {Bradaschia}, {Brady}, {Branchesi}, {Brau}, {Briant}, {Brillet}, {Brinkmann},
  {Brisson}, {Brockill}, {Broida}, {Brooks}, {Brown}, {Brown}, {Brunett},
  {Buchanan}, {Buikema}, {Bulik}, {Bulten}, {Buonanno}, {Buskulic}, {Buy},
  {Byer}, {Cabero}, {Cadonati}, {Cagnoli}, {Cahillane}, {Calder{\'o}n
  Bustillo}, {Callister}, {Calloni}, {Camp}, {Canepa}, {Canizares}, {Cannon},
  {Cao}, {Cao}, {Capano}, {Capocasa}, {Carbognani}, {Caride}, {Carney},
  {Casanueva Diaz}, {Casentini}, {Caudill}, {Cavagli{\`a}}, {Cavalier},
  {Cavalieri}, {Cella}, {Cepeda}, {Cerd{\'a}-Dur{\'a}n}, {Cerretani},
  {Cesarini}, {Chamberlin}, {Chan}, {Chao}, {Charlton}, {Chase},
  {Chassande-Mottin}, {Chatterjee}, {Chatziioannou}, {Cheeseboro}, {Chen},
  {Chen}, {Chen}, {Cheng}, {Chia}, {Chincarini}, {Chiummo}, {Chmiel}, {Cho},
  {Cho}, {Chow}, {Christensen}, {Chu}, {Chua}, {Chua}, {Chung}, {Chung},
  {Ciani}, {Ciolfi}, {Cirelli}, {Cirone}, {Clara}, {Clark}, {Clearwater},
  {Cleva}, {Cocchieri}, {Coccia}, {Cohadon}, {Cohen}, {Colla}, {Collette},
  {Cominsky}, {Constancio}, {Conti}, {Cooper}, {Corban}, {Corbitt},
  {Cordero-Carri{\'o}n}, {Corley}, {Cornish}, {Corsi}, {Cortese}, {Costa},
  {Coughlin}, {Coughlin}, {Coulon}, {Countryman}, {Couvares}, {Covas}, {Cowan},
  {Coward}, {Cowart}, {Coyne}, {Coyne}, {Creighton}, {Creighton}, {Cripe},
  {Crowder}, {Cullen}, {Cumming}, {Cunningham}, {Cuoco}, {Dal Canton},
  {D{\'a}lya}, {Danilishin}, {D'Antonio}, {Danzmann}, {Dasgupta}, {Da Silva
  Costa}, {Dattilo}, {Dave}, {Davier}, {Davis}, {Daw}, {Day}, {De}, {DeBra},
  {Degallaix}, {De Laurentis}, {Del{\'e}glise}, {Del Pozzo}, {Demos}, {Denker},
  {Dent}, {De Pietri}, {Dergachev}, {De Rosa}, {DeRosa}, {De Rossi}, {DeSalvo},
  {de Varona}, {Devenson}, {Dhurandhar}, {D{\'\i}az}, {Di Fiore}, {Di
  Giovanni}, {Di Girolamo}, {Di Lieto}, {Di Pace}, {Di Palma}, {Di Renzo},
  {Doctor}, {Dolique}, {Donovan}, {Dooley}, {Doravari}, {Dorrington},
  {Douglas}, {Dovale {\'A}lvarez}, {Downes}, {Drago}, {Dreissigacker},
  {Driggers}, {Du}, {Ducrot}, {Dupej}, {Dwyer}, {Edo}, {Edwards}, {Effler},
  {Ehrens}, {Eichholz}, {Eikenberry}, {Eisenstein}, {Essick}, {Estevez},
  {Etienne}, {Etzel}, {Evans}, {Evans}, {Factourovich}, {Fafone}, {Fair},
  {Fairhurst}, {Fan}, {Farinon}, {Farr}, {Farr}, {Fauchon-Jones}, {Favata},
  {Fays}, {Fee}, {Fehrmann}, {Feicht}, {Fejer}, {Fernandez-Galiana},
  {Ferrante}, {Ferreira}, {Ferrini}, {Fidecaro}, {Finstad}, {Fiori},
  {Fiorucci}, {Fishbach}, {Fisher}, {Fitz-Axen}, {Flaminio}, {Fletcher},
  {Fong}, {Font}, {Forsyth}, {Forsyth}, {Fournier}, {Frasca}, {Frasconi},
  {Frei}, {Freise}, {Frey}, {Frey}, {Fries}, {Fritschel}, {Frolov}, {Fulda},
  {Fyffe}, {Gabbard}, {Gadre}, {Gaebel}, {Gair}, {Gammaitoni}, {Ganija},
  {Gaonkar}, {Garcia-Quiros}, {Garufi}, {Gateley}, {Gaudio}, {Gaur},
  {Gayathri}, {Gehrels}, {Gemme}, {Genin}, {Gennai}, {George}, {George},
  {Gergely}, {Germain}, {Ghonge}, {Ghosh}, {Ghosh}, {Ghosh}, {Giaime},
  {Giardina}, {Giazotto}, {Gill}, {Glover}, {Goetz}, {Goetz}, {Gomes},
  {Goncharov}, {Gonz{\'a}lez}, {Gonzalez Castro}, {Gopakumar}, {Gorodetsky},
  {Gossan}, {Gosselin}, {Gouaty}, {Grado}, {Graef}, {Granata}, {Grant}, {Gras},
  {Gray}, {Greco}, {Green}, {Gretarsson}, {Griswold}, {Groot}, {Grote},
  {Grunewald}, {Gruning}, {Guidi}, {Guo}, {Gupta}, {Gupta}, {Gushwa},
  {Gustafson}, {Gustafson}, {Halim}, {Hall}, {Hall}, {Hamilton}, {Hammond},
  {Haney}, {Hanke}, {Hanks}, {Hanna}, {Hannam}, {Hannuksela}, {Hanson},
  {Hardwick}, {Harms}, {Harry}, {Harry}, {Hart}, {Haster}, {Haughian}, {Healy},
  {Heidmann}, {Heintze}, {Heitmann}, {Hello}, {Hemming}, {Hendry}, {Heng},
  {Hennig}, {Heptonstall}, {Heurs}, {Hild}, {Hinderer}, {Hoak}, {Hofman},
  {Holt}, {Holz}, {Hopkins}, {Horst}, {Hough}, {Houston}, {Howell}, {Hreibi},
  {Hu}, {Huerta}, {Huet}, {Hughey}, {Husa}, {Huttner}, {Huynh-Dinh}, {Indik},
  {Inta}, {Intini}, {Isa}, {Isac}, {Isi}, {Iyer}, {Izumi}, {Jacqmin}, {Jani},
  {Jaranowski}, {Jawahar}, {Jim{\'e}nez-Forteza}, {Johnson}, {Jones}, {Jones},
  {Jonker}, {Ju}, {Junker}, {Kalaghatgi}, {Kalogera}, {Kamai}, {Kandhasamy},
  {Kang}, {Kanner}, {Kapadia}, {Karki}, {Karvinen}, {Kasprzack}, {Katolik},
  {Katsavounidis}, {Katzman}, {Kaufer}, {Kawabe}, {K{\'e}f{\'e}lian}, {Keitel},
  {Kemball}, {Kennedy}, {Kent}, {Key}, {Khalili}, {Khan}, {Khan}, {Khan},
  {Khazanov}, {Kijbunchoo}, {Kim}, {Kim}, {Kim}, {Kim}, {Kim}, {Kim},
  {Kimbrell}, {King}, {King}, {Kinley-Hanlon}, {Kirchhoff}, {Kissel},
  {Kleybolte}, {Klimenko}, {Knowles}, {Koch}, {Koehlenbeck}, {Koley},
  {Kondrashov}, {Kontos}, {Korobko}, {Korth}, {Kowalska}, {Kozak},
  {Kr{\"a}mer}, {Kringel}, {Krishnan}, {Kr{\'o}lak}, {Kuehn}, {Kumar}, {Kumar},
  {Kumar}, {Kuo}, {Kutynia}, {Kwang}, {Lackey}, {Lai}, {Landry}, {Lang},
  {Lange}, {Lantz}, {Lanza}, {Larson}, {Lartaux-Vollard}, {Lasky}, {Laxen},
  {Lazzarini}, {Lazzaro}, {Leaci}, {Leavey}, {Lee}, {Lee}, {Lee}, {Lee}, {Lee},
  {Lehmann}, {Lenon}, {Leonardi}, {Leroy}, {Letendre}, {Levin}, {Li}, {Linker},
  {Littenberg}, {Liu}, {Lo}, {Lockerbie}, {London}, {Lord}, {Lorenzini},
  {Loriette}, {Lormand}, {Losurdo}, {Lough}, {Lousto}, {Lovelace}, {L{\"u}ck},
  {Lumaca}, {Lundgren}, {Lynch}, {Ma}, {Macas}, {Macfoy}, {Machenschalk},
  {MacInnis}, {Macleod}, {Maga{\~n}a Hernandez}, {Maga{\~n}a-Sandoval},
  {Maga{\~n}a Zertuche}, {Magee}, {Majorana}, {Maksimovic}, {Man}, {Mandic},
  {Mangano}, {Mansell}, {Manske}, {Mantovani}, {Marchesoni}, {Marion},
  {M{\'a}rka}, {M{\'a}rka}, {Markakis}, {Markosyan}, {Markowitz}, {Maros},
  {Marquina}, {Marsh}, {Martelli}, {Martellini}, {Martin}, {Martin},
  {Martynov}, {Mason}, {Massera}, {Masserot}, {Massinger}, {Masso-Reid},
  {Mastrogiovanni}, {Matas}, {Matichard}, {Matone}, {Mavalvala}, {Mazumder},
  {McCarthy}, {McClelland}, {McCormick}, {McCuller}, {McGuire}, {McIntyre},
  {McIver}, {McManus}, {McNeill}, {McRae}, {McWilliams}, {Meacher}, {Meadors},
  {Mehmet}, {Meidam}, {Mejuto-Villa}, {Melatos}, {Mendell}, {Mercer}, {Merilh},
  {Merzougui}, {Meshkov}, {Messenger}, {Messick}, {Metzdorff}, {Meyers},
  {Miao}, {Michel}, {Middleton}, {Mikhailov}, {Milano}, {Miller}, {Miller},
  {Miller}, {Millhouse}, {Milovich-Goff}, {Minazzoli}, {Minenkov}, {Ming},
  {Mishra}, {Mitra}, {Mitrofanov}, {Mitselmakher}, {Mittleman}, {Moffa},
  {Moggi}, {Mogushi}, {Mohan}, {Mohapatra}, {Montani}, {Moore}, {Moraru},
  {Moreno}, {Morriss}, {Mours}, {Mow-Lowry}, {Mueller}, {Muir}, {Mukherjee},
  {Mukherjee}, {Mukherjee}, {Mukund}, {Mullavey}, {Munch}, {Mu{\~n}iz},
  {Muratore}, {Murray}, {Napier}, {Nardecchia}, {Naticchioni}, {Nayak},
  {Neilson}, {Nelemans}, {Nelson}, {Nery}, {Neunzert}, {Nevin}, {Newport},
  {Newton}, {Ng}, {Nguyen}, {Nguyen}, {Nichols}, {Nielsen}, {Nissanke}, {Nitz},
  {Noack}, {Nocera}, {Nolting}, {North}, {Nuttall}, {Oberling}, {O'Dea},
  {Ogin}, {Oh}, {Oh}, {Ohme}, {Okada}, {Oliver}, {Oppermann}, {Oram},
  {O'Reilly}, {Ormiston}, {Ortega}, {O'Shaughnessy}, {Ossokine}, {Ottaway},
  {Overmier}, {Owen}, {Pace}, {Page}, {Page}, {Pai}, {Pai}, {Palamos},
  {Palashov}, {Palomba}, {Pal-Singh}, {Pan}, {Pan}, {Pang}, {Pang}, {Pankow},
  {Pannarale}, {Pant}, {Paoletti}, {Paoli}, {Papa}, {Parida}, {Parker},
  {Pascucci}, {Pasqualetti}, {Passaquieti}, {Passuello}, {Patil}, {Patricelli},
  {Pearlstone}, {Pedraza}, {Pedurand}, {Pekowsky}, {Pele}, {Penn}, {Perez},
  {Perreca}, {Perri}, {Pfeiffer}, {Phelps}, {Piccinni}, {Pichot},
  {Piergiovanni}, {Pierro}, {Pillant}, {Pinard}, {Pinto}, {Pirello}, {Pitkin},
  {Poe}, {Poggiani}, {Popolizio}, {Porter}, {Post}, {Powell}, {Prasad},
  {Pratt}, {Pratten}, {Predoi}, {Prestegard}, {Price}, {Prijatelj}, {Principe},
  {Privitera}, {Prodi}, {Prokhorov}, {Puncken}, {Punturo}, {Puppo},
  {P{\"u}rrer}, {Qi}, {Quetschke}, {Quintero}, {Quitzow-James}, {Raab},
  {Rabeling}, {Radkins}, {Raffai}, {Raja}, {Rajan}, {Rajbhandari}, {Rakhmanov},
  {Ramirez}, {Ramos-Buades}, {Rapagnani}, {Raymond}, {Razzano}, {Read},
  {Regimbau}, {Rei}, {Reid}, {Reitze}, {Ren}, {Reyes}, {Ricci}, {Ricker},
  {Rieger}, {Riles}, {Rizzo}, {Robertson}, {Robie}, {Robinet}, {Rocchi},
  {Rolland}, {Rollins}, {Roma}, {Romano}, {Romel}, {Romie}, {Rosi{\'n}ska},
  {Ross}, {Rowan}, {R{\"u}diger}, {Ruggi}, {Rutins}, {Ryan}, {Sachdev},
  {Sadecki}, {Sadeghian}, {Sakellariadou}, {Salconi}, {Saleem}, {Salemi},
  {Samajdar}, {Sammut}, {Sampson}, {Sanchez}, {Sanchez}, {Sanchis-Gual},
  {Sandberg}, {Sanders}, {Sassolas}, {Sathyaprakash}, {Saulson}, {Sauter},
  {Savage}, {Sawadsky}, {Schale}, {Scheel}, {Scheuer}, {Schmidt}, {Schmidt},
  {Schnabel}, {Schofield}, {Sch{\"o}nbeck}, {Schreiber}, {Schuette}, {Schulte},
  {Schutz}, {Schwalbe}, {Scott}, {Scott}, {Seidel}, {Sellers}, {Sengupta},
  {Sentenac}, {Sequino}, {Sergeev}, {Shaddock}, {Shaffer}, {Shah}, {Shahriar},
  {Shaner}, {Shao}, {Shapiro}, {Shawhan}, {Sheperd}, {Shoemaker}, {Shoemaker},
  {Siellez}, {Siemens}, {Sieniawska}, {Sigg}, {Silva}, {Singer}, {Singh},
  {Singhal}, {Sintes}, {Slagmolen}, {Smith}, {Smith}, {Smith}, {Somala}, {Son},
  {Sonnenberg}, {Sorazu}, {Sorrentino}, {Souradeep}, {Spencer}, {Srivastava},
  {Staats}, {Staley}, {Steinke}, {Steinlechner}, {Steinlechner}, {Steinmeyer},
  {Stevenson}, {Stone}, {Stops}, {Strain}, {Stratta}, {Strigin}, {Strunk},
  {Sturani}, {Stuver}, {Summerscales}, {Sun}, {Sunil}, {Suresh}, {Sutton},
  {Swinkels}, {Szczepa{\'n}czyk}, {Tacca}, {Tait}, {Talbot}, {Talukder},
  {Tanner}, {T{\'a}pai}, {Taracchini}, {Tasson}, {Taylor}, {Taylor}, {Tewari},
  {Theeg}, {Thies}, {Thomas}, {Thomas}, {Thomas}, {Thorne}, {Thorne}, {Thrane},
  {Tiwari}, {Tiwari}, {Tokmakov}, {Toland}, {Tonelli}, {Tornasi},
  {Torres-Forn{\'e}}, {Torrie}, {T{\"o}yr{\"a}}, {Travasso}, {Traylor},
  {Trinastic}, {Tringali}, {Trozzo}, {Tsang}, {Tse}, {Tso}, {Tsukada}, {Tsuna},
  {Tuyenbayev}, {Ueno}, {Ugolini}, {Unnikrishnan}, {Urban}, {Usman},
  {Vahlbruch}, {Vajente}, {Valdes}, {van Bakel}, {van Beuzekom}, {van den
  Brand}, {Van Den Broeck}, {Vander-Hyde}, {van der Schaaf}, {van Heijningen},
  {van Veggel}, {Vardaro}, {Varma}, {Vass}, {Vas{\'u}th}, {Vecchio},
  {Vedovato}, {Veitch}, {Veitch}, {Venkateswara}, {Venugopalan}, {Verkindt},
  {Vetrano}, {Vicer{\'e}}, {Viets}, {Vinciguerra}, {Vine}, {Vinet}, {Vitale},
  {Vo}, {Vocca}, {Vorvick}, {Vyatchanin}, {Wade}, {Wade}, {Wade}, {Walet},
  {Walker}, {Wallace}, {Walsh}, {Wang}, {Wang}, {Wang}, {Wang}, {Wang}, {Ward},
  {Warner}, {Was}, {Watchi}, {Weaver}, {Wei}, {Weinert}, {Weinstein}, {Weiss},
  {Wen}, {Wessel}, {Wessels}, {Westerweck}, {Westphal}, {Wette}, {Whelan},
  {Whitcomb}, {Whiting}, {Whittle}, {Wilken}, {Williams}, {Williams},
  {Williamson}, {Willis}, {Willke}, {Wimmer}, {Winkler}, {Wipf}, {Wittel},
  {Woan}, {Woehler}, {Wofford}, {Wong}, {Worden}, {Wright}, {Wu}, {Wysocki},
  {Xiao}, {Yamamoto}, {Yancey}, {Yang}, {Yap}, {Yazback}, {Yu}, {Yu}, {Yvert},
  {Zadrozny}, {Zanolin}, {Zelenova}, {Zendri}, {Zevin}, {Zhang}, {Zhang},
  {Zhang}, {Zhang}, {Zhao}, {Zhou}, {Zhou}, {Zhu}, {Zhu}, {Zimmerman},
  {Zucker}, {Zweizig}, {LIGO Scientific Collaboration}, {Virgo Collaboration},
  {Wilson-Hodge}, {Bissaldi}, {Blackburn}, {Briggs}, {Burns}, {Cleveland},
  {Connaughton}, {Gibby}, {Giles}, {Goldstein}, {Hamburg}, {Jenke}, {Hui},
  {Kippen}, {Kocevski}, {McBreen}, {Meegan}, {Paciesas}, {Poolakkil}, {Preece},
  {Racusin}, {Roberts}, {Stanbro}, {Veres}, {von Kienlin}, {GBM}, {Savchenko},
  {Ferrigno}, {Kuulkers}, {Bazzano}, {Bozzo}, {Brandt}, {Chenevez},
  {Courvoisier}, {Diehl}, {Domingo}, {Hanlon}, {Jourdain}, {Laurent}, {Lebrun},
  {Lutovinov}, {Martin-Carrillo}, {Mereghetti}, {Natalucci}, {Rodi}, {Roques},
  {Sunyaev}, {Ubertini}, {INTEGRAL}, {Aartsen}, {Ackermann}, {Adams},
  {Aguilar}, {Ahlers}, {Ahrens}, {Samarai}, {Altmann}, {Andeen}, {Anderson},
  {Ansseau}, {Anton}, {Arg{\"u}elles}, {Auffenberg}, {Axani}, {Bagherpour},
  {Bai}, {Barron}, {Barwick}, {Baum}, {Bay}, {Beatty}, {Becker Tjus},
  {Bernardini}, {Besson}, {Binder}, {Bindig}, {Blaufuss}, {Blot}, {Bohm},
  {B{\"o}rner}, {Bos}, {Bose}, {B{\"o}ser}, {Botner}, {Bourbeau}, {Bourbeau},
  {Bradascio}, {Braun}, {Brayeur}, {Brenzke}, {Bretz}, {Bron},
  {Brostean-Kaiser}, {Burgman}, {Carver}, {Casey}, {Casier}, {Cheung},
  {Chirkin}, {Christov}, {Clark}, {Classen}, {Coenders}, {Collin}, {Conrad},
  {Cowen}, {Cross}, {Day}, {de Andr{\'e}}, {De Clercq}, {DeLaunay},
  {Dembinski}, {De Ridder}, {Desiati}, {de Vries}, {de Wasseige}, {de With},
  {DeYoung}, {D{\'\i}az-V{\'e}lez}, {di Lorenzo}, {Dujmovic}, {Dumm},
  {Dunkman}, {Dvorak}, {Eberhardt}, {Ehrhardt}, {Eichmann}, {Eller}, {Evenson},
  {Fahey}, {Fazely}, {Felde}, {Filimonov}, {Finley}, {Flis}, {Franckowiak},
  {Friedman}, {Fuchs}, {Gaisser}, {Gallagher}, {Gerhardt}, {Ghorbani}, {Giang},
  {Glauch}, {Gl{\"u}senkamp}, {Goldschmidt}, {Gonzalez}, {Grant}, {Griffith},
  {Haack}, {Hallgren}, {Halzen}, {Hanson}, {Hebecker}, {Heereman}, {Helbing},
  {Hellauer}, {Hickford}, {Hignight}, {Hill}, {Hoffman}, {Hoffmann},
  {Hokanson-Fasig}, {Hoshina}, {Huang}, {Huber}, {Hultqvist}, {H{\"u}nnefeld},
  {In}, {Ishihara}, {Jacobi}, {Japaridze}, {Jeong}, {Jero}, {Jones},
  {Kalaczynski}, {Kang}, {Kappes}, {Karg}, {Karle}, {Kauer}, {Keivani},
  {Kelley}, {Kheirandish}, {Kim}, {Kim}, {Kintscher}, {Kiryluk}, {Kittler},
  {Klein}, {Kohnen}, {Koirala}, {Kolanoski}, {K{\"o}pke}, {Kopper}, {Kopper},
  {Koschinsky}, {Koskinen}, {Kowalski}, {Krings}, {Kroll}, {Kr{\"u}ckl},
  {Kunnen}, {Kunwar}, {Kurahashi}, {Kuwabara}, {Kyriacou}, {Labare},
  {Lanfranchi}, {Larson}, {Lauber}, {Lesiak-Bzdak}, {Leuermann}, {Liu}, {Lu},
  {L{\"u}nemann}, {Luszczak}, {Madsen}, {Maggi}, {Mahn}, {Mancina}, {Maruyama},
  {Mase}, {Maunu}, {McNally}, {Meagher}, {Medici}, {Meier}, {Menne}, {Merino},
  {Meures}, {Miarecki}, {Micallef}, {Moment{\'e}}, {Montaruli}, {Moore},
  {Moulai}, {Nahnhauer}, {Nakarmi}, {Naumann}, {Neer}, {Niederhausen},
  {Nowicki}, {Nygren}, {Obertacke Pollmann}, {Olivas}, {O'Murchadha},
  {Palczewski}, {Pandya}, {Pankova}, {Peiffer}, {Pepper}, {P{\'e}rez de los
  Heros}, {Pieloth}, {Pinat}, {Price}, {Przybylski}, {Raab}, {R{\"a}del},
  {Rameez}, {Rawlins}, {Rea}, {Reimann}, {Relethford}, {Relich}, {Resconi},
  {Rhode}, {Richman}, {Robertson}, {Rongen}, {Rott}, {Ruhe}, {Ryckbosch},
  {Rysewyk}, {S{\"a}lzer}, {Sanchez Herrera}, {Sandrock}, {Sandroos},
  {Santander}, {Sarkar}, {Sarkar}, {Satalecka}, {Schlunder}, {Schmidt},
  {Schneider}, {Schoenen}, {Sch{\"o}neberg}, {Schumacher}, {Seckel},
  {Seunarine}, {Soedingrekso}, {Soldin}, {Song}, {Spiczak}, {Spiering},
  {Stachurska}, {Stamatikos}, {Stanev}, {Stasik}, {Stettner}, {Steuer},
  {Stezelberger}, {Stokstad}, {St{\"o}ssl}, {Strotjohann}, {Stuttard},
  {Sullivan}, {Sutherland}, {Taboada}, {Tatar}, {Tenholt}, {Ter-Antonyan},
  {Terliuk}, {Te{\v{s}}i{\'c}}, {Tilav}, {Toale}, {Tobin}, {Toscano}, {Tosi},
  {Tselengidou}, {Tung}, {Turcati}, {Turley}, {Ty}, {Unger}, {Usner},
  {Vandenbroucke}, {Van Driessche}, {van Eijndhoven}, {Vanheule}, {van Santen},
  {Vehring}, {Vogel}, {Vraeghe}, {Walck}, {Wallace}, {Wallraff}, {Wandler},
  {Wandkowsky}, {Waza}, {Weaver}, {Weiss}, {Wendt}, {Werthebach}, {Whelan},
  {Wiebe}, {Wiebusch}, {Wille}, {Williams}, {Wills}, {Wolf}, {Wood}, {Woolsey},
  {Woschnagg}, {Xu}, {Xu}, {Xu}, {Yanez}, {Yodh}, {Yoshida}, {Yuan}, {Zoll},
  {IceCube Collaboration}, {Balasubramanian}, {Mate}, {Bhalerao},
  {Bhattacharya}, {Vibhute}, {Dewangan}, {Rao}, {Vadawale}, {AstroSat Cadmium
  Zinc Telluride Imager Team}, {Svinkin}, {Hurley}, {Aptekar}, {Frederiks},
  {Golenetskii}, {Kozlova}, {Lysenko}, {Oleynik}, {Tsvetkova}, {Ulanov},
  {Cline}, {IPN Collaboration}, {Li}, {Xiong}, {Zhang}, {Lu}, {Song}, {Cao},
  {Chang}, {Chen}, {Chen}, {Chen}, {Chen}, {Chen}, {Chen}, {Cui}, {Cui},
  {Deng}, {Dong}, {Du}, {Fu}, {Gao}, {Gao}, {Gao}, {Ge}, {Gu}, {Guan}, {Guo},
  {Han}, {Hu}, {Huang}, {Huo}, {Jia}, {Jiang}, {Jiang}, {Jin}, {Jin}, {Li},
  {Li}, {Li}, {Li}, {Li}, {Li}, {Li}, {Li}, {Li}, {Li}, {Li}, {Liang}, {Liao},
  {Liu}, {Liu}, {Liu}, {Liu}, {Liu}, {Liu}, {Liu}, {Lu}, {Lu}, {Luo}, {Ma},
  {Meng}, {Nang}, {Nie}, {Ou}, {Qu}, {Sai}, {Sun}, {Tan}, {Tao}, {Tao}, {Tuo},
  {Wang}, {Wang}, {Wang}, {Wang}, {Wang}, {Wen}, {Wu}, {Wu}, {Xiao}, {Xu},
  {Xu}, {Yan}, {Yang}, {Yang}, {Yang}, {Zhang}, {Zhang}, {Zhang}, {Zhang},
  {Zhang}, {Zhang}, {Zhang}, {Zhang}, {Zhang}, {Zhang}, {Zhang}, {Zhang},
  {Zhang}, {Zhang}, {Zhang}, {Zhang}, {Zhang}, {Zhang}, {Zhao}, {Zhao}, {Zhao},
  {Zheng}, {Zhu}, {Zhu}, {Zou}, {Insight-HXMT Collaboration}, {Albert},
  {Andr{\'e}}, {Anghinolfi}, {Ardid}, {Aubert}, {Aublin}, {Avgitas}, {Baret},
  {Barrios-Mart{\'\i}}, {Basa}, {Belhorma}, {Bertin}, {Biagi}, {Bormuth},
  {Bourret}, {Bouwhuis}, {Br{\^a}nza{\c{s}}}, {Bruijn}, {Brunner}, {Busto},
  {Capone}, {Caramete}, {Carr}, {Celli}, {Cherkaoui El Moursli}, {Chiarusi},
  {Circella}, {Coelho}, {Coleiro}, {Coniglione}, {Costantini}, {Coyle},
  {Creusot}, {D{\'\i}az}, {Deschamps}, {De Bonis}, {Distefano}, {Di Palma},
  {Domi}, {Donzaud}, {Dornic}, {Drouhin}, {Eberl}, {El Bojaddaini}, {El
  Khayati}, {Els{\"a}sser}, {Enzenh{\"o}fer}, {Ettahiri}, {Fassi}, {Felis},
  {Fusco}, {Gay}, {Giordano}, {Glotin}, {Gr{\'e}goire}, {Ruiz}, {Graf},
  {Hallmann}, {van Haren}, {Heijboer}, {Hello}, {Hern{\'a}ndez-Rey},
  {H{\"o}ssl}, {Hofest{\"a}dt}, {Hugon}, {Illuminati}, {James}, {de Jong},
  {Jongen}, {Kadler}, {Kalekin}, {Katz}, {Kiessling}, {Kouchner}, {Kreter},
  {Kreykenbohm}, {Kulikovskiy}, {Lachaud}, {Lahmann}, {Lef{\`e}vre}, {Leonora},
  {Lotze}, {Loucatos}, {Marcelin}, {Margiotta}, {Marinelli},
  {Mart{\'\i}nez-Mora}, {Mele}, {Melis}, {Michael}, {Migliozzi}, {Moussa},
  {Navas}, {Nezri}, {Organokov}, {P{\u{a}}v{\u{a}}la{\c{s}}}, {Pellegrino},
  {Perrina}, {Piattelli}, {Popa}, {Pradier}, {Quinn}, {Racca}, {Riccobene},
  {S{\'a}nchez-Losa}, {Salda{\~n}a}, {Salvadori}, {Samtleben}, {Sanguineti},
  {Sapienza}, {Sieger}, {Spurio}, {Stolarczyk}, {Taiuti}, {Tayalati},
  {Trovato}, {Turpin}, {T{\"o}nnis}, {Vallage}, {Van Elewyck}, {Versari},
  {Vivolo}, {Vizzoca}, {Wilms}, {Zornoza}, {Z{\'u}{\~n}iga}, {ANTARES
  Collaboration}, {Beardmore}, {Breeveld}, {Burrows}, {Cenko}, {Cusumano},
  {D'A{\`\i}}, {de Pasquale}, {Emery}, {Evans}, {Giommi}, {Gronwall}, {Kennea},
  {Krimm}, {Kuin}, {Lien}, {Marshall}, {Melandri}, {Nousek}, {Oates},
  {Osborne}, {Pagani}, {Page}, {Palmer}, {Perri}, {Siegel}, {Sbarufatti},
  {Tagliaferri}, {Tohuvavohu}, {Swift Collaboration}, {Tavani}, {Verrecchia},
  {Bulgarelli}, {Evangelista}, {Pacciani}, {Feroci}, {Pittori}, {Giuliani},
  {Del Monte}, {Donnarumma}, {Argan}, {Trois}, {Ursi}, {Cardillo}, {Piano},
  {Longo}, {Lucarelli}, {Munar-Adrover}, {Fuschino}, {Labanti}, {Marisaldi},
  {Minervini}, {Fioretti}, {Parmiggiani}, {Gianotti}, {Trifoglio}, {Di Persio},
  {Antonelli}, {Barbiellini}, {Caraveo}, {Cattaneo}, {Costa}, {Colafrancesco},
  {D'Amico}, {Ferrari}, {Morselli}, {Paoletti}, {Picozza}, {Pilia}, {Rappoldi},
  {Soffitta}, {Vercellone}, {AGILE Team}, {Foley}, {Coulter}, {Kilpatrick},
  {Drout}, {Piro}, {Shappee}, {Siebert}, {Simon}, {Ulloa}, {Kasen}, {Madore},
  {Murguia-Berthier}, {Pan}, {Prochaska}, {Ramirez-Ruiz}, {Rest},
  {Rojas-Bravo}, {1M2H Team}, {Berger}, {Soares-Santos}, {Annis}, {Alexander},
  {Allam}, {Balbinot}, {Blanchard}, {Brout}, {Butler}, {Chornock}, {Cook},
  {Cowperthwaite}, {Diehl}, {Drlica-Wagner}, {Drout}, {Durret}, {Eftekhari},
  {Finley}, {Fong}, {Frieman}, {Fryer}, {Garc{\'\i}a-Bellido}, {Gruendl},
  {Hartley}, {Herner}, {Kessler}, {Lin}, {Lopes}, {Louren{\c{c}}o}, {Margutti},
  {Marshall}, {Matheson}, {Medina}, {Metzger}, {Mu{\~n}oz}, {Muir}, {Nicholl},
  {Nugent}, {Palmese}, {Paz-Chinch{\'o}n}, {Quataert}, {Sako}, {Sauseda},
  {Schlegel}, {Scolnic}, {Secco}, {Smith}, {Sobreira}, {Villar}, {Vivas},
  {Wester}, {Williams}, {Yanny}, {Zenteno}, {Zhang}, {Abbott}, {Banerji},
  {Bechtol}, {Benoit-L{\'e}vy}, {Bertin}, {Brooks}, {Buckley-Geer}, {Burke},
  {Capozzi}, {Carnero Rosell}, {Carrasco Kind}, {Castander}, {Crocce}, {Cunha},
  {D'Andrea}, {da Costa}, {Davis}, {DePoy}, {Desai}, {Dietrich}, {Eifler},
  {Fernandez}, {Flaugher}, {Fosalba}, {Gaztanaga}, {Gerdes}, {Giannantonio},
  {Goldstein}, {Gruen}, {Gschwend}, {Gutierrez}, {Honscheid}, {James},
  {Jeltema}, {Johnson}, {Johnson}, {Kent}, {Krause}, {Kron}, {Kuehn}, {Lahav},
  {Lima}, {Maia}, {March}, {Martini}, {McMahon}, {Menanteau}, {Miller},
  {Miquel}, {Mohr}, {Nichol}, {Ogando}, {Plazas}, {Romer}, {Roodman}, {Rykoff},
  {Sanchez}, {Scarpine}, {Schindler}, {Schubnell}, {Sevilla-Noarbe}, {Sheldon},
  {Smith}, {Smith}, {Stebbins}, {Suchyta}, {Swanson}, {Tarle}, {Thomas},
  {Troxel}, {Tucker}, {Vikram}, {Walker}, {Wechsler}, {Weller}, {Carlin},
  {Gill}, {Li}, {Marriner}, {Neilsen}, {Dark Energy Camera GW-EM
  Collaboration}, {DES Collaboration}, {Haislip}, {Kouprianov}, {Reichart},
  {Sand}, {Tartaglia}, {Valenti}, {Yang}, {DLT40 Collaboration}, {Benetti},
  {Brocato}, {Campana}, {Cappellaro}, {Covino}, {D'Avanzo}, {D'Elia}, {Getman},
  {Ghirlanda}, {Ghisellini}, {Limatola}, {Nicastro}, {Palazzi}, {Pian},
  {Piranomonte}, {Possenti}, {Rossi}, {Salafia}, {Tomasella}, {Amati},
  {Antonelli}, {Bernardini}, {Bufano}, {Capaccioli}, {Casella}, {Dadina}, {De
  Cesare}, {Di Paola}, {Giuffrida}, {Giunta}, {Israel}, {Lisi}, {Maiorano},
  {Mapelli}, {Masetti}, {Pescalli}, {Pulone}, {Salvaterra}, {Schipani},
  {Spera}, {Stamerra}, {Stella}, {Testa}, {Turatto}, {Vergani}, {Aresu},
  {Bachetti}, {Buffa}, {Burgay}, {Buttu}, {Caria}, {Carretti}, {Casasola},
  {Castangia}, {Carboni}, {Casu}, {Concu}, {Corongiu}, {Deiana}, {Egron},
  {Fara}, {Gaudiomonte}, {Gusai}, {Ladu}, {Loru}, {Leurini}, {Marongiu},
  {Melis}, {Melis}, {Migoni}, {Milia}, {Navarrini}, {Orlati}, {Ortu}, {Palmas},
  {Pellizzoni}, {Perrodin}, {Pisanu}, {Poppi}, {Righini}, {Saba}, {Serra},
  {Serrau}, {Stagni}, {Surcis}, {Vacca}, {Vargiu}, {Hunt}, {Jin}, {Klose},
  {Kouveliotou}, {Mazzali}, {M{\o}ller}, {Nava}, {Piran}, {Selsing}, {Vergani},
  {Wiersema}, {Toma}, {Higgins}, {Mundell}, {di Serego Alighieri}, {G{\'o}tz},
  {Gao}, {Gomboc}, {Kaper}, {Kobayashi}, {Kopac}, {Mao}, {Starling}, {Steele},
  {van der Horst}, {GRAWITA: GRAvitational Wave Inaf TeAm}, {Acero}, {Atwood},
  {Baldini}, {Barbiellini}, {Bastieri}, {Berenji}, {Bellazzini}, {Bissaldi},
  {Blandford}, {Bloom}, {Bonino}, {Bottacini}, {Bregeon}, {Buehler}, {Buson},
  {Cameron}, {Caputo}, {Caraveo}, {Cavazzuti}, {Chekhtman}, {Cheung}, {Chiang},
  {Ciprini}, {Cohen-Tanugi}, {Cominsky}, {Costantin}, {Cuoco}, {D'Ammando}, {de
  Palma}, {Digel}, {Di Lalla}, {Di Mauro}, {Di Venere}, {Dubois}, {Fegan},
  {Focke}, {Franckowiak}, {Fukazawa}, {Funk}, {Fusco}, {Gargano}, {Gasparrini},
  {Giglietto}, {Giordano}, {Giroletti}, {Glanzman}, {Green}, {Grondin},
  {Guillemot}, {Guiriec}, {Harding}, {Horan}, {J{\'o}hannesson}, {Kamae},
  {Kensei}, {Kuss}, {La Mura}, {Latronico}, {Lemoine-Goumard}, {Longo},
  {Loparco}, {Lovellette}, {Lubrano}, {Magill}, {Maldera}, {Manfreda},
  {Mazziotta}, {McEnery}, {Meyer}, {Michelson}, {Mirabal}, {Monzani},
  {Moretti}, {Morselli}, {Moskalenko}, {Negro}, {Nuss}, {Ojha}, {Omodei},
  {Orienti}, {Orlando}, {Palatiello}, {Paliya}, {Paneque}, {Pesce-Rollins},
  {Piron}, {Porter}, {Principe}, {Rain{\`o}}, {Rando}, {Razzano}, {Razzaque},
  {Reimer}, {Reimer}, {Reposeur}, {Rochester}, {Saz Parkinson}, {Sgr{\`o}},
  {Siskind}, {Spada}, {Spandre}, {Suson}, {Takahashi}, {Tanaka}, {Thayer},
  {Thayer}, {Thompson}, {Tibaldo}, {Torres}, {Torresi}, {Troja}, {Venters},
  {Vianello}, {Zaharijas}, {Fermi Large Area Telescope Collaboration},
  {Allison}, {Bannister}, {Dobie}, {Kaplan}, {Lenc}, {Lynch}, {Murphy},
  {Sadler}, {Australia Telescope Compact Array}, {Hotan}, {James}, {Oslowski},
  {Raja}, {Shannon}, {Whiting}, {Australian SKA Pathfinder}, {Arcavi},
  {Howell}, {McCully}, {Hosseinzadeh}, {Hiramatsu}, {Poznanski}, {Barnes},
  {Zaltzman}, {Vasylyev}, {Maoz}, {Las Cumbres Observatory Group}, {Cooke},
  {Bailes}, {Wolf}, {Deller}, {Lidman}, {Wang}, {Gendre}, {Andreoni}, {Ackley},
  {Pritchard}, {Bessell}, {Chang}, {M{\"o}ller}, {Onken}, {Scalzo},
  {Ridden-Harper}, {Sharp}, {Tucker}, {Farrell}, {Elmer}, {Johnston},
  {Venkatraman Krishnan}, {Keane}, {Green}, {Jameson}, {Hu}, {Ma}, {Sun}, {Wu},
  {Wang}, {Shang}, {Hu}, {Ashley}, {Yuan}, {Li}, {Tao}, {Zhu}, {Zhang},
  {Suntzeff}, {Zhou}, {Yang}, {Orange}, {Morris}, {Cucchiara}, {Giblin},
  {Klotz}, {Staff}, {Thierry}, {Schmidt}, {OzGrav}, {(Deeper}, {Wider},
  {program}, {AST3}, {CAASTRO Collaborations}, {Tanvir}, {Levan}, {Cano}, {de
  Ugarte-Postigo}, {Gonz{\'a}lez-Fern{\'a}ndez}, {Greiner}, {Hjorth}, {Irwin},
  {Kr{\"u}hler}, {Mandel}, {Milvang-Jensen}, {O'Brien}, {Rol}, {Rosetti},
  {Rosswog}, {Rowlinson}, {Steeghs}, {Th{\"o}ne}, {Ulaczyk}, {Watson}, {Bruun},
  {Cutter}, {Figuera Jaimes}, {Fujii}, {Fruchter}, {Gompertz}, {Jakobsson},
  {Hodosan}, {J{\`e}rgensen}, {Kangas}, {Kann}, {Rabus}, {Schr{\o}der},
  {Stanway}, {Wijers}, {VINROUGE Collaboration}, {Lipunov}, {Gorbovskoy},
  {Kornilov}, {Tyurina}, {Balanutsa}, {Kuznetsov}, {Vlasenko}, {Podesta},
  {Lopez}, {Podesta}, {Levato}, {Saffe}, {Mallamaci}, {Budnev}, {Gress},
  {Kuvshinov}, {Gorbunov}, {Vladimirov}, {Zimnukhov}, {Gabovich}, {Yurkov},
  {Sergienko}, {Rebolo}, {Serra-Ricart}, {Tlatov}, {Ishmuhametova}, {MASTER
  Collaboration}, {Abe}, {Aoki}, {Aoki}, {Asakura}, {Baar}, {Barway}, {Bond},
  {Doi}, {Finet}, {Fujiyoshi}, {Furusawa}, {Honda}, {Itoh}, {Kanda},
  {Kawabata}, {Kawabata}, {Kim}, {Koshida}, {Kuroda}, {Lee}, {Liu},
  {Matsubayashi}, {Miyazaki}, {Morihana}, {Morokuma}, {Motohara}, {Murata},
  {Nagai}, {Nagashima}, {Nagayama}, {Nakaoka}, {Nakata}, {Ohsawa}, {Ohshima},
  {Ohta}, {Okita}, {Saito}, {Saito}, {Sako}, {Sekiguchi}, {Sumi}, {Tajitsu},
  {Takahashi}, {Takayama}, {Tamura}, {Tanaka}, {Tanaka}, {Terai}, {Tominaga},
  {Tristram}, {Uemura}, {Utsumi}, {Yamaguchi}, {Yasuda}, {Yoshida}, {Zenko},
  {J-GEM}, {Adams}, {Anupama}, {Bally}, {Barway}, {Bellm}, {Blagorodnova},
  {Cannella}, {Chandra}, {Chatterjee}, {Clarke}, {Cobb}, {Cook}, {Copperwheat},
  {De}, {Emery}, {Feindt}, {Foster}, {Fox}, {Frail}, {Fremling}, {Frohmaier},
  {Garcia}, {Ghosh}, {Giacintucci}, {Goobar}, {Gottlieb}, {Grefenstette},
  {Hallinan}, {Harrison}, {Heida}, {Helou}, {Ho}, {Horesh}, {Hotokezaka}, {Ip},
  {Itoh}, {Jacobs}, {Jencson}, {Kasen}, {Kasliwal}, {Kassim}, {Kim}, {Kiran},
  {Kuin}, {Kulkarni}, {Kupfer}, {Lau}, {Madsen}, {Mazzali}, {Miller},
  {Miyasaka}, {Mooley}, {Myers}, {Nakar}, {Ngeow}, {Nugent}, {Ofek},
  {Palliyaguru}, {Pavana}, {Perley}, {Peters}, {Pike}, {Piran}, {Qi}, {Quimby},
  {Rana}, {Rosswog}, {Rusu}, {Sadler}, {Van Sistine}, {Sollerman}, {Xu}, {Yan},
  {Yatsu}, {Yu}, {Zhang}, {Zhao}, {GROWTH}, {JAGWAR}, {Caltech-NRAO},
  {TTU-NRAO}, {NuSTAR Collaborations}, {Chambers}, {Huber}, {Schultz},
  {Bulger}, {Flewelling}, {Magnier}, {Lowe}, {Wainscoat}, {Waters}, {Willman},
  {Pan-STARRS}, {Ebisawa}, {Hanyu}, {Harita}, {Hashimoto}, {Hidaka}, {Hori},
  {Ishikawa}, {Isobe}, {Iwakiri}, {Kawai}, {Kawai}, {Kawamuro}, {Kawase},
  {Kitaoka}, {Makishima}, {Matsuoka}, {Mihara}, {Morita}, {Morita}, {Nakahira},
  {Nakajima}, {Nakamura}, {Negoro}, {Oda}, {Sakamaki}, {Sasaki}, {Serino},
  {Shidatsu}, {Shimomukai}, {Sugawara}, {Sugita}, {Sugizaki}, {Tachibana},
  {Takao}, {Tanimoto}, {Tomida}, {Tsuboi}, {Tsunemi}, {Ueda}, {Ueno}, {Yamada},
  {Yamaoka}, {Yamauchi}, {Yatabe}, {Yoneyama}, {Yoshii}, {MAXI Team}, {Coward},
  {Crisp}, {Macpherson}, {Andreoni}, {Laugier}, {Noysena}, {Klotz}, {Gendre},
  {Thierry}, {Turpin}, {Consortium}, {Im}, {Choi}, {Kim}, {Yoon}, {Lim}, {Lee},
  {Lee}, {Kim}, {Ko}, {Joe}, {Kwon}, {Kim}, {Lim}, {Choi}, {KU Collaboration},
  {Fynbo}, {Malesani}, {Xu}, {Optical Telescope}, {Smartt}, {Jerkstrand},
  {Kankare}, {Sim}, {Fraser}, {Inserra}, {Maguire}, {Leloudas}, {Magee},
  {Shingles}, {Smith}, {Young}, {Kotak}, {Gal-Yam}, {Lyman}, {Homan},
  {Agliozzo}, {Anderson}, {Angus}, {Ashall}, {Barbarino}, {Bauer}, {Berton},
  {Botticella}, {Bulla}, {Cannizzaro}, {Cartier}, {Cikota}, {Clark}, {De Cia},
  {Della Valle}, {Dennefeld}, {Dessart}, {Dimitriadis}, {Elias-Rosa}, {Firth},
  {Fl{\"o}rs}, {Frohmaier}, {Galbany}, {Gonz{\'a}lez-Gait{\'a}n}, {Gromadzki},
  {Guti{\'e}rrez}, {Hamanowicz}, {Harmanen}, {Heintz}, {Hernandez}, {Hodgkin},
  {Hook}, {Izzo}, {James}, {Jonker}, {Kerzendorf}, {Kostrzewa-Rutkowska},
  {Kromer}, {Kuncarayakti}, {Lawrence}, {Manulis}, {Mattila}, {McBrien},
  {M{\"u}ller}, {Nordin}, {O'Neill}, {Onori}, {Palmerio}, {Pastorello},
  {Patat}, {Pignata}, {Podsiadlowski}, {Razza}, {Reynolds}, {Roy}, {Ruiter},
  {Rybicki}, {Salmon}, {Pumo}, {Prentice}, {Seitenzahl}, {Smith}, {Sollerman},
  {Sullivan}, {Szegedi}, {Taddia}, {Taubenberger}, {Terreran}, {Van Soelen},
  {Vos}, {Walton}, {Wright}, {Wyrzykowski}, {Yaron}, {pre=''(''>ePESSTO},
  {Chen}, {Kr{\"u}hler}, {Schady}, {Wiseman}, {Greiner}, {Rau}, {Schweyer},
  {Klose}, {Nicuesa Guelbenzu}, {GROND}, {Palliyaguru}, {Tech University},
  {Shara}, {Williams}, {Vaisanen}, {Potter}, {Romero Colmenero}, {Crawford},
  {Buckley}, {Mao}, {SALT Group}, {D{\'\i}az}, {Macri}, {Garc{\'\i}a Lambas},
  {Mendes de Oliveira}, {Nilo Castell{\'o}n}, {Ribeiro}, {S{\'a}nchez},
  {Schoenell}, {Abramo}, {Akras}, {Alcaniz}, {Artola}, {Beroiz}, {Bonoli},
  {Cabral}, {Camuccio}, {Chavushyan}, {Coelho}, {Colazo}, {Costa-Duarte},
  {Cuevas Larenas}, {Dom{\'\i}nguez Romero}, {Dultzin}, {Fern{\'a}ndez},
  {Garc{\'\i}a}, {Girardini}, {Gon{\c{c}}alves}, {Gon{\c{c}}alves}, {Gurovich},
  {Jim{\'e}nez-Teja}, {Kanaan}, {Lares}, {Lopes de Oliveira}, {L{\'o}pez-Cruz},
  {Melia}, {Molino}, {Padilla}, {Pe{\~n}uela}, {Placco}, {Qui{\~n}ones},
  {Ram{\'\i}rez Rivera}, {Renzi}, {Riguccini}, {R{\'\i}os-L{\'o}pez},
  {Rodriguez}, {Sampedro}, {Schneiter}, {Sodr{\'e}}, {Starck}, {Torres-Flores},
  {Tornatore}, {Zadrozny}, {Castillo}, {TOROS: Transient Robotic Observatory of
  South Collaboration}, {Castro-Tirado}, {Tello}, {Hu}, {Zhang}, {Cunniffe},
  {Castell{\'o}n}, {Hiriart}, {Caballero-Garc{\'\i}a}, {Jel{\'\i}nek},
  {Kub{\'a}nek}, {P{\'e}rez del Pulgar}, {Park}, {Jeong}, {Castro Cer{\'o}n},
  {Pandey}, {Yock}, {Querel}, {Fan}, {Wang}, {BOOTES Collaboration},
  {Beardsley}, {Brown}, {Crosse}, {Emrich}, {Franzen}, {Gaensler}, {Horsley},
  {Johnston-Hollitt}, {Kenney}, {Morales}, {Pallot}, {Sokolowski}, {Steele},
  {Tingay}, {Trott}, {Walker}, {Wayth}, {Williams}, {Wu}, {Murchison Widefield
  Array}, {Yoshida}, {Sakamoto}, {Kawakubo}, {Yamaoka}, {Takahashi}, {Asaoka},
  {Ozawa}, {Torii}, {Shimizu}, {Tamura}, {Ishizaki}, {Cherry}, {Ricciarini},
  {Penacchioni}, {Marrocchesi}, {CALET Collaboration}, {Pozanenko}, {Volnova},
  {Mazaeva}, {Minaev}, {Krugov}, {Kusakin}, {Reva}, {Moskvitin}, {Rumyantsev},
  {Inasaridze}, {Klunko}, {Tungalag}, {Schmalz}, {Burhonov}, {IKI-GW Follow-up
  Collaboration}, {Abdalla}, {Abramowski}, {Aharonian}, {Ait Benkhali},
  {Ang{\"u}ner}, {Arakawa}, {Arrieta}, {Aubert}, {Backes}, {Balzer}, {Barnard},
  {Becherini}, {Becker Tjus}, {Berge}, {Bernhard}, {Bernl{\"o}hr}, {Blackwell},
  {B{\"o}ttcher}, {Boisson}, {Bolmont}, {Bonnefoy}, {Bordas}, {Bregeon},
  {Brun}, {Brun}, {Bryan}, {B{\"u}chele}, {Bulik}, {Capasso}, {Caroff},
  {Carosi}, {Casanova}, {Cerruti}, {Chakraborty}, {Chaves}, {Chen},
  {Chevalier}, {Colafrancesco}, {Condon}, {Conrad}, {Davids}, {Decock}, {Deil},
  {Devin}, {deWilt}, {Dirson}, {Djannati-Ata{\"\i}}, {Donath}, {O'C. Drury},
  {Dutson}, {Dyks}, {Edwards}, {Egberts}, {Emery}, {Ernenwein}, {Eschbach},
  {Farnier}, {Fegan}, {Fernandes}, {Fiasson}, {Fontaine}, {Funk},
  {F{\"u}ssling}, {Gabici}, {Gallant}, {Garrigoux}, {Gat{\'e}}, {Giavitto},
  {Giebels}, {Glawion}, {Glicenstein}, {Gottschall}, {Grondin}, {Hahn},
  {Haupt}, {Hawkes}, {Heinzelmann}, {Henri}, {Hermann}, {Hinton}, {Hofmann},
  {Hoischen}, {Holch}, {Holler}, {Horns}, {Ivascenko}, {Iwasaki},
  {Jacholkowska}, {Jamrozy}, {Jankowsky}, {Jankowsky}, {Jingo}, {Jouvin},
  {Jung-Richardt}, {Kastendieck}, {Katarzy{\'n}ski}, {Katsuragawa},
  {Kerszberg}, {Khangulyan}, {Kh{\'e}lifi}, {King}, {Klepser}, {Klochkov},
  {Klu{\'z}niak}, {Komin}, {Kosack}, {Krakau}, {Kraus}, {Kr{\"u}ger}, {Laffon},
  {Lamanna}, {Lau}, {Lees}, {Lefaucheur}, {Lemi{\`e}re}, {Lemoine-Goumard},
  {Lenain}, {Leser}, {Lohse}, {Lorentz}, {Liu}, {Lypova}, {Malyshev},
  {Marandon}, {Marcowith}, {Mariaud}, {Marx}, {Maurin}, {Maxted}, {Mayer},
  {Meintjes}, {Meyer}, {Mitchell}, {Moderski}, {Mohamed}, {Mohrmann},
  {Mor{\r{a}}}, {Moulin}, {Murach}, {Nakashima}, {de Naurois}, {Ndiyavala},
  {Niederwanger}, {Niemiec}, {Oakes}, {O'Brien}, {Odaka}, {Ohm}, {Ostrowski},
  {Oya}, {Padovani}, {Panter}, {Parsons}, {Pekeur}, {Pelletier}, {Perennes},
  {Petrucci}, {Peyaud}, {Piel}, {Pita}, {Poireau}, {Poon}, {Prokhorov},
  {Prokoph}, {P{\"u}hlhofer}, {Punch}, {Quirrenbach}, {Raab}, {Rauth},
  {Reimer}, {Reimer}, {Renaud}, {de los Reyes}, {Rieger}, {Rinchiuso},
  {Romoli}, {Rowell}, {Rudak}, {Rulten}, {Sahakian}, {Saito}, {Sanchez},
  {Santangelo}, {Sasaki}, {Schlickeiser}, {Sch{\"u}ssler}, {Schulz},
  {Schwanke}, {Schwemmer}, {Seglar-Arroyo}, {Settimo}, {Seyffert}, {Shafi},
  {Shilon}, {Shiningayamwe}, {Simoni}, {Sol}, {Spanier}, {Spir-Jacob},
  {Stawarz}, {Steenkamp}, {Stegmann}, {Steppa}, {Sushch}, {Takahashi},
  {Tavernet}, {Tavernier}, {Taylor}, {Terrier}, {Tibaldo}, {Tiziani},
  {Tluczykont}, {Trichard}, {Tsirou}, {Tsuji}, {Tuffs}, {Uchiyama}, {van der
  Walt}, {van Eldik}, {van Rensburg}, {van Soelen}, {Vasileiadis}, {Veh},
  {Venter}, {Viana}, {Vincent}, {Vink}, {Voisin}, {V{\"o}lk}, {Vuillaume},
  {Wadiasingh}, {Wagner}, {Wagner}, {Wagner}, {White}, {Wierzcholska},
  {Willmann}, {W{\"o}rnlein}, {Wouters}, {Yang}, {Zaborov}, {Zacharias},
  {Zanin}, {Zdziarski}, {Zech}, {Zefi}, {Ziegler}, {Zorn}, {Zywucka},
  {H.~E.~S.~S. Collaboration}, {Fender}, {Broderick}, {Rowlinson}, {Wijers},
  {Stewart}, {ter Veen}, {Shulevski}, {LOFAR Collaboration}, {Kavic},
  {Simonetti}, {League}, {Tsai}, {Obenberger}, {Nathaniel}, {Taylor}, {Dowell},
  {Liebling}, {Estes}, {Lippert}, {Sharma}, {Vincent}, {Farella}, {Wavelength
  Array}, {Abeysekara}, {Albert}, {Alfaro}, {Alvarez}, {Arceo},
  {Arteaga-Vel{\'a}zquez}, {Avila Rojas}, {Ayala Solares}, {Barber}, {Becerra
  Gonzalez}, {Becerril}, {Belmont-Moreno}, {BenZvi}, {Berley}, {Bernal},
  {Braun}, {Brisbois}, {Caballero-Mora}, {Capistr{\'a}n}, {Carrami{\~n}ana},
  {Casanova}, {Castillo}, {Cotti}, {Cotzomi}, {Couti{\~n}o de Le{\'o}n}, {De
  Le{\'o}n}, {De la Fuente}, {Diaz Hernandez}, {Dichiara}, {Dingus},
  {DuVernois}, {D{\'\i}az-V{\'e}lez}, {Ellsworth}, {Engel},
  {Enr{\'\i}quez-Rivera}, {Fiorino}, {Fleischhack}, {Fraija},
  {Garc{\'\i}a-Gonz{\'a}lez}, {Garfias}, {Gerhardt}, {Gonz{\~o}lez Mu{\~n}oz},
  {Gonz{\'a}lez}, {Goodman}, {Hampel-Arias}, {Harding}, {Hernandez},
  {Hernandez-Almada}, {Hona}, {H{\"u}ntemeyer}, {Iriarte}, {Jardin-Blicq},
  {Joshi}, {Kaufmann}, {Kieda}, {Lara}, {Lauer}, {Lennarz}, {Le{\'o}n Vargas},
  {Linnemann}, {Longinotti}, {Raya}, {Luna-Garc{\'\i}a}, {L{\'o}pez-Coto},
  {Malone}, {Marinelli}, {Martinez}, {Martinez-Castellanos},
  {Mart{\'\i}nez-Castro}, {Mart{\'\i}nez-Huerta}, {Matthews},
  {Miranda-Romagnoli}, {Moreno}, {Mostaf{\'a}}, {Nellen}, {Newbold}, {Nisa},
  {Noriega-Papaqui}, {Pelayo}, {Pretz}, {P{\'e}rez-P{\'e}rez}, {Ren}, {Rho},
  {Rivi{\`e}re}, {Rosa-Gonz{\'a}lez}, {Rosenberg}, {Ruiz-Velasco}, {Salazar},
  {Salesa Greus}, {Sandoval}, {Schneider}, {Schoorlemmer}, {Sinnis}, {Smith},
  {Springer}, {Surajbali}, {Tibolla}, {Tollefson}, {Torres}, {Ukwatta},
  {Weisgarber}, {Westerhoff}, {Wisher}, {Wood}, {Yapici}, {Yodh}, {Younk},
  {Zhou}, {{\'A}lvarez}, {HAWC Collaboration}, {Aab}, {Abreu}, {Aglietta},
  {Albuquerque}, {Albury}, {Allekotte}, {Almela}, {Alvarez Castillo},
  {Alvarez-Mu{\~n}iz}, {Anastasi}, {Anchordoqui}, {Andrada}, {Andringa},
  {Aramo}, {Arsene}, {Asorey}, {Assis}, {Avila}, {Badescu}, {Balaceanu},
  {Barbato}, {Barreira Luz}, {Becker}, {Bellido}, {Berat}, {Bertaina},
  {Bertou}, {Biermann}, {Biteau}, {Blaess}, {Blanco}, {Blazek}, {Bleve},
  {Boh{\'a}{\v{c}}ov{\'a}}, {Bonifazi}, {Borodai}, {Botti}, {Brack}, {Brancus},
  {Bretz}, {Bridgeman}, {Briechle}, {Buchholz}, {Bueno}, {Buitink}, {Buscemi},
  {Caballero-Mora}, {Caccianiga}, {Cancio}, {Canfora}, {Caruso}, {Castellina},
  {Catalani}, {Cataldi}, {Cazon}, {Chavez}, {Chinellato}, {Chudoba}, {Clay},
  {Cobos Cerutti}, {Colalillo}, {Coleman}, {Collica}, {Coluccia},
  {Concei{\c{c}}{\~a}o}, {Consolati}, {Contreras}, {Cooper}, {Coutu},
  {Covault}, {Cronin}, {D'Amico}, {Daniel}, {Dasso}, {Daumiller}, {Dawson},
  {Day}, {de Almeida}, {de Jong}, {De Mauro}, {de Mello Neto}, {De Mitri}, {de
  Oliveira}, {de Souza}, {Debatin}, {Deligny}, {D{\'\i}az Castro}, {Diogo},
  {Dobrigkeit}, {D'Olivo}, {Dorosti}, {Dos Anjos}, {Dova}, {Dundovic}, {Ebr},
  {Engel}, {Erdmann}, {Erfani}, {Escobar}, {Espadanal}, {Etchegoyen}, {Falcke},
  {Farmer}, {Farrar}, {Fauth}, {Fazzini}, {Feldbusch}, {Fenu}, {Fick},
  {Figueira}, {Filip{\v{c}}i{\v{c}}}, {Freire}, {Fujii}, {Fuster},
  {Ga{\"\i}or}, {Garc{\'\i}a}, {Gat{\'e}}, {Gemmeke}, {Gherghel-Lascu}, {Ghia},
  {Giaccari}, {Giammarchi}, {Giller}, {G{\l}as}, {Glaser}, {Golup}, {G{\'o}mez
  Berisso}, {G{\'o}mez Vitale}, {Gonz{\'a}lez}, {Gorgi}, {Gottowik}, {Grillo},
  {Grubb}, {Guarino}, {Guedes}, {Halliday}, {Hampel}, {Hansen}, {Harari},
  {Harrison}, {Harvey}, {Haungs}, {Hebbeker}, {Heck}, {Heimann}, {Herve},
  {Hill}, {Hojvat}, {Holt}, {Homola}, {H{\"o}randel}, {Horvath},
  {Hrabovsk{\'y}}, {Huege}, {Hulsman}, {Insolia}, {Isar}, {Jandt}, {Johnsen},
  {Josebachuili}, {Jurysek}, {K{\"a}{\"a}p{\"a}}, {Kampert}, {Keilhauer},
  {Kemmerich}, {Kemp}, {Kieckhafer}, {Klages}, {Kleifges}, {Kleinfeller},
  {Krause}, {Krohm}, {Kuempel}, {Kukec Mezek}, {Kunka}, {Kuotb Awad}, {Lago},
  {LaHurd}, {Lang}, {Lauscher}, {Legumina}, {Leigui de Oliveira},
  {Letessier-Selvon}, {Lhenry-Yvon}, {Link}, {Lo Presti}, {Lopes}, {L{\'o}pez},
  {L{\'o}pez Casado}, {Lorek}, {Luce}, {Lucero}, {Malacari}, {Mallamaci},
  {Mandat}, {Mantsch}, {Mariazzi}, {Maris}, {Marsella}, {Martello}, {Martinez},
  {Mart{\'\i}nez Bravo}, {Mas{\'\i}as Meza}, {Mathes}, {Mathys}, {Matthews},
  {Matthiae}, {Mayotte}, {Mazur}, {Medina}, {Medina-Tanco}, {Melo},
  {Menshikov}, {Merenda}, {Michal}, {Micheletti}, {Middendorf}, {Miramonti},
  {Mitrica}, {Mockler}, {Mollerach}, {Montanet}, {Morello}, {Morlino},
  {M{\"u}ller}, {M{\"u}ller}, {Muller}, {M{\"u}ller}, {Mussa}, {Naranjo},
  {Nguyen}, {Niculescu-Oglinzanu}, {Niechciol}, {Niemietz}, {Niggemann},
  {Nitz}, {Nosek}, {Novotny}, {No{\v{z}}ka}, {N{\'u}{\~n}ez}, {Oikonomou},
  {Olinto}, {Palatka}, {Pallotta}, {Papenbreer}, {Parente}, {Parra}, {Paul},
  {Pech}, {Pedreira}, {P{\c{e}}kala}, {Pe{\~n}a-Rodriguez}, {Pereira},
  {Perlin}, {Perrone}, {Peters}, {Petrera}, {Phuntsok}, {Pierog}, {Pimenta},
  {Pirronello}, {Platino}, {Plum}, {Poh}, {Porowski}, {Prado}, {Privitera},
  {Prouza}, {Quel}, {Querchfeld}, {Quinn}, {Ramos-Pollan}, {Rautenberg},
  {Ravignani}, {Ridky}, {Riehn}, {Risse}, {Ristori}, {Rizi}, {Rodrigues de
  Carvalho}, {Rodriguez Fernandez}, {Rodriguez Rojo}, {Roncoroni}, {Roth},
  {Roulet}, {Rovero}, {Ruehl}, {Saffi}, {Saftoiu}, {Salamida}, {Salazar},
  {Saleh}, {Salina}, {S{\'a}nchez}, {Sanchez-Lucas}, {Santos}, {Santos},
  {Sarazin}, {Sarmento}, {Sarmiento-Cano}, {Sato}, {Schauer}, {Scherini},
  {Schieler}, {Schimp}, {Schmidt}, {Scholten}, {Schov{\'a}nek}, {Schr{\"o}der},
  {Schr{\"o}der}, {Schulz}, {Schumacher}, {Sciutto}, {Segreto}, {Shadkam},
  {Shellard}, {Sigl}, {Silli}, {{\v{S}}m{\'\i}da}, {Snow}, {Sommers},
  {Sonntag}, {Soriano}, {Squartini}, {Stanca}, {Stani{\v{c}}}, {Stasielak},
  {Stassi}, {Stolpovskiy}, {Strafella}, {Streich}, {Suarez},
  {Suarez-Dur{\'a}n}, {Sudholz}, {Suomij{\"a}rvi}, {Supanitsky},
  {{\v{S}}up{\'\i}k}, {Swain}, {Szadkowski}, {Taboada}, {Taborda},
  {Timmermans}, {Todero Peixoto}, {Tomankova}, {Tom{\'e}}, {Torralba Elipe},
  {Travnicek}, {Trini}, {Tueros}, {Ulrich}, {Unger}, {Urban}, {Vald{\'e}s
  Galicia}, {Vali{\~n}o}, {Valore}, {van Aar}, {van Bodegom}, {van den Berg},
  {van Vliet}, {Varela}, {Vargas C{\'a}rdenas}, {V{\'a}zquez}, {Veberi{\v{c}}},
  {Ventura}, {Vergara Quispe}, {Verzi}, {Vicha}, {Villase{\~n}or}, {Vorobiov},
  {Wahlberg}, {Wainberg}, {Walz}, {Watson}, {Weber}, {Weindl}, {Wiede{\'n}ski},
  {Wiencke}, {Wilczy{\'n}ski}, {Wirtz}, {Wittkowski}, {Wundheiler}, {Yang},
  {Yushkov}, {Zas}, {Zavrtanik}, {Zavrtanik}, {Zepeda}, {Zimmermann},
  {Ziolkowski}, {Zong}, {Zuccarello}, {Pierre Auger Collaboration}, {Kim},
  {Schulze}, {Bauer}, {Corral-Santana}, {de Gregorio-Monsalvo},
  {Gonz{\'a}lez-L{\'o}pez}, {Hartmann}, {Ishwara-Chandra}, {Mart{\'\i}n},
  {Mehner}, {Misra}, {Micha{\l}owski}, {Resmi}, {ALMA Collaboration}, {Paragi},
  {Agudo}, {An}, {Beswick}, {Casadio}, {Frey}, {Jonker}, {Kettenis}, {Marcote},
  {Moldon}, {Szomoru}, {van Langevelde}, {Yang}, {Euro VLBI Team}, {Cwiek},
  {Cwiok}, {Czyrkowski}, {Dabrowski}, {Kasprowicz}, {Mankiewicz}, {Nawrocki},
  {Opiela}, {Piotrowski}, {Wrochna}, {Zaremba}, {Zarnecki}, {Pi of Sky
  Collaboration}, {Haggard}, {Nynka}, {Ruan}, {Chandra Team at McGill
  University}, {Bland}, {Booler}, {Devillepoix}, {de Gois}, {Hancock}, {Howie},
  {Paxman}, {Sansom}, {Towner}, {Desert Fireball Network}, {Tonry}, {Coughlin},
  {Stubbs}, {Denneau}, {Heinze}, {Stalder}, {Weiland}, {ATLAS}, {Eatough},
  {Kramer}, {Kraus}, {Time Resolution Universe Survey}, {Troja}, {Piro},
  {Becerra Gonz{\'a}lez}, {Butler}, {Fox}, {Khandrika}, {Kutyrev}, {Lee},
  {Ricci}, {Ryan}, {S{\'a}nchez-Ram{\'\i}rez}, {Veilleux}, {Watson},
  {Wieringa}, {Burgess}, {van Eerten}, {Fontes}, {Fryer}, {Korobkin},
  {Wollaeger}, {RIMAS}, {RATIR}, {Camilo}, {Foley}, {Goedhart}, {Makhathini},
  {Oozeer}, {Smirnov}, {Fender}, {Woudt}, and {South
  Africa/MeerKAT}}]{2017ApJ...848L..12A}
{Abbott} BP, {Abbott} R, {Abbott} TD, {Acernese} F, {Ackley} K, {Adams} C,
  {Adams} T, {Addesso} P, {Adhikari} RX, {Adya} VB, et~al. (2017{\natexlab{b}})
  {Multi-messenger Observations of a Binary Neutron Star Merger}. \apjl
  848(2):L12. \doi{10.3847/2041-8213/aa91c9}.
  {\href{https://arxiv.org/abs/1710.05833}{{arXiv:1710.05833}}} {[astro-ph.HE]}

\bibitem[{{Abbott} et~al.(2018{\natexlab{a}}){Abbott}, {Abbott}, {Abbott},
  {Acernese}, {Ackley}, {Adams}, {Adams}, {Addesso}, {Adhikari}, {Adya},
  {Affeldt}, and {et al.}}]{2018PhRvL.121p1101A}
{Abbott} BP, {Abbott} R, {Abbott} TD, {Acernese} F, {Ackley} K, {Adams} C,
  {Adams} T, {Addesso} P, {Adhikari} RX, {Adya} VB, et~al. (2018{\natexlab{a}})
  {GW170817: Measurements of Neutron Star Radii and Equation of State}. \prl
  121(16):161101. \doi{10.1103/PhysRevLett.121.161101}.
  {\href{https://arxiv.org/abs/1805.11581}{{arXiv:1805.11581}}} {[gr-qc]}

\bibitem[{{Abbott} et~al.(2018{\natexlab{b}}){Abbott}, {Abbott}, {Abbott},
  {Acernese}, {Ackley}, and et~al.}]{2018PhRvL.120i1101A}
{Abbott} BP, {Abbott} R, {Abbott} TD, {Acernese} F, {Ackley} K, et~al
  (2018{\natexlab{b}}) {GW170817: Implications for the Stochastic
  Gravitational-Wave Background from Compact Binary Coalescences}. \prl
  120(9):091101. \doi{10.1103/PhysRevLett.120.091101}.
  {\href{https://arxiv.org/abs/1710.05837}{{arXiv:1710.05837}}} {[gr-qc]}

\bibitem[{{Abbott} et~al.(2019){Abbott}, {Abbott}, {Abbott}, {Abraham},
  {Acernese}, {Ackley}, {Adams}, {Adhikari}, {Adya}, {Affeldt}, {Agathos},
  {Agatsuma}, {Aggarwal}, {Aguiar}, {Aiello}, {Ain}, and
  ...}]{2019PhRvX...9c1040A}
{Abbott} BP, {Abbott} R, {Abbott} TD, {Abraham} S, {Acernese} F, {Ackley} K,
  {Adams} C, {Adhikari} RX, {Adya} VB, {Affeldt} C, et~al. (2019) {GWTC-1: A
  Gravitational-Wave Transient Catalog of Compact Binary Mergers Observed by
  LIGO and Virgo during the First and Second Observing Runs}. Physical Review X
  9(3):031040. \doi{10.1103/PhysRevX.9.031040}.
  {\href{https://arxiv.org/abs/1811.12907}{{arXiv:1811.12907}}} {[astro-ph.HE]}

\bibitem[{{Abbott} et~al.(2020{\natexlab{a}}){Abbott}, {Abbott}, {Abbott},
  {Abraham}, {Acernese}, {Ackley}, {Adams}, {Adhikari}, {Adya}, {Affeldt},
  {Agathos}, {Agatsuma}, {Aggarwal}, and {et~al.}}]{2020ApJ...892L...3A}
{Abbott} BP, {Abbott} R, {Abbott} TD, {Abraham} S, {Acernese} F, {Ackley} K,
  {Adams} C, {Adhikari} RX, {Adya} VB, {Affeldt} C, et~al. (2020{\natexlab{a}})
  {GW190425: Observation of a Compact Binary Coalescence with Total Mass
  {\ensuremath{\sim}} 3.4 M$_{{\ensuremath{\odot}}}$}. \apjl 892(1):L3.
  \doi{10.3847/2041-8213/ab75f5}.
  {\href{https://arxiv.org/abs/2001.01761}{{arXiv:2001.01761}}} {[astro-ph.HE]}

\bibitem[{{Abbott} et~al.(2020{\natexlab{b}}){Abbott}, {Abbott}, {Abraham},
  {Acernese}, {Ackley}, {Adams}, {Adhikari}, {Adya}, {Affeldt}, {Agathos},
  {Agatsuma}, {Aggarwal}, {Aguiar}, {Aich}, {Aiello}, {Ain}, {Ajith}, and
  ...}]{2020ApJ...896L..44A}
{Abbott} R, {Abbott} TD, {Abraham} S, {Acernese} F, {Ackley} K, {Adams} C,
  {Adhikari} RX, {Adya} VB, {Affeldt} C, {Agathos} M, et~al.
  (2020{\natexlab{b}}) {GW190814: Gravitational Waves from the Coalescence of a
  23 Solar Mass Black Hole with a 2.6 Solar Mass Compact Object}. \apjl
  896(2):L44. \doi{10.3847/2041-8213/ab960f}.
  {\href{https://arxiv.org/abs/2006.12611}{{arXiv:2006.12611}}} {[astro-ph.HE]}

\bibitem[{{Abbott} et~al.(2020{\natexlab{c}}){Abbott}, {Abbott}, {Abraham},
  {Acernese}, {Ackley}, {Adams}, {Adhikari}, {Adya}, {Affeldt}, {Agathos},
  {Agatsuma}, {Aggarwal}, {Aguiar}, {Aich}, {Aiello}, {Ain}, {Ajith}, and
  et~al.}]{2020PhRvL.125j1102A}
{Abbott} R, {Abbott} TD, {Abraham} S, {Acernese} F, {Ackley} K, {Adams} C,
  {Adhikari} RX, {Adya} VB, {Affeldt} C, {Agathos} M, et~al.
  (2020{\natexlab{c}}) {GW190521: A Binary Black Hole Merger with a Total Mass
  of 150 M$_{\odot}$}. \prl 125(10):101102.
  \doi{10.1103/PhysRevLett.125.101102}

\bibitem[{{Abbott} et~al.(2020{\natexlab{d}}){Abbott}, {Abbott}, {Abraham},
  {Acernese}, {Ackley}, {Adams}, {Adhikari}, {Adya}, {Affeldt}, {Agathos},
  {Agatsuma}, {Aggarwal}, {Aguiar}, {Aich}, {Aiello}, {Ain}, {Ajith}, and
  et~al.}]{2020ApJ...900L..13A}
{Abbott} R, {Abbott} TD, {Abraham} S, {Acernese} F, {Ackley} K, {Adams} C,
  {Adhikari} RX, {Adya} VB, {Affeldt} C, {Agathos} M, et~al.
  (2020{\natexlab{d}}) {Properties and Astrophysical Implications of the 150
  M$_{\odot}$ Binary Black Hole Merger GW190521}. \apjl 900(1):L13.
  \doi{10.3847/2041-8213/aba493}

\bibitem[{{Abbott} et~al.(2021){Abbott}, {Abbott}, {Abraham}, {Acernese},
  {Ackley}, {Adams}, {Adams}, {Adhikari}, {Adya}, {Affeldt}, {Agathos},
  {Agatsuma}, {Aggarwal}, {Aguiar}, {Aiello}, {Ain}, {Ajith}, {Akcay}, {Allen},
  {Allocca}, {Altin}, {Amato}, {Anand}, {Ananyeva}, {Anderson}, {Anderson},
  {Angelova}, {Ansoldi}, {Antelis}, {Antier}, {Appert}, {Arai}, {Araya},
  {Areeda}, {Ar{\`e}ne}, {Arnaud}, {Aronson}, {Arun}, {Asali}, {Ascenzi},
  {Ashton}, {Aston}, {Astone}, {Aubin}, {Aufmuth}, {AultONeal}, {Austin},
  {Avendano}, {Babak}, {Badaracco}, {Bader}, {Bae}, {Baer}, {Bagnasco},
  {Baird}, {Ball}, {Ballardin}, {Ballmer}, {Bals}, {Balsamo}, {Baltus},
  {Banagiri}, {Bankar}, {Bankar}, {Barayoga}, {Barbieri}, {Barish}, {Barker},
  {Barneo}, {Barnum}, {Barone}, {Barr}, {Barsotti}, {Barsuglia}, {Barta},
  {Bartlett}, {Bartos}, {Bassiri}, {Basti}, {Bawaj}, {Bayley}, {Bazzan},
  {Becher}, {B{\'e}csy}, {Bedakihale}, {Bejger}, {Belahcene}, {Beniwal},
  {Benjamin}, {Bennett}, {Bentley}, {Bergamin}, {Berger}, {Bergmann},
  {Bernuzzi}, {Berry}, {Bersanetti}, {Bertolini}, {Betzwieser}, {Bhandare},
  {Bhandari}, {Bhattacharjee}, {Bidler}, {Bilenko}, {Billingsley}, {Birney},
  {Birnholtz}, {Biscans}, {Bischi}, {Biscoveanu}, {Bisht}, {Bitossi},
  {Bizouard}, {Blackburn}, {Blackman}, {Blair}, {Blair}, {Blair}, {Blanch},
  {Bobba}, {Bode}, {Boer}, {Boetzel}, {Bogaert}, {Boldrini}, {Bondu},
  {Bonilla}, {Bonnand}, {Booker}, {Boom}, {Bork}, {Boschi}, {Bose},
  {Bossilkov}, {Boudart}, {Bouffanais}, {Bozzi}, {Bradaschia}, {Brady},
  {Bramley}, {Branchesi}, {Brau}, {Breschi}, {Briant}, {Briggs}, {Brighenti},
  {Brillet}, {Brinkmann}, {Brockill}, {Brooks}, {Brooks}, {Brown}, {Brunett},
  {Bruno}, {Bruntz}, {Buikema}, {Bulik}, {Bulten}, {Buonanno}, {Buscicchio},
  {Buskulic}, {Byer}, {Cabero}, {Cadonati}, {Caesar}, {Cagnoli}, {Cahillane},
  {Calder{\'o}n Bustillo}, {Callaghan}, {Callister}, {Calloni}, {Camp},
  {Canepa}, {Cannon}, {Cao}, {Cao}, {Carapella}, {Carbognani}, {Carney},
  {Carpinelli}, {Carullo}, {Carver}, {Casanueva Diaz}, {Casentini}, {Caudill},
  {Cavagli{\`a}}, {Cavalier}, {Cavalieri}, {Cella}, {Cerd{\'a}-Dur{\'a}n},
  {Cesarini}, {Chaibi}, {Chakravarti}, {Chan}, {Chan}, {Chandra}, {Chanial},
  {Chao}, {Charlton}, {Chase}, {Chassande-Mottin}, {Chatterjee},
  {Chattopadhyay}, {Chaturvedi}, {Chatziioannou}, {Chen}, {Chen}, {Chen},
  {Chen}, {Cheng}, {Cheong}, {Chia}, {Chiadini}, {Chierici}, {Chincarini},
  {Chiummo}, {Cho}, {Cho}, {Cho}, {Choate}, {Christensen}, {Chu}, {Chua},
  {Chung}, {Chung}, {Ciani}, {Ciecielag}, {Cie{\'s}lar}, {Cifaldi}, {Ciobanu},
  {Ciolfi}, {Cipriano}, {Cirone}, {Clara}, {Clark}, {Clark}, {Clarke},
  {Clearwater}, {Clesse}, {Cleva}, {Coccia}, {Cohadon}, {Cohen}, {Colleoni},
  {Collette}, {Collins}, {Colpi}, {Constancio}, {Conti}, {Cooper}, {Corban},
  {Corbitt}, {Cordero-Carri{\'o}n}, {Corezzi}, {Corley}, {Cornish}, {Corre},
  {Corsi}, {Cortese}, {Costa}, {Cotesta}, {Coughlin}, {Coughlin}, {Coulon},
  {Countryman}, {Cousins}, {Couvares}, {Covas}, {Coward}, {Cowart}, {Coyne},
  {Coyne}, {Creighton}, {Creighton}, {Croquette}, {Crowder}, {Cudell},
  {Cullen}, {Cumming}, {Cummings}, {Cunningham}, {Cuoco}, {Cury{\l}o},
  {Canton}, {D{\'a}lya}, {Dana}, {DaneshgaranBajastani}, {D'Angelo}, {Danila},
  {Danilishin}, {D'Antonio}, {Danzmann}, {Darsow-Fromm}, {Dasgupta}, {Datrier},
  {Dattilo}, {Dave}, {Davier}, {Davies}, {Davis}, {Daw}, {Dean}, {DeBra},
  {Deenadayalan}, {Degallaix}, {De Laurentis}, {Del{\'e}glise}, {Del Favero},
  {De Lillo}, {De Lillo}, {Del Pozzo}, {DeMarchi}, {De Matteis}, {D'Emilio},
  {Demos}, {Denker}, {Dent}, {Depasse}, {De Pietri}, {De Rosa}, {De Rossi},
  {DeSalvo}, {de Varona}, {Dhurandhar}, {D{\'\i}az}, {Diaz-Ortiz}, {Didio},
  {Dietrich}, {Di Fiore}, {DiFronzo}, {Di Giorgio}, {Di Giovanni}, {Di
  Giovanni}, {Di Girolamo}, {Di Lieto}, {Ding}, {Di Pace}, {Di Palma}, {Di
  Renzo}, {Divakarla}, {Dmitriev}, {Doctor}, {D'Onofrio}, {Donovan}, {Dooley},
  {Doravari}, {Dorrington}, {Downes}, {Drago}, {Driggers}, {Du}, {Ducoin},
  {Dupej}, {Durante}, {D'Urso}, {Duverne}, {Dwyer}, {Easter}, {Eddolls},
  {Edelman}, {Edo}, {Edy}, {Effler}, {Eichholz}, {Eikenberry}, {Eisenmann},
  {Eisenstein}, {Ejlli}, {Errico}, {Essick}, {Estell{\'e}s}, {Estevez},
  {Etienne}, {Etzel}, {Evans}, {Evans}, {Ewing}, {Fafone}, {Fair}, {Fairhurst},
  {Fan}, {Farah}, {Farinon}, {Farr}, {Farr}, {Fauchon-Jones}, {Favata}, {Fays},
  {Fazio}, {Feicht}, {Fejer}, {Feng}, {Fenyvesi}, {Ferguson},
  {Fernandez-Galiana}, {Ferrante}, {Ferreira}, {Fidecaro}, {Figura}, {Fiori},
  {Fiorucci}, {Fishbach}, {Fisher}, {Fishner}, {Fittipaldi}, {Fitz-Axen},
  {Fiumara}, {Flaminio}, {Floden}, {Flynn}, {Fong}, {Font}, {Forsyth},
  {Fournier}, {Frasca}, {Frasconi}, {Frei}, {Freise}, {Frey}, {Frey},
  {Fritschel}, {Frolov}, {Fronz{\'e}}, {Fulda}, {Fyffe}, {Gabbard}, {Gadre},
  {Gaebel}, {Gair}, {Gais}, {Galaudage}, {Gamba}, {Ganapathy}, {Ganguly},
  {Gaonkar}, {Garaventa}, {Garc{\'\i}a-Quir{\'o}s}, {Garufi}, {Gateley},
  {Gaudio}, {Gayathri}, {Gemme}, {Gennai}, {George}, {George}, {George},
  {Gergely}, {Ghonge}, {Ghosh}, {Ghosh}, {Ghosh}, {Giacomazzo}, {Giacoppo},
  {Giaime}, {Giardina}, {Gibson}, {Gier}, {Gill}, {Giri}, {Glanzer}, {Gleckl},
  {Godwin}, {Goetz}, {Goetz}, {Gohlke}, {Goncharov}, {Gonz{\'a}lez},
  {Gopakumar}, {Gossan}, {Gosselin}, {Gouaty}, {Grace}, {Grado}, {Granata},
  {Granata}, {Grant}, {Gras}, {Grassia}, {Gray}, {Gray}, {Greco}, {Green},
  {Green}, {Gretarsson}, {Griggs}, {Grignani}, {Grimaldi}, {Grimes}, {Grimm},
  {Grote}, {Grunewald}, {Gruning}, {Guerrero}, {Guidi}, {Guimaraes},
  {Guix{\'e}}, {Gulati}, {Guo}, {Gupta}, {Gupta}, {Gupta}, {Gustafson},
  {Gustafson}, {Guzman}, {Haegel}, {Halim}, {Hall}, {Hamilton}, {Hammond},
  {Haney}, {Hanke}, {Hanks}, {Hanna}, {Hannam}, {Hannuksela}, {Hannuksela},
  {Hansen}, {Hansen}, {Hanson}, {Harder}, {Hardwick}, {Haris}, {Harms},
  {Harry}, {Harry}, {Hartwig}, {Hasskew}, {Haster}, {Haughian}, {Hayes},
  {Healy}, {Heidmann}, {Heintze}, {Heinze}, {Heinzel}, {Heitmann}, {Hellman},
  {Hello}, {Helmling-Cornell}, {Hemming}, {Hendry}, {Heng}, {Hennes}, {Hennig},
  {Hennig}, {Hernandez Vivanco}, {Heurs}, {Hild}, {Hill}, {Hines}, {Hochheim},
  {Hofgard}, {Hofman}, {Hohmann}, {Holgado}, {Holland}, {Hollows}, {Holmes},
  {Holt}, {Holz}, {Hopkins}, {Horst}, {Hough}, {Howell}, {Hoy}, {Hoyland},
  {Huang}, {H{\"u}bner}, {Huddart}, {Huerta}, {Hughey}, {Hui}, {Husa},
  {Huttner}, {Hutzler}, {Huxford}, {Huynh-Dinh}, {Idzkowski}, {Iess},
  {Imperato}, {Inchauspe}, {Ingram}, {Intini}, {Isi}, {Iyer},
  {JaberianHamedan}, {Jacqmin}, {Jadhav}, {Jadhav}, {James}, {Jani},
  {Janssens}, {Janthalur}, {Jaranowski}, {Jariwala}, {Jaume}, {Jenkins},
  {Jeunon}, {Jiang}, {Johns}, {Johnson-McDaniel}, {Jones}, {Jones}, {Jones},
  {Jones}, {Jones}, {Jonker}, {Ju}, {Junker}, {Kalaghatgi}, {Kalogera},
  {Kamai}, {Kandhasamy}, {Kang}, {Kanner}, {Kapadia}, {Kapasi}, {Karathanasis},
  {Karki}, {Kashyap}, {Kasprzack}, {Kastaun}, {Katsanevas}, {Katsavounidis},
  {Katzman}, {Kawabe}, {K{\'e}f{\'e}lian}, {Keitel}, {Key}, {Khadka},
  {Khalili}, {Khan}, {Khan}, {Khazanov}, {Khetan}, {Khursheed}, {Kijbunchoo},
  {Kim}, {Kim}, {Kim}, {Kim}, {Kim}, {Kim}, {Kimball}, {King}, {Kinley-Hanlon},
  {Kirchhoff}, {Kissel}, {Kleybolte}, {Klimenko}, {Knowles}, {Knyazev}, {Koch},
  {Koehlenbeck}, {Koekoek}, {Koley}, {Kolstein}, {Komori}, {Kondrashov},
  {Kontos}, {Koper}, {Korobko}, {Korth}, {Kovalam}, {Kozak}, {Kr{\"a}mer},
  {Kringel}, {Krishnendu}, {Kr{\'o}lak}, {Kuehn}, {Kumar}, {Kumar}, {Kumar},
  {Kumar}, {Kuns}, {Kwang}, {Lackey}, {Laghi}, {Lalande}, {Lam}, {Lamberts},
  {Landry}, {Lane}, {Lang}, {Lange}, {Lantz}, {Lanza}, {La Rosa},
  {Lartaux-Vollard}, {Lasky}, {Laxen}, {Lazzarini}, {Lazzaro}, {Leaci},
  {Leavey}, {Lecoeuche}, {Lee}, {Lee}, {Lee}, {Lee}, {Lehmann}, {Leon},
  {Leroy}, {Letendre}, {Levin}, {Li}, {Li}, {Li}, {Li}, {Li}, {Linde},
  {Linker}, {Linley}, {Littenberg}, {Liu}, {Liu}, {Llorens-Monteagudo}, {Lo},
  {Lockwood}, {London}, {Longo}, {Lorenzini}, {Loriette}, {Lormand}, {Losurdo},
  {Lough}, {Lousto}, {Lovelace}, {L{\"u}ck}, {Lumaca}, {Lundgren}, {Ma},
  {Macas}, {MacInnis}, {Macleod}, {MacMillan}, {Macquet}, {Maga{\~n}a
  Hernandez}, {Maga{\~n}a-Sandoval}, {Magazz{\`u}}, {Magee}, {Majorana},
  {Maksimovic}, {Maliakal}, {Malik}, {Man}, {Mandic}, {Mangano}, {Mansell},
  {Manske}, {Mantovani}, {Mapelli}, {Marchesoni}, {Marion}, {M{\'a}rka},
  {M{\'a}rka}, {Markakis}, {Markosyan}, {Markowitz}, {Maros}, {Marquina},
  {Marsat}, {Martelli}, {Martin}, {Martin}, {Martinez}, {Martinez}, {Martynov},
  {Masalehdan}, {Mason}, {Massera}, {Masserot}, {Massinger}, {Masso-Reid},
  {Mastrogiovanni}, {Matas}, {Mateu-Lucena}, {Matichard}, {Matiushechkina},
  {Mavalvala}, {Maynard}, {McCann}, {McCarthy}, {McClelland}, {McCormick},
  {McCuller}, {McGuire}, {McIsaac}, {McIver}, {McManus}, {McRae}, {McWilliams},
  {Meacher}, {Meadors}, {Mehmet}, {Mehta}, {Melatos}, {Melchor}, {Mendell},
  {Menendez-Vazquez}, {Mercer}, {Mereni}, {Merfeld}, {Merilh}, {Merritt},
  {Merzougui}, {Meshkov}, {Messenger}, {Messick}, {Metzdorff}, {Meyers},
  {Meylahn}, {Mhaske}, {Miani}, {Miao}, {Michaloliakos}, {Michel}, {Middleton},
  {Milano}, {Miller}, {Millhouse}, {Mills}, {Milotti}, {Milovich-Goff},
  {Minazzoli}, {Minenkov}, {Mir}, {Mishkin}, {Mishra}, {Mistry}, {Mitra},
  {Mitrofanov}, {Mitselmakher}, {Mittleman}, {Mo}, {Mogushi}, {Mohapatra},
  {Mohite}, {Molina}, {Molina-Ruiz}, {Mondin}, {Montani}, {Moore}, {Moraru},
  {Morawski}, {Moreno}, {Morisaki}, {Mours}, {Mow-Lowry}, {Mozzon},
  {Muciaccia}, {Mukherjee}, {Mukherjee}, {Mukherjee}, {Mukherjee}, {Mukund},
  {Mullavey}, {Munch}, {Mu{\~n}iz}, {Murray}, {Nadji}, {Nagar}, {Nardecchia},
  {Naticchioni}, {Nayak}, {Neil}, {Neilson}, {Nelemans}, {Nelson}, {Nery},
  {Neunzert}, {Nitz}, {Ng}, {Ng}, {Nguyen}, {Nguyen}, {Nguyen}, {Nichols},
  {Nissanke}, {Nocera}, {Noh}, {North}, {Nothard}, {Nuttall}, {Oberling},
  {O'Brien}, {O'Dell}, {Oganesyan}, {Ogin}, {Oh}, {Oh}, {Ohme}, {Ohta},
  {Okada}, {Olivetto}, {Oppermann}, {Oram}, {O'Reilly}, {Ormiston}, {Ortega},
  {O'Shaughnessy}, {Ossokine}, {Osthelder}, {Ottaway}, {Overmier}, {Owen},
  {Pace}, {Pagano}, {Page}, {Pagliaroli}, {Pai}, {Pai}, {Palamos}, {Palashov},
  {Palomba}, {Pan}, {Panda}, {Pang}, {Pankow}, {Pannarale}, {Pant}, {Paoletti},
  {Paoli}, {Paolone}, {Parker}, {Pascucci}, {Pasqualetti}, {Passaquieti},
  {Passuello}, {Patel}, {Patricelli}, {Payne}, {Pechsiri}, {Pedraza},
  {Pegoraro}, {Pele}, {Penn}, {Perego}, {Perez}, {P{\'e}rigois}, {Perreca},
  {Perri{\`e}s}, {Petermann}, {Petterson}, {Pfeiffer}, {Pham}, {Phukon},
  {Piccinni}, {Pichot}, {Piendibene}, {Piergiovanni}, {Pierini}, {Pierro},
  {Pillant}, {Pilo}, {Pinard}, {Pinto}, {Piotrzkowski}, {Pirello}, {Pitkin},
  {Placidi}, {Plastino}, {Pluchar}, {Poggiani}, {Polini}, {Pong}, {Ponrathnam},
  {Popolizio}, {Porter}, {Poverman}, {Powell}, {Pracchia}, {Prajapati},
  {Prasai}, {Prasanna}, {Pratten}, {Prestegard}, {Principe}, {Prodi},
  {Prokhorov}, {Prosposito}, {Prudenzi}, {Puecher}, {Punturo}, {Puosi},
  {Puppo}, {P{\"u}rrer}, {Qi}, {Quetschke}, {Quinonez}, {Quitzow-James},
  {Raab}, {Raaijmakers}, {Radkins}, {Radulesco}, {Raffai}, {Rafferty}, {Rail},
  {Raja}, {Rajan}, {Rajbhandari}, {Rakhmanov}, {Ramirez}, {Ramirez},
  {Ramos-Buades}, {Rana}, {Rao}, {Rapagnani}, {Rapol}, {Ratto}, {Raymond},
  {Razzano}, {Read}, {Regimbau}, {Rei}, {Reid}, {Reitze}, {Rettegno}, {Ricci},
  {Richardson}, {Richardson}, {Richardson}, {Ricker}, {Riemenschneider},
  {Riles}, {Rizzo}, {Robertson}, {Robinet}, {Rocchi}, {Rocha}, {Rodriguez},
  {Rodriguez-Soto}, {Rolland}, {Rollins}, {Roma}, {Romanelli}, {Romano},
  {Romel}, {Romero}, {Romero-Shaw}, {Romie}, {Ronchini}, {Rose}, {Rose},
  {Rose}, {Rosell}, {Rosi{\'n}ska}, {Rosofsky}, {Ross}, {Rowan}, {Rowlinson},
  {Roy}, {Roy}, {Ruggi}, {Ryan}, {Sachdev}, {Sadecki}, {Sadiq},
  {Sakellariadou}, {Salafia}, {Salconi}, {Saleem}, {Samajdar}, {Sanchez},
  {Sanchez}, {Sanchez}, {Sanchis-Gual}, {Sanders}, {Sandles}, {Santiago},
  {Santos}, {Saravanan}, {Sarin}, {Sassolas}, {Sathyaprakash}, {Sauter},
  {Savage}, {Savant}, {Sawant}, {Sayah}, {Schaetzl}, {Schale}, {Scheel},
  {Scheuer}, {Schindler-Tyka}, {Schmidt}, {Schnabel}, {Schofield},
  {Sch{\"o}nbeck}, {Schreiber}, {Schulte}, {Schutz}, {Schwarm}, {Schwartz},
  {Scott}, {Scott}, {Seglar-Arroyo}, {Seidel}, {Sellers}, {Sengupta},
  {Sennett}, {Sentenac}, {Sequino}, {Sergeev}, {Setyawati}, {Shaffer},
  {Shahriar}, {Sharifi}, {Sharma}, {Sharma}, {Shawhan}, {Shen}, {Shikauchi},
  {Shink}, {Shoemaker}, {Shoemaker}, {Shukla}, {ShyamSundar}, {Sieniawska},
  {Sigg}, {Singer}, {Singh}, {Singh}, {Singha}, {Singhal}, {Sintes}, {Sipala},
  {Skliris}, {Slagmolen}, {Slaven-Blair}, {Smetana}, {Smith}, {Smith},
  {Somala}, {Son}, {Soni}, {Soni}, {Sorazu}, {Sordini}, {Sorrentino},
  {Sorrentino}, {Soulard}, {Souradeep}, {Sowell}, {Spencer}, {Spera},
  {Srivastava}, {Srivastava}, {Staats}, {Stachie}, {Steer}, {Steinhoff},
  {Steinke}, {Steinlechner}, {Steinlechner}, {Steinmeyer}, {Stevenson},
  {Stolle-McAllister}, {Stops}, {Stover}, {Strain}, {Stratta}, {Strunk},
  {Sturani}, {Stuver}, {S{\"u}dbeck}, {Sudhagar}, {Sudhir}, {Suh},
  {Summerscales}, {Sun}, {Sun}, {Sunil}, {Sur}, {Suresh}, {Sutton}, {Swinkels},
  {Szczepa{\'n}czyk}, {Tacca}, {Tait}, {Talbot}, {Tanasijczuk}, {Tanner},
  {Tao}, {Tapia}, {Tapia San Martin}, {Tasson}, {Taylor}, {Tenorio},
  {Terkowski}, {Thirugnanasambandam}, {Thomas}, {Thomas}, {Thomas}, {Thompson},
  {Thondapu}, {Thorne}, {Thrane}, {Tiwari}, {Tiwari}, {Tiwari}, {Toland},
  {Tolley}, {Tonelli}, {Tornasi}, {Torres-Forn{\'e}}, {Torrie}, {e Melo},
  {T{\"o}yr{\"a}}, {Tran}, {Trapananti}, {Travasso}, {Traylor}, {Tringali},
  {Tripathee}, {Trovato}, {Trudeau}, {Tsai}, {Tsang}, {Tse}, {Tso}, {Tsukada},
  {Tsuna}, {Tsutsui}, {Turconi}, {Ubhi}, {Udall}, {Ueno}, {Ugolini},
  {Unnikrishnan}, {Urban}, {Usman}, {Utina}, {Vahlbruch}, {Vajente}, {Vajpeyi},
  {Valdes}, {Valentini}, {Valsan}, {van Bakel}, {van Beuzekom}, {van den
  Brand}, {Van Den Broeck}, {Vander-Hyde}, {van der Schaaf}, {van Heijningen},
  {Vardaro}, {Vargas}, {Varma}, {Vass}, {Vas{\'u}th}, {Vecchio}, {Vedovato},
  {Veitch}, {Veitch}, {Venkateswara}, {Venneberg}, {Venugopalan}, {Verkindt},
  {Verma}, {Veske}, {Vetrano}, {Vicer{\'e}}, {Viets}, {Vijaykumar},
  {Villa-Ortega}, {Vinet}, {Vitale}, {Vo}, {Vocca}, {Vorvick}, {Vyatchanin},
  {Wade}, {Wade}, {Wade}, {Walet}, {Walker}, {Wallace}, {Wallace}, {Walsh},
  {Wang}, {Wang}, {Wang}, {Wang}, {Ward}, {Warner}, {Was}, {Washington},
  {Watchi}, {Weaver}, {Wei}, {Weinert}, {Weinstein}, {Weiss}, {Wellmann},
  {Wen}, {We{\ss}els}, {Westhouse}, {Wette}, {Whelan}, {White}, {White},
  {Whiting}, {Whittle}, {Wilken}, {Williams}, {Williams}, {Williamson},
  {Willis}, {Willke}, {Wilson}, {Wimmer}, {Winkler}, {Wipf}, {Woan}, {Woehler},
  {Wofford}, {Wong}, {Wrangel}, {Wright}, {Wu}, {Wysocki}, {Xiao}, {Yamamoto},
  {Yang}, {Yang}, {Yang}, {Yap}, {Yeeles}, {Yoon}, {Yu}, {Yu}, {Yuen},
  {Zadrozny}, {Zanolin}, {Zelenova}, {Zendri}, {Zevin}, {Zhang}, {Zhang},
  {Zhang}, {Zhang}, {Zhao}, {Zhao}, {Zheng}, {Zhou}, {Zhou}, {Zhu},
  {Zimmerman}, {Zlochower}, {Zucker}, {Zweizig}, {LIGO Scientific
  Collaboration}, and {Virgo Collaboration}}]{2020arXiv201014527A}
{Abbott} R, {Abbott} TD, {Abraham} S, {Acernese} F, {Ackley} K, {Adams} A,
  {Adams} C, {Adhikari} RX, {Adya} VB, {Affeldt} C, et~al. (2021) {GWTC-2:
  Compact Binary Coalescences Observed by LIGO and Virgo during the First Half
  of the Third Observing Run}. Physical Review X 11(2):021053.
  \doi{10.1103/PhysRevX.11.021053}.
  {\href{https://arxiv.org/abs/2010.14527}{{arXiv:2010.14527}}} {[gr-qc]}

\bibitem[{{Abel} et~al.(2002){Abel}, {Bryan}, and
  {Norman}}]{2002Sci...295...93A}
{Abel} T, {Bryan} GL, {Norman} ML (2002) {The Formation of the First Star in
  the Universe}. Science 295:93--98. \doi{10.1126/science.1063991}.
  {\href{https://arxiv.org/abs/arXiv:astro-ph/0112088}{{arXiv:astro-ph/0112088}}}

\bibitem[{{Abuter} et~al.(2018){Abuter}, {Amorim}, {Anugu}, {Baub{\"o}ck},
  {Benisty}, {Berger}, {Blind}, {Bonnet}, {Brandner}, {Buron}, {Collin},
  {Chapron}, {Cl{\'e}net}, {Coud{\'e} Du Foresto}, {de Zeeuw}, {Deen},
  {Delplancke-Str{\"o}bele}, {Dembet}, {Dexter}, {Duvert}, {Eckart},
  {Eisenhauer}, {Finger}, {F{\"o}rster Schreiber}, {F{\'e}dou}, {Garcia},
  {Garcia Lopez}, {Gao}, {Gendron}, {Genzel}, {Gillessen}, {Gordo}, {Habibi},
  {Haubois}, {Haug}, {Hau{\ss}mann}, {Henning}, {Hippler}, {Horrobin},
  {Hubert}, {Hubin}, {Jimenez Rosales}, {Jochum}, {Jocou}, {Kaufer}, {Kellner},
  {Kendrew}, {Kervella}, {Kok}, {Kulas}, {Lacour}, {Lapeyr{\`e}re}, {Lazareff},
  {Le Bouquin}, {L{\'e}na}, {Lippa}, {Lenzen}, {M{\'e}rand}, {M{\"u}ler},
  {Neumann}, {Ott}, {Palanca}, {Paumard}, {Pasquini}, {Perraut}, {Perrin},
  {Pfuhl}, {Plewa}, {Rabien}, {Ram{\'\i}rez}, {Ramos}, {Rau},
  {Rodr{\'\i}guez-Coira}, {Rohloff}, {Rousset}, {Sanchez-Bermudez},
  {Scheithauer}, {Sch{\"o}ller}, {Schuler}, {Spyromilio}, {Straub},
  {Straubmeier}, {Sturm}, {Tacconi}, {Tristram}, {Vincent}, {von Fellenberg},
  {Wank}, {Waisberg}, {Widmann}, {Wieprecht}, {Wiest}, {Wiezorrek}, {Woillez},
  {Yazici}, {Ziegler}, and {Zins}}]{2018A&A...615L..15G}
{Abuter} R, {Amorim} A, {Anugu} N, {Baub{\"o}ck} M, {Benisty} M, {Berger} JP,
  {Blind} N, {Bonnet} H, {Brandner} W, {Buron} A, et~al. (2018) {Detection of
  the gravitational redshift in the orbit of the star S2 near the Galactic
  centre massive black hole}. \aap 615:L15. \doi{10.1051/0004-6361/201833718}.
  {\href{https://arxiv.org/abs/1807.09409}{{arXiv:1807.09409}}} {[astro-ph.GA]}

\bibitem[{{Abuter} et~al.(2020){Abuter}, {Amorim}, {Baub{\"o}ck}, {Berger},
  {Bonnet}, {Brand ner}, {Cardoso}, {Cl{\'e}net}, {de Zeeuw}, {Dexter},
  {Eckart}, {Eisenhauer}, {F{\"o}rster Schreiber}, {Garcia}, {Gao}, {Gendron},
  {Genzel}, {Gillessen}, {Habibi}, {Haubois}, {Henning}, {Hippler}, {Horrobin},
  {Jim{\'e}nez-Rosales}, {Jochum}, {Jocou}, {Kaufer}, {Kervella}, {Lacour},
  {Lapeyr{\`e}re}, {Le Bouquin}, {L{\'e}na}, {Nowak}, {Ott}, {Paumard},
  {Perraut}, {Perrin}, {Pfuhl}, {Rodr{\'\i}guez-Coira}, {Shangguan},
  {Scheithauer}, {Stadler}, {Straub}, {Straubmeier}, {Sturm}, {Tacconi},
  {Vincent}, {von Fellenberg}, {Waisberg}, {Widmann}, {Wieprecht}, {Wiezorrek},
  {Woillez}, {Yazici}, and {Zins}}]{2020A&A...636L...5G}
{Abuter} R, {Amorim} A, {Baub{\"o}ck} M, {Berger} JP, {Bonnet} H, {Brand ner}
  W, {Cardoso} V, {Cl{\'e}net} Y, {de Zeeuw} PT, {Dexter} J, et~al. (2020)
  {Detection of the Schwarzschild precession in the orbit of the star S2 near
  the Galactic centre massive black hole}. \aap 636:L5.
  \doi{10.1051/0004-6361/202037813}.
  {\href{https://arxiv.org/abs/2004.07187}{{arXiv:2004.07187}}} {[astro-ph.GA]}

\bibitem[{Acernese et~al.(2015)}]{AdvVirgo}
Acernese F, et~al. (2015) {Advanced Virgo: a second-generation interferometric
  gravitational wave detector}. Class Quant Grav 32(2):024001.
  \doi{10.1088/0264-9381/32/2/024001}.
  {\href{https://arxiv.org/abs/1408.3978}{{arXiv:1408.3978}}} {[gr-qc]}

\bibitem[{{Ackermann} et~al.(2018){Ackermann}, {Schawinski}, {Zhang}, {Weigel},
  and {Turp}}]{2018MNRAS.479..415A}
{Ackermann} S, {Schawinski} K, {Zhang} C, {Weigel} AK, {Turp} MD (2018) {Using
  transfer learning to detect galaxy mergers}. \mnras 479(1):415--425.
  \doi{10.1093/mnras/sty1398}.
  {\href{https://arxiv.org/abs/1805.10289}{{arXiv:1805.10289}}} {[astro-ph.IM]}

\bibitem[{{Adams} and {Cornish}(2014)}]{2014PhRvD..89b2001A}
{Adams} MR, {Cornish} NJ (2014) {Detecting a stochastic gravitational wave
  background in the presence of a galactic foreground and instrument noise}.
  \prd 89(2):022001. \doi{10.1103/PhysRevD.89.022001}.
  {\href{https://arxiv.org/abs/1307.4116}{{arXiv:1307.4116}}} {[gr-qc]}

\bibitem[{{Adams} et~al.(2012){Adams}, {Cornish}, and
  {Littenberg}}]{2012PhRvD..86l4032A}
{Adams} MR, {Cornish} NJ, {Littenberg} TB (2012) {Astrophysical model selection
  in gravitational wave astronomy}. \prd 86(12):124032.
  \doi{10.1103/PhysRevD.86.124032}.
  {\href{https://arxiv.org/abs/1209.6286}{{arXiv:1209.6286}}} {[gr-qc]}

\bibitem[{{Addison} et~al.(2019){Addison}, {Gracia-Linares}, {Laguna}, and
  {Larson}}]{2019GReGr..51...38A}
{Addison} E, {Gracia-Linares} M, {Laguna} P, {Larson} SL (2019) {Busting up
  binaries: encounters between compact binaries and a supermassive black hole}.
  General Relativity and Gravitation 51(3):38. \doi{10.1007/s10714-019-2523-4}

\bibitem[{Adri\'an-Mart\'{\i}nez et~al.(2016)}]{2016PhRvD..9312010A}
Adri\'an-Mart\'{\i}nez S, et~al. (2016) High-energy neutrino follow-up search
  of gravitational wave event gw150914 with antares and icecube. Phys Rev D
  93:122010. \doi{10.1103/PhysRevD.93.122010},
  \urlprefix\url{https://link.aps.org/doi/10.1103/PhysRevD.93.122010}

\bibitem[{{Agarwal} et~al.(2012){Agarwal}, {Khochfar}, {Johnson}, {Neistein},
  {Dalla Vecchia}, and {Livio}}]{2012MNRAS.425.2854A}
{Agarwal} B, {Khochfar} S, {Johnson} JL, {Neistein} E, {Dalla Vecchia} C,
  {Livio} M (2012) {Ubiquitous seeding of supermassive black holes by direct
  collapse}. \mnras 425(4):2854--2871. \doi{10.1111/j.1365-2966.2012.21651.x}.
  {\href{https://arxiv.org/abs/1205.6464}{{arXiv:1205.6464}}} {[astro-ph.CO]}

\bibitem[{{Agarwal} et~al.(2014){Agarwal}, {Dalla Vecchia}, {Johnson},
  {Khochfar}, and {Paardekooper}}]{2014MNRAS.443..648A}
{Agarwal} B, {Dalla Vecchia} C, {Johnson} JL, {Khochfar} S, {Paardekooper} JP
  (2014) {The First Billion Years project: birthplaces of direct collapse black
  holes}. \mnras 443(1):648--657. \doi{10.1093/mnras/stu1112}.
  {\href{https://arxiv.org/abs/1403.5267}{{arXiv:1403.5267}}} {[astro-ph.CO]}

\bibitem[{Ahn et~al.(2017)Ahn, Seth, den Brok, Strader, Baumgardt, van~den
  Bosch, Chilingarian, Frank, Hilker, and McDermid}]{Ahn:2017lzs}
Ahn CP, Seth AC, den Brok M, Strader J, Baumgardt H, van~den Bosch R,
  Chilingarian I, Frank M, Hilker M, McDermid R (2017) {Detection of
  Supermassive Black Holes in Two Virgo Ultracompact Dwarf Galaxies}. \apj
  839:72. \doi{10.3847/1538-4357/aa6972}.
  {\href{https://arxiv.org/abs/{arXiv:1703.09221
  [astro-ph.GA]}}{{{arXiv:1703.09221 [astro-ph.GA]}}}}

\bibitem[{{Aird} et~al.(2018){Aird}, {Coil}, and
  {Georgakakis}}]{2018MNRAS.474.1225A}
{Aird} J, {Coil} AL, {Georgakakis} A (2018) {X-rays across the galaxy
  population - II. The distribution of AGN accretion rates as a function of
  stellar mass and redshift.} \mnras 474:1225--1249.
  \doi{10.1093/mnras/stx2700}.
  {\href{https://arxiv.org/abs/1705.01132}{{arXiv:1705.01132}}} {[astro-ph.HE]}

\bibitem[{{Akiyama} et~al.(2019){Akiyama}, {Alberdi}, {Alef}, {Asada},
  {Azulay}, {Baczko}, {Ball}, {Balokovi{\'c}}, {Barrett}, {Bintley},
  {Blackburn}, {Boland}, {Bouman}, {Bower}, {Bremer}, {Brinkerink},
  {Brissenden}, {Britzen}, {Broderick}, {Broguiere}, {Bronzwaer}, {Byun},
  {Carlstrom}, {Chael}, {Chan}, {Chatterjee}, {Chatterjee}, {Chen}, {Chen},
  {Cho}, {Christian}, {Conway}, {Cordes}, {Crew}, {Cui}, {Davelaar}, {De
  Laurentis}, {Deane}, {Dempsey}, {Desvignes}, {Dexter}, {Doeleman}, {Eatough},
  {Falcke}, {Fish}, {Fomalont}, {Fraga-Encinas}, {Freeman}, {Friberg}, {Fromm},
  {G{\'o}mez}, {Galison}, {Gammie}, {Garc{\'\i}a}, {Gentaz}, {Georgiev},
  {Goddi}, {Gold}, {Gu}, {Gurwell}, {Hada}, {Hecht}, {Hesper}, {Ho}, {Ho},
  {Honma}, {Huang}, {Huang}, {Hughes}, {Ikeda}, {Inoue}, {Issaoun}, {James},
  {Jannuzi}, {Janssen}, {Jeter}, {Jiang}, {Johnson}, {Jorstad}, {Jung},
  {Karami}, {Karuppusamy}, {Kawashima}, {Keating}, {Kettenis}, {Kim}, {Kim},
  {Kim}, {Kino}, {Koay}, {Koch}, {Koyama}, {Kramer}, {Kramer}, {Krichbaum},
  {Kuo}, {Lauer}, {Lee}, {Li}, {Li}, {Lindqvist}, {Liu}, {Liuzzo}, {Lo},
  {Lobanov}, {Loinard}, {Lonsdale}, {Lu}, {MacDonald}, {Mao}, {Markoff},
  {Marrone}, {Marscher}, {Mart{\'\i}-Vidal}, {Matsushita}, {Matthews},
  {Medeiros}, {Menten}, {Mizuno}, {Mizuno}, {Moran}, {Moriyama},
  {Moscibrodzka}, {M{\"u}ller}, {Nagai}, {Nagar}, {Nakamura}, {Narayan},
  {Narayanan}, {Natarajan}, {Neri}, {Ni}, {Noutsos}, {Okino}, {Olivares},
  {Ortiz-Le{\'o}n}, {Oyama}, {{\"O}zel}, {Palumbo}, {Patel}, {Pen}, {Pesce},
  {Pi{\'e}tu}, {Plambeck}, {PopStefanija}, {Porth}, {Prather},
  {Preciado-L{\'o}pez}, {Psaltis}, {Pu}, {Ramakrishnan}, {Rao}, {Rawlings},
  {Raymond}, {Rezzolla}, {Ripperda}, {Roelofs}, {Rogers}, {Ros}, {Rose},
  {Roshanineshat}, {Rottmann}, {Roy}, {Ruszczyk}, {Ryan}, {Rygl},
  {S{\'a}nchez}, {S{\'a}nchez-Arguelles}, {Sasada}, {Savolainen}, {Schloerb},
  {Schuster}, {Shao}, {Shen}, {Small}, {Sohn}, {SooHoo}, {Tazaki}, {Tiede},
  {Tilanus}, {Titus}, {Toma}, {Torne}, {Trent}, {Trippe}, {Tsuda}, {van
  Bemmel}, {van Langevelde}, {van Rossum}, {Wagner}, {Wardle}, {Weintroub},
  {Wex}, {Wharton}, {Wielgus}, {Wong}, {Wu}, {Young}, {Young}, {Younsi},
  {Yuan}, {Yuan}, {Zensus}, {Zhao}, {Zhao}, {Zhu}, {Algaba}, {Allardi},
  {Amestica}, {Anczarski}, {Bach}, {Baganoff}, {Beaudoin}, {Benson},
  {Berthold}, {Blanchard}, {Blundell}, {Bustamente}, {Cappallo},
  {Castillo-Dom{\'\i}nguez}, {Chang}, {Chang}, {Chang}, {Chen}, {Chilson},
  {Chuter}, {C{\'o}rdova Rosado}, {Coulson}, {Crawford}, {Crowley}, {David},
  {Derome}, {Dexter}, {Dornbusch}, {Dudevoir}, {Dzib}, {Eckart}, {Eckert},
  {Erickson}, {Everett}, {Faber}, {Farah}, {Fath}, {Folkers}, {Forbes},
  {Freund}, {G{\'o}mez-Ruiz}, {Gale}, {Gao}, {Geertsema}, {Graham}, {Greer},
  {Grosslein}, {Gueth}, {Haggard}, {Halverson}, {Han}, {Han}, {Hao},
  {Hasegawa}, {Henning}, {Hern{\'a}ndez-G{\'o}mez}, {Herrero-Illana},
  {Heyminck}, {Hirota}, {Hoge}, {Huang}, {Impellizzeri}, {Jiang}, {Kamble},
  {Keisler}, {Kimura}, {Kono}, {Kubo}, {Kuroda}, {Lacasse}, {Laing}, {Leitch},
  {Li}, {Lin}, {Liu}, {Liu}, {Lu}, {Marson}, {Martin-Cocher}, {Massingill},
  {Matulonis}, {McColl}, {McWhirter}, {Messias}, {Meyer-Zhao}, {Michalik},
  {Monta{\~n}a}, {Montgomerie}, {Mora-Klein}, {Muders}, {Nadolski}, {Navarro},
  {Neilsen}, {Nguyen}, {Nishioka}, {Norton}, {Nowak}, {Nystrom}, {Ogawa},
  {Oshiro}, {Oyama}, {Parsons}, {Paine}, {Pe{\~n}alver}, {Phillips}, {Poirier},
  {Pradel}, {Primiani}, {Raffin}, {Rahlin}, {Reiland}, {Risacher}, {Ruiz},
  {S{\'a}ez-Mada{\'\i}n}, {Sassella}, {Schellart}, {Shaw}, {Silva}, {Shiokawa},
  {Smith}, {Snow}, {Souccar}, {Sousa}, {Sridharan}, {Srinivasan}, {Stahm},
  {Stark}, {Story}, {Timmer}, {Vertatschitsch}, {Walther}, {Wei}, {Whitehorn},
  {Whitney}, {Woody}, {Wouterloot}, {Wright}, {Yamaguchi}, {Yu}, {Zeballos},
  {Zhang}, and {Ziurys}}]{2019ApJ...875L...1E}
{Akiyama} K, {Alberdi} A, {Alef} W, {Asada} K, {Azulay} R, {Baczko} AK, {Ball}
  D, {Balokovi{\'c}} M, {Barrett} J, {Bintley} D, et~al. (2019) {First M87
  Event Horizon Telescope Results. I. The Shadow of the Supermassive Black
  Hole}. \apjl 875(1):L1. \doi{10.3847/2041-8213/ab0ec7}.
  {\href{https://arxiv.org/abs/1906.11238}{{arXiv:1906.11238}}} {[astro-ph.GA]}

\bibitem[{{Alexander} and {Hickox}(2012)}]{2012NewAR..56...93A}
{Alexander} DM, {Hickox} RC (2012) {What drives the growth of black holes?}
  \nar 56(4):93--121. \doi{10.1016/j.newar.2011.11.003}.
  {\href{https://arxiv.org/abs/1112.1949}{{arXiv:1112.1949}}} {[astro-ph.GA]}

\bibitem[{{Alexander} et~al.(2020){Alexander}, {van Velzen}, {Horesh}, and
  {Zauderer}}]{2020SSRv..216...81A}
{Alexander} KD, {van Velzen} S, {Horesh} A, {Zauderer} BA (2020) {Radio
  Properties of Tidal Disruption Events}. \ssr 216(5):81.
  \doi{10.1007/s11214-020-00702-w}.
  {\href{https://arxiv.org/abs/2006.01159}{{arXiv:2006.01159}}} {[astro-ph.HE]}

\bibitem[{{Alexander} et~al.(2008){Alexander}, {Armitage}, and
  {Cuadra}}]{2008MNRAS.389.1655A}
{Alexander} RD, {Armitage} PJ, {Cuadra} J (2008) {Binary formation and mass
  function variations in fragmenting discs with short cooling times}. \mnras
  389(4):1655--1664. \doi{10.1111/j.1365-2966.2008.13706.x}.
  {\href{https://arxiv.org/abs/0807.1731}{{arXiv:0807.1731}}} {[astro-ph]}

\bibitem[{{Alexander}(2017)}]{2017ARA&A..55...17A}
{Alexander} T (2017) {Stellar Dynamics and Stellar Phenomena Near a Massive
  Black Hole}. \araa 55(1):17--57. \doi{10.1146/annurev-astro-091916-055306}.
  {\href{https://arxiv.org/abs/1701.04762}{{arXiv:1701.04762}}} {[astro-ph.GA]}

\bibitem[{{Alexander} and {Bar-Or}(2017)}]{2017NatAs...1E.147A}
{Alexander} T, {Bar-Or} B (2017) {A universal minimal mass scale for
  present-day central black holes}. Nature Astronomy 1:0147.
  \doi{10.1038/s41550-017-0147}.
  {\href{https://arxiv.org/abs/1701.00415}{{arXiv:1701.00415}}} {[astro-ph.GA]}

\bibitem[{{Alexander} and {Natarajan}(2014)}]{Alexander_2014}
{Alexander} T, {Natarajan} P (2014) {Rapid growth of seed black holes in the
  early universe by supra-exponential accretion}. Science 345(6202):1330--1333.
  \doi{10.1126/science.1251053}.
  {\href{https://arxiv.org/abs/1408.1718}{{arXiv:1408.1718}}} {[astro-ph.GA]}

\bibitem[{{Ali} et~al.(2012){Ali}, {Christensen}, {Meyer}, and
  {R{\"o}ver}}]{2012CQGra..29n5014A}
{Ali} A, {Christensen} N, {Meyer} R, {R{\"o}ver} C (2012) {Bayesian inference
  on EMRI signals using low frequency approximations}. Classical and Quantum
  Gravity 29(14):145014. \doi{10.1088/0264-9381/29/14/145014}.
  {\href{https://arxiv.org/abs/1301.0455}{{arXiv:1301.0455}}} {[gr-qc]}

\bibitem[{{Ali-Ha{\"\i}moud} and {Kamionkowski}(2017)}]{2017PhRvD..95d3534A}
{Ali-Ha{\"\i}moud} Y, {Kamionkowski} M (2017) {Cosmic microwave background
  limits on accreting primordial black holes}. \prd 95(4):043534.
  \doi{10.1103/PhysRevD.95.043534}.
  {\href{https://arxiv.org/abs/1612.05644}{{arXiv:1612.05644}}} {[astro-ph.CO]}

\bibitem[{Ali-Ha\"\i{}moud et~al.(2017)Ali-Ha\"\i{}moud, Kovetz, and
  Kamionkowski}]{Ali-Haimoud:2017rtz}
Ali-Ha\"\i{}moud Y, Kovetz ED, Kamionkowski M (2017) {Merger rate of primordial
  black-hole binaries}. Phys Rev D 96(12):123523.
  \doi{10.1103/PhysRevD.96.123523}.
  {\href{https://arxiv.org/abs/1709.06576}{{arXiv:1709.06576}}} {[astro-ph.CO]}

\bibitem[{{Allen} et~al.(2012){Allen}, {Anderson}, {Brady}, {Brown}, and
  {Creighton}}]{2012PhRvD..85l2006A}
{Allen} B, {Anderson} WG, {Brady} PR, {Brown} DA, {Creighton} JDE (2012)
  {FINDCHIRP: An algorithm for detection of gravitational waves from
  inspiraling compact binaries}. \prd 85(12):122006.
  \doi{10.1103/PhysRevD.85.122006}.
  {\href{https://arxiv.org/abs/gr-qc/0509116}{{arXiv:gr-qc/0509116}}} {[gr-qc]}

\bibitem[{{Almeida} et~al.(2019){Almeida}, {de Almeida}, {Damineli},
  {Rodrigues}, {Castro}, {Ferreira Lopes}, {Jablonski}, {do Nascimento}, and
  {Pereira}}]{2019AJ....157..150A}
{Almeida} LA, {de Almeida} L, {Damineli} A, {Rodrigues} CV, {Castro} M,
  {Ferreira Lopes} CE, {Jablonski} F, {do Nascimento} J J~D, {Pereira} MG
  (2019) {Orbital Period Variation of KIC 10544976: Applegate Mechanism versus
  Light Travel Time Effect}. \aj 157(4):150. \doi{10.3847/1538-3881/ab0963}.
  {\href{https://arxiv.org/abs/1903.09637}{{arXiv:1903.09637}}} {[astro-ph.SR]}

\bibitem[{Aloni et~al.(2017)Aloni, Blum, and Flauger}]{Blum:2016cjs}
Aloni D, Blum K, Flauger R (2017) {Cosmic microwave background constraints on
  primordial black hole dark matter}. JCAP 05:017.
  \doi{10.1088/1475-7516/2017/05/017}.
  {\href{https://arxiv.org/abs/1612.06811}{{arXiv:1612.06811}}} {[astro-ph.CO]}

\bibitem[{{Amaro-Seoane}(2018{\natexlab{a}})}]{2018PhRvD..98f3018A}
{Amaro-Seoane} P (2018{\natexlab{a}}) {Detecting intermediate-mass ratio
  inspirals from the ground and space}. \prd 98(6):063018.
  \doi{10.1103/PhysRevD.98.063018}.
  {\href{https://arxiv.org/abs/1807.03824}{{arXiv:1807.03824}}} {[astro-ph.HE]}

\bibitem[{{Amaro-Seoane}(2018{\natexlab{b}})}]{2018LRR....21....4A}
{Amaro-Seoane} P (2018{\natexlab{b}}) {Relativistic dynamics and extreme mass
  ratio inspirals}. Living Reviews in Relativity 21(1):4.
  \doi{10.1007/s41114-018-0013-8}.
  {\href{https://arxiv.org/abs/1205.5240}{{arXiv:1205.5240}}} {[astro-ph.CO]}

\bibitem[{{Amaro-Seoane}(2019)}]{2019PhRvD..99l3025A}
{Amaro-Seoane} P (2019) {Extremely large mass-ratio inspirals}. \prd
  99(12):123025. \doi{10.1103/PhysRevD.99.123025}.
  {\href{https://arxiv.org/abs/1903.10871}{{arXiv:1903.10871}}} {[astro-ph.GA]}

\bibitem[{{Amaro-Seoane}(2020)}]{2020arXiv201103059A}
{Amaro-Seoane} P (2020) {The gravitational capture of compact objects by
  massive black holes}. arXiv e-prints arXiv:2011.03059.
  {\href{https://arxiv.org/abs/2011.03059}{{arXiv:2011.03059}}} {[gr-qc]}

\bibitem[{{Amaro-Seoane} and {Freitag}(2006)}]{Amaro:2006imbh}
{Amaro-Seoane} P, {Freitag} M (2006) {Intermediate-Mass Black Holes in
  Colliding Clusters: Implications for Lower Frequency Gravitational-Wave
  Astronomy}. \apjl 653:L53--L56. \doi{10.1086/510405}.
  {\href{https://arxiv.org/abs/arXiv:astro-ph/0610478}{{arXiv:astro-ph/0610478}}}

\bibitem[{{Amaro-Seoane} and {Preto}(2011)}]{2011CQGra..28i4017A}
{Amaro-Seoane} P, {Preto} M (2011) {The impact of realistic models of mass
  segregation on the event rate of extreme-mass ratio inspirals and cusp
  re-growth}. Classical and Quantum Gravity 28(9):094017.
  \doi{10.1088/0264-9381/28/9/094017}.
  {\href{https://arxiv.org/abs/1010.5781}{{arXiv:1010.5781}}} {[astro-ph.CO]}

\bibitem[{{Amaro-Seoane} and {Santamar{\'\i}a}(2010)}]{2010ApJ...722.1197A}
{Amaro-Seoane} P, {Santamar{\'\i}a} L (2010) {Detection of IMBHs with
  Ground-based Gravitational Wave Observatories: A Biography of a Binary of
  Black Holes, from Birth to Death}. \apj 722(2):1197--1206.
  \doi{10.1088/0004-637X/722/2/1197}.
  {\href{https://arxiv.org/abs/0910.0254}{{arXiv:0910.0254}}} {[astro-ph.CO]}

\bibitem[{{Amaro-Seoane} et~al.(2007){Amaro-Seoane}, {Gair}, {Freitag},
  {Miller}, {Mandel}, {Cutler}, and {Babak}}]{2007CQGra..24R.113A}
{Amaro-Seoane} P, {Gair} JR, {Freitag} M, {Miller} MC, {Mandel} I, {Cutler} CJ,
  {Babak} S (2007) {TOPICAL REVIEW: Intermediate and extreme mass-ratio
  inspirals{\textemdash}astrophysics, science applications and detection using
  LISA}. Classical and Quantum Gravity 24(17):R113--R169.
  \doi{10.1088/0264-9381/24/17/R01}.
  {\href{https://arxiv.org/abs/astro-ph/0703495}{{arXiv:astro-ph/0703495}}}
  {[astro-ph]}

\bibitem[{{Amaro-Seoane} et~al.(2010{\natexlab{a}}){Amaro-Seoane}, {Barranco},
  {Bernal}, and {Rezzolla}}]{2010JCAP...11..002A}
{Amaro-Seoane} P, {Barranco} J, {Bernal} A, {Rezzolla} L (2010{\natexlab{a}})
  {Constraining scalar fields with stellar kinematics and collisional dark
  matter}. \jcap 2010(11):002. \doi{10.1088/1475-7516/2010/11/002}.
  {\href{https://arxiv.org/abs/1009.0019}{{arXiv:1009.0019}}} {[astro-ph.CO]}

\bibitem[{{Amaro-Seoane} et~al.(2010{\natexlab{b}}){Amaro-Seoane}, {Sesana},
  {Hoffman}, {Benacquista}, {Eichhorn}, {Makino}, and
  {Spurzem}}]{2010MNRAS.402.2308A}
{Amaro-Seoane} P, {Sesana} A, {Hoffman} L, {Benacquista} M, {Eichhorn} C,
  {Makino} J, {Spurzem} R (2010{\natexlab{b}}) {Triplets of supermassive black
  holes: astrophysics, gravitational waves and detection}. \mnras
  402(4):2308--2320. \doi{10.1111/j.1365-2966.2009.16104.x}.
  {\href{https://arxiv.org/abs/0910.1587}{{arXiv:0910.1587}}} {[astro-ph.CO]}

\bibitem[{{Amaro-Seoane} et~al.(2012){Amaro-Seoane}, {Brem}, {Cuadra}, and
  {Armitage}}]{2012ApJ...744L..20A}
{Amaro-Seoane} P, {Brem} P, {Cuadra} J, {Armitage} PJ (2012) {The Butterfly
  Effect in the Extreme-mass Ratio Inspiral Problem}. \apjl 744(2):L20.
  \doi{10.1088/2041-8205/744/2/L20}.
  {\href{https://arxiv.org/abs/1108.5174}{{arXiv:1108.5174}}} {[astro-ph.CO]}

\bibitem[{{Amaro-Seoane} et~al.(2013){Amaro-Seoane}, {Sopuerta}, and
  {Freitag}}]{2013MNRAS.429.3155A}
{Amaro-Seoane} P, {Sopuerta} CF, {Freitag} MD (2013) {The role of the
  supermassive black hole spin in the estimation of the EMRI event rate}.
  \mnras 429(4):3155--3165. \doi{10.1093/mnras/sts572}.
  {\href{https://arxiv.org/abs/1205.4713}{{arXiv:1205.4713}}} {[astro-ph.CO]}

\bibitem[{{Amaro-Seoane} et~al.(2015){Amaro-Seoane}, {Gair}, {Pound}, {Hughes},
  and {Sopuerta}}]{2015JPhCS.610a2002A}
{Amaro-Seoane} P, {Gair} JR, {Pound} A, {Hughes} SA, {Sopuerta} CF (2015)
  {Research Update on Extreme-Mass-Ratio Inspirals}. In: Journal of Physics
  Conference Series. Journal of Physics Conference Series, vol 610. p 012002.
  \doi{10.1088/1742-6596/610/1/012002}.
  {\href{https://arxiv.org/abs/1410.0958}{{arXiv:1410.0958}}} {[astro-ph.CO]}

\bibitem[{{Amaro-Seoane} et~al.(2017){Amaro-Seoane}, {Audley}, {Babak},
  {Baker}, {Barausse}, {Bender}, {Berti}, {Binetruy}, {Born}, {Bortoluzzi},
  {Camp}, {Caprini}, {Cardoso}, {Colpi}, {Conklin}, {Cornish}, {Cutler},
  {Danzmann}, {Dolesi}, {Ferraioli}, {Ferroni}, {Fitzsimons}, {Gair}, {Gesa
  Bote}, {Giardini}, {Gibert}, {Grimani}, {Halloin}, {Heinzel}, {Hertog},
  {Hewitson}, {Holley-Bockelmann}, {Hollington}, {Hueller}, {Inchauspe},
  {Jetzer}, {Karnesis}, {Killow}, {Klein}, {Klipstein}, {Korsakova}, {Larson},
  {Livas}, {Lloro}, {Man}, {Mance}, {Martino}, {Mateos}, {McKenzie},
  {McWilliams}, {Miller}, {Mueller}, {Nardini}, {Nelemans}, {Nofrarias},
  {Petiteau}, {Pivato}, {Plagnol}, {Porter}, {Reiche}, {Robertson},
  {Robertson}, {Rossi}, {Russano}, {Schutz}, {Sesana}, {Shoemaker}, {Slutsky},
  {Sopuerta}, {Sumner}, {Tamanini}, {Thorpe}, {Troebs}, {Vallisneri},
  {Vecchio}, {Vetrugno}, {Vitale}, {Volonteri}, {Wanner}, {Ward}, {Wass},
  {Weber}, {Ziemer}, and {Zweifel}}]{2017arXiv170200786A}
{Amaro-Seoane} P, {Audley} H, {Babak} S, {Baker} J, {Barausse} E, {Bender} P,
  {Berti} E, {Binetruy} P, {Born} M, {Bortoluzzi} D, et~al. (2017) {Laser
  Interferometer Space Antenna}. arXiv e-prints arXiv:1702.00786.
  {\href{https://arxiv.org/abs/1702.00786}{{arXiv:1702.00786}}} {[astro-ph.IM]}

\bibitem[{{Amendola} et~al.(2013){Amendola}, {Appleby}, {Bacon}, {Baker},
  {Baldi}, {Bartolo}, {Blanchard}, {Bonvin}, {Borgani}, {Branchini}, {Burrage},
  {Camera}, {Carbone}, {Casarini}, {Cropper}, {de Rham}, {Di Porto}, {Ealet},
  {Ferreira}, {Finelli}, {Garc{\'\i}a-Bellido}, {Giannantonio}, {Guzzo},
  {Heavens}, {Heisenberg}, {Heymans}, {Hoekstra}, {Hollenstein}, {Holmes},
  {Horst}, {Jahnke}, {Kitching}, {Koivisto}, {Kunz}, {La Vacca}, {March},
  {Majerotto}, {Markovic}, {Marsh}, {Marulli}, {Massey}, {Mellier}, {Mota},
  {Nunes}, {Percival}, {Pettorino}, {Porciani}, {Quercellini}, {Read},
  {Rinaldi}, {Sapone}, {Scaramella}, {Skordis}, {Simpson}, {Taylor}, {Thomas},
  {Trotta}, {Verde}, {Vernizzi}, {Vollmer}, {Wang}, {Weller}, and
  {Zlosnik}}]{2013LRR....16....6A}
{Amendola} L, {Appleby} S, {Bacon} D, {Baker} T, {Baldi} M, {Bartolo} N,
  {Blanchard} A, {Bonvin} C, {Borgani} S, {Branchini} E, et~al. (2013)
  {Cosmology and Fundamental Physics with the Euclid Satellite}. Living Reviews
  in Relativity 16(1):6. \doi{10.12942/lrr-2013-6}.
  {\href{https://arxiv.org/abs/1206.1225}{{arXiv:1206.1225}}} {[astro-ph.CO]}

\bibitem[{{Anderson} et~al.(2005){Anderson}, {Haggard}, {Homer}, {Joshi},
  {Margon}, {Silvestri}, {Szkody}, and {et~ al.}}]{2005AJ....130.2230A}
{Anderson} SF, {Haggard} D, {Homer} L, {Joshi} NR, {Margon} B, {Silvestri} NM,
  {Szkody} P, {et~ al} (2005) {Ultracompact AM Canum Venaticorum Binaries from
  the Sloan Digital Sky Survey: Three Candidates Plus the First Confirmed
  Eclipsing System}. \aj 130(5):2230--2236. \doi{10.1086/491587}.
  {\href{https://arxiv.org/abs/astro-ph/0506730}{{arXiv:astro-ph/0506730}}}
  {[astro-ph]}

\bibitem[{{Andersson}(2019)}]{2019gwa..book.....A}
{Andersson} N (2019) {Gravitational-Wave Astronomy: Exploring the Dark Side of
  the Universe}. Oxford University Press.
  \doi{10.1093/oso/9780198568032.001.0001/oso-9780198568032}

\bibitem[{{Andrews} and {Zezas}(2019)}]{2019MNRAS.486.3213A}
{Andrews} JJ, {Zezas} A (2019) {Double neutron star formation: merger times,
  systemic velocities, and travel distances}. \mnras 486(3):3213--3227.
  \doi{10.1093/mnras/stz1066}.
  {\href{https://arxiv.org/abs/1904.06137}{{arXiv:1904.06137}}} {[astro-ph.HE]}

\bibitem[{{Andrews} et~al.(2020){Andrews}, {Breivik}, {Pankow}, {D'Orazio}, and
  {Safarzadeh}}]{2020ApJ...892L...9A}
{Andrews} JJ, {Breivik} K, {Pankow} C, {D'Orazio} DJ, {Safarzadeh} M (2020)
  {LISA and the Existence of a Fast-merging Double Neutron Star Formation
  Channel}. \apjl 892(1):L9. \doi{10.3847/2041-8213/ab5b9a}.
  {\href{https://arxiv.org/abs/1910.13436}{{arXiv:1910.13436}}} {[astro-ph.HE]}

\bibitem[{{Angl{\'e}s-Alc{\'a}zar} et~al.(2017){Angl{\'e}s-Alc{\'a}zar},
  {Dav{\'e}}, {Faucher-Gigu{\`e}re}, {{\"O}zel}, and
  {Hopkins}}]{2017MNRAS.464.2840A}
{Angl{\'e}s-Alc{\'a}zar} D, {Dav{\'e}} R, {Faucher-Gigu{\`e}re} CA, {{\"O}zel}
  F, {Hopkins} PF (2017) {Gravitational torque-driven black hole growth and
  feedback in cosmological simulations}. \mnras 464(3):2840--2853.
  \doi{10.1093/mnras/stw2565}.
  {\href{https://arxiv.org/abs/1603.08007}{{arXiv:1603.08007}}} {[astro-ph.GA]}

\bibitem[{{Annulli} et~al.(2020){Annulli}, {Cardoso}, and
  {Vicente}}]{2020arXiv200703700A}
{Annulli} L, {Cardoso} V, {Vicente} R (2020) {Stirred and shaken: Dynamical
  behavior of boson stars and dark matter cores}. Physics Letters B 811:135944.
  \doi{10.1016/j.physletb.2020.135944}.
  {\href{https://arxiv.org/abs/2007.03700}{{arXiv:2007.03700}}} {[astro-ph.HE]}

\bibitem[{{Antoni} et~al.(2019){Antoni}, {MacLeod}, and
  {Ramirez-Ruiz}}]{2019ApJ...884...22A}
{Antoni} A, {MacLeod} M, {Ramirez-Ruiz} E (2019) {The Evolution of Binaries in
  a Gaseous Medium: Three-dimensional Simulations of Binary
  Bondi-Hoyle-Lyttleton Accretion}. \apj 884(1):22.
  \doi{10.3847/1538-4357/ab3466}.
  {\href{https://arxiv.org/abs/1901.07572}{{arXiv:1901.07572}}} {[astro-ph.HE]}

\bibitem[{{Antonini}(2013)}]{2013ApJ...763...62A}
{Antonini} F (2013) {Origin and Growth of Nuclear Star Clusters around Massive
  Black Holes}. \apj 763(1):62. \doi{10.1088/0004-637X/763/1/62}.
  {\href{https://arxiv.org/abs/1207.6589}{{arXiv:1207.6589}}} {[astro-ph.GA]}

\bibitem[{{Antonini}(2014)}]{2014ApJ...794..106A}
{Antonini} F (2014) {On the Distribution of Stellar Remnants around Massive
  Black Holes: Slow Mass Segregation, Star Cluster Inspirals, and Correlated
  Orbits}. \apj 794(2):106. \doi{10.1088/0004-637X/794/2/106}.
  {\href{https://arxiv.org/abs/1402.4865}{{arXiv:1402.4865}}} {[astro-ph.GA]}

\bibitem[{{Antonini} and {Gieles}(2020)}]{2020MNRAS.492.2936A}
{Antonini} F, {Gieles} M (2020) {Population synthesis of black hole binary
  mergers from star clusters}. \mnras 492(2):2936--2954.
  \doi{10.1093/mnras/stz3584}.
  {\href{https://arxiv.org/abs/1906.11855}{{arXiv:1906.11855}}} {[astro-ph.HE]}

\bibitem[{{Antonini} and {Merritt}(2012)}]{2012ApJ...745...83A}
{Antonini} F, {Merritt} D (2012) {Dynamical Friction around Supermassive Black
  Holes}. \apj 745(1):83. \doi{10.1088/0004-637X/745/1/83}.
  {\href{https://arxiv.org/abs/1108.1163}{{arXiv:1108.1163}}} {[astro-ph.GA]}

\bibitem[{{Antonini} and {Merritt}(2013)}]{2013ApJ...763L..10A}
{Antonini} F, {Merritt} D (2013) {Relativity and the Evolution of the Galactic
  Center S-star orbits}. \apjl 763(1):L10. \doi{10.1088/2041-8205/763/1/L10}.
  {\href{https://arxiv.org/abs/1211.4594}{{arXiv:1211.4594}}} {[astro-ph.GA]}

\bibitem[{{Antonini} and {Perets}(2012)}]{2012ApJ...757...27A}
{Antonini} F, {Perets} HB (2012) {Secular Evolution of Compact Binaries near
  Massive Black Holes: Gravitational Wave Sources and Other Exotica}. \apj
  757(1):27. \doi{10.1088/0004-637X/757/1/27}.
  {\href{https://arxiv.org/abs/1203.2938}{{arXiv:1203.2938}}} {[astro-ph.GA]}

\bibitem[{{Antonini} et~al.(2016){Antonini}, {Chatterjee}, {Rodriguez},
  {Morscher}, {Pattabiraman}, {Kalogera}, and {Rasio}}]{2016ApJ...816...65A}
{Antonini} F, {Chatterjee} S, {Rodriguez} CL, {Morscher} M, {Pattabiraman} B,
  {Kalogera} V, {Rasio} FA (2016) {Black Hole Mergers and Blue Stragglers from
  Hierarchical Triples Formed in Globular Clusters}. \apj 816(2):65.
  \doi{10.3847/0004-637X/816/2/65}.
  {\href{https://arxiv.org/abs/1509.05080}{{arXiv:1509.05080}}} {[astro-ph.GA]}

\bibitem[{{Antonini} et~al.(2017){Antonini}, {Toonen}, and
  {Hamers}}]{2017ApJ...841...77A}
{Antonini} F, {Toonen} S, {Hamers} AS (2017) {Binary Black Hole Mergers from
  Field Triples: Properties, Rates, and the Impact of Stellar Evolution}. \apj
  841(2):77. \doi{10.3847/1538-4357/aa6f5e}.
  {\href{https://arxiv.org/abs/1703.06614}{{arXiv:1703.06614}}} {[astro-ph.GA]}

\bibitem[{{Antonini} et~al.(2018){Antonini}, {Rodriguez}, {Petrovich}, and
  {Fischer}}]{2018MNRAS.480L..58A}
{Antonini} F, {Rodriguez} CL, {Petrovich} C, {Fischer} CL (2018) {Precessional
  dynamics of black hole triples: binary mergers with near-zero effective
  spin}. \mnras 480(1):L58--L62. \doi{10.1093/mnrasl/sly126}.
  {\href{https://arxiv.org/abs/1711.07142}{{arXiv:1711.07142}}} {[astro-ph.HE]}

\bibitem[{{Antonini} et~al.(2019){Antonini}, {Gieles}, and
  {Gualandris}}]{2019MNRAS.486.5008A}
{Antonini} F, {Gieles} M, {Gualandris} A (2019) {Black hole growth through
  hierarchical black hole mergers in dense star clusters: implications for
  gravitational wave detections}. \mnras 486(4):5008--5021.
  \doi{10.1093/mnras/stz1149}.
  {\href{https://arxiv.org/abs/1811.03640}{{arXiv:1811.03640}}} {[astro-ph.HE]}

\bibitem[{{Apostolatos} et~al.(1994){Apostolatos}, {Cutler}, {Sussman}, and
  {Thorne}}]{1994PhRvD..49.6274A}
{Apostolatos} TA, {Cutler} C, {Sussman} GJ, {Thorne} KS (1994) {Spin-induced
  orbital precession and its modulation of the gravitational waveforms from
  merging binaries}. \prd 49(12):6274--6297. \doi{10.1103/PhysRevD.49.6274}

\bibitem[{{Apostolatos} et~al.(2009){Apostolatos}, {Lukes-Gerakopoulos}, and
  {Contopoulos}}]{2009PhRvL.103k1101A}
{Apostolatos} TA, {Lukes-Gerakopoulos} G, {Contopoulos} G (2009) {How to
  Observe a Non-Kerr Spacetime Using Gravitational Waves}. \prl 103(11):111101.
  \doi{10.1103/PhysRevLett.103.111101}.
  {\href{https://arxiv.org/abs/0906.0093}{{arXiv:0906.0093}}} {[gr-qc]}

\bibitem[{{Applegate} and {Shaham}(1994)}]{1994ApJ...436..312A}
{Applegate} JH, {Shaham} J (1994) {Orbital Period Variability in the Eclipsing
  Pulsar Binary PSR B1957+20: Evidence for a Tidally Powered Star}. \apj
  436:312. \doi{10.1086/174906}

\bibitem[{{Arca Sedda}(2020{\natexlab{a}})}]{2020ApJ...891...47A}
{Arca Sedda} M (2020{\natexlab{a}}) {Birth, Life, and Death of Black Hole
  Binaries around Supermassive Black Holes: Dynamical Evolution of
  Gravitational Wave Sources}. \apj 891(1):47. \doi{10.3847/1538-4357/ab723b}.
  {\href{https://arxiv.org/abs/2002.04037}{{arXiv:2002.04037}}} {[astro-ph.GA]}

\bibitem[{{Arca Sedda}(2020{\natexlab{b}})}]{2020CmPhy...3...43A}
{Arca Sedda} M (2020{\natexlab{b}}) {Dissecting the properties of neutron
  star-black hole mergers originating in dense star clusters}. Communications
  Physics 3(1):43. \doi{10.1038/s42005-020-0310-x}.
  {\href{https://arxiv.org/abs/2003.02279}{{arXiv:2003.02279}}} {[astro-ph.GA]}

\bibitem[{{Arca-Sedda} and {Capuzzo-Dolcetta}(2014)}]{2014MNRAS.444.3738A}
{Arca-Sedda} M, {Capuzzo-Dolcetta} R (2014) {The globular cluster migratory
  origin of nuclear star clusters}. \mnras 444(4):3738--3755.
  \doi{10.1093/mnras/stu1683}.
  {\href{https://arxiv.org/abs/1405.7593}{{arXiv:1405.7593}}} {[astro-ph.GA]}

\bibitem[{{Arca-Sedda} and {Capuzzo-Dolcetta}(2017)}]{2017MNRAS.464.3060A}
{Arca-Sedda} M, {Capuzzo-Dolcetta} R (2017) {Lack of nuclear clusters in dwarf
  spheroidal galaxies: implications for massive black holes formation and the
  cusp/core problem}. \mnras 464(3):3060--3070. \doi{10.1093/mnras/stw2483}.
  {\href{https://arxiv.org/abs/1611.01088}{{arXiv:1611.01088}}} {[astro-ph.GA]}

\bibitem[{{Arca-Sedda} and {Capuzzo-Dolcetta}(2019)}]{2019MNRAS.483..152A}
{Arca-Sedda} M, {Capuzzo-Dolcetta} R (2019) {The MEGaN project II.
  Gravitational waves from intermediate-mass and binary black holes around a
  supermassive black hole}. \mnras 483(1):152--171.
  \doi{10.1093/mnras/sty3096}.
  {\href{https://arxiv.org/abs/1709.05567}{{arXiv:1709.05567}}} {[astro-ph.GA]}

\bibitem[{{Arca-Sedda} and {Gualandris}(2018)}]{2018MNRAS.477.4423A}
{Arca-Sedda} M, {Gualandris} A (2018) {Gravitational wave sources from
  inspiralling globular clusters in the Galactic Centre and similar
  environments}. \mnras 477(4):4423--4442. \doi{10.1093/mnras/sty922}.
  {\href{https://arxiv.org/abs/1804.06116}{{arXiv:1804.06116}}} {[astro-ph.GA]}

\bibitem[{{Arca Sedda} and {Mastrobuono-Battisti}(2019)}]{2019arXiv190605864A}
{Arca Sedda} M, {Mastrobuono-Battisti} A (2019) {Mergers of globular clusters
  in the Galactic disc: intermediate mass black hole coalescence and
  implications for gravitational waves detection}. arXiv e-prints
  arXiv:1906.05864.
  {\href{https://arxiv.org/abs/1906.05864}{{arXiv:1906.05864}}} {[astro-ph.GA]}

\bibitem[{{Arca-Sedda} et~al.(2015){Arca-Sedda}, {Capuzzo-Dolcetta},
  {Antonini}, and {Seth}}]{2015ApJ...806..220A}
{Arca-Sedda} M, {Capuzzo-Dolcetta} R, {Antonini} F, {Seth} A (2015) {Henize
  2-10: The Ongoing Formation of a Nuclear Star Cluster around a Massive Black
  Hole}. \apj 806(2):220. \doi{10.1088/0004-637X/806/2/220}.
  {\href{https://arxiv.org/abs/1501.04567}{{arXiv:1501.04567}}} {[astro-ph.GA]}

\bibitem[{{Arca Sedda} et~al.(2018){Arca Sedda}, {Askar}, and
  {Giersz}}]{2018MNRAS.479.4652A}
{Arca Sedda} M, {Askar} A, {Giersz} M (2018) {MOCCA-Survey Database - I.
  Unravelling black hole subsystems in globular clusters}. \mnras
  479(4):4652--4664. \doi{10.1093/mnras/sty1859}.
  {\href{https://arxiv.org/abs/1801.00795}{{arXiv:1801.00795}}} {[astro-ph.GA]}

\bibitem[{{Arca Sedda} et~al.(2019{\natexlab{a}}){Arca Sedda}, {Askar}, and
  {Giersz}}]{2019arXiv190500902A}
{Arca Sedda} M, {Askar} A, {Giersz} M (2019{\natexlab{a}}) {MOCCA-SURVEY
  Database I. Intermediate mass black holes in Milky Way globular clusters and
  their connection to supermassive black holes}. arXiv e-prints
  arXiv:1905.00902.
  {\href{https://arxiv.org/abs/1905.00902}{{arXiv:1905.00902}}} {[astro-ph.GA]}

\bibitem[{{Arca Sedda} et~al.(2019{\natexlab{b}}){Arca Sedda}, {Berczik},
  {Capuzzo-Dolcetta}, {Fragione}, {Sobolenko}, and
  {Spurzem}}]{2019MNRAS.484..520A}
{Arca Sedda} M, {Berczik} P, {Capuzzo-Dolcetta} R, {Fragione} G, {Sobolenko} M,
  {Spurzem} R (2019{\natexlab{b}}) {Supermassive black holes coalescence
  mediated by massive perturbers: implications for gravitational waves emission
  and nuclear cluster formation}. \mnras 484(1):520--542.
  \doi{10.1093/mnras/sty3458}.
  {\href{https://arxiv.org/abs/1712.05810}{{arXiv:1712.05810}}} {[astro-ph.GA]}

\bibitem[{{Arca Sedda} et~al.(2020{\natexlab{a}}){Arca Sedda}, {Berry}, {Jani},
  {Amaro-Seoane}, {Auclair}, {Baird}, {Baker}, {Berti}, {Breivik}, {Burrows},
  {Caprini}, {Chen}, {Doneva}, {Ezquiaga}, {Saavik Ford}, {Katz}, {Kolkowitz},
  {McKernan}, {Mueller}, {Nardini}, {Pikovski}, {Rajendran}, {Sesana}, {Shao},
  {Tamanini}, {Vartanyan}, {Warburton}, {Witek}, {Wong}, and
  {Zevin}}]{2019arXiv190811375A}
{Arca Sedda} M, {Berry} CPL, {Jani} K, {Amaro-Seoane} P, {Auclair} P, {Baird}
  J, {Baker} T, {Berti} E, {Breivik} K, {Burrows} A, et~al.
  (2020{\natexlab{a}}) {The missing link in gravitational-wave astronomy:
  discoveries waiting in the decihertz range}. Classical and Quantum Gravity
  37(21):215011. \doi{10.1088/1361-6382/abb5c1}.
  {\href{https://arxiv.org/abs/1908.11375}{{arXiv:1908.11375}}} {[gr-qc]}

\bibitem[{{Arca Sedda} et~al.(2020{\natexlab{b}}){Arca Sedda}, {Mapelli},
  {Spera}, {Benacquista}, and {Giacobbo}}]{2020ApJ...894..133A}
{Arca Sedda} M, {Mapelli} M, {Spera} M, {Benacquista} M, {Giacobbo} N
  (2020{\natexlab{b}}) {Fingerprints of Binary Black Hole Formation Channels
  Encoded in the Mass and Spin of Merger Remnants}. \apj 894(2):133.
  \doi{10.3847/1538-4357/ab88b2}.
  {\href{https://arxiv.org/abs/2003.07409}{{arXiv:2003.07409}}} {[astro-ph.GA]}

\bibitem[{{Arca Sedda} et~al.(2021{\natexlab{a}}){Arca Sedda}, {Amaro Seoane},
  and {Chen}}]{2020arXiv200713746A}
{Arca Sedda} M, {Amaro Seoane} P, {Chen} X (2021{\natexlab{a}}) {Merging
  stellar and intermediate-mass black holes in dense clusters: implications for
  LIGO, LISA, and the next generation of gravitational wave detectors}. \aap
  652:A54. \doi{10.1051/0004-6361/202037785}.
  {\href{https://arxiv.org/abs/2007.13746}{{arXiv:2007.13746}}} {[astro-ph.GA]}

\bibitem[{{Arca Sedda} et~al.(2021{\natexlab{b}}){Arca Sedda}, {Li}, and
  {Kocsis}}]{2018arXiv180506458A}
{Arca Sedda} M, {Li} G, {Kocsis} B (2021{\natexlab{b}}) {Order in the chaos.
  Eccentric black hole binary mergers in triples formed via strong
  binary-binary scatterings}. \aap 650:A189. \doi{10.1051/0004-6361/202038795}.
  {\href{https://arxiv.org/abs/1805.06458}{{arXiv:1805.06458}}} {[astro-ph.HE]}

\bibitem[{{Arca-Sedda} et~al.(2021){Arca-Sedda}, {Rizzuto}, {Naab}, {Ostriker},
  {Giersz}, and {Spurzem}}]{2021ApJ...920..128A}
{Arca-Sedda} M, {Rizzuto} FP, {Naab} T, {Ostriker} J, {Giersz} M, {Spurzem} R
  (2021) {Breaching the Limit: Formation of GW190521-like and IMBH Mergers in
  Young Massive Clusters}. \apj 920(2):128. \doi{10.3847/1538-4357/ac1419}.
  {\href{https://arxiv.org/abs/2105.07003}{{arXiv:2105.07003}}} {[astro-ph.GA]}

\bibitem[{{Armitage} and {Natarajan}(2002)}]{2002ApJ...567L...9A}
{Armitage} PJ, {Natarajan} P (2002) {Accretion during the Merger of
  Supermassive Black Holes}. \apjl 567(1):L9--L12. \doi{10.1086/339770}.
  {\href{https://arxiv.org/abs/astro-ph/0201318}{{arXiv:astro-ph/0201318}}}
  {[astro-ph]}

\bibitem[{{Arnason} et~al.(2021){Arnason}, {Papei}, {Barmby}, {Bahramian}, and
  {Gorski}}]{2021MNRAS.502.5455A}
{Arnason} RM, {Papei} H, {Barmby} P, {Bahramian} A, {Gorski} MD (2021)
  {Distances to Galactic X-ray binaries with Gaia DR2}. \mnras
  502(4):5455--5470. \doi{10.1093/mnras/stab345}.
  {\href{https://arxiv.org/abs/2102.02615}{{arXiv:2102.02615}}} {[astro-ph.HE]}

\bibitem[{{Arnold}(1978)}]{1978mmcm.book.....A}
{Arnold} VI (1978) {Mathematical methods of classical mechanics}. Springer -
  Verlag

\bibitem[{{Artale} et~al.(2019){Artale}, {Mapelli}, {Giacobbo}, {Sabha},
  {Spera}, {Santoliquido}, and {Bressan}}]{2019MNRAS.487.1675A}
{Artale} MC, {Mapelli} M, {Giacobbo} N, {Sabha} NB, {Spera} M, {Santoliquido}
  F, {Bressan} A (2019) {Host galaxies of merging compact objects: mass, star
  formation rate, metallicity, and colours}. \mnras 487(2):1675--1688.
  \doi{10.1093/mnras/stz1382}.
  {\href{https://arxiv.org/abs/1903.00083}{{arXiv:1903.00083}}} {[astro-ph.GA]}

\bibitem[{{Artymowicz} et~al.(1993){Artymowicz}, {Lin}, and
  {Wampler}}]{1993ApJ...409..592A}
{Artymowicz} P, {Lin} DNC, {Wampler} EJ (1993) {Star Trapping and Metallicity
  Enrichment in Quasars and Active Galactic Nuclei}. \apj 409:592.
  \doi{10.1086/172690}

\bibitem[{{Arzoumanian} et~al.(2020{\natexlab{a}}){Arzoumanian}, {Baker},
  {Blumer}, {B{\'e}csy}, {Brazier}, {Brook}, {Burke-Spolaor}, {Chatterjee},
  {Chen}, {Cordes}, {Cornish}, {Crawford}, {Cromartie}, {Decesar}, {Demorest},
  {Dolch}, {Ellis}, {Ferrara}, {Fiore}, {Fonseca}, {Garver-Daniels}, {Gentile},
  {Good}, {Hazboun}, {Holgado}, {Islo}, {Jennings}, {Jones}, {Kaiser},
  {Kaplan}, {Kelley}, {Key}, {Laal}, {Lam}, {Lazio}, {Lorimer}, {Luo}, {Lynch},
  {Madison}, {McLaughlin}, {Mingarelli}, {Ng}, {Nice}, {Pennucci}, {Pol},
  {Ransom}, {Ray}, {Shapiro-Albert}, {Siemens}, {Simon}, {Spiewak}, {Stairs},
  {Stinebring}, {Stovall}, {Sun}, {Swiggum}, {Taylor}, {Turner}, {Vallisneri},
  {Vigeland}, {Witt}, and {Nanograv Collaboration}}]{2020arXiv200904496A}
{Arzoumanian} Z, {Baker} PT, {Blumer} H, {B{\'e}csy} B, {Brazier} A, {Brook}
  PR, {Burke-Spolaor} S, {Chatterjee} S, {Chen} S, {Cordes} JM, et~al.
  (2020{\natexlab{a}}) {The NANOGrav 12.5 yr Data Set: Search for an Isotropic
  Stochastic Gravitational-wave Background}. \apjl 905(2):L34.
  \doi{10.3847/2041-8213/abd401}.
  {\href{https://arxiv.org/abs/2009.04496}{{arXiv:2009.04496}}} {[astro-ph.HE]}

\bibitem[{{Arzoumanian} et~al.(2020{\natexlab{b}}){Arzoumanian}, {Baker},
  {Brazier}, {Brook}, {Burke-Spolaor}, {B{\'e}csy}, {Charisi}, {Chatterjee},
  {Cordes}, {Cornish}, {Crawford}, {Cromartie}, {Crowter}, {Decesar},
  {Demorest}, {Dolch}, {Elliott}, {Ellis}, {Ferdman}, {Ferrara}, {Fonseca},
  {Garver-Daniels}, {Gentile}, {Good}, {Hazboun}, {Islo}, {Jennings}, {Jones},
  {Kaiser}, {Kaplan}, {Kelley}, {Key}, {Lam}, {Lazio}, {Levin}, {Luo}, {Lynch},
  {Madison}, {McLaughlin}, {Mingarelli}, {Ng}, {Nice}, {Pennucci}, {Pol},
  {Ransom}, {Ray}, {Shapiro-Albert}, {Siemens}, {Simon}, {Spiewak}, {Stairs},
  {Stinebring}, {Stovall}, {Swiggum}, {Taylor}, {Vallisneri}, {Vigeland},
  {Witt}, {Zhu}, and {NANOGrav Collaboration}}]{2020ApJ...900..102A}
{Arzoumanian} Z, {Baker} PT, {Brazier} A, {Brook} PR, {Burke-Spolaor} S,
  {B{\'e}csy} B, {Charisi} M, {Chatterjee} S, {Cordes} JM, {Cornish} NJ, et~al.
  (2020{\natexlab{b}}) {Multimessenger Gravitational-wave Searches with Pulsar
  Timing Arrays: Application to 3C 66B Using the NANOGrav 11-year Data Set}.
  \apj 900(2):102. \doi{10.3847/1538-4357/ababa1}.
  {\href{https://arxiv.org/abs/2005.07123}{{arXiv:2005.07123}}} {[astro-ph.GA]}

\bibitem[{Arzoumanian et~al.(2020)}]{NANOgrav:2020bcs}
Arzoumanian Z, et~al. (2020) {The NANOGrav 12.5 yr Data Set: Search for an
  Isotropic Stochastic Gravitational-wave Background}. Astrophys J Lett
  905(2):L34. \doi{10.3847/2041-8213/abd401}.
  {\href{https://arxiv.org/abs/2009.04496}{{arXiv:2009.04496}}} {[astro-ph.HE]}

\bibitem[{{Askar} et~al.(2018){Askar}, {Arca Sedda}, and
  {Giersz}}]{2018MNRAS.478.1844A}
{Askar} A, {Arca Sedda} M, {Giersz} M (2018) {MOCCA-SURVEY Database I: Galactic
  globular clusters harbouring a black hole subsystem}. \mnras
  478(2):1844--1854. \doi{10.1093/mnras/sty1186}.
  {\href{https://arxiv.org/abs/1802.05284}{{arXiv:1802.05284}}} {[astro-ph.GA]}

\bibitem[{{Askar} et~al.(2021){Askar}, {Davies}, and
  {Church}}]{2020arXiv200604922A}
{Askar} A, {Davies} MB, {Church} RP (2021) {Formation of supermassive black
  holes in galactic nuclei - I. Delivering seed intermediate-mass black holes
  in massive stellar clusters}. \mnras 502(2):2682--2700.
  \doi{10.1093/mnras/stab113}.
  {\href{https://arxiv.org/abs/2006.04922}{{arXiv:2006.04922}}} {[astro-ph.GA]}

\bibitem[{{Athanassoula}(2002)}]{2002ASPC..275..141A}
{Athanassoula} E (2002) {Formation and Evolution of Bars in Disc Galaxies}. In:
  {Athanassoula} E, {Bosma} A, {Mujica} R (eds) Disks of Galaxies: Kinematics,
  Dynamics and Peturbations. Astronomical Society of the Pacific Conference
  Series, vol 275. pp 141--152.
  {\href{https://arxiv.org/abs/astro-ph/0209438}{{arXiv:astro-ph/0209438}}}
  {[astro-ph]}

\bibitem[{{Auchettl} et~al.(2017){Auchettl}, {Guillochon}, and
  {Ramirez-Ruiz}}]{2017ApJ...838..149A}
{Auchettl} K, {Guillochon} J, {Ramirez-Ruiz} E (2017) {New Physical Insights
  about Tidal Disruption Events from a Comprehensive Observational Inventory at
  X-Ray Wavelengths}. \apj 838(2):149. \doi{10.3847/1538-4357/aa633b}.
  {\href{https://arxiv.org/abs/1611.02291}{{arXiv:1611.02291}}} {[astro-ph.HE]}

\bibitem[{Auclair et~al.(to appear)}]{CosWGwhitepaper2021}
Auclair P, et~al. (to appear) {Cosmology with the Laser Interferometer Space
  Antenna}. Cosmology with the Laser Interferometer Space Antenna, to appear

\bibitem[{{Ba{\~n}ados} et~al.(2014){Ba{\~n}ados}, {Venemans}, {Morganson},
  {Decarli}, {Walter}, {Chambers}, {Rix}, {Farina}, {Fan}, {Jiang}, {McGreer},
  {De Rosa}, {Simcoe}, {Wei{\ss}}, {Price}, {Morgan}, {Burgett}, {Greiner},
  {Kaiser}, {Kudritzki}, {Magnier}, {Metcalfe}, {Stubbs}, {Sweeney}, {Tonry},
  {Wainscoat}, and {Waters}}]{2014AJ....148...14B}
{Ba{\~n}ados} E, {Venemans} BP, {Morganson} E, {Decarli} R, {Walter} F,
  {Chambers} KC, {Rix} HW, {Farina} EP, {Fan} X, {Jiang} L, et~al. (2014)
  {Discovery of Eight z \raisebox{-0.5ex}\textasciitilde 6 Quasars from
  Pan-STARRS1}. \aj 148(1):14. \doi{10.1088/0004-6256/148/1/14}.
  {\href{https://arxiv.org/abs/1405.3986}{{arXiv:1405.3986}}} {[astro-ph.GA]}

\bibitem[{{Ba{\~n}ados} et~al.(2016){Ba{\~n}ados}, {Venemans}, {Decarli},
  {Farina}, {Mazzucchelli}, {Walter}, {Fan}, {Stern}, {Schlafly}, {Chambers},
  {Rix}, {Jiang}, {McGreer}, {Simcoe}, {Wang}, {Yang}, {Morganson}, {De Rosa},
  {Greiner}, {Balokovi{\'c}}, {Burgett}, {Cooper}, {Draper}, {Flewelling},
  {Hodapp}, {Jun}, {Kaiser}, {Kudritzki}, {Magnier}, {Metcalfe}, {Miller},
  {Schindler}, {Tonry}, {Wainscoat}, {Waters}, and
  {Yang}}]{2016ApJS..227...11B}
{Ba{\~n}ados} E, {Venemans} BP, {Decarli} R, {Farina} EP, {Mazzucchelli} C,
  {Walter} F, {Fan} X, {Stern} D, {Schlafly} E, {Chambers} KC, et~al. (2016)
  {The Pan-STARRS1 Distant z > 5.6 Quasar Survey: More than 100 Quasars within
  the First Gyr of the Universe}. \apjs 227(1):11.
  \doi{10.3847/0067-0049/227/1/11}.
  {\href{https://arxiv.org/abs/1608.03279}{{arXiv:1608.03279}}} {[astro-ph.GA]}

\bibitem[{{Ba{\~n}ados} et~al.(2018{\natexlab{a}}){Ba{\~n}ados}, {Connor},
  {Stern}, {Mulchaey}, {Fan}, {Decarli}, {Farina}, {Mazzucchelli}, {Venemans},
  {Walter}, {Wang}, and {Yang}}]{2018ApJ...856L..25B}
{Ba{\~n}ados} E, {Connor} T, {Stern} D, {Mulchaey} J, {Fan} X, {Decarli} R,
  {Farina} EP, {Mazzucchelli} C, {Venemans} BP, {Walter} F, et~al.
  (2018{\natexlab{a}}) {Chandra X-Rays from the Redshift 7.54 Quasar ULAS
  J1342+0928}. \apjl 856(2):L25. \doi{10.3847/2041-8213/aab61e}.
  {\href{https://arxiv.org/abs/1803.08105}{{arXiv:1803.08105}}} {[astro-ph.GA]}

\bibitem[{{Ba{\~n}ados} et~al.(2018{\natexlab{b}}){Ba{\~n}ados}, {Venemans},
  {Mazzucchelli}, {Farina}, {Walter}, {Wang}, {Decarli}, {Stern}, {Fan},
  {Davies}, {Hennawi}, {Simcoe}, {Turner}, {Rix}, {Yang}, {Kelson}, {Rudie},
  and {Winters}}]{2018Natur.553..473B}
{Ba{\~n}ados} E, {Venemans} BP, {Mazzucchelli} C, {Farina} EP, {Walter} F,
  {Wang} F, {Decarli} R, {Stern} D, {Fan} X, {Davies} FB, et~al.
  (2018{\natexlab{b}}) {An 800-million-solar-mass black hole in a significantly
  neutral Universe at a redshift of 7.5}. \nat 553(7689):473--476.
  \doi{10.1038/nature25180}.
  {\href{https://arxiv.org/abs/1712.01860}{{arXiv:1712.01860}}} {[astro-ph.GA]}

\bibitem[{{Ba{\~n}ados} et~al.(2019){Ba{\~n}ados}, {Novak}, {Neeleman},
  {Walter}, {Decarli}, {Venemans}, {Mazzucchelli}, {Carilli}, {Wang}, {Fan},
  {Farina}, and {Rix}}]{2019ApJ...881L..23B}
{Ba{\~n}ados} E, {Novak} M, {Neeleman} M, {Walter} F, {Decarli} R, {Venemans}
  BP, {Mazzucchelli} C, {Carilli} C, {Wang} F, {Fan} X, et~al. (2019) {The z =
  7.54 Quasar ULAS J1342+0928 Is Hosted by a Galaxy Merger}. \apjl 881(1):L23.
  \doi{10.3847/2041-8213/ab3659}.
  {\href{https://arxiv.org/abs/1909.00027}{{arXiv:1909.00027}}} {[astro-ph.GA]}

\bibitem[{{Babak} et~al.(2007){Babak}, {Fang}, {Gair}, {Glampedakis}, and
  {Hughes}}]{2007PhRvD..75b4005B}
{Babak} S, {Fang} H, {Gair} JR, {Glampedakis} K, {Hughes} SA (2007) {``Kludge''
  gravitational waveforms for a test-body orbiting a Kerr black hole}. \prd
  75(2):024005. \doi{10.1103/PhysRevD.75.024005}.
  {\href{https://arxiv.org/abs/gr-qc/0607007}{{arXiv:gr-qc/0607007}}} {[gr-qc]}

\bibitem[{{Babak} et~al.(2008{\natexlab{a}}){Babak}, {Baker}, {Benacquista},
  {Cornish}, {Crowder}, {Cutler}, {Larson}, {Littenberg}, {Porter},
  {Vallisneri}, {Vecchio}, {data challenge task force}, {Auger}, {Barack},
  {B{\l}aut}, {Bloomer}, {Brown}, {Christensen}, {Clark}, {Fairhurst}, {Gair},
  {Halloin}, {Hendry}, {Jimenez}, {Kr{\'o}lak}, {Mand el}, {Messenger},
  {Meyer}, {Mohanty}, {Nayak}, {Petiteau}, {Pitkin}, {Plagnol}, {Prix},
  {Robinson}, {Roever}, {Savov}, {Stroeer}, {Toher}, {Veitch}, {Vinet}, {Wen},
  {Whelan}, {Woan}, and {Challenge-2 participants}}]{2008CQGra..25k4037B}
{Babak} S, {Baker} JG, {Benacquista} MJ, {Cornish} NJ, {Crowder} J, {Cutler} C,
  {Larson} SL, {Littenberg} TB, {Porter} EK, {Vallisneri} M, et~al.
  (2008{\natexlab{a}}) {Report on the second Mock LISA data challenge}.
  Classical and Quantum Gravity 25(11):114037.
  \doi{10.1088/0264-9381/25/11/114037}.
  {\href{https://arxiv.org/abs/0711.2667}{{arXiv:0711.2667}}} {[gr-qc]}

\bibitem[{{Babak} et~al.(2008{\natexlab{b}}){Babak}, {Baker}, {Benacquista},
  {Cornish}, {Crowder}, {Larson}, {Plagnol}, {Porter}, {Vallisneri}, {Vecchio},
  {Data Challenge Task Force}, {Arnaud}, {Barack}, {B{\l}aut}, {Cutler},
  {Fairhurst}, {Gair}, {Gong}, {Harry}, {Khurana}, {Kr{\'o}lak}, {Mandel},
  {Prix}, {Sathyaprakash}, {Savov}, {Shang}, {Trias}, {Veitch}, {Wang}, {Wen},
  {Whelan}, and {Challenge-1B participants}}]{2008CQGra..25r4026B}
{Babak} S, {Baker} JG, {Benacquista} MJ, {Cornish} NJ, {Crowder} J, {Larson}
  SL, {Plagnol} E, {Porter} EK, {Vallisneri} M, {Vecchio} A, et~al.
  (2008{\natexlab{b}}) {The Mock LISA Data Challenges: from Challenge 1B to
  Challenge 3}. Classical and Quantum Gravity 25(18):184026.
  \doi{10.1088/0264-9381/25/18/184026}.
  {\href{https://arxiv.org/abs/0806.2110}{{arXiv:0806.2110}}} {[gr-qc]}

\bibitem[{{Babak} et~al.(2009){Babak}, {Gair}, and
  {Porter}}]{2009CQGra..26m5004B}
{Babak} S, {Gair} JR, {Porter} EK (2009) {An algorithm for the detection of
  extreme mass ratio inspirals in LISA data}. Classical and Quantum Gravity
  26(13):135004. \doi{10.1088/0264-9381/26/13/135004}.
  {\href{https://arxiv.org/abs/0902.4133}{{arXiv:0902.4133}}} {[gr-qc]}

\bibitem[{{Babak} et~al.(2010){Babak}, {Baker}, {Benacquista}, {Cornish},
  {Larson}, {Mandel}, {McWilliams}, {Petiteau}, {Porter}, {Robinson},
  {Vallisneri}, {Vecchio}, {Data Challenge Task Force}, {Adams}, {Arnaud},
  {B{\l}aut}, {Bridges}, {Cohen}, {Cutler}, {Feroz}, {Gair}, {Graff}, {Hobson},
  {Shapiro Key}, {Kr{\'o}lak}, {Lasenby}, {Prix}, {Shang}, {Trias}, {Veitch},
  {Whelan}, and {participants}}]{2010CQGra..27h4009B}
{Babak} S, {Baker} JG, {Benacquista} MJ, {Cornish} NJ, {Larson} SL, {Mandel} I,
  {McWilliams} ST, {Petiteau} A, {Porter} EK, {Robinson} EL, et~al. (2010) {The
  Mock LISA Data Challenges: from challenge 3 to challenge 4}. Classical and
  Quantum Gravity 27(8):084009. \doi{10.1088/0264-9381/27/8/084009}.
  {\href{https://arxiv.org/abs/0912.0548}{{arXiv:0912.0548}}} {[gr-qc]}

\bibitem[{{Babak} et~al.(2017){Babak}, {Gair}, {Sesana}, {Barausse},
  {Sopuerta}, {Berry}, {Berti}, {Amaro-Seoane}, {Petiteau}, and
  {Klein}}]{2017PhRvD..95j3012B}
{Babak} S, {Gair} J, {Sesana} A, {Barausse} E, {Sopuerta} CF, {Berry} CPL,
  {Berti} E, {Amaro-Seoane} P, {Petiteau} A, {Klein} A (2017) {Science with the
  space-based interferometer LISA. V. Extreme mass-ratio inspirals}. \prd
  95(10):103012. \doi{10.1103/PhysRevD.95.103012}.
  {\href{https://arxiv.org/abs/1703.09722}{{arXiv:1703.09722}}} {[gr-qc]}

\bibitem[{{Babak} et~al.(2021){Babak}, {Hewitson}, and
  {Petiteau}}]{2021arXiv210801167B}
{Babak} S, {Hewitson} M, {Petiteau} A (2021) {LISA Sensitivity and SNR
  Calculations}. arXiv e-prints arXiv:2108.01167.
  {\href{https://arxiv.org/abs/2108.01167}{{arXiv:2108.01167}}} {[astro-ph.IM]}

\bibitem[{{Bade} et~al.(1996){Bade}, {Komossa}, and
  {Dahlem}}]{1996A&A...309L..35B}
{Bade} N, {Komossa} S, {Dahlem} M (1996) {Detection of an extremely soft X-ray
  outburst in the HII-like nucleus of NGC 5905.} \aap 309:L35--L38

\bibitem[{{Baghi} et~al.(2019){Baghi}, {Thorpe}, {Slutsky}, {Baker}, {Canton},
  {Korsakova}, and {Karnesis}}]{2019PhRvD.100b2003B}
{Baghi} Q, {Thorpe} JI, {Slutsky} J, {Baker} J, {Canton} TD, {Korsakova} N,
  {Karnesis} N (2019) {Gravitational-wave parameter estimation with gaps in
  LISA: A Bayesian data augmentation method}. \prd 100(2):022003.
  \doi{10.1103/PhysRevD.100.022003}.
  {\href{https://arxiv.org/abs/1907.04747}{{arXiv:1907.04747}}} {[gr-qc]}

\bibitem[{{Baghi} et~al.(2022){Baghi}, {Korsakova}, {Slutsky}, {Castelli},
  {Karnesis}, and {Bayle}}]{2022PhRvD.105d2002B}
{Baghi} Q, {Korsakova} N, {Slutsky} J, {Castelli} E, {Karnesis} N, {Bayle} JB
  (2022) {Detection and characterization of instrumental transients in LISA
  Pathfinder and their projection to LISA}. \prd 105(4):042002.
  \doi{10.1103/PhysRevD.105.042002}.
  {\href{https://arxiv.org/abs/2112.07490}{{arXiv:2112.07490}}} {[gr-qc]}

\bibitem[{{Bahcall} and {Wolf}(1976)}]{1976ApJ...209..214B}
{Bahcall} JN, {Wolf} RA (1976) {Star distribution around a massive black hole
  in a globular cluster.} \apj 209:214--232. \doi{10.1086/154711}

\bibitem[{{Bahcall}(1988)}]{1988ARA&A..26..631B}
{Bahcall} NA (1988) {Large-scale structure in the universe indicated by galaxy
  clusters.} \araa 26:631--686. \doi{10.1146/annurev.aa.26.090188.003215}

\bibitem[{{Bahramian} et~al.(2017){Bahramian}, {Heinke}, {Tudor},
  {Miller-Jones}, {Bogdanov}, {Maccarone}, {Knigge}, {Sivakoff}, {Chomiuk},
  {Strader}, {Garcia}, and {Kallman}}]{2017MNRAS.467.2199B}
{Bahramian} A, {Heinke} CO, {Tudor} V, {Miller-Jones} JCA, {Bogdanov} S,
  {Maccarone} TJ, {Knigge} C, {Sivakoff} GR, {Chomiuk} L, {Strader} J, et~al.
  (2017) {The ultracompact nature of the black hole candidate X-ray binary 47
  Tuc X9}. \mnras 467(2):2199--2216. \doi{10.1093/mnras/stx166}.
  {\href{https://arxiv.org/abs/1702.02167}{{arXiv:1702.02167}}} {[astro-ph.HE]}

\bibitem[{{Baker} et~al.(2006){Baker}, {Centrella}, {Choi}, {Koppitz}, and {van
  Meter}}]{2006PhRvL..96k1102B}
{Baker} JG, {Centrella} J, {Choi} DI, {Koppitz} M, {van Meter} J (2006)
  {Gravitational-Wave Extraction from an Inspiraling Configuration of Merging
  Black Holes}. \prl 96(11):111102. \doi{10.1103/PhysRevLett.96.111102}.
  {\href{https://arxiv.org/abs/gr-qc/0511103}{{arXiv:gr-qc/0511103}}} {[gr-qc]}

\bibitem[{{Baldassare} et~al.(2015){Baldassare}, {Reines}, {Gallo}, and
  {Greene}}]{2015ApJ...809L..14B}
{Baldassare} VF, {Reines} AE, {Gallo} E, {Greene} JE (2015) {A
  {\ensuremath{\sim}}50,000 M$_{\odot}$ Solar Mass Black Hole in the Nucleus of
  RGG 118}. \apjl 809(1):L14. \doi{10.1088/2041-8205/809/1/L14}.
  {\href{https://arxiv.org/abs/1506.07531}{{arXiv:1506.07531}}} {[astro-ph.GA]}

\bibitem[{{Baldassare} et~al.(2020){Baldassare}, {Dickey}, {Geha}, and
  {Reines}}]{2020ApJ...898L...3B}
{Baldassare} VF, {Dickey} C, {Geha} M, {Reines} AE (2020) {Populating the
  Low-mass End of the $M_{\rm BH}-\sigma_*$ Relation}. \apjl 898(1):L3.
  \doi{10.3847/2041-8213/aba0c1}.
  {\href{https://arxiv.org/abs/2006.15150}{{arXiv:2006.15150}}} {[astro-ph.GA]}

\bibitem[{{Baldry} et~al.(2004){Baldry}, {Glazebrook}, {Brinkmann},
  {Ivezi{\'c}}, {Lupton}, {Nichol}, and {Szalay}}]{2004ApJ...600..681B}
{Baldry} IK, {Glazebrook} K, {Brinkmann} J, {Ivezi{\'c}} {\v{Z}}, {Lupton} RH,
  {Nichol} RC, {Szalay} AS (2004) {Quantifying the Bimodal Color-Magnitude
  Distribution of Galaxies}. \apj 600(2):681--694. \doi{10.1086/380092}.
  {\href{https://arxiv.org/abs/astro-ph/0309710}{{arXiv:astro-ph/0309710}}}
  {[astro-ph]}

\bibitem[{{Baldry} et~al.(2012){Baldry}, {Driver}, {Loveday}, {Taylor},
  {Kelvin}, {Liske}, {Norberg}, {Robotham}, {Brough}, {Hopkins}, {Bamford},
  {Peacock}, {Bland-Hawthorn}, {Conselice}, {Croom}, {Jones}, {Parkinson},
  {Popescu}, {Prescott}, {Sharp}, and {Tuffs}}]{2012MNRAS.421..621B}
{Baldry} IK, {Driver} SP, {Loveday} J, {Taylor} EN, {Kelvin} LS, {Liske} J,
  {Norberg} P, {Robotham} ASG, {Brough} S, {Hopkins} AM, et~al. (2012) {Galaxy
  And Mass Assembly (GAMA): the galaxy stellar mass function at z < 0.06}.
  \mnras 421(1):621--634. \doi{10.1111/j.1365-2966.2012.20340.x}.
  {\href{https://arxiv.org/abs/1111.5707}{{arXiv:1111.5707}}} {[astro-ph.CO]}

\bibitem[{{Banerjee}(2017)}]{2017MNRAS.467..524B}
{Banerjee} S (2017) {Stellar-mass black holes in young massive and open stellar
  clusters and their role in gravitational-wave generation}. \mnras
  467(1):524--539. \doi{10.1093/mnras/stw3392}.
  {\href{https://arxiv.org/abs/1611.09357}{{arXiv:1611.09357}}} {[astro-ph.HE]}

\bibitem[{{Banerjee}(2018)}]{Banerjee2018}
{Banerjee} S (2018) {Stellar-mass black holes in young massive and open stellar
  clusters and their role in gravitational-wave generation - II}. \mnras
  473(1):909--926. \doi{10.1093/mnras/stx2347}.
  {\href{https://arxiv.org/abs/1707.00922}{{arXiv:1707.00922}}} {[astro-ph.HE]}

\bibitem[{{Banerjee}(2020)}]{Banerjee2020}
{Banerjee} S (2020) {LISA sources from young massive and open stellar
  clusters}. \prd 102(10):103002. \doi{10.1103/PhysRevD.102.103002}.
  {\href{https://arxiv.org/abs/2006.14587}{{arXiv:2006.14587}}} {[astro-ph.HE]}

\bibitem[{{Banerjee}(2021)}]{2021MNRAS.500.3002B}
{Banerjee} S (2021) {Stellar-mass black holes in young massive and open stellar
  clusters - IV. Updated stellar-evolutionary and black hole spin models and
  comparisons with the LIGO-Virgo O1/O2 merger-event data}. \mnras
  500(3):3002--3026. \doi{10.1093/mnras/staa2392}.
  {\href{https://arxiv.org/abs/2004.07382}{{arXiv:2004.07382}}} {[astro-ph.HE]}

\bibitem[{{Banerjee} et~al.(2010){Banerjee}, {Baumgardt}, and
  {Kroupa}}]{2010MNRAS.402..371B}
{Banerjee} S, {Baumgardt} H, {Kroupa} P (2010) {Stellar-mass black holes in
  star clusters: implications for gravitational wave radiation}. \mnras
  402(1):371--380. \doi{10.1111/j.1365-2966.2009.15880.x}.
  {\href{https://arxiv.org/abs/0910.3954}{{arXiv:0910.3954}}} {[astro-ph.SR]}

\bibitem[{{Banerjee} et~al.(2020){Banerjee}, {Belczynski}, {Fryer}, {Berczik},
  {Hurley}, {Spurzem}, and {Wang}}]{2020A&A...639A..41B}
{Banerjee} S, {Belczynski} K, {Fryer} CL, {Berczik} P, {Hurley} JR, {Spurzem}
  R, {Wang} L (2020) {BSE versus StarTrack: Implementations of new wind,
  remnant-formation, and natal-kick schemes in NBODY7 and their astrophysical
  consequences}. \aap 639:A41. \doi{10.1051/0004-6361/201935332}.
  {\href{https://arxiv.org/abs/1902.07718}{{arXiv:1902.07718}}} {[astro-ph.SR]}

\bibitem[{{Bansal} et~al.(2017){Bansal}, {Taylor}, {Peck}, {Zavala}, and
  {Romani}}]{2017ApJ...843...14B}
{Bansal} K, {Taylor} GB, {Peck} AB, {Zavala} RT, {Romani} RW (2017)
  {Constraining the Orbit of the Supermassive Black Hole Binary 0402+379}. \apj
  843(1):14. \doi{10.3847/1538-4357/aa74e1}.
  {\href{https://arxiv.org/abs/1705.08556}{{arXiv:1705.08556}}} {[astro-ph.GA]}

\bibitem[{{Bar-Or} and {Alexander}(2014)}]{2014CQGra..31x4003B}
{Bar-Or} B, {Alexander} T (2014) {The statistical mechanics of relativistic
  orbits around a massive black hole}. Classical and Quantum Gravity
  31(24):244003. \doi{10.1088/0264-9381/31/24/244003}.
  {\href{https://arxiv.org/abs/1404.0351}{{arXiv:1404.0351}}} {[astro-ph.GA]}

\bibitem[{{Bar-Or} and {Alexander}(2016)}]{2016ApJ...820..129B}
{Bar-Or} B, {Alexander} T (2016) {Steady-state Relativistic Stellar Dynamics
  Around a Massive Black hole}. \apj 820(2):129.
  \doi{10.3847/0004-637X/820/2/129}.
  {\href{https://arxiv.org/abs/1508.01390}{{arXiv:1508.01390}}} {[astro-ph.GA]}

\bibitem[{{Bar-Or} and {Fouvry}(2018)}]{2018ApJ...860L..23B}
{Bar-Or} B, {Fouvry} JB (2018) {Scalar Resonant Relaxation of Stars around a
  Massive Black Hole}. \apjl 860(2):L23. \doi{10.3847/2041-8213/aac88e}.
  {\href{https://arxiv.org/abs/1802.08890}{{arXiv:1802.08890}}} {[astro-ph.GA]}

\bibitem[{{Bar-Or} et~al.(2013){Bar-Or}, {Kupi}, and
  {Alexander}}]{2013ApJ...764...52B}
{Bar-Or} B, {Kupi} G, {Alexander} T (2013) {Stellar Energy Relaxation around a
  Massive Black Hole}. \apj 764(1):52. \doi{10.1088/0004-637X/764/1/52}.
  {\href{https://arxiv.org/abs/1209.4594}{{arXiv:1209.4594}}} {[astro-ph.GA]}

\bibitem[{{Barack} and {Cutler}(2004{\natexlab{a}})}]{2004PhRvD..70l2002B}
{Barack} L, {Cutler} C (2004{\natexlab{a}}) {Confusion noise from LISA capture
  sources}. \prd 70(12):122002. \doi{10.1103/PhysRevD.70.122002}.
  {\href{https://arxiv.org/abs/gr-qc/0409010}{{arXiv:gr-qc/0409010}}} {[gr-qc]}

\bibitem[{{Barack} and {Cutler}(2004{\natexlab{b}})}]{2004PhRvD..69h2005B}
{Barack} L, {Cutler} C (2004{\natexlab{b}}) {LISA capture sources: Approximate
  waveforms, signal-to-noise ratios, and parameter estimation accuracy}. \prd
  69(8):082005. \doi{10.1103/PhysRevD.69.082005}.
  {\href{https://arxiv.org/abs/gr-qc/0310125}{{arXiv:gr-qc/0310125}}} {[gr-qc]}

\bibitem[{{Barack} and {Pound}(2019)}]{2019RPPh...82a6904B}
{Barack} L, {Pound} A (2019) {Self-force and radiation reaction in general
  relativity}. Reports on Progress in Physics 82(1):016904.
  \doi{10.1088/1361-6633/aae552}.
  {\href{https://arxiv.org/abs/1805.10385}{{arXiv:1805.10385}}} {[gr-qc]}

\bibitem[{{Barai} et~al.(2016){Barai}, {Murante}, {Borgani}, {Gaspari},
  {Granato}, {Monaco}, and {Ragone-Figueroa}}]{2016MNRAS.461.1548B}
{Barai} P, {Murante} G, {Borgani} S, {Gaspari} M, {Granato} GL, {Monaco} P,
  {Ragone-Figueroa} C (2016) {Kinetic AGN feedback effects on cluster cool
  cores simulated using SPH}. \mnras 461(2):1548--1567.
  \doi{10.1093/mnras/stw1389}.
  {\href{https://arxiv.org/abs/1605.08051}{{arXiv:1605.08051}}} {[astro-ph.GA]}

\bibitem[{{Barausse}(2012)}]{2012MNRAS.423.2533B}
{Barausse} E (2012) {The evolution of massive black holes and their spins in
  their galactic hosts}. \mnras 423(3):2533--2557.
  \doi{10.1111/j.1365-2966.2012.21057.x}.
  {\href{https://arxiv.org/abs/1201.5888}{{arXiv:1201.5888}}} {[astro-ph.CO]}

\bibitem[{{Barausse} and {Rezzolla}(2008)}]{2008PhRvD..77j4027B}
{Barausse} E, {Rezzolla} L (2008) {Influence of the hydrodynamic drag from an
  accretion torus on extreme mass-ratio inspirals}. \prd 77(10):104027.
  \doi{10.1103/PhysRevD.77.104027}.
  {\href{https://arxiv.org/abs/0711.4558}{{arXiv:0711.4558}}} {[gr-qc]}

\bibitem[{{Barausse} and {Rezzolla}(2009)}]{2009ApJ...704L..40B}
{Barausse} E, {Rezzolla} L (2009) {Predicting the Direction of the Final Spin
  from the Coalescence of Two Black Holes}. \apjl 704(1):L40--L44.
  \doi{10.1088/0004-637X/704/1/L40}.
  {\href{https://arxiv.org/abs/0904.2577}{{arXiv:0904.2577}}} {[gr-qc]}

\bibitem[{{Barausse} et~al.(2007){Barausse}, {Rezzolla}, {Petroff}, and
  {Ansorg}}]{2007PhRvD..75f4026B}
{Barausse} E, {Rezzolla} L, {Petroff} D, {Ansorg} M (2007) {Gravitational waves
  from extreme mass ratio inspirals in nonpure Kerr spacetimes}. \prd
  75(6):064026. \doi{10.1103/PhysRevD.75.064026}.
  {\href{https://arxiv.org/abs/gr-qc/0612123}{{arXiv:gr-qc/0612123}}} {[gr-qc]}

\bibitem[{{Barausse} et~al.(2014){Barausse}, {Cardoso}, and
  {Pani}}]{2014PhRvD..89j4059B}
{Barausse} E, {Cardoso} V, {Pani} P (2014) {Can environmental effects spoil
  precision gravitational-wave astrophysics?} \prd 89(10):104059.
  \doi{10.1103/PhysRevD.89.104059}.
  {\href{https://arxiv.org/abs/1404.7149}{{arXiv:1404.7149}}} {[gr-qc]}

\bibitem[{{Barausse} et~al.(2016){Barausse}, {Yunes}, and
  {Chamberlain}}]{2016PhRvL.116x1104B}
{Barausse} E, {Yunes} N, {Chamberlain} K (2016) {Theory-Agnostic Constraints on
  Black-Hole Dipole Radiation with Multiband Gravitational-Wave Astrophysics}.
  \prl 116(24):241104. \doi{10.1103/PhysRevLett.116.241104}.
  {\href{https://arxiv.org/abs/1603.04075}{{arXiv:1603.04075}}} {[gr-qc]}

\bibitem[{{Barausse} et~al.(2020{\natexlab{a}}){Barausse}, {Berti}, {Hertog},
  {Hughes}, {Jetzer}, {Pani}, {Sotiriou}, {Tamanini}, {Witek}, {Yagi}, {Yunes},
  {Abdelsalhin}, {Achucarro}, {van Aelst}, {Afshordi}, {Akcay}, {Annulli},
  {Arun}, {Ayuso}, {Baibhav}, {Baker}, {Bantilan}, {Barreiro},
  {Barrera-Hinojosa}, {Bartolo}, {Baumann}, {Belgacem}, {Bellini}, {Bellomo},
  {Ben-Dayan}, {Bena}, {Benkel}, {Bergshoefs}, {Bernard}, {Bernuzzi},
  {Bertacca}, {Besancon}, {Beutler}, {Beyer}, {Bhagwat}, {Bicak}, {Biondini},
  {Bize}, {Blas}, {Boehmer}, {Boller}, {Bonga}, {Bonvin}, {Bosso}, {Bozzola},
  {Brax}, {Breitbach}, {Brito}, {Bruni}, {Br{\"u}gmann}, {Bulten}, {Buonanno},
  {Burko}, {Burrage}, {Cabral}, {Calcagni}, {Caprini},
  {C{\'a}rdenas-Avenda{\~n}o}, {Celoria}, {Chatziioannou}, {Chernoff},
  {Clough}, {Coates}, {Comelli}, {Comp{\`e}re}, {Croon}, {Cruces}, {Cusin},
  {Dalang}, {Danielsson}, {Das}, {Datta}, {de Boer}, {De Luca}, {De Rham},
  {Desjacques}, {Destounis}, {Di Filippo}, {Dima}, {Dimastrogiovanni}, {Dolan},
  {Doneva}, {Duque}, {Durrer}, {East}, {Easther}, {Elley}, {Ellis}, {Emparan},
  {Ezquiaga}, {Fairbairn}, {Fairhurst}, {Farmer}, {Fasiello}, {Ferrari},
  {Ferreira}, {Ficarra}, {Figueras}, {Fisenko}, {Foffa}, {Franchini},
  {Franciolini}, {Fransen}, {Frauendiener}, {Frusciante}, {Fujita}, {Gair},
  {Ganz}, {Garcia}, {Garcia-Bellido}, {Garriga}, {Geiger}, {Geng}, {Gergely},
  {Germani}, {Gerosa}, {Giddings}, {Gourgoulhon}, {Grand clement}, {Graziani},
  {Gualtieri}, {Haggard}, {Haino}, {Halburd}, {Han}, {Hawken}, {Hees}, {Heng},
  {Hennig}, {Herdeiro}, {Hervik}, {Holten}, {Hoyle}, {Hu}, {Hull}, {Ikeda},
  {Isi}, {Jenkins}, {Juli{\'e}}, {Kajfasz}, {Kalaghatgi}, {Kaloper},
  {Kamionkowski}, {Karas}, {Kastha}, {Keresztes}, {Kidder}, {Kimpson}, {Klein},
  {Klioner}, {Kokkotas}, {Kolesova}, {Kolkowitz}, {Kopp}, {Koyama},
  {Krishnendu}, {Kroon}, {Kunz}, {Lahav}, {Landragin}, {Lang}, {Le
  Poncin-Lafitte}, {Lemos}, {Li}, {Liberati}, {Liguori}, {Lin}, {Liu}, {Lobo},
  {Loll}, {Lombriser}, {Lovelace}, {Macedo}, {Madge}, {Maggio}, {Maggiore},
  {Marassi}, {Marcoccia}, {Markakis}, {Martens}, {Martinovic}, {Martins},
  {Maselli}, {Mastrogiovanni}, {Matarrese}, {Matas}, {Mavromatos}, {Mazumdar},
  {Meerburg}, {Megias}, {Miller}, {Mimoso}, {Mittnacht}, {Montero}, {Moore},
  {Martin-Moruno}, {Musco}, {Nakano}, {Nampalliwar}, {Nardini}, {Nielsen},
  {Nov{\'a}k}, {Nunes}, {Okounkova}, {Oliveri}, {Oppizzi}, {Orlando}, {Oshita},
  {Pappas}, {Paschalidis}, {Peiris}, {Peloso}, {Perkins}, {Pettorino},
  {Pikovski}, {Pilo}, {Podolsky}, {Pontzen}, {Prabhat}, {Pratten}, {Prokopec},
  {Prouza}, {Qi}, {Raccanelli}, {Rajantie}, {Randall}, {Raposo}, {Raymond},
  {Renaux-Petel}, {Ricciardone}, {Riotto}, {Robson}, {Roest}, {Rollo},
  {Rosofsky}, {Ruan}, {Rubiera-Garc{\'\i}a}, {Ruiz}, {Rusu}, {Sabatie}, {Sago},
  {Sakellariadou}, {Saltas}, {Sberna}, {Sathyaprakash}, {Scheel}, {Schmidt},
  {Schutz}, {Schwaller}, {Shao}, {Shapiro}, {Shoemaker}, {Silva}, {Simpson},
  {Sopuerta}, {Spallicci}, {Stefanek}, {Stein}, {Stergioulas}, {Stott},
  {Sutton}, {Svarc}, {Tagoshi}, {Tahamtan}, {Takeda}, {Tanaka}, {Tantilian},
  {Tasinato}, {Tattersall}, {Teukolsky}, {Tiec}, {Theureau}, {Trodden},
  {Tolley}, {Toubiana}, {Traykova}, {Tsokaros}, {Unal}, {Unnikrishnan},
  {Vagenas}, {Valageas}, {Vallisneri}, {Van den Brand}, {Van den Broeck}, {van
  de Meent}, {Vanhove}, {Varma}, {Veitch}, {Vercnocke}, {Verde}, {Vernieri},
  {Vernizzi}, {Vicente}, {Vidotto}, {Visser}, {Vlah}, {Vretinaris},
  {V{\"o}lkel}, {Wang}, {Wang}, {Werner}, {Westernacher}, {Weygaert},
  {Wiltshire}, {Wiseman}, {Wolf}, {Wu}, {Yamada}, {Yang}, {Yi}, {Yue}, {Yvon},
  {Zilh{\~a}o}, {Zimmerman}, and {Zumalacarregui}}]{2020GReGr..52...81B}
{Barausse} E, {Berti} E, {Hertog} T, {Hughes} SA, {Jetzer} P, {Pani} P,
  {Sotiriou} TP, {Tamanini} N, {Witek} H, {Yagi} K, et~al. (2020{\natexlab{a}})
  {Prospects for fundamental physics with LISA}. General Relativity and
  Gravitation 52(8):81. \doi{10.1007/s10714-020-02691-1}.
  {\href{https://arxiv.org/abs/2001.09793}{{arXiv:2001.09793}}} {[gr-qc]}

\bibitem[{{Barausse} et~al.(2020{\natexlab{b}}){Barausse}, {Dvorkin},
  {Tremmel}, {Volonteri}, and {Bonetti}}]{2020ApJ...904...16B}
{Barausse} E, {Dvorkin} I, {Tremmel} M, {Volonteri} M, {Bonetti} M
  (2020{\natexlab{b}}) {Massive Black Hole Merger Rates: The Effect of
  Kiloparsec Separation Wandering and Supernova Feedback}. \apj 904(1):16.
  \doi{10.3847/1538-4357/abba7f}.
  {\href{https://arxiv.org/abs/2006.03065}{{arXiv:2006.03065}}} {[astro-ph.GA]}

\bibitem[{{Barclay} et~al.(2011){Barclay}, {Ramsay}, {Hakala}, {Napiwotzki},
  {Nelemans}, {Potter}, and {Todd}}]{2011MNRAS.413.2696B}
{Barclay} T, {Ramsay} G, {Hakala} P, {Napiwotzki} R, {Nelemans} G, {Potter} S,
  {Todd} I (2011) {Stellar variability on time-scales of minutes: results from
  the first 5 yr of the Rapid Temporal Survey}. \mnras 413(4):2696--2708.
  \doi{10.1111/j.1365-2966.2011.18345.x}.
  {\href{https://arxiv.org/abs/1101.2445}{{arXiv:1101.2445}}} {[astro-ph.GA]}

\bibitem[{{Barcons} et~al.(2015){Barcons}, {Nandra}, {Barret}, {den Herder},
  {Fabian}, {Piro}, {Watson}, and {the Athena Team}}]{2015JPhCS.610a2008B}
{Barcons} X, {Nandra} K, {Barret} D, {den Herder} JW, {Fabian} AC, {Piro} L,
  {Watson} MG, {the Athena Team} (2015) {Athena: the X-ray observatory to study
  the hot and energetic Universe}. In: Journal of Physics Conference Series.
  Journal of Physics Conference Series, vol 610. p 012008.
  \doi{10.1088/1742-6596/610/1/012008}

\bibitem[{{Bardeen}(1970)}]{1970Natur.226...64B}
{Bardeen} JM (1970) {Kerr Metric Black Holes}. \nat 226(5240):64--65.
  \doi{10.1038/226064a0}

\bibitem[{{Bardeen} and {Petterson}(1975)}]{1975ApJ...195L..65B}
{Bardeen} JM, {Petterson} JA (1975) {The Lense-Thirring Effect and Accretion
  Disks around Kerr Black Holes}. \apjl 195:L65. \doi{10.1086/181711}

\bibitem[{{Barrow} et~al.(2018){Barrow}, {Aykutalp}, and
  {Wise}}]{2018NatAs...2..987B}
{Barrow} KSS, {Aykutalp} A, {Wise} JH (2018) {Observational signatures of
  massive black hole formation in the early Universe}. Nature Astronomy
  2:987--994. \doi{10.1038/s41550-018-0569-y}.
  {\href{https://arxiv.org/abs/1809.03526}{{arXiv:1809.03526}}} {[astro-ph.GA]}

\bibitem[{{Barstow} et~al.(2014){Barstow}, {Casewell}, {Catalan},
  {Copperwheat}, {Gaensicke}, {Garcia-Berro}, {Hambly}, {Heber}, {Holberg},
  {Isern}, {Jeffery}, {Jordan}, {Lawrie}, {Lynas-Gray}, {Maccarone}, {Marsh},
  {Parsons}, {Silvotti}, {Subasavage}, {Torres}, and
  {Wheatley}}]{2014arXiv1407.6163B}
{Barstow} MA, {Casewell} SL, {Catalan} S, {Copperwheat} C, {Gaensicke} B,
  {Garcia-Berro} E, {Hambly} N, {Heber} U, {Holberg} J, {Isern} J, et~al.
  (2014) {White paper: Gaia and the end states of stellar evolution}. arXiv
  e-prints arXiv:1407.6163.
  {\href{https://arxiv.org/abs/1407.6163}{{arXiv:1407.6163}}} {[astro-ph.SR]}

\bibitem[{{Bartko} et~al.(2009){Bartko}, {Martins}, {Fritz}, {Genzel}, {Levin},
  {Perets}, {Paumard}, {Nayakshin}, {Gerhard}, {Alexander}, {Dodds-Eden},
  {Eisenhauer}, {Gillessen}, {Mascetti}, {Ott}, {Perrin}, {Pfuhl}, {Reid},
  {Rouan}, {Sternberg}, and {Trippe}}]{2009ApJ...697.1741B}
{Bartko} H, {Martins} F, {Fritz} TK, {Genzel} R, {Levin} Y, {Perets} HB,
  {Paumard} T, {Nayakshin} S, {Gerhard} O, {Alexander} T, et~al. (2009)
  {Evidence for Warped Disks of Young Stars in the Galactic Center}. \apj
  697(2):1741--1763. \doi{10.1088/0004-637X/697/2/1741}.
  {\href{https://arxiv.org/abs/0811.3903}{{arXiv:0811.3903}}} {[astro-ph]}

\bibitem[{Bartolo et~al.(2019)Bartolo, De~Luca, Franciolini, Peloso, Racco, and
  Riotto}]{Bartolo:2018rku}
Bartolo N, De~Luca V, Franciolini G, Peloso M, Racco D, Riotto A (2019)
  {Testing primordial black holes as dark matter with LISA}. Phys Rev D
  99(10):103521. \doi{10.1103/PhysRevD.99.103521}.
  {\href{https://arxiv.org/abs/1810.12224}{{arXiv:1810.12224}}} {[astro-ph.CO]}

\bibitem[{{Bartos} et~al.(2017){Bartos}, {Kocsis}, {Haiman}, and
  {M{\'a}rka}}]{2017ApJ...835..165B}
{Bartos} I, {Kocsis} B, {Haiman} Z, {M{\'a}rka} S (2017) {Rapid and Bright
  Stellar-mass Binary Black Hole Mergers in Active Galactic Nuclei}. \apj
  835(2):165. \doi{10.3847/1538-4357/835/2/165}.
  {\href{https://arxiv.org/abs/1602.03831}{{arXiv:1602.03831}}} {[astro-ph.HE]}

\bibitem[{{Baruteau} and {Masset}(2013)}]{2013LNP...861..201B}
{Baruteau} C, {Masset} F (2013) {Recent Developments in Planet Migration
  Theory}, vol 861, p 201. \doi{10.1007/978-3-642-32961-6_6}

\bibitem[{{Baruteau} et~al.(2011){Baruteau}, {Cuadra}, and
  {Lin}}]{2011ApJ...726...28B}
{Baruteau} C, {Cuadra} J, {Lin} DNC (2011) {Binaries Migrating in a Gaseous
  Disk: Where are the Galactic Center Binaries?} \apj 726(1):28.
  \doi{10.1088/0004-637X/726/1/28}.
  {\href{https://arxiv.org/abs/1011.0360}{{arXiv:1011.0360}}} {[astro-ph.GA]}

\bibitem[{{Bassini} et~al.(2019){Bassini}, {Rasia}, {Borgani},
  {Ragone-Figueroa}, {Biffi}, {Dolag}, {Gaspari}, {Granato}, {Murante},
  {Taffoni}, and {Tornatore}}]{2019A&A...630A.144B}
{Bassini} L, {Rasia} E, {Borgani} S, {Ragone-Figueroa} C, {Biffi} V, {Dolag} K,
  {Gaspari} M, {Granato} GL, {Murante} G, {Taffoni} G, et~al. (2019) {Black
  hole mass of central galaxies and cluster mass correlation in cosmological
  hydro-dynamical simulations}. \aap 630:A144.
  \doi{10.1051/0004-6361/201935383}.
  {\href{https://arxiv.org/abs/1903.03142}{{arXiv:1903.03142}}} {[astro-ph.GA]}

\bibitem[{{Bastian} et~al.(2010){Bastian}, {Covey}, and
  {Meyer}}]{2010ARA&A..48..339B}
{Bastian} N, {Covey} KR, {Meyer} MR (2010) {A Universal Stellar Initial Mass
  Function? A Critical Look at Variations}. \araa 48:339--389.
  \doi{10.1146/annurev-astro-082708-101642}.
  {\href{https://arxiv.org/abs/1001.2965}{{arXiv:1001.2965}}} {[astro-ph.GA]}

\bibitem[{{Baumgardt}(2009)}]{2009gcgg.book..387B}
{Baumgardt} H (2009) {Dissolution of Globular Clusters}, p 387.
  \doi{10.1007/978-3-540-76961-3_89}

\bibitem[{{Baumgardt} et~al.(2018){Baumgardt}, {Amaro-Seoane}, and
  {Sch{\"o}del}}]{2018A&A...609A..28B}
{Baumgardt} H, {Amaro-Seoane} P, {Sch{\"o}del} R (2018) {The distribution of
  stars around the Milky Way's central black hole. III. Comparison with
  simulations}. \aap 609:A28. \doi{10.1051/0004-6361/201730462}.
  {\href{https://arxiv.org/abs/1701.03818}{{arXiv:1701.03818}}} {[astro-ph.GA]}

\bibitem[{{Bavera} et~al.(2020){Bavera}, {Fragos}, {Qin}, {Zapartas},
  {Neijssel}, {Mandel}, {Batta}, {Gaebel}, {Kimball}, and
  {Stevenson}}]{2020A&A...635A..97B}
{Bavera} SS, {Fragos} T, {Qin} Y, {Zapartas} E, {Neijssel} CJ, {Mandel} I,
  {Batta} A, {Gaebel} SM, {Kimball} C, {Stevenson} S (2020) {The origin of spin
  in binary black holes. Predicting the distributions of the main observables
  of Advanced LIGO}. \aap 635:A97. \doi{10.1051/0004-6361/201936204}.
  {\href{https://arxiv.org/abs/1906.12257}{{arXiv:1906.12257}}} {[astro-ph.HE]}

\bibitem[{{Bavera} et~al.(2021){Bavera}, {Fragos}, {Zevin}, {Berry},
  {Marchant}, {Andrews}, {Coughlin}, {Dotter}, {Kovlakas}, {Misra},
  {Serra-Perez}, {Qin}, {Rocha}, {Rom{\'a}n-Garza}, {Tran}, and
  {Zapartas}}]{2020arXiv201016333B}
{Bavera} SS, {Fragos} T, {Zevin} M, {Berry} CPL, {Marchant} P, {Andrews} JJ,
  {Coughlin} S, {Dotter} A, {Kovlakas} K, {Misra} D, et~al. (2021) {The impact
  of mass-transfer physics on the observable properties of field binary black
  hole populations}. \aap 647:A153. \doi{10.1051/0004-6361/202039804}.
  {\href{https://arxiv.org/abs/2010.16333}{{arXiv:2010.16333}}} {[astro-ph.HE]}

\bibitem[{{Begelman} et~al.(1980){Begelman}, {Blandford}, and
  {Rees}}]{1980Natur.287..307B}
{Begelman} MC, {Blandford} RD, {Rees} MJ (1980) {Massive black hole binaries in
  active galactic nuclei}. \nat 287(5780):307--309. \doi{10.1038/287307a0}

\bibitem[{{Bekenstein}(1973)}]{1973ApJ...183..657B}
{Bekenstein} JD (1973) {Gravitational-Radiation Recoil and Runaway Black
  Holes}. \apj 183:657--664. \doi{10.1086/152255}

\bibitem[{{Belczynski} et~al.(2002){Belczynski}, {Kalogera}, and
  {Bulik}}]{2002ApJ...572..407B}
{Belczynski} K, {Kalogera} V, {Bulik} T (2002) {A Comprehensive Study of Binary
  Compact Objects as Gravitational Wave Sources: Evolutionary Channels, Rates,
  and Physical Properties}. \apj 572(1):407--431. \doi{10.1086/340304}.
  {\href{https://arxiv.org/abs/astro-ph/0111452}{{arXiv:astro-ph/0111452}}}
  {[astro-ph]}

\bibitem[{{Belczynski} et~al.(2008){Belczynski}, {Kalogera}, {Rasio}, {Taam},
  {Zezas}, {Bulik}, {Maccarone}, and {Ivanova}}]{2008ApJS..174..223B}
{Belczynski} K, {Kalogera} V, {Rasio} FA, {Taam} RE, {Zezas} A, {Bulik} T,
  {Maccarone} TJ, {Ivanova} N (2008) {Compact Object Modeling with the
  StarTrack Population Synthesis Code}. \apjs 174(1):223--260.
  \doi{10.1086/521026}.
  {\href{https://arxiv.org/abs/astro-ph/0511811}{{arXiv:astro-ph/0511811}}}
  {[astro-ph]}

\bibitem[{{Belczynski} et~al.(2010{\natexlab{a}}){Belczynski}, {Benacquista},
  and {Bulik}}]{2010ApJ...725..816B}
{Belczynski} K, {Benacquista} M, {Bulik} T (2010{\natexlab{a}}) {Double Compact
  Objects as Low-frequency Gravitational Wave Sources}. \apj 725(1):816--823.
  \doi{10.1088/0004-637X/725/1/816}.
  {\href{https://arxiv.org/abs/0811.1602}{{arXiv:0811.1602}}} {[astro-ph]}

\bibitem[{{Belczynski} et~al.(2010{\natexlab{b}}){Belczynski}, {Dominik},
  {Bulik}, {O'Shaughnessy}, {Fryer}, and {Holz}}]{2010ApJ...715L.138B}
{Belczynski} K, {Dominik} M, {Bulik} T, {O'Shaughnessy} R, {Fryer} C, {Holz} DE
  (2010{\natexlab{b}}) {The Effect of Metallicity on the Detection Prospects
  for Gravitational Waves}. \apjl 715(2):L138--L141.
  \doi{10.1088/2041-8205/715/2/L138}.
  {\href{https://arxiv.org/abs/1004.0386}{{arXiv:1004.0386}}} {[astro-ph.HE]}

\bibitem[{{Belczynski} et~al.(2012){Belczynski}, {Bulik}, and
  {Fryer}}]{2012arXiv1208.2422B}
{Belczynski} K, {Bulik} T, {Fryer} CL (2012) {High Mass X-ray Binaries: Future
  Evolution and Fate}. arXiv e-prints arXiv:1208.2422.
  {\href{https://arxiv.org/abs/1208.2422}{{arXiv:1208.2422}}} {[astro-ph.HE]}

\bibitem[{{Belczynski} et~al.(2013){Belczynski}, {Bulik}, {Mandel},
  {Sathyaprakash}, {Zdziarski}, and {Miko{\l}ajewska}}]{2013ApJ...764...96B}
{Belczynski} K, {Bulik} T, {Mandel} I, {Sathyaprakash} BS, {Zdziarski} AA,
  {Miko{\l}ajewska} J (2013) {Cyg X-3: A Galactic Double Black Hole or
  Black-hole-Neutron-star Progenitor}. \apj 764(1):96.
  \doi{10.1088/0004-637X/764/1/96}.
  {\href{https://arxiv.org/abs/1209.2658}{{arXiv:1209.2658}}} {[astro-ph.HE]}

\bibitem[{{Belczynski} et~al.(2014){Belczynski}, {Buonanno}, {Cantiello},
  {Fryer}, {Holz}, {Mandel}, {Miller}, and {Walczak}}]{Belczynski:2014VMS}
{Belczynski} K, {Buonanno} A, {Cantiello} M, {Fryer} CL, {Holz} DE, {Mandel} I,
  {Miller} MC, {Walczak} M (2014) {The Formation and Gravitational-wave
  Detection of Massive Stellar Black Hole Binaries}. \apj 789:120.
  \doi{10.1088/0004-637X/789/2/120}.
  {\href{https://arxiv.org/abs/1403.0677}{{arXiv:1403.0677}}} {[astro-ph.HE]}

\bibitem[{{Belczynski} et~al.(2016{\natexlab{a}}){Belczynski}, {Holz}, {Bulik},
  and {O'Shaughnessy}}]{2016Natur.534..512B}
{Belczynski} K, {Holz} DE, {Bulik} T, {O'Shaughnessy} R (2016{\natexlab{a}})
  {The first gravitational-wave source from the isolated evolution of two stars
  in the 40-100 solar mass range}. \nat 534(7608):512--515.
  \doi{10.1038/nature18322}.
  {\href{https://arxiv.org/abs/1602.04531}{{arXiv:1602.04531}}} {[astro-ph.HE]}

\bibitem[{{Belczynski} et~al.(2016{\natexlab{b}}){Belczynski}, {Repetto},
  {Holz}, {O'Shaughnessy}, {Bulik}, {Berti}, {Fryer}, and
  {Dominik}}]{2016ApJ...819..108B}
{Belczynski} K, {Repetto} S, {Holz} DE, {O'Shaughnessy} R, {Bulik} T, {Berti}
  E, {Fryer} C, {Dominik} M (2016{\natexlab{b}}) {Compact Binary Merger Rates:
  Comparison with LIGO/Virgo Upper Limits}. \apj 819(2):108.
  \doi{10.3847/0004-637X/819/2/108}.
  {\href{https://arxiv.org/abs/1510.04615}{{arXiv:1510.04615}}} {[astro-ph.HE]}

\bibitem[{{Belczynski} et~al.(2020){Belczynski}, {Klencki}, {Fields}, {Olejak},
  {Berti}, {Meynet}, {Fryer}, {Holz}, {O'Shaughnessy}, {Brown}, {Bulik},
  {Leung}, {Nomoto}, {Madau}, {Hirschi}, {Kaiser}, {Jones}, {Mondal}
  et~al.}]{2020A&A...636A.104B}
{Belczynski} K, {Klencki} J, {Fields} CE, {Olejak} A, {Berti} E, {Meynet} G,
  {Fryer} CL, {Holz} DE, {O'Shaughnessy} R, {Brown} DA, et~al. (2020)
  {Evolutionary roads leading to low effective spins, high black hole masses,
  and O1/O2 rates for LIGO/Virgo binary black holes}. \aap 636:A104.
  \doi{10.1051/0004-6361/201936528}.
  {\href{https://arxiv.org/abs/1706.07053}{{arXiv:1706.07053}}} {[astro-ph.HE]}

\bibitem[{{Bell} et~al.(2017){Bell}, {Gianninas}, {Hermes}, {Winget}, {Kilic},
  {Montgomery}, {Castanheira}, {Vanderbosch}, {Winget}, and
  {Brown}}]{2017ApJ...835..180B}
{Bell} KJ, {Gianninas} A, {Hermes} JJ, {Winget} DE, {Kilic} M, {Montgomery} MH,
  {Castanheira} BG, {Vanderbosch} Z, {Winget} KI, {Brown} WR (2017) {Pruning
  The ELM Survey: Characterizing Candidate Low-mass White Dwarfs through
  Photometric Variability}. \apj 835(2):180. \doi{10.3847/1538-4357/835/2/180}.
  {\href{https://arxiv.org/abs/1612.06390}{{arXiv:1612.06390}}} {[astro-ph.SR]}

\bibitem[{{Bellm} et~al.(2019){Bellm}, {Kulkarni}, {Graham}, {Dekany}, {Smith},
  {Riddle}, {Masci}, and {et~al.}}]{2019PASP..131a8002B}
{Bellm} EC, {Kulkarni} SR, {Graham} MJ, {Dekany} R, {Smith} RM, {Riddle} R,
  {Masci} FJ, {et~al} (2019) {The Zwicky Transient Facility: System Overview,
  Performance, and First Results}. \pasp 131(995):018002.
  \doi{10.1088/1538-3873/aaecbe}.
  {\href{https://arxiv.org/abs/1902.01932}{{arXiv:1902.01932}}} {[astro-ph.IM]}

\bibitem[{{Belloni} et~al.(2016){Belloni}, {Giersz}, {Askar}, {Leigh}, and
  {Hypki}}]{2016MNRAS.462.2950B}
{Belloni} D, {Giersz} M, {Askar} A, {Leigh} N, {Hypki} A (2016) {MOCCA-SURVEY
  database I. Accreting white dwarf binary systems in globular clusters - I.
  Cataclysmic variables - present-day population}. \mnras 462(3):2950--2969.
  \doi{10.1093/mnras/stw1841}.
  {\href{https://arxiv.org/abs/1607.07619}{{arXiv:1607.07619}}} {[astro-ph.GA]}

\bibitem[{{Bellovary} et~al.(2010){Bellovary}, {Governato}, {Quinn}, {Wadsley},
  {Shen}, and {Volonteri}}]{2010ApJ...721L.148B}
{Bellovary} JM, {Governato} F, {Quinn} TR, {Wadsley} J, {Shen} S, {Volonteri} M
  (2010) {Wandering Black Holes in Bright Disk Galaxy Halos}. \apjl
  721(2):L148--L152. \doi{10.1088/2041-8205/721/2/L148}.
  {\href{https://arxiv.org/abs/1008.5147}{{arXiv:1008.5147}}} {[astro-ph.CO]}

\bibitem[{{Bellovary} et~al.(2016){Bellovary}, {Mac Low}, {McKernan}, and
  {Ford}}]{2016ApJ...819L..17B}
{Bellovary} JM, {Mac Low} MM, {McKernan} B, {Ford} KES (2016) {Migration Traps
  in Disks around Supermassive Black Holes}. \apjl 819(2):L17.
  \doi{10.3847/2041-8205/819/2/L17}.
  {\href{https://arxiv.org/abs/1511.00005}{{arXiv:1511.00005}}} {[astro-ph.GA]}

\bibitem[{{Bellovary} et~al.(2019){Bellovary}, {Cleary}, {Munshi}, {Tremmel},
  {Christensen}, {Brooks}, and {Quinn}}]{2019MNRAS.482.2913B}
{Bellovary} JM, {Cleary} CE, {Munshi} F, {Tremmel} M, {Christensen} CR,
  {Brooks} A, {Quinn} TR (2019) {Multimessenger signatures of massive black
  holes in dwarf galaxies}. \mnras 482(3):2913--2923.
  \doi{10.1093/mnras/sty2842}.
  {\href{https://arxiv.org/abs/1806.00471}{{arXiv:1806.00471}}} {[astro-ph.GA]}

\bibitem[{{Belokurov} et~al.(2006){Belokurov}, {Evans}, {Irwin}, {Hewett}, and
  {Wilkinson}}]{2006ApJ...637L..29B}
{Belokurov} V, {Evans} NW, {Irwin} MJ, {Hewett} PC, {Wilkinson} MI (2006) {The
  Discovery of Tidal Tails around the Globular Cluster NGC 5466}. \apjl
  637(1):L29--L32. \doi{10.1086/500362}.
  {\href{https://arxiv.org/abs/astro-ph/0511767}{{arXiv:astro-ph/0511767}}}
  {[astro-ph]}

\bibitem[{{Belotsky} et~al.(2014){Belotsky}, {Dmitriev}, {Esipova}, {Gani},
  {Grobov}, {Khlopov}, {Kirillov}, {Rubin}, and
  {Svadkovsky}}]{2014MPLA...2940005B}
{Belotsky} KM, {Dmitriev} AE, {Esipova} EA, {Gani} VA, {Grobov} AV, {Khlopov}
  MY, {Kirillov} AA, {Rubin} SG, {Svadkovsky} IV (2014) {Signatures of
  primordial black hole dark matter}. Modern Physics Letters A 29(37):1440005.
  \doi{10.1142/S0217732314400057}.
  {\href{https://arxiv.org/abs/1410.0203}{{arXiv:1410.0203}}} {[astro-ph.CO]}

\bibitem[{{Benacquista} and {Holley-Bockelmann}(2006)}]{2006ApJ...645..589B}
{Benacquista} M, {Holley-Bockelmann} K (2006) {Consequences of Disk Scale
  Height on LISA Confusion Noise from Close White Dwarf Binaries}. \apj
  645(1):589--596. \doi{10.1086/504024}.
  {\href{https://arxiv.org/abs/astro-ph/0504135}{{arXiv:astro-ph/0504135}}}
  {[astro-ph]}

\bibitem[{{Benacquista}(2011)}]{2011ApJ...740L..54B}
{Benacquista} MJ (2011) {Tidal Perturbations to the Gravitational Inspiral of
  J0651+2844}. \apjl 740(2):L54. \doi{10.1088/2041-8205/740/2/L54}.
  {\href{https://arxiv.org/abs/1109.2744}{{arXiv:1109.2744}}} {[astro-ph.SR]}

\bibitem[{{Bender} and {Hils}(1997)}]{1997CQGra..14.1439B}
{Bender} PL, {Hils} D (1997) {Confusion noise level due to galactic and
  extragalactic binaries}. Classical and Quantum Gravity 14(6):1439--1444.
  \doi{10.1088/0264-9381/14/6/008}

\bibitem[{{Bender} et~al.(2005){Bender}, {Kormendy}, {Bower}, {Green},
  {Thomas}, {Danks}, {Gull}, {Hutchings}, {Joseph}, {Kaiser}, {Lauer},
  {Nelson}, {Richstone}, {Weistrop}, and {Woodgate}}]{2005ApJ...631..280B}
{Bender} R, {Kormendy} J, {Bower} G, {Green} R, {Thomas} J, {Danks} AC, {Gull}
  T, {Hutchings} JB, {Joseph} CL, {Kaiser} ME, et~al. (2005) {HST STIS
  Spectroscopy of the Triple Nucleus of M31: Two Nested Disks in Keplerian
  Rotation around a Supermassive Black Hole}. \apj 631(1):280--300.
  \doi{10.1086/432434}.
  {\href{https://arxiv.org/abs/astro-ph/0509839}{{arXiv:astro-ph/0509839}}}
  {[astro-ph]}

\bibitem[{{Beniamini} and {Piran}(2016)}]{2016MNRAS.456.4089B}
{Beniamini} P, {Piran} T (2016) {Formation of double neutron star systems as
  implied by observations}. \mnras 456(4):4089--4099.
  \doi{10.1093/mnras/stv2903}.
  {\href{https://arxiv.org/abs/1510.03111}{{arXiv:1510.03111}}} {[astro-ph.HE]}

\bibitem[{{Benvenuto} et~al.(2014){Benvenuto}, {De Vito}, and
  {Horvath}}]{2014ApJ...786L...7B}
{Benvenuto} OG, {De Vito} MA, {Horvath} JE (2014) {Understanding the Evolution
  of Close Binary Systems with Radio Pulsars}. \apjl 786(1):L7.
  \doi{10.1088/2041-8205/786/1/L7}.
  {\href{https://arxiv.org/abs/1402.7338}{{arXiv:1402.7338}}} {[astro-ph.SR]}

\bibitem[{{Berczik} et~al.(2005){Berczik}, {Merritt}, and
  {Spurzem}}]{2005ApJ...633..680B}
{Berczik} P, {Merritt} D, {Spurzem} R (2005) {Long-Term Evolution of Massive
  Black Hole Binaries. II. Binary Evolution in Low-Density Galaxies}. \apj
  633(2):680--687. \doi{10.1086/491598}.
  {\href{https://arxiv.org/abs/astro-ph/0507260}{{arXiv:astro-ph/0507260}}}
  {[astro-ph]}

\bibitem[{{Berczik} et~al.(2006){Berczik}, {Merritt}, {Spurzem}, and
  {Bischof}}]{2006ApJ...642L..21B}
{Berczik} P, {Merritt} D, {Spurzem} R, {Bischof} HP (2006) {Efficient Merger of
  Binary Supermassive Black Holes in Nonaxisymmetric Galaxies}. \apjl
  642(1):L21--L24. \doi{10.1086/504426}.
  {\href{https://arxiv.org/abs/astro-ph/0601698}{{arXiv:astro-ph/0601698}}}
  {[astro-ph]}

\bibitem[{{Berczik} et~al.(2011){Berczik}, {Nitadori}, {Zhong}, {Spurzem},
  {Hamada}, {Wang}, {Berentzen}, {Veles}, and {Ge}}]{2011hpc..conf....8B}
{Berczik} P, {Nitadori} K, {Zhong} S, {Spurzem} R, {Hamada} T, {Wang} X,
  {Berentzen} I, {Veles} A, {Ge} W (2011) {High performance massively parallel
  direct N-body simulations on large GPU clusters.} In: International
  conference on High Performance Computing. pp 8--18

\bibitem[{{Berczik} et~al.(2013){Berczik}, {Spurzem}, {Wang}, {Zhong}, and
  {Huang}}]{2013hpc..conf...52B}
{Berczik} P, {Spurzem} R, {Wang} L, {Zhong} S, {Huang} S (2013) {Up to 700k GPU
  cores, Kepler, and the Exascale future for simulations of star clusters
  around black holes.} In: Third International Conference ''High Performance
  Computing. pp 52--59.
  {\href{https://arxiv.org/abs/1312.1789}{{arXiv:1312.1789}}} {[astro-ph.IM]}

\bibitem[{{Berrier} et~al.(2013){Berrier}, {Davis}, {Kennefick}, {Kennefick},
  {Seigar}, {Barrows}, {Hartley}, {Shields}, {Bentz}, and
  {Lacy}}]{2013ApJ...769..132B}
{Berrier} JC, {Davis} BL, {Kennefick} D, {Kennefick} JD, {Seigar} MS, {Barrows}
  RS, {Hartley} M, {Shields} D, {Bentz} MC, {Lacy} CHS (2013) {Further Evidence
  for a Supermassive Black Hole Mass-Pitch Angle Relation}. \apj 769(2):132.
  \doi{10.1088/0004-637X/769/2/132}.
  {\href{https://arxiv.org/abs/1304.4937}{{arXiv:1304.4937}}} {[astro-ph.GA]}

\bibitem[{{Berry} et~al.(2019){Berry}, {Hughes}, {Sopuerta}, {Chua},
  {Heffernan}, {Holley-Bockelmann}, {Mihaylov}, {Miller}, and
  {Sesana}}]{2019BAAS...51c..42B}
{Berry} C, {Hughes} S, {Sopuerta} C, {Chua} A, {Heffernan} A,
  {Holley-Bockelmann} K, {Mihaylov} D, {Miller} C, {Sesana} A (2019) {The
  unique potential of extreme mass-ratio inspirals for gravitational-wave
  astronomy}. \baas 51(3):42.
  {\href{https://arxiv.org/abs/1903.03686}{{arXiv:1903.03686}}} {[astro-ph.HE]}

\bibitem[{{Berry} and {Gair}(2013{\natexlab{a}})}]{2013MNRAS.435.3521B}
{Berry} CPL, {Gair} JR (2013{\natexlab{a}}) {Expectations for
  extreme-mass-ratio bursts from the Galactic Centre}. \mnras
  435(4):3521--3540. \doi{10.1093/mnras/stt1543}.
  {\href{https://arxiv.org/abs/1307.7276}{{arXiv:1307.7276}}} {[astro-ph.HE]}

\bibitem[{{Berry} and {Gair}(2013{\natexlab{b}})}]{2013MNRAS.433.3572B}
{Berry} CPL, {Gair} JR (2013{\natexlab{b}}) {Extreme-mass-ratio-bursts from
  extragalactic sources}. \mnras 433(4):3572--3583. \doi{10.1093/mnras/stt990}.
  {\href{https://arxiv.org/abs/1306.0774}{{arXiv:1306.0774}}} {[astro-ph.HE]}

\bibitem[{{Berry} and {Gair}(2013{\natexlab{c}})}]{2013MNRAS.429..589B}
{Berry} CPL, {Gair} JR (2013{\natexlab{c}}) {Observing the Galaxy's massive
  black hole with gravitational wave bursts}. \mnras 429(1):589--612.
  \doi{10.1093/mnras/sts360}.
  {\href{https://arxiv.org/abs/1210.2778}{{arXiv:1210.2778}}} {[astro-ph.HE]}

\bibitem[{{Berry} et~al.(2016){Berry}, {Cole}, {Ca{\~n}izares}, and
  {Gair}}]{2016PhRvD..94l4042B}
{Berry} CPL, {Cole} RH, {Ca{\~n}izares} P, {Gair} JR (2016) {Importance of
  transient resonances in extreme-mass-ratio inspirals}. \prd 94(12):124042.
  \doi{10.1103/PhysRevD.94.124042}.
  {\href{https://arxiv.org/abs/1608.08951}{{arXiv:1608.08951}}} {[gr-qc]}

\bibitem[{{Berti} and {Volonteri}(2008)}]{2008ApJ...684..822B}
{Berti} E, {Volonteri} M (2008) {Cosmological Black Hole Spin Evolution by
  Mergers and Accretion}. \apj 684(2):822--828. \doi{10.1086/590379}.
  {\href{https://arxiv.org/abs/0802.0025}{{arXiv:0802.0025}}} {[astro-ph]}

\bibitem[{{Berti} et~al.(2012){Berti}, {Kesden}, and
  {Sperhake}}]{2012PhRvD..85l4049B}
{Berti} E, {Kesden} M, {Sperhake} U (2012) {Effects of post-Newtonian spin
  alignment on the distribution of black-hole recoils}. \prd 85(12):124049.
  \doi{10.1103/PhysRevD.85.124049}.
  {\href{https://arxiv.org/abs/1203.2920}{{arXiv:1203.2920}}} {[astro-ph.HE]}

\bibitem[{Berti et~al.(2018)Berti, Yagi, Yang, and Yunes}]{Berti:2018vdi}
Berti E, Yagi K, Yang H, Yunes N (2018) {Extreme Gravity Tests with
  Gravitational Waves from Compact Binary Coalescences: (II) Ringdown}. Gen Rel
  Grav 50(5):49. \doi{10.1007/s10714-018-2372-6}.
  {\href{https://arxiv.org/abs/1801.03587}{{arXiv:1801.03587}}} {[gr-qc]}

\bibitem[{{Bertone} and {Merritt}(2005)}]{2005PhRvD..72j3502B}
{Bertone} G, {Merritt} D (2005) {Time-dependent models for dark matter at the
  galactic center}. \prd 72(10):103502. \doi{10.1103/PhysRevD.72.103502}.
  {\href{https://arxiv.org/abs/astro-ph/0501555}{{arXiv:astro-ph/0501555}}}
  {[astro-ph]}

\bibitem[{{Bertone} et~al.(2005){Bertone}, {Zentner}, and
  {Silk}}]{2005PhRvD..72j3517B}
{Bertone} G, {Zentner} AR, {Silk} J (2005) {New signature of dark matter
  annihilations: Gamma rays from intermediate-mass black holes}. \prd
  72(10):103517. \doi{10.1103/PhysRevD.72.103517}.
  {\href{https://arxiv.org/abs/astro-ph/0509565}{{arXiv:astro-ph/0509565}}}
  {[astro-ph]}

\bibitem[{{Beuermann} et~al.(2012){Beuermann}, {Dreizler}, {Hessman}, and
  {Deller}}]{2012A&A...543A.138B}
{Beuermann} K, {Dreizler} S, {Hessman} FV, {Deller} J (2012) {The quest for
  companions to post-common envelope binaries. III. A reexamination of
  <ASTROBJ>HW Virginis</ASTROBJ>}. \aap 543:A138.
  \doi{10.1051/0004-6361/201219391}.
  {\href{https://arxiv.org/abs/1206.3080}{{arXiv:1206.3080}}} {[astro-ph.SR]}

\bibitem[{{Bhattacharya} and {van den Heuvel}(1991)}]{1991PhR...203....1B}
{Bhattacharya} D, {van den Heuvel} EPJ (1991) {Formation and evolution of
  binary and millisecond radio pulsars}. \physrep 203(1-2):1--124.
  \doi{10.1016/0370-1573(91)90064-S}

\bibitem[{{Biava} et~al.(2019){Biava}, {Colpi}, {Capelo}, {Bonetti},
  {Volonteri}, {Tamfal}, {Mayer}, and {Sesana}}]{2019MNRAS.487.4985B}
{Biava} N, {Colpi} M, {Capelo} PR, {Bonetti} M, {Volonteri} M, {Tamfal} T,
  {Mayer} L, {Sesana} A (2019) {The lifetime of binary black holes in
  S{\'e}rsic galaxy models}. \mnras 487(4):4985--4994.
  \doi{10.1093/mnras/stz1614}.
  {\href{https://arxiv.org/abs/1903.05682}{{arXiv:1903.05682}}} {[astro-ph.GA]}

\bibitem[{{Bildsten} and {Cutler}(1992)}]{1992ApJ...400..175B}
{Bildsten} L, {Cutler} C (1992) {Tidal Interactions of Inspiraling Compact
  Binaries}. \apj 400:175. \doi{10.1086/171983}

\bibitem[{{Bildsten} et~al.(1992){Bildsten}, {Salpeter}, and
  {Wasserman}}]{1992ApJ...384..143B}
{Bildsten} L, {Salpeter} EE, {Wasserman} I (1992) {The Fate of Accreted CNO
  Elements in Neutron Star Atmospheres: X-Ray Bursts and Gamma-Ray Lines}. \apj
  384:143. \doi{10.1086/170860}

\bibitem[{{Bildsten} et~al.(2007){Bildsten}, {Shen}, {Weinberg}, and
  {Nelemans}}]{2007ApJ...662L..95B}
{Bildsten} L, {Shen} KJ, {Weinberg} NN, {Nelemans} G (2007) {Faint
  Thermonuclear Supernovae from AM Canum Venaticorum Binaries}. \apjl
  662(2):L95--L98. \doi{10.1086/519489}.
  {\href{https://arxiv.org/abs/astro-ph/0703578}{{arXiv:astro-ph/0703578}}}
  {[astro-ph]}

\bibitem[{{Binggeli} et~al.(2000){Binggeli}, {Barazza}, and
  {Jerjen}}]{2000A&A...359..447B}
{Binggeli} B, {Barazza} F, {Jerjen} H (2000) {Off-center nuclei in dwarf
  elliptical galaxies}. \aap 359:447--456

\bibitem[{{Binney} and {Tremaine}(1987)}]{1987gady.book.....B}
{Binney} J, {Tremaine} S (1987) {Galactic dynamics}. Princeton Univeristy Press

\bibitem[{Bird et~al.(2016)Bird, Cholis, Mu\~noz, Ali-Ha\"\i{}moud,
  Kamionkowski, Kovetz, Raccanelli, and Riess}]{Bird:2016dcv}
Bird S, Cholis I, Mu\~noz JB, Ali-Ha\"\i{}moud Y, Kamionkowski M, Kovetz ED,
  Raccanelli A, Riess AG (2016) {Did LIGO detect dark matter?} Phys Rev Lett
  116(20):201301. \doi{10.1103/PhysRevLett.116.201301}.
  {\href{https://arxiv.org/abs/1603.00464}{{arXiv:1603.00464}}} {[astro-ph.CO]}

\bibitem[{{Bitsch} and {Kley}(2010)}]{2010A&A...523A..30B}
{Bitsch} B, {Kley} W (2010) {Orbital evolution of eccentric planets in
  radiative discs}. \aap 523:A30. \doi{10.1051/0004-6361/201014414}.
  {\href{https://arxiv.org/abs/1008.2656}{{arXiv:1008.2656}}} {[astro-ph.EP]}

\bibitem[{{Bitsch} and {Kley}(2011)}]{2011A&A...530A..41B}
{Bitsch} B, {Kley} W (2011) {Evolution of inclined planets in three-dimensional
  radiative discs}. \aap 530:A41. \doi{10.1051/0004-6361/201016179}.
  {\href{https://arxiv.org/abs/1104.2408}{{arXiv:1104.2408}}} {[astro-ph.EP]}

\bibitem[{{Blaauw}(1961)}]{1961BAN....15..265B}
{Blaauw} A (1961) {On the origin of the O- and B-type stars with high
  velocities (the ``run-away'' stars), and some related problems}. \bain 15:265

\bibitem[{{Blaes} et~al.(2002){Blaes}, {Lee}, and
  {Socrates}}]{2002ApJ...578..775B}
{Blaes} O, {Lee} MH, {Socrates} A (2002) {The Kozai Mechanism and the Evolution
  of Binary Supermassive Black Holes}. \apj 578(2):775--786.
  \doi{10.1086/342655}.
  {\href{https://arxiv.org/abs/astro-ph/0203370}{{arXiv:astro-ph/0203370}}}
  {[astro-ph]}

\bibitem[{{Blanchet}(2014)}]{2014LRR....17....2B}
{Blanchet} L (2014) {Gravitational Radiation from Post-Newtonian Sources and
  Inspiralling Compact Binaries}. Living Reviews in Relativity 17(1):2.
  \doi{10.12942/lrr-2014-2}.
  {\href{https://arxiv.org/abs/1310.1528}{{arXiv:1310.1528}}} {[gr-qc]}

\bibitem[{{Blanchet}(2019)}]{2019IJMPD..2830011B}
{Blanchet} L (2019) {Analytic approximations in GR and gravitational waves}.
  International Journal of Modern Physics D 28(6):1930011-144.
  \doi{10.1142/S0218271819300118}.
  {\href{https://arxiv.org/abs/1812.07490}{{arXiv:1812.07490}}} {[gr-qc]}

\bibitem[{{Blanchet} et~al.(2005){Blanchet}, {Qusailah}, and
  {Will}}]{2005ApJ...635..508B}
{Blanchet} L, {Qusailah} MSS, {Will} CM (2005) {Gravitational Recoil of
  Inspiraling Black Hole Binaries to Second Post-Newtonian Order}. \apj
  635(1):508--515. \doi{10.1086/497332}.
  {\href{https://arxiv.org/abs/astro-ph/0507692}{{arXiv:astro-ph/0507692}}}
  {[astro-ph]}

\bibitem[{{Blandford} and {Znajek}(1977)}]{1977MNRAS.179..433B}
{Blandford} RD, {Znajek} RL (1977) {Electromagnetic extraction of energy from
  Kerr black holes.} \mnras 179:433--456. \doi{10.1093/mnras/179.3.433}

\bibitem[{{Blecha} and {Loeb}(2008)}]{2008MNRAS.390.1311B}
{Blecha} L, {Loeb} A (2008) {Effects of gravitational-wave recoil on the
  dynamics and growth of supermassive black holes}. \mnras 390(4):1311--1325.
  \doi{10.1111/j.1365-2966.2008.13790.x}.
  {\href{https://arxiv.org/abs/0805.1420}{{arXiv:0805.1420}}} {[astro-ph]}

\bibitem[{{Blecha} et~al.(2011){Blecha}, {Cox}, {Loeb}, and
  {Hernquist}}]{2011MNRAS.412.2154B}
{Blecha} L, {Cox} TJ, {Loeb} A, {Hernquist} L (2011) {Recoiling black holes in
  merging galaxies: relationship to active galactic nucleus lifetimes,
  starbursts and the M$_{BH}$-{\ensuremath{\sigma}}$_{*}$ relation}. \mnras
  412(4):2154--2182. \doi{10.1111/j.1365-2966.2010.18042.x}.
  {\href{https://arxiv.org/abs/1009.4940}{{arXiv:1009.4940}}} {[astro-ph.CO]}

\bibitem[{{Blecha} et~al.(2016){Blecha}, {Sijacki}, {Kelley}, {Torrey},
  {Vogelsberger}, {Nelson}, {Springel}, {Snyder}, and
  {Hernquist}}]{2016MNRAS.456..961B}
{Blecha} L, {Sijacki} D, {Kelley} LZ, {Torrey} P, {Vogelsberger} M, {Nelson} D,
  {Springel} V, {Snyder} G, {Hernquist} L (2016) {Recoiling black holes:
  prospects for detection and implications of spin alignment}. \mnras
  456(1):961--989. \doi{10.1093/mnras/stv2646}.
  {\href{https://arxiv.org/abs/1508.01524}{{arXiv:1508.01524}}} {[astro-ph.GA]}

\bibitem[{{Blelly} et~al.(2022){Blelly}, {Bobin}, and
  {Moutarde}}]{2022MNRAS.509.5902B}
{Blelly} A, {Bobin} J, {Moutarde} H (2022) {Sparse data inpainting for the
  recovery of Galactic-binary gravitational wave signals from gapped data}.
  \mnras 509(4):5902--5917. \doi{10.1093/mnras/stab3314}.
  {\href{https://arxiv.org/abs/2104.05250}{{arXiv:2104.05250}}} {[gr-qc]}

\bibitem[{{Bloom} et~al.(1999){Bloom}, {Sigurdsson}, and
  {Pols}}]{1999MNRAS.305..763B}
{Bloom} JS, {Sigurdsson} S, {Pols} OR (1999) {The spatial distribution of
  coalescing neutron star binaries: implications for gamma-ray bursts}. \mnras
  305(4):763--769. \doi{10.1046/j.1365-8711.1999.02437.x}.
  {\href{https://arxiv.org/abs/astro-ph/9805222}{{arXiv:astro-ph/9805222}}}
  {[astro-ph]}

\bibitem[{{Bobrick} et~al.(2017){Bobrick}, {Davies}, and
  {Church}}]{2017MNRAS.467.3556B}
{Bobrick} A, {Davies} MB, {Church} RP (2017) {Mass transfer in white
  dwarf-neutron star binaries}. \mnras 467(3):3556--3575.
  \doi{10.1093/mnras/stx312}.
  {\href{https://arxiv.org/abs/1702.02377}{{arXiv:1702.02377}}} {[astro-ph.HE]}

\bibitem[{{Boco} et~al.(2019){Boco}, {Lapi}, {Goswami}, {Perrotta},
  {Baccigalupi}, and {Danese}}]{2019ApJ...881..157B}
{Boco} L, {Lapi} A, {Goswami} S, {Perrotta} F, {Baccigalupi} C, {Danese} L
  (2019) {Merging Rates of Compact Binaries in Galaxies: Perspectives for
  Gravitational Wave Detections}. \apj 881(2):157.
  \doi{10.3847/1538-4357/ab328e}.
  {\href{https://arxiv.org/abs/1907.06841}{{arXiv:1907.06841}}} {[astro-ph.GA]}

\bibitem[{{Boekholt} et~al.(2018){Boekholt}, {Schleicher}, {Fellhauer},
  {Klessen}, {Reinoso}, {Stutz}, and {Haemmerl{\'e}}}]{2018MNRAS.476..366B}
{Boekholt} TCN, {Schleicher} DRG, {Fellhauer} M, {Klessen} RS, {Reinoso} B,
  {Stutz} AM, {Haemmerl{\'e}} L (2018) {Formation of massive seed black holes
  via collisions and accretion}. \mnras 476(1):366--380.
  \doi{10.1093/mnras/sty208}.
  {\href{https://arxiv.org/abs/1801.05841}{{arXiv:1801.05841}}} {[astro-ph.GA]}

\bibitem[{{Bogd{\'a}n} et~al.(2012){Bogd{\'a}n}, {Forman}, {Zhuravleva},
  {Mihos}, {Kraft}, {Harding}, {Guo}, {Li}, {Churazov}, {Vikhlinin}, {Nulsen},
  {Schindler}, and {Jones}}]{2012ApJ...753..140B}
{Bogd{\'a}n} {\'A}, {Forman} WR, {Zhuravleva} I, {Mihos} JC, {Kraft} RP,
  {Harding} P, {Guo} Q, {Li} Z, {Churazov} E, {Vikhlinin} A, et~al. (2012)
  {Exploring the Unusually High Black-hole-to-bulge Mass Ratios in NGC 4342 and
  NGC 4291: The Asynchronous Growth of Bulges and Black Holes}. \apj
  753(2):140. \doi{10.1088/0004-637X/753/2/140}.
  {\href{https://arxiv.org/abs/1203.1641}{{arXiv:1203.1641}}} {[astro-ph.CO]}

\bibitem[{{Bogdanovi{\'c}} et~al.(2007){Bogdanovi{\'c}}, {Reynolds}, and
  {Miller}}]{2007ApJ...661L.147B}
{Bogdanovi{\'c}} T, {Reynolds} CS, {Miller} MC (2007) {Alignment of the Spins
  of Supermassive Black Holes Prior to Coalescence}. \apjl 661(2):L147--L150.
  \doi{10.1086/518769}.
  {\href{https://arxiv.org/abs/astro-ph/0703054}{{arXiv:astro-ph/0703054}}}
  {[astro-ph]}

\bibitem[{{Bon} et~al.(2016){Bon}, {Zucker}, {Netzer}, {Marziani}, {Bon},
  {Jovanovi{\'c}}, {Shapovalova}, {Komossa}, {Gaskell}, {Popovi{\'c}},
  {Britzen}, {Chavushyan}, {Burenkov}, {Sergeev}, {La Mura}, {Vald{\'e}s}, and
  {Stalevski}}]{2016ApJS..225...29B}
{Bon} E, {Zucker} S, {Netzer} H, {Marziani} P, {Bon} N, {Jovanovi{\'c}} P,
  {Shapovalova} AI, {Komossa} S, {Gaskell} CM, {Popovi{\'c}} L{\v{C}}, et~al.
  (2016) {Evidence for Periodicity in 43 year-long Monitoring of NGC 5548}.
  \apjs 225(2):29. \doi{10.3847/0067-0049/225/2/29}.
  {\href{https://arxiv.org/abs/1606.04606}{{arXiv:1606.04606}}} {[astro-ph.HE]}

\bibitem[{{Bondi}(1952)}]{1952MNRAS.112..195B}
{Bondi} H (1952) {On spherically symmetrical accretion}. \mnras 112:195.
  \doi{10.1093/mnras/112.2.195}

\bibitem[{{Bondi} and {Hoyle}(1944)}]{1944MNRAS.104..273B}
{Bondi} H, {Hoyle} F (1944) {On the mechanism of accretion by stars}. \mnras
  104:273. \doi{10.1093/mnras/104.5.273}

\bibitem[{{Bonetti} and {Sesana}(2020)}]{2020arXiv200714403B}
{Bonetti} M, {Sesana} A (2020) {Gravitational wave background from extreme mass
  ratio inspirals}. \prd 102(10):103023. \doi{10.1103/PhysRevD.102.103023}.
  {\href{https://arxiv.org/abs/2007.14403}{{arXiv:2007.14403}}} {[astro-ph.GA]}

\bibitem[{{Bonetti} et~al.(2018){Bonetti}, {Haardt}, {Sesana}, and
  {Barausse}}]{2018MNRAS.477.3910B}
{Bonetti} M, {Haardt} F, {Sesana} A, {Barausse} E (2018) {Post-Newtonian
  evolution of massive black hole triplets in galactic nuclei - II. Survey of
  the parameter space}. \mnras 477(3):3910--3926. \doi{10.1093/mnras/sty896}.
  {\href{https://arxiv.org/abs/1709.06088}{{arXiv:1709.06088}}} {[astro-ph.GA]}

\bibitem[{{Bonetti} et~al.(2019){Bonetti}, {Sesana}, {Haardt}, {Barausse}, and
  {Colpi}}]{2019MNRAS.486.4044B}
{Bonetti} M, {Sesana} A, {Haardt} F, {Barausse} E, {Colpi} M (2019)
  {Post-Newtonian evolution of massive black hole triplets in galactic nuclei -
  IV. Implications for LISA}. \mnras 486(3):4044--4060.
  \doi{10.1093/mnras/stz903}.
  {\href{https://arxiv.org/abs/1812.01011}{{arXiv:1812.01011}}} {[astro-ph.GA]}

\bibitem[{{Bonetti} et~al.(2020{\natexlab{a}}){Bonetti}, {Bortolas}, {Lupi},
  {Dotti}, and {Raimundo}}]{2020MNRAS.494.3053B}
{Bonetti} M, {Bortolas} E, {Lupi} A, {Dotti} M, {Raimundo} SI
  (2020{\natexlab{a}}) {Dynamical friction-driven orbital circularization in
  rotating discs: a semi-analytical description}. \mnras 494(2):3053--3059.
  \doi{10.1093/mnras/staa964}.
  {\href{https://arxiv.org/abs/2002.04621}{{arXiv:2002.04621}}} {[astro-ph.GA]}

\bibitem[{{Bonetti} et~al.(2020{\natexlab{b}}){Bonetti}, {Rasskazov}, {Sesana},
  {Dotti}, {Haardt}, {Leigh}, {Arca Sedda}, {Fragione}, and
  {Rossi}}]{2020MNRAS.493L.114B}
{Bonetti} M, {Rasskazov} A, {Sesana} A, {Dotti} M, {Haardt} F, {Leigh} NWC,
  {Arca Sedda} M, {Fragione} G, {Rossi} E (2020{\natexlab{b}}) {On the
  eccentricity evolution of massive black hole binaries in stellar
  backgrounds}. \mnras 493(1):L114--L119. \doi{10.1093/mnrasl/slaa018}.
  {\href{https://arxiv.org/abs/2001.02231}{{arXiv:2001.02231}}} {[astro-ph.GA]}

\bibitem[{{Bonetti} et~al.(2021){Bonetti}, {Bortolas}, {Lupi}, and
  {Dotti}}]{2021MNRAS.502.3554B}
{Bonetti} M, {Bortolas} E, {Lupi} A, {Dotti} M (2021) {Dynamical evolution of
  massive perturbers in realistic multicomponent galaxy models I:
  implementation and validation}. \mnras 502(3):3554--3568.
  \doi{10.1093/mnras/stab222}.
  {\href{https://arxiv.org/abs/2010.08555}{{arXiv:2010.08555}}} {[astro-ph.GA]}

\bibitem[{{Bonfini} et~al.(2018){Bonfini}, {Bitsakis}, {Zezas}, {Duc},
  {Iodice}, {Gonz{\'a}lez-Mart{\'\i}n}, {Bruzual}, and {Gonz{\'a}lez
  Sanoja}}]{2018MNRAS.473L..94B}
{Bonfini} P, {Bitsakis} T, {Zezas} A, {Duc} PA, {Iodice} E,
  {Gonz{\'a}lez-Mart{\'\i}n} O, {Bruzual} G, {Gonz{\'a}lez Sanoja} AJ (2018)
  {Connecting traces of galaxy evolution: the missing core mass-morphological
  fine structure relation}. \mnras 473(1):L94--L100.
  \doi{10.1093/mnrasl/slx169}.
  {\href{https://arxiv.org/abs/1710.05025}{{arXiv:1710.05025}}} {[astro-ph.GA]}

\bibitem[{{Bonga} et~al.(2019){Bonga}, {Yang}, and
  {Hughes}}]{2019PhRvL.123j1103B}
{Bonga} B, {Yang} H, {Hughes} SA (2019) {Tidal Resonance in Extreme Mass-Ratio
  Inspirals}. \prl 123(10):101103. \doi{10.1103/PhysRevLett.123.101103}.
  {\href{https://arxiv.org/abs/1905.00030}{{arXiv:1905.00030}}} {[gr-qc]}

\bibitem[{{Bonnor} and {Rotenberg}(1961)}]{1961RSPSA.265..109B}
{Bonnor} WB, {Rotenberg} MA (1961) {Transport of Momentum by Gravitational
  Waves: The Linear Approximation}. Proceedings of the Royal Society of London
  Series A 265(1320):109--116. \doi{10.1098/rspa.1961.0226}

\bibitem[{{Bonoli} et~al.(2014){Bonoli}, {Mayer}, and
  {Callegari}}]{2014MNRAS.437.1576B}
{Bonoli} S, {Mayer} L, {Callegari} S (2014) {Massive black hole seeds born via
  direct gas collapse in galaxy mergers: their properties, statistics and
  environment}. \mnras 437(2):1576--1592. \doi{10.1093/mnras/stt1990}.
  {\href{https://arxiv.org/abs/1211.3752}{{arXiv:1211.3752}}} {[astro-ph.CO]}

\bibitem[{{Bonoli} et~al.(2016){Bonoli}, {Mayer}, {Kazantzidis}, {Madau},
  {Bellovary}, and {Governato}}]{2016MNRAS.459.2603B}
{Bonoli} S, {Mayer} L, {Kazantzidis} S, {Madau} P, {Bellovary} J, {Governato} F
  (2016) {Black hole starvation and bulge evolution in a Milky Way-like
  galaxy}. \mnras 459(3):2603--2617. \doi{10.1093/mnras/stw694}.
  {\href{https://arxiv.org/abs/1508.07328}{{arXiv:1508.07328}}} {[astro-ph.GA]}

\bibitem[{{Bortolas} and {Mapelli}(2019)}]{2019MNRAS.485.2125B}
{Bortolas} E, {Mapelli} M (2019) {Can supernova kicks trigger EMRIs in the
  Galactic Centre?} \mnras 485(2):2125--2138. \doi{10.1093/mnras/stz440}.
  {\href{https://arxiv.org/abs/1902.04581}{{arXiv:1902.04581}}} {[astro-ph.GA]}

\bibitem[{{Bortolas} et~al.(2016){Bortolas}, {Gualandris}, {Dotti}, {Spera},
  and {Mapelli}}]{2016MNRAS.461.1023B}
{Bortolas} E, {Gualandris} A, {Dotti} M, {Spera} M, {Mapelli} M (2016)
  {Brownian motion of massive black hole binaries and the final parsec
  problem}. \mnras 461(1):1023--1031. \doi{10.1093/mnras/stw1372}.
  {\href{https://arxiv.org/abs/1606.06728}{{arXiv:1606.06728}}} {[astro-ph.GA]}

\bibitem[{{Bortolas} et~al.(2018{\natexlab{a}}){Bortolas}, {Gualandris},
  {Dotti}, and {Read}}]{2018MNRAS.477.2310B}
{Bortolas} E, {Gualandris} A, {Dotti} M, {Read} JI (2018{\natexlab{a}}) {The
  influence of massive black hole binaries on the morphology of merger
  remnants}. \mnras 477(2):2310--2325. \doi{10.1093/mnras/sty775}.
  {\href{https://arxiv.org/abs/1710.04658}{{arXiv:1710.04658}}} {[astro-ph.GA]}

\bibitem[{{Bortolas} et~al.(2018{\natexlab{b}}){Bortolas}, {Mapelli}, and
  {Spera}}]{2018MNRAS.474.1054B}
{Bortolas} E, {Mapelli} M, {Spera} M (2018{\natexlab{b}}) {Star cluster
  disruption by a massive black hole binary}. \mnras 474(1):1054--1064.
  \doi{10.1093/mnras/stx2795}.
  {\href{https://arxiv.org/abs/1710.09418}{{arXiv:1710.09418}}} {[astro-ph.GA]}

\bibitem[{{Bortolas} et~al.(2020){Bortolas}, {Capelo}, {Zana}, {Mayer},
  {Bonetti}, {Dotti}, {Davies}, and {Madau}}]{2020MNRAS.498.3601B}
{Bortolas} E, {Capelo} PR, {Zana} T, {Mayer} L, {Bonetti} M, {Dotti} M,
  {Davies} MB, {Madau} P (2020) {Global torques and stochasticity as the
  drivers of massive black hole pairing in the young Universe}. \mnras
  498(3):3601--3615. \doi{10.1093/mnras/staa2628}.
  {\href{https://arxiv.org/abs/2005.02409}{{arXiv:2005.02409}}} {[astro-ph.GA]}

\bibitem[{{Bortolas} et~al.(2021){Bortolas}, {Franchini}, {Bonetti}, and
  {Sesana}}]{2021ApJ...918L..15B}
{Bortolas} E, {Franchini} A, {Bonetti} M, {Sesana} A (2021) {The Competing
  Effect of Gas and Stars in the Evolution of Massive Black Hole Binaries}.
  \apjl 918(1):L15. \doi{10.3847/2041-8213/ac1c0c}.
  {\href{https://arxiv.org/abs/2108.13436}{{arXiv:2108.13436}}} {[astro-ph.HE]}

\bibitem[{{Bortolas} et~al.(2022){Bortolas}, {Bonetti}, {Dotti}, {Lupi},
  {Capelo}, {Mayer}, and {Sesana}}]{2022MNRAS.512.3365B}
{Bortolas} E, {Bonetti} M, {Dotti} M, {Lupi} A, {Capelo} PR, {Mayer} L,
  {Sesana} A (2022) {The role of bars on the dynamical-friction-driven inspiral
  of massive objects}. \mnras 512(3):3365--3382. \doi{10.1093/mnras/stac645}.
  {\href{https://arxiv.org/abs/2103.07486}{{arXiv:2103.07486}}} {[astro-ph.GA]}

\bibitem[{{Bourne} and {Sijacki}(2017)}]{2017MNRAS.472.4707B}
{Bourne} MA, {Sijacki} D (2017) {AGN jet feedback on a moving mesh: cocoon
  inflation, gas flows and turbulence}. \mnras 472(4):4707--4735.
  \doi{10.1093/mnras/stx2269}.
  {\href{https://arxiv.org/abs/1705.07900}{{arXiv:1705.07900}}} {[astro-ph.GA]}

\bibitem[{{Bowen} et~al.(2018){Bowen}, {Mewes}, {Campanelli}, {Noble},
  {Krolik}, and {Zilh{\~a}o}}]{2018ApJ...853L..17B}
{Bowen} DB, {Mewes} V, {Campanelli} M, {Noble} SC, {Krolik} JH, {Zilh{\~a}o} M
  (2018) {Quasi-periodic Behavior of Mini-disks in Binary Black Holes
  Approaching Merger}. \apjl 853(1):L17. \doi{10.3847/2041-8213/aaa756}.
  {\href{https://arxiv.org/abs/1712.05451}{{arXiv:1712.05451}}} {[astro-ph.HE]}

\bibitem[{{Bowen} et~al.(2019){Bowen}, {Mewes}, {Noble}, {Avara}, {Campanelli},
  and {Krolik}}]{2019ApJ...879...76B}
{Bowen} DB, {Mewes} V, {Noble} SC, {Avara} M, {Campanelli} M, {Krolik} JH
  (2019) {Quasi-periodicity of Supermassive Binary Black Hole Accretion
  Approaching Merger}. \apj 879(2):76. \doi{10.3847/1538-4357/ab2453}.
  {\href{https://arxiv.org/abs/1904.12048}{{arXiv:1904.12048}}} {[astro-ph.HE]}

\bibitem[{Bowen et~al.(2017)}]{2017ApJ...838...42B}
Bowen DB, et~al. (2017) Relativistic dynamics and mass exchange in binary black
  hole mini-disks. ApJ 838(1):42. \doi{10.3847/1538-4357/aa63f3},
  \urlprefix\url{https://doi.org/10.3847\%2F1538-4357\%2Faa63f3}

\bibitem[{{Bower} et~al.(2006){Bower}, {Benson}, {Malbon}, {Helly}, {Frenk},
  {Baugh}, {Cole}, and {Lacey}}]{2006MNRAS.370..645B}
{Bower} RG, {Benson} AJ, {Malbon} R, {Helly} JC, {Frenk} CS, {Baugh} CM, {Cole}
  S, {Lacey} CG (2006) {Breaking the hierarchy of galaxy formation}. \mnras
  370(2):645--655. \doi{10.1111/j.1365-2966.2006.10519.x}.
  {\href{https://arxiv.org/abs/astro-ph/0511338}{{arXiv:astro-ph/0511338}}}
  {[astro-ph]}

\bibitem[{{Boylan-Kolchin} et~al.(2004){Boylan-Kolchin}, {Ma}, and
  {Quataert}}]{2004ApJ...613L..37B}
{Boylan-Kolchin} M, {Ma} CP, {Quataert} E (2004) {Core Formation in Galactic
  Nuclei due to Recoiling Black Holes}. \apjl 613(1):L37--L40.
  \doi{10.1086/425073}.
  {\href{https://arxiv.org/abs/astro-ph/0407488}{{arXiv:astro-ph/0407488}}}
  {[astro-ph]}

\bibitem[{{Bray} and {Eldridge}(2016)}]{2016MNRAS.461.3747B}
{Bray} JC, {Eldridge} JJ (2016) {Neutron star kicks and their relationship to
  supernovae ejecta mass}. \mnras 461(4):3747--3759.
  \doi{10.1093/mnras/stw1275}.
  {\href{https://arxiv.org/abs/1605.09529}{{arXiv:1605.09529}}} {[astro-ph.HE]}

\bibitem[{{Breedt} et~al.(2017){Breedt}, {Steeghs}, {Marsh}, {Gentile Fusillo},
  {Tremblay}, {Green}, {De Pasquale}, {Hermes}, {G{\"a}nsicke}, {Parsons},
  {Bours}, {Longa-Pe{\~n}a}, and {Rebassa-Mansergas}}]{2017MNRAS.468.2910B}
{Breedt} E, {Steeghs} D, {Marsh} TR, {Gentile Fusillo} NP, {Tremblay} PE,
  {Green} M, {De Pasquale} S, {Hermes} JJ, {G{\"a}nsicke} BT, {Parsons} SG,
  et~al. (2017) {Using large spectroscopic surveys to test the double
  degenerate model for Type Ia supernovae}. \mnras 468(3):2910--2922.
  \doi{10.1093/mnras/stx430}.
  {\href{https://arxiv.org/abs/1702.05117}{{arXiv:1702.05117}}} {[astro-ph.SR]}

\bibitem[{{Breen} and {Heggie}(2013)}]{2013MNRAS.432.2779B}
{Breen} PG, {Heggie} DC (2013) {Dynamical evolution of black hole subsystems in
  idealized star clusters}. \mnras 432(4):2779--2797.
  \doi{10.1093/mnras/stt628}.
  {\href{https://arxiv.org/abs/1304.3401}{{arXiv:1304.3401}}} {[astro-ph.GA]}

\bibitem[{{Bregman} and {Alexander}(2012)}]{2012ApJ...748...63B}
{Bregman} M, {Alexander} T (2012) {The Torquing of Circumnuclear Accretion
  Disks by Stars and the Evolution of Massive Black Holes}. \apj 748(1):63.
  \doi{10.1088/0004-637X/748/1/63}.
  {\href{https://arxiv.org/abs/1109.5384}{{arXiv:1109.5384}}} {[astro-ph.GA]}

\bibitem[{{Breivik} et~al.(2016){Breivik}, {Rodriguez}, {Larson}, {Kalogera},
  and {Rasio}}]{2016ApJ...830L..18B}
{Breivik} K, {Rodriguez} CL, {Larson} SL, {Kalogera} V, {Rasio} FA (2016)
  {Distinguishing between Formation Channels for Binary Black Holes with LISA}.
  \apjl 830(1):L18. \doi{10.3847/2041-8205/830/1/L18}.
  {\href{https://arxiv.org/abs/1606.09558}{{arXiv:1606.09558}}} {[astro-ph.GA]}

\bibitem[{{Breivik} et~al.(2017){Breivik}, {Chatterjee}, and
  {Larson}}]{2017ApJ...850L..13B}
{Breivik} K, {Chatterjee} S, {Larson} SL (2017) {Revealing Black Holes with
  Gaia}. \apjl 850(1):L13. \doi{10.3847/2041-8213/aa97d5}.
  {\href{https://arxiv.org/abs/1710.04657}{{arXiv:1710.04657}}} {[astro-ph.SR]}

\bibitem[{{Breivik} et~al.(2018){Breivik}, {Kremer}, {Bueno}, {Larson},
  {Coughlin}, and {Kalogera}}]{2018ApJ...854L...1B}
{Breivik} K, {Kremer} K, {Bueno} M, {Larson} SL, {Coughlin} S, {Kalogera} V
  (2018) {Characterizing Accreting Double White Dwarf Binaries with the Laser
  Interferometer Space Antenna and Gaia}. \apjl 854(1):L1.
  \doi{10.3847/2041-8213/aaaa23}.
  {\href{https://arxiv.org/abs/1710.08370}{{arXiv:1710.08370}}} {[astro-ph.SR]}

\bibitem[{{Breivik} et~al.(2020{\natexlab{a}}){Breivik}, {Coughlin}, {Zevin},
  {Rodriguez}, {Kremer}, {Ye}, {Andrews}, {Kurkowski}, {Digman}, {Larson}, and
  {Rasio}}]{2020ApJ...898...71B}
{Breivik} K, {Coughlin} S, {Zevin} M, {Rodriguez} CL, {Kremer} K, {Ye} CS,
  {Andrews} JJ, {Kurkowski} M, {Digman} MC, {Larson} SL, et~al.
  (2020{\natexlab{a}}) {COSMIC Variance in Binary Population Synthesis}. \apj
  898(1):71. \doi{10.3847/1538-4357/ab9d85}.
  {\href{https://arxiv.org/abs/1911.00903}{{arXiv:1911.00903}}} {[astro-ph.HE]}

\bibitem[{{Breivik} et~al.(2020{\natexlab{b}}){Breivik}, {Mingarelli}, and
  {Larson}}]{2020ApJ...901....4B}
{Breivik} K, {Mingarelli} CMF, {Larson} SL (2020{\natexlab{b}}) {Constraining
  Galactic Structure with the LISA White Dwarf Foreground}. \apj 901(1):4.
  \doi{10.3847/1538-4357/abab99}.
  {\href{https://arxiv.org/abs/1912.02200}{{arXiv:1912.02200}}} {[astro-ph.GA]}

\bibitem[{{Brem} et~al.(2013){Brem}, {Amaro-Seoane}, and
  {Spurzem}}]{2013MNRAS.434.2999B}
{Brem} P, {Amaro-Seoane} P, {Spurzem} R (2013) {Relativistic mergers of compact
  binaries in clusters: the fingerprint of the spin}. \mnras 434(4):2999--3007.
  \doi{10.1093/mnras/stt1220}.
  {\href{https://arxiv.org/abs/1302.3135}{{arXiv:1302.3135}}} {[astro-ph.CO]}

\bibitem[{{Brem} et~al.(2014){Brem}, {Amaro-Seoane}, and
  {Sopuerta}}]{2014MNRAS.437.1259B}
{Brem} P, {Amaro-Seoane} P, {Sopuerta} CF (2014) {Blocking low-eccentricity
  EMRIs: a statistical direct-summation N-body study of the Schwarzschild
  barrier}. \mnras 437(2):1259--1267. \doi{10.1093/mnras/stt1948}.
  {\href{https://arxiv.org/abs/1211.5601}{{arXiv:1211.5601}}} {[astro-ph.CO]}

\bibitem[{{Brenneman} and {Reynolds}(2006)}]{2006ApJ...652.1028B}
{Brenneman} LW, {Reynolds} CS (2006) {Constraining Black Hole Spin via X-Ray
  Spectroscopy}. \apj 652(2):1028--1043. \doi{10.1086/508146}.
  {\href{https://arxiv.org/abs/astro-ph/0608502}{{arXiv:astro-ph/0608502}}}
  {[astro-ph]}

\bibitem[{{Brenneman} et~al.(2011){Brenneman}, {Reynolds}, {Nowak}, {Reis},
  {Trippe}, {Fabian}, {Iwasawa}, {Lee}, {Miller}, {Mushotzky}, {Nand ra}, and
  {Volonteri}}]{2011ApJ...736..103B}
{Brenneman} LW, {Reynolds} CS, {Nowak} MA, {Reis} RC, {Trippe} M, {Fabian} AC,
  {Iwasawa} K, {Lee} JC, {Miller} JM, {Mushotzky} RF, et~al. (2011) {The Spin
  of the Supermassive Black Hole in NGC 3783}. \apj 736(2):103.
  \doi{10.1088/0004-637X/736/2/103}.
  {\href{https://arxiv.org/abs/1104.1172}{{arXiv:1104.1172}}} {[astro-ph.HE]}

\bibitem[{{Brough} et~al.(2020){Brough}, {Collins}, {Demarco}, {Ferguson},
  {Galaz}, {Holwerda}, {Martinez-Lombilla}, {Mihos}, and
  {Montes}}]{2020arXiv200111067B}
{Brough} S, {Collins} C, {Demarco} R, {Ferguson} HC, {Galaz} G, {Holwerda} B,
  {Martinez-Lombilla} C, {Mihos} C, {Montes} M (2020) {The Vera Rubin
  Observatory Legacy Survey of Space and Time and the Low Surface Brightness
  Universe}. arXiv e-prints arXiv:2001.11067.
  {\href{https://arxiv.org/abs/2001.11067}{{arXiv:2001.11067}}} {[astro-ph.GA]}

\bibitem[{{Brown} et~al.(2010){Brown}, {Kilic}, {Allende Prieto}, and
  {Kenyon}}]{2010ApJ...723.1072B}
{Brown} WR, {Kilic} M, {Allende Prieto} C, {Kenyon} SJ (2010) {The ELM Survey.
  I. A Complete Sample of Extremely Low-mass White Dwarfs}. \apj
  723(2):1072--1081. \doi{10.1088/0004-637X/723/2/1072}.
  {\href{https://arxiv.org/abs/1011.3050}{{arXiv:1011.3050}}} {[astro-ph.GA]}

\bibitem[{{Brown} et~al.(2011){Brown}, {Kilic}, {Hermes}, {Allende Prieto},
  {Kenyon}, and {Winget}}]{2011ApJ...737L..23B}
{Brown} WR, {Kilic} M, {Hermes} JJ, {Allende Prieto} C, {Kenyon} SJ, {Winget}
  DE (2011) {A 12 Minute Orbital Period Detached White Dwarf Eclipsing Binary}.
  \apjl 737(1):L23. \doi{10.1088/2041-8205/737/1/L23}.
  {\href{https://arxiv.org/abs/1107.2389}{{arXiv:1107.2389}}} {[astro-ph.GA]}

\bibitem[{{Brown} et~al.(2016){Brown}, {Kilic}, {Kenyon}, and
  {Gianninas}}]{2016ApJ...824...46B}
{Brown} WR, {Kilic} M, {Kenyon} SJ, {Gianninas} A (2016) {Most Double
  Degenerate Low-mass White Dwarf Binaries Merge}. \apj 824(1):46.
  \doi{10.3847/0004-637X/824/1/46}.
  {\href{https://arxiv.org/abs/1604.04269}{{arXiv:1604.04269}}} {[astro-ph.SR]}

\bibitem[{{Brown} et~al.(2020{\natexlab{a}}){Brown}, {Kilic}, {B{\'e}dard},
  {Kosakowski}, and {Bergeron}}]{2020ApJ...892L..35B}
{Brown} WR, {Kilic} M, {B{\'e}dard} A, {Kosakowski} A, {Bergeron} P
  (2020{\natexlab{a}}) {A 1201 s Orbital Period Detached Binary: The First
  Double Helium Core White Dwarf LISA Verification Binary}. \apjl 892(2):L35.
  \doi{10.3847/2041-8213/ab8228}.
  {\href{https://arxiv.org/abs/2004.00641}{{arXiv:2004.00641}}} {[astro-ph.SR]}

\bibitem[{{Brown} et~al.(2020{\natexlab{b}}){Brown}, {Kilic}, {Kosakowski},
  {Andrews}, {Heinke}, {Ag{\"u}eros}, {Camilo}, {Gianninas}, {Hermes}, and
  {Kenyon}}]{2020ApJ...889...49B}
{Brown} WR, {Kilic} M, {Kosakowski} A, {Andrews} JJ, {Heinke} CO, {Ag{\"u}eros}
  MA, {Camilo} F, {Gianninas} A, {Hermes} JJ, {Kenyon} SJ (2020{\natexlab{b}})
  {The ELM Survey. VIII. Ninety-eight Double White Dwarf Binaries}. \apj
  889(1):49. \doi{10.3847/1538-4357/ab63cd}.
  {\href{https://arxiv.org/abs/2002.00064}{{arXiv:2002.00064}}} {[astro-ph.SR]}

\bibitem[{{Br{\"u}gmann} et~al.(2008){Br{\"u}gmann}, {Gonz{\'a}lez}, {Hannam},
  {Husa}, and {Sperhake}}]{2008PhRvD..77l4047B}
{Br{\"u}gmann} B, {Gonz{\'a}lez} JA, {Hannam} M, {Husa} S, {Sperhake} U (2008)
  {Exploring black hole superkicks}. \prd 77(12):124047.
  \doi{10.1103/PhysRevD.77.124047}.
  {\href{https://arxiv.org/abs/0707.0135}{{arXiv:0707.0135}}} {[gr-qc]}

\bibitem[{{B{\"u}ning} and {Ritter}(2004)}]{2004A&A...423..281B}
{B{\"u}ning} A, {Ritter} H (2004) {Long-term evolution of compact binaries with
  irradiation feedback}. \aap 423:281--299. \doi{10.1051/0004-6361:20035678}.
  {\href{https://arxiv.org/abs/astro-ph/0403306}{{arXiv:astro-ph/0403306}}}
  {[astro-ph]}

\bibitem[{{Buonanno} and {Damour}(1999)}]{1999PhRvD..59h4006B}
{Buonanno} A, {Damour} T (1999) {Effective one-body approach to general
  relativistic two-body dynamics}. \prd 59(8):084006.
  \doi{10.1103/PhysRevD.59.084006}.
  {\href{https://arxiv.org/abs/gr-qc/9811091}{{arXiv:gr-qc/9811091}}} {[gr-qc]}

\bibitem[{{Buonanno} and {Damour}(2000)}]{2000PhRvD..62f4015B}
{Buonanno} A, {Damour} T (2000) {Transition from inspiral to plunge in binary
  black hole coalescences}. \prd 62(6):064015.
  \doi{10.1103/PhysRevD.62.064015}.
  {\href{https://arxiv.org/abs/gr-qc/0001013}{{arXiv:gr-qc/0001013}}} {[gr-qc]}

\bibitem[{{Buonanno} et~al.(2009){Buonanno}, {Iyer}, {Ochsner}, {Pan}, and
  {Sathyaprakash}}]{2009PhRvD..80h4043B}
{Buonanno} A, {Iyer} BR, {Ochsner} E, {Pan} Y, {Sathyaprakash} BS (2009)
  {Comparison of post-Newtonian templates for compact binary inspiral signals
  in gravitational-wave detectors}. \prd 80(8):084043.
  \doi{10.1103/PhysRevD.80.084043}.
  {\href{https://arxiv.org/abs/0907.0700}{{arXiv:0907.0700}}} {[gr-qc]}

\bibitem[{{Burdge} et~al.(2019{\natexlab{a}}){Burdge}, {Coughlin}, {Fuller},
  {Kupfer}, {Bellm}, {Bildsten}, {Graham}, and {et~ al.}}]{2019Natur.571..528B}
{Burdge} KB, {Coughlin} MW, {Fuller} J, {Kupfer} T, {Bellm} EC, {Bildsten} L,
  {Graham} MJ, {et~ al} (2019{\natexlab{a}}) {General relativistic orbital
  decay in a seven-minute-orbital-period eclipsing binary system}. \nat
  571(7766):528--531. \doi{10.1038/s41586-019-1403-0}.
  {\href{https://arxiv.org/abs/1907.11291}{{arXiv:1907.11291}}} {[astro-ph.SR]}

\bibitem[{{Burdge} et~al.(2019{\natexlab{b}}){Burdge}, {Fuller}, {Phinney},
  {van Roestel}, {Claret}, {Cukanovaite}, {Gentile Fusillo}, {Coughlin},
  {Kaplan}, {Kupfer}, {Tremblay}, {Dekany}, {Duev}, {Feeney}, {Riddle},
  {Kulkarni}, and {Prince}}]{2019ApJ...886L..12B}
{Burdge} KB, {Fuller} J, {Phinney} ES, {van Roestel} J, {Claret} A,
  {Cukanovaite} E, {Gentile Fusillo} NP, {Coughlin} MW, {Kaplan} DL, {Kupfer}
  T, et~al. (2019{\natexlab{b}}) {Orbital Decay in a 20 Minute Orbital Period
  Detached Binary with a Hydrogen-poor Low-mass White Dwarf}. \apjl 886(1):L12.
  \doi{10.3847/2041-8213/ab53e5}.
  {\href{https://arxiv.org/abs/1910.11389}{{arXiv:1910.11389}}} {[astro-ph.SR]}

\bibitem[{{Burdge} et~al.(2020{\natexlab{a}}){Burdge}, {Coughlin}, {Fuller},
  {Kaplan}, {Kulkarni}, {Marsh}, {Bellm}, {Dekany}, {Duev}, {Graham},
  {Mahabal}, {Masci}, {Laher}, {Riddle}, {Soumagnac}, and
  {Prince}}]{2020arXiv201003555B}
{Burdge} KB, {Coughlin} MW, {Fuller} J, {Kaplan} DL, {Kulkarni} SR, {Marsh} TR,
  {Bellm} EC, {Dekany} RG, {Duev} DA, {Graham} MJ, et~al. (2020{\natexlab{a}})
  {An 8.8 Minute Orbital Period Eclipsing Detached Double White Dwarf Binary}.
  \apjl 905(1):L7. \doi{10.3847/2041-8213/abca91}.
  {\href{https://arxiv.org/abs/2010.03555}{{arXiv:2010.03555}}} {[astro-ph.SR]}

\bibitem[{{Burdge} et~al.(2020{\natexlab{b}}){Burdge}, {Prince}, {Fuller},
  {Kaplan}, {Marsh}, and {et~al.}}]{2020ApJ...905...32B}
{Burdge} KB, {Prince} TA, {Fuller} J, {Kaplan} DL, {Marsh} TR, {et~al}
  (2020{\natexlab{b}}) {A Systematic Search of Zwicky Transient Facility Data
  for Ultracompact Binary LISA-detectable Gravitational-wave Sources}. \apj
  905(1):32. \doi{10.3847/1538-4357/abc261}.
  {\href{https://arxiv.org/abs/2009.02567}{{arXiv:2009.02567}}} {[astro-ph.SR]}

\bibitem[{{Burke-Spolaor} et~al.(2018){Burke-Spolaor}, {Blecha}, {Bogdanovic},
  {Comerford}, {Lazio}, {Liu}, {Maccarone}, {Pesce}, {Shen}, and
  {Taylor}}]{2018arXiv180804368B}
{Burke-Spolaor} S, {Blecha} L, {Bogdanovic} T, {Comerford} JM, {Lazio} TJW,
  {Liu} X, {Maccarone} TJ, {Pesce} D, {Shen} Y, {Taylor} G (2018) {The
  Next-Generation Very Large Array: Supermassive Black Hole Pairs and
  Binaries}. arXiv e-prints arXiv:1808.04368.
  {\href{https://arxiv.org/abs/1808.04368}{{arXiv:1808.04368}}} {[astro-ph.GA]}

\bibitem[{{Burke-Spolaor} et~al.(2019){Burke-Spolaor}, {Taylor}, {Charisi},
  {Dolch}, {Hazboun}, {Holgado}, {Kelley}, {Lazio}, {Madison}, {McMann},
  {Mingarelli}, {Rasskazov}, {Siemens}, {Simon}, and
  {Smith}}]{2019A&ARv..27....5B}
{Burke-Spolaor} S, {Taylor} SR, {Charisi} M, {Dolch} T, {Hazboun} JS, {Holgado}
  AM, {Kelley} LZ, {Lazio} TJW, {Madison} DR, {McMann} N, et~al. (2019) {The
  astrophysics of nanohertz gravitational waves}. \aapr 27(1):5.
  \doi{10.1007/s00159-019-0115-7}.
  {\href{https://arxiv.org/abs/1811.08826}{{arXiv:1811.08826}}} {[astro-ph.HE]}

\bibitem[{{Burningham}(2018)}]{2018haex.bookE.118B}
{Burningham} B (2018) {Large-Scale Searches for Brown Dwarfs and Free-Floating
  Planets}, p 118. \doi{10.1007/978-3-319-55333-7_118}

\bibitem[{{Buscicchio} et~al.(2021){Buscicchio}, {Klein}, {Roebber}, {Moore},
  {Gerosa}, {Finch}, and {Vecchio}}]{2021PhRvD.104d4065B}
{Buscicchio} R, {Klein} A, {Roebber} E, {Moore} CJ, {Gerosa} D, {Finch} E,
  {Vecchio} A (2021) {Bayesian parameter estimation of stellar-mass black-hole
  binaries with LISA}. \prd 104(4):044065. \doi{10.1103/PhysRevD.104.044065}.
  {\href{https://arxiv.org/abs/2106.05259}{{arXiv:2106.05259}}} {[astro-ph.HE]}

\bibitem[{{Bustamante} and {Springel}(2019)}]{2019MNRAS.490.4133B}
{Bustamante} S, {Springel} V (2019) {Spin evolution and feedback of
  supermassive black holes in cosmological simulations}. \mnras
  490(3):4133--4153. \doi{10.1093/mnras/stz2836}.
  {\href{https://arxiv.org/abs/1902.04651}{{arXiv:1902.04651}}} {[astro-ph.GA]}

\bibitem[{{Byrd} et~al.(1986){Byrd}, {Valtonen}, {Sundelius}, and
  {Valtaoja}}]{1986A&A...166...75B}
{Byrd} GG, {Valtonen} MJ, {Sundelius} B, {Valtaoja} L (1986) {Tidal triggering
  of Seyfert galaxies and quasars : perturbed galaxy disk models versus
  observations.} \aap 166:75--82

\bibitem[{Byrnes et~al.(2019)Byrnes, Cole, and Patil}]{Byrnes:2018txb}
Byrnes CT, Cole PS, Patil SP (2019) {Steepest growth of the power spectrum and
  primordial black holes}. JCAP 06:028. \doi{10.1088/1475-7516/2019/06/028}.
  {\href{https://arxiv.org/abs/1811.11158}{{arXiv:1811.11158}}} {[astro-ph.CO]}

\bibitem[{{Caballero}(2009)}]{2009A&A...507..251C}
{Caballero} JA (2009) {Reaching the boundary between stellar kinematic groups
  and very wide binaries. The Washington double stars with the widest angular
  separations}. \aap 507(1):251--259. \doi{10.1051/0004-6361/200912596}.
  {\href{https://arxiv.org/abs/0908.2761}{{arXiv:0908.2761}}} {[astro-ph.SR]}

\bibitem[{Cai et~al.(2019)Cai, Pi, and Sasaki}]{Cai:2018dig}
Cai Rg, Pi S, Sasaki M (2019) {Gravitational Waves Induced by non-Gaussian
  Scalar Perturbations}. Phys Rev Lett 122(20):201101.
  \doi{10.1103/PhysRevLett.122.201101}.
  {\href{https://arxiv.org/abs/1810.11000}{{arXiv:1810.11000}}} {[astro-ph.CO]}

\bibitem[{{Calder{\'o}n Bustillo} et~al.(2018){Calder{\'o}n Bustillo}, {Clark},
  {Laguna}, and {Shoemaker}}]{2018PhRvL.121s1102C}
{Calder{\'o}n Bustillo} J, {Clark} JA, {Laguna} P, {Shoemaker} D (2018)
  {Tracking Black Hole Kicks from Gravitational-Wave Observations}. \prl
  121(19):191102. \doi{10.1103/PhysRevLett.121.191102}.
  {\href{https://arxiv.org/abs/1806.11160}{{arXiv:1806.11160}}} {[gr-qc]}

\bibitem[{{Callegari} et~al.(2009){Callegari}, {Mayer}, {Kazantzidis}, {Colpi},
  {Governato}, {Quinn}, and {Wadsley}}]{2009ApJ...696L..89C}
{Callegari} S, {Mayer} L, {Kazantzidis} S, {Colpi} M, {Governato} F, {Quinn} T,
  {Wadsley} J (2009) {Pairing of Supermassive Black Holes in Unequal-Mass
  Galaxy Mergers}. \apjl 696(1):L89--L92. \doi{10.1088/0004-637X/696/1/L89}.
  {\href{https://arxiv.org/abs/0811.0615}{{arXiv:0811.0615}}} {[astro-ph]}

\bibitem[{{Callegari} et~al.(2011){Callegari}, {Kazantzidis}, {Mayer}, {Colpi},
  {Bellovary}, {Quinn}, and {Wadsley}}]{2011ApJ...729...85C}
{Callegari} S, {Kazantzidis} S, {Mayer} L, {Colpi} M, {Bellovary} JM, {Quinn}
  T, {Wadsley} J (2011) {Growing Massive Black Hole Pairs in Minor Mergers of
  Disk Galaxies}. \apj 729(2):85. \doi{10.1088/0004-637X/729/2/85}.
  {\href{https://arxiv.org/abs/1002.1712}{{arXiv:1002.1712}}} {[astro-ph.CO]}

\bibitem[{{Callister} et~al.(2016){Callister}, {Sammut}, {Qiu}, {Mandel}, and
  {Thrane}}]{Callister:2016}
{Callister} T, {Sammut} L, {Qiu} S, {Mandel} I, {Thrane} E (2016) {Limits of
  Astrophysics with Gravitational-Wave Backgrounds}. Physical Review X
  6(3):031018. \doi{10.1103/PhysRevX.6.031018}.
  {\href{https://arxiv.org/abs/1604.02513}{{arXiv:1604.02513}}} {[gr-qc]}

\bibitem[{{Cameron} et~al.(2020){Cameron}, {Champion}, {Bailes},
  {Balakrishnan}, {Barr}, {Bassa}, {Bates}, {Bhandari}, {Bhat}, {Burgay},
  {Burke-Spolaor}, {Flynn}, {Jameson}, {Johnston}, {Keith}, {Kramer}, {Levin},
  {Lyne}, {Ng}, {Petroff}, {Possenti}, {Smith}, {Stappers}, {van Straten},
  {Tiburzi}, and {Wu}}]{2020MNRAS.493.1063C}
{Cameron} AD, {Champion} DJ, {Bailes} M, {Balakrishnan} V, {Barr} ED, {Bassa}
  CG, {Bates} S, {Bhandari} S, {Bhat} NDR, {Burgay} M, et~al. (2020) {The High
  Time Resolution Universe Pulsar Survey - XVI. Discovery and timing of 40
  pulsars from the southern Galactic plane}. \mnras 493(1):1063--1087.
  \doi{10.1093/mnras/staa039}.
  {\href{https://arxiv.org/abs/2001.01823}{{arXiv:2001.01823}}} {[astro-ph.HE]}

\bibitem[{{Campanelli} et~al.(2006){Campanelli}, {Lousto}, {Marronetti}, and
  {Zlochower}}]{2006PhRvL..96k1101C}
{Campanelli} M, {Lousto} CO, {Marronetti} P, {Zlochower} Y (2006) {Accurate
  Evolutions of Orbiting Black-Hole Binaries without Excision}. \prl
  96(11):111101. \doi{10.1103/PhysRevLett.96.111101}.
  {\href{https://arxiv.org/abs/gr-qc/0511048}{{arXiv:gr-qc/0511048}}} {[gr-qc]}

\bibitem[{{Campanelli} et~al.(2007{\natexlab{a}}){Campanelli}, {Lousto},
  {Zlochower}, and {Merritt}}]{2007ApJ...659L...5C}
{Campanelli} M, {Lousto} C, {Zlochower} Y, {Merritt} D (2007{\natexlab{a}})
  {Large Merger Recoils and Spin Flips from Generic Black Hole Binaries}. \apjl
  659(1):L5--L8. \doi{10.1086/516712}.
  {\href{https://arxiv.org/abs/gr-qc/0701164}{{arXiv:gr-qc/0701164}}} {[gr-qc]}

\bibitem[{{Campanelli} et~al.(2007{\natexlab{b}}){Campanelli}, {Lousto},
  {Zlochower}, and {Merritt}}]{2007PhRvL..98w1102C}
{Campanelli} M, {Lousto} CO, {Zlochower} Y, {Merritt} D (2007{\natexlab{b}})
  {Maximum Gravitational Recoil}. \prl 98(23):231102.
  \doi{10.1103/PhysRevLett.98.231102}.
  {\href{https://arxiv.org/abs/gr-qc/0702133}{{arXiv:gr-qc/0702133}}} {[gr-qc]}

\bibitem[{{Cantiello} et~al.(2021){Cantiello}, {Jermyn}, and
  {Lin}}]{2020arXiv200903936C}
{Cantiello} M, {Jermyn} AS, {Lin} DNC (2021) {Stellar Evolution in AGN Disks}.
  \apj 910(2):94. \doi{10.3847/1538-4357/abdf4f}.
  {\href{https://arxiv.org/abs/2009.03936}{{arXiv:2009.03936}}} {[astro-ph.SR]}

\bibitem[{{Capelo} et~al.(2015){Capelo}, {Volonteri}, {Dotti}, {Bellovary},
  {Mayer}, and {Governato}}]{2015MNRAS.447.2123C}
{Capelo} PR, {Volonteri} M, {Dotti} M, {Bellovary} JM, {Mayer} L, {Governato} F
  (2015) {Growth and activity of black holes in galaxy mergers with varying
  mass ratios}. \mnras 447(3):2123--2143. \doi{10.1093/mnras/stu2500}.
  {\href{https://arxiv.org/abs/1409.0004}{{arXiv:1409.0004}}} {[astro-ph.GA]}

\bibitem[{{Cappelluti} et~al.(2022){Cappelluti}, {Hasinger}, and
  {Natarajan}}]{2022ApJ...926..205C}
{Cappelluti} N, {Hasinger} G, {Natarajan} P (2022) {Exploring the High-redshift
  PBH-{\ensuremath{\Lambda}}CDM Universe: Early Black Hole Seeding, the First
  Stars and Cosmic Radiation Backgrounds}. \apj 926(2):205.
  \doi{10.3847/1538-4357/ac332d}.
  {\href{https://arxiv.org/abs/2109.08701}{{arXiv:2109.08701}}} {[astro-ph.CO]}

\bibitem[{{Caprini} et~al.(2019){Caprini}, {Figueroa}, {Flauger}, {Nardini},
  {Peloso}, {Pieroni}, {Ricciardone}, and {Tasinato}}]{2019JCAP...11..017C}
{Caprini} C, {Figueroa} DG, {Flauger} R, {Nardini} G, {Peloso} M, {Pieroni} M,
  {Ricciardone} A, {Tasinato} G (2019) {Reconstructing the spectral shape of a
  stochastic gravitational wave background with LISA}. \jcap 2019(11):017.
  \doi{10.1088/1475-7516/2019/11/017}.
  {\href{https://arxiv.org/abs/1906.09244}{{arXiv:1906.09244}}} {[astro-ph.CO]}

\bibitem[{{Capuzzo-Dolcetta}(1993)}]{1993ApJ...415..616C}
{Capuzzo-Dolcetta} R (1993) {The Evolution of the Globular Cluster System in a
  Triaxial Galaxy: Can a Galactic Nucleus Form by Globular Cluster Capture?}
  \apj 415:616. \doi{10.1086/173189}.
  {\href{https://arxiv.org/abs/astro-ph/9301006}{{arXiv:astro-ph/9301006}}}
  {[astro-ph]}

\bibitem[{{Capuzzo-Dolcetta} et~al.(2013){Capuzzo-Dolcetta}, {Spera}, and
  {Punzo}}]{2013JCoPh.236..580C}
{Capuzzo-Dolcetta} R, {Spera} M, {Punzo} D (2013) {A fully parallel, high
  precision, N-body code running on hybrid computing platforms}. Journal of
  Computational Physics 236:580--593. \doi{10.1016/j.jcp.2012.11.013}.
  {\href{https://arxiv.org/abs/1207.2367}{{arXiv:1207.2367}}} {[astro-ph.IM]}

\bibitem[{{C{\'a}rdenas-Avenda{\~n}o} et~al.(2018){C{\'a}rdenas-Avenda{\~n}o},
  {Gutierrez}, {Pach{\'o}n}, and {Yunes}}]{2018CQGra..35p5010C}
{C{\'a}rdenas-Avenda{\~n}o} A, {Gutierrez} AF, {Pach{\'o}n} LA, {Yunes} N
  (2018) {The exact dynamical Chern-Simons metric for a spinning black hole
  possesses a fourth constant of motion: a dynamical-systems-based conjecture}.
  Classical and Quantum Gravity 35(16):165010. \doi{10.1088/1361-6382/aad06f}.
  {\href{https://arxiv.org/abs/1804.04002}{{arXiv:1804.04002}}} {[gr-qc]}

\bibitem[{{Cardoso} and {Maselli}(2020)}]{2019arXiv190905870C}
{Cardoso} V, {Maselli} A (2020) {Constraints on the astrophysical environment
  of binaries with gravitational-wave observations}. \aap 644:A147.
  \doi{10.1051/0004-6361/202037654}.
  {\href{https://arxiv.org/abs/1909.05870}{{arXiv:1909.05870}}} {[astro-ph.HE]}

\bibitem[{{Cardoso} et~al.(2021){Cardoso}, {Duque}, and
  {Khanna}}]{2021PhRvD.103h1501C}
{Cardoso} V, {Duque} F, {Khanna} G (2021) {Gravitational tuning forks and
  hierarchical triple systems}. \prd 103(8):L081501.
  \doi{10.1103/PhysRevD.103.L081501}.
  {\href{https://arxiv.org/abs/2101.01186}{{arXiv:2101.01186}}} {[gr-qc]}

\bibitem[{{Cardoso} et~al.(2022){Cardoso}, {Destounis}, {Duque}, {Macedo}, and
  {Maselli}}]{2022PhRvD.105f1501C}
{Cardoso} V, {Destounis} K, {Duque} F, {Macedo} RP, {Maselli} A (2022) {Black
  holes in galaxies: Environmental impact on gravitational-wave generation and
  propagation}. \prd 105(6):L061501. \doi{10.1103/PhysRevD.105.L061501}.
  {\href{https://arxiv.org/abs/2109.00005}{{arXiv:2109.00005}}} {[gr-qc]}

\bibitem[{{Carlberg} et~al.(1996){Carlberg}, {Yee}, {Ellingson}, {Abraham},
  {Gravel}, {Morris}, and {Pritchet}}]{1996ApJ...462...32C}
{Carlberg} RG, {Yee} HKC, {Ellingson} E, {Abraham} R, {Gravel} P, {Morris} S,
  {Pritchet} CJ (1996) {Galaxy Cluster Virial Masses and Omega}. \apj 462:32.
  \doi{10.1086/177125}.
  {\href{https://arxiv.org/abs/astro-ph/9509034}{{arXiv:astro-ph/9509034}}}
  {[astro-ph]}

\bibitem[{{Carr} and {K{\"u}hnel}(2020)}]{2020ARNPS..7050520C}
{Carr} B, {K{\"u}hnel} F (2020) {Primordial Black Holes as Dark Matter: Recent
  Developments}. Annual Review of Nuclear and Particle Science 70(1):annurev.
  \doi{10.1146/annurev-nucl-050520-125911}.
  {\href{https://arxiv.org/abs/2006.02838}{{arXiv:2006.02838}}} {[astro-ph.CO]}

\bibitem[{{Carr} et~al.(2021){Carr}, {Kohri}, {Sendouda}, and
  {Yokoyama}}]{2020arXiv200212778C}
{Carr} B, {Kohri} K, {Sendouda} Y, {Yokoyama} J (2021) {Constraints on
  primordial black holes}. Reports on Progress in Physics 84(11):116902.
  \doi{10.1088/1361-6633/ac1e31}.
  {\href{https://arxiv.org/abs/2002.12778}{{arXiv:2002.12778}}} {[astro-ph.CO]}

\bibitem[{{Carson} and {Yagi}(2020)}]{2020CQGra..37bLT01C}
{Carson} Z, {Yagi} K (2020) {Multi-band gravitational wave tests of general
  relativity}. Classical and Quantum Gravity 37(2):02LT01.
  \doi{10.1088/1361-6382/ab5c9a}.
  {\href{https://arxiv.org/abs/1905.13155}{{arXiv:1905.13155}}} {[gr-qc]}

\bibitem[{{Carter}(1968)}]{1968PhRv..174.1559C}
{Carter} B (1968) {Global Structure of the Kerr Family of Gravitational
  Fields}. Physical Review 174(5):1559--1571. \doi{10.1103/PhysRev.174.1559}

\bibitem[{{Carter} and {Luminet}(1983)}]{1983A&A...121...97C}
{Carter} B, {Luminet} JP (1983) {Tidal compression of a star by a large black
  hole. I Mechanical evolution and nuclear energy release by proton capture}.
  \aap 121(1):97--113

\bibitem[{{Carter} et~al.(2013){Carter}, {Marsh}, {Steeghs}, {Groot},
  {Nelemans}, {Levitan}, {Rau}, {Copperwheat}, {Kupfer}, and
  {Roelofs}}]{2013MNRAS.429.2143C}
{Carter} PJ, {Marsh} TR, {Steeghs} D, {Groot} PJ, {Nelemans} G, {Levitan} D,
  {Rau} A, {Copperwheat} CM, {Kupfer} T, {Roelofs} GHA (2013) {A search for the
  hidden population of AM CVn binaries in the Sloan Digital Sky Survey}. \mnras
  429(3):2143--2160. \doi{10.1093/mnras/sts485}.
  {\href{https://arxiv.org/abs/1211.6439}{{arXiv:1211.6439}}} {[astro-ph.SR]}

\bibitem[{{{\c{C}}atmabacak} et~al.(2022){{\c{C}}atmabacak}, {Feldmann},
  {Angl{\'e}s-Alc{\'a}zar}, {Faucher-Gigu{\`e}re}, {Hopkins}, and
  {Kere{\v{s}}}}]{2020arXiv200712185C}
{{\c{C}}atmabacak} O, {Feldmann} R, {Angl{\'e}s-Alc{\'a}zar} D,
  {Faucher-Gigu{\`e}re} CA, {Hopkins} PF, {Kere{\v{s}}} D (2022) {Black
  hole-galaxy scaling relations in FIRE: the importance of black hole location
  and mergers}. \mnras 511(1):506--535. \doi{10.1093/mnras/stac040}

\bibitem[{{Cembranos} et~al.(2012){Cembranos}, {de la Cruz-Dombriz}, and
  {Montes N{\'u}{\~n}ez}}]{2012JCAP...04..021C}
{Cembranos} JAR, {de la Cruz-Dombriz} A, {Montes N{\'u}{\~n}ez} B (2012)
  {Gravitational collapse in f(R) theories}. JCAP 2012(4):021.
  \doi{10.1088/1475-7516/2012/04/021}.
  {\href{https://arxiv.org/abs/1201.1289}{{arXiv:1201.1289}}} {[gr-qc]}

\bibitem[{{Cenci} et~al.(2020){Cenci}, {Sala}, {Lupi}, {Capelo}, and
  {Dotti}}]{2020MNRAS.500.3719C}
{Cenci} E, {Sala} L, {Lupi} A, {Capelo} PR, {Dotti} M (2020) {Black hole spin
  evolution in warped accretion discs}. \mnras 500(3):3719--3727.
  \doi{10.1093/mnras/staa3449}.
  {\href{https://arxiv.org/abs/2011.06596}{{arXiv:2011.06596}}} {[astro-ph.GA]}

\bibitem[{{Centrella} et~al.(2010){Centrella}, {Baker}, {Kelly}, and {van
  Meter}}]{2010RvMP...82.3069C}
{Centrella} J, {Baker} JG, {Kelly} BJ, {van Meter} JR (2010) {Black-hole
  binaries, gravitational waves, and numerical relativity}. Reviews of Modern
  Physics 82(4):3069--3119. \doi{10.1103/RevModPhys.82.3069}.
  {\href{https://arxiv.org/abs/1010.5260}{{arXiv:1010.5260}}} {[gr-qc]}

\bibitem[{{Cerioli} et~al.(2016){Cerioli}, {Lodato}, and
  {Price}}]{2016MNRAS.457..939C}
{Cerioli} A, {Lodato} G, {Price} DJ (2016) {Gas squeezing during the merger of
  a supermassive black hole binary}. \mnras 457(1):939--948.
  \doi{10.1093/mnras/stw034}.
  {\href{https://arxiv.org/abs/1601.03776}{{arXiv:1601.03776}}} {[astro-ph.HE]}

\bibitem[{{Ceverino} et~al.(2010){Ceverino}, {Dekel}, and
  {Bournaud}}]{2010MNRAS.404.2151C}
{Ceverino} D, {Dekel} A, {Bournaud} F (2010) {High-redshift clumpy discs and
  bulges in cosmological simulations}. \mnras 404(4):2151--2169.
  \doi{10.1111/j.1365-2966.2010.16433.x}.
  {\href{https://arxiv.org/abs/0907.3271}{{arXiv:0907.3271}}} {[astro-ph.CO]}

\bibitem[{{Chabrier}(2003)}]{2003PASP..115..763C}
{Chabrier} G (2003) {Galactic Stellar and Substellar Initial Mass Function}.
  \pasp 115(809):763--795. \doi{10.1086/376392}.
  {\href{https://arxiv.org/abs/astro-ph/0304382}{{arXiv:astro-ph/0304382}}}
  {[astro-ph]}

\bibitem[{{Chamandy} et~al.(2018){Chamandy}, {Frank}, {Blackman},
  {Carroll-Nellenback}, {Liu}, {Tu}, {Nordhaus}, {Chen}, and
  {Peng}}]{2018MNRAS.480.1898C}
{Chamandy} L, {Frank} A, {Blackman} EG, {Carroll-Nellenback} J, {Liu} B, {Tu}
  Y, {Nordhaus} J, {Chen} Z, {Peng} B (2018) {Accretion in common envelope
  evolution}. \mnras 480(2):1898--1911. \doi{10.1093/mnras/sty1950}.
  {\href{https://arxiv.org/abs/1805.03607}{{arXiv:1805.03607}}} {[astro-ph.SR]}

\bibitem[{{Chandrasekhar}(1943)}]{1943ApJ....97..255C}
{Chandrasekhar} S (1943) {Dynamical Friction. I. General Considerations: the
  Coefficient of Dynamical Friction.} \apj 97:255. \doi{10.1086/144517}

\bibitem[{{Chang} et~al.(2010){Chang}, {Strubbe}, {Menou}, and
  {Quataert}}]{2010MNRAS.407.2007C}
{Chang} P, {Strubbe} LE, {Menou} K, {Quataert} E (2010) {Fossil gas and the
  electromagnetic precursor of supermassive binary black hole mergers}. \mnras
  407(3):2007--2016. \doi{10.1111/j.1365-2966.2010.17056.x}.
  {\href{https://arxiv.org/abs/0906.0825}{{arXiv:0906.0825}}} {[astro-ph.HE]}

\bibitem[{{Chapon} et~al.(2013){Chapon}, {Mayer}, and
  {Teyssier}}]{2013MNRAS.429.3114C}
{Chapon} D, {Mayer} L, {Teyssier} R (2013) {Hydrodynamics of galaxy mergers
  with supermassive black holes: is there a last parsec problem?} \mnras
  429(4):3114--3122. \doi{10.1093/mnras/sts568}.
  {\href{https://arxiv.org/abs/1110.6086}{{arXiv:1110.6086}}} {[astro-ph.GA]}

\bibitem[{{Charisi} et~al.(2016){Charisi}, {Bartos}, {Haiman}, {Price-Whelan},
  {Graham}, {Bellm}, {Laher}, and {M{\'a}rka}}]{2016MNRAS.463.2145C}
{Charisi} M, {Bartos} I, {Haiman} Z, {Price-Whelan} AM, {Graham} MJ, {Bellm}
  EC, {Laher} RR, {M{\'a}rka} S (2016) {A population of short-period variable
  quasars from PTF as supermassive black hole binary candidates}. \mnras
  463(2):2145--2171. \doi{10.1093/mnras/stw1838}.
  {\href{https://arxiv.org/abs/1604.01020}{{arXiv:1604.01020}}} {[astro-ph.GA]}

\bibitem[{{Charisi} et~al.(2018){Charisi}, {Haiman}, {Schiminovich}, and
  {D'Orazio}}]{2018MNRAS.476.4617C}
{Charisi} M, {Haiman} Z, {Schiminovich} D, {D'Orazio} DJ (2018) {Testing the
  relativistic Doppler boost hypothesis for supermassive black hole binary
  candidates}. \mnras 476(4):4617--4628. \doi{10.1093/mnras/sty516}.
  {\href{https://arxiv.org/abs/1801.06189}{{arXiv:1801.06189}}} {[astro-ph.GA]}

\bibitem[{{Chatterjee} et~al.(2003){Chatterjee}, {Hernquist}, and
  {Loeb}}]{2003ApJ...592...32C}
{Chatterjee} P, {Hernquist} L, {Loeb} A (2003) {Effects of Wandering on the
  Coalescence of Black Hole Binaries in Galactic Centers}. \apj 592(1):32--41.
  \doi{10.1086/375552}.
  {\href{https://arxiv.org/abs/astro-ph/0302573}{{arXiv:astro-ph/0302573}}}
  {[astro-ph]}

\bibitem[{Chaty(2022)}]{10.1088/2514-3433/ac595f}
Chaty S (2022) Accreting Binaries. 2514-3433, IOP Publishing.
  \doi{10.1088/2514-3433/ac595f},
  \urlprefix\url{https://dx.doi.org/10.1088/2514-3433/ac595f}

\bibitem[{{Chen} et~al.(2013){Chen}, {Chen}, {Tauris}, and
  {Han}}]{2013ApJ...775...27C}
{Chen} HL, {Chen} X, {Tauris} TM, {Han} Z (2013) {Formation of Black Widows and
  Redbacks{\textemdash}Two Distinct Populations of Eclipsing Binary Millisecond
  Pulsars}. \apj 775(1):27. \doi{10.1088/0004-637X/775/1/27}.
  {\href{https://arxiv.org/abs/1308.4107}{{arXiv:1308.4107}}} {[astro-ph.SR]}

\bibitem[{{Chen} et~al.(2020{\natexlab{a}}){Chen}, {Liu}, and
  {Wang}}]{2020ApJ...900L...8C}
{Chen} WC, {Liu} DD, {Wang} B (2020{\natexlab{a}}) {Detectability of
  Ultra-compact X-Ray Binaries as LISA Sources}. \apjl 900(1):L8.
  \doi{10.3847/2041-8213/abae66}.
  {\href{https://arxiv.org/abs/2008.05143}{{arXiv:2008.05143}}} {[astro-ph.HE]}

\bibitem[{{Chen} and {Amaro-Seoane}(2014)}]{2014ApJ...786L..14C}
{Chen} X, {Amaro-Seoane} P (2014) {A Rapidly Evolving Region in the Galactic
  Center: Why S-stars Thermalize and More Massive Stars are Missing}. \apjl
  786(2):L14. \doi{10.1088/2041-8205/786/2/L14}.
  {\href{https://arxiv.org/abs/1401.6456}{{arXiv:1401.6456}}} {[astro-ph.GA]}

\bibitem[{{Chen} and {Han}(2018)}]{2018CmPhy...1...53C}
{Chen} X, {Han} WB (2018) {Extreme-mass-ratio inspirals produced by tidal
  capture of binary black holes}. Communications Physics 1(1):53.
  \doi{10.1038/s42005-018-0053-0}.
  {\href{https://arxiv.org/abs/1801.05780}{{arXiv:1801.05780}}} {[astro-ph.HE]}

\bibitem[{{Chen} and {Han}(2008)}]{2008MNRAS.387.1416C}
{Chen} X, {Han} Z (2008) {Mass transfer from a giant star to a main-sequence
  companion and its contribution to long-orbital-period blue stragglers}.
  \mnras 387(4):1416--1430. \doi{10.1111/j.1365-2966.2008.13334.x}.
  {\href{https://arxiv.org/abs/0804.2294}{{arXiv:0804.2294}}} {[astro-ph]}

\bibitem[{{Chen} and {Liu}(2013)}]{2013ApJ...762...95C}
{Chen} X, {Liu} FK (2013) {Is There an Intermediate Massive Black Hole in the
  Galactic Center: Imprints on the Stellar Tidal-disruption Rate}. \apj
  762(2):95. \doi{10.1088/0004-637X/762/2/95}.
  {\href{https://arxiv.org/abs/1211.4609}{{arXiv:1211.4609}}} {[astro-ph.GA]}

\bibitem[{{Chen} and {Zhang}(2022)}]{2022arXiv220608104C}
{Chen} X, {Zhang} Z (2022) {Binaries Wandering Around Supermassive Black Holes
  Due To Gravito-electromagnetism}. arXiv e-prints arXiv:2206.08104.
  {\href{https://arxiv.org/abs/2206.08104}{{arXiv:2206.08104}}} {[astro-ph.HE]}

\bibitem[{{Chen} et~al.(2019{\natexlab{a}}){Chen}, {Li}, and
  {Cao}}]{2019MNRAS.485L.141C}
{Chen} X, {Li} S, {Cao} Z (2019{\natexlab{a}}) {Mass-redshift degeneracy for
  the gravitational-wave sources in the vicinity of supermassive black holes}.
  \mnras 485(1):L141--L145. \doi{10.1093/mnrasl/slz046}.
  {\href{https://arxiv.org/abs/1703.10543}{{arXiv:1703.10543}}} {[astro-ph.HE]}

\bibitem[{{Chen} et~al.(2020{\natexlab{b}}){Chen}, {Yu}, and
  {Lu}}]{2020ApJ...897...86C}
{Chen} Y, {Yu} Q, {Lu} Y (2020{\natexlab{b}}) {Dynamical Evolution of Cosmic
  Supermassive Binary Black Holes and Their Gravitational-wave Radiation}. \apj
  897(1):86. \doi{10.3847/1538-4357/ab9594}.
  {\href{https://arxiv.org/abs/2005.10818}{{arXiv:2005.10818}}} {[astro-ph.HE]}

\bibitem[{{Chen} et~al.(2020{\natexlab{c}}){Chen}, {Liu}, {Liao}, {Holgado},
  {Guo}, {Gruendl}, {Morganson}, {Shen}, {Zhang}, {Abbott}, {Aguena}, {Allam},
  {Avila}, {Bertin}, {Bhargava}, {Brooks}, {Burke}, {Carnero Rosell},
  {Carollo}, {Carrasco Kind}, {Carretero}, {Costanzi}, {da Costa}, {Davis}, {De
  Vicente}, {Desai}, {Diehl}, {Doel}, {Everett}, {Flaugher}, {Friedel},
  {Frieman}, {Garc{\'\i}a-Bellido}, {Gaztanaga}, {Glazebrook}, {Gruen},
  {Gutierrez}, {Hinton}, {Hollowood}, {James}, {Kim}, {Kuehn}, {Kuropatkin},
  {Lewis}, {Lidman}, {Lima}, {Maia}, {March}, {Marshall}, {Menanteau},
  {Miquel}, {Palmese}, {Paz-Chinch{\'o}n}, {Plazas}, {Sanchez}, {Schubnell},
  {Serrano}, {Sevilla-Noarbe}, {Smith}, {Suchyta}, {Swanson}, {Tarle},
  {Tucker}, {Norbert Varga}, and {Walker}}]{2020arXiv200812329C}
{Chen} YC, {Liu} X, {Liao} WT, {Holgado} AM, {Guo} H, {Gruendl} RA, {Morganson}
  E, {Shen} Y, {Zhang} K, {Abbott} TMC, et~al. (2020{\natexlab{c}}) {Candidate
  periodically variable quasars from the Dark Energy Survey and the Sloan
  Digital Sky Survey}. \mnras 499(2):2245--2264. \doi{10.1093/mnras/staa2957}.
  {\href{https://arxiv.org/abs/2008.12329}{{arXiv:2008.12329}}} {[astro-ph.HE]}

\bibitem[{{Chen} et~al.(2019{\natexlab{b}}){Chen}, {Huang}, and
  {Huang}}]{2019ApJ...871...97C}
{Chen} ZC, {Huang} F, {Huang} QG (2019{\natexlab{b}}) {Stochastic
  Gravitational-wave Background from Binary Black Holes and Binary Neutron
  Stars and Implications for LISA}. \apj 871(1):97.
  \doi{10.3847/1538-4357/aaf581}.
  {\href{https://arxiv.org/abs/1809.10360}{{arXiv:1809.10360}}} {[gr-qc]}

\bibitem[{{Chiaberge} et~al.(2017){Chiaberge}, {Ely}, {Meyer},
  {Georganopoulos}, {Marinucci}, {Bianchi}, {Tremblay}, {Hilbert}, {Kotyla},
  {Capetti}, {Baum}, {Macchetto}, {Miley}, {O'Dea}, {Perlman}, {Sparks}, and
  {Norman}}]{2017A&A...600A..57C}
{Chiaberge} M, {Ely} JC, {Meyer} ET, {Georganopoulos} M, {Marinucci} A,
  {Bianchi} S, {Tremblay} GR, {Hilbert} B, {Kotyla} JP, {Capetti} A, et~al.
  (2017) {The puzzling case of the radio-loud QSO 3C 186: a gravitational wave
  recoiling black hole in a young radio source?} \aap 600:A57.
  \doi{10.1051/0004-6361/201629522}.
  {\href{https://arxiv.org/abs/1611.05501}{{arXiv:1611.05501}}} {[astro-ph.GA]}

\bibitem[{{Chicone} et~al.(2005){Chicone}, {Mashhoon}, and
  {Punsly}}]{2005PhLA..343....1C}
{Chicone} C, {Mashhoon} B, {Punsly} B (2005) {Relativistic motion of spinning
  particles in a gravitational field}. Physics Letters A 343(1-3):1--7.
  \doi{10.1016/j.physleta.2005.05.072}.
  {\href{https://arxiv.org/abs/gr-qc/0504146}{{arXiv:gr-qc/0504146}}} {[gr-qc]}

\bibitem[{{Chilingarian} et~al.(2018){Chilingarian}, {Katkov}, {Zolotukhin},
  {Grishin}, {Beletsky}, {Boutsia}, and {Osip}}]{2018ApJ...863....1C}
{Chilingarian} IV, {Katkov} IY, {Zolotukhin} IY, {Grishin} KA, {Beletsky} Y,
  {Boutsia} K, {Osip} DJ (2018) {A Population of Bona Fide Intermediate-mass
  Black Holes Identified as Low-luminosity Active Galactic Nuclei}. \apj
  863(1):1. \doi{10.3847/1538-4357/aad184}.
  {\href{https://arxiv.org/abs/1805.01467}{{arXiv:1805.01467}}} {[astro-ph.GA]}

\bibitem[{Chluba et~al.(2012)Chluba, Erickcek, and Ben-Dayan}]{Chluba:2012we}
Chluba J, Erickcek AL, Ben-Dayan I (2012) {Probing the inflaton: Small-scale
  power spectrum constraints from measurements of the CMB energy spectrum}.
  Astrophys J 758:76. \doi{10.1088/0004-637X/758/2/76}.
  {\href{https://arxiv.org/abs/1203.2681}{{arXiv:1203.2681}}} {[astro-ph.CO]}

\bibitem[{{Choi} et~al.(2012){Choi}, {Ostriker}, {Naab}, and
  {Johansson}}]{2012ApJ...754..125C}
{Choi} E, {Ostriker} JP, {Naab} T, {Johansson} PH (2012) {Radiative and
  Momentum-based Mechanical Active Galactic Nucleus Feedback in a
  Three-dimensional Galaxy Evolution Code}. \apj 754(2):125.
  \doi{10.1088/0004-637X/754/2/125}.
  {\href{https://arxiv.org/abs/1205.2082}{{arXiv:1205.2082}}} {[astro-ph.GA]}

\bibitem[{{Chon} and {Omukai}(2020)}]{2020MNRAS.494.2851C}
{Chon} S, {Omukai} K (2020) {Supermassive star formation via super competitive
  accretion in slightly metal-enriched clouds}. \mnras
  \doi{10.1093/mnras/staa863}.
  {\href{https://arxiv.org/abs/2001.06491}{{arXiv:2001.06491}}} {[astro-ph.GA]}

\bibitem[{{Chon} et~al.(2018){Chon}, {Hosokawa}, and
  {Yoshida}}]{2018MNRAS.475.4104C}
{Chon} S, {Hosokawa} T, {Yoshida} N (2018) {Radiation hydrodynamics simulations
  of the formation of direct-collapse supermassive stellar systems}. \mnras
  475(3):4104--4121. \doi{10.1093/mnras/sty086}.
  {\href{https://arxiv.org/abs/1711.05262}{{arXiv:1711.05262}}} {[astro-ph.GA]}

\bibitem[{{Chruslinska} and {Nelemans}(2019)}]{2019MNRAS.488.5300C}
{Chruslinska} M, {Nelemans} G (2019) {Metallicity of stars formed throughout
  the cosmic history based on the observational properties of star-forming
  galaxies}. \mnras 488(4):5300--5326. \doi{10.1093/mnras/stz2057}.
  {\href{https://arxiv.org/abs/1907.11243}{{arXiv:1907.11243}}} {[astro-ph.GA]}

\bibitem[{{Chruslinska} et~al.(2018){Chruslinska}, {Belczynski}, {Klencki}, and
  {Benacquista}}]{2018MNRAS.474.2937C}
{Chruslinska} M, {Belczynski} K, {Klencki} J, {Benacquista} M (2018) {Double
  neutron stars: merger rates revisited}. \mnras 474(3):2937--2958.
  \doi{10.1093/mnras/stx2923}.
  {\href{https://arxiv.org/abs/1708.07885}{{arXiv:1708.07885}}} {[astro-ph.HE]}

\bibitem[{{Chruslinska} et~al.(2019){Chruslinska}, {Nelemans}, and
  {Belczynski}}]{2019MNRAS.482.5012C}
{Chruslinska} M, {Nelemans} G, {Belczynski} K (2019) {The influence of the
  distribution of cosmic star formation at different metallicities on the
  properties of merging double compact objects}. \mnras 482(4):5012--5017.
  \doi{10.1093/mnras/sty3087}.
  {\href{https://arxiv.org/abs/1811.03565}{{arXiv:1811.03565}}} {[astro-ph.HE]}

\bibitem[{{Chru{\'s}li{\'n}ska} et~al.(2020){Chru{\'s}li{\'n}ska},
  {Je{\v{r}}{\'a}bkov{\'a}}, {Nelemans}, and {Yan}}]{2020A&A...636A..10C}
{Chru{\'s}li{\'n}ska} M, {Je{\v{r}}{\'a}bkov{\'a}} T, {Nelemans} G, {Yan} Z
  (2020) {The effect of the environment-dependent IMF on the formation and
  metallicities of stars over the cosmic history}. \aap 636:A10.
  \doi{10.1051/0004-6361/202037688}.
  {\href{https://arxiv.org/abs/2002.11122}{{arXiv:2002.11122}}} {[astro-ph.GA]}

\bibitem[{{Chua} and {Cutler}(2021)}]{2021arXiv210914254C}
{Chua} AJK, {Cutler} CJ (2021) {Non-local parameter degeneracy in the intrinsic
  space of gravitational-wave signals from extreme-mass-ratio inspirals}. arXiv
  e-prints arXiv:2109.14254.
  {\href{https://arxiv.org/abs/2109.14254}{{arXiv:2109.14254}}} {[gr-qc]}

\bibitem[{{Chua} and {Vallisneri}(2020)}]{2020PhRvL.124d1102C}
{Chua} AJK, {Vallisneri} M (2020) {Learning Bayesian Posteriors with Neural
  Networks for Gravitational-Wave Inference}. \prl 124(4):041102.
  \doi{10.1103/PhysRevLett.124.041102}.
  {\href{https://arxiv.org/abs/1909.05966}{{arXiv:1909.05966}}} {[gr-qc]}

\bibitem[{{Chua} et~al.(2017){Chua}, {Moore}, and {Gair}}]{2017PhRvD..96d4005C}
{Chua} AJK, {Moore} CJ, {Gair} JR (2017) {Augmented kludge waveforms for
  detecting extreme-mass-ratio inspirals}. \prd 96(4):044005.
  \doi{10.1103/PhysRevD.96.044005}.
  {\href{https://arxiv.org/abs/1705.04259}{{arXiv:1705.04259}}} {[gr-qc]}

\bibitem[{{Church} et~al.(2006){Church}, {Bush}, {Tout}, and
  {Davies}}]{2006MNRAS.372..715C}
{Church} RP, {Bush} SJ, {Tout} CA, {Davies} MB (2006) {Detailed models of the
  binary pulsars J1141-6545 and B2303+46}. \mnras 372(2):715--727.
  \doi{10.1111/j.1365-2966.2006.10897.x}

\bibitem[{{Church} et~al.(2017){Church}, {Strader}, {Davies}, and
  {Bobrick}}]{2017ApJ...851L...4C}
{Church} RP, {Strader} J, {Davies} MB, {Bobrick} A (2017) {Formation
  Constraints Indicate a Black Hole Accretor in 47 Tuc X9}. \apjl 851(1):L4.
  \doi{10.3847/2041-8213/aa9aeb}.
  {\href{https://arxiv.org/abs/1801.00796}{{arXiv:1801.00796}}} {[astro-ph.HE]}

\bibitem[{{Civano} et~al.(2012){Civano}, {Elvis}, {Lanzuisi}, {Aldcroft},
  {Trichas}, {Bongiorno}, {Brusa}, {Blecha}, {Comastri}, {Loeb}, {Salvato},
  {Fruscione}, {Koekemoer}, {Komossa}, {Gilli}, {Mainieri}, {Piconcelli}, and
  {Vignali}}]{2012ApJ...752...49C}
{Civano} F, {Elvis} M, {Lanzuisi} G, {Aldcroft} T, {Trichas} M, {Bongiorno} A,
  {Brusa} M, {Blecha} L, {Comastri} A, {Loeb} A, et~al. (2012) {Chandra
  High-resolution observations of CID-42, a Candidate Recoiling Supermassive
  Black Hole}. \apj 752(1):49. \doi{10.1088/0004-637X/752/1/49}.
  {\href{https://arxiv.org/abs/1205.0815}{{arXiv:1205.0815}}} {[astro-ph.CO]}

\bibitem[{{Clark} et~al.(2011{\natexlab{a}}){Clark}, {Glover}, {Klessen}, and
  {Bromm}}]{2011ApJ...727..110C}
{Clark} PC, {Glover} SCO, {Klessen} RS, {Bromm} V (2011{\natexlab{a}})
  {Gravitational Fragmentation in Turbulent Primordial Gas and the Initial Mass
  Function of Population III Stars}. \apj 727:110.
  \doi{10.1088/0004-637X/727/2/110}.
  {\href{https://arxiv.org/abs/1006.1508}{{arXiv:1006.1508}}} {[astro-ph.GA]}

\bibitem[{{Clark} et~al.(2011{\natexlab{b}}){Clark}, {Glover}, {Smith},
  {Greif}, {Klessen}, and {Bromm}}]{2011Sci...331.1040C}
{Clark} PC, {Glover} SCO, {Smith} RJ, {Greif} TH, {Klessen} RS, {Bromm} V
  (2011{\natexlab{b}}) {The Formation and Fragmentation of Disks Around
  Primordial Protostars}. Science 331:1040--. \doi{10.1126/science.1198027}.
  {\href{https://arxiv.org/abs/1101.5284}{{arXiv:1101.5284}}} {[astro-ph.CO]}

\bibitem[{{Clayton} et~al.(2017){Clayton}, {Podsiadlowski}, {Ivanova}, and
  {Justham}}]{2017MNRAS.470.1788C}
{Clayton} M, {Podsiadlowski} P, {Ivanova} N, {Justham} S (2017) {Episodic mass
  ejections from common-envelope objects}. \mnras 470(2):1788--1808.
  \doi{10.1093/mnras/stx1290}.
  {\href{https://arxiv.org/abs/1705.08457}{{arXiv:1705.08457}}} {[astro-ph.SR]}

\bibitem[{{Clesse} and {Garc{\'\i}a-Bellido}(2015)}]{2015PhRvD..92b3524C}
{Clesse} S, {Garc{\'\i}a-Bellido} J (2015) {Massive primordial black holes from
  hybrid inflation as dark matter and the seeds of galaxies}. \prd
  92(2):023524. \doi{10.1103/PhysRevD.92.023524}.
  {\href{https://arxiv.org/abs/1501.07565}{{arXiv:1501.07565}}} {[astro-ph.CO]}

\bibitem[{Clesse and Garc\'\i{}a-Bellido(2017)}]{Clesse:2016vqa}
Clesse S, Garc\'\i{}a-Bellido J (2017) {The clustering of massive Primordial
  Black Holes as Dark Matter: measuring their mass distribution with Advanced
  LIGO}. Phys Dark Univ 15:142--147. \doi{10.1016/j.dark.2016.10.002}.
  {\href{https://arxiv.org/abs/1603.05234}{{arXiv:1603.05234}}} {[astro-ph.CO]}

\bibitem[{{Cohn} and {Kulsrud}(1978)}]{1978ApJ...226.1087C}
{Cohn} H, {Kulsrud} RM (1978) {The stellar distribution around a black hole:
  numerical integration of the Fokker-Planck equation.} \apj 226:1087--1108.
  \doi{10.1086/156685}

\bibitem[{{Colin} et~al.(2017){Colin}, {Mohayaee}, {Rameez}, and
  {Sarkar}}]{2017MNRAS.471.1045C}
{Colin} J, {Mohayaee} R, {Rameez} M, {Sarkar} S (2017) {High-redshift radio
  galaxies and divergence from the CMB dipole}. \mnras 471(1):1045--1055.
  \doi{10.1093/mnras/stx1631}.
  {\href{https://arxiv.org/abs/1703.09376}{{arXiv:1703.09376}}} {[astro-ph.CO]}

\bibitem[{{Colpi}(2014)}]{2014SSRv..183..189C}
{Colpi} M (2014) {Massive Binary Black Holes in Galactic Nuclei and Their Path
  to Coalescence}. \ssr 183(1-4):189--221. \doi{10.1007/s11214-014-0067-1}.
  {\href{https://arxiv.org/abs/1407.3102}{{arXiv:1407.3102}}} {[astro-ph.GA]}

\bibitem[{{Comastri} et~al.(2015){Comastri}, {Gilli}, {Marconi}, {Risaliti},
  and {Salvati}}]{2015A&A...574L..10C}
{Comastri} A, {Gilli} R, {Marconi} A, {Risaliti} G, {Salvati} M (2015) {Mass
  without radiation: Heavily obscured AGNs, the X-ray background, and the black
  hole mass density}. \aap 574:L10. \doi{10.1051/0004-6361/201425496}.
  {\href{https://arxiv.org/abs/1501.03620}{{arXiv:1501.03620}}} {[astro-ph.GA]}

\bibitem[{{Comerford} and {Izzard}(2020)}]{2020MNRAS.498.2957C}
{Comerford} TAF, {Izzard} RG (2020) {Estimating the outcomes of common envelope
  evolution in triple stellar systems}. \mnras 498(2):2957--2967.
  \doi{10.1093/mnras/staa2539}.
  {\href{https://arxiv.org/abs/2008.09671}{{arXiv:2008.09671}}} {[astro-ph.SR]}

\bibitem[{{Comerford} et~al.(2019){Comerford}, {Izzard}, {Booth}, and
  {Rosotti}}]{2019MNRAS.490.5196C}
{Comerford} TAF, {Izzard} RG, {Booth} RA, {Rosotti} G (2019)
  {Bondi-Hoyle-Lyttleton accretion by binary stars}. \mnras 490(4):5196--5209.
  \doi{10.1093/mnras/stz2977}.
  {\href{https://arxiv.org/abs/1910.13353}{{arXiv:1910.13353}}} {[astro-ph.SR]}

\bibitem[{{Connors} et~al.(1980){Connors}, {Piran}, and
  {Stark}}]{1980ApJ...235..224C}
{Connors} PA, {Piran} T, {Stark} RF (1980) {Polarization features of X-ray
  radiation emitted near black holes.} \apj 235:224--244. \doi{10.1086/157627}

\bibitem[{{Consolandi}(2016)}]{2016A&A...595A..67C}
{Consolandi} G (2016) {Automated bar detection in local disk galaxies from the
  SDSS. The colors of bars}. \aap 595:A67. \doi{10.1051/0004-6361/201629115}.
  {\href{https://arxiv.org/abs/1607.05563}{{arXiv:1607.05563}}} {[astro-ph.GA]}

\bibitem[{{Contenta} et~al.(2018){Contenta}, {Balbinot}, {Petts}, {Read},
  {Gieles}, {Collins}, {Pe{\~n}arrubia}, {Delorme}, and
  {Gualandris}}]{2018MNRAS.476.3124C}
{Contenta} F, {Balbinot} E, {Petts} JA, {Read} JI, {Gieles} M, {Collins} MLM,
  {Pe{\~n}arrubia} J, {Delorme} M, {Gualandris} A (2018) {Probing dark matter
  with star clusters: a dark matter core in the ultra-faint dwarf Eridanus II}.
  \mnras 476(3):3124--3136. \doi{10.1093/mnras/sty424}.
  {\href{https://arxiv.org/abs/1705.01820}{{arXiv:1705.01820}}} {[astro-ph.GA]}

\bibitem[{{Contopoulos}(2002)}]{2002ocda.book.....C}
{Contopoulos} G (2002) {Order and chaos in dynamical astronomy}. Springer

\bibitem[{{Contopoulos} et~al.(2011){Contopoulos}, {Lukes-Gerakopoulos}, and
  {Apostolatos}}]{2011IJBC...21.2261C}
{Contopoulos} G, {Lukes-Gerakopoulos} G, {Apostolatos} TA (2011) {Orbits in a
  Non-Kerr Dynamical System}. International Journal of Bifurcation and Chaos
  21(8):2261. \doi{10.1142/S0218127411029768}.
  {\href{https://arxiv.org/abs/1108.5057}{{arXiv:1108.5057}}} {[gr-qc]}

\bibitem[{{Copperwheat} et~al.(2011){Copperwheat}, {Marsh}, {Littlefair},
  {Dhillon}, {Ramsay}, {Drake}, {G{\"a}nsicke}, {Groot}, {Hakala}, {Koester},
  {Nelemans}, {Roelofs}, {Southworth}, {Steeghs}, and
  {Tulloch}}]{2011MNRAS.410.1113C}
{Copperwheat} CM, {Marsh} TR, {Littlefair} SP, {Dhillon} VS, {Ramsay} G,
  {Drake} AJ, {G{\"a}nsicke} BT, {Groot} PJ, {Hakala} P, {Koester} D, et~al.
  (2011) {SDSS J0926+3624: the shortest period eclipsing binary star}. \mnras
  410(2):1113--1129. \doi{10.1111/j.1365-2966.2010.17508.x}.
  {\href{https://arxiv.org/abs/1008.1907}{{arXiv:1008.1907}}} {[astro-ph.SR]}

\bibitem[{{Cornish} and {Robson}(2017)}]{2017JPhCS.840a2024C}
{Cornish} N, {Robson} T (2017) {Galactic binary science with the new LISA
  design}. In: Journal of Physics Conference Series. Journal of Physics
  Conference Series, vol 840. p 012024. \doi{10.1088/1742-6596/840/1/012024}.
  {\href{https://arxiv.org/abs/1703.09858}{{arXiv:1703.09858}}} {[astro-ph.IM]}

\bibitem[{{Cornish}(2011)}]{2011CQGra..28i4016C}
{Cornish} NJ (2011) {Detection strategies for extreme mass ratio inspirals}.
  Classical and Quantum Gravity 28(9):094016.
  \doi{10.1088/0264-9381/28/9/094016}.
  {\href{https://arxiv.org/abs/0804.3323}{{arXiv:0804.3323}}} {[gr-qc]}

\bibitem[{{Cornish} and {Crowder}(2005)}]{2005PhRvD..72d3005C}
{Cornish} NJ, {Crowder} J (2005) {LISA data analysis using Markov chain Monte
  Carlo methods}. \prd 72(4):043005. \doi{10.1103/PhysRevD.72.043005}.
  {\href{https://arxiv.org/abs/gr-qc/0506059}{{arXiv:gr-qc/0506059}}}
  {[astro-ph]}

\bibitem[{{Cornish} and {Larson}(2003)}]{2003PhRvD..67j3001C}
{Cornish} NJ, {Larson} SL (2003) {LISA data analysis: Source identification and
  subtraction}. \prd 67(10):103001. \doi{10.1103/PhysRevD.67.103001}.
  {\href{https://arxiv.org/abs/astro-ph/0301548}{{arXiv:astro-ph/0301548}}}
  {[astro-ph]}

\bibitem[{{Cornish} and {Littenberg}(2007)}]{2007PhRvD..76h3006C}
{Cornish} NJ, {Littenberg} TB (2007) {Tests of Bayesian model selection
  techniques for gravitational wave astronomy}. \prd 76(8):083006.
  \doi{10.1103/PhysRevD.76.083006}.
  {\href{https://arxiv.org/abs/0704.1808}{{arXiv:0704.1808}}} {[gr-qc]}

\bibitem[{{Cornish} and {Shuman}(2020)}]{2020PhRvD.101l4008C}
{Cornish} NJ, {Shuman} K (2020) {Black hole hunting with LISA}. \prd
  101(12):124008. \doi{10.1103/PhysRevD.101.124008}.
  {\href{https://arxiv.org/abs/2005.03610}{{arXiv:2005.03610}}} {[gr-qc]}

\bibitem[{{Corrales} et~al.(2019){Corrales}, {Mills}, {Heinz}, and
  {Williger}}]{2019ApJ...874..155C}
{Corrales} L, {Mills} BS, {Heinz} S, {Williger} GM (2019) {The X-Ray Variable
  Sky as Seen by MAXI: The Future of Dust-echo Tomography with Bright Galactic
  X-Ray Bursts}. \apj 874(2):155. \doi{10.3847/1538-4357/ab0c9b}.
  {\href{https://arxiv.org/abs/1903.08299}{{arXiv:1903.08299}}} {[astro-ph.HE]}

\bibitem[{{Corrales} et~al.(2010){Corrales}, {Haiman}, and
  {MacFadyen}}]{2010MNRAS.404..947C}
{Corrales} LR, {Haiman} Z, {MacFadyen} A (2010) {Hydrodynamical response of a
  circumbinary gas disc to black hole recoil and mass loss}. \mnras
  404(2):947--962. \doi{10.1111/j.1365-2966.2010.16324.x}.
  {\href{https://arxiv.org/abs/0910.0014}{{arXiv:0910.0014}}} {[astro-ph.HE]}

\bibitem[{{Coughlin} et~al.(2018){Coughlin}, {Dietrich}, {Doctor}, {Kasen},
  {Coughlin}, {Jerkstrand}, {Leloudas}, {McBrien}, {Metzger}, {O'Shaughnessy},
  and {Smartt}}]{2018MNRAS.480.3871C}
{Coughlin} MW, {Dietrich} T, {Doctor} Z, {Kasen} D, {Coughlin} S, {Jerkstrand}
  A, {Leloudas} G, {McBrien} O, {Metzger} BD, {O'Shaughnessy} R, et~al. (2018)
  {Constraints on the neutron star equation of state from AT2017gfo using
  radiative transfer simulations}. \mnras 480(3):3871--3878.
  \doi{10.1093/mnras/sty2174}.
  {\href{https://arxiv.org/abs/1805.09371}{{arXiv:1805.09371}}} {[astro-ph.HE]}

\bibitem[{{Cresswell} et~al.(2007){Cresswell}, {Dirksen}, {Kley}, and
  {Nelson}}]{2007A&A...473..329C}
{Cresswell} P, {Dirksen} G, {Kley} W, {Nelson} RP (2007) {On the evolution of
  eccentric and inclined protoplanets embedded in protoplanetary disks}. \aap
  473(1):329--342. \doi{10.1051/0004-6361:20077666}.
  {\href{https://arxiv.org/abs/0707.2225}{{arXiv:0707.2225}}} {[astro-ph]}

\bibitem[{{Cromartie} et~al.(2020){Cromartie}, {Fonseca}, {Ransom}, {Demorest},
  {Arzoumanian}, {Blumer}, {Brook}, {DeCesar}, {Dolch}, {Ellis}, {Ferdman},
  {Ferrara}, and {et al.}}]{2020NatAs...4...72C}
{Cromartie} HT, {Fonseca} E, {Ransom} SM, {Demorest} PB, {Arzoumanian} Z,
  {Blumer} H, {Brook} PR, {DeCesar} ME, {Dolch} T, {Ellis} JA, et~al. (2020)
  {Relativistic Shapiro delay measurements of an extremely massive millisecond
  pulsar}. Nature Astronomy 4:72--76. \doi{10.1038/s41550-019-0880-2}.
  {\href{https://arxiv.org/abs/1904.06759}{{arXiv:1904.06759}}} {[astro-ph.HE]}

\bibitem[{{Croton} et~al.(2006){Croton}, {Springel}, {White}, {De Lucia},
  {Frenk}, {Gao}, {Jenkins}, {Kauffmann}, {Navarro}, and
  {Yoshida}}]{2006MNRAS.365...11C}
{Croton} DJ, {Springel} V, {White} SDM, {De Lucia} G, {Frenk} CS, {Gao} L,
  {Jenkins} A, {Kauffmann} G, {Navarro} JF, {Yoshida} N (2006) {The many lives
  of active galactic nuclei: cooling flows, black holes and the luminosities
  and colours of galaxies}. \mnras 365(1):11--28.
  \doi{10.1111/j.1365-2966.2005.09675.x}.
  {\href{https://arxiv.org/abs/astro-ph/0508046}{{arXiv:astro-ph/0508046}}}
  {[astro-ph]}

\bibitem[{{Cuadra} et~al.(2009){Cuadra}, {Armitage}, {Alexander}, and
  {Begelman}}]{2009MNRAS.393.1423C}
{Cuadra} J, {Armitage} PJ, {Alexander} RD, {Begelman} MC (2009) {Massive black
  hole binary mergers within subparsec scale gas discs}. \mnras
  393(4):1423--1432. \doi{10.1111/j.1365-2966.2008.14147.x}.
  {\href{https://arxiv.org/abs/0809.0311}{{arXiv:0809.0311}}} {[astro-ph]}

\bibitem[{{Cunningham} et~al.(2020){Cunningham}, {Garavito-Camargo}, {Deason},
  {Johnston}, {Erkal}, {Laporte}, {Besla}, {Luger}, and
  {Sanderson}}]{2020ApJ...898....4C}
{Cunningham} EC, {Garavito-Camargo} N, {Deason} AJ, {Johnston} KV, {Erkal} D,
  {Laporte} CFP, {Besla} G, {Luger} R, {Sanderson} RE (2020) {Quantifying the
  Stellar Halo's Response to the LMC's Infall with Spherical Harmonics}. \apj
  898(1):4. \doi{10.3847/1538-4357/ab9b88}.
  {\href{https://arxiv.org/abs/2006.08621}{{arXiv:2006.08621}}} {[astro-ph.GA]}

\bibitem[{{Cusin} et~al.(2020){Cusin}, {Dvorkin}, {Pitrou}, and
  {Uzan}}]{2020MNRAS.493L...1C}
{Cusin} G, {Dvorkin} I, {Pitrou} C, {Uzan} JP (2020) {Stochastic gravitational
  wave background anisotropies in the mHz band: astrophysical dependencies}.
  \mnras 493(1):L1--L5. \doi{10.1093/mnrasl/slz182}.
  {\href{https://arxiv.org/abs/1904.07757}{{arXiv:1904.07757}}} {[astro-ph.CO]}

\bibitem[{{Cutler} and {Flanagan}(1994)}]{1994PhRvD..49.2658C}
{Cutler} C, {Flanagan} {\'E}E (1994) {Gravitational waves from merging compact
  binaries: How accurately can one extract the binary's parameters from the
  inspiral waveform?} \prd 49(6):2658--2697. \doi{10.1103/PhysRevD.49.2658}.
  {\href{https://arxiv.org/abs/gr-qc/9402014}{{arXiv:gr-qc/9402014}}} {[gr-qc]}

\bibitem[{{Cutler} et~al.(2019){Cutler}, {Berti}, {Holley-Bockelmann}, {Jani},
  {Kovetz}, {Larson}, {Littenberg}, {McWilliams}, {Mueller}, {Randall},
  {Schnittman}, {Shoemaker}, {Vallisneri}, {Vitale}, and
  {Wong}}]{2019BAAS...51c.109C}
{Cutler} C, {Berti} E, {Holley-Bockelmann} K, {Jani} K, {Kovetz} ED, {Larson}
  SL, {Littenberg} T, {McWilliams} ST, {Mueller} G, {Randall} L, et~al. (2019)
  {What can we learn from multi-band observations of black hole binaries?}
  \baas 51(3):109.
  {\href{https://arxiv.org/abs/1903.04069}{{arXiv:1903.04069}}} {[astro-ph.HE]}

\bibitem[{{Dabringhausen} et~al.(2009){Dabringhausen}, {Kroupa}, and
  {Baumgardt}}]{2009MNRAS.394.1529D}
{Dabringhausen} J, {Kroupa} P, {Baumgardt} H (2009) {A top-heavy stellar
  initial mass function in starbursts as an explanation for the high
  mass-to-light ratios of ultra-compact dwarf galaxies}. \mnras
  394(3):1529--1543. \doi{10.1111/j.1365-2966.2009.14425.x}.
  {\href{https://arxiv.org/abs/0901.0915}{{arXiv:0901.0915}}} {[astro-ph.GA]}

\bibitem[{{Dage} et~al.(2019){Dage}, {Zepf}, {Bahramian}, {Strader},
  {Maccarone}, {Peacock}, {Kundu}, {Steele}, and {Britt}}]{2019MNRAS.489.4783D}
{Dage} KC, {Zepf} SE, {Bahramian} A, {Strader} J, {Maccarone} TJ, {Peacock} MB,
  {Kundu} A, {Steele} MM, {Britt} CT (2019) {Slow decline and rise of the broad
  [O III] emission line in globular cluster black hole candidate RZ2109}.
  \mnras 489(4):4783--4790. \doi{10.1093/mnras/stz2514}.
  {\href{https://arxiv.org/abs/1909.02683}{{arXiv:1909.02683}}} {[astro-ph.HE]}

\bibitem[{{Dai} et~al.(2018){Dai}, {McKinney}, {Roth}, {Ramirez-Ruiz}, and
  {Miller}}]{2018ApJ...859L..20D}
{Dai} L, {McKinney} JC, {Roth} N, {Ramirez-Ruiz} E, {Miller} MC (2018) {A
  Unified Model for Tidal Disruption Events}. \apjl 859(2):L20.
  \doi{10.3847/2041-8213/aab429}.
  {\href{https://arxiv.org/abs/1803.03265}{{arXiv:1803.03265}}} {[astro-ph.HE]}

\bibitem[{{Dal Canton} et~al.(2019){Dal Canton}, Mangiagli, Noble, Schnittman,
  Ptak, Klein, Sesana, and Camp}]{2019ApJ...886..146D}
{Dal Canton} T, Mangiagli A, Noble SC, Schnittman J, Ptak A, Klein A, Sesana A,
  Camp J (2019) Detectability of modulated x-rays from {LISA}'s supermassive
  black hole mergers. ApJ 886(2):146. \doi{10.3847/1538-4357/ab505a},
  \urlprefix\url{https://doi.org/10.3847\%2F1538-4357\%2Fab505a}

\bibitem[{{Dall'Osso} and {Rossi}(2013)}]{2013MNRAS.428..518D}
{Dall'Osso} S, {Rossi} EM (2013) {Tidal torque induced by orbital decay in
  compact object binaries}. \mnras 428(1):518--531. \doi{10.1093/mnras/sts037}.
  {\href{https://arxiv.org/abs/1203.3440}{{arXiv:1203.3440}}} {[astro-ph.HE]}

\bibitem[{{Dall'Osso} and {Rossi}(2014)}]{2014MNRAS.443.1057D}
{Dall'Osso} S, {Rossi} EM (2014) {Constraining white dwarf viscosity through
  tidal heating in detached binary systems}. \mnras 443(2):1057--1064.
  \doi{10.1093/mnras/stu901}.
  {\href{https://arxiv.org/abs/1308.1664}{{arXiv:1308.1664}}} {[astro-ph.HE]}

\bibitem[{{Daly}(2011)}]{2011MNRAS.414.1253D}
{Daly} RA (2011) {Estimates of black hole spin properties of 55 sources}.
  \mnras 414(2):1253--1262. \doi{10.1111/j.1365-2966.2011.18452.x}.
  {\href{https://arxiv.org/abs/1103.0940}{{arXiv:1103.0940}}} {[astro-ph.CO]}

\bibitem[{{Damour} and {Gopakumar}(2006)}]{2006PhRvD..73l4006D}
{Damour} T, {Gopakumar} A (2006) {Gravitational recoil during binary black hole
  coalescence using the effective one body approach}. \prd 73(12):124006.
  \doi{10.1103/PhysRevD.73.124006}.
  {\href{https://arxiv.org/abs/gr-qc/0602117}{{arXiv:gr-qc/0602117}}} {[gr-qc]}

\bibitem[{{Dan} et~al.(2011){Dan}, {Rosswog}, {Guillochon}, and
  {Ramirez-Ruiz}}]{2011ApJ...737...89D}
{Dan} M, {Rosswog} S, {Guillochon} J, {Ramirez-Ruiz} E (2011) {Prelude to A
  Double Degenerate Merger: The Onset of Mass Transfer and Its Impact on
  Gravitational Waves and Surface Detonations}. \apj 737(2):89.
  \doi{10.1088/0004-637X/737/2/89}.
  {\href{https://arxiv.org/abs/1101.5132}{{arXiv:1101.5132}}} {[astro-ph.HE]}

\bibitem[{{D'Angelo} et~al.(2006){D'Angelo}, {Lubow}, and
  {Bate}}]{2006ApJ...652.1698D}
{D'Angelo} G, {Lubow} SH, {Bate} MR (2006) {Evolution of Giant Planets in
  Eccentric Disks}. \apj 652(2):1698--1714. \doi{10.1086/508451}.
  {\href{https://arxiv.org/abs/astro-ph/0608355}{{arXiv:astro-ph/0608355}}}
  {[astro-ph]}

\bibitem[{{Danielski} and {Tamanini}(2020)}]{2020arXiv200707010D}
{Danielski} C, {Tamanini} N (2020) {Will Gravitational Waves Discover the First
  Extra-Galactic Planetary System?} arXiv e-prints arXiv:2007.07010.
  {\href{https://arxiv.org/abs/2007.07010}{{arXiv:2007.07010}}} {[astro-ph.IM]}

\bibitem[{{Danielski} et~al.(2019){Danielski}, {Korol}, {Tamanini}, and
  {Rossi}}]{2019A&A...632A.113D}
{Danielski} C, {Korol} V, {Tamanini} N, {Rossi} EM (2019) {Circumbinary
  exoplanets and brown dwarfs with the Laser Interferometer Space Antenna}.
  \aap 632:A113. \doi{10.1051/0004-6361/201936729}.
  {\href{https://arxiv.org/abs/1910.05414}{{arXiv:1910.05414}}} {[astro-ph.EP]}

\bibitem[{d'Ascoli et~al.(2018)}]{2018ApJ...865..140D}
d'Ascoli S, et~al. (2018) Electromagnetic emission from supermassive binary
  black holes approaching merger. ApJ 865:140. \doi{10.3847/1538-4357/aad8b4}.
  {\href{https://arxiv.org/abs/1806.05697}{{arXiv:1806.05697}}} {[astro-ph.HE]}

\bibitem[{{Datta} et~al.(2020){Datta}, {Gupta}, {Kastha}, {Arun}, and
  {Sathyaprakash}}]{2020arXiv200612137D}
{Datta} S, {Gupta} A, {Kastha} S, {Arun} KG, {Sathyaprakash} BS (2020) {Tests
  of general relativity using multiband observations of intermediate mass
  binary black hole mergers}. arXiv e-prints arXiv:2006.12137.
  {\href{https://arxiv.org/abs/2006.12137}{{arXiv:2006.12137}}} {[gr-qc]}

\bibitem[{{Dav{\'e}} et~al.(2019){Dav{\'e}}, {Angl{\'e}s-Alc{\'a}zar},
  {Narayanan}, {Li}, {Rafieferantsoa}, and {Appleby}}]{2019MNRAS.486.2827D}
{Dav{\'e}} R, {Angl{\'e}s-Alc{\'a}zar} D, {Narayanan} D, {Li} Q,
  {Rafieferantsoa} MH, {Appleby} S (2019) {SIMBA: Cosmological simulations with
  black hole growth and feedback}. \mnras 486(2):2827--2849.
  \doi{10.1093/mnras/stz937}.
  {\href{https://arxiv.org/abs/1901.10203}{{arXiv:1901.10203}}} {[astro-ph.GA]}

\bibitem[{{Davies} and {Lin}(2020)}]{2020MNRAS.tmp.2525D}
{Davies} MB, {Lin} DNC (2020) {Making massive stars in the Galactic Centre via
  accretion onto low-mass stars within an accretion disc}. \mnras
  \doi{10.1093/mnras/staa2590}.
  {\href{https://arxiv.org/abs/2008.10033}{{arXiv:2008.10033}}} {[astro-ph.GA]}

\bibitem[{{Davies} et~al.(2011){Davies}, {Miller}, and
  {Bellovary}}]{2011ApJ...740L..42D}
{Davies} MB, {Miller} MC, {Bellovary} JM (2011) {Supermassive Black Hole
  Formation Via Gas Accretion in Nuclear Stellar Clusters}. \apjl 740(2):L42.
  \doi{10.1088/2041-8205/740/2/L42}.
  {\href{https://arxiv.org/abs/1106.5943}{{arXiv:1106.5943}}} {[astro-ph.CO]}

\bibitem[{{Davies} et~al.(2007){Davies}, {M{\"u}ller S{\'a}nchez}, {Genzel},
  {Tacconi}, {Hicks}, {Friedrich}, and {Sternberg}}]{2007ApJ...671.1388D}
{Davies} RI, {M{\"u}ller S{\'a}nchez} F, {Genzel} R, {Tacconi} LJ, {Hicks} EKS,
  {Friedrich} S, {Sternberg} A (2007) {A Close Look at Star Formation around
  Active Galactic Nuclei}. \apj 671(2):1388--1412. \doi{10.1086/523032}.
  {\href{https://arxiv.org/abs/0704.1374}{{arXiv:0704.1374}}} {[astro-ph]}

\bibitem[{{Davis} et~al.(2017){Davis}, {Graham}, and
  {Seigar}}]{2017MNRAS.471.2187D}
{Davis} BL, {Graham} AW, {Seigar} MS (2017) {Updating the (supermassive black
  hole mass)-(spiral arm pitch angle) relation: a strong correlation for
  galaxies with pseudobulges}. \mnras 471(2):2187--2203.
  \doi{10.1093/mnras/stx1794}.
  {\href{https://arxiv.org/abs/1707.04001}{{arXiv:1707.04001}}} {[astro-ph.GA]}

\bibitem[{{Davis} et~al.(2018){Davis}, {Graham}, and
  {Cameron}}]{2018ApJ...869..113D}
{Davis} BL, {Graham} AW, {Cameron} E (2018) {Black Hole Mass Scaling Relations
  for Spiral Galaxies. II. M $_{BH}$-M $_{*,tot}$ and M $_{BH}$-M $_{*,disk}$}.
  \apj 869(2):113. \doi{10.3847/1538-4357/aae820}.
  {\href{https://arxiv.org/abs/1810.04888}{{arXiv:1810.04888}}} {[astro-ph.GA]}

\bibitem[{{Davis} et~al.(2019{\natexlab{a}}){Davis}, {Graham}, and
  {Cameron}}]{2019ApJ...873...85D}
{Davis} BL, {Graham} AW, {Cameron} E (2019{\natexlab{a}}) {Black Hole Mass
  Scaling Relations for Spiral Galaxies. I. M $_{BH}$-M $_{*,sph}$}. \apj
  873(1):85. \doi{10.3847/1538-4357/aaf3b8}.
  {\href{https://arxiv.org/abs/1810.04887}{{arXiv:1810.04887}}} {[astro-ph.GA]}

\bibitem[{{Davis} et~al.(2019{\natexlab{b}}){Davis}, {Graham}, and
  {Combes}}]{2019ApJ...877...64D}
{Davis} BL, {Graham} AW, {Combes} F (2019{\natexlab{b}}) {A Consistent Set of
  Empirical Scaling Relations for Spiral Galaxies: The (v $_{max}$, M
  $_{oM}$)-({\ensuremath{\sigma}} $_{0}$, M $_{BH}$, {\ensuremath{\phi}})
  Relations}. \apj 877(1):64. \doi{10.3847/1538-4357/ab1aa4}.
  {\href{https://arxiv.org/abs/1901.06509}{{arXiv:1901.06509}}} {[astro-ph.GA]}

\bibitem[{{Dayal} and {Ferrara}(2018)}]{2018PhR...780....1D}
{Dayal} P, {Ferrara} A (2018) {Early galaxy formation and its large-scale
  effects}. \physrep 780:1--64. \doi{10.1016/j.physrep.2018.10.002}.
  {\href{https://arxiv.org/abs/1809.09136}{{arXiv:1809.09136}}} {[astro-ph.GA]}

\bibitem[{{Dayal} et~al.(2019){Dayal}, {Rossi}, {Shiralilou}, {Piana},
  {Choudhury}, and {Volonteri}}]{2019MNRAS.486.2336D}
{Dayal} P, {Rossi} EM, {Shiralilou} B, {Piana} O, {Choudhury} TR, {Volonteri} M
  (2019) {The hierarchical assembly of galaxies and black holes in the first
  billion years: predictions for the era of gravitational wave astronomy}.
  \mnras 486(2):2336--2350. \doi{10.1093/mnras/stz897}.
  {\href{https://arxiv.org/abs/1810.11033}{{arXiv:1810.11033}}} {[astro-ph.GA]}

\bibitem[{{De} et~al.(2020){De}, {MacLeod}, {Everson}, {Antoni}, {Mandel}, and
  {Ramirez-Ruiz}}]{2020ApJ...897..130D}
{De} S, {MacLeod} M, {Everson} RW, {Antoni} A, {Mandel} I, {Ramirez-Ruiz} E
  (2020) {Common Envelope Wind Tunnel: The Effects of Binary Mass Ratio and
  Implications for the Accretion-driven Growth of LIGO Binary Black Holes}.
  \apj 897(2):130. \doi{10.3847/1538-4357/ab9ac6}.
  {\href{https://arxiv.org/abs/1910.13333}{{arXiv:1910.13333}}} {[astro-ph.SR]}

\bibitem[{De~Luca et~al.(2020)De~Luca, Franciolini, Pani, and
  Riotto}]{DeLuca:2020fpg}
De~Luca V, Franciolini G, Pani P, Riotto A (2020) {Constraints on Primordial
  Black Holes: the Importance of Accretion}. Phys Rev D 102(4):043505.
  \doi{10.1103/PhysRevD.102.043505}.
  {\href{https://arxiv.org/abs/2003.12589}{{arXiv:2003.12589}}} {[astro-ph.CO]}

\bibitem[{{De Luca} et~al.(2020){De Luca}, {Franciolini}, and
  {Riotto}}]{2020arXiv200908268D}
{De Luca} V, {Franciolini} G, {Riotto} A (2020) {NANOGrav Hints to Primordial
  Black Holes as Dark Matter}. arXiv e-prints arXiv:2009.08268.
  {\href{https://arxiv.org/abs/2009.08268}{{arXiv:2009.08268}}} {[astro-ph.CO]}

\bibitem[{{De Marco} et~al.(2011){De Marco}, {Passy}, {Moe}, {Herwig}, {Mac
  Low}, and {Paxton}}]{2011MNRAS.411.2277D}
{De Marco} O, {Passy} JC, {Moe} M, {Herwig} F, {Mac Low} MM, {Paxton} B (2011)
  {On the {\ensuremath{\alpha}} formalism for the common envelope interaction}.
  \mnras 411(4):2277--2292. \doi{10.1111/j.1365-2966.2010.17891.x}.
  {\href{https://arxiv.org/abs/1010.4374}{{arXiv:1010.4374}}} {[astro-ph.SR]}

\bibitem[{{de Mink} et~al.(2009){de Mink}, {Cantiello}, {Langer}, {Pols},
  {Brott}, and {Yoon}}]{2009A&A...497..243D}
{de Mink} SE, {Cantiello} M, {Langer} N, {Pols} OR, {Brott} I, {Yoon} SC (2009)
  {Rotational mixing in massive binaries. Detached short-period systems}. \aap
  497(1):243--253. \doi{10.1051/0004-6361/200811439}.
  {\href{https://arxiv.org/abs/0902.1751}{{arXiv:0902.1751}}} {[astro-ph.SR]}

\bibitem[{{De Rosa} et~al.(2019{\natexlab{a}}){De Rosa}, {Uttley}, {Gou},
  {Liu}, {Bambi}, {Barret}, {Belloni}, {Berti}, {Bianchi}, {Caiazzo},
  {Casella}, {Feroci}, {Ferrari}, {Gualtieri}, {Heyl}, {Ingram}, {Karas}, {Lu},
  {Luo}, {Matt}, {Motta}, {Neilsen}, {Pani}, {Santangelo}, {Shu}, {Wang},
  {Wang}, {Xue}, {Xu}, {Yuan}, {Yuan}, {Zhang}, {Zhang}, {Agudo}, {Amati},
  {Andersson}, {Baglio}, {Bakala}, {Baykal}, {Bhattacharyya}, {Bombaci},
  {Bucciantini}, {Capitanio}, {Ciolfi}, {Cui}, {D'Ammand o}, {Dauser}, {Del
  Santo}, {De Marco}, {Di Salvo}, {Done}, {Dov{\v{c}}iak}, {Fabian}, {Falanga},
  {Gambino}, {Gendre}, {Grinberg}, {Heger}, {Homan}, {Iaria}, {Jiang}, {Jin},
  {Koerding}, {Linares}, {Liu}, {Maccarone}, {Malzac}, {Manousakis}, {Marin},
  {Marinucci}, {Mehdipour}, {M{\'e}ndez}, {Migliari}, {Miller}, {Miniutti},
  {Nardini}, {O'Brien}, {Osborne}, {Petrucci}, {Possenti}, {Riggio},
  {Rodriguez}, {Sanna}, {Shao}, {Sobolewska}, {Sramkova}, {Stevens}, {Stiele},
  {Stratta}, {Stuchlik}, {Svoboda}, {Tamburini}, {Tauris}, {Tombesi}, {Torok},
  {Urbanec}, {Vincent}, {Wu}, {Yuan}, {in't Zand }, {Zdziarski}, and
  {Zhou}}]{2019SCPMA..6229504D}
{De Rosa} A, {Uttley} P, {Gou} L, {Liu} Y, {Bambi} C, {Barret} D, {Belloni} T,
  {Berti} E, {Bianchi} S, {Caiazzo} I, et~al. (2019{\natexlab{a}}) {Accretion
  in strong field gravity with eXTP}. Science China Physics, Mechanics, and
  Astronomy 62(2):29504. \doi{10.1007/s11433-018-9297-0}.
  {\href{https://arxiv.org/abs/1812.04022}{{arXiv:1812.04022}}} {[astro-ph.HE]}

\bibitem[{{De Rosa} et~al.(2019{\natexlab{b}}){De Rosa}, {Vignali},
  {Bogdanovi{\'c}}, {Capelo}, {Charisi}, {Dotti}, {Husemann}, {Lusso}, {Mayer},
  {Paragi}, {Runnoe}, {Sesana}, {Steinborn}, {Bianchi}, {Colpi}, {del Valle},
  {Frey}, {Gab{\'a}nyi}, {Giustini}, {Guainazzi}, {Haiman}, {Herrera Ruiz},
  {Herrero-Illana}, {Iwasawa}, {Komossa}, {Lena}, {Loiseau}, {Perez-Torres},
  {Piconcelli}, and {Volonteri}}]{2019NewAR..8601525D}
{De Rosa} A, {Vignali} C, {Bogdanovi{\'c}} T, {Capelo} PR, {Charisi} M, {Dotti}
  M, {Husemann} B, {Lusso} E, {Mayer} L, {Paragi} Z, et~al.
  (2019{\natexlab{b}}) {The quest for dual and binary supermassive black holes:
  A multi-messenger view}. \nar 86:101525. \doi{10.1016/j.newar.2020.101525}.
  {\href{https://arxiv.org/abs/2001.06293}{{arXiv:2001.06293}}} {[astro-ph.GA]}

\bibitem[{{de Val-Borro} et~al.(2017){de Val-Borro}, {Karovska}, {Sasselov},
  and {Stone}}]{2017MNRAS.468.3408D}
{de Val-Borro} M, {Karovska} M, {Sasselov} DD, {Stone} JM (2017)
  {Three-dimensional hydrodynamical models of wind and outburst-related
  accretion in symbiotic systems}. \mnras 468(3):3408--3417.
  \doi{10.1093/mnras/stx684}.
  {\href{https://arxiv.org/abs/1704.03460}{{arXiv:1704.03460}}} {[astro-ph.SR]}

\bibitem[{{Deane} et~al.(2014){Deane}, {Paragi}, {Jarvis}, {Coriat},
  {Bernardi}, {Fender}, {Frey}, {Heywood}, {Kl{\"o}ckner}, {Grainge}, and
  {Rumsey}}]{2014Natur.511...57D}
{Deane} RP, {Paragi} Z, {Jarvis} MJ, {Coriat} M, {Bernardi} G, {Fender} RP,
  {Frey} S, {Heywood} I, {Kl{\"o}ckner} HR, {Grainge} K, et~al. (2014) {A
  close-pair binary in a distant triple supermassive black hole system}. \nat
  511(7507):57--60. \doi{10.1038/nature13454}.
  {\href{https://arxiv.org/abs/1406.6365}{{arXiv:1406.6365}}} {[astro-ph.GA]}

\bibitem[{{Decarli} et~al.(2013){Decarli}, {Dotti}, {Fumagalli}, {Tsalmantza},
  {Montuori}, {Lusso}, {Hogg}, and {Prochaska}}]{2013MNRAS.433.1492D}
{Decarli} R, {Dotti} M, {Fumagalli} M, {Tsalmantza} P, {Montuori} C, {Lusso} E,
  {Hogg} DW, {Prochaska} JX (2013) {The nature of massive black hole binary
  candidates - I. Spectral properties and evolution}. \mnras 433(2):1492--1504.
  \doi{10.1093/mnras/stt831}.
  {\href{https://arxiv.org/abs/1305.4941}{{arXiv:1305.4941}}} {[astro-ph.CO]}

\bibitem[{{Decarli} et~al.(2014){Decarli}, {Dotti}, {Mazzucchelli}, {Montuori},
  and {Volonteri}}]{2014MNRAS.445.1558D}
{Decarli} R, {Dotti} M, {Mazzucchelli} C, {Montuori} C, {Volonteri} M (2014)
  {New insights on the recoiling/binary black hole candidate J0927+2943 via
  molecular gas observations}. \mnras 445(2):1558--1566.
  \doi{10.1093/mnras/stu1810}.
  {\href{https://arxiv.org/abs/1409.1585}{{arXiv:1409.1585}}} {[astro-ph.GA]}

\bibitem[{{Decarli} et~al.(2018){Decarli}, {Walter}, {Venemans}, {Ba{\~n}ados},
  {Bertoldi}, {Carilli}, {Fan}, {Farina}, {Mazzucchelli}, {Riechers}, {Rix},
  {Strauss}, {Wang}, and {Yang}}]{2018ApJ...854...97D}
{Decarli} R, {Walter} F, {Venemans} BP, {Ba{\~n}ados} E, {Bertoldi} F,
  {Carilli} C, {Fan} X, {Farina} EP, {Mazzucchelli} C, {Riechers} D, et~al.
  (2018) {An ALMA [C II] Survey of 27 Quasars at z > 5.94}. \apj 854(2):97.
  \doi{10.3847/1538-4357/aaa5aa}.
  {\href{https://arxiv.org/abs/1801.02641}{{arXiv:1801.02641}}} {[astro-ph.GA]}

\bibitem[{{Decarli} et~al.(2020){Decarli}, {Aravena}, {Boogaard}, {Carilli},
  {Gonz{\'a}lez-L{\'o}pez}, {Walter}, {Cortes}, {Cox}, {da Cunha}, {Daddi},
  {D{\'\i}az-Santos}, {Hodge}, {Inami}, {Neeleman}, {Novak}, {Oesch},
  {Popping}, {Riechers}, {Smail}, {Uzgil}, {van der Werf}, {Wagg}, and
  {Weiss}}]{2020ApJ...902..110D}
{Decarli} R, {Aravena} M, {Boogaard} L, {Carilli} C, {Gonz{\'a}lez-L{\'o}pez}
  J, {Walter} F, {Cortes} PC, {Cox} P, {da Cunha} E, {Daddi} E, et~al. (2020)
  {The ALMA Spectroscopic Survey in the Hubble Ultra Deep Field: Multiband
  Constraints on Line-luminosity Functions and the Cosmic Density of Molecular
  Gas}. \apj 902(2):110. \doi{10.3847/1538-4357/abaa3b}.
  {\href{https://arxiv.org/abs/2009.10744}{{arXiv:2009.10744}}} {[astro-ph.GA]}

\bibitem[{{DeGraf} and {Sijacki}(2020)}]{2020MNRAS.491.4973D}
{DeGraf} C, {Sijacki} D (2020) {Cosmological simulations of massive black hole
  seeds: predictions for next-generation electromagnetic and gravitational wave
  observations}. \mnras 491(4):4973--4992. \doi{10.1093/mnras/stz3309}.
  {\href{https://arxiv.org/abs/1906.11271}{{arXiv:1906.11271}}} {[astro-ph.GA]}

\bibitem[{{DeGraf} et~al.(2020){DeGraf}, {Sijacki}, {Di Matteo},
  {Holley-Bockelmann}, {Snyder}, and {Springel}}]{2020arXiv201200775D}
{DeGraf} C, {Sijacki} D, {Di Matteo} T, {Holley-Bockelmann} K, {Snyder} G,
  {Springel} V (2020) {Morphological evolution of supermassive black hole
  merger hosts and multimessenger signatures}. arXiv e-prints arXiv:2012.00775.
  {\href{https://arxiv.org/abs/2012.00775}{{arXiv:2012.00775}}} {[astro-ph.GA]}

\bibitem[{{del Valle} and {Volonteri}(2018)}]{2018MNRAS.480..439D}
{del Valle} L, {Volonteri} M (2018) {The effect of AGN feedback on the
  migration time-scale of supermassive black holes binaries}. \mnras
  480(1):439--450. \doi{10.1093/mnras/sty1815}.
  {\href{https://arxiv.org/abs/1807.03844}{{arXiv:1807.03844}}} {[astro-ph.GA]}

\bibitem[{{del Valle} et~al.(2015){del Valle}, {Escala}, {Maureira-Fredes},
  {Molina}, {Cuadra}, and {Amaro-Seoane}}]{2015ApJ...811...59D}
{del Valle} L, {Escala} A, {Maureira-Fredes} C, {Molina} J, {Cuadra} J,
  {Amaro-Seoane} P (2015) {Supermassive Black Holes in a Star-forming Gaseous
  Circumnuclear Disk}. \apj 811(1):59. \doi{10.1088/0004-637X/811/1/59}.
  {\href{https://arxiv.org/abs/1503.01664}{{arXiv:1503.01664}}} {[astro-ph.GA]}

\bibitem[{{Deloye} and {Taam}(2006)}]{2006ApJ...649L..99D}
{Deloye} CJ, {Taam} RE (2006) {The Turn-On of Mass Transfer in AM CVn Binaries:
  Implications for RX J0806+1527 and RX J1914+2456}. \apjl 649(2):L99--L102.
  \doi{10.1086/508372}.
  {\href{https://arxiv.org/abs/astro-ph/0608442}{{arXiv:astro-ph/0608442}}}
  {[astro-ph]}

\bibitem[{{Deme} et~al.(2020{\natexlab{a}}){Deme}, {Hoang}, {Naoz}, and
  {Kocsis}}]{2020ApJ...901..125D}
{Deme} B, {Hoang} BM, {Naoz} S, {Kocsis} B (2020{\natexlab{a}}) {Detecting
  Kozai-Lidov Imprints on the Gravitational Waves of Intermediate-mass Black
  Holes in Galactic Nuclei}. \apj 901(2):125. \doi{10.3847/1538-4357/abafa3}.
  {\href{https://arxiv.org/abs/2005.03677}{{arXiv:2005.03677}}} {[astro-ph.HE]}

\bibitem[{{Deme} et~al.(2020{\natexlab{b}}){Deme}, {Meiron}, and
  {Kocsis}}]{2020ApJ...892..130D}
{Deme} B, {Meiron} Y, {Kocsis} B (2020{\natexlab{b}}) {Intermediate-mass Black
  Holes' Effects on Compact Object Binaries}. \apj 892(2):130.
  \doi{10.3847/1538-4357/ab7921}.
  {\href{https://arxiv.org/abs/1909.04678}{{arXiv:1909.04678}}} {[astro-ph.GA]}

\bibitem[{{Derdzinski} et~al.(2020){Derdzinski}, {D'Orazio}, {Duffell},
  {Haiman}, and {Macfadyen}}]{2020arXiv200511333D}
{Derdzinski} A, {D'Orazio} D, {Duffell} P, {Haiman} Z, {Macfadyen} A (2020)
  {Evolution of gas disc-embedded intermediate mass ratio inspirals in the LISA
  band}. arXiv e-prints arXiv:2005.11333.
  {\href{https://arxiv.org/abs/2005.11333}{{arXiv:2005.11333}}} {[astro-ph.HE]}

\bibitem[{{Derdzinski} et~al.(2019){Derdzinski}, {D'Orazio}, {Duffell},
  {Haiman}, and {MacFadyen}}]{2019MNRAS.486.2754D}
{Derdzinski} AM, {D'Orazio} D, {Duffell} P, {Haiman} Z, {MacFadyen} A (2019)
  {Probing gas disc physics with LISA: simulations of an intermediate mass
  ratio inspiral in an accretion disc}. \mnras 486(2):2754--2765.
  \doi{10.1093/mnras/stz1026}.
  {\href{https://arxiv.org/abs/1810.03623}{{arXiv:1810.03623}}} {[astro-ph.HE]}

\bibitem[{{DESI Collaboration} et~al.(2016){DESI Collaboration}, {Aghamousa},
  {Aguilar}, {Ahlen}, {Alam}, {Allen}, {Allende Prieto}, {Annis}, {Bailey},
  {Balland}, {Ballester}, {Baltay}, {Beaufore}, {Bebek}, {Beers}, {Bell},
  {Bernal}, {Besuner}, {Beutler}, {Blake}, {Bleuler}, {Blomqvist}, {Blum},
  {Bolton}, {Briceno}, {Brooks}, {Brownstein}, {Buckley-Geer}, {Burden},
  {Burtin}, {Busca}, {Cahn}, {Cai}, {Cardiel-Sas}, {Carlberg}, {Carton},
  {Casas}, {Castander}, {Cervantes-Cota}, {Claybaugh}, {Close}, {Coker},
  {Cole}, {Comparat}, {Cooper}, {Cousinou}, {Crocce}, {Cuby}, {Cunningham},
  {Davis}, {Dawson}, {de la Macorra}, {De Vicente}, {Delubac}, {Derwent},
  {Dey}, {Dhungana}, {Ding}, {Doel}, {Duan}, {Ealet}, {Edelstein},
  {Eftekharzadeh}, {Eisenstein}, {Elliott}, {Escoffier}, {Evatt}, {Fagrelius},
  {Fan}, {Fanning}, {Farahi}, {Farihi}, {Favole}, {Feng}, {Fernandez},
  {Findlay}, {Finkbeiner}, {Fitzpatrick}, {Flaugher}, {Flender}, {Font-Ribera},
  {Forero-Romero}, {Fosalba}, {Frenk}, {Fumagalli}, {Gaensicke}, {Gallo},
  {Garcia-Bellido}, {Gaztanaga}, {Pietro Gentile Fusillo}, {Gerard},
  {Gershkovich}, {Giannantonio}, {Gillet}, {Gonzalez-de-Rivera},
  {Gonzalez-Perez}, {Gott}, {Graur}, {Gutierrez}, {Guy}, {Habib}, {Heetderks},
  {Heetderks}, {Heitmann}, {Hellwing}, {Herrera}, {Ho}, {Holland}, {Honscheid},
  {Huff}, {Hutchinson}, {Huterer}, {Hwang}, {Illa Laguna}, {Ishikawa},
  {Jacobs}, {Jeffrey}, {Jelinsky}, {Jennings}, {Jiang}, {Jimenez}, {Johnson},
  {Joyce}, {Jullo}, {Juneau}, {Kama}, {Karcher}, {Karkar}, {Kehoe}, {Kennamer},
  {Kent}, {Kilbinger}, {Kim}, {Kirkby}, {Kisner}, {Kitanidis}, {Kneib},
  {Koposov}, {Kovacs}, {Koyama}, {Kremin}, {Kron}, {Kronig}, {Kueter-Young},
  {Lacey}, {Lafever}, {Lahav}, {Lambert}, {Lampton}, {Landriau}, {Lang},
  {Lauer}, {Le Goff}, {Le Guillou}, {Le Van Suu}, {Lee}, {Lee}, {Leitner},
  {Lesser}, {Levi}, {L'Huillier}, {Li}, {Liang}, {Lin}, {Linder}, {Loebman},
  {Luki{\'c}}, {Ma}, {MacCrann}, {Magneville}, {Makarem}, {Manera}, {Manser},
  {Marshall}, {Martini}, {Massey}, {Matheson}, {McCauley}, {McDonald},
  {McGreer}, {Meisner}, {Metcalfe}, {Miller}, {Miquel}, {Moustakas}, {Myers},
  {Naik}, {Newman}, {Nichol}, {Nicola}, {Nicolati da Costa}, {Nie}, {Niz},
  {Norberg}, {Nord}, {Norman}, {Nugent}, {O'Brien}, {Oh}, {Olsen}, {Padilla},
  {Padmanabhan}, {Padmanabhan}, {Palanque-Delabrouille}, {Palmese},
  {Pappalardo}, {P{\^a}ris}, {Park}, {Patej}, {Peacock}, {Peiris}, {Peng},
  {Percival}, {Perruchot}, {Pieri}, {Pogge}, {Pollack}, {Poppett}, {Prada},
  {Prakash}, {Probst}, {Rabinowitz}, {Raichoor}, {Ree}, {Refregier}, {Regal},
  {Reid}, {Reil}, {Rezaie}, {Rockosi}, {Roe}, {Ronayette}, {Roodman}, {Ross},
  {Ross}, {Rossi}, {Rozo}, {Ruhlmann-Kleider}, {Rykoff}, {Sabiu}, {Samushia},
  {Sanchez}, {Sanchez}, {Schlegel}, {Schneider}, {Schubnell}, {Secroun},
  {Seljak}, {Seo}, {Serrano}, {Shafieloo}, {Shan}, {Sharples}, {Sholl},
  {Shourt}, {Silber}, {Silva}, {Sirk}, {Slosar}, {Smith}, {Smoot}, {Som},
  {Song}, {Sprayberry}, {Staten}, {Stefanik}, {Tarle}, {Sien Tie}, {Tinker},
  {Tojeiro}, {Valdes}, {Valenzuela}, {Valluri}, {Vargas-Magana}, {Verde},
  {Walker}, {Wang}, {Wang}, {Weaver}, {Weaverdyck}, {Wechsler}, {Weinberg},
  {White}, {Yang}, {Yeche}, {Zhang}, {Zhao}, {Zheng}, {Zhou}, {Zhou}, {Zhu},
  {Zou}, and {Zu}}]{2016arXiv161100036D}
{DESI Collaboration}, {Aghamousa} A, {Aguilar} J, {Ahlen} S, {Alam} S, {Allen}
  LE, {Allende Prieto} C, {Annis} J, {Bailey} S, {Balland} C, et~al. (2016)
  {The DESI Experiment Part I: Science,Targeting, and Survey Design}. arXiv
  e-prints arXiv:1611.00036.
  {\href{https://arxiv.org/abs/1611.00036}{{arXiv:1611.00036}}} {[astro-ph.IM]}

\bibitem[{{Dessart} et~al.(2006){Dessart}, {Burrows}, {Ott}, {Livne}, {Yoon},
  and {Langer}}]{2006ApJ...644.1063D}
{Dessart} L, {Burrows} A, {Ott} CD, {Livne} E, {Yoon} SC, {Langer} N (2006)
  {Multidimensional Simulations of the Accretion-induced Collapse of White
  Dwarfs to Neutron Stars}. \apj 644(2):1063--1084. \doi{10.1086/503626}.
  {\href{https://arxiv.org/abs/astro-ph/0601603}{{arXiv:astro-ph/0601603}}}
  {[astro-ph]}

\bibitem[{{Destounis} and {Kokkotas}(2021)}]{2021PhRvD.104f4023D}
{Destounis} K, {Kokkotas} KD (2021) {Gravitational-wave glitches: Resonant
  islands and frequency jumps in nonintegrable extreme-mass-ratio inspirals}.
  \prd 104(6):064023. \doi{10.1103/PhysRevD.104.064023}.
  {\href{https://arxiv.org/abs/2108.02782}{{arXiv:2108.02782}}} {[gr-qc]}

\bibitem[{Destounis et~al.(2020)Destounis, Suvorov, and
  Kokkotas}]{PhysRevD.102.064041}
Destounis K, Suvorov AG, Kokkotas KD (2020) Testing spacetime symmetry through
  gravitational waves from extreme-mass-ratio inspirals. Phys Rev D 102:064041.
  \doi{10.1103/PhysRevD.102.064041},
  \urlprefix\url{https://link.aps.org/doi/10.1103/PhysRevD.102.064041}

\bibitem[{{Destounis} et~al.(2021){Destounis}, {Suvorov}, and
  {Kokkotas}}]{2021PhRvL.126n1102D}
{Destounis} K, {Suvorov} AG, {Kokkotas} KD (2021) {Gravitational Wave Glitches
  in Chaotic Extreme-Mass-Ratio Inspirals}. \prl 126(14):141102.
  \doi{10.1103/PhysRevLett.126.141102}.
  {\href{https://arxiv.org/abs/2103.05643}{{arXiv:2103.05643}}} {[gr-qc]}

\bibitem[{{Desvignes} et~al.(2016){Desvignes}, {Caballero}, {Lentati},
  {Verbiest}, {Champion}, {Stappers}, {Janssen}, {Lazarus}, {Os{\l}owski},
  {Babak}, {Bassa}, {Brem}, {Burgay}, {Cognard}, {Gair}, {Graikou},
  {Guillemot}, {Hessels}, {Jessner}, {Jordan}, {Karuppusamy}, {Kramer},
  {Lassus}, {Lazaridis}, {Lee}, {Liu}, {Lyne}, {McKee}, {Mingarelli},
  {Perrodin}, {Petiteau}, {Possenti}, {Purver}, {Rosado}, {Sanidas}, {Sesana},
  {Shaifullah}, {Smits}, {Taylor}, {Theureau}, {Tiburzi}, {van Haasteren}, and
  {Vecchio}}]{2016MNRAS.458.3341D}
{Desvignes} G, {Caballero} RN, {Lentati} L, {Verbiest} JPW, {Champion} DJ,
  {Stappers} BW, {Janssen} GH, {Lazarus} P, {Os{\l}owski} S, {Babak} S, et~al.
  (2016) {High-precision timing of 42 millisecond pulsars with the European
  Pulsar Timing Array}. \mnras 458(3):3341--3380. \doi{10.1093/mnras/stw483}.
  {\href{https://arxiv.org/abs/1602.08511}{{arXiv:1602.08511}}} {[astro-ph.HE]}

\bibitem[{{Devecchi} and {Volonteri}(2009)}]{2009ApJ...694..302D}
{Devecchi} B, {Volonteri} M (2009) {Formation of the First Nuclear Clusters and
  Massive Black Holes at High Redshift}. \apj 694(1):302--313.
  \doi{10.1088/0004-637X/694/1/302}.
  {\href{https://arxiv.org/abs/0810.1057}{{arXiv:0810.1057}}} {[astro-ph]}

\bibitem[{{Dewdney} et~al.(2009){Dewdney}, {Hall}, {Schilizzi}, and
  {Lazio}}]{2009IEEEP..97.1482D}
{Dewdney} PE, {Hall} PJ, {Schilizzi} RT, {Lazio} TJLW (2009) {The Square
  Kilometre Array}. IEEE Proceedings 97(8):1482--1496.
  \doi{10.1109/JPROC.2009.2021005}

\bibitem[{{Dewi} and {Pols}(2003)}]{2003MNRAS.344..629D}
{Dewi} JDM, {Pols} OR (2003) {The late stages of evolution of helium
  star-neutron star binaries and the formation of double neutron star systems}.
  \mnras 344(2):629--643. \doi{10.1046/j.1365-8711.2003.06844.x}.
  {\href{https://arxiv.org/abs/astro-ph/0306066}{{arXiv:astro-ph/0306066}}}
  {[astro-ph]}

\bibitem[{{Dewi} et~al.(2005){Dewi}, {Podsiadlowski}, and
  {Pols}}]{2005MNRAS.363L..71D}
{Dewi} JDM, {Podsiadlowski} P, {Pols} OR (2005) {The spin period-eccentricity
  relation of double neutron stars: evidence for weak supernova kicks?} \mnras
  363(1):L71--L75. \doi{10.1111/j.1745-3933.2005.00085.x}.
  {\href{https://arxiv.org/abs/astro-ph/0507628}{{arXiv:astro-ph/0507628}}}
  {[astro-ph]}

\bibitem[{{Di Carlo} et~al.(2019){Di Carlo}, {Giacobbo}, {Mapelli}, {Pasquato},
  {Spera}, {Wang}, and {Haardt}}]{2019MNRAS.487.2947D}
{Di Carlo} UN, {Giacobbo} N, {Mapelli} M, {Pasquato} M, {Spera} M, {Wang} L,
  {Haardt} F (2019) {Merging black holes in young star clusters}. \mnras
  487(2):2947--2960. \doi{10.1093/mnras/stz1453}.
  {\href{https://arxiv.org/abs/1901.00863}{{arXiv:1901.00863}}} {[astro-ph.HE]}

\bibitem[{{Di Carlo} et~al.(2020{\natexlab{a}}){Di Carlo}, {Mapelli},
  {Bouffanais}, {Giacobbo}, {Santoliquido}, {Bressan}, {Spera}, and
  {Haardt}}]{2020MNRAS.497.1043D}
{Di Carlo} UN, {Mapelli} M, {Bouffanais} Y, {Giacobbo} N, {Santoliquido} F,
  {Bressan} Ar, {Spera} M, {Haardt} F (2020{\natexlab{a}}) {Binary black holes
  in the pair instability mass gap}. \mnras 497(1):1043--1049.
  \doi{10.1093/mnras/staa1997}.
  {\href{https://arxiv.org/abs/1911.01434}{{arXiv:1911.01434}}} {[astro-ph.HE]}

\bibitem[{{Di Carlo} et~al.(2020{\natexlab{b}}){Di Carlo}, {Mapelli},
  {Giacobbo}, {Spera}, {Bouffanais}, {Rastello}, {Santoliquido}, {Pasquato},
  {Ballone}, {Trani}, {Torniamenti}, and {Haardt}}]{2020MNRAS.498..495D}
{Di Carlo} UN, {Mapelli} M, {Giacobbo} N, {Spera} M, {Bouffanais} Y, {Rastello}
  S, {Santoliquido} F, {Pasquato} M, {Ballone} Ar, {Trani} AA, et~al.
  (2020{\natexlab{b}}) {Binary black holes in young star clusters: the impact
  of metallicity}. \mnras 498(1):495--506. \doi{10.1093/mnras/staa2286}.
  {\href{https://arxiv.org/abs/2004.09525}{{arXiv:2004.09525}}} {[astro-ph.HE]}

\bibitem[{{Di Carlo} et~al.(2021){Di Carlo}, {Mapelli}, {Pasquato}, {Rastello},
  {Ballone}, {Dall'Amico}, {Giacobbo}, {Iorio}, {Spera}, {Torniamenti}, and
  {Haardt}}]{2021MNRAS.507.5132D}
{Di Carlo} UN, {Mapelli} M, {Pasquato} M, {Rastello} S, {Ballone} A,
  {Dall'Amico} M, {Giacobbo} N, {Iorio} G, {Spera} M, {Torniamenti} S, et~al.
  (2021) {Intermediate-mass black holes from stellar mergers in young star
  clusters}. \mnras 507(4):5132--5143. \doi{10.1093/mnras/stab2390}.
  {\href{https://arxiv.org/abs/2105.01085}{{arXiv:2105.01085}}} {[astro-ph.GA]}

\bibitem[{{Di Cintio} et~al.(2017){Di Cintio}, {Tremmel}, {Governato},
  {Pontzen}, {Zavala}, {Bastidas Fry}, {Brooks}, and
  {Vogelsberger}}]{2017MNRAS.469.2845D}
{Di Cintio} A, {Tremmel} M, {Governato} F, {Pontzen} A, {Zavala} J, {Bastidas
  Fry} Ae, {Brooks} A, {Vogelsberger} M (2017) {A rumble in the dark:
  signatures of self-interacting dark matter in supermassive black hole
  dynamics and galaxy density profiles}. \mnras 469(3):2845--2854.
  \doi{10.1093/mnras/stx1043}.
  {\href{https://arxiv.org/abs/1701.04410}{{arXiv:1701.04410}}} {[astro-ph.GA]}

\bibitem[{{Di Matteo} et~al.(2005){Di Matteo}, {Springel}, and
  {Hernquist}}]{2005Natur.433..604D}
{Di Matteo} T, {Springel} V, {Hernquist} L (2005) {Energy input from quasars
  regulates the growth and activity of black holes and their host galaxies}.
  \nat 433(7026):604--607. \doi{10.1038/nature03335}.
  {\href{https://arxiv.org/abs/astro-ph/0502199}{{arXiv:astro-ph/0502199}}}
  {[astro-ph]}

\bibitem[{{Dieball} et~al.(2005){Dieball}, {Knigge}, {Zurek}, {Shara}, {Long},
  {Charles}, {Hannikainen}, and {van Zyl}}]{2005ApJ...634L.105D}
{Dieball} A, {Knigge} C, {Zurek} DR, {Shara} MM, {Long} KS, {Charles} PA,
  {Hannikainen} DC, {van Zyl} L (2005) {An Ultracompact X-Ray Binary in the
  Globular Cluster M15 (NGC 7078)}. \apjl 634(1):L105--L108.
  \doi{10.1086/498712}.
  {\href{https://arxiv.org/abs/astro-ph/0510430}{{arXiv:astro-ph/0510430}}}
  {[astro-ph]}

\bibitem[{{Dijkstra} et~al.(2008){Dijkstra}, {Haiman}, {Mesinger}, and
  {Wyithe}}]{2008MNRAS.391.1961D}
{Dijkstra} M, {Haiman} Z, {Mesinger} A, {Wyithe} JSB (2008) {Fluctuations in
  the high-redshift Lyman-Werner background: close halo pairs as the origin of
  supermassive black holes}. \mnras 391:1961--1972.
  \doi{10.1111/j.1365-2966.2008.14031.x}.
  {\href{https://arxiv.org/abs/0810.0014}{{arXiv:0810.0014}}}

\bibitem[{{Dijkstra} et~al.(2014){Dijkstra}, {Ferrara}, and
  {Mesinger}}]{2014MNRAS.442.2036D}
{Dijkstra} M, {Ferrara} A, {Mesinger} A (2014) {Feedback-regulated supermassive
  black hole seed formation}. \mnras 442(3):2036--2047.
  \doi{10.1093/mnras/stu1007}.
  {\href{https://arxiv.org/abs/1405.6743}{{arXiv:1405.6743}}} {[astro-ph.GA]}

\bibitem[{{Dittmann} and {Miller}(2020)}]{2020MNRAS.493.3732D}
{Dittmann} AJ, {Miller} MC (2020) {Star formation in accretion discs and SMBH
  growth}. \mnras 493(3):3732--3743. \doi{10.1093/mnras/staa463}.
  {\href{https://arxiv.org/abs/1911.08685}{{arXiv:1911.08685}}} {[astro-ph.HE]}

\bibitem[{Domcke et~al.(2017)Domcke, Muia, Pieroni, and
  Witkowski}]{Domcke:2017fix}
Domcke V, Muia F, Pieroni M, Witkowski LT (2017) {PBH dark matter from axion
  inflation}. JCAP 07:048. \doi{10.1088/1475-7516/2017/07/048}.
  {\href{https://arxiv.org/abs/1704.03464}{{arXiv:1704.03464}}} {[astro-ph.CO]}

\bibitem[{{Dominik} et~al.(2013){Dominik}, {Belczynski}, {Fryer}, {Holz},
  {Berti}, {Bulik}, {Mand el}, and {O'Shaughnessy}}]{2013ApJ...779...72D}
{Dominik} M, {Belczynski} K, {Fryer} C, {Holz} DE, {Berti} E, {Bulik} T, {Mand
  el} I, {O'Shaughnessy} R (2013) {Double Compact Objects. II. Cosmological
  Merger Rates}. \apj 779(1):72. \doi{10.1088/0004-637X/779/1/72}.
  {\href{https://arxiv.org/abs/1308.1546}{{arXiv:1308.1546}}} {[astro-ph.HE]}

\bibitem[{{D'Orazio} and {Di Stefano}(2018)}]{2018MNRAS.474.2975D}
{D'Orazio} DJ, {Di Stefano} R (2018) {Periodic self-lensing from accreting
  massive black hole binaries}. \mnras 474(3):2975--2986.
  \doi{10.1093/mnras/stx2936}.
  {\href{https://arxiv.org/abs/1707.02335}{{arXiv:1707.02335}}} {[astro-ph.HE]}

\bibitem[{{D'Orazio} and {Loeb}(2018)}]{2018ApJ...863..185D}
{D'Orazio} DJ, {Loeb} A (2018) {Repeated Imaging of Massive Black Hole Binary
  Orbits with Millimeter Interferometry: Measuring Black Hole Masses and the
  Hubble Constant}. \apj 863(2):185. \doi{10.3847/1538-4357/aad413}.
  {\href{https://arxiv.org/abs/1712.02362}{{arXiv:1712.02362}}} {[astro-ph.HE]}

\bibitem[{{D'Orazio} and {Samsing}(2018)}]{2018MNRAS.481.4775D}
{D'Orazio} DJ, {Samsing} J (2018) {Black hole mergers from globular clusters
  observable by LISA II. Resolved eccentric sources and the gravitational wave
  background}. \mnras 481(4):4775--4785. \doi{10.1093/mnras/sty2568}.
  {\href{https://arxiv.org/abs/1805.06194}{{arXiv:1805.06194}}} {[astro-ph.HE]}

\bibitem[{{D'Orazio} et~al.(2013){D'Orazio}, {Haiman}, and
  {MacFadyen}}]{2013MNRAS.436.2997D}
{D'Orazio} DJ, {Haiman} Z, {MacFadyen} A (2013) {Accretion into the central
  cavity of a circumbinary disc}. \mnras 436(4):2997--3020.
  \doi{10.1093/mnras/stt1787}.
  {\href{https://arxiv.org/abs/1210.0536}{{arXiv:1210.0536}}} {[astro-ph.GA]}

\bibitem[{{D'Orazio} et~al.(2015){D'Orazio}, {Haiman}, and
  {Schiminovich}}]{2015Natur.525..351D}
{D'Orazio} DJ, {Haiman} Z, {Schiminovich} D (2015) {Relativistic boost as the
  cause of periodicity in a massive black-hole binary candidate}. \nat
  525(7569):351--353. \doi{10.1038/nature15262}.
  {\href{https://arxiv.org/abs/1509.04301}{{arXiv:1509.04301}}} {[astro-ph.HE]}

\bibitem[{{D'Orazio} et~al.(2016){D'Orazio}, {Haiman}, {Duffell}, {MacFadyen},
  and {Farris}}]{2016MNRAS.459.2379D}
{D'Orazio} DJ, {Haiman} Z, {Duffell} P, {MacFadyen} A, {Farris} B (2016) {A
  transition in circumbinary accretion discs at a binary mass ratio of 1:25}.
  \mnras 459(3):2379--2393. \doi{10.1093/mnras/stw792}.
  {\href{https://arxiv.org/abs/1512.05788}{{arXiv:1512.05788}}} {[astro-ph.HE]}

\bibitem[{{Dosopoulou} and {Antonini}(2017)}]{2017ApJ...840...31D}
{Dosopoulou} F, {Antonini} F (2017) {Dynamical Friction and the Evolution of
  Supermassive Black Hole Binaries: The Final Hundred-parsec Problem}. \apj
  840(1):31. \doi{10.3847/1538-4357/aa6b58}.
  {\href{https://arxiv.org/abs/1611.06573}{{arXiv:1611.06573}}} {[astro-ph.GA]}

\bibitem[{{Dotti} et~al.(2006){Dotti}, {Colpi}, and
  {Haardt}}]{2006MNRAS.367..103D}
{Dotti} M, {Colpi} M, {Haardt} F (2006) {Laser Interferometer Space Antenna
  double black holes: dynamics in gaseous nuclear discs}. \mnras
  367(1):103--112. \doi{10.1111/j.1365-2966.2005.09956.x}.
  {\href{https://arxiv.org/abs/astro-ph/0509813}{{arXiv:astro-ph/0509813}}}
  {[astro-ph]}

\bibitem[{{Dotti} et~al.(2007){Dotti}, {Colpi}, {Haardt}, and
  {Mayer}}]{2007MNRAS.379..956D}
{Dotti} M, {Colpi} M, {Haardt} F, {Mayer} L (2007) {Supermassive black hole
  binaries in gaseous and stellar circumnuclear discs: orbital dynamics and gas
  accretion}. \mnras 379(3):956--962. \doi{10.1111/j.1365-2966.2007.12010.x}.
  {\href{https://arxiv.org/abs/astro-ph/0612505}{{arXiv:astro-ph/0612505}}}
  {[astro-ph]}

\bibitem[{{Dotti} et~al.(2010){Dotti}, {Volonteri}, {Perego}, {Colpi},
  {Ruszkowski}, and {Haardt}}]{2010MNRAS.402..682D}
{Dotti} M, {Volonteri} M, {Perego} A, {Colpi} M, {Ruszkowski} M, {Haardt} F
  (2010) {Dual black holes in merger remnants - II. Spin evolution and
  gravitational recoil}. \mnras 402(1):682--690.
  \doi{10.1111/j.1365-2966.2009.15922.x}.
  {\href{https://arxiv.org/abs/0910.5729}{{arXiv:0910.5729}}} {[astro-ph.HE]}

\bibitem[{{Dotti} et~al.(2012){Dotti}, {Sesana}, and
  {Decarli}}]{2012AdAst2012E...3D}
{Dotti} M, {Sesana} A, {Decarli} R (2012) {Massive Black Hole Binaries:
  Dynamical Evolution and Observational Signatures}. Advances in Astronomy
  2012:940568. \doi{10.1155/2012/940568}.
  {\href{https://arxiv.org/abs/1111.0664}{{arXiv:1111.0664}}} {[astro-ph.CO]}

\bibitem[{{Dotti} et~al.(2013){Dotti}, {Colpi}, {Pallini}, {Perego}, and
  {Volonteri}}]{2013ApJ...762...68D}
{Dotti} M, {Colpi} M, {Pallini} S, {Perego} A, {Volonteri} M (2013) {On the
  Orientation and Magnitude of the Black Hole Spin in Galactic Nuclei}. \apj
  762(2):68. \doi{10.1088/0004-637X/762/2/68}.
  {\href{https://arxiv.org/abs/1211.4871}{{arXiv:1211.4871}}} {[astro-ph.CO]}

\bibitem[{{Dovciak} et~al.(2013){Dovciak}, {Matt}, {Bianchi}, {Boller},
  {Brenneman}, {Bursa}, {D'Ai}, {di Salvo}, {de Marco}, {Goosmann}, {Karas},
  {Iwasawa}, {Kara}, {Miller}, {Miniutti}, {Papadakis}, {Petrucci}, {Ponti},
  {Porquet}, {Reynolds}, {Risaliti}, {Rozanska}, {Zampieri}, {Zezas}, and
  {Young}}]{2013arXiv1306.2331D}
{Dovciak} M, {Matt} G, {Bianchi} S, {Boller} T, {Brenneman} L, {Bursa} M,
  {D'Ai} A, {di Salvo} T, {de Marco} B, {Goosmann} R, et~al. (2013) {The Hot
  and Energetic Universe: The close environments of supermassive black holes}.
  arXiv e-prints arXiv:1306.2331.
  {\href{https://arxiv.org/abs/1306.2331}{{arXiv:1306.2331}}} {[astro-ph.HE]}

\bibitem[{{Dov{\v{c}}iak} et~al.(2008){Dov{\v{c}}iak}, {Muleri}, {Goosmann},
  {Karas}, and {Matt}}]{2008MNRAS.391...32D}
{Dov{\v{c}}iak} M, {Muleri} F, {Goosmann} RW, {Karas} V, {Matt} G (2008)
  {Thermal disc emission from a rotating black hole: X-ray polarization
  signatures}. \mnras 391(1):32--38. \doi{10.1111/j.1365-2966.2008.13872.x}.
  {\href{https://arxiv.org/abs/0809.0418}{{arXiv:0809.0418}}} {[astro-ph]}

\bibitem[{{Drasco}(2009)}]{2009PhRvD..79j4016D}
{Drasco} S (2009) {Verifying black hole orbits with gravitational
  spectroscopy}. \prd 79(10):104016. \doi{10.1103/PhysRevD.79.104016}.
  {\href{https://arxiv.org/abs/0711.4644}{{arXiv:0711.4644}}} {[gr-qc]}

\bibitem[{{du Buisson} et~al.(2020){du Buisson}, {Marchant}, {Podsiadlowski},
  {Kobayashi}, {Abdalla}, {Taylor}, {Mandel}, {de Mink}, {Moriya}, and
  {Langer}}]{2020arXiv200211630D}
{du Buisson} L, {Marchant} P, {Podsiadlowski} P, {Kobayashi} C, {Abdalla} FB,
  {Taylor} P, {Mandel} I, {de Mink} SE, {Moriya} TJ, {Langer} N (2020) {Cosmic
  Rates of Black Hole Mergers and Pair-Instability Supernovae from Chemically
  Homogeneous Binary Evolution}. arXiv e-prints arXiv:2002.11630.
  {\href{https://arxiv.org/abs/2002.11630}{{arXiv:2002.11630}}} {[astro-ph.HE]}

\bibitem[{{Dubois} et~al.(2012){Dubois}, {Devriendt}, {Slyz}, and
  {Teyssier}}]{2012MNRAS.420.2662D}
{Dubois} Y, {Devriendt} J, {Slyz} A, {Teyssier} R (2012) {Self-regulated growth
  of supermassive black holes by a dual jet-heating active galactic nucleus
  feedback mechanism: methods, tests and implications for cosmological
  simulations}. \mnras 420(3):2662--2683.
  \doi{10.1111/j.1365-2966.2011.20236.x}.
  {\href{https://arxiv.org/abs/1108.0110}{{arXiv:1108.0110}}} {[astro-ph.CO]}

\bibitem[{{Dubois} et~al.(2013){Dubois}, {Pichon}, {Devriendt}, {Silk},
  {Haehnelt}, {Kimm}, and {Slyz}}]{2013MNRAS.428.2885D}
{Dubois} Y, {Pichon} C, {Devriendt} J, {Silk} J, {Haehnelt} M, {Kimm} T, {Slyz}
  A (2013) {Blowing cold flows away: the impact of early AGN activity on the
  formation of a brightest cluster galaxy progenitor}. \mnras
  428(4):2885--2900. \doi{10.1093/mnras/sts224}.
  {\href{https://arxiv.org/abs/1206.5838}{{arXiv:1206.5838}}} {[astro-ph.CO]}

\bibitem[{{Dubois} et~al.(2014{\natexlab{a}}){Dubois}, {Pichon}, {Welker}, {Le
  Borgne}, {Devriendt}, {Laigle}, {Codis}, {Pogosyan}, {Arnouts}, {Benabed},
  {Bertin}, {Blaizot}, {Bouchet}, {Cardoso}, {Colombi}, {de Lapparent},
  {Desjacques}, {Gavazzi}, {Kassin}, {Kimm}, {McCracken}, {Milliard},
  {Peirani}, {Prunet}, {Rouberol}, {Silk}, {Slyz}, {Sousbie}, {Teyssier},
  {Tresse}, {Treyer}, {Vibert}, and {Volonteri}}]{2014MNRAS.444.1453D}
{Dubois} Y, {Pichon} C, {Welker} C, {Le Borgne} D, {Devriendt} J, {Laigle} C,
  {Codis} S, {Pogosyan} D, {Arnouts} S, {Benabed} K, et~al.
  (2014{\natexlab{a}}) {Dancing in the dark: galactic properties trace spin
  swings along the cosmic web}. \mnras 444(2):1453--1468.
  \doi{10.1093/mnras/stu1227}.
  {\href{https://arxiv.org/abs/1402.1165}{{arXiv:1402.1165}}} {[astro-ph.CO]}

\bibitem[{{Dubois} et~al.(2014{\natexlab{b}}){Dubois}, {Volonteri}, and
  {Silk}}]{2014MNRAS.440.1590D}
{Dubois} Y, {Volonteri} M, {Silk} J (2014{\natexlab{b}}) {Black hole evolution
  - III. Statistical properties of mass growth and spin evolution using
  large-scale hydrodynamical cosmological simulations}. \mnras
  440(2):1590--1606. \doi{10.1093/mnras/stu373}.
  {\href{https://arxiv.org/abs/1304.4583}{{arXiv:1304.4583}}} {[astro-ph.CO]}

\bibitem[{{Dubois} et~al.(2015){Dubois}, {Volonteri}, {Silk}, {Devriendt},
  {Slyz}, and {Teyssier}}]{2015MNRAS.452.1502D}
{Dubois} Y, {Volonteri} M, {Silk} J, {Devriendt} J, {Slyz} A, {Teyssier} R
  (2015) {Black hole evolution - I. Supernova-regulated black hole growth}.
  \mnras 452(2):1502--1518. \doi{10.1093/mnras/stv1416}.
  {\href{https://arxiv.org/abs/1504.00018}{{arXiv:1504.00018}}} {[astro-ph.GA]}

\bibitem[{{Dubois} et~al.(2020){Dubois}, {Beckmann}, {Bournaud}, {Choi},
  {Devriendt}, {Jackson}, {Kaviraj}, {Kimm}, {Kraljic}, {Laigle}, {Martin},
  {Park}, {Peirani}, {Pichon}, {Volonteri}, and {Yi}}]{2020arXiv200910578D}
{Dubois} Y, {Beckmann} R, {Bournaud} F, {Choi} H, {Devriendt} J, {Jackson} R,
  {Kaviraj} S, {Kimm} T, {Kraljic} K, {Laigle} C, et~al. (2020) {Introducing
  the NewHorizon simulation: galaxy properties with resolved internal dynamics
  across cosmic time}. arXiv e-prints arXiv:2009.10578.
  {\href{https://arxiv.org/abs/2009.10578}{{arXiv:2009.10578}}} {[astro-ph.GA]}

\bibitem[{Duechting(2004)}]{Duechting:2004dk}
Duechting N (2004) {Supermassive black holes from primordial black hole seeds}.
  Phys Rev D 70:064015. \doi{10.1103/PhysRevD.70.064015}.
  {\href{https://arxiv.org/abs/astro-ph/0406260}{{arXiv:astro-ph/0406260}}}

\bibitem[{{Duez} and {Zlochower}(2019)}]{2019RPPh...82a6902D}
{Duez} MD, {Zlochower} Y (2019) {Numerical relativity of compact binaries in
  the 21st century}. Reports on Progress in Physics 82(1):016902.
  \doi{10.1088/1361-6633/aadb16}.
  {\href{https://arxiv.org/abs/1808.06011}{{arXiv:1808.06011}}} {[gr-qc]}

\bibitem[{{Duffell}(2015)}]{2015ApJ...806..182D}
{Duffell} PC (2015) {Halting Migration: Numerical Calculations of Corotation
  Torques in the Weakly Nonlinear Regime}. \apj 806(2):182.
  \doi{10.1088/0004-637X/806/2/182}.
  {\href{https://arxiv.org/abs/1412.8092}{{arXiv:1412.8092}}} {[astro-ph.EP]}

\bibitem[{{Duffell} et~al.(2019){Duffell}, {D'Orazio}, {Derdzinski}, {Haiman},
  {MacFadyen}, {Rosen}, and {Zrake}}]{2019arXiv191105506D}
{Duffell} PC, {D'Orazio} D, {Derdzinski} A, {Haiman} Z, {MacFadyen} A, {Rosen}
  AL, {Zrake} J (2019) {Circumbinary Disks: Accretion and Torque as a Function
  of Mass Ratio and Disk Viscosity}. arXiv e-prints arXiv:1911.05506.
  {\href{https://arxiv.org/abs/1911.05506}{{arXiv:1911.05506}}} {[astro-ph.SR]}

\bibitem[{{Dullo}(2019)}]{2019ApJ...886...80D}
{Dullo} BT (2019) {The Most Massive Galaxies with Large Depleted Cores:
  Structural Parameter Relations and Black Hole Masses}. \apj 886(2):80.
  \doi{10.3847/1538-4357/ab4d4f}.
  {\href{https://arxiv.org/abs/1910.10240}{{arXiv:1910.10240}}} {[astro-ph.GA]}

\bibitem[{{Dullo} and {Graham}(2013)}]{2013ApJ...768...36D}
{Dullo} BT, {Graham} AW (2013) {Central Stellar Mass Deficits in the Bulges of
  Local Lenticular Galaxies, and the Connection with Compact z
  \raisebox{-0.5ex}\textasciitilde 1.5 Galaxies}. \apj 768(1):36.
  \doi{10.1088/0004-637X/768/1/36}.
  {\href{https://arxiv.org/abs/1303.1273}{{arXiv:1303.1273}}} {[astro-ph.CO]}

\bibitem[{{Dullo} and {Graham}(2014)}]{2014MNRAS.444.2700D}
{Dullo} BT, {Graham} AW (2014) {Depleted cores, multicomponent fits, and
  structural parameter relations for luminous early-type galaxies}. \mnras
  444(3):2700--2722. \doi{10.1093/mnras/stu1590}.
  {\href{https://arxiv.org/abs/1310.5867}{{arXiv:1310.5867}}} {[astro-ph.CO]}

\bibitem[{{Duncan} and {Lissauer}(1998)}]{1998Icar..134..303D}
{Duncan} MJ, {Lissauer} JJ (1998) {The Effects of Post-Main-Sequence Solar Mass
  Loss on the Stability of Our Planetary System}. \icarus 134(2):303--310.
  \doi{10.1006/icar.1998.5962}

\bibitem[{{Dunhill} et~al.(2013){Dunhill}, {Alexander}, and
  {Armitage}}]{2013MNRAS.428.3072D}
{Dunhill} AC, {Alexander} RD, {Armitage} PJ (2013) {A limit on eccentricity
  growth from global 3D simulations of disc-planet interactions}. \mnras
  428(4):3072--3082. \doi{10.1093/mnras/sts254}.
  {\href{https://arxiv.org/abs/1210.6035}{{arXiv:1210.6035}}} {[astro-ph.EP]}

\bibitem[{{Dunlop} et~al.(2021){Dunlop}, {Abraham}, {Ashby}, {Bagley}, {Best},
  {Bongiorno}, {Bouwens}, {Bowler}, {Brammer}, {Bremer}, {Calabro'}, {Carnall},
  {Castellano}, {Cirasuolo}, {Conselice}, {Cullen}, {Dave}, {Dayal}, {Dekel},
  {Dickinson}, {Duncan}, {Elbaz}, {Ellis}, {Ferguson}, {Ferrara},
  {Finkelstein}, {Fontana}, {Furlanetto}, {Fynbo}, {Gallerani}, {Gardner},
  {Giavalisco}, {Grazian}, {Grogin}, {Harikane}, {Hopkins}, {Ilbert},
  {Illingworth}, {Juneau}, {Jung}, {Kartaltepe}, {Kassin}, {Kauffmann},
  {Khochfar}, {Kirkpatrick}, {Kocevski}, {Koekemoer}, {Labbe}, {Laporte},
  {Larson}, {Lucas}, {Magee}, {Mason}, {McCracken}, {McLeod}, {McLure},
  {Merlin}, {Mesinger}, {Milvang-Jensen}, {Newman}, {Oesch}, {Ouchi},
  {Pacifici}, {Papovich}, {Peacock}, {Peeples}, {Pentericci}, {Perez-Gonzalez},
  {Pirzkal}, {Pope}, {Pye}, {Reddy}, {Robertson}, {Salvato}, {Santini},
  {Schaerer}, {Shapley}, {Simons}, {Smit}, {Smith}, {Snyder}, {Somerville},
  {Stanway}, {Stefanon}, {Tasca}, {Tikkanen}, {Tresse}, {Trump}, {Whitaker},
  {Wilkins}, {Wright}, {Wyithe}, {van Dokkum}, and {van der
  Werf}}]{2021jwst.prop.1837D}
{Dunlop} JS, {Abraham} RG, {Ashby} MLN, {Bagley} M, {Best} PN, {Bongiorno} A,
  {Bouwens} R, {Bowler} RAA, {Brammer} G, {Bremer} M, et~al. (2021) {PRIMER:
  Public Release IMaging for Extragalactic Research}. JWST Proposal. Cycle 1,
  ID. \#1837

\bibitem[{{Dunn} et~al.(2018){Dunn}, {Bellovary}, {Holley-Bockelmann},
  {Christensen}, and {Quinn}}]{2018ApJ...861...39D}
{Dunn} G, {Bellovary} J, {Holley-Bockelmann} K, {Christensen} C, {Quinn} T
  (2018) {Sowing Black Hole Seeds: Direct Collapse Black Hole Formation with
  Realistic Lyman-Werner Radiation in Cosmological Simulations}. \apj
  861(1):39. \doi{10.3847/1538-4357/aac7c2}.
  {\href{https://arxiv.org/abs/1803.01007}{{arXiv:1803.01007}}} {[astro-ph.GA]}

\bibitem[{{Dunn} et~al.(2020){Dunn}, {Holley-Bockelmann}, and
  {Bellovary}}]{2020ApJ...896...72D}
{Dunn} G, {Holley-Bockelmann} K, {Bellovary} J (2020) {The Role of
  Gravitational Recoil in the Assembly of Massive Black Hole Seeds}. \apj
  896(1):72. \doi{10.3847/1538-4357/ab7cd2}.
  {\href{https://arxiv.org/abs/2002.04740}{{arXiv:2002.04740}}} {[astro-ph.GA]}

\bibitem[{{Dvorkin} et~al.(2016{\natexlab{a}}){Dvorkin}, {Uzan}, {Vangioni},
  and {Silk}}]{2016PhRvD..94j3011D}
{Dvorkin} I, {Uzan} JP, {Vangioni} E, {Silk} J (2016{\natexlab{a}}) {Synthetic
  model of the gravitational wave background from evolving binary compact
  objects}. \prd 94(10):103011. \doi{10.1103/PhysRevD.94.103011}.
  {\href{https://arxiv.org/abs/1607.06818}{{arXiv:1607.06818}}} {[astro-ph.HE]}

\bibitem[{{Dvorkin} et~al.(2016{\natexlab{b}}){Dvorkin}, {Vangioni}, {Silk},
  {Uzan}, and {Olive}}]{2016MNRAS.461.3877D}
{Dvorkin} I, {Vangioni} E, {Silk} J, {Uzan} JP, {Olive} KA (2016{\natexlab{b}})
  {Metallicity-constrained merger rates of binary black holes and the
  stochastic gravitational wave background}. \mnras 461(4):3877--3885.
  \doi{10.1093/mnras/stw1477}.
  {\href{https://arxiv.org/abs/1604.04288}{{arXiv:1604.04288}}} {[astro-ph.HE]}

\bibitem[{{Eatough} et~al.(2013){Eatough}, {Kramer}, {Lyne}, and
  {Keith}}]{2013MNRAS.431..292E}
{Eatough} RP, {Kramer} M, {Lyne} AG, {Keith} MJ (2013) {A coherent acceleration
  search of the Parkes multibeam pulsar survey - techniques and the discovery
  and timing of 16 pulsars}. \mnras 431(1):292--307.
  \doi{10.1093/mnras/stt161}.
  {\href{https://arxiv.org/abs/1301.6346}{{arXiv:1301.6346}}} {[astro-ph.IM]}

\bibitem[{{Ebisuzaki} et~al.(1991){Ebisuzaki}, {Makino}, and
  {Okumura}}]{1991Natur.354..212E}
{Ebisuzaki} T, {Makino} J, {Okumura} SK (1991) {Merging of two galaxies with
  central black holes}. \nat 354(6350):212--214. \doi{10.1038/354212a0}

\bibitem[{{Ebisuzaki} et~al.(2001){Ebisuzaki}, {Makino}, {Tsuru}, {Funato},
  {Portegies Zwart}, {Hut}, {McMillan}, {Matsushita}, {Matsumoto}, and
  {Kawabe}}]{2001ApJ...562L..19E}
{Ebisuzaki} T, {Makino} J, {Tsuru} TG, {Funato} Y, {Portegies Zwart} S, {Hut}
  P, {McMillan} S, {Matsushita} S, {Matsumoto} H, {Kawabe} R (2001) {Missing
  Link Found? The ``Runaway'' Path to Supermassive Black Holes}. \apjl
  562(1):L19--L22. \doi{10.1086/338118}.
  {\href{https://arxiv.org/abs/astro-ph/0106252}{{arXiv:astro-ph/0106252}}}
  {[astro-ph]}

\bibitem[{{Eda} et~al.(2013){Eda}, {Itoh}, {Kuroyanagi}, and
  {Silk}}]{2013PhRvL.110v1101E}
{Eda} K, {Itoh} Y, {Kuroyanagi} S, {Silk} J (2013) {New Probe of Dark-Matter
  Properties: Gravitational Waves from an Intermediate-Mass Black Hole Embedded
  in a Dark-Matter Minispike}. \prl 110(22):221101.
  \doi{10.1103/PhysRevLett.110.221101}.
  {\href{https://arxiv.org/abs/1301.5971}{{arXiv:1301.5971}}} {[gr-qc]}

\bibitem[{{Edlund} et~al.(2005){Edlund}, {Tinto}, {Kr{\'o}lak}, and
  {Nelemans}}]{2005PhRvD..71l2003E}
{Edlund} JA, {Tinto} M, {Kr{\'o}lak} A, {Nelemans} G (2005) {White-dwarf
  white-dwarf galactic background in the LISA data}. \prd 71(12):122003.
  \doi{10.1103/PhysRevD.71.122003}.
  {\href{https://arxiv.org/abs/gr-qc/0504112}{{arXiv:gr-qc/0504112}}} {[gr-qc]}

\bibitem[{{Edwards} and {Patton}(2012)}]{2012MNRAS.425..287E}
{Edwards} LOV, {Patton} DR (2012) {Close companions to brightest cluster
  galaxies: support for minor mergers and downsizing}. \mnras 425(1):287--295.
  \doi{10.1111/j.1365-2966.2012.21457.x}.
  {\href{https://arxiv.org/abs/1206.1612}{{arXiv:1206.1612}}} {[astro-ph.CO]}

\bibitem[{{Eggleton}(2006)}]{2006epbm.book.....E}
{Eggleton} P (2006) {Evolutionary Processes in Binary and Multiple Stars}

\bibitem[{{Eggleton}(1983)}]{1983ApJ...268..368E}
{Eggleton} PP (1983) {Aproximations to the radii of Roche lobes.} \apj
  268:368--369. \doi{10.1086/160960}

\bibitem[{{Eilon} et~al.(2009){Eilon}, {Kupi}, and
  {Alexander}}]{2009ApJ...698..641E}
{Eilon} E, {Kupi} G, {Alexander} T (2009) {The Efficiency of Resonant
  Relaxation Around a Massive Black Hole}. \apj 698(1):641--647.
  \doi{10.1088/0004-637X/698/1/641}.
  {\href{https://arxiv.org/abs/0807.1430}{{arXiv:0807.1430}}} {[astro-ph]}

\bibitem[{{Elbert} et~al.(2018){Elbert}, {Bullock}, and
  {Kaplinghat}}]{2018MNRAS.473.1186E}
{Elbert} OD, {Bullock} JS, {Kaplinghat} M (2018) {Counting black holes: The
  cosmic stellar remnant population and implications for LIGO}. \mnras
  473(1):1186--1194. \doi{10.1093/mnras/stx1959}.
  {\href{https://arxiv.org/abs/1703.02551}{{arXiv:1703.02551}}} {[astro-ph.GA]}

\bibitem[{{Eldridge} and {Stanway}(2016)}]{2016MNRAS.462.3302E}
{Eldridge} JJ, {Stanway} ER (2016) {BPASS predictions for binary black hole
  mergers}. \mnras 462(3):3302--3313. \doi{10.1093/mnras/stw1772}.
  {\href{https://arxiv.org/abs/1602.03790}{{arXiv:1602.03790}}} {[astro-ph.HE]}

\bibitem[{{Enoki} et~al.(2005){Enoki}, {Inoue}, {Nagashima}, and
  {Sugiyama}}]{2005ARAOJ...7...34E}
{Enoki} M, {Inoue} KT, {Nagashima} M, {Sugiyama} N (2005) {Gravitational waves
  from coalescing supermassive black hole binaries in a hierarchical galaxy
  formation model}. Annual Report of the National Astronomical Observatory of
  Japan 7:34.
  {\href{https://arxiv.org/abs/astro-ph/0502529}{{arXiv:astro-ph/0502529}}}
  {[astro-ph]}

\bibitem[{{Eracleous} et~al.(2012){Eracleous}, {Boroson}, {Halpern}, and
  {Liu}}]{2012ApJS..201...23E}
{Eracleous} M, {Boroson} TA, {Halpern} JP, {Liu} J (2012) {A Large Systematic
  Search for Close Supermassive Binary and Rapidly Recoiling Black Holes}.
  \apjs 201(2):23. \doi{10.1088/0067-0049/201/2/23}.
  {\href{https://arxiv.org/abs/1106.2952}{{arXiv:1106.2952}}} {[astro-ph.CO]}

\bibitem[{{Espaillat} et~al.(2005){Espaillat}, {Patterson}, {Warner}, and
  {Woudt}}]{2005PASP..117..189E}
{Espaillat} C, {Patterson} J, {Warner} B, {Woudt} P (2005) {The Helium-rich
  Cataclysmic Variable ES Ceti}. \pasp 117(828):189--198. \doi{10.1086/427959}.
  {\href{https://arxiv.org/abs/astro-ph/0412068}{{arXiv:astro-ph/0412068}}}
  {[astro-ph]}

\bibitem[{{Euclid Collaboration} et~al.(2019){Euclid Collaboration}, {Barnett},
  {Warren}, {Mortlock}, {Cuby}, {Conselice}, {Hewett}, {Willott}, {Auricchio},
  {Balaguera-Antol{\'\i}nez}, {Baldi}, {Bardelli}, {Bellagamba}, {Bender},
  {Biviano}, {Bonino}, {Bozzo}, {Branchini}, {Brescia}, {Brinchmann},
  {Burigana}, {Camera}, {Capobianco}, {Carbone}, {Carretero}, {Carvalho},
  {Castander}, {Castellano}, {Cavuoti}, {Cimatti}, {Cl{\'e}dassou}, {Congedo},
  {Conversi}, {Copin}, {Corcione}, {Coupon}, {Courtois}, {Cropper}, {Da Silva},
  {Duncan}, {Dusini}, {Ealet}, {Farrens}, {Fosalba}, {Fotopoulou},
  {Fourmanoit}, {Frailis}, {Fumana}, {Galeotta}, {Garilli}, {Gillard},
  {Gillis}, {Graci{\'a}-Carpio}, {Grupp}, {Hoekstra}, {Hormuth}, {Israel},
  {Jahnke}, {Kermiche}, {Kilbinger}, {Kirkpatrick}, {Kitching}, {Kohley},
  {Kubik}, {Kunz}, {Kurki-Suonio}, {Laureijs}, {Ligori}, {Lilje}, {Lloro},
  {Maiorano}, {Mansutti}, {Marggraf}, {Martinet}, {Marulli}, {Massey}, {Mauri},
  {Medinaceli}, {Mei}, {Mellier}, {Metcalf}, {Metge}, {Meylan}, {Moresco},
  {Moscardini}, {Munari}, {Neissner}, {Niemi}, {Nutma}, {Padilla}, {Paltani},
  {Pasian}, {Paykari}, {Percival}, {Pettorino}, {Polenta}, {Poncet},
  {Pozzetti}, {Raison}, {Renzi}, {Rhodes}, {Rix}, {Romelli}, {Roncarelli},
  {Rossetti}, {Saglia}, {Sapone}, {Scaramella}, {Schneider}, {Scottez},
  {Secroun}, {Serrano}, {Sirri}, {Stanco}, {Sureau}, {Tallada-Cresp{\'\i}},
  {Tavagnacco}, {Taylor}, {Tenti}, {Tereno}, {Toledo-Moreo}, {Torradeflot},
  {Valenziano}, {Vassallo}, {Wang}, {Zacchei}, {Zamorani}, {Zoubian}, and
  {Zucca}}]{2019A&A...631A..85E}
{Euclid Collaboration}, {Barnett} R, {Warren} SJ, {Mortlock} DJ, {Cuby} JG,
  {Conselice} C, {Hewett} PC, {Willott} CJ, {Auricchio} N,
  {Balaguera-Antol{\'\i}nez} A, et~al. (2019) {Euclid preparation. V. Predicted
  yield of redshift 7 < z < 9 quasars from the wide survey}. \aap 631:A85.
  \doi{10.1051/0004-6361/201936427}.
  {\href{https://arxiv.org/abs/1908.04310}{{arXiv:1908.04310}}} {[astro-ph.GA]}

\bibitem[{{Event Horizon Telescope Collaboration} et~al.(2019){Event Horizon
  Telescope Collaboration}, {Akiyama}, {Alberdi}, {Alef}, {Asada}, {Azulay},
  {Baczko}, {Ball}, {Balokovi{\'c}}, {Barrett}, {Bintley}, {Blackburn},
  {Boland}, {Bouman}, {Bower}, {Bremer}, {Brinkerink}, {Brissenden}, {Britzen},
  {Broderick}, {Broguiere}, {Bronzwaer}, {Byun}, {Carlstrom}, {Chael}, {Chan},
  {Chatterjee}, {Chatterjee}, {Chen}, {Chen}, {Cho}, {Christian}, {Conway},
  {Cordes}, {Crew}, {Cui}, {Davelaar}, {De Laurentis}, {Deane}, {Dempsey},
  {Desvignes}, {Dexter}, {Doeleman}, {Eatough}, {Falcke}, {Fish}, {Fomalont},
  {Fraga-Encinas}, {Friberg}, {Fromm}, {G{\'o}mez}, {Galison}, {Gammie},
  {Garc{\'\i}a}, {Gentaz}, {Georgiev}, {Goddi}, {Gold}, {Gu}, {Gurwell},
  {Hada}, {Hecht}, {Hesper}, {Ho}, {Ho}, {Honma}, {Huang}, {Huang}, {Hughes},
  {Ikeda}, {Inoue}, {Issaoun}, {James}, {Jannuzi}, {Janssen}, {Jeter}, {Jiang},
  {Johnson}, {Jorstad}, {Jung}, {Karami}, {Karuppusamy}, {Kawashima},
  {Keating}, {Kettenis}, {Kim}, {Kim}, {Kim}, {Kino}, {Koay}, {Koch}, {Koyama},
  {Kramer}, {Kramer}, {Krichbaum}, {Kuo}, {Lauer}, {Lee}, {Li}, {Li},
  {Lindqvist}, {Liu}, {Liuzzo}, {Lo}, {Lobanov}, {Loinard}, {Lonsdale}, {Lu},
  {MacDonald}, {Mao}, {Markoff}, {Marrone}, {Marscher}, {Mart{\'\i}-Vidal},
  {Matsushita}, {Matthews}, {Medeiros}, {Menten}, {Mizuno}, {Mizuno}, {Moran},
  {Moriyama}, {Moscibrodzka}, {M{\"u}ller}, {Nagai}, {Nagar}, {Nakamura},
  {Narayan}, {Narayanan}, {Natarajan}, {Neri}, {Ni}, {Noutsos}, {Okino},
  {Olivares}, {Ortiz-Le{\'o}n}, {Oyama}, {{\"O}zel}, {Palumbo}, {Patel}, {Pen},
  {Pesce}, {Pi{\'e}tu}, {Plambeck}, {PopStefanija}, {Porth}, {Prather},
  {Preciado-L{\'o}pez}, {Psaltis}, {Pu}, {Ramakrishnan}, {Rao}, {Rawlings},
  {Raymond}, {Rezzolla}, {Ripperda}, {Roelofs}, {Rogers}, {Ros}, {Rose},
  {Roshanineshat}, {Rottmann}, {Roy}, {Ruszczyk}, {Ryan}, {Rygl},
  {S{\'a}nchez}, {S{\'a}nchez-Arguelles}, {Sasada}, {Savolainen}, {Schloerb},
  {Schuster}, {Shao}, {Shen}, {Small}, {Sohn}, {SooHoo}, {Tazaki}, {Tiede},
  {Tilanus}, {Titus}, {Toma}, {Torne}, {Trent}, {Trippe}, {Tsuda}, {van
  Bemmel}, {van Langevelde}, {van Rossum}, {Wagner}, {Wardle}, {Weintroub},
  {Wex}, {Wharton}, {Wielgus}, {Wong}, {Wu}, {Young}, {Young}, {Younsi},
  {Yuan}, {Yuan}, {Zensus}, {Zhao}, {Zhao}, {Zhu}, {Algaba}, {Allardi},
  {Amestica}, {Bach}, {Beaudoin}, {Benson}, {Berthold}, {Blanchard},
  {Blundell}, {Bustamente}, {Cappallo}, {Castillo-Dom{\'\i}nguez}, {Chang},
  {Chang}, {Chang}, {Chen}, {Chilson}, {Chuter}, {C{\'o}rdova Rosado},
  {Coulson}, {Crawford}, {Crowley}, {David}, {Derome}, {Dexter}, {Dornbusch},
  {Dudevoir}, {Dzib}, {Eckert}, {Erickson}, {Everett}, {Faber}, {Farah},
  {Fath}, {Folkers}, {Forbes}, {Freund}, {G{\'o}mez-Ruiz}, {Gale}, {Gao},
  {Geertsema}, {Graham}, {Greer}, {Grosslein}, {Gueth}, {Halverson}, {Han},
  {Han}, {Hao}, {Hasegawa}, {Henning}, {Hern{\'a}ndez-G{\'o}mez},
  {Herrero-Illana}, {Heyminck}, {Hirota}, {Hoge}, {Huang}, {Impellizzeri},
  {Jiang}, {Kamble}, {Keisler}, {Kimura}, {Kono}, {Kubo}, {Kuroda}, {Lacasse},
  {Laing}, {Leitch}, {Li}, {Lin}, {Liu}, {Liu}, {Lu}, {Marson},
  {Martin-Cocher}, {Massingill}, {Matulonis}, {McColl}, {McWhirter}, {Messias},
  {Meyer-Zhao}, {Michalik}, {Monta{\~n}a}, {Montgomerie}, {Mora-Klein},
  {Muders}, {Nadolski}, {Navarro}, {Nguyen}, {Nishioka}, {Norton}, {Nystrom},
  {Ogawa}, {Oshiro}, {Oyama}, {Padin}, {Parsons}, {Paine}, {Pe{\~n}alver},
  {Phillips}, {Poirier}, {Pradel}, {Primiani}, {Raffin}, {Rahlin}, {Reiland},
  {Risacher}, {Ruiz}, {S{\'a}ez-Mada{\'\i}n}, {Sassella}, {Schellart}, {Shaw},
  {Silva}, {Shiokawa}, {Smith}, {Snow}, {Souccar}, {Sousa}, {Sridharan},
  {Srinivasan}, {Stahm}, {Stark}, {Story}, {Timmer}, {Vertatschitsch},
  {Walther}, {Wei}, {Whitehorn}, {Whitney}, {Woody}, {Wouterloot}, {Wright},
  {Yamaguchi}, {Yu}, {Zeballos}, and {Ziurys}}]{2019ApJ...875L...2E}
{Event Horizon Telescope Collaboration}, {Akiyama} K, {Alberdi} A, {Alef} W,
  {Asada} K, {Azulay} R, {Baczko} AK, {Ball} D, {Balokovi{\'c}} M, {Barrett} J,
  et~al. (2019) {First M87 Event Horizon Telescope Results. II. Array and
  Instrumentation}. \apjl 875(1):L2. \doi{10.3847/2041-8213/ab0c96}.
  {\href{https://arxiv.org/abs/1906.11239}{{arXiv:1906.11239}}} {[astro-ph.IM]}

\bibitem[{{Everson} et~al.(2020){Everson}, {MacLeod}, {De}, {Macias}, and
  {Ramirez-Ruiz}}]{2020ApJ...899...77E}
{Everson} RW, {MacLeod} M, {De} S, {Macias} P, {Ramirez-Ruiz} E (2020) {Common
  Envelope Wind Tunnel: Range of Applicability and Self-similarity in Realistic
  Stellar Envelopes}. \apj 899(1):77. \doi{10.3847/1538-4357/aba75c}.
  {\href{https://arxiv.org/abs/2006.07471}{{arXiv:2006.07471}}} {[astro-ph.SR]}

\bibitem[{{Fabian}(2012)}]{2012ARA&A..50..455F}
{Fabian} AC (2012) {Observational Evidence of Active Galactic Nuclei Feedback}.
  \araa 50:455--489. \doi{10.1146/annurev-astro-081811-125521}.
  {\href{https://arxiv.org/abs/1204.4114}{{arXiv:1204.4114}}} {[astro-ph.CO]}

\bibitem[{{Fabian} et~al.(1975){Fabian}, {Pringle}, and
  {Rees}}]{1975MNRAS.172P..15F}
{Fabian} AC, {Pringle} JE, {Rees} MJ (1975) {Tidal capture formation of binary
  systems and X-ray sources in globular clusters.} \mnras 172:15.
  \doi{10.1093/mnras/172.1.15P}

\bibitem[{{Fabian} et~al.(2000){Fabian}, {Iwasawa}, {Reynolds}, and
  {Young}}]{2000PASP..112.1145F}
{Fabian} AC, {Iwasawa} K, {Reynolds} CS, {Young} AJ (2000) {Broad Iron Lines in
  Active Galactic Nuclei}. \pasp 112(775):1145--1161. \doi{10.1086/316610}.
  {\href{https://arxiv.org/abs/astro-ph/0004366}{{arXiv:astro-ph/0004366}}}
  {[astro-ph]}

\bibitem[{{Fabj} et~al.(2020){Fabj}, {Nasim}, {Caban}, {Ford}, {McKernan}, and
  {Bellovary}}]{2020arXiv200611229F}
{Fabj} G, {Nasim} SS, {Caban} F, {Ford} KES, {McKernan} B, {Bellovary} JM
  (2020) {Aligning Nuclear Cluster Orbits with an Active Galactic Nucleus
  Accretion Disk}. MNRAS 499(2):arXiv:2006.11229. \doi{10.1093/mnras/staa3004}.
  {\href{https://arxiv.org/abs/2006.11229}{{arXiv:2006.11229}}} {[astro-ph.GA]}

\bibitem[{{Fabrycky} and {Tremaine}(2007)}]{2007ApJ...669.1298F}
{Fabrycky} D, {Tremaine} S (2007) {Shrinking Binary and Planetary Orbits by
  Kozai Cycles with Tidal Friction}. \apj 669(2):1298--1315.
  \doi{10.1086/521702}.
  {\href{https://arxiv.org/abs/0705.4285}{{arXiv:0705.4285}}} {[astro-ph]}

\bibitem[{Fairhurst(2009)}]{Fairhurst:2009}
Fairhurst S (2009) Triangulation of gravitational wave sources with a network
  of detectors. New Journal of Physics 11:123006

\bibitem[{{Fakhouri} et~al.(2010){Fakhouri}, {Ma}, and
  {Boylan-Kolchin}}]{2010MNRAS.406.2267F}
{Fakhouri} O, {Ma} CP, {Boylan-Kolchin} M (2010) {The merger rates and mass
  assembly histories of dark matter haloes in the two Millennium simulations}.
  \mnras 406(4):2267--2278. \doi{10.1111/j.1365-2966.2010.16859.x}.
  {\href{https://arxiv.org/abs/1001.2304}{{arXiv:1001.2304}}} {[astro-ph.CO]}

\bibitem[{{Falta} et~al.(2011){Falta}, {Fisher}, and
  {Khanna}}]{2011PhRvL.106t1103F}
{Falta} D, {Fisher} R, {Khanna} G (2011) {Gravitational Wave Emission from the
  Single-Degenerate Channel of Type Ia Supernovae}. \prl 106(20):201103.
  \doi{10.1103/PhysRevLett.106.201103}.
  {\href{https://arxiv.org/abs/1011.6387}{{arXiv:1011.6387}}} {[astro-ph.HE]}

\bibitem[{{Fan} et~al.(2003){Fan}, {Strauss}, {Schneider}, {Becker}, {White},
  {Haiman}, {Gregg}, {Pentericci}, {Grebel}, {Narayanan}, {Loh}, {Richards},
  {Gunn}, {Lupton}, {Knapp}, {Ivezi{\'c}}, {Brandt}, {Collinge}, {Hao},
  {Harbeck}, {Prada}, {Schaye}, {Strateva}, {Zakamska}, {Anderson},
  {Brinkmann}, {Bahcall}, {Lamb}, {Okamura}, {Szalay}, and
  {York}}]{2003AJ....125.1649F}
{Fan} X, {Strauss} MA, {Schneider} DP, {Becker} RH, {White} RL, {Haiman} Z,
  {Gregg} M, {Pentericci} L, {Grebel} EK, {Narayanan} VK, et~al. (2003) {A
  Survey of z>5.7 Quasars in the Sloan Digital Sky Survey. II. Discovery of
  Three Additional Quasars at z>6}. \aj 125(4):1649--1659.
  \doi{10.1086/368246}.
  {\href{https://arxiv.org/abs/astro-ph/0301135}{{arXiv:astro-ph/0301135}}}
  {[astro-ph]}

\bibitem[{{Fan} et~al.(2006){Fan}, {Carilli}, and
  {Keating}}]{2006ARA&A..44..415F}
{Fan} X, {Carilli} CL, {Keating} B (2006) {Observational Constraints on Cosmic
  Reionization}. \araa 44(1):415--462.
  \doi{10.1146/annurev.astro.44.051905.092514}.
  {\href{https://arxiv.org/abs/astro-ph/0602375}{{arXiv:astro-ph/0602375}}}
  {[astro-ph]}

\bibitem[{{Fan} et~al.(2019){Fan}, {Barth}, {Banados}, {De Rosa}, {Decarli},
  {Eilers}, {Farina}, {Greene}, {Habouzit}, {Jiang}, {Jun}, {Koekemoer},
  {Malhotra}, {Mazzucchelli}, {Pacucci}, {Rhoads}, {Riechers}, {Rigby}, {Shen},
  {Simcoe}, {Stern}, {Strauss}, {Treu}, {Venemans}, {Vestergaard}, {Volonteri},
  {Walter}, {Yang}, and {Wang}}]{2019BAAS...51c.121F}
{Fan} X, {Barth} A, {Banados} E, {De Rosa} G, {Decarli} R, {Eilers} AC,
  {Farina} EP, {Greene} J, {Habouzit} M, {Jiang} L, et~al. (2019) {The First
  Luminous Quasars and Their Host Galaxies}. \baas 51(3):121.
  {\href{https://arxiv.org/abs/1903.04078}{{arXiv:1903.04078}}} {[astro-ph.GA]}

\bibitem[{{Farihi} et~al.(2005){Farihi}, {Becklin}, and
  {Zuckerman}}]{2005ApJS..161..394F}
{Farihi} J, {Becklin} EE, {Zuckerman} B (2005) {Low-Luminosity Companions to
  White Dwarfs}. \apjs 161(2):394--428. \doi{10.1086/444362}.
  {\href{https://arxiv.org/abs/astro-ph/0506017}{{arXiv:astro-ph/0506017}}}
  {[astro-ph]}

\bibitem[{{Farihi} et~al.(2009){Farihi}, {Jura}, and
  {Zuckerman}}]{2009ApJ...694..805F}
{Farihi} J, {Jura} M, {Zuckerman} B (2009) {Infrared Signatures of Disrupted
  Minor Planets at White Dwarfs}. \apj 694(2):805--819.
  \doi{10.1088/0004-637X/694/2/805}.
  {\href{https://arxiv.org/abs/0901.0973}{{arXiv:0901.0973}}} {[astro-ph.EP]}

\bibitem[{{Farmer} and {Phinney}(2003)}]{2003MNRAS.346.1197F}
{Farmer} AJ, {Phinney} ES (2003) {The gravitational wave background from
  cosmological compact binaries}. \mnras 346(4):1197--1214.
  \doi{10.1111/j.1365-2966.2003.07176.x}.
  {\href{https://arxiv.org/abs/astro-ph/0304393}{{arXiv:astro-ph/0304393}}}
  {[astro-ph]}

\bibitem[{{Farmer} et~al.(2019){Farmer}, {Renzo}, {de Mink}, {Marchant}, and
  {Justham}}]{2019ApJ...887...53F}
{Farmer} R, {Renzo} M, {de Mink} SE, {Marchant} P, {Justham} S (2019) {Mind the
  Gap: The Location of the Lower Edge of the Pair-instability Supernova Black
  Hole Mass Gap}. \apj 887(1):53. \doi{10.3847/1538-4357/ab518b}.
  {\href{https://arxiv.org/abs/1910.12874}{{arXiv:1910.12874}}} {[astro-ph.SR]}

\bibitem[{{Farr} et~al.(2017){Farr}, {Stevenson}, {Miller}, {Mandel}, {Farr},
  and {Vecchio}}]{Farr:2017}
{Farr} WM, {Stevenson} S, {Miller} MC, {Mandel} I, {Farr} B, {Vecchio} A (2017)
  {Distinguishing spin-aligned and isotropic black hole populations with
  gravitational waves}. \nat 548:426--429. \doi{10.1038/nature23453}.
  {\href{https://arxiv.org/abs/1706.01385}{{arXiv:1706.01385}}} {[astro-ph.HE]}

\bibitem[{{Farris} et~al.(2010){Farris}, {Liu}, and
  {Shapiro}}]{2010PhRvD..81h4008F}
{Farris} BD, {Liu} YT, {Shapiro} SL (2010) {Binary black hole mergers in
  gaseous environments: ``Binary Bondi`` and ``binary Bondi-Hoyle-Lyttleton''
  accretion}. \prd 81(8):084008. \doi{10.1103/PhysRevD.81.084008}.
  {\href{https://arxiv.org/abs/0912.2096}{{arXiv:0912.2096}}} {[astro-ph.HE]}

\bibitem[{{Farris} et~al.(2012){Farris}, {Gold}, {Paschalidis}, {Etienne}, and
  {Shapiro}}]{2012PhRvL.109v1102F}
{Farris} BD, {Gold} R, {Paschalidis} V, {Etienne} ZB, {Shapiro} SL (2012)
  {Binary Black-Hole Mergers in Magnetized Disks: Simulations in Full General
  Relativity}. \prl 109(22):221102. \doi{10.1103/PhysRevLett.109.221102}.
  {\href{https://arxiv.org/abs/1207.3354}{{arXiv:1207.3354}}} {[astro-ph.HE]}

\bibitem[{{Farris} et~al.(2015{\natexlab{a}}){Farris}, {Duffell}, {MacFadyen},
  and {Haiman}}]{2015MNRAS.447L..80F}
{Farris} BD, {Duffell} P, {MacFadyen} AI, {Haiman} Z (2015{\natexlab{a}})
  {Binary black hole accretion during inspiral and merger.} \mnras
  447:L80--L84. \doi{10.1093/mnrasl/slu184}.
  {\href{https://arxiv.org/abs/1409.5124}{{arXiv:1409.5124}}} {[astro-ph.HE]}

\bibitem[{{Farris} et~al.(2015{\natexlab{b}}){Farris}, {Duffell}, {MacFadyen},
  and {Haiman}}]{2015MNRAS.446L..36F}
{Farris} BD, {Duffell} P, {MacFadyen} AI, {Haiman} Z (2015{\natexlab{b}})
  {Characteristic signatures in the thermal emission from accreting binary
  black holes.} \mnras 446:L36--L40. \doi{10.1093/mnrasl/slu160}.
  {\href{https://arxiv.org/abs/1406.0007}{{arXiv:1406.0007}}} {[astro-ph.HE]}

\bibitem[{{Farrow} et~al.(2019){Farrow}, {Zhu}, and
  {Thrane}}]{2019ApJ...876...18F}
{Farrow} N, {Zhu} XJ, {Thrane} E (2019) {The Mass Distribution of Galactic
  Double Neutron Stars}. \apj 876(1):18. \doi{10.3847/1538-4357/ab12e3}.
  {\href{https://arxiv.org/abs/1902.03300}{{arXiv:1902.03300}}} {[astro-ph.HE]}

\bibitem[{{Faulkner}(1971)}]{1971ApJ...170L..99F}
{Faulkner} J (1971) {Ultrashort-Period Binaries, Gravitational Radiation, and
  Mass Transfer. I. The Standard Model, with Applications to WZ Sagittae and Z
  Camelopardalis}. \apjl 170:L99. \doi{10.1086/180848}

\bibitem[{{Feng} and {Soria}(2011)}]{2011NewAR..55..166F}
{Feng} H, {Soria} R (2011) {Ultraluminous X-ray sources in the Chandra and
  XMM-Newton era}. \nar 55(5):166--183. \doi{10.1016/j.newar.2011.08.002}.
  {\href{https://arxiv.org/abs/1109.1610}{{arXiv:1109.1610}}} {[astro-ph.HE]}

\bibitem[{{Fern{\'a}ndez} and {Metzger}(2013)}]{2013ApJ...763..108F}
{Fern{\'a}ndez} R, {Metzger} BD (2013) {Nuclear Dominated Accretion Flows in
  Two Dimensions. I. Torus Evolution with Parametric Microphysics}. \apj
  763(2):108. \doi{10.1088/0004-637X/763/2/108}.
  {\href{https://arxiv.org/abs/1209.2712}{{arXiv:1209.2712}}} {[astro-ph.HE]}

\bibitem[{{Fernandez} et~al.(2014){Fernandez}, {Bryan}, {Haiman}, and
  {Li}}]{2014MNRAS.439.3798F}
{Fernandez} R, {Bryan} GL, {Haiman} Z, {Li} M (2014) {H$_{2}$ suppression with
  shocking inflows: testing a pathway for supermassive black hole formation}.
  \mnras 439:3798--3807. \doi{10.1093/mnras/stu230}.
  {\href{https://arxiv.org/abs/1401.5803}{{arXiv:1401.5803}}} {[astro-ph.CO]}

\bibitem[{{Ferrara} et~al.(2014){Ferrara}, {Salvadori}, {Yue}, and
  {Schleicher}}]{2014MNRAS.443.2410F}
{Ferrara} A, {Salvadori} S, {Yue} B, {Schleicher} D (2014) {Initial mass
  function of intermediate-mass black hole seeds}. \mnras 443(3):2410--2425.
  \doi{10.1093/mnras/stu1280}.
  {\href{https://arxiv.org/abs/1406.6685}{{arXiv:1406.6685}}} {[astro-ph.GA]}

\bibitem[{{Ferrarese} and {Merritt}(2000)}]{2000ApJ...539L...9F}
{Ferrarese} L, {Merritt} D (2000) {A Fundamental Relation between Supermassive
  Black Holes and Their Host Galaxies}. \apjl 539(1):L9--L12.
  \doi{10.1086/312838}.
  {\href{https://arxiv.org/abs/astro-ph/0006053}{{arXiv:astro-ph/0006053}}}
  {[astro-ph]}

\bibitem[{{Ferrario} et~al.(2015){Ferrario}, {de Martino}, and
  {G{\"a}nsicke}}]{2015SSRv..191..111F}
{Ferrario} L, {de Martino} D, {G{\"a}nsicke} BT (2015) {Magnetic White Dwarfs}.
  \ssr 191(1-4):111--169. \doi{10.1007/s11214-015-0152-0}.
  {\href{https://arxiv.org/abs/1504.08072}{{arXiv:1504.08072}}} {[astro-ph.SR]}

\bibitem[{{Fiacconi} et~al.(2013){Fiacconi}, {Mayer}, {Ro{\v{s}}kar}, and
  {Colpi}}]{2013ApJ...777L..14F}
{Fiacconi} D, {Mayer} L, {Ro{\v{s}}kar} R, {Colpi} M (2013) {Massive Black Hole
  Pairs in Clumpy, Self-gravitating Circumnuclear Disks: Stochastic Orbital
  Decay}. \apjl 777(1):L14. \doi{10.1088/2041-8205/777/1/L14}.
  {\href{https://arxiv.org/abs/1307.0822}{{arXiv:1307.0822}}} {[astro-ph.CO]}

\bibitem[{{Fiacconi} et~al.(2017){Fiacconi}, {Mayer}, {Madau}, {Lupi}, {Dotti},
  and {Haardt}}]{2017MNRAS.467.4080F}
{Fiacconi} D, {Mayer} L, {Madau} P, {Lupi} A, {Dotti} M, {Haardt} F (2017)
  {Young and turbulent: the early life of massive galaxy progenitors}. \mnras
  467(4):4080--4100. \doi{10.1093/mnras/stx335}.
  {\href{https://arxiv.org/abs/1609.09499}{{arXiv:1609.09499}}} {[astro-ph.GA]}

\bibitem[{{Fiacconi} et~al.(2018){Fiacconi}, {Sijacki}, and
  {Pringle}}]{2018MNRAS.477.3807F}
{Fiacconi} D, {Sijacki} D, {Pringle} JE (2018) {Galactic nuclei evolution with
  spinning black holes: method and implementation}. \mnras 477(3):3807--3835.
  \doi{10.1093/mnras/sty893}.
  {\href{https://arxiv.org/abs/1712.00023}{{arXiv:1712.00023}}} {[astro-ph.GA]}

\bibitem[{{Fink} et~al.(2010){Fink}, {R{\"o}pke}, {Hillebrandt}, {Seitenzahl},
  {Sim}, and {Kromer}}]{2010A&A...514A..53F}
{Fink} M, {R{\"o}pke} FK, {Hillebrandt} W, {Seitenzahl} IR, {Sim} SA, {Kromer}
  M (2010) {Double-detonation sub-Chandrasekhar supernovae: can minimum helium
  shell masses detonate the core?} \aap 514:A53.
  \doi{10.1051/0004-6361/200913892}.
  {\href{https://arxiv.org/abs/1002.2173}{{arXiv:1002.2173}}} {[astro-ph.SR]}

\bibitem[{{Finkelstein} et~al.(2021){Finkelstein}, {Papovich}, {Pirzkal},
  {Bagley}, {Berg}, {Castellano}, {Chavez Ortiz}, {Chworowsky}, {Dave},
  {Dickinson}, {Estrada-Carpenter}, {Faber}, {Ferguson}, {Fontana},
  {Giavalisco}, {Grazian}, {Grogin}, {Jaskot}, {Jung}, {Kartaltepe}, {Kewley},
  {Kirkpatrick}, {Kocevski}, {Larson}, {Leung}, {Lotz}, {Mantha}, {Matharu},
  {McCarron}, {McIntosh}, {Natarajan}, {Pentericci}, {Ravindranath},
  {Rodriguez-Gomez}, {Rothberg}, {Ryan}, {Simons}, {Snyder}, {Somerville},
  {Trump}, {Wilkins}, and {Yung}}]{2021jwst.prop.2079F}
{Finkelstein} SL, {Papovich} C, {Pirzkal} N, {Bagley} M, {Berg} D, {Castellano}
  M, {Chavez Ortiz} OA, {Chworowsky} K, {Dave} R, {Dickinson} M, et~al. (2021)
  {The Webb Deep Extragalactic Exploratory Public (WDEEP) Survey: Feedback in
  Low-Mass Galaxies from Cosmic Dawn to Dusk}. JWST Proposal. Cycle 1, ID.
  \#2079

\bibitem[{{Fiore} et~al.(2009){Fiore}, {Puccetti}, {Brusa}, {Salvato},
  {Zamorani}, {Aldcroft}, {Aussel}, {Brunner}, {Capak}, {Cappelluti}, {Civano},
  {Comastri}, {Elvis}, {Feruglio}, {Finoguenov}, {Fruscione}, {Gilli},
  {Hasinger}, {Koekemoer}, {Kartaltepe}, {Ilbert}, {Impey}, {Le Floc'h},
  {Lilly}, {Mainieri}, {Martinez-Sansigre}, {McCracken}, {Menci}, {Merloni},
  {Miyaji}, {Sanders}, {Sargent}, {Schinnerer}, {Scoville}, {Silverman},
  {Smolcic}, {Steffen}, {Santini}, {Taniguchi}, {Thompson}, {Trump}, {Vignali},
  {Urry}, and {Yan}}]{2009ApJ...693..447F}
{Fiore} F, {Puccetti} S, {Brusa} M, {Salvato} M, {Zamorani} G, {Aldcroft} T,
  {Aussel} H, {Brunner} H, {Capak} P, {Cappelluti} N, et~al. (2009) {Chasing
  Highly Obscured QSOs in the COSMOS Field}. \apj 693(1):447--462.
  \doi{10.1088/0004-637X/693/1/447}.
  {\href{https://arxiv.org/abs/0810.0720}{{arXiv:0810.0720}}} {[astro-ph]}

\bibitem[{{Fitchett}(1983)}]{1983MNRAS.203.1049F}
{Fitchett} MJ (1983) {The influence of gravitational wave momentum losses on
  the centre of mass motion of a Newtonian binay system.} \mnras
  203:1049--1062. \doi{10.1093/mnras/203.4.1049}

\bibitem[{{Flanagan} and {Hinderer}(2008)}]{2008PhRvD..77b1502F}
{Flanagan} {\'E}{\'E}, {Hinderer} T (2008) {Constraining neutron-star tidal
  Love numbers with gravitational-wave detectors}. \prd 77(2):021502.
  \doi{10.1103/PhysRevD.77.021502}.
  {\href{https://arxiv.org/abs/0709.1915}{{arXiv:0709.1915}}} {[astro-ph]}

\bibitem[{{Flanagan} and {Hinderer}(2012)}]{2012PhRvL.109g1102F}
{Flanagan} {\'E}{\'E}, {Hinderer} T (2012) {Transient Resonances in the
  Inspirals of Point Particles into Black Holes}. \prl 109(7):071102.
  \doi{10.1103/PhysRevLett.109.071102}.
  {\href{https://arxiv.org/abs/1009.4923}{{arXiv:1009.4923}}} {[gr-qc]}

\bibitem[{{Fluri} et~al.(2019){Fluri}, {Kacprzak}, {Lucchi}, {Refregier},
  {Amara}, {Hofmann}, and {Schneider}}]{2019PhRvD.100f3514F}
{Fluri} J, {Kacprzak} T, {Lucchi} A, {Refregier} A, {Amara} A, {Hofmann} T,
  {Schneider} A (2019) {Cosmological constraints with deep learning from
  KiDS-450 weak lensing maps}. \prd 100(6):063514.
  \doi{10.1103/PhysRevD.100.063514}.
  {\href{https://arxiv.org/abs/1906.03156}{{arXiv:1906.03156}}} {[astro-ph.CO]}

\bibitem[{{Fontaine} et~al.(2011){Fontaine}, {Brassard}, {Green}, {Charpinet},
  {Dufour}, {Hubeny}, {Steeghs}, {Aerts}, {Randall}, {Bergeron}, {Guvenen},
  {O'Malley}, {Van Grootel}, {{\O}stensen}, {Bloemen}, {Silvotti}, {Howell},
  {Baran}, {Kepler}, {Marsh}, {Montgomery}, {Oreiro}, {Provencal}, {Telting},
  {Winget}, {Zima}, {Christensen-Dalsgaard}, and
  {Kjeldsen}}]{2011ApJ...726...92F}
{Fontaine} G, {Brassard} P, {Green} EM, {Charpinet} S, {Dufour} P, {Hubeny} I,
  {Steeghs} D, {Aerts} C, {Randall} SK, {Bergeron} P, et~al. (2011) {Discovery
  of a New AM CVn System with the Kepler Satellite}. \apj 726(2):92.
  \doi{10.1088/0004-637X/726/2/92}

\bibitem[{{Fontecilla} et~al.(2017){Fontecilla}, {Chen}, and
  {Cuadra}}]{2017MNRAS.468L..50F}
{Fontecilla} C, {Chen} X, {Cuadra} J (2017) {A second decoupling between
  merging binary black holes and the inner disc - impact on the electromagnetic
  counterpart}. \mnras 468(1):L50--L54. \doi{10.1093/mnrasl/slw258}.
  {\href{https://arxiv.org/abs/1610.09382}{{arXiv:1610.09382}}} {[astro-ph.HE]}

\bibitem[{{Ford} et~al.(2000){Ford}, {Kozinsky}, and
  {Rasio}}]{2000ApJ...535..385F}
{Ford} EB, {Kozinsky} B, {Rasio} FA (2000) {Secular Evolution of Hierarchical
  Triple Star Systems}. \apj 535(1):385--401. \doi{10.1086/308815}

\bibitem[{{Fouvry} et~al.(2019){Fouvry}, {Bar-Or}, and
  {Chavanis}}]{2019ApJ...883..161F}
{Fouvry} JB, {Bar-Or} B, {Chavanis} PH (2019) {Vector Resonant Relaxation of
  Stars around a Massive Black Hole}. \apj 883(2):161.
  \doi{10.3847/1538-4357/ab2f78}.
  {\href{https://arxiv.org/abs/1812.07053}{{arXiv:1812.07053}}} {[astro-ph.GA]}

\bibitem[{{Fragione}(2022)}]{2022arXiv220205618F}
{Fragione} G (2022) {Mergers of supermassive and intermediate-mass black holes
  in galactic nuclei from disruptions of star clusters}. arXiv e-prints
  arXiv:2202.05618.
  {\href{https://arxiv.org/abs/2202.05618}{{arXiv:2202.05618}}} {[astro-ph.HE]}

\bibitem[{{Fragione} and {Kocsis}(2019)}]{2019MNRAS.486.4781F}
{Fragione} G, {Kocsis} B (2019) {Black hole mergers from quadruples}. \mnras
  486(4):4781--4789. \doi{10.1093/mnras/stz1175}.
  {\href{https://arxiv.org/abs/1903.03112}{{arXiv:1903.03112}}} {[astro-ph.GA]}

\bibitem[{{Fragione} and {Loeb}(2019)}]{2019MNRAS.486.4443F}
{Fragione} G, {Loeb} A (2019) {Black hole-neutron star mergers from triples}.
  \mnras 486(3):4443--4450. \doi{10.1093/mnras/stz1131}.
  {\href{https://arxiv.org/abs/1903.10511}{{arXiv:1903.10511}}} {[astro-ph.GA]}

\bibitem[{{Fragione} and {Silk}(2020)}]{2020MNRAS.tmp.2017F}
{Fragione} G, {Silk} J (2020) {Repeated mergers and ejection of black holes
  within nuclear star clusters}. \mnras \doi{10.1093/mnras/staa2629}.
  {\href{https://arxiv.org/abs/2006.01867}{{arXiv:2006.01867}}} {[astro-ph.GA]}

\bibitem[{{Fragione} et~al.(2018){Fragione}, {Ginsburg}, and
  {Kocsis}}]{2018ApJ...856...92F}
{Fragione} G, {Ginsburg} I, {Kocsis} B (2018) {Gravitational Waves and
  Intermediate-mass Black Hole Retention in Globular Clusters}. \apj 856(2):92.
  \doi{10.3847/1538-4357/aab368}.
  {\href{https://arxiv.org/abs/1711.00483}{{arXiv:1711.00483}}} {[astro-ph.GA]}

\bibitem[{{Fragione} et~al.(2019){Fragione}, {Grishin}, {Leigh}, {Perets}, and
  {Perna}}]{2019MNRAS.488...47F}
{Fragione} G, {Grishin} E, {Leigh} NWC, {Perets} HB, {Perna} R (2019) {Black
  hole and neutron star mergers in galactic nuclei}. \mnras 488(1):47--63.
  \doi{10.1093/mnras/stz1651}.
  {\href{https://arxiv.org/abs/1811.10627}{{arXiv:1811.10627}}} {[astro-ph.GA]}

\bibitem[{{Fragione} et~al.(2020){Fragione}, {Martinez}, {Kremer},
  {Chatterjee}, {Rodriguez}, {Ye}, {Weatherford}, {Naoz}, and
  {Rasio}}]{2020ApJ...900...16F}
{Fragione} G, {Martinez} MAS, {Kremer} K, {Chatterjee} S, {Rodriguez} CL, {Ye}
  CS, {Weatherford} NC, {Naoz} S, {Rasio} FA (2020) {Demographics of Triple
  Systems in Dense Star Clusters}. \apj 900(1):16.
  \doi{10.3847/1538-4357/aba89b}.
  {\href{https://arxiv.org/abs/2007.11605}{{arXiv:2007.11605}}} {[astro-ph.GA]}

\bibitem[{{Fragos} et~al.(2019){Fragos}, {Andrews}, {Ramirez-Ruiz}, {Meynet},
  {Kalogera}, {Taam}, and {Zezas}}]{2019ApJ...883L..45F}
{Fragos} T, {Andrews} JJ, {Ramirez-Ruiz} E, {Meynet} G, {Kalogera} V, {Taam}
  RE, {Zezas} A (2019) {The Complete Evolution of a Neutron-star Binary through
  a Common Envelope Phase Using 1D Hydrodynamic Simulations}. \apjl 883(2):L45.
  \doi{10.3847/2041-8213/ab40d1}.
  {\href{https://arxiv.org/abs/1907.12573}{{arXiv:1907.12573}}} {[astro-ph.HE]}

\bibitem[{{Fragos} et~al.(2022){Fragos}, {Andrews}, {Bavera}, {Berry},
  {Coughlin}, {Dotter}, {Giri}, {Kalogera}, {Katsaggelos}, {Kovlakas},
  {Lalvani}, {Misra}, {Srivastava}, {Qin}, {Rocha}, {Roman-Garza}, {Serra},
  {Stahle}, {Sun}, {Teng}, {Trajcevski}, {Hai Tran}, {Xing}, {Zapartas}, and
  {Zevin}}]{2022arXiv220205892F}
{Fragos} T, {Andrews} JJ, {Bavera} SS, {Berry} CPL, {Coughlin} S, {Dotter} A,
  {Giri} P, {Kalogera} V, {Katsaggelos} A, {Kovlakas} K, et~al. (2022)
  {POSYDON: A General-Purpose Population Synthesis Code with Detailed
  Binary-Evolution Simulations}. arXiv e-prints arXiv:2202.05892.
  {\href{https://arxiv.org/abs/2202.05892}{{arXiv:2202.05892}}} {[astro-ph.SR]}

\bibitem[{{Franchini} et~al.(2021){Franchini}, {Sesana}, and
  {Dotti}}]{2021MNRAS.507.1458F}
{Franchini} A, {Sesana} A, {Dotti} M (2021) {Circumbinary disc self-gravity
  governing supermassive black hole binary mergers}. \mnras 507(1):1458--1467.
  \doi{10.1093/mnras/stab2234}.
  {\href{https://arxiv.org/abs/2106.13253}{{arXiv:2106.13253}}} {[astro-ph.HE]}

\bibitem[{{Franchini} et~al.(2022){Franchini}, {Lupi}, and
  {Sesana}}]{2022ApJ...929L..13F}
{Franchini} A, {Lupi} A, {Sesana} A (2022) {Resolving Massive Black Hole Binary
  Evolution via Adaptive Particle Splitting}. \apjl 929(1):L13.
  \doi{10.3847/2041-8213/ac63a2}.
  {\href{https://arxiv.org/abs/2201.05619}{{arXiv:2201.05619}}} {[astro-ph.HE]}

\bibitem[{{Frank} and {Rees}(1976)}]{1976MNRAS.176..633F}
{Frank} J, {Rees} MJ (1976) {Effects of massive black holes on dense stellar
  systems.} \mnras 176:633--647. \doi{10.1093/mnras/176.3.633}

\bibitem[{{Frank} et~al.(2002){Frank}, {King}, and
  {Raine}}]{2002apa..book.....F}
{Frank} J, {King} A, {Raine} DJ (2002) {Accretion Power in Astrophysics: Third
  Edition}

\bibitem[{{Freitag}(2001)}]{2001CQGra..18.4033F}
{Freitag} M (2001) {Monte Carlo cluster simulations to determine the rate of
  compact star inspiralling to a central galactic black hole}. Classical and
  Quantum Gravity 18(19):4033--4038. \doi{10.1088/0264-9381/18/19/309}.
  {\href{https://arxiv.org/abs/astro-ph/0107193}{{arXiv:astro-ph/0107193}}}
  {[astro-ph]}

\bibitem[{{Freitag} et~al.(2006{\natexlab{a}}){Freitag}, {G{\"u}rkan}, and
  {Rasio}}]{2006MNRAS.368..141F}
{Freitag} M, {G{\"u}rkan} MA, {Rasio} FA (2006{\natexlab{a}}) {Runaway
  collisions in young star clusters - II. Numerical results}. \mnras
  368(1):141--161. \doi{10.1111/j.1365-2966.2006.10096.x}.
  {\href{https://arxiv.org/abs/astro-ph/0503130}{{arXiv:astro-ph/0503130}}}
  {[astro-ph]}

\bibitem[{{Freitag} et~al.(2006{\natexlab{b}}){Freitag}, {Rasio}, and
  {Baumgardt}}]{2006MNRAS.368..121F}
{Freitag} M, {Rasio} FA, {Baumgardt} H (2006{\natexlab{b}}) {Runaway collisions
  in young star clusters - I. Methods and tests}. \mnras 368(1):121--140.
  \doi{10.1111/j.1365-2966.2006.10095.x}.
  {\href{https://arxiv.org/abs/astro-ph/0503129}{{arXiv:astro-ph/0503129}}}
  {[astro-ph]}

\bibitem[{{French} et~al.(2016){French}, {Arcavi}, and
  {Zabludoff}}]{2016ApJ...818L..21F}
{French} KD, {Arcavi} I, {Zabludoff} A (2016) {Tidal Disruption Events Prefer
  Unusual Host Galaxies}. \apjl 818(1):L21. \doi{10.3847/2041-8205/818/1/L21}.
  {\href{https://arxiv.org/abs/1601.04705}{{arXiv:1601.04705}}} {[astro-ph.GA]}

\bibitem[{{Fruchter} et~al.(1988){Fruchter}, {Stinebring}, and
  {Taylor}}]{1988Natur.333..237F}
{Fruchter} AS, {Stinebring} DR, {Taylor} JH (1988) {A millisecond pulsar in an
  eclipsing binary}. \nat 333(6170):237--239. \doi{10.1038/333237a0}

\bibitem[{{Fryer} et~al.(2012){Fryer}, {Belczynski}, {Wiktorowicz}, {Dominik},
  {Kalogera}, and {Holz}}]{2012ApJ...749...91F}
{Fryer} CL, {Belczynski} K, {Wiktorowicz} G, {Dominik} M, {Kalogera} V, {Holz}
  DE (2012) {Compact Remnant Mass Function: Dependence on the Explosion
  Mechanism and Metallicity}. \apj 749(1):91. \doi{10.1088/0004-637X/749/1/91}.
  {\href{https://arxiv.org/abs/1110.1726}{{arXiv:1110.1726}}} {[astro-ph.SR]}

\bibitem[{{Fujita}(2015)}]{2015PTEP.2015c3E01F}
{Fujita} R (2015) {Gravitational waves from a particle in circular orbits
  around a rotating black hole to the 11th post-Newtonian order}. Progress of
  Theoretical and Experimental Physics 2015(3):033E01.
  \doi{10.1093/ptep/ptv012}.
  {\href{https://arxiv.org/abs/1412.5689}{{arXiv:1412.5689}}} {[gr-qc]}

\bibitem[{{Fuller} and {Lai}(2012)}]{2012MNRAS.421..426F}
{Fuller} J, {Lai} D (2012) {Dynamical tides in compact white dwarf binaries:
  tidal synchronization and dissipation}. \mnras 421(1):426--445.
  \doi{10.1111/j.1365-2966.2011.20320.x}.
  {\href{https://arxiv.org/abs/1108.4910}{{arXiv:1108.4910}}} {[astro-ph.SR]}

\bibitem[{{Gabbard} et~al.(2018){Gabbard}, {Williams}, {Hayes}, and
  {Messenger}}]{2018PhRvL.120n1103G}
{Gabbard} H, {Williams} M, {Hayes} F, {Messenger} C (2018) {Matching Matched
  Filtering with Deep Networks for Gravitational-Wave Astronomy}. \prl
  120(14):141103. \doi{10.1103/PhysRevLett.120.141103}.
  {\href{https://arxiv.org/abs/1712.06041}{{arXiv:1712.06041}}} {[astro-ph.IM]}

\bibitem[{{Gaia Collaboration} et~al.(2018{\natexlab{a}}){Gaia Collaboration},
  {Brown}, {Vallenari}, {Prusti}, {de Bruijne}, {Babusiaux}, {Bailer-Jones},
  {Biermann}, {Evans}, {Eyer}, {Jansen}, {Jordi}, and
  {et~al.}}]{2018A&A...616A...1G}
{Gaia Collaboration}, {Brown} AGA, {Vallenari} A, {Prusti} T, {de Bruijne} JHJ,
  {Babusiaux} C, {Bailer-Jones} CAL, {Biermann} M, {Evans} DW, {Eyer} L, et~al.
  (2018{\natexlab{a}}) {Gaia Data Release 2. Summary of the contents and survey
  properties}. \aap 616:A1. \doi{10.1051/0004-6361/201833051}.
  {\href{https://arxiv.org/abs/1804.09365}{{arXiv:1804.09365}}} {[astro-ph.GA]}

\bibitem[{{Gaia Collaboration} et~al.(2018{\natexlab{b}}){Gaia Collaboration},
  {Katz}, {Antoja}, {Romero-G{\'o}mez}, {Drimmel}, {Reyl{\'e}}, {Seabroke},
  {Soubiran}, {Babusiaux}, {Di Matteo}, {Figueras}, {Poggio}, {Robin}, {Evans},
  {Brown}, {Vallenari}, {Prusti}, {de Bruijne}, {Bailer-Jones}, {Biermann},
  {Eyer}, {Jansen}, {Jordi}, {Klioner}, {Lammers}, {Lindegren}, {Luri},
  {Mignard}, {Panem}, {Pourbaix}, {Randich}, {Sartoretti}, {Siddiqui}, {van
  Leeuwen}, {Walton}, {Arenou}, {Bastian}, {Cropper}, {Lattanzi}, {Bakker},
  {Cacciari}, {Casta n}, {Chaoul}, {Cheek}, {De Angeli}, {Fabricius}, {Guerra},
  {Holl}, {Masana}, {Messineo}, {Mowlavi}, {Nienartowicz}, {Panuzzo},
  {Portell}, {Riello}, {Tanga}, {Th{\'e}venin}, {Gracia-Abril}, {Comoretto},
  {Garcia-Reinaldos}, {Teyssier}, {Altmann}, {Andrae}, {Audard},
  {Bellas-Velidis}, {Benson}, {Berthier}, {Blomme}, {Burgess}, {Busso},
  {Carry}, {Cellino}, {Clementini}, {Clotet}, {Creevey}, {Davidson}, {De
  Ridder}, {Delchambre}, {Dell'Oro}, {Ducourant},
  {Fern{\'a}ndez-Hern{\'a}ndez}, {Fouesneau}, {Fr{\'e}mat}, {Galluccio},
  {Garc{\'\i}a-Torres}, {Gonz{\'a}lez-N{\'u}{\~n}ez}, {Gonz{\'a}lez-Vidal},
  {Gosset}, {Guy}, {Halbwachs}, {Hambly}, {Harrison}, {Hern{\'a}ndez},
  {Hestroffer}, {Hodgkin}, {Hutton}, {Jasniewicz}, {Jean-Antoine-Piccolo},
  {Jordan}, {Korn}, {Krone-Martins}, {Lanzafame}, {Lebzelter}, {L{\"o}ffler},
  {Manteiga}, {Marrese}, {Mart{\'\i}n-Fleitas}, {Moitinho}, {Mora}, {Muinonen},
  {Osinde}, {Pancino}, {Pauwels}, {Petit}, {Recio-Blanco}, {Richards},
  {Rimoldini}, {Sarro}, {Siopis}, {Smith}, {Sozzetti}, {S{\"u}veges}, {Torra},
  {van Reeven}, {Abbas}, {Abreu Aramburu}, {Accart}, {Aerts}, {Altavilla},
  {{\'A}lvarez}, {Alvarez}, {Alves}, {Anderson}, {Andrei}, {Anglada Varela},
  {Antiche}, {Arcay}, {Astraatmadja}, {Bach}, {Baker},
  {Balaguer-N{\'u}{\~n}ez}, {Balm}, {Barache}, {Barata}, {Barbato}, {Barblan},
  {Barklem}, {Barrado}, {Barros}, {Barstow}, {Bartholom{\'e} Mu{\~n}oz},
  {Bassilana}, {Becciani}, {Bellazzini}, {Berihuete}, {Bertone}, {Bianchi},
  {Bienaym{\'e}}, {Blanco-Cuaresma}, {Boch}, {Boeche}, {Bombrun}, {Borrachero},
  {Bossini}, {Bouquillon}, {Bourda}, {Bragaglia}, {Bramante}, {Breddels},
  {Bressan}, {Brouillet}, {Br{\"u}semeister}, {Brugaletta}, {Bucciarelli},
  {Burlacu}, {Busonero}, {Butkevich}, {Buzzi}, {Caffau}, {Cancelliere},
  {Cannizzaro}, {Cantat-Gaudin}, {Carballo}, {Carlucci}, {Carrasco},
  {Casamiquela}, {Castellani}, {Castro-Ginard}, {Charlot}, {Chemin},
  {Chiavassa}, {Cocozza}, {Costigan}, {Cowell}, {Crifo}, {Crosta}, {Crowley},
  {Cuypers}, {Dafonte}, {Damerdji}, {Dapergolas}, {David}, {David}, {de
  Laverny}, {De Luise}, {De March}, {de Souza}, {de Torres}, {Debosscher}, {del
  Pozo}, {Delbo}, {Delgado}, {Delgado}, {Diakite}, {Diener}, {Distefano},
  {Dolding}, {Drazinos}, {Dur{\'a}n}, {Edvardsson}, {Enke}, {Eriksson},
  {Esquej}, {Eynard Bontemps}, {Fabre}, {Fabrizio}, {Faigler}, {Falc a},
  {Farr{\`a}s Casas}, {Federici}, {Fedorets}, {Fernique}, {Filippi},
  {Findeisen}, {Fonti}, {Fraile}, {Fraser}, {Fr{\'e}zouls}, {Gai}, {Galleti},
  {Garabato}, {Garc{\'\i}a-Sedano}, {Garofalo}, {Garralda}, {Gavel}, {Gavras},
  {Gerssen}, {Geyer}, {Giacobbe}, {Gilmore}, {Girona}, {Giuffrida}, {Glass},
  {Gomes}, {Granvik}, {Gueguen}, {Guerrier}, {Guiraud}, {Guti{\'e}}, {Haigron},
  {Hatzidimitriou}, {Hauser}, {Haywood}, {Heiter}, {Helmi}, {Heu}, {Hilger},
  {Hobbs}, {Hofmann}, {Holland}, {Huckle}, {Hypki}, {Icardi}, {Jan{\ss}en},
  {Jevardat de Fombelle}, {Jonker}, {Juh{\'a}sz}, {Julbe}, {Karampelas},
  {Kewley}, {Klar}, {Kochoska}, {Kohley}, {Kolenberg}, {Kontizas}, {Kontizas},
  {Koposov}, {Kordopatis}, {Kostrzewa-Rutkowska}, {Koubsky}, {Lambert},
  {Lanza}, {Lasne}, {Lavigne}, {Le Fustec}, {Le Poncin-Lafitte}, {Lebreton},
  {Leccia}, {Leclerc}, {Lecoeur-Taibi}, {Lenhardt}, {Leroux}, {Liao}, {Licata},
  {Lindstr{\o}m}, {Lister}, {Livanou}, {Lobel}, {L{\'o}pez}, {Managau}, {Mann},
  {Mantelet}, {Marchal}, {Marchant}, {Marconi}, {Marinoni}, {Marschalk{\'o}},
  {Marshall}, {Martino}, {Marton}, {Mary}, {Massari}, {Matijevi{\v{c}}},
  {Mazeh}, {McMillan}, {Messina}, {Michalik}, {Millar}, {Molina}, {Molinaro},
  {Moln{\'a}r}, {Montegriffo}, {Mor}, {Morbidelli}, {Morel}, {Morris},
  {Mulone}, {Muraveva}, {Musella}, {Nelemans}, {Nicastro}, {Noval},
  {O'Mullane}, {Ord{\'e}novic}, {Ord{\'o}{\~n}ez-Blanco}, {Osborne}, {Pagani},
  {Pagano}, {Pailler}, {Palacin}, {Palaversa}, {Panahi}, {Pawlak},
  {Piersimoni}, {Pineau}, {Plachy}, {Plum}, {Poujoulet}, {Pr{\v{s}}a},
  {Pulone}, {Racero}, {Ragaini}, {Rambaux}, {Ramos-Lerate}, {Regibo}, {Riclet},
  {Ripepi}, {Riva}, {Rivard}, {Rixon}, {Roegiers}, {Roelens}, {Rowell},
  {Royer}, {Ruiz-Dern}, {Sadowski}, {Sagrist{\`a} Sell{\'e}s}, {Sahlmann},
  {Salgado}, {Salguero}, {Sanna}, {Santana-Ros}, {Sarasso}, {Savietto},
  {Schultheis}, {Sciacca}, {Segol}, {Segovia}, {S{\'e}gransan}, {Shih},
  {Siltala}, {Silva}, {Smart}, {Smith}, {Solano}, {Solitro}, {Sordo}, {Soria
  Nieto}, {Souchay}, {Spagna}, {Spoto}, {Stampa}, {Steele},
  {Steidelm{\"u}ller}, {Stephenson}, {Stoev}, {Suess}, {Surdej}, {Szabados},
  {Szegedi-Elek}, {Tapiador}, {Taris}, {Tauran}, {Taylor}, {Teixeira},
  {Terrett}, {Teyssandier}, {Thuillot}, {Titarenko}, {Torra Clotet}, {Turon},
  {Ulla}, {Utrilla}, {Uzzi}, {Vaillant}, {Valentini}, {Valette}, {van Elteren},
  {Van Hemelryck}, {van Leeuwen}, {Vaschetto}, {Vecchiato}, {Veljanoski},
  {Viala}, {Vicente}, {Vogt}, {von Essen}, {Voss}, {Votruba}, {Voutsinas},
  {Walmsley}, {Weiler}, {Wertz}, {Wevers}, {Wyrzykowski}, {Yoldas},
  {{\v{Z}}erjal}, {Ziaeepour}, {Zorec}, {Zschocke}, {Zucker}, {Zurbach}, and
  {Zwitter}}]{2018A&A...616A..11G}
{Gaia Collaboration}, {Katz} D, {Antoja} T, {Romero-G{\'o}mez} M, {Drimmel} R,
  {Reyl{\'e}} C, {Seabroke} GM, {Soubiran} C, {Babusiaux} C, {Di Matteo} P,
  et~al. (2018{\natexlab{b}}) {Gaia Data Release 2. Mapping the Milky Way disc
  kinematics}. \aap 616:A11. \doi{10.1051/0004-6361/201832865}.
  {\href{https://arxiv.org/abs/1804.09380}{{arXiv:1804.09380}}} {[astro-ph.GA]}

\bibitem[{{Gaia Collaboration} et~al.(2020){Gaia Collaboration}, {Brown},
  {Vallenari}, {Prusti}, {de Bruijne}, {Babusiaux}, and
  {Biermann}}]{2020arXiv201201533G}
{Gaia Collaboration}, {Brown} AGA, {Vallenari} A, {Prusti} T, {de Bruijne} JHJ,
  {Babusiaux} C, {Biermann} M (2020) {Gaia Early Data Release 3: Summary of the
  contents and survey properties}. arXiv e-prints arXiv:2012.01533.
  {\href{https://arxiv.org/abs/2012.01533}{{arXiv:2012.01533}}} {[astro-ph.GA]}

\bibitem[{{Gair} and {Jones}(2007)}]{2007CQGra..24.1145G}
{Gair} J, {Jones} G (2007) {Detecting extreme mass ratio inspiral events in
  LISA data using the hierarchical algorithm for clusters and ridges (HACR)}.
  Classical and Quantum Gravity 24(5):1145--1168.
  \doi{10.1088/0264-9381/24/5/007}.
  {\href{https://arxiv.org/abs/gr-qc/0610046}{{arXiv:gr-qc/0610046}}} {[gr-qc]}

\bibitem[{{Gair} et~al.(2004){Gair}, {Barack}, {Creighton}, {Cutler}, {Larson},
  {Phinney}, and {Vallisneri}}]{2004CQGra..21S1595G}
{Gair} JR, {Barack} L, {Creighton} T, {Cutler} C, {Larson} SL, {Phinney} ES,
  {Vallisneri} M (2004) {Event rate estimates for LISA extreme mass ratio
  capture sources}. Classical and Quantum Gravity 21(20):S1595--S1606.
  \doi{10.1088/0264-9381/21/20/003}.
  {\href{https://arxiv.org/abs/gr-qc/0405137}{{arXiv:gr-qc/0405137}}} {[gr-qc]}

\bibitem[{{Gair} et~al.(2008{\natexlab{a}}){Gair}, {Mandel}, and
  {Wen}}]{2008CQGra..25r4031G}
{Gair} JR, {Mandel} I, {Wen} L (2008{\natexlab{a}}) {Improved time frequency
  analysis of extreme-mass-ratio inspiral signals in mock LISA data}. Classical
  and Quantum Gravity 25(18):184031. \doi{10.1088/0264-9381/25/18/184031}.
  {\href{https://arxiv.org/abs/0804.1084}{{arXiv:0804.1084}}} {[gr-qc]}

\bibitem[{{Gair} et~al.(2008{\natexlab{b}}){Gair}, {Mandel}, and
  {Wen}}]{2008JPhCS.122a2037G}
{Gair} JR, {Mandel} I, {Wen} L (2008{\natexlab{b}}) {Time-frequency analysis of
  extreme-mass-ratio inspiral signals in mock LISA data}. In: Journal of
  Physics Conference Series. Journal of Physics Conference Series, vol 122. p
  012037. \doi{10.1088/1742-6596/122/1/012037}.
  {\href{https://arxiv.org/abs/0710.5250}{{arXiv:0710.5250}}} {[gr-qc]}

\bibitem[{{Gair} et~al.(2008{\natexlab{c}}){Gair}, {Porter}, {Babak}, and
  {Barack}}]{2008CQGra..25r4030G}
{Gair} JR, {Porter} E, {Babak} S, {Barack} L (2008{\natexlab{c}}) {A
  constrained Metropolis Hastings search for EMRIs in the Mock LISA Data
  Challenge 1B}. Classical and Quantum Gravity 25(18):184030.
  \doi{10.1088/0264-9381/25/18/184030}.
  {\href{https://arxiv.org/abs/0804.3322}{{arXiv:0804.3322}}} {[gr-qc]}

\bibitem[{{Gair} et~al.(2011){Gair}, {Mandel}, {Miller}, and
  {Volonteri}}]{2011GReGr..43..485G}
{Gair} JR, {Mandel} I, {Miller} MC, {Volonteri} M (2011) {Exploring
  intermediate and massive black-hole binaries with the Einstein Telescope}.
  General Relativity and Gravitation 43(2):485--518.
  \doi{10.1007/s10714-010-1104-3}.
  {\href{https://arxiv.org/abs/0907.5450}{{arXiv:0907.5450}}} {[astro-ph.CO]}

\bibitem[{{Galaudage} et~al.(2021){Galaudage}, {Adamcewicz}, {Zhu},
  {Stevenson}, and {Thrane}}]{2021ApJ...909L..19G}
{Galaudage} S, {Adamcewicz} C, {Zhu} XJ, {Stevenson} S, {Thrane} E (2021)
  {Heavy Double Neutron Stars: Birth, Midlife, and Death}. \apjl 909(2):L19.
  \doi{10.3847/2041-8213/abe7f6}.
  {\href{https://arxiv.org/abs/2011.01495}{{arXiv:2011.01495}}} {[astro-ph.HE]}

\bibitem[{{Gallego-Cano} et~al.(2018){Gallego-Cano}, {Sch{\"o}del}, {Dong},
  {Nogueras-Lara}, {Gallego-Calvente}, {Amaro-Seoane}, and
  {Baumgardt}}]{2018A&A...609A..26G}
{Gallego-Cano} E, {Sch{\"o}del} R, {Dong} H, {Nogueras-Lara} F,
  {Gallego-Calvente} AT, {Amaro-Seoane} P, {Baumgardt} H (2018) {The
  distribution of stars around the Milky Way's central black hole. I. Deep star
  counts}. \aap 609:A26. \doi{10.1051/0004-6361/201730451}.
  {\href{https://arxiv.org/abs/1701.03816}{{arXiv:1701.03816}}} {[astro-ph.GA]}

\bibitem[{{Gammie} et~al.(2004){Gammie}, {Shapiro}, and
  {McKinney}}]{2004ApJ...602..312G}
{Gammie} CF, {Shapiro} SL, {McKinney} JC (2004) {Black Hole Spin Evolution}.
  \apj 602(1):312--319. \doi{10.1086/380996}.
  {\href{https://arxiv.org/abs/astro-ph/0310886}{{arXiv:astro-ph/0310886}}}
  {[astro-ph]}

\bibitem[{{Gandhi} et~al.(2019){Gandhi}, {Rao}, {Johnson}, {Paice}, and
  {Maccarone}}]{2019MNRAS.485.2642G}
{Gandhi} P, {Rao} A, {Johnson} MAC, {Paice} JA, {Maccarone} TJ (2019) {Gaia
  Data Release 2 distances and peculiar velocities for Galactic black hole
  transients}. \mnras 485(2):2642--2655. \doi{10.1093/mnras/stz438}.
  {\href{https://arxiv.org/abs/1804.11349}{{arXiv:1804.11349}}} {[astro-ph.HE]}

\bibitem[{{G{\"a}nsicke} et~al.(2006){G{\"a}nsicke}, {Marsh}, {Southworth}, and
  {Rebassa-Mansergas}}]{2006Sci...314.1908G}
{G{\"a}nsicke} BT, {Marsh} TR, {Southworth} J, {Rebassa-Mansergas} A (2006) {A
  Gaseous Metal Disk Around a White Dwarf}. Science 314(5807):1908.
  \doi{10.1126/science.1135033}.
  {\href{https://arxiv.org/abs/astro-ph/0612697}{{arXiv:astro-ph/0612697}}}
  {[astro-ph]}

\bibitem[{{G{\"a}nsicke} et~al.(2019){G{\"a}nsicke}, {Schreiber}, {Toloza},
  {Gentile Fusillo}, {Koester}, and {Manser}}]{2019Natur.576...61G}
{G{\"a}nsicke} BT, {Schreiber} MR, {Toloza} O, {Gentile Fusillo} NP, {Koester}
  D, {Manser} CJ (2019) {Accretion of a giant planet onto a white dwarf star}.
  \nat 576(7785):61--64. \doi{10.1038/s41586-019-1789-8}.
  {\href{https://arxiv.org/abs/1912.01611}{{arXiv:1912.01611}}} {[astro-ph.EP]}

\bibitem[{{Garavito-Camargo} et~al.(2019){Garavito-Camargo}, {Besla},
  {Laporte}, {Johnston}, {G{\'o}mez}, and {Watkins}}]{2019ApJ...884...51G}
{Garavito-Camargo} N, {Besla} G, {Laporte} CFP, {Johnston} KV, {G{\'o}mez} FA,
  {Watkins} LL (2019) {Hunting for the Dark Matter Wake Induced by the Large
  Magellanic Cloud}. \apj 884(1):51. \doi{10.3847/1538-4357/ab32eb}.
  {\href{https://arxiv.org/abs/1902.05089}{{arXiv:1902.05089}}} {[astro-ph.GA]}

\bibitem[{{Garavito-Camargo} et~al.(2020){Garavito-Camargo}, {Besla},
  {Laporte}, {Price-Whelan}, {Cunningham}, {Johnston}, {Weinberg}, and
  {Gomez}}]{2020arXiv201000816G}
{Garavito-Camargo} N, {Besla} G, {Laporte} CFP, {Price-Whelan} AM, {Cunningham}
  EC, {Johnston} KV, {Weinberg} MD, {Gomez} FA (2020) {Quantifying the impact
  of the Large Magellanic Cloud on the structure of the Milky Way's dark matter
  halo using Basis Function Expansions}. arXiv e-prints arXiv:2010.00816.
  {\href{https://arxiv.org/abs/2010.00816}{{arXiv:2010.00816}}} {[astro-ph.GA]}

\bibitem[{{Garc{\'\i}a} et~al.(2016){Garc{\'\i}a}, {Fabian}, {Kallman},
  {Dauser}, {Parker}, {McClintock}, {Steiner}, and
  {Wilms}}]{2016MNRAS.462..751G}
{Garc{\'\i}a} JA, {Fabian} AC, {Kallman} TR, {Dauser} T, {Parker} ML,
  {McClintock} JE, {Steiner} JF, {Wilms} J (2016) {The effects of high density
  on the X-ray spectrum reflected from accretion discs around black holes}.
  \mnras 462(1):751--760. \doi{10.1093/mnras/stw1696}.
  {\href{https://arxiv.org/abs/1603.05259}{{arXiv:1603.05259}}} {[astro-ph.HE]}

\bibitem[{Garcia-Bellido et~al.(2016)Garcia-Bellido, Peloso, and
  Unal}]{Garcia-Bellido:2016dkw}
Garcia-Bellido J, Peloso M, Unal C (2016) {Gravitational waves at
  interferometer scales and primordial black holes in axion inflation}. JCAP
  12:031. \doi{10.1088/1475-7516/2016/12/031}.
  {\href{https://arxiv.org/abs/1610.03763}{{arXiv:1610.03763}}} {[astro-ph.CO]}

\bibitem[{Garcia-Bellido et~al.(2017)Garcia-Bellido, Peloso, and
  Unal}]{Garcia-Bellido:2017aan}
Garcia-Bellido J, Peloso M, Unal C (2017) {Gravitational Wave signatures of
  inflationary models from Primordial Black Hole Dark Matter}. JCAP 09:013.
  \doi{10.1088/1475-7516/2017/09/013}.
  {\href{https://arxiv.org/abs/1707.02441}{{arXiv:1707.02441}}} {[astro-ph.CO]}

\bibitem[{{Gardner} et~al.(2006){Gardner}, {Mather}, {Clampin}, {Doyon},
  {Greenhouse}, {Hammel}, {Hutchings}, {Jakobsen}, {Lilly}, {Long}, {Lunine},
  {McCaughrean}, {Mountain}, {Nella}, {Rieke}, {Rieke}, {Rix}, {Smith},
  {Sonneborn}, {Stiavelli}, {Stockman}, {Windhorst}, and
  {Wright}}]{2006SSRv..123..485G}
{Gardner} JP, {Mather} JC, {Clampin} M, {Doyon} R, {Greenhouse} MA, {Hammel}
  HB, {Hutchings} JB, {Jakobsen} P, {Lilly} SJ, {Long} KS, et~al. (2006) {The
  James Webb Space Telescope}. \ssr 123(4):485--606.
  \doi{10.1007/s11214-006-8315-7}.
  {\href{https://arxiv.org/abs/astro-ph/0606175}{{arXiv:astro-ph/0606175}}}
  {[astro-ph]}

\bibitem[{{Gaskin} et~al.(2019){Gaskin}, {Swartz}, {Vikhlinin}, {{\"O}zel},
  {Gelmis}, {Arenberg}, {Bandler}, {Bautz}, {Civitani}, {Dominguez}, {Eckart},
  {Falcone}, {Figueroa-Feliciano}, {Freeman}, {G{\"u}nther}, {Havey},
  {Heilmann}, {Kilaru}, {Kraft}, {McCarley}, {McEntaffer}, {Pareschi},
  {Purcell}, {Reid}, {Schattenburg}, {Schwartz}, {Schwartz}, {Tananbaum},
  {Tremblay}, {Zhang}, and {Zuhone}}]{2019JATIS...5b1001G}
{Gaskin} JA, {Swartz} DA, {Vikhlinin} A, {{\"O}zel} F, {Gelmis} KE, {Arenberg}
  JW, {Bandler} SR, {Bautz} MW, {Civitani} MM, {Dominguez} A, et~al. (2019)
  {Lynx X-Ray Observatory: an overview}. Journal of Astronomical Telescopes,
  Instruments, and Systems 5:021001. \doi{10.1117/1.JATIS.5.2.021001}

\bibitem[{{Gaspari} et~al.(2012){Gaspari}, {Ruszkowski}, and
  {Sharma}}]{2012ApJ...746...94G}
{Gaspari} M, {Ruszkowski} M, {Sharma} P (2012) {Cause and Effect of Feedback:
  Multiphase Gas in Cluster Cores Heated by AGN Jets}. \apj 746(1):94.
  \doi{10.1088/0004-637X/746/1/94}.
  {\href{https://arxiv.org/abs/1110.6063}{{arXiv:1110.6063}}} {[astro-ph.CO]}

\bibitem[{{Gaspari} et~al.(2013){Gaspari}, {Ruszkowski}, and
  {Oh}}]{2013MNRAS.432.3401G}
{Gaspari} M, {Ruszkowski} M, {Oh} SP (2013) {Chaotic cold accretion on to black
  holes}. \mnras 432(4):3401--3422. \doi{10.1093/mnras/stt692}.
  {\href{https://arxiv.org/abs/1301.3130}{{arXiv:1301.3130}}} {[astro-ph.CO]}

\bibitem[{{Gaspari} et~al.(2015){Gaspari}, {Brighenti}, and
  {Temi}}]{2015A&A...579A..62G}
{Gaspari} M, {Brighenti} F, {Temi} P (2015) {Chaotic cold accretion on to black
  holes in rotating atmospheres}. \aap 579:A62.
  \doi{10.1051/0004-6361/201526151}.
  {\href{https://arxiv.org/abs/1407.7531}{{arXiv:1407.7531}}} {[astro-ph.GA]}

\bibitem[{{Gaspari} et~al.(2017){Gaspari}, {Temi}, and
  {Brighenti}}]{2017MNRAS.466..677G}
{Gaspari} M, {Temi} P, {Brighenti} F (2017) {Raining on black holes and massive
  galaxies: the top-down multiphase condensation model}. \mnras
  466(1):677--704. \doi{10.1093/mnras/stw3108}.
  {\href{https://arxiv.org/abs/1608.08216}{{arXiv:1608.08216}}} {[astro-ph.GA]}

\bibitem[{{Gaspari} et~al.(2019){Gaspari}, {Eckert}, {Ettori}, {Tozzi},
  {Bassini}, {Rasia}, {Brighenti}, {Sun}, {Borgani}, {Johnson}, {Tremblay},
  {Stone}, {Temi}, {Yang}, {Tombesi}, and {Cappi}}]{2019ApJ...884..169G}
{Gaspari} M, {Eckert} D, {Ettori} S, {Tozzi} P, {Bassini} L, {Rasia} E,
  {Brighenti} F, {Sun} M, {Borgani} S, {Johnson} SD, et~al. (2019) {The X-Ray
  Halo Scaling Relations of Supermassive Black Holes}. \apj 884(2):169.
  \doi{10.3847/1538-4357/ab3c5d}.
  {\href{https://arxiv.org/abs/1904.10972}{{arXiv:1904.10972}}} {[astro-ph.GA]}

\bibitem[{{Gaspari} et~al.(2020){Gaspari}, {Tombesi}, and
  {Cappi}}]{2020NatAs...4...10G}
{Gaspari} M, {Tombesi} F, {Cappi} M (2020) {Linking macro-, meso- and
  microscales in multiphase AGN feeding and feedback}. Nature Astronomy
  4:10--13. \doi{10.1038/s41550-019-0970-1}.
  {\href{https://arxiv.org/abs/2001.04985}{{arXiv:2001.04985}}} {[astro-ph.GA]}

\bibitem[{{Gatti} et~al.(2015){Gatti}, {Lamastra}, {Menci}, {Bongiorno}, and
  {Fiore}}]{2015A&A...576A..32G}
{Gatti} M, {Lamastra} A, {Menci} N, {Bongiorno} A, {Fiore} F (2015) {Physical
  properties of AGN host galaxies as a probe of supermassive black hole feeding
  mechanisms}. \aap 576:A32. \doi{10.1051/0004-6361/201425094}.
  {\href{https://arxiv.org/abs/1412.7660}{{arXiv:1412.7660}}} {[astro-ph.GA]}

\bibitem[{{Gavazzi} et~al.(2015){Gavazzi}, {Consolandi}, {Dotti}, {Fanali},
  {Fossati}, {Fumagalli}, {Viscardi}, {Savorgnan}, {Boselli}, {Guti{\'e}rrez},
  {Hern{\'a}ndez Toledo}, {Giovanelli}, and {Haynes}}]{2015A&A...580A.116G}
{Gavazzi} G, {Consolandi} G, {Dotti} M, {Fanali} R, {Fossati} M, {Fumagalli} M,
  {Viscardi} E, {Savorgnan} G, {Boselli} A, {Guti{\'e}rrez} L, et~al. (2015)
  {H{\ensuremath{\alpha}}3: an H{\ensuremath{\alpha}} imaging survey of HI
  selected galaxies from ALFALFA. VI. The role of bars in quenching star
  formation from z = 3 to the present epoch}. \aap 580:A116.
  \doi{10.1051/0004-6361/201425351}.
  {\href{https://arxiv.org/abs/1505.07836}{{arXiv:1505.07836}}} {[astro-ph.GA]}

\bibitem[{{Ge} et~al.(2020){Ge}, {Webbink}, and {Han}}]{2020ApJS..249....9G}
{Ge} H, {Webbink} RF, {Han} Z (2020) {The Thermal Equilibrium Mass-loss Model
  and Its Applications in Binary Evolution}. \apjs 249(1):9.
  \doi{10.3847/1538-4365/ab98f6}.
  {\href{https://arxiv.org/abs/2006.00774}{{arXiv:2006.00774}}} {[astro-ph.SR]}

\bibitem[{{Gebhardt} et~al.(2000){Gebhardt}, {Bender}, {Bower}, {Dressler},
  {Faber}, {Filippenko}, {Green}, {Grillmair}, {Ho}, {Kormendy}, {Lauer},
  {Magorrian}, {Pinkney}, {Richstone}, and {Tremaine}}]{2000ApJ...539L..13G}
{Gebhardt} K, {Bender} R, {Bower} G, {Dressler} A, {Faber} SM, {Filippenko} AV,
  {Green} R, {Grillmair} C, {Ho} LC, {Kormendy} J, et~al. (2000) {A
  Relationship between Nuclear Black Hole Mass and Galaxy Velocity Dispersion}.
  \apjl 539(1):L13--L16. \doi{10.1086/312840}.
  {\href{https://arxiv.org/abs/astro-ph/0006289}{{arXiv:astro-ph/0006289}}}
  {[astro-ph]}

\bibitem[{{Geha} et~al.(2013){Geha}, {Brown}, {Tumlinson}, {Kalirai}, {Simon},
  {Kirby}, {Vand enBerg}, {Mu{\~n}oz}, {Avila}, {Guhathakurta}, and
  {Ferguson}}]{2013ApJ...771...29G}
{Geha} M, {Brown} TM, {Tumlinson} J, {Kalirai} JS, {Simon} JD, {Kirby} EN,
  {Vand enBerg} DA, {Mu{\~n}oz} RR, {Avila} RJ, {Guhathakurta} P, et~al. (2013)
  {The Stellar Initial Mass Function of Ultra-faint Dwarf Galaxies: Evidence
  for IMF Variations with Galactic Environment}. \apj 771(1):29.
  \doi{10.1088/0004-637X/771/1/29}.
  {\href{https://arxiv.org/abs/1304.7769}{{arXiv:1304.7769}}} {[astro-ph.CO]}

\bibitem[{{Geier} et~al.(2013){Geier}, {Marsh}, {Wang}, {Dunlap}, {Barlow},
  {Schaffenroth}, {Chen}, {Irrgang}, {Maxted}, and {et
  al.}}]{2013A&A...554A..54G}
{Geier} S, {Marsh} TR, {Wang} B, {Dunlap} B, {Barlow} BN, {Schaffenroth} V,
  {Chen} X, {Irrgang} A, {Maxted} PFL, {et al} (2013) {A progenitor binary and
  an ejected mass donor remnant of faint type Ia supernovae}. \aap 554:A54.
  \doi{10.1051/0004-6361/201321395}.
  {\href{https://arxiv.org/abs/1304.4452}{{arXiv:1304.4452}}} {[astro-ph.SR]}

\bibitem[{{Gendreau} et~al.(2016){Gendreau}, {Arzoumanian}, {Adkins}, {Albert},
  {Anders}, {Aylward}, {Baker}, {Balsamo}, {Bamford}, {Benegalrao}, {Berry},
  {Bhalwani}, {Black}, {Blaurock}, {Bronke}, {Brown}, {Budinoff}, and {et
  al}}]{2016SPIE.9905E..1HG}
{Gendreau} KC, {Arzoumanian} Z, {Adkins} PW, {Albert} CL, {Anders} JF,
  {Aylward} AT, {Baker} CL, {Balsamo} ER, {Bamford} WA, {Benegalrao} SS, et~al.
  (2016) {The Neutron star Interior Composition Explorer (NICER): design and
  development}. In: {den Herder} JWA, {Takahashi} T, {Bautz} M (eds) Space
  Telescopes and Instrumentation 2016: Ultraviolet to Gamma Ray. Society of
  Photo-Optical Instrumentation Engineers (SPIE) Conference Series, vol 9905. p
  99051H. \doi{10.1117/12.2231304}

\bibitem[{{Generozov} and {Madigan}(2020)}]{2020ApJ...896..137G}
{Generozov} A, {Madigan} AM (2020) {The Hills Mechanism and the Galactic Center
  S-stars}. \apj 896(2):137. \doi{10.3847/1538-4357/ab94bc}.
  {\href{https://arxiv.org/abs/2002.10547}{{arXiv:2002.10547}}} {[astro-ph.GA]}

\bibitem[{{Generozov} et~al.(2018){Generozov}, {Stone}, {Metzger}, and
  {Ostriker}}]{2018MNRAS.478.4030G}
{Generozov} A, {Stone} NC, {Metzger} BD, {Ostriker} JP (2018) {An overabundance
  of black hole X-ray binaries in the Galactic Centre from tidal captures}.
  \mnras 478(3):4030--4051. \doi{10.1093/mnras/sty1262}.
  {\href{https://arxiv.org/abs/1804.01543}{{arXiv:1804.01543}}} {[astro-ph.HE]}

\bibitem[{{Genzel} et~al.(2010){Genzel}, {Eisenhauer}, and
  {Gillessen}}]{2010RvMP...82.3121G}
{Genzel} R, {Eisenhauer} F, {Gillessen} S (2010) {The Galactic Center massive
  black hole and nuclear star cluster}. Reviews of Modern Physics
  82(4):3121--3195. \doi{10.1103/RevModPhys.82.3121}.
  {\href{https://arxiv.org/abs/1006.0064}{{arXiv:1006.0064}}} {[astro-ph.GA]}

\bibitem[{{Georgakarakos}(2008)}]{2008CeMDA.100..151G}
{Georgakarakos} N (2008) {Stability criteria for hierarchical triple systems}.
  Celestial Mechanics and Dynamical Astronomy 100(2):151--168.
  \doi{10.1007/s10569-007-9109-2}.
  {\href{https://arxiv.org/abs/1408.5431}{{arXiv:1408.5431}}} {[astro-ph.EP]}

\bibitem[{{George} et~al.(2018){George}, {Shen}, and
  {Huerta}}]{2018PhRvD..97j1501G}
{George} D, {Shen} H, {Huerta} EA (2018) {Classification and unsupervised
  clustering of LIGO data with Deep Transfer Learning}. \prd 97(10):101501.
  \doi{10.1103/PhysRevD.97.101501}.
  {\href{https://arxiv.org/abs/1706.07446}{{arXiv:1706.07446}}} {[gr-qc]}

\bibitem[{{Gergely} et~al.(2010){Gergely}, {Biermann}, and
  {Caramete}}]{2010CQGra..27s4009G}
{Gergely} L{\'A}, {Biermann} PL, {Caramete} LI (2010) {Supermassive black hole
  spin-flip during the inspiral}. Classical and Quantum Gravity 27(19):194009.
  \doi{10.1088/0264-9381/27/19/194009}.
  {\href{https://arxiv.org/abs/1005.2287}{{arXiv:1005.2287}}} {[astro-ph.CO]}

\bibitem[{{Gerosa} and {Kesden}(2016)}]{2016PhRvD..93l4066G}
{Gerosa} D, {Kesden} M (2016) {precession: Dynamics of spinning black-hole
  binaries with python}. \prd 93(12):124066. \doi{10.1103/PhysRevD.93.124066}.
  {\href{https://arxiv.org/abs/1605.01067}{{arXiv:1605.01067}}} {[astro-ph.HE]}

\bibitem[{{Gerosa} and {Moore}(2016)}]{2016PhRvL.117a1101G}
{Gerosa} D, {Moore} CJ (2016) {Black Hole Kicks as New Gravitational Wave
  Observables}. \prl 117(1):011101. \doi{10.1103/PhysRevLett.117.011101}.
  {\href{https://arxiv.org/abs/1606.04226}{{arXiv:1606.04226}}} {[gr-qc]}

\bibitem[{{Gerosa} and {Sesana}(2015)}]{2015MNRAS.446...38G}
{Gerosa} D, {Sesana} A (2015) {Missing black holes in brightest cluster
  galaxies as evidence for the occurrence of superkicks in nature}. \mnras
  446(1):38--55. \doi{10.1093/mnras/stu2049}.
  {\href{https://arxiv.org/abs/1405.2072}{{arXiv:1405.2072}}} {[astro-ph.GA]}

\bibitem[{{Gerosa} et~al.(2013){Gerosa}, {Kesden}, {Berti}, {O'Shaughnessy},
  and {Sperhake}}]{2013PhRvD..87j4028G}
{Gerosa} D, {Kesden} M, {Berti} E, {O'Shaughnessy} R, {Sperhake} U (2013)
  {Resonant-plane locking and spin alignment in stellar-mass black-hole
  binaries: A diagnostic of compact-binary formation}. \prd 87(10):104028.
  \doi{10.1103/PhysRevD.87.104028}.
  {\href{https://arxiv.org/abs/1302.4442}{{arXiv:1302.4442}}} {[gr-qc]}

\bibitem[{{Gerosa} et~al.(2015{\natexlab{a}}){Gerosa}, {Kesden}, {Sperhake},
  {Berti}, and {O'Shaughnessy}}]{2015PhRvD..92f4016G}
{Gerosa} D, {Kesden} M, {Sperhake} U, {Berti} E, {O'Shaughnessy} R
  (2015{\natexlab{a}}) {Multi-timescale analysis of phase transitions in
  precessing black-hole binaries}. \prd 92(6):064016.
  \doi{10.1103/PhysRevD.92.064016}.
  {\href{https://arxiv.org/abs/1506.03492}{{arXiv:1506.03492}}} {[gr-qc]}

\bibitem[{{Gerosa} et~al.(2015{\natexlab{b}}){Gerosa}, {Veronesi}, {Lodato},
  and {Rosotti}}]{2015MNRAS.451.3941G}
{Gerosa} D, {Veronesi} B, {Lodato} G, {Rosotti} G (2015{\natexlab{b}}) {Spin
  alignment and differential accretion in merging black hole binaries}. \mnras
  451(4):3941--3954. \doi{10.1093/mnras/stv1214}.
  {\href{https://arxiv.org/abs/1503.06807}{{arXiv:1503.06807}}} {[astro-ph.GA]}

\bibitem[{{Gerosa} et~al.(2018{\natexlab{a}}){Gerosa}, {Berti},
  {O'Shaughnessy}, {Belczynski}, {Kesden}, {Wysocki}, and
  {Gladysz}}]{2018PhRvD..98h4036G}
{Gerosa} D, {Berti} E, {O'Shaughnessy} R, {Belczynski} K, {Kesden} M, {Wysocki}
  D, {Gladysz} W (2018{\natexlab{a}}) {Spin orientations of merging black holes
  formed from the evolution of stellar binaries}. \prd 98(8):084036.
  \doi{10.1103/PhysRevD.98.084036}.
  {\href{https://arxiv.org/abs/1808.02491}{{arXiv:1808.02491}}} {[astro-ph.HE]}

\bibitem[{{Gerosa} et~al.(2018{\natexlab{b}}){Gerosa}, {H{\'e}bert}, and
  {Stein}}]{2018PhRvD..97j4049G}
{Gerosa} D, {H{\'e}bert} F, {Stein} LC (2018{\natexlab{b}}) {Black-hole kicks
  from numerical-relativity surrogate models}. \prd 97(10):104049.
  \doi{10.1103/PhysRevD.97.104049}.
  {\href{https://arxiv.org/abs/1802.04276}{{arXiv:1802.04276}}} {[gr-qc]}

\bibitem[{{Gerosa} et~al.(2019){Gerosa}, {Ma}, {Wong}, {Berti},
  {O'Shaughnessy}, {Chen}, and {Belczynski}}]{2019PhRvD..99j3004G}
{Gerosa} D, {Ma} S, {Wong} KWK, {Berti} E, {O'Shaughnessy} R, {Chen} Y,
  {Belczynski} K (2019) {Multiband gravitational-wave event rates and stellar
  physics}. \prd 99(10):103004. \doi{10.1103/PhysRevD.99.103004}.
  {\href{https://arxiv.org/abs/1902.00021}{{arXiv:1902.00021}}} {[astro-ph.HE]}

\bibitem[{{Gerosa} et~al.(2020){Gerosa}, {Rosotti}, and
  {Barbieri}}]{2020MNRAS.496.3060G}
{Gerosa} D, {Rosotti} G, {Barbieri} R (2020) {The Bardeen-Petterson effect in
  accreting supermassive black hole binaries: a systematic approach}. \mnras
  496(3):3060--3075. \doi{10.1093/mnras/staa1693}.
  {\href{https://arxiv.org/abs/2004.02894}{{arXiv:2004.02894}}} {[astro-ph.GA]}

\bibitem[{{Giacobbo} and {Mapelli}(2018)}]{2018MNRAS.480.2011G}
{Giacobbo} N, {Mapelli} M (2018) {The progenitors of compact-object binaries:
  impact of metallicity, common envelope and natal kicks}. \mnras
  480(2):2011--2030. \doi{10.1093/mnras/sty1999}.
  {\href{https://arxiv.org/abs/1806.00001}{{arXiv:1806.00001}}} {[astro-ph.HE]}

\bibitem[{{Giacobbo} and {Mapelli}(2020)}]{2020ApJ...891..141G}
{Giacobbo} N, {Mapelli} M (2020) {Revising Natal Kick Prescriptions in
  Population Synthesis Simulations}. \apj 891(2):141.
  \doi{10.3847/1538-4357/ab7335}.
  {\href{https://arxiv.org/abs/1909.06385}{{arXiv:1909.06385}}} {[astro-ph.HE]}

\bibitem[{{Giacobbo} et~al.(2018){Giacobbo}, {Mapelli}, and
  {Spera}}]{2018MNRAS.474.2959G}
{Giacobbo} N, {Mapelli} M, {Spera} M (2018) {Merging black hole binaries: the
  effects of progenitor's metallicity, mass-loss rate and Eddington factor}.
  \mnras 474(3):2959--2974. \doi{10.1093/mnras/stx2933}.
  {\href{https://arxiv.org/abs/1711.03556}{{arXiv:1711.03556}}} {[astro-ph.SR]}

\bibitem[{{Giacomazzo} et~al.(2012){Giacomazzo}, {Baker}, {Miller}, {Reynolds},
  and {van Meter}}]{2012ApJ...752L..15G}
{Giacomazzo} B, {Baker} JG, {Miller} MC, {Reynolds} CS, {van Meter} JR (2012)
  {General Relativistic Simulations of Magnetized Plasmas around Merging
  Supermassive Black Holes}. \apjl 752(1):L15.
  \doi{10.1088/2041-8205/752/1/L15}.
  {\href{https://arxiv.org/abs/1203.6108}{{arXiv:1203.6108}}} {[astro-ph.HE]}

\bibitem[{{Giammichele} et~al.(2012){Giammichele}, {Bergeron}, and
  {Dufour}}]{2012ApJS..199...29G}
{Giammichele} N, {Bergeron} P, {Dufour} P (2012) {Know Your Neighborhood: A
  Detailed Model Atmosphere Analysis of Nearby White Dwarfs}. \apjs 199(2):29.
  \doi{10.1088/0067-0049/199/2/29}.
  {\href{https://arxiv.org/abs/1202.5581}{{arXiv:1202.5581}}} {[astro-ph.SR]}

\bibitem[{{Giblin} et~al.(2012){Giblin}, {Price}, {Siemens}, and
  {Vlcek}}]{2012JCAP...11..006G}
{Giblin} J John~T, {Price} LR, {Siemens} X, {Vlcek} B (2012) {Gravitational
  waves from global second order phase transitions}. \jcap 2012(11):006.
  \doi{10.1088/1475-7516/2012/11/006}.
  {\href{https://arxiv.org/abs/1111.4014}{{arXiv:1111.4014}}} {[astro-ph.CO]}

\bibitem[{{Gieles} and {Bastian}(2008)}]{2008A&A...482..165G}
{Gieles} M, {Bastian} N (2008) {An alternative method to study star cluster
  disruption}. \aap 482(1):165--171. \doi{10.1051/0004-6361:20078909}.
  {\href{https://arxiv.org/abs/0802.3387}{{arXiv:0802.3387}}} {[astro-ph]}

\bibitem[{{Giersz} et~al.(2013){Giersz}, {Heggie}, {Hurley}, and
  {Hypki}}]{2013MNRAS.431.2184G}
{Giersz} M, {Heggie} DC, {Hurley} JR, {Hypki} A (2013) {MOCCA code for star
  cluster simulations - II. Comparison with N-body simulations}. \mnras
  431(3):2184--2199. \doi{10.1093/mnras/stt307}.
  {\href{https://arxiv.org/abs/1112.6246}{{arXiv:1112.6246}}} {[astro-ph.GA]}

\bibitem[{{Giersz} et~al.(2015){Giersz}, {Leigh}, {Hypki}, {L{\"u}tzgendorf},
  and {Askar}}]{2015MNRAS.454.3150G}
{Giersz} M, {Leigh} N, {Hypki} A, {L{\"u}tzgendorf} N, {Askar} A (2015) {MOCCA
  code for star cluster simulations - IV. A new scenario for intermediate mass
  black hole formation in globular clusters}. \mnras 454(3):3150--3165.
  \doi{10.1093/mnras/stv2162}.
  {\href{https://arxiv.org/abs/1506.05234}{{arXiv:1506.05234}}} {[astro-ph.GA]}

\bibitem[{{Giersz} et~al.(2019){Giersz}, {Askar}, {Wang}, {Hypki}, {Leveque},
  and {Spurzem}}]{2019MNRAS.487.2412G}
{Giersz} M, {Askar} A, {Wang} L, {Hypki} A, {Leveque} A, {Spurzem} R (2019)
  {MOCCA survey data base- I. Dissolution of tidally filling star clusters
  harbouring black hole subsystems}. \mnras 487(2):2412--2423.
  \doi{10.1093/mnras/stz1460}.
  {\href{https://arxiv.org/abs/1904.01227}{{arXiv:1904.01227}}} {[astro-ph.GA]}

\bibitem[{{Gillessen} et~al.(2017){Gillessen}, {Plewa}, {Eisenhauer}, {Sari},
  {Waisberg}, {Habibi}, {Pfuhl}, {George}, {Dexter}, {von Fellenberg}, {Ott},
  and {Genzel}}]{2017ApJ...837...30G}
{Gillessen} S, {Plewa} PM, {Eisenhauer} F, {Sari} R, {Waisberg} I, {Habibi} M,
  {Pfuhl} O, {George} E, {Dexter} J, {von Fellenberg} S, et~al. (2017) {An
  Update on Monitoring Stellar Orbits in the Galactic Center}. \apj 837(1):30.
  \doi{10.3847/1538-4357/aa5c41}.
  {\href{https://arxiv.org/abs/1611.09144}{{arXiv:1611.09144}}} {[astro-ph.GA]}

\bibitem[{{Gilli} et~al.(2005){Gilli}, {Daddi}, {Zamorani}, {Tozzi}, {Borgani},
  {Bergeron}, {Giacconi}, {Hasinger}, {Mainieri}, {Norman}, {Rosati},
  {Szokoly}, and {Zheng}}]{2005A&A...430..811G}
{Gilli} R, {Daddi} E, {Zamorani} G, {Tozzi} P, {Borgani} S, {Bergeron} J,
  {Giacconi} R, {Hasinger} G, {Mainieri} V, {Norman} C, et~al. (2005) {The
  spatial clustering of X-ray selected AGN and galaxies in the Chandra Deep
  Field South and North}. \aap 430:811--825. \doi{10.1051/0004-6361:20041375}.
  {\href{https://arxiv.org/abs/astro-ph/0409759}{{arXiv:astro-ph/0409759}}}
  {[astro-ph]}

\bibitem[{{Gilli} et~al.(2022){Gilli}, {Norman}, {Calura}, {Vito}, {Decarli},
  {Marchesi}, {Iwasawa}, {Comastri}, {Lanzuisi}, {Pozzi}, {D'Amato}, {Vignali},
  {Brusa}, {Mignoli}, and {Cox}}]{2022arXiv220603508G}
{Gilli} R, {Norman} C, {Calura} F, {Vito} F, {Decarli} R, {Marchesi} S,
  {Iwasawa} K, {Comastri} A, {Lanzuisi} G, {Pozzi} F, et~al. (2022)
  {Supermassive Black Holes at High Redshift are Expected to be Obscured by
  their Massive Host Galaxies' Inter Stellar Medium}. arXiv e-prints
  arXiv:2206.03508.
  {\href{https://arxiv.org/abs/2206.03508}{{arXiv:2206.03508}}} {[astro-ph.GA]}

\bibitem[{{Ginat} et~al.(2020){Ginat}, {Glanz}, {Perets}, {Grishin}, and
  {Desjacques}}]{2020MNRAS.493.4861G}
{Ginat} YB, {Glanz} H, {Perets} HB, {Grishin} E, {Desjacques} V (2020)
  {Gravitational waves from in-spirals of compact objects in binary
  common-envelope evolution}. \mnras 493(4):4861--4867.
  \doi{10.1093/mnras/staa465}.
  {\href{https://arxiv.org/abs/1903.11072}{{arXiv:1903.11072}}} {[astro-ph.SR]}

\bibitem[{{Girardi} et~al.(1996){Girardi}, {Fadda}, {Giuricin}, {Mardirossian},
  {Mezzetti}, and {Biviano}}]{1996ApJ...457...61G}
{Girardi} M, {Fadda} D, {Giuricin} G, {Mardirossian} F, {Mezzetti} M, {Biviano}
  A (1996) {Velocity Dispersions and X-Ray Temperatures of Galaxy Clusters}.
  \apj 457:61. \doi{10.1086/176711}.
  {\href{https://arxiv.org/abs/astro-ph/9507031}{{arXiv:astro-ph/9507031}}}
  {[astro-ph]}

\bibitem[{{Glanz} and {Perets}(2021)}]{2021MNRAS.500.1921G}
{Glanz} H, {Perets} HB (2021) {Simulations of common envelope evolution in
  triple systems: circumstellar case}. \mnras 500(2):1921--1932.
  \doi{10.1093/mnras/staa3242}.
  {\href{https://arxiv.org/abs/2004.00020}{{arXiv:2004.00020}}} {[astro-ph.SR]}

\bibitem[{{Gnedin} et~al.(2014){Gnedin}, {Ostriker}, and
  {Tremaine}}]{2014ApJ...785...71G}
{Gnedin} OY, {Ostriker} JP, {Tremaine} S (2014) {Co-evolution of Galactic
  Nuclei and Globular Cluster Systems}. \apj 785(1):71.
  \doi{10.1088/0004-637X/785/1/71}.
  {\href{https://arxiv.org/abs/1308.0021}{{arXiv:1308.0021}}} {[astro-ph.CO]}

\bibitem[{{Gnocchi} et~al.(2019){Gnocchi}, {Maselli}, {Abdelsalhin},
  {Giacobbo}, and {Mapelli}}]{2019PhRvD.100f4024G}
{Gnocchi} G, {Maselli} A, {Abdelsalhin} T, {Giacobbo} N, {Mapelli} M (2019)
  {Bounding alternative theories of gravity with multiband GW observations}.
  \prd 100(6):064024. \doi{10.1103/PhysRevD.100.064024}.
  {\href{https://arxiv.org/abs/1905.13460}{{arXiv:1905.13460}}} {[gr-qc]}

\bibitem[{{Goicovic} et~al.(2016){Goicovic}, {Cuadra}, {Sesana}, {Stasyszyn},
  {Amaro-Seoane}, and {Tanaka}}]{2016MNRAS.455.1989G}
{Goicovic} FG, {Cuadra} J, {Sesana} A, {Stasyszyn} F, {Amaro-Seoane} P,
  {Tanaka} TL (2016) {Infalling clouds on to supermassive black hole binaries -
  I. Formation of discs, accretion and gas dynamics}. \mnras 455(2):1989--2003.
  \doi{10.1093/mnras/stv2470}.
  {\href{https://arxiv.org/abs/1507.05596}{{arXiv:1507.05596}}} {[astro-ph.HE]}

\bibitem[{{Goicovic} et~al.(2017){Goicovic}, {Sesana}, {Cuadra}, and
  {Stasyszyn}}]{2017MNRAS.472..514G}
{Goicovic} FG, {Sesana} A, {Cuadra} J, {Stasyszyn} F (2017) {Infalling clouds
  on to supermassive black hole binaries - II. Binary evolution and the final
  parsec problem}. \mnras 472(1):514--531. \doi{10.1093/mnras/stx1996}.
  {\href{https://arxiv.org/abs/1602.01966}{{arXiv:1602.01966}}} {[astro-ph.HE]}

\bibitem[{{Goicovic} et~al.(2018){Goicovic}, {Maureira-Fredes}, {Sesana},
  {Amaro-Seoane}, and {Cuadra}}]{2018MNRAS.479.3438G}
{Goicovic} FG, {Maureira-Fredes} C, {Sesana} A, {Amaro-Seoane} P, {Cuadra} J
  (2018) {Accretion of clumpy cold gas onto massive black hole binaries: a
  possible fast route to binary coalescence}. \mnras 479(3):3438--3455.
  \doi{10.1093/mnras/sty1709}.
  {\href{https://arxiv.org/abs/1801.04937}{{arXiv:1801.04937}}} {[astro-ph.HE]}

\bibitem[{{Gokhale} et~al.(2007){Gokhale}, {Peng}, and
  {Frank}}]{2007ApJ...655.1010G}
{Gokhale} V, {Peng} Xa, {Frank} J (2007) {Evolution of Close White Dwarf
  Binaries}. \apj 655(2):1010--1024. \doi{10.1086/510119}.
  {\href{https://arxiv.org/abs/astro-ph/0610919}{{arXiv:astro-ph/0610919}}}
  {[astro-ph]}

\bibitem[{{Gold} et~al.(2014{\natexlab{a}}){Gold}, {Paschalidis}, {Etienne},
  {Shapiro}, and {Pfeiffer}}]{2014PhRvD..89f4060G}
{Gold} R, {Paschalidis} V, {Etienne} ZB, {Shapiro} SL, {Pfeiffer} HP
  (2014{\natexlab{a}}) {Accretion disks around binary black holes of unequal
  mass: General relativistic magnetohydrodynamic simulations near decoupling}.
  \prd 89(6):064060. \doi{10.1103/PhysRevD.89.064060}.
  {\href{https://arxiv.org/abs/1312.0600}{{arXiv:1312.0600}}} {[astro-ph.HE]}

\bibitem[{{Gold} et~al.(2014{\natexlab{b}}){Gold}, {Paschalidis}, {Ruiz},
  {Shapiro}, {Etienne}, and {Pfeiffer}}]{2014PhRvD..90j4030G}
{Gold} R, {Paschalidis} V, {Ruiz} M, {Shapiro} SL, {Etienne} ZB, {Pfeiffer} HP
  (2014{\natexlab{b}}) {Accretion disks around binary black holes of unequal
  mass: General relativistic MHD simulations of postdecoupling and merger}.
  \prd 90(10):104030. \doi{10.1103/PhysRevD.90.104030}.
  {\href{https://arxiv.org/abs/1410.1543}{{arXiv:1410.1543}}} {[astro-ph.GA]}

\bibitem[{{Gondolo} and {Silk}(1999)}]{1999PhRvL..83.1719G}
{Gondolo} P, {Silk} J (1999) {Dark Matter Annihilation at the Galactic Center}.
  \prl 83(9):1719--1722. \doi{10.1103/PhysRevLett.83.1719}.
  {\href{https://arxiv.org/abs/astro-ph/9906391}{{arXiv:astro-ph/9906391}}}
  {[astro-ph]}

\bibitem[{{Gonz{\'a}lez} et~al.(2021){Gonz{\'a}lez}, {Kremer}, {Chatterjee},
  {Fragione}, {Rodriguez}, {Weatherford}, {Ye}, and
  {Rasio}}]{2021ApJ...908L..29G}
{Gonz{\'a}lez} E, {Kremer} K, {Chatterjee} S, {Fragione} G, {Rodriguez} CL,
  {Weatherford} NC, {Ye} CS, {Rasio} FA (2021) {Intermediate-mass Black Holes
  from High Massive-star Binary Fractions in Young Star Clusters}. \apjl
  908(2):L29. \doi{10.3847/2041-8213/abdf5b}.
  {\href{https://arxiv.org/abs/2012.10497}{{arXiv:2012.10497}}} {[astro-ph.HE]}

\bibitem[{{Gonz{\'a}lez} et~al.(2007{\natexlab{a}}){Gonz{\'a}lez}, {Hannam},
  {Sperhake}, {Br{\"u}gmann}, and {Husa}}]{2007PhRvL..98w1101G}
{Gonz{\'a}lez} JA, {Hannam} M, {Sperhake} U, {Br{\"u}gmann} B, {Husa} S
  (2007{\natexlab{a}}) {Supermassive Recoil Velocities for Binary Black-Hole
  Mergers with Antialigned Spins}. \prl 98(23):231101.
  \doi{10.1103/PhysRevLett.98.231101}.
  {\href{https://arxiv.org/abs/gr-qc/0702052}{{arXiv:gr-qc/0702052}}} {[gr-qc]}

\bibitem[{{Gonz{\'a}lez} et~al.(2007{\natexlab{b}}){Gonz{\'a}lez}, {Sperhake},
  {Br{\"u}gmann}, {Hannam}, and {Husa}}]{2007PhRvL..98i1101G}
{Gonz{\'a}lez} JA, {Sperhake} U, {Br{\"u}gmann} B, {Hannam} M, {Husa} S
  (2007{\natexlab{b}}) {Maximum Kick from Nonspinning Black-Hole Binary
  Inspiral}. \prl 98(9):091101. \doi{10.1103/PhysRevLett.98.091101}.
  {\href{https://arxiv.org/abs/gr-qc/0610154}{{arXiv:gr-qc/0610154}}} {[gr-qc]}

\bibitem[{{Gonz{\'a}lez Delgado} et~al.(2008){Gonz{\'a}lez Delgado},
  {P{\'e}rez}, {Cid Fernandes}, and {Schmitt}}]{2008AJ....135..747G}
{Gonz{\'a}lez Delgado} RM, {P{\'e}rez} E, {Cid Fernandes} R, {Schmitt} H (2008)
  {HST/WFPC2 Imaging of the Circumnuclear Structure of Low-Luminosity Active
  Galactic Nuclei. I. Data and Nuclear Morphology}. \aj 135(3):747--765.
  \doi{10.1088/0004-6256/135/3/747}.
  {\href{https://arxiv.org/abs/0710.4450}{{arXiv:0710.4450}}} {[astro-ph]}

\bibitem[{{Goodman}(2003)}]{2003MNRAS.339..937G}
{Goodman} J (2003) {Self-gravity and quasi-stellar object discs}. \mnras
  339(4):937--948. \doi{10.1046/j.1365-8711.2003.06241.x}.
  {\href{https://arxiv.org/abs/astro-ph/0201001}{{arXiv:astro-ph/0201001}}}
  {[astro-ph]}

\bibitem[{{Goodman} and {Hut}(1993)}]{1993ApJ...403..271G}
{Goodman} J, {Hut} P (1993) {Binary--Single-Star Scattering. V. Steady State
  Binary Distribution in a Homogeneous Static Background of Single Stars}. \apj
  403:271. \doi{10.1086/172200}

\bibitem[{{Goodman} and {Tan}(2004)}]{2004ApJ...608..108G}
{Goodman} J, {Tan} JC (2004) {Supermassive Stars in Quasar Disks}. \apj
  608(1):108--118. \doi{10.1086/386360}.
  {\href{https://arxiv.org/abs/astro-ph/0307361}{{arXiv:astro-ph/0307361}}}
  {[astro-ph]}

\bibitem[{{G{\"o}tberg} et~al.(2020){G{\"o}tberg}, {Korol}, {Lamberts},
  {Kupfer}, {Breivik}, {Ludwig}, and {Drout}}]{2020arXiv200607382G}
{G{\"o}tberg} Y, {Korol} V, {Lamberts} A, {Kupfer} T, {Breivik} K, {Ludwig} B,
  {Drout} MR (2020) {Stars stripped in binaries -- the living gravitational
  wave sources}. arXiv e-prints arXiv:2006.07382.
  {\href{https://arxiv.org/abs/2006.07382}{{arXiv:2006.07382}}} {[astro-ph.SR]}

\bibitem[{{Goulding} et~al.(2018){Goulding}, {Greene}, {Bezanson}, {Greco},
  {Johnson}, {Leauthaud}, {Matsuoka}, {Medezinski}, and
  {Price-Whelan}}]{2018PASJ...70S..37G}
{Goulding} AD, {Greene} JE, {Bezanson} R, {Greco} J, {Johnson} S, {Leauthaud}
  A, {Matsuoka} Y, {Medezinski} E, {Price-Whelan} AM (2018) {Galaxy
  interactions trigger rapid black hole growth: An unprecedented view from the
  Hyper Suprime-Cam survey}. \pasj 70:S37. \doi{10.1093/pasj/psx135}.
  {\href{https://arxiv.org/abs/1706.07436}{{arXiv:1706.07436}}} {[astro-ph.GA]}

\bibitem[{{Gourgoulhon} et~al.(2019){Gourgoulhon}, {Le Tiec}, {Vincent}, and
  {Warburton}}]{2019A&A...627A..92G}
{Gourgoulhon} E, {Le Tiec} A, {Vincent} FH, {Warburton} N (2019) {Gravitational
  waves from bodies orbiting the Galactic center black hole and their
  detectability by LISA}. \aap 627:A92. \doi{10.1051/0004-6361/201935406}.
  {\href{https://arxiv.org/abs/1903.02049}{{arXiv:1903.02049}}} {[gr-qc]}

\bibitem[{{Governato} et~al.(2010){Governato}, {Brook}, {Mayer}, {Brooks},
  {Rhee}, {Wadsley}, {Jonsson}, {Willman}, {Stinson}, {Quinn}, and
  {Madau}}]{2010Natur.463..203G}
{Governato} F, {Brook} C, {Mayer} L, {Brooks} A, {Rhee} G, {Wadsley} J,
  {Jonsson} P, {Willman} B, {Stinson} G, {Quinn} T, et~al. (2010) {Bulgeless
  dwarf galaxies and dark matter cores from supernova-driven outflows}. \nat
  463(7278):203--206. \doi{10.1038/nature08640}.
  {\href{https://arxiv.org/abs/0911.2237}{{arXiv:0911.2237}}} {[astro-ph.CO]}

\bibitem[{{Gow} et~al.(2020){Gow}, {Byrnes}, {Cole}, and
  {Young}}]{2020arXiv200803289G}
{Gow} AD, {Byrnes} CT, {Cole} PS, {Young} S (2020) {The power spectrum on small
  scales: Robust constraints and comparing PBH methodologies}. arXiv e-prints
  arXiv:2008.03289.
  {\href{https://arxiv.org/abs/2008.03289}{{arXiv:2008.03289}}} {[astro-ph.CO]}

\bibitem[{{Go{\'z}dziewski} et~al.(2015){Go{\'z}dziewski}, {S{\l}owikowska},
  {Dimitrov}, {Krzeszowski}, {Zejmo}, {Kanbach}, {Burwitz}, {Rau}, {Irawati},
  {Richichi}, {Gawro{\'n}ski}, {Nowak}, {Nasiroglu}, and
  {Kubicki}}]{2015MNRAS.448.1118G}
{Go{\'z}dziewski} K, {S{\l}owikowska} A, {Dimitrov} D, {Krzeszowski} K, {Zejmo}
  M, {Kanbach} G, {Burwitz} V, {Rau} A, {Irawati} P, {Richichi} A, et~al.
  (2015) {The HU Aqr planetary system hypothesis revisited}. \mnras
  448(2):1118--1136. \doi{10.1093/mnras/stu2728}.
  {\href{https://arxiv.org/abs/1412.5899}{{arXiv:1412.5899}}} {[astro-ph.EP]}

\bibitem[{{Graham}(2004)}]{2004ApJ...613L..33G}
{Graham} AW (2004) {Core Depletion from Coalescing Supermassive Black Holes}.
  \apjl 613(1):L33--L36. \doi{10.1086/424928}.
  {\href{https://arxiv.org/abs/astro-ph/0503177}{{arXiv:astro-ph/0503177}}}
  {[astro-ph]}

\bibitem[{{Graham}(2016{\natexlab{a}})}]{2016IAUS..312..269G}
{Graham} AW (2016{\natexlab{a}}) {Black hole and nuclear cluster scaling
  relations: M $_{bh}$ \raisebox{-0.5ex}\textasciitilde M $_{nc}$
  $^{2.7+/-0.7}$}. Proceedings of the International Astronomical Union
  10(S312):269--273. \doi{10.1017/S1743921315008017}.
  {\href{https://arxiv.org/abs/1412.5715}{{arXiv:1412.5715}}} {[astro-ph.GA]}

\bibitem[{{Graham}(2016{\natexlab{b}})}]{2016ASSL..418..263G}
{Graham} AW (2016{\natexlab{b}}) {Galaxy Bulges and Their Massive Black Holes:
  A Review}, vol 418, p 263. \doi{10.1007/978-3-319-19378-6_11}

\bibitem[{{Graham} and {Driver}(2007)}]{2007ApJ...655...77G}
{Graham} AW, {Driver} SP (2007) {A Log-Quadratic Relation for Predicting
  Supermassive Black Hole Masses from the Host Bulge S{\'e}rsic Index}. \apj
  655(1):77--87. \doi{10.1086/509758}.
  {\href{https://arxiv.org/abs/astro-ph/0607378}{{arXiv:astro-ph/0607378}}}
  {[astro-ph]}

\bibitem[{{Graham} and {Scott}(2015)}]{2015ApJ...798...54G}
{Graham} AW, {Scott} N (2015) {The (Black Hole)-bulge Mass Scaling Relation at
  Low Masses}. \apj 798(1):54. \doi{10.1088/0004-637X/798/1/54}.
  {\href{https://arxiv.org/abs/1412.3091}{{arXiv:1412.3091}}} {[astro-ph.GA]}

\bibitem[{{Graham} and {Soria}(2019)}]{2019MNRAS.484..794G}
{Graham} AW, {Soria} R (2019) {Expected intermediate-mass black holes in the
  Virgo cluster - I. Early-type galaxies}. \mnras 484(1):794--813.
  \doi{10.1093/mnras/sty3398}.
  {\href{https://arxiv.org/abs/1812.01231}{{arXiv:1812.01231}}} {[astro-ph.HE]}

\bibitem[{{Graham} and {Spitler}(2009)}]{2009MNRAS.397.2148G}
{Graham} AW, {Spitler} LR (2009) {Quantifying the coexistence of massive black
  holes and dense nuclear star clusters}. \mnras 397(4):2148--2162.
  \doi{10.1111/j.1365-2966.2009.15118.x}.
  {\href{https://arxiv.org/abs/0907.5250}{{arXiv:0907.5250}}} {[astro-ph.CO]}

\bibitem[{{Graham} et~al.(2003){Graham}, {Erwin}, {Trujillo}, and {Asensio
  Ramos}}]{2003AJ....125.2951G}
{Graham} AW, {Erwin} P, {Trujillo} I, {Asensio Ramos} A (2003) {A New Empirical
  Model for the Structural Analysis of Early-Type Galaxies, and A Critical
  Review of the Nuker Model}. \aj 125(6):2951--2963. \doi{10.1086/375320}.
  {\href{https://arxiv.org/abs/astro-ph/0306023}{{arXiv:astro-ph/0306023}}}
  {[astro-ph]}

\bibitem[{{Graham} et~al.(2007){Graham}, {Driver}, {Allen}, and
  {Liske}}]{2007MNRAS.378..198G}
{Graham} AW, {Driver} SP, {Allen} PD, {Liske} J (2007) {The Millennium Galaxy
  Catalogue: the local supermassive black hole mass function in early- and
  late-type galaxies}. \mnras 378(1):198--210.
  \doi{10.1111/j.1365-2966.2007.11770.x}.
  {\href{https://arxiv.org/abs/0704.0316}{{arXiv:0704.0316}}} {[astro-ph]}

\bibitem[{{Graham} et~al.(2019){Graham}, {Soria}, and
  {Davis}}]{2019MNRAS.484..814G}
{Graham} AW, {Soria} R, {Davis} BL (2019) {Expected intermediate-mass black
  holes in the Virgo cluster - II. Late-type galaxies}. \mnras 484(1):814--831.
  \doi{10.1093/mnras/sty3068}.
  {\href{https://arxiv.org/abs/1811.03232}{{arXiv:1811.03232}}} {[astro-ph.GA]}

\bibitem[{{Graham} et~al.(2015){Graham}, {Djorgovski}, {Stern}, {Drake},
  {Mahabal}, {Donalek}, {Glikman}, {Larson}, and
  {Christensen}}]{2015MNRAS.453.1562G}
{Graham} MJ, {Djorgovski} SG, {Stern} D, {Drake} AJ, {Mahabal} AA, {Donalek} C,
  {Glikman} E, {Larson} S, {Christensen} E (2015) {A systematic search for
  close supermassive black hole binaries in the Catalina Real-time Transient
  Survey}. \mnras 453(2):1562--1576. \doi{10.1093/mnras/stv1726}.
  {\href{https://arxiv.org/abs/1507.07603}{{arXiv:1507.07603}}} {[astro-ph.GA]}

\bibitem[{{Graham} et~al.(2020){Graham}, {Ford}, {McKernan}, {Ross}, {Stern},
  {Burdge}, {Coughlin}, {Djorgovski}, {Drake}, {Duev}, {Kasliwal}, {Mahabal},
  {van Velzen}, {Belecki}, {Bellm}, {Burruss}, {Cenko}, and
  ...}]{2020PhRvL.124y1102G}
{Graham} MJ, {Ford} KES, {McKernan} B, {Ross} NP, {Stern} D, {Burdge} K,
  {Coughlin} M, {Djorgovski} SG, {Drake} AJ, {Duev} D, et~al. (2020) {Candidate
  Electromagnetic Counterpart to the Binary Black Hole Merger
  Gravitational-Wave Event S190521g$^{*}$}. \prl 124(25):251102.
  \doi{10.1103/PhysRevLett.124.251102}.
  {\href{https://arxiv.org/abs/2006.14122}{{arXiv:2006.14122}}} {[astro-ph.HE]}

\bibitem[{{Graziani}(2019)}]{2019Physi...1..412G}
{Graziani} L (2019) {Hunting for Dwarf Galaxies Hosting the Formation and
  Coalescence of Compact Binaries}. Physics 1(3):412--429.
  \doi{10.3390/physics1030030}

\bibitem[{{Graziani} et~al.(2015){Graziani}, {Salvadori}, {Schneider},
  {Kawata}, {de Bennassuti}, and {Maselli}}]{2015MNRAS.449.3137G}
{Graziani} L, {Salvadori} S, {Schneider} R, {Kawata} D, {de Bennassuti} M,
  {Maselli} A (2015) {Galaxy formation with radiative and chemical feedback}.
  \mnras 449(3):3137--3148. \doi{10.1093/mnras/stv494}.
  {\href{https://arxiv.org/abs/1502.07344}{{arXiv:1502.07344}}} {[astro-ph.GA]}

\bibitem[{{Graziani} et~al.(2017){Graziani}, {de Bennassuti}, {Schneider},
  {Kawata}, and {Salvadori}}]{2017MNRAS.469.1101G}
{Graziani} L, {de Bennassuti} M, {Schneider} R, {Kawata} D, {Salvadori} S
  (2017) {The history of the dark and luminous side of Milky Way-like
  progenitors}. \mnras 469(1):1101--1116. \doi{10.1093/mnras/stx900}.
  {\href{https://arxiv.org/abs/1704.02983}{{arXiv:1704.02983}}} {[astro-ph.GA]}

\bibitem[{{Graziani} et~al.(2020){Graziani}, {Schneider}, {Marassi}, {Del
  Pozzo}, {Mapelli}, and {Giacobbo}}]{2020MNRAS.495L..81G}
{Graziani} L, {Schneider} R, {Marassi} S, {Del Pozzo} W, {Mapelli} M,
  {Giacobbo} N (2020) {Cosmic archaeology with massive stellar black hole
  binaries}. \mnras 495(1):L81--L85. \doi{10.1093/mnrasl/slaa063}.
  {\href{https://arxiv.org/abs/2004.03603}{{arXiv:2004.03603}}} {[astro-ph.GA]}

\bibitem[{{Green} et~al.(2012){Green}, {Schechter}, {Baltay}, {Bean},
  {Bennett}, {Brown}, {Conselice}, {Donahue}, {Fan}, {Gaudi}, {Hirata},
  {Kalirai}, {Lauer}, {Nichol}, {Padmanabhan}, {Perlmutter}, {Rauscher},
  {Rhodes}, {Roellig}, {Stern}, {Sumi}, {Tanner}, {Wang}, {Weinberg}, {Wright},
  {Gehrels}, {Sambruna}, {Traub}, {Anderson}, {Cook}, {Garnavich},
  {Hillenbrand}, {Ivezic}, {Kerins}, {Lunine}, {McDonald}, {Penny}, {Phillips},
  {Rieke}, {Riess}, {van der Marel}, {Barry}, {Cheng}, {Content}, {Cutri},
  {Goullioud}, {Grady}, {Helou}, {Jackson}, {Kruk}, {Melton}, {Peddie},
  {Rioux}, and {Seiffert}}]{2012arXiv1208.4012G}
{Green} J, {Schechter} P, {Baltay} C, {Bean} R, {Bennett} D, {Brown} R,
  {Conselice} C, {Donahue} M, {Fan} X, {Gaudi} BS, et~al. (2012) {Wide-Field
  InfraRed Survey Telescope (WFIRST) Final Report}. arXiv e-prints
  arXiv:1208.4012. {\href{https://arxiv.org/abs/1208.4012}{{arXiv:1208.4012}}}
  {[astro-ph.IM]}

\bibitem[{{Green} et~al.(2018{\natexlab{a}}){Green}, {Hermes}, {Marsh},
  {Steeghs}, {Bell}, {Littlefair}, {Parsons}, {Dennihy}, {Fuchs}, {Reding},
  {Kaiser}, {Ashley}, {Breedt}, {Dhillon}, {Gentile Fusillo}, {Kerry}, and
  {Sahman}}]{2018MNRAS.477.5646G}
{Green} MJ, {Hermes} JJ, {Marsh} TR, {Steeghs} DTH, {Bell} KJ, {Littlefair} SP,
  {Parsons} SG, {Dennihy} E, {Fuchs} JT, {Reding} JS, et~al.
  (2018{\natexlab{a}}) {A 15.7-minAM CVn binary discovered in K2}. \mnras
  477(4):5646--5656. \doi{10.1093/mnras/sty1032}.
  {\href{https://arxiv.org/abs/1804.07138}{{arXiv:1804.07138}}} {[astro-ph.SR]}

\bibitem[{{Green} et~al.(2018{\natexlab{b}}){Green}, {Marsh}, {Steeghs},
  {Kupfer}, {Ashley}, {Bloemen}, and {et~ al.}}]{2018MNRAS.476.1663G}
{Green} MJ, {Marsh} TR, {Steeghs} DTH, {Kupfer} T, {Ashley} RP, {Bloemen} S,
  {et~ al} (2018{\natexlab{b}}) {High-speed photometry of Gaia14aae: an
  eclipsing AM CVn that challenges formation models}. \mnras 476(2):1663--1679.
  \doi{10.1093/mnras/sty299}.
  {\href{https://arxiv.org/abs/1802.00499}{{arXiv:1802.00499}}} {[astro-ph.SR]}

\bibitem[{{Green} and {Gair}(2020)}]{2020arXiv200803312G}
{Green} SR, {Gair} J (2020) {Complete parameter inference for GW150914 using
  deep learning}. arXiv e-prints arXiv:2008.03312.
  {\href{https://arxiv.org/abs/2008.03312}{{arXiv:2008.03312}}} {[astro-ph.IM]}

\bibitem[{{Greene}(2012)}]{2012NatCo...3.1304G}
{Greene} JE (2012) {Low-mass black holes as the remnants of primordial black
  hole formation}. Nature Communications 3:1304. \doi{10.1038/ncomms2314}.
  {\href{https://arxiv.org/abs/1211.7082}{{arXiv:1211.7082}}} {[astro-ph.CO]}

\bibitem[{{Greene} et~al.(2019){Greene}, {Strader}, and
  {Ho}}]{2019arXiv191109678G}
{Greene} JE, {Strader} J, {Ho} LC (2019) {Intermediate-Mass Black Holes}. arXiv
  e-prints arXiv:1911.09678.
  {\href{https://arxiv.org/abs/1911.09678}{{arXiv:1911.09678}}} {[astro-ph.GA]}

\bibitem[{{Grefenstette} et~al.(2014){Grefenstette}, {Harrison}, {Boggs},
  {Reynolds}, {Fryer}, {Madsen}, {Wik}, and {et~al.}}]{2014Natur.506..339G}
{Grefenstette} BW, {Harrison} FA, {Boggs} SE, {Reynolds} SP, {Fryer} CL,
  {Madsen} KK, {Wik} DR, {et~al} (2014) {Asymmetries in core-collapse
  supernovae from maps of radioactive $^{44}$Ti in CassiopeiaA}. \nat
  506(7488):339--342. \doi{10.1038/nature12997}.
  {\href{https://arxiv.org/abs/1403.4978}{{arXiv:1403.4978}}} {[astro-ph.HE]}

\bibitem[{Gregely and Biermann(2009)}]{ApJ.697.1621}
Gregely LA, Biermann PL (2009) The spin flip phenomenon and supermassive black
  hole binary systems. \apj 697:1621--1633. \doi{10.1088/0004-637X/697/2/1621},
  \urlprefix\url{https://doi.org/10.1088-0004-637X/697/2/1621}

\bibitem[{{Greiner} et~al.(2000){Greiner}, {Schwarz}, {Zharikov}, and
  {Orio}}]{2000A&A...362L..25G}
{Greiner} J, {Schwarz} R, {Zharikov} S, {Orio} M (2000) {RX J1420.4+5334 -
  another tidal disruption event?} \aap 362:L25--L28.
  {\href{https://arxiv.org/abs/astro-ph/0009430}{{arXiv:astro-ph/0009430}}}
  {[astro-ph]}

\bibitem[{{Grindlay} et~al.(1995){Grindlay}, {Cool}, {Callanan}, {Bailyn},
  {Cohn}, and {Lugger}}]{1995ApJ...455L..47G}
{Grindlay} JE, {Cool} AM, {Callanan} PJ, {Bailyn} CD, {Cohn} HN, {Lugger} PM
  (1995) {Spectroscopic Identification of Probable Cataclysmic Variables in the
  Globular Cluster NGC 6397}. \apjl 455:L47. \doi{10.1086/309806}

\bibitem[{{Grover} et~al.(2014){Grover}, {Fairhurst}, {Farr}
  et~al.}]{Grover:2013}
{Grover} K, {Fairhurst} S, {Farr} BF, et~al. (2014) {Comparison of
  gravitational wave detector network sky localization approximations}. \prd
  89(4):042004. \doi{10.1103/PhysRevD.89.042004}.
  {\href{https://arxiv.org/abs/1310.7454}{{arXiv:1310.7454}}} {[gr-qc]}

\bibitem[{{Grupe} et~al.(1999){Grupe}, {Thomas}, and
  {Leighly}}]{1999A&A...350L..31G}
{Grupe} D, {Thomas} HC, {Leighly} KM (1999) {RX J1624.9+7554: a new X-ray
  transient AGN}. \aap 350:L31--L34.
  {\href{https://arxiv.org/abs/astro-ph/9909101}{{arXiv:astro-ph/9909101}}}
  {[astro-ph]}

\bibitem[{{Gruzinov} et~al.(2020){Gruzinov}, {Levin}, and
  {Matzner}}]{2020MNRAS.492.2755G}
{Gruzinov} A, {Levin} Y, {Matzner} CD (2020) {Negative dynamical friction on
  compact objects moving through dense gas}. \mnras 492(2):2755--2761.
  \doi{10.1093/mnras/staa013}.
  {\href{https://arxiv.org/abs/1906.01186}{{arXiv:1906.01186}}} {[astro-ph.HE]}

\bibitem[{{Gualandris} and {Merritt}(2008)}]{2008ApJ...678..780G}
{Gualandris} A, {Merritt} D (2008) {Ejection of Supermassive Black Holes from
  Galaxy Cores}. \apj 678(2):780--797. \doi{10.1086/586877}.
  {\href{https://arxiv.org/abs/0708.0771}{{arXiv:0708.0771}}} {[astro-ph]}

\bibitem[{{Gualandris} and {Merritt}(2009)}]{2009ApJ...705..361G}
{Gualandris} A, {Merritt} D (2009) {Perturbations of Intermediate-mass Black
  Holes on Stellar Orbits in the Galactic Center}. \apj 705(1):361--371.
  \doi{10.1088/0004-637X/705/1/361}.
  {\href{https://arxiv.org/abs/0905.4514}{{arXiv:0905.4514}}} {[astro-ph.GA]}

\bibitem[{{Gualandris} et~al.(2012){Gualandris}, {Dotti}, and
  {Sesana}}]{2012MNRAS.420L..38G}
{Gualandris} A, {Dotti} M, {Sesana} A (2012) {Massive black hole binary plane
  reorientation in rotating stellar systems}. \mnras 420(1):L38--L42.
  \doi{10.1111/j.1745-3933.2011.01188.x}.
  {\href{https://arxiv.org/abs/1109.3707}{{arXiv:1109.3707}}} {[astro-ph.GA]}

\bibitem[{{Gualandris} et~al.(2017){Gualandris}, {Read}, {Dehnen}, and
  {Bortolas}}]{2017MNRAS.464.2301G}
{Gualandris} A, {Read} JI, {Dehnen} W, {Bortolas} E (2017) {Collisionless
  loss-cone refilling: there is no final parsec problem}. \mnras
  464(2):2301--2310. \doi{10.1093/mnras/stw2528}.
  {\href{https://arxiv.org/abs/1609.09383}{{arXiv:1609.09383}}} {[astro-ph.GA]}

\bibitem[{{Gualtieri} et~al.(2008){Gualtieri}, {Berti}, {Cardoso}, and
  {Sperhake}}]{2008PhRvD..78d4024G}
{Gualtieri} L, {Berti} E, {Cardoso} V, {Sperhake} U (2008) {Transformation of
  the multipolar components of gravitational radiation under rotations and
  boosts}. \prd 78(4):044024. \doi{10.1103/PhysRevD.78.044024}.
  {\href{https://arxiv.org/abs/0805.1017}{{arXiv:0805.1017}}} {[gr-qc]}

\bibitem[{{Guillochon} et~al.(2009){Guillochon}, {Ramirez-Ruiz}, {Rosswog}, and
  {Kasen}}]{2009ApJ...705..844G}
{Guillochon} J, {Ramirez-Ruiz} E, {Rosswog} S, {Kasen} D (2009)
  {Three-dimensional Simulations of Tidally Disrupted Solar-type Stars and the
  Observational Signatures of Shock Breakout}. \apj 705(1):844--853.
  \doi{10.1088/0004-637X/705/1/844}.
  {\href{https://arxiv.org/abs/0811.1370}{{arXiv:0811.1370}}} {[astro-ph]}

\bibitem[{{Guillochon} et~al.(2010){Guillochon}, {Dan}, {Ramirez-Ruiz}, and
  {Rosswog}}]{2010ApJ...709L..64G}
{Guillochon} J, {Dan} M, {Ramirez-Ruiz} E, {Rosswog} S (2010) {Surface
  Detonations in Double Degenerate Binary Systems Triggered by Accretion Stream
  Instabilities}. \apjl 709(1):L64--L69. \doi{10.1088/2041-8205/709/1/L64}.
  {\href{https://arxiv.org/abs/0911.0416}{{arXiv:0911.0416}}} {[astro-ph.HE]}

\bibitem[{{G{\"u}ltekin} et~al.(2009){G{\"u}ltekin}, {Richstone}, {Gebhardt},
  {Lauer}, {Tremaine}, {Aller}, {Bender}, {Dressler}, {Faber}, {Filippenko},
  {Green}, {Ho}, {Kormendy}, {Magorrian}, {Pinkney}, and
  {Siopis}}]{2009ApJ...698..198G}
{G{\"u}ltekin} K, {Richstone} DO, {Gebhardt} K, {Lauer} TR, {Tremaine} S,
  {Aller} MC, {Bender} R, {Dressler} A, {Faber} SM, {Filippenko} AV, et~al.
  (2009) {The M-{\ensuremath{\sigma}} and M-L Relations in Galactic Bulges, and
  Determinations of Their Intrinsic Scatter}. \apj 698(1):198--221.
  \doi{10.1088/0004-637X/698/1/198}.
  {\href{https://arxiv.org/abs/0903.4897}{{arXiv:0903.4897}}} {[astro-ph.GA]}

\bibitem[{{Gunn} and {Ostriker}(1970)}]{1970ApJ...160..979G}
{Gunn} JE, {Ostriker} JP (1970) {On the Nature of Pulsars. III. Analysis of
  Observations}. \apj 160:979. \doi{10.1086/150487}

\bibitem[{{Guo} et~al.(2019){Guo}, {Liu}, {Shen}, {Loeb}, {Monroe}, and
  {Prochaska}}]{2019MNRAS.482.3288G}
{Guo} H, {Liu} X, {Shen} Y, {Loeb} A, {Monroe} T, {Prochaska} JX (2019)
  {Constraining sub-parsec binary supermassive black holes in quasars with
  multi-epoch spectroscopy - III. Candidates from continued radial velocity
  tests}. \mnras 482(3):3288--3307. \doi{10.1093/mnras/sty2920}.
  {\href{https://arxiv.org/abs/1809.04610}{{arXiv:1809.04610}}} {[astro-ph.GA]}

\bibitem[{{G{\"u}rkan} et~al.(2004){G{\"u}rkan}, {Freitag}, and
  {Rasio}}]{2004ApJ...604..632G}
{G{\"u}rkan} MA, {Freitag} M, {Rasio} FA (2004) {Formation of Massive Black
  Holes in Dense Star Clusters. I. Mass Segregation and Core Collapse}. \apj
  604(2):632--652. \doi{10.1086/381968}.
  {\href{https://arxiv.org/abs/astro-ph/0308449}{{arXiv:astro-ph/0308449}}}
  {[astro-ph]}

\bibitem[{{G{\"u}rkan} et~al.(2006){G{\"u}rkan}, {Fregeau}, and
  {Rasio}}]{2006ApJ...640L..39G}
{G{\"u}rkan} MA, {Fregeau} JM, {Rasio} FA (2006) {Massive Black Hole Binaries
  from Collisional Runaways}. \apjl 640(1):L39--L42. \doi{10.1086/503295}.
  {\href{https://arxiv.org/abs/astro-ph/0512642}{{arXiv:astro-ph/0512642}}}
  {[astro-ph]}

\bibitem[{{Habets}(1986)}]{1986A&A...165...95H}
{Habets} GMHJ (1986) {The evolution of a single and a binary helium star of 2.5
  solar mass up to neon ignition.} \aap 165:95--109

\bibitem[{{Habouzit} et~al.(2016){Habouzit}, {Volonteri}, {Latif}, {Dubois},
  and {Peirani}}]{2016MNRAS.463..529H}
{Habouzit} M, {Volonteri} M, {Latif} M, {Dubois} Y, {Peirani} S (2016) {On the
  number density of `direct collapse' black hole seeds}. \mnras
  463(1):529--540. \doi{10.1093/mnras/stw1924}.
  {\href{https://arxiv.org/abs/1601.00557}{{arXiv:1601.00557}}} {[astro-ph.GA]}

\bibitem[{{Habouzit} et~al.(2017){Habouzit}, {Volonteri}, and
  {Dubois}}]{2017MNRAS.468.3935H}
{Habouzit} M, {Volonteri} M, {Dubois} Y (2017) {Blossoms from black hole seeds:
  properties and early growth regulated by supernova feedback}. \mnras
  468(4):3935--3948. \doi{10.1093/mnras/stx666}.
  {\href{https://arxiv.org/abs/1605.09394}{{arXiv:1605.09394}}} {[astro-ph.GA]}

\bibitem[{{Hachisu} et~al.(1999){Hachisu}, {Kato}, and
  {Nomoto}}]{1999ApJ...522..487H}
{Hachisu} I, {Kato} M, {Nomoto} K (1999) {A Wide Symbiotic Channel to Type IA
  Supernovae}. \apj 522(1):487--503. \doi{10.1086/307608}.
  {\href{https://arxiv.org/abs/astro-ph/9902304}{{arXiv:astro-ph/9902304}}}
  {[astro-ph]}

\bibitem[{{Haehnelt}(1994)}]{1994MNRAS.269..199H}
{Haehnelt} MG (1994) {Low-Frequency Gravitational Waves from Supermassive
  Black-Holes}. \mnras 269:199. \doi{10.1093/mnras/269.1.199}.
  {\href{https://arxiv.org/abs/astro-ph/9405032}{{arXiv:astro-ph/9405032}}}
  {[astro-ph]}

\bibitem[{{Haehnelt} et~al.(1998){Haehnelt}, {Natarajan}, and
  {Rees}}]{1998MNRAS.300..817H}
{Haehnelt} MG, {Natarajan} P, {Rees} MJ (1998) {High-redshift galaxies, their
  active nuclei and central black holes}. \mnras 300(3):817--827.
  \doi{10.1046/j.1365-8711.1998.01951.x}.
  {\href{https://arxiv.org/abs/astro-ph/9712259}{{arXiv:astro-ph/9712259}}}
  {[astro-ph]}

\bibitem[{{Haemmerl{\'e}} et~al.(2018){Haemmerl{\'e}}, {Woods}, {Klessen},
  {Heger}, and {Whalen}}]{2018MNRAS.474.2757H}
{Haemmerl{\'e}} L, {Woods} TE, {Klessen} RS, {Heger} A, {Whalen} DJ (2018) {The
  evolution of supermassive Population III stars}. \mnras 474(2):2757--2773.
  \doi{10.1093/mnras/stx2919}.
  {\href{https://arxiv.org/abs/1705.09301}{{arXiv:1705.09301}}} {[astro-ph.SR]}

\bibitem[{{Haemmerl{\'e}} et~al.(2021){Haemmerl{\'e}}, {Klessen}, {Mayer}, and
  {Zwick}}]{Haemmerle_et_al_2021}
{Haemmerl{\'e}} L, {Klessen} RS, {Mayer} L, {Zwick} L (2021) {Maximum accretion
  rate of supermassive stars}. \aap 652(L7):L7--L11.
  \doi{10.1051/0004-6361/202141376}.
  {\href{https://arxiv.org/abs/arXiv:2105.13373}{{arXiv:arXiv:2105.13373}}}
  {[astro-ph.SR]}

\bibitem[{{Haiman}(2004)}]{2004ApJ...613...36H}
{Haiman} Z (2004) {Constraints from Gravitational Recoil on the Growth of
  Supermassive Black Holes at High Redshift}. \apj 613(1):36--40.
  \doi{10.1086/422910}.
  {\href{https://arxiv.org/abs/astro-ph/0404196}{{arXiv:astro-ph/0404196}}}
  {[astro-ph]}

\bibitem[{{Haiman}(2017)}]{2017PhRvD..96b3004H}
{Haiman} Z (2017) {Electromagnetic chirp of a compact binary black hole: A
  phase template for the gravitational wave inspiral}. \prd 96(2):023004.
  \doi{10.1103/PhysRevD.96.023004}.
  {\href{https://arxiv.org/abs/1705.06765}{{arXiv:1705.06765}}} {[astro-ph.HE]}

\bibitem[{{Haiman} and {Loeb}(2001)}]{2001ApJ...552..459H}
{Haiman} Z, {Loeb} A (2001) {What Is the Highest Plausible Redshift of Luminous
  Quasars?} \apj 552(2):459--463. \doi{10.1086/320586}.
  {\href{https://arxiv.org/abs/astro-ph/0011529}{{arXiv:astro-ph/0011529}}}
  {[astro-ph]}

\bibitem[{{Haiman} et~al.(2009){Haiman}, {Kocsis}, and
  {Menou}}]{2009ApJ...700.1952H}
{Haiman} Z, {Kocsis} B, {Menou} K (2009) {The Population of Viscosity- and
  Gravitational Wave-driven Supermassive Black Hole Binaries Among Luminous
  Active Galactic Nuclei}. \apj 700(2):1952--1969.
  \doi{10.1088/0004-637X/700/2/1952}.
  {\href{https://arxiv.org/abs/0904.1383}{{arXiv:0904.1383}}} {[astro-ph.CO]}

\bibitem[{{Halpern} and {Holt}(1992)}]{1992Natur.357..222H}
{Halpern} JP, {Holt} SS (1992) {Discovery of soft X-ray pulsations from the
  {\ensuremath{\gamma}}-ray source Geminga}. \nat 357(6375):222--224.
  \doi{10.1038/357222a0}

\bibitem[{{Hamers}(2017)}]{2017MNRAS.466.4107H}
{Hamers} AS (2017) {On the formation of hot and warm Jupiters via secular
  high-eccentricity migration in stellar triples}. \mnras 466(4):4107--4120.
  \doi{10.1093/mnras/stx035}.
  {\href{https://arxiv.org/abs/1701.01733}{{arXiv:1701.01733}}} {[astro-ph.EP]}

\bibitem[{{Hamers} and {Portegies Zwart}(2016)}]{2016MNRAS.462L..84H}
{Hamers} AS, {Portegies Zwart} SF (2016) {White dwarf pollution by planets in
  stellar binaries}. \mnras 462(1):L84--L87. \doi{10.1093/mnrasl/slw134}.
  {\href{https://arxiv.org/abs/1607.01397}{{arXiv:1607.01397}}} {[astro-ph.EP]}

\bibitem[{{Hamers} and {Safarzadeh}(2020)}]{2020ApJ...898...99H}
{Hamers} AS, {Safarzadeh} M (2020) {Was GW190412 Born from a Hierarchical 3 + 1
  Quadruple Configuration?} \apj 898(2):99. \doi{10.3847/1538-4357/ab9b27}.
  {\href{https://arxiv.org/abs/2005.03045}{{arXiv:2005.03045}}} {[astro-ph.HE]}

\bibitem[{{Hamers} et~al.(2013){Hamers}, {Pols}, {Claeys}, and
  {Nelemans}}]{2013MNRAS.430.2262H}
{Hamers} AS, {Pols} OR, {Claeys} JSW, {Nelemans} G (2013) {Population synthesis
  of triple systems in the context of mergers of carbon-oxygen white dwarfs}.
  \mnras 430(3):2262--2280. \doi{10.1093/mnras/stt046}.
  {\href{https://arxiv.org/abs/1301.1469}{{arXiv:1301.1469}}} {[astro-ph.SR]}

\bibitem[{{Hamers} et~al.(2014){Hamers}, {Portegies Zwart}, and
  {Merritt}}]{2014MNRAS.443..355H}
{Hamers} AS, {Portegies Zwart} SF, {Merritt} D (2014) {Relativistic dynamics of
  stars near a supermassive black hole}. \mnras 443(1):355--387.
  \doi{10.1093/mnras/stu1126}.
  {\href{https://arxiv.org/abs/1406.2846}{{arXiv:1406.2846}}} {[astro-ph.GA]}

\bibitem[{{Hamilton} and {Heinemann}(2020)}]{2020arXiv201114812H}
{Hamilton} C, {Heinemann} T (2020) {Noise and waves: a unified kinetic theory
  for stellar systems}. arXiv e-prints arXiv:2011.14812.
  {\href{https://arxiv.org/abs/2011.14812}{{arXiv:2011.14812}}} {[astro-ph.GA]}

\bibitem[{{Hamilton} and {Rafikov}(2019{\natexlab{a}})}]{2019ApJ...881L..13H}
{Hamilton} C, {Rafikov} RR (2019{\natexlab{a}}) {Compact Object Binary Mergers
  Driven By Cluster Tides: A New Channel for LIGO/Virgo Gravitational-wave
  Events}. \apjl 881(1):L13. \doi{10.3847/2041-8213/ab3468}.
  {\href{https://arxiv.org/abs/1907.00994}{{arXiv:1907.00994}}} {[astro-ph.GA]}

\bibitem[{{Hamilton} and {Rafikov}(2019{\natexlab{b}})}]{2019MNRAS.488.5489H}
{Hamilton} C, {Rafikov} RR (2019{\natexlab{b}}) {Secular dynamics of binaries
  in stellar clusters - I. General formulation and dependence on cluster
  potential}. \mnras 488(4):5489--5511. \doi{10.1093/mnras/stz1730}.
  {\href{https://arxiv.org/abs/1902.01344}{{arXiv:1902.01344}}} {[astro-ph.GA]}

\bibitem[{{Han} and {Chen}(2019)}]{2019MNRAS.485L..29H}
{Han} WB, {Chen} X (2019) {Testing general relativity using binary
  extreme-mass-ratio inspirals}. \mnras 485(1):L29--L33.
  \doi{10.1093/mnrasl/slz021}.
  {\href{https://arxiv.org/abs/1801.07060}{{arXiv:1801.07060}}} {[gr-qc]}

\bibitem[{{Han} and {Fan}(2018)}]{2018ApJ...856...82H}
{Han} WB, {Fan} XL (2018) {Determining the Nature of White Dwarfs from
  Low-frequency Gravitational Waves}. \apj 856(1):82.
  \doi{10.3847/1538-4357/aab03c}.
  {\href{https://arxiv.org/abs/1711.08628}{{arXiv:1711.08628}}} {[astro-ph.HE]}

\bibitem[{{Han} et~al.(2020){Han}, {Zhong}, {Chen}, and
  {Xin}}]{2020MNRAS.498L..61H}
{Han} WB, {Zhong} XY, {Chen} X, {Xin} S (2020) {Very extreme mass-ratio bursts
  in the Galaxy and neighbouring galaxies in relation to space-borne
  detectors}. \mnras 498(1):L61--L65. \doi{10.1093/mnrasl/slaa115}.
  {\href{https://arxiv.org/abs/2004.04016}{{arXiv:2004.04016}}} {[gr-qc]}

\bibitem[{Hannuksela et~al.(2019)Hannuksela, Wong, Brito, Berti, and
  Li}]{Hannuksela:2018izj}
Hannuksela OA, Wong KW, Brito R, Berti E, Li TG (2019) {Probing the existence
  of ultralight bosons with a single gravitational-wave measurement}. Nature
  Astron 3(5):447--451. \doi{10.1038/s41550-019-0712-4}.
  {\href{https://arxiv.org/abs/1804.09659}{{arXiv:1804.09659}}} {[astro-ph.HE]}

\bibitem[{{Hannuksela} et~al.(2020){Hannuksela}, {Ng}, and
  {Li}}]{2020PhRvD.102j3022H}
{Hannuksela} OA, {Ng} KCY, {Li} TGF (2020) {Extreme dark matter tests with
  extreme mass ratio inspirals}. \prd 102(10):103022.
  \doi{10.1103/PhysRevD.102.103022}.
  {\href{https://arxiv.org/abs/1906.11845}{{arXiv:1906.11845}}} {[astro-ph.CO]}

\bibitem[{{Hansen} and {Milosavljevi{\'c}}(2003)}]{2003ApJ...593L..77H}
{Hansen} BMS, {Milosavljevi{\'c}} M (2003) {The Need for a Second Black Hole at
  the Galactic Center}. \apjl 593(2):L77--L80. \doi{10.1086/378182}.
  {\href{https://arxiv.org/abs/astro-ph/0306074}{{arXiv:astro-ph/0306074}}}
  {[astro-ph]}

\bibitem[{{Hansen}(1972)}]{1972PhRvD...5.1021H}
{Hansen} RO (1972) {Post-Newtonian Gravitational Radiation from Point Masses in
  a Hyperbolic Kepler Orbit}. \prd 5(4):1021--1023.
  \doi{10.1103/PhysRevD.5.1021}

\bibitem[{{Harko} et~al.(2010){Harko}, {Kov{\'a}cs}, and
  {Lobo}}]{2010CQGra..27j5010H}
{Harko} T, {Kov{\'a}cs} Z, {Lobo} FSN (2010) {Thin accretion disk signatures in
  dynamical Chern-Simons-modified gravity}. Classical and Quantum Gravity
  27(10):105010. \doi{10.1088/0264-9381/27/10/105010}.
  {\href{https://arxiv.org/abs/0909.1267}{{arXiv:0909.1267}}} {[gr-qc]}

\bibitem[{{Hartnett} and {Luiten}(2011)}]{2011RvMP...83....1H}
{Hartnett} JG, {Luiten} AN (2011) {Colloquium: Comparison of astrophysical and
  terrestrial frequency standards}. Reviews of Modern Physics 83(1):1--9.
  \doi{10.1103/RevModPhys.83.1}.
  {\href{https://arxiv.org/abs/1004.0115}{{arXiv:1004.0115}}} {[astro-ph.IM]}

\bibitem[{{Haster} et~al.(2016{\natexlab{a}}){Haster}, {Antonini}, {Kalogera},
  and {Mand el}}]{2016ApJ...832..192H}
{Haster} CJ, {Antonini} F, {Kalogera} V, {Mand el} I (2016{\natexlab{a}})
  {N-Body Dynamics of Intermediate Mass-ratio Inspirals in Star Clusters}. \apj
  832(2):192. \doi{10.3847/0004-637X/832/2/192}.
  {\href{https://arxiv.org/abs/1606.07097}{{arXiv:1606.07097}}} {[astro-ph.HE]}

\bibitem[{{Haster} et~al.(2016{\natexlab{b}}){Haster}, {Wang}, {Berry},
  {Stevenson}, {Veitch}, and {Mandel}}]{2016MNRAS.457.4499H}
{Haster} CJ, {Wang} Z, {Berry} CPL, {Stevenson} S, {Veitch} J, {Mandel} I
  (2016{\natexlab{b}}) {Inference on gravitational waves from coalescences of
  stellar-mass compact objects and intermediate-mass black holes}. \mnras
  457(4):4499--4506. \doi{10.1093/mnras/stw233}.
  {\href{https://arxiv.org/abs/1511.01431}{{arXiv:1511.01431}}} {[astro-ph.HE]}

\bibitem[{{He} and {Petrovich}(2018)}]{2018MNRAS.474...20H}
{He} MY, {Petrovich} C (2018) {On the stability and collisions in triple
  stellar systems}. \mnras 474(1):20--31. \doi{10.1093/mnras/stx2718}.
  {\href{https://arxiv.org/abs/1710.04698}{{arXiv:1710.04698}}} {[astro-ph.SR]}

\bibitem[{{Heath} and {Nixon}(2020)}]{2020A&A...641A..64H}
{Heath} RM, {Nixon} CJ (2020) {On the orbital evolution of binaries with
  circumbinary discs}. \aap 641:A64. \doi{10.1051/0004-6361/202038548}.
  {\href{https://arxiv.org/abs/2007.11592}{{arXiv:2007.11592}}} {[astro-ph.HE]}

\bibitem[{{Heber}(2016)}]{2016PASP..128h2001H}
{Heber} U (2016) {Hot Subluminous Stars}. \pasp 128(966):082001.
  \doi{10.1088/1538-3873/128/966/082001}.
  {\href{https://arxiv.org/abs/1604.07749}{{arXiv:1604.07749}}} {[astro-ph.SR]}

\bibitem[{{Heckman} and {Best}(2014)}]{2014ARA&A..52..589H}
{Heckman} TM, {Best} PN (2014) {The Coevolution of Galaxies and Supermassive
  Black Holes: Insights from Surveys of the Contemporary Universe}. \araa
  52:589--660. \doi{10.1146/annurev-astro-081913-035722}.
  {\href{https://arxiv.org/abs/1403.4620}{{arXiv:1403.4620}}} {[astro-ph.GA]}

\bibitem[{{Heger} et~al.(2003){Heger}, {Fryer}, {Woosley}, {Langer}, and
  {Hartmann}}]{2003ApJ...591..288H}
{Heger} A, {Fryer} CL, {Woosley} SE, {Langer} N, {Hartmann} DH (2003) {How
  Massive Single Stars End Their Life}. \apj 591:288--300.
  \doi{10.1086/375341}.
  {\href{https://arxiv.org/abs/arXiv:astro-ph/0212469}{{arXiv:astro-ph/0212469}}}

\bibitem[{{Heggie} and {Hut}(2003)}]{2003gmbp.book.....H}
{Heggie} D, {Hut} P (2003) {The Gravitational Million-Body Problem: A
  Multidisciplinary Approach to Star Cluster Dynamics}

\bibitem[{{Heggie}(1975)}]{1975MNRAS.173..729H}
{Heggie} DC (1975) {Binary evolution in stellar dynamics.} \mnras 173:729--787.
  \doi{10.1093/mnras/173.3.729}

\bibitem[{{Hein{\"a}m{\"a}ki}(2001)}]{2001A&A...371..795H}
{Hein{\"a}m{\"a}ki} P (2001) {Symmetry of black hole ejections in mergers of
  galaxies}. \aap 371:795--805. \doi{10.1051/0004-6361:20010460}

\bibitem[{{Heinke} et~al.(2003){Heinke}, {Grindlay}, {Lugger}, {Cohn},
  {Edmonds}, {Lloyd}, and {Cool}}]{2003ApJ...598..501H}
{Heinke} CO, {Grindlay} JE, {Lugger} PM, {Cohn} HN, {Edmonds} PD, {Lloyd} DA,
  {Cool} AM (2003) {Analysis of the Quiescent Low-Mass X-Ray Binary Population
  in Galactic Globular Clusters}. \apj 598(1):501--515. \doi{10.1086/378885}.
  {\href{https://arxiv.org/abs/astro-ph/0305445}{{arXiv:astro-ph/0305445}}}
  {[astro-ph]}

\bibitem[{{Heinke} et~al.(2013){Heinke}, {Ivanova}, {Engel}, {Pavlovskii},
  {Sivakoff}, {Cartwright}, and {Gladstone}}]{2013ApJ...768..184H}
{Heinke} CO, {Ivanova} N, {Engel} MC, {Pavlovskii} K, {Sivakoff} GR,
  {Cartwright} TF, {Gladstone} JC (2013) {Galactic Ultracompact X-Ray Binaries:
  Disk Stability and Evolution}. \apj 768(2):184.
  \doi{10.1088/0004-637X/768/2/184}.
  {\href{https://arxiv.org/abs/1303.5864}{{arXiv:1303.5864}}} {[astro-ph.HE]}

\bibitem[{{Hellings} and {Downs}(1983)}]{1983ApJ...265L..39H}
{Hellings} RW, {Downs} GS (1983) {Upper limits on the isotopic gravitational
  radiation background frompulsar timing analysis.} \apjl 265:L39--L42.
  \doi{10.1086/183954}

\bibitem[{{Hennawi} et~al.(2015){Hennawi}, {Prochaska}, {Cantalupo}, and
  {Arrigoni-Battaia}}]{2015Sci...348..779H}
{Hennawi} JF, {Prochaska} JX, {Cantalupo} S, {Arrigoni-Battaia} F (2015)
  {Quasar quartet embedded in giant nebula reveals rare massive structure in
  distant universe}. Science 348(6236):779--783. \doi{10.1126/science.aaa5397}.
  {\href{https://arxiv.org/abs/1505.03786}{{arXiv:1505.03786}}} {[astro-ph.GA]}

\bibitem[{{Hermes} et~al.(2012){Hermes}, {Kilic}, {Brown}, {Winget}, {Allende
  Prieto}, {Gianninas}, {Mukadam}, {Cabrera-Lavers}, and
  {Kenyon}}]{2012ApJ...757L..21H}
{Hermes} JJ, {Kilic} M, {Brown} WR, {Winget} DE, {Allende Prieto} C,
  {Gianninas} A, {Mukadam} AS, {Cabrera-Lavers} A, {Kenyon} SJ (2012) {Rapid
  Orbital Decay in the 12.75-minute Binary White Dwarf J0651+2844}. \apjl
  757(2):L21. \doi{10.1088/2041-8205/757/2/L21}.
  {\href{https://arxiv.org/abs/1208.5051}{{arXiv:1208.5051}}} {[astro-ph.SR]}

\bibitem[{{Hickox} et~al.(2009){Hickox}, {Jones}, {Forman}, {Murray},
  {Kochanek}, {Eisenstein}, {Jannuzi}, {Dey}, {Brown}, {Stern}, {Eisenhardt},
  {Gorjian}, {Brodwin}, {Narayan}, {Cool}, {Kenter}, {Caldwell}, and
  {Anderson}}]{2009ApJ...696..891H}
{Hickox} RC, {Jones} C, {Forman} WR, {Murray} SS, {Kochanek} CS, {Eisenstein}
  D, {Jannuzi} BT, {Dey} A, {Brown} MJI, {Stern} D, et~al. (2009) {Host
  Galaxies, Clustering, Eddington Ratios, and Evolution of Radio, X-Ray, and
  Infrared-Selected AGNs}. \apj 696(1):891--919.
  \doi{10.1088/0004-637X/696/1/891}.
  {\href{https://arxiv.org/abs/0901.4121}{{arXiv:0901.4121}}} {[astro-ph.GA]}

\bibitem[{{Hicks} et~al.(2020){Hicks}, {Wells}, {Norman}, {Wise}, {Smith}, and
  {O'Shea}}]{2020arXiv200905499H}
{Hicks} W, {Wells} A, {Norman} ML, {Wise} JH, {Smith} BD, {O'Shea} BW (2020)
  {External Enrichment of Minihalos by the First Supernovae}. arXiv e-prints
  arXiv:2009.05499.
  {\href{https://arxiv.org/abs/2009.05499}{{arXiv:2009.05499}}} {[astro-ph.GA]}

\bibitem[{{Hild} et~al.(2011)}]{Hild2011}
{Hild} S, et~al. (2011) {Sensitivity studies for third-generation gravitational
  wave observatories}. Classical and Quantum Gravity 28(9):094013.
  \doi{10.1088/0264-9381/28/9/094013}.
  {\href{https://arxiv.org/abs/1012.0908}{{arXiv:1012.0908}}} {[gr-qc]}

\bibitem[{{Hilditch}(2001)}]{2001icbs.book.....H}
{Hilditch} RW (2001) {An Introduction to Close Binary Stars}

\bibitem[{{Hills}(1976)}]{1976MNRAS.175P...1H}
{Hills} JG (1976) {The formation of binaries containing black holes by the
  exchange of companions and the X-ray sources in globular clusters.} \mnras
  175:1P--4P. \doi{10.1093/mnras/175.1.1P}

\bibitem[{{Hills} and {Fullerton}(1980)}]{1980AJ.....85.1281H}
{Hills} JG, {Fullerton} LW (1980) {Computer simulations of close encounters
  between single stars and hard binaries}. \aj 85:1281--1291.
  \doi{10.1086/112798}

\bibitem[{{Hils} et~al.(1990){Hils}, {Bender}, and
  {Webbink}}]{1990ApJ...360...75H}
{Hils} D, {Bender} PL, {Webbink} RF (1990) {Gravitational Radiation from the
  Galaxy}. \apj 360:75. \doi{10.1086/169098}

\bibitem[{{Hinderer} and {Flanagan}(2008)}]{2008PhRvD..78f4028H}
{Hinderer} T, {Flanagan} {\'E}{\'E} (2008) {Two-timescale analysis of extreme
  mass ratio inspirals in Kerr spacetime: Orbital motion}. \prd 78(6):064028.
  \doi{10.1103/PhysRevD.78.064028}.
  {\href{https://arxiv.org/abs/0805.3337}{{arXiv:0805.3337}}} {[gr-qc]}

\bibitem[{{Hirano} et~al.(2014){Hirano}, {Hosokawa}, {Yoshida}, {Umeda},
  {Omukai}, {Chiaki}, and {Yorke}}]{2014ApJ...781...60H}
{Hirano} S, {Hosokawa} T, {Yoshida} N, {Umeda} H, {Omukai} K, {Chiaki} G,
  {Yorke} HW (2014) {One Hundred First Stars: Protostellar Evolution and the
  Final Masses}. \apj 781(2):60. \doi{10.1088/0004-637X/781/2/60}.
  {\href{https://arxiv.org/abs/1308.4456}{{arXiv:1308.4456}}} {[astro-ph.CO]}

\bibitem[{{Hirschmann} et~al.(2014){Hirschmann}, {Dolag}, {Saro}, {Bachmann},
  {Borgani}, and {Burkert}}]{2014MNRAS.442.2304H}
{Hirschmann} M, {Dolag} K, {Saro} A, {Bachmann} L, {Borgani} S, {Burkert} A
  (2014) {Cosmological simulations of black hole growth: AGN luminosities and
  downsizing}. \mnras 442(3):2304--2324. \doi{10.1093/mnras/stu1023}.
  {\href{https://arxiv.org/abs/1308.0333}{{arXiv:1308.0333}}} {[astro-ph.CO]}

\bibitem[{{Hjellming} and {Webbink}(1987)}]{1987ApJ...318..794H}
{Hjellming} MS, {Webbink} RF (1987) {Thresholds for Rapid Mass Transfer in
  Binary System. I. Polytropic Models}. \apj 318:794. \doi{10.1086/165412}

\bibitem[{{Hoang} et~al.(2018){Hoang}, {Naoz}, {Kocsis}, {Rasio}, and
  {Dosopoulou}}]{2018ApJ...856..140H}
{Hoang} BM, {Naoz} S, {Kocsis} B, {Rasio} FA, {Dosopoulou} F (2018) {Black Hole
  Mergers in Galactic Nuclei Induced by the Eccentric Kozai-Lidov Effect}. \apj
  856(2):140. \doi{10.3847/1538-4357/aaafce}.
  {\href{https://arxiv.org/abs/1706.09896}{{arXiv:1706.09896}}} {[astro-ph.HE]}

\bibitem[{{Hoang} et~al.(2019){Hoang}, {Naoz}, {Kocsis}, {Farr}, and
  {McIver}}]{2019ApJ...875L..31H}
{Hoang} BM, {Naoz} S, {Kocsis} B, {Farr} WM, {McIver} J (2019) {Detecting
  Supermassive Black Hole-induced Binary Eccentricity Oscillations with LISA}.
  \apjl 875(2):L31. \doi{10.3847/2041-8213/ab14f7}.
  {\href{https://arxiv.org/abs/1903.00134}{{arXiv:1903.00134}}} {[astro-ph.HE]}

\bibitem[{{Hobbs} et~al.(2012){Hobbs}, {Power}, {Nayakshin}, and
  {King}}]{2012MNRAS.421.3443H}
{Hobbs} A, {Power} C, {Nayakshin} S, {King} AR (2012) {Modelling supermassive
  black hole growth: towards an improved sub-grid prescription}. \mnras
  421(4):3443--3449. \doi{10.1111/j.1365-2966.2012.20563.x}.
  {\href{https://arxiv.org/abs/1202.4725}{{arXiv:1202.4725}}} {[astro-ph.IM]}

\bibitem[{{Hobbs} et~al.(2005){Hobbs}, {Lorimer}, {Lyne}, and
  {Kramer}}]{2005MNRAS.360..974H}
{Hobbs} G, {Lorimer} DR, {Lyne} AG, {Kramer} M (2005) {A statistical study of
  233 pulsar proper motions}. \mnras 360(3):974--992.
  \doi{10.1111/j.1365-2966.2005.09087.x}.
  {\href{https://arxiv.org/abs/astro-ph/0504584}{{arXiv:astro-ph/0504584}}}
  {[astro-ph]}

\bibitem[{{Hoffman} and {Loeb}(2007)}]{2007MNRAS.377..957H}
{Hoffman} L, {Loeb} A (2007) {Dynamics of triple black hole systems in
  hierarchically merging massive galaxies}. \mnras 377(3):957--976.
  \doi{10.1111/j.1365-2966.2007.11694.x}.
  {\href{https://arxiv.org/abs/astro-ph/0612517}{{arXiv:astro-ph/0612517}}}
  {[astro-ph]}

\bibitem[{{Hofmann} et~al.(2016){Hofmann}, {Barausse}, and
  {Rezzolla}}]{2016ApJ...825L..19H}
{Hofmann} F, {Barausse} E, {Rezzolla} L (2016) {The Final Spin from Binary
  Black Holes in Quasi-circular Orbits}. \apjl 825(2):L19.
  \doi{10.3847/2041-8205/825/2/L19}.
  {\href{https://arxiv.org/abs/1605.01938}{{arXiv:1605.01938}}} {[gr-qc]}

\bibitem[{{Holley-Bockelmann} and {Khan}(2015)}]{2015ApJ...810..139H}
{Holley-Bockelmann} K, {Khan} FM (2015) {Galaxy Rotation and Rapid Supermassive
  Binary Coalescence}. \apj 810(2):139. \doi{10.1088/0004-637X/810/2/139}.
  {\href{https://arxiv.org/abs/1505.06203}{{arXiv:1505.06203}}} {[astro-ph.GA]}

\bibitem[{{Holley-Bockelmann} et~al.(2008){Holley-Bockelmann}, {G{\"u}ltekin},
  {Shoemaker}, and {Yunes}}]{2008ApJ...686..829H}
{Holley-Bockelmann} K, {G{\"u}ltekin} K, {Shoemaker} D, {Yunes} N (2008)
  {Gravitational Wave Recoil and the Retention of Intermediate-Mass Black
  Holes}. \apj 686(2):829--837. \doi{10.1086/591218}.
  {\href{https://arxiv.org/abs/0707.1334}{{arXiv:0707.1334}}} {[astro-ph]}

\bibitem[{{Hong} and {Lee}(2015)}]{2015MNRAS.448..754H}
{Hong} J, {Lee} HM (2015) {Black hole binaries in galactic nuclei and
  gravitational wave sources}. \mnras 448(1):754--770.
  \doi{10.1093/mnras/stv035}.
  {\href{https://arxiv.org/abs/1501.02717}{{arXiv:1501.02717}}} {[astro-ph.GA]}

\bibitem[{{Hong} et~al.(2020){Hong}, {Askar}, {Giersz}, {Hypki}, and
  {Yoon}}]{2020MNRAS.tmp.2099H}
{Hong} J, {Askar} A, {Giersz} M, {Hypki} A, {Yoon} SJ (2020) {MOCCA-SURVEY
  Database I: Binary Black Hole Mergers from Globular Clusters with
  Intermediate Mass Black Holes}. \mnras \doi{10.1093/mnras/staa2677}.
  {\href{https://arxiv.org/abs/2008.10823}{{arXiv:2008.10823}}} {[astro-ph.HE]}

\bibitem[{{Hopkins} and {Quataert}(2010)}]{2010MNRAS.407.1529H}
{Hopkins} PF, {Quataert} E (2010) {How do massive black holes get their gas?}
  \mnras 407(3):1529--1564. \doi{10.1111/j.1365-2966.2010.17064.x}.
  {\href{https://arxiv.org/abs/0912.3257}{{arXiv:0912.3257}}} {[astro-ph.CO]}

\bibitem[{{Hopkins} et~al.(2007){Hopkins}, {Richards}, and
  {Hernquist}}]{2007ApJ...654..731H}
{Hopkins} PF, {Richards} GT, {Hernquist} L (2007) {An Observational
  Determination of the Bolometric Quasar Luminosity Function}. \apj
  654(2):731--753. \doi{10.1086/509629}.
  {\href{https://arxiv.org/abs/astro-ph/0605678}{{arXiv:astro-ph/0605678}}}
  {[astro-ph]}

\bibitem[{{Hopkins} et~al.(2014){Hopkins}, {Kere{\v{s}}}, {O{\~n}orbe},
  {Faucher-Gigu{\`e}re}, {Quataert}, {Murray}, and
  {Bullock}}]{2014MNRAS.445..581H}
{Hopkins} PF, {Kere{\v{s}}} D, {O{\~n}orbe} J, {Faucher-Gigu{\`e}re} CA,
  {Quataert} E, {Murray} N, {Bullock} JS (2014) {Galaxies on FIRE (Feedback In
  Realistic Environments): stellar feedback explains cosmologically inefficient
  star formation}. \mnras 445(1):581--603. \doi{10.1093/mnras/stu1738}.
  {\href{https://arxiv.org/abs/1311.2073}{{arXiv:1311.2073}}} {[astro-ph.CO]}

\bibitem[{{Hopkins} et~al.(2018){Hopkins}, {Wetzel}, {Kere{\v{s}}},
  {Faucher-Gigu{\`e}re}, {Quataert}, {Boylan-Kolchin}, {Murray}, {Hayward},
  {Garrison-Kimmel}, {Hummels}, {Feldmann}, {Torrey}, {Ma},
  {Angl{\'e}s-Alc{\'a}zar}, {Su}, {Orr}, {Schmitz}, {Escala}, {Sanderson},
  {Grudi{\'c}}, {Hafen}, {Kim}, {Fitts}, {Bullock}, {Wheeler}, {Chan},
  {Elbert}, and {Narayanan}}]{2018MNRAS.480..800H}
{Hopkins} PF, {Wetzel} A, {Kere{\v{s}}} D, {Faucher-Gigu{\`e}re} CA, {Quataert}
  E, {Boylan-Kolchin} M, {Murray} N, {Hayward} CC, {Garrison-Kimmel} S,
  {Hummels} C, et~al. (2018) {FIRE-2 simulations: physics versus numerics in
  galaxy formation}. \mnras 480(1):800--863. \doi{10.1093/mnras/sty1690}.
  {\href{https://arxiv.org/abs/1702.06148}{{arXiv:1702.06148}}} {[astro-ph.GA]}

\bibitem[{{Hopman} and {Alexander}(2006)}]{2006ApJ...645.1152H}
{Hopman} C, {Alexander} T (2006) {Resonant Relaxation near a Massive Black
  Hole: The Stellar Distribution and Gravitational Wave Sources}. \apj
  645(2):1152--1163. \doi{10.1086/504400}.
  {\href{https://arxiv.org/abs/astro-ph/0601161}{{arXiv:astro-ph/0601161}}}
  {[astro-ph]}

\bibitem[{{Hopman} et~al.(2007){Hopman}, {Freitag}, and
  {Larson}}]{2007MNRAS.378..129H}
{Hopman} C, {Freitag} M, {Larson} SL (2007) {Gravitational wave bursts from the
  Galactic massive black hole}. \mnras 378(1):129--136.
  \doi{10.1111/j.1365-2966.2007.11758.x}.
  {\href{https://arxiv.org/abs/astro-ph/0612337}{{arXiv:astro-ph/0612337}}}
  {[astro-ph]}

\bibitem[{{Horton} et~al.(2020){Horton}, {Hardcastle}, {Read}, and
  {Krause}}]{2020MNRAS.493.3911H}
{Horton} MA, {Hardcastle} MJ, {Read} SC, {Krause} MGH (2020) {A Markov chain
  Monte Carlo approach for measurement of jet precession in radio-loud active
  galactic nuclei}. \mnras 493(3):3911--3919. \doi{10.1093/mnras/staa429}.
  {\href{https://arxiv.org/abs/2002.04966}{{arXiv:2002.04966}}} {[astro-ph.GA]}

\bibitem[{{Howitt} et~al.(2020){Howitt}, {Stevenson}, {Vigna-G{\'o}mez},
  {Justham}, {Ivanova}, {Woods}, {Neijssel}, and
  {Mandel}}]{2020MNRAS.492.3229H}
{Howitt} G, {Stevenson} S, {Vigna-G{\'o}mez} Ar, {Justham} S, {Ivanova} N,
  {Woods} TE, {Neijssel} CJ, {Mandel} I (2020) {Luminous Red Novae: population
  models and future prospects}. \mnras 492(3):3229--3240.
  \doi{10.1093/mnras/stz3542}.
  {\href{https://arxiv.org/abs/1912.07771}{{arXiv:1912.07771}}} {[astro-ph.HE]}

\bibitem[{{Hoyle} and {Lyttleton}(1939)}]{1939PCPS...35..405H}
{Hoyle} F, {Lyttleton} RA (1939) {The effect of interstellar matter on climatic
  variation}. Proceedings of the Cambridge Philosophical Society 35(3):405.
  \doi{10.1017/S0305004100021150}

\bibitem[{{Hu} et~al.(2020){Hu}, {D'Orazio}, {Haiman}, {Smith}, {Snios},
  {Charisi}, and {Di Stefano}}]{2020MNRAS.495.4061H}
{Hu} BX, {D'Orazio} DJ, {Haiman} Z, {Smith} KL, {Snios} B, {Charisi} M, {Di
  Stefano} R (2020) {Spikey: self-lensing flares from eccentric SMBH binaries}.
  \mnras 495(4):4061--4070. \doi{10.1093/mnras/staa1312}.
  {\href{https://arxiv.org/abs/1910.05348}{{arXiv:1910.05348}}} {[astro-ph.HE]}

\bibitem[{{Hu} et~al.(2018){Hu}, {Li}, {Wang}, {Feng}, {Zhou}, {Hu}, {Hu},
  {Mei}, and {Shao}}]{2018CQGra..35i5008H}
{Hu} XC, {Li} XH, {Wang} Y, {Feng} WF, {Zhou} MY, {Hu} YM, {Hu} SC, {Mei} JW,
  {Shao} CG (2018) {Fundamentals of the orbit and response for TianQin}.
  Classical and Quantum Gravity 35(9):095008. \doi{10.1088/1361-6382/aab52f}.
  {\href{https://arxiv.org/abs/1803.03368}{{arXiv:1803.03368}}} {[gr-qc]}

\bibitem[{{Huang} and {Wu}(2014)}]{2014PhRvD..89l4034H}
{Huang} G, {Wu} X (2014) {Dynamics of the post-Newtonian circular restricted
  three-body problem with compact objects}. \prd 89(12):124034.
  \doi{10.1103/PhysRevD.89.124034}

\bibitem[{{Huang} et~al.(2020){Huang}, {Hu}, {Korol}, {Li}, {Liang}, {Lu},
  {Wang}, {Yu}, and {Mei}}]{2020arXiv200507889H}
{Huang} SJ, {Hu} YM, {Korol} V, {Li} PC, {Liang} ZC, {Lu} Y, {Wang} HT, {Yu} S,
  {Mei} J (2020) {Science with the TianQin Observatory: Preliminary results on
  Galactic double white dwarf binaries}. arXiv e-prints arXiv:2005.07889.
  {\href{https://arxiv.org/abs/2005.07889}{{arXiv:2005.07889}}} {[astro-ph.HE]}

\bibitem[{{Hughes} and {Blandford}(2003)}]{2003ApJ...585L.101H}
{Hughes} SA, {Blandford} RD (2003) {Black Hole Mass and Spin Coevolution by
  Mergers}. \apjl 585(2):L101--L104. \doi{10.1086/375495}.
  {\href{https://arxiv.org/abs/astro-ph/0208484}{{arXiv:astro-ph/0208484}}}
  {[astro-ph]}

\bibitem[{{Hui} et~al.(2018){Hui}, {Wu}, {Han}, {Kong}, and
  {Tam}}]{2018ApJ...864...30H}
{Hui} CY, {Wu} K, {Han} Q, {Kong} AKH, {Tam} PHT (2018) {On the Orbital
  Properties of Millisecond Pulsar Binaries}. \apj 864(1):30.
  \doi{10.3847/1538-4357/aad5ec}.
  {\href{https://arxiv.org/abs/1807.09001}{{arXiv:1807.09001}}} {[astro-ph.HE]}

\bibitem[{{Hui} et~al.(2017){Hui}, {Ostriker}, {Tremaine}, and
  {Witten}}]{2017PhRvD..95d3541H}
{Hui} L, {Ostriker} JP, {Tremaine} S, {Witten} E (2017) {Ultralight scalars as
  cosmological dark matter}. \prd 95(4):043541.
  \doi{10.1103/PhysRevD.95.043541}.
  {\href{https://arxiv.org/abs/1610.08297}{{arXiv:1610.08297}}} {[astro-ph.CO]}

\bibitem[{{Hulse} and {Taylor}(1975)}]{1975ApJ...195L..51H}
{Hulse} RA, {Taylor} JH (1975) {Discovery of a pulsar in a binary system.}
  \apjl 195:L51--L53. \doi{10.1086/181708}

\bibitem[{{Hurley} et~al.(2002){Hurley}, {Tout}, and
  {Pols}}]{2002MNRAS.329..897H}
{Hurley} JR, {Tout} CA, {Pols} OR (2002) {Evolution of binary stars and the
  effect of tides on binary populations}. \mnras 329(4):897--928.
  \doi{10.1046/j.1365-8711.2002.05038.x}.
  {\href{https://arxiv.org/abs/astro-ph/0201220}{{arXiv:astro-ph/0201220}}}
  {[astro-ph]}

\bibitem[{{Hut}(1981)}]{1981A&A....99..126H}
{Hut} P (1981) {Tidal evolution in close binary systems.} \aap 99:126--140

\bibitem[{{Hut} and {Bahcall}(1983)}]{1983ApJ...268..319H}
{Hut} P, {Bahcall} JN (1983) {Binary-single star scattering. I - Numerical
  experiments for equal masses}. \apj 268:319--341. \doi{10.1086/160956}

\bibitem[{{Hut} and {Paczynski}(1984)}]{1984ApJ...284..675H}
{Hut} P, {Paczynski} B (1984) {Effects of encoounters with field stars on the
  evolution of low-mass semidetached binaries.} \apj 284:675--684.
  \doi{10.1086/162450}

\bibitem[{{Hut} et~al.(1992){Hut}, {McMillan}, {Goodman}, {Mateo}, {Phinney},
  {Pryor}, {Richer}, {Verbunt}, and {Weinberg}}]{1992PASP..104..981H}
{Hut} P, {McMillan} S, {Goodman} J, {Mateo} M, {Phinney} ES, {Pryor} C,
  {Richer} HB, {Verbunt} F, {Weinberg} M (1992) {Binaries in Globular
  Clusters}. \pasp 104:981. \doi{10.1086/133085}

\bibitem[{{Hypki} and {Giersz}(2013)}]{2013MNRAS.429.1221H}
{Hypki} A, {Giersz} M (2013) {MOCCA code for star cluster simulations - I. Blue
  stragglers, first results}. \mnras 429(2):1221--1243.
  \doi{10.1093/mnras/sts415}.
  {\href{https://arxiv.org/abs/1207.6700}{{arXiv:1207.6700}}} {[astro-ph.GA]}

\bibitem[{{Iaconi} et~al.(2018){Iaconi}, {De Marco}, {Passy}, and
  {Staff}}]{2018MNRAS.477.2349I}
{Iaconi} R, {De Marco} O, {Passy} JC, {Staff} J (2018) {The effect of binding
  energy and resolution in simulations of the common envelope binary
  interaction}. \mnras 477(2):2349--2365. \doi{10.1093/mnras/sty794}.
  {\href{https://arxiv.org/abs/1706.09786}{{arXiv:1706.09786}}} {[astro-ph.SR]}

\bibitem[{{Iben} et~al.(1998){Iben}, {Tutukov}, and
  {Fedorova}}]{1998ApJ...503..344I}
{Iben} J Icko, {Tutukov} AV, {Fedorova} ArV (1998) {On the Luminosity of White
  Dwarfs in Close Binaries Merging under the Influence of Gravitational Wave
  Radiation}. \apj 503(1):344--349. \doi{10.1086/305972}

\bibitem[{{Inayoshi} et~al.(2016){Inayoshi}, {Haiman}, and
  {Ostriker}}]{2016MNRAS.459.3738I}
{Inayoshi} K, {Haiman} Z, {Ostriker} JP (2016) {Hyper-Eddington accretion flows
  on to massive black holes}. \mnras 459:3738--3755.
  \doi{10.1093/mnras/stw836}.
  {\href{https://arxiv.org/abs/1511.02116}{{arXiv:1511.02116}}} {[astro-ph.HE]}

\bibitem[{{Inayoshi} et~al.(2017{\natexlab{a}}){Inayoshi}, {Hirai}, {Kinugawa},
  and {Hotokezaka}}]{2017MNRAS.468.5020I}
{Inayoshi} K, {Hirai} R, {Kinugawa} T, {Hotokezaka} K (2017{\natexlab{a}})
  {Formation pathway of Population III coalescing binary black holes through
  stable mass transfer}. \mnras 468(4):5020--5032. \doi{10.1093/mnras/stx757}.
  {\href{https://arxiv.org/abs/1701.04823}{{arXiv:1701.04823}}} {[astro-ph.HE]}

\bibitem[{{Inayoshi} et~al.(2017{\natexlab{b}}){Inayoshi}, {Tamanini},
  {Caprini}, and {Haiman}}]{2017PhRvD..96f3014I}
{Inayoshi} K, {Tamanini} N, {Caprini} C, {Haiman} Z (2017{\natexlab{b}})
  {Probing stellar binary black hole formation in galactic nuclei via the
  imprint of their center of mass acceleration on their gravitational wave
  signal}. \prd 96(6):063014. \doi{10.1103/PhysRevD.96.063014}.
  {\href{https://arxiv.org/abs/1702.06529}{{arXiv:1702.06529}}} {[astro-ph.HE]}

\bibitem[{{Inayoshi} et~al.(2019){Inayoshi}, {Visbal}, and
  {Haiman}}]{2019arXiv191105791I}
{Inayoshi} K, {Visbal} E, {Haiman} Z (2019) {The Assembly of the First Massive
  Black Holes}. arXiv e-prints arXiv:1911.05791.
  {\href{https://arxiv.org/abs/1911.05791}{{arXiv:1911.05791}}} {[astro-ph.GA]}

\bibitem[{Inman and Ali-Ha\"\i{}moud(2019)}]{Inman:2019wvr}
Inman D, Ali-Ha\"\i{}moud Y (2019) {Early structure formation in primordial
  black hole cosmologies}. Phys Rev D 100(8):083528.
  \doi{10.1103/PhysRevD.100.083528}.
  {\href{https://arxiv.org/abs/1907.08129}{{arXiv:1907.08129}}} {[astro-ph.CO]}

\bibitem[{Inomata and Nakama(2019)}]{Inomata:2018epa}
Inomata K, Nakama T (2019) {Gravitational waves induced by scalar perturbations
  as probes of the small-scale primordial spectrum}. Phys Rev D 99(4):043511.
  \doi{10.1103/PhysRevD.99.043511}.
  {\href{https://arxiv.org/abs/1812.00674}{{arXiv:1812.00674}}} {[astro-ph.CO]}

\bibitem[{{Inomata} et~al.(2017){Inomata}, {Kawasaki}, {Mukaida}, {Tada}, and
  {Yanagida}}]{2017PhRvD..95l3510I}
{Inomata} K, {Kawasaki} M, {Mukaida} K, {Tada} Y, {Yanagida} TT (2017)
  {Inflationary primordial black holes for the LIGO gravitational wave events
  and pulsar timing array experiments}. \prd 95(12):123510.
  \doi{10.1103/PhysRevD.95.123510}.
  {\href{https://arxiv.org/abs/1611.06130}{{arXiv:1611.06130}}} {[astro-ph.CO]}

\bibitem[{{in't Zand} et~al.(2005){in't Zand}, {Cumming}, {van der Sluys},
  {Verbunt}, and {Pols}}]{2005A&A...441..675I}
{in't Zand} JJM, {Cumming} A, {van der Sluys} MV, {Verbunt} F, {Pols} OR (2005)
  {On the possibility of a helium white dwarf donor in the presumed
  ultracompact binary 2S 0918-549}. \aap 441(2):675--684.
  \doi{10.1051/0004-6361:20053002}.
  {\href{https://arxiv.org/abs/astro-ph/0506666}{{arXiv:astro-ph/0506666}}}
  {[astro-ph]}

\bibitem[{{Iorio}(2012)}]{2012GReGr..44..719I}
{Iorio} L (2012) {General relativistic spin-orbit and spin-spin effects on the
  motion of rotating particles in an external gravitational field}. General
  Relativity and Gravitation 44(3):719--736. \doi{10.1007/s10714-011-1302-7}.
  {\href{https://arxiv.org/abs/1012.5622}{{arXiv:1012.5622}}} {[gr-qc]}

\bibitem[{{Ishibashi} and {Fabian}(2016)}]{2016MNRAS.463.1291I}
{Ishibashi} W, {Fabian} AC (2016) {AGN-starburst evolutionary connection: a
  physical interpretation based on radiative feedback}. \mnras
  463(2):1291--1296. \doi{10.1093/mnras/stw2063}.
  {\href{https://arxiv.org/abs/1609.08963}{{arXiv:1609.08963}}} {[astro-ph.GA]}

\bibitem[{{Isoyama} et~al.(2018){Isoyama}, {Nakano}, and
  {Nakamura}}]{2018PTEP.2018g3E01I}
{Isoyama} S, {Nakano} H, {Nakamura} T (2018) {Multiband gravitational-wave
  astronomy: Observing binary inspirals with a decihertz detector, B-DECIGO}.
  PTEP 2018(7):073E01. \doi{10.1093/ptep/pty078}.
  {\href{https://arxiv.org/abs/1802.06977}{{arXiv:1802.06977}}} {[gr-qc]}

\bibitem[{{Istrate} et~al.(2014{\natexlab{a}}){Istrate}, {Tauris}, and
  {Langer}}]{2014A&A...571A..45I}
{Istrate} AG, {Tauris} TM, {Langer} N (2014{\natexlab{a}}) {The formation of
  low-mass helium white dwarfs orbiting pulsars . Evolution of low-mass X-ray
  binaries below the bifurcation period}. \aap 571:A45.
  \doi{10.1051/0004-6361/201424680}.
  {\href{https://arxiv.org/abs/1410.5470}{{arXiv:1410.5470}}} {[astro-ph.SR]}

\bibitem[{{Istrate} et~al.(2014{\natexlab{b}}){Istrate}, {Tauris}, {Langer},
  and {Antoniadis}}]{2014A&A...571L...3I}
{Istrate} AG, {Tauris} TM, {Langer} N, {Antoniadis} J (2014{\natexlab{b}}) {The
  timescale of low-mass proto-helium white dwarf evolution}. \aap 571:L3.
  \doi{10.1051/0004-6361/201424681}.
  {\href{https://arxiv.org/abs/1410.5471}{{arXiv:1410.5471}}} {[astro-ph.SR]}

\bibitem[{{Istrate} et~al.(2016){Istrate}, {Marchant}, {Tauris}, {Langer},
  {Stancliffe}, and {Grassitelli}}]{2016A&A...595A..35I}
{Istrate} AG, {Marchant} P, {Tauris} TM, {Langer} N, {Stancliffe} RJ,
  {Grassitelli} L (2016) {Models of low-mass helium white dwarfs including
  gravitational settling, thermal and chemical diffusion, and rotational
  mixing}. \aap 595:A35. \doi{10.1051/0004-6361/201628874}.
  {\href{https://arxiv.org/abs/1606.04947}{{arXiv:1606.04947}}} {[astro-ph.SR]}

\bibitem[{{Ivanov} et~al.(2015){Ivanov}, {Papaloizou}, {Paardekooper}, and
  {Polnarev}}]{2015A&A...576A..29I}
{Ivanov} PB, {Papaloizou} JCB, {Paardekooper} SJ, {Polnarev} AG (2015) {The
  evolution of a binary in a retrograde circular orbit embedded in an accretion
  disk}. \aap 576:A29. \doi{10.1051/0004-6361/201424359}.
  {\href{https://arxiv.org/abs/1410.3250}{{arXiv:1410.3250}}} {[astro-ph.HE]}

\bibitem[{{Ivanova} et~al.(2003){Ivanova}, {Belczynski}, {Kalogera}, {Rasio},
  and {Taam}}]{2003ApJ...592..475I}
{Ivanova} N, {Belczynski} K, {Kalogera} V, {Rasio} FA, {Taam} RE (2003) {The
  Role of Helium Stars in the Formation of Double Neutron Stars}. \apj
  592(1):475--485. \doi{10.1086/375578}.
  {\href{https://arxiv.org/abs/astro-ph/0210267}{{arXiv:astro-ph/0210267}}}
  {[astro-ph]}

\bibitem[{{Ivanova} et~al.(2006){Ivanova}, {Heinke}, {Rasio}, {Taam},
  {Belczynski}, and {Fregeau}}]{2006MNRAS.372.1043I}
{Ivanova} N, {Heinke} CO, {Rasio} FA, {Taam} RE, {Belczynski} K, {Fregeau} J
  (2006) {Formation and evolution of compact binaries in globular clusters - I.
  Binaries with white dwarfs}. \mnras 372(3):1043--1059.
  \doi{10.1111/j.1365-2966.2006.10876.x}.
  {\href{https://arxiv.org/abs/astro-ph/0604085}{{arXiv:astro-ph/0604085}}}
  {[astro-ph]}

\bibitem[{{Ivanova} et~al.(2013){Ivanova}, {Justham}, {Chen}, {De Marco},
  {Fryer}, {Gaburov}, {Ge}, {Glebbeek}, {Han}, {Li}, {Lu}, {Marsh},
  {Podsiadlowski}, {Potter}, {Soker}, {Taam}, {Tauris}, {van den Heuvel}, and
  {Webbink}}]{2013A&ARv..21...59I}
{Ivanova} N, {Justham} S, {Chen} X, {De Marco} O, {Fryer} CL, {Gaburov} E, {Ge}
  H, {Glebbeek} E, {Han} Z, {Li} XD, et~al. (2013) {Common envelope evolution:
  where we stand and how we can move forward}. \aapr 21:59.
  \doi{10.1007/s00159-013-0059-2}.
  {\href{https://arxiv.org/abs/1209.4302}{{arXiv:1209.4302}}} {[astro-ph.HE]}

\bibitem[{{Ivanova} et~al.(2020){Ivanova}, {Justham}, and
  {Ricker}}]{2020cee..book.....I}
{Ivanova} N, {Justham} S, {Ricker} P (2020) {Common Envelope Evolution}.
  \doi{10.1088/2514-3433/abb6f0}

\bibitem[{{Ivezi{\'c}} et~al.(2019){Ivezi{\'c}}, {Kahn}, {Tyson}, {Abel},
  {Acosta}, {Allsman}, {Alonso}, {AlSayyad}, {Anderson}, {Andrew}, {Angel},
  {Angeli}, {Ansari}, {Antilogus}, {Araujo}, {Armstrong}, {Arndt}, {Astier},
  {Aubourg}, {Auza}, {Axelrod}, {Bard}, {Barr}, {Barrau}, {Bartlett}, {Bauer},
  {Bauman}, {Baumont}, {Bechtol}, {Bechtol}, {Becker}, {Becla}, {Beldica},
  {Bellavia}, {Bianco}, {Biswas}, {Blanc}, {Blazek}, {Blandford}, {Bloom},
  {Bogart}, {Bond}, {Booth}, {Borgland}, {Borne}, {Bosch}, {Boutigny},
  {Brackett}, {Bradshaw}, {Brandt}, {Brown}, {Bullock}, {Burchat}, {Burke},
  {Cagnoli}, {Calabrese}, {Callahan}, {Callen}, {Carlin}, {Carlson},
  {Chandrasekharan}, {Charles-Emerson}, {Chesley}, {Cheu}, {Chiang}, {Chiang},
  {Chirino}, {Chow}, {Ciardi}, {Claver}, {Cohen-Tanugi}, {Cockrum}, {Coles},
  {Connolly}, {Cook}, {Cooray}, {Covey}, {Cribbs}, {Cui}, {Cutri}, {Daly},
  {Daniel}, {Daruich}, {Daubard}, {Daues}, {Dawson}, {Delgado}, {Dellapenna},
  {de Peyster}, {de Val-Borro}, {Digel}, {Doherty}, {Dubois},
  {Dubois-Felsmann}, {Durech}, {Economou}, {Eifler}, {Eracleous}, {Emmons},
  {Fausti Neto}, {Ferguson}, {Figueroa}, {Fisher-Levine}, {Focke}, {Foss},
  {Frank}, {Freemon}, {Gangler}, {Gawiser}, {Geary}, {Gee}, {Geha}, {Gessner},
  {Gibson}, {Gilmore}, {Glanzman}, {Glick}, {Goldina}, {Goldstein}, {Goodenow},
  {Graham}, {Gressler}, {Gris}, {Guy}, {Guyonnet}, {Haller}, {Harris},
  {Hascall}, {Haupt}, {Hernandez}, {Herrmann}, {Hileman}, {Hoblitt}, {Hodgson},
  {Hogan}, {Howard}, {Huang}, {Huffer}, {Ingraham}, {Innes}, {Jacoby}, {Jain},
  {Jammes}, {Jee}, {Jenness}, {Jernigan}, {Jevremovi{\'c}}, {Johns}, {Johnson},
  {Johnson}, {Jones}, {Juramy-Gilles}, {Juri{\'c}}, {Kalirai}, {Kallivayalil},
  {Kalmbach}, {Kantor}, {Karst}, {Kasliwal}, {Kelly}, {Kessler}, {Kinnison},
  {Kirkby}, {Knox}, {Kotov}, {Krabbendam}, {Krughoff}, {Kub{\'a}nek},
  {Kuczewski}, {Kulkarni}, {Ku}, {Kurita}, {Lage}, {Lambert}, {Lange},
  {Langton}, {Le Guillou}, {Levine}, {Liang}, {Lim}, {Lintott}, {Long},
  {Lopez}, {Lotz}, {Lupton}, {Lust}, {MacArthur}, {Mahabal}, {Mandelbaum},
  {Markiewicz}, {Marsh}, {Marshall}, {Marshall}, {May}, {McKercher}, {McQueen},
  {Meyers}, {Migliore}, {Miller}, {Mills}, {Miraval}, {Moeyens}, {Moolekamp},
  {Monet}, {Moniez}, {Monkewitz}, {Montgomery}, {Morrison}, {Mueller},
  {Muller}, {Mu{\~n}oz Arancibia}, {Neill}, {Newbry}, {Nief}, {Nomerotski},
  {Nordby}, {O'Connor}, {Oliver}, {Olivier}, {Olsen}, {O'Mullane}, {Ortiz},
  {Osier}, {Owen}, {Pain}, {Palecek}, {Parejko}, {Parsons}, {Pease},
  {Peterson}, {Peterson}, {Petravick}, {Libby Petrick}, {Petry},
  {Pierfederici}, {Pietrowicz}, {Pike}, {Pinto}, {Plante}, {Plate}, {Plutchak},
  {Price}, {Prouza}, {Radeka}, {Rajagopal}, {Rasmussen}, {Regnault}, {Reil},
  {Reiss}, {Reuter}, {Ridgway}, {Riot}, {Ritz}, {Robinson}, {Roby}, {Roodman},
  {Rosing}, {Roucelle}, {Rumore}, {Russo}, {Saha}, {Sassolas}, {Schalk},
  {Schellart}, {Schindler}, {Schmidt}, {Schneider}, {Schneider}, {Schoening},
  {Schumacher}, {Schwamb}, {Sebag}, {Selvy}, {Sembroski}, {Seppala}, {Serio},
  {Serrano}, {Shaw}, {Shipsey}, {Sick}, {Silvestri}, {Slater}, {Smith},
  {Smith}, {Sobhani}, {Soldahl}, {Storrie-Lombardi}, {Stover}, {Strauss},
  {Street}, {Stubbs}, {Sullivan}, {Sweeney}, {Swinbank}, {Szalay}, {Takacs},
  {Tether}, {Thaler}, {Thayer}, {Thomas}, {Thornton}, {Thukral}, {Tice},
  {Trilling}, {Turri}, {Van Berg}, {Vanden Berk}, {Vetter}, {Virieux},
  {Vucina}, {Wahl}, {Walkowicz}, {Walsh}, {Walter}, {Wang}, {Wang}, {Warner},
  {Wiecha}, {Willman}, {Winters}, {Wittman}, {Wolff}, {Wood-Vasey}, {Wu},
  {Xin}, {Yoachim}, and {Zhan}}]{2019ApJ...873..111I}
{Ivezi{\'c}} {\v{Z}}, {Kahn} SM, {Tyson} JA, {Abel} B, {Acosta} E, {Allsman} R,
  {Alonso} D, {AlSayyad} Y, {Anderson} SF, {Andrew} J, et~al. (2019) {LSST:
  From Science Drivers to Reference Design and Anticipated Data Products}. \apj
  873(2):111. \doi{10.3847/1538-4357/ab042c}.
  {\href{https://arxiv.org/abs/0805.2366}{{arXiv:0805.2366}}} {[astro-ph]}

\bibitem[{{Izquierdo-Villalba} et~al.(2020){Izquierdo-Villalba}, {Bonoli},
  {Dotti}, {Sesana}, {Rosas-Guevara}, and {Spinoso}}]{2020MNRAS.495.4681I}
{Izquierdo-Villalba} D, {Bonoli} S, {Dotti} M, {Sesana} A, {Rosas-Guevara} Y,
  {Spinoso} D (2020) {From galactic nuclei to the halo outskirts: tracing
  supermassive black holes across cosmic history and environments}. \mnras
  495(4):4681--4706. \doi{10.1093/mnras/staa1399}.
  {\href{https://arxiv.org/abs/2001.10548}{{arXiv:2001.10548}}} {[astro-ph.GA]}

\bibitem[{{Izzard}(2004)}]{2004PhDT........45I}
{Izzard} RG (2004) {Nucleosynthesis In Binary Stars}. PhD thesis, University of
  Cambridge

\bibitem[{{Izzard} et~al.(2006){Izzard}, {Dray}, {Karakas}, {Lugaro}, and
  {Tout}}]{2006AA...460..565I}
{Izzard} RG, {Dray} LM, {Karakas} AI, {Lugaro} M, {Tout} CA (2006) {Population
  nucleosynthesis in single and binary stars. I. Model}. \aap 460(2):565--572.
  \doi{10.1051/0004-6361:20066129}

\bibitem[{{Izzard} et~al.(2009){Izzard}, {Glebbeek}, {Stancliffe}, and
  {Pols}}]{2009AA...508.1359I}
{Izzard} RG, {Glebbeek} E, {Stancliffe} RJ, {Pols} OR (2009) {Population
  synthesis of binary carbon-enhanced metal-poor stars}. \aap
  508(3):1359--1374. \doi{10.1051/0004-6361/200912827}.
  {\href{https://arxiv.org/abs/0910.2158}{{arXiv:0910.2158}}} {[astro-ph.SR]}

\bibitem[{{Jani} et~al.(2019){Jani}, {Shoemaker}, and
  {Cutler}}]{2020NatAs...4..260J}
{Jani} K, {Shoemaker} D, {Cutler} C (2019) {Detectability of intermediate-mass
  black holes in multiband gravitational wave astronomy}. Nature Astronomy
  4:260--265. \doi{10.1038/s41550-019-0932-7}.
  {\href{https://arxiv.org/abs/1908.04985}{{arXiv:1908.04985}}} {[gr-qc]}

\bibitem[{{Janka}(2012)}]{2012ARNPS..62..407J}
{Janka} HT (2012) {Explosion Mechanisms of Core-Collapse Supernovae}. Annual
  Review of Nuclear and Particle Science 62(1):407--451.
  \doi{10.1146/annurev-nucl-102711-094901}.
  {\href{https://arxiv.org/abs/1206.2503}{{arXiv:1206.2503}}} {[astro-ph.SR]}

\bibitem[{{Janka}(2017)}]{2017ApJ...837...84J}
{Janka} HT (2017) {Neutron Star Kicks by the Gravitational Tug-boat Mechanism
  in Asymmetric Supernova Explosions: Progenitor and Explosion Dependence}.
  \apj 837(1):84. \doi{10.3847/1538-4357/aa618e}.
  {\href{https://arxiv.org/abs/1611.07562}{{arXiv:1611.07562}}} {[astro-ph.HE]}

\bibitem[{{Jha} et~al.(2019){Jha}, {Maguire}, and
  {Sullivan}}]{2019NatAs...3..706J}
{Jha} SW, {Maguire} K, {Sullivan} M (2019) {Observational properties of
  thermonuclear supernovae}. Nature Astronomy 3:706--716.
  \doi{10.1038/s41550-019-0858-0}.
  {\href{https://arxiv.org/abs/1908.02303}{{arXiv:1908.02303}}} {[astro-ph.HE]}

\bibitem[{{Jia} and {Li}(2016)}]{2016ApJ...830..153J}
{Jia} K, {Li} XD (2016) {Evolution of Low-mass X-Ray Binaries: The Effect of
  Donor Evaporation}. \apj 830(2):153. \doi{10.3847/0004-637X/830/2/153}.
  {\href{https://arxiv.org/abs/1608.01076}{{arXiv:1608.01076}}} {[astro-ph.HE]}

\bibitem[{{Jiang} and {Blaes}(2020)}]{2020ApJ...900...25J}
{Jiang} YF, {Blaes} O (2020) {Opacity-driven Convection and Variability in
  Accretion Disks around Supermassive Black Holes}. \apj 900(1):25.
  \doi{10.3847/1538-4357/aba4b7}.
  {\href{https://arxiv.org/abs/2006.08657}{{arXiv:2006.08657}}} {[astro-ph.HE]}

\bibitem[{{Jiang} et~al.(2014){Jiang}, {Stone}, and
  {Davis}}]{2014ApJ...796..106J}
{Jiang} YF, {Stone} JM, {Davis} SW (2014) {A Global Three-dimensional Radiation
  Magneto-hydrodynamic Simulation of Super-Eddington Accretion Disks}. \apj
  796(2):106. \doi{10.1088/0004-637X/796/2/106}.
  {\href{https://arxiv.org/abs/1410.0678}{{arXiv:1410.0678}}} {[astro-ph.HE]}

\bibitem[{{Jiang} et~al.(2016){Jiang}, {Davis}, and
  {Stone}}]{2016ApJ...827...10J}
{Jiang} YF, {Davis} SW, {Stone} JM (2016) {Iron Opacity Bump Changes the
  Stability and Structure of Accretion Disks in Active Galactic Nuclei}. \apj
  827(1):10. \doi{10.3847/0004-637X/827/1/10}.
  {\href{https://arxiv.org/abs/1601.06836}{{arXiv:1601.06836}}} {[astro-ph.HE]}

\bibitem[{{Jiang} et~al.(2019){Jiang}, {Blaes}, {Stone}, and
  {Davis}}]{2019ApJ...885..144J}
{Jiang} YF, {Blaes} O, {Stone} JM, {Davis} SW (2019) {Global Radiation
  Magnetohydrodynamic Simulations of sub-Eddington Accretion Disks around
  Supermassive Black Holes}. \apj 885(2):144. \doi{10.3847/1538-4357/ab4a00}.
  {\href{https://arxiv.org/abs/1904.01674}{{arXiv:1904.01674}}} {[astro-ph.HE]}

\bibitem[{{Johannsen}(2013)}]{2013PhRvD..88d4002J}
{Johannsen} T (2013) {Regular black hole metric with three constants of
  motion}. \prd 88(4):044002. \doi{10.1103/PhysRevD.88.044002}.
  {\href{https://arxiv.org/abs/1501.02809}{{arXiv:1501.02809}}} {[gr-qc]}

\bibitem[{{Johnson} and {Haardt}(2016)}]{2016PASA...33....7J}
{Johnson} JL, {Haardt} F (2016) {The Early Growth of the First Black Holes}.
  \pasa 33:e007. \doi{10.1017/pasa.2016.4}.
  {\href{https://arxiv.org/abs/1601.05473}{{arXiv:1601.05473}}} {[astro-ph.GA]}

\bibitem[{{Johnson} et~al.(2019){Johnson}, {Haworth}, {Pesce}, {Palumbo},
  {Blackburn}, {Akiyama}, {Boroson}, {Bouman}, {Farah}, {Fish}, {Honma},
  {Kawashima}, {Kino}, {Raymond}, {Silver}, {Weintroub}, {Wielgus}, {Doeleman},
  {Kauffmann}, {Keating}, {Krichbaum}, {Loinard}, {Narayanan}, {Doi}, {James},
  {Marrone}, {Mizuno}, and {Nagai}}]{2019BAAS...51g.235J}
{Johnson} M, {Haworth} K, {Pesce} DW, {Palumbo} DCM, {Blackburn} L, {Akiyama}
  K, {Boroson} D, {Bouman} KL, {Farah} JR, {Fish} VL, et~al. (2019) {Studying
  black holes on horizon scales with space-VLBI}. In: Bulletin of the American
  Astronomical Society. vol~51. p 235.
  {\href{https://arxiv.org/abs/1909.01405}{{arXiv:1909.01405}}} {[astro-ph.IM]}

\bibitem[{{Jones} et~al.(2013){Jones}, {Hirschi}, {Nomoto}, {Fischer},
  {Timmes}, {Herwig}, {Paxton}, {Toki}, {Suzuki}, {Mart{\'\i}nez-Pinedo},
  {Lam}, and {Bertolli}}]{2013ApJ...772..150J}
{Jones} S, {Hirschi} R, {Nomoto} K, {Fischer} T, {Timmes} FX, {Herwig} F,
  {Paxton} B, {Toki} H, {Suzuki} T, {Mart{\'\i}nez-Pinedo} G, et~al. (2013)
  {Advanced Burning Stages and Fate of 8-10 M $_{\odot}$ Stars}. \apj
  772(2):150. \doi{10.1088/0004-637X/772/2/150}.
  {\href{https://arxiv.org/abs/1306.2030}{{arXiv:1306.2030}}} {[astro-ph.SR]}

\bibitem[{{Ju} et~al.(2013){Ju}, {Greene}, {Rafikov}, {Bickerton}, and
  {Badenes}}]{2013ApJ...777...44J}
{Ju} W, {Greene} JE, {Rafikov} RR, {Bickerton} SJ, {Badenes} C (2013) {Search
  for Supermassive Black Hole Binaries in the Sloan Digital Sky Survey
  Spectroscopic Sample}. \apj 777(1):44. \doi{10.1088/0004-637X/777/1/44}.
  {\href{https://arxiv.org/abs/1306.4987}{{arXiv:1306.4987}}} {[astro-ph.CO]}

\bibitem[{{Juett} et~al.(2001){Juett}, {Psaltis}, and
  {Chakrabarty}}]{2001ApJ...560L..59J}
{Juett} AM, {Psaltis} D, {Chakrabarty} D (2001) {Ultracompact X-Ray Binaries
  with Neon-rich Degenerate Donors}. \apjl 560(1):L59--L63.
  \doi{10.1086/324225}.
  {\href{https://arxiv.org/abs/astro-ph/0108102}{{arXiv:astro-ph/0108102}}}
  {[astro-ph]}

\bibitem[{{Just} et~al.(2012){Just}, {Yurin}, {Makukov}, {Berczik}, {Omarov},
  {Spurzem}, and {Vilkoviskij}}]{2012ApJ...758...51J}
{Just} A, {Yurin} D, {Makukov} M, {Berczik} P, {Omarov} C, {Spurzem} R,
  {Vilkoviskij} EY (2012) {Enhanced Accretion Rates of Stars on Supermassive
  Black Holes by Star-Disk Interactions in Galactic Nuclei}. \apj 758(1):51.
  \doi{10.1088/0004-637X/758/1/51}.
  {\href{https://arxiv.org/abs/1208.4954}{{arXiv:1208.4954}}} {[astro-ph.CO]}

\bibitem[{{Kahabka} and {van den Heuvel}(1997)}]{1997ARA&A..35...69K}
{Kahabka} P, {van den Heuvel} EPJ (1997) {Luminous Supersoft X-Ray Sources}.
  \araa 35:69--100. \doi{10.1146/annurev.astro.35.1.69}

\bibitem[{Kalaja et~al.(2019)Kalaja, Bellomo, Bartolo, Bertacca, Matarrese,
  Musco, Raccanelli, and Verde}]{Kalaja:2019uju}
Kalaja A, Bellomo N, Bartolo N, Bertacca D, Matarrese S, Musco I, Raccanelli A,
  Verde L (2019) {From Primordial Black Holes Abundance to Primordial Curvature
  Power Spectrum (and back)}. JCAP 10:031. \doi{10.1088/1475-7516/2019/10/031}.
  {\href{https://arxiv.org/abs/1908.03596}{{arXiv:1908.03596}}} {[astro-ph.CO]}

\bibitem[{{Kalfountzou} et~al.(2017){Kalfountzou}, {Santos Lleo}, and
  {Trichas}}]{2017ApJ...851L..15K}
{Kalfountzou} E, {Santos Lleo} M, {Trichas} M (2017) {SDSS J1056+5516: A Triple
  AGN or an SMBH Recoil Candidate?} \apjl 851(1):L15.
  \doi{10.3847/2041-8213/aa9b2d}.
  {\href{https://arxiv.org/abs/1712.03909}{{arXiv:1712.03909}}} {[astro-ph.GA]}

\bibitem[{{Kalogera} et~al.(2007){Kalogera}, {Belczynski}, {Kim},
  {O'Shaughnessy}, and {Willems}}]{2007PhR...442...75K}
{Kalogera} V, {Belczynski} K, {Kim} C, {O'Shaughnessy} R, {Willems} B (2007)
  {Formation of double compact objects}. \physrep 442(1-6):75--108.
  \doi{10.1016/j.physrep.2007.02.008}.
  {\href{https://arxiv.org/abs/astro-ph/0612144}{{arXiv:astro-ph/0612144}}}
  {[astro-ph]}

\bibitem[{{Karnesis} et~al.(2019){Karnesis}, {Lilley}, and
  {Petiteau}}]{2019arXiv190609027K}
{Karnesis} N, {Lilley} M, {Petiteau} A (2019) {Assessing the detectability of a
  Stochastic Gravitational Wave Background with LISA, using an excess of power
  approach}. arXiv e-prints arXiv:1906.09027.
  {\href{https://arxiv.org/abs/1906.09027}{{arXiv:1906.09027}}} {[astro-ph.IM]}

\bibitem[{{Karnesis} et~al.(2021){Karnesis}, {Babak}, {Pieroni}, {Cornish}, and
  {Littenberg}}]{2021PhRvD.104d3019K}
{Karnesis} N, {Babak} S, {Pieroni} M, {Cornish} N, {Littenberg} T (2021)
  {Characterization of the stochastic signal originating from compact binary
  populations as measured by LISA}. \prd 104(4):043019.
  \doi{10.1103/PhysRevD.104.043019}.
  {\href{https://arxiv.org/abs/2103.14598}{{arXiv:2103.14598}}} {[astro-ph.IM]}

\bibitem[{{Kato} and {Hachisu}(1999)}]{1999ApJ...513L..41K}
{Kato} M, {Hachisu} I (1999) {A New Estimation of Mass Accumulation Efficiency
  in Helium Shell Flashestoward Type IA Supernova Explosions}. \apjl
  513(1):L41--L44. \doi{10.1086/311893}.
  {\href{https://arxiv.org/abs/astro-ph/9901080}{{arXiv:astro-ph/9901080}}}
  {[astro-ph]}

\bibitem[{{Katz} et~al.(2015){Katz}, {Sijacki}, and
  {Haehnelt}}]{2015MNRAS.451.2352K}
{Katz} H, {Sijacki} D, {Haehnelt} MG (2015) {Seeding high-redshift QSOs by
  collisional runaway in primordial star clusters}. \mnras 451(3):2352--2369.
  \doi{10.1093/mnras/stv1048}.
  {\href{https://arxiv.org/abs/1502.03448}{{arXiv:1502.03448}}} {[astro-ph.GA]}

\bibitem[{{Katz} et~al.(2020){Katz}, {Kelley}, {Dosopoulou}, {Berry}, {Blecha},
  and {Larson}}]{2020MNRAS.491.2301K}
{Katz} ML, {Kelley} LZ, {Dosopoulou} F, {Berry} S, {Blecha} L, {Larson} SL
  (2020) {Probing massive black hole binary populations with LISA}. \mnras
  491(2):2301--2317. \doi{10.1093/mnras/stz3102}.
  {\href{https://arxiv.org/abs/1908.05779}{{arXiv:1908.05779}}} {[astro-ph.HE]}

\bibitem[{{Kauffmann} and {Haehnelt}(2000)}]{2000MNRAS.311..576K}
{Kauffmann} G, {Haehnelt} M (2000) {A unified model for the evolution of
  galaxies and quasars}. \mnras 311(3):576--588.
  \doi{10.1046/j.1365-8711.2000.03077.x}.
  {\href{https://arxiv.org/abs/astro-ph/9906493}{{arXiv:astro-ph/9906493}}}
  {[astro-ph]}

\bibitem[{{Kauffmann} et~al.(1993){Kauffmann}, {White}, and
  {Guiderdoni}}]{1993MNRAS.264..201K}
{Kauffmann} G, {White} SDM, {Guiderdoni} B (1993) {The formation and evolution
  of galaxies within merging dark matter haloes.} \mnras 264:201--218.
  \doi{10.1093/mnras/264.1.201}

\bibitem[{{Kavanagh} et~al.(2020){Kavanagh}, {Nichols}, {Bertone}, and
  {Gaggero}}]{2020PhRvD.102h3006K}
{Kavanagh} BJ, {Nichols} DA, {Bertone} G, {Gaggero} D (2020) {Detecting dark
  matter around black holes with gravitational waves: Effects of dark-matter
  dynamics on the gravitational waveform}. \prd 102(8):083006.
  \doi{10.1103/PhysRevD.102.083006}.
  {\href{https://arxiv.org/abs/2002.12811}{{arXiv:2002.12811}}} {[gr-qc]}

\bibitem[{{Kawamura} et~al.(2011){Kawamura}, {Ando}, {Seto}, {Sato},
  {Nakamura}, {Tsubono}, {Kand a}, {Tanaka}, {Yokoyama}, {Funaki}, and
  et~al.}]{2011CQGra..28i4011K}
{Kawamura} S, {Ando} M, {Seto} N, {Sato} S, {Nakamura} T, {Tsubono} K, {Kand a}
  N, {Tanaka} T, {Yokoyama} J, {Funaki} I, et~al. (2011) {The Japanese space
  gravitational wave antenna: DECIGO}. CQGra 28(9):094011.
  \doi{10.1088/0264-9381/28/9/094011}

\bibitem[{{Kawamura} et~al.(2020){Kawamura}, {Ando}, {Seto}, {Sato}, {Musha},
  {Kawano}, {Yokoyama}, {Tanaka}, {Ioka}, {Akutsu}, and {et
  al.}}]{2020arXiv200613545K}
{Kawamura} S, {Ando} M, {Seto} N, {Sato} S, {Musha} M, {Kawano} I, {Yokoyama}
  J, {Tanaka} T, {Ioka} K, {Akutsu} T, et~al. (2020) {Current status of space
  gravitational wave antenna DECIGO and B-DECIGO}. arXiv e-prints
  arXiv:2006.13545.
  {\href{https://arxiv.org/abs/2006.13545}{{arXiv:2006.13545}}} {[gr-qc]}

\bibitem[{{Kawanaka} et~al.(2017){Kawanaka}, {Yamaguchi}, {Piran}, and
  {Bulik}}]{2017IAUS..324...41K}
{Kawanaka} N, {Yamaguchi} M, {Piran} T, {Bulik} T (2017) {Prospects for the
  Discovery of Black Hole Binaries without Mass Accretion with Gaia}. In:
  {Gomboc} A (ed) New Frontiers in Black Hole Astrophysics. IAU Symposium, vol
  324. pp 41--42. \doi{10.1017/S1743921316012606}

\bibitem[{{Kawka} et~al.(2020){Kawka}, {Vennes}, and
  {Ferrario}}]{2020MNRAS.491L..40K}
{Kawka} A, {Vennes} S, {Ferrario} L (2020) {An ancient double degenerate merger
  in the Milky Way halo}. \mnras 491(1):L40--L45. \doi{10.1093/mnrasl/slz165}.
  {\href{https://arxiv.org/abs/1910.13053}{{arXiv:1910.13053}}} {[astro-ph.SR]}

\bibitem[{{Keane} et~al.(2015){Keane}, {Bhattacharyya}, {Kramer}, {Stappers},
  {Keane}, {Bhattacharyya}, {Kramer}, {Stappers}, and {et~
  al.}}]{2015aska.confE..40K}
{Keane} E, {Bhattacharyya} B, {Kramer} M, {Stappers} B, {Keane} EF,
  {Bhattacharyya} B, {Kramer} M, {Stappers} BW, {et~ al} (2015) {A Cosmic
  Census of Radio Pulsars with the SKA}. In: Advancing Astrophysics with the
  Square Kilometre Array (AASKA14). p~40.
  {\href{https://arxiv.org/abs/1501.00056}{{arXiv:1501.00056}}} {[astro-ph.IM]}

\bibitem[{{Kelley} et~al.(2019){Kelley}, {Charisi}, {Burke-Spolaor}, {Simon},
  {Blecha}, {Bogdanovic}, {Colpi}, {Comerford}, {D'Orazio}, {Dotti},
  {Eracleous}, {Graham}, {Greene}, {Haiman}, {Holley-Bockelmann}, {Kara},
  {Kelly}, {Komossa}, {Larson}, {Liu}, {Ma}, {Noble}, {Paschalidis}, {Rafikov},
  {Ravi}, {Runnoe}, {Sesana}, {Stern}, {Strauss}, {U}, {Volonteri}, and
  {Nanograv Collaboration}}]{2019BAAS...51c.490K}
{Kelley} L, {Charisi} M, {Burke-Spolaor} S, {Simon} J, {Blecha} L, {Bogdanovic}
  T, {Colpi} M, {Comerford} J, {D'Orazio} D, {Dotti} M, et~al. (2019)
  {Multi-Messenger Astrophysics With Pulsar Timing Arrays}. \baas 51(3):490.
  {\href{https://arxiv.org/abs/1903.07644}{{arXiv:1903.07644}}} {[astro-ph.HE]}

\bibitem[{{Kelley} et~al.(2017{\natexlab{a}}){Kelley}, {Blecha}, and
  {Hernquist}}]{2017MNRAS.464.3131K}
{Kelley} LZ, {Blecha} L, {Hernquist} L (2017{\natexlab{a}}) {Massive black hole
  binary mergers in dynamical galactic environments}. \mnras 464(3):3131--3157.
  \doi{10.1093/mnras/stw2452}.
  {\href{https://arxiv.org/abs/1606.01900}{{arXiv:1606.01900}}} {[astro-ph.HE]}

\bibitem[{{Kelley} et~al.(2017{\natexlab{b}}){Kelley}, {Blecha}, {Hernquist},
  {Sesana}, and {Taylor}}]{2017MNRAS.471.4508K}
{Kelley} LZ, {Blecha} L, {Hernquist} L, {Sesana} A, {Taylor} SR
  (2017{\natexlab{b}}) {The gravitational wave background from massive black
  hole binaries in Illustris: spectral features and time to detection with
  pulsar timing arrays}. \mnras 471(4):4508--4526. \doi{10.1093/mnras/stx1638}.
  {\href{https://arxiv.org/abs/1702.02180}{{arXiv:1702.02180}}} {[astro-ph.HE]}

\bibitem[{{Kelley} et~al.(2018){Kelley}, {Blecha}, {Hernquist}, {Sesana}, and
  {Taylor}}]{2018MNRAS.477..964K}
{Kelley} LZ, {Blecha} L, {Hernquist} L, {Sesana} A, {Taylor} SR (2018) {Single
  sources in the low-frequency gravitational wave sky: properties and time to
  detection by pulsar timing arrays}. \mnras 477(1):964--976.
  \doi{10.1093/mnras/sty689}.
  {\href{https://arxiv.org/abs/1711.00075}{{arXiv:1711.00075}}} {[astro-ph.HE]}

\bibitem[{{Kelly} et~al.(2017){Kelly}, {Baker}, {Etienne}, {Giacomazzo}, and
  {Schnittman}}]{2017PhRvD..96l3003K}
{Kelly} BJ, {Baker} JG, {Etienne} ZB, {Giacomazzo} B, {Schnittman} J (2017)
  {Prompt electromagnetic transients from binary black hole mergers}. \prd
  96(12):123003. \doi{10.1103/PhysRevD.96.123003}.
  {\href{https://arxiv.org/abs/1710.02132}{{arXiv:1710.02132}}} {[astro-ph.HE]}

\bibitem[{{Kennedy} et~al.(2016){Kennedy}, {Meiron}, {Shukirgaliyev},
  {Panamarev}, {Berczik}, {Just}, and {Spurzem}}]{Kennedy16}
{Kennedy} GF, {Meiron} Y, {Shukirgaliyev} B, {Panamarev} T, {Berczik} P, {Just}
  A, {Spurzem} R (2016) {Star-disc interaction in galactic nuclei: orbits and
  rates of accreted stars}. \mnras 460(1):240--255. \doi{10.1093/mnras/stw908}.
  {\href{https://arxiv.org/abs/1604.05309}{{arXiv:1604.05309}}} {[astro-ph.GA]}

\bibitem[{{Kerr} et~al.(2020){Kerr}, {Reardon}, {Hobbs}, {Shannon},
  {Manchester}, {Dai}, {Russell}, {Zhang}, {van Straten}, {Os{\l}owski},
  {Parthasarathy}, {Spiewak}, {Bailes}, {Bhat}, {Cameron}, {Coles}, {Dempsey},
  {Deng}, {Goncharov}, {Kaczmarek}, {Keith}, {Lasky}, {Lower}, {Preisig},
  {Sarkissian}, {Toomey}, {Wang}, {Wang}, {Zhang}, and
  {Zhu}}]{2020PASA...37...20K}
{Kerr} M, {Reardon} DJ, {Hobbs} G, {Shannon} RM, {Manchester} RN, {Dai} S,
  {Russell} CJ, {Zhang} S, {van Straten} W, {Os{\l}owski} S, et~al. (2020) {The
  Parkes Pulsar Timing Array project: second data release}. \pasa 37:e020.
  \doi{10.1017/pasa.2020.11}.
  {\href{https://arxiv.org/abs/2003.09780}{{arXiv:2003.09780}}} {[astro-ph.IM]}

\bibitem[{{Kesden} et~al.(2010{\natexlab{a}}){Kesden}, {Sperhake}, and
  {Berti}}]{2010PhRvD..81h4054K}
{Kesden} M, {Sperhake} U, {Berti} E (2010{\natexlab{a}}) {Final spins from the
  merger of precessing binary black holes}. \prd 81(8):084054.
  \doi{10.1103/PhysRevD.81.084054}.
  {\href{https://arxiv.org/abs/1002.2643}{{arXiv:1002.2643}}} {[astro-ph.GA]}

\bibitem[{{Kesden} et~al.(2010{\natexlab{b}}){Kesden}, {Sperhake}, and
  {Berti}}]{2010ApJ...715.1006K}
{Kesden} M, {Sperhake} U, {Berti} E (2010{\natexlab{b}}) {Relativistic
  Suppression of Black Hole Recoils}. \apj 715(2):1006--1011.
  \doi{10.1088/0004-637X/715/2/1006}.
  {\href{https://arxiv.org/abs/1003.4993}{{arXiv:1003.4993}}} {[astro-ph.CO]}

\bibitem[{{Kesden} et~al.(2015){Kesden}, {Gerosa}, {O'Shaughnessy}, {Berti},
  and {Sperhake}}]{2015PhRvL.114h1103K}
{Kesden} M, {Gerosa} D, {O'Shaughnessy} R, {Berti} E, {Sperhake} U (2015)
  {Effective Potentials and Morphological Transitions for Binary Black Hole
  Spin Precession}. \prl 114(8):081103. \doi{10.1103/PhysRevLett.114.081103}.
  {\href{https://arxiv.org/abs/1411.0674}{{arXiv:1411.0674}}} {[gr-qc]}

\bibitem[{{Khan} et~al.(2018{\natexlab{a}}){Khan}, {Paschalidis}, {Ruiz}, and
  {Shapiro}}]{2018PhRvD..97d4036K}
{Khan} A, {Paschalidis} V, {Ruiz} M, {Shapiro} SL (2018{\natexlab{a}}) {Disks
  around merging binary black holes: From GW150914 to supermassive black
  holes}. \prd 97(4):044036. \doi{10.1103/PhysRevD.97.044036}.
  {\href{https://arxiv.org/abs/1801.02624}{{arXiv:1801.02624}}} {[astro-ph.HE]}

\bibitem[{{Khan} et~al.(2011){Khan}, {Just}, and
  {Merritt}}]{2011ApJ...732...89K}
{Khan} FM, {Just} A, {Merritt} D (2011) {Efficient Merger of Binary
  Supermassive Black Holes in Merging Galaxies}. \apj 732(2):89.
  \doi{10.1088/0004-637X/732/2/89}.
  {\href{https://arxiv.org/abs/1103.0272}{{arXiv:1103.0272}}} {[astro-ph.CO]}

\bibitem[{{Khan} et~al.(2012){Khan}, {Preto}, {Berczik}, {Berentzen}, {Just},
  and {Spurzem}}]{2012ApJ...749..147K}
{Khan} FM, {Preto} M, {Berczik} P, {Berentzen} I, {Just} A, {Spurzem} R (2012)
  {Mergers of Unequal-mass Galaxies: Supermassive Black Hole Binary Evolution
  and Structure of Merger Remnants}. \apj 749(2):147.
  \doi{10.1088/0004-637X/749/2/147}.
  {\href{https://arxiv.org/abs/1202.2124}{{arXiv:1202.2124}}} {[astro-ph.CO]}

\bibitem[{{Khan} et~al.(2016){Khan}, {Fiacconi}, {Mayer}, {Berczik}, and
  {Just}}]{2016ApJ...828...73K}
{Khan} FM, {Fiacconi} D, {Mayer} L, {Berczik} P, {Just} A (2016) {Swift
  Coalescence of Supermassive Black Holes in Cosmological Mergers of Massive
  Galaxies}. \apj 828(2):73. \doi{10.3847/0004-637X/828/2/73}.
  {\href{https://arxiv.org/abs/1604.00015}{{arXiv:1604.00015}}} {[astro-ph.GA]}

\bibitem[{{Khan} et~al.(2018{\natexlab{b}}){Khan}, {Berczik}, and
  {Just}}]{2018A&A...615A..71K}
{Khan} FM, {Berczik} P, {Just} A (2018{\natexlab{b}}) {Gravitational wave
  driven mergers and coalescence time of supermassive black holes}. \aap
  615:A71. \doi{10.1051/0004-6361/201730489}.
  {\href{https://arxiv.org/abs/1803.11394}{{arXiv:1803.11394}}} {[astro-ph.GA]}

\bibitem[{{Khan} et~al.(2018{\natexlab{c}}){Khan}, {Capelo}, {Mayer}, and
  {Berczik}}]{2018ApJ...868...97K}
{Khan} FM, {Capelo} PR, {Mayer} L, {Berczik} P (2018{\natexlab{c}}) {Dynamical
  Evolution and Merger Timescales of LISA Massive Black Hole Binaries in Disk
  Galaxy Mergers}. \apj 868(2):97. \doi{10.3847/1538-4357/aae77b}.
  {\href{https://arxiv.org/abs/1807.11004}{{arXiv:1807.11004}}} {[astro-ph.GA]}

\bibitem[{{Khan} et~al.(2020){Khan}, {Mirza}, and
  {Holley-Bockelmann}}]{2020MNRAS.492..256K}
{Khan} FM, {Mirza} MA, {Holley-Bockelmann} K (2020) {Inward bound: the
  incredible journey of massive black holes as they pair and merge - I. The
  effect of mass ratio in flattened rotating galactic nuclei}. \mnras
  492(1):256--267. \doi{10.1093/mnras/stz3360}.
  {\href{https://arxiv.org/abs/1911.07946}{{arXiv:1911.07946}}} {[astro-ph.GA]}

\bibitem[{{Kidder}(1995)}]{1995PhRvD..52..821K}
{Kidder} LE (1995) {Coalescing binary systems of compact objects to
  (post)$^{5/2}$-Newtonian order. V. Spin effects}. \prd 52(2):821--847.
  \doi{10.1103/PhysRevD.52.821}.
  {\href{https://arxiv.org/abs/gr-qc/9506022}{{arXiv:gr-qc/9506022}}} {[gr-qc]}

\bibitem[{{Kilic} et~al.(2011){Kilic}, {Brown}, {Hermes}, {Allende Prieto},
  {Kenyon}, {Winget}, and {Winget}}]{2011MNRAS.418L.157K}
{Kilic} M, {Brown} WR, {Hermes} JJ, {Allende Prieto} C, {Kenyon} SJ, {Winget}
  DE, {Winget} KI (2011) {SDSS J163030.58+423305.8: a 40-min orbital period
  detached white dwarf binary}. \mnras 418(1):L157--L161.
  \doi{10.1111/j.1745-3933.2011.01165.x}.
  {\href{https://arxiv.org/abs/1109.6339}{{arXiv:1109.6339}}} {[astro-ph.GA]}

\bibitem[{{Kilic} et~al.(2012){Kilic}, {Brown}, {Allende Prieto}, {Kenyon},
  {Heinke}, {Ag{\"u}eros}, and {Kleinman}}]{2012ApJ...751..141K}
{Kilic} M, {Brown} WR, {Allende Prieto} C, {Kenyon} SJ, {Heinke} CO,
  {Ag{\"u}eros} MA, {Kleinman} SJ (2012) {The ELM Survey. IV. 24 White Dwarf
  Merger Systems}. \apj 751(2):141. \doi{10.1088/0004-637X/751/2/141}.
  {\href{https://arxiv.org/abs/1204.0028}{{arXiv:1204.0028}}} {[astro-ph.GA]}

\bibitem[{{Kilic} et~al.(2014){Kilic}, {Brown}, {Gianninas}, {Hermes}, {Allende
  Prieto}, and {Kenyon}}]{2014MNRAS.444L...1K}
{Kilic} M, {Brown} WR, {Gianninas} A, {Hermes} JJ, {Allende Prieto} C, {Kenyon}
  SJ (2014) {A new gravitational wave verification source.} \mnras 444:L1--L5.
  \doi{10.1093/mnrasl/slu093}.
  {\href{https://arxiv.org/abs/1406.3346}{{arXiv:1406.3346}}} {[astro-ph.SR]}

\bibitem[{{Kilic} et~al.(2017){Kilic}, {Brown}, {Gianninas}, {Curd}, {Bell},
  and {Allende Prieto}}]{2017MNRAS.471.4218K}
{Kilic} M, {Brown} WR, {Gianninas} A, {Curd} B, {Bell} KJ, {Allende Prieto} C
  (2017) {A Gemini snapshot survey for double degenerates}. \mnras
  471(4):4218--4227. \doi{10.1093/mnras/stx1886}.
  {\href{https://arxiv.org/abs/1707.08948}{{arXiv:1707.08948}}} {[astro-ph.SR]}

\bibitem[{{Kim} et~al.(2003){Kim}, {Kalogera}, and
  {Lorimer}}]{2003ApJ...584..985K}
{Kim} C, {Kalogera} V, {Lorimer} DR (2003) {The Probability Distribution of
  Binary Pulsar Coalescence Rates. I. Double Neutron Star Systems in the
  Galactic Field}. \apj 584(2):985--995. \doi{10.1086/345740}.
  {\href{https://arxiv.org/abs/astro-ph/0207408}{{arXiv:astro-ph/0207408}}}
  {[astro-ph]}

\bibitem[{{Kim} et~al.(2017){Kim}, {Yoon}, {Privon}, {Evans}, {Harvey},
  {Stierwalt}, and {Kim}}]{2017ApJ...840...71K}
{Kim} DC, {Yoon} I, {Privon} GC, {Evans} AS, {Harvey} D, {Stierwalt} S, {Kim}
  JH (2017) {A Potential Recoiling Supermassive Black Hole, CXO
  J101527.2+625911}. \apj 840(2):71. \doi{10.3847/1538-4357/aa6030}.
  {\href{https://arxiv.org/abs/1704.05549}{{arXiv:1704.05549}}} {[astro-ph.GA]}

\bibitem[{{Kimpson} et~al.(2019{\natexlab{a}}){Kimpson}, {Wu}, and
  {Zane}}]{2019MNRAS.486..360K}
{Kimpson} T, {Wu} K, {Zane} S (2019{\natexlab{a}}) {Pulsar timing in extreme
  mass ratio binaries: a general relativistic approach}. \mnras
  486(1):360--377. \doi{10.1093/mnras/stz845}.
  {\href{https://arxiv.org/abs/1903.08258}{{arXiv:1903.08258}}} {[astro-ph.HE]}

\bibitem[{{Kimpson} et~al.(2019{\natexlab{b}}){Kimpson}, {Wu}, and
  {Zane}}]{2019MNRAS.484.2411K}
{Kimpson} T, {Wu} K, {Zane} S (2019{\natexlab{b}}) {Spatial dispersion of light
  rays propagating through a plasma in Kerr space-time}. \mnras
  484(2):2411--2419. \doi{10.1093/mnras/stz138}.
  {\href{https://arxiv.org/abs/1901.03733}{{arXiv:1901.03733}}} {[astro-ph.HE]}

\bibitem[{{Kimpson} et~al.(2020{\natexlab{a}}){Kimpson}, {Wu}, and
  {Zane}}]{2020MNRAS.495..600K}
{Kimpson} T, {Wu} K, {Zane} S (2020{\natexlab{a}}) {Gravitational burst
  radiation from pulsars in the Galactic centre and stellar clusters}. \mnras
  495(1):600--613. \doi{10.1093/mnras/staa1259}.
  {\href{https://arxiv.org/abs/2005.02053}{{arXiv:2005.02053}}} {[astro-ph.HE]}

\bibitem[{{Kimpson} et~al.(2020{\natexlab{b}}){Kimpson}, {Wu}, and
  {Zane}}]{2020MNRAS.497.5421K}
{Kimpson} T, {Wu} K, {Zane} S (2020{\natexlab{b}}) {Orbital spin dynamics of a
  millisecond pulsar around a massive BH with a general mass quadrupole}.
  \mnras 497(4):5421--5431. \doi{10.1093/mnras/staa2103}.
  {\href{https://arxiv.org/abs/2007.05219}{{arXiv:2007.05219}}} {[astro-ph.HE]}

\bibitem[{{King}(2003)}]{2003ApJ...596L..27K}
{King} A (2003) {Black Holes, Galaxy Formation, and the
  M$_{BH}$-{\ensuremath{\sigma}} Relation}. \apjl 596(1):L27--L29.
  \doi{10.1086/379143}.
  {\href{https://arxiv.org/abs/astro-ph/0308342}{{arXiv:astro-ph/0308342}}}
  {[astro-ph]}

\bibitem[{{King} and {Pringle}(2006)}]{2006MNRAS.373L..90K}
{King} AR, {Pringle} JE (2006) {Growing supermassive black holes by chaotic
  accretion}. \mnras 373(1):L90--L92. \doi{10.1111/j.1745-3933.2006.00249.x}.
  {\href{https://arxiv.org/abs/astro-ph/0609598}{{arXiv:astro-ph/0609598}}}
  {[astro-ph]}

\bibitem[{{King} et~al.(1990){King}, {Frank}, and
  {Whitehurst}}]{1990MNRAS.244..731K}
{King} AR, {Frank} J, {Whitehurst} R (1990) {Synchronous rotation in AM
  Herculis systems-I. Equilibrium configurations.} \mnras 244:731

\bibitem[{{King} et~al.(2008){King}, {Pringle}, and
  {Hofmann}}]{2008MNRAS.385.1621K}
{King} AR, {Pringle} JE, {Hofmann} JA (2008) {The evolution of black hole mass
  and spin in active galactic nuclei}. \mnras 385(3):1621--1627.
  \doi{10.1111/j.1365-2966.2008.12943.x}.
  {\href{https://arxiv.org/abs/0801.1564}{{arXiv:0801.1564}}} {[astro-ph]}

\bibitem[{{Kiseleva} et~al.(1994){Kiseleva}, {Eggleton}, and
  {Orlov}}]{1994MNRAS.270..936K}
{Kiseleva} LG, {Eggleton} PP, {Orlov} VV (1994) {Instability of close triple
  systems with coplanar initial doubly circular motion.} \mnras 270:936--946.
  \doi{10.1093/mnras/270.4.936}

\bibitem[{{Kiseleva} et~al.(1998){Kiseleva}, {Eggleton}, and
  {Mikkola}}]{1998MNRAS.300..292K}
{Kiseleva} LG, {Eggleton} PP, {Mikkola} S (1998) {Tidal friction in triple
  stars}. \mnras 300(1):292--302. \doi{10.1046/j.1365-8711.1998.01903.x}

\bibitem[{{Kiuchi} and {Maeda}(2004)}]{2004PhRvD..70f4036K}
{Kiuchi} K, {Maeda} KI (2004) {Gravitational waves from a chaotic dynamical
  system}. \prd 70(6):064036. \doi{10.1103/PhysRevD.70.064036}.
  {\href{https://arxiv.org/abs/gr-qc/0404124}{{arXiv:gr-qc/0404124}}} {[gr-qc]}

\bibitem[{{Kiuchi} et~al.(2011){Kiuchi}, {Shibata}, {Montero}, and
  {Font}}]{2011PhRvL.106y1102K}
{Kiuchi} K, {Shibata} M, {Montero} PJ, {Font} JA (2011) {Gravitational Waves
  from the Papaloizou-Pringle Instability in Black-Hole-Torus Systems}. \prl
  106(25):251102. \doi{10.1103/PhysRevLett.106.251102}.
  {\href{https://arxiv.org/abs/1105.5035}{{arXiv:1105.5035}}} {[astro-ph.HE]}

\bibitem[{{Klein} et~al.(2016){Klein}, {Barausse}, {Sesana}, {Petiteau},
  {Berti}, {Babak}, {Gair}, {Aoudia}, {Hinder}, {Ohme}, and
  {Wardell}}]{2016PhRvD..93b4003K}
{Klein} A, {Barausse} E, {Sesana} A, {Petiteau} A, {Berti} E, {Babak} S, {Gair}
  J, {Aoudia} S, {Hinder} I, {Ohme} F, et~al. (2016) {Science with the
  space-based interferometer eLISA: Supermassive black hole binaries}. \prd
  93(2):024003. \doi{10.1103/PhysRevD.93.024003}.
  {\href{https://arxiv.org/abs/1511.05581}{{arXiv:1511.05581}}} {[gr-qc]}

\bibitem[{{Klencki} et~al.(2018){Klencki}, {Moe}, {Gladysz}, {Chruslinska},
  {Holz}, and {Belczynski}}]{2018A&A...619A..77K}
{Klencki} J, {Moe} M, {Gladysz} W, {Chruslinska} M, {Holz} DE, {Belczynski} K
  (2018) {Impact of inter-correlated initial binary parameters on double black
  hole and neutron star mergers}. \aap 619:A77.
  \doi{10.1051/0004-6361/201833025}.
  {\href{https://arxiv.org/abs/1808.07889}{{arXiv:1808.07889}}} {[astro-ph.HE]}

\bibitem[{{Knigge} et~al.(2011){Knigge}, {Baraffe}, and
  {Patterson}}]{2011ApJS..194...28K}
{Knigge} C, {Baraffe} I, {Patterson} J (2011) {The Evolution of Cataclysmic
  Variables as Revealed by Their Donor Stars}. \apjs 194(2):28.
  \doi{10.1088/0067-0049/194/2/28}.
  {\href{https://arxiv.org/abs/1102.2440}{{arXiv:1102.2440}}} {[astro-ph.SR]}

\bibitem[{{Kobayashi} et~al.(2004){Kobayashi}, {Laguna}, {Phinney}, and
  {M{\'e}sz{\'a}ros}}]{2004ApJ...615..855K}
{Kobayashi} S, {Laguna} P, {Phinney} ES, {M{\'e}sz{\'a}ros} P (2004)
  {Gravitational Waves and X-Ray Signals from Stellar Disruption by a Massive
  Black Hole}. \apj 615(2):855--865. \doi{10.1086/424684}.
  {\href{https://arxiv.org/abs/astro-ph/0404173}{{arXiv:astro-ph/0404173}}}
  {[astro-ph]}

\bibitem[{{Kochanek}(1992)}]{1992ApJ...398..234K}
{Kochanek} CS (1992) {Coalescing Binary Neutron Stars}. \apj 398:234.
  \doi{10.1086/171851}

\bibitem[{{Kochanek} et~al.(1990){Kochanek}, {Shapiro}, {Teukolsky}, and
  {Chernoff}}]{1990ApJ...358...81K}
{Kochanek} CS, {Shapiro} SL, {Teukolsky} SA, {Chernoff} DF (1990)
  {Gravitational Radiation from Colliding Clusters: Newtonian Simulations in
  Three Dimensions}. \apj 358:81. \doi{10.1086/168964}

\bibitem[{{Kocsis} and {Tremaine}(2015)}]{2015MNRAS.448.3265K}
{Kocsis} B, {Tremaine} S (2015) {A numerical study of vector resonant
  relaxation}. \mnras 448(4):3265--3296. \doi{10.1093/mnras/stv057}.
  {\href{https://arxiv.org/abs/1406.1178}{{arXiv:1406.1178}}} {[astro-ph.GA]}

\bibitem[{{Kocsis} et~al.(2006){Kocsis}, {G{\'a}sp{\'a}r}, and
  {M{\'a}rka}}]{2006ApJ...648..411K}
{Kocsis} B, {G{\'a}sp{\'a}r} ME, {M{\'a}rka} S (2006) {Detection Rate Estimates
  of Gravity Waves Emitted during Parabolic Encounters of Stellar Black Holes
  in Globular Clusters}. \apj 648(1):411--429. \doi{10.1086/505641}.
  {\href{https://arxiv.org/abs/astro-ph/0603441}{{arXiv:astro-ph/0603441}}}
  {[astro-ph]}

\bibitem[{{Kocsis} et~al.(2011){Kocsis}, {Yunes}, and
  {Loeb}}]{2011PhRvD..84b4032K}
{Kocsis} B, {Yunes} N, {Loeb} A (2011) {Observable signatures of extreme
  mass-ratio inspiral black hole binaries embedded in thin accretion disks}.
  \prd 84(2):024032. \doi{10.1103/PhysRevD.84.024032}.
  {\href{https://arxiv.org/abs/1104.2322}{{arXiv:1104.2322}}} {[astro-ph.GA]}

\bibitem[{{Koester} et~al.(2014){Koester}, {G{\"a}nsicke}, and
  {Farihi}}]{2014A&A...566A..34K}
{Koester} D, {G{\"a}nsicke} BT, {Farihi} J (2014) {The frequency of planetary
  debris around young white dwarfs}. \aap 566:A34.
  \doi{10.1051/0004-6361/201423691}.
  {\href{https://arxiv.org/abs/1404.2617}{{arXiv:1404.2617}}} {[astro-ph.SR]}

\bibitem[{{Kohri} and {Terada}(2020)}]{2020arXiv200911853K}
{Kohri} K, {Terada} T (2020) {Solar-Mass Primordial Black Holes Explain
  NANOGrav Hint of Gravitational Waves}. arXiv e-prints arXiv:2009.11853.
  {\href{https://arxiv.org/abs/2009.11853}{{arXiv:2009.11853}}} {[astro-ph.CO]}

\bibitem[{{Kolb}(1993)}]{1993A&A...271..149K}
{Kolb} U (1993) {A model for the intrinsic population of cataclysmic
  variables}. \aap 271:149

\bibitem[{{Koliopanos} et~al.(2017){Koliopanos}, {Ciambur}, {Graham}, {Webb},
  {Coriat}, {Mutlu-Pakdil}, {Davis}, {Godet}, {Barret}, and
  {Seigar}}]{2017A&A...601A..20K}
{Koliopanos} F, {Ciambur} BC, {Graham} AW, {Webb} NA, {Coriat} M,
  {Mutlu-Pakdil} B, {Davis} BL, {Godet} O, {Barret} D, {Seigar} MS (2017)
  {Searching for intermediate-mass black holes in galaxies with low-luminosity
  AGN: a multiple-method approach}. \aap 601:A20.
  \doi{10.1051/0004-6361/201630061}.
  {\href{https://arxiv.org/abs/1612.06794}{{arXiv:1612.06794}}} {[astro-ph.GA]}

\bibitem[{{Koliopanos} et~al.(2020){Koliopanos}, {Peault}, {Vasilopoulos}, and
  {Webb}}]{2020arXiv200100716K}
{Koliopanos} F, {Peault} M, {Vasilopoulos} G, {Webb} N (2020) {The chemical
  composition of the accretion disk and donor star in Ultra Compact X-ray
  Binaries: A comprehensive X-ray analysis}. arXiv e-prints arXiv:2001.00716.
  {\href{https://arxiv.org/abs/2001.00716}{{arXiv:2001.00716}}} {[astro-ph.HE]}

\bibitem[{{Kollmeier} et~al.(2017){Kollmeier}, {Zasowski}, {Rix}, {Johns},
  {Anderson}, {Drory}, {Johnson}, {Pogge}, {Bird}, {Blanc}, {Brownstein},
  {Crane}, {De Lee}, {Klaene}, {Kreckel}, {MacDonald}, {Merloni}, {Ness},
  {O'Brien}, {Sanchez-Gallego}, {Sayres}, {Shen}, {Thakar}, {Tkachenko},
  {Aerts}, {Blanton}, {Eisenstein}, {Holtzman}, {Maoz}, {Nandra}, {Rockosi},
  {Weinberg}, {Bovy}, {Casey}, {Chaname}, {Clerc}, {Conroy}, {Eracleous},
  {G{\"a}nsicke}, {Hekker}, {Horne}, {Kauffmann}, {McQuinn}, {Pellegrini},
  {Schinnerer}, {Schlafly}, {Schwope}, {Seibert}, {Teske}, and {van
  Saders}}]{2017arXiv171103234K}
{Kollmeier} JA, {Zasowski} G, {Rix} HW, {Johns} M, {Anderson} SF, {Drory} N,
  {Johnson} JA, {Pogge} RW, {Bird} JC, {Blanc} GA, et~al. (2017) {SDSS-V:
  Pioneering Panoptic Spectroscopy}. arXiv e-prints arXiv:1711.03234.
  {\href{https://arxiv.org/abs/1711.03234}{{arXiv:1711.03234}}} {[astro-ph.GA]}

\bibitem[{{Komossa}(2012)}]{2012AdAst2012E..14K}
{Komossa} S (2012) {Recoiling Black Holes: Electromagnetic Signatures,
  Candidates, and Astrophysical Implications}. Advances in Astronomy
  2012:364973. \doi{10.1155/2012/364973}.
  {\href{https://arxiv.org/abs/1202.1977}{{arXiv:1202.1977}}} {[astro-ph.CO]}

\bibitem[{{Komossa} and {Greiner}(1999)}]{1999A&A...349L..45K}
{Komossa} S, {Greiner} J (1999) {Discovery of a giant and luminous X-ray
  outburst from the optically inactive galaxy pair RX J1242.6-1119}. \aap
  349:L45--L48.
  {\href{https://arxiv.org/abs/astro-ph/9908216}{{arXiv:astro-ph/9908216}}}
  {[astro-ph]}

\bibitem[{{Komossa} et~al.(2008){Komossa}, {Zhou}, and
  {Lu}}]{2008ApJ...678L..81K}
{Komossa} S, {Zhou} H, {Lu} H (2008) {A Recoiling Supermassive Black Hole in
  the Quasar SDSS J092712.65+294344.0?} \apjl 678(2):L81. \doi{10.1086/588656}.
  {\href{https://arxiv.org/abs/0804.4585}{{arXiv:0804.4585}}} {[astro-ph]}

\bibitem[{{Konstantinidis} et~al.(2013){Konstantinidis}, {Amaro-Seoane}, and
  {Kokkotas}}]{2013A&A...557A.135K}
{Konstantinidis} S, {Amaro-Seoane} P, {Kokkotas} KD (2013) {Investigating the
  retention of intermediate-mass black holes in star clusters using N-body
  simulations}. \aap 557:A135. \doi{10.1051/0004-6361/201219620}.
  {\href{https://arxiv.org/abs/1108.5175}{{arXiv:1108.5175}}} {[astro-ph.CO]}

\bibitem[{{Kormendy}(2004)}]{2004cbhg.symp....1K}
{Kormendy} J (2004) {The Stellar-Dynamical Search for Supermassive Black Holes
  in Galactic Nuclei}. In: {Ho} LC (ed) Coevolution of Black Holes and
  Galaxies. p~1.
  {\href{https://arxiv.org/abs/astro-ph/0306353}{{arXiv:astro-ph/0306353}}}
  {[astro-ph]}

\bibitem[{{Kormendy} and {Ho}(2013)}]{2013ARA&A..51..511K}
{Kormendy} J, {Ho} LC (2013) {Coevolution (Or Not) of Supermassive Black Holes
  and Host Galaxies}. \araa 51(1):511--653.
  \doi{10.1146/annurev-astro-082708-101811}.
  {\href{https://arxiv.org/abs/1304.7762}{{arXiv:1304.7762}}} {[astro-ph.CO]}

\bibitem[{{Kormendy} and {Richstone}(1995)}]{1995ARA&A..33..581K}
{Kormendy} J, {Richstone} D (1995) {Inward Bound---The Search For Supermassive
  Black Holes In Galactic Nuclei}. \araa 33:581.
  \doi{10.1146/annurev.aa.33.090195.003053}

\bibitem[{{Korol} and {Safarzadeh}(2021)}]{2021MNRAS.502.5576K}
{Korol} V, {Safarzadeh} M (2021) {How can LISA probe a population of
  GW190425-like binary neutron stars in the Milky Way?} \mnras
  502(4):5576--5583. \doi{10.1093/mnras/stab310}.
  {\href{https://arxiv.org/abs/2012.03070}{{arXiv:2012.03070}}} {[astro-ph.HE]}

\bibitem[{{Korol} et~al.(2017){Korol}, {Rossi}, {Groot}, {Nelemans}, {Toonen},
  and {Brown}}]{2017MNRAS.470.1894K}
{Korol} V, {Rossi} EM, {Groot} PJ, {Nelemans} G, {Toonen} S, {Brown} AGA (2017)
  {Prospects for detection of detached double white dwarf binaries with Gaia,
  LSST and LISA}. \mnras 470(2):1894--1910. \doi{10.1093/mnras/stx1285}.
  {\href{https://arxiv.org/abs/1703.02555}{{arXiv:1703.02555}}} {[astro-ph.HE]}

\bibitem[{{Korol} et~al.(2018){Korol}, {Koop}, and
  {Rossi}}]{2018ApJ...866L..20K}
{Korol} V, {Koop} O, {Rossi} EM (2018) {Detectability of Double White Dwarfs in
  the Local Group with LISA}. \apjl 866(2):L20. \doi{10.3847/2041-8213/aae587}.
  {\href{https://arxiv.org/abs/1808.05959}{{arXiv:1808.05959}}} {[astro-ph.HE]}

\bibitem[{{Korol} et~al.(2019){Korol}, {Rossi}, and
  {Barausse}}]{2019MNRAS.483.5518K}
{Korol} V, {Rossi} EM, {Barausse} E (2019) {A multimessenger study of the Milky
  Way's stellar disc and bulge with LISA, Gaia, and LSST}. \mnras
  483(4):5518--5533. \doi{10.1093/mnras/sty3440}.
  {\href{https://arxiv.org/abs/1806.03306}{{arXiv:1806.03306}}} {[astro-ph.GA]}

\bibitem[{{Korol} et~al.(2020){Korol}, {Toonen}, {Klein}, {Belokurov},
  {Vincenzo}, {Buscicchio}, {Gerosa}, {Moore}, {Roebber}, {Rossi}, and
  {Vecchio}}]{2020A&A...638A.153K}
{Korol} V, {Toonen} S, {Klein} A, {Belokurov} V, {Vincenzo} F, {Buscicchio} R,
  {Gerosa} D, {Moore} CJ, {Roebber} E, {Rossi} EM, et~al. (2020) {Populations
  of double white dwarfs in Milky Way satellites and their detectability with
  LISA}. \aap 638:A153. \doi{10.1051/0004-6361/202037764}.
  {\href{https://arxiv.org/abs/2002.10462}{{arXiv:2002.10462}}} {[astro-ph.GA]}

\bibitem[{{Korol} et~al.(2021){Korol}, {Belokurov}, {Moore}, and
  {Toonen}}]{2021MNRAS.502L..55K}
{Korol} V, {Belokurov} V, {Moore} CJ, {Toonen} S (2021) {Weighing Milky Way
  satellites with LISA}. \mnras 502(1):L55--L60. \doi{10.1093/mnrasl/slab003}.
  {\href{https://arxiv.org/abs/2010.05918}{{arXiv:2010.05918}}} {[astro-ph.GA]}

\bibitem[{{Koss} et~al.(2014){Koss}, {Blecha}, {Mushotzky}, {Hung}, {Veilleux},
  {Trakhtenbrot}, {Schawinski}, {Stern}, {Smith}, {Li}, {Man}, {Filippenko},
  {Mauerhan}, {Stanek}, and {Sanders}}]{2014MNRAS.445..515K}
{Koss} M, {Blecha} L, {Mushotzky} R, {Hung} CL, {Veilleux} S, {Trakhtenbrot} B,
  {Schawinski} K, {Stern} D, {Smith} N, {Li} Y, et~al. (2014) {SDSS1133: an
  unusually persistent transient in a nearby dwarf galaxy}. \mnras
  445(1):515--527. \doi{10.1093/mnras/stu1673}.
  {\href{https://arxiv.org/abs/1401.6798}{{arXiv:1401.6798}}} {[astro-ph.GA]}

\bibitem[{{Kostov} et~al.(2016){Kostov}, {Moore}, {Tamayo}, {Jayawardhana}, and
  {Rinehart}}]{2016ApJ...832..183K}
{Kostov} VB, {Moore} K, {Tamayo} D, {Jayawardhana} R, {Rinehart} SA (2016)
  {Tatooine{\textquoteright}s Future: The Eccentric Response of
  Kepler{\textquoteright}s Circumbinary Planets to Common-envelope Evolution of
  Their Host Stars}. \apj 832(2):183. \doi{10.3847/0004-637X/832/2/183}.
  {\href{https://arxiv.org/abs/1610.03436}{{arXiv:1610.03436}}} {[astro-ph.EP]}

\bibitem[{{Kozai}(1962)}]{1962AJ.....67..591K}
{Kozai} Y (1962) {Secular perturbations of asteroids with high inclination and
  eccentricity}. \aj 67:591--598. \doi{10.1086/108790}

\bibitem[{{Kramer} et~al.(2020){Kramer}, {Schneider}, {Ohlmann}, {Geier},
  {Schaffenroth}, {Pakmor}, and {Roepke}}]{2020arXiv200700019K}
{Kramer} M, {Schneider} FRN, {Ohlmann} ST, {Geier} S, {Schaffenroth} V,
  {Pakmor} R, {Roepke} FK (2020) {Formation of sdB-stars via common envelope
  ejection by substellar companions}. arXiv e-prints arXiv:2007.00019.
  {\href{https://arxiv.org/abs/2007.00019}{{arXiv:2007.00019}}} {[astro-ph.SR]}

\bibitem[{{Kremer} et~al.(2017){Kremer}, {Breivik}, {Larson}, and
  {Kalogera}}]{2017ApJ...846...95K}
{Kremer} K, {Breivik} K, {Larson} SL, {Kalogera} V (2017) {Accreting Double
  White Dwarf Binaries: Implications for LISA}. \apj 846(2):95.
  \doi{10.3847/1538-4357/aa8557}.
  {\href{https://arxiv.org/abs/1707.01104}{{arXiv:1707.01104}}} {[astro-ph.HE]}

\bibitem[{{Kremer} et~al.(2018{\natexlab{a}}){Kremer}, {Chatterjee}, {Breivik},
  {Rodriguez}, {Larson}, and {Rasio}}]{2018PhRvL.120s1103K}
{Kremer} K, {Chatterjee} S, {Breivik} K, {Rodriguez} CL, {Larson} SL, {Rasio}
  FA (2018{\natexlab{a}}) {LISA Sources in Milky Way Globular Clusters}. \prl
  120(19):191103. \doi{10.1103/PhysRevLett.120.191103}.
  {\href{https://arxiv.org/abs/1802.05661}{{arXiv:1802.05661}}} {[astro-ph.HE]}

\bibitem[{{Kremer} et~al.(2018{\natexlab{b}}){Kremer}, {Chatterjee},
  {Rodriguez}, and {Rasio}}]{2018ApJ...852...29K}
{Kremer} K, {Chatterjee} S, {Rodriguez} CL, {Rasio} FA (2018{\natexlab{b}})
  {Accreting Black Hole Binaries in Globular Clusters}. \apj 852(1):29.
  \doi{10.3847/1538-4357/aa99df}.
  {\href{https://arxiv.org/abs/1709.05444}{{arXiv:1709.05444}}} {[astro-ph.HE]}

\bibitem[{{Kremer} et~al.(2019{\natexlab{a}}){Kremer}, {Chatterjee}, {Ye},
  {Rodriguez}, and {Rasio}}]{2019ApJ...871...38K}
{Kremer} K, {Chatterjee} S, {Ye} CS, {Rodriguez} CL, {Rasio} FA
  (2019{\natexlab{a}}) {How Initial Size Governs Core Collapse in Globular
  Clusters}. \apj 871(1):38. \doi{10.3847/1538-4357/aaf646}.
  {\href{https://arxiv.org/abs/1808.02204}{{arXiv:1808.02204}}} {[astro-ph.GA]}

\bibitem[{{Kremer} et~al.(2019{\natexlab{b}}){Kremer}, {Rodriguez},
  {Amaro-Seoane}, {Breivik}, {Chatterjee}, {Katz}, {Larson}, {Rasio},
  {Samsing}, {Ye}, and {Zevin}}]{2019PhRvD..99f3003K}
{Kremer} K, {Rodriguez} CL, {Amaro-Seoane} P, {Breivik} K, {Chatterjee} S,
  {Katz} ML, {Larson} SL, {Rasio} FA, {Samsing} J, {Ye} CS, et~al.
  (2019{\natexlab{b}}) {Post-Newtonian dynamics in dense star clusters: Binary
  black holes in the LISA band}. \prd 99(6):063003.
  \doi{10.1103/PhysRevD.99.063003}.
  {\href{https://arxiv.org/abs/1811.11812}{{arXiv:1811.11812}}} {[astro-ph.HE]}

\bibitem[{{Kremer} et~al.(2020{\natexlab{a}}){Kremer}, {Spera}, {Becker},
  {Chatterjee}, {Di Carlo}, {Fragione}, {Rodriguez}, {Ye}, and
  {Rasio}}]{2020arXiv200610771K}
{Kremer} K, {Spera} M, {Becker} D, {Chatterjee} S, {Di Carlo} UN, {Fragione} G,
  {Rodriguez} CL, {Ye} CS, {Rasio} FA (2020{\natexlab{a}}) {Populating the
  upper black hole mass gap through stellar collisions in dense star clusters}.
  arXiv e-prints arXiv:2006.10771.
  {\href{https://arxiv.org/abs/2006.10771}{{arXiv:2006.10771}}} {[astro-ph.HE]}

\bibitem[{{Kremer} et~al.(2020{\natexlab{b}}){Kremer}, {Ye}, {Rui},
  {Weatherford}, {Chatterjee}, {Fragione}, {Rodriguez}, {Spera}, and
  {Rasio}}]{2020ApJS..247...48K}
{Kremer} K, {Ye} CS, {Rui} NZ, {Weatherford} NC, {Chatterjee} S, {Fragione} G,
  {Rodriguez} CL, {Spera} M, {Rasio} FA (2020{\natexlab{b}}) {Modeling Dense
  Star Clusters in the Milky Way and Beyond with the CMC Cluster Catalog}.
  \apjs 247(2):48. \doi{10.3847/1538-4365/ab7919}.
  {\href{https://arxiv.org/abs/1911.00018}{{arXiv:1911.00018}}} {[astro-ph.HE]}

\bibitem[{{Krolik} et~al.(2019){Krolik}, {Volonteri}, {Dubois}, and
  {Devriendt}}]{2019ApJ...879..110K}
{Krolik} JH, {Volonteri} M, {Dubois} Y, {Devriendt} J (2019) Population
  estimates for electromagnetically distinguishable supermassive binary black
  holes. \apj 879(2):110. \doi{10.3847/1538-4357/ab24c9}.
  {\href{https://arxiv.org/abs/1905.10450}{{arXiv:1905.10450}}} {[astro-ph.GA]}

\bibitem[{{Kroupa}(2001)}]{2001MNRAS.322..231K}
{Kroupa} P (2001) {On the variation of the initial mass function}. \mnras
  322(2):231--246. \doi{10.1046/j.1365-8711.2001.04022.x}.
  {\href{https://arxiv.org/abs/astro-ph/0009005}{{arXiv:astro-ph/0009005}}}
  {[astro-ph]}

\bibitem[{{Kruckow} et~al.(2016){Kruckow}, {Tauris}, {Langer}, {Sz{\'e}csi},
  {Marchant}, and {Podsiadlowski}}]{2016A&A...596A..58K}
{Kruckow} MU, {Tauris} TM, {Langer} N, {Sz{\'e}csi} D, {Marchant} P,
  {Podsiadlowski} P (2016) {Common-envelope ejection in massive binary stars.
  Implications for the progenitors of GW150914 and GW151226}. \aap 596:A58.
  \doi{10.1051/0004-6361/201629420}.
  {\href{https://arxiv.org/abs/1610.04417}{{arXiv:1610.04417}}} {[astro-ph.SR]}

\bibitem[{{Kruckow} et~al.(2018){Kruckow}, {Tauris}, {Langer}, {Kramer}, and
  {Izzard}}]{2018MNRAS.481.1908K}
{Kruckow} MU, {Tauris} TM, {Langer} N, {Kramer} M, {Izzard} RG (2018)
  {Progenitors of gravitational wave mergers: binary evolution with the stellar
  grid-based code COMBINE}. \mnras 481(2):1908--1949.
  \doi{10.1093/mnras/sty2190}.
  {\href{https://arxiv.org/abs/1801.05433}{{arXiv:1801.05433}}} {[astro-ph.SR]}

\bibitem[{{Krumholz}(2014)}]{2014PhR...539...49K}
{Krumholz} MR (2014) {The big problems in star formation: The star formation
  rate, stellar clustering, and the initial mass function}. \physrep
  539:49--134. \doi{10.1016/j.physrep.2014.02.001}.
  {\href{https://arxiv.org/abs/1402.0867}{{arXiv:1402.0867}}} {[astro-ph.GA]}

\bibitem[{{Kuhlen} et~al.(2012){Kuhlen}, {Vogelsberger}, and
  {Angulo}}]{2012PDU.....1...50K}
{Kuhlen} M, {Vogelsberger} M, {Angulo} R (2012) {Numerical simulations of the
  dark universe: State of the art and the next decade}. Physics of the Dark
  Universe 1(1-2):50--93. \doi{10.1016/j.dark.2012.10.002}.
  {\href{https://arxiv.org/abs/1209.5745}{{arXiv:1209.5745}}} {[astro-ph.CO]}

\bibitem[{{Kuin} et~al.(2019){Kuin}, {Wu}, {Oates}, {Lien}, {Emery}, {Kennea},
  {de Pasquale}, {Han}, {Brown}, {Tohuvavohu}, {Breeveld}, {Burrows}, {Cenko},
  {Campana}, {Levan}, {Markwardt}, {Osborne}, {Page}, {Page}, {Sbarufatti},
  {Siegel}, and {Troja}}]{2019MNRAS.487.2505K}
{Kuin} NPM, {Wu} K, {Oates} S, {Lien} A, {Emery} S, {Kennea} JA, {de Pasquale}
  M, {Han} Q, {Brown} PJ, {Tohuvavohu} A, et~al. (2019) {Swift spectra of
  AT2018cow: a white dwarf tidal disruption event?} \mnras 487(2):2505--2521.
  \doi{10.1093/mnras/stz053}.
  {\href{https://arxiv.org/abs/1808.08492}{{arXiv:1808.08492}}} {[astro-ph.HE]}

\bibitem[{{Kulkarni} and {Loeb}(2012)}]{2012MNRAS.422.1306K}
{Kulkarni} G, {Loeb} A (2012) {Formation of galactic nuclei with multiple
  supermassive black holes at high redshifts}. \mnras 422(2):1306--1323.
  \doi{10.1111/j.1365-2966.2012.20699.x}.
  {\href{https://arxiv.org/abs/1107.0517}{{arXiv:1107.0517}}} {[astro-ph.CO]}

\bibitem[{{Kulkarni} et~al.(2007){Kulkarni}, {Ofek}, {Rau}, {Cenko},
  {Soderberg}, {Fox}, {Gal-Yam}, {Capak}, {Moon}, {Li}, {Filippenko}, {Egami},
  {Kartaltepe}, and {Sanders}}]{2007Natur.447..458K}
{Kulkarni} SR, {Ofek} EO, {Rau} A, {Cenko} SB, {Soderberg} AM, {Fox} DB,
  {Gal-Yam} A, {Capak} PL, {Moon} DS, {Li} W, et~al. (2007) {An unusually
  brilliant transient in the galaxy M85}. \nat 447(7143):458--460.
  \doi{10.1038/nature05822}.
  {\href{https://arxiv.org/abs/0705.3668}{{arXiv:0705.3668}}} {[astro-ph]}

\bibitem[{{Kumamoto} et~al.(2019){Kumamoto}, {Fujii}, and
  {Tanikawa}}]{2019MNRAS.486.3942K}
{Kumamoto} J, {Fujii} MS, {Tanikawa} A (2019) {Gravitational-wave emission from
  binary black holes formed in open clusters}. \mnras 486(3):3942--3950.
  \doi{10.1093/mnras/stz1068}.
  {\href{https://arxiv.org/abs/1811.06726}{{arXiv:1811.06726}}} {[astro-ph.HE]}

\bibitem[{{Kumamoto} et~al.(2020){Kumamoto}, {Fujii}, and
  {Tanikawa}}]{2020MNRAS.495.4268K}
{Kumamoto} J, {Fujii} MS, {Tanikawa} A (2020) {Merger rate density of binary
  black holes formed in open clusters}. \mnras 495(4):4268--4278.
  \doi{10.1093/mnras/staa1440}.
  {\href{https://arxiv.org/abs/2001.10690}{{arXiv:2001.10690}}} {[astro-ph.HE]}

\bibitem[{{Kupfer} et~al.(2015){Kupfer}, {Groot}, {Bloemen}, {Levitan},
  {Steeghs}, {Marsh}, {Rutten}, {Nelemans}, {Prince}, {F{\"u}rst}, and
  {Geier}}]{2015MNRAS.453..483K}
{Kupfer} T, {Groot} PJ, {Bloemen} S, {Levitan} D, {Steeghs} D, {Marsh} TR,
  {Rutten} RGM, {Nelemans} G, {Prince} TA, {F{\"u}rst} F, et~al. (2015)
  {Phase-resolved spectroscopy and Kepler photometry of the ultracompact AM CVn
  binary SDSS J190817.07+394036.4}. \mnras 453(1):483--496.
  \doi{10.1093/mnras/stv1609}.
  {\href{https://arxiv.org/abs/1507.03926}{{arXiv:1507.03926}}} {[astro-ph.SR]}

\bibitem[{{Kupfer} et~al.(2017){Kupfer}, {Ramsay}, {van Roestel}, {Brooks},
  {MacFarlane}, {Toma}, {Groot}, {Woudt}, {Bildsten}, and {et
  al.}}]{2017ApJ...851...28K}
{Kupfer} T, {Ramsay} G, {van Roestel} J, {Brooks} J, {MacFarlane} SA, {Toma} R,
  {Groot} PJ, {Woudt} PA, {Bildsten} L, {et al} (2017) {The OmegaWhite Survey
  for Short-period Variable Stars. V. Discovery of an Ultracompact Hot Subdwarf
  Binary with a Compact Companion in a 44-minute Orbit}. \apj 851(1):28.
  \doi{10.3847/1538-4357/aa9522}.
  {\href{https://arxiv.org/abs/1710.07287}{{arXiv:1710.07287}}} {[astro-ph.SR]}

\bibitem[{{Kupfer} et~al.(2018){Kupfer}, {Korol}, {Shah}, {Nelemans}, {Marsh},
  {Ramsay}, {Groot}, {Steeghs}, and {Rossi}}]{2018MNRAS.480..302K}
{Kupfer} T, {Korol} V, {Shah} S, {Nelemans} G, {Marsh} TR, {Ramsay} G, {Groot}
  PJ, {Steeghs} DTH, {Rossi} EM (2018) {LISA verification binaries with updated
  distances from Gaia Data Release 2}. \mnras 480(1):302--309.
  \doi{10.1093/mnras/sty1545}.
  {\href{https://arxiv.org/abs/1805.00482}{{arXiv:1805.00482}}} {[astro-ph.SR]}

\bibitem[{{Kupfer} et~al.(2020{\natexlab{a}}){Kupfer}, {Bauer}, {Burdge},
  {Roestel}, {Bellm}, {Fuller}, {Hermes}, {Marsh}, and {et
  al.}}]{2020ApJ...898L..25K}
{Kupfer} T, {Bauer} EB, {Burdge} KB, {Roestel} Jv, {Bellm} EC, {Fuller} J,
  {Hermes} J, {Marsh} TR, {et al} (2020{\natexlab{a}}) {A New Class of Roche
  Lobe-filling Hot Subdwarf Binaries}. \apjl 898(1):L25.
  \doi{10.3847/2041-8213/aba3c2}.
  {\href{https://arxiv.org/abs/2007.05349}{{arXiv:2007.05349}}} {[astro-ph.SR]}

\bibitem[{{Kupfer} et~al.(2020{\natexlab{b}}){Kupfer}, {Bauer}, {Marsh}, {van
  Roestel}, {Bellm}, {Burdge}, {Coughlin}, {Fuller}, and {et
  al.}}]{2020ApJ...891...45K}
{Kupfer} T, {Bauer} EB, {Marsh} TR, {van Roestel} J, {Bellm} EC, {Burdge} KB,
  {Coughlin} MW, {Fuller} J, {et al} (2020{\natexlab{b}}) {The First
  Ultracompact Roche Lobe-Filling Hot Subdwarf Binary}. \apj 891(1):45.
  \doi{10.3847/1538-4357/ab72ff}.
  {\href{https://arxiv.org/abs/2002.01485}{{arXiv:2002.01485}}} {[astro-ph.SR]}

\bibitem[{{Kupfer} et~al.(2021){Kupfer}, {Prince}, {van Roestel}, {Bellm},
  {Bildsten}, {Coughlin}, {Drake}, {Graham}, {Klein}, {Kulkarni}, {Masci},
  {Walters}, {Andreoni}, {Biswas}, {Bradshaw}, {Duev}, {Dekany}, {Guidry},
  {Hermes}, {Laher}, and {Riddle}}]{2021MNRAS.505.1254K}
{Kupfer} T, {Prince} TA, {van Roestel} J, {Bellm} EC, {Bildsten} L, {Coughlin}
  MW, {Drake} AJ, {Graham} MJ, {Klein} C, {Kulkarni} SR, et~al. (2021) {Year 1
  of the ZTF high-cadence Galactic plane survey: strategy, goals, and early
  results on new single-mode hot subdwarf B-star pulsators}. \mnras
  505(1):1254--1267. \doi{10.1093/mnras/stab1344}.
  {\href{https://arxiv.org/abs/2105.02758}{{arXiv:2105.02758}}} {[astro-ph.SR]}

\bibitem[{{Kupi} et~al.(2006){Kupi}, {Amaro-Seoane}, and
  {Spurzem}}]{2006MNRAS.371L..45K}
{Kupi} G, {Amaro-Seoane} P, {Spurzem} R (2006) {Dynamics of compact object
  clusters: a post-Newtonian study}. \mnras 371(1):L45--L49.
  \doi{10.1111/j.1745-3933.2006.00205.x}.
  {\href{https://arxiv.org/abs/astro-ph/0602125}{{arXiv:astro-ph/0602125}}}
  {[astro-ph]}

\bibitem[{{Kuroda} et~al.(2020){Kuroda}, {Arcones}, {Takiwaki}, and
  {Kotake}}]{2020ApJ...896..102K}
{Kuroda} T, {Arcones} A, {Takiwaki} T, {Kotake} K (2020) {Magnetorotational
  Explosion of a Massive Star Supported by Neutrino Heating in General
  Relativistic Three-dimensional Simulations}. \apj 896(2):102.
  \doi{10.3847/1538-4357/ab9308}.
  {\href{https://arxiv.org/abs/2003.02004}{{arXiv:2003.02004}}} {[astro-ph.HE]}

\bibitem[{{Kuruwita} et~al.(2016){Kuruwita}, {Staff}, and {De
  Marco}}]{2016MNRAS.461..486K}
{Kuruwita} RL, {Staff} J, {De Marco} O (2016) {Considerations on the role of
  fall-back discs in the final stages of the common envelope binary
  interaction}. \mnras 461(1):486--496. \doi{10.1093/mnras/stw1414}.
  {\href{https://arxiv.org/abs/1606.04635}{{arXiv:1606.04635}}} {[astro-ph.SR]}

\bibitem[{{Kushnir} et~al.(2013){Kushnir}, {Katz}, {Dong}, {Livne}, and
  {Fern{\'a}ndez}}]{2013ApJ...778L..37K}
{Kushnir} D, {Katz} B, {Dong} S, {Livne} E, {Fern{\'a}ndez} R (2013) {Head-on
  Collisions of White Dwarfs in Triple Systems Could Explain Type Ia
  Supernovae}. \apjl 778(2):L37. \doi{10.1088/2041-8205/778/2/L37}.
  {\href{https://arxiv.org/abs/1303.1180}{{arXiv:1303.1180}}} {[astro-ph.HE]}

\bibitem[{{Kuulkers} et~al.(2003){Kuulkers}, {den Hartog}, {in't Zand},
  {Verbunt}, {Harris}, and {Cocchi}}]{2003A&A...399..663K}
{Kuulkers} E, {den Hartog} PR, {in't Zand} JJM, {Verbunt} FWM, {Harris} WE,
  {Cocchi} M (2003) {Photospheric radius expansion X-ray bursts as standard
  candles}. \aap 399:663--680. \doi{10.1051/0004-6361:20021781}.
  {\href{https://arxiv.org/abs/astro-ph/0212028}{{arXiv:astro-ph/0212028}}}
  {[astro-ph]}

\bibitem[{{Kyutoku} and {Seto}(2016)}]{2016MNRAS.462.2177K}
{Kyutoku} K, {Seto} N (2016) {Concise estimate of the expected number of
  detections for stellar-mass binary black holes by eLISA}. \mnras
  462(2):2177--2183. \doi{10.1093/mnras/stw1767}.
  {\href{https://arxiv.org/abs/1606.02298}{{arXiv:1606.02298}}} {[astro-ph.HE]}

\bibitem[{{Kyutoku} et~al.(2019){Kyutoku}, {Nishino}, and
  {Seto}}]{2019MNRAS.483.2615K}
{Kyutoku} K, {Nishino} Y, {Seto} N (2019) {How to detect the shortest period
  binary pulsars in the era of LISA}. \mnras 483(2):2615--2620.
  \doi{10.1093/mnras/sty3322}.
  {\href{https://arxiv.org/abs/1812.02177}{{arXiv:1812.02177}}} {[astro-ph.HE]}

\bibitem[{{Lacey} and {Cole}(1993)}]{1993MNRAS.262..627L}
{Lacey} C, {Cole} S (1993) {Merger rates in hierarchical models of galaxy
  formation}. \mnras 262(3):627--649. \doi{10.1093/mnras/262.3.627}

\bibitem[{{Lacy} et~al.(2015){Lacy}, {Ridgway}, {Sajina}, {Petric}, {Gates},
  {Urrutia}, and {Storrie-Lombardi}}]{2015ApJ...802..102L}
{Lacy} M, {Ridgway} SE, {Sajina} A, {Petric} AO, {Gates} EL, {Urrutia} T,
  {Storrie-Lombardi} LJ (2015) {The Spitzer Mid-infrared AGN Survey. II. The
  Demographics and Cosmic Evolution of the AGN Population}. \apj 802(2):102.
  \doi{10.1088/0004-637X/802/2/102}.
  {\href{https://arxiv.org/abs/1501.04118}{{arXiv:1501.04118}}} {[astro-ph.GA]}

\bibitem[{{Lai} and {Shapiro}(1995)}]{1995ApJ...443..705L}
{Lai} D, {Shapiro} SL (1995) {Hydrodynamics of Coalescing Binary Neutron Stars:
  Ellipsoidal Treatment}. \apj 443:705. \doi{10.1086/175562}.
  {\href{https://arxiv.org/abs/astro-ph/9408054}{{arXiv:astro-ph/9408054}}}
  {[astro-ph]}

\bibitem[{{Lai} et~al.(1994){Lai}, {Rasio}, and
  {Shapiro}}]{1994ApJ...437..742L}
{Lai} D, {Rasio} FA, {Shapiro} SL (1994) {Hydrodynamics of Rotating Stars and
  Close Binary Interactions: Compressible Ellipsoid Models}. \apj 437:742.
  \doi{10.1086/175036}.
  {\href{https://arxiv.org/abs/astro-ph/9404031}{{arXiv:astro-ph/9404031}}}
  {[astro-ph]}

\bibitem[{{Lai} et~al.(2001){Lai}, {Chernoff}, and
  {Cordes}}]{2001ApJ...549.1111L}
{Lai} D, {Chernoff} DF, {Cordes} JM (2001) {Pulsar Jets: Implications for
  Neutron Star Kicks and Initial Spins}. \apj 549(2):1111--1118.
  \doi{10.1086/319455}.
  {\href{https://arxiv.org/abs/astro-ph/0007272}{{arXiv:astro-ph/0007272}}}
  {[astro-ph]}

\bibitem[{{Lamberts} et~al.(2016){Lamberts}, {Garrison-Kimmel}, {Clausen}, and
  {Hopkins}}]{2016MNRAS.463L..31L}
{Lamberts} A, {Garrison-Kimmel} S, {Clausen} DR, {Hopkins} PF (2016) {When and
  where did GW150914 form?} \mnras 463(1):L31--L35.
  \doi{10.1093/mnrasl/slw152}.
  {\href{https://arxiv.org/abs/1605.08783}{{arXiv:1605.08783}}} {[astro-ph.HE]}

\bibitem[{{Lamberts} et~al.(2018){Lamberts}, {Garrison-Kimmel}, {Hopkins},
  {Quataert}, {Bullock}, {Faucher-Gigu{\`e}re}, {Wetzel}, {Kere{\v{s}}},
  {Drango}, and {Sand erson}}]{2018MNRAS.480.2704L}
{Lamberts} A, {Garrison-Kimmel} S, {Hopkins} PF, {Quataert} E, {Bullock} JS,
  {Faucher-Gigu{\`e}re} CA, {Wetzel} A, {Kere{\v{s}}} D, {Drango} K, {Sand
  erson} RE (2018) {Predicting the binary black hole population of the Milky
  Way with cosmological simulations}. \mnras 480(2):2704--2718.
  \doi{10.1093/mnras/sty2035}.
  {\href{https://arxiv.org/abs/1801.03099}{{arXiv:1801.03099}}} {[astro-ph.GA]}

\bibitem[{{Lamberts} et~al.(2019){Lamberts}, {Blunt}, {Littenberg},
  {Garrison-Kimmel}, {Kupfer}, and {Sanderson}}]{2019MNRAS.490.5888L}
{Lamberts} A, {Blunt} S, {Littenberg} TB, {Garrison-Kimmel} S, {Kupfer} T,
  {Sanderson} RE (2019) {Predicting the LISA white dwarf binary population in
  the Milky Way with cosmological simulations}. \mnras 490(4):5888--5903.
  \doi{10.1093/mnras/stz2834}.
  {\href{https://arxiv.org/abs/1907.00014}{{arXiv:1907.00014}}} {[astro-ph.HE]}

\bibitem[{{Lasky} et~al.(2016){Lasky}, {Mingarelli}, {Smith}
  et~al.}]{Lasky:2016}
{Lasky} PD, {Mingarelli} CMF, {Smith} TL, et~al. (2016) {Gravitational-Wave
  Cosmology across 29 Decades in Frequency}. Physical Review X 6(1):011035.
  \doi{10.1103/PhysRevX.6.011035}.
  {\href{https://arxiv.org/abs/1511.05994}{{arXiv:1511.05994}}}

\bibitem[{{Lattimer} and {Prakash}(2016)}]{2016PhR...621..127L}
{Lattimer} JM, {Prakash} M (2016) {The equation of state of hot, dense matter
  and neutron stars}. \physrep 621:127--164.
  \doi{10.1016/j.physrep.2015.12.005}.
  {\href{https://arxiv.org/abs/1512.07820}{{arXiv:1512.07820}}} {[astro-ph.SR]}

\bibitem[{{Lau} et~al.(2020){Lau}, {Mandel}, {Vigna-G{\'o}mez}, {Neijssel},
  {Stevenson}, and {Sesana}}]{2020MNRAS.492.3061L}
{Lau} MYM, {Mandel} I, {Vigna-G{\'o}mez} A, {Neijssel} CJ, {Stevenson} S,
  {Sesana} A (2020) {Detecting double neutron stars with LISA}. \mnras
  492(3):3061--3072. \doi{10.1093/mnras/staa002}.
  {\href{https://arxiv.org/abs/1910.12422}{{arXiv:1910.12422}}} {[astro-ph.HE]}

\bibitem[{{Law-Smith} et~al.(2020){Law-Smith}, {Everson}, {Ramirez-Ruiz}, {de
  Mink}, {van Son}, {G{\"o}tberg}, {Zellmann}, {Vigna-G{\'o}mez}, {Renzo},
  {Wu}, {Schr{\o}der}, {Foley}, and {Hutchinson-Smith}}]{2020arXiv201106630L}
{Law-Smith} JAP, {Everson} RW, {Ramirez-Ruiz} E, {de Mink} SE, {van Son} LAC,
  {G{\"o}tberg} Y, {Zellmann} S, {Vigna-G{\'o}mez} A, {Renzo} M, {Wu} S, et~al.
  (2020) {Successful Common Envelope Ejection and Binary Neutron Star Formation
  in 3D Hydrodynamics}. arXiv e-prints arXiv:2011.06630.
  {\href{https://arxiv.org/abs/2011.06630}{{arXiv:2011.06630}}} {[astro-ph.HE]}

\bibitem[{{Lee}(1995)}]{1995MNRAS.272..605L}
{Lee} HM (1995) {Evolution of galactic nuclei with 10-M\_ black holes}. \mnras
  272(3):605--617. \doi{10.1093/mnras/272.3.605}.
  {\href{https://arxiv.org/abs/astro-ph/9409073}{{arXiv:astro-ph/9409073}}}
  {[astro-ph]}

\bibitem[{{Leigh} and {Geller}(2013)}]{2013MNRAS.432.2474L}
{Leigh} NWC, {Geller} AM (2013) {The dynamical significance of triple star
  systems in star clusters}. \mnras 432(3):2474--2479.
  \doi{10.1093/mnras/stt617}.
  {\href{https://arxiv.org/abs/1304.2775}{{arXiv:1304.2775}}} {[astro-ph.SR]}

\bibitem[{{Leigh} et~al.(2013){Leigh}, {B{\"o}ker}, {Maccarone}, and
  {Perets}}]{2013MNRAS.429.2997L}
{Leigh} NWC, {B{\"o}ker} T, {Maccarone} TJ, {Perets} HB (2013) {Gas depletion
  in primordial globular clusters due to accretion on to stellar-mass black
  holes}. \mnras 429(4):2997--3006. \doi{10.1093/mnras/sts554}.
  {\href{https://arxiv.org/abs/1212.1461}{{arXiv:1212.1461}}} {[astro-ph.SR]}

\bibitem[{{Leigh} et~al.(2014){Leigh}, {L{\"u}tzgendorf}, {Geller},
  {Maccarone}, {Heinke}, and {Sesana}}]{2014MNRAS.444...29L}
{Leigh} NWC, {L{\"u}tzgendorf} N, {Geller} AM, {Maccarone} TJ, {Heinke} C,
  {Sesana} A (2014) {On the coexistence of stellar-mass and intermediate-mass
  black holes in globular clusters}. \mnras 444(1):29--42.
  \doi{10.1093/mnras/stu1437}.
  {\href{https://arxiv.org/abs/1407.4459}{{arXiv:1407.4459}}} {[astro-ph.SR]}

\bibitem[{{Leitao} et~al.(2012){Leitao}, {M{\'e}gevand}, and
  {S{\'a}nchez}}]{2012JCAP...10..024L}
{Leitao} L, {M{\'e}gevand} A, {S{\'a}nchez} AD (2012) {Gravitational waves from
  the electroweak phase transition}. \jcap 2012(10):024.
  \doi{10.1088/1475-7516/2012/10/024}.
  {\href{https://arxiv.org/abs/1205.3070}{{arXiv:1205.3070}}} {[astro-ph.CO]}

\bibitem[{{Lena} et~al.(2014){Lena}, {Robinson}, {Marconi}, {Axon}, {Capetti},
  {Merritt}, and {Batcheldor}}]{2014ApJ...795..146L}
{Lena} D, {Robinson} A, {Marconi} A, {Axon} DJ, {Capetti} A, {Merritt} D,
  {Batcheldor} D (2014) {Recoiling Supermassive Black Holes: A Search in the
  Nearby Universe}. \apj 795(2):146. \doi{10.1088/0004-637X/795/2/146}.
  {\href{https://arxiv.org/abs/1409.3976}{{arXiv:1409.3976}}} {[astro-ph.GA]}

\bibitem[{{Lenon} et~al.(2020){Lenon}, {Nitz}, and
  {Brown}}]{2020MNRAS.497.1966L}
{Lenon} AK, {Nitz} AH, {Brown} DA (2020) {Measuring the eccentricity of
  GW170817 and GW190425}. \mnras 497(2):1966--1971.
  \doi{10.1093/mnras/staa2120}.
  {\href{https://arxiv.org/abs/2005.14146}{{arXiv:2005.14146}}} {[astro-ph.HE]}

\bibitem[{{Leung} et~al.(2020){Leung}, {Leaman}, {van de Ven}, and
  {Battaglia}}]{2020MNRAS.493..320L}
{Leung} GYC, {Leaman} R, {van de Ven} G, {Battaglia} G (2020) {A dwarf-dwarf
  merger and dark matter core as a solution to the globular cluster problems in
  the Fornax dSph}. \mnras 493(1):320--336. \doi{10.1093/mnras/stz3017}.
  {\href{https://arxiv.org/abs/1911.09167}{{arXiv:1911.09167}}} {[astro-ph.GA]}

\bibitem[{{Leveque} et~al.(2022){Leveque}, {Giersz}, {Arca-Sedda}, and
  {Askar}}]{2022MNRAS.514.5751L}
{Leveque} A, {Giersz} M, {Arca-Sedda} M, {Askar} A (2022) {MOCCA-survey data
  base: extra galactic globular clusters - II. Milky Way and Andromeda}. \mnras
  514(4):5751--5766. \doi{10.1093/mnras/stac1694}

\bibitem[{{Levin} et~al.(2013){Levin}, {Bailes}, {Barsdell}, {Bates}, {Bhat},
  {Burgay}, {Burke-Spolaor}, {Champion}, {Coster}, {D'Amico}, {Jameson},
  {Johnston}, {Keith}, {Kramer}, {Milia}, {Ng}, {Possenti}, {Stappers},
  {Thornton}, and {van Straten}}]{2013MNRAS.434.1387L}
{Levin} L, {Bailes} M, {Barsdell} BR, {Bates} SD, {Bhat} NDR, {Burgay} M,
  {Burke-Spolaor} S, {Champion} DJ, {Coster} P, {D'Amico} N, et~al. (2013) {The
  High Time Resolution Universe Pulsar Survey -VIII. The Galactic millisecond
  pulsar population}. \mnras 434(2):1387--1397. \doi{10.1093/mnras/stt1103}.
  {\href{https://arxiv.org/abs/1306.4190}{{arXiv:1306.4190}}} {[astro-ph.SR]}

\bibitem[{{Levin}(2007)}]{2007MNRAS.374..515L}
{Levin} Y (2007) {Starbursts near supermassive black holes: young stars in the
  Galactic Centre, and gravitational waves in LISA band}. \mnras
  374(2):515--524. \doi{10.1111/j.1365-2966.2006.11155.x}.
  {\href{https://arxiv.org/abs/astro-ph/0603583}{{arXiv:astro-ph/0603583}}}
  {[astro-ph]}

\bibitem[{{Levin} and {Beloborodov}(2003)}]{2003ApJ...590L..33L}
{Levin} Y, {Beloborodov} AM (2003) {Stellar Disk in the Galactic Center: A
  Remnant of a Dense Accretion Disk?} \apjl 590(1):L33--L36.
  \doi{10.1086/376675}.
  {\href{https://arxiv.org/abs/astro-ph/0303436}{{arXiv:astro-ph/0303436}}}
  {[astro-ph]}

\bibitem[{{Levine} et~al.(2010){Levine}, {Gnedin}, and
  {Hamilton}}]{2010ApJ...716.1386L}
{Levine} R, {Gnedin} NY, {Hamilton} AJS (2010) {Measuring Gas Accretion and
  Angular Momentum Near Simulated Supermassive Black Holes}. \apj
  716(2):1386--1396. \doi{10.1088/0004-637X/716/2/1386}.
  {\href{https://arxiv.org/abs/1004.3785}{{arXiv:1004.3785}}} {[astro-ph.CO]}

\bibitem[{{Levitan} et~al.(2014){Levitan}, {Kupfer}, {Groot}, {Margon},
  {Prince}, {Kulkarni}, {Hallinan}, {Harding}, {Kyne}, {Laher}, {Ofek},
  {Rutten}, {Sesar}, and {Surace}}]{2014ApJ...785..114L}
{Levitan} D, {Kupfer} T, {Groot} PJ, {Margon} B, {Prince} TA, {Kulkarni} SR,
  {Hallinan} G, {Harding} LK, {Kyne} G, {Laher} R, et~al. (2014) {PTF1
  J191905.19+481506.2{\textemdash}a Partially Eclipsing AM CVn System
  Discovered in the Palomar Transient Factory}. \apj 785(2):114.
  \doi{10.1088/0004-637X/785/2/114}.
  {\href{https://arxiv.org/abs/1402.7129}{{arXiv:1402.7129}}} {[astro-ph.SR]}

\bibitem[{{Lezhnin} and {Vasiliev}(2016)}]{2016ApJ...831...84L}
{Lezhnin} K, {Vasiliev} E (2016) {Tidal Disruption Rates in Non-spherical
  Galactic Nuclei Formed by Galaxy Mergers}. \apj 831(1):84.
  \doi{10.3847/0004-637X/831/1/84}.
  {\href{https://arxiv.org/abs/1609.00009}{{arXiv:1609.00009}}} {[astro-ph.GA]}

\bibitem[{{Li} et~al.(2020{\natexlab{a}}){Li}, {Bogdanovi{\'c}}, and
  {Ballantyne}}]{2020ApJ...896..113L}
{Li} K, {Bogdanovi{\'c}} T, {Ballantyne} DR (2020{\natexlab{a}}) {Pairing of
  Massive Black Holes in Merger Galaxies Driven by Dynamical Friction}. \apj
  896(2):113. \doi{10.3847/1538-4357/ab93c6}.
  {\href{https://arxiv.org/abs/2006.08520}{{arXiv:2006.08520}}} {[astro-ph.GA]}

\bibitem[{{Li} et~al.(2020{\natexlab{b}}){Li}, {Bogdanovic}, and
  {Ballantyne}}]{2020arXiv200702051L}
{Li} K, {Bogdanovic} T, {Ballantyne} DR (2020{\natexlab{b}}) {The Pairing
  Probability of Massive Black Holes in Merger Galaxies in the Presence of
  Radiative Feedback}. arXiv e-prints arXiv:2007.02051.
  {\href{https://arxiv.org/abs/2007.02051}{{arXiv:2007.02051}}} {[astro-ph.GA]}

\bibitem[{{Li} et~al.(2019){Li}, {Wu}, and {Singh}}]{2019MNRAS.485.1053L}
{Li} KJ, {Wu} K, {Singh} D (2019) {Spin dynamics of a millisecond pulsar
  orbiting closely around a massive black hole}. \mnras 485(1):1053--1066.
  \doi{10.1093/mnras/stz389}.
  {\href{https://arxiv.org/abs/1902.03146}{{arXiv:1902.03146}}} {[astro-ph.HE]}

\bibitem[{{Li} et~al.(2009){Li}, {Narayan}, and
  {McClintock}}]{2009ApJ...691..847L}
{Li} LX, {Narayan} R, {McClintock} JE (2009) {Inferring the Inclination of a
  Black Hole Accretion Disk from Observations of its Polarized Continuum
  Radiation}. \apj 691(1):847--865. \doi{10.1088/0004-637X/691/1/847}.
  {\href{https://arxiv.org/abs/0809.0866}{{arXiv:0809.0866}}} {[astro-ph]}

\bibitem[{{Li} et~al.(2017){Li}, {Liu}, {Berczik}, and
  {Spurzem}}]{2017ApJ...834..195L}
{Li} S, {Liu} FK, {Berczik} P, {Spurzem} R (2017) {Boosted Tidal Disruption by
  Massive Black Hole Binaries During Galaxy Mergers from the View of N-Body
  Simulation}. \apj 834(2):195. \doi{10.3847/1538-4357/834/2/195}.
  {\href{https://arxiv.org/abs/1509.00158}{{arXiv:1509.00158}}} {[astro-ph.GA]}

\bibitem[{{Li} et~al.(2018){Li}, {Simon}, {Kuehn}, {Pace}, {Erkal}, {Bechtol},
  {Yanny}, {Drlica-Wagner}, {Marshall}, {Lidman}, {Balbinot}, {Carollo},
  {Jenkins}, {Mart{\'\i}nez-V{\'a}zquez}, {Shipp}, {Stringer}, {Vivas},
  {Walker}, {Wechsler}, {Abdalla}, {Allam}, {Annis}, {Avila}, {Bertin},
  {Brooks}, {Buckley-Geer}, {Burke}, {Carnero Rosell}, {Carrasco Kind},
  {Carretero}, {Cunha}, {D'Andrea}, {da Costa}, {Davis}, {De Vicente}, {Doel},
  {Eifler}, {Evrard}, {Flaugher}, {Frieman}, {Garc{\'\i}a-Bellido},
  {Gaztanaga}, {Gerdes}, {Gruen}, {Gruendl}, {Gschwend}, {Gutierrez},
  {Hartley}, {Hollowood}, {Honscheid}, {James}, {Krause}, {Maia}, {March},
  {Menanteau}, {Miquel}, {Plazas}, {Sanchez}, {Santiago}, {Scarpine},
  {Schindler}, {Schubnell}, {Sevilla-Noarbe}, {Smith}, {Smith},
  {Soares-Santos}, {Sobreira}, {Suchyta}, {Swanson}, {Tarle}, {Tucker}, and
  {DES Collaboration}}]{2018ApJ...866...22L}
{Li} TS, {Simon} JD, {Kuehn} K, {Pace} AB, {Erkal} D, {Bechtol} K, {Yanny} B,
  {Drlica-Wagner} A, {Marshall} JL, {Lidman} C, et~al. (2018) {The First
  Tidally Disrupted Ultra-faint Dwarf Galaxy?: A Spectroscopic Analysis of the
  Tucana III Stream}. \apj 866(1):22. \doi{10.3847/1538-4357/aadf91}.
  {\href{https://arxiv.org/abs/1804.07761}{{arXiv:1804.07761}}} {[astro-ph.GA]}

\bibitem[{{Li} et~al.(2020{\natexlab{c}}){Li}, {Ni}, {Croft}, {Di Matteo},
  {Bird}, and {Feng}}]{2020arXiv201006608L}
{Li} Y, {Ni} Y, {Croft} RAC, {Di Matteo} T, {Bird} S, {Feng} Y
  (2020{\natexlab{c}}) {AI-assisted super-resolution cosmological simulations}.
  arXiv e-prints arXiv:2010.06608.
  {\href{https://arxiv.org/abs/2010.06608}{{arXiv:2010.06608}}} {[astro-ph.CO]}

\bibitem[{{Lidov}(1962)}]{1962P&SS....9..719L}
{Lidov} ML (1962) {The evolution of orbits of artificial satellites of planets
  under the action of gravitational perturbations of external bodies}. \planss
  9(10):719--759. \doi{10.1016/0032-0633(62)90129-0}

\bibitem[{{Lim} and {Rodriguez}(2020)}]{2020PhRvD.102f4033L}
{Lim} H, {Rodriguez} CL (2020) {Relativistic three-body effects in hierarchical
  triples}. \prd 102(6):064033. \doi{10.1103/PhysRevD.102.064033}.
  {\href{https://arxiv.org/abs/2001.03654}{{arXiv:2001.03654}}} {[astro-ph.HE]}

\bibitem[{{Lippai} et~al.(2008){Lippai}, {Frei}, and
  {Haiman}}]{2008ApJ...676L...5L}
{Lippai} Z, {Frei} Z, {Haiman} Z (2008) {Prompt Shocks in the Gas Disk around a
  Recoiling Supermassive Black Hole Binary}. \apjl 676(1):L5.
  \doi{10.1086/587034}.
  {\href{https://arxiv.org/abs/0801.0739}{{arXiv:0801.0739}}} {[astro-ph]}

\bibitem[{{Liptai} and {Price}(2019)}]{2019MNRAS.485..819L}
{Liptai} D, {Price} DJ (2019) {General relativistic smoothed particle
  hydrodynamics}. \mnras 485(1):819--842. \doi{10.1093/mnras/stz111}.
  {\href{https://arxiv.org/abs/1901.08064}{{arXiv:1901.08064}}} {[astro-ph.IM]}

\bibitem[{{Lipunov} et~al.(1996){Lipunov}, {Postnov}, and
  {Prokhorov}}]{1996smbs.book.....L}
{Lipunov} VM, {Postnov} KA, {Prokhorov} ME (1996) {The scenario machine: Binary
  star population synthesis}

\bibitem[{{Lipunov} et~al.(2009){Lipunov}, {Postnov}, {Prokhorov}, and
  {Bogomazov}}]{2009ARep...53..915L}
{Lipunov} VM, {Postnov} KA, {Prokhorov} ME, {Bogomazov} AI (2009) {Description
  of the ``Scenario Machine''}. Astronomy Reports 53(10):915--940.
  \doi{10.1134/S1063772909100047}.
  {\href{https://arxiv.org/abs/0704.1387}{{arXiv:0704.1387}}} {[astro-ph]}

\bibitem[{{LISA Pathfinder Collaboration}(2022)}]{2022arXiv220511938L}
{LISA Pathfinder Collaboration} (2022) {Transient acceleration events in LISA
  Pathfinder: properties and possible physical origin}. arXiv e-prints
  arXiv:2205.11938.
  {\href{https://arxiv.org/abs/2205.11938}{{arXiv:2205.11938}}} {[astro-ph.IM]}

\bibitem[{{Littenberg} and {Cornish}(2019)}]{2019ApJ...881L..43L}
{Littenberg} TB, {Cornish} NJ (2019) {Prospects for Gravitational Wave
  Measurement of ZTF J1539+5027}. \apjl 881(2):L43.
  \doi{10.3847/2041-8213/ab385f}.
  {\href{https://arxiv.org/abs/1908.00678}{{arXiv:1908.00678}}} {[astro-ph.IM]}

\bibitem[{{Littenberg} et~al.(2013){Littenberg}, {Larson}, {Nelemans}, and
  {Cornish}}]{2013MNRAS.429.2361L}
{Littenberg} TB, {Larson} SL, {Nelemans} G, {Cornish} NJ (2013) {Prospects for
  observing ultracompact binaries with space-based gravitational wave
  interferometers and optical telescopes}. \mnras 429(3):2361--2365.
  \doi{10.1093/mnras/sts507}.
  {\href{https://arxiv.org/abs/1207.4848}{{arXiv:1207.4848}}} {[astro-ph.IM]}

\bibitem[{{Littenberg} et~al.(2020){Littenberg}, {Cornish}, {Lackeos}, and
  {Robson}}]{2020PhRvD.101l3021L}
{Littenberg} TB, {Cornish} NJ, {Lackeos} K, {Robson} T (2020) {Global analysis
  of the gravitational wave signal from Galactic binaries}. \prd
  101(12):123021. \doi{10.1103/PhysRevD.101.123021}.
  {\href{https://arxiv.org/abs/2004.08464}{{arXiv:2004.08464}}} {[gr-qc]}

\bibitem[{{Liu} and {Lai}(2017)}]{2017ApJ...846L..11L}
{Liu} B, {Lai} D (2017) {Spin-Orbit Misalignment of Merging Black Hole Binaries
  with Tertiary Companions}. \apjl 846(1):L11. \doi{10.3847/2041-8213/aa8727}.
  {\href{https://arxiv.org/abs/1706.02309}{{arXiv:1706.02309}}} {[astro-ph.HE]}

\bibitem[{{Liu} and {Lai}(2018)}]{2018ApJ...863...68L}
{Liu} B, {Lai} D (2018) {Black Hole and Neutron Star Binary Mergers in Triple
  Systems: Merger Fraction and Spin-Orbit Misalignment}. \apj 863(1):68.
  \doi{10.3847/1538-4357/aad09f}.
  {\href{https://arxiv.org/abs/1805.03202}{{arXiv:1805.03202}}} {[astro-ph.HE]}

\bibitem[{{Liu} et~al.(2020){Liu}, {Shao}, {Zhao}, and
  {Gao}}]{2020MNRAS.496..182L}
{Liu} C, {Shao} L, {Zhao} J, {Gao} Y (2020) {Multiband observation of
  LIGO/Virgo binary black hole mergers in the gravitational-wave transient
  catalog GWTC-1}. MNRAS 496(1):182--196. \doi{10.1093/mnras/staa1512}.
  {\href{https://arxiv.org/abs/2004.12096}{{arXiv:2004.12096}}} {[astro-ph.HE]}

\bibitem[{{Liu} et~al.(2010){Liu}, {Han}, {Zhang}, and
  {Zhang}}]{2010ApJ...719.1546L}
{Liu} J, {Han} Z, {Zhang} F, {Zhang} Y (2010) {A Comprehensive Study of Close
  Double White Dwarfs as Gravitational Wave Sources: Evolutionary Channels,
  Birth Rates, and Physical Properties}. \apj 719(2):1546--1552.
  \doi{10.1088/0004-637X/719/2/1546}

\bibitem[{{Liu} et~al.(2019{\natexlab{a}}){Liu}, {Zhang}, {Howard}, {Bai},
  {Lu}, {Soria}, {Justham}, {Li}, {Zheng}, and ...}]{2019Natur.575..618L}
{Liu} J, {Zhang} H, {Howard} AW, {Bai} Z, {Lu} Y, {Soria} R, {Justham} S, {Li}
  X, {Zheng} Z,  (2019{\natexlab{a}}) {A wide star-black-hole binary system
  from radial-velocity measurements}. \nat 575(7784):618--621.
  \doi{10.1038/s41586-019-1766-2}.
  {\href{https://arxiv.org/abs/1911.11989}{{arXiv:1911.11989}}} {[astro-ph.SR]}

\bibitem[{{Liu} et~al.(2019{\natexlab{b}}){Liu}, {Gezari}, {Ayers}, {Burgett},
  {Chambers}, {Hodapp}, {Huber}, {Kudritzki}, {Metcalfe}, {Tonry}, {Wainscoat},
  and {Waters}}]{2019ApJ...884...36L}
{Liu} T, {Gezari} S, {Ayers} M, {Burgett} W, {Chambers} K, {Hodapp} K, {Huber}
  ME, {Kudritzki} RP, {Metcalfe} N, {Tonry} J, et~al. (2019{\natexlab{b}})
  {Supermassive Black Hole Binary Candidates from the Pan-STARRS1 Medium Deep
  Survey}. \apj 884(1):36. \doi{10.3847/1538-4357/ab40cb}.
  {\href{https://arxiv.org/abs/1906.08315}{{arXiv:1906.08315}}} {[astro-ph.HE]}

\bibitem[{{Liu} et~al.(2014){Liu}, {Shen}, {Bian}, {Loeb}, and
  {Tremaine}}]{2014ApJ...789..140L}
{Liu} X, {Shen} Y, {Bian} F, {Loeb} A, {Tremaine} S (2014) {Constraining
  Sub-parsec Binary Supermassive Black Holes in Quasars with Multi-epoch
  Spectroscopy. II. The Population with Kinematically Offset Broad Balmer
  Emission Lines}. \apj 789(2):140. \doi{10.1088/0004-637X/789/2/140}.
  {\href{https://arxiv.org/abs/1312.6694}{{arXiv:1312.6694}}} {[astro-ph.CO]}

\bibitem[{Liu et~al.(2007)Liu, Shapiro, and Stephens}]{Liu:2007cf}
Liu YT, Shapiro SL, Stephens BC (2007) {Magnetorotational collapse of very
  massive stars in full general relativity}. Phys Rev D 76:084017.
  \doi{10.1103/PhysRevD.76.084017}.
  {\href{https://arxiv.org/abs/0706.2360}{{arXiv:0706.2360}}} {[astro-ph]}

\bibitem[{{Livio} and {Soker}(1984)}]{1984MNRAS.208..763L}
{Livio} M, {Soker} N (1984) {Star-planet systems as possible progenitors of
  cataclysmic binaries.} \mnras 208:763--781. \doi{10.1093/mnras/208.4.763}

\bibitem[{{Lodato} and {Gerosa}(2013)}]{2013MNRAS.429L..30L}
{Lodato} G, {Gerosa} D (2013) {Black hole mergers: do gas discs lead to spin
  alignment?} \mnras 429:L30--L34. \doi{10.1093/mnrasl/sls018}.
  {\href{https://arxiv.org/abs/1211.0284}{{arXiv:1211.0284}}} {[astro-ph.CO]}

\bibitem[{{Lodato} and {Natarajan}(2006)}]{2006MNRAS.371.1813L}
{Lodato} G, {Natarajan} P (2006) {Supermassive black hole formation during the
  assembly of pre-galactic discs}. \mnras 371(4):1813--1823.
  \doi{10.1111/j.1365-2966.2006.10801.x}.
  {\href{https://arxiv.org/abs/astro-ph/0606159}{{arXiv:astro-ph/0606159}}}
  {[astro-ph]}

\bibitem[{{Lodato} and {Rossi}(2011)}]{2011MNRAS.410..359L}
{Lodato} G, {Rossi} EM (2011) {Multiband light curves of tidal disruption
  events}. \mnras 410(1):359--367. \doi{10.1111/j.1365-2966.2010.17448.x}.
  {\href{https://arxiv.org/abs/1008.4589}{{arXiv:1008.4589}}} {[astro-ph.CO]}

\bibitem[{{Lodato} et~al.(2009){Lodato}, {Nayakshin}, {King}, and
  {Pringle}}]{Lodato09}
{Lodato} G, {Nayakshin} S, {King} AR, {Pringle} JE (2009) {Black hole mergers:
  can gas discs solve the `final parsec' problem?} \mnras 398(3):1392--1402.
  \doi{10.1111/j.1365-2966.2009.15179.x}.
  {\href{https://arxiv.org/abs/0906.0737}{{arXiv:0906.0737}}} {[astro-ph.CO]}

\bibitem[{{Lops} et~al.(2022){Lops}, {Izquierdo-Villalba}, {Colpi}, {Bonoli},
  {Sesana}, and {Mangiagli}}]{2022arXiv220710683L}
{Lops} G, {Izquierdo-Villalba} D, {Colpi} M, {Bonoli} S, {Sesana} A,
  {Mangiagli} A (2022) {Galaxy fields of LISA massive black hole mergers in a
  simulated Universe}. arXiv e-prints arXiv:2207.10683.
  {\href{https://arxiv.org/abs/2207.10683}{{arXiv:2207.10683}}} {[astro-ph.GA]}

\bibitem[{{Lousto} and {Zlochower}(2008)}]{2008PhRvD..77d4028L}
{Lousto} CO, {Zlochower} Y (2008) {Further insight into gravitational recoil}.
  \prd 77(4):044028. \doi{10.1103/PhysRevD.77.044028}.
  {\href{https://arxiv.org/abs/0708.4048}{{arXiv:0708.4048}}} {[gr-qc]}

\bibitem[{{Lousto} and {Zlochower}(2011)}]{2011PhRvL.107w1102L}
{Lousto} CO, {Zlochower} Y (2011) {Hangup Kicks: Still Larger Recoils by
  Partial Spin-Orbit Alignment of Black-Hole Binaries}. \prl 107(23):231102.
  \doi{10.1103/PhysRevLett.107.231102}.
  {\href{https://arxiv.org/abs/1108.2009}{{arXiv:1108.2009}}} {[gr-qc]}

\bibitem[{{Lousto} and {Zlochower}(2013)}]{2013PhRvD..87h4027L}
{Lousto} CO, {Zlochower} Y (2013) {Nonlinear gravitational recoil from the
  mergers of precessing black-hole binaries}. \prd 87(8):084027.
  \doi{10.1103/PhysRevD.87.084027}.
  {\href{https://arxiv.org/abs/1211.7099}{{arXiv:1211.7099}}} {[gr-qc]}

\bibitem[{{Lousto} et~al.(2010){Lousto}, {Campanelli}, {Zlochower}, and
  {Nakano}}]{2010CQGra..27k4006L}
{Lousto} CO, {Campanelli} M, {Zlochower} Y, {Nakano} H (2010) {Remnant masses,
  spins and recoils from the merger of generic black hole binaries}. Classical
  and Quantum Gravity 27(11):114006. \doi{10.1088/0264-9381/27/11/114006}.
  {\href{https://arxiv.org/abs/0904.3541}{{arXiv:0904.3541}}} {[gr-qc]}

\bibitem[{{Lousto} et~al.(2012){Lousto}, {Zlochower}, {Dotti}, and
  {Volonteri}}]{2012PhRvD..85h4015L}
{Lousto} CO, {Zlochower} Y, {Dotti} M, {Volonteri} M (2012) {Gravitational
  recoil from accretion-aligned black-hole binaries}. \prd 85(8):084015.
  \doi{10.1103/PhysRevD.85.084015}.
  {\href{https://arxiv.org/abs/1201.1923}{{arXiv:1201.1923}}} {[gr-qc]}

\bibitem[{{Lovelace} et~al.(2011){Lovelace}, {Scheel}, and
  {Szil{\'a}gyi}}]{2011PhRvD..83b4010L}
{Lovelace} G, {Scheel} MA, {Szil{\'a}gyi} B (2011) {Simulating merging binary
  black holes with nearly extremal spins}. \prd 83(2):024010.
  \doi{10.1103/PhysRevD.83.024010}.
  {\href{https://arxiv.org/abs/1010.2777}{{arXiv:1010.2777}}} {[gr-qc]}

\bibitem[{{Loveridge} et~al.(2011){Loveridge}, {van der Sluys}, and
  {Kalogera}}]{2011ApJ...743...49L}
{Loveridge} AJ, {van der Sluys} MV, {Kalogera} V (2011) {Analytical Expressions
  for the Envelope Binding Energy of Giants as a Function of Basic Stellar
  Parameters}. \apj 743(1):49. \doi{10.1088/0004-637X/743/1/49}.
  {\href{https://arxiv.org/abs/1009.5400}{{arXiv:1009.5400}}} {[astro-ph.SR]}

\bibitem[{{LSST Science Collaboration} et~al.(2009){LSST Science
  Collaboration}, {Abell}, {Allison}, {Anderson}, {Andrew}, {Angel}, {Armus},
  {Arnett}, {Asztalos}, {Axelrod}, {Bailey}, {Ballantyne}, {Bankert},
  {Barkhouse}, {Barr}, {Barrientos}, {Barth}, {Bartlett}, {Becker}, {Becla},
  {Beers}, {Bernstein}, {Biswas}, {Blanton}, {Bloom}, {Bochanski}, {Boeshaar},
  {Borne}, {Bradac}, {Brandt}, {Bridge}, {Brown}, {Brunner}, {Bullock},
  {Burgasser}, {Burge}, {Burke}, {Cargile}, {Chand rasekharan}, {Chartas},
  {Chesley}, {Chu}, {Cinabro}, {Claire}, {Claver}, {Clowe}, {Connolly}, {Cook},
  {Cooke}, {Cooray}, {Covey}, {Culliton}, {de Jong}, {de Vries}, {Debattista},
  {Delgado}, {Dell'Antonio}, {Dhital}, {Di Stefano}, {Dickinson}, {Dilday},
  {Djorgovski}, {Dobler}, {Donalek}, {Dubois-Felsmann}, {Durech},
  {Eliasdottir}, {Eracleous}, {Eyer}, {Falco}, {Fan}, {Fassnacht}, {Ferguson},
  {Fernandez}, {Fields}, {Finkbeiner}, {Figueroa}, {Fox}, {Francke}, {Frank},
  {Frieman}, {Fromenteau}, {Furqan}, {Galaz}, {Gal-Yam}, {Garnavich},
  {Gawiser}, {Geary}, {Gee}, {Gibson}, {Gilmore}, {Grace}, {Green}, {Gressler},
  {Grillmair}, {Habib}, {Haggerty}, {Hamuy}, {Harris}, {Hawley}, {Heavens},
  {Hebb}, {Henry}, {Hileman}, {Hilton}, {Hoadley}, {Holberg}, {Holman},
  {Howell}, {Infante}, {Ivezic}, {Jacoby}, {Jain}, {R}, {Jedicke}, {Jee},
  {Garrett Jernigan}, {Jha}, {Johnston}, {Jones}, {Juric}, {Kaasalainen},
  {Styliani}, {Kafka}, {Kahn}, {Kaib}, {Kalirai}, {Kantor}, {Kasliwal},
  {Keeton}, {Kessler}, {Knezevic}, {Kowalski}, {Krabbendam}, {Krughoff},
  {Kulkarni}, {Kuhlman}, {Lacy}, {Lepine}, {Liang}, {Lien}, {Lira}, {Long},
  {Lorenz}, {Lotz}, {Lupton}, {Lutz}, {Macri}, {Mahabal}, {Mandelbaum},
  {Marshall}, {May}, {McGehee}, {Meadows}, {Meert}, {Milani}, {Miller},
  {Miller}, {Mills}, {Minniti}, {Monet}, {Mukadam}, {Nakar}, {Neill}, {Newman},
  {Nikolaev}, {Nordby}, {O'Connor}, {Oguri}, {Oliver}, {Olivier}, {Olsen},
  {Olsen}, {Olszewski}, {Oluseyi}, {Padilla}, {Parker}, {Pepper}, {Peterson},
  {Petry}, {Pinto}, {Pizagno}, {Popescu}, {Prsa}, {Radcka}, {Raddick},
  {Rasmussen}, {Rau}, {Rho}, {Rhoads}, {Richards}, {Ridgway}, {Robertson},
  {Roskar}, {Saha}, {Sarajedini}, {Scannapieco}, {Schalk}, {Schindler},
  {Schmidt}, {Schmidt}, {Schneider}, {Schumacher}, {Scranton}, {Sebag},
  {Seppala}, {Shemmer}, {Simon}, {Sivertz}, {Smith}, {Allyn Smith}, {Smith},
  {Spitz}, {Stanford}, {Stassun}, {Strader}, {Strauss}, {Stubbs}, {Sweeney},
  {Szalay}, {Szkody}, {Takada}, {Thorman}, {Trilling}, {Trimble}, {Tyson}, {Van
  Berg}, {Vand en Berk}, {VanderPlas}, {Verde}, {Vrsnak}, {Walkowicz}, {Wand
  elt}, {Wang}, {Wang}, {Warner}, {Wechsler}, {West}, {Wiecha}, {Williams},
  {Willman}, {Wittman}, {Wolff}, {Wood-Vasey}, {Wozniak}, {Young}, {Zentner},
  and {Zhan}}]{2009arXiv0912.0201L}
{LSST Science Collaboration}, {Abell} PA, {Allison} J, {Anderson} SF, {Andrew}
  JR, {Angel} JRP, {Armus} L, {Arnett} D, {Asztalos} SJ, {Axelrod} TS, et~al.
  (2009) {LSST Science Book, Version 2.0}. arXiv e-prints arXiv:0912.0201.
  {\href{https://arxiv.org/abs/0912.0201}{{arXiv:0912.0201}}} {[astro-ph.IM]}

\bibitem[{{Lu} et~al.(2020){Lu}, {Beniamini}, and
  {Bonnerot}}]{2020arXiv200910082L}
{Lu} W, {Beniamini} P, {Bonnerot} C (2020) {On the formation of GW190814}.
  arXiv e-prints arXiv:2009.10082.
  {\href{https://arxiv.org/abs/2009.10082}{{arXiv:2009.10082}}} {[astro-ph.HE]}

\bibitem[{{Lukes-Gerakopoulos} and {Witzany}(2021)}]{2021hgwa.bookE..42L}
{Lukes-Gerakopoulos} G, {Witzany} V (2021) {Nonlinear Effects in EMRI Dynamics
  and Their Imprints on Gravitational Waves}. In: Handbook of Gravitational
  Wave Astronomy. p~42. \doi{10.1007/978-981-15-4702-7\_42-1}

\bibitem[{{Lukes-Gerakopoulos} et~al.(2010){Lukes-Gerakopoulos}, {Apostolatos},
  and {Contopoulos}}]{2010PhRvD..81l4005L}
{Lukes-Gerakopoulos} G, {Apostolatos} TA, {Contopoulos} G (2010) {Observable
  signature of a background deviating from the Kerr metric}. \prd
  81(12):124005. \doi{10.1103/PhysRevD.81.124005}.
  {\href{https://arxiv.org/abs/1003.3120}{{arXiv:1003.3120}}} {[gr-qc]}

\bibitem[{{Lukes-Gerakopoulos} et~al.(2014){Lukes-Gerakopoulos}, {Seyrich}, and
  {Kunst}}]{2014PhRvD..90j4019L}
{Lukes-Gerakopoulos} G, {Seyrich} J, {Kunst} D (2014) {Investigating spinning
  test particles: Spin supplementary conditions and the Hamiltonian formalism}.
  \prd 90(10):104019. \doi{10.1103/PhysRevD.90.104019}.
  {\href{https://arxiv.org/abs/1409.4314}{{arXiv:1409.4314}}} {[gr-qc]}

\bibitem[{{Luo} et~al.(2016){Luo}, {Chen}, {Duan}, {Gong}, {Hu}, {Ji}, {Liu},
  {Mei}, {Milyukov}, {Sazhin}, {Shao}, {Toth}, {Tu}, {Wang}, {Wang}, {Yeh},
  {Zhan}, {Zhang}, {Zharov}, and {Zhou}}]{2016CQGra..33c5010L}
{Luo} J, {Chen} LS, {Duan} HZ, {Gong} YG, {Hu} S, {Ji} J, {Liu} Q, {Mei} J,
  {Milyukov} V, {Sazhin} M, et~al. (2016) {TianQin: a space-borne gravitational
  wave detector}. Classical and Quantum Gravity 33(3):035010.
  \doi{10.1088/0264-9381/33/3/035010}.
  {\href{https://arxiv.org/abs/1512.02076}{{arXiv:1512.02076}}} {[astro-ph.IM]}

\bibitem[{{Lupi} et~al.(2016){Lupi}, {Haardt}, {Dotti}, {Fiacconi}, {Mayer},
  and {Madau}}]{2016MNRAS.456.2993L}
{Lupi} A, {Haardt} F, {Dotti} M, {Fiacconi} D, {Mayer} L, {Madau} P (2016)
  {Growing massive black holes through supercritical accretion of stellar-mass
  seeds}. \mnras 456:2993--3003. \doi{10.1093/mnras/stv2877}.
  {\href{https://arxiv.org/abs/1512.02651}{{arXiv:1512.02651}}}

\bibitem[{{Luyten}(1949)}]{1949ApJ...109..528L}
{Luyten} WJ (1949) {An Atlas of Identification Charts of White Dwarfs.} \apj
  109:528. \doi{10.1086/145156}

\bibitem[{{Lyne} et~al.(2004){Lyne}, {Burgay}, {Kramer}, {Possenti},
  {Manchester}, {Camilo}, {McLaughlin}, {Lorimer}, {D'Amico}, {Joshi},
  {Reynolds}, and {Freire}}]{2004Sci...303.1153L}
{Lyne} AG, {Burgay} M, {Kramer} M, {Possenti} A, {Manchester} RN, {Camilo} F,
  {McLaughlin} MA, {Lorimer} DR, {D'Amico} N, {Joshi} BC, et~al. (2004) {A
  Double-Pulsar System: A Rare Laboratory for Relativistic Gravity and Plasma
  Physics}. Science 303(5661):1153--1157. \doi{10.1126/science.1094645}.
  {\href{https://arxiv.org/abs/astro-ph/0401086}{{arXiv:astro-ph/0401086}}}
  {[astro-ph]}

\bibitem[{{Ma} et~al.(2021){Ma}, {Hopkins}, {Ma}, {Angl{\'e}s-Alc{\'a}zar},
  {Faucher-Gigu{\`e}re}, and {Kelley}}]{2021MNRAS.508.1973M}
{Ma} L, {Hopkins} PF, {Ma} X, {Angl{\'e}s-Alc{\'a}zar} D, {Faucher-Gigu{\`e}re}
  CA, {Kelley} LZ (2021) {Seeds don't sink: even massive black hole 'seeds'
  cannot migrate to galaxy centres efficiently}. \mnras 508(2):1973--1985.
  \doi{10.1093/mnras/stab2713}.
  {\href{https://arxiv.org/abs/2101.02727}{{arXiv:2101.02727}}} {[astro-ph.GA]}

\bibitem[{{Maccarone} et~al.(2007){Maccarone}, {Kundu}, {Zepf}, and
  {Rhode}}]{2007Natur.445..183M}
{Maccarone} TJ, {Kundu} A, {Zepf} SE, {Rhode} KL (2007) {A black hole in a
  globular cluster}. \nat 445(7124):183--185. \doi{10.1038/nature05434}.
  {\href{https://arxiv.org/abs/astro-ph/0701310}{{arXiv:astro-ph/0701310}}}
  {[astro-ph]}

\bibitem[{{MacFadyen} and {Milosavljevi{\'c}}(2008)}]{2008ApJ...672...83M}
{MacFadyen} AI, {Milosavljevi{\'c}} M (2008) {An Eccentric Circumbinary
  Accretion Disk and the Detection of Binary Massive Black Holes}. \apj
  672(1):83--93. \doi{10.1086/523869}.
  {\href{https://arxiv.org/abs/astro-ph/0607467}{{arXiv:astro-ph/0607467}}}
  {[astro-ph]}

\bibitem[{{Macfarlane} et~al.(2015){Macfarlane}, {Toma}, {Ramsay}, {Groot},
  {Woudt}, {Drew}, {Barentsen}, and {Eisl{\"o}ffel}}]{2015MNRAS.454..507M}
{Macfarlane} SA, {Toma} R, {Ramsay} G, {Groot} PJ, {Woudt} PA, {Drew} JE,
  {Barentsen} G, {Eisl{\"o}ffel} J (2015) {The OmegaWhite survey for
  short-period variable stars - I. Overview and first results}. \mnras
  454(1):507--530. \doi{10.1093/mnras/stv1989}.
  {\href{https://arxiv.org/abs/1508.06277}{{arXiv:1508.06277}}} {[astro-ph.SR]}

\bibitem[{Mack et~al.(2007)Mack, Ostriker, and Ricotti}]{Mack:2006gz}
Mack KJ, Ostriker JP, Ricotti M (2007) {Growth of structure seeded by
  primordial black holes}. Astrophys J 665:1277--1287. \doi{10.1086/518998}.
  {\href{https://arxiv.org/abs/astro-ph/0608642}{{arXiv:astro-ph/0608642}}}

\bibitem[{{Mackey} et~al.(2007){Mackey}, {Wilkinson}, {Davies}, and
  {Gilmore}}]{2007MNRAS.379L..40M}
{Mackey} AD, {Wilkinson} MI, {Davies} MB, {Gilmore} GF (2007) {The effect of
  stellar-mass black holes on the structural evolution of massive star
  clusters}. \mnras 379(1):L40--L44. \doi{10.1111/j.1745-3933.2007.00330.x}.
  {\href{https://arxiv.org/abs/0704.2494}{{arXiv:0704.2494}}} {[astro-ph]}

\bibitem[{{Mackey} et~al.(2008){Mackey}, {Wilkinson}, {Davies}, and
  {Gilmore}}]{2008MNRAS.386...65M}
{Mackey} AD, {Wilkinson} MI, {Davies} MB, {Gilmore} GF (2008) {Black holes and
  core expansion in massive star clusters}. \mnras 386(1):65--95.
  \doi{10.1111/j.1365-2966.2008.13052.x}.
  {\href{https://arxiv.org/abs/0802.0513}{{arXiv:0802.0513}}} {[astro-ph]}

\bibitem[{MacLeod and Hogan(2008)}]{MacLeod:2007jd}
MacLeod CL, Hogan CJ (2008) {Precision of Hubble constant derived using black
  hole binary absolute distances and statistical redshift information}. Phys
  Rev D 77:043512. \doi{10.1103/PhysRevD.77.043512}.
  {\href{https://arxiv.org/abs/0712.0618}{{arXiv:0712.0618}}} {[astro-ph]}

\bibitem[{{MacLeod} and {Lin}(2020)}]{2020ApJ...889...94M}
{MacLeod} M, {Lin} DNC (2020) {The Effect of Star-Disk Interactions on Highly
  Eccentric Stellar Orbits in Active Galactic Nuclei: A Disk Loss Cone and
  Implications for Stellar Tidal Disruption Events}. \apj 889(2):94.
  \doi{10.3847/1538-4357/ab64db}.
  {\href{https://arxiv.org/abs/1909.09645}{{arXiv:1909.09645}}} {[astro-ph.SR]}

\bibitem[{{MacLeod} and {Ramirez-Ruiz}(2015)}]{2015ApJ...798L..19M}
{MacLeod} M, {Ramirez-Ruiz} E (2015) {On the Accretion-fed Growth of Neutron
  Stars during Common Envelope}. \apjl 798(1):L19.
  \doi{10.1088/2041-8205/798/1/L19}.
  {\href{https://arxiv.org/abs/1410.5421}{{arXiv:1410.5421}}} {[astro-ph.SR]}

\bibitem[{{MacLeod} et~al.(2016{\natexlab{a}}){MacLeod}, {Guillochon},
  {Ramirez-Ruiz}, {Kasen}, and {Rosswog}}]{2016ApJ...819....3M}
{MacLeod} M, {Guillochon} J, {Ramirez-Ruiz} E, {Kasen} D, {Rosswog} S
  (2016{\natexlab{a}}) {Optical Thermonuclear Transients from Tidal Compression
  of White Dwarfs as Tracers of the Low End of the Massive Black Hole Mass
  Function}. \apj 819(1):3. \doi{10.3847/0004-637X/819/1/3}.
  {\href{https://arxiv.org/abs/1508.02399}{{arXiv:1508.02399}}} {[astro-ph.HE]}

\bibitem[{{MacLeod} et~al.(2016{\natexlab{b}}){MacLeod}, {Trenti}, and
  {Ramirez-Ruiz}}]{2016ApJ...819...70M}
{MacLeod} M, {Trenti} M, {Ramirez-Ruiz} E (2016{\natexlab{b}}) {The Close
  Stellar Companions to Intermediate-mass Black Holes}. \apj 819(1):70.
  \doi{10.3847/0004-637X/819/1/70}.
  {\href{https://arxiv.org/abs/1508.07000}{{arXiv:1508.07000}}} {[astro-ph.HE]}

\bibitem[{{MacLeod} et~al.(2017){MacLeod}, {Antoni}, {Murguia-Berthier},
  {Macias}, and {Ramirez-Ruiz}}]{2017ApJ...838...56M}
{MacLeod} M, {Antoni} A, {Murguia-Berthier} A, {Macias} P, {Ramirez-Ruiz} E
  (2017) {Common Envelope Wind Tunnel: Coefficients of Drag and Accretion in a
  Simplified Context for Studying Flows around Objects Embedded within Stellar
  Envelopes}. \apj 838(1):56. \doi{10.3847/1538-4357/aa6117}.
  {\href{https://arxiv.org/abs/1704.02372}{{arXiv:1704.02372}}} {[astro-ph.SR]}

\bibitem[{{Madau} and {Dickinson}(2014)}]{2014ARA&A..52..415M}
{Madau} P, {Dickinson} M (2014) {Cosmic Star-Formation History}. \araa
  52:415--486. \doi{10.1146/annurev-astro-081811-125615}.
  {\href{https://arxiv.org/abs/1403.0007}{{arXiv:1403.0007}}} {[astro-ph.CO]}

\bibitem[{{Madau} and {Fragos}(2017)}]{2017ApJ...840...39M}
{Madau} P, {Fragos} T (2017) {Radiation Backgrounds at Cosmic Dawn: X-Rays from
  Compact Binaries}. \apj 840(1):39. \doi{10.3847/1538-4357/aa6af9}.
  {\href{https://arxiv.org/abs/1606.07887}{{arXiv:1606.07887}}} {[astro-ph.GA]}

\bibitem[{{Madau} and {Rees}(2001)}]{2001ApJ...551L..27M}
{Madau} P, {Rees} MJ (2001) {Massive Black Holes as Population III Remnants}.
  \apjl 551(1):L27--L30. \doi{10.1086/319848}.
  {\href{https://arxiv.org/abs/astro-ph/0101223}{{arXiv:astro-ph/0101223}}}
  {[astro-ph]}

\bibitem[{{Madej} et~al.(2013){Madej}, {Jonker}, {Groot}, {van Haaften},
  {Nelemans}, and {Maccarone}}]{2013MNRAS.429.2986M}
{Madej} OK, {Jonker} PG, {Groot} PJ, {van Haaften} LM, {Nelemans} G,
  {Maccarone} TJ (2013) {Time-resolved X-Shooter spectra and RXTE light curves
  of the ultra-compact X-ray binary candidate 4U 0614+091}. \mnras
  429(4):2986--2996. \doi{10.1093/mnras/sts550}.
  {\href{https://arxiv.org/abs/1212.0862}{{arXiv:1212.0862}}} {[astro-ph.HE]}

\bibitem[{{Madigan} et~al.(2011){Madigan}, {Hopman}, and
  {Levin}}]{2011ApJ...738...99M}
{Madigan} AM, {Hopman} C, {Levin} Y (2011) {Secular Stellar Dynamics near a
  Massive Black Hole}. \apj 738(1):99. \doi{10.1088/0004-637X/738/1/99}.
  {\href{https://arxiv.org/abs/1010.1535}{{arXiv:1010.1535}}} {[astro-ph.GA]}

\bibitem[{{Maggiore}(2000)}]{2000gr.qc.....8027M}
{Maggiore} M (2000) {Stochastic backgrounds of gravitational waves}. arXiv
  e-prints gr-qc/0008027.
  {\href{https://arxiv.org/abs/gr-qc/0008027}{{arXiv:gr-qc/0008027}}}
  {[astro-ph]}

\bibitem[{{Magorrian} et~al.(1998){Magorrian}, {Tremaine}, {Richstone},
  {Bender}, {Bower}, {Dressler}, {Faber}, {Gebhardt}, {Green}, {Grillmair},
  {Kormendy}, and {Lauer}}]{1998AJ....115.2285M}
{Magorrian} J, {Tremaine} S, {Richstone} D, {Bender} R, {Bower} G, {Dressler}
  A, {Faber} SM, {Gebhardt} K, {Green} R, {Grillmair} C, et~al. (1998) {The
  Demography of Massive Dark Objects in Galaxy Centers}. \aj 115(6):2285--2305.
  \doi{10.1086/300353}.
  {\href{https://arxiv.org/abs/astro-ph/9708072}{{arXiv:astro-ph/9708072}}}
  {[astro-ph]}

\bibitem[{{Maguire} et~al.(2020){Maguire}, {Eracleous}, {Jonker}, {MacLeod},
  and {Rosswog}}]{2020SSRv..216...39M}
{Maguire} K, {Eracleous} M, {Jonker} PG, {MacLeod} M, {Rosswog} S (2020) {Tidal
  Disruptions of White Dwarfs: Theoretical Models and Observational Prospects}.
  \ssr 216(3):39. \doi{10.1007/s11214-020-00661-2}.
  {\href{https://arxiv.org/abs/2004.00146}{{arXiv:2004.00146}}} {[astro-ph.HE]}

\bibitem[{{Maio} et~al.(2013){Maio}, {Dotti}, {Petkova}, {Perego}, and
  {Volonteri}}]{2013ApJ...767...37M}
{Maio} U, {Dotti} M, {Petkova} M, {Perego} A, {Volonteri} M (2013) {Effects of
  Circumnuclear Disk Gas Evolution on the Spin of Central Black Holes}. \apj
  767(1):37. \doi{10.1088/0004-637X/767/1/37}.
  {\href{https://arxiv.org/abs/1203.1877}{{arXiv:1203.1877}}} {[astro-ph.HE]}

\bibitem[{{Makino} and {Funato}(2004)}]{2004ApJ...602...93M}
{Makino} J, {Funato} Y (2004) {Evolution of Massive Black Hole Binaries}. \apj
  602(1):93--102. \doi{10.1086/380917}.
  {\href{https://arxiv.org/abs/astro-ph/0307327}{{arXiv:astro-ph/0307327}}}
  {[astro-ph]}

\bibitem[{{Manchester} et~al.(2005){Manchester}, {Hobbs}, {Teoh}, and
  {Hobbs}}]{2005AJ....129.1993M}
{Manchester} RN, {Hobbs} GB, {Teoh} A, {Hobbs} M (2005) {The Australia
  Telescope National Facility Pulsar Catalogue}. \aj 129(4):1993--2006.
  \doi{10.1086/428488}.
  {\href{https://arxiv.org/abs/astro-ph/0412641}{{arXiv:astro-ph/0412641}}}
  {[astro-ph]}

\bibitem[{{Mandel}(2016)}]{2016MNRAS.456..578M}
{Mandel} I (2016) {Estimates of black hole natal kick velocities from
  observations of low-mass X-ray binaries}. \mnras 456(1):578--581.
  \doi{10.1093/mnras/stv2733}.
  {\href{https://arxiv.org/abs/1510.03871}{{arXiv:1510.03871}}} {[astro-ph.HE]}

\bibitem[{{Mandel} and {de Mink}(2016)}]{2016MNRAS.458.2634M}
{Mandel} I, {de Mink} SE (2016) {Merging binary black holes formed through
  chemically homogeneous evolution in short-period stellar binaries}. \mnras
  458(3):2634--2647. \doi{10.1093/mnras/stw379}.
  {\href{https://arxiv.org/abs/1601.00007}{{arXiv:1601.00007}}} {[astro-ph.HE]}

\bibitem[{{Mandel} and {M{\"u}ller}(2020)}]{2020arXiv200608360M}
{Mandel} I, {M{\"u}ller} B (2020) {Simple recipes for compact remnant masses
  and natal kicks}. arXiv e-prints arXiv:2006.08360.
  {\href{https://arxiv.org/abs/2006.08360}{{arXiv:2006.08360}}} {[astro-ph.HE]}

\bibitem[{{Mandel} et~al.(2008){Mandel}, {Brown}, {Gair}, and
  {Miller}}]{Mandel:2008}
{Mandel} I, {Brown} DA, {Gair} JR, {Miller} MC (2008) {Rates and
  Characteristics of Intermediate Mass Ratio Inspirals Detectable by Advanced
  LIGO}. \apj 681:1431--1447. \doi{10.1086/588246}.
  {\href{https://arxiv.org/abs/0705.0285}{{arXiv:0705.0285}}}

\bibitem[{{Mandel} et~al.(2018){Mandel}, {Sesana}, and {Vecchio}}]{Mandel:2017}
{Mandel} I, {Sesana} A, {Vecchio} A (2018) {The astrophysical science case for
  a decihertz gravitational-wave detector}. Classical and Quantum Gravity
  35(5):054004. \doi{10.1088/1361-6382/aaa7e0}.
  {\href{https://arxiv.org/abs/1710.11187}{{arXiv:1710.11187}}} {[astro-ph.HE]}

\bibitem[{{Mandic} et~al.(2012){Mandic}, {Thrane}, {Giampanis}, and
  {Regimbau}}]{Mandic:2012}
{Mandic} V, {Thrane} E, {Giampanis} S, {Regimbau} T (2012) {Parameter
  Estimation in Searches for the Stochastic Gravitational-Wave Background}.
  Physical Review Letters 109(17):171102. \doi{10.1103/PhysRevLett.109.171102}.
  {\href{https://arxiv.org/abs/1209.3847}{{arXiv:1209.3847}}} {[astro-ph.CO]}

\bibitem[{{Mangiagli} et~al.(2019){Mangiagli}, {Klein}, {Sesana}, {Barausse},
  and {Colpi}}]{2019PhRvD..99f4056M}
{Mangiagli} A, {Klein} A, {Sesana} A, {Barausse} E, {Colpi} M (2019)
  {Post-Newtonian phase accuracy requirements for stellar black hole binaries
  with LISA}. \prd 99(6):064056. \doi{10.1103/PhysRevD.99.064056}.
  {\href{https://arxiv.org/abs/1811.01805}{{arXiv:1811.01805}}} {[gr-qc]}

\bibitem[{{Mangiagli} et~al.(2020){Mangiagli}, {Klein}, {Bonetti}, {Katz},
  {Sesana}, {Volonteri}, {Colpi}, {Marsat}, and {Babak}}]{2020PhRvD.102h4056M}
{Mangiagli} A, {Klein} A, {Bonetti} M, {Katz} ML, {Sesana} A, {Volonteri} M,
  {Colpi} M, {Marsat} S, {Babak} S (2020) {Observing the inspiral of coalescing
  massive black hole binaries with LISA in the era of multimessenger
  astrophysics}. \prd 102(8):084056. \doi{10.1103/PhysRevD.102.084056}.
  {\href{https://arxiv.org/abs/2006.12513}{{arXiv:2006.12513}}} {[astro-ph.HE]}

\bibitem[{{Mangiagli} et~al.(2022){Mangiagli}, {Caprini}, {Volonteri},
  {Marsat}, {Vergani}, {Tamanini}, and {Inchausp{\'e}}}]{2022arXiv220710678M}
{Mangiagli} A, {Caprini} C, {Volonteri} M, {Marsat} S, {Vergani} S, {Tamanini}
  N, {Inchausp{\'e}} H (2022) {Massive black hole binaries in LISA:
  multimessenger prospects and electromagnetic counterparts}. arXiv e-prints
  arXiv:2207.10678.
  {\href{https://arxiv.org/abs/2207.10678}{{arXiv:2207.10678}}} {[astro-ph.HE]}

\bibitem[{{Mannerkoski} et~al.(2019){Mannerkoski}, {Johansson}, {Pihajoki},
  {Rantala}, and {Naab}}]{2019ApJ...887...35M}
{Mannerkoski} M, {Johansson} PH, {Pihajoki} P, {Rantala} A, {Naab} T (2019)
  {Gravitational Waves from the Inspiral of Supermassive Black Holes in
  Galactic-scale Simulations}. \apj 887(1):35. \doi{10.3847/1538-4357/ab52f9}.
  {\href{https://arxiv.org/abs/1909.01373}{{arXiv:1909.01373}}} {[astro-ph.GA]}

\bibitem[{{Mannerkoski} et~al.(2021){Mannerkoski}, {Johansson}, {Rantala},
  {Naab}, and {Liao}}]{2021ApJ...912L..20M}
{Mannerkoski} M, {Johansson} PH, {Rantala} A, {Naab} T, {Liao} S (2021)
  {Resolving the Complex Evolution of a Supermassive Black Hole Triplet in a
  Cosmological Simulation}. \apjl 912(2):L20. \doi{10.3847/2041-8213/abf9a5}.
  {\href{https://arxiv.org/abs/2103.16254}{{arXiv:2103.16254}}} {[astro-ph.GA]}

\bibitem[{{Manser} et~al.(2016){Manser}, {G{\"a}nsicke}, {Marsh}, {Veras},
  {Koester}, {Breedt}, {Pala}, {Parsons}, and
  {Southworth}}]{2016MNRAS.455.4467M}
{Manser} CJ, {G{\"a}nsicke} BT, {Marsh} TR, {Veras} D, {Koester} D, {Breedt} E,
  {Pala} AF, {Parsons} SG, {Southworth} J (2016) {Doppler imaging of the
  planetary debris disc at the white dwarf SDSS J122859.93+104032.9}. \mnras
  455(4):4467--4478. \doi{10.1093/mnras/stv2603}.
  {\href{https://arxiv.org/abs/1511.02230}{{arXiv:1511.02230}}} {[astro-ph.SR]}

\bibitem[{{Manser} et~al.(2019){Manser}, {G{\"a}nsicke}, {Eggl}, {Hollands},
  {Izquierdo}, {Koester}, {Landstreet}, {Lyra}, {Marsh}, {Meru}, {Mustill},
  {Rodr{\'\i}guez-Gil}, {Toloza}, {Veras}, {Wilson}, {Burleigh}, {Davies},
  {Farihi}, {Gentile Fusillo}, {de Martino}, {Parsons}, {Quirrenbach}, {Raddi},
  {Reffert}, {Del Santo}, {Schreiber}, {Silvotti}, {Toonen}, {Villaver},
  {Wyatt}, {Xu}, and {Portegies Zwart}}]{2019Sci...364...66M}
{Manser} CJ, {G{\"a}nsicke} BT, {Eggl} S, {Hollands} M, {Izquierdo} P,
  {Koester} D, {Landstreet} JD, {Lyra} W, {Marsh} TR, {Meru} F, et~al. (2019)
  {A planetesimal orbiting within the debris disc around a white dwarf star}.
  Science 364(6435):66--69. \doi{10.1126/science.aat5330}.
  {\href{https://arxiv.org/abs/1904.02163}{{arXiv:1904.02163}}} {[astro-ph.EP]}

\bibitem[{{Mapelli}(2016)}]{2016MNRAS.459.3432M}
{Mapelli} M (2016) {Massive black hole binaries from runaway collisions: the
  impact of metallicity}. \mnras 459(4):3432--3446. \doi{10.1093/mnras/stw869}.
  {\href{https://arxiv.org/abs/1604.03559}{{arXiv:1604.03559}}} {[astro-ph.GA]}

\bibitem[{{Mapelli} and {Giacobbo}(2018)}]{2018MNRAS.479.4391M}
{Mapelli} M, {Giacobbo} N (2018) {The cosmic merger rate of neutron stars and
  black holes}. \mnras 479(4):4391--4398. \doi{10.1093/mnras/sty1613}.
  {\href{https://arxiv.org/abs/1806.04866}{{arXiv:1806.04866}}} {[astro-ph.HE]}

\bibitem[{{Mapelli} et~al.(2012){Mapelli}, {Ripamonti}, {Vecchio}, {Graham},
  and {Gualandris}}]{2012A&A...542A.102M}
{Mapelli} M, {Ripamonti} E, {Vecchio} A, {Graham} AW, {Gualandris} A (2012) {A
  cosmological view of extreme mass-ratio inspirals in nuclear star clusters}.
  \aap 542:A102. \doi{10.1051/0004-6361/201118444}.
  {\href{https://arxiv.org/abs/1205.2702}{{arXiv:1205.2702}}} {[astro-ph.CO]}

\bibitem[{{Mapelli} et~al.(2017){Mapelli}, {Giacobbo}, {Ripamonti}, and
  {Spera}}]{2017MNRAS.472.2422M}
{Mapelli} M, {Giacobbo} N, {Ripamonti} E, {Spera} M (2017) {The cosmic merger
  rate of stellar black hole binaries from the Illustris simulation}. \mnras
  472(2):2422--2435. \doi{10.1093/mnras/stx2123}.
  {\href{https://arxiv.org/abs/1708.05722}{{arXiv:1708.05722}}} {[astro-ph.GA]}

\bibitem[{{Mapelli} et~al.(2020){Mapelli}, {Santoliquido}, {Bouffanais}, {Arca
  Sedda}, {Giacobbo}, {Artale}, and {Ballone}}]{2020arXiv200715022M}
{Mapelli} M, {Santoliquido} F, {Bouffanais} Y, {Arca Sedda} M, {Giacobbo} N,
  {Artale} MC, {Ballone} A (2020) {Hierarchical mergers in young, globular and
  nuclear star clusters: black hole masses and merger rates}. arXiv e-prints
  arXiv:2007.15022.
  {\href{https://arxiv.org/abs/2007.15022}{{arXiv:2007.15022}}} {[astro-ph.HE]}

\bibitem[{{Marassi} et~al.(2011){Marassi}, {Schneider}, {Corvino}, {Ferrari},
  and {Portegies Zwart}}]{2011PhRvD..84l4037M}
{Marassi} S, {Schneider} R, {Corvino} G, {Ferrari} V, {Portegies Zwart} S
  (2011) {Imprint of the merger and ring-down on the gravitational wave
  background from black hole binaries coalescence}. \prd 84(12):124037.
  \doi{10.1103/PhysRevD.84.124037}.
  {\href{https://arxiv.org/abs/1111.6125}{{arXiv:1111.6125}}} {[astro-ph.CO]}

\bibitem[{{Marassi} et~al.(2019){Marassi}, {Graziani}, {Ginolfi}, {Schneider},
  {Mapelli}, {Spera}, and {Alparone}}]{2019MNRAS.484.3219M}
{Marassi} S, {Graziani} L, {Ginolfi} M, {Schneider} R, {Mapelli} M, {Spera} M,
  {Alparone} M (2019) {Evolution of dwarf galaxies hosting GW150914-like
  events}. \mnras 484(3):3219--3232. \doi{10.1093/mnras/stz170}.
  {\href{https://arxiv.org/abs/1901.04494}{{arXiv:1901.04494}}} {[astro-ph.GA]}

\bibitem[{{Marchant} et~al.(2016){Marchant}, {Langer}, {Podsiadlowski},
  {Tauris}, and {Moriya}}]{2016A&A...588A..50M}
{Marchant} P, {Langer} N, {Podsiadlowski} P, {Tauris} TM, {Moriya} TJ (2016) {A
  new route towards merging massive black holes}. \aap 588:A50.
  \doi{10.1051/0004-6361/201628133}.
  {\href{https://arxiv.org/abs/1601.03718}{{arXiv:1601.03718}}} {[astro-ph.SR]}

\bibitem[{{Marconi} et~al.(2004){Marconi}, {Risaliti}, {Gilli}, {Hunt},
  {Maiolino}, and {Salvati}}]{2004MNRAS.351..169M}
{Marconi} A, {Risaliti} G, {Gilli} R, {Hunt} LK, {Maiolino} R, {Salvati} M
  (2004) {Local supermassive black holes, relics of active galactic nuclei and
  the X-ray background}. \mnras 351(1):169--185.
  \doi{10.1111/j.1365-2966.2004.07765.x}.
  {\href{https://arxiv.org/abs/astro-ph/0311619}{{arXiv:astro-ph/0311619}}}
  {[astro-ph]}

\bibitem[{{Mardling} and {Aarseth}(1999)}]{1999ASIC..522..385M}
{Mardling} R, {Aarseth} S (1999) {Dynamics and Stability of Three-Body
  Systems}. In: {Steves} BA, {Roy} AE (eds) The Dynamics of Small Bodies in the
  Solar System, A Major Key to Solar System Studies. NATO Advanced Study
  Institute (ASI) Series C, vol 522. p 385

\bibitem[{{Marelli} et~al.(2015){Marelli}, {Mignani}, {De Luca}, {Saz
  Parkinson}, {Salvetti}, {Den Hartog}, and {Wolff}}]{2015ApJ...802...78M}
{Marelli} M, {Mignani} RP, {De Luca} A, {Saz Parkinson} PM, {Salvetti} D, {Den
  Hartog} PR, {Wolff} MT (2015) {Radio-quiet and Radio-loud Pulsars: Similar in
  Gamma-Rays but Different in X-Rays}. \apj 802(2):78.
  \doi{10.1088/0004-637X/802/2/78}.
  {\href{https://arxiv.org/abs/1501.06215}{{arXiv:1501.06215}}} {[astro-ph.HE]}

\bibitem[{{Margalit} and {Metzger}(2019)}]{2019ApJ...880L..15M}
{Margalit} B, {Metzger} BD (2019) {The Multi-messenger Matrix: The Future of
  Neutron Star Merger Constraints on the Nuclear Equation of State}. \apjl
  880(1):L15. \doi{10.3847/2041-8213/ab2ae2}.
  {\href{https://arxiv.org/abs/1904.11995}{{arXiv:1904.11995}}} {[astro-ph.HE]}

\bibitem[{{Marinacci} et~al.(2019){Marinacci}, {Sales}, {Vogelsberger},
  {Torrey}, and {Springel}}]{2019MNRAS.489.4233M}
{Marinacci} F, {Sales} LV, {Vogelsberger} M, {Torrey} P, {Springel} V (2019)
  {Simulating the interstellar medium and stellar feedback on a moving mesh:
  implementation and isolated galaxies}. \mnras 489(3):4233--4260.
  \doi{10.1093/mnras/stz2391}.
  {\href{https://arxiv.org/abs/1905.08806}{{arXiv:1905.08806}}} {[astro-ph.GA]}

\bibitem[{{Marronetti} et~al.(2008){Marronetti}, {Tichy}, {Br{\"u}gmann},
  {Gonz{\'a}lez}, and {Sperhake}}]{2008PhRvD..77f4010M}
{Marronetti} P, {Tichy} W, {Br{\"u}gmann} B, {Gonz{\'a}lez} J, {Sperhake} U
  (2008) {High-spin binary black hole mergers}. \prd 77(6):064010.
  \doi{10.1103/PhysRevD.77.064010}.
  {\href{https://arxiv.org/abs/0709.2160}{{arXiv:0709.2160}}} {[gr-qc]}

\bibitem[{{Marsh} and {Steeghs}(2002)}]{2002MNRAS.331L...7M}
{Marsh} TR, {Steeghs} D (2002) {V407 Vul: a direct impact accretor}. \mnras
  331(1):L7--L11. \doi{10.1046/j.1365-8711.2002.05346.x}.
  {\href{https://arxiv.org/abs/astro-ph/0201309}{{arXiv:astro-ph/0201309}}}
  {[astro-ph]}

\bibitem[{{Marsh} et~al.(1995){Marsh}, {Dhillon}, and
  {Duck}}]{1995MNRAS.275..828M}
{Marsh} TR, {Dhillon} VS, {Duck} SR (1995) {Low-Mass White Dwarfs Need Friends
  - Five New Double-Degenerate Close Binary Stars}. \mnras 275:828.
  \doi{10.1093/mnras/275.3.828}

\bibitem[{{Marsh} et~al.(2004){Marsh}, {Nelemans}, and
  {Steeghs}}]{2004MNRAS.350..113M}
{Marsh} TR, {Nelemans} G, {Steeghs} D (2004) {Mass transfer between double
  white dwarfs}. \mnras 350(1):113--128.
  \doi{10.1111/j.1365-2966.2004.07564.x}.
  {\href{https://arxiv.org/abs/astro-ph/0312577}{{arXiv:astro-ph/0312577}}}
  {[astro-ph]}

\bibitem[{{Martinez} et~al.(2020){Martinez}, {Fragione}, {Kremer},
  {Chatterjee}, {Rodriguez}, {Samsing}, {Ye}, {Weatherford}, {Zevin}, {Naoz},
  and {Rasio}}]{2020arXiv200908468M}
{Martinez} MAS, {Fragione} G, {Kremer} K, {Chatterjee} S, {Rodriguez} CL,
  {Samsing} J, {Ye} CS, {Weatherford} NC, {Zevin} M, {Naoz} S, et~al. (2020)
  {Black Hole Mergers from Hierarchical Triples in Dense Star Clusters}. arXiv
  e-prints arXiv:2009.08468.
  {\href{https://arxiv.org/abs/2009.08468}{{arXiv:2009.08468}}} {[astro-ph.GA]}

\bibitem[{{Martinez-Valpuesta} et~al.(2016){Martinez-Valpuesta}, {Aguerri}, and
  {Gonz{\'a}lez-Garc{\'\i}a}}]{2016Galax...4....7M}
{Martinez-Valpuesta} I, {Aguerri} J, {Gonz{\'a}lez-Garc{\'\i}a} C (2016)
  {Characterization of Bars Induced by Interactions}. Galaxies 4(2):7.
  \doi{10.3390/galaxies4020007}

\bibitem[{{Marulli} et~al.(2008){Marulli}, {Bonoli}, {Branchini}, {Moscardini},
  and {Springel}}]{2008MNRAS.385.1846M}
{Marulli} F, {Bonoli} S, {Branchini} E, {Moscardini} L, {Springel} V (2008)
  {Modelling the cosmological co-evolution of supermassive black holes and
  galaxies - I. BH scaling relations and the AGN luminosity function}. \mnras
  385(4):1846--1858. \doi{10.1111/j.1365-2966.2008.12988.x}.
  {\href{https://arxiv.org/abs/0711.2053}{{arXiv:0711.2053}}} {[astro-ph]}

\bibitem[{{Masci} et~al.(2019){Masci}, {Laher}, {Rusholme}, {Shupe}, {Groom},
  {Surace}, {Jackson}, {Monkewitz}, {Beck}, {Flynn}, {Terek}, {Landry},
  {Hacopians}, {Desai}, {Howell}, {Brooke}, {Imel}, {Wachter}, {Ye}, {Lin},
  {Cenko}, {Cunningham}, {Rebbapragada}, {Bue}, {Miller}, {Mahabal}, {Bellm},
  {Patterson}, {Juri{\'c}}, {Golkhou}, {Ofek}, {Walters}, {Graham}, {Kasliwal},
  {Dekany}, {Kupfer}, {Burdge}, {Cannella}, {Barlow}, {Van Sistine}, {Giomi},
  {Fremling}, {Blagorodnova}, {Levitan}, {Riddle}, {Smith}, {Helou}, {Prince},
  and {Kulkarni}}]{2019PASP..131a8003M}
{Masci} FJ, {Laher} RR, {Rusholme} B, {Shupe} DL, {Groom} S, {Surace} J,
  {Jackson} E, {Monkewitz} S, {Beck} R, {Flynn} D, et~al. (2019) {The Zwicky
  Transient Facility: Data Processing, Products, and Archive}. \pasp
  131(995):018003. \doi{10.1088/1538-3873/aae8ac}.
  {\href{https://arxiv.org/abs/1902.01872}{{arXiv:1902.01872}}} {[astro-ph.IM]}

\bibitem[{{Mashian} and {Loeb}(2017)}]{2017MNRAS.470.2611M}
{Mashian} N, {Loeb} A (2017) {Hunting black holes with Gaia}. \mnras
  470(3):2611--2616. \doi{10.1093/mnras/stx1410}.
  {\href{https://arxiv.org/abs/1704.03455}{{arXiv:1704.03455}}} {[astro-ph.HE]}

\bibitem[{{Mastrobuono-Battisti} et~al.(2014){Mastrobuono-Battisti}, {Perets},
  and {Loeb}}]{2014ApJ...796...40M}
{Mastrobuono-Battisti} A, {Perets} HB, {Loeb} A (2014) {Effects of Intermediate
  Mass Black Holes on Nuclear Star Clusters}. \apj 796(1):40.
  \doi{10.1088/0004-637X/796/1/40}.
  {\href{https://arxiv.org/abs/1403.3094}{{arXiv:1403.3094}}} {[astro-ph.GA]}

\bibitem[{{Mathis}(2019)}]{2019EAS....82....5M}
{Mathis} S (2019) {Tidal dissipation in stars and giant planets: Jean-Paul
  Zahn's pioneering work and legacy}. In: EAS Publications Series. EAS
  Publications Series, vol~82. pp 5--33. \doi{10.1051/eas/1982002}

\bibitem[{{Matsubayashi} et~al.(2004){Matsubayashi}, {Shinkai}, and
  {Ebisuzaki}}]{2004ApJ...614..864M}
{Matsubayashi} T, {Shinkai} Ha, {Ebisuzaki} T (2004) {Gravitational Waves from
  Merging Intermediate-Mass Black Holes}. \apj 614(2):864--868.
  \doi{10.1086/423796}

\bibitem[{{Matsubayashi} et~al.(2007){Matsubayashi}, {Makino}, and
  {Ebisuzaki}}]{2007ApJ...656..879M}
{Matsubayashi} T, {Makino} J, {Ebisuzaki} T (2007) {Orbital Evolution of an
  IMBH in the Galactic Nucleus with a Massive Central Black Hole}. \apj
  656(2):879--896. \doi{10.1086/510344}.
  {\href{https://arxiv.org/abs/astro-ph/0511782}{{arXiv:astro-ph/0511782}}}
  {[astro-ph]}

\bibitem[{{Matsuoka} et~al.(2019){Matsuoka}, {Iwasawa}, {Onoue}, {Kashikawa},
  {Strauss}, {Lee}, {Imanishi}, {Nagao}, {Akiyama}, {Asami}, {Bosch},
  {Furusawa}, {Goto}, {Gunn}, {Harikane}, {Ikeda}, {Izumi}, {Kawaguchi},
  {Kato}, {Kikuta}, {Kohno}, {Komiyama}, {Koyama}, {Lupton}, {Minezaki},
  {Miyazaki}, {Murayama}, {Niida}, {Nishizawa}, {Noboriguchi}, {Oguri}, {Ono},
  {Ouchi}, {Price}, {Sameshima}, {Schulze}, {Silverman}, {Sugiyama}, {Tait},
  {Takada}, {Takata}, {Tanaka}, {Tang}, {Toba}, {Utsumi}, {Wang}, and
  {Yamashita}}]{2019ApJ...883..183M}
{Matsuoka} Y, {Iwasawa} K, {Onoue} M, {Kashikawa} N, {Strauss} MA, {Lee} CH,
  {Imanishi} M, {Nagao} T, {Akiyama} M, {Asami} N, et~al. (2019) {Subaru High-z
  Exploration of Low-luminosity Quasars (SHELLQs). X. Discovery of 35 Quasars
  and Luminous Galaxies at 5.7 {\ensuremath{\leq}} z {\ensuremath{\leq}} 7.0}.
  \apj 883(2):183. \doi{10.3847/1538-4357/ab3c60}.
  {\href{https://arxiv.org/abs/1908.07910}{{arXiv:1908.07910}}} {[astro-ph.GA]}

\bibitem[{{Maureira-Fredes} et~al.(2018){Maureira-Fredes}, {Goicovic},
  {Amaro-Seoane}, and {Sesana}}]{2018MNRAS.478.1726M}
{Maureira-Fredes} C, {Goicovic} FG, {Amaro-Seoane} P, {Sesana} A (2018)
  {Accretion of clumpy cold gas on to massive black hole binaries: the
  challenging formation of extended circumbinary structures}. \mnras
  478(2):1726--1748. \doi{10.1093/mnras/sty1105}.
  {\href{https://arxiv.org/abs/1801.06179}{{arXiv:1801.06179}}} {[astro-ph.HE]}

\bibitem[{{Maxted} et~al.(2000{\natexlab{a}}){Maxted}, {Marsh}, and
  {Moran}}]{2000MNRAS.319..305M}
{Maxted} PFL, {Marsh} TR, {Moran} CKJ (2000{\natexlab{a}}) {Radial velocity
  measurements of white dwarfs}. \mnras 319(1):305--317.
  \doi{10.1046/j.1365-8711.2000.03840.x}.
  {\href{https://arxiv.org/abs/astro-ph/0007129}{{arXiv:astro-ph/0007129}}}
  {[astro-ph]}

\bibitem[{{Maxted} et~al.(2000{\natexlab{b}}){Maxted}, {Marsh}, {Moran}, and
  {Han}}]{2000MNRAS.314..334M}
{Maxted} PFL, {Marsh} TR, {Moran} CKJ, {Han} Z (2000{\natexlab{b}}) {The triple
  degenerate star WD 1704+481}. \mnras 314(2):334--337.
  \doi{10.1046/j.1365-8711.2000.03343.x}.
  {\href{https://arxiv.org/abs/astro-ph/0001212}{{arXiv:astro-ph/0001212}}}
  {[astro-ph]}

\bibitem[{{Maxted} et~al.(2002){Maxted}, {Marsh}, and
  {Moran}}]{2002MNRAS.332..745M}
{Maxted} PFL, {Marsh} TR, {Moran} CKJ (2002) {The mass ratio distribution of
  short-period double degenerate stars}. \mnras 332(3):745--753.
  \doi{10.1046/j.1365-8711.2002.05368.x}.
  {\href{https://arxiv.org/abs/astro-ph/0201411}{{arXiv:astro-ph/0201411}}}
  {[astro-ph]}

\bibitem[{{Maxted} et~al.(2006){Maxted}, {Napiwotzki}, {Dobbie}, and
  {Burleigh}}]{2006Natur.442..543M}
{Maxted} PFL, {Napiwotzki} R, {Dobbie} PD, {Burleigh} MR (2006) {Survival of a
  brown dwarf after engulfment by a red giant star}. \nat 442(7102):543--545.
  \doi{10.1038/nature04987}.
  {\href{https://arxiv.org/abs/astro-ph/0608054}{{arXiv:astro-ph/0608054}}}
  {[astro-ph]}

\bibitem[{{Mayer}(2013)}]{2013CQGra..30x4008M}
{Mayer} L (2013) {Massive black hole binaries in gas-rich galaxy mergers;
  multiple regimes of orbital decay and interplay with gas inflows}. Classical
  and Quantum Gravity 30(24):244008. \doi{10.1088/0264-9381/30/24/244008}.
  {\href{https://arxiv.org/abs/1308.0431}{{arXiv:1308.0431}}} {[astro-ph.CO]}

\bibitem[{{Mayer}(2017)}]{2017JPhCS.840a2025M}
{Mayer} L (2017) {Multiple regimes and coalescence timescales for massive black
  hole pairs; the critical role of galaxy formation physics}. In: Journal of
  Physics Conference Series. Journal of Physics Conference Series, vol 840. p
  012025. \doi{10.1088/1742-6596/840/1/012025}.
  {\href{https://arxiv.org/abs/1703.00661}{{arXiv:1703.00661}}} {[astro-ph.GA]}

\bibitem[{{Mayer} and {Wadsley}(2004)}]{2004MNRAS.347..277M}
{Mayer} L, {Wadsley} J (2004) {The formation and evolution of bars in low
  surface brightness galaxies with cold dark matter haloes}. \mnras
  347(1):277--294. \doi{10.1111/j.1365-2966.2004.07202.x}.
  {\href{https://arxiv.org/abs/astro-ph/0303239}{{arXiv:astro-ph/0303239}}}
  {[astro-ph]}

\bibitem[{{Mayer} et~al.(2007){Mayer}, {Kazantzidis}, {Madau}, {Colpi},
  {Quinn}, and {Wadsley}}]{2007Sci...316.1874M}
{Mayer} L, {Kazantzidis} S, {Madau} P, {Colpi} M, {Quinn} T, {Wadsley} J (2007)
  {Rapid Formation of Supermassive Black Hole Binaries in Galaxy Mergers with
  Gas}. Science 316(5833):1874. \doi{10.1126/science.1141858}.
  {\href{https://arxiv.org/abs/0706.1562}{{arXiv:0706.1562}}} {[astro-ph]}

\bibitem[{{Mayer} et~al.(2010){Mayer}, {Kazantzidis}, {Escala}, and
  {Callegari}}]{2010Natur.466.1082M}
{Mayer} L, {Kazantzidis} S, {Escala} A, {Callegari} S (2010) {Direct formation
  of supermassive black holes via multi-scale gas inflows in galaxy mergers}.
  \nat 466(7310):1082--1084. \doi{10.1038/nature09294}.
  {\href{https://arxiv.org/abs/0912.4262}{{arXiv:0912.4262}}} {[astro-ph.CO]}

\bibitem[{{Mayer} et~al.(2015){Mayer}, {Fiacconi}, {Bonoli}, {Quinn}, {Ro{\v
  s}kar}, {Shen}, and {Wadsley}}]{2015ApJ...810...51M}
{Mayer} L, {Fiacconi} D, {Bonoli} S, {Quinn} T, {Ro{\v s}kar} R, {Shen} S,
  {Wadsley} J (2015) {Direct Formation of Supermassive Black Holes in
  Metal-enriched Gas at the Heart of High-redshift Galaxy Mergers}. \apj
  810:51. \doi{10.1088/0004-637X/810/1/51}.
  {\href{https://arxiv.org/abs/1411.5683}{{arXiv:1411.5683}}}

\bibitem[{{Mazeh} and {Shaham}(1979)}]{1979A&A....77..145M}
{Mazeh} T, {Shaham} J (1979) {The orbital evolution of close triple systems:
  the binary eccentricity.} \aap 77:145

\bibitem[{{McClintock} and {Remillard}(2006)}]{2006csxs.book..157M}
{McClintock} JE, {Remillard} RA (2006) {Black hole binaries}, vol~39, pp
  157--213

\bibitem[{{McClintock} et~al.(2011){McClintock}, {Narayan}, {Davis}, {Gou},
  {Kulkarni}, {Orosz}, {Penna}, {Remillard}, and
  {Steiner}}]{2011CQGra..28k4009M}
{McClintock} JE, {Narayan} R, {Davis} SW, {Gou} L, {Kulkarni} A, {Orosz} JA,
  {Penna} RF, {Remillard} RA, {Steiner} JF (2011) {Measuring the spins of
  accreting black holes}. Classical and Quantum Gravity 28(11):114009.
  \doi{10.1088/0264-9381/28/11/114009}.
  {\href{https://arxiv.org/abs/1101.0811}{{arXiv:1101.0811}}} {[astro-ph.HE]}

\bibitem[{{McConnell} and {Ma}(2013)}]{2013ApJ...764..184M}
{McConnell} NJ, {Ma} CP (2013) {Revisiting the Scaling Relations of Black Hole
  Masses and Host Galaxy Properties}. \apj 764(2):184.
  \doi{10.1088/0004-637X/764/2/184}.
  {\href{https://arxiv.org/abs/1211.2816}{{arXiv:1211.2816}}} {[astro-ph.CO]}

\bibitem[{{McConnell} et~al.(2011){McConnell}, {Ma}, {Gebhardt}, {Wright},
  {Murphy}, {Lauer}, {Graham}, and {Richstone}}]{2011Natur.480..215M}
{McConnell} NJ, {Ma} CP, {Gebhardt} K, {Wright} SA, {Murphy} JD, {Lauer} TR,
  {Graham} JR, {Richstone} DO (2011) {Two ten-billion-solar-mass black holes at
  the centres of giant elliptical galaxies}. \nat 480(7376):215--218.
  \doi{10.1038/nature10636}.
  {\href{https://arxiv.org/abs/1112.1078}{{arXiv:1112.1078}}} {[astro-ph.CO]}

\bibitem[{{McGee} et~al.(2020){McGee}, {Sesana}, and
  {Vecchio}}]{2020NatAs...4...26M}
{McGee} S, {Sesana} A, {Vecchio} A (2020) {Linking gravitational waves and
  X-ray phenomena with joint LISA and Athena observations}. Nature Astronomy
  4:26--31. \doi{10.1038/s41550-019-0969-7}.
  {\href{https://arxiv.org/abs/1811.00050}{{arXiv:1811.00050}}} {[astro-ph.HE]}

\bibitem[{{McGee}(2013)}]{2013MNRAS.436.2708M}
{McGee} SL (2013) {The strong environmental dependence of black hole scaling
  relations}. \mnras 436(3):2708--2721. \doi{10.1093/mnras/stt1769}.
  {\href{https://arxiv.org/abs/1302.6237}{{arXiv:1302.6237}}} {[astro-ph.CO]}

\bibitem[{{McKernan} and {Ford}(2015)}]{2015MNRAS.452L...1M}
{McKernan} B, {Ford} KES (2015) {Detection of radial velocity shifts due to
  black hole binaries near merger.} \mnras 452:L1--L5.
  \doi{10.1093/mnrasl/slv076}.
  {\href{https://arxiv.org/abs/1505.04120}{{arXiv:1505.04120}}} {[astro-ph.HE]}

\bibitem[{{McKernan} et~al.(2012){McKernan}, {Ford}, {Lyra}, and
  {Perets}}]{2012MNRAS.425..460M}
{McKernan} B, {Ford} KES, {Lyra} W, {Perets} HB (2012) {Intermediate mass black
  holes in AGN discs - I. Production and growth}. \mnras 425(1):460--469.
  \doi{10.1111/j.1365-2966.2012.21486.x}.
  {\href{https://arxiv.org/abs/1206.2309}{{arXiv:1206.2309}}} {[astro-ph.GA]}

\bibitem[{{McKernan} et~al.(2013){McKernan}, {Ford}, {Kocsis}, and
  {Haiman}}]{2013MNRAS.432.1468M}
{McKernan} B, {Ford} KES, {Kocsis} B, {Haiman} Z (2013) {Ripple effects and
  oscillations in the broad Fe K{\ensuremath{\alpha}} line as a probe of
  massive black hole mergers}. \mnras 432(2):1468--1482.
  \doi{10.1093/mnras/stt567}.
  {\href{https://arxiv.org/abs/1303.7206}{{arXiv:1303.7206}}} {[astro-ph.HE]}

\bibitem[{{McKernan} et~al.(2014){McKernan}, {Ford}, {Kocsis}, {Lyra}, and
  {Winter}}]{2014MNRAS.441..900M}
{McKernan} B, {Ford} KES, {Kocsis} B, {Lyra} W, {Winter} LM (2014)
  {Intermediate-mass black holes in AGN discs - II. Model predictions and
  observational constraints}. \mnras 441(1):900--909.
  \doi{10.1093/mnras/stu553}.
  {\href{https://arxiv.org/abs/1403.6433}{{arXiv:1403.6433}}} {[astro-ph.GA]}

\bibitem[{{McKernan} et~al.(2019){McKernan}, {Ford}, {Bartos}, {Graham},
  {Lyra}, {Marka}, {Marka}, {Ross}, {Stern}, and {Yang}}]{2019ApJ...884L..50M}
{McKernan} B, {Ford} KES, {Bartos} I, {Graham} MJ, {Lyra} W, {Marka} S, {Marka}
  Z, {Ross} NP, {Stern} D, {Yang} Y (2019) {Ram-pressure Stripping of a Kicked
  Hill Sphere: Prompt Electromagnetic Emission from the Merger of Stellar Mass
  Black Holes in an AGN Accretion Disk}. \apjl 884(2):L50.
  \doi{10.3847/2041-8213/ab4886}.
  {\href{https://arxiv.org/abs/1907.03746}{{arXiv:1907.03746}}} {[astro-ph.HE]}

\bibitem[{{McKernan} et~al.(2020{\natexlab{a}}){McKernan}, {Ford}, and
  {O'Shaughnessy}}]{2020MNRAS.498.4088M}
{McKernan} B, {Ford} KES, {O'Shaughnessy} R (2020{\natexlab{a}}) {Black hole,
  neutron star, and white dwarf merger rates in AGN discs}. \mnras
  498(3):4088--4094. \doi{10.1093/mnras/staa2681}.
  {\href{https://arxiv.org/abs/2002.00046}{{arXiv:2002.00046}}} {[astro-ph.HE]}

\bibitem[{{McKernan} et~al.(2020{\natexlab{b}}){McKernan}, {Ford},
  {O'Shaugnessy}, and {Wysocki}}]{2020MNRAS.494.1203M}
{McKernan} B, {Ford} KES, {O'Shaugnessy} R, {Wysocki} D (2020{\natexlab{b}})
  {Monte Carlo simulations of black hole mergers in AGN discs: Low
  {\ensuremath{\chi}}$_{eff}$ mergers and predictions for LIGO}. \mnras
  494(1):1203--1216. \doi{10.1093/mnras/staa740}.
  {\href{https://arxiv.org/abs/1907.04356}{{arXiv:1907.04356}}} {[astro-ph.HE]}

\bibitem[{{McKinney} et~al.(2015){McKinney}, {Dai}, and
  {Avara}}]{2015MNRAS.454L...6M}
{McKinney} JC, {Dai} L, {Avara} MJ (2015) {Efficiency of super-Eddington
  magnetically-arrested accretion}. \mnras 454(1):L6--L10.
  \doi{10.1093/mnrasl/slv115}.
  {\href{https://arxiv.org/abs/1508.02433}{{arXiv:1508.02433}}} {[astro-ph.HE]}

\bibitem[{{McMillan} et~al.(1991){McMillan}, {Hut}, and
  {Makino}}]{1991ApJ...372..111M}
{McMillan} S, {Hut} P, {Makino} J (1991) {Star Cluster Evolution with
  Primordial Binaries. II. Detailed Analysis}. \apj 372:111.
  \doi{10.1086/169958}

\bibitem[{{McNeill} et~al.(2020){McNeill}, {Mardling}, and
  {M{\"u}ller}}]{2020MNRAS.491.3000M}
{McNeill} LO, {Mardling} RA, {M{\"u}ller} B (2020) {Gravitational waves from
  dynamical tides in white-dwarf binaries}. \mnras 491(2):3000--3012.
  \doi{10.1093/mnras/stz3215}.
  {\href{https://arxiv.org/abs/1901.09045}{{arXiv:1901.09045}}} {[astro-ph.HE]}

\bibitem[{{Mei} et~al.(2020){Mei}, {Bai}, {Bao}, {Barausse}, {Cai}, {Canuto},
  {Cao}, {Chen}, and {et~al.}}]{2020arXiv200810332M}
{Mei} J, {Bai} YZ, {Bao} J, {Barausse} E, {Cai} L, {Canuto} E, {Cao} B, {Chen}
  WM, {et~al} (2020) {The TianQin project: current progress on science and
  technology}. arXiv e-prints arXiv:2008.10332.
  {\href{https://arxiv.org/abs/2008.10332}{{arXiv:2008.10332}}} {[gr-qc]}

\bibitem[{{Meiron} et~al.(2017){Meiron}, {Kocsis}, and
  {Loeb}}]{2017ApJ...834..200M}
{Meiron} Y, {Kocsis} B, {Loeb} A (2017) {Detecting Triple Systems with
  Gravitational Wave Observations}. \apj 834(2):200.
  \doi{10.3847/1538-4357/834/2/200}.
  {\href{https://arxiv.org/abs/1604.02148}{{arXiv:1604.02148}}} {[astro-ph.HE]}

\bibitem[{{Melvin} et~al.(2014){Melvin}, {Masters}, {Lintott}, {Nichol},
  {Simmons}, {Bamford}, {Casteels}, {Cheung}, {Edmondson}, {Fortson},
  {Schawinski}, {Skibba}, {Smith}, and {Willett}}]{2014MNRAS.438.2882M}
{Melvin} T, {Masters} K, {Lintott} C, {Nichol} RC, {Simmons} B, {Bamford} SP,
  {Casteels} KRV, {Cheung} E, {Edmondson} EM, {Fortson} L, et~al. (2014)
  {Galaxy Zoo: an independent look at the evolution of the bar fraction over
  the last eight billion years from HST-COSMOS}. \mnras 438(4):2882--2897.
  \doi{10.1093/mnras/stt2397}.
  {\href{https://arxiv.org/abs/1401.3334}{{arXiv:1401.3334}}} {[astro-ph.GA]}

\bibitem[{{Memmesheimer} et~al.(2004){Memmesheimer}, {Gopakumar}, and
  {Sch{\"a}fer}}]{2004PhRvD..70j4011M}
{Memmesheimer} RM, {Gopakumar} A, {Sch{\"a}fer} G (2004) {Third post-Newtonian
  accurate generalized quasi-Keplerian parametrization for compact binaries in
  eccentric orbits}. \prd 70(10):104011. \doi{10.1103/PhysRevD.70.104011}.
  {\href{https://arxiv.org/abs/gr-qc/0407049}{{gr-qc/0407049}}}

\bibitem[{{Menci} et~al.(2014){Menci}, {Gatti}, {Fiore}, and
  {Lamastra}}]{2014A&A...569A..37M}
{Menci} N, {Gatti} M, {Fiore} F, {Lamastra} A (2014) {Triggering active
  galactic nuclei in hierarchical galaxy formation: disk instability vs.
  interactions}. \aap 569:A37. \doi{10.1051/0004-6361/201424217}.
  {\href{https://arxiv.org/abs/1406.7740}{{arXiv:1406.7740}}} {[astro-ph.GA]}

\bibitem[{{Menou} et~al.(2008){Menou}, {Haiman}, and
  {Kocsis}}]{2008NewAR..51..884M}
{Menou} K, {Haiman} Z, {Kocsis} B (2008) {Cosmological physics with black holes
  (and possibly white dwarfs)}. \nar 51(10-12):884--890.
  \doi{10.1016/j.newar.2008.03.020}.
  {\href{https://arxiv.org/abs/0803.3627}{{arXiv:0803.3627}}} {[astro-ph]}

\bibitem[{{Merloni} and {Heinz}(2008)}]{2008MNRAS.388.1011M}
{Merloni} A, {Heinz} S (2008) {A synthesis model for AGN evolution:
  supermassive black holes growth and feedback modes}. \mnras
  388(3):1011--1030. \doi{10.1111/j.1365-2966.2008.13472.x}.
  {\href{https://arxiv.org/abs/0805.2499}{{arXiv:0805.2499}}} {[astro-ph]}

\bibitem[{{Merritt}(2001)}]{2001ApJ...556..245M}
{Merritt} D (2001) {Brownian Motion of a Massive Binary}. \apj 556(1):245--264.
  \doi{10.1086/321550}.
  {\href{https://arxiv.org/abs/astro-ph/0012264}{{arXiv:astro-ph/0012264}}}
  {[astro-ph]}

\bibitem[{{Merritt}(2004)}]{2004PhRvL..92t1304M}
{Merritt} D (2004) {Evolution of the Dark Matter Distribution at the Galactic
  Center}. \prl 92(20):201304. \doi{10.1103/PhysRevLett.92.201304}.
  {\href{https://arxiv.org/abs/astro-ph/0311594}{{arXiv:astro-ph/0311594}}}
  {[astro-ph]}

\bibitem[{{Merritt}(2013)}]{2013degn.book.....M}
{Merritt} D (2013) {Dynamics and Evolution of Galactic Nuclei}

\bibitem[{{Merritt}(2015)}]{2015ApJ...804...52M}
{Merritt} D (2015) {Gravitational Encounters and the Evolution of Galactic
  Nuclei. I. Method}. \apj 804(1):52. \doi{10.1088/0004-637X/804/1/52}.
  {\href{https://arxiv.org/abs/1505.07516}{{arXiv:1505.07516}}} {[astro-ph.GA]}

\bibitem[{{Merritt} and {Milosavljevi{\'c}}(2005)}]{2005LRR.....8....8M}
{Merritt} D, {Milosavljevi{\'c}} M (2005) {Massive Black Hole Binary
  Evolution}. Living Reviews in Relativity 8:8. \doi{10.12942/lrr-2005-8}.
  {\href{https://arxiv.org/abs/astro-ph/0410364}{{arXiv:astro-ph/0410364}}}
  {[astro-ph]}

\bibitem[{{Merritt} and {Poon}(2004)}]{2004ApJ...606..788M}
{Merritt} D, {Poon} MY (2004) {Chaotic Loss Cones and Black Hole Fueling}. \apj
  606(2):788--798. \doi{10.1086/382497}.
  {\href{https://arxiv.org/abs/astro-ph/0302296}{{arXiv:astro-ph/0302296}}}
  {[astro-ph]}

\bibitem[{{Merritt} and {Vasiliev}(2011)}]{2011ApJ...726...61M}
{Merritt} D, {Vasiliev} E (2011) {Orbits Around Black Holes in Triaxial
  Nuclei}. \apj 726(2):61. \doi{10.1088/0004-637X/726/2/61}.
  {\href{https://arxiv.org/abs/1005.0040}{{arXiv:1005.0040}}} {[astro-ph.GA]}

\bibitem[{{Merritt} et~al.(2002){Merritt}, {Milosavljevi{\'c}}, {Verde}, and
  {Jimenez}}]{2002PhRvL..88s1301M}
{Merritt} D, {Milosavljevi{\'c}} M, {Verde} L, {Jimenez} R (2002) {Dark Matter
  Spikes and Annihilation Radiation from the Galactic Center}. \prl
  88(19):191301. \doi{10.1103/PhysRevLett.88.191301}.
  {\href{https://arxiv.org/abs/astro-ph/0201376}{{arXiv:astro-ph/0201376}}}
  {[astro-ph]}

\bibitem[{{Merritt} et~al.(2004){Merritt}, {Milosavljevi{\'c}}, {Favata},
  {Hughes}, and {Holz}}]{2004ApJ...607L...9M}
{Merritt} D, {Milosavljevi{\'c}} M, {Favata} M, {Hughes} SA, {Holz} DE (2004)
  {Consequences of Gravitational Radiation Recoil}. \apjl 607(1):L9--L12.
  \doi{10.1086/421551}.
  {\href{https://arxiv.org/abs/astro-ph/0402057}{{arXiv:astro-ph/0402057}}}
  {[astro-ph]}

\bibitem[{{Merritt} et~al.(2011){Merritt}, {Alexander}, {Mikkola}, and
  {Will}}]{2011PhRvD..84d4024M}
{Merritt} D, {Alexander} T, {Mikkola} S, {Will} CM (2011) {Stellar dynamics of
  extreme-mass-ratio inspirals}. \prd 84(4):044024.
  \doi{10.1103/PhysRevD.84.044024}.
  {\href{https://arxiv.org/abs/1102.3180}{{arXiv:1102.3180}}} {[astro-ph.CO]}

\bibitem[{{Metzger}(2012)}]{2012MNRAS.419..827M}
{Metzger} BD (2012) {Nuclear-dominated accretion and subluminous supernovae
  from the merger of a white dwarf with a neutron star or black hole}. \mnras
  419(1):827--840. \doi{10.1111/j.1365-2966.2011.19747.x}.
  {\href{https://arxiv.org/abs/1105.6096}{{arXiv:1105.6096}}} {[astro-ph.HE]}

\bibitem[{{Metzger}(2019)}]{2019LRR....23....1M}
{Metzger} BD (2019) {Kilonovae}. Living Reviews in Relativity 23(1):1.
  \doi{10.1007/s41114-019-0024-0}.
  {\href{https://arxiv.org/abs/1910.01617}{{arXiv:1910.01617}}} {[astro-ph.HE]}

\bibitem[{{Metzger} and {Stone}(2016)}]{2016MNRAS.461..948M}
{Metzger} BD, {Stone} NC (2016) {A bright year for tidal disruptions}. \mnras
  461(1):948--966. \doi{10.1093/mnras/stw1394}.
  {\href{https://arxiv.org/abs/1506.03453}{{arXiv:1506.03453}}} {[astro-ph.HE]}

\bibitem[{{Mezcua}(2017)}]{2017IJMPD..2630021M}
{Mezcua} M (2017) {Observational evidence for intermediate-mass black holes}.
  International Journal of Modern Physics D 26(11):1730021.
  \doi{10.1142/S021827181730021X}.
  {\href{https://arxiv.org/abs/1705.09667}{{arXiv:1705.09667}}} {[astro-ph.GA]}

\bibitem[{{Mezcua} and {Dom{\'\i}nguez
  S{\'a}nchez}(2020)}]{2020ApJ...898L..30M}
{Mezcua} M, {Dom{\'\i}nguez S{\'a}nchez} H (2020) {Hidden AGNs in Dwarf
  Galaxies Revealed by MaNGA: Light Echoes, Off-nuclear Wanderers, and a New
  Broad-line AGN}. \apjl 898(2):L30. \doi{10.3847/2041-8213/aba199}.
  {\href{https://arxiv.org/abs/2007.08527}{{arXiv:2007.08527}}} {[astro-ph.GA]}

\bibitem[{{Mezcua} et~al.(2016){Mezcua}, {Civano}, {Fabbiano}, {Miyaji}, and
  {Marchesi}}]{2016ApJ...817...20M}
{Mezcua} M, {Civano} F, {Fabbiano} G, {Miyaji} T, {Marchesi} S (2016) {A
  Population of Intermediate-mass Black Holes in Dwarf Starburst Galaxies Up to
  Redshift=1.5}. \apj 817(1):20. \doi{10.3847/0004-637X/817/1/20}.
  {\href{https://arxiv.org/abs/1511.05844}{{arXiv:1511.05844}}} {[astro-ph.GA]}

\bibitem[{{Mezcua} et~al.(2018){Mezcua}, {Civano}, {Marchesi}, {Suh},
  {Fabbiano}, and {Volonteri}}]{2018MNRAS.478.2576M}
{Mezcua} M, {Civano} F, {Marchesi} S, {Suh} H, {Fabbiano} G, {Volonteri} M
  (2018) {Intermediate-mass black holes in dwarf galaxies out to redshift
  {\ensuremath{\sim}}2.4 in the Chandra COSMOS-Legacy Survey}. \mnras
  478(2):2576--2591. \doi{10.1093/mnras/sty1163}.
  {\href{https://arxiv.org/abs/1802.01567}{{arXiv:1802.01567}}} {[astro-ph.GA]}

\bibitem[{{Michaely} and {Perets}(2014)}]{2014ApJ...794..122M}
{Michaely} E, {Perets} HB (2014) {Secular Dynamics in Hierarchical Three-body
  Systems with Mass Loss and Mass Transfer}. \apj 794(2):122.
  \doi{10.1088/0004-637X/794/2/122}.
  {\href{https://arxiv.org/abs/1406.3035}{{arXiv:1406.3035}}} {[astro-ph.SR]}

\bibitem[{{Mikkola}(1983)}]{1983MNRAS.203.1107M}
{Mikkola} S (1983) {Encounters of binaries. I - Equal energies}. \mnras
  203:1107--1121. \doi{10.1093/mnras/203.4.1107}

\bibitem[{{Mikkola}(1984)}]{1984MNRAS.207..115M}
{Mikkola} S (1984) {Encounters of binaries. II - Unequal energies}. \mnras
  207:115--126. \doi{10.1093/mnras/207.1.115}

\bibitem[{{Mikkola} and {Valtonen}(1990)}]{1990ApJ...348..412M}
{Mikkola} S, {Valtonen} MJ (1990) {The Slingshot Ejections in Merging
  Galaxies}. \apj 348:412. \doi{10.1086/168250}

\bibitem[{{Mikkola} and {Valtonen}(1992)}]{1992MNRAS.259..115M}
{Mikkola} S, {Valtonen} MJ (1992) {Evolution of binaries in the field of light
  particles and the problem of two black holes}. \mnras 259(1):115--120.
  \doi{10.1093/mnras/259.1.115}

\bibitem[{{Miller} et~al.(2016){Miller}, {Wardell}, and
  {Pound}}]{2016PhRvD..94j4018M}
{Miller} J, {Wardell} B, {Pound} A (2016) {Second-order perturbation theory:
  The problem of infinite mode coupling}. \prd 94(10):104018.
  \doi{10.1103/PhysRevD.94.104018}.
  {\href{https://arxiv.org/abs/1608.06783}{{arXiv:1608.06783}}} {[gr-qc]}

\bibitem[{{Miller}(2007)}]{2007ARA&A..45..441M}
{Miller} JM (2007) {Relativistic X-Ray Lines from the Inner Accretion Disks
  Around Black Holes}. \araa 45(1):441--479.
  \doi{10.1146/annurev.astro.45.051806.110555}.
  {\href{https://arxiv.org/abs/0705.0540}{{arXiv:0705.0540}}} {[astro-ph]}

\bibitem[{{Miller}(2005)}]{2005ApJ...618..426M}
{Miller} MC (2005) {Probing General Relativity with Mergers of Supermassive and
  Intermediate-Mass Black Holes}. \apj 618(1):426--431. \doi{10.1086/425910}.
  {\href{https://arxiv.org/abs/astro-ph/0409331}{{arXiv:astro-ph/0409331}}}
  {[astro-ph]}

\bibitem[{{Miller} and {Colbert}(2004)}]{MillerColbert:2004}
{Miller} MC, {Colbert} EJM (2004) {Intermediate-Mass Black Holes}.
  International Journal of Modern Physics D 13:1--64.
  \doi{10.1142/S0218271804004426}.
  {\href{https://arxiv.org/abs/arXiv:astro-ph/0308402}{{arXiv:astro-ph/0308402}}}

\bibitem[{{Miller} and {Davies}(2012)}]{2012ApJ...755...81M}
{Miller} MC, {Davies} MB (2012) {An Upper Limit to the Velocity Dispersion of
  Relaxed Stellar Systems without Massive Black Holes}. \apj 755(1):81.
  \doi{10.1088/0004-637X/755/1/81}.
  {\href{https://arxiv.org/abs/1206.6167}{{arXiv:1206.6167}}} {[astro-ph.GA]}

\bibitem[{{Miller} and {Hamilton}(2002{\natexlab{a}})}]{2002ApJ...576..894M}
{Miller} MC, {Hamilton} DP (2002{\natexlab{a}}) {Four-Body Effects in Globular
  Cluster Black Hole Coalescence}. \apj 576(2):894--898. \doi{10.1086/341788}.
  {\href{https://arxiv.org/abs/astro-ph/0202298}{{arXiv:astro-ph/0202298}}}
  {[astro-ph]}

\bibitem[{{Miller} and {Hamilton}(2002{\natexlab{b}})}]{2002MNRAS.330..232C}
{Miller} MC, {Hamilton} DP (2002{\natexlab{b}}) {Production of
  intermediate-mass black holes in globular clusters}. \mnras 330(1):232--240.
  \doi{10.1046/j.1365-8711.2002.05112.x}.
  {\href{https://arxiv.org/abs/astro-ph/0106188}{{arXiv:astro-ph/0106188}}}
  {[astro-ph]}

\bibitem[{{Miller} and {Krolik}(2013)}]{2013ApJ...774...43M}
{Miller} MC, {Krolik} JH (2013) {Alignment of Supermassive Black Hole Binary
  Orbits and Spins}. \apj 774(1):43. \doi{10.1088/0004-637X/774/1/43}.
  {\href{https://arxiv.org/abs/1307.6569}{{arXiv:1307.6569}}} {[astro-ph.HE]}

\bibitem[{{Miller} and {Lauburg}(2009)}]{2009ApJ...692..917M}
{Miller} MC, {Lauburg} VM (2009) {Mergers of Stellar-Mass Black Holes in
  Nuclear Star Clusters}. \apj 692(1):917--923.
  \doi{10.1088/0004-637X/692/1/917}.
  {\href{https://arxiv.org/abs/0804.2783}{{arXiv:0804.2783}}} {[astro-ph]}

\bibitem[{{Miller} et~al.(2005){Miller}, {Freitag}, {Hamilton}, and
  {Lauburg}}]{2005ApJ...631L.117M}
{Miller} MC, {Freitag} M, {Hamilton} DP, {Lauburg} VM (2005) {Binary Encounters
  with Supermassive Black Holes: Zero-Eccentricity LISA Events}. \apjl
  631(2):L117--L120. \doi{10.1086/497335}.
  {\href{https://arxiv.org/abs/astro-ph/0507133}{{arXiv:astro-ph/0507133}}}
  {[astro-ph]}

\bibitem[{{Miller} et~al.(2019){Miller}, {Lamb}, {Dittmann}, {Bogdanov},
  {Arzoumanian}, {Gendreau}, {Guillot}, {Harding}, {Ho}, {Lattimer}, {Ludlam},
  and {et al.}}]{2019ApJ...887L..24M}
{Miller} MC, {Lamb} FK, {Dittmann} AJ, {Bogdanov} S, {Arzoumanian} Z,
  {Gendreau} KC, {Guillot} S, {Harding} AK, {Ho} WCG, {Lattimer} JM, et~al.
  (2019) {PSR J0030+0451 Mass and Radius from NICER Data and Implications for
  the Properties of Neutron Star Matter}. \apjl 887(1):L24.
  \doi{10.3847/2041-8213/ab50c5}.
  {\href{https://arxiv.org/abs/1912.05705}{{arXiv:1912.05705}}} {[astro-ph.HE]}

\bibitem[{{Miller-Jones} et~al.(2015){Miller-Jones}, {Strader}, {Heinke},
  {Maccarone}, {van den Berg}, {Knigge}, {Chomiuk}, {Noyola}, {Russell},
  {Seth}, and {Sivakoff}}]{2015MNRAS.453.3918M}
{Miller-Jones} JCA, {Strader} J, {Heinke} CO, {Maccarone} TJ, {van den Berg} M,
  {Knigge} C, {Chomiuk} L, {Noyola} E, {Russell} TD, {Seth} AC, et~al. (2015)
  {Deep radio imaging of 47 Tuc identifies the peculiar X-ray source X9 as a
  new black hole candidate}. \mnras 453(4):3918--3931.
  \doi{10.1093/mnras/stv1869}.
  {\href{https://arxiv.org/abs/1509.02579}{{arXiv:1509.02579}}} {[astro-ph.HE]}

\bibitem[{{Miller-Jones} et~al.(2021){Miller-Jones}, {Bahramian}, {Orosz},
  {Mandel}, {Gou}, {Maccarone}, {Neijssel}, {Zhao}, {Zi{\'o}{\l}kowski},
  {Reid}, {Uttley}, {Zheng}, {Byun}, {Dodson}, {Grinberg}, {Jung}, {Kim},
  {Marcote}, {Markoff}, {Rioja}, {Rushton}, {Russell}, {Sivakoff}, {Tetarenko},
  {Tudose}, and {Wilms}}]{2021Sci...371.1046M}
{Miller-Jones} JCA, {Bahramian} A, {Orosz} JA, {Mandel} I, {Gou} L, {Maccarone}
  TJ, {Neijssel} CJ, {Zhao} X, {Zi{\'o}{\l}kowski} J, {Reid} MJ, et~al. (2021)
  {Cygnus X-1 contains a 21{\textendash}solar mass black
  hole{\textemdash}Implications for massive star winds}. Science
  371(6533):1046--1049. \doi{10.1126/science.abb3363}.
  {\href{https://arxiv.org/abs/2102.09091}{{arXiv:2102.09091}}} {[astro-ph.HE]}

\bibitem[{{Milosavljevi{\'c}} and {Merritt}(2001)}]{2001ApJ...563...34M}
{Milosavljevi{\'c}} M, {Merritt} D (2001) {Formation of Galactic Nuclei}. \apj
  563(1):34--62. \doi{10.1086/323830}.
  {\href{https://arxiv.org/abs/astro-ph/0103350}{{arXiv:astro-ph/0103350}}}
  {[astro-ph]}

\bibitem[{{Milosavljevi{\'c}} and {Merritt}(2003)}]{2003ApJ...596..860M}
{Milosavljevi{\'c}} M, {Merritt} D (2003) {Long-Term Evolution of Massive Black
  Hole Binaries}. \apj 596(2):860--878. \doi{10.1086/378086}.
  {\href{https://arxiv.org/abs/astro-ph/0212459}{{arXiv:astro-ph/0212459}}}
  {[astro-ph]}

\bibitem[{{Milosavljevi{\'c}} and {Phinney}(2005)}]{2005ApJ...622L..93M}
{Milosavljevi{\'c}} M, {Phinney} ES (2005) {The Afterglow of Massive Black Hole
  Coalescence}. \apjl 622(2):L93--L96. \doi{10.1086/429618}.
  {\href{https://arxiv.org/abs/astro-ph/0410343}{{arXiv:astro-ph/0410343}}}
  {[astro-ph]}

\bibitem[{{Mingarelli} et~al.(2017){Mingarelli}, {Lazio}, {Sesana}, {Greene},
  {Ellis}, {Ma}, {Croft}, {Burke-Spolaor}, and {Taylor}}]{2017NatAs...1..886M}
{Mingarelli} CMF, {Lazio} TJW, {Sesana} A, {Greene} JE, {Ellis} JA, {Ma} CP,
  {Croft} S, {Burke-Spolaor} S, {Taylor} SR (2017) {The local nanohertz
  gravitational-wave landscape from supermassive black hole binaries}. Nature
  Astronomy 1:886--892. \doi{10.1038/s41550-017-0299-6}.
  {\href{https://arxiv.org/abs/1708.03491}{{arXiv:1708.03491}}} {[astro-ph.GA]}

\bibitem[{{Mirabel}(2017)}]{2017NewAR..78....1M}
{Mirabel} F (2017) {The formation of stellar black holes}. \nar 78:1--15.
  \doi{10.1016/j.newar.2017.04.002}

\bibitem[{{Miralda-Escud{\'e}} and {Kollmeier}(2005)}]{2005ApJ...619...30M}
{Miralda-Escud{\'e}} J, {Kollmeier} JA (2005) {Star Captures by Quasar
  Accretion Disks: A Possible Explanation of the M-{\ensuremath{\sigma}}
  Relation}. \apj 619(1):30--40. \doi{10.1086/426467}.
  {\href{https://arxiv.org/abs/astro-ph/0310717}{{arXiv:astro-ph/0310717}}}
  {[astro-ph]}

\bibitem[{{Mirza} et~al.(2017){Mirza}, {Tahir}, {Khan}, {Holley-Bockelmann},
  {Baig}, {Berczik}, and {Chishtie}}]{2017MNRAS.470..940M}
{Mirza} MA, {Tahir} A, {Khan} FM, {Holley-Bockelmann} H, {Baig} AM, {Berczik}
  P, {Chishtie} F (2017) {Galaxy rotation and supermassive black hole binary
  evolution}. \mnras 470(1):940--947. \doi{10.1093/mnras/stx1248}.
  {\href{https://arxiv.org/abs/1704.03490}{{arXiv:1704.03490}}} {[astro-ph.GA]}

\bibitem[{{Misra} et~al.(2020){Misra}, {Fragos}, {Tauris}, {Zapartas}, and
  {Aguilera-Dena}}]{2020A&A...642A.174M}
{Misra} D, {Fragos} T, {Tauris} TM, {Zapartas} E, {Aguilera-Dena} DR (2020)
  {The origin of pulsating ultra-luminous X-ray sources: Low- and
  intermediate-mass X-ray binaries containing neutron star accretors}. \aap
  642:A174. \doi{10.1051/0004-6361/202038070}.
  {\href{https://arxiv.org/abs/2004.01205}{{arXiv:2004.01205}}} {[astro-ph.HE]}

\bibitem[{{Montuori} et~al.(2011){Montuori}, {Dotti}, {Colpi}, {Decarli}, and
  {Haardt}}]{2011MNRAS.412...26M}
{Montuori} C, {Dotti} M, {Colpi} M, {Decarli} R, {Haardt} F (2011) {Search for
  sub-parsec massive binary black holes through line diagnosis}. \mnras
  412(1):26--32. \doi{10.1111/j.1365-2966.2010.17888.x}.
  {\href{https://arxiv.org/abs/1010.4303}{{arXiv:1010.4303}}} {[astro-ph.CO]}

\bibitem[{{Montuori} et~al.(2012){Montuori}, {Dotti}, {Haardt}, {Colpi}, and
  {Decarli}}]{2012MNRAS.425.1633M}
{Montuori} C, {Dotti} M, {Haardt} F, {Colpi} M, {Decarli} R (2012) {Search for
  sub-parsec massive binary black holes through line diagnosis - II}. \mnras
  425(3):1633--1639. \doi{10.1111/j.1365-2966.2012.21530.x}.
  {\href{https://arxiv.org/abs/1207.0813}{{arXiv:1207.0813}}} {[astro-ph.CO]}

\bibitem[{{Moody} et~al.(2019){Moody}, {Shi}, and
  {Stone}}]{2019ApJ...875...66M}
{Moody} MSL, {Shi} JM, {Stone} JM (2019) {Hydrodynamic Torques in Circumbinary
  Accretion Disks}. \apj 875(1):66. \doi{10.3847/1538-4357/ab09ee}.
  {\href{https://arxiv.org/abs/1903.00008}{{arXiv:1903.00008}}} {[astro-ph.HE]}

\bibitem[{{Moore}(1994)}]{1994Natur.370..629M}
{Moore} B (1994) {Evidence against dissipation-less dark matter from
  observations of galaxy haloes}. \nat 370(6491):629--631.
  \doi{10.1038/370629a0}

\bibitem[{{Moore} et~al.(2015){Moore}, {Cole}, and
  {Berry}}]{2015CQGra..32a5014M}
{Moore} CJ, {Cole} RH, {Berry} CPL (2015) {Gravitational-wave sensitivity
  curves}. Classical and Quantum Gravity 32(1):015014.
  \doi{10.1088/0264-9381/32/1/015014}.
  {\href{https://arxiv.org/abs/1408.0740}{{arXiv:1408.0740}}} {[gr-qc]}

\bibitem[{{Moore} et~al.(2017){Moore}, {Mihaylov}, {Lasenby}, and
  {Gilmore}}]{2017PhRvL.119z1102M}
{Moore} CJ, {Mihaylov} DP, {Lasenby} A, {Gilmore} G (2017) {Astrometric Search
  Method for Individually Resolvable Gravitational Wave Sources with Gaia}.
  \prl 119(26):261102. \doi{10.1103/PhysRevLett.119.261102}.
  {\href{https://arxiv.org/abs/1707.06239}{{arXiv:1707.06239}}} {[astro-ph.IM]}

\bibitem[{{Moore} et~al.(2019){Moore}, {Gerosa}, and
  {Klein}}]{2019MNRAS.488L..94M}
{Moore} CJ, {Gerosa} D, {Klein} A (2019) {Are stellar-mass black-hole binaries
  too quiet for LISA?} \mnras 488(1):L94--L98. \doi{10.1093/mnrasl/slz104}.
  {\href{https://arxiv.org/abs/1905.11998}{{arXiv:1905.11998}}} {[astro-ph.HE]}

\bibitem[{{Moran} et~al.(2014){Moran}, {Shahinyan}, {Sugarman}, {V{\'e}lez},
  and {Eracleous}}]{2014AJ....148..136M}
{Moran} EC, {Shahinyan} K, {Sugarman} HR, {V{\'e}lez} DO, {Eracleous} M (2014)
  {Black Holes At the Centers of Nearby Dwarf Galaxies}. \aj 148(6):136.
  \doi{10.1088/0004-6256/148/6/136}.
  {\href{https://arxiv.org/abs/1408.4451}{{arXiv:1408.4451}}} {[astro-ph.GA]}

\bibitem[{{Morawski} et~al.(2018){Morawski}, {Giersz}, {Askar}, and
  {Belczynski}}]{2018MNRAS.481.2168M}
{Morawski} J, {Giersz} M, {Askar} A, {Belczynski} K (2018) {MOCCA-SURVEY
  Database I: Assessing GW kick retention fractions for BH-BH mergers in
  globular clusters}. \mnras 481(2):2168--2179. \doi{10.1093/mnras/sty2401}.
  {\href{https://arxiv.org/abs/1802.01192}{{arXiv:1802.01192}}} {[astro-ph.GA]}

\bibitem[{{Morris}(1993)}]{1993ApJ...408..496M}
{Morris} M (1993) {Massive Star Formation near the Galactic Center and the Fate
  of the Stellar Remnants}. \apj 408:496. \doi{10.1086/172607}

\bibitem[{{Morscher} et~al.(2013){Morscher}, {Umbreit}, {Farr}, and
  {Rasio}}]{2013ApJ...763L..15M}
{Morscher} M, {Umbreit} S, {Farr} WM, {Rasio} FA (2013) {Retention of
  Stellar-mass Black Holes in Globular Clusters}. \apjl 763(1):L15.
  \doi{10.1088/2041-8205/763/1/L15}.
  {\href{https://arxiv.org/abs/1211.3372}{{arXiv:1211.3372}}} {[astro-ph.GA]}

\bibitem[{{Morscher} et~al.(2015){Morscher}, {Pattabiraman}, {Rodriguez},
  {Rasio}, and {Umbreit}}]{2015ApJ...800....9M}
{Morscher} M, {Pattabiraman} B, {Rodriguez} C, {Rasio} FA, {Umbreit} S (2015)
  {The Dynamical Evolution of Stellar Black Holes in Globular Clusters}. \apj
  800(1):9. \doi{10.1088/0004-637X/800/1/9}.
  {\href{https://arxiv.org/abs/1409.0866}{{arXiv:1409.0866}}} {[astro-ph.GA]}

\bibitem[{{Mortlock} et~al.(2011){Mortlock}, {Warren}, {Venemans}, {Patel},
  {Hewett}, {McMahon}, {Simpson}, {Theuns}, {Gonz{\'a}les-Solares}, {Adamson},
  {Dye}, {Hambly}, {Hirst}, {Irwin}, {Kuiper}, {Lawrence}, and
  {R{\"o}ttgering}}]{2011Natur.474..616M}
{Mortlock} DJ, {Warren} SJ, {Venemans} BP, {Patel} M, {Hewett} PC, {McMahon}
  RG, {Simpson} C, {Theuns} T, {Gonz{\'a}les-Solares} EA, {Adamson} A, et~al.
  (2011) {A luminous quasar at a redshift of z = 7.085}. \nat
  474(7353):616--619. \doi{10.1038/nature10159}.
  {\href{https://arxiv.org/abs/1106.6088}{{arXiv:1106.6088}}} {[astro-ph.CO]}

\bibitem[{M\"osta et~al.(2010)}]{2010PhRvD..81f4017M}
M\"osta P, et~al. (2010) Vacuum electromagnetic counterparts of binary
  black-hole mergers. Phys Rev D 81:064017. \doi{10.1103/PhysRevD.81.064017},
  \urlprefix\url{https://link.aps.org/doi/10.1103/PhysRevD.81.064017}

\bibitem[{M\"{o}sta et~al.(2012)}]{2012ApJ...749L..32M}
M\"{o}sta P, et~al. (2012) On the detectability of dual jets from binary black
  holes. ApJ 749(2):L32. \doi{10.1088/2041-8205/749/2/l32},
  \urlprefix\url{https://doi.org/10.1088\%2F2041-8205\%2F749\%2F2\%2Fl32}

\bibitem[{{Motl} et~al.(2007){Motl}, {Frank}, {Tohline}, and
  {D'Souza}}]{2007ApJ...670.1314M}
{Motl} PM, {Frank} J, {Tohline} JE, {D'Souza} MCR (2007) {The Stability of
  Double White Dwarf Binaries Undergoing Direct-Impact Accretion}. \apj
  670(2):1314--1325. \doi{10.1086/522076}.
  {\href{https://arxiv.org/abs/astro-ph/0702388}{{arXiv:astro-ph/0702388}}}
  {[astro-ph]}

\bibitem[{{Moxon} and {Flanagan}(2018)}]{2018PhRvD..97j5001M}
{Moxon} J, {Flanagan} {\'E} (2018) {Radiation-reaction force on a small charged
  body to second order}. \prd 97(10):105001. \doi{10.1103/PhysRevD.97.105001}.
  {\href{https://arxiv.org/abs/1711.05212}{{arXiv:1711.05212}}} {[gr-qc]}

\bibitem[{{Mu{\~n}oz} et~al.(2019){Mu{\~n}oz}, {Miranda}, and
  {Lai}}]{2019ApJ...871...84M}
{Mu{\~n}oz} DJ, {Miranda} R, {Lai} D (2019) {Hydrodynamics of Circumbinary
  Accretion: Angular Momentum Transfer and Binary Orbital Evolution}. \apj
  871(1):84. \doi{10.3847/1538-4357/aaf867}.
  {\href{https://arxiv.org/abs/1810.04676}{{arXiv:1810.04676}}} {[astro-ph.HE]}

\bibitem[{{Mu{\~n}oz} et~al.(2020){Mu{\~n}oz}, {Lai}, {Kratter}, and {Mirand
  a}}]{2020ApJ...889..114M}
{Mu{\~n}oz} DJ, {Lai} D, {Kratter} K, {Mirand a} R (2020) {Circumbinary
  Accretion from Finite and Infinite Disks}. \apj 889(2):114.
  \doi{10.3847/1538-4357/ab5d33}.
  {\href{https://arxiv.org/abs/1910.04763}{{arXiv:1910.04763}}} {[astro-ph.HE]}

\bibitem[{{M{\"u}ller} et~al.(2019){M{\"u}ller}, {Tauris}, {Heger}, {Banerjee},
  {Qian}, {Powell}, {Chan}, {Gay}, and {Langer}}]{2019MNRAS.484.3307M}
{M{\"u}ller} B, {Tauris} TM, {Heger} A, {Banerjee} P, {Qian} YZ, {Powell} J,
  {Chan} C, {Gay} DW, {Langer} N (2019) {Three-dimensional simulations of
  neutrino-driven core-collapse supernovae from low-mass single and binary star
  progenitors}. \mnras 484(3):3307--3324. \doi{10.1093/mnras/stz216}.
  {\href{https://arxiv.org/abs/1811.05483}{{arXiv:1811.05483}}} {[astro-ph.SR]}

\bibitem[{{Munna}(2020)}]{2020arXiv200810622M}
{Munna} C (2020) {Analytic post-Newtonian expansion of the energy and angular
  momentum radiated to infinity by eccentric-orbit non-spinning
  extreme-mass-ratio inspirals to 19PN}. arXiv e-prints arXiv:2008.10622.
  {\href{https://arxiv.org/abs/2008.10622}{{arXiv:2008.10622}}} {[gr-qc]}

\bibitem[{{Muratov} and {Gnedin}(2010)}]{2010ApJ...718.1266M}
{Muratov} AL, {Gnedin} OY (2010) {Modeling the Metallicity Distribution of
  Globular Clusters}. \apj 718(2):1266--1288.
  \doi{10.1088/0004-637X/718/2/1266}.
  {\href{https://arxiv.org/abs/1002.1325}{{arXiv:1002.1325}}} {[astro-ph.GA]}

\bibitem[{{Murguia-Berthier} et~al.(2020){Murguia-Berthier}, {Batta}, {Janiuk},
  {Ramirez-Ruiz}, {Mandel}, {Noble}, and {Everson}}]{2020ApJ...901L..24M}
{Murguia-Berthier} A, {Batta} A, {Janiuk} A, {Ramirez-Ruiz} E, {Mandel} I,
  {Noble} SC, {Everson} RW (2020) {On the Maximum Stellar Rotation to form a
  Black Hole without an Accompanying Luminous Transient}. \apjl 901(2):L24.
  \doi{10.3847/2041-8213/abb818}.
  {\href{https://arxiv.org/abs/2005.10212}{{arXiv:2005.10212}}} {[astro-ph.HE]}

\bibitem[{{Murphy} et~al.(2018){Murphy}, {Bolatto}, {Chatterjee}, {Casey},
  {Chomiuk}, {Dale}, {de Pater}, {Dickinson}, {Francesco}, {Hallinan},
  {Isella}, {Kohno}, {Kulkarni}, {Lang}, {Lazio}, {Leroy}, {Loinard},
  {Maccarone}, {Matthews}, {Osten}, {Reid}, {Riechers}, {Sakai}, {Walter}, and
  {Wilner}}]{2018ASPC..517....3M}
{Murphy} EJ, {Bolatto} A, {Chatterjee} S, {Casey} CM, {Chomiuk} L, {Dale} D,
  {de Pater} I, {Dickinson} M, {Francesco} JD, {Hallinan} G, et~al. (2018) {The
  ngVLA Science Case and Associated Science Requirements}. In: {Murphy} E (ed)
  Science with a Next Generation Very Large Array. Astronomical Society of the
  Pacific Conference Series, vol 517. p~3.
  {\href{https://arxiv.org/abs/1810.07524}{{arXiv:1810.07524}}} {[astro-ph.IM]}

\bibitem[{{Murray} and {Dermott}(1999)}]{1999ssd..book.....M}
{Murray} CD, {Dermott} SF (1999) {Solar system dynamics}

\bibitem[{{Mushotzky}(2018)}]{2018SPIE10699E..29M}
{Mushotzky} R (2018) {AXIS: a probe class next generation high angular
  resolution x-ray imaging satellite}. In: Space Telescopes and Instrumentation
  2018: Ultraviolet to Gamma Ray. Society of Photo-Optical Instrumentation
  Engineers (SPIE) Conference Series, vol 10699. p 1069929.
  \doi{10.1117/12.2310003}.
  {\href{https://arxiv.org/abs/1807.02122}{{arXiv:1807.02122}}} {[astro-ph.HE]}

\bibitem[{{Mushotzky} et~al.(2019){Mushotzky}, {Aird}, {Barger}, {Cappelluti},
  {Chartas}, {Corrales}, {Eufrasio}, {Fabian}, {Falcone}, {Gallo}, {Gilli},
  {Grant}, {Hardcastle}, {Hodges-Kluck}, {Kara}, {Koss}, {Li}, {Lisse},
  {Loewenstein}, {Markevitch}, {Meyer}, {Miller}, {Mulchaey}, {Petre}, {Ptak},
  {Reynolds}, {Russell}, {Safi-Harb}, {Smith}, {Snios}, {Tombesi}, {Valencic},
  {Walker}, {Williams}, {Winter}, {Yamaguchi}, {Zhang}, {Arenberg}, {Brandt},
  {Burrows}, {Georganopoulos}, {Miller}, {Norman}, and
  {Rosati}}]{2019BAAS...51g.107M}
{Mushotzky} R, {Aird} J, {Barger} AJ, {Cappelluti} N, {Chartas} G, {Corrales}
  L, {Eufrasio} R, {Fabian} AC, {Falcone} AD, {Gallo} E, et~al. (2019) {The
  Advanced X-ray Imaging Satellite}. In: Bulletin of the American Astronomical
  Society. vol~51. p 107.
  {\href{https://arxiv.org/abs/1903.04083}{{arXiv:1903.04083}}} {[astro-ph.HE]}

\bibitem[{{Nakama} et~al.(2016){Nakama}, {Suyama}, and
  {Yokoyama}}]{2016PhRvD..94j3522N}
{Nakama} T, {Suyama} T, {Yokoyama} J (2016) {Supermassive black holes formed by
  direct collapse of inflationary perturbations}. \prd 94(10):103522.
  \doi{10.1103/PhysRevD.94.103522}.
  {\href{https://arxiv.org/abs/1609.02245}{{arXiv:1609.02245}}} {[gr-qc]}

\bibitem[{Nakama et~al.(2018)Nakama, Carr, and Silk}]{Nakama:2017xvq}
Nakama T, Carr B, Silk J (2018) {Limits on primordial black holes from $\mu$
  distortions in cosmic microwave background}. Phys Rev D 97(4):043525.
  \doi{10.1103/PhysRevD.97.043525}.
  {\href{https://arxiv.org/abs/1710.06945}{{arXiv:1710.06945}}} {[astro-ph.CO]}

\bibitem[{{Nandez} and {Ivanova}(2016)}]{2016MNRAS.460.3992N}
{Nandez} JLA, {Ivanova} N (2016) {Common envelope events with low-mass giants:
  understanding the energy budget}. \mnras 460(4):3992--4002.
  \doi{10.1093/mnras/stw1266}.
  {\href{https://arxiv.org/abs/1606.04922}{{arXiv:1606.04922}}} {[astro-ph.SR]}

\bibitem[{{Nandez} et~al.(2015){Nandez}, {Ivanova}, and
  {Lombardi}}]{2015MNRAS.450L..39N}
{Nandez} JLA, {Ivanova} N, {Lombardi} JCJ (2015) {Recombination energy in
  double white dwarf formation.} \mnras 450:L39--L43.
  \doi{10.1093/mnrasl/slv043}.
  {\href{https://arxiv.org/abs/1503.02750}{{arXiv:1503.02750}}} {[astro-ph.SR]}

\bibitem[{{Nandra} et~al.(1997){Nandra}, {George}, {Mushotzky}, {Turner}, and
  {Yaqoob}}]{1997ApJ...477..602N}
{Nandra} K, {George} IM, {Mushotzky} RF, {Turner} TJ, {Yaqoob} T (1997) {ASCA
  Observations of Seyfert 1 Galaxies. II. Relativistic Iron
  K{\ensuremath{\alpha}} Emission}. \apj 477(2):602--622. \doi{10.1086/303721}.
  {\href{https://arxiv.org/abs/astro-ph/9606169}{{arXiv:astro-ph/9606169}}}
  {[astro-ph]}

\bibitem[{{Nandra} et~al.(2013){Nandra}, {Barret}, {Barcons}, {Fabian}, {den
  Herder}, {Piro}, {Watson}, {Adami}, {Aird}, {Afonso}, {Alexander},
  {Argiroffi}, {Amati}, {Arnaud}, {Atteia}, {Audard}, {Badenes}, {Ballet},
  {Ballo}, {Bamba}, {Bhardwaj}, {Stefano Battistelli}, {Becker}, {De Becker},
  {Behar}, {Bianchi}, {Biffi}, {B{\^\i}rzan}, {Bocchino}, {Bogdanov}, {Boirin},
  {Boller}, {Borgani}, {Borm}, {Bouch{\'e}}, {Bourdin}, {Bower}, {Braito},
  {Branchini}, {Branduardi-Raymont}, {Bregman}, {Brenneman}, {Brightman},
  {Br{\"u}ggen}, {Buchner}, {Bulbul}, {Brusa}, {Bursa}, {Caccianiga},
  {Cackett}, {Campana}, {Cappelluti}, {Cappi}, {Carrera}, {Ceballos},
  {Christensen}, {Chu}, {Churazov}, {Clerc}, {Corbel}, {Corral}, {Comastri},
  {Costantini}, {Croston}, {Dadina}, {D'Ai}, {Decourchelle}, {Della Ceca},
  {Dennerl}, {Dolag}, {Done}, {Dovciak}, {Drake}, {Eckert}, {Edge}, {Ettori},
  {Ezoe}, {Feigelson}, {Fender}, {Feruglio}, {Finoguenov}, {Fiore}, {Galeazzi},
  {Gallagher}, {Gandhi}, {Gaspari}, {Gastaldello}, {Georgakakis},
  {Georgantopoulos}, {Gilfanov}, {Gitti}, {Gladstone}, {Goosmann}, {Gosset},
  {Grosso}, {Guedel}, {Guerrero}, {Haberl}, {Hardcastle}, {Heinz}, {Alonso
  Herrero}, {Herv{\'e}}, {Holmstrom}, {Iwasawa}, {Jonker}, {Kaastra}, {Kara},
  {Karas}, {Kastner}, {King}, {Kosenko}, {Koutroumpa}, {Kraft}, {Kreykenbohm},
  {Lallement}, {Lanzuisi}, {Lee}, {Lemoine-Goumard}, {Lobban}, {Lodato},
  {Lovisari}, {Lotti}, {McCharthy}, {McNamara}, {Maggio}, {Maiolino}, {De
  Marco}, {de Martino}, {Mateos}, {Matt}, {Maughan}, {Mazzotta}, {Mendez},
  {Merloni}, {Micela}, {Miceli}, {Mignani}, {Miller}, {Miniutti}, {Molendi},
  {Montez}, {Moretti}, {Motch}, {Naz{\'e}}, {Nevalainen}, {Nicastro}, {Nulsen},
  {Ohashi}, {O'Brien}, {Osborne}, {Oskinova}, {Pacaud}, {Paerels}, {Page},
  {Papadakis}, {Pareschi}, {Petre}, {Petrucci}, {Piconcelli}, {Pillitteri},
  {Pinto}, {de Plaa}, {Pointecouteau}, {Ponman}, {Ponti}, {Porquet}, {Pounds},
  {Pratt}, {Predehl}, {Proga}, {Psaltis}, {Rafferty}, {Ramos-Ceja}, {Ranalli},
  {Rasia}, {Rau}, {Rauw}, {Rea}, {Read}, {Reeves}, {Reiprich}, {Renaud},
  {Reynolds}, {Risaliti}, {Rodriguez}, {Rodriguez Hidalgo}, {Roncarelli},
  {Rosario}, {Rossetti}, {Rozanska}, {Rovilos}, {Salvaterra}, {Salvato}, {Di
  Salvo}, {Sanders}, {Sanz-Forcada}, {Schawinski}, {Schaye}, {Schwope},
  {Sciortino}, {Severgnini}, {Shankar}, {Sijacki}, {Sim}, {Schmid}, {Smith},
  {Steiner}, {Stelzer}, {Stewart}, {Strohmayer}, {Str{\"u}der}, {Sun}, {Takei},
  {Tatischeff}, {Tiengo}, {Tombesi}, {Trinchieri}, {Tsuru}, {Ud-Doula},
  {Ursino}, {Valencic}, {Vanzella}, {Vaughan}, {Vignali}, {Vink}, {Vito},
  {Volonteri}, {Wang}, {Webb}, {Willingale}, {Wilms}, {Wise}, {Worrall},
  {Young}, {Zampieri}, {In't Zand}, {Zane}, {Zezas}, {Zhang}, and
  {Zhuravleva}}]{2013arXiv1306.2307N}
{Nandra} K, {Barret} D, {Barcons} X, {Fabian} A, {den Herder} JW, {Piro} L,
  {Watson} M, {Adami} C, {Aird} J, {Afonso} JM, et~al. (2013) {The Hot and
  Energetic Universe: A White Paper presenting the science theme motivating the
  Athena+ mission}. arXiv e-prints arXiv:1306.2307.
  {\href{https://arxiv.org/abs/1306.2307}{{arXiv:1306.2307}}} {[astro-ph.HE]}

\bibitem[{{Naoz}(2016)}]{2016ARA&A..54..441N}
{Naoz} S (2016) {The Eccentric Kozai-Lidov Effect and Its Applications}. \araa
  54:441--489. \doi{10.1146/annurev-astro-081915-023315}.
  {\href{https://arxiv.org/abs/1601.07175}{{arXiv:1601.07175}}} {[astro-ph.EP]}

\bibitem[{{Naoz} et~al.(2020){Naoz}, {Will}, {Ramirez-Ruiz}, {Hees}, {Ghez},
  and {Do}}]{2020ApJ...888L...8N}
{Naoz} S, {Will} CM, {Ramirez-Ruiz} E, {Hees} A, {Ghez} AM, {Do} T (2020) {A
  Hidden Friend for the Galactic Center Black Hole, Sgr A*}. \apjl 888(1):L8.
  \doi{10.3847/2041-8213/ab5e3b}.
  {\href{https://arxiv.org/abs/1912.04910}{{arXiv:1912.04910}}} {[astro-ph.GA]}

\bibitem[{{Napiwotzki} et~al.(2020){Napiwotzki}, {Karl}, {Lisker},
  {Catal{\'a}n}, {Drechsel}, {Heber}, {Homeier}, {Koester}, {Leibundgut},
  {Marsh}, {Moehler}, {Nelemans}, {Reimers}, {Renzini}, {Str{\"o}er}, and
  {Yungelson}}]{2020A&A...638A.131N}
{Napiwotzki} R, {Karl} CA, {Lisker} T, {Catal{\'a}n} S, {Drechsel} H, {Heber}
  U, {Homeier} D, {Koester} D, {Leibundgut} B, {Marsh} TR, et~al. (2020) {The
  ESO supernovae type Ia progenitor survey (SPY). The radial velocities of 643
  DA white dwarfs}. \aap 638:A131. \doi{10.1051/0004-6361/201629648}.
  {\href{https://arxiv.org/abs/1906.10977}{{arXiv:1906.10977}}} {[astro-ph.SR]}

\bibitem[{{Narayan} and {Yi}(1994)}]{1994ApJ...428L..13N}
{Narayan} R, {Yi} I (1994) {Advection-dominated Accretion: A Self-similar
  Solution}. \apjl 428:L13. \doi{10.1086/187381}.
  {\href{https://arxiv.org/abs/astro-ph/9403052}{{arXiv:astro-ph/9403052}}}
  {[astro-ph]}

\bibitem[{{Narayan} et~al.(1992){Narayan}, {Paczynski}, and
  {Piran}}]{1992ApJ...395L..83N}
{Narayan} R, {Paczynski} B, {Piran} T (1992) {Gamma-Ray Bursts as the Death
  Throes of Massive Binary Stars}. \apjl 395:L83. \doi{10.1086/186493}.
  {\href{https://arxiv.org/abs/astro-ph/9204001}{{arXiv:astro-ph/9204001}}}
  {[astro-ph]}

\bibitem[{{Natarajan}(2020)}]{2020arXiv200909156N}
{Natarajan} P (2020) {A new channel to form IMBHs throughout cosmic time}.
  arXiv e-prints arXiv:2009.09156.
  {\href{https://arxiv.org/abs/2009.09156}{{arXiv:2009.09156}}} {[astro-ph.GA]}

\bibitem[{{Natarajan} et~al.(2017){Natarajan}, {Pacucci}, {Ferrara}, {Agarwal},
  {Ricarte}, {Zackrisson}, and {Cappelluti}}]{2017ApJ...838..117N}
{Natarajan} P, {Pacucci} F, {Ferrara} A, {Agarwal} B, {Ricarte} A, {Zackrisson}
  E, {Cappelluti} N (2017) {Unveiling the First Black Holes With
  JWST:Multi-wavelength Spectral Predictions}. \apj 838(2):117.
  \doi{10.3847/1538-4357/aa6330}.
  {\href{https://arxiv.org/abs/1610.05312}{{arXiv:1610.05312}}} {[astro-ph.GA]}

\bibitem[{{Nayakshin} et~al.(2007){Nayakshin}, {Cuadra}, and
  {Springel}}]{2007MNRAS.379...21N}
{Nayakshin} S, {Cuadra} J, {Springel} V (2007) {Simulations of star formation
  in a gaseous disc around Sgr A* - a failed active galactic nucleus}. \mnras
  379(1):21--33. \doi{10.1111/j.1365-2966.2007.11938.x}.
  {\href{https://arxiv.org/abs/astro-ph/0701141}{{arXiv:astro-ph/0701141}}}
  {[astro-ph]}

\bibitem[{{Negri} and {Volonteri}(2017)}]{2017MNRAS.467.3475N}
{Negri} A, {Volonteri} M (2017) {Black hole feeding and feedback: the physics
  inside the `sub-grid'}. \mnras 467(3):3475--3492. \doi{10.1093/mnras/stx362}.
  {\href{https://arxiv.org/abs/1610.04753}{{arXiv:1610.04753}}} {[astro-ph.GA]}

\bibitem[{{Neijssel} et~al.(2019){Neijssel}, {Vigna-G{\'o}mez}, {Stevenson},
  {Barrett}, {Gaebel}, {Broekgaarden}, {de Mink}, {Sz{\'e}csi}, {Vinciguerra},
  and {Mandel}}]{2019MNRAS.490.3740N}
{Neijssel} CJ, {Vigna-G{\'o}mez} A, {Stevenson} S, {Barrett} JW, {Gaebel} SM,
  {Broekgaarden} FS, {de Mink} SE, {Sz{\'e}csi} D, {Vinciguerra} S, {Mandel} I
  (2019) {The effect of the metallicity-specific star formation history on
  double compact object mergers}. \mnras 490(3):3740--3759.
  \doi{10.1093/mnras/stz2840}.
  {\href{https://arxiv.org/abs/1906.08136}{{arXiv:1906.08136}}} {[astro-ph.SR]}

\bibitem[{{Nelemans} and {Tauris}(1998)}]{1998A&A...335L..85N}
{Nelemans} G, {Tauris} TM (1998) {Formation of undermassive single white dwarfs
  and the influence of planets on late stellar evolution}. \aap 335:L85--L88.
  {\href{https://arxiv.org/abs/astro-ph/9806011}{{arXiv:astro-ph/9806011}}}
  {[astro-ph]}

\bibitem[{{Nelemans} et~al.(2000){Nelemans}, {Verbunt}, {Yungelson}, and
  {Portegies Zwart}}]{2000A&A...360.1011N}
{Nelemans} G, {Verbunt} F, {Yungelson} LR, {Portegies Zwart} SF (2000)
  {Reconstructing the evolution of double helium white dwarfs: envelope loss
  without spiral-in}. \aap 360:1011--1018.
  {\href{https://arxiv.org/abs/astro-ph/0006216}{{arXiv:astro-ph/0006216}}}
  {[astro-ph]}

\bibitem[{{Nelemans} et~al.(2001{\natexlab{a}}){Nelemans}, {Portegies Zwart},
  {Verbunt}, and {Yungelson}}]{2001A&A...368..939N}
{Nelemans} G, {Portegies Zwart} SF, {Verbunt} F, {Yungelson} LR
  (2001{\natexlab{a}}) {Population synthesis for double white dwarfs. II.
  Semi-detached systems: AM CVn stars}. \aap 368:939--949.
  \doi{10.1051/0004-6361:20010049}.
  {\href{https://arxiv.org/abs/astro-ph/0101123}{{arXiv:astro-ph/0101123}}}
  {[astro-ph]}

\bibitem[{{Nelemans} et~al.(2001{\natexlab{b}}){Nelemans}, {Yungelson}, and
  {Portegies Zwart}}]{2001A&A...375..890N}
{Nelemans} G, {Yungelson} LR, {Portegies Zwart} SF (2001{\natexlab{b}}) {The
  gravitational wave signal from the Galactic disk population of binaries
  containing two compact objects}. \aap 375:890--898.
  \doi{10.1051/0004-6361:20010683}.
  {\href{https://arxiv.org/abs/astro-ph/0105221}{{arXiv:astro-ph/0105221}}}
  {[astro-ph]}

\bibitem[{{Nelemans} et~al.(2001{\natexlab{c}}){Nelemans}, {Yungelson},
  {Portegies Zwart}, and {Verbunt}}]{2001A&A...365..491N}
{Nelemans} G, {Yungelson} LR, {Portegies Zwart} SF, {Verbunt} F
  (2001{\natexlab{c}}) {Population synthesis for double white dwarfs . I. Close
  detached systems}. \aap 365:491--507. \doi{10.1051/0004-6361:20000147}.
  {\href{https://arxiv.org/abs/astro-ph/0010457}{{arXiv:astro-ph/0010457}}}
  {[astro-ph]}

\bibitem[{{Nelemans} et~al.(2004{\natexlab{a}}){Nelemans}, {Jonker}, {Marsh},
  and {van der Klis}}]{2004MNRAS.348L...7N}
{Nelemans} G, {Jonker} PG, {Marsh} TR, {van der Klis} M (2004{\natexlab{a}})
  {Optical spectra of the carbon-oxygen accretion discs in the ultra-compact
  X-ray binaries 4U 0614+09, 4U 1543-624 and 2S 0918-549}. \mnras
  348(1):L7--L11. \doi{10.1111/j.1365-2966.2004.07486.x}.
  {\href{https://arxiv.org/abs/astro-ph/0312008}{{arXiv:astro-ph/0312008}}}
  {[astro-ph]}

\bibitem[{{Nelemans} et~al.(2004{\natexlab{b}}){Nelemans}, {Yungelson}, and
  {Portegies Zwart}}]{2004MNRAS.349..181N}
{Nelemans} G, {Yungelson} LR, {Portegies Zwart} SF (2004{\natexlab{b}})
  {Short-period AM CVn systems as optical, X-ray and gravitational-wave
  sources}. \mnras 349(1):181--192. \doi{10.1111/j.1365-2966.2004.07479.x}.
  {\href{https://arxiv.org/abs/astro-ph/0312193}{{arXiv:astro-ph/0312193}}}
  {[astro-ph]}

\bibitem[{{Nelemans} et~al.(2005){Nelemans}, {Napiwotzki}, {Karl}, {Marsh},
  {Voss}, {Roelofs}, {Izzard}, {Montgomery}, {Reerink}, {Christlieb}, and
  {Reimers}}]{2005A&A...440.1087N}
{Nelemans} G, {Napiwotzki} R, {Karl} C, {Marsh} TR, {Voss} B, {Roelofs} G,
  {Izzard} RG, {Montgomery} M, {Reerink} T, {Christlieb} N, et~al. (2005)
  {Binaries discovered by the SPYproject. IV. Five single-lined DA double white
  dwarfs}. \aap 440(3):1087--1095. \doi{10.1051/0004-6361:20053174}.
  {\href{https://arxiv.org/abs/astro-ph/0506231}{{arXiv:astro-ph/0506231}}}
  {[astro-ph]}

\bibitem[{{Nelemans} et~al.(2006){Nelemans}, {Jonker}, and
  {Steeghs}}]{2006MNRAS.370..255N}
{Nelemans} G, {Jonker} PG, {Steeghs} D (2006) {Optical spectroscopy of
  (candidate) ultracompact X-ray binaries: constraints on the composition of
  the donor stars}. \mnras 370(1):255--262.
  \doi{10.1111/j.1365-2966.2006.10496.x}.
  {\href{https://arxiv.org/abs/astro-ph/0604597}{{arXiv:astro-ph/0604597}}}
  {[astro-ph]}

\bibitem[{{Nelemans} et~al.(2010){Nelemans}, {Yungelson}, {van der Sluys}, and
  {Tout}}]{2010MNRAS.401.1347N}
{Nelemans} G, {Yungelson} LR, {van der Sluys} MV, {Tout} CA (2010) {The
  chemical composition of donors in AM CVn stars and ultracompact X-ray
  binaries: observational tests of their formation}. \mnras 401(2):1347--1359.
  \doi{10.1111/j.1365-2966.2009.15731.x}.
  {\href{https://arxiv.org/abs/0909.3376}{{arXiv:0909.3376}}} {[astro-ph.SR]}

\bibitem[{{Nelson} et~al.(1986){Nelson}, {Rappaport}, and
  {Joss}}]{1986ApJ...304..231N}
{Nelson} LA, {Rappaport} SA, {Joss} PC (1986) {The Evolution of Ultrashort
  Period Binary Systems}. \apj 304:231. \doi{10.1086/164156}

\bibitem[{{Neumayer} et~al.(2020){Neumayer}, {Seth}, and
  {B{\"o}ker}}]{2020A&ARv..28....4N}
{Neumayer} N, {Seth} A, {B{\"o}ker} T (2020) {Nuclear star clusters}. \aapr
  28(1):4. \doi{10.1007/s00159-020-00125-0}.
  {\href{https://arxiv.org/abs/2001.03626}{{arXiv:2001.03626}}} {[astro-ph.GA]}

\bibitem[{{Nevin} et~al.(2019){Nevin}, {Blecha}, {Comerford}, and
  {Greene}}]{2019ApJ...872...76N}
{Nevin} R, {Blecha} L, {Comerford} J, {Greene} J (2019) {Accurate
  Identification of Galaxy Mergers with Imaging}. \apj 872(1):76.
  \doi{10.3847/1538-4357/aafd34}.
  {\href{https://arxiv.org/abs/1901.01975}{{arXiv:1901.01975}}} {[astro-ph.GA]}

\bibitem[{{Nguyen} and {Bogdanovi{\'c}}(2016)}]{2016ApJ...828...68N}
{Nguyen} K, {Bogdanovi{\'c}} T (2016) {Emission Signatures from Sub-parsec
  Binary Supermassive Black Holes. I. Diagnostic Power of Broad Emission
  Lines}. \apj 828(2):68. \doi{10.3847/0004-637X/828/2/68}.
  {\href{https://arxiv.org/abs/1605.09389}{{arXiv:1605.09389}}} {[astro-ph.HE]}

\bibitem[{{Nguyen} et~al.(2019){Nguyen}, {Bogdanovi{\'c}}, {Runnoe},
  {Eracleous}, {Sigurdsson}, and {Boroson}}]{2019ApJ...870...16N}
{Nguyen} K, {Bogdanovi{\'c}} T, {Runnoe} JC, {Eracleous} M, {Sigurdsson} S,
  {Boroson} T (2019) {Emission Signatures from Sub-parsec Binary Supermassive
  Black Holes. II. Effect of Accretion Disk Wind on Broad Emission Lines}. \apj
  870(1):16. \doi{10.3847/1538-4357/aaeff0}.
  {\href{https://arxiv.org/abs/1807.09782}{{arXiv:1807.09782}}} {[astro-ph.HE]}

\bibitem[{{Nguyen} et~al.(2020){Nguyen}, {Bogdanovi{\'c}}, {Runnoe},
  {Eracleous}, {Sigurdsson}, and {Boroson}}]{2020ApJ...894..105N}
{Nguyen} K, {Bogdanovi{\'c}} T, {Runnoe} JC, {Eracleous} M, {Sigurdsson} S,
  {Boroson} T (2020) {Emission Signatures from Subparsec Binary Supermassive
  Black Holes. III. Comparison of Models with Observations}. \apj 894(2):105.
  \doi{10.3847/1538-4357/ab88b5}.
  {\href{https://arxiv.org/abs/1908.01799}{{arXiv:1908.01799}}} {[astro-ph.HE]}

\bibitem[{{Nishizawa} et~al.(2016){Nishizawa}, {Berti}, {Klein}, and
  {Sesana}}]{2016PhRvD..94f4020N}
{Nishizawa} A, {Berti} E, {Klein} A, {Sesana} A (2016) {eLISA eccentricity
  measurements as tracers of binary black hole formation}. \prd 94(6):064020.
  \doi{10.1103/PhysRevD.94.064020}.
  {\href{https://arxiv.org/abs/1605.01341}{{arXiv:1605.01341}}} {[gr-qc]}

\bibitem[{{Nishizawa} et~al.(2017){Nishizawa}, {Sesana}, {Berti}, and
  {Klein}}]{2017MNRAS.465.4375N}
{Nishizawa} A, {Sesana} A, {Berti} E, {Klein} A (2017) {Constraining stellar
  binary black hole formation scenarios with eLISA eccentricity measurements}.
  \mnras 465(4):4375--4380. \doi{10.1093/mnras/stw2993}.
  {\href{https://arxiv.org/abs/1606.09295}{{arXiv:1606.09295}}} {[astro-ph.HE]}

\bibitem[{{Nissanke} et~al.(2012){Nissanke}, {Vallisneri}, {Nelemans}, and
  {Prince}}]{2012ApJ...758..131N}
{Nissanke} S, {Vallisneri} M, {Nelemans} G, {Prince} TA (2012)
  {Gravitational-wave Emission from Compact Galactic Binaries}. \apj
  758(2):131. \doi{10.1088/0004-637X/758/2/131}.
  {\href{https://arxiv.org/abs/1201.4613}{{arXiv:1201.4613}}} {[astro-ph.GA]}

\bibitem[{{Nitadori} and {Aarseth}(2012)}]{2012MNRAS.424..545N}
{Nitadori} K, {Aarseth} SJ (2012) {Accelerating NBODY6 with graphics processing
  units}. \mnras 424(1):545--552. \doi{10.1111/j.1365-2966.2012.21227.x}.
  {\href{https://arxiv.org/abs/1205.1222}{{arXiv:1205.1222}}} {[astro-ph.IM]}

\bibitem[{{Noble} et~al.(2012){Noble}, {Mundim}, {Nakano}, {Krolik},
  {Campanelli}, {Zlochower}, and {Yunes}}]{2012ApJ...755...51N}
{Noble} SC, {Mundim} BC, {Nakano} H, {Krolik} JH, {Campanelli} M, {Zlochower}
  Y, {Yunes} N (2012) {Circumbinary Magnetohydrodynamic Accretion into
  Inspiraling Binary Black Holes}. \apj 755(1):51.
  \doi{10.1088/0004-637X/755/1/51}.
  {\href{https://arxiv.org/abs/1204.1073}{{arXiv:1204.1073}}} {[astro-ph.HE]}

\bibitem[{{Nomoto} et~al.(2007){Nomoto}, {Saio}, {Kato}, and
  {Hachisu}}]{2007ApJ...663.1269N}
{Nomoto} K, {Saio} H, {Kato} M, {Hachisu} I (2007) {Thermal Stability of White
  Dwarfs Accreting Hydrogen-rich Matter and Progenitors of Type Ia Supernovae}.
  \apj 663(2):1269--1276. \doi{10.1086/518465}.
  {\href{https://arxiv.org/abs/astro-ph/0603351}{{arXiv:astro-ph/0603351}}}
  {[astro-ph]}

\bibitem[{{Noutsos} et~al.(2012){Noutsos}, {Kramer}, {Carr}, and
  {Johnston}}]{2012MNRAS.423.2736N}
{Noutsos} A, {Kramer} M, {Carr} P, {Johnston} S (2012) {Pulsar spin-velocity
  alignment: further results and discussion}. \mnras 423(3):2736--2752.
  \doi{10.1111/j.1365-2966.2012.21083.x}.
  {\href{https://arxiv.org/abs/1205.2305}{{arXiv:1205.2305}}} {[astro-ph.GA]}

\bibitem[{{Novikov} and {Thorne}(1973)}]{1973blho.conf..343N}
{Novikov} ID, {Thorne} KS (1973) {Astrophysics of black holes.} In: Black Holes
  (Les Astres Occlus). pp 343--450

\bibitem[{{Obergaulinger} and {Aloy}(2020)}]{2020MNRAS.492.4613O}
{Obergaulinger} M, {Aloy} M{\'A} (2020) {Magnetorotational core collapse of
  possible GRB progenitors - I. Explosion mechanisms}. \mnras
  492(4):4613--4634. \doi{10.1093/mnras/staa096}.
  {\href{https://arxiv.org/abs/1909.01105}{{arXiv:1909.01105}}} {[astro-ph.HE]}

\bibitem[{{Oesch} et~al.(2021){Oesch}, {Bouwens}, {Brammer}, {Chisholm},
  {Fudamoto}, {Illingworth}, {Kerutt}, {Labbe}, {Magee}, {Marchesini},
  {Maseda}, {Mason}, {Naidu}, {Nelson}, {Qin}, {Reddy}, {Schaerer}, {Shapley},
  {Shivaei}, {Smit}, {Whitaker}, {Wuyts}, {Wyithe}, and {van
  Dokkum}}]{2021jwst.prop.1895O}
{Oesch} P, {Bouwens} R, {Brammer} G, {Chisholm} J, {Fudamoto} Y, {Illingworth}
  GD, {Kerutt} J, {Labbe} I, {Magee} DK, {Marchesini} D, et~al. (2021) {FRESCO:
  The First Reionization Epoch Spectroscopic COmplete Survey}. JWST Proposal.
  Cycle 1, ID. \#1895

\bibitem[{{Ogilvie}(2013)}]{2013MNRAS.429..613O}
{Ogilvie} GI (2013) {Tides in rotating barotropic fluid bodies: the
  contribution of inertial waves and the role of internal structure}. \mnras
  429(1):613--632. \doi{10.1093/mnras/sts362}.
  {\href{https://arxiv.org/abs/1211.0837}{{arXiv:1211.0837}}} {[astro-ph.EP]}

\bibitem[{{Ogilvie}(2014)}]{2014ARA&A..52..171O}
{Ogilvie} GI (2014) {Tidal Dissipation in Stars and Giant Planets}. \araa
  52:171--210. \doi{10.1146/annurev-astro-081913-035941}.
  {\href{https://arxiv.org/abs/1406.2207}{{arXiv:1406.2207}}} {[astro-ph.SR]}

\bibitem[{{Ogiya} et~al.(2020){Ogiya}, {Hahn}, {Mingarelli}, and
  {Volonteri}}]{2020MNRAS.493.3676O}
{Ogiya} G, {Hahn} O, {Mingarelli} CMF, {Volonteri} M (2020) {Accelerated
  orbital decay of supermassive black hole binaries in merging nuclear star
  clusters}. \mnras 493(3):3676--3689. \doi{10.1093/mnras/staa444}.
  {\href{https://arxiv.org/abs/1911.11526}{{arXiv:1911.11526}}} {[astro-ph.GA]}

\bibitem[{{Oh} et~al.(2015){Oh}, {Hunter}, {Brinks}, {Elmegreen}, {Schruba},
  {Walter}, {Rupen}, {Young}, {Simpson}, {Johnson}, {Herrmann}, {Ficut-Vicas},
  {Cigan}, {Heesen}, {Ashley}, and {Zhang}}]{2015AJ....149..180O}
{Oh} SH, {Hunter} DA, {Brinks} E, {Elmegreen} BG, {Schruba} A, {Walter} F,
  {Rupen} MP, {Young} LM, {Simpson} CE, {Johnson} MC, et~al. (2015)
  {High-resolution Mass Models of Dwarf Galaxies from LITTLE THINGS}. \aj
  149(6):180. \doi{10.1088/0004-6256/149/6/180}.
  {\href{https://arxiv.org/abs/1502.01281}{{arXiv:1502.01281}}} {[astro-ph.GA]}

\bibitem[{{Oh} and {Haiman}(2003)}]{2003MNRAS.346..456O}
{Oh} SP, {Haiman} Z (2003) {Fossil H II regions: self-limiting star formation
  at high redshift}. \mnras 346(2):456--472.
  \doi{10.1046/j.1365-2966.2003.07103.x}.
  {\href{https://arxiv.org/abs/astro-ph/0307135}{{arXiv:astro-ph/0307135}}}
  {[astro-ph]}

\bibitem[{{Ohlmann}(2016)}]{2016PhDT........74O}
{Ohlmann} ST (2016) {Hydrodynamics of the Common Envelope Phase in Binary
  Stellar Evolution.} PhD thesis, -

\bibitem[{{Ohlmann} et~al.(2016{\natexlab{a}}){Ohlmann}, {R{\"o}pke}, {Pakmor},
  and {Springel}}]{2016ApJ...816L...9O}
{Ohlmann} ST, {R{\"o}pke} FK, {Pakmor} R, {Springel} V (2016{\natexlab{a}})
  {Hydrodynamic Moving-mesh Simulations of the Common Envelope Phase in Binary
  Stellar Systems}. \apjl 816(1):L9. \doi{10.3847/2041-8205/816/1/L9}.
  {\href{https://arxiv.org/abs/1512.04529}{{arXiv:1512.04529}}} {[astro-ph.SR]}

\bibitem[{{Ohlmann} et~al.(2016{\natexlab{b}}){Ohlmann}, {R{\"o}pke}, {Pakmor},
  {Springel}, and {M{\"u}ller}}]{2016MNRAS.462L.121O}
{Ohlmann} ST, {R{\"o}pke} FK, {Pakmor} R, {Springel} V, {M{\"u}ller} E
  (2016{\natexlab{b}}) {Magnetic field amplification during the common envelope
  phase}. \mnras 462(1):L121--L125. \doi{10.1093/mnrasl/slw144}.
  {\href{https://arxiv.org/abs/1607.05996}{{arXiv:1607.05996}}} {[astro-ph.SR]}

\bibitem[{{Oka} et~al.(2017){Oka}, {Tsujimoto}, {Iwata}, {Nomura}, and
  {Takekawa}}]{2017NatAs...1..709O}
{Oka} T, {Tsujimoto} S, {Iwata} Y, {Nomura} M, {Takekawa} S (2017)
  {Millimetre-wave emission from an intermediate-mass black hole candidate in
  the Milky Way}. Nature Astronomy 1:709--712. \doi{10.1038/s41550-017-0224-z}.
  {\href{https://arxiv.org/abs/1707.07603}{{arXiv:1707.07603}}} {[astro-ph.GA]}

\bibitem[{{O'Leary} et~al.(2021){O'Leary}, {Moster}, {Naab}, and
  {Somerville}}]{2021MNRAS.501.3215O}
{O'Leary} JA, {Moster} BP, {Naab} T, {Somerville} RS (2021) {EMERGE: empirical
  predictions of galaxy merger rates since z {\ensuremath{\sim}} 6}. \mnras
  501(3):3215--3237. \doi{10.1093/mnras/staa3746}.
  {\href{https://arxiv.org/abs/2001.02687}{{arXiv:2001.02687}}}

\bibitem[{{Omukai} and {Inutsuka}(2002)}]{2002MNRAS.332...59O}
{Omukai} K, {Inutsuka} Si (2002) {An upper limit on the mass of a primordial
  star due to the formation of an Hii region: the effect of ionizing radiation
  force}. \mnras 332(1):59--64. \doi{10.1046/j.1365-8711.2002.05276.x}.
  {\href{https://arxiv.org/abs/astro-ph/0112345}{{arXiv:astro-ph/0112345}}}
  {[astro-ph]}

\bibitem[{{Omukai} and {Palla}(2003)}]{2003ApJ...589..677O}
{Omukai} K, {Palla} F (2003) {Formation of the First Stars by Accretion}. \apj
  589(2):677--687. \doi{10.1086/374810}.
  {\href{https://arxiv.org/abs/astro-ph/0302345}{{arXiv:astro-ph/0302345}}}
  {[astro-ph]}

\bibitem[{{Omukai} et~al.(2008){Omukai}, {Schneider}, and
  {Haiman}}]{2008ApJ...686..801O}
{Omukai} K, {Schneider} R, {Haiman} Z (2008) {Can Supermassive Black Holes Form
  in Metal-enriched High-Redshift Protogalaxies?} \apj 686(2):801--814.
  \doi{10.1086/591636}.
  {\href{https://arxiv.org/abs/0804.3141}{{arXiv:0804.3141}}} {[astro-ph]}

\bibitem[{{O'Shaughnessy} et~al.(2017){O'Shaughnessy}, {Bellovary}, {Brooks},
  {Shen}, {Governato}, and {Christensen}}]{2017MNRAS.464.2831O}
{O'Shaughnessy} R, {Bellovary} JM, {Brooks} A, {Shen} S, {Governato} F,
  {Christensen} CR (2017) {The effects of host galaxy properties on merging
  compact binaries detectable by LIGO}. \mnras 464(3):2831--2839.
  \doi{10.1093/mnras/stw2550}.
  {\href{https://arxiv.org/abs/1609.06715}{{arXiv:1609.06715}}} {[astro-ph.GA]}

\bibitem[{{O'Shea} et~al.(2015){O'Shea}, {Wise}, {Xu}, and
  {Norman}}]{2015ApJ...807L..12O}
{O'Shea} BW, {Wise} JH, {Xu} H, {Norman} ML (2015) {Probing the Ultraviolet
  Luminosity Function of the Earliest Galaxies with the Renaissance
  Simulations}. \apjl 807:L12. \doi{10.1088/2041-8205/807/1/L12}.
  {\href{https://arxiv.org/abs/1503.01110}{{arXiv:1503.01110}}}

\bibitem[{{Ossokine} et~al.(2020){Ossokine}, {Buonanno}, {Marsat}, {Cotesta},
  {Babak}, {Dietrich}, {Haas}, {Hinder}, {Pfeiffer}, {P{\"u}rrer}, {Woodford},
  {Boyle}, {Kidder}, {Scheel}, and {Szil{\'a}gyi}}]{2020PhRvD.102d4055O}
{Ossokine} S, {Buonanno} A, {Marsat} S, {Cotesta} R, {Babak} S, {Dietrich} T,
  {Haas} R, {Hinder} I, {Pfeiffer} HP, {P{\"u}rrer} M, et~al. (2020)
  {Multipolar effective-one-body waveforms for precessing binary black holes:
  Construction and validation}. \prd 102(4):044055.
  \doi{10.1103/PhysRevD.102.044055}.
  {\href{https://arxiv.org/abs/2004.09442}{{arXiv:2004.09442}}} {[gr-qc]}

\bibitem[{{Ostriker}(1999)}]{1999ApJ...513..252O}
{Ostriker} EC (1999) {Dynamical Friction in a Gaseous Medium}. \apj
  513(1):252--258. \doi{10.1086/306858}.
  {\href{https://arxiv.org/abs/astro-ph/9810324}{{arXiv:astro-ph/9810324}}}
  {[astro-ph]}

\bibitem[{{Owen}(1996)}]{1996PhRvD..53.6749O}
{Owen} BJ (1996) {Search templates for gravitational waves from inspiraling
  binaries: Choice of template spacing}. \prd 53(12):6749--6761.
  \doi{10.1103/PhysRevD.53.6749}.
  {\href{https://arxiv.org/abs/gr-qc/9511032}{{arXiv:gr-qc/9511032}}} {[gr-qc]}

\bibitem[{{{\"O}zel} and {Freire}(2016)}]{2016ARA&A..54..401O}
{{\"O}zel} F, {Freire} P (2016) {Masses, Radii, and the Equation of State of
  Neutron Stars}. \araa 54:401--440. \doi{10.1146/annurev-astro-081915-023322}.
  {\href{https://arxiv.org/abs/1603.02698}{{arXiv:1603.02698}}} {[astro-ph.HE]}

\bibitem[{{Pacucci} and {Loeb}(2020)}]{2020ApJ...895...95P}
{Pacucci} F, {Loeb} A (2020) {Separating Accretion and Mergers in the Cosmic
  Growth of Black Holes with X-Ray and Gravitational-wave Observations}. \apj
  895(2):95. \doi{10.3847/1538-4357/ab886e}.
  {\href{https://arxiv.org/abs/2004.07246}{{arXiv:2004.07246}}} {[astro-ph.GA]}

\bibitem[{{Pacucci} et~al.(2015){Pacucci}, {Ferrara}, {Volonteri}, and
  {Dubus}}]{2015MNRAS.454.3771P}
{Pacucci} F, {Ferrara} A, {Volonteri} M, {Dubus} G (2015) {Shining in the dark:
  the spectral evolution of the first black holes}. \mnras 454(4):3771--3777.
  \doi{10.1093/mnras/stv2196}.
  {\href{https://arxiv.org/abs/1506.05299}{{arXiv:1506.05299}}} {[astro-ph.HE]}

\bibitem[{{Pacucci} et~al.(2017){Pacucci}, {Natarajan}, {Volonteri},
  {Cappelluti}, and {Urry}}]{2017ApJ...850L..42P}
{Pacucci} F, {Natarajan} P, {Volonteri} M, {Cappelluti} N, {Urry} CM (2017)
  {Conditions for Optimal Growth of Black Hole Seeds}. \apjl 850(2):L42.
  \doi{10.3847/2041-8213/aa9aea}.
  {\href{https://arxiv.org/abs/1710.09375}{{arXiv:1710.09375}}} {[astro-ph.GA]}

\bibitem[{{Pacucci} et~al.(2018){Pacucci}, {Loeb}, {Mezcua}, and
  {Mart{\'\i}n-Navarro}}]{2018ApJ...864L...6P}
{Pacucci} F, {Loeb} A, {Mezcua} M, {Mart{\'\i}n-Navarro} I (2018) {Glimmering
  in the Dark: Modeling the Low-mass End of the M
  $_{{\ensuremath{\bullet}}}$-{\ensuremath{\sigma}} Relation and of the Quasar
  Luminosity Function}. \apjl 864(1):L6. \doi{10.3847/2041-8213/aad8b2}.
  {\href{https://arxiv.org/abs/1808.09452}{{arXiv:1808.09452}}} {[astro-ph.GA]}

\bibitem[{{Paczy{\'n}ski}(1967)}]{1967AcA....17..287P}
{Paczy{\'n}ski} B (1967) {Gravitational Waves and the Evolution of Close
  Binaries}. \actaa 17:287

\bibitem[{{Paczynski}(1976)}]{1976IAUS...73...75P}
{Paczynski} B (1976) {Common Envelope Binaries}. In: {Eggleton} P, {Mitton} S,
  {Whelan} J (eds) Structure and Evolution of Close Binary Systems. IAUS,
  vol~73. p~75

\bibitem[{{Paczynski}(1986)}]{1986ApJ...308L..43P}
{Paczynski} B (1986) {Gamma-ray bursters at cosmological distances}. \apjl
  308:L43--L46. \doi{10.1086/184740}

\bibitem[{{Paczy{\'n}ski} and {Sienkiewicz}(1972)}]{1972AcA....22...73P}
{Paczy{\'n}ski} B, {Sienkiewicz} R (1972) {Evolution of Close Binaries VIII.
  Mass Exchange on the Dynamical Time Scale}. \actaa 22:73--91

\bibitem[{{Padmanabhan} and {Loeb}(2020)}]{2020JCAP...11..055P}
{Padmanabhan} H, {Loeb} A (2020) {Constraining the host galaxy halos of massive
  black holes from LISA event rates}. \jcap 2020(11):055.
  \doi{10.1088/1475-7516/2020/11/055}.
  {\href{https://arxiv.org/abs/2007.12710}{{arXiv:2007.12710}}} {[astro-ph.CO]}

\bibitem[{{Pakmor} et~al.(2010){Pakmor}, {Kromer}, {R{\"o}pke}, {Sim},
  {Ruiter}, and {Hillebrandt}}]{2010Natur.463...61P}
{Pakmor} R, {Kromer} M, {R{\"o}pke} FK, {Sim} SA, {Ruiter} AJ, {Hillebrandt} W
  (2010) {Sub-luminous type Ia supernovae from the mergers of equal-mass white
  dwarfs with mass \raisebox{-0.5ex}\textasciitilde 0.9M$_{solar}$}. \nat
  463(7277):61--64. \doi{10.1038/nature08642}.
  {\href{https://arxiv.org/abs/0911.0926}{{arXiv:0911.0926}}} {[astro-ph.HE]}

\bibitem[{{Pakmor} et~al.(2012){Pakmor}, {Kromer}, {Taubenberger}, {Sim},
  {R{\"o}pke}, and {Hillebrandt}}]{2012ApJ...747L..10P}
{Pakmor} R, {Kromer} M, {Taubenberger} S, {Sim} SA, {R{\"o}pke} FK,
  {Hillebrandt} W (2012) {Normal Type Ia Supernovae from Violent Mergers of
  White Dwarf Binaries}. \apjl 747(1):L10. \doi{10.1088/2041-8205/747/1/L10}.
  {\href{https://arxiv.org/abs/1201.5123}{{arXiv:1201.5123}}} {[astro-ph.HE]}

\bibitem[{{Pala} et~al.(2018){Pala}, {Schmidtobreick}, {Tappert},
  {G{\"a}nsicke}, and {Mehner}}]{2018MNRAS.481.2523P}
{Pala} AF, {Schmidtobreick} L, {Tappert} C, {G{\"a}nsicke} BT, {Mehner} A
  (2018) {The cataclysmic variable QZ Lib: a period bouncer}. \mnras
  481(2):2523--2535. \doi{10.1093/mnras/sty2434}.
  {\href{https://arxiv.org/abs/1809.02135}{{arXiv:1809.02135}}} {[astro-ph.SR]}

\bibitem[{Palenzuela et~al.(2010{\natexlab{a}})Palenzuela, Garrett, Lehner, and
  Liebling}]{2010PhRvD..82d4045P}
Palenzuela C, Garrett T, Lehner L, Liebling SL (2010{\natexlab{a}})
  Magnetospheres of black hole systems in force-free plasma. Phys Rev D
  82:044045. \doi{10.1103/PhysRevD.82.044045},
  \urlprefix\url{https://link.aps.org/doi/10.1103/PhysRevD.82.044045}

\bibitem[{Palenzuela et~al.(2010{\natexlab{b}})Palenzuela, Lehner, and
  Yoshida}]{2010PhRvD..81h4007P}
Palenzuela C, Lehner L, Yoshida S (2010{\natexlab{b}}) Understanding possible
  electromagnetic counterparts to loud gravitational wave events: Binary black
  hole effects on electromagnetic fields. Phys Rev D 81:084007.
  \doi{10.1103/PhysRevD.81.084007},
  \urlprefix\url{https://link.aps.org/doi/10.1103/PhysRevD.81.084007}

\bibitem[{Palenzuela et~al.(2009)}]{2009PhRvL.103h1101P}
Palenzuela C, et~al. (2009) Binary black holes' effects on electromagnetic
  fields. Phys Rev Lett 103:081101. \doi{10.1103/PhysRevLett.103.081101},
  \urlprefix\url{https://link.aps.org/doi/10.1103/PhysRevLett.103.081101}

\bibitem[{Palenzuela et~al.(2010{\natexlab{c}})}]{2010Sci...329..927P}
Palenzuela C, et~al. (2010{\natexlab{c}}) Dual jets from binary black holes.
  Science 329:927--930. \doi{10.1126/science.1191766},
  \urlprefix\url{http://doi.org/10.1126/science.1191766}

\bibitem[{{Panamarev} et~al.(2018){Panamarev}, {Shukirgaliyev}, and
  {Meiron}}]{Bekdaulet2018}
{Panamarev} T, {Shukirgaliyev} B, {Meiron} Y (2018) {Star–disc interaction in
  galactic nuclei: formation of a central stellar disc}. \mnras

\bibitem[{{Panamarev} et~al.(2019){Panamarev}, {Just}, {Spurzem}, {Berczik},
  {Wang}, and {Arca Sedda}}]{2019MNRAS.484.3279P}
{Panamarev} T, {Just} A, {Spurzem} R, {Berczik} P, {Wang} L, {Arca Sedda} M
  (2019) {Direct N-body simulation of the Galactic centre}. \mnras
  484(3):3279--3290. \doi{10.1093/mnras/stz208}.
  {\href{https://arxiv.org/abs/1805.02153}{{arXiv:1805.02153}}} {[astro-ph.GA]}

\bibitem[{{Papadopoulos} and {Kokkotas}(2018)}]{2018CQGra..35r5014P}
{Papadopoulos} GO, {Kokkotas} KD (2018) {Preserving Kerr symmetries in deformed
  spacetimes}. Classical and Quantum Gravity 35(18):185014.
  \doi{10.1088/1361-6382/aad7f4}.
  {\href{https://arxiv.org/abs/1807.08594}{{arXiv:1807.08594}}} {[gr-qc]}

\bibitem[{{Paragi} et~al.(2015){Paragi}, {Godfrey}, {Reynolds}, {Rioja},
  {Deller}, {Zhang}, {Gurvits}, {Bietenholz}, {Szomoru}, {Bignall}, {Boven},
  {Charlot}, {Dodson}, {Frey}, {Garrett}, {Imai}, {Lobanov}, {Reid}, {Ros},
  {van Langevelde}, {Zensus}, {Zheng}, {Alberdi}, {Agudo}, {An}, {Argo},
  {Beswick}, {Biggs}, {Brunthaler}, {Campbell}, {Cimo}, {Colomer}, {Corbel},
  {Conway}, {Cseh}, {Deane}, {Falcke}, {Gawronski}, {Gaylard}, {Giovannini},
  {Giroletti}, {Goddi}, {Goedhart}, {G{\'o}mez}, {Gunn}, {Kharb}, {Kloeckner},
  {Koerding}, {Kovalev}, {Kunert-Bajraszewska}, {Lindqvist}, {Lister},
  {Mantovani}, {Marti-Vidal}, {Mezcua}, {McKean}, {Middelberg}, {Miller-Jones},
  {Moldon}, {Muxlow}, {O'Brien}, {Perez-Torres}, {Pogrebenko}, {Quick},
  {Rushton}, {Schilizzi}, {Smirnov}, {Sohn}, {Surcis}, {Taylor}, {Tingay},
  {Tudose}, {van der Horst}, {van Leeuwen}, {Venturi}, {Vermeulen},
  {Vlemmings}, {de Witt}, {Wucknitz}, {Yang}, {Gab{\"a}nyi}, and
  {Jung}}]{2015aska.confE.143P}
{Paragi} Z, {Godfrey} L, {Reynolds} C, {Rioja} MJ, {Deller} A, {Zhang} B,
  {Gurvits} L, {Bietenholz} M, {Szomoru} A, {Bignall} HE, et~al. (2015) {Very
  Long Baseline Interferometry with the SKA}. In: Advancing Astrophysics with
  the Square Kilometre Array (AASKA14). p 143.
  {\href{https://arxiv.org/abs/1412.5971}{{arXiv:1412.5971}}} {[astro-ph.IM]}

\bibitem[{{Pardo} et~al.(2016){Pardo}, {Goulding}, {Greene}, {Somerville},
  {Gallo}, {Hickox}, {Miller}, {Reines}, and {Silverman}}]{2016ApJ...831..203P}
{Pardo} K, {Goulding} AD, {Greene} JE, {Somerville} RS, {Gallo} E, {Hickox} RC,
  {Miller} BP, {Reines} AE, {Silverman} JD (2016) {X-Ray Detected Active
  Galactic Nuclei in Dwarf Galaxies at 0 < z < 1}. \apj 831(2):203.
  \doi{10.3847/0004-637X/831/2/203}.
  {\href{https://arxiv.org/abs/1603.01622}{{arXiv:1603.01622}}} {[astro-ph.GA]}

\bibitem[{{Park} and {Bogdanovi{\'c}}(2017)}]{2017ApJ...838..103P}
{Park} K, {Bogdanovi{\'c}} T (2017) {Gaseous Dynamical Friction in Presence of
  Black Hole Radiative Feedback}. \apj 838(2):103.
  \doi{10.3847/1538-4357/aa65ce}.
  {\href{https://arxiv.org/abs/1701.00526}{{arXiv:1701.00526}}} {[astro-ph.GA]}

\bibitem[{{Park} and {Bogdanovi{\'c}}(2019)}]{2019ApJ...883..209P}
{Park} K, {Bogdanovi{\'c}} T (2019) {Erratum: {\textquotedblleft}Gaseous
  Dynamical Friction in Presence of Black Hole Radiative
  Feedback{\textquotedblright} (<A
  href=``http://doi.org/10.3847/1538-4357/aa65ce''>2017, ApJ, 838, 103</A>)}.
  \apj 883(2):209. \doi{10.3847/1538-4357/ab3f30}

\bibitem[{{Paschalidis} and {Stergioulas}(2017)}]{2017LRR....20....7P}
{Paschalidis} V, {Stergioulas} N (2017) {Rotating stars in relativity}. Living
  Reviews in Relativity 20(1):7. \doi{10.1007/s41114-017-0008-x}.
  {\href{https://arxiv.org/abs/1612.03050}{{arXiv:1612.03050}}} {[astro-ph.HE]}

\bibitem[{{Paschalidis} et~al.(2009){Paschalidis}, {MacLeod}, {Baumgarte}, and
  {Shapiro}}]{2009PhRvD..80b4006P}
{Paschalidis} V, {MacLeod} M, {Baumgarte} TW, {Shapiro} SL (2009) {Merger of
  white dwarf-neutron star binaries: Prelude to hydrodynamic simulations in
  general relativity}. \prd 80(2):024006. \doi{10.1103/PhysRevD.80.024006}.
  {\href{https://arxiv.org/abs/0910.5719}{{arXiv:0910.5719}}} {[astro-ph.HE]}

\bibitem[{{Paschalidis} et~al.(2021){Paschalidis}, {Bright}, {Ruiz}, and
  {Gold}}]{2021arXiv210206712P}
{Paschalidis} V, {Bright} J, {Ruiz} M, {Gold} R (2021) {Minidisk dynamics in
  accreting, spinning black hole binaries: Simulations in full general
  relativity}. arXiv e-prints, In press in ApJ Lett arXiv:2102.06712.
  {\href{https://arxiv.org/abs/2102.06712}{{arXiv:2102.06712}}} {[astro-ph.HE]}

\bibitem[{{Passy} et~al.(2012{\natexlab{a}}){Passy}, {De Marco}, {Fryer},
  {Herwig}, {Diehl}, {Oishi}, {Mac Low}, {Bryan}, and
  {Rockefeller}}]{2012ApJ...744...52P}
{Passy} JC, {De Marco} O, {Fryer} CL, {Herwig} F, {Diehl} S, {Oishi} JS, {Mac
  Low} MM, {Bryan} GL, {Rockefeller} G (2012{\natexlab{a}}) {Simulating the
  Common Envelope Phase of a Red Giant Using Smoothed-particle Hydrodynamics
  and Uniform-grid Codes}. \apj 744(1):52. \doi{10.1088/0004-637X/744/1/52}.
  {\href{https://arxiv.org/abs/1107.5072}{{arXiv:1107.5072}}} {[astro-ph.SR]}

\bibitem[{{Passy} et~al.(2012{\natexlab{b}}){Passy}, {Herwig}, and
  {Paxton}}]{2012ApJ...760...90P}
{Passy} JC, {Herwig} F, {Paxton} B (2012{\natexlab{b}}) {The Response of Giant
  Stars to Dynamical-timescale Mass Loss}. \apj 760(1):90.
  \doi{10.1088/0004-637X/760/1/90}.
  {\href{https://arxiv.org/abs/1111.4202}{{arXiv:1111.4202}}} {[astro-ph.SR]}

\bibitem[{{Pavl{\'\i}k} et~al.(2018){Pavl{\'\i}k}, {Je{\v{r}}{\'a}bkov{\'a}},
  {Kroupa}, and {Baumgardt}}]{2018A&A...617A..69P}
{Pavl{\'\i}k} V, {Je{\v{r}}{\'a}bkov{\'a}} T, {Kroupa} P, {Baumgardt} H (2018)
  {The black hole retention fraction in star clusters}. \aap 617:A69.
  \doi{10.1051/0004-6361/201832919}.
  {\href{https://arxiv.org/abs/1806.05192}{{arXiv:1806.05192}}} {[astro-ph.GA]}

\bibitem[{{Pavlovskii} and {Ivanova}(2015)}]{2015MNRAS.449.4415P}
{Pavlovskii} K, {Ivanova} N (2015) {Mass transfer from giant donors}. \mnras
  449(4):4415--4427. \doi{10.1093/mnras/stv619}.
  {\href{https://arxiv.org/abs/1410.5109}{{arXiv:1410.5109}}} {[astro-ph.SR]}

\bibitem[{{Pearson} et~al.(2019){Pearson}, {Wang}, {Trayford}, {Petrillo}, and
  {van der Tak}}]{2019A&A...626A..49P}
{Pearson} WJ, {Wang} L, {Trayford} JW, {Petrillo} CE, {van der Tak} FFS (2019)
  {Identifying galaxy mergers in observations and simulations with deep
  learning}. \aap 626:A49. \doi{10.1051/0004-6361/201935355}.
  {\href{https://arxiv.org/abs/1902.10626}{{arXiv:1902.10626}}} {[astro-ph.GA]}

\bibitem[{{Pei{\ss}ker} et~al.(2020){Pei{\ss}ker}, {Eckart}, {Zaja{\v{c}}ek},
  {Ali}, and {Parsa}}]{2020ApJ...899...50P}
{Pei{\ss}ker} F, {Eckart} A, {Zaja{\v{c}}ek} M, {Ali} B, {Parsa} M (2020) {S62
  and S4711: Indications of a Population of Faint Fast-moving Stars inside the
  S2 Orbit{\textemdash}S4711 on a 7.6 yr Orbit around Sgr A*}. \apj 899(1):50.
  \doi{10.3847/1538-4357/ab9c1c}.
  {\href{https://arxiv.org/abs/2008.04764}{{arXiv:2008.04764}}} {[astro-ph.GA]}

\bibitem[{{Perera} et~al.(2019){Perera}, {DeCesar}, {Demorest}, {Kerr},
  {Lentati}, {Nice}, {Os{\l}owski}, {Ransom}, {Keith}, {Arzoumanian}, {Bailes},
  {Baker}, {Bassa}, {Bhat}, {Brazier}, {Burgay}, {Burke-Spolaor}, {Caballero},
  {Champion}, {Chatterjee}, {Chen}, {Cognard}, {Cordes}, {Crowter}, {Dai},
  {Desvignes}, {Dolch}, {Ferdman}, {Ferrara}, {Fonseca}, {Goldstein},
  {Graikou}, {Guillemot}, {Hazboun}, {Hobbs}, {Hu}, {Islo}, {Janssen},
  {Karuppusamy}, {Kramer}, {Lam}, {Lee}, {Liu}, {Luo}, {Lyne}, {Manchester},
  {McKee}, {McLaughlin}, {Mingarelli}, {Parthasarathy}, {Pennucci}, {Perrodin},
  {Possenti}, {Reardon}, {Russell}, {Sanidas}, {Sesana}, {Shaifullah},
  {Shannon}, {Siemens}, {Simon}, {Spiewak}, {Stairs}, {Stappers}, {Swiggum},
  {Taylor}, {Theureau}, {Tiburzi}, {Vallisneri}, {Vecchio}, {Wang}, {Zhang},
  {Zhang}, {Zhu}, and {Zhu}}]{2019MNRAS.490.4666P}
{Perera} BBP, {DeCesar} ME, {Demorest} PB, {Kerr} M, {Lentati} L, {Nice} DJ,
  {Os{\l}owski} S, {Ransom} SM, {Keith} MJ, {Arzoumanian} Z, et~al. (2019) {The
  International Pulsar Timing Array: second data release}. \mnras
  490(4):4666--4687. \doi{10.1093/mnras/stz2857}.
  {\href{https://arxiv.org/abs/1909.04534}{{arXiv:1909.04534}}} {[astro-ph.HE]}

\bibitem[{{Peres}(1962)}]{1962PhRv..128.2471P}
{Peres} A (1962) {Classical Radiation Recoil}. Physical Review
  128(5):2471--2475. \doi{10.1103/PhysRev.128.2471}

\bibitem[{{Perets} and {Alexander}(2008)}]{2008ApJ...677..146P}
{Perets} HB, {Alexander} T (2008) {Massive Perturbers and the Efficient Merger
  of Binary Massive Black Holes}. \apj 677(1):146--159. \doi{10.1086/527525}.
  {\href{https://arxiv.org/abs/0705.2123}{{arXiv:0705.2123}}} {[astro-ph]}

\bibitem[{{Perets} et~al.(2007){Perets}, {Hopman}, and
  {Alexander}}]{2007ApJ...656..709P}
{Perets} HB, {Hopman} C, {Alexander} T (2007) {Massive Perturber-driven
  Interactions between Stars and a Massive Black Hole}. \apj 656(2):709--720.
  \doi{10.1086/510377}.
  {\href{https://arxiv.org/abs/astro-ph/0606443}{{arXiv:astro-ph/0606443}}}
  {[astro-ph]}

\bibitem[{{Perets} et~al.(2009){Perets}, {Gualandris}, {Kupi}, {Merritt}, and
  {Alexander}}]{2009ApJ...702..884P}
{Perets} HB, {Gualandris} A, {Kupi} G, {Merritt} D, {Alexander} T (2009)
  {Dynamical Evolution of the Young Stars in the Galactic Center: N-body
  Simulations of the S-Stars}. \apj 702(2):884--889.
  \doi{10.1088/0004-637X/702/2/884}.
  {\href{https://arxiv.org/abs/0903.2912}{{arXiv:0903.2912}}} {[astro-ph.GA]}

\bibitem[{{P{\'e}rigois} et~al.(2021){P{\'e}rigois}, {Belczynski}, {Bulik}, and
  {Regimbau}}]{2020arXiv200804890P}
{P{\'e}rigois} C, {Belczynski} C, {Bulik} T, {Regimbau} T (2021) {StarTrack
  predictions of the stochastic gravitational-wave background from compact
  binary mergers}. \prd 103(4):043002. \doi{10.1103/PhysRevD.103.043002}.
  {\href{https://arxiv.org/abs/2008.04890}{{arXiv:2008.04890}}} {[astro-ph.CO]}

\bibitem[{{Perley} et~al.(2019){Perley}, {Mazzali}, {Yan}, {Cenko}, {Gezari},
  {Taggart}, {Blagorodnova}, {Fremling}, {Mockler}, {Singh}, {Tominaga},
  {Tanaka}, {Watson}, {Ahumada}, {Anupama}, {Ashall}, {Becerra}, {Bersier},
  {Bhalerao}, {Bloom}, {Butler}, {Copperwheat}, {Coughlin}, {De}, {Drake},
  {Duev}, {Frederick}, {Gonz{\'a}lez}, {Goobar}, {Heida}, {Ho}, {Horst},
  {Hung}, {Itoh}, {Jencson}, {Kasliwal}, {Kawai}, {Khanam}, {Kulkarni},
  {Kumar}, {Kumar}, {Kutyrev}, {Lee}, {Maeda}, {Mahabal}, {Murata}, {Neill},
  {Ngeow}, {Penprase}, {Pian}, {Quimby}, {Ramirez-Ruiz}, {Richer},
  {Rom{\'a}n-Z{\'u}{\~n}iga}, {Sahu}, {Srivastav}, {Socia}, {Sollerman},
  {Tachibana}, {Taddia}, {Tinyanont}, {Troja}, {Ward}, {Wee}, and
  {Yu}}]{2019MNRAS.484.1031P}
{Perley} DA, {Mazzali} PA, {Yan} L, {Cenko} SB, {Gezari} S, {Taggart} K,
  {Blagorodnova} N, {Fremling} C, {Mockler} B, {Singh} A, et~al. (2019) {The
  fast, luminous ultraviolet transient AT2018cow: extreme supernova, or
  disruption of a star by an intermediate-mass black hole?} \mnras
  484(1):1031--1049. \doi{10.1093/mnras/sty3420}.
  {\href{https://arxiv.org/abs/1808.00969}{{arXiv:1808.00969}}} {[astro-ph.HE]}

\bibitem[{{Perpiny{\`a}-Vall{\`e}s} et~al.(2019){Perpiny{\`a}-Vall{\`e}s},
  {Rebassa-Mansergas}, {G{\"a}nsicke}, {Toonen}, {Hermes}, {Gentile Fusillo},
  and {Tremblay}}]{2019MNRAS.483..901P}
{Perpiny{\`a}-Vall{\`e}s} M, {Rebassa-Mansergas} A, {G{\"a}nsicke} BT, {Toonen}
  S, {Hermes} JJ, {Gentile Fusillo} NP, {Tremblay} PE (2019) {Discovery of the
  first resolved triple white dwarf}. \mnras 483(1):901--907.
  \doi{10.1093/mnras/sty3149}.
  {\href{https://arxiv.org/abs/1811.07752}{{arXiv:1811.07752}}} {[astro-ph.SR]}

\bibitem[{{Peschken} and {{\L}okas}(2019)}]{2019MNRAS.483.2721P}
{Peschken} N, {{\L}okas} EL (2019) {Tidally induced bars in Illustris
  galaxies}. \mnras 483(2):2721--2735. \doi{10.1093/mnras/sty3277}.
  {\href{https://arxiv.org/abs/1804.06241}{{arXiv:1804.06241}}} {[astro-ph.GA]}

\bibitem[{{Pestoni} et~al.(2021){Pestoni}, {Bortolas}, {Capelo}, and
  {Mayer}}]{2021MNRAS.500.4628P}
{Pestoni} B, {Bortolas} E, {Capelo} PR, {Mayer} L (2021) {Generation of
  gravitational waves and tidal disruptions in clumpy galaxies}. \mnras
  500(4):4628--4638. \doi{10.1093/mnras/staa3496}.
  {\href{https://arxiv.org/abs/2011.02488}{{arXiv:2011.02488}}} {[astro-ph.GA]}

\bibitem[{{Peters}(1964{\natexlab{a}})}]{1964PhRv..136.1224P}
{Peters} PC (1964{\natexlab{a}}) {Gravitational Radiation and the Motion of Two
  Point Masses}. Physical Review 136(4B):1224--1232.
  \doi{10.1103/PhysRev.136.B1224}

\bibitem[{{Peters}(1964{\natexlab{b}})}]{1964PhDT........51P}
{Peters} PC (1964{\natexlab{b}}) {Gravitational radiation and the motion of two
  point masses}. PhD thesis, California Institute of Technology

\bibitem[{{Peters} and {Mathews}(1963)}]{1963PhRv..131..435P}
{Peters} PC, {Mathews} J (1963) {Gravitational Radiation from Point Masses in a
  Keplerian Orbit}. Physical Review 131(1):435--440.
  \doi{10.1103/PhysRev.131.435}

\bibitem[{{Petrovich} and {Antonini}(2017)}]{2017ApJ...846..146P}
{Petrovich} C, {Antonini} F (2017) {Greatly Enhanced Merger Rates of
  Compact-object Binaries in Non-spherical Nuclear Star Clusters}. \apj
  846(2):146. \doi{10.3847/1538-4357/aa8628}.
  {\href{https://arxiv.org/abs/1705.05848}{{arXiv:1705.05848}}} {[astro-ph.HE]}

\bibitem[{{Pezzulli} et~al.(2016){Pezzulli}, {Valiante}, and
  {Schneider}}]{2016MNRAS.458.3047P}
{Pezzulli} E, {Valiante} R, {Schneider} R (2016) {Super-Eddington growth of the
  first black holes}. \mnras 458(3):3047--3059. \doi{10.1093/mnras/stw505}.
  {\href{https://arxiv.org/abs/1603.00475}{{arXiv:1603.00475}}} {[astro-ph.GA]}

\bibitem[{{Pezzulli} et~al.(2017){Pezzulli}, {Volonteri}, {Schneider}, and
  {Valiante}}]{2017MNRAS.471..589P}
{Pezzulli} E, {Volonteri} M, {Schneider} R, {Valiante} R (2017) {The
  sustainable growth of the first black holes}. \mnras 471(1):589--595.
  \doi{10.1093/mnras/stx1640}.
  {\href{https://arxiv.org/abs/1706.06592}{{arXiv:1706.06592}}} {[astro-ph.GA]}

\bibitem[{{Pfister} et~al.(2017){Pfister}, {Lupi}, {Capelo}, {Volonteri},
  {Bellovary}, and {Dotti}}]{2017MNRAS.471.3646P}
{Pfister} H, {Lupi} A, {Capelo} PR, {Volonteri} M, {Bellovary} JM, {Dotti} M
  (2017) {The birth of a supermassive black hole binary}. \mnras
  471(3):3646--3656. \doi{10.1093/mnras/stx1853}.
  {\href{https://arxiv.org/abs/1706.04010}{{arXiv:1706.04010}}} {[astro-ph.GA]}

\bibitem[{{Pfister} et~al.(2019{\natexlab{a}}){Pfister}, {Bar-Or}, {Volonteri},
  {Dubois}, and {Capelo}}]{2019MNRAS.488L..29P}
{Pfister} H, {Bar-Or} B, {Volonteri} M, {Dubois} Y, {Capelo} PR
  (2019{\natexlab{a}}) {Tidal disruption event rates in galaxy merger
  remnants}. \mnras 488(1):L29--L34. \doi{10.1093/mnrasl/slz091}.
  {\href{https://arxiv.org/abs/1903.09124}{{arXiv:1903.09124}}} {[astro-ph.GA]}

\bibitem[{{Pfister} et~al.(2019{\natexlab{b}}){Pfister}, {Volonteri}, {Dubois},
  {Dotti}, and {Colpi}}]{2019MNRAS.486..101P}
{Pfister} H, {Volonteri} M, {Dubois} Y, {Dotti} M, {Colpi} M
  (2019{\natexlab{b}}) {The erratic dynamical life of black hole seeds in
  high-redshift galaxies}. \mnras 486(1):101--111. \doi{10.1093/mnras/stz822}.
  {\href{https://arxiv.org/abs/1902.01297}{{arXiv:1902.01297}}} {[astro-ph.GA]}

\bibitem[{{Pfister} et~al.(2020){Pfister}, {Dotti}, {Laigle}, {Dubois}, and
  {Volonteri}}]{2020MNRAS.493..922P}
{Pfister} H, {Dotti} M, {Laigle} C, {Dubois} Y, {Volonteri} M (2020) {Real
  galaxy mergers from galaxy pair catalogues}. \mnras 493(1):922--929.
  \doi{10.1093/mnras/staa227}.
  {\href{https://arxiv.org/abs/2001.02461}{{arXiv:2001.02461}}} {[astro-ph.GA]}

\bibitem[{{Pfister} et~al.(2021{\natexlab{a}}){Pfister}, {Dai}, {Volonteri},
  {Auchettl}, {Trebitsch}, and {Ramirez-Ruiz}}]{2021MNRAS.500.3944P}
{Pfister} H, {Dai} JL, {Volonteri} M, {Auchettl} K, {Trebitsch} M,
  {Ramirez-Ruiz} E (2021{\natexlab{a}}) {Tidal disruption events in the first
  billion years of a galaxy}. \mnras 500(3):3944--3956.
  \doi{10.1093/mnras/staa3471}.
  {\href{https://arxiv.org/abs/2006.06565}{{arXiv:2006.06565}}} {[astro-ph.GA]}

\bibitem[{{Pfister} et~al.(2021{\natexlab{b}}){Pfister}, {Toscani}, {Wong},
  {Lixin Dai}, {Lodato}, and {Rossi}}]{2021arXiv210305883P}
{Pfister} H, {Toscani} M, {Wong} THT, {Lixin Dai} J, {Lodato} G, {Rossi} EM
  (2021{\natexlab{b}}) {Observable gravitational waves from tidal disruption
  events and their electromagnetic counterpart}. arXiv e-prints
  arXiv:2103.05883.
  {\href{https://arxiv.org/abs/2103.05883}{{arXiv:2103.05883}}} {[astro-ph.HE]}

\bibitem[{{Pflueger} et~al.(2018){Pflueger}, {Nguyen}, {Bogdanovi{\'c}},
  {Eracleous}, {Runnoe}, {Sigurdsson}, and {Boroson}}]{2018ApJ...861...59P}
{Pflueger} BJ, {Nguyen} K, {Bogdanovi{\'c}} T, {Eracleous} M, {Runnoe} JC,
  {Sigurdsson} S, {Boroson} T (2018) {Likelihood for Detection of Subparsec
  Supermassive Black Hole Binaries in Spectroscopic Surveys}. \apj 861(1):59.
  \doi{10.3847/1538-4357/aaca2c}.
  {\href{https://arxiv.org/abs/1803.02368}{{arXiv:1803.02368}}} {[astro-ph.HE]}

\bibitem[{{Phillips} and {Podsiadlowski}(2002)}]{2002MNRAS.337..431P}
{Phillips} SN, {Podsiadlowski} P (2002) {Irradiation pressure effects in close
  binary systems}. \mnras 337(2):431--444.
  \doi{10.1046/j.1365-8711.2002.05886.x}.
  {\href{https://arxiv.org/abs/astro-ph/0109304}{{arXiv:astro-ph/0109304}}}
  {[astro-ph]}

\bibitem[{{Phinney}(1989)}]{1989IAUS..136..543P}
{Phinney} ES (1989) {Manifestations of a Massive Black Hole in the Galactic
  Center}. In: {Morris} M (ed) The Center of the Galaxy. IAU Symposium, vol
  136. p 543

\bibitem[{{Phinney}(1991)}]{1991ApJ...380L..17P}
{Phinney} ES (1991) {The Rate of Neutron Star Binary Mergers in the Universe:
  Minimal Predictions for Gravity Wave Detectors}. \apjl 380:L17.
  \doi{10.1086/186163}

\bibitem[{{Piana} et~al.(2021){Piana}, {Dayal}, {Volonteri}, and
  {Choudhury}}]{2021MNRAS.500.2146P}
{Piana} O, {Dayal} P, {Volonteri} M, {Choudhury} TR (2021) {The mass assembly
  of high-redshift black holes}. \mnras 500(2):2146--2158.
  \doi{10.1093/mnras/staa3363}

\bibitem[{{Pieroni} and {Barausse}(2020)}]{2020JCAP...07..021P}
{Pieroni} M, {Barausse} E (2020) {Foreground cleaning and template-free
  stochastic background extraction for LISA}. \jcap 2020(7):021.
  \doi{10.1088/1475-7516/2020/07/021}.
  {\href{https://arxiv.org/abs/2004.01135}{{arXiv:2004.01135}}} {[astro-ph.CO]}

\bibitem[{{Pillepich} et~al.(2019){Pillepich}, {Nelson}, {Springel}, {Pakmor},
  {Torrey}, {Weinberger}, {Vogelsberger}, {Marinacci}, {Genel}, {van der Wel},
  and {Hernquist}}]{2019MNRAS.490.3196P}
{Pillepich} A, {Nelson} D, {Springel} V, {Pakmor} R, {Torrey} P, {Weinberger}
  R, {Vogelsberger} M, {Marinacci} F, {Genel} S, {van der Wel} A, et~al. (2019)
  {First results from the TNG50 simulation: the evolution of stellar and
  gaseous discs across cosmic time}. \mnras 490(3):3196--3233.
  \doi{10.1093/mnras/stz2338}.
  {\href{https://arxiv.org/abs/1902.05553}{{arXiv:1902.05553}}} {[astro-ph.GA]}

\bibitem[{{Piovano} et~al.(2020){Piovano}, {Maselli}, and
  {Pani}}]{2020PhRvD.102b4041P}
{Piovano} GA, {Maselli} A, {Pani} P (2020) {Extreme mass ratio inspirals with
  spinning secondary: A detailed study of equatorial circular motion}. \prd
  102(2):024041. \doi{10.1103/PhysRevD.102.024041}.
  {\href{https://arxiv.org/abs/2004.02654}{{arXiv:2004.02654}}} {[gr-qc]}

\bibitem[{{Pipino} et~al.(2014){Pipino}, {Cibinel}, {Tacchella}, {Carollo},
  {Lilly}, {Miniati}, {Silverman}, {van Gorkom}, and
  {Finoguenov}}]{2014ApJ...797..127P}
{Pipino} A, {Cibinel} A, {Tacchella} S, {Carollo} CM, {Lilly} SJ, {Miniati} F,
  {Silverman} JD, {van Gorkom} JH, {Finoguenov} A (2014) {The Zurich
  Environmental Study (ZENS) of Galaxies in Groups along the Cosmic Web. V.
  Properties and Frequency of Merging Satellites and Centrals in Different
  Environments}. \apj 797(2):127. \doi{10.1088/0004-637X/797/2/127}.
  {\href{https://arxiv.org/abs/1409.8298}{{arXiv:1409.8298}}} {[astro-ph.GA]}

\bibitem[{{Piran} et~al.(2015){Piran}, {Svirski}, {Krolik}, {Cheng}, and
  {Shiokawa}}]{2015ApJ...806..164P}
{Piran} T, {Svirski} G, {Krolik} J, {Cheng} RM, {Shiokawa} H (2015) {Disk
  Formation Versus Disk Accretion{\textemdash}What Powers Tidal Disruption
  Events?} \apj 806(2):164. \doi{10.1088/0004-637X/806/2/164}.
  {\href{https://arxiv.org/abs/1502.05792}{{arXiv:1502.05792}}} {[astro-ph.HE]}

\bibitem[{{Piro}(2011)}]{2011ApJ...740L..53P}
{Piro} AL (2011) {Tidal Interactions in Merging White Dwarf Binaries}. \apjl
  740(2):L53. \doi{10.1088/2041-8205/740/2/L53}.
  {\href{https://arxiv.org/abs/1108.3110}{{arXiv:1108.3110}}} {[astro-ph.SR]}

\bibitem[{{Piro}(2012)}]{2012ApJ...755...80P}
{Piro} AL (2012) {Magnetic Interactions in Coalescing Neutron Star Binaries}.
  \apj 755(1):80. \doi{10.1088/0004-637X/755/1/80}.
  {\href{https://arxiv.org/abs/1205.6482}{{arXiv:1205.6482}}} {[astro-ph.HE]}

\bibitem[{{Piro} et~al.(2021){Piro}, {Ahlers}, {Coleiro}, {Colpi}, {de O{\~n}a
  Wilhelmi}, {Guainazzi}, {Jonker}, {Mc Namara}, {Nichols}, {O'Brien}, {Troja},
  {Vink}, {Aird}, {Amati}, {Anand}, {Bozzo}, {Carrera}, {Fabian}, {Fryer},
  {Hall}, {Korobkin}, {Korol}, {Mangiagli}, {Mart{\'\i}nez-N{\'u}{\~n}ez},
  {Nissanke}, {Osborne}, {Padovani}, {Rossi}, {Ryan}, {Sesana}, {Stratta},
  {Tanvir}, and {van Eerten}}]{2021arXiv211015677P}
{Piro} L, {Ahlers} M, {Coleiro} A, {Colpi} M, {de O{\~n}a Wilhelmi} E,
  {Guainazzi} M, {Jonker} PG, {Mc Namara} P, {Nichols} DA, {O'Brien} P, et~al.
  (2021) {Multi-messenger-Athena Synergy White Paper}. arXiv e-prints
  arXiv:2110.15677.
  {\href{https://arxiv.org/abs/2110.15677}{{arXiv:2110.15677}}} {[astro-ph.HE]}

\bibitem[{{Podsiadlowski}(1991)}]{1991Natur.350..136P}
{Podsiadlowski} P (1991) {Irradiation-driven mass transfer in low-mass X-ray
  binaries}. \nat 350(6314):136--138. \doi{10.1038/350136a0}

\bibitem[{{Podsiadlowski} et~al.(2002){Podsiadlowski}, {Rappaport}, and
  {Pfahl}}]{2002ApJ...565.1107P}
{Podsiadlowski} P, {Rappaport} S, {Pfahl} ED (2002) {Evolutionary Sequences for
  Low- and Intermediate-Mass X-Ray Binaries}. \apj 565(2):1107--1133.
  \doi{10.1086/324686}.
  {\href{https://arxiv.org/abs/astro-ph/0107261}{{arXiv:astro-ph/0107261}}}
  {[astro-ph]}

\bibitem[{{Podsiadlowski} et~al.(2003){Podsiadlowski}, {Rappaport}, and
  {Han}}]{2003MNRAS.341..385P}
{Podsiadlowski} P, {Rappaport} S, {Han} Z (2003) {On the formation and
  evolution of black hole binaries}. \mnras 341(2):385--404.
  \doi{10.1046/j.1365-8711.2003.06464.x}.
  {\href{https://arxiv.org/abs/astro-ph/0207153}{{arXiv:astro-ph/0207153}}}
  {[astro-ph]}

\bibitem[{{Podsiadlowski} et~al.(2004){Podsiadlowski}, {Langer}, {Poelarends},
  {Rappaport}, {Heger}, and {Pfahl}}]{2004ApJ...612.1044P}
{Podsiadlowski} P, {Langer} N, {Poelarends} AJT, {Rappaport} S, {Heger} A,
  {Pfahl} E (2004) {The Effects of Binary Evolution on the Dynamics of Core
  Collapse and Neutron Star Kicks}. \apj 612(2):1044--1051.
  \doi{10.1086/421713}.
  {\href{https://arxiv.org/abs/astro-ph/0309588}{{arXiv:astro-ph/0309588}}}
  {[astro-ph]}

\bibitem[{{Poggianti} et~al.(2017){Poggianti}, {Jaff{\'e}}, {Moretti},
  {Gullieuszik}, {Radovich}, {Tonnesen}, {Fritz}, {Bettoni}, {Vulcani},
  {Fasano}, {Bellhouse}, {Hau}, and {Omizzolo}}]{2017Natur.548..304P}
{Poggianti} BM, {Jaff{\'e}} YL, {Moretti} A, {Gullieuszik} M, {Radovich} M,
  {Tonnesen} S, {Fritz} J, {Bettoni} D, {Vulcani} B, {Fasano} G, et~al. (2017)
  {Ram-pressure feeding of supermassive black holes}. \nat 548(7667):304--309.
  \doi{10.1038/nature23462}.
  {\href{https://arxiv.org/abs/1708.09036}{{arXiv:1708.09036}}} {[astro-ph.GA]}

\bibitem[{{Poisson} and {Will}(1995)}]{PoissonWill:1995}
{Poisson} E, {Will} CM (1995) {Gravitational waves from inspiraling compact
  binaries: Parameter estimation using second-post-Newtonian waveforms}. \prd
  52:848--855.
  {\href{https://arxiv.org/abs/arXiv:gr-qc/9502040}{{arXiv:gr-qc/9502040}}}

\bibitem[{{Poisson} et~al.(2011){Poisson}, {Pound}, and
  {Vega}}]{2011LRR....14....7P}
{Poisson} E, {Pound} A, {Vega} I (2011) {The Motion of Point Particles in
  Curved Spacetime}. Living Reviews in Relativity 14(1):7.
  \doi{10.12942/lrr-2011-7}.
  {\href{https://arxiv.org/abs/1102.0529}{{arXiv:1102.0529}}} {[gr-qc]}

\bibitem[{{Pol} et~al.(2020){Pol}, {McLaughlin}, {Lorimer}, and
  {Garver-Daniels}}]{2020arXiv201004151P}
{Pol} N, {McLaughlin} M, {Lorimer} DR, {Garver-Daniels} N (2020) {On the
  detectability of ultra-compact binary pulsar systems}. arXiv e-prints
  arXiv:2010.04151.
  {\href{https://arxiv.org/abs/2010.04151}{{arXiv:2010.04151}}} {[astro-ph.HE]}

\bibitem[{{Popham} and {Gammie}(1998)}]{1998ApJ...504..419P}
{Popham} R, {Gammie} CF (1998) {Advection-dominated Accretion Flows in the Kerr
  Metric. II. Steady State Global Solutions}. \apj 504(1):419--430.
  \doi{10.1086/306054}.
  {\href{https://arxiv.org/abs/astro-ph/9802321}{{arXiv:astro-ph/9802321}}}
  {[astro-ph]}

\bibitem[{{Portegies Zwart}(2013)}]{2013MNRAS.429L..45P}
{Portegies Zwart} S (2013) {Planet-mediated precision reconstruction of the
  evolution of the cataclysmic variable HU Aquarii.} \mnras 429:L45--L49.
  \doi{10.1093/mnrasl/sls022}.
  {\href{https://arxiv.org/abs/1210.5540}{{arXiv:1210.5540}}} {[astro-ph.EP]}

\bibitem[{{Portegies Zwart} and {McMillan}(2000)}]{2000ApJ...528L..17P}
{Portegies Zwart} SF, {McMillan} SLW (2000) {Black Hole Mergers in the
  Universe}. \apjl 528(1):L17--L20. \doi{10.1086/312422}.
  {\href{https://arxiv.org/abs/astro-ph/9910061}{{arXiv:astro-ph/9910061}}}
  {[astro-ph]}

\bibitem[{{Portegies Zwart} and {McMillan}(2002)}]{2002ApJ...576..899P}
{Portegies Zwart} SF, {McMillan} SLW (2002) {The Runaway Growth of
  Intermediate-Mass Black Holes in Dense Star Clusters}. \apj 576(2):899--907.
  \doi{10.1086/341798}.
  {\href{https://arxiv.org/abs/astro-ph/0201055}{{arXiv:astro-ph/0201055}}}
  {[astro-ph]}

\bibitem[{{Portegies Zwart} and {Verbunt}(1996)}]{1996AA...309..179P}
{Portegies Zwart} SF, {Verbunt} F (1996) {Population synthesis of high-mass
  binaries.} \aap 309:179--196

\bibitem[{{Portegies Zwart} and {Yungelson}(1998)}]{1998A&A...332..173P}
{Portegies Zwart} SF, {Yungelson} LR (1998) {Formation and evolution of binary
  neutron stars}. \aap 332:173--188.
  {\href{https://arxiv.org/abs/astro-ph/9710347}{{arXiv:astro-ph/9710347}}}
  {[astro-ph]}

\bibitem[{{Portegies Zwart} et~al.(2004){Portegies Zwart}, {Baumgardt}, {Hut},
  {Makino}, and {McMillan}}]{2004Natur.428..724P}
{Portegies Zwart} SF, {Baumgardt} H, {Hut} P, {Makino} J, {McMillan} SLW (2004)
  {Formation of massive black holes through runaway collisions in dense young
  star clusters}. \nat 428(6984):724--726. \doi{10.1038/nature02448}.
  {\href{https://arxiv.org/abs/astro-ph/0402622}{{arXiv:astro-ph/0402622}}}
  {[astro-ph]}

\bibitem[{{Portegies Zwart} et~al.(2006){Portegies Zwart}, {Baumgardt},
  {McMillan}, {Makino}, {Hut}, and {Ebisuzaki}}]{2006ApJ...641..319P}
{Portegies Zwart} SF, {Baumgardt} H, {McMillan} SLW, {Makino} J, {Hut} P,
  {Ebisuzaki} T (2006) {The Ecology of Star Clusters and Intermediate-Mass
  Black Holes in the Galactic Bulge}. \apj 641(1):319--326.
  \doi{10.1086/500361}.
  {\href{https://arxiv.org/abs/astro-ph/0511397}{{arXiv:astro-ph/0511397}}}
  {[astro-ph]}

\bibitem[{{Postnov} and {Yungelson}(2014)}]{2014LRR....17....3P}
{Postnov} KA, {Yungelson} LR (2014) {The Evolution of Compact Binary Star
  Systems}. Living Reviews in Relativity 17(1):3. \doi{10.12942/lrr-2014-3}.
  {\href{https://arxiv.org/abs/1403.4754}{{arXiv:1403.4754}}} {[astro-ph.HE]}

\bibitem[{{Pound}(2014)}]{2014PhRvD..90h4039P}
{Pound} A (2014) {Conservative effect of the second-order gravitational
  self-force on quasicircular orbits in Schwarzschild spacetime}. \prd
  90(8):084039. \doi{10.1103/PhysRevD.90.084039}.
  {\href{https://arxiv.org/abs/1404.1543}{{arXiv:1404.1543}}} {[gr-qc]}

\bibitem[{{Pound}(2017)}]{2017PhRvD..95j4056P}
{Pound} A (2017) {Nonlinear gravitational self-force: Second-order equation of
  motion}. \prd 95(10):104056. \doi{10.1103/PhysRevD.95.104056}.
  {\href{https://arxiv.org/abs/1703.02836}{{arXiv:1703.02836}}} {[gr-qc]}

\bibitem[{{Pound} and {Miller}(2014)}]{2014PhRvD..89j4020P}
{Pound} A, {Miller} J (2014) {Practical, covariant puncture for second-order
  self-force calculations}. \prd 89(10):104020.
  \doi{10.1103/PhysRevD.89.104020}.
  {\href{https://arxiv.org/abs/1403.1843}{{arXiv:1403.1843}}} {[gr-qc]}

\bibitem[{{Pound} et~al.(2020){Pound}, {Wardell}, {Warburton}, and
  {Miller}}]{2020PhRvL.124b1101P}
{Pound} A, {Wardell} B, {Warburton} N, {Miller} J (2020) {Second-Order
  Self-Force Calculation of Gravitational Binding Energy in Compact Binaries}.
  \prl 124(2):021101. \doi{10.1103/PhysRevLett.124.021101}.
  {\href{https://arxiv.org/abs/1908.07419}{{arXiv:1908.07419}}} {[gr-qc]}

\bibitem[{{Poveda} et~al.(1994){Poveda}, {Herrera}, {Allen}, {Cordero}, and
  {Lavalley}}]{1994RMxAA..28...43P}
{Poveda} A, {Herrera} MA, {Allen} C, {Cordero} G, {Lavalley} C (1994)
  {Statistical studies of visual double and multiple stars. II. A catalogue of
  nearby wide binary and multiple systems.} \rmxaa 28:43--89

\bibitem[{{Power} et~al.(2010){Power}, {Baugh}, and
  {Lacey}}]{2010MNRAS.406...43P}
{Power} C, {Baugh} CM, {Lacey} CG (2010) {The redshift evolution of the mass
  function of cold gas in hierarchical galaxy formation models}. \mnras
  406(1):43--59. \doi{10.1111/j.1365-2966.2010.16481.x}.
  {\href{https://arxiv.org/abs/0908.1396}{{arXiv:0908.1396}}} {[astro-ph.CO]}

\bibitem[{{Predehl} et~al.(2010){Predehl}, {Andritschke}, {B{\"o}hringer},
  {Bornemann}, {Br{\"a}uninger}, {Brunner}, {Brusa}, {Burkert}, {Burwitz},
  {Cappelluti}, {Churazov}, {Dennerl}, {Eder}, {Elbs}, {Freyberg}, {Friedrich},
  {F{\"u}rmetz}, {Gaida}, {H{\"a}lker}, {Hartner}, {Hasinger}, {Hermann},
  {Huber}, {Kendziorra}, {von Kienlin}, {Kink}, {Kreykenbohm}, {Lamer},
  {Lapchov}, {Lehmann}, {Meidinger}, {Mican}, {Mohr}, {M{\"u}hlegger},
  {M{\"u}ller}, {Nandra}, {Pavlinsky}, {Pfeffermann}, {Reiprich}, {Robrade},
  {Roh{\'e}}, {Santangelo}, {Sch{\"a}chner}, {Schanz}, {Schmid}, {Schmitt},
  {Schreib}, {Schrey}, {Schwope}, {Steinmetz}, {Str{\"u}der}, {Sunyaev},
  {Tenzer}, {Tiedemann}, {Vongehr}, and {Wilms}}]{2010SPIE.7732E..0UP}
{Predehl} P, {Andritschke} R, {B{\"o}hringer} H, {Bornemann} W,
  {Br{\"a}uninger} H, {Brunner} H, {Brusa} M, {Burkert} W, {Burwitz} V,
  {Cappelluti} N, et~al. (2010) {eROSITA on SRG}. In: {Arnaud} M, {Murray} SS,
  {Takahashi} T (eds) Space Telescopes and Instrumentation 2010: Ultraviolet to
  Gamma Ray. Society of Photo-Optical Instrumentation Engineers (SPIE)
  Conference Series, vol 7732. p 77320U. \doi{10.1117/12.856577}.
  {\href{https://arxiv.org/abs/1001.2502}{{arXiv:1001.2502}}} {[astro-ph.CO]}

\bibitem[{{Press} and {Schechter}(1974)}]{1974ApJ...187..425P}
{Press} WH, {Schechter} P (1974) {Formation of Galaxies and Clusters of
  Galaxies by Self-Similar Gravitational Condensation}. \apj 187:425--438.
  \doi{10.1086/152650}

\bibitem[{{Preto}(2010)}]{2010arXiv1005.4048A}
{Preto} M (2010) {Gravitational Waves Notes, Issue \#3 : ``Stellar cusps in
  galactic nuclei - How stars distribute around a massive black hole''}. arXiv
  e-prints arXiv:1005.4048.
  {\href{https://arxiv.org/abs/1005.4048}{{arXiv:1005.4048}}} {[astro-ph.CO]}

\bibitem[{{Preto} et~al.(2011){Preto}, {Berentzen}, {Berczik}, and
  {Spurzem}}]{2011ApJ...732L..26P}
{Preto} M, {Berentzen} I, {Berczik} P, {Spurzem} R (2011) {Fast Coalescence of
  Massive Black Hole Binaries from Mergers of Galactic Nuclei: Implications for
  Low-frequency Gravitational-wave Astrophysics}. \apjl 732(2):L26.
  \doi{10.1088/2041-8205/732/2/L26}.
  {\href{https://arxiv.org/abs/1102.4855}{{arXiv:1102.4855}}} {[astro-ph.GA]}

\bibitem[{{Pretorius}(2005)}]{2005PhRvL..95l1101P}
{Pretorius} F (2005) {Evolution of Binary Black-Hole Spacetimes}. \prl
  95(12):121101. \doi{10.1103/PhysRevLett.95.121101}.
  {\href{https://arxiv.org/abs/gr-qc/0507014}{{arXiv:gr-qc/0507014}}} {[gr-qc]}

\bibitem[{{Provencal} et~al.(1997){Provencal}, {Winget}, {Nather}, {Robinson},
  {Clemens}, {Bradley}, {Claver}, {Kleinman}, {Grauer}, {Hine}, {Ferrario},
  {Warner}, {Vauclair}, {Chevreton}, {Kepler}, {Wood}, and
  {Henry}}]{1997ApJ...480..383P}
{Provencal} JL, {Winget} DE, {Nather} RE, {Robinson} EL, {Clemens} JC,
  {Bradley} PA, {Claver} CF, {Kleinman} SJ, {Grauer} AD, {Hine} BP, et~al.
  (1997) {Whole Earth Telescope Observations of the Helium Interacting Binary
  PG 1346+082 (CR Bootis)}. \apj 480(1):383--394. \doi{10.1086/303971}

\bibitem[{{Prust} and {Chang}(2019)}]{2019MNRAS.486.5809P}
{Prust} LJ, {Chang} P (2019) {Common envelope evolution on a moving mesh}.
  \mnras 486(4):5809--5818. \doi{10.1093/mnras/stz1219}.
  {\href{https://arxiv.org/abs/1904.09256}{{arXiv:1904.09256}}} {[astro-ph.SR]}

\bibitem[{{Punturo} et~al.(2010){Punturo}, {Abernathy}, {Acernese}, {Allen},
  {Andersson}, {Arun}, {Barone}, {Barr}, {Barsuglia}, {Beker}, {Beveridge},
  {Birindelli}, {Bose}, {Bosi}, {Braccini}, {Bradaschia}, {Bulik}, {Calloni},
  {Cella}, {Chassande Mottin}, {Chelkowski}, {Chincarini}, {Clark}, {Coccia},
  {Colacino}, {Colas}, {Cumming}, {Cunningham}, {Cuoco}, {Danilishin},
  {Danzmann}, {De Luca}, {De Salvo}, {Dent}, {Derosa}, {Di Fiore}, {Di
  Virgilio}, {Doets}, {Fafone}, {Falferi}, {Flaminio}, {Franc}, {Frasconi},
  {Freise}, {Fulda}, {Gair}, {Gemme}, {Gennai}, {Giazotto}, {Glampedakis},
  {Granata}, {Grote}, {Guidi}, {Hammond}, {Hannam}, {Harms}, {Heinert},
  {Hendry}, {Heng}, {Hennes}, {Hild}, {Hough}, {Husa}, {Huttner}, {Jones},
  {Khalili}, {Kokeyama}, {Kokkotas}, {Krishnan}, {Lorenzini}, {L{\"u}ck},
  {Majorana}, {Mandel}, {Mandic}, {Martin}, {Michel}, {Minenkov}, {Morgado},
  {Mosca}, {Mours}, {M{\"u}ller-Ebhardt}, {Murray}, {Nawrodt}, {Nelson},
  {Oshaughnessy}, {Ott}, {Palomba}, {Paoli}, {Parguez}, {Pasqualetti},
  {Passaquieti}, {Passuello}, {Pinard}, {Poggiani}, {Popolizio}, {Prato},
  {Puppo}, {Rabeling}, {Rapagnani}, {Read}, {Regimbau}, {Rehbein}, {Reid},
  {Rezzolla}, {Ricci}, {Richard}, {Rocchi}, {Rowan}, {R{\"u}diger}, {Sassolas},
  {Sathyaprakash}, {Schnabel}, {Schwarz}, {Seidel}, {Sintes}, {Somiya},
  {Speirits}, {Strain}, {Strigin}, {Sutton}, {Tarabrin}, {van den Brand}, {van
  Leewen}, {van Veggel}, {van den Broeck}, {Vecchio}, {Veitch}, {Vetrano},
  {Vicere}, {Vyatchanin}, {Willke}, {Woan}, {Wolfango}, and
  {Yamamoto}}]{2010CQGra..27h4007P}
{Punturo} M, {Abernathy} M, {Acernese} F, {Allen} B, {Andersson} N, {Arun} K,
  {Barone} F, {Barr} B, {Barsuglia} M, {Beker} M, et~al. (2010) {The third
  generation of gravitational wave observatories and their science reach}.
  Classical and Quantum Gravity 27(8):084007.
  \doi{10.1088/0264-9381/27/8/084007}

\bibitem[{{Qin} et~al.(2018){Qin}, {Fragos}, {Meynet}, {Andrews},
  {S{\o}rensen}, and {Song}}]{2018A&A...616A..28Q}
{Qin} Y, {Fragos} T, {Meynet} G, {Andrews} J, {S{\o}rensen} M, {Song} HF (2018)
  {The spin of the second-born black hole in coalescing binary black holes}.
  \aap 616:A28. \doi{10.1051/0004-6361/201832839}.
  {\href{https://arxiv.org/abs/1802.05738}{{arXiv:1802.05738}}} {[astro-ph.SR]}

\bibitem[{{Quinlan}(1996)}]{1996NewA....1...35Q}
{Quinlan} GD (1996) {The dynamical evolution of massive black hole binaries I.
  Hardening in a fixed stellar background}. \na 1(1):35--56.
  \doi{10.1016/S1384-1076(96)00003-6}.
  {\href{https://arxiv.org/abs/astro-ph/9601092}{{arXiv:astro-ph/9601092}}}
  {[astro-ph]}

\bibitem[{{Quinlan} and {Hernquist}(1997)}]{1997NewA....2..533Q}
{Quinlan} GD, {Hernquist} L (1997) {The dynamical evolution of massive black
  hole binaries {\textemdash} II. Self-consistent N-body integrations}. \na
  2(6):533--554. \doi{10.1016/S1384-1076(97)00039-0}.
  {\href{https://arxiv.org/abs/astro-ph/9706298}{{arXiv:astro-ph/9706298}}}
  {[astro-ph]}

\bibitem[{{Quinlan} and {Shapiro}(1989)}]{1989ApJ...343..725Q}
{Quinlan} GD, {Shapiro} SL (1989) {Dynamical Evolution of Dense Clusters of
  Compact Stars}. \apj 343:725. \doi{10.1086/167745}

\bibitem[{{Raaijmakers} et~al.(2020){Raaijmakers}, {Greif}, {Riley},
  {Hinderer}, {Hebeler}, {Schwenk}, {Watts}, {Nissanke}, {Guillot}, {Lattimer},
  and {Ludlam}}]{2020ApJ...893L..21R}
{Raaijmakers} G, {Greif} SK, {Riley} TE, {Hinderer} T, {Hebeler} K, {Schwenk}
  A, {Watts} AL, {Nissanke} S, {Guillot} S, {Lattimer} JM, et~al. (2020)
  {Constraining the Dense Matter Equation of State with Joint Analysis of NICER
  and LIGO/Virgo Measurements}. \apjl 893(1):L21.
  \doi{10.3847/2041-8213/ab822f}.
  {\href{https://arxiv.org/abs/1912.11031}{{arXiv:1912.11031}}} {[astro-ph.HE]}

\bibitem[{{Raffai} et~al.(2016){Raffai}, {Haiman}, and
  {Frei}}]{2016MNRAS.455..484R}
{Raffai} P, {Haiman} Z, {Frei} Z (2016) {A statistical method to search for
  recoiling supermassive black holes in active galactic nuclei}. \mnras
  455(1):484--492. \doi{10.1093/mnras/stv2371}.
  {\href{https://arxiv.org/abs/1509.02075}{{arXiv:1509.02075}}} {[astro-ph.GA]}

\bibitem[{{Ragusa} et~al.(2016){Ragusa}, {Lodato}, and
  {Price}}]{2016MNRAS.460.1243R}
{Ragusa} E, {Lodato} G, {Price} DJ (2016) {Suppression of the accretion rate in
  thin discs around binary black holes}. \mnras 460(2):1243--1253.
  \doi{10.1093/mnras/stw1081}.
  {\href{https://arxiv.org/abs/1605.01730}{{arXiv:1605.01730}}} {[astro-ph.HE]}

\bibitem[{Raidal et~al.(2019)Raidal, Spethmann, Vaskonen, and
  Veerm\"ae}]{Raidal:2018bbj}
Raidal M, Spethmann C, Vaskonen V, Veerm\"ae H (2019) {Formation and Evolution
  of Primordial Black Hole Binaries in the Early Universe}. JCAP 02:018.
  \doi{10.1088/1475-7516/2019/02/018}.
  {\href{https://arxiv.org/abs/1812.01930}{{arXiv:1812.01930}}} {[astro-ph.CO]}

\bibitem[{{Ramsay} and {Hakala}(2005)}]{2005MNRAS.360..314R}
{Ramsay} G, {Hakala} P (2005) {RApid Temporal Survey (RATS) - I. Overview and
  first results}. \mnras 360(1):314--321.
  \doi{10.1111/j.1365-2966.2005.09035.x}.
  {\href{https://arxiv.org/abs/astro-ph/0503138}{{arXiv:astro-ph/0503138}}}
  {[astro-ph]}

\bibitem[{{Ramsay} et~al.(2005){Ramsay}, {Hakala}, {Marsh}, {Nelemans},
  {Steeghs}, and {Cropper}}]{2005A&A...440..675R}
{Ramsay} G, {Hakala} P, {Marsh} T, {Nelemans} G, {Steeghs} D, {Cropper} M
  (2005) {XMM-Newton observations of AM CVn binaries}. \aap 440(2):675--681.
  \doi{10.1051/0004-6361:20052950}.
  {\href{https://arxiv.org/abs/astro-ph/0505549}{{arXiv:astro-ph/0505549}}}
  {[astro-ph]}

\bibitem[{{Ramsay} et~al.(2006){Ramsay}, {Groot}, {Marsh}, {Nelemans},
  {Steeghs}, and {Hakala}}]{2006A&A...457..623R}
{Ramsay} G, {Groot} PJ, {Marsh} T, {Nelemans} G, {Steeghs} D, {Hakala} P (2006)
  {XMM-Newton observations of AM CVn binaries: V396 Hya and SDSS J1240-01}.
  \aap 457(2):623--627. \doi{10.1051/0004-6361:20065491}.
  {\href{https://arxiv.org/abs/astro-ph/0607178}{{arXiv:astro-ph/0607178}}}
  {[astro-ph]}

\bibitem[{{Ramsay} et~al.(2018){Ramsay}, {Green}, {Marsh}, {Kupfer}, {Breedt},
  {Korol}, {Groot}, {Knigge}, {Nelemans}, {Steeghs}, {Woudt}, and
  {Aungwerojwit}}]{2018A&A...620A.141R}
{Ramsay} G, {Green} MJ, {Marsh} TR, {Kupfer} T, {Breedt} E, {Korol} V, {Groot}
  PJ, {Knigge} C, {Nelemans} G, {Steeghs} D, et~al. (2018) {Physical properties
  of AM CVn stars: New insights from Gaia DR2}. \aap 620:A141.
  \doi{10.1051/0004-6361/201834261}.
  {\href{https://arxiv.org/abs/1810.06548}{{arXiv:1810.06548}}} {[astro-ph.SR]}

\bibitem[{{Randall} and {Xianyu}(2019{\natexlab{a}})}]{2019arXiv190702283R}
{Randall} L, {Xianyu} ZZ (2019{\natexlab{a}}) {Eccentricity Without Measuring
  Eccentricity: Discriminating Among Stellar Mass Black Hole Binary Formation
  Channels}. arXiv e-prints arXiv:1907.02283.
  {\href{https://arxiv.org/abs/1907.02283}{{arXiv:1907.02283}}} {[astro-ph.HE]}

\bibitem[{{Randall} and {Xianyu}(2019{\natexlab{b}})}]{2019arXiv190208604R}
{Randall} L, {Xianyu} ZZ (2019{\natexlab{b}}) {Observing Eccentricity
  Oscillations of Binary Black Holes in LISA}. arXiv e-prints arXiv:1902.08604.
  {\href{https://arxiv.org/abs/1902.08604}{{arXiv:1902.08604}}} {[astro-ph.HE]}

\bibitem[{{Randall} et~al.(2021){Randall}, {Shelest}, and
  {Xianyu}}]{2021arXiv210316030R}
{Randall} L, {Shelest} A, {Xianyu} ZZ (2021) {An Efficient Signal to Noise
  Approximation for Eccentric Inspiraling Binaries}. arXiv e-prints
  arXiv:2103.16030.
  {\href{https://arxiv.org/abs/2103.16030}{{arXiv:2103.16030}}} {[astro-ph.HE]}

\bibitem[{{Ransom} et~al.(2019){Ransom}, {Brazier}, {Chatterjee}, {Cohen},
  {Cordes}, {DeCesar}, {Demorest}, {Hazboun}, {Lam}, {Lynch}, {McLaughlin},
  {Ransom}, {Siemens}, {Taylor}, and {Vigeland}}]{2019BAAS...51g.195R}
{Ransom} S, {Brazier} A, {Chatterjee} S, {Cohen} T, {Cordes} JM, {DeCesar} ME,
  {Demorest} PB, {Hazboun} JS, {Lam} MT, {Lynch} RS, et~al. (2019) {The
  NANOGrav Program for Gravitational Waves and Fundamental Physics}. In:
  Bulletin of the American Astronomical Society. vol~51. p 195.
  {\href{https://arxiv.org/abs/1908.05356}{{arXiv:1908.05356}}} {[astro-ph.IM]}

\bibitem[{{Ransom} et~al.(2014){Ransom}, {Stairs}, {Archibald}, {Hessels},
  {Kaplan}, {van Kerkwijk}, and {et~al.}}]{2014Natur.505..520R}
{Ransom} SM, {Stairs} IH, {Archibald} AM, {Hessels} JWT, {Kaplan} DL, {van
  Kerkwijk} MH, {et~al} (2014) {A millisecond pulsar in a stellar triple
  system}. \nat 505(7484):520--524. \doi{10.1038/nature12917}.
  {\href{https://arxiv.org/abs/1401.0535}{{arXiv:1401.0535}}} {[astro-ph.SR]}

\bibitem[{{Rantala} et~al.(2017){Rantala}, {Pihajoki}, {Johansson}, {Naab},
  {Lah{\'e}n}, and {Sawala}}]{2017ApJ...840...53R}
{Rantala} A, {Pihajoki} P, {Johansson} PH, {Naab} T, {Lah{\'e}n} N, {Sawala} T
  (2017) {Post-Newtonian Dynamical Modeling of Supermassive Black Holes in
  Galactic-scale Simulations}. \apj 840(1):53. \doi{10.3847/1538-4357/aa6d65}.
  {\href{https://arxiv.org/abs/1611.07028}{{arXiv:1611.07028}}} {[astro-ph.GA]}

\bibitem[{{Rantala} et~al.(2018){Rantala}, {Johansson}, {Naab}, {Thomas}, and
  {Frigo}}]{2018ApJ...864..113R}
{Rantala} A, {Johansson} PH, {Naab} T, {Thomas} J, {Frigo} M (2018) {The
  Formation of Extremely Diffuse Galaxy Cores by Merging Supermassive Black
  Holes}. \apj 864(2):113. \doi{10.3847/1538-4357/aada47}.
  {\href{https://arxiv.org/abs/1805.10295}{{arXiv:1805.10295}}} {[astro-ph.GA]}

\bibitem[{{Rappaport} et~al.(1982){Rappaport}, {Joss}, and
  {Webbink}}]{1982ApJ...254..616R}
{Rappaport} S, {Joss} PC, {Webbink} RF (1982) {The evolution of highly compact
  binary stellar systems.} \apj 254:616--640. \doi{10.1086/159772}

\bibitem[{{Rappaport} et~al.(2021){Rappaport}, {Vanderburg}, {Schwab}, and
  {Nelson}}]{2021arXiv210412083R}
{Rappaport} S, {Vanderburg} A, {Schwab} J, {Nelson} L (2021) {Minimum Orbital
  Periods of H-Rich Bodies}. arXiv e-prints arXiv:2104.12083.
  {\href{https://arxiv.org/abs/2104.12083}{{arXiv:2104.12083}}} {[astro-ph.SR]}

\bibitem[{{Rasio} and {Livio}(1996)}]{1996ApJ...471..366R}
{Rasio} FA, {Livio} M (1996) {On the Formation and Evolution of Common Envelope
  Systems}. \apj 471:366. \doi{10.1086/177975}.
  {\href{https://arxiv.org/abs/astro-ph/9511054}{{arXiv:astro-ph/9511054}}}
  {[astro-ph]}

\bibitem[{{Raskin} et~al.(2009){Raskin}, {Timmes}, {Scannapieco}, {Diehl}, and
  {Fryer}}]{2009MNRAS.399L.156R}
{Raskin} C, {Timmes} FX, {Scannapieco} E, {Diehl} S, {Fryer} C (2009) {On Type
  Ia supernovae from the collisions of two white dwarfs}. \mnras
  399(1):L156--L159. \doi{10.1111/j.1745-3933.2009.00743.x}.
  {\href{https://arxiv.org/abs/0907.3915}{{arXiv:0907.3915}}} {[astro-ph.SR]}

\bibitem[{{Rasskazov} and {Merritt}(2017)}]{2017ApJ...837..135R}
{Rasskazov} A, {Merritt} D (2017) {Evolution of Binary Supermassive Black Holes
  in Rotating Nuclei}. \apj 837(2):135. \doi{10.3847/1538-4357/aa6188}.
  {\href{https://arxiv.org/abs/1610.08555}{{arXiv:1610.08555}}} {[astro-ph.GA]}

\bibitem[{{Rasskazov} et~al.(2019){Rasskazov}, {Fragione}, {Leigh}, {Tagawa},
  {Sesana}, {Price-Whelan}, and {Rossi}}]{2019ApJ...878...17R}
{Rasskazov} A, {Fragione} G, {Leigh} NWC, {Tagawa} H, {Sesana} A,
  {Price-Whelan} A, {Rossi} EM (2019) {Hypervelocity Stars from a Supermassive
  Black Hole-Intermediate-mass Black Hole Binary}. \apj 878(1):17.
  \doi{10.3847/1538-4357/ab1c5d}.
  {\href{https://arxiv.org/abs/1810.12354}{{arXiv:1810.12354}}} {[astro-ph.GA]}

\bibitem[{{Rastello} et~al.(2019){Rastello}, {Amaro-Seoane}, {Arca-Sedda},
  {Capuzzo-Dolcetta}, {Fragione}, and {Tosta e Melo}}]{2019MNRAS.483.1233R}
{Rastello} S, {Amaro-Seoane} P, {Arca-Sedda} M, {Capuzzo-Dolcetta} R,
  {Fragione} G, {Tosta e Melo} I (2019) {Stellar black hole binary mergers in
  open clusters}. \mnras 483(1):1233--1246. \doi{10.1093/mnras/sty3193}.
  {\href{https://arxiv.org/abs/1811.10628}{{arXiv:1811.10628}}} {[astro-ph.GA]}

\bibitem[{{Rastello} et~al.(2020){Rastello}, {Mapelli}, {Di Carlo}, {Giacobbo},
  {Santoliquido}, {Spera}, {Ballone}, and {Iorio}}]{2020MNRAS.497.1563R}
{Rastello} S, {Mapelli} M, {Di Carlo} UN, {Giacobbo} N, {Santoliquido} F,
  {Spera} M, {Ballone} A, {Iorio} G (2020) {Dynamics of black hole-neutron star
  binaries in young star clusters}. \mnras 497(2):1563--1570.
  \doi{10.1093/mnras/staa2018}.
  {\href{https://arxiv.org/abs/2003.02277}{{arXiv:2003.02277}}} {[astro-ph.HE]}

\bibitem[{{Rauch} and {Tremaine}(1996)}]{1996NewA....1..149R}
{Rauch} KP, {Tremaine} S (1996) {Resonant relaxation in stellar systems}. \na
  1(2):149--170. \doi{10.1016/S1384-1076(96)00012-7}.
  {\href{https://arxiv.org/abs/astro-ph/9603018}{{arXiv:astro-ph/9603018}}}
  {[astro-ph]}

\bibitem[{{Ravenhall} and {Pethick}(1994)}]{1994ApJ...424..846R}
{Ravenhall} DG, {Pethick} CJ (1994) {Neutron Star Moments of Inertia}. \apj
  424:846. \doi{10.1086/173935}

\bibitem[{{Ray} et~al.(2018){Ray}, {Arzoumanian}, {Brandt}, {Burns},
  {Chakrabarty}, {Feroci}, {Gendreau}, {Gevin}, {Hernanz}, {Jenke}, {Kenyon},
  {G{\'a}lvez}, {Maccarone}, {Okajima}, {Remillard}, {Schanne}, {Tenzer},
  {Vacchi}, {Wilson-Hodge}, {Winter}, {Zane}, {Ballantyne}, {Bozzo},
  {Brenneman}, {Cackett}, {De Rosa}, {Goldstein}, {Hartmann}, {McDonald},
  {Stevens}, {Tomsick}, {Watts}, {Wood}, and {Zoghbi}}]{2018SPIE10699E..19R}
{Ray} PS, {Arzoumanian} Z, {Brandt} S, {Burns} E, {Chakrabarty} D, {Feroci} M,
  {Gendreau} KC, {Gevin} O, {Hernanz} M, {Jenke} P, et~al. (2018) {STROBE-X: a
  probe-class mission for x-ray spectroscopy and timing on timescales from
  microseconds to years}. In: {den Herder} JWA, {Nikzad} S, {Nakazawa} K (eds)
  Space Telescopes and Instrumentation 2018: Ultraviolet to Gamma Ray. Society
  of Photo-Optical Instrumentation Engineers (SPIE) Conference Series, vol
  10699. p 1069919. \doi{10.1117/12.2312257}.
  {\href{https://arxiv.org/abs/1807.01179}{{arXiv:1807.01179}}} {[astro-ph.IM]}

\bibitem[{{Ray} et~al.(2019){Ray}, {Arzoumanian}, {Ballantyne}, {Bozzo},
  {Brandt}, {Brenneman}, {Chakrabarty}, {Christophersen}, {DeRosa}, {Feroci},
  {Gendreau}, {Goldstein}, {Hartmann}, and {et al.}}]{2019arXiv190303035R}
{Ray} PS, {Arzoumanian} Z, {Ballantyne} D, {Bozzo} E, {Brandt} S, {Brenneman}
  L, {Chakrabarty} D, {Christophersen} M, {DeRosa} A, {Feroci} M, et~al. (2019)
  {STROBE-X: X-ray Timing and Spectroscopy on Dynamical Timescales from
  Microseconds to Years}. arXiv e-prints arXiv:1903.03035.
  {\href{https://arxiv.org/abs/1903.03035}{{arXiv:1903.03035}}} {[astro-ph.IM]}

\bibitem[{{Razzano} and {Cuoco}(2018)}]{2018CQGra..35i5016R}
{Razzano} M, {Cuoco} E (2018) {Image-based deep learning for classification of
  noise transients in gravitational wave detectors}. Classical and Quantum
  Gravity 35(9):095016. \doi{10.1088/1361-6382/aab793}.
  {\href{https://arxiv.org/abs/1803.09933}{{arXiv:1803.09933}}} {[gr-qc]}

\bibitem[{{Rebassa-Mansergas} et~al.(2007){Rebassa-Mansergas}, {G{\"a}nsicke},
  {Rodr{\'\i}guez-Gil}, {Schreiber}, and {Koester}}]{2007MNRAS.382.1377R}
{Rebassa-Mansergas} A, {G{\"a}nsicke} BT, {Rodr{\'\i}guez-Gil} P, {Schreiber}
  MR, {Koester} D (2007) {Post-common-envelope binaries from SDSS - I. 101
  white dwarf main-sequence binaries with multiple Sloan Digital Sky Survey
  spectroscopy}. \mnras 382(4):1377--1393.
  \doi{10.1111/j.1365-2966.2007.12288.x}.
  {\href{https://arxiv.org/abs/0707.4107}{{arXiv:0707.4107}}} {[astro-ph]}

\bibitem[{{Rebassa-Mansergas} et~al.(2019){Rebassa-Mansergas}, {Toonen},
  {Korol}, and {Torres}}]{2019MNRAS.482.3656R}
{Rebassa-Mansergas} A, {Toonen} S, {Korol} V, {Torres} S (2019) {Where are the
  double-degenerate progenitors of Type Ia supernovae?} \mnras
  482(3):3656--3668. \doi{10.1093/mnras/sty2965}.
  {\href{https://arxiv.org/abs/1809.07158}{{arXiv:1809.07158}}} {[astro-ph.SR]}

\bibitem[{{Redmount} and {Rees}(1989)}]{1989ComAp..14..165R}
{Redmount} IH, {Rees} MJ (1989) {Gravitational-radiation rocket effects and
  galactic structure.} Comments on Astrophysics 14:165

\bibitem[{{Rees}(1988)}]{1988Natur.333..523R}
{Rees} MJ (1988) {Tidal disruption of stars by black holes of
  {}10$^{6}$-{}10$^{8}$ solar masses in nearby galaxies}. \nat
  333(6173):523--528. \doi{10.1038/333523a0}

\bibitem[{{Regan} et~al.(2017){Regan}, {Visbal}, {Wise}, , {Haiman},
  {Johansson}, and {Bryan}}]{2017NatAs...1E..75R}
{Regan} JA, {Visbal} E, {Wise} JH, , {Haiman} Z, {Johansson} PH, {Bryan} GL
  (2017) {Rapid formation of massive black holes in close proximity to
  embryonic protogalaxies}. Nature Astronomy 1:0075.
  \doi{10.1038/s41550-017-0075}

\bibitem[{{Regan} et~al.(2019){Regan}, {Downes}, {Volonteri}, {Beckmann},
  {Lupi}, {Trebitsch}, and {Dubois}}]{2019MNRAS.486.3892R}
{Regan} JA, {Downes} TP, {Volonteri} M, {Beckmann} R, {Lupi} A, {Trebitsch} M,
  {Dubois} Y (2019) {Super-Eddington accretion and feedback from the first
  massive seed black holes}. \mnras 486(3):3892--3906.
  \doi{10.1093/mnras/stz1045}.
  {\href{https://arxiv.org/abs/1811.04953}{{arXiv:1811.04953}}} {[astro-ph.GA]}

\bibitem[{{Regan} et~al.(2020{\natexlab{a}}){Regan}, {Haiman}, {Wise},
  {O'Shea}, and {Norman}}]{2020OJAp....3E...9R}
{Regan} JA, {Haiman} Z, {Wise} JH, {O'Shea} BW, {Norman} ML
  (2020{\natexlab{a}}) {Massive Star Formation in Metal-Enriched Haloes at High
  Redshift}. The Open Journal of Astrophysics 3:E9.
  \doi{10.21105/astro.2006.14625}.
  {\href{https://arxiv.org/abs/2006.14625}{{arXiv:2006.14625}}} {[astro-ph.GA]}

\bibitem[{{Regan} et~al.(2020{\natexlab{b}}){Regan}, {Wise}, {O'Shea}, and
  {Norman}}]{2020MNRAS.492.3021R}
{Regan} JA, {Wise} JH, {O'Shea} BW, {Norman} ML (2020{\natexlab{b}}) {The
  emergence of the first star-free atomic cooling haloes in the Universe}.
  \mnras 492(2):3021--3031. \doi{10.1093/mnras/staa035}.
  {\href{https://arxiv.org/abs/1908.02823}{{arXiv:1908.02823}}} {[astro-ph.GA]}

\bibitem[{{Regimbau}(2011)}]{2011RAA....11..369R}
{Regimbau} T (2011) {The astrophysical gravitational wave stochastic
  background}. Research in Astronomy and Astrophysics 11(4):369--390.
  \doi{10.1088/1674-4527/11/4/001}.
  {\href{https://arxiv.org/abs/1101.2762}{{arXiv:1101.2762}}} {[astro-ph.CO]}

\bibitem[{{Reichardt} et~al.(2019){Reichardt}, {De Marco}, {Iaconi}, {Tout},
  and {Price}}]{2019MNRAS.484..631R}
{Reichardt} TA, {De Marco} O, {Iaconi} R, {Tout} CA, {Price} DJ (2019)
  {Extending common envelope simulations from Roche lobe overflow to the
  nebular phase}. \mnras 484(1):631--647. \doi{10.1093/mnras/sty3485}.
  {\href{https://arxiv.org/abs/1809.02297}{{arXiv:1809.02297}}} {[astro-ph.SR]}

\bibitem[{{Reid} and {Brunthaler}(2004)}]{2004ApJ...616..872R}
{Reid} MJ, {Brunthaler} A (2004) {The Proper Motion of Sagittarius A*. II. The
  Mass of Sagittarius A*}. \apj 616(2):872--884. \doi{10.1086/424960}.
  {\href{https://arxiv.org/abs/astro-ph/0408107}{{arXiv:astro-ph/0408107}}}
  {[astro-ph]}

\bibitem[{{Reines} and {Volonteri}(2015)}]{2015ApJ...813...82R}
{Reines} AE, {Volonteri} M (2015) {Relations between Central Black Hole Mass
  and Total Galaxy Stellar Mass in the Local Universe}. \apj 813(2):82.
  \doi{10.1088/0004-637X/813/2/82}.
  {\href{https://arxiv.org/abs/1508.06274}{{arXiv:1508.06274}}} {[astro-ph.GA]}

\bibitem[{{Reines} et~al.(2013){Reines}, {Greene}, and
  {Geha}}]{2013ApJ...775..116R}
{Reines} AE, {Greene} JE, {Geha} M (2013) {Dwarf Galaxies with Optical
  Signatures of Active Massive Black Holes}. \apj 775:116.
  \doi{10.1088/0004-637X/775/2/116}.
  {\href{https://arxiv.org/abs/1308.0328}{{arXiv:1308.0328}}} {[astro-ph.CO]}

\bibitem[{{Reines} et~al.(2020){Reines}, {Condon}, {Darling}, and
  {Greene}}]{2020ApJ...888...36R}
{Reines} AE, {Condon} JJ, {Darling} J, {Greene} JE (2020) {A New Sample of
  (Wandering) Massive Black Holes in Dwarf Galaxies from High-resolution Radio
  Observations}. \apj 888(1):36. \doi{10.3847/1538-4357/ab4999}.
  {\href{https://arxiv.org/abs/1909.04670}{{arXiv:1909.04670}}} {[astro-ph.GA]}

\bibitem[{{Reinoso} et~al.(2018){Reinoso}, {Schleicher}, {Fellhauer},
  {Klessen}, and {Boekholt}}]{2018A&A...614A..14R}
{Reinoso} B, {Schleicher} DRG, {Fellhauer} M, {Klessen} RS, {Boekholt} TCN
  (2018) {Collisions in primordial star clusters. Formation pathway for
  intermediate mass black holes}. \aap 614:A14.
  \doi{10.1051/0004-6361/201732224}.
  {\href{https://arxiv.org/abs/1801.05891}{{arXiv:1801.05891}}} {[astro-ph.GA]}

\bibitem[{{Remillard} and {McClintock}(2006)}]{2006ARA&A..44...49R}
{Remillard} RA, {McClintock} JE (2006) {X-Ray Properties of Black-Hole
  Binaries}. \araa 44(1):49--92. \doi{10.1146/annurev.astro.44.051905.092532}.
  {\href{https://arxiv.org/abs/astro-ph/0606352}{{arXiv:astro-ph/0606352}}}
  {[astro-ph]}

\bibitem[{{Remmen} and {Wu}(2013)}]{2013MNRAS.430.1940R}
{Remmen} GN, {Wu} K (2013) {Complex orbital dynamics of a double neutron star
  system revolving around a massive black hole}. \mnras 430(3):1940--1955.
  \doi{10.1093/mnras/stt023}.
  {\href{https://arxiv.org/abs/1301.2836}{{arXiv:1301.2836}}} {[astro-ph.HE]}

\bibitem[{{Remus} et~al.(2012){Remus}, {Mathis}, and
  {Zahn}}]{2012A&A...544A.132R}
{Remus} F, {Mathis} S, {Zahn} JP (2012) {The equilibrium tide in stars and
  giant planets. I. The coplanar case}. \aap 544:A132.
  \doi{10.1051/0004-6361/201118160}.
  {\href{https://arxiv.org/abs/1205.3536}{{arXiv:1205.3536}}} {[astro-ph.SR]}

\bibitem[{{Renzo} et~al.(2021){Renzo}, {Callister}, {Chatziioannou}, {van Son},
  {Mingarelli}, {Cantiello}, {Ford}, {McKernan}, and
  {Ashton}}]{2021arXiv210200078R}
{Renzo} M, {Callister} T, {Chatziioannou} K, {van Son} LAC, {Mingarelli} CMF,
  {Cantiello} M, {Ford} KES, {McKernan} B, {Ashton} G (2021) {Prospects of
  gravitational-waves detections from common-envelope evolution with LISA}.
  arXiv e-prints arXiv:2102.00078.
  {\href{https://arxiv.org/abs/2102.00078}{{arXiv:2102.00078}}} {[astro-ph.SR]}

\bibitem[{{Reynolds}(2014)}]{2014SSRv..183..277R}
{Reynolds} CS (2014) {Measuring Black Hole Spin Using X-Ray Reflection
  Spectroscopy}. \ssr 183(1-4):277--294. \doi{10.1007/s11214-013-0006-6}.
  {\href{https://arxiv.org/abs/1302.3260}{{arXiv:1302.3260}}} {[astro-ph.HE]}

\bibitem[{{Reynolds}(2019)}]{2019NatAs...3...41R}
{Reynolds} CS (2019) {Observing black holes spin}. Nature Astronomy 3:41--47.
  \doi{10.1038/s41550-018-0665-z}.
  {\href{https://arxiv.org/abs/1903.11704}{{arXiv:1903.11704}}} {[astro-ph.HE]}

\bibitem[{{Reynolds} and {Nowak}(2003)}]{2003PhR...377..389R}
{Reynolds} CS, {Nowak} MA (2003) {Fluorescent iron lines as a probe of
  astrophysical black hole systems}. \physrep 377(6):389--466.
  \doi{10.1016/S0370-1573(02)00584-7}.
  {\href{https://arxiv.org/abs/astro-ph/0212065}{{arXiv:astro-ph/0212065}}}
  {[astro-ph]}

\bibitem[{{Rezzolla} et~al.(2008){Rezzolla}, {Barausse}, {Dorband}, {Pollney},
  {Reisswig}, {Seiler}, and {Husa}}]{2008PhRvD..78d4002R}
{Rezzolla} L, {Barausse} E, {Dorband} EN, {Pollney} D, {Reisswig} C, {Seiler}
  J, {Husa} S (2008) {Final spin from the coalescence of two black holes}. \prd
  78(4):044002. \doi{10.1103/PhysRevD.78.044002}.
  {\href{https://arxiv.org/abs/0712.3541}{{arXiv:0712.3541}}} {[gr-qc]}

\bibitem[{{Ricarte} and {Natarajan}(2018{\natexlab{a}})}]{2018MNRAS.474.1995R}
{Ricarte} A, {Natarajan} P (2018{\natexlab{a}}) {Exploring SMBH assembly with
  semi-analytic modelling}. \mnras 474(2):1995--2011.
  \doi{10.1093/mnras/stx2851}.
  {\href{https://arxiv.org/abs/1710.11532}{{arXiv:1710.11532}}} {[astro-ph.HE]}

\bibitem[{{Ricarte} and {Natarajan}(2018{\natexlab{b}})}]{2018MNRAS.481.3278R}
{Ricarte} A, {Natarajan} P (2018{\natexlab{b}}) {The observational signatures
  of supermassive black hole seeds}. \mnras 481(3):3278--3292.
  \doi{10.1093/mnras/sty2448}.
  {\href{https://arxiv.org/abs/1809.01177}{{arXiv:1809.01177}}} {[astro-ph.GA]}

\bibitem[{{Ricarte} et~al.(2020){Ricarte}, {Tremmel}, {Natarajan}, and
  {Quinn}}]{2020ApJ...895L...8R}
{Ricarte} A, {Tremmel} M, {Natarajan} P, {Quinn} T (2020) {A Link between Ram
  Pressure Stripping and Active Galactic Nuclei}. \apjl 895(1):L8.
  \doi{10.3847/2041-8213/ab9022}.
  {\href{https://arxiv.org/abs/2003.05950}{{arXiv:2003.05950}}} {[astro-ph.GA]}

\bibitem[{{Ricker} and {Taam}(2012)}]{2012ApJ...746...74R}
{Ricker} PM, {Taam} RE (2012) {An AMR Study of the Common-envelope Phase of
  Binary Evolution}. \apj 746(1):74. \doi{10.1088/0004-637X/746/1/74}.
  {\href{https://arxiv.org/abs/1107.3889}{{arXiv:1107.3889}}} {[astro-ph.SR]}

\bibitem[{{Riley} et~al.(2019){Riley}, {Watts}, {Bogdanov}, {Ray}, {Ludlam},
  {Guillot}, {Arzoumanian}, {Baker}, {Bilous}, {Chakrabarty}, {Gendreau},
  {Harding}, {Ho}, {Lattimer}, {Morsink}, and
  {Strohmayer}}]{2019ApJ...887L..21R}
{Riley} TE, {Watts} AL, {Bogdanov} S, {Ray} PS, {Ludlam} RM, {Guillot} S,
  {Arzoumanian} Z, {Baker} CL, {Bilous} AV, {Chakrabarty} D, et~al. (2019) {A
  NICER View of PSR J0030+0451: Millisecond Pulsar Parameter Estimation}. \apjl
  887(1):L21. \doi{10.3847/2041-8213/ab481c}.
  {\href{https://arxiv.org/abs/1912.05702}{{arXiv:1912.05702}}} {[astro-ph.HE]}

\bibitem[{{Rivera Sandoval} et~al.(2018){Rivera Sandoval}, {van den Berg},
  {Heinke}, {Cohn}, {Lugger}, {Anderson}, {Cool}, {Edmonds}, {Wijnands},
  {Ivanova}, and {Grindlay}}]{2018MNRAS.475.4841R}
{Rivera Sandoval} LE, {van den Berg} M, {Heinke} CO, {Cohn} HN, {Lugger} PM,
  {Anderson} J, {Cool} AM, {Edmonds} PD, {Wijnands} R, {Ivanova} N, et~al.
  (2018) {New cataclysmic variables and other exotic binaries in the globular
  cluster 47 Tucanae$^{*}$}. \mnras 475(4):4841--4867.
  \doi{10.1093/mnras/sty058}.
  {\href{https://arxiv.org/abs/1705.07100}{{arXiv:1705.07100}}} {[astro-ph.SR]}

\bibitem[{{Rizzuto} et~al.(2020){Rizzuto}, {Naab}, {Spurzem}, {Giersz},
  {Ostriker}, {Stone}, {Wang}, {Berczik}, and {Rampp}}]{2020arXiv200809571R}
{Rizzuto} FP, {Naab} T, {Spurzem} R, {Giersz} M, {Ostriker} JP, {Stone} NC,
  {Wang} L, {Berczik} P, {Rampp} M (2020) {Intermediate Mass Black Hole
  Formation in compact Young Massive Star Clusters}. arXiv e-prints
  arXiv:2008.09571.
  {\href{https://arxiv.org/abs/2008.09571}{{arXiv:2008.09571}}} {[astro-ph.GA]}

\bibitem[{{Rizzuto} et~al.(2021){Rizzuto}, {Naab}, {Spurzem}, {Giersz},
  {Ostriker}, {Stone}, {Wang}, {Berczik}, and {Rampp}}]{2021MNRAS.501.5257R}
{Rizzuto} FP, {Naab} T, {Spurzem} R, {Giersz} M, {Ostriker} JP, {Stone} NC,
  {Wang} L, {Berczik} P, {Rampp} M (2021) {Intermediate mass black hole
  formation in compact young massive star clusters}. \mnras 501(4):5257--5273.
  \doi{10.1093/mnras/staa3634}.
  {\href{https://arxiv.org/abs/2008.09571}{{arXiv:2008.09571}}} {[astro-ph.GA]}

\bibitem[{{Rizzuto} et~al.(2022){Rizzuto}, {Naab}, {Spurzem}, {Arca-Sedda},
  {Giersz}, {Ostriker}, and {Banerjee}}]{2022MNRAS.512..884R}
{Rizzuto} FP, {Naab} T, {Spurzem} R, {Arca-Sedda} M, {Giersz} M, {Ostriker} JP,
  {Banerjee} S (2022) {Black hole mergers in compact star clusters and massive
  black hole formation beyond the mass gap}. \mnras 512(1):884--898.
  \doi{10.1093/mnras/stac231}.
  {\href{https://arxiv.org/abs/2108.11457}{{arXiv:2108.11457}}} {[astro-ph.GA]}

\bibitem[{{Roberts} et~al.(2015){Roberts}, {Saripalli}, and
  {Subrahmanyan}}]{2015ApJ...810L...6R}
{Roberts} DH, {Saripalli} L, {Subrahmanyan} R (2015) {The Abundance of X-shaped
  Radio Sources: Implications for the Gravitational Wave Background}. \apjl
  810(1):L6. \doi{10.1088/2041-8205/810/1/L6}.
  {\href{https://arxiv.org/abs/1503.02021}{{arXiv:1503.02021}}} {[astro-ph.GA]}

\bibitem[{{Roberts}(2013)}]{2013IAUS..291..127R}
{Roberts} MSE (2013) {Surrounded by spiders! New black widows and redbacks in
  the Galactic field}. In: {van Leeuwen} J (ed) Neutron Stars and Pulsars:
  Challenges and Opportunities after 80 years. IAU Symposium, vol 291. pp
  127--132. \doi{10.1017/S174392131202337X}.
  {\href{https://arxiv.org/abs/1210.6903}{{arXiv:1210.6903}}} {[astro-ph.HE]}

\bibitem[{{Robinson} et~al.(2008){Robinson}, {Romano}, and
  {Vecchio}}]{2008CQGra..25r4019R}
{Robinson} EL, {Romano} JD, {Vecchio} A (2008) {Search for a stochastic
  gravitational-wave signal in the second round of the Mock LISA Data
  Challenges}. Classical and Quantum Gravity 25(18):184019.
  \doi{10.1088/0264-9381/25/18/184019}.
  {\href{https://arxiv.org/abs/0804.4144}{{arXiv:0804.4144}}} {[gr-qc]}

\bibitem[{{Robson} et~al.(2018){Robson}, {Cornish}, {Tamanini}, and
  {Toonen}}]{2018PhRvD..98f4012R}
{Robson} T, {Cornish} NJ, {Tamanini} N, {Toonen} S (2018) {Detecting
  hierarchical stellar systems with LISA}. \prd 98(6):064012.
  \doi{10.1103/PhysRevD.98.064012}.
  {\href{https://arxiv.org/abs/1806.00500}{{arXiv:1806.00500}}} {[gr-qc]}

\bibitem[{{Robson} et~al.(2019){Robson}, {Cornish}, and
  {Liu}}]{2019CQGra..36j5011R}
{Robson} T, {Cornish} NJ, {Liu} C (2019) {The construction and use of LISA
  sensitivity curves}. Classical and Quantum Gravity 36(10):105011.
  \doi{10.1088/1361-6382/ab1101}.
  {\href{https://arxiv.org/abs/1803.01944}{{arXiv:1803.01944}}} {[astro-ph.HE]}

\bibitem[{{Rodriguez} et~al.(2006){Rodriguez}, {Taylor}, {Zavala}, {Peck},
  {Pollack}, and {Romani}}]{2006ApJ...646...49R}
{Rodriguez} C, {Taylor} GB, {Zavala} RT, {Peck} AB, {Pollack} LK, {Romani} RW
  (2006) {A Compact Supermassive Binary Black Hole System}. \apj 646(1):49--60.
  \doi{10.1086/504825}.
  {\href{https://arxiv.org/abs/astro-ph/0604042}{{arXiv:astro-ph/0604042}}}
  {[astro-ph]}

\bibitem[{{Rodriguez} and {Antonini}(2018)}]{2018ApJ...863....7R}
{Rodriguez} CL, {Antonini} F (2018) {A Triple Origin for the Heavy and Low-spin
  Binary Black Holes Detected by LIGO/VIRGO}. \apj 863(1):7.
  \doi{10.3847/1538-4357/aacea4}.
  {\href{https://arxiv.org/abs/1805.08212}{{arXiv:1805.08212}}} {[astro-ph.HE]}

\bibitem[{{Rodriguez} et~al.(2019){Rodriguez}, {Zevin}, {Amaro-Seoane},
  {Chatterjee}, {Kremer}, {Rasio}, and {Ye}}]{2019PhRvD.100d3027R}
{Rodriguez} CL, {Zevin} M, {Amaro-Seoane} P, {Chatterjee} S, {Kremer} K,
  {Rasio} FA, {Ye} CS (2019) {Black holes: The next
  generation{\textemdash}repeated mergers in dense star clusters and their
  gravitational-wave properties}. \prd 100(4):043027.
  \doi{10.1103/PhysRevD.100.043027}.
  {\href{https://arxiv.org/abs/1906.10260}{{arXiv:1906.10260}}} {[astro-ph.HE]}

\bibitem[{{Roebber} et~al.(2020){Roebber}, {Buscicchio}, {Vecchio}, {Moore},
  {Klein}, {Korol}, {Toonen}, {Gerosa}, {Goldstein}, {Gaebel}, and
  {Woods}}]{2020ApJ...894L..15R}
{Roebber} E, {Buscicchio} R, {Vecchio} A, {Moore} CJ, {Klein} A, {Korol} V,
  {Toonen} S, {Gerosa} D, {Goldstein} J, {Gaebel} SM, et~al. (2020) {Milky Way
  Satellites Shining Bright in Gravitational Waves}. \apjl 894(2):L15.
  \doi{10.3847/2041-8213/ab8ac9}.
  {\href{https://arxiv.org/abs/2002.10465}{{arXiv:2002.10465}}} {[astro-ph.GA]}

\bibitem[{{Roedig} and {Sesana}(2014)}]{2014MNRAS.439.3476R}
{Roedig} C, {Sesana} A (2014) {Migration of massive black hole binaries in
  self-gravitating discs: retrograde versus prograde}. \mnras
  439(4):3476--3489. \doi{10.1093/mnras/stu194}.
  {\href{https://arxiv.org/abs/1307.6283}{{arXiv:1307.6283}}} {[astro-ph.HE]}

\bibitem[{{Roedig} et~al.(2011){Roedig}, {Dotti}, {Sesana}, {Cuadra}, and
  {Colpi}}]{2011MNRAS.415.3033R}
{Roedig} C, {Dotti} M, {Sesana} A, {Cuadra} J, {Colpi} M (2011) {Limiting
  eccentricity of subparsec massive black hole binaries surrounded by
  self-gravitating gas discs}. \mnras 415(4):3033--3041.
  \doi{10.1111/j.1365-2966.2011.18927.x}.
  {\href{https://arxiv.org/abs/1104.3868}{{arXiv:1104.3868}}} {[astro-ph.CO]}

\bibitem[{{Roedig} et~al.(2012){Roedig}, {Sesana}, {Dotti}, {Cuadra},
  {Amaro-Seoane}, and {Haardt}}]{2012A&A...545A.127R}
{Roedig} C, {Sesana} A, {Dotti} M, {Cuadra} J, {Amaro-Seoane} P, {Haardt} F
  (2012) {Evolution of binary black holes in self gravitating discs. Dissecting
  the torques}. \aap 545:A127. \doi{10.1051/0004-6361/201219986}.
  {\href{https://arxiv.org/abs/1202.6063}{{arXiv:1202.6063}}} {[astro-ph.CO]}

\bibitem[{{Roedig} et~al.(2014){Roedig}, {Krolik}, and
  {Miller}}]{2014ApJ...785..115R}
{Roedig} C, {Krolik} JH, {Miller} MC (2014) Observational signatures of binary
  supermassive black holes. \apj 785(2):115. \doi{10.1088/0004-637X/785/2/115}.
  {\href{https://arxiv.org/abs/1402.7098}{{arXiv:1402.7098}}} {[astro-ph.HE]}

\bibitem[{{Roelofs} et~al.(2006){Roelofs}, {Groot}, {Nelemans}, {Marsh}, and
  {Steeghs}}]{2006MNRAS.371.1231R}
{Roelofs} GHA, {Groot} PJ, {Nelemans} G, {Marsh} TR, {Steeghs} D (2006)
  {Kinematics of the ultracompact helium accretor AM Canum Venaticorum}. \mnras
  371(3):1231--1242. \doi{10.1111/j.1365-2966.2006.10718.x}.
  {\href{https://arxiv.org/abs/astro-ph/0606327}{{arXiv:astro-ph/0606327}}}
  {[astro-ph]}

\bibitem[{{Roelofs} et~al.(2007{\natexlab{a}}){Roelofs}, {Groot}, {Benedict},
  {McArthur}, {Steeghs}, {Morales-Rueda}, {Marsh}, and
  {Nelemans}}]{2007ApJ...666.1174R}
{Roelofs} GHA, {Groot} PJ, {Benedict} GF, {McArthur} BE, {Steeghs} D,
  {Morales-Rueda} L, {Marsh} TR, {Nelemans} G (2007{\natexlab{a}}) {Hubble
  Space Telescope Parallaxes of AM CVn Stars and Astrophysical Consequences}.
  \apj 666(2):1174--1188. \doi{10.1086/520491}.
  {\href{https://arxiv.org/abs/0705.3855}{{arXiv:0705.3855}}} {[astro-ph]}

\bibitem[{{Roelofs} et~al.(2007{\natexlab{b}}){Roelofs}, {Groot}, {Nelemans},
  {Marsh}, and {Steeghs}}]{2007MNRAS.379..176R}
{Roelofs} GHA, {Groot} PJ, {Nelemans} G, {Marsh} TR, {Steeghs} D
  (2007{\natexlab{b}}) {On the orbital periods of the AM CVn stars HP Librae
  and V803 Centauri}. \mnras 379(1):176--182.
  \doi{10.1111/j.1365-2966.2007.11931.x}.
  {\href{https://arxiv.org/abs/0705.0402}{{arXiv:0705.0402}}} {[astro-ph]}

\bibitem[{{Roelofs} et~al.(2007{\natexlab{c}}){Roelofs}, {Nelemans}, and
  {Groot}}]{2007MNRAS.382..685R}
{Roelofs} GHA, {Nelemans} G, {Groot} PJ (2007{\natexlab{c}}) {The population of
  AM CVn stars from the Sloan Digital Sky Survey}. \mnras 382(2):685--692.
  \doi{10.1111/j.1365-2966.2007.12451.x}.
  {\href{https://arxiv.org/abs/0709.2951}{{arXiv:0709.2951}}} {[astro-ph]}

\bibitem[{{Roelofs} et~al.(2010){Roelofs}, {Rau}, {Marsh}, {Steeghs}, {Groot},
  and {Nelemans}}]{2010ApJ...711L.138R}
{Roelofs} GHA, {Rau} A, {Marsh} TR, {Steeghs} D, {Groot} PJ, {Nelemans} G
  (2010) {Spectroscopic Evidence for a 5.4 Minute Orbital Period in HM Cancri}.
  \apjl 711(2):L138--L142. \doi{10.1088/2041-8205/711/2/L138}.
  {\href{https://arxiv.org/abs/1003.0658}{{arXiv:1003.0658}}} {[astro-ph.SR]}

\bibitem[{{Romano} and {Cornish}(2017)}]{2017LRR....20....2R}
{Romano} JD, {Cornish} NJ (2017) {Detection methods for stochastic
  gravitational-wave backgrounds: a unified treatment}. Living Reviews in
  Relativity 20(1):2. \doi{10.1007/s41114-017-0004-1}.
  {\href{https://arxiv.org/abs/1608.06889}{{arXiv:1608.06889}}} {[gr-qc]}

\bibitem[{{Romano-D{\'\i}az} et~al.(2008){Romano-D{\'\i}az}, {Shlosman},
  {Heller}, and {Hoffman}}]{2008ApJ...687L..13R}
{Romano-D{\'\i}az} E, {Shlosman} I, {Heller} C, {Hoffman} Y (2008) {Disk
  Evolution and Bar Triggering Driven by Interactions with Dark Matter
  Substructure}. \apjl 687(1):L13. \doi{10.1086/593168}.
  {\href{https://arxiv.org/abs/0809.2785}{{arXiv:0809.2785}}} {[astro-ph]}

\bibitem[{{Romero-Shaw} et~al.(2019){Romero-Shaw}, {Lasky}, and
  {Thrane}}]{RomeroShaw:2019}
{Romero-Shaw} IM, {Lasky} PD, {Thrane} E (2019) {Searching for eccentricity:
  signatures of dynamical formation in the first gravitational-wave transient
  catalogue of LIGO and Virgo}. \mnras 490(4):5210--5216.
  \doi{10.1093/mnras/stz2996}.
  {\href{https://arxiv.org/abs/1909.05466}{{arXiv:1909.05466}}} {[astro-ph.HE]}

\bibitem[{{Rosado} et~al.(2015){Rosado}, {Sesana}, and
  {Gair}}]{2015MNRAS.451.2417R}
{Rosado} PA, {Sesana} A, {Gair} J (2015) {Expected properties of the first
  gravitational wave signal detected with pulsar timing arrays}. \mnras
  451(3):2417--2433. \doi{10.1093/mnras/stv1098}.
  {\href{https://arxiv.org/abs/1503.04803}{{arXiv:1503.04803}}} {[astro-ph.HE]}

\bibitem[{{Rosas-Guevara} et~al.(2016){Rosas-Guevara}, {Bower}, {Schaye},
  {McAlpine}, {Dalla Vecchia}, {Frenk}, {Schaller}, and
  {Theuns}}]{2016MNRAS.462..190R}
{Rosas-Guevara} Y, {Bower} RG, {Schaye} J, {McAlpine} S, {Dalla Vecchia} C,
  {Frenk} CS, {Schaller} M, {Theuns} T (2016) {Supermassive black holes in the
  EAGLE Universe. Revealing the observables of their growth}. \mnras
  462(1):190--205. \doi{10.1093/mnras/stw1679}.
  {\href{https://arxiv.org/abs/1604.00020}{{arXiv:1604.00020}}} {[astro-ph.GA]}

\bibitem[{{Rosas-Guevara} et~al.(2015){Rosas-Guevara}, {Bower}, {Schaye},
  {Furlong}, {Frenk}, {Booth}, {Crain}, {Dalla Vecchia}, {Schaller}, and
  {Theuns}}]{2015MNRAS.454.1038R}
{Rosas-Guevara} YM, {Bower} RG, {Schaye} J, {Furlong} M, {Frenk} CS, {Booth}
  CM, {Crain} RA, {Dalla Vecchia} C, {Schaller} M, {Theuns} T (2015) {The
  impact of angular momentum on black hole accretion rates in simulations of
  galaxy formation}. \mnras 454(1):1038--1057. \doi{10.1093/mnras/stv2056}.
  {\href{https://arxiv.org/abs/1312.0598}{{arXiv:1312.0598}}} {[astro-ph.CO]}

\bibitem[{{Rosenthal}(2006)}]{2006PhRvD..73d4034R}
{Rosenthal} E (2006) {Construction of the second-order gravitational
  perturbations produced by a compact object}. \prd 73(4):044034.
  \doi{10.1103/PhysRevD.73.044034}.
  {\href{https://arxiv.org/abs/gr-qc/0602066}{{arXiv:gr-qc/0602066}}} {[gr-qc]}

\bibitem[{{Ross} et~al.(2009){Ross}, {Shen}, {Strauss}, {Vanden Berk},
  {Connolly}, {Richards}, {Schneider}, {Weinberg}, {Hall}, {Bahcall}, and
  {Brunner}}]{2009ApJ...697.1634R}
{Ross} NP, {Shen} Y, {Strauss} MA, {Vanden Berk} DE, {Connolly} AJ, {Richards}
  GT, {Schneider} DP, {Weinberg} DH, {Hall} PB, {Bahcall} NA, et~al. (2009)
  {Clustering of Low-redshift (z <= 2.2) Quasars from the Sloan Digital Sky
  Survey}. \apj 697(2):1634--1655. \doi{10.1088/0004-637X/697/2/1634}.
  {\href{https://arxiv.org/abs/0903.3230}{{arXiv:0903.3230}}} {[astro-ph.CO]}

\bibitem[{{Rossi} et~al.(2010){Rossi}, {Lodato}, {Armitage}, {Pringle}, and
  {King}}]{2010MNRAS.401.2021R}
{Rossi} EM, {Lodato} G, {Armitage} PJ, {Pringle} JE, {King} AR (2010) {Black
  hole mergers: the first light}. \mnras 401(3):2021--2035.
  \doi{10.1111/j.1365-2966.2009.15802.x}.
  {\href{https://arxiv.org/abs/0910.0002}{{arXiv:0910.0002}}} {[astro-ph.HE]}

\bibitem[{{Rossi} et~al.(2020){Rossi}, {Stone}, {Law-Smith}, {MacLeod},
  {Lodato}, {Dai}, and {Mand el}}]{2020arXiv200512528R}
{Rossi} EM, {Stone} NC, {Law-Smith} JAP, {MacLeod} M, {Lodato} G, {Dai} JL,
  {Mand el} I (2020) {The Process of Stellar Tidal Disruption by Supermassive
  Black Holes. The first pericenter passage}. arXiv e-prints arXiv:2005.12528.
  {\href{https://arxiv.org/abs/2005.12528}{{arXiv:2005.12528}}} {[astro-ph.HE]}

\bibitem[{{Rosswog} et~al.(2008){Rosswog}, {Ramirez-Ruiz}, and
  {Hix}}]{2008ApJ...679.1385R}
{Rosswog} S, {Ramirez-Ruiz} E, {Hix} WR (2008) {Atypical Thermonuclear
  Supernovae from Tidally Crushed White Dwarfs}. \apj 679(2):1385--1389.
  \doi{10.1086/528738}.
  {\href{https://arxiv.org/abs/0712.2513}{{arXiv:0712.2513}}} {[astro-ph]}

\bibitem[{{Rosswog} et~al.(2009{\natexlab{a}}){Rosswog}, {Kasen}, {Guillochon},
  and {Ramirez-Ruiz}}]{2009ApJ...705L.128R}
{Rosswog} S, {Kasen} D, {Guillochon} J, {Ramirez-Ruiz} E (2009{\natexlab{a}})
  {Collisions of White Dwarfs as a New Progenitor Channel for Type Ia
  Supernovae}. \apjl 705(2):L128--L132. \doi{10.1088/0004-637X/705/2/L128}.
  {\href{https://arxiv.org/abs/0907.3196}{{arXiv:0907.3196}}} {[astro-ph.HE]}

\bibitem[{{Rosswog} et~al.(2009{\natexlab{b}}){Rosswog}, {Ramirez-Ruiz}, and
  {Hix}}]{2009ApJ...695..404R}
{Rosswog} S, {Ramirez-Ruiz} E, {Hix} WR (2009{\natexlab{b}}) {Tidal Disruption
  and Ignition of White Dwarfs by Moderately Massive Black Holes}. \apj
  695(1):404--419. \doi{10.1088/0004-637X/695/1/404}.
  {\href{https://arxiv.org/abs/0808.2143}{{arXiv:0808.2143}}} {[astro-ph]}

\bibitem[{{Roupas} et~al.(2017){Roupas}, {Kocsis}, and
  {Tremaine}}]{2017ApJ...842...90R}
{Roupas} Z, {Kocsis} B, {Tremaine} S (2017) {Isotropic-Nematic Phase
  Transitions in Gravitational Systems}. \apj 842(2):90.
  \doi{10.3847/1538-4357/aa7141}.
  {\href{https://arxiv.org/abs/1701.03271}{{arXiv:1701.03271}}} {[astro-ph.GA]}

\bibitem[{{Ro{\v{s}}kar} et~al.(2015){Ro{\v{s}}kar}, {Fiacconi}, {Mayer},
  {Kazantzidis}, {Quinn}, and {Wadsley}}]{2015MNRAS.449..494R}
{Ro{\v{s}}kar} R, {Fiacconi} D, {Mayer} L, {Kazantzidis} S, {Quinn} TR,
  {Wadsley} J (2015) {Orbital decay of supermassive black hole binaries in
  clumpy multiphase merger remnants}. \mnras 449(1):494--505.
  \doi{10.1093/mnras/stv312}.
  {\href{https://arxiv.org/abs/1406.4505}{{arXiv:1406.4505}}} {[astro-ph.GA]}

\bibitem[{{Ruan} et~al.(2018){Ruan}, {Guo}, {Cai}, and
  {Zhang}}]{2018arXiv180709495R}
{Ruan} WH, {Guo} ZK, {Cai} RG, {Zhang} YZ (2018) {Taiji Program:
  Gravitational-Wave Sources}. arXiv e-prints arXiv:1807.09495.
  {\href{https://arxiv.org/abs/1807.09495}{{arXiv:1807.09495}}} {[gr-qc]}

\bibitem[{{Ruan} et~al.(2021){Ruan}, {Liu}, {Guo}, {Wu}, and
  {Cai}}]{2021Resea202114164R}
{Ruan} WH, {Liu} C, {Guo} ZK, {Wu} YL, {Cai} RG (2021) {The LISA-Taiji Network:
  Precision Localization of Coalescing Massive Black Hole Binaries}. Research
  2021:6014164. \doi{10.34133/2021/6014164}

\bibitem[{{Rubbo} et~al.(2006){Rubbo}, {Holley-Bockelmann}, and
  {Finn}}]{2006ApJ...649L..25R}
{Rubbo} LJ, {Holley-Bockelmann} K, {Finn} LS (2006) {Event Rate for Extreme
  Mass Ratio Burst Signals in the Laser Interferometer Space Antenna Band}.
  \apjl 649(1):L25--L28. \doi{10.1086/508326}

\bibitem[{{Rueda} et~al.(2019){Rueda}, {Ruffini}, {Wang}, {Bianco},
  {Blanco-Iglesias}, {Karlica}, {Lor{\'e}n-Aguilar}, {Moradi}, and
  {Sahakyan}}]{2019JCAP...03..044R}
{Rueda} JA, {Ruffini} R, {Wang} Y, {Bianco} CL, {Blanco-Iglesias} JM, {Karlica}
  M, {Lor{\'e}n-Aguilar} P, {Moradi} R, {Sahakyan} N (2019) {Electromagnetic
  emission of white dwarf binary mergers}. \jcap 2019(3):044.
  \doi{10.1088/1475-7516/2019/03/044}.
  {\href{https://arxiv.org/abs/1807.07905}{{arXiv:1807.07905}}} {[astro-ph.HE]}

\bibitem[{{Ruel} et~al.(2014){Ruel}, {Bazin}, {Bayliss}, {Brodwin}, {Foley},
  {Stalder}, {Aird}, {Armstrong}, {Ashby}, {Bautz}, {Benson}, {Bleem},
  {Bocquet}, {Carlstrom}, {Chang}, {Chapman}, {Cho}, {Clocchiatti}, {Crawford},
  {Crites}, {de Haan}, {Desai}, {Dobbs}, {Dudley}, {Forman}, {George},
  {Gladders}, {Gonzalez}, {Halverson}, {Harrington}, {High}, {Holder},
  {Holzapfel}, {Hrubes}, {Jones}, {Joy}, {Keisler}, {Knox}, {Lee}, {Leitch},
  {Liu}, {Lueker}, {Luong-Van}, {Mantz}, {Marrone}, {McDonald}, {McMahon},
  {Mehl}, {Meyer}, {Mocanu}, {Mohr}, {Montroy}, {Murray}, {Natoli},
  {Nurgaliev}, {Padin}, {Plagge}, {Pryke}, {Reichardt}, {Rest}, {Ruhl},
  {Saliwanchik}, {Saro}, {Sayre}, {Schaffer}, {Shaw}, {Shirokoff}, {Song},
  {{\v{S}}uhada}, {Spieler}, {Stanford}, {Staniszewski}, {Starsk}, {Story},
  {Stubbs}, {van Engelen}, {Vanderlinde}, {Vieira}, {Vikhlinin}, {Williamson},
  {Zahn}, and {Zenteno}}]{2014ApJ...792...45R}
{Ruel} J, {Bazin} G, {Bayliss} M, {Brodwin} M, {Foley} RJ, {Stalder} B, {Aird}
  KA, {Armstrong} R, {Ashby} MLN, {Bautz} M, et~al. (2014) {Optical
  Spectroscopy and Velocity Dispersions of Galaxy Clusters from the SPT-SZ
  Survey}. \apj 792(1):45. \doi{10.1088/0004-637X/792/1/45}.
  {\href{https://arxiv.org/abs/1311.4953}{{arXiv:1311.4953}}} {[astro-ph.CO]}

\bibitem[{{Ruiter}(2020)}]{2020IAUS..357....1R}
{Ruiter} AJ (2020) {Type Ia supernova sub-classes and progenitor origin}. IAU
  Symposium 357:1--15. \doi{10.1017/S1743921320000587}.
  {\href{https://arxiv.org/abs/2001.02947}{{arXiv:2001.02947}}} {[astro-ph.SR]}

\bibitem[{{Ruiter} et~al.(2009){Ruiter}, {Belczynski}, {Benacquista}, and
  {Holley-Bockelmann}}]{2009ApJ...693..383R}
{Ruiter} AJ, {Belczynski} K, {Benacquista} M, {Holley-Bockelmann} K (2009) {The
  Contribution of Halo White Dwarf Binaries to the Laser Interferometer Space
  Antenna Signal}. \apj 693(1):383--387. \doi{10.1088/0004-637X/693/1/383}

\bibitem[{{Ruiter} et~al.(2010){Ruiter}, {Belczynski}, {Benacquista}, {Larson},
  and {Williams}}]{2010ApJ...717.1006R}
{Ruiter} AJ, {Belczynski} K, {Benacquista} M, {Larson} SL, {Williams} G (2010)
  {The LISA Gravitational Wave Foreground: A Study of Double White Dwarfs}.
  \apj 717(2):1006--1021. \doi{10.1088/0004-637X/717/2/1006}.
  {\href{https://arxiv.org/abs/0705.3272}{{arXiv:0705.3272}}} {[astro-ph]}

\bibitem[{{Ruiter} et~al.(2019){Ruiter}, {Ferrario}, {Belczynski},
  {Seitenzahl}, {Crocker}, and {Karakas}}]{2019MNRAS.484..698R}
{Ruiter} AJ, {Ferrario} L, {Belczynski} K, {Seitenzahl} IR, {Crocker} RM,
  {Karakas} AI (2019) {On the formation of neutron stars via accretion-induced
  collapse in binaries}. \mnras 484(1):698--711. \doi{10.1093/mnras/stz001}.
  {\href{https://arxiv.org/abs/1802.02437}{{arXiv:1802.02437}}} {[astro-ph.SR]}

\bibitem[{{Runnoe} et~al.(2015){Runnoe}, {Eracleous}, {Mathes}, {Pennell},
  {Boroson}, {Sigur{\dh}sson}, {Bogdanovi{\'c}}, {Halpern}, and
  {Liu}}]{2015ApJS..221....7R}
{Runnoe} JC, {Eracleous} M, {Mathes} G, {Pennell} A, {Boroson} T,
  {Sigur{\dh}sson} S, {Bogdanovi{\'c}} T, {Halpern} JP, {Liu} J (2015) {A Large
  Systematic Search for Close Supermassive Binary and Rapidly Recoiling Black
  Holes. II. Continued Spectroscopic Monitoring and Optical Flux Variability}.
  \apjs 221(1):7. \doi{10.1088/0067-0049/221/1/7}.
  {\href{https://arxiv.org/abs/1509.02575}{{arXiv:1509.02575}}} {[astro-ph.GA]}

\bibitem[{{Runnoe} et~al.(2017){Runnoe}, {Eracleous}, {Pennell}, {Mathes},
  {Boroson}, {Sigur{\dh}sson}, {Bogdanovi{\'c}}, {Halpern}, {Liu}, and
  {Brown}}]{2017MNRAS.468.1683R}
{Runnoe} JC, {Eracleous} M, {Pennell} A, {Mathes} G, {Boroson} T,
  {Sigur{\dh}sson} S, {Bogdanovi{\'c}} T, {Halpern} JP, {Liu} J, {Brown} S
  (2017) {A large systematic search for close supermassive binary and rapidly
  recoiling black holes - III. Radial velocity variations}. \mnras
  468(2):1683--1702. \doi{10.1093/mnras/stx452}.
  {\href{https://arxiv.org/abs/1702.05465}{{arXiv:1702.05465}}} {[astro-ph.GA]}

\bibitem[{{Ryu} et~al.(2018){Ryu}, {Perna}, {Haiman}, {Ostriker}, and
  {Stone}}]{2018MNRAS.473.3410R}
{Ryu} T, {Perna} R, {Haiman} Z, {Ostriker} JP, {Stone} NC (2018) {Interactions
  between multiple supermassive black holes in galactic nuclei: a solution to
  the final parsec problem}. \mnras 473(3):3410--3433.
  \doi{10.1093/mnras/stx2524}.
  {\href{https://arxiv.org/abs/1709.06501}{{arXiv:1709.06501}}} {[astro-ph.GA]}

\bibitem[{{Ryu} et~al.(2020){Ryu}, {Krolik}, {Piran}, and
  {Noble}}]{2020ApJ...904...99R}
{Ryu} T, {Krolik} J, {Piran} T, {Noble} SC (2020) {Tidal Disruptions of
  Main-sequence Stars. II. Simulation Methodology and Stellar Mass Dependence
  of the Character of Full Tidal Disruptions}. \apj 904(2):99.
  \doi{10.3847/1538-4357/abb3cd}.
  {\href{https://arxiv.org/abs/2001.03502}{{arXiv:2001.03502}}} {[astro-ph.HE]}

\bibitem[{{Sadeghian} et~al.(2013){Sadeghian}, {Ferrer}, and
  {Will}}]{2013PhRvD..88f3522S}
{Sadeghian} L, {Ferrer} F, {Will} CM (2013) {Dark-matter distributions around
  massive black holes: A general relativistic analysis}. \prd 88(6):063522.
  \doi{10.1103/PhysRevD.88.063522}.
  {\href{https://arxiv.org/abs/1305.2619}{{arXiv:1305.2619}}} {[astro-ph.GA]}

\bibitem[{{Safarzadeh} et~al.(2019){Safarzadeh}, {Ramirez-Ruiz}, {Andrews},
  {Macias}, {Fragos}, and {Scannapieco}}]{2019ApJ...872..105S}
{Safarzadeh} M, {Ramirez-Ruiz} E, {Andrews} JJ, {Macias} P, {Fragos} T,
  {Scannapieco} E (2019) {r-process Enrichment of the Ultra-faint Dwarf
  Galaxies by Fast-merging Double-neutron Stars}. \apj 872(1):105.
  \doi{10.3847/1538-4357/aafe0e}.
  {\href{https://arxiv.org/abs/1810.04176}{{arXiv:1810.04176}}} {[astro-ph.HE]}

\bibitem[{{Safarzadeh} et~al.(2020{\natexlab{a}}){Safarzadeh}, {Hamers},
  {Loeb}, and {Berger}}]{2020ApJ...888L...3S}
{Safarzadeh} M, {Hamers} AS, {Loeb} A, {Berger} E (2020{\natexlab{a}})
  {Formation and Merging of Mass Gap Black Holes in Gravitational-wave Merger
  Events from Wide Hierarchical Quadruple Systems}. \apjl 888(1):L3.
  \doi{10.3847/2041-8213/ab5dc8}.
  {\href{https://arxiv.org/abs/1911.04495}{{arXiv:1911.04495}}} {[astro-ph.HE]}

\bibitem[{{Safarzadeh} et~al.(2020{\natexlab{b}}){Safarzadeh}, {Ramirez-Ruiz},
  and {Berger}}]{2020ApJ...900...13S}
{Safarzadeh} M, {Ramirez-Ruiz} E, {Berger} E (2020{\natexlab{b}}) {Does
  GW190425 Require an Alternative Formation Pathway than a Fast-merging
  Channel?} \apj 900(1):13. \doi{10.3847/1538-4357/aba596}.
  {\href{https://arxiv.org/abs/2001.04502}{{arXiv:2001.04502}}} {[astro-ph.HE]}

\bibitem[{{Saffer} et~al.(1998){Saffer}, {Livio}, and
  {Yungelson}}]{1998ApJ...502..394S}
{Saffer} RA, {Livio} M, {Yungelson} LR (1998) {Close Binary White Dwarf
  Systems: Numerous New Detections and Their Interpretation}. \apj
  502(1):394--407. \doi{10.1086/305907}.
  {\href{https://arxiv.org/abs/astro-ph/9802356}{{arXiv:astro-ph/9802356}}}
  {[astro-ph]}

\bibitem[{{Sago} and {Fujita}(2015)}]{2015PTEP.2015g3E03S}
{Sago} N, {Fujita} R (2015) {Calculation of radiation reaction effect on
  orbital parameters in Kerr spacetime}. Progress of Theoretical and
  Experimental Physics 2015(7):073E03. \doi{10.1093/ptep/ptv092}.
  {\href{https://arxiv.org/abs/1505.01600}{{arXiv:1505.01600}}} {[gr-qc]}

\bibitem[{{Sahu} et~al.(2019{\natexlab{a}}){Sahu}, {Graham}, and
  {Davis}}]{2019ApJ...876..155S}
{Sahu} N, {Graham} AW, {Davis} BL (2019{\natexlab{a}}) {Black Hole Mass Scaling
  Relations for Early-type Galaxies. I. M $_{BH}$-M $_{*,}$ $_{sph}$ and M
  $_{BH}$-M $_{*,gal}$}. \apj 876(2):155. \doi{10.3847/1538-4357/ab0f32}.
  {\href{https://arxiv.org/abs/1903.04738}{{arXiv:1903.04738}}} {[astro-ph.GA]}

\bibitem[{{Sahu} et~al.(2019{\natexlab{b}}){Sahu}, {Graham}, and
  {Davis}}]{2019ApJ...887...10S}
{Sahu} N, {Graham} AW, {Davis} BL (2019{\natexlab{b}}) {Revealing Hidden
  Substructures in the M $_{BH}$-{\ensuremath{\sigma}} Diagram, and Refining
  the Bend in the L-{\ensuremath{\sigma}} Relation}. \apj 887(1):10.
  \doi{10.3847/1538-4357/ab50b7}.
  {\href{https://arxiv.org/abs/1908.06838}{{arXiv:1908.06838}}} {[astro-ph.GA]}

\bibitem[{{Sahu} et~al.(2020){Sahu}, {Graham}, and
  {Davis}}]{2020ApJ...903...97S}
{Sahu} N, {Graham} AW, {Davis} BL (2020) {Defining the (Black Hole)-Spheroid
  Connection with the Discovery of Morphology-dependent Substructure in the
  M$_{BH}$-n$_{sph}$ and M$_{BH}$-R$_{e,sph}$ Diagrams: New Tests for Advanced
  Theories and Realistic Simulations}. \apj 903(2):97.
  \doi{10.3847/1538-4357/abb675}.
  {\href{https://arxiv.org/abs/2101.04895}{{arXiv:2101.04895}}} {[astro-ph.GA]}

\bibitem[{Saito and Yokoyama(2009)}]{Saito:2008jc}
Saito R, Yokoyama J (2009) {Gravitational wave background as a probe of the
  primordial black hole abundance}. Phys Rev Lett 102:161101.
  \doi{10.1103/PhysRevLett.102.161101}, [Erratum: Phys.Rev.Lett. 107, 069901
  (2011)]. {\href{https://arxiv.org/abs/0812.4339}{{arXiv:0812.4339}}}
  {[astro-ph]}

\bibitem[{{Sakurai} et~al.(2016){Sakurai}, {Vorobyov}, {Hosokawa}, {Yoshida},
  {Omukai}, and {Yorke}}]{2016MNRAS.459.1137S}
{Sakurai} Y, {Vorobyov} EI, {Hosokawa} T, {Yoshida} N, {Omukai} K, {Yorke} HW
  (2016) {Supermassive star formation via episodic accretion: protostellar disc
  instability and radiative feedback efficiency}. \mnras 459:1137--1145.
  \doi{10.1093/mnras/stw637}.
  {\href{https://arxiv.org/abs/1511.06080}{{arXiv:1511.06080}}} {[astro-ph.SR]}

\bibitem[{{Sakurai} et~al.(2017){Sakurai}, {Yoshida}, {Fujii}, and
  {Hirano}}]{2017MNRAS.472.1677S}
{Sakurai} Y, {Yoshida} N, {Fujii} MS, {Hirano} S (2017) {Formation of
  intermediate-mass black holes through runaway collisions in the first star
  clusters}. \mnras 472(2):1677--1684. \doi{10.1093/mnras/stx2044}.
  {\href{https://arxiv.org/abs/1704.06130}{{arXiv:1704.06130}}} {[astro-ph.GA]}

\bibitem[{{Sakurai} et~al.(2019){Sakurai}, {Yoshida}, and
  {Fujii}}]{2019MNRAS.484.4665S}
{Sakurai} Y, {Yoshida} N, {Fujii} MS (2019) {Growth of intermediate mass black
  holes by tidal disruption events in the first star clusters}. \mnras
  484(4):4665--4677. \doi{10.1093/mnras/stz315}.
  {\href{https://arxiv.org/abs/1810.01985}{{arXiv:1810.01985}}} {[astro-ph.GA]}

\bibitem[{{Sakurai} et~al.(2020){Sakurai}, {Haiman}, and
  {Inayoshi}}]{2020arXiv200902629S}
{Sakurai} Y, {Haiman} Z, {Inayoshi} K (2020) {Supermassive star formation in a
  massive cloud with H2 molecules}. arXiv e-prints arXiv:2009.02629.
  {\href{https://arxiv.org/abs/2009.02629}{{arXiv:2009.02629}}} {[astro-ph.GA]}

\bibitem[{{Sala} et~al.(2021){Sala}, {Cenci}, {Capelo}, {Lupi}, and
  {Dotti}}]{2020MNRAS.tmp.3351S}
{Sala} L, {Cenci} E, {Capelo} PR, {Lupi} A, {Dotti} M (2021) {Non-isotropic
  feedback from accreting spinning black holes}. \mnras 500(4):4788--4800.
  \doi{10.1093/mnras/staa3552}.
  {\href{https://arxiv.org/abs/2011.06606}{{arXiv:2011.06606}}} {[astro-ph.GA]}

\bibitem[{{Saladino} et~al.(2018){Saladino}, {Pols}, {van der Helm},
  {Pelupessy}, and {Portegies Zwart}}]{2018A&A...618A..50S}
{Saladino} MI, {Pols} OR, {van der Helm} E, {Pelupessy} I, {Portegies Zwart} S
  (2018) {Gone with the wind: the impact of wind mass transfer on the orbital
  evolution of AGB binary systems}. \aap 618:A50.
  \doi{10.1051/0004-6361/201832967}.
  {\href{https://arxiv.org/abs/1805.03208}{{arXiv:1805.03208}}} {[astro-ph.SR]}

\bibitem[{{Saladino} et~al.(2019){Saladino}, {Pols}, and
  {Abate}}]{2019A&A...626A..68S}
{Saladino} MI, {Pols} OR, {Abate} C (2019) {Slowly, slowly in the wind. 3D
  hydrodynamical simulations of wind mass transfer and angular-momentum loss in
  AGB binary systems}. \aap 626:A68. \doi{10.1051/0004-6361/201834598}.
  {\href{https://arxiv.org/abs/1903.04515}{{arXiv:1903.04515}}} {[astro-ph.SR]}

\bibitem[{{Salcido} et~al.(2016){Salcido}, {Bower}, {Theuns}, {McAlpine},
  {Schaller}, {Crain}, {Schaye}, and {Regan}}]{2016MNRAS.463..870S}
{Salcido} J, {Bower} RG, {Theuns} T, {McAlpine} S, {Schaller} M, {Crain} RA,
  {Schaye} J, {Regan} J (2016) {Music from the heavens - gravitational waves
  from supermassive black hole mergers in the EAGLE simulations}. \mnras
  463(1):870--885. \doi{10.1093/mnras/stw2048}.
  {\href{https://arxiv.org/abs/1601.06156}{{arXiv:1601.06156}}} {[astro-ph.GA]}

\bibitem[{{Samsing}(2018)}]{2018PhRvD..97j3014S}
{Samsing} J (2018) {Eccentric black hole mergers forming in globular clusters}.
  \prd 97(10):103014. \doi{10.1103/PhysRevD.97.103014}.
  {\href{https://arxiv.org/abs/1711.07452}{{arXiv:1711.07452}}} {[astro-ph.HE]}

\bibitem[{{Samsing} and {D'Orazio}(2018)}]{2018MNRAS.481.5445S}
{Samsing} J, {D'Orazio} DJ (2018) {Black Hole Mergers From Globular Clusters
  Observable by LISA I: Eccentric Sources Originating From Relativistic N-body
  Dynamics}. \mnras 481(4):5445--5450. \doi{10.1093/mnras/sty2334}.
  {\href{https://arxiv.org/abs/1804.06519}{{arXiv:1804.06519}}} {[astro-ph.HE]}

\bibitem[{{Samsing} and {D'Orazio}(2019)}]{Samsing2019a}
{Samsing} J, {D'Orazio} DJ (2019) {How post-Newtonian dynamics shape the
  distribution of stationary binary black hole LISA sources in nearby globular
  clusters}. \prd 99(6):063006. \doi{10.1103/PhysRevD.99.063006}.
  {\href{https://arxiv.org/abs/1807.08864}{{arXiv:1807.08864}}} {[astro-ph.HE]}

\bibitem[{{Samsing} and {Hotokezaka}(2020)}]{2020arXiv200609744S}
{Samsing} J, {Hotokezaka} K (2020) {Populating the Black Hole Mass Gaps In
  Stellar Clusters: General Relations and Upper Limits}. arXiv e-prints
  arXiv:2006.09744.
  {\href{https://arxiv.org/abs/2006.09744}{{arXiv:2006.09744}}} {[astro-ph.HE]}

\bibitem[{{Samsing} et~al.(2014){Samsing}, {MacLeod}, and
  {Ramirez-Ruiz}}]{2014ApJ...784...71S}
{Samsing} J, {MacLeod} M, {Ramirez-Ruiz} E (2014) {The Formation of Eccentric
  Compact Binary Inspirals and the Role of Gravitational Wave Emission in
  Binary-Single Stellar Encounters}. \apj 784(1):71.
  \doi{10.1088/0004-637X/784/1/71}.
  {\href{https://arxiv.org/abs/1308.2964}{{arXiv:1308.2964}}} {[astro-ph.HE]}

\bibitem[{{Sand} et~al.(2020){Sand}, {Ohlmann}, {Schneider}, {Pakmor}, and
  {Roepke}}]{2020arXiv200711000S}
{Sand} C, {Ohlmann} ST, {Schneider} FRN, {Pakmor} R, {Roepke} FK (2020)
  {Common-envelope evolution with an asymptotic giant branch star}. arXiv
  e-prints arXiv:2007.11000.
  {\href{https://arxiv.org/abs/2007.11000}{{arXiv:2007.11000}}} {[astro-ph.SR]}

\bibitem[{{Sanders}(2013)}]{2013JApA...34...81S}
{Sanders} GH (2013) {The Thirty Meter Telescope (TMT): An International
  Observatory}. Journal of Astrophysics and Astronomy 34(2):81--86.
  \doi{10.1007/s12036-013-9169-5}

\bibitem[{{Sanders} et~al.(2018){Sanders}, {Evans}, and
  {Dehnen}}]{2018MNRAS.478.3879S}
{Sanders} JL, {Evans} NW, {Dehnen} W (2018) {Tidal disruption of dwarf
  spheroidal galaxies: the strange case of Crater II}. \mnras
  478(3):3879--3889. \doi{10.1093/mnras/sty1278}.
  {\href{https://arxiv.org/abs/1802.09537}{{arXiv:1802.09537}}} {[astro-ph.GA]}

\bibitem[{{Sandquist} et~al.(2000){Sandquist}, {Taam}, and
  {Burkert}}]{2000ApJ...533..984S}
{Sandquist} EL, {Taam} RE, {Burkert} A (2000) {On the Formation of Helium
  Double Degenerate Stars and Pre-Cataclysmic Variables}. \apj 533(2):984--997.
  \doi{10.1086/308687}.
  {\href{https://arxiv.org/abs/astro-ph/9912243}{{arXiv:astro-ph/9912243}}}
  {[astro-ph]}

\bibitem[{{Santamar{\'{\i}}a} et~al.(2010){Santamar{\'{\i}}a}, {Ohme}, {Ajith},
  {Br{\"u}gmann}, {Dorband}, {Hannam}, {Husa}, {M{\"o}sta}, {Pollney},
  {Reisswig}, {Robinson}, {Seiler}, and {Krishnan}}]{Santamaria10}
{Santamar{\'{\i}}a} L, {Ohme} F, {Ajith} P, {Br{\"u}gmann} B, {Dorband} N,
  {Hannam} M, {Husa} S, {M{\"o}sta} P, {Pollney} D, {Reisswig} C, et~al. (2010)
  {Matching post-Newtonian and numerical relativity waveforms: Systematic
  errors and a new phenomenological model for nonprecessing black hole
  binaries}. \prd 82(6):064016. \doi{10.1103/PhysRevD.82.064016}.
  {\href{https://arxiv.org/abs/1005.3306}{{arXiv:1005.3306}}} {[gr-qc]}

\bibitem[{{Santoliquido} et~al.(2020{\natexlab{a}}){Santoliquido}, {Mapelli},
  {Bouffanais}, {Giacobbo}, {Di Carlo}, {Rastello}, {Artale}, and
  {Ballone}}]{2020ApJ...898..152S}
{Santoliquido} F, {Mapelli} M, {Bouffanais} Y, {Giacobbo} N, {Di Carlo} UN,
  {Rastello} S, {Artale} MC, {Ballone} A (2020{\natexlab{a}}) {The Cosmic
  Merger Rate Density Evolution of Compact Binaries Formed in Young Star
  Clusters and in Isolated Binaries}. \apj 898(2):152.
  \doi{10.3847/1538-4357/ab9b78}.
  {\href{https://arxiv.org/abs/2004.09533}{{arXiv:2004.09533}}} {[astro-ph.HE]}

\bibitem[{{Santoliquido} et~al.(2020{\natexlab{b}}){Santoliquido}, {Mapelli},
  {Giacobbo}, {Bouffanais}, and {Artale}}]{2020arXiv200903911S}
{Santoliquido} F, {Mapelli} M, {Giacobbo} N, {Bouffanais} Y, {Artale} MC
  (2020{\natexlab{b}}) {The cosmic merger rate density of compact objects:
  impact of star formation, metallicity, initial mass function and binary
  evolution}. arXiv e-prints arXiv:2009.03911.
  {\href{https://arxiv.org/abs/2009.03911}{{arXiv:2009.03911}}} {[astro-ph.HE]}

\bibitem[{{Sanyal} et~al.(2015){Sanyal}, {Grassitelli}, {Langer}, and
  {Bestenlehner}}]{2015A&A...580A..20S}
{Sanyal} D, {Grassitelli} L, {Langer} N, {Bestenlehner} JM (2015) {Massive
  main-sequence stars evolving at the Eddington limit}. \aap 580:A20.
  \doi{10.1051/0004-6361/201525945}.
  {\href{https://arxiv.org/abs/1506.02997}{{arXiv:1506.02997}}} {[astro-ph.SR]}

\bibitem[{Sasaki et~al.(2016)Sasaki, Suyama, Tanaka, and
  Yokoyama}]{Sasaki:2016jop}
Sasaki M, Suyama T, Tanaka T, Yokoyama S (2016) {Primordial Black Hole Scenario
  for the Gravitational-Wave Event GW150914}. Phys Rev Lett 117(6):061101.
  \doi{10.1103/PhysRevLett.117.061101}, [Erratum: Phys.Rev.Lett. 121, 059901
  (2018)]. {\href{https://arxiv.org/abs/1603.08338}{{arXiv:1603.08338}}}
  {[astro-ph.CO]}

\bibitem[{{Sato} et~al.(2017){Sato}, {Kawamura}, {Ando}, {Nakamura}, {Tsubono},
  {Araya}, and {et~al.}}]{2017JPhCS.840a2010S}
{Sato} S, {Kawamura} S, {Ando} M, {Nakamura} T, {Tsubono} K, {Araya} A, {et~al}
  (2017) {The status of DECIGO}. In: Journal of Physics Conference Series.
  Journal of Physics Conference Series, vol 840. p 012010.
  \doi{10.1088/1742-6596/840/1/012010}

\bibitem[{{Sawai} et~al.(2008){Sawai}, {Kotake}, and
  {Yamada}}]{2008ApJ...672..465S}
{Sawai} H, {Kotake} K, {Yamada} S (2008) {Numerical Simulations of Equatorially
  Asymmetric Magnetized Supernovae: Formation of Magnetars and Their Kicks}.
  \apj 672(1):465--478. \doi{10.1086/523624}.
  {\href{https://arxiv.org/abs/0709.1795}{{arXiv:0709.1795}}} {[astro-ph]}

\bibitem[{{Saxton} et~al.(2020){Saxton}, {Komossa}, {Auchettl}, and
  {Jonker}}]{2020SSRv..216...85S}
{Saxton} R, {Komossa} S, {Auchettl} K, {Jonker} PG (2020) {X-Ray Properties of
  TDEs}. \ssr 216(5):85. \doi{10.1007/s11214-020-00708-4}

\bibitem[{{Sayeb} et~al.(2020){Sayeb}, {Blecha}, {Kelley}, {Gerosa}, {Kesden},
  and {Thomas}}]{2020arXiv200606647S}
{Sayeb} M, {Blecha} L, {Kelley} LZ, {Gerosa} D, {Kesden} M, {Thomas} J (2020)
  {Massive black hole binary inspiral and spin evolution in a cosmological
  framework}. arXiv e-prints arXiv:2006.06647.
  {\href{https://arxiv.org/abs/2006.06647}{{arXiv:2006.06647}}} {[astro-ph.GA]}

\bibitem[{{Sberna} et~al.(2020){Sberna}, {Toubiana}, and
  {Miller}}]{2020arXiv201005974S}
{Sberna} L, {Toubiana} A, {Miller} MC (2020) {Golden galactic binaries for
  LISA: mass-transferring white dwarf black hole binaries}. arXiv e-prints
  arXiv:2010.05974.
  {\href{https://arxiv.org/abs/2010.05974}{{arXiv:2010.05974}}} {[astro-ph.SR]}

\bibitem[{{Schaye} et~al.(2015){Schaye}, {Crain}, {Bower}, {Furlong},
  {Schaller}, {Theuns}, {Dalla Vecchia}, {Frenk}, {McCarthy}, {Helly},
  {Jenkins}, {Rosas-Guevara}, {White}, {Baes}, {Booth}, {Camps}, {Navarro},
  {Qu}, {Rahmati}, {Sawala}, {Thomas}, and {Trayford}}]{2015MNRAS.446..521S}
{Schaye} J, {Crain} RA, {Bower} RG, {Furlong} M, {Schaller} M, {Theuns} T,
  {Dalla Vecchia} C, {Frenk} CS, {McCarthy} IG, {Helly} JC, et~al. (2015) {The
  EAGLE project: simulating the evolution and assembly of galaxies and their
  environments}. \mnras 446(1):521--554. \doi{10.1093/mnras/stu2058}.
  {\href{https://arxiv.org/abs/1407.7040}{{arXiv:1407.7040}}} {[astro-ph.GA]}

\bibitem[{{Scheck} et~al.(2006){Scheck}, {Kifonidis}, {Janka}, and
  {M{\"u}ller}}]{2006A&A...457..963S}
{Scheck} L, {Kifonidis} K, {Janka} HT, {M{\"u}ller} E (2006) {Multidimensional
  supernova simulations with approximative neutrino transport. I. Neutron star
  kicks and the anisotropy of neutrino-driven explosions in two spatial
  dimensions}. \aap 457(3):963--986. \doi{10.1051/0004-6361:20064855}.
  {\href{https://arxiv.org/abs/astro-ph/0601302}{{arXiv:astro-ph/0601302}}}
  {[astro-ph]}

\bibitem[{{Scheel} et~al.(2009){Scheel}, {Boyle}, {Chu}, {Kidder}, {Matthews},
  and {Pfeiffer}}]{2009PhRvD..79b4003S}
{Scheel} MA, {Boyle} M, {Chu} T, {Kidder} LE, {Matthews} KD, {Pfeiffer} HP
  (2009) {High-accuracy waveforms for binary black hole inspiral, merger, and
  ringdown}. \prd 79(2):024003. \doi{10.1103/PhysRevD.79.024003}.
  {\href{https://arxiv.org/abs/0810.1767}{{arXiv:0810.1767}}} {[gr-qc]}

\bibitem[{{Schleicher} and {Dreizler}(2014)}]{2014A&A...563A..61S}
{Schleicher} DRG, {Dreizler} S (2014) {Planet formation from the ejecta of
  common envelopes}. \aap 563:A61. \doi{10.1051/0004-6361/201322860}.
  {\href{https://arxiv.org/abs/1312.3479}{{arXiv:1312.3479}}} {[astro-ph.EP]}

\bibitem[{{Schleicher} et~al.(2022){Schleicher}, {Reinoso}, {Latif}, {Klessen},
  {Vergara}, {Das}, {Alister}, {D{\'\i}az}, and {Solar}}]{2022MNRAS.512.6192S}
{Schleicher} DRG, {Reinoso} B, {Latif} M, {Klessen} RS, {Vergara} MZC, {Das} A,
  {Alister} P, {D{\'\i}az} VB, {Solar} PA (2022) {Origin of supermassive black
  holes in massive metal-poor protoclusters}. \mnras 512(4):6192--6200.
  \doi{10.1093/mnras/stac926}.
  {\href{https://arxiv.org/abs/2204.02361}{{arXiv:2204.02361}}} {[astro-ph.GA]}

\bibitem[{{Schneider} et~al.(2001){Schneider}, {Ferrari}, {Matarrese}, and
  {Portegies Zwart}}]{2001MNRAS.324..797S}
{Schneider} R, {Ferrari} V, {Matarrese} S, {Portegies Zwart} SF (2001)
  {Low-frequency gravitational waves from cosmological compact binaries}.
  \mnras 324(4):797--810. \doi{10.1046/j.1365-8711.2001.04217.x}.
  {\href{https://arxiv.org/abs/astro-ph/0002055}{{arXiv:astro-ph/0002055}}}
  {[astro-ph]}

\bibitem[{{Schneider} et~al.(2017){Schneider}, {Graziani}, {Marassi}, {Spera},
  {Mapelli}, {Alparone}, and {Bennassuti}}]{2017MNRAS.471L.105S}
{Schneider} R, {Graziani} L, {Marassi} S, {Spera} M, {Mapelli} M, {Alparone} M,
  {Bennassuti} Md (2017) {The formation and coalescence sites of the first
  gravitational wave events}. \mnras 471(1):L105--L109.
  \doi{10.1093/mnrasl/slx118}.
  {\href{https://arxiv.org/abs/1705.06781}{{arXiv:1705.06781}}} {[astro-ph.GA]}

\bibitem[{{Schnittman}(2007)}]{2007ApJ...667L.133S}
{Schnittman} JD (2007) {Retaining Black Holes with Very Large Recoil
  Velocities}. \apjl 667(2):L133--L136. \doi{10.1086/522203}.
  {\href{https://arxiv.org/abs/0706.1548}{{arXiv:0706.1548}}} {[astro-ph]}

\bibitem[{{Schnittman} and {Buonanno}(2007)}]{2007ApJ...662L..63S}
{Schnittman} JD, {Buonanno} A (2007) {The Distribution of Recoil Velocities
  from Merging Black Holes}. \apjl 662(2):L63--L66. \doi{10.1086/519309}.
  {\href{https://arxiv.org/abs/astro-ph/0702641}{{arXiv:astro-ph/0702641}}}
  {[astro-ph]}

\bibitem[{{Schnittman} and {Krolik}(2008)}]{2008ApJ...684..835S}
{Schnittman} JD, {Krolik} JH (2008) {The Infrared Afterglow of Supermassive
  Black Hole Mergers}. \apj 684(2):835--844. \doi{10.1086/590363}.
  {\href{https://arxiv.org/abs/0802.3556}{{arXiv:0802.3556}}} {[astro-ph]}

\bibitem[{{Schnittman} and {Krolik}(2009)}]{2009ApJ...701.1175S}
{Schnittman} JD, {Krolik} JH (2009) {X-ray Polarization from Accreting Black
  Holes: The Thermal State}. \apj 701(2):1175--1187.
  \doi{10.1088/0004-637X/701/2/1175}.
  {\href{https://arxiv.org/abs/0902.3982}{{arXiv:0902.3982}}} {[astro-ph.HE]}

\bibitem[{{Sch{\"o}del} et~al.(2018){Sch{\"o}del}, {Gallego-Cano}, {Dong},
  {Nogueras-Lara}, {Gallego-Calvente}, {Amaro-Seoane}, and
  {Baumgardt}}]{2018A&A...609A..27S}
{Sch{\"o}del} R, {Gallego-Cano} E, {Dong} H, {Nogueras-Lara} F,
  {Gallego-Calvente} AT, {Amaro-Seoane} P, {Baumgardt} H (2018) {The
  distribution of stars around the Milky Way's central black hole. II. Diffuse
  light from sub-giants and dwarfs}. \aap 609:A27.
  \doi{10.1051/0004-6361/201730452}.
  {\href{https://arxiv.org/abs/1701.03817}{{arXiv:1701.03817}}} {[astro-ph.GA]}

\bibitem[{{Schr{\"o}der} and {Smith}(2008)}]{2008MNRAS.386..155S}
{Schr{\"o}der} KP, {Smith} RC (2008) {Distant future of the Sun and Earth
  revisited}. \mnras 386(1):155--163. \doi{10.1111/j.1365-2966.2008.13022.x}.
  {\href{https://arxiv.org/abs/0801.4031}{{arXiv:0801.4031}}} {[astro-ph]}

\bibitem[{{Scott} and {Graham}(2013)}]{2013ApJ...763...76S}
{Scott} N, {Graham} AW (2013) {Updated Mass Scaling Relations for Nuclear Star
  Clusters and a Comparison to Supermassive Black Holes}. \apj 763(2):76.
  \doi{10.1088/0004-637X/763/2/76}.
  {\href{https://arxiv.org/abs/1205.5338}{{arXiv:1205.5338}}} {[astro-ph.CO]}

\bibitem[{{Scrimgeour} et~al.(2016){Scrimgeour}, {Davis}, {Blake},
  {Staveley-Smith}, {Magoulas}, {Springob}, {Beutler}, {Colless}, {Johnson},
  {Jones}, {Koda}, {Lucey}, {Ma}, {Mould}, and {Poole}}]{2016MNRAS.455..386S}
{Scrimgeour} MI, {Davis} TM, {Blake} C, {Staveley-Smith} L, {Magoulas} C,
  {Springob} CM, {Beutler} F, {Colless} M, {Johnson} A, {Jones} DH, et~al.
  (2016) {The 6dF Galaxy Survey: bulk flows on 50-70 h$^{-1}$ Mpc scales}.
  \mnras 455(1):386--401. \doi{10.1093/mnras/stv2146}.
  {\href{https://arxiv.org/abs/1511.06930}{{arXiv:1511.06930}}} {[astro-ph.CO]}

\bibitem[{{Secunda} et~al.(2019){Secunda}, {Bellovary}, {Mac Low}, {Ford},
  {McKernan}, {Leigh}, {Lyra}, and {S{\'a}ndor}}]{2019ApJ...878...85S}
{Secunda} A, {Bellovary} J, {Mac Low} MM, {Ford} KES, {McKernan} B, {Leigh}
  NWC, {Lyra} W, {S{\'a}ndor} Z (2019) {Orbital Migration of Interacting
  Stellar Mass Black Holes in Disks around Supermassive Black Holes}. \apj
  878(2):85. \doi{10.3847/1538-4357/ab20ca}.
  {\href{https://arxiv.org/abs/1807.02859}{{arXiv:1807.02859}}} {[astro-ph.HE]}

\bibitem[{{Secunda} et~al.(2020{\natexlab{a}}){Secunda}, {Bellovary}, {Mac
  Low}, {Ford}, {McKernan}, {Leigh}, {Lyra}, {S{\'a}ndor}, and
  {Adorno}}]{2020arXiv200411936S}
{Secunda} A, {Bellovary} J, {Mac Low} MM, {Ford} KES, {McKernan} B, {Leigh}
  NWC, {Lyra} W, {S{\'a}ndor} Z, {Adorno} JI (2020{\natexlab{a}}) {Orbital
  Migration of Interacting Stellar Mass Black Holes in Disks around
  Supermassive Black Holes II. Spins and Incoming Objects}. arXiv e-prints
  arXiv:2004.11936.
  {\href{https://arxiv.org/abs/2004.11936}{{arXiv:2004.11936}}} {[astro-ph.HE]}

\bibitem[{{Secunda} et~al.(2020{\natexlab{b}}){Secunda}, {Hernandez},
  {Goodman}, {Leigh}, {McKernan}, {Ford}, and {Adorno}}]{2020arXiv200903910S}
{Secunda} A, {Hernandez} B, {Goodman} J, {Leigh} NWC, {McKernan} B, {Ford} KES,
  {Adorno} JI (2020{\natexlab{b}}) {Evolution of Retrograde Orbiters in an AGN
  Disk}. arXiv e-prints arXiv:2009.03910.
  {\href{https://arxiv.org/abs/2009.03910}{{arXiv:2009.03910}}} {[astro-ph.HE]}

\bibitem[{{Seigar} et~al.(2008){Seigar}, {Kennefick}, {Kennefick}, and
  {Lacy}}]{2008ApJ...678L..93S}
{Seigar} MS, {Kennefick} D, {Kennefick} J, {Lacy} CHS (2008) {Discovery of a
  Relationship between Spiral Arm Morphology and Supermassive Black Hole Mass
  in Disk Galaxies}. \apjl 678(2):L93. \doi{10.1086/588727}.
  {\href{https://arxiv.org/abs/0804.0773}{{arXiv:0804.0773}}} {[astro-ph]}

\bibitem[{{Seitenzahl} et~al.(2013){Seitenzahl}, {Cescutti}, {R{\"o}pke},
  {Ruiter}, and {Pakmor}}]{2013A&A...559L...5S}
{Seitenzahl} IR, {Cescutti} G, {R{\"o}pke} FK, {Ruiter} AJ, {Pakmor} R (2013)
  {Solar abundance of manganese: a case for near Chandrasekhar-mass Type Ia
  supernova progenitors}. \aap 559:L5. \doi{10.1051/0004-6361/201322599}.
  {\href{https://arxiv.org/abs/1309.2397}{{arXiv:1309.2397}}} {[astro-ph.SR]}

\bibitem[{{Seitenzahl} et~al.(2015){Seitenzahl}, {Herzog}, {Ruiter},
  {Marquardt}, {Ohlmann}, and {R{\"o}pke}}]{2015PhRvD..92l4013S}
{Seitenzahl} IR, {Herzog} M, {Ruiter} AJ, {Marquardt} K, {Ohlmann} ST,
  {R{\"o}pke} FK (2015) {Neutrino and gravitational wave signal of a
  delayed-detonation model of type Ia supernovae}. \prd 92(12):124013.
  \doi{10.1103/PhysRevD.92.124013}.
  {\href{https://arxiv.org/abs/1511.02542}{{arXiv:1511.02542}}} {[astro-ph.SR]}

\bibitem[{{Sellwood}(2014)}]{2014RvMP...86....1S}
{Sellwood} JA (2014) {Secular evolution in disk galaxies}. Reviews of Modern
  Physics 86(1):1--46. \doi{10.1103/RevModPhys.86.1}.
  {\href{https://arxiv.org/abs/1310.0403}{{arXiv:1310.0403}}} {[astro-ph.GA]}

\bibitem[{{Sengar} et~al.(2017){Sengar}, {Tauris}, {Langer}, and
  {Istrate}}]{2017MNRAS.470L...6S}
{Sengar} R, {Tauris} TM, {Langer} N, {Istrate} AG (2017) {Novel modelling of
  ultracompact X-ray binary evolution - stable mass transfer from white dwarfs
  to neutron stars}. \mnras 470(1):L6--L10. \doi{10.1093/mnrasl/slx064}.
  {\href{https://arxiv.org/abs/1704.08260}{{arXiv:1704.08260}}} {[astro-ph.SR]}

\bibitem[{Serpico et~al.(2020)Serpico, Poulin, Inman, and
  Kohri}]{Serpico:2020ehh}
Serpico PD, Poulin V, Inman D, Kohri K (2020) {Cosmic microwave background
  bounds on primordial black holes including dark matter halo accretion}. Phys
  Rev Res 2(2):023204. \doi{10.1103/PhysRevResearch.2.023204}.
  {\href{https://arxiv.org/abs/2002.10771}{{arXiv:2002.10771}}} {[astro-ph.CO]}

\bibitem[{{Sesana}(2007)}]{2007MNRAS.382L...6S}
{Sesana} A (2007) {Extreme recoils: impact on the detection of gravitational
  waves from massive black hole binaries}. \mnras 382(1):L6--L10.
  \doi{10.1111/j.1745-3933.2007.00375.x}.
  {\href{https://arxiv.org/abs/0707.4677}{{arXiv:0707.4677}}} {[astro-ph]}

\bibitem[{{Sesana}(2016)}]{2016PhRvL.116w1102S}
{Sesana} A (2016) {Prospects for Multiband Gravitational-Wave Astronomy after
  GW150914}. \prl 116(23):231102. \doi{10.1103/PhysRevLett.116.231102}.
  {\href{https://arxiv.org/abs/1602.06951}{{arXiv:1602.06951}}} {[gr-qc]}

\bibitem[{{Sesana}(2017)}]{2017JPhCS.840a2018S}
{Sesana} A (2017) {Multi-band gravitational wave astronomy: science with joint
  space- and ground-based observations of black hole binaries}. In: Journal of
  Physics Conference Series. Journal of Physics Conference Series, vol 840. p
  012018. \doi{10.1088/1742-6596/840/1/012018}.
  {\href{https://arxiv.org/abs/1702.04356}{{arXiv:1702.04356}}} {[astro-ph.HE]}

\bibitem[{{Sesana} and {Khan}(2015)}]{2015MNRAS.454L..66S}
{Sesana} A, {Khan} FM (2015) {Scattering experiments meet N-body - I. A
  practical recipe for the evolution of massive black hole binaries in stellar
  environments}. \mnras 454(1):L66--L70. \doi{10.1093/mnrasl/slv131}.
  {\href{https://arxiv.org/abs/1505.02062}{{arXiv:1505.02062}}} {[astro-ph.GA]}

\bibitem[{{Sesana} et~al.(2005){Sesana}, {Haardt}, {Madau}, and
  {Volonteri}}]{2005ApJ...623...23S}
{Sesana} A, {Haardt} F, {Madau} P, {Volonteri} M (2005) {The Gravitational Wave
  Signal from Massive Black Hole Binaries and Its Contribution to the LISA Data
  Stream}. \apj 623(1):23--30. \doi{10.1086/428492}.
  {\href{https://arxiv.org/abs/astro-ph/0409255}{{arXiv:astro-ph/0409255}}}
  {[astro-ph]}

\bibitem[{{Sesana} et~al.(2006){Sesana}, {Haardt}, and
  {Madau}}]{2006ApJ...651..392S}
{Sesana} A, {Haardt} F, {Madau} P (2006) {Interaction of Massive Black Hole
  Binaries with Their Stellar Environment. I. Ejection of Hypervelocity Stars}.
  \apj 651(1):392--400. \doi{10.1086/507596}.
  {\href{https://arxiv.org/abs/astro-ph/0604299}{{arXiv:astro-ph/0604299}}}
  {[astro-ph]}

\bibitem[{{Sesana} et~al.(2008{\natexlab{a}}){Sesana}, {Haardt}, and
  {Madau}}]{2008ApJ...686..432S}
{Sesana} A, {Haardt} F, {Madau} P (2008{\natexlab{a}}) {Interaction of Massive
  Black Hole Binaries with Their Stellar Environment. III. Scattering of Bound
  Stars}. \apj 686(1):432--447. \doi{10.1086/590651}.
  {\href{https://arxiv.org/abs/0710.4301}{{arXiv:0710.4301}}} {[astro-ph]}

\bibitem[{{Sesana} et~al.(2008{\natexlab{b}}){Sesana}, {Vecchio}, {Eracleous},
  and {Sigurdsson}}]{2008MNRAS.391..718S}
{Sesana} A, {Vecchio} A, {Eracleous} M, {Sigurdsson} S (2008{\natexlab{b}})
  {Observing white dwarfs orbiting massive black holes in the gravitational
  wave and electro-magnetic window}. \mnras 391(2):718--726.
  \doi{10.1111/j.1365-2966.2008.13904.x}.
  {\href{https://arxiv.org/abs/0806.0624}{{arXiv:0806.0624}}} {[astro-ph]}

\bibitem[{{Sesana} et~al.(2009){Sesana}, {Vecchio}, and
  {Volonteri}}]{2009MNRAS.394.2255S}
{Sesana} A, {Vecchio} A, {Volonteri} M (2009) {Gravitational waves from
  resolvable massive black hole binary systems and observations with Pulsar
  Timing Arrays}. \mnras 394(4):2255--2265.
  \doi{10.1111/j.1365-2966.2009.14499.x}.
  {\href{https://arxiv.org/abs/0809.3412}{{arXiv:0809.3412}}} {[astro-ph]}

\bibitem[{{Sesana} et~al.(2011{\natexlab{a}}){Sesana}, {Gair}, {Berti}, and
  {Volonteri}}]{2011PhRvD..83d4036S}
{Sesana} A, {Gair} J, {Berti} E, {Volonteri} M (2011{\natexlab{a}})
  {Reconstructing the massive black hole cosmic history through gravitational
  waves}. \prd 83(4):044036. \doi{10.1103/PhysRevD.83.044036}.
  {\href{https://arxiv.org/abs/1011.5893}{{arXiv:1011.5893}}} {[astro-ph.CO]}

\bibitem[{{Sesana} et~al.(2011{\natexlab{b}}){Sesana}, {Gualandris}, and
  {Dotti}}]{2011MNRAS.415L..35S}
{Sesana} A, {Gualandris} A, {Dotti} M (2011{\natexlab{b}}) {Massive black hole
  binary eccentricity in rotating stellar systems}. \mnras 415(1):L35--L39.
  \doi{10.1111/j.1745-3933.2011.01073.x}.
  {\href{https://arxiv.org/abs/1105.0670}{{arXiv:1105.0670}}} {[astro-ph.GA]}

\bibitem[{{Sesana} et~al.(2012){Sesana}, {Roedig}, {Reynolds}, and
  {Dotti}}]{2012MNRAS.420..860S}
{Sesana} A, {Roedig} C, {Reynolds} MT, {Dotti} M (2012) Multimessenger
  astronomy with pulsar timing and x-ray observations of massive black hole
  binaries. MNRAS 420(1):860--877. \doi{10.1111/j.1365-2966.2011.20097.x}.
  {\href{https://arxiv.org/abs/1107.2927}{{arXiv:1107.2927}}} {[astro-ph.CO]}

\bibitem[{{Sesana} et~al.(2014){Sesana}, {Barausse}, {Dotti}, and
  {Rossi}}]{2014ApJ...794..104S}
{Sesana} A, {Barausse} E, {Dotti} M, {Rossi} EM (2014) {Linking the Spin
  Evolution of Massive Black Holes to Galaxy Kinematics}. \apj 794(2):104.
  \doi{10.1088/0004-637X/794/2/104}.
  {\href{https://arxiv.org/abs/1402.7088}{{arXiv:1402.7088}}} {[astro-ph.CO]}

\bibitem[{{Sesana} et~al.(2020){Sesana}, {Lamberts}, and
  {Petiteau}}]{2020MNRAS.494L..75S}
{Sesana} A, {Lamberts} A, {Petiteau} A (2020) {Finding binary black holes in
  the Milky Way with LISA}. \mnras 494(1):L75--L80.
  \doi{10.1093/mnrasl/slaa039}.
  {\href{https://arxiv.org/abs/1912.07627}{{arXiv:1912.07627}}} {[astro-ph.GA]}

\bibitem[{{Sesana} et~al.(2021){Sesana}, {Korsakova}, {Arca Sedda}, {Baibhav},
  {Barausse}, {Barke}, {Berti}, {Bonetti}, {Capelo}, {Caprini},
  {Garcia-Bellido}, {Haiman}, {Jani}, {Jennrich}, {Johansson}, {Khan}, {Korol},
  {Lamberts}, {Lupi}, {Mangiagli}, {Mayer}, {Nardini}, {Pacucci}, {Petiteau},
  {Raccanelli}, {Rajendran}, {Regan}, {Shao}, {Spallicci}, {Tamanini},
  {Volonteri}, {Warburton}, {Wong}, and {Zumalacarregui}}]{2019arXiv190811391S}
{Sesana} A, {Korsakova} N, {Arca Sedda} M, {Baibhav} V, {Barausse} E, {Barke}
  S, {Berti} E, {Bonetti} M, {Capelo} PR, {Caprini} C, et~al. (2021) {Unveiling
  the gravitational universe at {\ensuremath{\mu}}-Hz frequencies}.
  Experimental Astronomy 51(3):1333--1383. \doi{10.1007/s10686-021-09709-9}.
  {\href{https://arxiv.org/abs/1908.11391}{{arXiv:1908.11391}}} {[astro-ph.IM]}

\bibitem[{{Seto}(2016)}]{2016MNRAS.460L...1S}
{Seto} N (2016) {Prospects of eLISA for detecting Galactic binary black holes
  similar to GW150914}. \mnras 460(1):L1--L4. \doi{10.1093/mnrasl/slw060}.
  {\href{https://arxiv.org/abs/1602.04715}{{arXiv:1602.04715}}} {[astro-ph.HE]}

\bibitem[{{Seto}(2019)}]{2019MNRAS.489.4513S}
{Seto} N (2019) {Search for neutron star binaries in the Local Group galaxies
  using LISA}. \mnras 489(4):4513--4519. \doi{10.1093/mnras/stz2439}.
  {\href{https://arxiv.org/abs/1909.01471}{{arXiv:1909.01471}}} {[astro-ph.HE]}

\bibitem[{{Severgnini} et~al.(2018){Severgnini}, {Cicone}, {Della Ceca},
  {Braito}, {Caccianiga}, {Ballo}, {Campana}, {Moretti}, {La Parola},
  {Vignali}, {Zaino}, {Matzeu}, and {Landoni}}]{2018MNRAS.479.3804S}
{Severgnini} P, {Cicone} C, {Della Ceca} R, {Braito} V, {Caccianiga} A, {Ballo}
  L, {Campana} S, {Moretti} A, {La Parola} V, {Vignali} C, et~al. (2018) {Swift
  data hint at a binary supermassive black hole candidate at sub-parsec
  separation}. \mnras 479(3):3804--3813. \doi{10.1093/mnras/sty1699}.
  {\href{https://arxiv.org/abs/1806.10150}{{arXiv:1806.10150}}} {[astro-ph.HE]}

\bibitem[{{Shah} and {Nelemans}(2014{\natexlab{a}})}]{2014ApJ...790..161S}
{Shah} S, {Nelemans} G (2014{\natexlab{a}}) {Constraining Parameters of
  White-dwarf Binaries Using Gravitational-wave and Electromagnetic
  Observations}. \apj 790(2):161. \doi{10.1088/0004-637X/790/2/161}.
  {\href{https://arxiv.org/abs/1406.3599}{{arXiv:1406.3599}}} {[astro-ph.SR]}

\bibitem[{{Shah} and {Nelemans}(2014{\natexlab{b}})}]{2014ApJ...791...76S}
{Shah} S, {Nelemans} G (2014{\natexlab{b}}) {Measuring Tides and Binary
  Parameters from Gravitational Wave Data and Eclipsing Timings of Detached
  White Dwarf Binaries}. \apj 791(2):76. \doi{10.1088/0004-637X/791/2/76}.
  {\href{https://arxiv.org/abs/1406.3603}{{arXiv:1406.3603}}} {[astro-ph.SR]}

\bibitem[{{Shah} et~al.(2012){Shah}, {van der Sluys}, and
  {Nelemans}}]{2012A&A...544A.153S}
{Shah} S, {van der Sluys} M, {Nelemans} G (2012) {Using electromagnetic
  observations to aid gravitational-wave parameter estimation of compact
  binaries observed with LISA}. \aap 544:A153.
  \doi{10.1051/0004-6361/201219309}.
  {\href{https://arxiv.org/abs/1207.6770}{{arXiv:1207.6770}}} {[astro-ph.IM]}

\bibitem[{{Shah} et~al.(2013){Shah}, {Nelemans}, and {van der
  Sluys}}]{2013A&A...553A..82S}
{Shah} S, {Nelemans} G, {van der Sluys} M (2013) {Using electromagnetic
  observations to aid gravitational-wave parameter estimation of compact
  binaries observed with LISA. II. The effect of knowing the sky position}.
  \aap 553:A82. \doi{10.1051/0004-6361/201321123}.
  {\href{https://arxiv.org/abs/1303.6116}{{arXiv:1303.6116}}} {[astro-ph.IM]}

\bibitem[{{Shakura} and {Sunyaev}(1973)}]{1973A&A....24..337S}
{Shakura} NI, {Sunyaev} RA (1973) {Reprint of 1973A\&A....24..337S. Black holes
  in binary systems. Observational appearance.} \aap 500:33--51

\bibitem[{{Shakura} and {Sunyaev}(1976)}]{1976MNRAS.175..613S}
{Shakura} NI, {Sunyaev} RA (1976) {A theory of the instability of disk
  accretion on to black holes and the variability of binary X-ray sources,
  galactic nuclei and quasars.} \mnras 175:613--632.
  \doi{10.1093/mnras/175.3.613}

\bibitem[{{Shankar}(2009)}]{2009NewAR..53...57S}
{Shankar} F (2009) {The demography of supermassive black holes: Growing
  monsters at the heart of galaxies}. \nar 53(4-6):57--77.
  \doi{10.1016/j.newar.2009.07.006}.
  {\href{https://arxiv.org/abs/0907.5213}{{arXiv:0907.5213}}} {[astro-ph.CO]}

\bibitem[{{Shankar} et~al.(2009){Shankar}, {Weinberg}, and
  {Miralda-Escud{\'e}}}]{2009ApJ...690...20S}
{Shankar} F, {Weinberg} DH, {Miralda-Escud{\'e}} J (2009) {Self-Consistent
  Models of the AGN and Black Hole Populations: Duty Cycles, Accretion Rates,
  and the Mean Radiative Efficiency}. \apj 690(1):20--41.
  \doi{10.1088/0004-637X/690/1/20}.
  {\href{https://arxiv.org/abs/0710.4488}{{arXiv:0710.4488}}} {[astro-ph]}

\bibitem[{{Shankar} et~al.(2013){Shankar}, {Weinberg}, and
  {Miralda-Escud{\'e}}}]{2013MNRAS.428..421S}
{Shankar} F, {Weinberg} DH, {Miralda-Escud{\'e}} J (2013) {Accretion-driven
  evolution of black holes: Eddington ratios, duty cycles and active galaxy
  fractions}. \mnras 428(1):421--446. \doi{10.1093/mnras/sts026}.
  {\href{https://arxiv.org/abs/1111.3574}{{arXiv:1111.3574}}} {[astro-ph.CO]}

\bibitem[{{Shannon} et~al.(2015){Shannon}, {Ravi}, {Lentati}, {Lasky}, {Hobbs},
  {Kerr}, {Manchester}, {Coles}, {Levin}, {Bailes}, {Bhat}, {Burke-Spolaor},
  {Dai}, {Keith}, {Os{\l}owski}, {Reardon}, {van Straten}, {Toomey}, {Wang},
  {Wen}, {Wyithe}, and {Zhu}}]{2015Sci...349.1522S}
{Shannon} RM, {Ravi} V, {Lentati} LT, {Lasky} PD, {Hobbs} G, {Kerr} M,
  {Manchester} RN, {Coles} WA, {Levin} Y, {Bailes} M, et~al. (2015)
  {Gravitational waves from binary supermassive black holes missing in pulsar
  observations}. Science 349(6255):1522--1525. \doi{10.1126/science.aab1910}.
  {\href{https://arxiv.org/abs/1509.07320}{{arXiv:1509.07320}}} {[astro-ph.CO]}

\bibitem[{{Shao} et~al.(2017){Shao}, {Sennett}, {Buonanno}, {Kramer}, and
  {Wex}}]{2017PhRvX...7d1025S}
{Shao} L, {Sennett} N, {Buonanno} A, {Kramer} M, {Wex} N (2017) {Constraining
  Nonperturbative Strong-Field Effects in Scalar-Tensor Gravity by Combining
  Pulsar Timing and Laser-Interferometer Gravitational-Wave Detectors}.
  Physical Review X 7(4):041025. \doi{10.1103/PhysRevX.7.041025}.
  {\href{https://arxiv.org/abs/1704.07561}{{arXiv:1704.07561}}} {[gr-qc]}

\bibitem[{{Shapiro} and {Marchant}(1978)}]{1978ApJ...225..603S}
{Shapiro} SL, {Marchant} AB (1978) {Star clusters containing massive, central
  black holes: Monte Carlo simulations in two-dimensional phase space.} \apj
  225:603--624. \doi{10.1086/156521}

\bibitem[{{Shappee} and {Thompson}(2013)}]{2013ApJ...766...64S}
{Shappee} BJ, {Thompson} TA (2013) {The Mass-loss-induced Eccentric Kozai
  Mechanism: A New Channel for the Production of Close Compact Object-Stellar
  Binaries}. \apj 766(1):64. \doi{10.1088/0004-637X/766/1/64}.
  {\href{https://arxiv.org/abs/1204.1053}{{arXiv:1204.1053}}} {[astro-ph.SR]}

\bibitem[{{Shemmer} et~al.(2004){Shemmer}, {Netzer}, {Maiolino}, {Oliva},
  {Croom}, {Corbett}, and {di Fabrizio}}]{2004ApJ...614..547S}
{Shemmer} O, {Netzer} H, {Maiolino} R, {Oliva} E, {Croom} S, {Corbett} E, {di
  Fabrizio} L (2004) {Near-Infrared Spectroscopy of High-Redshift Active
  Galactic Nuclei. I. A Metallicity-Accretion Rate Relationship}. \apj
  614(2):547--557. \doi{10.1086/423607}.
  {\href{https://arxiv.org/abs/astro-ph/0406559}{{arXiv:astro-ph/0406559}}}
  {[astro-ph]}

\bibitem[{{Shen}(2015)}]{2015ApJ...805L...6S}
{Shen} KJ (2015) {Every Interacting Double White Dwarf Binary May Merge}. \apjl
  805(1):L6. \doi{10.1088/2041-8205/805/1/L6}.
  {\href{https://arxiv.org/abs/1502.05052}{{arXiv:1502.05052}}} {[astro-ph.SR]}

\bibitem[{{Shen} et~al.(2020){Shen}, {Hopkins}, {Faucher-Gigu{\`e}re},
  {Alexander}, {Richards}, {Ross}, and {Hickox}}]{2020MNRAS.495.3252S}
{Shen} X, {Hopkins} PF, {Faucher-Gigu{\`e}re} CA, {Alexander} DM, {Richards}
  GT, {Ross} NP, {Hickox} RC (2020) {The bolometric quasar luminosity function
  at z = 0-7}. \mnras 495(3):3252--3275. \doi{10.1093/mnras/staa1381}.
  {\href{https://arxiv.org/abs/2001.02696}{{arXiv:2001.02696}}} {[astro-ph.GA]}

\bibitem[{{Shen} et~al.(2013){Shen}, {Liu}, {Loeb}, and
  {Tremaine}}]{2013ApJ...775...49S}
{Shen} Y, {Liu} X, {Loeb} A, {Tremaine} S (2013) {Constraining Sub-parsec
  Binary Supermassive Black Holes in Quasars with Multi-epoch Spectroscopy. I.
  The General Quasar Population}. \apj 775(1):49.
  \doi{10.1088/0004-637X/775/1/49}.
  {\href{https://arxiv.org/abs/1306.4330}{{arXiv:1306.4330}}} {[astro-ph.CO]}

\bibitem[{{Sheth} et~al.(2008){Sheth}, {Elmegreen}, {Elmegreen}, {Capak},
  {Abraham}, {Athanassoula}, {Ellis}, {Mobasher}, {Salvato}, {Schinnerer},
  {Scoville}, {Spalsbury}, {Strubbe}, {Carollo}, {Rich}, and
  {West}}]{2008ApJ...675.1141S}
{Sheth} K, {Elmegreen} DM, {Elmegreen} BG, {Capak} P, {Abraham} RG,
  {Athanassoula} E, {Ellis} RS, {Mobasher} B, {Salvato} M, {Schinnerer} E,
  et~al. (2008) {Evolution of the Bar Fraction in COSMOS: Quantifying the
  Assembly of the Hubble Sequence}. \apj 675(2):1141--1155.
  \doi{10.1086/524980}.
  {\href{https://arxiv.org/abs/0710.4552}{{arXiv:0710.4552}}} {[astro-ph]}

\bibitem[{Shi and Krolik(2016)}]{2016ApJ...832...22S}
Shi JM, Krolik JH (2016) How bright are the gaps in circumbinary disk systems?
  ApJ 832(1):22. \doi{10.3847/0004-637x/832/1/22},
  \urlprefix\url{https://doi.org/10.3847\%2F0004-637x\%2F832\%2F1\%2F22}

\bibitem[{{Shibata} and {Taniguchi}(2011)}]{2011LRR....14....6S}
{Shibata} M, {Taniguchi} K (2011) {Coalescence of Black Hole-Neutron Star
  Binaries}. Living Reviews in Relativity 14(1):6. \doi{10.12942/lrr-2011-6}

\bibitem[{{Shibata} et~al.(2016){Shibata}, {Sekiguchi}, {Uchida}, and
  {Umeda}}]{2016PhRvD..94b1501S}
{Shibata} M, {Sekiguchi} Y, {Uchida} H, {Umeda} H (2016) {Gravitational waves
  from supermassive stars collapsing to a supermassive black hole}. \prd
  94(2):021501. \doi{10.1103/PhysRevD.94.021501}.
  {\href{https://arxiv.org/abs/1606.07147}{{arXiv:1606.07147}}} {[astro-ph.HE]}

\bibitem[{{Shiber} et~al.(2019){Shiber}, {Iaconi}, {De Marco}, and
  {Soker}}]{2019MNRAS.488.5615S}
{Shiber} S, {Iaconi} R, {De Marco} O, {Soker} N (2019) {Companion-launched jets
  and their effect on the dynamics of common envelope interaction simulations}.
  \mnras 488(4):5615--5632. \doi{10.1093/mnras/stz2013}.
  {\href{https://arxiv.org/abs/1902.03931}{{arXiv:1902.03931}}} {[astro-ph.SR]}

\bibitem[{{Shibuya} et~al.(2016){Shibuya}, {Ouchi}, {Kubo}, and
  {Harikane}}]{2016ApJ...821...72S}
{Shibuya} T, {Ouchi} M, {Kubo} M, {Harikane} Y (2016) {Morphologies of
  \raisebox{-0.5ex}\textasciitilde190,000 Galaxies at z = 0-10 Revealed with
  HST Legacy Data. II. Evolution of Clumpy Galaxies}. \apj 821(2):72.
  \doi{10.3847/0004-637X/821/2/72}.
  {\href{https://arxiv.org/abs/1511.07054}{{arXiv:1511.07054}}} {[astro-ph.GA]}

\bibitem[{{Shields} and {Bonning}(2008)}]{2008ApJ...682..758S}
{Shields} GA, {Bonning} EW (2008) {Powerful Flares from Recoiling Black Holes
  in Quasars}. \apj 682(2):758--766. \doi{10.1086/589427}.
  {\href{https://arxiv.org/abs/0802.3873}{{arXiv:0802.3873}}} {[astro-ph]}

\bibitem[{{Shiokawa} et~al.(2015){Shiokawa}, {Krolik}, {Cheng}, {Piran}, and
  {Noble}}]{2015ApJ...804...85S}
{Shiokawa} H, {Krolik} JH, {Cheng} RM, {Piran} T, {Noble} SC (2015) {General
  Relativistic Hydrodynamic Simulation of Accretion Flow from a Stellar Tidal
  Disruption}. \apj 804(2):85. \doi{10.1088/0004-637X/804/2/85}.
  {\href{https://arxiv.org/abs/1501.04365}{{arXiv:1501.04365}}} {[astro-ph.HE]}

\bibitem[{{Shirakata} et~al.(2019){Shirakata}, {Okamoto}, {Kawaguchi},
  {Nagashima}, {Ishiyama}, {Makiya}, {Kobayashi}, {Enoki}, {Oogi}, and
  {Okoshi}}]{2019MNRAS.482.4846S}
{Shirakata} H, {Okamoto} T, {Kawaguchi} T, {Nagashima} M, {Ishiyama} T,
  {Makiya} R, {Kobayashi} MAR, {Enoki} M, {Oogi} T, {Okoshi} K (2019) {The New
  Numerical Galaxy Catalogue ({\ensuremath{\nu}}$^{2}$GC): properties of active
  galactic nuclei and their host galaxies}. \mnras 482(4):4846--4873.
  \doi{10.1093/mnras/sty2958}.
  {\href{https://arxiv.org/abs/1802.02169}{{arXiv:1802.02169}}} {[astro-ph.GA]}

\bibitem[{{Shlosman} and {Begelman}(1989)}]{1989ApJ...341..685S}
{Shlosman} I, {Begelman} MC (1989) {Evolution of Self-Gravitating Accretion
  Disks in Active Galactic Nuclei}. \apj 341:685. \doi{10.1086/167526}

\bibitem[{{Shore} et~al.(1994){Shore}, {Livio}, and {van den Heuvel}}]{slv94}
{Shore} SN, {Livio} M, {van den Heuvel} EPJ (1994) {Interacting binaries}

\bibitem[{{Shuman} and {Cornish}(2022)}]{2022PhRvD.105f4055S}
{Shuman} KJ, {Cornish} NJ (2022) {Massive black hole binaries and where to find
  them with dual detector networks}. \prd 105(6):064055.
  \doi{10.1103/PhysRevD.105.064055}.
  {\href{https://arxiv.org/abs/2105.02943}{{arXiv:2105.02943}}} {[gr-qc]}

\bibitem[{{Siemens} et~al.(2007){Siemens}, {Mandic}, and
  {Creighton}}]{2007PhRvL..98k1101S}
{Siemens} X, {Mandic} V, {Creighton} J (2007) {Gravitational-Wave Stochastic
  Background from Cosmic Strings}. \prl 98(11):111101.
  \doi{10.1103/PhysRevLett.98.111101}.
  {\href{https://arxiv.org/abs/astro-ph/0610920}{{arXiv:astro-ph/0610920}}}
  {[astro-ph]}

\bibitem[{{Sigurdsson} and {Phinney}(1993)}]{1993ApJ...415..631S}
{Sigurdsson} S, {Phinney} ES (1993) {Binary--Single Star Interactions in
  Globular Clusters}. \apj 415:631. \doi{10.1086/173190}

\bibitem[{{Sigurdsson} and {Rees}(1997)}]{1997MNRAS.284..318S}
{Sigurdsson} S, {Rees} MJ (1997) {Capture of stellar mass compact objects by
  massive black holes in galactic cusps}. \mnras 284(2):318--326.
  \doi{10.1093/mnras/284.2.318}.
  {\href{https://arxiv.org/abs/astro-ph/9608093}{{arXiv:astro-ph/9608093}}}
  {[astro-ph]}

\bibitem[{{Sigurdsson} et~al.(2003){Sigurdsson}, {Richer}, {Hansen}, {Stairs},
  and {Thorsett}}]{2003Sci...301..193S}
{Sigurdsson} S, {Richer} HB, {Hansen} BM, {Stairs} IH, {Thorsett} SE (2003) {A
  Young White Dwarf Companion to Pulsar B1620-26: Evidence for Early Planet
  Formation}. Science 301(5630):193--196. \doi{10.1126/science.1086326}.
  {\href{https://arxiv.org/abs/astro-ph/0307339}{{arXiv:astro-ph/0307339}}}
  {[astro-ph]}

\bibitem[{{Sijacki} et~al.(2007){Sijacki}, {Springel}, {Di Matteo}, and
  {Hernquist}}]{2007MNRAS.380..877S}
{Sijacki} D, {Springel} V, {Di Matteo} T, {Hernquist} L (2007) {A unified model
  for AGN feedback in cosmological simulations of structure formation}. \mnras
  380(3):877--900. \doi{10.1111/j.1365-2966.2007.12153.x}.
  {\href{https://arxiv.org/abs/0705.2238}{{arXiv:0705.2238}}} {[astro-ph]}

\bibitem[{{Sijacki} et~al.(2011){Sijacki}, {Springel}, and
  {Haehnelt}}]{2011MNRAS.414.3656S}
{Sijacki} D, {Springel} V, {Haehnelt} MG (2011) {Gravitational recoils of
  supermassive black holes in hydrodynamical simulations of gas-rich galaxies}.
  \mnras 414:3656--3670. \doi{10.1111/j.1365-2966.2011.18666.x}.
  {\href{https://arxiv.org/abs/1008.3313}{{arXiv:1008.3313}}}

\bibitem[{{Sijacki} et~al.(2015){Sijacki}, {Vogelsberger}, {Genel}, {Springel},
  {Torrey}, {Snyder}, {Nelson}, and {Hernquist}}]{2015MNRAS.452..575S}
{Sijacki} D, {Vogelsberger} M, {Genel} S, {Springel} V, {Torrey} P, {Snyder}
  GF, {Nelson} D, {Hernquist} L (2015) {The Illustris simulation: the evolving
  population of black holes across cosmic time}. \mnras 452(1):575--596.
  \doi{10.1093/mnras/stv1340}.
  {\href{https://arxiv.org/abs/1408.6842}{{arXiv:1408.6842}}} {[astro-ph.GA]}

\bibitem[{{Silsbee} and {Tremaine}(2017)}]{2017ApJ...836...39S}
{Silsbee} K, {Tremaine} S (2017) {Lidov-Kozai Cycles with Gravitational
  Radiation: Merging Black Holes in Isolated Triple Systems}. \apj 836(1):39.
  \doi{10.3847/1538-4357/aa5729}.
  {\href{https://arxiv.org/abs/1608.07642}{{arXiv:1608.07642}}} {[astro-ph.HE]}

\bibitem[{{Simmons} et~al.(2014){Simmons}, {Melvin}, {Lintott}, {Masters},
  {Willett}, {Keel}, {Smethurst}, {Cheung}, {Nichol}, {Schawinski},
  {Rutkowski}, {Kartaltepe}, {Bell}, {Casteels}, {Conselice}, {Almaini},
  {Ferguson}, {Fortson}, {Hartley}, {Kocevski}, {Koekemoer}, {McIntosh},
  {Mortlock}, {Newman}, {Ownsworth}, {Bamford}, {Dahlen}, {Faber},
  {Finkelstein}, {Fontana}, {Galametz}, {Grogin}, {Gr{\"u}tzbauch}, {Guo},
  {H{\"a}u{\ss}ler}, {Jek}, {Kaviraj}, {Lucas}, {Peth}, {Salvato}, {Wiklind},
  and {Wuyts}}]{2014MNRAS.445.3466S}
{Simmons} BD, {Melvin} T, {Lintott} C, {Masters} KL, {Willett} KW, {Keel} WC,
  {Smethurst} RJ, {Cheung} E, {Nichol} RC, {Schawinski} K, et~al. (2014)
  {Galaxy Zoo: CANDELS barred discs and bar fractions}. \mnras
  445(4):3466--3474. \doi{10.1093/mnras/stu1817}.
  {\href{https://arxiv.org/abs/1409.1214}{{arXiv:1409.1214}}} {[astro-ph.GA]}

\bibitem[{{Simon} and {Burke-Spolaor}(2016)}]{2016ApJ...826...11S}
{Simon} J, {Burke-Spolaor} S (2016) {Constraints on Black Hole/Host Galaxy
  Co-evolution and Binary Stalling Using Pulsar Timing Arrays}. \apj 826(1):11.
  \doi{10.3847/0004-637X/826/1/11}.
  {\href{https://arxiv.org/abs/1603.06577}{{arXiv:1603.06577}}} {[astro-ph.GA]}

\bibitem[{{Singh} et~al.(2014){Singh}, {Wu}, and {Sarty}}]{2014MNRAS.441..800S}
{Singh} D, {Wu} K, {Sarty} GE (2014) {Fast spinning pulsars as probes of
  massive black holes' gravity}. \mnras 441(1):800--808.
  \doi{10.1093/mnras/stu614}.
  {\href{https://arxiv.org/abs/1403.7171}{{arXiv:1403.7171}}} {[astro-ph.HE]}

\bibitem[{{Sippel} and {Hurley}(2013)}]{2013MNRAS.430L..30S}
{Sippel} AC, {Hurley} JR (2013) {Multiple stellar-mass black holes in globular
  clusters: theoretical confirmation.} \mnras 430:L30--L34.
  \doi{10.1093/mnrasl/sls044}.
  {\href{https://arxiv.org/abs/1211.6608}{{arXiv:1211.6608}}} {[astro-ph.GA]}

\bibitem[{Sirko and Goodman(2003)}]{SG}
Sirko E, Goodman J (2003) {Spectral energy distributions of marginally
  self-gravitating quasi-stellar object discs}. MNRAS 341(2):501--508.
  \doi{10.1046/j.1365-8711.2003.06431.x},
  \urlprefix\url{http://dx.doi.org/10.1046/j.1365-8711.2003.06431.x}

\bibitem[{{S{\k{a}}dowski} and {Gaspari}(2017)}]{2017MNRAS.468.1398S}
{S{\k{a}}dowski} A, {Gaspari} M (2017) {Kinetic and radiative power from
  optically thin accretion flows}. \mnras 468(2):1398--1404.
  \doi{10.1093/mnras/stx543}.
  {\href{https://arxiv.org/abs/1701.07033}{{arXiv:1701.07033}}} {[astro-ph.HE]}

\bibitem[{{S{\k{a}}dowski} and {Narayan}(2016)}]{2016MNRAS.456.3929S}
{S{\k{a}}dowski} A, {Narayan} R (2016) {Three-dimensional simulations of
  supercritical black hole accretion discs - luminosities, photon trapping and
  variability}. \mnras 456(4):3929--3947. \doi{10.1093/mnras/stv2941}.
  {\href{https://arxiv.org/abs/1509.03168}{{arXiv:1509.03168}}} {[astro-ph.HE]}

\bibitem[{{Skillman} et~al.(1999){Skillman}, {Patterson}, {Kemp}, {Harvey},
  {Fried}, {Retter}, {Lipkin}, and {Vanmunster}}]{1999PASP..111.1281S}
{Skillman} DR, {Patterson} J, {Kemp} J, {Harvey} DA, {Fried} RE, {Retter} A,
  {Lipkin} Y, {Vanmunster} T (1999) {Superhumps in Cataclysmic Binaries. XVII.
  AM Canum Venaticorum}. \pasp 111(764):1281--1291. \doi{10.1086/316437}

\bibitem[{{Smarr} and {Blandford}(1976)}]{1976ApJ...207..574S}
{Smarr} LL, {Blandford} R (1976) {The binary pulsar: physical processes,
  possible companions, and evolutionary histories.} \apj 207:574--588.
  \doi{10.1086/154524}

\bibitem[{{Smith} et~al.(2015){Smith}, {Wise}, {O'Shea}, {Norman}, and
  {Khochfar}}]{2015MNRAS.452.2822S}
{Smith} BD, {Wise} JH, {O'Shea} BW, {Norman} ML, {Khochfar} S (2015) {The first
  Population II stars formed in externally enriched mini-haloes}. \mnras
  452:2822--2836. \doi{10.1093/mnras/stv1509}.
  {\href{https://arxiv.org/abs/1504.07639}{{arXiv:1504.07639}}}

\bibitem[{{Smith} et~al.(2018){Smith}, {Regan}, {Downes}, {Norman}, {O'Shea},
  and {Wise}}]{2018MNRAS.480.3762S}
{Smith} BD, {Regan} JA, {Downes} TP, {Norman} ML, {O'Shea} BW, {Wise} JH (2018)
  {The growth of black holes from Population III remnants in the Renaissance
  simulations}. \mnras 480(3):3762--3773. \doi{10.1093/mnras/sty2103}.
  {\href{https://arxiv.org/abs/1804.06477}{{arXiv:1804.06477}}} {[astro-ph.GA]}

\bibitem[{{Smith} and {Caldwell}(2019)}]{2019PhRvD.100j4055S}
{Smith} TL, {Caldwell} RR (2019) {LISA for cosmologists: Calculating the
  signal-to-noise ratio for stochastic and deterministic sources}. \prd
  100(10):104055. \doi{10.1103/PhysRevD.100.104055}.
  {\href{https://arxiv.org/abs/1908.00546}{{arXiv:1908.00546}}} {[astro-ph.CO]}

\bibitem[{{Snyder} et~al.(2019){Snyder}, {Rodriguez-Gomez}, {Lotz}, {Torrey},
  {Quirk}, {Hernquist}, {Vogelsberger}, and {Freeman}}]{2019MNRAS.486.3702S}
{Snyder} GF, {Rodriguez-Gomez} V, {Lotz} JM, {Torrey} P, {Quirk} ACN,
  {Hernquist} L, {Vogelsberger} M, {Freeman} PE (2019) {Automated distant
  galaxy merger classifications from Space Telescope images using the Illustris
  simulation}. \mnras 486(3):3702--3720. \doi{10.1093/mnras/stz1059}.
  {\href{https://arxiv.org/abs/1809.02136}{{arXiv:1809.02136}}} {[astro-ph.GA]}

\bibitem[{{Soberman} et~al.(1997){Soberman}, {Phinney}, and {van den
  Heuvel}}]{1997A&A...327..620S}
{Soberman} GE, {Phinney} ES, {van den Heuvel} EPJ (1997) {Stability criteria
  for mass transfer in binary stellar evolution.} \aap 327:620--635.
  {\href{https://arxiv.org/abs/astro-ph/9703016}{{arXiv:astro-ph/9703016}}}
  {[astro-ph]}

\bibitem[{{Soker} and {Tylenda}(2003)}]{2003ApJ...582L.105S}
{Soker} N, {Tylenda} R (2003) {Main-Sequence Stellar Eruption Model for V838
  Monocerotis}. \apjl 582(2):L105--L108. \doi{10.1086/367759}.
  {\href{https://arxiv.org/abs/astro-ph/0210463}{{arXiv:astro-ph/0210463}}}
  {[astro-ph]}

\bibitem[{{Solanki} et~al.(2021){Solanki}, {Kupfer}, {Blaes}, {Breedt}, and
  {Scaringi}}]{2021MNRAS.500.1222S}
{Solanki} S, {Kupfer} T, {Blaes} O, {Breedt} E, {Scaringi} S (2021)
  {Periodicities in the K2 light curve of HP Librae}. \mnras 500(1):1222--1230.
  \doi{10.1093/mnras/staa3240}.
  {\href{https://arxiv.org/abs/2010.09754}{{arXiv:2010.09754}}} {[astro-ph.HE]}

\bibitem[{{Solheim}(2010)}]{2010PASP..122.1133S}
{Solheim} JE (2010) {AM CVn Stars: Status and Challenges}. \pasp 122(896):1133.
  \doi{10.1086/656680}

\bibitem[{{Soltan}(1982)}]{1982MNRAS.200..115S}
{Soltan} A (1982) {Masses of quasars.} \mnras 200:115--122.
  \doi{10.1093/mnras/200.1.115}

\bibitem[{{Somerville} et~al.(2008){Somerville}, {Hopkins}, {Cox}, {Robertson},
  and {Hernquist}}]{2008MNRAS.391..481S}
{Somerville} RS, {Hopkins} PF, {Cox} TJ, {Robertson} BE, {Hernquist} L (2008)
  {A semi-analytic model for the co-evolution of galaxies, black holes and
  active galactic nuclei}. \mnras 391(2):481--506.
  \doi{10.1111/j.1365-2966.2008.13805.x}.
  {\href{https://arxiv.org/abs/0808.1227}{{arXiv:0808.1227}}} {[astro-ph]}

\bibitem[{{Souza Lima} et~al.(2017){Souza Lima}, {Mayer}, {Capelo}, and
  {Bellovary}}]{2017ApJ...838...13S}
{Souza Lima} R, {Mayer} L, {Capelo} PR, {Bellovary} JM (2017) {The Pairing of
  Accreting Massive Black Holes in Multiphase Circumnuclear Disks: the
  Interplay Between Radiative Cooling, Star Formation, and Feedback Processes}.
  \apj 838(1):13. \doi{10.3847/1538-4357/aa5d19}.
  {\href{https://arxiv.org/abs/1610.01600}{{arXiv:1610.01600}}} {[astro-ph.GA]}

\bibitem[{{Souza Lima} et~al.(2020){Souza Lima}, {Mayer}, {Capelo}, {Bortolas},
  and {Quinn}}]{2020ApJ...899..126S}
{Souza Lima} R, {Mayer} L, {Capelo} PR, {Bortolas} E, {Quinn} TR (2020) {The
  Erratic Path to Coalescence of LISA Massive Black Hole Binaries in
  Subparsec-resolution Simulations of Smooth Circumnuclear Gas Disks}. \apj
  899(2):126. \doi{10.3847/1538-4357/aba624}.
  {\href{https://arxiv.org/abs/2003.13789}{{arXiv:2003.13789}}} {[astro-ph.GA]}

\bibitem[{{Spera} and {Mapelli}(2017)}]{2017MNRAS.470.4739S}
{Spera} M, {Mapelli} M (2017) {Very massive stars, pair-instability supernovae
  and intermediate-mass black holes with the sevn code}. \mnras
  470(4):4739--4749. \doi{10.1093/mnras/stx1576}.
  {\href{https://arxiv.org/abs/1706.06109}{{arXiv:1706.06109}}} {[astro-ph.SR]}

\bibitem[{{Spera} et~al.(2015){Spera}, {Mapelli}, and
  {Bressan}}]{2015MNRAS.451.4086S}
{Spera} M, {Mapelli} M, {Bressan} A (2015) {The mass spectrum of compact
  remnants from the PARSEC stellar evolution tracks}. \mnras 451(4):4086--4103.
  \doi{10.1093/mnras/stv1161}.
  {\href{https://arxiv.org/abs/1505.05201}{{arXiv:1505.05201}}} {[astro-ph.SR]}

\bibitem[{{Spergel} et~al.(2015){Spergel}, {Gehrels}, {Baltay}, {Bennett},
  {Breckinridge}, {Donahue}, {Dressler}, {Gaudi}, {Greene}, {Guyon}, {Hirata},
  {Kalirai}, {Kasdin}, {Macintosh}, {Moos}, {Perlmutter}, {Postman},
  {Rauscher}, {Rhodes}, {Wang}, {Weinberg}, {Benford}, {Hudson}, {Jeong},
  {Mellier}, {Traub}, {Yamada}, {Capak}, {Colbert}, {Masters}, {Penny},
  {Savransky}, {Stern}, {Zimmerman}, {Barry}, {Bartusek}, {Carpenter}, {Cheng},
  {Content}, {Dekens}, {Demers}, {Grady}, {Jackson}, {Kuan}, {Kruk}, {Melton},
  {Nemati}, {Parvin}, {Poberezhskiy}, {Peddie}, {Ruffa}, {Wallace}, {Whipple},
  {Wollack}, and {Zhao}}]{2015arXiv150303757S}
{Spergel} D, {Gehrels} N, {Baltay} C, {Bennett} D, {Breckinridge} J, {Donahue}
  M, {Dressler} A, {Gaudi} BS, {Greene} T, {Guyon} O, et~al. (2015) {Wide-Field
  InfrarRed Survey Telescope-Astrophysics Focused Telescope Assets WFIRST-AFTA
  2015 Report}. arXiv e-prints arXiv:1503.03757.
  {\href{https://arxiv.org/abs/1503.03757}{{arXiv:1503.03757}}} {[astro-ph.IM]}

\bibitem[{{Sperhake}(2015)}]{2015CQGra..32l4011S}
{Sperhake} U (2015) {The numerical relativity breakthrough for binary black
  holes}. Classical and Quantum Gravity 32(12):124011.
  \doi{10.1088/0264-9381/32/12/124011}.
  {\href{https://arxiv.org/abs/1411.3997}{{arXiv:1411.3997}}} {[gr-qc]}

\bibitem[{{Spitzer}(1969)}]{1969ApJ...158L.139S}
{Spitzer} J Lyman (1969) {Equipartition and the Formation of Compact Nuclei in
  Spherical Stellar Systems}. \apjl 158:L139. \doi{10.1086/180451}

\bibitem[{{Spitzer} and {Schwarzschild}(1951)}]{1951ApJ...114..385S}
{Spitzer} J Lyman, {Schwarzschild} M (1951) {The Possible Influence of
  Interstellar Clouds on Stellar Velocities.} \apj 114:385.
  \doi{10.1086/145478}

\bibitem[{{Spitzer}(1987)}]{1987degc.book.....S}
{Spitzer} L (1987) {Dynamical evolution of globular clusters}

\bibitem[{{Springel} et~al.(2001){Springel}, {White}, {Tormen}, and
  {Kauffmann}}]{2001MNRAS.328..726S}
{Springel} V, {White} SDM, {Tormen} G, {Kauffmann} G (2001) {Populating a
  cluster of galaxies - I. Results at [formmu2]z=0}. \mnras 328(3):726--750.
  \doi{10.1046/j.1365-8711.2001.04912.x}.
  {\href{https://arxiv.org/abs/astro-ph/0012055}{{arXiv:astro-ph/0012055}}}
  {[astro-ph]}

\bibitem[{{Springel} et~al.(2005{\natexlab{a}}){Springel}, {Di Matteo}, and
  {Hernquist}}]{2005ApJ...620L..79S}
{Springel} V, {Di Matteo} T, {Hernquist} L (2005{\natexlab{a}}) {Black Holes in
  Galaxy Mergers: The Formation of Red Elliptical Galaxies}. \apjl
  620(2):L79--L82. \doi{10.1086/428772}.
  {\href{https://arxiv.org/abs/astro-ph/0409436}{{arXiv:astro-ph/0409436}}}
  {[astro-ph]}

\bibitem[{{Springel} et~al.(2005{\natexlab{b}}){Springel}, {Di Matteo}, and
  {Hernquist}}]{2005MNRAS.361..776S}
{Springel} V, {Di Matteo} T, {Hernquist} L (2005{\natexlab{b}}) {Modelling
  feedback from stars and black holes in galaxy mergers}. \mnras
  361(3):776--794. \doi{10.1111/j.1365-2966.2005.09238.x}.
  {\href{https://arxiv.org/abs/astro-ph/0411108}{{arXiv:astro-ph/0411108}}}
  {[astro-ph]}

\bibitem[{{Sridhar} and {Touma}(2016)}]{2016MNRAS.458.4143S}
{Sridhar} S, {Touma} JR (2016) {Stellar dynamics around a massive black hole -
  II. Resonant relaxation}. \mnras 458(4):4143--4161.
  \doi{10.1093/mnras/stw543}.
  {\href{https://arxiv.org/abs/1509.02401}{{arXiv:1509.02401}}} {[astro-ph.GA]}

\bibitem[{{Stadel}(2001)}]{2001PhDT........21S}
{Stadel} JG (2001) {Cosmological N-body simulations and their analysis}. PhD
  thesis, UNIVERSITY OF WASHINGTON

\bibitem[{{Stanway} and {Eldridge}(2018)}]{2018MNRAS.479...75S}
{Stanway} ER, {Eldridge} JJ (2018) {Re-evaluating old stellar populations}.
  \mnras 479(1):75--93. \doi{10.1093/mnras/sty1353}.
  {\href{https://arxiv.org/abs/1805.08784}{{arXiv:1805.08784}}} {[astro-ph.GA]}

\bibitem[{{Steele} et~al.(2014){Steele}, {Zepf}, {Maccarone}, {Kundu}, {Rhode},
  and {Salzer}}]{2014ApJ...785..147S}
{Steele} MM, {Zepf} SE, {Maccarone} TJ, {Kundu} A, {Rhode} KL, {Salzer} JJ
  (2014) {Composition of an Emission Line System in Black Hole Host Globular
  Cluster RZ2109}. \apj 785(2):147. \doi{10.1088/0004-637X/785/2/147}.
  {\href{https://arxiv.org/abs/1403.2784}{{arXiv:1403.2784}}} {[astro-ph.GA]}

\bibitem[{{Stella}(1987)}]{1987vgex.conf..157S}
{Stella} L (1987) {A 685 second orbital period from the X-ray source 4U 1820-30
  in the globular cluster NGC 6624.} In: Variability of Galactic and
  Extragalactic X-ray Sources. pp 157--164

\bibitem[{{Stella} et~al.(1987){Stella}, {Priedhorsky}, and
  {White}}]{1987ApJ...312L..17S}
{Stella} L, {Priedhorsky} W, {White} NE (1987) {The Discovery of a Second
  Orbital Period from the X-Ray Source 4U 1820-30 in the Globular Cluster NGC
  6624}. \apjl 312:L17. \doi{10.1086/184811}

\bibitem[{{Stephan} et~al.(2016){Stephan}, {Naoz}, {Ghez}, {Witzel},
  {Sitarski}, {Do}, and {Kocsis}}]{2016MNRAS.460.3494S}
{Stephan} AP, {Naoz} S, {Ghez} AM, {Witzel} G, {Sitarski} BN, {Do} T, {Kocsis}
  B (2016) {Merging binaries in the Galactic Center: the eccentric Kozai-Lidov
  mechanism with stellar evolution}. \mnras 460(4):3494--3504.
  \doi{10.1093/mnras/stw1220}.
  {\href{https://arxiv.org/abs/1603.02709}{{arXiv:1603.02709}}} {[astro-ph.SR]}

\bibitem[{{Stephan} et~al.(2018){Stephan}, {Naoz}, and
  {Gaudi}}]{2018AJ....156..128S}
{Stephan} AP, {Naoz} S, {Gaudi} BS (2018) {A-type Stars, the Destroyers of
  Worlds: The Lives and Deaths of Jupiters in Evolving Stellar Binaries}. \aj
  156(3):128. \doi{10.3847/1538-3881/aad6e5}.
  {\href{https://arxiv.org/abs/1806.04145}{{arXiv:1806.04145}}} {[astro-ph.SR]}

\bibitem[{{Stephan} et~al.(2019){Stephan}, {Naoz}, {Ghez}, {Morris}, {Ciurlo},
  {Do}, {Breivik}, {Coughlin}, and {Rodriguez}}]{2019ApJ...878...58S}
{Stephan} AP, {Naoz} S, {Ghez} AM, {Morris} MR, {Ciurlo} A, {Do} T, {Breivik}
  K, {Coughlin} S, {Rodriguez} CL (2019) {The Fate of Binaries in the Galactic
  Center: The Mundane and the Exotic}. \apj 878(1):58.
  \doi{10.3847/1538-4357/ab1e4d}.
  {\href{https://arxiv.org/abs/1903.00010}{{arXiv:1903.00010}}} {[astro-ph.SR]}

\bibitem[{{Stephan} et~al.(2020){Stephan}, {Naoz}, {Gaudi}, and
  {Salas}}]{2020ApJ...889...45S}
{Stephan} AP, {Naoz} S, {Gaudi} BS, {Salas} JM (2020) {Eating Planets for Lunch
  and Dinner: Signatures of Planet Consumption by Evolving Stars}. \apj
  889(1):45. \doi{10.3847/1538-4357/ab5b00}.
  {\href{https://arxiv.org/abs/1909.05259}{{arXiv:1909.05259}}} {[astro-ph.SR]}

\bibitem[{{Stevenson} et~al.(2017{\natexlab{a}}){Stevenson}, {Berry}, and
  {Mandel}}]{Stevenson:2017spin}
{Stevenson} S, {Berry} CPL, {Mandel} I (2017{\natexlab{a}}) {Hierarchical
  analysis of gravitational-wave measurements of binary black hole spin-orbit
  misalignments}. \mnras 471:2801--2811. \doi{10.1093/mnras/stx1764}.
  {\href{https://arxiv.org/abs/1703.06873}{{arXiv:1703.06873}}} {[astro-ph.HE]}

\bibitem[{{Stevenson} et~al.(2017{\natexlab{b}}){Stevenson}, {Vigna-G{\'o}mez},
  {Mandel}, {Barrett}, {Neijssel}, {Perkins}, and {de
  Mink}}]{2017NatCo...814906S}
{Stevenson} S, {Vigna-G{\'o}mez} A, {Mandel} I, {Barrett} JW, {Neijssel} CJ,
  {Perkins} D, {de Mink} SE (2017{\natexlab{b}}) {Formation of the first three
  gravitational-wave observations through isolated binary evolution}. Nature
  Communications 8:14906. \doi{10.1038/ncomms14906}.
  {\href{https://arxiv.org/abs/1704.01352}{{arXiv:1704.01352}}} {[astro-ph.HE]}

\bibitem[{{Stone} et~al.(2013){Stone}, {Sari}, and
  {Loeb}}]{2013MNRAS.435.1809S}
{Stone} N, {Sari} R, {Loeb} A (2013) {Consequences of strong compression in
  tidal disruption events}. \mnras 435(3):1809--1824.
  \doi{10.1093/mnras/stt1270}.
  {\href{https://arxiv.org/abs/1210.3374}{{arXiv:1210.3374}}} {[astro-ph.HE]}

\bibitem[{{Stone} and {van Velzen}(2016)}]{2016ApJ...825L..14S}
{Stone} NC, {van Velzen} S (2016) {An Enhanced Rate of Tidal Disruptions in the
  Centrally Overdense E+A Galaxy NGC 3156}. \apjl 825(1):L14.
  \doi{10.3847/2041-8205/825/1/L14}.
  {\href{https://arxiv.org/abs/1604.02056}{{arXiv:1604.02056}}} {[astro-ph.GA]}

\bibitem[{{Stone} et~al.(2017{\natexlab{a}}){Stone}, {K{\"u}pper}, and
  {Ostriker}}]{2017MNRAS.467.4180S}
{Stone} NC, {K{\"u}pper} AHW, {Ostriker} JP (2017{\natexlab{a}}) {Formation of
  massive black holes in galactic nuclei: runaway tidal encounters}. \mnras
  467(4):4180--4199. \doi{10.1093/mnras/stx097}.
  {\href{https://arxiv.org/abs/1606.01909}{{arXiv:1606.01909}}} {[astro-ph.GA]}

\bibitem[{{Stone} et~al.(2017{\natexlab{b}}){Stone}, {Metzger}, and
  {Haiman}}]{2017MNRAS.464..946S}
{Stone} NC, {Metzger} BD, {Haiman} Z (2017{\natexlab{b}}) {Assisted inspirals
  of stellar mass black holes embedded in AGN discs: solving the `final au
  problem'}. \mnras 464(1):946--954. \doi{10.1093/mnras/stw2260}.
  {\href{https://arxiv.org/abs/1602.04226}{{arXiv:1602.04226}}} {[astro-ph.GA]}

\bibitem[{{Stone} et~al.(2018){Stone}, {Generozov}, {Vasiliev}, and
  {Metzger}}]{2018MNRAS.480.5060S}
{Stone} NC, {Generozov} A, {Vasiliev} E, {Metzger} BD (2018) {The delay time
  distribution of tidal disruption flares}. \mnras 480(4):5060--5077.
  \doi{10.1093/mnras/sty2045}.
  {\href{https://arxiv.org/abs/1709.00423}{{arXiv:1709.00423}}} {[astro-ph.GA]}

\bibitem[{{Strader} et~al.(2012){Strader}, {Chomiuk}, {Maccarone},
  {Miller-Jones}, and {Seth}}]{2012Natur.490...71S}
{Strader} J, {Chomiuk} L, {Maccarone} TJ, {Miller-Jones} JCA, {Seth} AC (2012)
  {Two stellar-mass black holes in the globular cluster M22}. \nat
  490(7418):71--73. \doi{10.1038/nature11490}.
  {\href{https://arxiv.org/abs/1210.0901}{{arXiv:1210.0901}}} {[astro-ph.HE]}

\bibitem[{{Stritzinger} et~al.(2020){Stritzinger}, {Taddia}, {Fraser},
  {Tauris}, {Contreras}, {Drybye}, {Galbany}, {Holmbo}, {Morrell},
  {Pastorello}, {Phillips}, {Pignata}, {Tartaglia}, {Suntzeff}, {Anais},
  {Ashall}, {Baron}, {Burns}, {Hoeflich}, {Hsiao}, {Karamehmetoglu}, {Moriya},
  {Bock}, {Campillay}, {Castell{\'o}n}, {Inserra}, {Gonz{\'a}lez}, {Marples},
  {Parker}, {Reichart}, {Torres-Robledo}, and {Young}}]{2020A&A...639A.104S}
{Stritzinger} MD, {Taddia} F, {Fraser} M, {Tauris} TM, {Contreras} C, {Drybye}
  S, {Galbany} L, {Holmbo} S, {Morrell} N, {Pastorello} A, et~al. (2020) {The
  Carnegie Supernova Project II. Observations of the luminous red nova AT
  2014ej}. \aap 639:A104. \doi{10.1051/0004-6361/202038019}.
  {\href{https://arxiv.org/abs/2005.00076}{{arXiv:2005.00076}}} {[astro-ph.HE]}

\bibitem[{{Stroeer} and {Vecchio}(2006)}]{2006CQGra..23S.809S}
{Stroeer} A, {Vecchio} A (2006) {The LISA verification binaries}. Classical and
  Quantum Gravity 23(19):S809--S817. \doi{10.1088/0264-9381/23/19/S19}.
  {\href{https://arxiv.org/abs/astro-ph/0605227}{{arXiv:astro-ph/0605227}}}
  {[astro-ph]}

\bibitem[{{Stroeer} et~al.(2006){Stroeer}, {Gair}, and
  {Vecchio}}]{2006AIPC..873..444S}
{Stroeer} A, {Gair} J, {Vecchio} A (2006) {Automatic Bayesian inference for
  LISA data analysis strategies}. In: {Merkovitz} SM, {Livas} JC (eds) Laser
  Interferometer Space Antenna: 6th International LISA Symposium. American
  Institute of Physics Conference Series, vol 873. pp 444--451.
  \doi{10.1063/1.2405082}.
  {\href{https://arxiv.org/abs/gr-qc/0609010}{{arXiv:gr-qc/0609010}}} {[gr-qc]}

\bibitem[{{Strohmayer}(2004)}]{2004ApJ...614..358S}
{Strohmayer} TE (2004) {Chandra Detection of the AM Canum Venaticorum Binary ES
  Ceti (KUV 01584-0939)}. \apj 614(1):358--362. \doi{10.1086/423615}.
  {\href{https://arxiv.org/abs/astro-ph/0405203}{{arXiv:astro-ph/0405203}}}
  {[astro-ph]}

\bibitem[{{Strohmayer}(2005)}]{2005ApJ...627..920S}
{Strohmayer} TE (2005) {Precision X-Ray Timing of RX J0806.3+1527 with Chandra:
  Evidence for Gravitational Radiation from an Ultracompact Binary}. \apj
  627(2):920--925. \doi{10.1086/430439}.
  {\href{https://arxiv.org/abs/astro-ph/0504150}{{arXiv:astro-ph/0504150}}}
  {[astro-ph]}

\bibitem[{{Strubbe} and {Quataert}(2009)}]{2009MNRAS.400.2070S}
{Strubbe} LE, {Quataert} E (2009) {Optical flares from the tidal disruption of
  stars by massive black holes}. \mnras 400(4):2070--2084.
  \doi{10.1111/j.1365-2966.2009.15599.x}.
  {\href{https://arxiv.org/abs/0905.3735}{{arXiv:0905.3735}}} {[astro-ph.CO]}

\bibitem[{Sun et~al.(2017)Sun, Paschalidis, Ruiz, and Shapiro}]{Sun:2017voo}
Sun L, Paschalidis V, Ruiz M, Shapiro SL (2017) {Magnetorotational Collapse of
  Supermassive Stars: Black Hole Formation, Gravitational Waves and Jets}. Phys
  Rev D 96(4):043006. \doi{10.1103/PhysRevD.96.043006}.
  {\href{https://arxiv.org/abs/1704.04502}{{arXiv:1704.04502}}} {[astro-ph.HE]}

\bibitem[{Sun et~al.(2018)Sun, Ruiz, and Shapiro}]{Sun:2018puk}
Sun L, Ruiz M, Shapiro SL (2018) {Simulating the Magnetorotational Collapse of
  Supermassive Stars: Incorporating Gas Pressure Perturbations and Different
  Rotation Profiles}. Phys Rev D 98(10):103008.
  \doi{10.1103/PhysRevD.98.103008}.
  {\href{https://arxiv.org/abs/1807.07970}{{arXiv:1807.07970}}} {[astro-ph.HE]}

\bibitem[{{Sutantyo}(1975)}]{1975A&A....44..227S}
{Sutantyo} W (1975) {The formation of globular cluster X-ray sources through
  neutron star - giant collisions.} \aap 44:227--230

\bibitem[{{Suvorov}(2021)}]{2021MNRAS.503.5495S}
{Suvorov} AG (2021) {Ultra-compact X-ray binaries as dual-line
  gravitational-wave sources}. \mnras 503(4):5495--5503.
  \doi{10.1093/mnras/stab825}.
  {\href{https://arxiv.org/abs/2103.09858}{{arXiv:2103.09858}}} {[astro-ph.HE]}

\bibitem[{{Syer} et~al.(1991){Syer}, {Clarke}, and
  {Rees}}]{1991MNRAS.250..505S}
{Syer} D, {Clarke} CJ, {Rees} MJ (1991) {Star-disc interactions near a massive
  black hole}. \mnras 250:505--512. \doi{10.1093/mnras/250.3.505}

\bibitem[{{Sz{\"o}lgy{\'e}n} and {Kocsis}(2018)}]{2018PhRvL.121j1101S}
{Sz{\"o}lgy{\'e}n} {\'A}, {Kocsis} B (2018) {Black Hole Disks in Galactic
  Nuclei}. \prl 121(10):101101. \doi{10.1103/PhysRevLett.121.101101}.
  {\href{https://arxiv.org/abs/1803.07090}{{arXiv:1803.07090}}} {[astro-ph.GA]}

\bibitem[{{Tacconi} et~al.(2018){Tacconi}, {Genzel}, {Saintonge}, {Combes},
  {Garc{\'\i}a-Burillo}, {Neri}, {Bolatto}, {Contini}, {F{\"o}rster Schreiber},
  {Lilly}, {Lutz}, {Wuyts}, {Accurso}, {Boissier}, {Boone}, {Bouch{\'e}},
  {Bournaud}, {Burkert}, {Carollo}, {Cooper}, {Cox}, {Feruglio}, {Freundlich},
  {Herrera-Camus}, {Juneau}, {Lippa}, {Naab}, {Renzini}, {Salome}, {Sternberg},
  {Tadaki}, {{\"U}bler}, {Walter}, {Weiner}, and {Weiss}}]{2018ApJ...853..179T}
{Tacconi} LJ, {Genzel} R, {Saintonge} A, {Combes} F, {Garc{\'\i}a-Burillo} S,
  {Neri} R, {Bolatto} A, {Contini} T, {F{\"o}rster Schreiber} NM, {Lilly} S,
  et~al. (2018) {PHIBSS: Unified Scaling Relations of Gas Depletion Time and
  Molecular Gas Fractions}. \apj 853(2):179. \doi{10.3847/1538-4357/aaa4b4}.
  {\href{https://arxiv.org/abs/1702.01140}{{arXiv:1702.01140}}} {[astro-ph.GA]}

\bibitem[{{Tagawa} et~al.(2020{\natexlab{a}}){Tagawa}, {Haiman}, and
  {Kocsis}}]{2020ApJ...898...25T}
{Tagawa} H, {Haiman} Z, {Kocsis} B (2020{\natexlab{a}}) {Formation and
  Evolution of Compact-object Binaries in AGN Disks}. \apj 898(1):25.
  \doi{10.3847/1538-4357/ab9b8c}.
  {\href{https://arxiv.org/abs/1912.08218}{{arXiv:1912.08218}}} {[astro-ph.GA]}

\bibitem[{{Tagawa} et~al.(2020{\natexlab{b}}){Tagawa}, {Haiman}, and
  {Kocsis}}]{2020ApJ...892...36T}
{Tagawa} H, {Haiman} Z, {Kocsis} B (2020{\natexlab{b}}) {Making a Supermassive
  Star by Stellar Bombardment}. \apj 892(1):36. \doi{10.3847/1538-4357/ab7922}.
  {\href{https://arxiv.org/abs/1909.10517}{{arXiv:1909.10517}}} {[astro-ph.GA]}

\bibitem[{{Tak{\'a}cs} and {Kocsis}(2018)}]{2018ApJ...856..113T}
{Tak{\'a}cs} {\'A}, {Kocsis} B (2018) {Isotropic-Nematic Phase Transitions in
  Gravitational Systems. II. Higher Order Multipoles}. \apj 856(2):113.
  \doi{10.3847/1538-4357/aab268}.
  {\href{https://arxiv.org/abs/1712.04449}{{arXiv:1712.04449}}} {[astro-ph.GA]}

\bibitem[{{Takekawa} et~al.(2019){Takekawa}, {Oka}, {Iwata}, {Tsujimoto}, and
  {Nomura}}]{2019ApJ...871L...1T}
{Takekawa} S, {Oka} T, {Iwata} Y, {Tsujimoto} S, {Nomura} M (2019) {Indication
  of Another Intermediate-mass Black Hole in the Galactic Center}. \apjl
  871(1):L1. \doi{10.3847/2041-8213/aafb07}.
  {\href{https://arxiv.org/abs/1812.10733}{{arXiv:1812.10733}}} {[astro-ph.GA]}

\bibitem[{{Takekawa} et~al.(2020){Takekawa}, {Oka}, {Iwata}, {Tsujimoto}, and
  {Nomura}}]{2020ApJ...890..167T}
{Takekawa} S, {Oka} T, {Iwata} Y, {Tsujimoto} S, {Nomura} M (2020) {The Fifth
  Candidate for an Intermediate-mass Black Hole in the Galactic Center}. \apj
  890(2):167. \doi{10.3847/1538-4357/ab6f6f}.
  {\href{https://arxiv.org/abs/2002.05173}{{arXiv:2002.05173}}} {[astro-ph.GA]}

\bibitem[{{Tamai} et~al.(2016){Tamai}, {Cirasuolo}, {Gonz{\'a}lez}, {Koehler},
  and {Tuti}}]{2016SPIE.9906E..0WT}
{Tamai} R, {Cirasuolo} M, {Gonz{\'a}lez} JC, {Koehler} B, {Tuti} M (2016) {The
  E-ELT program status}. In: {Hall} HJ, {Gilmozzi} R, {Marshall} HK (eds)
  Ground-based and Airborne Telescopes VI. Society of Photo-Optical
  Instrumentation Engineers (SPIE) Conference Series, vol 9906. p 99060W.
  \doi{10.1117/12.2232690}

\bibitem[{{Tamanini} and {Danielski}(2019)}]{2019NatAs...3..858T}
{Tamanini} N, {Danielski} C (2019) {The gravitational-wave detection of
  exoplanets orbiting white dwarf binaries using LISA}. Nature Astronomy
  3:858--866. \doi{10.1038/s41550-019-0807-y}.
  {\href{https://arxiv.org/abs/1812.04330}{{arXiv:1812.04330}}} {[astro-ph.EP]}

\bibitem[{{Tamanini} et~al.(2016){Tamanini}, {Caprini}, {Barausse}, {Sesana},
  {Klein}, and {Petiteau}}]{2016JCAP...04..002T}
{Tamanini} N, {Caprini} C, {Barausse} E, {Sesana} A, {Klein} A, {Petiteau} A
  (2016) {Science with the space-based interferometer eLISA. III: probing the
  expansion of the universe using gravitational wave standard sirens}. \jcap
  2016(4):002. \doi{10.1088/1475-7516/2016/04/002}.
  {\href{https://arxiv.org/abs/1601.07112}{{arXiv:1601.07112}}} {[astro-ph.CO]}

\bibitem[{{Tamanini} et~al.(2020){Tamanini}, {Klein}, {Bonvin}, {Barausse}, and
  {Caprini}}]{2020PhRvD.101f3002T}
{Tamanini} N, {Klein} A, {Bonvin} C, {Barausse} E, {Caprini} C (2020) {Peculiar
  acceleration of stellar-origin black hole binaries: Measurement and biases
  with LISA}. \prd 101(6):063002. \doi{10.1103/PhysRevD.101.063002}.
  {\href{https://arxiv.org/abs/1907.02018}{{arXiv:1907.02018}}} {[astro-ph.IM]}

\bibitem[{{Tamburello} et~al.(2015){Tamburello}, {Mayer}, {Shen}, and
  {Wadsley}}]{2015MNRAS.453.2490T}
{Tamburello} V, {Mayer} L, {Shen} S, {Wadsley} J (2015) {A lower fragmentation
  mass scale in high-redshift galaxies and its implications on giant clumps: a
  systematic numerical study}. \mnras 453(3):2490--2514.
  \doi{10.1093/mnras/stv1695}.
  {\href{https://arxiv.org/abs/1412.3319}{{arXiv:1412.3319}}} {[astro-ph.GA]}

\bibitem[{{Tamburello} et~al.(2017{\natexlab{a}}){Tamburello}, {Capelo},
  {Mayer}, {Bellovary}, and {Wadsley}}]{2017MNRAS.464.2952T}
{Tamburello} V, {Capelo} PR, {Mayer} L, {Bellovary} JM, {Wadsley} JW
  (2017{\natexlab{a}}) {Supermassive black hole pairs in clumpy galaxies at
  high redshift: delayed binary formation and concurrent mass growth}. \mnras
  464(3):2952--2962. \doi{10.1093/mnras/stw2561}.
  {\href{https://arxiv.org/abs/1603.00021}{{arXiv:1603.00021}}} {[astro-ph.GA]}

\bibitem[{{Tamburello} et~al.(2017{\natexlab{b}}){Tamburello}, {Rahmati},
  {Mayer}, {Cava}, {Dessauges-Zavadsky}, and {Schaerer}}]{2017MNRAS.468.4792T}
{Tamburello} V, {Rahmati} A, {Mayer} L, {Cava} A, {Dessauges-Zavadsky} M,
  {Schaerer} D (2017{\natexlab{b}}) {Clumpy galaxies seen in H
  {\ensuremath{\alpha}}: inflated observed clump properties due to limited
  spatial resolution and sensitivity}. \mnras 468(4):4792--4800.
  \doi{10.1093/mnras/stx784}.
  {\href{https://arxiv.org/abs/1610.05304}{{arXiv:1610.05304}}} {[astro-ph.GA]}

\bibitem[{{Tamfal} et~al.(2018){Tamfal}, {Capelo}, {Kazantzidis}, {Mayer},
  {Potter}, {Stadel}, and {Widrow}}]{2018ApJ...864L..19T}
{Tamfal} T, {Capelo} PR, {Kazantzidis} S, {Mayer} L, {Potter} D, {Stadel} J,
  {Widrow} LM (2018) {Formation of LISA Black Hole Binaries in Merging Dwarf
  Galaxies: The Imprint of Dark Matter}. \apjl 864(1):L19.
  \doi{10.3847/2041-8213/aada4b}.
  {\href{https://arxiv.org/abs/1806.11112}{{arXiv:1806.11112}}} {[astro-ph.GA]}

\bibitem[{{Tamfal} et~al.(2021){Tamfal}, {Mayer}, {Quinn}, {Capelo},
  {Kazantzidis}, {Babul}, and {Potter}}]{2020arXiv200713763T}
{Tamfal} T, {Mayer} L, {Quinn} TR, {Capelo} PR, {Kazantzidis} S, {Babul} A,
  {Potter} D (2021) {Revisiting Dynamical Friction: The Role of Global Modes
  and Local Wakes}. \apj 916(1):55. \doi{10.3847/1538-4357/ac0627}.
  {\href{https://arxiv.org/abs/2007.13763}{{arXiv:2007.13763}}} {[astro-ph.GA]}

\bibitem[{{Tan} et~al.(1991){Tan}, {Morgan}, {Lewin}, {Penninx}, {van der
  Klis}, {van Paradijs}, {Makishima}, {Inoue}, {Dotani}, and
  {Mitsuda}}]{1991ApJ...374..291T}
{Tan} J, {Morgan} E, {Lewin} WHG, {Penninx} W, {van der Klis} M, {van Paradijs}
  J, {Makishima} K, {Inoue} H, {Dotani} T, {Mitsuda} K (1991) {Changes in the
  11 Minute Period of 4U 1820-30}. \apj 374:291. \doi{10.1086/170118}

\bibitem[{{Tanaka} and {Ward}(2004)}]{2004ApJ...602..388T}
{Tanaka} H, {Ward} WR (2004) {Three-dimensional Interaction between a Planet
  and an Isothermal Gaseous Disk. II. Eccentricity Waves and Bending Waves}.
  \apj 602(1):388--395. \doi{10.1086/380992}

\bibitem[{{Tanaka} et~al.(2002){Tanaka}, {Takeuchi}, and
  {Ward}}]{2002ApJ...565.1257T}
{Tanaka} H, {Takeuchi} T, {Ward} WR (2002) {Three-Dimensional Interaction
  between a Planet and an Isothermal Gaseous Disk. I. Corotation and Lindblad
  Torques and Planet Migration}. \apj 565(2):1257--1274. \doi{10.1086/324713}

\bibitem[{{Tanaka} and {Haiman}(2009)}]{2009ApJ...696.1798T}
{Tanaka} T, {Haiman} Z (2009) {The Assembly of Supermassive Black Holes at High
  Redshifts}. \apj 696(2):1798--1822. \doi{10.1088/0004-637X/696/2/1798}.
  {\href{https://arxiv.org/abs/0807.4702}{{arXiv:0807.4702}}} {[astro-ph]}

\bibitem[{{Tang} et~al.(2017){Tang}, {MacFadyen}, and
  {Haiman}}]{2017MNRAS.469.4258T}
{Tang} Y, {MacFadyen} A, {Haiman} Z (2017) {On the orbital evolution of
  supermassive black hole binaries with circumbinary accretion discs}. \mnras
  469(4):4258--4267. \doi{10.1093/mnras/stx1130}.
  {\href{https://arxiv.org/abs/1703.03913}{{arXiv:1703.03913}}} {[astro-ph.HE]}

\bibitem[{Tang et~al.(2018)}]{2018MNRAS.476.2249T}
Tang Y, et~al. (2018) The late inspiral of supermassive black hole binaries
  with circumbinary gas discs in the lisa band. MNRAS 476(2):2249--2257.
  \doi{10.1093/mnras/sty423},
  \urlprefix\url{https://doi.org/10.1093/mnras/sty423}.
  {\href{https://arxiv.org/abs/http://oup.prod.sis.lan/mnras/article-pdf/476/2/2249/24353233/sty423.pdf}{{http://oup.prod.sis.lan/mnras/article-pdf/476/2/2249/24353233/sty423.pdf}}}

\bibitem[{{Taracchini} et~al.(2014){Taracchini}, {Buonanno}, {Pan}, {Hinderer},
  {Boyle}, {Hemberger}, {Kidder}, {Lovelace}, {Mrou{\'e}}, {Pfeiffer},
  {Scheel}, {Szil{\'a}gyi}, {Taylor}, and {Zenginoglu}}]{2014PhRvD..89f1502T}
{Taracchini} A, {Buonanno} A, {Pan} Y, {Hinderer} T, {Boyle} M, {Hemberger} DA,
  {Kidder} LE, {Lovelace} G, {Mrou{\'e}} AH, {Pfeiffer} HP, et~al. (2014)
  {Effective-one-body model for black-hole binaries with generic mass ratios
  and spins}. \prd 89(6):061502. \doi{10.1103/PhysRevD.89.061502}.
  {\href{https://arxiv.org/abs/1311.2544}{{arXiv:1311.2544}}} {[gr-qc]}

\bibitem[{{Tashiro} et~al.(2018){Tashiro}, {Maejima}, {Toda}, {Kelley},
  {Reichenthal}, {Lobell}, {Petre}, {Guainazzi}, {Costantini}, {Edison},
  {Fujimoto}, {Grim}, {Hayashida}, {den Herder}, {Ishisaki}, {Paltani},
  {Matsushita}, {Mori}, {Sneiderman}, {Takei}, {Terada}, {Tomida}, {Akamatsu},
  {Angelini}, {Arai}, {Awaki}, {Babyk}, {Bamba}, {Barfknecht}, {Barnstable},
  {Bialas}, {Blagojevic}, {Bonafede}, {Brambora}, {Brenneman}, {Brown},
  {Brown}, {Burns}, {Canavan}, {Carnahan}, {Chiao}, {Comber}, {Corrales}, {de
  Vries}, {Dercksen}, {Diaz-Trigo}, {Dillard}, {DiPirro}, {Done}, {Dotani},
  {Ebisawa}, {Eckart}, {Enoto}, {Ezoe}, {Ferrigno}, {Fukazawa}, {Fujita},
  {Furuzawa}, {Gallo}, {Graham}, {Gu}, {Hagino}, {Hamaguchi}, {Hatsukade},
  {Hawes}, {Hayashi}, {Hegarty}, {Hell}, {Hiraga}, {Hodges-Kluck}, {Holland},
  {Hornschemeier}, {Hoshino}, {Ichinohe}, {Iizuka}, {Ishibashi}, {Ishida},
  {Ishikawa}, {Ishimura}, {James}, {Kallman}, {Kara}, {Katsuda}, {Kenyon},
  {Kilbourne}, {Kimball}, {Kitaguti}, {Kitamoto}, {Kobayashi}, {Kohmura},
  {Koyama}, {Kubota}, {Leutenegger}, {Lockard}, {Loewenstein}, {Maeda},
  {Marbley}, {Markevitch}, {Matsumoto}, {Matsuzaki}, {McCammon}, {McNamara},
  {Miko}, {Miller}, {Miller}, {Minesugi}, {Mitsuishi}, {Mizuno}, {Mori},
  {Mukai}, {Murakami}, {Mushotzky}, {Nakajima}, {Nakamura}, {Nakashima},
  {Nakazawa}, {Natsukari}, {Nigo}, {Nishioka}, {Nobukawa}, {Nobukawa}, {Noda},
  {Odaka}, {Ogawa}, {Ohashi}, {Ohno}, {Ohta}, {Okajima}, {Okamoto}, {Onizuka},
  {Ota}, {Ozaki}, {Plucinsky}, {Porter}, {Pottschmidt}, {Sato}, {Sato},
  {Sawada}, {Seta}, {Shelton}, {Shibano}, {Shida}, {Shidatsu}, {Shirron},
  {Simionescu}, {Smith}, {Someya}, {Soong}, {Suagawara}, {Szymkowiak},
  {Takahashi}, {Tamagawa}, {Tamura}, {Tanaka}, {Terashima}, {Tsuboi},
  {Tsujimoto}, {Tsunemi}, {Tsuru}, {Uchida}, {Uchiyama}, {Ueda}, {Uno},
  {Walsh}, {Watanabe}, {Williams}, {Wolfs}, {Wright}, {Yamada}, {Yamaguchi},
  {Yamaoka}, {Yamasaki}, {Yamauchi}, {Yamauchi}, {Yanagase}, {Yaqoob},
  {Yasuda}, {Yoshioka}, {Zabala}, and {Irina}}]{2018SPIE10699E..22T}
{Tashiro} M, {Maejima} H, {Toda} K, {Kelley} R, {Reichenthal} L, {Lobell} J,
  {Petre} R, {Guainazzi} M, {Costantini} E, {Edison} M, et~al. (2018) {Concept
  of the X-ray Astronomy Recovery Mission}. In: {den Herder} JWA, {Nikzad} S,
  {Nakazawa} K (eds) Space Telescopes and Instrumentation 2018: Ultraviolet to
  Gamma Ray. Society of Photo-Optical Instrumentation Engineers (SPIE)
  Conference Series, vol 10699. p 1069922. \doi{10.1117/12.2309455}

\bibitem[{{Tauris}(2018)}]{2018PhRvL.121m1105T}
{Tauris} TM (2018) {Disentangling Coalescing Neutron-Star-White-Dwarf Binaries
  for LISA}. \prl 121(13):131105. \doi{10.1103/PhysRevLett.121.131105}.
  {\href{https://arxiv.org/abs/1809.03504}{{arXiv:1809.03504}}} {[astro-ph.SR]}

\bibitem[{{Tauris} and {Dewi}(2001)}]{2001A&A...369..170T}
{Tauris} TM, {Dewi} JDM (2001) {Research Note On the binding energy parameter
  of common envelope evolution. Dependency on the definition of the stellar
  core boundary during spiral-in}. \aap 369:170--173.
  \doi{10.1051/0004-6361:20010099}.
  {\href{https://arxiv.org/abs/astro-ph/0101530}{{arXiv:astro-ph/0101530}}}
  {[astro-ph]}

\bibitem[{{Tauris} and {Savonije}(1999)}]{1999A&A...350..928T}
{Tauris} TM, {Savonije} GJ (1999) {Formation of millisecond pulsars. I.
  Evolution of low-mass X-ray binaries with P\_orb> 2 days}. \aap 350:928--944.
  {\href{https://arxiv.org/abs/astro-ph/9909147}{{arXiv:astro-ph/9909147}}}
  {[astro-ph]}

\bibitem[{{Tauris} and {Sennels}(2000)}]{2000A&A...355..236T}
{Tauris} TM, {Sennels} T (2000) {Formation of the binary pulsars PSR B2303+46
  and PSR J1141-6545. Young neutron stars with old white dwarf companions}.
  \aap 355:236--244.
  {\href{https://arxiv.org/abs/astro-ph/9909149}{{arXiv:astro-ph/9909149}}}
  {[astro-ph]}

\bibitem[{{Tauris} and {van den Heuvel}(2014)}]{2014ApJ...781L..13T}
{Tauris} TM, {van den Heuvel} EPJ (2014) {Formation of the Galactic Millisecond
  Pulsar Triple System PSR J0337+1715{\textemdash}A Neutron Star with Two
  Orbiting White Dwarfs}. \apjl 781(1):L13. \doi{10.1088/2041-8205/781/1/L13}.
  {\href{https://arxiv.org/abs/1401.0941}{{arXiv:1401.0941}}} {[astro-ph.SR]}

\bibitem[{{Tauris} and {van den Heuvel}(2023)}]{TvdH23}
{Tauris} TM, {van den Heuvel} EPJ (2023) {Physics of Binary Star Evolution.
  From Stars to X-ray Binaries and Gravitational Wave Sources}. Princeton
  University Press

\bibitem[{{Tauris} et~al.(2000){Tauris}, {van den Heuvel}, and
  {Savonije}}]{2000ApJ...530L..93T}
{Tauris} TM, {van den Heuvel} EPJ, {Savonije} GJ (2000) {Formation of
  Millisecond Pulsars with Heavy White Dwarf Companions:Extreme Mass Transfer
  on Subthermal Timescales}. \apjl 530(2):L93--L96. \doi{10.1086/312496}.
  {\href{https://arxiv.org/abs/astro-ph/0001013}{{arXiv:astro-ph/0001013}}}
  {[astro-ph]}

\bibitem[{{Tauris} et~al.(2012){Tauris}, {Langer}, and
  {Kramer}}]{2012MNRAS.425.1601T}
{Tauris} TM, {Langer} N, {Kramer} M (2012) {Formation of millisecond pulsars
  with CO white dwarf companions - II. Accretion, spin-up, true ages and
  comparison to MSPs with He white dwarf companions}. \mnras 425(3):1601--1627.
  \doi{10.1111/j.1365-2966.2012.21446.x}.
  {\href{https://arxiv.org/abs/1206.1862}{{arXiv:1206.1862}}} {[astro-ph.SR]}

\bibitem[{{Tauris} et~al.(2013){Tauris}, {Langer}, {Moriya}, {Podsiadlowski},
  {Yoon}, and {Blinnikov}}]{2013ApJ...778L..23T}
{Tauris} TM, {Langer} N, {Moriya} TJ, {Podsiadlowski} P, {Yoon} SC, {Blinnikov}
  SI (2013) {Ultra-stripped Type Ic Supernovae from Close Binary Evolution}.
  \apjl 778(2):L23. \doi{10.1088/2041-8205/778/2/L23}.
  {\href{https://arxiv.org/abs/1310.6356}{{arXiv:1310.6356}}} {[astro-ph.SR]}

\bibitem[{{Tauris} et~al.(2015){Tauris}, {Langer}, and
  {Podsiadlowski}}]{2015MNRAS.451.2123T}
{Tauris} TM, {Langer} N, {Podsiadlowski} P (2015) {Ultra-stripped supernovae:
  progenitors and fate}. \mnras 451(2):2123--2144. \doi{10.1093/mnras/stv990}.
  {\href{https://arxiv.org/abs/1505.00270}{{arXiv:1505.00270}}} {[astro-ph.SR]}

\bibitem[{{Tauris} et~al.(2017){Tauris}, {Kramer}, {Freire}, {Wex}, {Janka},
  {Langer}, {Podsiadlowski}, {Bozzo}, {Chaty}, {Kruckow}, {van den Heuvel},
  {Antoniadis}, {Breton}, and {Champion}}]{2017ApJ...846..170T}
{Tauris} TM, {Kramer} M, {Freire} PCC, {Wex} N, {Janka} HT, {Langer} N,
  {Podsiadlowski} P, {Bozzo} E, {Chaty} S, {Kruckow} MU, et~al. (2017)
  {Formation of Double Neutron Star Systems}. \apj 846(2):170.
  \doi{10.3847/1538-4357/aa7e89}.
  {\href{https://arxiv.org/abs/1706.09438}{{arXiv:1706.09438}}} {[astro-ph.HE]}

\bibitem[{{Tavani} and {Brookshaw}(1992)}]{1992Natur.356..320T}
{Tavani} M, {Brookshaw} L (1992) {The origin of planets orbiting millisecond
  pulsars}. \nat 356(6367):320--322. \doi{10.1038/356320a0}

\bibitem[{{Taylor} and {Kobayashi}(2014)}]{2014MNRAS.442.2751T}
{Taylor} P, {Kobayashi} C (2014) {Seeding black holes in cosmological
  simulations}. \mnras 442(3):2751--2767. \doi{10.1093/mnras/stu983}.
  {\href{https://arxiv.org/abs/1405.4194}{{arXiv:1405.4194}}} {[astro-ph.GA]}

\bibitem[{{Taylor} et~al.(2019){Taylor}, {Burke-Spolaor}, {Baker}, {Charisi},
  {Islo}, {Kelley}, {Madison}, {Simon}, {Vigeland}, and {Nanograv
  Collaboration}}]{2019BAAS...51c.336T}
{Taylor} S, {Burke-Spolaor} S, {Baker} PT, {Charisi} M, {Islo} K, {Kelley} LZ,
  {Madison} DR, {Simon} J, {Vigeland} S, {Nanograv Collaboration} (2019)
  {Supermassive Black-hole Demographics \&Environments With Pulsar Timing
  Arrays}. \baas 51(3):336.
  {\href{https://arxiv.org/abs/1903.08183}{{arXiv:1903.08183}}} {[astro-ph.GA]}

\bibitem[{{Taylor} et~al.(2016){Taylor}, {Vallisneri}, {Ellis}, {Mingarelli},
  {Lazio}, and {van Haasteren}}]{2016ApJ...819L...6T}
{Taylor} SR, {Vallisneri} M, {Ellis} JA, {Mingarelli} CMF, {Lazio} TJW, {van
  Haasteren} R (2016) {Are We There Yet? Time to Detection of Nanohertz
  Gravitational Waves Based on Pulsar-timing Array Limits}. \apjl 819(1):L6.
  \doi{10.3847/2041-8205/819/1/L6}.
  {\href{https://arxiv.org/abs/1511.05564}{{arXiv:1511.05564}}} {[astro-ph.IM]}

\bibitem[{{Taylor} et~al.(2017){Taylor}, {Simon}, and
  {Sampson}}]{2017PhRvL.118r1102T}
{Taylor} SR, {Simon} J, {Sampson} L (2017) {Constraints on the Dynamical
  Environments of Supermassive Black-Hole Binaries Using Pulsar-Timing Arrays}.
  \prl 118(18):181102. \doi{10.1103/PhysRevLett.118.181102}.
  {\href{https://arxiv.org/abs/1612.02817}{{arXiv:1612.02817}}} {[astro-ph.GA]}

\bibitem[{{Tazzari} and {Lodato}(2015)}]{2015MNRAS.449.1118T}
{Tazzari} M, {Lodato} G (2015) {Estimating the fossil disc mass during
  supermassive black hole mergers: the importance of torque implementation}.
  \mnras 449(1):1118--1128. \doi{10.1093/mnras/stv352}.
  {\href{https://arxiv.org/abs/1502.05046}{{arXiv:1502.05046}}} {[astro-ph.HE]}

\bibitem[{{Terrazas} et~al.(2016){Terrazas}, {Bell}, {Henriques}, {White},
  {Cattaneo}, and {Woo}}]{2016ApJ...830L..12T}
{Terrazas} BA, {Bell} EF, {Henriques} BMB, {White} SDM, {Cattaneo} A, {Woo} J
  (2016) {Quiescence Correlates Strongly with Directly Measured Black Hole Mass
  in Central Galaxies}. \apjl 830(1):L12. \doi{10.3847/2041-8205/830/1/L12}.
  {\href{https://arxiv.org/abs/1609.07141}{{arXiv:1609.07141}}} {[astro-ph.GA]}

\bibitem[{{Teyssandier} and {Ogilvie}(2019)}]{2019EAS....82..415T}
{Teyssandier} J, {Ogilvie} G (2019) {Growth of eccentricity in planet-disc
  interactions}. In: EAS Publications Series. EAS Publications Series, vol~82.
  pp 415--422. \doi{10.1051/eas/1982036}

\bibitem[{{Teyssandier} and {Ogilvie}(2016)}]{2016MNRAS.458.3221T}
{Teyssandier} J, {Ogilvie} GI (2016) {Growth of eccentric modes in disc-planet
  interactions}. \mnras 458(3):3221--3247. \doi{10.1093/mnras/stw521}.
  {\href{https://arxiv.org/abs/1603.00653}{{arXiv:1603.00653}}} {[astro-ph.EP]}

\bibitem[{{The LIGO Scientific Collaboration} et~al.(2020){The LIGO Scientific
  Collaboration}, {the Virgo Collaboration}, {Abbott}, {Abbott}, {Abraham},
  {Acernese}, {Ackley}, {Adams}, {Adams}, {Adhikari}, {Adya}, {Affeldt}, and
  {et~al.}}]{2020arXiv201014533T}
{The LIGO Scientific Collaboration}, {the Virgo Collaboration}, {Abbott} R,
  {Abbott} TD, {Abraham} S, {Acernese} F, {Ackley} K, {Adams} A, {Adams} C,
  {Adhikari} RX, et~al. (2020) {Population Properties of Compact Objects from
  the Second LIGO-Virgo Gravitational-Wave Transient Catalog}. arXiv e-prints
  arXiv:2010.14533.
  {\href{https://arxiv.org/abs/2010.14533}{{arXiv:2010.14533}}} {[astro-ph.HE]}

\bibitem[{{The Lynx Team}(2018)}]{2018arXiv180909642T}
{The Lynx Team} (2018) {The Lynx Mission Concept Study Interim Report}. arXiv
  e-prints arXiv:1809.09642.
  {\href{https://arxiv.org/abs/1809.09642}{{arXiv:1809.09642}}} {[astro-ph.IM]}

\bibitem[{{Thompson}(2011)}]{2011ApJ...741...82T}
{Thompson} TA (2011) {Accelerating Compact Object Mergers in Triple Systems
  with the Kozai Resonance: A Mechanism for ``Prompt'' Type Ia Supernovae,
  Gamma-Ray Bursts, and Other Exotica}. \apj 741(2):82.
  \doi{10.1088/0004-637X/741/2/82}.
  {\href{https://arxiv.org/abs/1011.4322}{{arXiv:1011.4322}}} {[astro-ph.HE]}

\bibitem[{Thompson et~al.(2005)Thompson, Quataert, and Murray}]{TQM}
Thompson TA, Quataert E, Murray N (2005) {Radiation
  Pressure{\textendash}supported Starburst Disks and Active Galactic Nucleus
  Fueling}. ApJ 630(1):167--185. \doi{10.1086/431923},
  \urlprefix\url{http://dx.doi.org/10.1086/431923}

\bibitem[{{Thompson} et~al.(2019){Thompson}, {Kochanek}, {Stanek}, {Badenes},
  {Post}, {Jayasinghe}, {Latham}, {Bieryla}, {Esquerdo}, {Berlind}, {Calkins},
  {Tayar}, {Lindegren}, {Johnson}, {Holoien}, {Auchettl}, and
  {Covey}}]{2019Sci...366..637T}
{Thompson} TA, {Kochanek} CS, {Stanek} KZ, {Badenes} C, {Post} RS, {Jayasinghe}
  T, {Latham} DW, {Bieryla} A, {Esquerdo} GA, {Berlind} P, et~al. (2019) {A
  noninteracting low-mass black hole-giant star binary system}. Science
  366(6465):637--640. \doi{10.1126/science.aau4005}.
  {\href{https://arxiv.org/abs/1806.02751}{{arXiv:1806.02751}}} {[astro-ph.HE]}

\bibitem[{{Thorne}(1974)}]{1974ApJ...191..507T}
{Thorne} KS (1974) {Disk-Accretion onto a Black Hole. II. Evolution of the
  Hole}. \apj 191:507--520. \doi{10.1086/152991}

\bibitem[{{Thorsett} et~al.(1999){Thorsett}, {Arzoumanian}, {Camilo}, and
  {Lyne}}]{1999ApJ...523..763T}
{Thorsett} SE, {Arzoumanian} Z, {Camilo} F, {Lyne} AG (1999) {The Triple Pulsar
  System PSR B1620-26 in M4}. \apj 523(2):763--770. \doi{10.1086/307771}.
  {\href{https://arxiv.org/abs/astro-ph/9903227}{{arXiv:astro-ph/9903227}}}
  {[astro-ph]}

\bibitem[{{Thrane} et~al.(2020){Thrane}, {Os{\l}owski}, and
  {Lasky}}]{2020MNRAS.493.5408T}
{Thrane} E, {Os{\l}owski} S, {Lasky} PD (2020) {Ultrarelativistic astrophysics
  using multimessenger observations of double neutron stars with LISA and the
  SKA}. \mnras 493(4):5408--5412. \doi{10.1093/mnras/staa593}.
  {\href{https://arxiv.org/abs/1910.12330}{{arXiv:1910.12330}}} {[astro-ph.HE]}

\bibitem[{{Tichy} and {Marronetti}(2007)}]{2007PhRvD..76f1502T}
{Tichy} W, {Marronetti} P (2007) {Binary black hole mergers: Large kicks for
  generic spin orientations}. \prd 76(6):061502.
  \doi{10.1103/PhysRevD.76.061502}.
  {\href{https://arxiv.org/abs/gr-qc/0703075}{{arXiv:gr-qc/0703075}}} {[gr-qc]}

\bibitem[{{Tiede} et~al.(2020){Tiede}, {Zrake}, {MacFadyen}, and
  {Haiman}}]{2020arXiv200509555T}
{Tiede} C, {Zrake} J, {MacFadyen} A, {Haiman} Z (2020) {Gas-driven inspiral of
  binaries in thin accretion disks}. arXiv e-prints arXiv:2005.09555.
  {\href{https://arxiv.org/abs/2005.09555}{{arXiv:2005.09555}}} {[astro-ph.GA]}

\bibitem[{{Timpano} et~al.(2006){Timpano}, {Rubbo}, and
  {Cornish}}]{2006PhRvD..73l2001T}
{Timpano} SE, {Rubbo} LJ, {Cornish} NJ (2006) {Characterizing the galactic
  gravitational wave background with LISA}. \prd 73(12):122001.
  \doi{10.1103/PhysRevD.73.122001}.
  {\href{https://arxiv.org/abs/gr-qc/0504071}{{arXiv:gr-qc/0504071}}} {[gr-qc]}

\bibitem[{{Tisserand} et~al.(2020){Tisserand}, {Clayton}, {Bessell}, {Welch},
  {Kamath}, {Wood}, {Wils}, {Wyrzykowski}, {Mr{\'o}z}, and
  {Udalski}}]{2020A&A...635A..14T}
{Tisserand} P, {Clayton} GC, {Bessell} MS, {Welch} DL, {Kamath} D, {Wood} PR,
  {Wils} P, {Wyrzykowski} {\L}, {Mr{\'o}z} P, {Udalski} A (2020) {A plethora of
  new R Coronae Borealis stars discovered from a dedicated spectroscopic
  follow-up survey}. \aap 635:A14. \doi{10.1051/0004-6361/201834410}.
  {\href{https://arxiv.org/abs/1809.01743}{{arXiv:1809.01743}}} {[astro-ph.SR]}

\bibitem[{{Tomsick} et~al.(2018){Tomsick}, {Parker}, {Garc{\'\i}a}, {Yamaoka},
  {Barret}, {Chiu}, {Clavel}, {Fabian}, {F{\"u}rst}, {Gandhi}, {Grinberg},
  {Miller}, {Pottschmidt}, and {Walton}}]{2018ApJ...855....3T}
{Tomsick} JA, {Parker} ML, {Garc{\'\i}a} JA, {Yamaoka} K, {Barret} D, {Chiu}
  JL, {Clavel} M, {Fabian} A, {F{\"u}rst} F, {Gandhi} P, et~al. (2018)
  {Alternative Explanations for Extreme Supersolar Iron Abundances Inferred
  from the Energy Spectrum of Cygnus X-1}. \apj 855(1):3.
  \doi{10.3847/1538-4357/aaaab1}.
  {\href{https://arxiv.org/abs/1801.07267}{{arXiv:1801.07267}}} {[astro-ph.HE]}

\bibitem[{{Toonen} and {Nelemans}(2013)}]{2013A&A...557A..87T}
{Toonen} S, {Nelemans} G (2013) {The effect of common-envelope evolution on the
  visible population of post-common-envelope binaries}. \aap 557:A87.
  \doi{10.1051/0004-6361/201321753}.
  {\href{https://arxiv.org/abs/1309.0327}{{arXiv:1309.0327}}} {[astro-ph.SR]}

\bibitem[{{Toonen} et~al.(2012){Toonen}, {Nelemans}, and {Portegies
  Zwart}}]{2012AA...546A..70T}
{Toonen} S, {Nelemans} G, {Portegies Zwart} S (2012) {Supernova Type Ia
  progenitors from merging double white dwarfs. Using a new population
  synthesis model}. \aap 546:A70. \doi{10.1051/0004-6361/201218966}.
  {\href{https://arxiv.org/abs/1208.6446}{{arXiv:1208.6446}}} {[astro-ph.HE]}

\bibitem[{{Toonen} et~al.(2014){Toonen}, {Claeys}, {Mennekens}, and
  {Ruiter}}]{2014A&A...562A..14T}
{Toonen} S, {Claeys} JSW, {Mennekens} N, {Ruiter} AJ (2014) {PopCORN: Hunting
  down the differences between binary population synthesis codes}. \aap
  562:A14. \doi{10.1051/0004-6361/201321576}.
  {\href{https://arxiv.org/abs/1311.6503}{{arXiv:1311.6503}}} {[astro-ph.SR]}

\bibitem[{{Toonen} et~al.(2016){Toonen}, {Hamers}, and {Portegies
  Zwart}}]{2016ComAC...3....6T}
{Toonen} S, {Hamers} A, {Portegies Zwart} S (2016) {The evolution of
  hierarchical triple star-systems}. Computational Astrophysics and Cosmology
  3(1):6. \doi{10.1186/s40668-016-0019-0}.
  {\href{https://arxiv.org/abs/1612.06172}{{arXiv:1612.06172}}} {[astro-ph.SR]}

\bibitem[{{Toonen} et~al.(2017){Toonen}, {Hollands}, {G{\"a}nsicke}, and
  {Boekholt}}]{2017A&A...602A..16T}
{Toonen} S, {Hollands} M, {G{\"a}nsicke} BT, {Boekholt} T (2017) {The binarity
  of the local white dwarf population}. \aap 602:A16.
  \doi{10.1051/0004-6361/201629978}.
  {\href{https://arxiv.org/abs/1703.06893}{{arXiv:1703.06893}}} {[astro-ph.SR]}

\bibitem[{{Toonen} et~al.(2018){Toonen}, {Perets}, and
  {Hamers}}]{2018A&A...610A..22T}
{Toonen} S, {Perets} HB, {Hamers} AS (2018) {Rate of WD-WD head-on collisions
  in isolated triples is too low to explain standard type Ia supernovae}. \aap
  610:A22. \doi{10.1051/0004-6361/201731874}.
  {\href{https://arxiv.org/abs/1709.00422}{{arXiv:1709.00422}}} {[astro-ph.HE]}

\bibitem[{{Toonen} et~al.(2020){Toonen}, {Portegies Zwart}, {Hamers}, and {Band
  opadhyay}}]{2020A&A...640A..16T}
{Toonen} S, {Portegies Zwart} S, {Hamers} AS, {Band opadhyay} D (2020) {The
  evolution of stellar triples. The most common evolutionary pathways}. \aap
  640:A16. \doi{10.1051/0004-6361/201936835}.
  {\href{https://arxiv.org/abs/2004.07848}{{arXiv:2004.07848}}} {[astro-ph.SR]}

\bibitem[{{Torres-Orjuela} et~al.(2020{\natexlab{a}}){Torres-Orjuela}, {Chen},
  and {Amaro-Seoane}}]{2020arXiv201015856T}
{Torres-Orjuela} A, {Chen} X, {Amaro-Seoane} P (2020{\natexlab{a}}) {Excitation
  of gravitational wave modes by a center-of-mass velocity of the source}.
  arXiv e-prints arXiv:2010.15856.
  {\href{https://arxiv.org/abs/2010.15856}{{arXiv:2010.15856}}} {[astro-ph.CO]}

\bibitem[{{Torres-Orjuela} et~al.(2020{\natexlab{b}}){Torres-Orjuela}, {Chen},
  and {Amaro-Seoane}}]{2020PhRvD.101h3028T}
{Torres-Orjuela} A, {Chen} X, {Amaro-Seoane} P (2020{\natexlab{b}}) {Phase
  shift of gravitational waves induced by aberration}. \prd 101(8):083028.
  \doi{10.1103/PhysRevD.101.083028}.
  {\href{https://arxiv.org/abs/2001.00721}{{arXiv:2001.00721}}} {[astro-ph.HE]}

\bibitem[{{Torres-Orjuela} et~al.(2021){Torres-Orjuela}, {Amaro Seoane},
  {Xuan}, {Chua}, {Rosell}, and {Chen}}]{2020arXiv201015842A}
{Torres-Orjuela} A, {Amaro Seoane} P, {Xuan} Z, {Chua} AJK, {Rosell} MJB,
  {Chen} X (2021) {Exciting Modes due to the Aberration of Gravitational Waves:
  Measurability for Extreme-Mass-Ratio Inspirals}. \prl 127(4):041102.
  \doi{10.1103/PhysRevLett.127.041102}.
  {\href{https://arxiv.org/abs/2010.15842}{{arXiv:2010.15842}}} {[gr-qc]}

\bibitem[{{Toscani} et~al.(2019){Toscani}, {Lodato}, and
  {Nealon}}]{2019MNRAS.489..699T}
{Toscani} M, {Lodato} G, {Nealon} R (2019) {Gravitational wave emission from
  unstable accretion discs in tidal disruption events}. \mnras 489(1):699--706.
  \doi{10.1093/mnras/stz2201}.
  {\href{https://arxiv.org/abs/1908.02969}{{arXiv:1908.02969}}} {[astro-ph.HE]}

\bibitem[{{Toscani} et~al.(2020){Toscani}, {Rossi}, and
  {Lodato}}]{2020MNRAS.498..507T}
{Toscani} M, {Rossi} EM, {Lodato} G (2020) {The gravitational wave background
  signal from tidal disruption events}. \mnras 498(1):507--516.
  \doi{10.1093/mnras/staa2290}.
  {\href{https://arxiv.org/abs/2007.13225}{{arXiv:2007.13225}}} {[astro-ph.HE]}

\bibitem[{{Toscani} et~al.(2022){Toscani}, {Lodato}, {Price}, and
  {Liptai}}]{2022MNRAS.510..992T}
{Toscani} M, {Lodato} G, {Price} DJ, {Liptai} D (2022) {Gravitational waves
  from tidal disruption events: an open and comprehensive catalog}. \mnras
  510(1):992--1001. \doi{10.1093/mnras/stab3384}.
  {\href{https://arxiv.org/abs/2111.05145}{{arXiv:2111.05145}}} {[astro-ph.HE]}

\bibitem[{{Toubiana} et~al.(2020{\natexlab{a}}){Toubiana}, {Marsat}, {Babak},
  {Barausse}, and {Baker}}]{2020PhRvD.101j4038T}
{Toubiana} A, {Marsat} S, {Babak} S, {Barausse} E, {Baker} J
  (2020{\natexlab{a}}) {Tests of general relativity with stellar-mass black
  hole binaries observed by LISA}. \prd 101(10):104038.
  \doi{10.1103/PhysRevD.101.104038}.
  {\href{https://arxiv.org/abs/2004.03626}{{arXiv:2004.03626}}} {[gr-qc]}

\bibitem[{{Toubiana} et~al.(2020{\natexlab{b}}){Toubiana}, {Sberna}, {Caputo},
  {Cusin}, {Marsat}, {Jani}, {Babak}, {Barausse}, {Caprini}, {Pani}, {Sesana},
  and {Tamanini}}]{2020arXiv201006056T}
{Toubiana} A, {Sberna} L, {Caputo} A, {Cusin} G, {Marsat} S, {Jani} K, {Babak}
  S, {Barausse} E, {Caprini} C, {Pani} P, et~al. (2020{\natexlab{b}})
  {Detectable environmental effects in GW190521-like black-hole binaries with
  LISA}. arXiv e-prints arXiv:2010.06056.
  {\href{https://arxiv.org/abs/2010.06056}{{arXiv:2010.06056}}} {[astro-ph.HE]}

\bibitem[{{Touma} et~al.(2009){Touma}, {Tremaine}, and
  {Kazandjian}}]{2009MNRAS.394.1085T}
{Touma} JR, {Tremaine} S, {Kazandjian} MV (2009) {Gauss's method for secular
  dynamics, softened}. \mnras 394(2):1085--1108.
  \doi{10.1111/j.1365-2966.2009.14409.x}.
  {\href{https://arxiv.org/abs/0811.2812}{{arXiv:0811.2812}}} {[astro-ph]}

\bibitem[{{Toyouchi} et~al.(2020){Toyouchi}, {Hosokawa}, {Sugimura}, and
  {Kuiper}}]{2020MNRAS.496.1909T}
{Toyouchi} D, {Hosokawa} T, {Sugimura} K, {Kuiper} R (2020) {Gaseous dynamical
  friction under radiative feedback: do intermediate-mass black holes speed up
  or down?} \mnras 496(2):1909--1921. \doi{10.1093/mnras/staa1338}.
  {\href{https://arxiv.org/abs/2002.08017}{{arXiv:2002.08017}}} {[astro-ph.GA]}

\bibitem[{{Trebitsch} et~al.(2020){Trebitsch}, {Dubois}, {Volonteri},
  {Pfister}, {Cadiou}, {Katz}, {Rosdahl}, {Kimm}, {Pichon}, {Beckmann},
  {Devriendt}, and {Slyz}}]{2020arXiv200204045T}
{Trebitsch} M, {Dubois} Y, {Volonteri} M, {Pfister} H, {Cadiou} C, {Katz} H,
  {Rosdahl} J, {Kimm} T, {Pichon} C, {Beckmann} RS, et~al. (2020) {The Obelisk
  simulation: galaxies contribute more than AGN to HI reionization of
  protoclusters}. arXiv e-prints arXiv:2002.04045.
  {\href{https://arxiv.org/abs/2002.04045}{{arXiv:2002.04045}}} {[astro-ph.GA]}

\bibitem[{{Tremaine} and {Weinberg}(1984)}]{1984MNRAS.209..729T}
{Tremaine} S, {Weinberg} MD (1984) {Dynamical friction in spherical systems.}
  \mnras 209:729--757. \doi{10.1093/mnras/209.4.729}

\bibitem[{{Tremaine} et~al.(2002){Tremaine}, {Gebhardt}, {Bender}, {Bower},
  {Dressler}, {Faber}, {Filippenko}, {Green}, {Grillmair}, {Ho}, {Kormendy},
  {Lauer}, {Magorrian}, {Pinkney}, and {Richstone}}]{2002ApJ...574..740T}
{Tremaine} S, {Gebhardt} K, {Bender} R, {Bower} G, {Dressler} A, {Faber} SM,
  {Filippenko} AV, {Green} R, {Grillmair} C, {Ho} LC, et~al. (2002) {The Slope
  of the Black Hole Mass versus Velocity Dispersion Correlation}. \apj
  574(2):740--753. \doi{10.1086/341002}.
  {\href{https://arxiv.org/abs/astro-ph/0203468}{{arXiv:astro-ph/0203468}}}
  {[astro-ph]}

\bibitem[{{Tremaine}(1976)}]{1976ApJ...203..345T}
{Tremaine} SD (1976) {The formation of the nuclei of galaxies. II. The local
  group.} \apj 203:345--351. \doi{10.1086/154085}

\bibitem[{{Tremaine} et~al.(1975){Tremaine}, {Ostriker}, and
  {Spitzer}}]{1975ApJ...196..407T}
{Tremaine} SD, {Ostriker} JP, {Spitzer} J L (1975) {The formation of the nuclei
  of galaxies. I. M31.} \apj 196:407--411. \doi{10.1086/153422}

\bibitem[{{Tremmel} et~al.(2015){Tremmel}, {Governato}, {Volonteri}, and
  {Quinn}}]{2015MNRAS.451.1868T}
{Tremmel} M, {Governato} F, {Volonteri} M, {Quinn} TR (2015) {Off the beaten
  path: a new approach to realistically model the orbital decay of supermassive
  black holes in galaxy formation simulations}. \mnras 451(2):1868--1874.
  \doi{10.1093/mnras/stv1060}.
  {\href{https://arxiv.org/abs/1501.07609}{{arXiv:1501.07609}}} {[astro-ph.GA]}

\bibitem[{{Tremmel} et~al.(2017){Tremmel}, {Karcher}, {Governato}, {Volonteri},
  {Quinn}, {Pontzen}, {Anderson}, and {Bellovary}}]{2017MNRAS.470.1121T}
{Tremmel} M, {Karcher} M, {Governato} F, {Volonteri} M, {Quinn} TR, {Pontzen}
  A, {Anderson} L, {Bellovary} J (2017) {The Romulus cosmological simulations:
  a physical approach to the formation, dynamics and accretion models of
  SMBHs}. \mnras 470(1):1121--1139. \doi{10.1093/mnras/stx1160}.
  {\href{https://arxiv.org/abs/1607.02151}{{arXiv:1607.02151}}} {[astro-ph.GA]}

\bibitem[{{Tremmel} et~al.(2018{\natexlab{a}}){Tremmel}, {Governato},
  {Volonteri}, {Pontzen}, and {Quinn}}]{2018ApJ...857L..22T}
{Tremmel} M, {Governato} F, {Volonteri} M, {Pontzen} A, {Quinn} TR
  (2018{\natexlab{a}}) {Wandering Supermassive Black Holes in Milky-Way-mass
  Halos}. \apjl 857(2):L22. \doi{10.3847/2041-8213/aabc0a}.
  {\href{https://arxiv.org/abs/1802.06783}{{arXiv:1802.06783}}} {[astro-ph.GA]}

\bibitem[{{Tremmel} et~al.(2018{\natexlab{b}}){Tremmel}, {Governato},
  {Volonteri}, {Quinn}, and {Pontzen}}]{2018MNRAS.475.4967T}
{Tremmel} M, {Governato} F, {Volonteri} M, {Quinn} TR, {Pontzen} A
  (2018{\natexlab{b}}) {Dancing to CHANGA: a self-consistent prediction for
  close SMBH pair formation time-scales following galaxy mergers}. \mnras
  475(4):4967--4977. \doi{10.1093/mnras/sty139}.
  {\href{https://arxiv.org/abs/1708.07126}{{arXiv:1708.07126}}} {[astro-ph.GA]}

\bibitem[{{Tr{\"u}mper} and {Sch{\"o}nfelder}(1973)}]{1973A&A....25..445T}
{Tr{\"u}mper} J, {Sch{\"o}nfelder} V (1973) {Distance Determination of Variable
  X-ray Sources}. \aap 25:445

\bibitem[{{Tsalmantza} et~al.(2011){Tsalmantza}, {Decarli}, {Dotti}, and
  {Hogg}}]{2011ApJ...738...20T}
{Tsalmantza} P, {Decarli} R, {Dotti} M, {Hogg} DW (2011) {A Systematic Search
  for Massive Black Hole Binaries in the Sloan Digital Sky Survey Spectroscopic
  Sample}. \apj 738(1):20. \doi{10.1088/0004-637X/738/1/20}.
  {\href{https://arxiv.org/abs/1106.1180}{{arXiv:1106.1180}}} {[astro-ph.CO]}

\bibitem[{{Tsizh} et~al.(2020){Tsizh}, {Novosyadlyj}, {Holovatch}, and
  {Libeskind}}]{2020MNRAS.495.1311T}
{Tsizh} M, {Novosyadlyj} B, {Holovatch} Y, {Libeskind} NI (2020) {Large-scale
  structures in the {\ensuremath{\Lambda}}CDM Universe: network analysis and
  machine learning}. \mnras 495(1):1311--1320. \doi{10.1093/mnras/staa1030}.
  {\href{https://arxiv.org/abs/1910.07868}{{arXiv:1910.07868}}} {[astro-ph.CO]}

\bibitem[{{Tso} et~al.(2019){Tso}, {Gerosa}, and {Chen}}]{2019PhRvD..99l4043T}
{Tso} R, {Gerosa} D, {Chen} Y (2019) {Optimizing LIGO with LISA forewarnings to
  improve black-hole spectroscopy}. \prd 99(12):124043.
  \doi{10.1103/PhysRevD.99.124043}.
  {\href{https://arxiv.org/abs/1807.00075}{{arXiv:1807.00075}}} {[gr-qc]}

\bibitem[{{Turk} et~al.(2009){Turk}, {Abel}, and
  {O'Shea}}]{2009Sci...325..601T}
{Turk} MJ, {Abel} T, {O'Shea} B (2009) {The Formation of Population III
  Binaries from Cosmological Initial Conditions}. Science 325:601--.
  \doi{10.1126/science.1173540}.
  {\href{https://arxiv.org/abs/0907.2919}{{arXiv:0907.2919}}} {[astro-ph.CO]}

\bibitem[{{Tutukov} and {Yungelson}(1993)}]{1993MNRAS.260..675T}
{Tutukov} AV, {Yungelson} LR (1993) {The merger rate of neutron star and black
  hole binaries.} \mnras 260:675--678. \doi{10.1093/mnras/260.3.675}

\bibitem[{{Tylenda} et~al.(2011){Tylenda}, {Hajduk}, {Kami{\'n}ski}, {Udalski},
  {Soszy{\'n}ski}, {Szyma{\'n}ski}, {Kubiak}, {Pietrzy{\'n}ski}, {Poleski},
  {Wyrzykowski}, and {Ulaczyk}}]{2011A&A...528A.114T}
{Tylenda} R, {Hajduk} M, {Kami{\'n}ski} T, {Udalski} A, {Soszy{\'n}ski} I,
  {Szyma{\'n}ski} MK, {Kubiak} M, {Pietrzy{\'n}ski} G, {Poleski} R,
  {Wyrzykowski} {\L}, et~al. (2011) {V1309 Scorpii: merger of a contact
  binary}. \aap 528:A114. \doi{10.1051/0004-6361/201016221}.
  {\href{https://arxiv.org/abs/1012.0163}{{arXiv:1012.0163}}} {[astro-ph.SR]}

\bibitem[{{Ullio} et~al.(2001){Ullio}, {Zhao}, and
  {Kamionkowski}}]{2001PhRvD..64d3504U}
{Ullio} P, {Zhao} H, {Kamionkowski} M (2001) {Dark-matter spike at the galactic
  center?} \prd 64(4):043504. \doi{10.1103/PhysRevD.64.043504}.
  {\href{https://arxiv.org/abs/astro-ph/0101481}{{arXiv:astro-ph/0101481}}}
  {[astro-ph]}

\bibitem[{{Ulmer}(1999)}]{1999ApJ...514..180U}
{Ulmer} A (1999) {Flares from the Tidal Disruption of Stars by Massive Black
  Holes}. \apj 514(1):180--187. \doi{10.1086/306909}

\bibitem[{{Umst{\"a}tter} et~al.(2005){Umst{\"a}tter}, {Christensen}, {Hendry},
  {Meyer}, {Simha}, {Veitch}, {Vigeland}, and {Woan}}]{2005PhRvD..72b2001U}
{Umst{\"a}tter} R, {Christensen} N, {Hendry} M, {Meyer} R, {Simha} V, {Veitch}
  J, {Vigeland} S, {Woan} G (2005) {Bayesian modeling of source confusion in
  LISA data}. \prd 72(2):022001. \doi{10.1103/PhysRevD.72.022001}.
  {\href{https://arxiv.org/abs/gr-qc/0506055}{{arXiv:gr-qc/0506055}}} {[gr-qc]}

\bibitem[{Unal(2019)}]{Unal:2018yaa}
Unal C (2019) {Imprints of Primordial Non-Gaussianity on Gravitational Wave
  Spectrum}. Phys Rev D 99(4):041301. \doi{10.1103/PhysRevD.99.041301}.
  {\href{https://arxiv.org/abs/1811.09151}{{arXiv:1811.09151}}} {[astro-ph.CO]}

\bibitem[{Unal and Loeb(2020)}]{Unal:2020add}
Unal C, Loeb A (2020) {On Spin Dependence of the Fundamental Plane of Black
  Hole Activity}. Mon Not Roy Astron Soc 495(1):278--284.
  \doi{10.1093/mnras/staa1119}.
  {\href{https://arxiv.org/abs/2002.11778}{{arXiv:2002.11778}}} {[astro-ph.HE]}

\bibitem[{{Unal} et~al.(2020){Unal}, {Pacucci}, and
  {Loeb}}]{2020arXiv201212790U}
{Unal} C, {Pacucci} F, {Loeb} A (2020) {Properties of Ultralight Bosons from
  Heavy Quasar Spins via Superradiance}. arXiv e-prints arXiv:2012.12790.
  {\href{https://arxiv.org/abs/2012.12790}{{arXiv:2012.12790}}} {[hep-ph]}

\bibitem[{{Valiante} et~al.(2016){Valiante}, {Schneider}, {Volonteri}, and
  {Omukai}}]{2016MNRAS.457.3356V}
{Valiante} R, {Schneider} R, {Volonteri} M, {Omukai} K (2016) {From the first
  stars to the first black holes}. \mnras 457(3):3356--3371.
  \doi{10.1093/mnras/stw225}.
  {\href{https://arxiv.org/abs/1601.07915}{{arXiv:1601.07915}}} {[astro-ph.GA]}

\bibitem[{{Valiante} et~al.(2017){Valiante}, {Agarwal}, {Habouzit}, and
  {Pezzulli}}]{2017PASA...34...31V}
{Valiante} R, {Agarwal} B, {Habouzit} M, {Pezzulli} E (2017) {On the Formation
  of the First Quasars}. \pasa 34:e031. \doi{10.1017/pasa.2017.25}.
  {\href{https://arxiv.org/abs/1703.03808}{{arXiv:1703.03808}}} {[astro-ph.GA]}

\bibitem[{{Valiante} et~al.(2018{\natexlab{a}}){Valiante}, {Schneider},
  {Graziani}, and {Zappacosta}}]{2018MNRAS.474.3825V}
{Valiante} R, {Schneider} R, {Graziani} L, {Zappacosta} L (2018{\natexlab{a}})
  Chasing the observational signatures of seed black holes at z > 7: candidate
  statistics. \mnras 474(3):3825--3834. \doi{10.1093/mnras/stx3028}.
  {\href{https://arxiv.org/abs/1801.08165}{{arXiv:1801.08165}}} {[astro-ph.GA]}

\bibitem[{{Valiante} et~al.(2018{\natexlab{b}}){Valiante}, {Schneider},
  {Zappacosta}, {Graziani}, {Pezzulli}, and {Volonteri}}]{2018MNRAS.476..407V}
{Valiante} R, {Schneider} R, {Zappacosta} L, {Graziani} L, {Pezzulli} E,
  {Volonteri} M (2018{\natexlab{b}}) Chasing the observational signatures of
  seed black holes at z > 7: candidate observability. \mnras 476(1):407--420.
  \doi{10.1093/mnras/sty213}

\bibitem[{{Valiante} et~al.(2021){Valiante}, {Colpi}, {Schneider}, {Mangiagli},
  {Bonetti}, {Cerini}, {Fairhurst}, {Haardt}, {Mills}, and
  {Sesana}}]{2021MNRAS.500.4095V}
{Valiante} R, {Colpi} M, {Schneider} R, {Mangiagli} A, {Bonetti} M, {Cerini} G,
  {Fairhurst} S, {Haardt} F, {Mills} C, {Sesana} A (2021) {Unveiling early
  black hole growth with multifrequency gravitational wave observations}.
  \mnras 500(3):4095--4109. \doi{10.1093/mnras/staa3395}.
  {\href{https://arxiv.org/abs/2010.15096}{{arXiv:2010.15096}}} {[astro-ph.GA]}

\bibitem[{{van de Meent}(2018)}]{2018PhRvD..97j4033V}
{van de Meent} M (2018) {Gravitational self-force on generic bound geodesics in
  Kerr spacetime}. \prd 97(10):104033. \doi{10.1103/PhysRevD.97.104033}.
  {\href{https://arxiv.org/abs/1711.09607}{{arXiv:1711.09607}}} {[gr-qc]}

\bibitem[{{van de Meent} and {Pfeiffer}(2020)}]{2020arXiv200612036V}
{van de Meent} M, {Pfeiffer} HP (2020) {Intermediate mass-ratio black hole
  binaries: Applicability of small mass-ratio perturbation theory}. arXiv
  e-prints arXiv:2006.12036.
  {\href{https://arxiv.org/abs/2006.12036}{{arXiv:2006.12036}}} {[gr-qc]}

\bibitem[{{van den Heuvel}(1976)}]{1976IAUS...73...35V}
{van den Heuvel} EPJ (1976) {Late Stages of Close Binary Systems}. In:
  {Eggleton} P, {Mitton} S, {Whelan} J (eds) Structure and Evolution of Close
  Binary Systems. IAU Symposium, vol~73. p~35

\bibitem[{{van den Heuvel}(2007)}]{2007AIPC..924..598V}
{van den Heuvel} EPJ (2007) {Double Neutron Stars: Evidence For Two Different
  Neutron-Star Formation Mechanisms}. In: {di Salvo} T, {Israel} GL, {Piersant}
  L, {Burderi} L, {Matt} G, {Tornambe} A, {Menna} MT (eds) The Multicolored
  Landscape of Compact Objects and Their Explosive Origins. American Institute
  of Physics Conference Series, vol 924. pp 598--606. \doi{10.1063/1.2774916}.
  {\href{https://arxiv.org/abs/0704.1215}{{arXiv:0704.1215}}} {[astro-ph]}

\bibitem[{{van den Heuvel}(2019)}]{2019IAUS..346....1V}
{van den Heuvel} EPJ (2019) {High-Mass X-ray Binaries: progenitors of double
  compact objects}. In: {Oskinova} LM, {Bozzo} E, {Bulik} T, {Gies} DR (eds)
  IAU Symposium. IAU Symposium, vol 346. pp 1--13.
  \doi{10.1017/S1743921319001315}.
  {\href{https://arxiv.org/abs/1901.06939}{{arXiv:1901.06939}}} {[astro-ph.HE]}

\bibitem[{{van den Heuvel} and {Tauris}(2020)}]{2020Sci...368.3282V}
{van den Heuvel} EPJ, {Tauris} TM (2020) {Comment on {\textquotedblleft}A
  noninteracting low-mass black hole-giant star binary
  system{\textquotedblright}}. Science 368(6491):eaba3282.
  \doi{10.1126/science.aba3282}.
  {\href{https://arxiv.org/abs/2005.04896}{{arXiv:2005.04896}}} {[astro-ph.SR]}

\bibitem[{{van den Heuvel} et~al.(2017){van den Heuvel}, {Portegies Zwart}, and
  {de Mink}}]{2017MNRAS.471.4256V}
{van den Heuvel} EPJ, {Portegies Zwart} SF, {de Mink} SE (2017) {Forming
  short-period Wolf-Rayet X-ray binaries and double black holes through stable
  mass transfer}. \mnras 471(4):4256--4264. \doi{10.1093/mnras/stx1430}.
  {\href{https://arxiv.org/abs/1701.02355}{{arXiv:1701.02355}}} {[astro-ph.SR]}

\bibitem[{{van der Sluys} et~al.(2005){van der Sluys}, {Verbunt}, and
  {Pols}}]{2005A&A...440..973V}
{van der Sluys} MV, {Verbunt} F, {Pols} OR (2005) {Reduced magnetic braking and
  the magnetic capture model for the formation of ultra-compact binaries}. \aap
  440(3):973--979. \doi{10.1051/0004-6361:20052696}.
  {\href{https://arxiv.org/abs/astro-ph/0506375}{{arXiv:astro-ph/0506375}}}
  {[astro-ph]}

\bibitem[{{van der Sluys} et~al.(2006){van der Sluys}, {Verbunt}, and
  {Pols}}]{2006A&A...460..209V}
{van der Sluys} MV, {Verbunt} F, {Pols} OR (2006) {Modelling the formation of
  double white dwarfs}. \aap 460(1):209--228. \doi{10.1051/0004-6361:20065066}.
  {\href{https://arxiv.org/abs/astro-ph/0610492}{{arXiv:astro-ph/0610492}}}
  {[astro-ph]}

\bibitem[{{van Haaften} et~al.(2012){van Haaften}, {Nelemans}, {Voss}, {Wood},
  and {Kuijpers}}]{2012A&A...537A.104V}
{van Haaften} LM, {Nelemans} G, {Voss} R, {Wood} MA, {Kuijpers} J (2012) {The
  evolution of ultracompact X-ray binaries}. \aap 537:A104.
  \doi{10.1051/0004-6361/201117880}.
  {\href{https://arxiv.org/abs/1111.5978}{{arXiv:1111.5978}}} {[astro-ph.SR]}

\bibitem[{{van Meter} et~al.(2010){van Meter}, {Miller}, {Baker}, {Boggs}, and
  {Kelly}}]{2010ApJ...719.1427V}
{van Meter} JR, {Miller} MC, {Baker} JG, {Boggs} WD, {Kelly} BJ (2010) {Test of
  a General Formula for Black Hole Gravitational Wave Kicks}. \apj
  719(2):1427--1432. \doi{10.1088/0004-637X/719/2/1427}.
  {\href{https://arxiv.org/abs/1003.3865}{{arXiv:1003.3865}}} {[astro-ph.HE]}

\bibitem[{{van Oirschot} et~al.(2014){van Oirschot}, {Nelemans}, {Toonen},
  {Pols}, {Brown}, {Helmi}, and {Portegies Zwart}}]{2014A&A...569A..42V}
{van Oirschot} P, {Nelemans} G, {Toonen} S, {Pols} O, {Brown} AGA, {Helmi} A,
  {Portegies Zwart} S (2014) {Binary white dwarfs in the halo of the Milky
  Way}. \aap 569:A42. \doi{10.1051/0004-6361/201424195}.
  {\href{https://arxiv.org/abs/1407.2405}{{arXiv:1407.2405}}} {[astro-ph.GA]}

\bibitem[{{van Putten}(2001)}]{2001PhRvL..87i1101V}
{van Putten} MH (2001) {Proposed Source of Gravitational Radiation from a Torus
  around a Black Hole}. \prl 87(9):091101. \doi{10.1103/PhysRevLett.87.091101}.
  {\href{https://arxiv.org/abs/astro-ph/0107007}{{arXiv:astro-ph/0107007}}}
  {[astro-ph]}

\bibitem[{{van Putten}(2002)}]{2002ApJ...575L..71V}
{van Putten} MHPM (2002) {LIGO/VIRGO Searches for Gravitational Radiation in
  Hypernovae}. \apjl 575(2):L71--L74. \doi{10.1086/342781}.
  {\href{https://arxiv.org/abs/astro-ph/0207242}{{arXiv:astro-ph/0207242}}}
  {[astro-ph]}

\bibitem[{{van Putten} et~al.(2019){van Putten}, {Levinson}, {Frontera},
  {Guidorzi}, {Amati}, and {Della Valle}}]{2019EPJP..134..537V}
{van Putten} MHPM, {Levinson} A, {Frontera} F, {Guidorzi} C, {Amati} L, {Della
  Valle} M (2019) {Prospects for multi-messenger extended emission from
  core-collapse supernovae in the Local Universe}. European Physical Journal
  Plus 134(10):537. \doi{10.1140/epjp/i2019-12932-3}.
  {\href{https://arxiv.org/abs/1709.04455}{{arXiv:1709.04455}}} {[astro-ph.HE]}

\bibitem[{{van Son} et~al.(2019){van Son}, {Barber}, {Bah{\'e}}, {Schaye},
  {Barnes}, {Crain}, {Kay}, {Theuns}, and {Dalla
  Vecchia}}]{2019MNRAS.485..396V}
{van Son} LAC, {Barber} C, {Bah{\'e}} YM, {Schaye} J, {Barnes} DJ, {Crain} RA,
  {Kay} ST, {Theuns} T, {Dalla Vecchia} C (2019) {Galaxies with monstrous black
  holes in galaxy cluster environments}. \mnras 485(1):396--407.
  \doi{10.1093/mnras/stz399}.
  {\href{https://arxiv.org/abs/1901.03156}{{arXiv:1901.03156}}} {[astro-ph.GA]}

\bibitem[{{van Velzen} et~al.(2020){van Velzen}, {Holoien}, {Onori}, {Hung},
  and {Arcavi}}]{2020SSRv..216..124V}
{van Velzen} S, {Holoien} TWS, {Onori} F, {Hung} T, {Arcavi} I (2020)
  {Optical-Ultraviolet Tidal Disruption Events}. \ssr 216(8):124.
  \doi{10.1007/s11214-020-00753-z}.
  {\href{https://arxiv.org/abs/2008.05461}{{arXiv:2008.05461}}} {[astro-ph.HE]}

\bibitem[{{van Wassenhove} et~al.(2010){van Wassenhove}, {Volonteri}, {Walker},
  and {Gair}}]{2010MNRAS.408.1139V}
{van Wassenhove} S, {Volonteri} M, {Walker} MG, {Gair} JR (2010) {Massive black
  holes lurking in Milky Way satellites}. \mnras 408(2):1139--1146.
  \doi{10.1111/j.1365-2966.2010.17189.x}.
  {\href{https://arxiv.org/abs/1001.5451}{{arXiv:1001.5451}}} {[astro-ph.CO]}

\bibitem[{{Van Wassenhove} et~al.(2014){Van Wassenhove}, {Capelo}, {Volonteri},
  {Dotti}, {Bellovary}, {Mayer}, and {Governato}}]{2014MNRAS.439..474V}
{Van Wassenhove} S, {Capelo} PR, {Volonteri} M, {Dotti} M, {Bellovary} JM,
  {Mayer} L, {Governato} F (2014) {Nuclear coups: dynamics of black holes in
  galaxy mergers}. \mnras 439(1):474--487. \doi{10.1093/mnras/stu024}.
  {\href{https://arxiv.org/abs/1310.7581}{{arXiv:1310.7581}}} {[astro-ph.CO]}

\bibitem[{{Vanderburg} et~al.(2015){Vanderburg}, {Johnson}, {Rappaport},
  {Bieryla}, {Irwin}, {Lewis}, {Kipping}, {Brown}, {Dufour}, {Ciardi}, {Angus},
  {Schaefer}, {Latham}, {Charbonneau}, {Beichman}, {Eastman}, {McCrady},
  {Wittenmyer}, and {Wright}}]{2015Natur.526..546V}
{Vanderburg} A, {Johnson} JA, {Rappaport} S, {Bieryla} A, {Irwin} J, {Lewis}
  JA, {Kipping} D, {Brown} WR, {Dufour} P, {Ciardi} DR, et~al. (2015) {A
  disintegrating minor planet transiting a white dwarf}. \nat
  526(7574):546--549. \doi{10.1038/nature15527}.
  {\href{https://arxiv.org/abs/1510.06387}{{arXiv:1510.06387}}} {[astro-ph.EP]}

\bibitem[{{Vanderburg} et~al.(2020){Vanderburg}, {Rappaport}, {Xu},
  {Crossfield}, {Becker}, {Gary}, {Murgas}, {Blouin}, {Kaye}, {Palle}, and
  {et~al.}}]{2020Natur.585..363V}
{Vanderburg} A, {Rappaport} SA, {Xu} S, {Crossfield} IJM, {Becker} JC, {Gary}
  B, {Murgas} F, {Blouin} S, {Kaye} TG, {Palle} E, et~al. (2020) {A giant
  planet candidate transiting a white dwarf}. \nat 585(7825):363--367.
  \doi{10.1038/s41586-020-2713-y}.
  {\href{https://arxiv.org/abs/2009.07282}{{arXiv:2009.07282}}} {[astro-ph.EP]}

\bibitem[{{Varma} et~al.(2019){Varma}, {Gerosa}, {Stein}, {H{\'e}bert}, and
  {Zhang}}]{2019PhRvL.122a1101V}
{Varma} V, {Gerosa} D, {Stein} LC, {H{\'e}bert} F, {Zhang} H (2019)
  {High-Accuracy Mass, Spin, and Recoil Predictions of Generic Black-Hole
  Merger Remnants}. \prl 122(1):011101. \doi{10.1103/PhysRevLett.122.011101}.
  {\href{https://arxiv.org/abs/1809.09125}{{arXiv:1809.09125}}} {[gr-qc]}

\bibitem[{{Varma} et~al.(2020){Varma}, {Isi}, and
  {Biscoveanu}}]{2020PhRvL.124j1104V}
{Varma} V, {Isi} M, {Biscoveanu} S (2020) {Extracting the Gravitational Recoil
  from Black Hole Merger Signals}. \prl 124(10):101104.
  \doi{10.1103/PhysRevLett.124.101104}.
  {\href{https://arxiv.org/abs/2002.00296}{{arXiv:2002.00296}}} {[gr-qc]}

\bibitem[{{Vasiliev}(2017)}]{2017ApJ...848...10V}
{Vasiliev} E (2017) {A New Fokker-Planck Approach for the Relaxation-driven
  Evolution of Galactic Nuclei}. \apj 848(1):10.
  \doi{10.3847/1538-4357/aa8cc8}.
  {\href{https://arxiv.org/abs/1709.04467}{{arXiv:1709.04467}}} {[astro-ph.GA]}

\bibitem[{{Vasiliev} et~al.(2015){Vasiliev}, {Antonini}, and
  {Merritt}}]{2015ApJ...810...49V}
{Vasiliev} E, {Antonini} F, {Merritt} D (2015) {The Final-parsec Problem in the
  Collisionless Limit}. \apj 810(1):49. \doi{10.1088/0004-637X/810/1/49}.
  {\href{https://arxiv.org/abs/1505.05480}{{arXiv:1505.05480}}} {[astro-ph.GA]}

\bibitem[{{Vaskonen} and {Veerm{\"a}e}(2020)}]{2020arXiv200907832V}
{Vaskonen} V, {Veerm{\"a}e} H (2020) {Did NANOGrav see a signal from primordial
  black hole formation?} arXiv e-prints arXiv:2009.07832.
  {\href{https://arxiv.org/abs/2009.07832}{{arXiv:2009.07832}}} {[astro-ph.CO]}

\bibitem[{{Vasudevan} et~al.(2016){Vasudevan}, {Fabian}, {Reynolds}, {Aird},
  {Dauser}, and {Gallo}}]{2016MNRAS.458.2012V}
{Vasudevan} RV, {Fabian} AC, {Reynolds} CS, {Aird} J, {Dauser} T, {Gallo} LC
  (2016) {A selection effect boosting the contribution from rapidly spinning
  black holes to the cosmic X-ray background}. \mnras 458(2):2012--2023.
  \doi{10.1093/mnras/stw363}.
  {\href{https://arxiv.org/abs/1506.01027}{{arXiv:1506.01027}}} {[astro-ph.HE]}

\bibitem[{{Vaughan} et~al.(2016){Vaughan}, {Uttley}, {Markowitz},
  {Huppenkothen}, {Middleton}, {Alston}, {Scargle}, and
  {Farr}}]{2016MNRAS.461.3145V}
{Vaughan} S, {Uttley} P, {Markowitz} AG, {Huppenkothen} D, {Middleton} MJ,
  {Alston} WN, {Scargle} JD, {Farr} WM (2016) {False periodicities in quasar
  time-domain surveys}. \mnras 461(3):3145--3152. \doi{10.1093/mnras/stw1412}.
  {\href{https://arxiv.org/abs/1606.02620}{{arXiv:1606.02620}}} {[astro-ph.IM]}

\bibitem[{{Vennes} et~al.(2012){Vennes}, {Kawka}, {O'Toole}, {N{\'e}meth}, and
  {Burton}}]{2012ApJ...759L..25V}
{Vennes} S, {Kawka} A, {O'Toole} SJ, {N{\'e}meth} P, {Burton} D (2012) {The
  Shortest Period sdB Plus White Dwarf Binary CD-30 11223 (GALEX J1411-3053)}.
  \apjl 759(1):L25. \doi{10.1088/2041-8205/759/1/L25}.
  {\href{https://arxiv.org/abs/1210.1512}{{arXiv:1210.1512}}} {[astro-ph.SR]}

\bibitem[{{Venturi} et~al.(2020){Venturi}, {Paragi}, {Lindqvist},
  {Bartkiewicz}, {Beswick}, {Bogdanovi{\'c}}, {Brisken}, {Charlot}, {Colomer},
  {Conway}, {Frey}, {Guirado}, {Gurvits}, {van Langevelde}, {Lobanov},
  {McKean}, {Morganti}, {Muxlow}, {P{\'e}rez-Torres}, {Rygl}, {Schulz},
  {Szomoru}, {de Vicente}, {An}, {Anglada}, {Argo}, {Azulay}, {van Bemmel},
  {Bocanegra}, {Boccardi}, {Castangia}, {Chibueze}, {Cim{\`o}}, {Climent},
  {Deane}, {Deller}, {Dodson}, {Duev}, {Etoka}, {Fenech}, {Gab{\'a}nyi},
  {Gabuzda}, {Garrett}, {Gawro{\'n}ski}, {Ghirlanda}, {Giroletti}, {Goddi},
  {G{\'o}mez}, {Gray}, {Greaves}, {Hessels}, {van der Horst}, {Hunter},
  {Laing}, {Vir Lal}, {Lambert}, {Loinard}, {Marcote}, {Merloni},
  {Miller-Jones}, {Molera Calv{\'e}s}, {Moscadelli}, {Olofsson}, {Petrov},
  {Pizzo}, {Possenti}, {Quiroga-Nu{\~n}ez}, {Reynolds}, {Richards}, {Rioja},
  {Sanna}, {Savolainen}, {Sbarrato}, {Spingola}, {Surcis}, {Trigilio},
  {Varenius}, {Vlemmings}, {van Velzen}, and {van der
  Walt}}]{2020arXiv200702347V}
{Venturi} T, {Paragi} Z, {Lindqvist} M, {Bartkiewicz} A, {Beswick} R,
  {Bogdanovi{\'c}} T, {Brisken} W, {Charlot} P, {Colomer} F, {Conway} J, et~al.
  (2020) {VLBI20-30: a scientific roadmap for the next decade -- The future of
  the European VLBI Network}. arXiv e-prints arXiv:2007.02347.
  {\href{https://arxiv.org/abs/2007.02347}{{arXiv:2007.02347}}} {[astro-ph.IM]}

\bibitem[{{Veras} et~al.(2018){Veras}, {Georgakarakos}, {G{\"a}nsicke}, and
  {Dobbs-Dixon}}]{2018MNRAS.481.2180V}
{Veras} D, {Georgakarakos} N, {G{\"a}nsicke} BT, {Dobbs-Dixon} I (2018)
  {Effects of non-Kozai mutual inclinations on two-planet system stability
  through all phases of stellar evolution}. \mnras 481(2):2180--2188.
  \doi{10.1093/mnras/sty2409}.
  {\href{https://arxiv.org/abs/1809.01157}{{arXiv:1809.01157}}} {[astro-ph.EP]}

\bibitem[{{Verbunt} et~al.(2017){Verbunt}, {Igoshev}, and
  {Cator}}]{2017A&A...608A..57V}
{Verbunt} F, {Igoshev} A, {Cator} E (2017) {The observed velocity distribution
  of young pulsars}. \aap 608:A57. \doi{10.1051/0004-6361/201731518}.
  {\href{https://arxiv.org/abs/1708.08281}{{arXiv:1708.08281}}} {[astro-ph.HE]}

\bibitem[{{Vick} and {Lai}(2019)}]{2019PhRvD.100f3001V}
{Vick} M, {Lai} D (2019) {Tidal effects in eccentric coalescing neutron star
  binaries}. \prd 100(6):063001. \doi{10.1103/PhysRevD.100.063001}.
  {\href{https://arxiv.org/abs/1906.08780}{{arXiv:1906.08780}}} {[astro-ph.HE]}

\bibitem[{{Vigna-G{\'o}mez} et~al.(2018){Vigna-G{\'o}mez}, {Neijssel},
  {Stevenson}, {Barrett}, {Belczynski}, {Justham}, {de Mink}, {M{\"u}ller},
  {Podsiadlowski}, {Renzo}, {Sz{\'e}csi}, and {Mandel}}]{2018MNRAS.481.4009V}
{Vigna-G{\'o}mez} A, {Neijssel} CJ, {Stevenson} S, {Barrett} JW, {Belczynski}
  K, {Justham} S, {de Mink} SE, {M{\"u}ller} B, {Podsiadlowski} P, {Renzo} M,
  et~al. (2018) {On the formation history of Galactic double neutron stars}.
  \mnras 481(3):4009--4029. \doi{10.1093/mnras/sty2463}.
  {\href{https://arxiv.org/abs/1805.07974}{{arXiv:1805.07974}}} {[astro-ph.SR]}

\bibitem[{{Vigna-G{\'o}mez} et~al.(2020){Vigna-G{\'o}mez}, {MacLeod},
  {Neijssel}, {Broekgaarden}, {Justham}, {Howitt}, {de Mink}, {Vinciguerra},
  and {Mandel}}]{2020arXiv200109829V}
{Vigna-G{\'o}mez} A, {MacLeod} M, {Neijssel} CJ, {Broekgaarden} FS, {Justham}
  S, {Howitt} G, {de Mink} SE, {Vinciguerra} S, {Mandel} I (2020)
  {Common-Envelope Episodes that lead to Double Neutron Star formation}. arXiv
  e-prints arXiv:2001.09829.
  {\href{https://arxiv.org/abs/2001.09829}{{arXiv:2001.09829}}} {[astro-ph.SR]}

\bibitem[{{Vila}(1971)}]{1971ApJ...168..217V}
{Vila} SC (1971) {Late Evolution of Close Binaries}. \apj 168:217.
  \doi{10.1086/151076}

\bibitem[{{Vinciguerra} et~al.(2020){Vinciguerra}, {Neijssel},
  {Vigna-G{\'o}mez}, {Mandel}, {Podsiadlowski}, {Maccarone}, {Nicholl},
  {Kingdon}, {Perry}, and {Salemi}}]{2020MNRAS.498.4705V}
{Vinciguerra} S, {Neijssel} CJ, {Vigna-G{\'o}mez} A, {Mandel} I,
  {Podsiadlowski} P, {Maccarone} TJ, {Nicholl} M, {Kingdon} S, {Perry} A,
  {Salemi} F (2020) {Be X-ray binaries in the SMC as indicators of
  mass-transfer efficiency}. \mnras 498(4):4705--4720.
  \doi{10.1093/mnras/staa2177}.
  {\href{https://arxiv.org/abs/2003.00195}{{arXiv:2003.00195}}} {[astro-ph.HE]}

\bibitem[{{Visbal} et~al.(2014){Visbal}, {Haiman}, and
  {Bryan}}]{2014MNRAS.445.1056V}
{Visbal} E, {Haiman} Z, {Bryan} GL (2014) {Direct collapse black hole formation
  from synchronized pairs of atomic cooling haloes}. \mnras 445(1):1056--1063.
  \doi{10.1093/mnras/stu1794}.
  {\href{https://arxiv.org/abs/1406.7020}{{arXiv:1406.7020}}} {[astro-ph.GA]}

\bibitem[{{Vitale}(2016)}]{2016PhRvL.117e1102V}
{Vitale} S (2016) {Multiband Gravitational-Wave Astronomy: Parameter Estimation
  and Tests of General Relativity with Space- and Ground-Based Detectors}. \prl
  117(5):051102. \doi{10.1103/PhysRevLett.117.051102}.
  {\href{https://arxiv.org/abs/1605.01037}{{arXiv:1605.01037}}} {[gr-qc]}

\bibitem[{{Vitale} et~al.(2017){Vitale}, {Lynch}, {Sturani}, and
  {Graff}}]{Vitale:2015}
{Vitale} S, {Lynch} R, {Sturani} R, {Graff} P (2017) {Use of gravitational
  waves to probe the formation channels of compact binaries}. Classical and
  Quantum Gravity 34(3):03LT01. \doi{10.1088/1361-6382/aa552e}.
  {\href{https://arxiv.org/abs/1503.04307}{{arXiv:1503.04307}}} {[gr-qc]}

\bibitem[{{Vitale} et~al.(2019){Vitale}, {Farr}, {Ng}, and
  {Rodriguez}}]{2019ApJ...886L...1V}
{Vitale} S, {Farr} WM, {Ng} KKY, {Rodriguez} CL (2019) {Measuring the Star
  Formation Rate with Gravitational Waves from Binary Black Holes}. \apjl
  886(1):L1. \doi{10.3847/2041-8213/ab50c0}.
  {\href{https://arxiv.org/abs/1808.00901}{{arXiv:1808.00901}}} {[astro-ph.HE]}

\bibitem[{{Vogelsberger} et~al.(2014){Vogelsberger}, {Genel}, {Springel},
  {Torrey}, {Sijacki}, {Xu}, {Snyder}, {Nelson}, and
  {Hernquist}}]{2014MNRAS.444.1518V}
{Vogelsberger} M, {Genel} S, {Springel} V, {Torrey} P, {Sijacki} D, {Xu} D,
  {Snyder} G, {Nelson} D, {Hernquist} L (2014) {Introducing the Illustris
  Project: simulating the coevolution of dark and visible matter in the
  Universe}. \mnras 444(2):1518--1547. \doi{10.1093/mnras/stu1536}.
  {\href{https://arxiv.org/abs/1405.2921}{{arXiv:1405.2921}}} {[astro-ph.CO]}

\bibitem[{{Voit} et~al.(2017){Voit}, {Meece}, {Li}, {O'Shea}, {Bryan}, and
  {Donahue}}]{2017ApJ...845...80V}
{Voit} GM, {Meece} G, {Li} Y, {O'Shea} BW, {Bryan} GL, {Donahue} M (2017) {A
  Global Model for Circumgalactic and Cluster-core Precipitation}. \apj
  845(1):80. \doi{10.3847/1538-4357/aa7d04}.
  {\href{https://arxiv.org/abs/1607.02212}{{arXiv:1607.02212}}} {[astro-ph.GA]}

\bibitem[{{Volonteri}(2007)}]{2007ApJ...663L...5V}
{Volonteri} M (2007) {Gravitational Recoil: Signatures on the Massive Black
  Hole Population}. \apjl 663(1):L5--L8. \doi{10.1086/519525}.
  {\href{https://arxiv.org/abs/astro-ph/0703180}{{arXiv:astro-ph/0703180}}}
  {[astro-ph]}

\bibitem[{{Volonteri}(2010)}]{2010A&ARv..18..279V}
{Volonteri} M (2010) {Formation of supermassive black holes}. \aapr
  18(3):279--315. \doi{10.1007/s00159-010-0029-x}.
  {\href{https://arxiv.org/abs/1003.4404}{{arXiv:1003.4404}}} {[astro-ph.CO]}

\bibitem[{{Volonteri} and {Madau}(2008)}]{2008ApJ...687L..57V}
{Volonteri} M, {Madau} P (2008) {Off-Nuclear AGNs as a Signature of Recoiling
  Massive Black Holes}. \apjl 687(2):L57. \doi{10.1086/593353}.
  {\href{https://arxiv.org/abs/0809.4007}{{arXiv:0809.4007}}} {[astro-ph]}

\bibitem[{{Volonteri} and {Natarajan}(2009)}]{2009MNRAS.400.1911V}
{Volonteri} M, {Natarajan} P (2009) {Journey to the
  M$_{BH}$-{\ensuremath{\sigma}} relation: the fate of low-mass black holes in
  the Universe}. \mnras 400(4):1911--1918.
  \doi{10.1111/j.1365-2966.2009.15577.x}.
  {\href{https://arxiv.org/abs/0903.2262}{{arXiv:0903.2262}}} {[astro-ph.CO]}

\bibitem[{{Volonteri} and {Perna}(2005)}]{2005MNRAS.358..913V}
{Volonteri} M, {Perna} R (2005) {Dynamical evolution of intermediate-mass black
  holes and their observable signatures in the nearby Universe}. \mnras
  358(3):913--922. \doi{10.1111/j.1365-2966.2005.08832.x}.
  {\href{https://arxiv.org/abs/astro-ph/0501345}{{arXiv:astro-ph/0501345}}}
  {[astro-ph]}

\bibitem[{{Volonteri} et~al.(2003{\natexlab{a}}){Volonteri}, {Haardt}, and
  {Madau}}]{2003ApJ...582..559V}
{Volonteri} M, {Haardt} F, {Madau} P (2003{\natexlab{a}}) {The Assembly and
  Merging History of Supermassive Black Holes in Hierarchical Models of Galaxy
  Formation}. \apj 582(2):559--573. \doi{10.1086/344675}.
  {\href{https://arxiv.org/abs/astro-ph/0207276}{{arXiv:astro-ph/0207276}}}
  {[astro-ph]}

\bibitem[{{Volonteri} et~al.(2003{\natexlab{b}}){Volonteri}, {Madau}, and
  {Haardt}}]{2003ApJ...593..661V}
{Volonteri} M, {Madau} P, {Haardt} F (2003{\natexlab{b}}) {The Formation of
  Galaxy Stellar Cores by the Hierarchical Merging of Supermassive Black
  Holes}. \apj 593(2):661--666. \doi{10.1086/376722}.
  {\href{https://arxiv.org/abs/astro-ph/0304389}{{arXiv:astro-ph/0304389}}}
  {[astro-ph]}

\bibitem[{{Volonteri} et~al.(2005){Volonteri}, {Madau}, {Quataert}, and
  {Rees}}]{2005ApJ...620...69V}
{Volonteri} M, {Madau} P, {Quataert} E, {Rees} MJ (2005) {The Distribution and
  Cosmic Evolution of Massive Black Hole Spins}. \apj 620(1):69--77.
  \doi{10.1086/426858}.
  {\href{https://arxiv.org/abs/astro-ph/0410342}{{arXiv:astro-ph/0410342}}}
  {[astro-ph]}

\bibitem[{{Volonteri} et~al.(2007){Volonteri}, {Sikora}, and
  {Lasota}}]{2007ApJ...667..704V}
{Volonteri} M, {Sikora} M, {Lasota} JP (2007) {Black Hole Spin and Galactic
  Morphology}. \apj 667(2):704--713. \doi{10.1086/521186}.
  {\href{https://arxiv.org/abs/0706.3900}{{arXiv:0706.3900}}} {[astro-ph]}

\bibitem[{{Volonteri} et~al.(2008{\natexlab{a}}){Volonteri}, {Haardt}, and
  {G{\"u}ltekin}}]{2008MNRAS.384.1387V}
{Volonteri} M, {Haardt} F, {G{\"u}ltekin} K (2008{\natexlab{a}}) {Compact
  massive objects in Virgo galaxies: the black hole population}. \mnras
  384(4):1387--1392. \doi{10.1111/j.1365-2966.2008.12911.x}.
  {\href{https://arxiv.org/abs/0710.5770}{{arXiv:0710.5770}}} {[astro-ph]}

\bibitem[{{Volonteri} et~al.(2008{\natexlab{b}}){Volonteri}, {Lodato}, and
  {Natarajan}}]{2008MNRAS.383.1079V}
{Volonteri} M, {Lodato} G, {Natarajan} P (2008{\natexlab{b}}) {The evolution of
  massive black hole seeds}. \mnras 383(3):1079--1088.
  \doi{10.1111/j.1365-2966.2007.12589.x}.
  {\href{https://arxiv.org/abs/0709.0529}{{arXiv:0709.0529}}} {[astro-ph]}

\bibitem[{{Volonteri} et~al.(2013){Volonteri}, {Sikora}, {Lasota}, and
  {Merloni}}]{2013ApJ...775...94V}
{Volonteri} M, {Sikora} M, {Lasota} JP, {Merloni} A (2013) {The Evolution of
  Active Galactic Nuclei and their Spins}. \apj 775(2):94.
  \doi{10.1088/0004-637X/775/2/94}.
  {\href{https://arxiv.org/abs/1210.1025}{{arXiv:1210.1025}}} {[astro-ph.HE]}

\bibitem[{{Volonteri} et~al.(2016){Volonteri}, {Dubois}, {Pichon}, and
  {Devriendt}}]{2016MNRAS.460.2979V}
{Volonteri} M, {Dubois} Y, {Pichon} C, {Devriendt} J (2016) {The cosmic
  evolution of massive black holes in the Horizon-AGN simulation}. \mnras
  460:2979--2996. \doi{10.1093/mnras/stw1123}.
  {\href{https://arxiv.org/abs/1602.01941}{{arXiv:1602.01941}}}

\bibitem[{{Volonteri} et~al.(2020){Volonteri}, {Pfister}, {Beckmann}, {Dubois},
  {Colpi}, {Conselice}, {Dotti}, {Martin}, {Jackson}, {Kraljic}, {Pichon},
  {Trebitsch}, {Yi}, {Devriendt}, and {Peirani}}]{2020MNRAS.498.2219V}
{Volonteri} M, {Pfister} H, {Beckmann} RS, {Dubois} Y, {Colpi} M, {Conselice}
  CJ, {Dotti} M, {Martin} G, {Jackson} R, {Kraljic} K, et~al. (2020) {Black
  hole mergers from dwarf to massive galaxies with the NewHorizon and
  Horizon-AGN simulations}. \mnras 498:2219. \doi{10.1093/mnras/staa2384}.
  {\href{https://arxiv.org/abs/2005.04902}{{arXiv:2005.04902}}} {[astro-ph.GA]}

\bibitem[{{Volonteri} et~al.(2021){Volonteri}, {Habouzit}, and
  {Colpi}}]{2021NatRP...3..732V}
{Volonteri} M, {Habouzit} M, {Colpi} M (2021) {The origins of massive black
  holes}. Nature Reviews Physics 3(11):732--743.
  \doi{10.1038/s42254-021-00364-9}.
  {\href{https://arxiv.org/abs/2110.10175}{{arXiv:2110.10175}}} {[astro-ph.GA]}

\bibitem[{{von Zeipel}(1910)}]{1910AN....183..345V}
{von Zeipel} H (1910) {Sur l'application des s{\'e}ries de M. Lindstedt {\`a}
  l'{\'e}tude du mouvement des com{\`e}tes p{\'e}riodiques}. Astronomische
  Nachrichten 183(22):345. \doi{10.1002/asna.19091832202}

\bibitem[{{Voss} and {Tauris}(2003)}]{2003MNRAS.342.1169V}
{Voss} R, {Tauris} TM (2003) {Galactic distribution of merging neutron stars
  and black holes - prospects for short gamma-ray burst progenitors and
  LIGO/VIRGO}. \mnras 342(4):1169--1184.
  \doi{10.1046/j.1365-8711.2003.06616.x}.
  {\href{https://arxiv.org/abs/astro-ph/0303227}{{arXiv:astro-ph/0303227}}}
  {[astro-ph]}

\bibitem[{{Wadsley} et~al.(2004){Wadsley}, {Stadel}, and
  {Quinn}}]{2004NewA....9..137W}
{Wadsley} JW, {Stadel} J, {Quinn} T (2004) {Gasoline: a flexible, parallel
  implementation of TreeSPH}. \na 9(2):137--158.
  \doi{10.1016/j.newast.2003.08.004}.
  {\href{https://arxiv.org/abs/astro-ph/0303521}{{arXiv:astro-ph/0303521}}}
  {[astro-ph]}

\bibitem[{{Wan} et~al.(2020){Wan}, {Lewis}, {Li}, {Simpson}, {Martell},
  {Zucker}, {Mould}, {Erkal}, {Pace}, {Mackey}, {Ji}, {Koposov}, {Kuehn},
  {Shipp}, {Balbinot}, {Bland -Hawthorn}, {Casey}, {Da Costa}, {Kafle},
  {Sharma}, and {De Silva}}]{2020Natur.583..768W}
{Wan} Z, {Lewis} GF, {Li} TS, {Simpson} JD, {Martell} SL, {Zucker} DB, {Mould}
  JR, {Erkal} D, {Pace} AB, {Mackey} D, et~al. (2020) {The tidal remnant of an
  unusually metal-poor globular cluster}. \nat 583(7818):768--770.
  \doi{10.1038/s41586-020-2483-6}.
  {\href{https://arxiv.org/abs/2007.14577}{{arXiv:2007.14577}}} {[astro-ph.GA]}

\bibitem[{{Wang} and {Han}(2012)}]{2012NewAR..56..122W}
{Wang} B, {Han} Z (2012) {Progenitors of type Ia supernovae}. \nar
  56(4):122--141. \doi{10.1016/j.newar.2012.04.001}.
  {\href{https://arxiv.org/abs/1204.1155}{{arXiv:1204.1155}}} {[astro-ph.SR]}

\bibitem[{{Wang} et~al.(2006){Wang}, {Lai}, and {Han}}]{2006ApJ...639.1007W}
{Wang} C, {Lai} D, {Han} JL (2006) {Neutron Star Kicks in Isolated and Binary
  Pulsars: Observational Constraints and Implications for Kick Mechanisms}.
  \apj 639(2):1007--1017. \doi{10.1086/499397}.
  {\href{https://arxiv.org/abs/astro-ph/0509484}{{arXiv:astro-ph/0509484}}}
  {[astro-ph]}

\bibitem[{{Wang} et~al.(2016{\natexlab{a}}){Wang}, {Jia}, and
  {Li}}]{2016RAA....16..126W}
{Wang} C, {Jia} K, {Li} XD (2016{\natexlab{a}}) {The binding energy parameter
  for common envelope evolution}. Research in Astronomy and Astrophysics
  16(8):126. \doi{10.1088/1674-4527/16/8/126}.
  {\href{https://arxiv.org/abs/1605.03668}{{arXiv:1605.03668}}} {[astro-ph.SR]}

\bibitem[{{Wang} et~al.(2021{\natexlab{a}}){Wang}, {Yang}, {Fan}, {Hennawi},
  {Barth}, {Banados}, {Bian}, {Boutsia}, {Connor}, {Davies}, {Decarli},
  {Eilers}, {Farina}, {Green}, {Jiang}, {Li}, {Mazzucchelli}, {Nanni},
  {Schindler}, {Venemans}, {Walter}, {Wu}, and {Yue}}]{2021arXiv210103179W}
{Wang} F, {Yang} J, {Fan} X, {Hennawi} JF, {Barth} AJ, {Banados} E, {Bian} F,
  {Boutsia} K, {Connor} T, {Davies} FB, et~al. (2021{\natexlab{a}}) {A Luminous
  Quasar at Redshift 7.642}. arXiv e-prints arXiv:2101.03179.
  {\href{https://arxiv.org/abs/2101.03179}{{arXiv:2101.03179}}} {[astro-ph.GA]}

\bibitem[{{Wang} et~al.(2021{\natexlab{b}}){Wang}, {Ni}, {Han}, {Xu}, and
  {Luo}}]{2021PhRvD.104b4012W}
{Wang} G, {Ni} WT, {Han} WB, {Xu} P, {Luo} Z (2021{\natexlab{b}}) {Alternative
  LISA-TAIJI networks}. \prd 104(2):024012. \doi{10.1103/PhysRevD.104.024012}.
  {\href{https://arxiv.org/abs/2105.00746}{{arXiv:2105.00746}}} {[gr-qc]}

\bibitem[{{Wang} and {Lai}(2020)}]{2020arXiv200908300W}
{Wang} JS, {Lai} D (2020) {Evolution of inspiralling neutron star binaries:
  effects of tidal interactions and orbital eccentricities}. arXiv e-prints
  arXiv:2009.08300.
  {\href{https://arxiv.org/abs/2009.08300}{{arXiv:2009.08300}}} {[astro-ph.HE]}

\bibitem[{{Wang} et~al.(2018){Wang}, {Peng}, {Wu}, and
  {Dai}}]{2018ApJ...868...19W}
{Wang} JS, {Peng} FK, {Wu} K, {Dai} ZG (2018) {Pre-merger Electromagnetic
  Counterparts of Binary Compact Stars}. \apj 868(1):19.
  \doi{10.3847/1538-4357/aae531}.
  {\href{https://arxiv.org/abs/1810.00170}{{arXiv:1810.00170}}} {[astro-ph.HE]}

\bibitem[{{Wang} et~al.(2015){Wang}, {Spurzem}, {Aarseth}, {Nitadori},
  {Berczik}, {Kouwenhoven}, and {Naab}}]{2015MNRAS.450.4070W}
{Wang} L, {Spurzem} R, {Aarseth} S, {Nitadori} K, {Berczik} P, {Kouwenhoven}
  MBN, {Naab} T (2015) {NBODY6++GPU: ready for the gravitational million-body
  problem}. \mnras 450(4):4070--4080. \doi{10.1093/mnras/stv817}.
  {\href{https://arxiv.org/abs/1504.03687}{{arXiv:1504.03687}}} {[astro-ph.IM]}

\bibitem[{{Wang} et~al.(2016{\natexlab{b}}){Wang}, {Spurzem}, {Aarseth},
  {Giersz}, {Askar}, {Berczik}, {Naab}, {Schadow}, and
  {Kouwenhoven}}]{2016MNRAS.458.1450W}
{Wang} L, {Spurzem} R, {Aarseth} S, {Giersz} M, {Askar} A, {Berczik} P, {Naab}
  T, {Schadow} R, {Kouwenhoven} MBN (2016{\natexlab{b}}) {The DRAGON
  simulations: globular cluster evolution with a million stars}. \mnras
  458(2):1450--1465. \doi{10.1093/mnras/stw274}.
  {\href{https://arxiv.org/abs/1602.00759}{{arXiv:1602.00759}}} {[astro-ph.SR]}

\bibitem[{{Wang} et~al.(2017){Wang}, {Greene}, {Ju}, {Rafikov}, {Ruan}, and
  {Schneider}}]{2017ApJ...834..129W}
{Wang} L, {Greene} JE, {Ju} W, {Rafikov} RR, {Ruan} JJ, {Schneider} DP (2017)
  {Searching for Binary Supermassive Black Holes via Variable Broad Emission
  Line Shifts: Low Binary Fraction}. \apj 834(2):129.
  \doi{10.3847/1538-4357/834/2/129}.
  {\href{https://arxiv.org/abs/1611.00039}{{arXiv:1611.00039}}} {[astro-ph.GA]}

\bibitem[{{Wang} et~al.(2012){Wang}, {Shang}, and
  {Babak}}]{2012PhRvD..86j4050W}
{Wang} Y, {Shang} Y, {Babak} S (2012) {Extreme mass ratio inspiral data
  analysis with a phenomenological waveform}. \prd 86(10):104050.
  \doi{10.1103/PhysRevD.86.104050}.
  {\href{https://arxiv.org/abs/1207.4956}{{arXiv:1207.4956}}} {[gr-qc]}

\bibitem[{{Ward}(1988)}]{1988Icar...73..330W}
{Ward} WR (1988) {On disk-planet interactions and orbital eccentricities}.
  \icarus 73(2):330--348. \doi{10.1016/0019-1035(88)90103-0}

\bibitem[{{Ward}(1997)}]{1997Icar..126..261W}
{Ward} WR (1997) {Protoplanet Migration by Nebula Tides}. \icarus
  126(2):261--281. \doi{10.1006/icar.1996.5647}

\bibitem[{{Warner}(1995)}]{1995Ap&SS.225..249W}
{Warner} B (1995) {The AM Canum Venaticorum Stars}. \apss 225(2):249--270.
  \doi{10.1007/BF00613240}

\bibitem[{{Watts}(2019)}]{2019AIPC.2127b0008W}
{Watts} AL (2019) {Constraining the neutron star equation of state using pulse
  profile modeling}. In: Xiamen-CUSTIPEN Workshop on the Equation of State of
  Dense Neutron-Rich Matter in the Era of Gravitational Wave Astronomy.
  American Institute of Physics Conference Series, vol 2127. p 020008.
  \doi{10.1063/1.5117798}.
  {\href{https://arxiv.org/abs/1904.07012}{{arXiv:1904.07012}}} {[astro-ph.HE]}

\bibitem[{{Webb} et~al.(2018){Webb}, {Leigh}, {Singh}, {Ford}, {McKernan}, and
  {Bellovary}}]{2018MNRAS.474.3835W}
{Webb} JJ, {Leigh} NWC, {Singh} A, {Ford} KES, {McKernan} B, {Bellovary} J
  (2018) {The evolution of kicked stellar-mass black holes in star cluster
  environments}. \mnras 474(3):3835--3846. \doi{10.1093/mnras/stx3024}.
  {\href{https://arxiv.org/abs/1711.09100}{{arXiv:1711.09100}}} {[astro-ph.GA]}

\bibitem[{{Webbink}(1979)}]{1979ApJ...227..178W}
{Webbink} RF (1979) {The evolution of low-mass close binary systems. VI.
  Population II W Ursae Majoris systems.} \apj 227:178--184.
  \doi{10.1086/156717}

\bibitem[{{Webbink}(1984)}]{1984ApJ...277..355W}
{Webbink} RF (1984) {Double white dwarfs as progenitors of R Coronae Borealis
  stars and type I supernovae.} \apj 277:355--360. \doi{10.1086/161701}

\bibitem[{{Webbink}(1985)}]{1985ibs..book...39W}
{Webbink} RF (1985) {Stellar evolution and binaries}, p~39

\bibitem[{{Webbink} and {Iben}(1988)}]{WebbinkIben1988}
{Webbink} RF, {Iben} IJ (1988) 2d Conf.on Faint Blue Stars, Schenectady: L.
  Davis Press. p 445

\bibitem[{{Weinberg}(1986)}]{1986ApJ...300...93W}
{Weinberg} MD (1986) {Orbital Decay of Satellite Galaxies in Spherical
  Systems}. \apj 300:93. \doi{10.1086/163785}

\bibitem[{{Weinberg}(1989)}]{1989MNRAS.239..549W}
{Weinberg} MD (1989) {Self-gravitating response of a spherical galaxy to
  sinking satellites}. \mnras 239:549--569. \doi{10.1093/mnras/239.2.549}

\bibitem[{{Weinberger} et~al.(2017){Weinberger}, {Springel}, {Hernquist},
  {Pillepich}, {Marinacci}, {Pakmor}, {Nelson}, {Genel}, {Vogelsberger},
  {Naiman}, and {Torrey}}]{2017MNRAS.465.3291W}
{Weinberger} R, {Springel} V, {Hernquist} L, {Pillepich} A, {Marinacci} F,
  {Pakmor} R, {Nelson} D, {Genel} S, {Vogelsberger} M, {Naiman} J, et~al.
  (2017) {Simulating galaxy formation with black hole driven thermal and
  kinetic feedback}. \mnras 465(3):3291--3308. \doi{10.1093/mnras/stw2944}.
  {\href{https://arxiv.org/abs/1607.03486}{{arXiv:1607.03486}}} {[astro-ph.GA]}

\bibitem[{{Weisberg} and {Huang}(2016)}]{2016ApJ...829...55W}
{Weisberg} JM, {Huang} Y (2016) {Relativistic Measurements from Timing the
  Binary Pulsar PSR B1913+16}. \apj 829(1):55.
  \doi{10.3847/0004-637X/829/1/55}.
  {\href{https://arxiv.org/abs/1606.02744}{{arXiv:1606.02744}}} {[astro-ph.HE]}

\bibitem[{{Weisskopf} et~al.(2016){Weisskopf}, {Ramsey}, {O'Dell}, {Tennant},
  {Elsner}, {Soffitta}, {Bellazzini}, {Costa}, {Kolodziejczak}, {Kaspi},
  {Muleri}, {Marshall}, {Matt}, and {Romani}}]{2016SPIE.9905E..17W}
{Weisskopf} MC, {Ramsey} B, {O'Dell} S, {Tennant} A, {Elsner} R, {Soffitta} P,
  {Bellazzini} R, {Costa} E, {Kolodziejczak} J, {Kaspi} V, et~al. (2016) {The
  Imaging X-ray Polarimetry Explorer (IXPE)}. In: {den Herder} JWA, {Takahashi}
  T, {Bautz} M (eds) Space Telescopes and Instrumentation 2016: Ultraviolet to
  Gamma Ray. Society of Photo-Optical Instrumentation Engineers (SPIE)
  Conference Series, vol 9905. p 990517. \doi{10.1117/12.2235240}

\bibitem[{{Wen}(2003)}]{2003ApJ...598..419W}
{Wen} L (2003) {On the Eccentricity Distribution of Coalescing Black Hole
  Binaries Driven by the Kozai Mechanism in Globular Clusters}. \apj
  598(1):419--430. \doi{10.1086/378794}.
  {\href{https://arxiv.org/abs/astro-ph/0211492}{{arXiv:astro-ph/0211492}}}
  {[astro-ph]}

\bibitem[{{Wen} and {Chen}(2010)}]{2010PhRvD..81h2001W}
{Wen} L, {Chen} Y (2010) {Geometrical expression for the angular resolution of
  a network of gravitational-wave detectors}. \prd 81(8):082001.
  \doi{10.1103/PhysRevD.81.082001}.
  {\href{https://arxiv.org/abs/1003.2504}{{arXiv:1003.2504}}} {[astro-ph.CO]}

\bibitem[{{Wen} and {Gair}(2005)}]{2005CQGra..22S.445W}
{Wen} L, {Gair} JR (2005) {Detecting extreme mass ratio inspirals with LISA
  using time frequency methods}. Classical and Quantum Gravity
  22(10):S445--S451. \doi{10.1088/0264-9381/22/10/041}.
  {\href{https://arxiv.org/abs/gr-qc/0502100}{{arXiv:gr-qc/0502100}}} {[gr-qc]}

\bibitem[{{Wesson} et~al.(2018){Wesson}, {Jones}, {Garc{\'\i}a-Rojas},
  {Boffin}, and {Corradi}}]{2018MNRAS.480.4589W}
{Wesson} R, {Jones} D, {Garc{\'\i}a-Rojas} J, {Boffin} HMJ, {Corradi} RLM
  (2018) {Confirmation of the link between central star binarity and extreme
  abundance discrepancy factors in planetary nebulae}. \mnras
  480(4):4589--4613. \doi{10.1093/mnras/sty1871}.
  {\href{https://arxiv.org/abs/1807.09272}{{arXiv:1807.09272}}} {[astro-ph.SR]}

\bibitem[{{Wetzel} et~al.(2016){Wetzel}, {Hopkins}, {Kim},
  {Faucher-Gigu{\`e}re}, {Kere{\v{s}}}, and {Quataert}}]{2016ApJ...827L..23W}
{Wetzel} AR, {Hopkins} PF, {Kim} Jh, {Faucher-Gigu{\`e}re} CA, {Kere{\v{s}}} D,
  {Quataert} E (2016) {Reconciling Dwarf Galaxies with
  {\ensuremath{\Lambda}}CDM Cosmology: Simulating a Realistic Population of
  Satellites around a Milky Way-mass Galaxy}. \apjl 827(2):L23.
  \doi{10.3847/2041-8205/827/2/L23}.
  {\href{https://arxiv.org/abs/1602.05957}{{arXiv:1602.05957}}} {[astro-ph.GA]}

\bibitem[{{Wevers} et~al.(2016){Wevers}, {Torres}, {Jonker}, {Wetuski},
  {Nelemans}, {Steeghs}, {Maccarone}, {Heinke}, {Hynes}, {Udalski},
  {Kostrzewa-Rutkowska}, {Groot}, {Gazer}, {Szyma{\'n}ski}, {Britt},
  {Wyrzykowski}, and {Poleski}}]{2016MNRAS.462L.106W}
{Wevers} T, {Torres} MAP, {Jonker} PG, {Wetuski} JD, {Nelemans} G, {Steeghs} D,
  {Maccarone} TJ, {Heinke} C, {Hynes} RI, {Udalski} A, et~al. (2016) {Discovery
  of a high state AM CVn binary in the Galactic Bulge Survey}. \mnras
  462(1):L106--L110. \doi{10.1093/mnrasl/slw141}.
  {\href{https://arxiv.org/abs/1607.04262}{{arXiv:1607.04262}}} {[astro-ph.HE]}

\bibitem[{{Wevers} et~al.(2019){Wevers}, {Stone}, {van Velzen}, {Jonker},
  {Hung}, {Auchettl}, {Gezari}, {Onori}, {Mata S{\'a}nchez},
  {Kostrzewa-Rutkowska}, and {Casares}}]{2019MNRAS.487.4136W}
{Wevers} T, {Stone} NC, {van Velzen} S, {Jonker} PG, {Hung} T, {Auchettl} K,
  {Gezari} S, {Onori} F, {Mata S{\'a}nchez} D, {Kostrzewa-Rutkowska} Z, et~al.
  (2019) {Black hole masses of tidal disruption event host galaxies II}. \mnras
  487(3):4136--4152. \doi{10.1093/mnras/stz1602}.
  {\href{https://arxiv.org/abs/1902.04077}{{arXiv:1902.04077}}} {[astro-ph.HE]}

\bibitem[{{Whalen} et~al.(2004){Whalen}, {Abel}, and
  {Norman}}]{2004ApJ...610...14W}
{Whalen} D, {Abel} T, {Norman} ML (2004) {Radiation Hydrodynamic Evolution of
  Primordial H II Regions}. \apj 610(1):14--22. \doi{10.1086/421548}.
  {\href{https://arxiv.org/abs/astro-ph/0310283}{{arXiv:astro-ph/0310283}}}
  {[astro-ph]}

\bibitem[{{Wild} et~al.(2010){Wild}, {Heckman}, and
  {Charlot}}]{2010MNRAS.405..933W}
{Wild} V, {Heckman} T, {Charlot} S (2010) {Timing the starburst-AGN
  connection}. \mnras 405(2):933--947. \doi{10.1111/j.1365-2966.2010.16536.x}.
  {\href{https://arxiv.org/abs/1002.3156}{{arXiv:1002.3156}}} {[astro-ph.CO]}

\bibitem[{{Wilhelm} et~al.(2020){Wilhelm}, {Korol}, {Rossi}, and
  {D'Onghia}}]{2020arXiv200311074W}
{Wilhelm} MJC, {Korol} V, {Rossi} EM, {D'Onghia} E (2020) {The Milky Way's bar
  structural properties from gravitational waves}. arXiv e-prints
  arXiv:2003.11074.
  {\href{https://arxiv.org/abs/2003.11074}{{arXiv:2003.11074}}} {[astro-ph.GA]}

\bibitem[{{Will}(2006)}]{2006LRR.....9....3W}
{Will} CM (2006) {The Confrontation between General Relativity and Experiment}.
  Living Reviews in Relativity 9(1):3. \doi{10.12942/lrr-2006-3}.
  {\href{https://arxiv.org/abs/gr-qc/0510072}{{arXiv:gr-qc/0510072}}} {[gr-qc]}

\bibitem[{{Will}(2014)}]{2014CQGra..31x4001W}
{Will} CM (2014) {Post-Newtonian effects in N-body dynamics: conserved
  quantities in hierarchical triple systems}. Classical and Quantum Gravity
  31(24):244001. \doi{10.1088/0264-9381/31/24/244001}.
  {\href{https://arxiv.org/abs/1404.7724}{{arXiv:1404.7724}}} {[astro-ph.GA]}

\bibitem[{{Willems} et~al.(2007){Willems}, {Kalogera}, {Vecchio}, {Ivanova},
  {Rasio}, {Fregeau}, and {Belczynski}}]{Willems2007}
{Willems} B, {Kalogera} V, {Vecchio} A, {Ivanova} N, {Rasio} FA, {Fregeau} JM,
  {Belczynski} K (2007) {Eccentric Double White Dwarfs as LISA Sources in
  Globular Clusters}. \apjl 665(1):L59--L62. \doi{10.1086/521049}.
  {\href{https://arxiv.org/abs/0705.4287}{{arXiv:0705.4287}}} {[astro-ph]}

\bibitem[{{Williams} et~al.(2018){Williams}, {Curtis-Lake}, {Hainline},
  {Chevallard}, {Robertson}, {Charlot}, {Endsley}, {Stark}, {Willmer},
  {Alberts}, {Amorin}, {Arribas}, {Baum}, {Bunker}, {Carniani}, {Crand all},
  {Egami}, {Eisenstein}, {Ferruit}, {Husemann}, {Maseda}, {Maiolino}, {Rawle},
  {Rieke}, {Smit}, {Tacchella}, and {Willott}}]{2018ApJS..236...33W}
{Williams} CC, {Curtis-Lake} E, {Hainline} KN, {Chevallard} J, {Robertson} BE,
  {Charlot} S, {Endsley} R, {Stark} DP, {Willmer} CNA, {Alberts} S, et~al.
  (2018) {The JWST Extragalactic Mock Catalog: Modeling Galaxy Populations from
  the UV through the Near-IR over 13 Billion Years of Cosmic History}. \apjs
  236(2):33. \doi{10.3847/1538-4365/aabcbb}.
  {\href{https://arxiv.org/abs/1802.05272}{{arXiv:1802.05272}}} {[astro-ph.GA]}

\bibitem[{{Wise} et~al.(2019){Wise}, {Regan}, {O'Shea}, {Norman}, {Downes}, and
  {Xu}}]{2019Natur.566...85W}
{Wise} JH, {Regan} JA, {O'Shea} BW, {Norman} ML, {Downes} TP, {Xu} H (2019)
  {Formation of massive black holes in rapidly growing pre-galactic gas
  clouds}. \nat 566(7742):85--88. \doi{10.1038/s41586-019-0873-4}.
  {\href{https://arxiv.org/abs/1901.07563}{{arXiv:1901.07563}}} {[astro-ph.GA]}

\bibitem[{{Wittor} and {Gaspari}(2020)}]{2020MNRAS.498.4983W}
{Wittor} D, {Gaspari} M (2020) {Dissecting the turbulent weather driven by
  mechanical AGN feedback}. \mnras 498(4):4983--5002.
  \doi{10.1093/mnras/staa2747}.
  {\href{https://arxiv.org/abs/2009.03344}{{arXiv:2009.03344}}} {[astro-ph.GA]}

\bibitem[{{Wong} et~al.(2019{\natexlab{a}}){Wong}, {Baibhav}, and
  {Berti}}]{2019MNRAS.488.5665W}
{Wong} KWK, {Baibhav} V, {Berti} E (2019{\natexlab{a}}) {Binary radial velocity
  measurements with space-based gravitational-wave detectors}. \mnras
  488(4):5665--5670. \doi{10.1093/mnras/stz2077}.
  {\href{https://arxiv.org/abs/1902.01402}{{arXiv:1902.01402}}} {[astro-ph.HE]}

\bibitem[{{Wong} et~al.(2019{\natexlab{b}}){Wong}, {Berti}, {Gabella}, and
  {Holley-Bockelmann}}]{2019MNRAS.483L..33W}
{Wong} KWK, {Berti} E, {Gabella} WE, {Holley-Bockelmann} K (2019{\natexlab{b}})
  {On the possibility of detecting ultrashort period exoplanets with LISA}.
  \mnras 483(1):L33--L36. \doi{10.1093/mnrasl/sly208}.
  {\href{https://arxiv.org/abs/1808.07055}{{arXiv:1808.07055}}} {[astro-ph.EP]}

\bibitem[{{Wongwathanarat} et~al.(2013){Wongwathanarat}, {Janka}, and
  {M{\"u}ller}}]{2013A&A...552A.126W}
{Wongwathanarat} A, {Janka} HT, {M{\"u}ller} E (2013) {Three-dimensional
  neutrino-driven supernovae: Neutron star kicks, spins, and asymmetric
  ejection of nucleosynthesis products}. \aap 552:A126.
  \doi{10.1051/0004-6361/201220636}.
  {\href{https://arxiv.org/abs/1210.8148}{{arXiv:1210.8148}}} {[astro-ph.HE]}

\bibitem[{{Woods} and {Ivanova}(2011)}]{2011ApJ...739L..48W}
{Woods} TE, {Ivanova} N (2011) {Can We Trust Models for Adiabatic Mass Loss?}
  \apjl 739(2):L48. \doi{10.1088/2041-8205/739/2/L48}.
  {\href{https://arxiv.org/abs/1108.2752}{{arXiv:1108.2752}}} {[astro-ph.SR]}

\bibitem[{{Woods} et~al.(2012){Woods}, {Ivanova}, {van der Sluys}, and
  {Chaichenets}}]{2012ApJ...744...12W}
{Woods} TE, {Ivanova} N, {van der Sluys} MV, {Chaichenets} S (2012) {On the
  Formation of Double White Dwarfs through Stable Mass Transfer and a Common
  Envelope}. \apj 744(1):12. \doi{10.1088/0004-637X/744/1/12}.
  {\href{https://arxiv.org/abs/1102.1039}{{arXiv:1102.1039}}} {[astro-ph.SR]}

\bibitem[{{Woods} et~al.(2019){Woods}, {Agarwal}, {Bromm}, {Bunker}, {Chen},
  {Chon}, {Ferrara}, {Glover}, {Haemmerl{\'e}}, {Haiman}, {Hartwig}, {Heger},
  {Hirano}, {Hosokawa}, {Inayoshi}, {Klessen}, {Kobayashi}, {Koliopanos},
  {Latif}, {Li}, {Mayer}, {Mezcua}, {Natarajan}, {Pacucci}, {Rees}, {Regan},
  {Sakurai}, {Salvadori}, {Schneider}, {Surace}, {Tanaka}, {Whalen}, and
  {Yoshida}}]{2019PASA...36...27W}
{Woods} TE, {Agarwal} B, {Bromm} V, {Bunker} A, {Chen} KJ, {Chon} S, {Ferrara}
  A, {Glover} SCO, {Haemmerl{\'e}} L, {Haiman} Z, et~al. (2019) {Titans of the
  early Universe: The Prato statement on the origin of the first supermassive
  black holes}. \pasa 36:e027. \doi{10.1017/pasa.2019.14}.
  {\href{https://arxiv.org/abs/1810.12310}{{arXiv:1810.12310}}} {[astro-ph.GA]}

\bibitem[{{Woosley}(1993)}]{1993ApJ...405..273W}
{Woosley} SE (1993) {Gamma-Ray Bursts from Stellar Mass Accretion Disks around
  Black Holes}. \apj 405:273. \doi{10.1086/172359}

\bibitem[{{Woosley} and {Heger}(2021)}]{2021arXiv210307933W}
{Woosley} SE, {Heger} A (2021) {The Pair-Instability Mass Gap for Black Holes}.
  arXiv e-prints arXiv:2103.07933.
  {\href{https://arxiv.org/abs/2103.07933}{{arXiv:2103.07933}}} {[astro-ph.SR]}

\bibitem[{{Woosley} et~al.(2007){Woosley}, {Blinnikov}, and
  {Heger}}]{2007Natur.450..390W}
{Woosley} SE, {Blinnikov} S, {Heger} A (2007) {Pulsational pair instability as
  an explanation for the most luminous supernovae}. \nat 450(7168):390--392.
  \doi{10.1038/nature06333}.
  {\href{https://arxiv.org/abs/0710.3314}{{arXiv:0710.3314}}} {[astro-ph]}

\bibitem[{{Worsley} et~al.(2005){Worsley}, {Fabian}, {Bauer}, {Alexand er},
  {Hasinger}, {Mateos}, {Brunner}, {Brand t}, and
  {Schneider}}]{2005MNRAS.357.1281W}
{Worsley} MA, {Fabian} AC, {Bauer} FE, {Alexand er} DM, {Hasinger} G, {Mateos}
  S, {Brunner} H, {Brand t} WN, {Schneider} DP (2005) {The unresolved hard
  X-ray background: the missing source population implied by the Chandra and
  XMM-Newton deep fields}. \mnras 357(4):1281--1287.
  \doi{10.1111/j.1365-2966.2005.08731.x}.
  {\href{https://arxiv.org/abs/astro-ph/0412266}{{arXiv:astro-ph/0412266}}}
  {[astro-ph]}

\bibitem[{{Wu} et~al.(2002){Wu}, {Cropper}, {Ramsay}, and
  {Sekiguchi}}]{2002MNRAS.331..221W}
{Wu} K, {Cropper} M, {Ramsay} G, {Sekiguchi} K (2002) {An electrically powered
  binary star?} \mnras 331(1):221--227. \doi{10.1046/j.1365-8711.2002.05190.x}.
  {\href{https://arxiv.org/abs/astro-ph/0111358}{{arXiv:astro-ph/0111358}}}
  {[astro-ph]}

\bibitem[{{Wu} et~al.(2015){Wu}, {Wang}, {Fan}, {Yi}, {Zuo}, {Bian}, {Jiang},
  {McGreer}, {Wang}, {Yang}, {Yang}, {Thompson}, and
  {Beletsky}}]{2015Natur.518..512W}
{Wu} XB, {Wang} F, {Fan} X, {Yi} W, {Zuo} W, {Bian} F, {Jiang} L, {McGreer} ID,
  {Wang} R, {Yang} J, et~al. (2015) {An ultraluminous quasar with a
  twelve-billion-solar-mass black hole at redshift 6.30}. \nat
  518(7540):512--515. \doi{10.1038/nature14241}.
  {\href{https://arxiv.org/abs/1502.07418}{{arXiv:1502.07418}}} {[astro-ph.GA]}

\bibitem[{{Xin} and {Haiman}(2021)}]{2021MNRAS.506.2408X}
{Xin} C, {Haiman} Z (2021) {Ultra-short-period massive black hole binary
  candidates in LSST as LISA 'verification binaries'}. \mnras
  506(2):2408--2417. \doi{10.1093/mnras/stab1856}.
  {\href{https://arxiv.org/abs/2105.00005}{{arXiv:2105.00005}}} {[astro-ph.HE]}

\bibitem[{{Xu} et~al.(2016){Xu}, {Norman}, {O'Shea}, and
  {Wise}}]{2016ApJ...823..140X}
{Xu} H, {Norman} ML, {O'Shea} BW, {Wise} JH (2016) {Late Pop III Star Formation
  During the Epoch of Reionization: Results from the Renaissance Simulations}.
  \apj 823:140. \doi{10.3847/0004-637X/823/2/140}.
  {\href{https://arxiv.org/abs/1604.03586}{{arXiv:1604.03586}}}

\bibitem[{{Xu} and {Lai}(2017)}]{2017PhRvD..96h3005X}
{Xu} W, {Lai} D (2017) {Resonant tidal excitation of oscillation modes in
  merging binary neutron stars: Inertial-gravity modes}. \prd 96(8):083005.
  \doi{10.1103/PhysRevD.96.083005}.
  {\href{https://arxiv.org/abs/1708.01839}{{arXiv:1708.01839}}} {[astro-ph.HE]}

\bibitem[{{Xu} and {Li}(2010)}]{2010ApJ...716..114X}
{Xu} XJ, {Li} XD (2010) {On the Binding Energy Parameter {\ensuremath{\lambda}}
  of Common Envelope Evolution}. \apj 716(1):114--121.
  \doi{10.1088/0004-637X/716/1/114}.
  {\href{https://arxiv.org/abs/1004.4957}{{arXiv:1004.4957}}} {[astro-ph.SR]}

\bibitem[{{Yagi} and {Seto}(2011)}]{2011PhRvD..83d4011Y}
{Yagi} K, {Seto} N (2011) {Detector configuration of DECIGO/BBO and
  identification of cosmological neutron-star binaries}. PhRvD 83(4):044011.
  \doi{10.1103/PhysRevD.83.044011}.
  {\href{https://arxiv.org/abs/1101.3940}{{arXiv:1101.3940}}} {[astro-ph.CO]}

\bibitem[{{Yagi} et~al.(2010){Yagi}, {Yoshida}, {Komiyama}, {Kashikawa},
  {Furusawa}, {Okamura}, {Graham}, {Miller}, {Carter}, {Mobasher}, and
  {Jogee}}]{2010AJ....140.1814Y}
{Yagi} M, {Yoshida} M, {Komiyama} Y, {Kashikawa} N, {Furusawa} H, {Okamura} S,
  {Graham} AW, {Miller} NA, {Carter} D, {Mobasher} B, et~al. (2010) {A Dozen
  New Galaxies Caught in the Act: Gas Stripping and Extended Emission Line
  Regions in the Coma Cluster}. \aj 140(6):1814--1829.
  \doi{10.1088/0004-6256/140/6/1814}.
  {\href{https://arxiv.org/abs/1005.3874}{{arXiv:1005.3874}}} {[astro-ph.CO]}

\bibitem[{{Yamaguchi} et~al.(2018){Yamaguchi}, {Kawanaka}, {Bulik}, and
  {Piran}}]{2018ApJ...861...21Y}
{Yamaguchi} MS, {Kawanaka} N, {Bulik} T, {Piran} T (2018) {Detecting Black Hole
  Binaries by Gaia}. \apj 861(1):21. \doi{10.3847/1538-4357/aac5ec}.
  {\href{https://arxiv.org/abs/1710.09839}{{arXiv:1710.09839}}} {[astro-ph.SR]}

\bibitem[{{Yang} et~al.(2020{\natexlab{a}}){Yang}, {Wang}, {Fan}, {Hennawi},
  {Davies}, {Yue}, {Banados}, {Wu}, {Venemans}, {Barth}, {Bian}, {Boutsia},
  {Decarli}, {Farina}, {Green}, {Jiang}, {Li}, {Mazzucchelli}, and
  {Walter}}]{2020ApJ...897L..14Y}
{Yang} J, {Wang} F, {Fan} X, {Hennawi} JF, {Davies} FB, {Yue} M, {Banados} E,
  {Wu} XB, {Venemans} B, {Barth} AJ, et~al. (2020{\natexlab{a}})
  {P{\={o}}niu{\={a}}'ena: A Luminous z = 7.5 Quasar Hosting a 1.5 Billion
  Solar Mass Black Hole}. \apjl 897(1):L14. \doi{10.3847/2041-8213/ab9c26}.
  {\href{https://arxiv.org/abs/2006.13452}{{arXiv:2006.13452}}} {[astro-ph.GA]}

\bibitem[{{Yang} et~al.(2019){Yang}, {Bartos}, {Gayathri}, {Ford}, {Haiman},
  {Klimenko}, {Kocsis}, {M{\'a}rka}, {M{\'a}rka}, {McKernan}, and
  {O'Shaughnessy}}]{2019PhRvL.123r1101Y}
{Yang} Y, {Bartos} I, {Gayathri} V, {Ford} KES, {Haiman} Z, {Klimenko} S,
  {Kocsis} B, {M{\'a}rka} S, {M{\'a}rka} Z, {McKernan} B, et~al. (2019)
  {Hierarchical Black Hole Mergers in Active Galactic Nuclei}. \prl
  123(18):181101. \doi{10.1103/PhysRevLett.123.181101}.
  {\href{https://arxiv.org/abs/1906.09281}{{arXiv:1906.09281}}} {[astro-ph.HE]}

\bibitem[{{Yang} et~al.(2020{\natexlab{b}}){Yang}, {Gayathri}, {Bartos},
  {Haiman}, {Safarzadeh}, and {Tagawa}}]{2020ApJ...901L..34Y}
{Yang} Y, {Gayathri} V, {Bartos} I, {Haiman} Z, {Safarzadeh} M, {Tagawa} H
  (2020{\natexlab{b}}) {Black Hole Formation in the Lower Mass Gap through
  Mergers and Accretion in AGN Disks}. \apjl 901(2):L34.
  \doi{10.3847/2041-8213/abb940}.
  {\href{https://arxiv.org/abs/2007.04781}{{arXiv:2007.04781}}} {[astro-ph.HE]}

\bibitem[{{Ye} et~al.(2019){Ye}, {Zhang}, {Zhou}, {Wang}, {Yuan}, {Gu}, {Ding},
  {Zhang}, {Mei}, and {Luo}}]{2019IJMPD..2850121Y}
{Ye} BB, {Zhang} X, {Zhou} MY, {Wang} Y, {Yuan} HM, {Gu} D, {Ding} Y, {Zhang}
  J, {Mei} J, {Luo} J (2019) {Optimizing orbits for TianQin}. International
  Journal of Modern Physics D 28(9):1950121. \doi{10.1142/S0218271819501219}

\bibitem[{{Ye} et~al.(2020){Ye}, {Fong}, {Kremer}, {Rodriguez}, {Chatterjee},
  {Fragione}, and {Rasio}}]{2020ApJ...888L..10Y}
{Ye} CS, {Fong} Wf, {Kremer} K, {Rodriguez} CL, {Chatterjee} S, {Fragione} G,
  {Rasio} FA (2020) {On the Rate of Neutron Star Binary Mergers from Globular
  Clusters}. \apjl 888(1):L10. \doi{10.3847/2041-8213/ab5dc5}.
  {\href{https://arxiv.org/abs/1910.10740}{{arXiv:1910.10740}}} {[astro-ph.HE]}

\bibitem[{{Yelda} et~al.(2014){Yelda}, {Ghez}, {Lu}, {Do}, {Meyer}, {Morris},
  and {Matthews}}]{2014ApJ...783..131Y}
{Yelda} S, {Ghez} AM, {Lu} JR, {Do} T, {Meyer} L, {Morris} MR, {Matthews} K
  (2014) {Properties of the Remnant Clockwise Disk of Young Stars in the
  Galactic Center}. \apj 783(2):131. \doi{10.1088/0004-637X/783/2/131}.
  {\href{https://arxiv.org/abs/1401.7354}{{arXiv:1401.7354}}} {[astro-ph.GA]}

\bibitem[{{Yi} et~al.(2021){Yi}, {Nelemans}, {Brinkerink},
  {Kostrzewa-Rutkowska}, {Timmer}, {Stoppa}, {Rossi}, and {Portegies
  Zwart}}]{2021arXiv210613662Y}
{Yi} SX, {Nelemans} G, {Brinkerink} C, {Kostrzewa-Rutkowska} Z, {Timmer} ST,
  {Stoppa} F, {Rossi} EM, {Portegies Zwart} SF (2021) {The Gravitational Wave
  Universe Toolbox: A software package to simulate observation of the
  Gravitational Wave Universe with different detectors}. arXiv e-prints
  arXiv:2106.13662.
  {\href{https://arxiv.org/abs/2106.13662}{{arXiv:2106.13662}}} {[astro-ph.HE]}

\bibitem[{{Yoshida} et~al.(2003){Yoshida}, {Abel}, {Hernquist}, and
  {Sugiyama}}]{2003ApJ...592..645Y}
{Yoshida} N, {Abel} T, {Hernquist} L, {Sugiyama} N (2003) {Simulations of Early
  Structure Formation: Primordial Gas Clouds}. \apj 592:645--663.
  \doi{10.1086/375810}.
  {\href{https://arxiv.org/abs/arXiv:astro-ph/0301645}{{arXiv:astro-ph/0301645}}}

\bibitem[{{Yu} et~al.(2020){Yu}, {Weinberg}, and
  {Fuller}}]{2020MNRAS.496.5482Y}
{Yu} H, {Weinberg} NN, {Fuller} J (2020) {Non-linear dynamical tides in white
  dwarf binaries}. \mnras 496(4):5482--5502. \doi{10.1093/mnras/staa1858}.
  {\href{https://arxiv.org/abs/2005.03058}{{arXiv:2005.03058}}} {[astro-ph.SR]}

\bibitem[{{Yu}(2002)}]{2002MNRAS.331..935Y}
{Yu} Q (2002) {Evolution of massive binary black holes}. \mnras
  331(4):935--958. \doi{10.1046/j.1365-8711.2002.05242.x}.
  {\href{https://arxiv.org/abs/astro-ph/0109530}{{arXiv:astro-ph/0109530}}}
  {[astro-ph]}

\bibitem[{{Yu} and {Tremaine}(2002)}]{2002MNRAS.335..965Y}
{Yu} Q, {Tremaine} S (2002) {Observational constraints on growth of massive
  black holes}. \mnras 335(4):965--976. \doi{10.1046/j.1365-8711.2002.05532.x}.
  {\href{https://arxiv.org/abs/astro-ph/0203082}{{arXiv:astro-ph/0203082}}}
  {[astro-ph]}

\bibitem[{{Yu} and {Tremaine}(2003)}]{2003ApJ...599.1129Y}
{Yu} Q, {Tremaine} S (2003) {Ejection of Hypervelocity Stars by the (Binary)
  Black Hole in the Galactic Center}. \apj 599(2):1129--1138.
  \doi{10.1086/379546}.
  {\href{https://arxiv.org/abs/astro-ph/0309084}{{arXiv:astro-ph/0309084}}}
  {[astro-ph]}

\bibitem[{{Yu} and {Jeffery}(2010)}]{2010A&A...521A..85Y}
{Yu} S, {Jeffery} CS (2010) {The gravitational wave signal from diverse
  populations of double white dwarf binaries in the Galaxy}. \aap 521:A85.
  \doi{10.1051/0004-6361/201014827}.
  {\href{https://arxiv.org/abs/1007.4267}{{arXiv:1007.4267}}} {[astro-ph.SR]}

\bibitem[{{Yu} et~al.(2021){Yu}, {Lu}, and {Jeffery}}]{2021MNRAS.503.2776Y}
{Yu} S, {Lu} Y, {Jeffery} CS (2021) {Orbital evolution of
  neutron-star-white-dwarf binaries by Roche lobe overflow and gravitational
  wave radiation}. \mnras 503(2):2776--2790. \doi{10.1093/mnras/stab626}.
  {\href{https://arxiv.org/abs/2103.01884}{{arXiv:2103.01884}}} {[astro-ph.HE]}

\bibitem[{{Yuan} and {Narayan}(2014)}]{2014ARA&A..52..529Y}
{Yuan} F, {Narayan} R (2014) {Hot Accretion Flows Around Black Holes}. \araa
  52:529--588. \doi{10.1146/annurev-astro-082812-141003}.
  {\href{https://arxiv.org/abs/1401.0586}{{arXiv:1401.0586}}} {[astro-ph.HE]}

\bibitem[{{Yue} and {Han}(2018)}]{2018PhRvD..97f4003Y}
{Yue} XJ, {Han} WB (2018) {Gravitational waves with dark matter minispikes: The
  combined effect}. \prd 97(6):064003. \doi{10.1103/PhysRevD.97.064003}.
  {\href{https://arxiv.org/abs/1711.09706}{{arXiv:1711.09706}}} {[gr-qc]}

\bibitem[{{Yue} et~al.(2019){Yue}, {Han}, and {Chen}}]{2019ApJ...874...34Y}
{Yue} XJ, {Han} WB, {Chen} X (2019) {Dark Matter: An Efficient Catalyst for
  Intermediate-mass-ratio-inspiral Events}. \apj 874(1):34.
  \doi{10.3847/1538-4357/ab06f6}.
  {\href{https://arxiv.org/abs/1802.03739}{{arXiv:1802.03739}}} {[gr-qc]}

\bibitem[{{Yunes} and {Berti}(2008)}]{2008PhRvD..77l4006Y}
{Yunes} N, {Berti} E (2008) {Accuracy of the post-Newtonian approximation:
  Optimal asymptotic expansion for quasicircular, extreme-mass ratio
  inspirals}. \prd 77(12):124006. \doi{10.1103/PhysRevD.77.124006}.
  {\href{https://arxiv.org/abs/0803.1853}{{arXiv:0803.1853}}} {[gr-qc]}

\bibitem[{{Yunes} et~al.(2011{\natexlab{a}}){Yunes}, {Kocsis}, {Loeb}, and
  {Haiman}}]{2011PhRvL.107q1103Y}
{Yunes} N, {Kocsis} B, {Loeb} A, {Haiman} Z (2011{\natexlab{a}}) {Imprint of
  Accretion Disk-Induced Migration on Gravitational Waves from Extreme Mass
  Ratio Inspirals}. \prl 107(17):171103. \doi{10.1103/PhysRevLett.107.171103}.
  {\href{https://arxiv.org/abs/1103.4609}{{arXiv:1103.4609}}} {[astro-ph.CO]}

\bibitem[{{Yunes} et~al.(2011{\natexlab{b}}){Yunes}, {Miller}, and
  {Thornburg}}]{2011PhRvD..83d4030Y}
{Yunes} N, {Miller} MC, {Thornburg} J (2011{\natexlab{b}}) {Effect of massive
  perturbers on extreme mass-ratio inspiral waveforms}. \prd 83(4):044030.
  \doi{10.1103/PhysRevD.83.044030}.
  {\href{https://arxiv.org/abs/1010.1721}{{arXiv:1010.1721}}} {[astro-ph.GA]}

\bibitem[{{Yungelson}(2008)}]{2008AstL...34..620Y}
{Yungelson} LR (2008) {Evolution of low-mass helium stars in semidetached
  binaries}. Astronomy Letters 34(9):620--634. \doi{10.1134/S1063773708090053}.
  {\href{https://arxiv.org/abs/0804.2780}{{arXiv:0804.2780}}} {[astro-ph]}

\bibitem[{{Yungelson} et~al.(2006){Yungelson}, {Lasota}, {Nelemans}, {Dubus},
  {van den Heuvel}, {Dewi}, and {Portegies Zwart}}]{2006A&A...454..559Y}
{Yungelson} LR, {Lasota} JP, {Nelemans} G, {Dubus} G, {van den Heuvel} EPJ,
  {Dewi} J, {Portegies Zwart} S (2006) {The origin and fate of short-period
  low-mass black-hole binaries}. \aap 454(2):559--569.
  \doi{10.1051/0004-6361:20064984}.
  {\href{https://arxiv.org/abs/astro-ph/0604434}{{arXiv:astro-ph/0604434}}}
  {[astro-ph]}

\bibitem[{{Zahn}(1966)}]{1966AnAp...29..313Z}
{Zahn} JP (1966) {Les mar{\'e}es dans une {\'e}toile double serr{\'e}e}.
  Annales d'Astrophysique 29:313

\bibitem[{{Zahn}(1977)}]{1977A&A....57..383Z}
{Zahn} JP (1977) {Reprint of 1977A\&A....57..383Z. Tidal friction in close
  binary stars.} \aap 500:121--132

\bibitem[{{Zana} et~al.(2018{\natexlab{a}}){Zana}, {Dotti}, {Capelo}, {Bonoli},
  {Haardt}, {Mayer}, and {Spinoso}}]{2018MNRAS.473.2608Z}
{Zana} T, {Dotti} M, {Capelo} PR, {Bonoli} S, {Haardt} F, {Mayer} L, {Spinoso}
  D (2018{\natexlab{a}}) {External versus internal triggers of bar formation in
  cosmological zoom-in simulations}. \mnras 473(2):2608--2621.
  \doi{10.1093/mnras/stx2503}.
  {\href{https://arxiv.org/abs/1705.02348}{{arXiv:1705.02348}}} {[astro-ph.GA]}

\bibitem[{{Zana} et~al.(2018{\natexlab{b}}){Zana}, {Dotti}, {Capelo}, {Mayer},
  {Haardt}, {Shen}, and {Bonoli}}]{2018MNRAS.479.5214Z}
{Zana} T, {Dotti} M, {Capelo} PR, {Mayer} L, {Haardt} F, {Shen} S, {Bonoli} S
  (2018{\natexlab{b}}) {Bar resilience to flybys in a cosmological framework}.
  \mnras 479(4):5214--5219. \doi{10.1093/mnras/sty1850}.
  {\href{https://arxiv.org/abs/1805.03658}{{arXiv:1805.03658}}} {[astro-ph.GA]}

\bibitem[{{Zelenka} et~al.(2020){Zelenka}, {Lukes-Gerakopoulos}, {Witzany}, and
  {Kop{\'a}{\v{c}}ek}}]{2020PhRvD.101b4037Z}
{Zelenka} O, {Lukes-Gerakopoulos} G, {Witzany} V, {Kop{\'a}{\v{c}}ek} O (2020)
  {Growth of resonances and chaos for a spinning test particle in the
  Schwarzschild background}. \prd 101(2):024037.
  \doi{10.1103/PhysRevD.101.024037}.
  {\href{https://arxiv.org/abs/1911.00414}{{arXiv:1911.00414}}} {[gr-qc]}

\bibitem[{{Zepf} et~al.(2008){Zepf}, {Stern}, {Maccarone}, {Kundu},
  {Kamionkowski}, {Rhode}, {Salzer}, {Ciardullo}, and
  {Gronwall}}]{2008ApJ...683L.139Z}
{Zepf} SE, {Stern} D, {Maccarone} TJ, {Kundu} A, {Kamionkowski} M, {Rhode} KL,
  {Salzer} JJ, {Ciardullo} R, {Gronwall} C (2008) {Very Broad [O III]
  {\ensuremath{\lambda}}{\ensuremath{\lambda}}4959, 5007 Emission from the NGC
  4472 Globular Cluster RZ 2109 and Implications for the Mass of Its Black Hole
  X-Ray Source}. \apjl 683(2):L139. \doi{10.1086/591937}.
  {\href{https://arxiv.org/abs/0805.2952}{{arXiv:0805.2952}}} {[astro-ph]}

\bibitem[{{Zevin} et~al.(2017){Zevin}, {Pankow}, {Rodriguez}, {Sampson},
  {Chase}, {Kalogera}, and {Rasio}}]{Zevin:2017}
{Zevin} M, {Pankow} C, {Rodriguez} CL, {Sampson} L, {Chase} E, {Kalogera} V,
  {Rasio} FA (2017) {Constraining Formation Models of Binary Black Holes with
  Gravitational-wave Observations}. \apj 846:82.
  \doi{10.3847/1538-4357/aa8408}.
  {\href{https://arxiv.org/abs/1704.07379}{{arXiv:1704.07379}}} {[astro-ph.HE]}

\bibitem[{{Zevin} et~al.(2019{\natexlab{a}}){Zevin}, {Kremer}, {Siegel},
  {Coughlin}, {Tsang}, {Berry}, and {Kalogera}}]{2019ApJ...886....4Z}
{Zevin} M, {Kremer} K, {Siegel} DM, {Coughlin} S, {Tsang} BTH, {Berry} CPL,
  {Kalogera} V (2019{\natexlab{a}}) {Can Neutron-star Mergers Explain the
  r-process Enrichment in Globular Clusters?} \apj 886(1):4.
  \doi{10.3847/1538-4357/ab498b}.
  {\href{https://arxiv.org/abs/1906.11299}{{arXiv:1906.11299}}} {[astro-ph.HE]}

\bibitem[{{Zevin} et~al.(2019{\natexlab{b}}){Zevin}, {Samsing}, {Rodriguez},
  {Haster}, and {Ramirez-Ruiz}}]{2019ApJ...871...91Z}
{Zevin} M, {Samsing} J, {Rodriguez} C, {Haster} CJ, {Ramirez-Ruiz} E
  (2019{\natexlab{b}}) {Eccentric Black Hole Mergers in Dense Star Clusters:
  The Role of Binary-Binary Encounters}. \apj 871(1):91.
  \doi{10.3847/1538-4357/aaf6ec}.
  {\href{https://arxiv.org/abs/1810.00901}{{arXiv:1810.00901}}} {[astro-ph.HE]}

\bibitem[{{Zhang} et~al.(2019){Zhang}, {Santangelo}, {Feroci}, {Xu}, {Lu},
  {Chen}, {Feng}, {Zhang}, {Brandt}, {Hernanz}, {Baldini}, {Bozzo}, {Campana},
  {De Rosa}, {Dong}, {Evangelista}, {Karas}, and {et
  al.}}]{2019SCPMA..6229502Z}
{Zhang} S, {Santangelo} A, {Feroci} M, {Xu} Y, {Lu} F, {Chen} Y, {Feng} H,
  {Zhang} S, {Brandt} S, {Hernanz} M, et~al. (2019) {The enhanced X-ray Timing
  and Polarimetry mission{\textemdash}eXTP}. Science China Physics, Mechanics,
  and Astronomy 62(2):29502. \doi{10.1007/s11433-018-9309-2}.
  {\href{https://arxiv.org/abs/1812.04020}{{arXiv:1812.04020}}} {[astro-ph.IM]}

\bibitem[{{Zhang} et~al.(2011){Zhang}, {Yunes}, and
  {Berti}}]{2011PhRvD..84b4029Z}
{Zhang} Z, {Yunes} N, {Berti} E (2011) {Accuracy of the post-Newtonian
  approximation. II. Optimal asymptotic expansion of the energy flux for
  quasicircular, extreme mass-ratio inspirals into a Kerr black hole}. \prd
  84(2):024029. \doi{10.1103/PhysRevD.84.024029}.
  {\href{https://arxiv.org/abs/1103.6041}{{arXiv:1103.6041}}} {[gr-qc]}

\bibitem[{{Zhao} and {Silk}(2005)}]{2005PhRvL..95a1301Z}
{Zhao} H, {Silk} J (2005) {Dark Minihalos with Intermediate Mass Black Holes}.
  \prl 95(1):011301. \doi{10.1103/PhysRevLett.95.011301}.
  {\href{https://arxiv.org/abs/astro-ph/0501625}{{arXiv:astro-ph/0501625}}}
  {[astro-ph]}

\bibitem[{{Zhu} and {Thrane}(2020)}]{2020ApJ...900..117Z}
{Zhu} XJ, {Thrane} E (2020) {Toward the Unambiguous Identification of
  Supermassive Binary Black Holes through Bayesian Inference}. \apj 900(2):117.
  \doi{10.3847/1538-4357/abac5a}.
  {\href{https://arxiv.org/abs/2004.10944}{{arXiv:2004.10944}}} {[astro-ph.HE]}

\bibitem[{{Zinn} and {West}(1984)}]{1984ApJS...55...45Z}
{Zinn} R, {West} MJ (1984) {The globular cluster system of the Galaxy. III.
  Measurements of radial velocity and metallicity for 60 clusters and a
  compilation of metallicities for 121 clusters.} \apjs 55:45--66.
  \doi{10.1086/190947}

\bibitem[{{Ziosi} et~al.(2014){Ziosi}, {Mapelli}, {Branchesi}, and
  {Tormen}}]{2014MNRAS.441.3703Z}
{Ziosi} BM, {Mapelli} M, {Branchesi} M, {Tormen} G (2014) {Dynamics of stellar
  black holes in young star clusters with different metallicities - II. Black
  hole-black hole binaries}. \mnras 441(4):3703--3717.
  \doi{10.1093/mnras/stu824}.
  {\href{https://arxiv.org/abs/1404.7147}{{arXiv:1404.7147}}} {[astro-ph.GA]}

\bibitem[{{Zorotovic} et~al.(2010){Zorotovic}, {Schreiber}, {G{\"a}nsicke}, and
  {Nebot G{\'o}mez-Mor{\'a}n}}]{2010A&A...520A..86Z}
{Zorotovic} M, {Schreiber} MR, {G{\"a}nsicke} BT, {Nebot G{\'o}mez-Mor{\'a}n} A
  (2010) {Post-common-envelope binaries from SDSS. IX: Constraining the
  common-envelope efficiency}. \aap 520:A86. \doi{10.1051/0004-6361/200913658}.
  {\href{https://arxiv.org/abs/1006.1621}{{arXiv:1006.1621}}} {[astro-ph.SR]}

\bibitem[{{Zorotovic} et~al.(2011){Zorotovic}, {Schreiber}, and
  {G{\"a}nsicke}}]{2011A&A...536A..42Z}
{Zorotovic} M, {Schreiber} MR, {G{\"a}nsicke} BT (2011) {Post common envelope
  binaries from SDSS. XI. The white dwarf mass distributions of CVs and
  pre-CVs}. \aap 536:A42. \doi{10.1051/0004-6361/201116626}.
  {\href{https://arxiv.org/abs/1108.4600}{{arXiv:1108.4600}}} {[astro-ph.SR]}

\bibitem[{{Zorotovic} et~al.(2014){Zorotovic}, {Schreiber},
  {Garc{\'\i}a-Berro}, {Camacho}, {Torres}, {Rebassa-Mansergas}, and
  {G{\"a}nsicke}}]{2014A&A...568A..68Z}
{Zorotovic} M, {Schreiber} MR, {Garc{\'\i}a-Berro} E, {Camacho} J, {Torres} S,
  {Rebassa-Mansergas} A, {G{\"a}nsicke} BT (2014) {Monte Carlo simulations of
  post-common-envelope white dwarf + main sequence binaries: The effects of
  including recombination energy}. \aap 568:A68.
  \doi{10.1051/0004-6361/201323039}.
  {\href{https://arxiv.org/abs/1407.3301}{{arXiv:1407.3301}}} {[astro-ph.SR]}

\bibitem[{{Zubovas} and {King}(2012)}]{2012MNRAS.426.2751Z}
{Zubovas} K, {King} AR (2012) {The M-{\ensuremath{\sigma}} relation in
  different environments}. \mnras 426(4):2751--2757.
  \doi{10.1111/j.1365-2966.2012.21845.x}.
  {\href{https://arxiv.org/abs/1208.1380}{{arXiv:1208.1380}}} {[astro-ph.GA]}

\bibitem[{{Zuckerman} et~al.(2010){Zuckerman}, {Melis}, {Klein}, {Koester}, and
  {Jura}}]{2010ApJ...722..725Z}
{Zuckerman} B, {Melis} C, {Klein} B, {Koester} D, {Jura} M (2010) {Ancient
  Planetary Systems are Orbiting a Large Fraction of White Dwarf Stars}. \apj
  722(1):725--736. \doi{10.1088/0004-637X/722/1/725}.
  {\href{https://arxiv.org/abs/1007.2252}{{arXiv:1007.2252}}} {[astro-ph.SR]}

\bibitem[{{Zurek} et~al.(2009){Zurek}, {Knigge}, {Maccarone}, {Dieball}, and
  {Long}}]{2009ApJ...699.1113Z}
{Zurek} DR, {Knigge} C, {Maccarone} TJ, {Dieball} A, {Long} KS (2009) {An
  Ultracompact X-Ray Binary in the Globular Cluster NGC 1851}. \apj
  699(2):1113--1118. \doi{10.1088/0004-637X/699/2/1113}.
  {\href{https://arxiv.org/abs/0905.0145}{{arXiv:0905.0145}}} {[astro-ph.HE]}

\bibitem[{{Zurek} et~al.(2016){Zurek}, {Knigge}, {Maccarone}, {Pooley},
  {Dieball}, {Long}, {Shara}, and {Sarajedini}}]{2016MNRAS.460.3660Z}
{Zurek} DR, {Knigge} C, {Maccarone} TJ, {Pooley} D, {Dieball} A, {Long} KS,
  {Shara} M, {Sarajedini} A (2016) {A far-ultraviolet variable with an
  18-minute period in the globular cluster NGC 1851}. \mnras 460(4):3660--3668.
  \doi{10.1093/mnras/stw1190}.
  {\href{https://arxiv.org/abs/1605.04827}{{arXiv:1605.04827}}} {[astro-ph.SR]}

\bibitem[{{Zwick} et~al.(2020){Zwick}, {Capelo}, {Bortolas}, {Mayer}, and
  {Amaro-Seoane}}]{2020MNRAS.495.2321Z}
{Zwick} L, {Capelo} PR, {Bortolas} E, {Mayer} L, {Amaro-Seoane} P (2020)
  {Improved gravitational radiation time-scales: significance for LISA and
  LIGO-Virgo sources}. \mnras 495(2):2321--2331. \doi{10.1093/mnras/staa1314}.
  {\href{https://arxiv.org/abs/1911.06024}{{arXiv:1911.06024}}} {[astro-ph.GA]}

\bibitem[{{Zwick} et~al.(2021){Zwick}, {Capelo}, {Bortolas},
  {V{\'a}zquez-Aceves}, {Mayer}, and {Amaro-Seoane}}]{2021arXiv210200015Z}
{Zwick} L, {Capelo} PR, {Bortolas} E, {V{\'a}zquez-Aceves} V, {Mayer} L,
  {Amaro-Seoane} P (2021) {Improved gravitational radiation time-scales II:
  Spin-orbit contributions and environmental perturbations}. \mnras
  506(1):1007--1018. \doi{10.1093/mnras/stab1818}.
  {\href{https://arxiv.org/abs/2102.00015}{{arXiv:2102.00015}}} {[astro-ph.GA]}

\end{thebibliography}
